


%
%


\def\famname{
 \textfont0=\textrm \scriptfont0=\scriptrm
 \scriptscriptfont0=\sscriptrm
 \textfont1=\textmi \scriptfont1=\scriptmi
 \scriptscriptfont1=\sscriptmi
 \textfont2=\textsy \scriptfont2=\scriptsy \scriptscriptfont2=\sscriptsy
 \textfont3=\textex \scriptfont3=\textex \scriptscriptfont3=\textex
 \textfont4=\textbf \scriptfont4=\scriptbf \scriptscriptfont4=\sscriptbf
 \skewchar\textmi='177 \skewchar\scriptmi='177
 \skewchar\sscriptmi='177
 \skewchar\textsy='60 \skewchar\scriptsy='60
 \skewchar\sscriptsy='60
 \def\rm{\fam0 \textrm} \def\bf{\fam4 \textbf}}
\def\sca#1{scaled\magstep#1} \def\scah{scaled\magstephalf} 
\def\twelvepoint{
 \font\textrm=cmr12 \font\scriptrm=cmr8 \font\sscriptrm=cmr6
 \font\textmi=cmmi12 \font\scriptmi=cmmi8 \font\sscriptmi=cmmi6 
 \font\textsy=cmsy10 \sca1 \font\scriptsy=cmsy8
 \font\sscriptsy=cmsy6
 \font\textex=cmex10 \sca1
 \font\textbf=cmbx12 \font\scriptbf=cmbx8 \font\sscriptbf=cmbx6
 \font\it=cmti12
 \font\sectfont=cmbx12 \sca1
 \font\sectmath=cmmib10 \sca2
 \font\sectsymb=cmbsy10 \sca2
 \font\refrm=cmr10 \scah \font\refit=cmti10 \scah
 \font\refbf=cmbx10 \scah
 \def\twelverm{\textrm} \def\twelveit{\it} \def\twelvebf{\textbf}
 \famname \textrm 
 \advance\voffset by .06in \advance\hoffset by .28in
 \normalbaselineskip=17.5pt plus 1pt \baselineskip=\normalbaselineskip
 \parindent=21pt
 \setbox\strutbox=\hbox{\vrule height10.5pt depth4pt width0pt}}


\catcode`@=11

{\catcode`\'=\active \def'{{}^\bgroup\prim@s}}

\def\screwcount{\alloc@0\count\countdef\insc@unt}   
\def\screwdimen{\alloc@1\dimen\dimendef\insc@unt} 
\def\screwbox{\alloc@4\box\chardef\insc@unt}

\catcode`@=12


\overfullrule=0pt			
\vsize=9in \hsize=6in
\lineskip=0pt				
\abovedisplayskip=1.2em plus.3em minus.9em 
\belowdisplayskip=1.2em plus.3em minus.9em	
\abovedisplayshortskip=0em plus.3em	
\belowdisplayshortskip=.7em plus.3em minus.4em	
\parindent=21pt
\setbox\strutbox=\hbox{\vrule height10.5pt depth4pt width0pt}
\def\makefootline{\baselineskip=30pt \line{\the\footline}}
\footline={\ifnum\count0=1 \hfil \else\hss\twelverm\folio\hss \fi}
\pageno=1


\def\put(#1,#2)#3{\screwdimen\unit  \unit=1in
	\vbox to0pt{\kern-#2\unit\hbox{\kern#1\unit
	\vbox{#3}}\vss}\nointerlineskip}


\def\\{\hfil\break}
\def\newpage{\vfill\eject}
\def\center{\leftskip=0pt plus 1fill \rightskip=\leftskip \parindent=0pt
 \def\textindent##1{\par\hangindent21pt\footrm\noindent\hskip21pt
 \llap{##1\enspace}\ignorespaces}\par}
\def\unnarrower{\leftskip=0pt \rightskip=\leftskip}


\def\sect#1\par{\par\ifdim\lastskip<\medskipamount
	\bigskip\medskip\goodbreak\else\nobreak\fi
	\noindent{\sectfont{#1}}\par\nobreak\medskip} 
\def\itemize#1 {\item{[#1]}}	
\def\vol#1 {{\refbf#1} }		 

\def\ref#1{\setbox0=\hbox{M}$\vbox to\ht0{}^{#1}$}


\def\NP #1 {{\refit Nucl. Phys.} {\refbf B{#1}} }
\def\PL #1 {{\refit Phys. Lett.} {\refbf{#1}} }
\def\PR #1 {{\refit Phys. Rev. Lett.} {\refbf{#1}} }
\def\PRD #1 {{\refit Phys. Rev.} {\refbf D{#1}} }


\hyphenation{pre-print}
\hyphenation{quan-ti-za-tion}

%
%

\def\on#1#2{{\buildrel{\mkern2.5mu#1\mkern-2.5mu}\over{#2}}}
\def\oonoo#1#2#3{\vbox{\ialign{##\crcr
	\hfil\hfil\hfil{$#3{#1}$}\hfil\crcr\noalign{\kern1pt\nointerlineskip}
	$#3{#2}$\crcr}}}
\def\oon#1#2{\mathchoice{\oonoo{#1}{#2}{\displaystyle}}
	{\oonoo{#1}{#2}{\textstyle}}{\oonoo{#1}{#2}{\scriptstyle}}
	{\oonoo{#1}{#2}{\scriptscriptstyle}}}
\def\dt#1{\oon{\hbox{\bf .}}{#1}}  
\def\ddt#1{\oon{\hbox{\bf .\kern-1pt.}}#1}    
\def\slap#1#2{\setbox0=\hbox{$#1{#2}$}
	#2\kern-\wd0{\hfuzz=1pt\hbox to\wd0{\hfil$#1{/}$\hfil}}}
\def\sla#1{\mathpalette\slap{#1}}                
\def\bop#1{\setbox0=\hbox{$#1M$}\mkern1.5mu
	\lower.02\ht0\vbox{\hrule height0pt depth.06\ht0
	\hbox{\vrule width.06\ht0 height.9\ht0 \kern.9\ht0
	\vrule width.06\ht0}\hrule height.06\ht0}\mkern1.5mu}
\def\bo{{\mathpalette\bop{}}}                        
\def~{\widetilde} 
\mathcode`\*="702A                  
\def\in{\relax\ifmmode\mathchar"3232\else{\refit in\/}\fi} 
\def\f#1#2{{\textstyle{#1\over#2}}}	   
\def\half{{\textstyle{1\over{\raise.1ex\hbox{$\scriptstyle{2}$}}}}}

\def\Gamma{\mathchar"0100}
\def\Delta{\mathchar"0101}
\def\Theta{\mathchar"0102}
\def\Lambda{\mathchar"0103}
\def\Xi{\mathchar"0104}
\def\Pi{\mathchar"0105}
\def\Sigma{\mathchar"0106}
\def\Upsilon{\mathchar"0107}
\def\Phi{\mathchar"0108}
\def\Psi{\mathchar"0109}
\def\Omega{\mathchar"010A}

\catcode`\^^?=13				    
\catcode128=13 \def €{\"A}                 
\catcode129=13 \def {\AA}                 
\catcode130=13 \def '{\c}           	   
\catcode131=13 \def ƒ{\'E}                   
\catcode132=13 \def "{\~N}                   
\catcode133=13 \def …{\"O}                 
\catcode134=13 \def †{\"U}                  
\catcode135=13 \def ‡{\'a}                  
\catcode136=13 \def ˆ{\`a}                   
\catcode137=13 \def ‰{\^a}                 
\catcode138=13 \def Š{\"a}                 
\catcode139=13 \def ‹{\~a}                   
\catcode140=13 \def Œ{\alpha}            
\catcode141=13 \def {\chi}                
\catcode142=13 \def Ž{\'e}                   
\catcode143=13 \def {\`e}                    
\catcode144=13 \def {\^e}                  
\catcode145=13 \def '{\"e}                
\catcode146=13 \def '{\'\i}                 
\catcode147=13 \def "{\`\i}                  
\catcode148=13 \def "{\^\i}                
\catcode149=13 \def •{\"\i}                
\catcode150=13 \def –{\~n}                  
\catcode151=13 \def —{\'o}                 
\catcode152=13 \def ˜{\`o}                  
\catcode153=13 \def ™{\^o}                
\catcode154=13 \def š{\"o}                 
\catcode155=13 \def ›{\~o}                  
\catcode156=13 \def œ{\'u}                  
\catcode157=13 \def {\`u}                  
\catcode158=13 \def ž{\^u}                
\catcode159=13 \def Ÿ{\"u}                
\catcode160=13 \def  {\tau}               
\catcode161=13 \mathchardef ¡="2203     
\catcode162=13 \def ¢{\oplus}           
\catcode163=13 \def £{\relax\ifmmode\to\else\itemize\fi} 
\catcode164=13 \def ¤{\subset}	  
\catcode165=13 \def ¥{\infty}           
\catcode166=13 \def ¦{\mp}                
\catcode167=13 \def §{\sigma}           
\catcode168=13 \def ¨{\rho}               
\catcode169=13 \def ©{\gamma}         
\catcode170=13 \def ª{\leftrightarrow} 
\catcode171=13 \def «{\relax\ifmmode\acute\else\expandafter\'\fi}
\catcode172=13 \def ¬{\relax\ifmmode\expandafter\ddt\else\expandafter\"\fi}
\catcode173=13 \def ­{\equiv}            
\catcode174=13 \def ®{\approx}          
\catcode175=13 \def ¯{\Omega}          
\catcode176=13 \def °{\otimes}          
\catcode177=13 \def ±{\ne}                 
\catcode178=13 \def ²{\le}                   
\catcode179=13 \def ³{\ge}                  
\catcode180=13 \def ´{\upsilon}          
\catcode181=13 \def µ{\mu}                
\catcode182=13 \def ¶{\delta}             
\catcode183=13 \def ·{\epsilon}          
\catcode184=13 \def ¸{\Pi}                  
\catcode185=13 \def ¹{\pi}                  
\catcode186=13 \def º{\beta}               
\catcode187=13 \def »{\partial}           
\catcode188=13 \def ¼{\nobreak\ }       
\catcode189=13 \def ½{\zeta}               
\catcode190=13 \def ¾{\sim}                 
\catcode191=13 \def ¿{\omega}           
\catcode192=13 \def À{\dt}                     
\catcode193=13 \def Á{\gets}                
\catcode194=13 \def Â{\lambda}           
\catcode195=13 \def Ã{\nu}                   
\catcode196=13 \def Ä{\phi}                  
\catcode197=13 \def Å{\xi}                     
\catcode198=13 \def Æ{\psi}                  
\catcode199=13 \def Ç{\int}                    
\catcode200=13 \def È{\oint}                 
\catcode201=13 \def É{\relax\ifmmode\cdot\else\vol\fi}    
\catcode202=13 \def Ê{\relax\ifmmode\,\else\thinspace\fi}
\catcode203=13 \def Ë{\`A}                      
\catcode204=13 \def Ì{\~A}                      
\catcode205=13 \def Í{\~O}                      
\catcode206=13 \def Î{\Theta}              
\catcode207=13 \def Ï{\theta}               
\catcode208=13 \def Ð{\relax\ifmmode\bar\else\expandafter\=\fi}
\catcode209=13 \def Ñ{\overline}             
\catcode210=13 \def Ò{\langle}               
\catcode211=13 \def Ó{\relax\ifmmode\{\else\ital\fi}      
\catcode212=13 \def Ô{\rangle}               
\catcode213=13 \def Õ{\}}                        
\catcode214=13 \def Ö{\sla}                      
\catcode215=13 \def ×{\relax\ifmmode\check\else\expandafter\v\fi}
\catcode216=13 \def Ø{\"y}                     
\catcode217=13 \def Ù{\"Y}  		    
\catcode218=13 \def Ú{\Leftarrow}       
\catcode219=13 \def Û{\Leftrightarrow}       
\catcode220=13 \def Ü{\relax\ifmmode\Rightarrow\else\sect\fi}
\catcode221=13 \def Ý{\sum}                  
\catcode222=13 \def Þ{\prod}                 
\catcode223=13 \def ß{\widehat}              
\catcode224=13 \def à{\pm}                     
\catcode225=13 \def á{\nabla}                
\catcode226=13 \def â{\quad}                 
\catcode227=13 \def ã{\in}               	
\catcode228=13 \def ä{\star}      	      
\catcode229=13 \def å{\sqrt}                   
\catcode230=13 \def æ{\^E}			
\catcode231=13 \def ç{\Upsilon}              
\catcode232=13 \def è{\"E}    	   	 
\catcode233=13 \def é{\`E}               	  
\catcode234=13 \def ê{\Sigma}                
\catcode235=13 \def ë{\Delta}                 
\catcode236=13 \def ì{\Phi}                     
\catcode237=13 \def í{\`I}        		   
\catcode238=13 \def î{\iota}        	     
\catcode239=13 \def ï{\Psi}                     
\catcode240=13 \def ð{\times}                  
\catcode241=13 \def ñ{\Lambda}             
\catcode242=13 \def ò{\cdots}                
\catcode243=13 \def ó{\^U}			
\catcode244=13 \def ô{\`U}    	              
\catcode245=13 \def õ{\bo}                       
\catcode246=13 \def ö{\relax\ifmmode\hat\else\expandafter\^\fi}
\catcode247=13 \def÷{\relax\ifmmode\tilde\else\expandafter\~\fi}
\catcode248=13 \def ø{\ll}                         
\catcode249=13 \def ù{\gg}                       
\catcode250=13 \def ú{\eta}                      
\catcode251=13 \def û{\kappa}                  
\catcode252=13 \def ü{\half}     		 
\catcode253=13 \def ý{\Gamma} 		
\catcode254=13 \def þ{\Xi}   			
\catcode255=13 \def ÿ{\relax\ifmmode{}^{\dagger}{}\else\dag\fi}


\def\ital#1Õ{{\it#1\/}}	     
\def\un#1{\relax\ifmmode\underline#1\else $\underline{\hbox{#1}}$
	\relax\fi}

\def\ron#1#2{{\buildrel{#1}\over{#2}}}	
\def\roonoo#1#2#3{\vbox{\ialign{##\crcr
	\hfil{$#3{#1}$}\hfil\crcr\noalign{\kern1pt\nointerlineskip}
	$#3{#2}$\crcr}}}
\def\roon#1#2{\mathchoice{\roonoo{#1}{#2}{\displaystyle}}
	{\roonoo{#1}{#2}{\textstyle}}{\roonoo{#1}{#2}{\scriptstyle}}
	{\roonoo{#1}{#2}{\scriptscriptstyle}}}
\def\rdt#1{\roon{\hbox{\bf .}}{#1}}  
\def\tdt#1{\oon{\hbox{\bf .\kern-1pt.\kern-1pt.}}#1}   
\def\({\eqno(}
\def\li{\eqalignno}
\def\refs{\sect{REFERENCES}\par\medskip \frenchspacing 
	\parskip=0pt \refrm \baselineskip=1.23em plus 1pt
	\def\ital##1Õ{{\refit##1\/}}}
\def\unrefs{\textrm \nonfrenchspacing 
	\parskip=\normalparskip \baselineskip=\normalbaselineskip
	\def\ital##1Õ{{\it##1\/}}}


\def\õ#1{
	\screwcount\num
	\num=1
	\screwdimen\downsy
	\downsy=-1.5ex
	\mkern-3.5mu
	õ
	\loop
	\ifnum\num<#1
	\llap{\raise\num\downsy\hbox{$õ$}}
	\advance\num by1
	\repeat}
\def\upõ#1#2{\screwcount\numup
	\numup=#1
	\advance\numup by-1
	\screwdimen\upsy
	\upsy=.75ex
	\mkern3.5mu
	\raise\numup\upsy\hbox{$#2$}}



\newcount\marknumber	\marknumber=1
\newcount\countdp \newcount\countwd \newcount\countht 

%
%
\ifx\pdfoutput\undefined
\def\rgboo#1{}
\input epsf
\def\fig#1{\epsfbox{#1.eps}}
\def\figscale#1#2{\epsfxsize=#2\epsfbox{#1.eps}}
\def\postscript#1{\special{" #1}}		
\postscript{
	/bd {bind def} bind def
	/fsd {findfont exch scalefont def} bd
	/sms {setfont moveto show} bd
	/ms {moveto show} bd
	/pdfmark where		
	{pop} {userdict /pdfmark /cleartomark load put} ifelse
	[ /PageMode /UseOutlines		
	/DOCVIEW pdfmark}
\def\bookmark#1#2{\postscript{		
	[ /Dest /MyDest\the\marknumber /View [ /XYZ null null null ] /DEST pdfmark
	[ /Title (#2) /Count #1 /Dest /MyDest\the\marknumber /OUT pdfmark}%
	\advance\marknumber by1}
\def\pdfklink#1#2{%
	\hskip-.25em\setbox0=\hbox{#1}%
		\countdp=\dp0 \countwd=\wd0 \countht=\ht0%
		\divide\countdp by65536 \divide\countwd by65536%
			\divide\countht by65536%
		\advance\countdp by1 \advance\countwd by1%
			\advance\countht by1%
		\def\linkdp{\the\countdp} \def\linkwd{\the\countwd}%
			\def\linkht{\the\countht}%
	\postscript{
		[ /Rect [ -1.5 -\linkdp.0 0\linkwd.0 0\linkht.5 ] 
		/Border [ 0 0 0 ]
		/Action << /Subtype /URI /URI (#2) >>
		/Subtype /Link
		/ANN pdfmark}{\rgb{1 0 0}{#1}}}
%
%
\else
\def\rgboo#1{\pdfliteral{#1 rg #1 RG}}
\def\fig#1{\pdfximage {#1.pdf}\pdfrefximage\pdflastximage}
\def\figscale#1#2{\pdfximage width#2 {#1.pdf}\pdfrefximage\pdflastximage}
\pdfcatalog{/PageMode /UseOutlines}		
\def\bookmark#1#2{
	\pdfdest num \marknumber xyz
	\pdfoutline goto num \marknumber count #1 {#2}
	\advance\marknumber by1}
\def\pdfklink#1#2{%
	\noindent\pdfstartlink user
		{/Subtype /Link
		/Border [ 0 0 0 ]
		/A << /S /URI /URI (#2) >>}{\rgb{1 0 0}{#1}}%
	\pdfendlink}
\fi

\def\rgbo#1#2{\rgboo{#1}#2\rgboo{0 0 0}}
\def\rgb#1#2{\mark{#1}\rgbo{#1}{#2}\mark{0 0 0}}
\def\pdflink#1{\pdfklink{#1}{#1}}
\def\xxxlink#1{\pdfklink{#1}{http://arXiv.org/abs/#1}}

\catcode`@=11

\def\wlog#1{}	


\def\makeheadline{\vbox to\z@{\vskip-36.5\p@
	\line{\vbox to8.5\p@{}\the\headline%
	\ifnum\pageno=\z@\rgboo{0 0 0}\else\rgboo{\topmark}\fi%
	}\vss}\nointerlineskip}
\headline={
	\ifnum\pageno=\z@
		\hfil
	\else
		\ifnum\pageno<\z@
			\ifodd\pageno
				\tenrm\romannumeral-\pageno\hfil\lefthead\hfil
			\else
				\tenrm\hfil\righthead\hfil\romannumeral-\pageno
			\fi
		\else
			\ifodd\pageno
				\tenrm\hfil\righthead\hfil\number\pageno
			\else
				\tenrm\number\pageno\hfil\lefthead\hfil
			\fi
		\fi
	\fi}

\catcode`@=12

\def\righthead{\hfil} \def\lefthead{\hfil}
\nopagenumbers


\def\chrulefill{\rgb{1 0 0}{\hrulefill}}
\def\cdotfill{\rgb{1 0 0}{\dotfill}}
\newcount\area	\area=1
\newcount\cross	\cross=1
\def\volume#1\par{\newpage\noindent{\biggest{\rgb{1 .5 0}{#1}}}
	\par\nobreak\bigskip\medskip\area=0}
\def\chapskip{\par\ifnum\area=0\bigskip\medskip\goodbreak
	\else\newpage\fi}
\def\chapy#1{\area=1\cross=0
	\xdef\lefthead{\rgbo{1 0 .5}{#1}}\vbox{\biggerer\offinterlineskip
	\line{\chrulefill¼\hphantom{\lefthead}\chrulefill}
	\line{\chrulefill¼\lefthead\chrulefill}}\par\nobreak\medskip}
\def\chap#1\par{\chapskip\bookmark3{#1}\chapy{#1}}
\def\sectskip{\par\ifnum\cross=0\bigskip\medskip\goodbreak
	\else\newpage\fi}
\def\secty#1{\cross=1
	\xdef\righthead{\rgbo{1 0 1}{#1}}\vbox{\bigger\offinterlineskip
	\line{\cdotfill¼\hphantom{\righthead}\cdotfill}
	\line{\cdotfill¼\righthead\cdotfill}}\par\nobreak\medskip}
\def\sect#1 #2\par{\sectskip\bookmark{#1}{#2}\secty{#2}}
\def\subsectskip{\par\ifdim\lastskip<\medskipamount
	\bigskip\medskip\goodbreak\else\nobreak\fi}
\def\subsecty#1{\noindent{\sectfont{\rgbo{.5 0 1}{#1}}}\par\nobreak\medskip}
\def\subsect#1\par{\subsectskip\bookmark0{#1}\subsecty{#1}}
\long\def\x#1 #2\par{\hangindent2\parindent%
\mark{0 0 1}\rgboo{0 0 1}{\bf Exercise #1}\\#2%
\par\rgboo{0 0 0}\mark{0 0 0}}
\def\refs{\bigskip\noindent{\bf \rgbo{0 .5 1}{REFERENCES}}\par\nobreak\medskip
	\frenchspacing \parskip=0pt \refrm \baselineskip=1.23em plus 1pt
	\def\ital##1Õ{{\refit##1\/}}}
\long\def\twocolumn#1#2{\hbox to\hsize{\vtop{\hsize=2.9in#1}
	\hfil\vtop{\hsize=2.9in #2}}}


\twelvepoint
\font\bigger=cmbx12 \sca2
\font\biggerer=cmb10 \sca5
\font\biggest=cmssdc10 scaled 3583
\font\subtitlefont=cmbsy10 \sca5
\font\suptitlefont=cmbsy10 scaled 3583
\def\subsubtitlefont{\bigger}
\font\subsubsubtitlefont=cmssbx10 \sca3
\font\small=cmr9


\def Ü{\relax\ifmmode\Rightarrow\else\expandafter\subsect\fi}
\def Û{\relax\ifmmode\Leftrightarrow\else\expandafter\sect\fi}
\def Ú{\relax\ifmmode\Leftarrow\else\expandafter\chap\fi}
\def\its{\item{$\bullet$}}
\def\itemize#1 {\item{\bf#1}}
\def\itemizze#1 {\itemitem{\bf#1}}
\def\itemutem{\par\indent\indent \hangindent3\parindent \textindent}
\def\itemizzze#1 {\itemutem{\bf#1}}
\def ª{\relax\ifmmode\leftrightarrow\else\itemizze\fi}
\def Á{\relax\ifmmode\gets\else\itemizzze\fi}

\def\tbt#1#2#3#4{\left({#1\atop#2}{#3\atop#4}\right)}
\def\tat#1#2#3#4{{\textstyle\left({#1\atop#2}{#3\atop#4}\right)}}
\def\astop#1#2{\scriptstyle#1\atop\scriptstyle#2}
\def\¢{\ominus}
\def\A{{\cal A}}  \def\B{{\cal B}}  \def\C{{\cal C}}  \def\D{{\cal D}}
\def\E{{\cal E}}  \def\F{{\cal F}}  \def\G{{\cal G}}  \def\H{{\cal H}}
\def\J{{\cal J}}  \let\Sll=\L  \def\L{{\cal L}}  \def\N{{\cal N}}  
\def\O{{\cal O}}  \def\P{{\cal P}}  \def\Q{{\cal Q}}  \def\R{{\cal R}}
\def\S{{\cal S}}  \def\T{{\cal T}}  \def\V{{\cal V}}  \def\W{{\cal W}}
\def\Z{{\cal Z}}
\def\Ä{\varphi}  \def\¿{\varpi}
\def\h{\hbar}
\def ò{\relax\ifmmode\cdots\else\dotfill\fi}
\def\dit{\hbox{\tt "}}


\def\cvrule{\rgbo{0 .5 1}{\vrule}}
\def\chrule{\rgbo{0 .5 1}{\hrule}}
\def\boxit#1{\leavevmode\thinspace\hbox{\cvrule\vtop{\vbox{\chrule%
	\vskip3pt\kern1pt\hbox{\vphantom{\bf/}\thinspace\thinspace%
	{\bf#1}\thinspace\thinspace}}\kern1pt\vskip3pt\chrule}\cvrule}%
	\thinspace}
\def\Boxit#1{\noindent\vbox{\chrule\hbox{\cvrule\kern3pt\vbox{
	\advance\hsize-7pt\vskip-\parskip\kern3pt\bf#1
	\hbox{\vrule height0pt depth\dp\strutbox width0pt}
	\kern3pt}\kern3pt\cvrule}\chrule}}


\def\boxeq#1{\boxit{${\displaystyle #1 }$}}          
\def\Boxeq#1{\Boxit{\vskip-.1in#1\vskip-.3in}}   


\def\today{\ifcase\month\or
 January\or February\or March\or April\or May\or June\or July\or
 August\or September\or October\or November\or December\fi
 \space\number\day, \number\year}

\parindent=20pt
\newskip\normalparskip	\normalparskip=.7\medskipamount
\parskip=\normalparskip	



\catcode`\|=\active \catcode`\<=\active \catcode`\>=\active 
\def|{\relax\ifmmode\delimiter"026A30C \else$\mathchar"026A$\fi}
\def<{\relax\ifmmode\mathchar"313C \else$\mathchar"313C$\fi}
\def>{\relax\ifmmode\mathchar"313E \else$\mathchar"313E$\fi}


\pageno=0\phantom{duh}\vfill

\centerline{\fig{title}}	\vskip1in{\center
\rgbo{0 0 1}
{{\suptitlefont W}\negthinspace\negthinspace{\subtitlefont ARREN}â{\suptitlefont S}{\subtitlefont IEGEL}}\vskip.5in
{\subsubtitlefont C. N. Yang Institute for Theoretical Physics\vskip-.02in
	State University of New York at Stony Brook\vskip-.02in
	Stony Brook, New York 11794-3840â¼USA}\vskip.5in
{\subsubsubtitlefont 
	\pdflink{mailto:siegel@insti.physics.sunysb.edu}\vskip-.02in
	\pdfklink{http://insti.physics.sunysb.edu/\~{}siegel/plan.html}
	{http://insti.physics.sunysb.edu/\noexpand~siegel/plan.html}
	}\par}

\newpage

\pageno=2

\parskip=0pt 
\baselineskip=13.3pt minus .2pt

ÚCONTENTS

\def\lefthead{\hfil}

\twocolumn{

{\bf\noindent Preface}ò23

}{

{\bf\noindent Some field theory texts}ò36

}

\vskip-.1in

Û0 PART ONE: SYMMETRY

\twocolumn{

{\bf\noindent I. Global}
	ªA. Coordinates
		Á1. Nonrelativityò39
		Á2. Fermionsò46
		Á3. Lie algebraò51
		Á4. Relativityò54
		Á5. Discrete: C, P, Tò65
		Á6. Conformalò68
	ªB. Indices
		Á1. Matricesò73
		Á2. Representationsò76
		Á3. Determinantsò81
		Á4. Classical groupsò84
		Á5. Tensor notationò86
	ªC. Representations
		Á1. More coordinatesò92
		Á2. Coordinate tensorsò94
		Á3. Young tableauxò99
		Á4. Color and flavorò101
		Á5. Covering groupsò107

\vskip6pt minus 4pt
{\bf\noindent II. Spin}
	ªA. Two components
		Á1. 3-vectorsò110
		Á2. Rotationsò114
		Á3. Spinorsò115
		Á4. Indicesò117
		Á5. Lorentzò120
		Á6. Diracò126
		Á7. Chirality/dualityò128
	ªB. Poincar«e
		Á1. Field equationsò131
		Á2. Examplesò134
		Á3. Solutionò137
		Á4. Massò141
		Á5. Foldy-Wouthuysenò144
		Á6. Twistorsò148
		Á7. Helicityò151
	ªC. Supersymmetry
		Á1. Algebraò156
		Á2. Supercoordinatesò157
		Á3. Supergroupsò160
		Á4. Superconformalò163
		Á5. Supertwistorsò164

}{

{\bf\noindent III. Local}
	ªA. Actions
		Á1. Generalò169
		Á2. Fermionsò174
		Á3. Fieldsò176
		Á4. Relativityò180
		Á5. Constrained systemsò186
	ªB. Particles
		Á1. Freeò191
		Á2. Gaugesò195
		Á3. Couplingò197
		Á4. Conservationò198
		Á5. Pair creationò201
	ªC. Yang-Mills
		Á1. Nonabelianò204
		Á2. Lightconeò208
		Á3. Plane wavesò212
		Á4. Self-dualityò213
		Á5. Twistorsò217
		Á6. Instantonsò220
		Á7. ADHMò224
		Á8. Monopolesò226

\vskip6pt minus 4pt
{\bf\noindent IV. Mixed}
	ªA. Hidden symmetry
		Á1. Spontaneous breakdownò232
		Á2. Sigma modelsò234
		Á3. Coset spaceò237
		Á4. Chiral symmetryò240
		Á5. St¬uckelbergò243
		Á6. Higgsò245
		Á7. Dilaton cosmologyò247
	ªB. Standard model
		Á1. Chromodynamicsò259
		Á2. Electroweakò264
		Á3. Familiesò267
		Á4. Grand Unified Theoriesò269
	ªC. Supersymmetry
		Á1. Chiralò275
		Á2. Actionsò277
		Á3. Covariant derivativesò280
		Á4. Prepotentialò282
		Á5. Gauge actionsò284
		Á6. Breakingò287
		Á7. Extendedò289

}\eject

Û0 PART TWO: QUANTA

\def\righthead{\hfil}

\twocolumn{

{\bf\noindent V. Quantization}
	ªA. General
		Á1. Path integralsò298
		Á2. Semiclassical expansionò303
		Á3. Propagatorsò307
		Á4. S-matricesò310
		Á5. Wick rotationò315
	ªB. Propagators
		Á1. Particlesò319
		Á2. Propertiesò322
		Á3. Generalizationsò326
		Á4. Wick rotationò329
	ªC. S-matrix
		Á1. Path integralsò334
		Á2. Graphsò339
		Á3. Semiclassical expansionò344
		Á4. Feynman rulesò349
		Á5. Semiclassical unitarityò355
		Á6. Cutting rulesò358
		Á7. Cross sectionsò361
		Á8. Singularitiesò366
		Á9. Group theoryò368

\vskip6pt minus 4pt
{\bf\noindent VI. Quantum gauge theory}
	ªA. Becchi-Rouet-Stora-Tyutin
		Á1. Hamiltonianò373
		Á2. Lagrangianò378
		Á3. Particlesò381
		Á4. Fieldsò382
	ªB. Gauges
		Á1. Radialò386
		Á2. Lorenzò389
		Á3. Massiveò391
		Á4. Gervais-Neveuò393
		Á5. Super Gervais-Neveuò396
		Á6. Spaceconeò399
		Á7. Superspaceconeò403
		Á8. Background-fieldò406
		Á9. Nielsen-Kalloshò412
		Á10. Super background-fieldò415
	ªC. Scattering
		Á1. Yang-Millsò419
		Á2. Recursionò423
		Á3. Fermionsò426
		Á4. Massesò429
		Á5. Supergraphsò435

}{

{\bf\noindent VII. Loops}
	ªA. General
		Á1. Dimensional renormaliz'nò440
		Á2. Momentum integrationò443
		Á3. Modified subtractionsò447
		Á4. Optical theoremò451
		Á5. Power countingò453
		Á6. Infrared divergencesò458
	ªB. Examples
		Á1. Tadpolesò462
		Á2. Effective potentialò465
		Á3. Dimensional transmut'nò468
		Á4. Massless propagatorsò470
		Á5. Bosonizationò473
		Á6. Massive propagatorsò478
		Á7. Renormalization groupò482
		Á8. Overlapping divergencesò485
	ªC. Resummation
		Á1. Improved perturbationò492
		Á2. Renormalonsò497
		Á3. Borelò500
		Á4. 1/N expansionò504

\vskip6pt minus 4pt
{\bf\noindent VIII. Gauge loops}
	ªA. Propagators
		Á1. Fermionò511
		Á2. Photonò514
		Á3. Gluonò515
		Á4. Grand Unified Theoriesò521
		Á5. Supermatterò524
		Á6. Supergluonò527
		Á7. Schwinger modelò531
	ªB. Low energy
		Á1. JWKBò537
		Á2. Axial anomalyò540
		Á3. Anomaly cancellationò544
		Á4. $¹^0 £ 2©$ò546
		Á5. Vertexò548
		Á6. Nonrelativistic JWKBò551
		Á7. Latticeò554
	ªC. High energy
		Á1. Conformal anomalyò561
		Á2. $e^+ e^- £$ hadronsò564
		Á3. Parton modelò566
		Á4. Maximal supersymmetryò573
		Á5. First quantizationò576

}\newpage

Û0 PART THREE: HIGHER SPIN

\twocolumn{

{\bf\noindent IX. General relativity}
	ªA. Actions
		Á1. Gauge invarianceò587
		Á2. Covariant derivativesò592
		Á3. Conditionsò598
		Á4. Integrationò601
		Á5. Gravityò605
		Á6. Energy-momentumò609
		Á7. Weyl scaleò611
	ªB. Gauges
		Á1. Lorenzò620
		Á2. Geodesicsò622
		Á3. Axialò625
		Á4. Radialò629
		Á5. Weyl scaleò633
	ªC. Curved spaces
		Á1. Self-dualityò638
		Á2. De Sitterò640
		Á3. Cosmologyò642
		Á4. Red shiftò645
		Á5. Schwarzschildò646
		Á6. Experimentsò654
		Á7. Black holesò660

\vskip6pt minus 4pt
{\bf\noindent X. Supergravity}
	ªA. Superspace
		Á1. Covariant derivativesò664
		Á2. Field strengthsò669
		Á3. Compensatorsò672
		Á4. Scale gaugesò675
	ªB. Actions
		Á1. Integrationò681
		Á2. Ectoplasmò684
		Á3. Component transform'nsò687
		Á4. Component approachò689
		Á5. Dualityò692
		Á6. Superhiggsò695
		Á7. No-scaleò698
	ªC. Higher dimensions
		Á1. Dirac spinorsò701
		Á2. Wick rotationò704
		Á3. Other spinsò708
		Á4. Supersymmetryò709
		Á5. Theoriesò713
		Á6. Reduction to D=4ò715

}{

{\bf\noindent XI. Strings}
	ªA. Generalities
		Á1. Regge theoryò724
		Á2. Topologyò728
		Á3. Classical mechanicsò733
		Á4. Typesò736
		Á5. T-dualityò740
		Á6. Dilatonò742
		Á7. Latticesò747
	ªB. Quantization
		Á1. Gaugesò756
		Á2. Quantum mechanicsò761
		Á3. Commutatorsò766
		Á4. Conformal transformat'nsò769
		Á5. Trialityò773
		Á6. Treesò778
		Á7. Ghostsò785
	ªC. Loops
		Á1. Partition functionò791
		Á2. Jacobi Theta functionò794
		Á3. Green functionò797
		Á4. Openò801
		Á5. Closedò806
		Á6. Superò810
		Á7. Anomaliesò814

\vskip6pt minus 4pt
{\bf\noindent XII. Mechanics}
	ªA. OSp(1,1|2)
		Á1. Lightconeò819
		Á2. Algebraò822
		Á3. Actionò826
		Á4. Spinorsò827
		Á5. Examplesò829
	ªB. IGL(1)
		Á1. Algebraò834
		Á2. Inner productò835
		Á3. Actionò837
		Á4. Solutionò840
		Á5. Spinorsò843
		Á6. Massesò844
		Á7. Background fieldsò845
		Á8. Stringsò847
		Á9. Relation to OSp(1,1|2)ò852
	ªC. Gauge fixing
		Á1. Antibracketò855
		Á2. ZJBVò858
		Á3. BRSTò862

}\vskip.17in minus.1in

\line{{\bf\noindent AfterMath}ò866}

\baselineskip=16.9pt plus 1pt

ÚOUTLINE

\def\lefthead{\hfil}

In this Outline we give a brief description of each item listed in
the Contents.  While the Contents and Index are quick ways to search, or
learn the general layout of the book, the Outline
gives more detail for the uninitiated.  (The PDF version also allows use of
the ``Find" command in PDF readers.)

\line{\chrulefill}
\vskip15pt
{\bf\noindent Preface}ò23\\
general remarks on style, organization, focus, content, use, differences from other texts,
etc.

{\bf\noindent Some field theory texts}ò36\\
recommended alternatives or supplements (but see Preface)

Û0 PART ONE: SYMMETRY

\noindent Relativistic quantum mechanics and classical field theory.
Poincar«e group = special relativity.
Enlarged spacetime symmetries: conformal and supersymmetry.
Equations of motion and actions for particles and fields/wave functions.  
Internal symmetries: global (classifying particles), local (field interactions).

\vskip10pt minus 3pt
{\bf\noindent I. Global}\\
Spacetime and internal symmetries.
	ªA. {\bf Coordinates}\\
	spacetime symmetries
		Á1. {\bf Nonrelativity}ò39\\
		Poisson bracket, Einstein summation convention, 
		Galilean symmetry (introductory example)
		Á2. {\bf Fermions}ò46\\
		statistics, anticommutator; anticommuting variables, differentiation, integration
		Á3. {\bf Lie algebra}ò51\\
		general structure of symmetries (including internal);
		Lie bracket, group, structure constants;
		brief summary of group theory
		Á4. {\bf Relativity}ò54\\
		Minkowski space, antiparticles, Lorentz and Poincar«e symmetries, proper time,
		Mandelstam variables, lightcone bases
		Á5. {\bf Discrete: C, P, T}ò65\\
		charge conjugation, parity, time reversal, 
		in classical mechanics and field theory;
		Levi-Civita tensor
		Á6. {\bf Conformal}ò68\\
		broken, but useful, enlargement of Poincar«e; projective lightcone
	ªB. {\bf Indices}\\
	easy way to group theory
		Á1. {\bf Matrices}ò73\\
		Hilbert-space notation
		Á2. {\bf Representations}ò76\\
		adjoint, Cartan metric, Dynkin index, Casimir, (pseudo)reality,
		direct sum and product
		Á3. {\bf Determinants}ò81\\
		with Levi-Civita tensors, Gaussian integrals; Pfaffian
		Á4. {\bf Classical groups}ò84\\
		and generalizations, via tensor methods
		Á5. {\bf Tensor notation}ò86\\
		index notation, simplest bases for simplest representations
	ªC. {\bf Representations}\\
	useful special cases
		Á1. {\bf More coordinates}ò92\\
		Dirac gamma matrices as coordinates for orthogonal groups
		Á2. {\bf Coordinate tensors}ò94\\
		formulations of coordinate transformations; differential forms
		Á3. {\bf Young tableaux}ò99\\
		pictures for representations, their symmetries, sizes, direct products
		Á4. {\bf Color and flavor}ò101\\
		symmetries of particles of Standard Model and observed light hadrons
		Á5. {\bf Covering groups}ò107\\
		relating spinors and vectors

\vskip10pt minus 3pt
{\bf\noindent II. Spin}\\
Extension of spacetime symmetry to include spin.
Field equations for field strengths of all spins.
Most efficient methods for Lorentz indices in 
QuantumChromoDynamics or pure Yang-Mills.
Supersymmetry relates bosons and fermions, also useful for QCD.
	ªA. {\bf Two components}\\
	2$ð$2 matrices describe the spacetime groups more easily (2<4)
		Á1. {\bf 3-vectors}ò110\\
		algebraic properties of 2$ð$2 matrices, vectors as quaternions
		Á2. {\bf Rotations}ò114\\
		in three (space) dimensions
		Á3. {\bf Spinors}ò115\\
		basis for spinor notation
		Á4. {\bf Indices}ò117\\
		review of spin in simpler notation: many indices instead of bigger;
		tensor notation avoids Clebsch-Gordan-Wigner coefficients
		Á5. {\bf Lorentz}ò120\\
		still 2$ð$2 matrices, but four dimensions; dotted and undotted indices;
		antisymmetric tensors; matrix identities
		Á6. {\bf Dirac}ò126\\
		example in free field theory; 4-component identities
		Á7. {\bf Chirality/duality}ò128\\
		chiral symmetry, simpler with two-component spinor indices;
		more examples; duality
	ªB. {\bf Poincar«e}\\
	relativistic solutions
		Á1. {\bf Field equations}ò131\\
		conformal group as unified way to all massless free equations
		Á2. {\bf Examples}ò134\\
		reproduction of familiar cases (Dirac and Maxwell equations)
		Á3. {\bf Solution}ò137\\
		proof; lightcone methods; transformations
		Á4. {\bf Mass}ò141\\
		dimensional reduction; St¬uckelberg formalism for vector in terms
		of massless vector + scalar
		Á5. {\bf Foldy-Wouthuysen}ò144\\
		an application, for arbitrary spin, from massless analog; 
		transformation to nonrelativistic + corrections;
		minimal electromagnetic coupling to spin 1/2; preparation for
		nonminimal coupling in chapter VIII for Lamb shift
		Á6. {\bf Twistors}ò148\\
		convenient and covariant method to solve massless equations;
		related to conformal invariance and self-duality;
		useful for QCD computations in chapter VI
		Á7. {\bf Helicity}ò151\\
		via twistors; Penrose transform
	ªC. {\bf Supersymmetry}\\
	symmetry relating fermions to bosons, generalizing translations;
	general properties, representations
		Á1. {\bf Algebra}ò156\\
		definition of supersymmetry; positive energy automatic
		Á2. {\bf Supercoordinates}ò157\\
		superspace includes anticommuting coordinates; 
		covariant derivatives\\ generalize spacetime derivatives
		Á3. {\bf Supergroups}ò160\\
		generalizing classical groups; supertrace, superdeterminant
		Á4. {\bf Superconformal}ò163\\
		also broken but useful, enlargement of supersymmetry, as classical 
		group
		Á5. {\bf Supertwistors}ò164\\
		massless representations of supersymmetry

\vskip10pt minus 3pt
{\bf\noindent III. Local}\\
Symmetries that act independently at each point in spacetime.
Basis of fundamental forces.
	ªA. {\bf Actions}\\
	for previous examples (spins 0, 1/2, 1)
		Á1. {\bf General}ò169\\
		action principle, variation, functional derivative, Lagrangians
		Á2. {\bf Fermions}ò174\\
		quantizing anticommuting quantities; spin
		Á3. {\bf Fields}ò176\\
		actions in nonrelativistic field theory, Hamiltonian and Lagrangian 
		densities
		Á4. {\bf Relativity}ò180\\
		relativistic particles and fields, charge conjugation, good ultraviolet 
		behavior,  general forces
		Á5. {\bf Constrained systems}ò186\\
		role of gauge invariance; first-order formalism; gauge fixing
	ªB. {\bf Particles}\\
	relativistic classical mechanics; useful later in understanding
	Feynman diagrams; simple example of local symmetry
		Á1. {\bf Free}ò191\\
		worldline metric, gauge invariance of actions
		Á2. {\bf Gauges}ò195\\
		gauge fixing, lightcone gauge
		Á3. {\bf Coupling}ò197\\
		external fields
		Á4. {\bf Conservation}ò198\\
		for classical particles; true vs.¼canonical energy
		Á5. {\bf Pair creation}ò201\\
		and annihilation, for classical particle and antiparticle
	ªC. {\bf Yang-Mills}\\
	self-coupling for spin 1; describes forces of Standard Model
		Á1. {\bf Nonabelian}ò204\\
		self-interactions; covariant derivatives, field strengths, Jacobi identities, 
		action
		Á2. {\bf Lightcone}ò208\\
		a unitary gauge; axial gauges; spin 1/2
		Á3. {\bf Plane waves}ò212\\
		simple exact solutions to interacting theory
		Á4. {\bf Self-duality}ò213\\
		and massive analog 
		Á5. {\bf Twistors}ò217\\
		useful for self-duality; lightcone gauge for solving self-duality
		Á6. {\bf Instantons}ò220\\
		nonperturbative self-dual solutions, via twistors; 't Hooft ansatz;
		Chern-Simons form
		Á7. {\bf ADHM}ò224\\
		general instanton solution of Atiyah, Drinfel'd, Hitchin, and Manin
		Á8. {\bf Monopoles}ò226\\
		more nonperturbative self-dual solutions, but static

\vskip10pt minus 3pt
{\bf\noindent IV. Mixed}\\
Global symmetries of interacting theories.
Gauge symmetry coupled to lower spins.
	ªA. {\bf Hidden symmetry}\\
	explicit and soft breaking, confinement
		Á1. {\bf Spontaneous breakdown}ò232\\
		method; Goldstone theorem of massless scalars
		Á2. {\bf Sigma models}ò234\\
		linear and nonlinear; low-energy theories of scalars
		Á3. {\bf Coset space}ò237\\
		general construction, using gauge invariance, for sigma models
		Á4. {\bf Chiral symmetry}ò240\\
		low-energy symmetry, quarks, pseudogoldstone boson, 
		Partially Con-\\ served Axial Current
		Á5. {\bf St¬uckelberg}ò243\\
		scalars generate mass for vectors; free case
		     Á6. {\bf Higgs}ò245\\
		     same for interactions; Gervais-Neveu model; unitary gauge
		     Á7. {\bf Dilaton cosmology}ò247\\
		     cosmology with gravity replaced by Goldstone boson of scale invariance
	ªB. {\bf Standard model}\\
	application to real world
		Á1. {\bf Chromodynamics}ò259\\
		strong interactions, using Yang-Mills; C and P
		Á2. {\bf Electroweak}ò264\\
		unification of electromagnetic and weak interactions, 
		using also Higgs
		Á3. {\bf Families}ò267\\
		including all known fundamental leptons; 
		Cabibbo-Kobayashi-Maskawa transformation;
		flavor-changing neutral currents
		Á4. {\bf Grand Unified Theories}ò269\\
		unification of all leptons and vector mesons
	ªC. {\bf Supersymmetry}\\
	superfield theory, using superspace;
	useful for solving problems of perturbation resummation (chapter VIII)
		Á1. {\bf Chiral}ò275\\
		simplest (``matter") multiplet
		Á2. {\bf Actions}ò277\\
		to introduce interactions; component expansion, superfield equations
		Á3. {\bf Covariant derivatives}ò280\\
		approach to gauge multiplet; vielbein, torsion; solution to Jacobi 
		identities
		Á4. {\bf Prepotential}ò282\\
		fundamental superfield for constructing covariant derivatives;
		solution to constraints, chiral representation
		Á5. {\bf Gauge actions}ò284\\
		for gauge and matter multiplets; Fayet-Iliopoulos term
		Á6. {\bf Breaking}ò287\\
		of supersymmetry; spurions
		Á7. {\bf Extended}ò289\\
		introduction to multiple supersymmetries; central charges

Û0 PART TWO: QUANTA

\noindent Quantum aspects of field theory.
Perturbation theory:  
expansions in loops, helicity, and internal symmetry.
Although some have
conjectured that nonperturbative approaches might solve
renormalization difficulties found in perturbation, all evidence indicates
these problems worsen instead in complete theory.

\vskip10pt minus 3pt
{\bf\noindent V. Quantization}\\
Quantization of classical theories by path integrals.
Backgrounds fields instead of sources exclusively:  
All uses of Feynman diagrams involve either
S-matrix or effective action, both of which require removal of
external propagators, equivalent to replacing sources with
fields.
	ªA. {\bf General}\\
	various properties of quantum physics in general context, 
	so these items need not be repeated in more specialized and
	complicated cases of field theory
		Á1. {\bf Path integrals}ò298\\
		Feynman's alternative to Heisenberg and Schr¬odinger methods;
		relation to canonical quantization; unitarity, causality
		Á2. {\bf Semiclassical expansion}ò303\\
		JWKB in path integral; free particle
		Á3. {\bf Propagators}ò307\\
		Green functions; solution to Schr¬odinger equation via path integrals
		Á4. {\bf S-matrices}ò310\\
		scattering, most common use of field theory; unitarity, causality
		Á5. {\bf Wick rotation}ò315\\
		imaginary time, to get Euclidean space, has important role
		in quantum mechanics
	ªB. {\bf Propagators}\\
	relativistic quantum mechanics, free quantum field theory
		Á1. {\bf Particles}ò319\\
		St¬uckelberg-Feynman propagator for spin 0; 
		covariant gauge, lightcone gauge
		Á2. {\bf Properties}ò322\\
		features, relations to classical Green functions, inner product
		Á3. {\bf Generalizations}ò326\\
		other spins, nature of quantum corrections
		Á4. {\bf Wick rotation}ò329\\
		its relativistic use, in mechanics and field theory
	ªC. {\bf S-matrix}\\
	path integration of field theory produces Feynman diagrams/graphs;
	simple examples from scalar theories
		Á1. {\bf Path integrals}ò334\\
		definition of initial/final states;
		generating functional of background fields;
		perturbative evaluation
		Á2. {\bf Graphs}ò339\\
		pictorial interpretation of perturbation theory;
		connected and one-\\ particle-irreducible parts, effective action
		Á3. {\bf Semiclassical expansion}ò344\\
		classical (tree) graphs give perturbative solution to classical field 
		equations
		Á4. {\bf Feynman rules}ò349\\
		collection of simplified steps from action to graphs,
		in Wick-rotated (Euclidean) momentum space
		Á5. {\bf Semiclassical unitarity}ò355\\
		properties of classical action needed for unitarity
		Á6. {\bf Cutting rules}ò358\\
		diagrammatic translation of unitarity and causality
		Á7. {\bf Cross sections}ò361\\
		scattering probabilities;
		differential cross sections; cut propagators
		Á8. {\bf Singularities}ò366\\
		relation of Landau singularities in momenta to classical mechanics
		Á9. {\bf Group theory}ò368\\
		quark line rules for easily dealing with group theory in graphs

\vskip10pt minus 3pt
{\bf\noindent VI. Quantum gauge theory}\\
Gauge fixing and more complicated vertices require additional methods.
	ªA. {\bf Becchi-Rouet-Stora-Tyutin}\\
	easiest way to gauge fix, with fermionic symmetry
	relating unphysical degrees of freedom; 
	unitarity clear by relating general gauges to unitary gauges;
	general discussion in framework of quantum physics and 
	canonical quantization, so field theory can be addressed
	covariantly with path integrals
		Á1. {\bf Hamiltonian}ò373\\
		canonical quantization, ghosts (unphysical states), 
		cohomology (physical states/operators)
		Á2. {\bf Lagrangian}ò378\\
		relation to Hamiltonian approach, Nakanishi-Lautrup auxiliary fields
		Á3. {\bf Particles}ò381\\
		first-quantization
		Á4. {\bf Fields}ò382\\
		Yang-Mills, unitary gauges
	ªB. {\bf Gauges}\\
	different gauges for different uses (else BRST wouldn't be necessary)
		Á1. {\bf Radial}ò386\\
		for particles in external fields
		Á2. {\bf Lorenz}ò389\\
		covariant class of gauges is simplest; 
		Landau, Fermi-Feynman gauges
		Á3. {\bf Massive}ò391\\
		Higgs requires modifications; 
		unitary and renormalizable gauges
		Á4. {\bf Gervais-Neveu}ò393\\
		special Lorenz gauges that simplify interactions,
		complex gauges similar to lightcone; anti-Gervais-Neveu
		Á5. {\bf Super Gervais-Neveu}ò396\\
		supersymmetry has interesting new features
		Á6. {\bf Spacecone}ò399\\
		general axial gauges;
		Wick rotation of lightcone,
		best gauge for trees;
		lightcone-based simplifications for covariant rules
		Á7. {\bf Superspacecone}ò403\\
		supersymmetric rules also useful for nonsupersymmetric
		theories (like QCD)
		Á8. {\bf Background-field}ò406\\
		class of gauges that simplifies BRST to ordinary gauge invariance
		for loops
		Á9. {\bf Nielsen-Kallosh}ò412\\
		methods for more general gauges
		Á10. {\bf Super background-field}ò415\\
		again new features for superfields;
		prepotentials only as potentials
	ªC. {\bf Scattering}\\
	applications to S-matrices
		Á1. {\bf Yang-Mills}ò419\\
		explicit tree graphs made easy; 4-gluon and 5-gluon trees evaluated
		Á2. {\bf Recursion}ò423\\
		methods for generalizations to arbitrary number of external lines
		Á3. {\bf Fermions}ò426\\
		similar simplifications for high-energy QCD
		Á4. {\bf Masses}ò429\\
		more complicated trees for massive theories;
		all 4-point tree amplitudes of QED, differential cross sections
		Á5. {\bf Supergraphs}ò435\\
		supersymmetric theories are simpler because of superspace;
		anticommuting integrals reduce to algebra of covariant derivatives;
		explicit locality of effective action in anticommuting coordinates implies 
		nonrenormalization theorems

\vskip10pt minus 3pt
{\bf\noindent VII. Loops}\\
General features of higher orders in perturbation theory due to
momentum integration.
	ªA. {\bf General}\\
	properties and methods
		Á1. {\bf Dimensional renormalization}ò440\\
		eliminating infinities;
		method (but not proof); dimensional regularization
		Á2. {\bf Momentum integration}ò443\\
		general method for performing integrals, Beta and Gamma functions
		Á3. {\bf Modified subtractions}ò447\\
		schemes: minimal (MS), modified minimal ($Ñ{\rm MS}$),
		momentum (MOM)
		Á4. {\bf Optical theorem}ò451\\
		unitarity applied to loops; decay rates
		Á5. {\bf Power counting}ò453\\
		how divergent, UV divergences, divergent terms, renormalizability,\\
		Furry's theorem
		Á6. {\bf Infrared divergences}ò458\\
		brief introduction to long-range infinities; soft and colinear divergences
	ªB. {\bf Examples}\\
	mostly one loop
		Á1. {\bf Tadpoles}ò462\\
		simplest examples: one external line, one or more loops,
		massless and massive
		Á2. {\bf Effective potential}ò465\\
		simplest application -- low energy; first-quantization
		Á3. {\bf Dimensional transmutation}ò468\\
		most important loop effect; massless theories can get mass
		Á4. {\bf Massless propagators}ò470\\
		next simplest examples
		Á5. {\bf Bosonization}ò473\\
		fermion fields from boson fields in D=2
		Á6. {\bf Massive propagators}ò478\\
		masses mean dimensional analysis is not as useful
		Á7. {\bf Renormalization group}ò482\\
		application of dimensional transmutation;
		running of couplings at high energy 
		Á8. {\bf Overlapping divergences}ò485\\
		two-loop complications (massless);
		renormalization of subdivergences
	ªC. {\bf Resummation}\\
	how good perturbation is
		Á1. {\bf Improved perturbation}ò492\\
		using renormalization group to resum
		Á2. {\bf Renormalons}ò497\\
		how good renormalization is;
		instantons and IR and UV renormalons create ambiguities
		tantamount to nonrenormalizability
		Á3. {\bf Borel}ò500\\
		improving resummation;
		ambiguities related to nonperturbative vacuum values of composite 
		fields
		Á4. {\bf 1/N expansion}ò504\\
		reorganization of resummation based on group theory;
		useful at finite orders of perturbation;
		related to string theory; Okubo-Zweig-Iizuka rule;
		a solution to instanton ambiguity

\vskip10pt minus 3pt
{\bf\noindent VIII. Gauge loops}\\
(Mostly) one-loop complications in gauge theories.
	ªA. {\bf Propagators}\\
	QED and QCD
		Á1. {\bf Fermion}ò511\\
		correction to fermion propagator from gauge field
		Á2. {\bf Photon}ò514\\
		correction to gauge propagator from matter
		Á3. {\bf Gluon}ò515\\
		correction to gauge propagator from self-interaction;
		total contribution to high-energy behavior
		Á4. {\bf Grand Unified Theories}ò521\\
		3 gauge couplings running to 1 at high energy
		Á5. {\bf Supermatter}ò524\\
		supergraphs at 1 loop
		Á6. {\bf Supergluon}ò527\\
		finite N=1 supersymmetric theories as solution to renormalon problem
		Á7. {\bf Schwinger model}ò531\\
		kinematic ``bound" states at one loop (quantum St¬uckelberg), 
		axial\\ anomaly
	ªB. {\bf Low energy}\\
	QED and anomaly effects
		Á1. {\bf JWKB}ò537\\
		first-quantized approach to 1 loop at low-enregy
		Á2. {\bf Axial anomaly}ò540\\
		classical symmetry broken at one loop
		Á3. {\bf Anomaly cancellation}ò544\\
		constraints on electroweak models
		Á4. {\bf $¹^0 £ 2©$}ò546\\
		application of uncanceled anomaly
		Á5. {\bf Vertex}ò548\\
		one-loop 3-point function in QED
		Á6. {\bf Nonrelativistic JWKB}ò551\\
		nonrelativistic form of effective action useful for finding Lamb shift 
		(including anomalous magnetic moment), using
		Foldy-Wouthuysen transformation
		Á7. {\bf Lattice}ò554\\
		lattices for nonperturbative QCD; regulator; no gauge fixing;
		problems with fermions;
		Wilson loop, confinement; nonuniversality
	ªC. {\bf High energy}\\
	brief introduction to perturbative QCD
		Á1. {\bf Conformal anomaly}ò561\\
		relation to asymptotic freedom
		Á2. {\bf $e^+ e^- £$ hadrons}ò564\\
		simplest QCD loop application; jets
		Á3. {\bf Parton model}ò566\\
		factorization and evolution;
		deep inelastic and Drell-Yan scattering
		Á4. {\bf Maximal supersymmetry}ò573\\
		3- \& 4-pt.¼amplitudes for N=4 supersymmetry
		Á5. {\bf First quantization}ò576\\
		making use of the worldline for loop calculations

Û0 PART THREE: HIGHER SPIN

\noindent General spins.  
Spin 2 must be included in any complete theory of nature.  
Higher spins are observed experimentally for bound states, 
but may be required also as fundamental fields.

\vskip10pt minus 3pt
{\bf\noindent IX. General relativity}\\
Treatment closely related to that applied to Yang-Mills,
super Yang-Mills, and supergravity.
Based on methods that can be applied directly to spinors, and
therefore to supergravity and superstrings.
Somewhat new, but simplest, methods of
calculating curvatures for purposes of solving the classical field equations.
	ªA. {\bf Actions}\\
	starting point for deriving field equations for gravity (and matter)
		Á1. {\bf Gauge invariance}ò587\\
		curved (spacetime) and flat (tangent space) indices;
		coordinate (spacetime) and local Lorentz (tangent space) symmetries
		Á2. {\bf Covariant derivatives}ò592\\
		gauge fields: vierbein (coordinate symmetry) and Lorentz connection;
		generalization of unit vectors used as basis in curvilinear coordinates;
		basis for deriving field strengths (torsion, curvature), matter coupling;
		Killing vectors (symmetries of solutions)
		Á3. {\bf Conditions}ò598\\
		gauges, constraints;
		Weyl tensor, Ricci tensor and scalar
		Á4. {\bf Integration}ò601\\
		measure, invariance, densities
		Á5. {\bf Gravity}ò605\\
		pure gravity, field equations
		Á6. {\bf Energy-momentum}ò609\\
		matter coupling; gravitational energy-momentum
		Á7. {\bf Weyl scale}ò611\\
		used later for cosmological and spherically symmetric solutions, 
		gauge fixing, field redefinitions, studying conformal properties,
		generalization to supergravity and strings
	ªB. {\bf Gauges}\\
	coordinate and other choices
		Á1. {\bf Lorenz}ò620\\
		globally Lorentz covariant gauges, de Donder gauge;
		perturbation, BRST
		Á2. {\bf Geodesics}ò622\\
		straight lines as solutions for particle equations of motion;
		dust
		Á3. {\bf Axial}ò625\\
		simplest unitary gauges; lightcone, Gaussian normal coordinates
		Á4. {\bf Radial}ò629\\
		gauges for external fields, Riemann normal coordinates;
		local inertial frame, parallel transport
		Á5. {\bf Weyl scale}ò633\\
		use of Weyl scale invariance to simplify gauges; dilaton;
		string gauge
	ªC. {\bf Curved spaces}\\
	solutions
		Á1. {\bf Self-duality}ò638\\
		simplest solutions, waves; lightcone gauge for self-duality
		Á2. {\bf De Sitter}ò640\\
		cosmological term and its vacuum, 
		spaces of constant curvature
		Á3. {\bf Cosmology}ò642\\
		the universe, Big Bang
		Á4. {\bf Red shift}ò645\\
		cosmological measurements:
		Hubble constant, deceleration parameter
		Á5. {\bf Schwarzschild}ò646\\
		spherical symmetry;
		applications of general methods for solving field equations;
		electromagnetism
		Á6. {\bf Experiments}ò654\\
		classic experimental tests: gravitational redshift, geodesics
		Á7. {\bf Black holes}ò660\\
		extrapolation of spherically symmetric solutions
		(Kruskal-Szekeres coordinates);
		gravitational collapse, event horizon, physical singularity

\vskip10pt minus 3pt
{\bf\noindent X. Supergravity}\\
Graviton and spin-3/2 particle (gravitino) from supersymmetry;
local supersymmetry.
	ªA. {\bf Superspace}\\
	simplest (yet complicated) method for general applications,
	especially quantum
		Á1. {\bf Covariant derivatives}ò664\\
		general starting point for gauge theories;
		R gauge symmetry;
		constraints, solution, prepotentials
		Á2. {\bf Field strengths}ò669\\
		generalization of curvatures;
		solution of Bianchi (Jacobi) identities
		Á3. {\bf Compensators}ò672\\
		more than one generalization of dilaton; 
		minimal coupling
		Á4. {\bf Scale gauges}ò675\\
		supersymmetric generalization of Weyl scale;
		transformations of field\\ strengths, compensators
	ªB. {\bf Actions}\\
	generalization from global supersymmetry, using superspace,
	components, and compensators
		Á1. {\bf Integration}ò681\\
		different measures, supergravity action, matter,
		first-order action
		Á2. {\bf Ectoplasm}ò684\\
		alternative method of integration,
		geared for components
		Á3. {\bf Component transformations}ò687\\
		derivation from superspace
		Á4. {\bf Component approach}ò689\\
		starting directly from components can be simpler
		for pure supergravity; second-order vs.¼first-order
		formulations
		Á5. {\bf Duality}ò692\\
		relations of different formulations, old minimal vs.¼new minimal
		Á6. {\bf Superhiggs}ò695\\
		mass to the gravitino; analysis using compensators
		Á7. {\bf No-scale}ò698\\
		simple models with naturally vanishing cosmological constant;
		sigma-model symmetries
	ªC. {\bf Higher dimensions}\\
	useful for superstring and other unifications; extended supergravity
		Á1. {\bf Dirac spinors}ò701\\
		spinors for general orthogonal groups; metrics; also useful for GUTs
		Á2. {\bf Wick rotation}ò704\\
		generalization to arbitrary signature; sigma matrices, Majorana spinors
		Á3. {\bf Other spins}ò708\\
		all types of fields that can appear in supergravity;
		restrictions on number of dimensions
		Á4. {\bf Supersymmetry}ò709\\
		supersymmetry in general dimensions; extended supersymmetry
		Á5. {\bf Theories}ò713\\
		examples of supersymmetric and supergravity theories
		Á6. {\bf Reduction to D=4}ò715\\
		D=4 theories from higher dimensions; 
		extended supergravities; S-duality

\vskip10pt minus 3pt
{\bf\noindent XI. Strings}\\
Approach to studying most important yet least understood property of QCD: 
confinement.
Other proposed methods have achieved explicit results for only low hadron energy.
String theory is also useful for field theory in general.
	ªA. {\bf Generalities}\\
	known string theories not suitable for describing hadrons
	quantitatively, but useful models of observed properties;
	qualitative features of general string theories,
	using dilaton and closed = open $°$ open
		Á1. {\bf Regge theory}ò724\\
		observed high-energy behavior of hadrons;
		bound states, s-t duality
		Á2. {\bf Topology}ò728\\
		loop simplifications from geometry;
		closed from open
		Á3. {\bf Classical mechanics}ò733\\
		action, string tension, Virasoro constraints,
		boundary conditions
		Á4. {\bf Types}ò736\\
		reality and group properties; twisting;
		generalizations for massless part of theory;
		supergravity theories appearing in superstrings;
		string types: heterotic, Types I and II
		Á5. {\bf T-duality}ò740\\
		strings unify massless antisymmetric tensors with gravity;
		transformations, O(D,D)
		Á6. {\bf Dilaton}ò742\\
		how it appears in strings and superstrings; 
		constraints on backgrounds; S-duality
		Á7. {\bf Lattices}ò747\\
		discretization of string worldsheet into sum of Feynman diagrams;
		alternative lattice theories relevant to string theory of hadrons
	ªB. {\bf Quantization}\\
	calculational methods; spectrum; tree graphs
		Á1. {\bf Gauges}ò756\\
		fixing 2D coordinates; conformal and lightcone gauges;
		 open vs.¼closed strings
		Á2. {\bf Quantum mechanics}ò761\\
		mode expansion; spectrum; ghosts
		Á3. {\bf Commutators}ò766\\
		commutators from propagators
		Á4. {\bf Conformal transformations}ò769\\
		consequences of conformal invariance of the worldsheet;
		vertex operators
		Á5. {\bf Triality}ò773\\
		bosonization relates physical fermions and bosons
		Á6. {\bf Trees}ò778\\
		interactions, path integral;
		Regge behavior, but not parton behavior
		Á7. {\bf Ghosts}ò785\\
		ghosts manifest conformal invariance
	ªC. {\bf Loops}\\
	quantum field theory corrections via first quantization
		Á1. {\bf Partition function}ò791\\
		general setup
		Á2. {\bf Jacobi Theta function}ò794\\
		functional properties for analyzing amplitudes
		Á3. {\bf Green function}ò797\\
		main part of path integral
		Á4. {\bf Open}ò801\\
		singularities, including generation of closed strings
		Á5. {\bf Closed}ò806\\
		singularities, modular invariance
		Á6. {\bf Super}ò810\\
		supersymmetry, cancellation of divergences
		Á7. {\bf Anomalies}ò814\\
		avoidance of potential problems with 10D supergravity

\vskip10pt minus 3pt
{\bf\noindent XII. Mechanics}\\
General derivation of free actions for any gauge
theory, based on adding equal numbers of 
commuting and anticommuting ghost dimensions.
Usual ghost fields appear as components of gauge fields in
anticommuting directions, as do necessary auxiliary fields like
determinant of metric tensor in gravity.
	ªA. {\bf OSp(1,1|2)}\\
	enlarged group of BRST, applied to first-quantization
		Á1. {\bf Lightcone}ò819\\
		BRST based on lightcone formulation of Poincar«e group
		Á2. {\bf Algebra}ò822\\
		add extra dimensions; nonminimal terms
		Á3. {\bf Action}ò826\\
		for general spin
		Á4. {\bf Spinors}ò827\\
		slight generalization for half-integer spin
		Á5. {\bf Examples}ò829\\
		specialization to usual known results:
		massless spins 0, $ü$, 1, $\f32$, 2
	ªB. {\bf IGL(1)}\\
	subalgebra is simpler and sufficient; 
	gauge fixing is automatic
		Á1. {\bf Algebra}ò834\\
		restriction from OSp(1,1|2)
		Á2. {\bf Inner product}ò835\\
		modified by restriction
		Á3. {\bf Action}ò837\\
		simpler form, but extra fields
		Á4. {\bf Solution}ò840\\
		of cohomology;
		proof of equivalence to lightcone (unitarity)
		Á5. {\bf Spinors}ò843\\
		modified action; cohomology
		Á6. {\bf Masses}ò844\\
		by dimensional reduction; examples: spins $ü$, 1
		Á7. {\bf Background fields}ò845\\
		as generalization of BRST operator;
		vertex operators in Yang-Mills
		Á8. {\bf Strings}ò847\\
		as special case; ghost structure from OSp;
		dilaton vs.¼physical scalar;
		heterotic string;
		vertex operators
		Á9. {\bf Relation to OSp(1,1|2)}ò852\\
		proof of equivalence
	ªC. {\bf Gauge fixing}\\
	Fermi-Feynman gauge is automatic
		Á1. {\bf Antibracket}ò855\\
		antifields and antibracket appear naturally from
		anticommuting coordinate, first-quantized ghost of
		Klein-Gordon equation
		Á2. {\bf ZJBV}ò858\\
		equivalence to Zinn-Justin-Batalin-Vilkovisky method
		Á3. {\bf BRST}ò862\\
		relation to field theory BRST

\vskip.5in minus.3in

\line{{\bf\noindent AfterMath}ò866}

\noindent Following the body of the text (and preceding the Index) is the AfterMath,
containing conventions and some of the more important equations.

\parskip=.3\medskipamount

Û7 PREFACE

\def\lefthead{\hfil}\def\righthead{\hfil}

ÜScientific method

Although there are many fine textbooks on quantum field theory, they all
have various shortcomings.  ÓInstinctÕ is claimed as a basis for most
discussions of quantum field theory, though clearly this topic is too
recent to affect evolution.  Their subjectivity more accurately identifies
this as ÓfashionÕ:  (1) The Óold-fashionedÕ approach justifies itself with
the instinct of ÓintuitionÕ.  However, anyone who remembers when they
first learned quantum mechanics or special relativity knows they are
counter-intuitive; quantum field theory is the synthesis of those two
topics.  Thus, the intuition in this case is probably just ÓhabitÕ:  Such an
approach is actually ÓhistoricalÕ or ÓtraditionalÕ, recounting the
chronological development of the subject.  Generally the first half (or
volume) is devoted to quantum electrodynamics, treated in the way it
was viewed in the 1950's, while the second half tells the story of
quantum chromodynamics, as it was understood in the 1970's.  Such a
``dualistic" approach is necessarily redundant, e.g., using canonical
quantization for QED but path-integral quantization for QCD, contrary to
scientific principles, which advocate applying the same ``unified"
methods to all theories.  While some teachers may feel more comfortable
by beginning a topic the way they first learned it, students may wonder
why the course didn't begin with the approach that they will wind up
using in the end.  Topics that are unfamiliar to the author's intuition are
often labeled as ``formal" (lacking substance) or even ``mathematical"
(devoid of physics).  Recent topics are usually treated there as advanced: 
The opposite is often true, since explanations simplify with time, as the
topic is better understood.  On the positive side, this approach generally
presents topics with better experimental verification.

(2) In contrast, the ÓfashionableÕ approach is described as being based on
the instinct of ÓbeautyÕ.  But this subjective beauty of ÓartÕ is not the
instinctive beauty of nature, and in science it is merely a consolation. 
Treatments based on this approach are usually found in review articles
rather than textbooks, due to the shorter life expectancy of the latest
fashion. On the other hand, this approach has more imagination than the
traditional one, and attempts to capture the future of the subject.

A related issue in the treatment of field theory is the relative importance
of ÓconceptsÕ vs.¼ÓcalculationsÕ:  (1) Some texts emphasize the concepts,
including those which have not proven of practical value, but were
considered motivational historically (in the traditional approach) or
currently (in the artistic approach).  However, many approaches that
were once considered at the forefront of research have faded into
oblivion not because they were proven wrong by experimental evidence
or lacked conceptual attractiveness, but because they were too complex
for calculation, or so vague they lacked predicitive ability.  Some methods
claimed total generality, which they used to prove theorems (though
sometimes without examples); but ultimately the only useful proofs of
theorems are by construction.  Often a dualistic, two-volume approach is
again advocated (and frequently the author writes only one of the two
volumes):  Like the traditional approach of QED volume + QCD volume,
some prefer concept volume + calculation volume.  Generally, this means
that gauge theory S-matrix calculations are omitted from the conceptual
field theory course, and left for a ``particle physics" course, or perhaps
an ``advanced field theory" course.  Unfortunately, the particle physics
course will find the specialized techniques of gauge theory too technical
to cover, while the advanced field theory course will frighten away many
students by its title alone.

(2) On the other hand, some authors express a desire to introduce
Feynman graphs as quickly as possible:  This suggests a lack of
appreciation of field theory outside of diagrammatics.  Many essential
aspects of field theory (such as symmetry breaking and the Higgs effect)
can be seen only from the action, and its analysis also leads to better
methods of applying perturbation theory than those obtained from a
fixed set of rules.  Also, functional equations are often simpler than
pictorial ones, especially when they are nonlinear in the fields.  The result
of over-emphasizing the calculations is a cookbook, of the kind familiar
from some lower-division undergraduate courses intended for physics
majors but designed for engineers. 

The best explanation of a theory is the one that fits the principles of
Óscientific methodÕ: simplicity, generality, and experimental verification. 
In this text we thus take a more ÓeconomicalÕ or ÓpragmaticÕ approach,
with methods based on efficiency and power.  Unattractiveness or
counter-intuitiveness of such methods become advantages, because they
force one to accept new and better ways of thinking about the subject: 
The efficiency of the method directs one to the underlying idea.  For
example, although some consider Einstein's original explanation of special
relativity in terms of relativistic trains and Lorentz transformations with
square roots as being more physical, the concept of Minkowski space
gave a much simpler explanation and deeper understanding that proved
more useful and led to generalization.  Many theories have ``miraculous
cancellations" when traditional methods are used, which led to new
methods (background field gauge, supergraphs, spacecone, etc.)¼that not
only incorporate the cancellations automatically (so that the ``zeros"
need not be calculated), but are built on the principles that explain them. 
We place an emphasis on such new concepts, as well as the calculational
methods that allow them to be compared with nature.  It is important not
to neglect one for the sake of the other, artificial and misleading to try to
separate them.

As a result, many of our explanations of the standard topics are new to
textbooks, and some are completely new.  For example: 
\item{(1)} We derive the
Foldy-Wouthuysen transformation by dimensional reduction from an
analogous one for the massless case (subsections IIB3,5).  
\item{(2)} Cosmology is discussed with just the dilaton instead of full general relativity (subsection IVA7).  With only some minor fudges, this is sufficient to fit the post-inflation universe to observations.
\item{(3)} We derive
the Feynman rules in terms of background fields rather than sources
(subsection VC1); this avoids the need for amputation of external lines for
S-matrices or effective actions, and is more useful for background-field
gauges.  
\item{(4)} We obtain the nonrelativistic QED effective action, used in
modern treatments of the Lamb shift (because it makes perturbation
easier than the older Bethe-Salpeter methods), by field redefinition of the
relativistic effective action (subsection VIIIB6), rather than fitting
parameters by comparing Feynman diagrams from the relativistic and
nonrelativistic actions.  (In general, manipulations in the action are easier
than in diagrams.)  
\item{(5)} We present two somewhat new methods for solving
for the covariant derivatives and
curvature in general relativity that are slightly easier than all
previous methods (subsections IXA2,A7,C5). 
 
\noindent There are also some completely
new topics, like: 
\item{(1)} the anti-Gervais-Neveu gauge, where spin in U(N)
Yang-Mills is treated in almost the same way as internal symmetry ---
with Chan-Paton factors (subsection VIB4); 
\item{(2)} the superspacecone gauge,
the simplest gauge for QCD (subsection VIB7); and 
\item{(3)} a new
``(almost-)first-order" superspace action for supergravity, analogous to
the one for super Yang-Mills (subsection XB1). 

We try to give the simplest possible calculational tools, not only for the
above reasons, but also so group theory (internal and spacetime) and
integrals can be performed with the least effort and memory.  Some
traditionalists may claim that the old methods are easy enough, but their
arguments are less convincing when the order of perturbation is
increased.  Even computer calculations are more efficient when left as a
last resort; and you can't see what's going on when the computer's
doing the calculating, so you don't gain any new understanding.  
We give examples of (and exercises on) these methods, but
not exhaustively.  We also include more recent topics (or those more
recently appreciated in the particle physics community) that might be
deemed non-introductory, but are commonly used, and are simple and
important enough to include at the earliest level.  For example, the
related topics of (unitary) lightcone gauge, twistors, and spinor helicity
are absent from all field theory texts, and as a result no such text
performs the calculation of as basic a diagram as the 4-gluon tree
amplitude.  Another missing topic is the relation of QCD to strings through
the random worldsheet lattice and large-color (1/N) expansion, which is
the only known method that might quantitatively describe its
high-energy nonperturbative behavior (bound states of arbitrarily large
mass). 

This text is meant to cover all the field theory every high energy theorist
should know, but not all that any particular theorist might need to know. 
It is not meant as an introduction to research, but as a preliminary to
such courses:  We try to fill in the cracks that often lie between standard
field theory courses and advanced specialized courses.  For example, we
have some discussion of string theory, but it is more oriented toward the
strong interactions, where it has some experimental justification, rather
than quantum gravity and unification, where its usefulness is still under
investigation.  We do not mention statistical mechanics, although many
of the field theory methods we discuss are useful there.  Also, we do not
discuss any experimental results in detail; phenomenology and analysis of
experiments deserve their own text.  We give and apply the methods of
calculation and discuss the qualitative features of the results, but do not
make a numerical comparison to nature:  We concentrate more on the
``forest" than the ``trees".

Unfortunately, our discussions of the (somewhat related) topics of
infrared-diver\-gence cancellation, Lamb shift, and the parton model are
sketchy, due to our inability to give fully satisfying treatments --- but
maybe in a later edition?

Unlike all previous texts on quantum field theory, this one is available for
free over the Internet (as usual, from arXiv.org and its mirrors), and
may be periodically updated.  Errata, additions, and other changes will be
posted on my web page at
 \pdfklink{http://insti.physics.sunysb.edu/\~{}siegel/plan.html}
	{http://insti.physics.sunysb.edu/\noexpand~siegel/plan.html}
 until enough are accumulated for a new edition.  Electronic distribution
has many advantages:
\its It's free.
\its It's available quickly and easily. You can download it from the arXive.org or its mirrors, just like preprints, without a trip to the library (where it may be checked out) or bookstore or waiting for an order from the publisher.  (If your connection is slow, download overnight.)  And it won't go ``out of print".
\its Download it at work, home, etc.\ (or carry it on a CD), rather than carrying a book or printing multiple copies.
\its Get updates just as quickly, rather than printing yet again.
\its It has the usual Web links, so you can get the referenced papers just as easily.
\its It has a separate ``outline" window containing a table of contents on which you can click to take the main window to that item.
\its You can electronically search (do a ``find" on) the text.
\its Easier to read on the computer screen (arbitrary magnification, etc.)
\its Save trees, ink, and space.
\its Theft is not a problem.
\its No wear or tear.
\its No paper cuts.
\its You can even add notes (far bigger than would fit in the margin) with various software programs.

ÜHighlights

The preceding Table of Contents lists the three parts of the text: 
Symmetry, Quanta, and Higher Spin. Each part is divided into four chapters,
each of which has three sections, divided further into subsections.  Each
section is followed by references to reviews and original papers. 
Exercises appear throughout the text, immediately following the items
they test:  This purposely disrupts the flow of the text, forcing the reader
to stop and think about what he has just learned.  These exercises are
interesting in their own right, and not just examples or memory tests. 
This is not a crime for homeworks and exams, which at least by graduate
school should be about more than just grades.

This text also differs from any other in most of the following ways:  
\item{(1)} We
place a greater emphasis on ÓmechanicsÕ in introducing some of the more
elementary physical concepts of field theory:  
\itemitem{(a)} Some basic ideas, such
as antiparticles, can be more simply understood already with classical
mechanics.  
\itemitem{(b)} Some interactions can also be treated through
first-quantization:  This is sufficient for evaluating certain tree and
one-loop graphs as particles in external fields.  Also, Schwinger
parameters can be understood from first-quantization:  They are useful
for  performing momentum integrals (reducing them to Gaussians),
studying the high-energy behavior of Feynman graphs, and finding their
singularities in a way that exposes their classical mechanics
interpretation.  
\itemitem{(c)} Quantum mechanics is very similar to free classical
field theory, by the usual ``semiclassical" correspondence (``duality") between
particles (mechanics) and waves (fields).  They use the same wave
equations, since the mechanics Hamiltonian or Becchi-Rouet-Stora-Tyutin
operator is the kinetic operator of the corresponding classical field
theory, so the free theories are equivalent.  In particular, (relativistic)
quantum mechanical BRST provides a simple explanation of the off-shell
degrees of freedom of general gauge theories, and introduces concepts
useful in string theory.  As in the nonrelativistic case, this treatment
starts directly with quantum mechanics, rather than by
\hbox{(first-)} quantization of a classical mechanical system.  Since
supersymmetry and strings are so important in present theoretical
research, it is useful to have a text that includes the field theory
concepts that are prerequisites to a course on these topics.  (For the
same reason, and because it can be treated so similarly to Yang-Mills, we
also discuss general relativity.)

\item{(2)} We also emphasize Óconformal invarianceÕ.  Although a badly broken
symmetry, the fact that it is larger than Poincar«e invariance makes it
useful in many ways:  
\itemitem{(a)} General classical theories can be described most
simply by first analyzing conformal theories, and then introducing mass
scales by various techniques.  This is particularly useful for the general
analysis of free theories, for finding solutions in gravity theories,
and for constructing actions for supergravity theories.  
\itemitem{(b)} Spontaneously broken conformal invariance produces the dilaton, which can be used in place of general relativity to describe cosmology.
\itemitem{(c)} Quantum theories that are well-defined within perturbation
theory are conformal (``scaling") at high energies.  (A possible exception
is string theories, but the supposedly well understood string theories that
are finite perturbatively have been discovered to be hard-to-quantize
membranes in disguise nonperturbatively.)  This makes methods based on
conformal invariance useful for finding classical solutions, as well as
studying the high-energy behavior of the quantum theory, and simplifying
the calculation of amplitudes.  
\itemitem{(d)} Theories whose conformal invariance is
not (further) broken by quantum corrections avoid certain problems at the
nonperturbative level.  Thus conformal theories ultimately may be
required for an unambiguous description of high-energy physics.

\item{(3)} We make extensive use of Ótwo-component (chiral) spinorsÕ, which are
ubiquitous in particle physics:  
\itemitem{(a)} The method of twistors (more recently
dubbed ``spinor helicity") greatly simplifies the Lorentz algebra in
Feynman diagrams for massless (or high-energy) particles with spin, and
it's now a standard in QCD.  (Twistors are also related to conformal
invariance and self-duality.)  On the other hand, most texts still struggle
with 4-component Dirac (rather than 2-component Weyl) spinor notation,
which requires gamma-matrix and Fierz identities, when discussing QCD
calculations.  
\itemitem{(b)} Chirality and duality are important concepts in all the
interactions:  Two-compo\-nent spinors were first found useful for weak
interactions in the days of 4-fermion interactions.  Chiral symmetry in
strong interactions has been important since the early days of pion
physics; the related topic of instantons (self-dual solutions) is simplified
by two-component notation, and general self-dual solutions are
expressed in terms of twistors.  Duality is simplest in two-component
spinor notation, even when applied to just the electromagnetic field.  
\itemitem{(c)}
Supersymmetry still has no convincing experimental verification (at least
not at the moment I'm typing this), but its theoretical properties promise
to solve many of the fundamental problems of quantum field theory.  
(Although there is no experimental evidence for supersymmetry, there is also no experimental evidence for the Higgs boson. They are equally important for predictability in particle physics, although for one this is seen in perturbation theory, while for the other only when attempting to resum it.)
It is an element of most of the proposed generalizations of the Standard
Model.  Chiral symmetry is built into supersymmetry, making
two-component spinors unavoidable.

\item{(4)} The topics are ÓorderedÕ in a more pedagogical manner:  
\itemitem{(a)}  Abelian
and nonabelian gauge theories are treated together using modern
techniques.  (Classical gravity is treated with the same methods.)  
\itemitem{(b)}
Classical Yang-Mills theory is discussed before any quantum field theory. 
This allows much of the physics, such as the Standard Model (which may
appeal to a wider audience), of which Yang-Mills is an essential part, to be
introduced earlier.  In particular, symmetries and mass generation in the
Standard Model appear already at the classical level, and can be seen
more easily from the action (classically) or effective action (quantum)
than from diagrams.  
\itemitem{(c)} Only the method of path integrals is used for
second-quantization.  Canonical quantization is more cumbersome and
hides Lorentz invariance, as has been emphasized even by Feynman when
he introduced his diagrams.  We thus avoid such spurious concepts as the
``Dirac sea", which supposedly explains positrons while being totally
inapplicable to bosons.  However, for quantum physics of general systems
or single particles, operator methods are more powerful than any type of
first-quantization of a classical system, and path integrals are mainly of
pedagogical interest.  We therefore ``review" quantum physics first,
discussing various properties (path integrals, S-matrices, unitarity, BRST,
etc.)¼in a general (but simpler) framework, so that these properties need
not be rederived for the special case of quantum field theory, for which
path-integral methods are then sufficient as well as preferable.

\item{(5)} ÓGauge fixingÕ is discussed in a way more general and efficient than
older methods:  
\itemitem{(a)} The best gauge for studying unitarity is the
(unitary) lightcone gauge.  This rarely appears in field theory (and gravity) texts, or is
treated only half way, missing the important explicit elimination of all
unphysical degrees of freedom.  
\itemitem{(b)} Ghosts are introduced by BRST
symmetry, which proves unitarity by showing equivalence of convenient
and manifestly covariant gauges to the manifestly unitary lightcone
gauge.  It can be applied directly to the classical action, avoiding the
explicit use of functional determinants of the older Faddeev-Popov
method.  It also allows direct introduction of more general gauges (again
at the classical level) through the use of Nakanishi-Lautrup fields (which
are omitted in older treatments of BRST), rather than the functional
averaging over Landau gauges required by the Faddeev-Popov method. 
\itemitem{(c)} For nonabelian gauge theories the background field gauge is a must. 
It makes the effective action gauge invariant, so Slavnov-Taylor
identities need not be applied to it.  Beta functions can be found from just
propagator corrections.

\item{(6)} ÓDimensional regularizationÕ is used exclusively (with the exception of
one-loop axial anomaly calculations):  
\itemitem{(a)} It is the only one that preserves
all possible symmetries, as well as being the only one practical enough
for higher-loop calculations.  
\itemitem{(b)} We also use it exclusively for infrared
regularization, allowing all divergences to be regularized with a single
regulator (in contrast, e.g., to the ÓthreeÕ regulators used for the
standard treatment of Lamb shift).  
\itemitem{(c)} It is good not only for
regularization, but renormalization (``dimensional renormalization").  For
example, the renormalization group is most simply described using
dimensional regularization methods.  More importantly, renormalization
itself is performed most simply by a minimal prescription implied by
dimensional regularization.  Unfortunately, many books, even among those
that use dimensional regularization, apply more complicated
renormalization procedures that require additional, finite
renormalizations as prescribed by Slavnov-Taylor identities.  This is a
needless duplication of effort that ignores the manifest gauge invariance
whose preservation led to the choice of dimensional regularization in the
first place.  By using dimensional renormalization, gauge theories are as
easy to treat as scalar theories:  BRST does not have to be applied to
amplitudes explicitly, since the dimensional regularization and
renormalization procedure preserves it.

\item{(7)} Perhaps the most fundamental omission in most field theory texts is
the expansion of QCD in the inverse of the Ónumber of colorsÕ:  
\itemitem{(a)} It
provides a gauge-invariant organization of graphs into subsets,
allowing simplifications of calculations at intermediate stages, and is
commonly used in QCD today.  
\itemitem{(b)} It is useful as a perturbation expansion,
whose experimental basis is the Okubo-Zweig-Iizuka rule.  
\itemitem{(c)} At the
nonperturbative level, it leads to a resummation of diagrams in a way
that can be associated with strings, suggesting an explanation of
confinement.

\item{(8)} Our treatment of gravity is closely related to that applied to Yang-Mills
theory, and differs from that of most texts on gravity:  
\itemitem{(a)} We emphasize
the action for deriving field equations for gravity (and matter), rather
than treating it as an afterthought.  
\itemitem{(b)} We make use of local (Weyl) scale
invariance for cosmological and spherically symmetric
solutions, gauge fixing, field redefinitions, and
studying conformal properties.  In particular, other texts neglect the
(unphysical) dilaton, which is crucial in such treatments (especially for
generalization to supergravity and strings).  
\itemitem{(c)}  While most gravity
texts leave spinors till the end, and treat them briefly, our discussion of
gravity is based on methods that can be applied directly to spinors, and
therefore to supergravity and superstrings.  
\itemitem{(d)} Our methods of
calculating curvatures for purposes of solving the classical field equations
are somewhat new, but probably the simplest, and are directly related to the
simplest methods for super Yang-Mills theory and supergravity.

ÜNotes for instructors

This text is intended for reference and as the basis for a course
on relativistic quantum field theory for second-year graduate students.  
The first two parts were repeatedly used for a one-year course
I taught at Stony Brook. 
(There is more there than can fit comfortably into one year,
so I skipped some subsections, but my choice varied.)
It also includes material I used for a one-semester relativity course, and for my third of a one-year string course, both of which I also gave several
times here --- I used most of the following:  
\vskip.1in
relativity: IA, B3, C2; IIA; IIIA-C5; IVA7; VIB1; IX; XIA3, A5-6, B1-2
\vskip.1in
strings: IIB1-2; VIIA2, B5, C4; VIIIB2, C4-5; XI; XIIA2, B1-3, B8
\vskip.1in
The prerequisites (for the quantum field theory
course) are the usual graduate courses in classical mechanics, classical
electrodynamics, and quantum mechanics.  For example, the student
should be familiar with Hamiltonians and Lagrangians, Lorentz
transformations for particles and electromagnetism, Green functions for
wave equations, SU(2) and spin, and Hilbert space.  Unfortunately, I find
that many second-year graduate students (especially many who got their
undergraduate training in the USA) still have only an undergraduate level
of understanding of the prerequisite topics, lacking a working knowledge
of action principles, commutators, creation and annihilation operators,
etc.  While most such topics are briefly reviewed here, they should be
learned elsewhere.
	
Generally students need to be prepared to begin research at the beginning of their third year.  This means they have to begin preparation for research in the middle of their second year, so standard courses for high-energy theorists, such as quantum field theory (and maybe even string theory), should already be finished by then.  This is rather difficult, considering that quantum field theory is usually considered a one-year course that follows one-year prerequisites.  The best solution would be to improve undergraduate courses, making them less repetitive, so first-semester graduate courses could be eliminated.  An easier fix would be to make graduate courses more efficient, or at least better coordinated and more modern.  For example:
\item{(1)} Sometimes relativistic quantum mechanics is taught in the second semester of quantum mechanics.  If this were done consistently, it wouldn't need to be treated in the quantum field theory course.
\item{(2)} The useful parts of classical electrodynamics are covered in the first semester.  (Do all physicists really need to learn wave guides?)  This is especially true if methods for solving wave equations (special functions, radiation, etc.)¼are not covered twice, once in quantum mechanics and once in electromagentism.  Furthermore, we now know (since the early 20th century) that electromagnetism is not the only useful classical field theory:  Why not have a one-year course on classical field theory, covering not only electromagnetism, but also Yang-Mills and general relativity?
\item{(3)} A lot of the important concepts in the Standard Model (especially the electroweak interactions) are essentially classical: spontaneous symmetry breaking, the Higgs effect, tree graphs, etc.  They could be covered as a third semester of classical field theory.
\item{(4)} Meanwhile, true quantum field theory (quantization, loops, etc.) could become a third semester of quantum theory, taken in parallel with the Standard Model.
\item{(5)} Much of string theory is mechanics, not field theory.  A string theory course could begin in the first semester of the second year (classical and statistical mechanics having been covered in the first year).

\noindent In summary, a curriculum for high-energy theorists could look something like...
$$ \vbox{\halign{\strut#&&â#\hfil\cr
sem. & Mechanics & Classical fields & Quantum \cr
\noalign{\hrule}
1 & Classical mechanics & Actions \& symmetries & Quantum theory \cr
2 & Statistical mechanics & Yang-Mills \& gravity & Solving wave equations \cr
3 & Strings & Standard Model & Quantum field theory \cr
}} $$
followed by more advanced courses (e.g., more quantum field theory or strings).  An alternative is to start quantum field theory in the second semester of the first year.

Unfortunately, in most places students start quantum field theory in their second year, having had little relativistic quantum mechanics and no Yang-Mills, so those subjects will comprise the first semester of the ``quantum" field theory course, while the true quantum field theory will wait till the second semester of that year.  

To fit these various scenarios, the ordering of the chapters is
somewhat flexible:  The ``flow" is indicated by the following ``3D" plot:

{\abovedisplayskip=-10pt
$$ \leftÓ\matrix{\hbox{lower spin} \cr \searrow¼\swarrow \cr
		\hbox{\small higher spin}\cr}\right.
	âââ\vcenter{\halign{\strut#â&&#â\cr
  & \multispan2 \hfil \bf classicalââ\hfil $\rightarrow$
	& \multispan2 \hfil \bf quantum \hfil \cr
  & \it symmetry & \it fieldsâââ& \it quantize & ââ\it loop \cr
  \bf Bose &I\hfil & III\hfil & \hfil V & \hfil VII \cr
  \hfil $\downarrow$ \hfil && \hfill {\small IX}â & {\small XI}\hfil & \cr
  && \hfil {\small X}â & {\small XII}\hfil & \cr
  \bf Fermi & II & IV \hfil & \hfil VI & \hfil VIII\cr}} $$}%
\noindent where the 3 dimensions are spin (``$j$"), quantization
(``$\h$"), and statistics (``$s$"):  The three independent flows are down
the page, to the right, and into the page.  (The third dimension has been
represented as perpendicular to the page, with ``higher spin" in smaller
type to indicate perspective, for legibility.)  To present these chapters in
the 1 dimension of time we have classified them as $j\h s$, but other
orderings are possible:

$$ \matrix{ j\h s : & \hbox{I II III IV V VI VII VIII IX X XI XII} \cr
	j s \h : & \hbox{I III V VII II IV VI VIII IX XI X XII} \cr
	\h j s : & \hbox{I II III IV IX X V VI XI XII VII VIII} \cr
	\h s j : & \hbox{I II III IX IV X V XI VI XII VII VIII} \cr
	s j \h : & \hbox{I III V VII IX XI II IV VI VIII X XII} \cr
	s \h j : & \hbox{I III IX V XI VII II IV X VI XII VIII} \cr} $$

\noindent
 (However, the spinor notation of II is used for discussing instantons in
III, so some rearrangement would be required, except in the $j\h s$, $\h
js$, and $\h sj$ cases.)  For example, the first half of the course can
cover all of the classical, and the second quantum, dividing Part Three
between them ($\h js$ or $\h sj$).  Another alternative ($js\h$) is a
one-semester course on quantum field theory, followed by a semester on
the Standard Model, and finishing with supergravity and strings.  Although
some of these (especially the first two) allow division of the course into
one-semester courses, this should not be used as an excuse to treat such
courses as complete:  Any particle physics student who was content to sit
through another entire year of quantum mechanics in graduate school
should be prepared to take at least a year of field theory.

ÜNotes for students

Field theory is a hard course.  (If you don't think so, name me a harder
one at this level.)  But you knew as an undergraduate that physics was a
hard major.  Students who plan to do research in field theory will find the
topic challenging; those with less enthusiasm for the topic may find it
overwhelming.  The main difference between field theory and lower
courses is that it is not set in stone:  There is much more variation in
style and content among field theory courses than, e.g., quantum
mechanics courses, since quantum mechanics (to the extent taught in
courses) was pretty much finished in the 1920's, while field theory is still
an active research topic, even though it has had many experimentally
confirmed results since the 1940's.  As a result, a field theory course has
the flavor of research:  There is no set of mathematically rigorous rules
to solve any problem.  Answers are not final, and should be treated as
questions:  One should not be satisfied with the solution of a problem, but
consider it as a first step toward generalization.  The student should not
expect to capture all the details of field theory the first time through,
since many of them are not yet fully understood by people who work in
the area.  (It is far more likely that instead you will discover details that
you missed in earlier courses.)  And one reminder:  The ÓonlyÕ reason for
lectures (including seminars and conferences) is for the attendees to ask
questions (and not just in private), and there are no stupid questions
(except for the infamous ``How many questions are on the exam?").  Only
half of teaching is the responsibility of the instructor.

Some students who have a good undergraduate background may want
to begin graduate school taking field theory.  That can be difficult, so
you should be sure you have a good understanding of most of the 
following topics:
\item{(1)}Classical mechanics: Hamiltonians, Lagrangians, actions;
Lorentz transformations; Poisson brackets
\item{(2)}Classical electrodynamics: Lagrangian for electromagnetism;
Lorentz transformations for electromagnetic fields, 4-vector potential,
4-vector Lorentz force law; Green functions
\item{(3)}Quantum mechanics: coupling to electromagnetism;
spin, SU(2), symmetries; Green functions for Schr¬odinger equation;
Hilbert space, commutators, Heisenberg and Schr¬odinger pictures;
creation and annihilation operators, statistics (bosons and fermions);
JWKB expansion

\noindent It is not necessary to be familiar with all these topics,
and most will be briefly reviewed, but if most of these topics are
not familiar then there will not be enough time to catch up.
A standard undergraduate education in these three courses is
ÓnotÕ enough.

ÜAcknowledgments

I thank everyone with whom I have discussed field theory, especially
Gordon Chalmers, Marc Grisaru, Marcelo Leite, Martin Ro×cek, Jack Smith,
George Sterman, and Peter van Nieuwenhuizen.  More generally, I thank
the human race, without whom this work would have been neither
possible nor necessary.

\vskip.1in\rightline{December 20, 1999}

ÜVersion 2

This update contains no new topics or subsections, but many small changes
(amounting to a 10\% increase in size):
corrections, improved explanations, examples, (20\% more) exercises, 
figures, references, cosmetics (including more color), and
an expanded Outline and AfterMath.  There are also a few small additions,
such as a more fundamental explanation of causality and unitarity in
quantum mechanics, and the use of Weyl scaling as a general method
for spherical (as well as cosmological) solutions to Einstein's equations.
It now TeX's with either ordinary TeX or pdftex.
(PDF figures can be created from PS with ghostscript; 
also available at my web site.)

\vskip.1in\rightline{September 19, 2002}

ÜVersion 3

The size has increased another 10\%.  I added a new subsection on cosmology using just the dilaton instead of general relativity, one on 1-loop graphs in N=4 Yang-Mills, another on first-quantization in gauge theory, and a whole section on string loops.  I fixed some EPS figures so they come out as bright as the PDF ones (and smaller files than before).  Also, more figures, exercises, references, etc.  (PDF can be created from EPS natively with Mac OS X now.)

\vskip.1in\rightline{August 23, 2005}

Û0 SOME FIELD THEORY TEXTS

\def\righthead{\hfil}

\def\texty #1\par#2\par{\bigskip\noindent{\sectfont{#1}}\\
	{#2}\medskip}

\frenchspacing \parskip=0pt

\texty Traditional, leaning toward concepts

Canonically quantize QED and calculate, then introduce path integrals

£1 S. Weinberg, ÓThe quantum theory of fieldsÕ, 3 v.¼(Cambridge
	University, 1995, 1996, 2000) 609+489+419 pp.:\\ 
	First volume QED; second volume contains many interesting
	topics; third volume supersymmetry (but no 2-component spinors,
	and only 11 pg.¼on supergraphs, non-gauge).  By one of the developers of the
	Standard Model.

£2 M. Kaku, ÓQuantum field theory: a modern introductionÕ (Oxford
	University, 1993) 785 pp.:\\
	Includes introduction to supergravity and superstrings.

£3 C. Itzykson and J.-B. Zuber, ÓQuantum field theoryÕ (McGraw-Hill,
	1980) 705 pp. (but with lots of {\small small print}):\\
	Emphasis on QED.

£4 N.N. Bogoliubov and D.V. Shirkov, ÓIntroduction to the theory of
	quantized fieldsÕ, 3rd ed. (Wiley, 1980) 620 pp.:\\
	Ahead of its time (1st English ed. 1959): early treatments of path 
	integrals, causality, background fields, and renormalization of general
	field theories; but before Yang-Mills and Higgs.

\texty Traditional, leaning toward calculations

Emphasis on Feynman diagrams

£5 M.E. Peskin and D.V. Schroeder, ÓAn introduction to quantum field theoryÕ\\
	(Perseus, 1995) 842 pp.:\\
	Comprehensive; style similar to Bjorken and Drell.

£6 B. de Wit and J. Smith, ÓField theory in particle physicsÕ, v. 1 (Elsevier
	Science, 1986) 490 pp.:\\
	No Yang-Mills or Higgs (but wait till v. 2, due any day now...).

£7 A.I. Akhiezer and V.B. Berestetskii, ÓQuantum electrodynamicsÕ (Wiley,
	1965) 868 pp.:\\
	Numerous examples of QED calculations.

£8 R.P. Feynman, ÓQuantum electrodynamics: a lecture note and reprint volumeÕ\\
	(Perseus, 1961) 198 pp.:\\
	Original treatment of quantum field theory as we know it today,
	but from mechanics; includes reprints of original articles (1949).
	
\newpage

\texty Modern, but somewhat specialized

Basics, plus thorough treatment of an advanced topic

£9 J. Zinn-Justin, ÓQuantum field theory and critical phenomenaÕ, 4th
	ed.¼(Clarendon, 2002) 1074 pp.:\\
	First 1/2 is basic text, with interesting treatments of many
	topics, but no S-matrix examples or discussion of cross sections;
	second 1/2 is statistical mechanics.

£10 G. Sterman, ÓAn introduction to quantum field theoryÕ (Cambridge
	University, 1993) 572 pp.:\\
	First 3/4 can be used as basic text, including S-matrix examples; last
	1/4 has extensive treatment of perturbative QCD, emphasizing
	factorization.  (Weak interactions are in an appendix.)

\texty Modern, but basic: few S-matrix examples

Should be supplemented with a ``QED/particle physics text"

£11 L.H. Ryder, ÓQuantum field theoryÕ, 2nd ed. (Cambridge University,
	1996) 487 pp.:\\
	Includes introduction to supersymmetry.

£12 D. Bailin and A. Love, ÓIntroduction to gauge field theoryÕ, 2nd ed.
	(Institute of Physics, 1993) 364 pp.:\\
	All the fundamentals.

£13 P. Ramond, ÓField theory: a modern primerÕ, 2nd ed.¼(Perseus, 1989) 329 pp.:\\
	Short text on QCD: no weak interactions.

\texty Advanced topics

For further reading; including brief reviews of some standard topics

£14 Theoretical Advanced Study Institute in Elementary Particle Physics
	(TASI) proceedings, University of Colorado, Boulder, CO (World
	Scientific):\\
	Annual collection of summer school lectures on recent research
	topics.

£15 W. Siegel, ÓIntroduction to string field theoryÕ (World Scientific, 1988), 
	\pdfklink{hep-\break th/0107094}{http://arXiv.org/abs/hep-th/0107094},
	244 pp.:\\
	Reviews lightcone, BRST, gravity, first-quantization, spinors, twistors,
	strings; besides, I like the author.

£16 S.J. Gates, Jr., M.T. Grisaru, M. Ro×cek, and W. Siegel, ÓSuperspace: or
	one thousand and one lessons in supersymmetryÕ
	(Benjamin/Cummings, 1983), \xxxlink{hep-th/0108200}, 548 pp.:\\
	Covers supersymmetry, spinor notation, lightcone, St¬uckelberg
	fields, gravity, Weyl scale, gauge fixing, background-field method,
	regularization, and anomalies; same author as previous, plus three
	other guys whose names sound familiar.

\unrefs

\volume PART ONE: SYMMETRY


The first four chapters present a one-semester course on ``classical field
theory".  Perhaps a more accurate description would be ``everything you
should know before learning quantum field theory".  

One of the most important and fundamental principles in physics is symmetry.  A symmetry is a transformation (a change of variables) under which the laws of nature do not change.  It places strong restrictions on what kinds of objects can exist, and how they can interact.  When dynamics are described by an action principle (Lagrangian, Hamiltonian, etc.), as required by quantum mechanics (but also useful classically), ÓcontinuousÕ symmetries are equivalent to conservation laws, which are the sole content of Newton's laws.
In particular, ÓlocalÕ (``gauge") symmetries, which allow independent
transformations at each coordinate point, are basic to all the
fundamental interactions:  All the fundamental forces are mediated by
particles described by Yang-Mills theory and its generalizations.  

From a practical
viewpoint, symmetry simplifies calculations by relating different solutions to
equations of motion, and allowing these equations to be written more
concisely by treating independent degrees of freedom as a single entity. 

Part One is basically a
study of global and local symmetries:  Classical dynamics represents only
a certain limit of quantum dynamics, and not the one usually emphasized,
but most of the symmetries of classical physics survive quantization.  The
phenomenon of symmetry breaking, and the related mechanisms of mass
generation, can also be seen at the classical level.  In perturbative
quantum field theory, classical field theory is simply the leading term in
the perturbation expansion.

Note that ``global" (time-, and usually space-independent) symmetries
can eliminate a variable, but not its time derivative.  For example,
translation invariance allows us to fix (i.e., eliminate) the position of the
center of mass of a system at some initial time, but not its time
derivative, which is just the total momentum, whose conservation is a
consequence of that same symmetry.  A local symmetry, being time
dependent, may allow the elimination of a variable at all times:  The
existence of this possibility depends on the dynamics, and will be
discussed later.

Of particular interest are ways in which symmetries can be made
manifest.  Frequently in the literature ``manifest" is used vacuously; a
``manifest symmetry" is an obvious one:  If you know the group, the
representation under consideration doesn't need to be stated, but can be
seen from just the notation.  (In fact, one of the main uses of index
notation is just to manifest the symmetry.)  Formulations where global
and local symmetries are manifest simplify calculations and their results,
as well as clarifying their meaning.

One of the main uses of manifest symmetry is rarely needing to explicitly
perform a specific symmetry transformation.  For example, one might
need to examine a relativistic problem in different Lorentz frames. 
Rather than starting with a description of the problem in one frame, and
then explicitly transforming to another, it is much simpler to start with a
manifestly covariant description, make one choice of frame, then make
another choice of frame.  One then never uses the messy square roots of
the familiar Lorentz contraction factors (although they may appear at the
end from kinematic constraints).  A more extreme example is the
corresponding situation for local symmetries, where such
transformations are intractable in general, and one always starts with
the manifestly covariant form.

ÚI. GLOBAL

In the first chapter we study symmetry in general, concentrating
primarily on spacetime symmetries, but also discussing general
properties that will have other applications in the following chapter.

Û6 A. COORDINATES

In this section we discuss the Poincar«e (and conformal) group as
coordinate transformations.  This is the simplest way to represent it on
the physical world.  In later sections we find general representations by
adding spin.

Ü1. Nonrelativity

We begin by reviewing some general properties of symmetries, including
as an example the symmetry group of nonrelativistic physics.  
Symmetries are the result of a redundant, but useful, description of a
theory.  (Note that here we refer to symmetries of a theory, not of a
solution to the theory.)  For example, translation invariance says that
only differences in position are measurable, not absolute position:  We
can't measure the position of the ``origin".  There are three ways to deal
with this:  
\item{(1)} Keep this invariance, and the corresponding
redundant variables, which allows all particles to be treated equally.  
\item{(2)} Choose an origin; i.e., make a ``choice of coordinates".  For
example, place an object at the origin; i.e., choose the position of an
object at a certain time to be the origin.  We can use
translational invariance to fix the position of any one particle at a given
time, but not the rest:  The differences in position are ``translationally
invariant".  
In this example, for N particles there are 3N
coordinates describing the particles, but still only 3 translations:  The
particles interact in the same 3-dimensional space.  
\item{(3)} Work only in terms of the differences of positions themselves as the
variables, allowing a symmetric treatment of the particles in terms of
translationally invariant variables:  However, in this example this would require applying
constraints on the variables, since there are 3N(N$-$1)/2 differences, of
which only 3(N$-$1) are independent.  

\noindent Although the last choice is most physical, the first is usually most
convenient:  The use of redundant variables, together with symmetry,
often gives a simpler description of a theory.  
We will find similar features later for
``local" invariances:  In general, the most convenient description of a
theory is with the invariance; the invariance can then be fixed, or
invariant combinations of variables used, appropriately for the particular
application.

\x IA1.1  Consider a system of objects labeled by the index $I$, each object located at position $x_I$.  (For simplicity, we can consider one spatial dimension, or just ignore an index labeling the different directions.)  Because of translational invariance
$$ x'_I = x_I +¶x $$
where $¶x$ is a constant independent of $I$, we are led to define new variables
$$ x_{IJ} ­ x_I - x_J $$
invariant under the above symmetry.  But these are not independent, satisfying
$$ x_{IJ} = -x_{JI},ââx_{IJ} +x_{JK} +x_{KI} = 0 $$
for all $I,J,K$.  Start with $x_{IJ}$ as fundamental instead, and show that the solution of these constraints is always in terms of some derived variables $x_I$ as in our original definition.  (Hint:  What happens if we define $x_1=0$?)  The appearance of a new invariance upon solving constraints in terms of new variables is common in physics: e.g., the gauge invariance of the potential upon solving the source-free half of Maxwell's equations.

Another example is
quantum mechanics, where the arbitrariness of the phase of the wave
function can be considered a symmetry:  Although quantum mechanics can
be reformulated in terms of phase-invariant probabilities, currents, or
density matrices instead of wave functions, and this can be useful for
some purposes of exposing physical properties, formulating and solving
the Schr¬odinger equation is simpler in terms of the wave function.  The
same applies to ``local" symmetries, where there is an independent
symmetry at each point of space and time:  For example, quarks and
gluons have a local ``color" symmetry, and are not (yet) observed
independently in nature, but are simpler objects in terms of which to
describe strong interactions than the observed hadrons (protons,
neutrons, etc.), which are described by color-invariant products of
quark/gluon wave functions, in the same way that probabilities are
phase-invariant products of wave functions.  

(Note that in quantum
mechanics there is a subtle distinction between observed and observer
that can obscure this symmetry if the observer is not invariant under it. 
This can always be avoided by choosing to define the observer as
invariant:  For example, the detection apparatus can be included as part
of the quantum mechanical system, while the observer can be defined as
some ``remote" recorder, who may be abstracted as even being
translationally invariant.  In practice we are less precise, and abstract
even the detection apparatus to be invariant:  For example, we describe
the scattering of particles in terms of the coordinates of only the
particles, and deal with the origin problem as above in terms of just
those coordinates.)

In the
Hamiltonian approach to mechanics, both symmetries and dynamics can
be expressed conveniently in terms of a ``bracket": the Poisson bracket
for classical mechanics, the commutator for quantum mechanics.  In this
formulation, the fundamental variables (operators) are some set of
coordinates and their canonically conjugate momenta, as functions of
time.  The (Heisenberg) operator approach to quantum mechanics then is
related to classical mechanics by identifying the semiclassical limit of the
commutator as the Poisson bracket:  For any functions $A$ and $B$ of $p$
and $q$, the quantum mechanical commutator is
$$ AB -BA = -i\h\left({»A\over »p_m}{»B\over »q^m} 
	- {»B\over »p_m}{»A\over »q^m}\right) +\O (\h^2 ) $$
 where all terms are generated by re-ordering.  (For example, if we define ``normal ordering" in $A$ and $B$ by putting all $q$'s to the left of all $p$'s, then doing so in the products will lead to an automatic cancellation of the ``classical" terms, with all the original $p$'s and $q$'s.)
In other words, the true classical limit of $AB-BA$ is zero, since
classically functions commute; thus the semiclassical limit is defined by
$$ \lim_{\h£0}\left[ {1\over \h}(AB-BA) \right] $$
 (which is really a derivative with respect to $\h$).  We therefore define
the bracket for the two cases by
$$ [A,B] ­\cases{ \displaystyle{ -i\left({»A\over »p_m}{»B\over »q^m} 
	- {»B\over »p_m}{»A\over »q^m}\right) } & semiclassically \cr
	AB -BA & \vrule height18pt depth0pt width0pt
	quantum mechanically \cr} $$
 The semiclassical definition of the bracket then can be applied to
classical physics (where it was originally discovered).  Classically $A$ and
$B$ are two arbitrary functions of the coordinates $q$ and momenta $p$;
in quantum mechanics they can be arbitrary operators.  We have included
an ``$i$" in the classical normalization so the two agree in the
semiclassical limit.  We generally use (natural/Planck) units $\h=1$, so
mass is measured as inverse length, etc.  (In fact, proposals have been made to fix the value of $\h$ by definition, and then determine the value of the kilogram by experimental apparatus such as the ``watt balance", rather than relying on a cylinder somewhere in Paris.)  When we do use an explicit $\h$,
it is a dimensionless parameter, and appears only for defining
Jeffries-Wentzel-Kramers-Brillouin (JWKB) expansions or (semi)classical
limits.

Our indices may appear either as subscripts or superscripts, with
preferences to be explained later:  For nonrelativistic purposes we treat
them the same.   We also use the Einstein summation convention, that any
repeated index in a product is summed over (``contracted"); usually we
contract a superscript with a subscript:
$$ A^m B_m ­ Ý_m A^m B_m $$
 The definition of the bracket is equivalent to using
$$ [p_m,q^n] = -i¶_m^n $$
 (where $¶_m^n$ is the ``Kronecker delta function": 1 if $m=n$, 0 if
$m±n$) together with the general properties of the bracket
$$ [A,B] = -[B,A],ââ[A,B]ÿ = -[Aÿ,Bÿ] $$
$$ [[A,B],C] +[[B,C],A] +[[C,A],B] = 0 $$
$$ [A,BC] = [A,B]C + B[A,C] $$
 The first set of identities exhibit the antisymmetry of the bracket; next
are the ``Jacobi identities".  In the last identity the ordering is important
only in the quantum mechanical case:  In general, the difference between
classical and quantum mechanics comes from the fact that in the
quantum case operator reordering after taking the commutator results in
multiple commutators.  

Infinitesimal symmetry transformations are then written as
$$ ¶A = i[G,A],ââA' = A +¶A $$
 where $G$ is the ``generator" of the transformation.  More explicitly,
infinitesimal generators will contain infinitesimal parameters:  For
example, for translations we have 
$$ G=·^i p_iâÜâ¶x^i = i[G,x^i] = ·^i,â¶p_i = 0 $$
 where $·^i$ are infinitesimal numbers.
 
As we'll see later (subsection IA3), the bracket of any two generators of infinitesimal transformations is also an infinitesimal transformation.  Thus, any symmetry group defines an algebra whose properties follow from the above general properties of the bracket.

The most evident physical symmetries are those involving spacetime.  For
nonrelativistic particles, these symmetries form the ``Galilean group": 
For the free particle, those infinitesimal transformations are linear
combinations of
$$ M=m,âP_i = p_i,âJ_{ij} = x_{[i}p_{j]} ­ x_i p_j -x_j p_i,â
	E = H = {p_i^2\over 2m},âV_i = mx_i -p_i t $$
 in terms of the position $x^i$ ($i=1,2,3$), momenta $p_i$, and
(nonvanishing) mass $m$, where $[ij]$ means to antisymmetrize in those
indices, by summing over all permutations (just two in this case), with
plus signs for even permutations and minus for odd.  (In three spatial
dimensions, one often writes $J_i=ü·_{ijk}J_{jk}$ to make $J$ into a
vector, where $·$ is totally antisymmetric in its indices and $·_{123}=1$.  This is a peculiarity of three dimensions, and will lose its utility
once we consider relativity in four spacetime dimensions.)  These
transformations are the space translations (momentum) $P$, rotations
(angular momentum --- just orbital for the spinless case) $J$, time
translations (energy) $E$, and velocity transformations (``Galilean
boosts") $V$.  (The mass $M$ is not normally associated with a symmetry,
and is not conserved relativistically.)  

\x IA1.2 Let's examine the Galilean group more closely.  Using just the
relations for $[x,p]$ and $[A,BC]$ (and the antisymmetry of the bracket): 
 ªa Find the action on $x_i$ of each kind of infinitesimal Galilean
transformation.
 ªb Show that the nonvanishing commutation relations for the generators
are
$$ [J_{ij},P_k] = i¶_{k[i}P_{j]},ââ[J_{ij},V_k] = i¶_{k[i}V_{j]},ââ
	[J_{ij},J^{kl}] = i¶^{[k}_{[i}J_{j]}{}^{l]} $$
$$ [P_i,V_j] = -i¶_{ij}M,ââ[H,V_i] = -iP_i $$

For more than one free particle, we introduce an $m$, $x^i$, and $p_i$ for
each particle (but the same $t$), and the generators are the sums over all
particles of the above expressions.  If the particles interact with each
other the expression for $H$ is modified, in such a way as to preserve the
commutation relations.  If the particles also interact with dynamical
fields, field-dependent terms must be added to the generators.  (External,
nondynamical fields break the invariance.  For example, a particle in a
Coulomb potential is not translation invariant since the potential is
centered about some point.)  

\x IA1.3  Show that the Hamiltonian
$$ H = \left( Ý_I {p_I^2\over 2m_I} \right) + U[(x_I -x_J)^2] $$
preserves the algebra of exercise IA1.2 for the Galilean group, where the other generators are modified only by summing over the index ``$I$Ê" labeling each particle.  (There are also implicit sums over the usual vector index ``$i$"; $U$ is a function of coordinate differences for ÓeachÕ $I$ and $J$.)

The rotations (or at least their ``orbital" parts) and space translations
are examples of coordinate transformations.  In general, generators of
coordinate transformations are of the form
$$ G = Â^i(x)p_iâÜâ¶Ä(x) = i[G,Ä] = Â^i »_i Ä $$
 where $»_i = »/»x^i$ and $Ä(x)$ is a ``scalar field" (or ``spin-0 wave
function"), a function of only the coordinates.

In classical mechanics, or quantum mechanics in the Heisenberg picture,
time development also can be expressed in terms of the Hamiltonian using
the bracket:
$$ {d\over dt}A = \left[{»\over »t}+iH,A\right] = {»\over »t}A +i[H,A] $$
 (The middle expression with the commutator of $»/»t$ makes sense only
in the quantum case, and is not defined for the Poisson bracket.)  Again,
this general relation is equivalent to the special cases, which in the
classical limit are Hamilton's equations of motion:
$$ {dq^m\over dt} = i[H,q^m] = {»H\over »p_m},ââ
	{dp_m\over dt} = i[H,p_m] = -{»H\over »q^m} $$
 The Hamiltonian has no explicit time dependence in the absence of
time-dependent nondyamical fields (external potentials whose time
dependence is fixed by hand, rather than by introducing the fields and
their conjugate variables into the Hamiltonian).  Consequently, time
development is itself a symmetry:  Time translations are generated by the
Hamiltonian; the $»/»t$ term in $d/dt$ term can be dropped when acting
on operators without explicit time dependence.

Invariance of the theory under a symmetry means that the equations of
motion are unchanged under the transformation:
$$ \left({dA\over dt}\right)' = {dA'\over dt} $$
 To apply our above translation of infinitesimal transformations into
bracket language, we define $¶(d/dt)$ by
$$ ¶\left({d\over dt}A\right) = ¶\left({d\over dt}\right)A +{d\over dt}¶A $$
 In the quantum case we can write
$$ ¶\left({d\over dt}\right) = 
	\left[ \left[ iG, {»\over »t}+iH\right],â\right] $$
 which follows from the Jacobi identity using $B=iG$ and $C=»/»t+iH$, and
inserting $A$ into the blank spaces of the commutators above.  (The
classical case can be treated similarly, except that the time derivatives
are not written as brackets.)  We then find that the generator of a
symmetry transformation is conserved (constant), since
$$ 0 = ¶\left({d\over dt}\right) = \left[-i{»G\over »t} -[G,H],â\right]
	= -i\left[{dG\over dt},â\right] $$

\x IA1.4 Show that the generators of the Galilean group are conserved:
 ªa Use the relation $d/dt=»/»t+i[H,¼]$ for the hamiltonian $H$ of a free
particle.  
 ªb Solve the equations of motion for $x(t)$ and $p(t)$ in terms of
initial conditions, and substitute into the expression for the generators to
give an independent derivation of their time independence.

Note that in the case where the Galilean symmetry persists for
interacting multiparticle systems, (total) mass is conserved.
In particular, invariance under translations and velocity transformations
implies mass conservation.

In the cases where time dependence is not involved, symmetries can be
treated in almost exactly the same way either classically or quantum
mechanically using the corresponding bracket (Poisson or commutator),
by using the properties that they have in common.  In particular, the fact
that a symmetry generator $G=Â^m(x)p_m$ is conserved means that we
can solve for a component of $p$ in terms of the constant $G$, and
substitute the result into the remaining equations of motion,
and that the conjugate to that component doesn't appear in $H$.  For
example, translation invariance of a potential in a particular direction
means that component of the momentum is a constant
($dp_1/dt=-»H/»q^1=0$), rotational invariance about some axis means
that component of angular momentum is a constant ($dJ/dt=-»H/»Ï=0$),
etc.

Ü2. Fermions

As we learned in our quantum mechanics course, two particles of the same
type are indistinguishable.  Furthermore, while an arbitrary number of 
ÓbosonsÕ (particles satisfying Bose-Einstein statistics) can each exist
in the same one-particle state, only one (or zero) ÓfermionsÕ can exist
in the same one-particle state.  (For example, we can have 
a state consisting of 17 photons
each of the same momentum and each of the same polarization,
and we can't tell which is which,
but we can only have 1 electron in such a state.)
In terms of wave functions, e.g., a 2-particle wave function, made from
1-particle wave functions of the form $Æ_i(x)$ (where $x$ labels the
spatial position and $i$ other properties), we conveniently define
$$ \li{ bosons: & âÆ_{ii'}(x,x') = +Æ_{i'i}(x',x) \cr
	fermions: & âÆ_{ii'}(x,x') = -Æ_{i'i}(x',x) \cr} $$
 For $x=x'$ and $i=i'$ the signs (which could be phases, but are chosen real
for convenience) are chosen so $Æ_{ii}(x,x)$ vanishes for fermions but not
necessarily for bosons, so no 2 fermions are in the same state.  For
other cases the relation avoids double counting for the 2 particles
being switched; the signs are arbitrary, but are chosen consistently with
the previous case so that the relation is local.
The symmetry of wave functions for bosons and antisymmetry for
fermions corresponds to operators that commute for bosons and
anticommute for fermions (or for properties associated with fermions).

As we know experimentally, and we will see follows from relativistic field
theory, particles with half-integral spins obey Fermi-Dirac statistics. 
Let's therefore consider the classical limit of fermions:  
This will prove useful later, when we define quantum field theory by quantizing classical field theory.  (A similar approach can be taken to the quantum mechanics of fermions, but is less useful, which is one reason why nonrelativistic quantum mechanics of spin $ü$ is usually done directly, without reference to the corresponding classical mechanics.)
This will lead to
generalizations of the concepts of brackets and coordinates.  Bosons 
(more generally, bosonic operators) obey
commutation relations, such as $[x,p]=i\h$; in the classical limit they just
commute.  Fermions obey anticommutation relations, such as $Ó½,½ÿÕ=\h$
for a single fermionic harmonic oscillator, where
$$ ÓA,BÕ = AB +BA $$
 is the anticommutator, expressed in terms of ``braces" $ÓÊ , Õ$" instead of the ``(square) brackets" $[Ê , ]$ used for commutators.  
 So, in the truly classical (not semiclassical) limit
they anticommute, $½½ÿ+½ÿ½=0$.  Actually, the simplest case is a single real
(hermitian) fermion:  Quantum mechanically, or semiclassically, we have
$$ \h = ÓÅ,ÅÕ = 2Å^2  $$
 while classically $Å^2=0$.  There is no analog for a single boson:
$[x,x]=x^2-x^2=0$.  This means that classical fermionic fields must be
``anticommuting":  Two such objects get a minus sign when pushed past
each other.  As a result, the product of two fermionic quantities is
bosonic, while fermionic times bosonic gives fermionic.

\x IA2.1 Show
$$ [B,C] = [A,D] = 0âÜâ[AB,CD] = üÓA,CÕ[B,D] +ü[A,C]ÓB,DÕ $$

Functions of anticommuting variables are simpler than functions of commuting variables in every way (algebra and calculus) ÓexceptÕ for keeping track of signs.
This is because Taylor expansions in anticommuting variables always terminate.
For instance, given a single
anticommuting variable $Æ$, we need to be able to Taylor expand
functions in $Æ$, e.g., to find a basis for the states.  We then have simply
$$ f(Æ) = a +bÆ $$
for constants $a$ and $b$, since $Æ^2=0$.  This generalizes in an obvious way to a function of many anticommuting variables:  For $N$ such variables, we have $2^N$ terms in the Taylor expansion, since any term can be either independent or first-order in each variable.

Note that $a$ has the same statistics as $f$, while $b$ has the opposite; thus functions of anticommuting variables will include some anticommuting coefficients.
In general, when Taylor expanding a function of anticommuting variables
we must preserve the statistics:  If we Taylor expand a quantity that is
defined to be commuting (bosonic), then the coefficients of even powers
of anticommuting variables will also be commuting, while the coefficients
of odd powers will be anticommuting (fermionic), to maintain the
commuting nature of that term (the product of the variables and
coefficient).  Similarly, when expanding an anticommuting quantity the
coefficients of even powers will also be anticommuting, while for odd
powers it will be commuting.

To work with wave functions that are functions of anticommuting
numbers, we must also understand calculus of anticommuting variables.
Since the Taylor expansion of a function terminates because $Æ^2=0$, as follows from anticommutativity, an
anticommuting derivative $»/»Æ$ must also satisfy
$$ \left({»\over »Æ}\right)^2 = 0 $$
 from either anticommutativity or the fact functions of $Æ$ terminate at
first order in $Æ$.  We also need a $Æ$ integral to define the inner product;
indefinite integration turns out to be enough.  The most important
property of the integral is integration by parts; then, when acting on any
function of $Æ$,
$$ ÇdƼ{»\over »Æ} = 0âÜâÇdÆ = {»\over »Æ} $$
 where the normalization is fixed for convenience.  This also implies
a definition of the ``(anticommuting) Dirac delta function",
$$ ¶(Æ) = Æ $$
 which satisfies
$$ ÇdÆ'¼¶(Æ'-Æ)f(Æ') = f(Æ) $$
 for any function $f$.  However, unlike the commuting case, we also have
 $$ ¶(-Æ) = -¶(Æ) $$

\x IA2.2 Prove this is the most general possibility for anticommuting
integration by considering action of integration and differentiation on the
most general function of $Æ$ (which has only two terms).

We can now consider operators that depend on both commuting ($Ä^m$)
and anticommuting ($Æ^µ$) classical variables, 
$$ ì^M = (Ä^m,Æ^µ) $$
 Classically they satisfy the ``graded" commutation relations
(anticommutation if both elements are fermionic, commutation
otherwise), not to be confused with the Poisson bracket,
$$ classicallyââ[ì^M,ì^NÕ = 0¼:ââ
	Ä^m Ä^n-Ä^n Ä^m = Ä^m Æ^Ã-Æ^Ã Ä^m = Æ^µ Æ^Ã+Æ^Ã Æ^µ = 0 $$
where we use mixed brackets (square and brace), the square one to the left to indicate the usual commutator unless both arguments are fermionic.
 This relation is then generalized to the graded quantum
mechanical commutator or Poisson bracket by
$$ [ì^M,ì^NÕ = \h ¯^{MN},ââ¯^{MN}¯_{PN} = ¶_P^M $$
 where $¯$ is constant, hermitian, and ``graded antisymmetric":
$$ ¯_{(MN]} = 0 :ââ¯_{(mn)} = ¯_{[µÃ]} = ¯_{mÃ} +¯_{Ãm} = 0 $$
 where $[µÃ]$ is the difference of the two orderings, as above, while
$(µÃ)$ is the sum.
For the standard normalization of canonically conjugate pairs of bosons
$$ Ä^m = Ä^{iŒ} = (q^i,p^i) $$
 and self-conjugate fermions, we choose
$$ ¯^{µÃ} = ¶^{µÃ};ââ¯^{iŒ,jº} = ¶^{ij}C^{Œº},ââC^{Œº} = \tat0{-i}i0 $$

Because of signs resulting from ordering anticommuting quantities, we
define derivatives unambiguously by their action from the left:
$$ {»\over »ì^M}ì^N = ¶_M^N $$
 The general Poisson bracket then can be written as
$$ semiclassicallyââ[A,BÕ ­ 
	-A{\onÁ»\over »ì^M}¯^{NM}{»\over »ì^N}B $$
 Since derivatives are normally defined to act from the left, there is a
minus sign from pushing the first derivative to the left if $A$ and that
particular component of $»/»ì^M$ are both fermionic.

\x IA2.3  Let's examine some properties of fermionic oscillators:
 ªa  For a single set of harmonic oscillators we have
$$ Óa,aÿÕ = 1,ââÓa,aÕ = Óaÿ,aÿÕ = 0 $$
 Show that the ``number operator" $aÿa$ has the property
$$ Óa,e^{i¹aÿa}Õ = 0 $$
 (Hint:  Since this system has only 2 states, the easiest way is to check
the action on those states.)
 ªb  Define eigenstates of the annihilation operator (``coherent states")
by
$$ a|½Ô = ½|½Ô $$
 where $½$ is anticommuting.  Show that this implies
$$ aÿ|½Ô = -{»\over »½}|½Ô,â|½Ô = e^{-½aÿ}|0Ô,âe^{-½'aÿ}|½Ô = |½+½'Ô,â
	x^{aÿa}|½Ô = |x½Ô, $$
$$ Ò½|½'Ô = e^{½*½'},â1= Çd½*d½¼e^{-½*½}|½ÔÒ½| $$
 Define wave functions in this space, $ï(½*)=Ò½|ïÔ$.  Taylor expand them
in $½*$, and compare this to the usual two-component representation
using $|0Ô$ and $aÿ|0Ô$ as a basis.
 ªc  Define the ``supertrace" by
$$ str(A) = Çd½*d½¼e^{-½*½}Ò½|A|½Ô $$
 Find the relation between any operator in this space and a 2$ð$2 matrix,
and find the expression for the supertrace in terms of this matrix.
 ªd For ÓtwoÕ sets of fermionic oscillators, we define
$$ Óa_1,aÿ_1Õ = Óa_2,aÿ_2Õ = 1,ââother¼Ó¼,¼Õ = 0 $$
 Show that the new operators
$$ ÷a_1 = a_1,ââ÷a_2 = e^{i¹aÿ_1 a_1}a_2 $$
 (and their Hermitian conjugates) are equivalent to the original ones
except that one set of the new oscillators ÓcommutesÕ (not anticommutes)
with the other ($[÷a_1,÷aÿ_2]=0$, etc.), even though each set satisfies the
same anticommutation relations with itself ($Ó÷a_1,÷aÿ_1Õ=1$, etc.).  Thus,
choice of statistics is relevant only for particles in the same state: at
most one fermion, but unlimited bosons.  (This change of oscillator basis
is called a ``Klein transformation".  It can be useful for discrete sets of
oscillators, but not for those labeled by a continuous parameter, because
of the discontinuity in the commutation relations when the two labels are
equal.)

\x IA2.4  Repeat exercise IA2.3 for the bosonic oscillator ($[a,aÿ]=1$), 
where the Hilbert space is infinite-dimensional, paying attention to 
signs, interchanging commutators with anticommutators where
necessary, etc.  Show
that the analog of part {\bf c} defines the ordinary trace.

Ü3. Lie algebra

Since the same symmetries can be expressed in terms of different kinds
of brackets for classical and quantum theories, it can be useful to work
with just those properties that the Poisson bracket and commutator have
in common, i.e., those that involve only the bracket of two operators, not
just their ordinary product:
$$ [ŒA+ºB,C] = Œ[A,C] +º[B,C]â\hbox{for numbers
	$Œ,º$ââ(distributivity)} $$
$$ [A,B] = -[B,A]ââ\hbox{(antisymmetry)} $$
$$ [A,[B,C]] +[B,[C,A]] +[C,[A,B]] = 0ââ\hbox{(Jacobi identity)} $$
 with similar expressions (differing only by signs) for anticommutators or
mixed commutators and anticommutators.

\x IA3.1  Find the generalizations of the Jacobi identity using also
anticommutators, corresponding to the cases where 2 or 3 of the objects
involved are considered as fermionic instead of bosonic.

These properties also give an abstract definition of a form of
multiplication, the ``Lie bracket", which defines a ``Lie algebra".  (The
first property is true of algebras in general.)  Other Lie brackets include
those defined by another, associative, form of multiplication, such as
matrix multiplication, or operator (infinite matrix) multiplication as in
quantum mechanics:  In those cases we can write $[A,B]=AB-BA$, and use
the usual properties of multiplication (distributivity and associativity) to
derive the properties of the Lie bracket.  (Another familiar example in
physics is the ``cross" product for three-vectors; however, this can also
be expressed in terms of matrix multiplication.)  The most important use
of Lie algebras for physics is for describing (continuous) infinitesimal
transformations, especially those describing symmetries.

\x IA3.2 Using only the commutation relations of the generators of the
Galilean group (exercise IA1.2), check all the Jacobi identities.

For describing transformations, we can also think of the bracket as a
derivative:  The ``Lie derivative" of $B$ with respect to $A$ is defined as
$$ \L_A B = [A,B] $$
 As a consequence of the properties of the Lie bracket, this derivative
satisfies the usual properties of a derivative, including the Leibniz (distributive) rule. 
(In fact, for coordinate transformations the Lie derivative is really a
derivative with respect to the coordinates.)

We can now define finite transformations by exponentiating infinitesimal
ones:
$$ A' ® (1 + i·\L_G)AâÜâA' = \lim_{·£0}(1 +i·\L_G)^{1/·} A = e^{i\L_G}A $$
 In cases where we have $[A,B]=AB-BA$, we can also write
$$ e^{i\L_G}A = e^{iG}A e^{-iG} $$
 This follows from replacing $G$ on both sides with $ŒG$ and taking the
derivative with respect to $Œ$, to see that both satisfy the same
differential equation with the same initial condition.  We then can
recognize this as the way transformations are performed in quantum
mechanics:  A linear transformation that preserves the Hilbert-space inner
product must be unitary, which means it can be written as the
exponential of an antihermitian operator.

Just as infinitesimal transformations define a Lie algebra with elements
$G$, finite ones define a ``Lie group" with elements 
$$ g = e^{iG} $$
 (or similarly with $\L_G$).  The multiplication law of two group elements follows from the fact the
product of two exponentials can be expressed in terms of multiple
commutators:
$$ e^A e^B = e^{A + B + ü[A,B] + ...} $$
 We now have the mathematical properties that define a group, namely:
\item{(1)} a product, so that for two group elements $g_1$ and $g_2$, we can
define $g_1 g_2$, which is another element of the group (closure), 
\item{(2)} an
identity element, so $g I = I g = g$, 
\item{(3)} an inverse, where $g g^{-1} =
g^{-1} g = I$, and 
\item{(4)} associativity, $g_1 (g_2 g_3) = (g_1 g_2) g_3$.  

\noindent In
this case the identity is $1=e^0$, while the inverse is $(e^A)^{-1}=e^{-A}$.

Thus two consecutive symmetry transformations will automatically involve Lie brackets of the generators of infinitesimal transformations.  In particular,
performing two consecutive infinitesimal transformations, followed by the inverse transformations in Óthe sameÕ order, gives their bracket:
$$ e^A e^B e^{-A} e^{-B} = exp(e^A B e^{-A}) e^{-B} ® e^{[A,B]} $$

Since the elements of a Lie algebra form a vector space (we can add them
and multiply by numbers), it's useful to define a basis:
$$ G = Œ^i G_iâÜâg = e^{iŒ^i G_i} $$
 The parameters $Œ^i$ then also give a set of coordinates for
the Lie group.  (Previously they were required to be infinitesimal, for
infinitesimal transformations; now they are finite, but may be periodic, as
determined by topological considerations that we will mostly ignore.) 
Now the multiplication rules for both the algebra and the group are given
by those of the basis:
$$ [G_i,G_j] = -if_{ij}{}^k G_k $$
 for the (``structure") constants $f_{ij}{}^k=-f_{ji}{}^k$, which define the
algebra/group (but are ambiguous up to a change of basis).  They satisfy
the Jacobi identity
$$ [[G_{[i},G_j],G_{k]}] = 0âÜâf_{[ij}{}^l f_{k]l}{}^m = 0 $$
 A familiar example is SO(3) (SU(2)), 3D rotations, where
$f_{ij}{}^k=·_{ijk}$ if we use $G_i=ü·_{ijk}J_{jk}$.

Another useful concept is a ``subgroup":  If some subset of the elements
of a group also form a group, that is called a ``subgroup" of the original
group.  In particular, for a Lie group the basis of that subgroup will be a
subset of some basis for the original group.  For example, for the Galilean
group $J_{ij}$ generate the rotation subgroup.

\x IA3.3  Let's examine the subgroup of the Galilean group describing
(spatial) coordinate transformations --- rotations and spatial translations:
 ªa  Show that the infinitesimal transformations are given by
$$ ¶x^i = x^j ·_j{}^i + ö·^i,ââ·_{ij} = - ·_{ji} $$
 where the $·$'s are constants.
 ªb  Exponentiate to find the finite transformations
$$ x'^i = x^j ñ_j{}^i +öñ^i $$
 ªc  Show that $ñ_i{}^j$ must satisfy
$$ ñ_i{}^k ñ_j{}^l ¶_{kl} = ¶_{ij} $$
 both to preserve the scalar product, and as a consequence of
exponentiating.  (Hint:  Use matrix notation, and find the equivalent
relation between $ñ$ and $ñ^{-1}$.)
 ªd  Show that the last equation implies $det¼ñ=à1$, while
exponentiating can give only $det¼ñ=1$ (since +1 can't change
continuously to $-1$).  What is the physical interpretation of a
transformation with $det¼ñ=-1$?  (Hint:  Consider a simple example.)

These results can be generalized to include anticommutators:  When some
of the basis elements $G_i$ are fermionic, the corresponding parameters
$Œ^i$ are anticommuting numbers, the structure constants are defined
by $[G_i,G_jÕ$, etc..  Then $G=Œ^i G_i$ is bosonic term by term, as is $g$, so
bosons transform into bosons and fermions into fermions, but Taylor
expansion in the $Œ$'s will have both bosonic and fermionic coefficients. 
(For example, for $¶A=·B$, if $A$ is bosonic, then so is $·B$, but if also $·$
is fermionic, then $B$ will also be fermionic.)

For some purposes it is more convenient to absorb the ``$i$" in the
infinitesimal transformation into the definition of the generator:
$$ G £ -iGâÜâ¶A = [G,A] = \L_G A,âg = e^G,â[G_i,G_j] = f_{ij}{}^k G_k $$
 This affects the reality properties of $G$:  In particular, if $g$ is unitary
($ggÿ=I$), as usually required in quantum mechanics, $g=e^{iG}$ makes $G$
hermitian ($G=Gÿ$), while $g=e^G$ makes $G$ antihermitian ($G=-Gÿ$).  In
some cases anithermiticity can be an advantage:  For example, for
translations we would then have $P_i=»_i$ and for rotations
$J_{ij}=x_{[i}»_{j]}$, which is more convenient since we know the $i$'s in
these (and any) coordinate transformations must cancel anyway.  On the
other hand, the U(1) transformations of electrodynamics (on the wave
function for a charged particle) are just phase transformations $g=e^{iÏ}$
(where $Ï$ is a real number), so clearly we want the explicit $i$; then the
only generator has the representation $G_i=1$.  In general we'll find that
for our purposes absorbing the $i$'s into the generators is more
convenient for just spacetime symmetries, while explicit $i$'s are more
convenient for internal symmetries.

Ü4. Relativity

The Hamiltonian approach singles out the time coordinate.  In relativistic
theories time can be treated on equal footing with space, and it is useful
to take advantage of this fact, so that the full Poincar«e invariance is
manifest.  So, we treat the time $t$ and spatial position $x^i$ together as
a four-vector (or D-vector in D$-$1 space and 1 time dimension)
$$ x^m = (x^0,x^i) = (t,x^i) $$
 where $m=0,1,...,3$ (or D$-1$), $i=1,2,3$.  Since the energy $E$ and
three-momentum $p^i$ are canonically conjugate to them,
$$ [p^i,x^j] = -i¶^{ij},ââ[E,t] = +i $$
 we define the 4-momentum as
$$ p^m = (E,p^i) = ú^{mn}p_n,ââp_m = ú_{mn}p^n;ââ
	[p^m,x^n] = -iú^{mn},ââ[p_m,x^n] = -i¶_m^n $$
 where we raise and lower indices with the ``Minkowski metric", in an
``orthonormal basis",
$$ ú_{mn} = \bordermatrix{ &0&1&2&3\cr 0&-1&0&0&0\cr 
	1&0&1&0&0\cr 2&0&0&1&0\cr 3&0&0&0&1\cr }
	âÜâp_0 = -p^0 = -E $$
 in four spacetime dimensions, with obvious generalizations to higher
dimensions.  (Sometimes the metric with signs $+---$ is used; we prefer
$-+++$ because it is more convenient for quantum calculations.
The numbers of positive and negative eigenvalues of an invertible
matrix is known as its ``signature".) 
Therefore, we now distinguish upper and lower indices in general:  At
least for position and momentum, the upper-indexed $x^m$ and $p^m$
have the usual physical interpretation (so $x_m$ and $p_m$ have extra
signs).  This is consistent with our previous nonrelativistic notation,
since 3-vector indices do not change sign upon raising or lowering.

Of course, we could have done that much nonrelativistically.  Relativity is
a symmetry of kinematics and dynamics:  In particular, a free, spinless,
relativistic particle is completely described by the constraint
$$ p^2 + m^2 = 0 $$
 where we define the covariant square
$$ p^2 = p^m p_m = p^m p^n ú_{mn} 
	= -(p^0)^2 +(p^1)^2 +(p^2)^2 +(p^3)^2 $$
 (The square of $p$ on the left should not be confused with 
the second component of $p$ on the right.) 
Our relativistic symmetry must leave this constraint invariant:  Thus the
metric defines the norm of a vector (and an invariant inner product). 
Therefore, to preserve Lorentz invariance it is important that we contract
only an upper index with a lower index.  For similar reasons, we have
$$ »_m = {»\over »x^m},ââ»_m x^n = ¶_m^n $$
 so quantum mechanically $p_m=-i»_m$.

We use (natural/Planck) units $c=1$ (where $c$ is the speed of light in a vacuum), so length and
duration are measured in the same units; $c$ then appears only as a
parameter for defining nonrelativistic expansions and limits.
For example, in astronomical units, $c$=1 light year/year.  In fact, the speed of light is no longer measured, but used to define the meter (since 1986) in terms of the second (itself defined by an atomic clock), as the distance light travels in a vacuum in exactly 1/299,792,458th of a second.  So, using metric system units for $c$ is no different than measuring land distance in miles and altitude in feet and writing $ds^2=dx^2+dy^2+b^2 dz^2$, where $b$=(1/5280)miles/foot is the slope of a line raised up $45^\circ$.  (As we mentioned in subsection IA1, similar remarks will soon apply to $\h$ and the kilogram, $\h=1$ being another natural/Planck unit.)

Unlike the positive-definite nonrelativistic norm of a 3-vector $V^i$, for
an arbitrary 4-vector $V^m$ we can have
$$ V^2 \leftÓ \matrix{ < \cr = \cr > \cr} \rightÕ 0:â\leftÓ \matrix{ 
	timelike\hfill \cr lightlike/null \cr spacelike\hfill \cr} \right. $$
 In particular, the 4-momentum is timelike for massive particles
($m^2>0$) and lightlike for massless ones (while ``tachyons", with
spacelike momenta and $m^2<0$, do not exist, for reasons that are most
clear from quantum field theory).
With respect to ``proper" Lorentz transformations, those that can be obtained continuously from the identity, we can further classify timelike and lightlike vectors as ``forward" and ``backward", since there is no way to continuously ``rotate" a vector from forward to backward without it being spacelike (``sideways"), so only spacelike vectors can have their time component change sign continuously.  

The quantum mechanics will be described later, but the result is that this
constraint can be used as the wave equation.  The main qualitative
distinction from the nonrelativistic case in the constraint
$$ \li{ nonrelativistic: &â-2mE + \vec pÊ{}^2 = 0 \cr
	relativistic: &â-E^2 +m^2 +\vec pÊ{}^2 = 0 \cr } $$
 is that the equation for the energy $E­p^0$ is now quadratic, and thus
has two solutions: 
$$ p^0 = à¿,ââ¿ = å{(p^i)^2+m^2} $$
 Later we'll see how the second solution is interpreted as an
``antiparticle".

The translations and Lorentz transformations make up the Poincar«e
group, the symmetry that defines special relativity.   (The Lorentz group
in D$-$1 space and 1 time dimension is the ``orthogonal" group
``O(D$-$1,1)".  The ``proper" Lorentz group ``SO(D$-$1,1)", where the ``S"
is for ``special", transforms the coordinates by a matrix whose
determinant is 1.  The Poincar«e group is ISO(D$-$1,1), where the ``I"
stands for ``inhomogeneous".)  For the spinless particle they are
generated by coordinate transformations $G_I=(P_a,J_{ab})$:
$$ P_a = p_a,âJ_{ab} = x_{[a}p_{b]} $$
 (where also $a,b=0,...,3$).  Then the fact that the physics of the free
particle is invariant under Poincar«e transformations is expressed as
$$ [P_a,p^2 +m^2] = [J_{ab},p^2 +m^2] = 0 $$
 Writing an arbitrary infinitesimal transformation as a linear combination
of the generators, we find
$$ ¶x^m = x^n ·_n{}^m + ö·^m,ââ·_{mn} = - ·_{nm} $$
 where the $·$'s are constants.  Note that antisymmetry of $·_{mn}$
does not imply antisymmetry of $·_m{}^n=·_{mp}ú^{pn}$, because of
additional signs.  (Similar remarks apply to $J_{ab}$.)  Exponentiating to
find the finite transformations, we have
$$ x'^m = x^n ñ_n{}^m +öñ^m,ââñ_m{}^p ñ_n{}^q ú_{pq} = ú_{mn} $$
 The same Lorentz transformations apply to $p^m$, but the translations
do not affect it.  The condition on $ñ$ follows from preservation of the
Minkowski norm (or inner product), but it is equivalent to the
antisymmetry of $·_m{}^n$ by exponentiating $ñ=e^·$ (compare
exercise IA3.3).

Since $dx^a p_a$ is invariant under the coordinate transformations
defined by the Poisson bracket (the chain rule, since effectively
$p_a¾»_a$), it follows that the Poincar«e invariance of $p^2$ is equivalent
to the invariance of the line element
$$ ds^2 = - dx^m dx^n ú_{mn} $$
 which defines the ``proper time" $s$.  Spacetime with this indefinite
metric is called ``Minkowski space", in contrast to the ``Euclidean space"
with positive definite metric used to describe nonrelativistic length
measured in just the three spatial dimensions.
(The signature of the metric is thus the numbers of space and time
dimensions.)

\x IA4.1  For general variables $(q^m,p_m)$ and generator $G$, 
show from the definition of the Poisson bracket that
$$ ¶(dq^m p_m) = -d\left( G - p_m{»G\over »p_m} \right) $$
 and that this vanishes for any coordinate transformation.

For the massive case, we also have
$$ p^a = m{dx^a\over ds} $$
 For the massless case $ds=0$:  Massless particles travel along lightlike
lines.  However, we can define a new parameter $ $ such that
$$ p^a = {dx^a\over d } $$
 is well-defined in the massless case.  In general, we then have
$$ s = m  $$
 While this fixes $ =s/m$ in the massive case, in the massless case it
instead restricts $s=0$.  Thus, proper time does not provide a useful
parametrization of the world line of a classical massless particle, while
$ $ does:  For any piece of such a line, $d $ is given in terms of (any
component of) $p^a$ and $dx^a$.  Later we'll see how this parameter
appears in relativistic classical mechanics, and is useful for quantum
mechanics and field theory.

\x IA4.2  Starting from the usual Lorentz force law for a ÓmassiveÕ particle
in terms of proper time $s$ (which doesn't apply to $m=0$), rewrite it in
terms of $ $ to find a form which can apply to $m=0$.

\x IA4.3  The relation between $x$ and $p$ is closely related to the
Poincar«e conservation laws:
 ªa Show that 
$$ dP_a = dJ_{ab} = 0âÜâp_{[a}dx_{b]} = 0 $$
 and use this to prove that conservation of $P$ and $J$ imply the
existence of a parameter $ $ such that $p^a=dx^a/d $.
 ªb Consider a multiparticle system (but still without spin) where some of
the particles can interact only when at the same point (i.e., by collision;
they act as free particles otherwise).  Define $P_a=Ý_I p^I_a$ and
$J_{ab}=Ý_I x^I_{[a}p^I_{b]}$ as the sum of the individual momenta 
and angular momenta (where we label the particle with ``$IÊ$").  
Show that momentum conservation implies angular momentum
conservation,
$$ ëP_a = 0âÜâëJ_{ab} = 0 $$
 where ``$ë$" refers to the change from before to after the collision(s).

Special relativity can also be stated as the fact that the only physically
observable quantities are those that are Poincar«e invariant.  (Other
objects, such as vectors, depend on the choice of reference frame.)  For
example, consider two spinless particles that interact by collision,
producing two spinless particles (which may differ from the originals). 
Consider just
the momenta.  (Quantum mechanically, this is a complete
description.)  All invariants
can be expressed in terms of the masses and the ``Mandelstam
variables" (not to be confused with time and proper time)
$$ s = -(p_1+p_2)^2,âât = -(p_1-p_3)^2,ââu = -(p_1-p_4)^2 $$
 where we have used momentum conservation, which shows that even
these three quantities are not independent:
$$ p_I^2 = -m_I^2,âp_1+p_2 = p_3+p_4âÜâs+t+u = Ý_{I=1}^4 m_I^2 $$
 (The explicit index now labels the particle, for the process 1+2$£$3+4.) 
The simplest reference frame to describe this interaction is the
center-of-mass frame (actually the center of momentum, where the two
3-momenta cancel).  In that Lorentz frame, using also rotational
invariance, momentum conservation, and the mass-shell conditions, the
momenta can be written in terms of these invariants as
$$ \li{ p_1 & = \f1{ås}(ü(s+m_1^2-m_2^2),Â_{12},0,0) \cr
	p_2 & = \f1{ås}(ü(s+m_2^2-m_1^2),-Â_{12},0,0) \cr
	p_3 & = \f1{ås}(ü(s+m_3^2-m_4^2),Â_{34}ÊcosÊÏ,Â_{34}ÊsinÊÏ,0) \cr
     p_4 & = \f1{ås}(ü(s+m_4^2-m_3^2),-Â_{34}ÊcosÊÏ,-Â_{34}ÊsinÊÏ,0) \cr} $$
$$ cos¼Ï = {s^2 +2st -(Ým_I^2)s +(m_1^2-m_2^2)(m_3^2-m_4^2)\over
	4Â_{12}Â_{34}} $$
$$ Â_{IJ}^2 = \f14 [s-(m_I+m_J)^2][s-(m_I-m_J)^2] $$
 The ``physical region" of momentum space is then given by
$s³(m_1+m_2)^2$ and $(m_3+m_4)^2$, and $|cos¼Ï|²1$.

\x IA4.4 Derive the above expressions for the momenta in terms of
invariants in the center-of-mass frame.

\x IA4.5 Find the conditions on $s,t$ and $u$ that define the physical
region in the case where all masses are equal.

For some purposes it will prove more convenient to use a ``lightcone
basis" 
$$ p^à = \f1{å2}(p^0àp^1)âÜâ
	ú_{mn} = \bordermatrix{ &+&-&2&3\cr +&0&-1&0&0\cr 
	-&-1&0&0&0\cr 2&0&0&1&0\cr 3&0&0&0&1\cr },â
	p^2 = -2p^+ p^- +(p^2)^2 +(p^3)^2 $$
 and similarly for the ``lightcone coordinates" $(x^à,x^2,x^3)$.
(``Lightcone" is an unfortunate but common misnomer, having nothing to
do with cones in most usages.)  In this basis the solution to the
mass-shell condition $p^2+m^2=0$ can be written as
$$ p^à = -p_¦ = {(p^i)^2+m^2\over 2p^¦} $$
 (where now $i=2,3$), which more closely resembles the nonrelativistic
expression.  (Note the change on indices $+ª-$ upon raising and
lowering.)  A special lightcone basis is the ``null basis",
$$ p^à = \f1{å2}(p^0àp^1),â
	p^t = \f1{å2}(p^2-ip^3),âÐp^t = \f1{å2}(p^2+ip^3) $$
$$ Üâú_{mn} = \bordermatrix{ &+&-&t&Ðt\cr +&0&-1&0&0\cr 
	-&-1&0&0&0\cr t&0&0&0&1\cr Ðt&0&0&1&0\cr },ââ
	p^2 = -2p^+ p^- +2p^t Ðp^t $$
 where the square of a vector is linear in each component.  (We often use
``$Ñ{\phantom{M}}$" to indicate complex conjugation.)

\x IA4.6  Show that for $p^2+m^2=0$ ($m^2³0$, $p^a±0$), the signs of
$p^+$ and $p^-$ are always the same as the sign of the canonical energy
$p^0$.

\x IA4.7 Consider the Poincar«e group in 1 extra space dimension (D space,
1 time) for a massless particle.  Interpret $p^+$ as the mass, and $p^-$ as
the energy.  
 ªa Show that the constraint $p^2=0$ gives the usual
ÓnonrelativisticÕ expression for the energy.  
 ªb Show that the subgroup of
the Poincar«e group generated by all generators that commute with $p^+$
is the Galilean group (in D$-1$ space and 1 time dimensions).  Now
nonrelativistic mass conservation is part of momentum conservation, and
all the Galilean transformations are coordinate transformations.  Also,
positivity of the mass is related to positivity of the energy (see
exercise IA4.4).

There are two standard examples of relativistic effects on geometry.  Without loss of generality we can consider 2 dimensions, by considering motion in just 1 spatial direction.
One example is called ``Lorentz-Fitzgerald contraction":  Consider a finite-sized
object moving with constant velocity.  In our 2D space, this looks like 2
parallel lines, representing the endpoints:
$$ \figscale{contract}{1.3in} $$
 (In higher dimensions, this represents a one-spatial-dimensional object,
like a thin ruler, moving in the direction of its length.)  If we were in the
``rest frame" of this object, the lines would be vertical.  In that frame,
there is a simple physical way to measure the length of the object:  Send
light from a clock sitting at one end to a mirror sitting at the other end,
and time how long it takes to make the round trip.  A clock measures
something physical, namely the proper time $T­Çå{ds^2}$ along its ``worldline" (the curve describing its history in spacetime).  Since $ds^2$ is by definition
the same in any frame, we can calculate this quantity in our frame.
$$ \figscale{contract2}{1.3in} $$
 In this 2D picture lightlike lines are always slanted at $à45^\circ$.  The 2
lines representing the ends of the object are (in this frame) $x=vt$ and
$x=L+vt$.  Some simple geometry then gives 
$$ T = {2L\overå{1-v^2}}âÜâL =  å{1-v^2}¼T/2 $$
 This means that the length $L$ we measure for the object is ÓshorterÕ
than the length $T/2$ measured in the object's rest frame by a factor
$å{1-v^2}<1$.  Unlike $T$, the $L$ we have defined is not a physical
property of the object:  It depends on both the object and our velocity
with respect to it.  There is a direct analogy for rotations:  We can easily
define an infinite strip of constant width in terms of 2 parallel lines (the
ends), where the width is ÓdefinedÕ by measuring along a line
perpendicular to the ends.  If we instead measure at an arbitrary angle to
the ends, we won't find the width, but the width times a factor depending
on that angle.

The most common point of confusion about relativity is that events that
are simultaneous in one reference frame are not simultaneous in another
(unless they are at the same place, in which case they are the same
event).  A frequent example is of this sort:  You have too much junk in
your garage, so your car won't fit anymore.  So your
spouse/roommate/whatever says, ``No problem, just drive it near the
speed of light, and it will Lorentz contract to fit."  So you try it, but in your frame inside the car you
find it is the garage that has contracted, so your car fits even worse.  The
real question is, ``What happens to the car when it stops?"  The answer
is, ``It depends on when the front end stops, and when the back end
stops."  You might expect that they stop at the same time.  That's
probably wrong, but assuming it's true, we have (at least) two
possibilities:  (1) They stop at the same time as measured in the garage's
reference frame.  Then the car fits.  However, in the car's frame (its initial
fast frame), the front end has stopped first, and the back end keeps going
until it smashes into the front enough to make it fit.  (2) They stop at the
same time in the car's frame.  In the garage's frame, the back end of
the car stops first, and the front end keeps going until it smashes out the
back of the garage.

The other standard example is ``time dilation":  Consider two clocks.  One
moves with constant velocity, so we choose the frame where it is at rest. 
The other moves at constant ÓspeedÕ in this frame, but it starts at the
position of the first clock, moves away, and then returns.  (It is
usually convenient to compare two clocks when they are at the same
point in space, since that makes it unambiguous that one is
reading the two clocks at the same time.) 

$$ \figscale{dilate}{.8in} $$
 A simple calculation shows that when the moving clock returns it
measures a time that is ÓshorterÕ by a factor of $å{1-v^2}$.  Of course,
this also has a Newtonian analog:  Curves between two given points are
longer than straight lines.  For relativity, straight lines are always the
ÓlongestÕ timelike curves because of the funny minus sign in the metric.

\x IA4.8  You are standing in the road, and a police car comes toward you, flashing
its lights at regular intervals.  It runs you down and keeps right on going,
as you watch it continue to flash its lights at you at the same intervals
(as measured by the clock in the car).  Treat this as a two-dimensional
problem (one space, one time), and approximate the car's velocity as
constant.  Draw the Minkowski-space picture (including you, the car, and
the light rays).  If the car moves at speed $v$ and flashes its lights at
intervals $t_0$ (as measured by the clock's car), at what intervals
(according to your watch) do you see the lights flashing when it is
approaching, and at what intervals as it is leaving?

Special relativity is so fundamental a part of physics that in some areas
of physics ÓeveryÕ experiment is more evidence for it, so that the many
early experimental tests of it are more of historical interest than
scientific.

The Galilean group is a symmetry of particles moving at speeds small compared to light, but electromagnetism is symmetric under the Poincar«e group (actually the conformal group).  This caused some confusion historically:  Since the two groups have only translations and rotations in common, it was assumed that nature was invariant under no velocity transformation (neither Galilean nor Lorentz boost).  In particular, the speed of light itself would seem to depend on the reference frame, since the laws of nature would be correct only in a ``rest frame".  To explain ``at rest with
respect to what," physicists invented something that is invariant under
rotations and space and time translations, but not velocity
transformations, and called this ``medium" for wave propagation the
``ether," probably because they were only semiconscious at the time. 
(The idea was supposed to be like sound traveling through the air,
although nobody had ever felt an ethereal wind.)

Many experiments were performed to test the existence of the ether, or
at least to show that the wave equation for light was correct only in
references frames at rest.  So as not to keep you in suspense, we first
tell you the general result was that the ether theory was wrong.  On the
contrary, one finds that the speed of light in a vacuum is measured as
$c$ in both of two reference frames that are moving at constant velocity
with respect to each other.  This means that electromagnetism is right
and Newtonian mechanics is wrong (or at least inaccurate), since Maxwell's equations are consistent with the speed of light being the same in all
frames, while Newtonian mechanics is not consistent with any speed
being the same in all frames.

The first such experiment was performed by A.A. Michelson and E.W.
Morley in 1887.  They measured the speed of light in various
directions at various times of year to try to detect the effect of the
Earth's motion around the sun.  They detected no differences, to an
accuracy of 1/6th the Earth's speed around the sun ($®10^{-4}c$).  (The
method was interferometry: seeing if a light beam split into
perpendicular paths of equal length interfered with itself.)

Another interesting experiment was performed in 1971 by J.C. Hafele and R. Keating,
who compared synchronized atomic clocks, one at rest with respect to the
Earth's surface, one carried by plane (a commercial airliner) west around
the world, one east.  Afterwards the clocks disagreed in a way predicted
by the relativistic effect of time dilation.

Probably the most convincing evidence of special relativity comes from
experiments related to atomic, nuclear, and particle physics.  In atoms the
speed of the electrons is of the order of the fine structure constant
($®$1/137) times $c$, and the corresponding effects on atomic energy
levels and such is typically of the order of the square of that ($®10^{-4}$),
well within the accuracy of such experiments.  In particle accelerators
(and also cosmic rays), various particles are accelerated to over 99\% $c$,
so relativistic effects are exaggerated to the point where particles act
more like light waves than Newtonian particles.  In nuclear physics the
relativistic relation between mass and energy is demonstrated by nuclear
decay where, unlike Newtonian mechanics, the sum of the (rest) masses is
not conserved; thus the atomic bomb provides a strong proof of special
relativity (although it seems like a rather extreme way to prove a point).

Ü5. Discrete: C, P, T

By considering only symmetries than can be obtained continuously from
the identity (Lie groups), we have missed some important symmetries:
those that reflect some of the coordinates.  It's sufficient to consider a
single reflection of a spacelike axis, and one of a timelike axis; all other
reflections can be obtained by combining these with the continuous
(``proper, orthochronous") Lorentz transformations.  (Spacelike and
timelike vectors can't be Lorentz transformed into each other, and
reflection of a lightlike axis won't preserve $p^2+m^2$.)  Also, the
reflection of one spatial axis can be combined with a $¹$ rotation about
that axis, resulting in reflection of all three spatial coordinates.  (Similar
generalizations hold for higher dimensions.  Note that the product of an
even number of reflections about different axes is a proper rotation;
thus, for even numbers of spatial dimensions reflections of all spatial
coordinates are proper rotations, even though the reflection of a single
axis is not.)  The reversal of the spatial coordinates is called ``parity (P)",
while that of the time coordinate is called ``time reversal" (``T"; actually,
for historical reasons, to be explained shortly, this is usually labeled
``CT".)  These transformations have the same effect on the momentum,
so that the definition of the Poisson bracket is also preserved.  These
``discrete" transformations, unlike the proper ones, are not symmetries
of nature (except in certain approximations):  The only exception is the
transformation that reflects all axes (``CPT").

While the metric $ú_{mn}$ is invariant under all Lorentz transformations
(by definition), the ``Levi-Civita tensor"
$$ ·_{mnpq}¼totally¼antisymmetric,ââ·_{0123} = -·^{0123} = 1 $$
 is invariant under only proper Lorentz transformations:  It has an
odd number of space indices and of time indices, so it changes sign under
parity or time reversal.  (More precisely, under P or T the Levi-Civita tensor 
does not suffer the expected sign change, since it's constant, so there is 
an ``extra" sign compared to the one expected for a tensor.)
Consequently, we can use it to define
``pseudotensors":  Given ``polar vectors", whose signs change as position
or momentum under improper Lorentz transformations, and scalars,
which are invariant, we can define ``axial vectors" and ``pseudoscalars"
as
$$ V_a = ·_{abcd}B^b C^c D^d,ââÄ = ·_{abcd}A^a B^b C^c D^d $$
 which get an extra sign change under such transformations (P or CT, but
not CPT).

There is another such ``discrete" transformation that is defined on phase
space, but which does not affect spacetime.  It changes the sign of all
components of the momentum, while leaving the spacetime coordinates
unchanged.  This transformation is called ``charge conjugation (C)",
and is also only an approximate symmetry in nature.  (Quantum
mechanically, complex conjugation of the position-space wave function
changes the sign of the momentum.)  Furthermore, it does not preserve
the Poisson bracket, but changes it by an overall sign.  (The misnomer
``CT" for time reversal follows historically from the fact that the
combination of reversing the time axis and charge conjugation preserves
the sign of the energy.)  The physical meaning of this transformation is
clear from the spacetime-momentum relation of relativistic classical
mechanics $p=m¼dx/ds$: It is proper-time reversal, changing the sign of
$s$.  The relation to charge follows from ``minimal coupling":  The
``covariant momentum" $m¼dx/ds=p+qA$ (for charge $q$) appears in the
constraint $(p+qA)^2+m^2=0$ in an electromagnetic background; $p£-p$
then has the same effect as $q£-q$.

In the previous subsection, we mentioned how negative energies were
associated with ``antiparticles".  Now we can better see the relation in
terms of charge conjugation.  Note that charge conjugation, since it only
changes the sign of $ $ but does not effect the coordinates, does not
change the path of the particle, but only how it is parametrized.  This is
also true in terms of momentum, since the velocity is given by $p^i/p^0$. 
Thus, the only observable property that is changed is charge; spacetime
properties (path, velocity, mass; also spin, as we'll see later) remain the
same.  Another way to say this is that charge conjugation commutes with
the Poincar«e group.  One way to identify an antiparticle is that it has all
the same kinematical properties (mass, spin) as the corresponding
particle, but opposite sign for internal quantum numbers (like charge). 
(Another way is pair creation and annihilation:  See subsection IIIB5
below.)  

All these transformations are summarized in the table:

$$ \vbox{\offinterlineskip
\hrule
\halign{ &\vrule#&\strut¼$#$¼\cr
height2pt&\omit&\hskip2pt\vrule&\omit&&\omit&&\omit&&\omit&\cr
& &\hskip2pt\vrule& C && CT¼ÊP && T¼ÊCP && PT¼ÊCPT &\cr 
height2pt&\omit&\hskip2pt\vrule&\omit&&\omit&&\omit&&\omit&\cr
\noalign{\hrule}
height2pt&\omit&\hskip2pt\vrule&\omit&&\omit&&\omit&&\omit&\cr
\noalign{\hrule}
height2pt&\omit&\hskip2pt\vrule&\omit&&\omit&&\omit&&\omit&\cr
& s &\hskip2pt\vrule& - && + \hfil + && - \hfil - && -ââ+¼ &\cr
height2pt&\omit&\hskip2pt\vrule&\omit&&\omit&&\omit&&\omit&\cr
\noalign{\hrule}
height2pt&\omit&\hskip2pt\vrule&\omit&&\omit&&\omit&&\omit&\cr
& t &\hskip2pt\vrule&  + && - \hfil + && - \hfil + && -ââ-¼ &\cr
& \vec x &\hskip2pt\vrule& + && + \hfil - && + \hfil - && -ââ-¼ &\cr
height2pt&\omit&\hskip2pt\vrule&\omit&&\omit&&\omit&&\omit&\cr
\noalign{\hrule}
height2pt&\omit&\hskip2pt\vrule&\omit&&\omit&&\omit&&\omit&\cr
& E &\hskip2pt\vrule& - && - \hfil + && + \hfil - && +ââ-¼ &\cr
& \vec p &\hskip2pt\vrule& - && + \hfil - && - \hfil + && +ââ-¼ &\cr
height2pt&\omit&\hskip2pt\vrule&\omit&&\omit&&\omit&&\omit&\cr
}\hrule} $$
 (The upper-left 3$ð$3 matrix contains the definitions, the rest is
implied.)  In terms of complex wave functions, we see that $C$ is just
complex conjugation (no effect on coordinates, but momentum and 
energy change sign because of the ``$i$" in the Fourier transform).
On the other hand, for $CT$ and $P$ there is no complex conjugation,
but changes in sign of the coordinates that are arguments of the
wave functions, and also on the corresponding indices --- 
the ``orbital" and ``spin" parts of these discrete transformations.
(For example, derivatives $»_a$ have sign changes because $x^a$ does,
so a vector wave function $Æ^a$ must have the same sign changes
on its indices for $»_a Æ^a$ to transform as a scalar.)  The other
transformations follow as products of these.

\x IA5.1  Find the effect of each of these 7 transformations on
wave functions that are: {\bf a} scalars, {\bf b} pseudoscalars,
{\bf c} vectors, {\bf d} axial vectors.

However, from the point of view of the ``particle" there ÓisÕ some kind of
kinematic change, since the proper time has changed sign:  If we think of
the mechanics of a particle as a one-dimensional theory in $ $ space (the
worldline), where $x( )$ (as well as any such variables describing spin or
internal symmetry) is a wave function or field on that space, then $ £- $
is T on that one-dimensional space.  (The fact we don't get CT can be
seen when we add additional variables:  For example, if we describe
internal U(N) symmetry in terms of creation and annihilation operators
$aÿ^i$ and $a_i$, then C mixes them on both the worldline and spacetime.
 So, on the worldline we have the ``pure" worldline geometric symmetry
CT times C = T.)  Thus, in terms of ``zeroth quantization",
$$ worldline¼T ª spacetime¼C $$
 On the other hand, ÓspacetimeÕ $P$ and $CT$ are simply internal
symmetries with respect to the worldline (as are proper, orthochronous
Poincar«e transformations).

Quantum mechanically, there is a good reason for particles of
negative energy:  They appear in complex-conjugate wave functions,
since $(e^{-i¿t})*=e^{+i¿t}$.  Since we always evaluate expressions
of the form $Òf|iÔ$, it is natural for energies of both signs to
appear.  

In classical field theory, we can identify a particle with its
antiparticle by requiring the field to be invariant under
charge conjugation:  For example, for a scalar field (spinless particle), we
have the reality condition
$$ Ä(x) = Ä*(x) $$
 or in momentum space, by Fourier transformation,
$$ ր(p) = [ր(-p)]* $$
 which implies the particle has charge zero (neutral).

Ü6. Conformal

Poincar«e transformations are the most general coordinate
transformations that preserve the mass condition $p^2+m^2=0$, but there is a
larger group, the ``conformal group", that preserves this constraint in the
massless case.  
Although conformal symmetry is not observed in nature, it is important in
all approaches to field theory:  
\item{(1)} First of all, it is useful in the
construction of free theories (see subsections IIB1-4 below).  All massive
fields can be described consistently in quantum field theory in terms of
coupling massless fields.  Massless theories are a subset of conformal
theories, and some conditions on massless theories can be found more
easily by finding the appropriate subset of those on conformal theories. 
This is related to the fact that the conformal group, unlike the Poincar«e
group, is ``simple":  It has no nontrivial subgroup that transforms into
itself under the rest of the group (like the way translations transform into
themselves under Lorentz transformations). 
\item{(2)} In interacting theories at
the classical level, conformal symmetry is also important in finding and
classifying solutions, since at least some parts of the action are
conformally invariant, so corresponding solutions are related by conformal
transformations (see subsections IIIC5-7).  Furthermore, it is often
convenient to treat arbitrary theories as broken conformal theories,
introducing fields with which the breaking is associated, and analyze the
conformal and conformal-breaking fields separately.  This is particularly
true for the case of gravity (see subsections IXA7,B5,C2-3,XA3-4,B5-7). 
\item{(3)} Within quantum field theory at the perturbative level, the only physical
quantum field theories are ones that are conformal at high energies (see
subsection VIIIC1). The quantum corrections to conformal invariance at
high energy are relatively simple.  
\item{(4)} Beyond perturbation theory, the
only quantum theories that are well defined may be just the ones whose
breaking of conformal invariance at low energy is only classical (see
subsections VIIC2-3,VIIIA5-6).  Furthermore, the largest possible
symmetry of a nontrivial S-matrix is conformal symmetry (or
superconformal symmetry if we include fermionic generators).  
\item{(5)}
Self-duality (a generalization of a condition that equates electric and
magnetic fields) is useful for finding solutions to classical field
equations as well as simplifying perturbation theory, and is closely
related to ``twistors" (see subsections IIB6-7,C5,IIIC4-7).  In general,
self-duality is related to conformal invariance:  For example, it can be
shown that the free conformal theories in arbitrary even dimensions are
just those with (on-mass-shell) field strengths on which self-duality can
be imposed.  (In arbitrary odd dimensions the free conformal theories are
just the scalar and spinor.)  

Transformations $Â$ that satisfy
$$ [Â^a(x)p_a,p^2] = ½(x)p^2 $$
 for some $½$ also preserve $p^2=0$, although they don't leave $p^2$
invariant.  Equivalently, we can look for coordinate transformations that
scale
$$ dx'^2 = Å(x)dx^2 $$

\x IA6.1  Find the conformal group explicitly in two dimensions, and show
it's infinite dimensional (not just the SO(2,2) described below).  (Hint:  Use
lightcone coordinates.)

This symmetry can be made manifest by starting with a space with one
extra space and time dimension:
$$ y^A = (y^a,y^+,y^-)âÜây^2 = y^A y^B ú_{AB} = (y^a)^2 -2y^+y^- $$
 where $(y^a)^2=y^a y^b ú_{ab}$ uses the usual
D-dimensional Minkowski-space metric $ú_{ab}$, and the two additional
dimensions have been written in a lightcone basis (not to be confused for
the similar basis that can be used for the Minkowski metric itself).  With
respect to this metric, the original SO(D$-$1,1) Lorentz symmetry has
been enlarged to SO(D,2).  This is the conformal group in D dimensions. 
However, rather than also preserving (D+2)-dimensional translation
invariance, we instead impose the constraint and invariance
$$ y^2 = 0,â¶y^A = ½(y)y^A $$
 This reduces the original space to the ``projective" (invariant under the
$½$ scaling) lightcone (which in this case really is a cone).  

These two conditions can be solved by
$$ y^A = ew^A,âw^A = (x^a,1,üx^a x_a) $$
 Projective invariance then means independence from $e$ ($y^+$), while
the lightcone condition has determined $y^-$.  $y^2=0$ implies $yÉdy=0$,
so the simplest conformal invariant is
$$ dy^2 = (edw+wde)^2 = e^2 dw^2 = e^2 dx^2 $$
 where we have used $w^2=0ÜwÉdw=0$.  This means any SO(D,2)
transformation on $y^A$ will simply scale $dx^2$, and scale $e^2$ in the
opposite way:
$$ dx'^2 = \left( {e^2 \over e'^2} \right) dx^2 $$
 in agreement with the previous definition of the conformal group.

The explicit form of conformal transformations on $x^a=y^a/y^+$ now
follows from their linear form on $y^A$, using the generators
$$ G^{AB} = y^{[A}r^{B]},ââ[r_A,y^B] = -i¶_A^B $$
 of SO(D,2) in terms of the momentum $r_A$ conjugate to $y^A$.  (These
are defined the same way as the Lorentz generators
$J^{ab}=x^{[a}p^{b]}$.)  For example, $G^{+-}$ just scales
$x^a$.  (Scale transformations are also known as ``dilatations", or just ``dilations".)  We can
also recognize $G^{+a}$ as generating translations on $x^a$.  The only
complicated transformations are generated by $G^{-a}$, known as
``conformal boosts" (acceleration transformations).  Since they commute
with each other (like translations), it's easy to exponentiate to find the
finite transformations:
$$ y' = e^G y,ââG = v_a y^{[-}»^{a]} $$
 for some constant D-vector $v^a$ (where $»_A­»/»y^A$).  Since the
conformal boosts act as ``lowering operators" for scale weight
($+£a£-$), only the first three terms in the exponential survive:
$$ Gy^- = 0,ââGy^a = v^a y^-,ââGy^+ = v^a y_aâÜ $$
$$ y'^- = y^-,¼y'^a = y^a +v^a y^-,¼y'^+ = y^+ +v^a y_a +üv^2 y^-âÜ $$
$$ x'^a = {x^a +üv^a x^2\over 1+vÉx +\f14 v^2 x^2} $$
 using $x^a=y^a/y^+$, $y^-/y^+=üx^2$.

\x IA6.2  Make the change of variables to $x^a=y^a/y^+$, $e=y^+$,
$z=üy^2$.  Express $r_A$ in terms of the momenta $(p_a,n,s)$ conjugate
to $(x^a,e,z)$.  Show that the conditions $y^2=y^A r_A=r^2=0$ become
$z=en=p^2=0$ in terms of the new variables.

\x IA6.3  Find the generator of infinitesimal conformal boosts in terms of
$x^a$ and $p_a$.

We actually have the full O(D,2) symmetry:  Besides the continuous
symmetries, and the discrete ones of SO(D$-$1,1), we have a second
``time" reversal (from our second time dimension):
$$ y^+ ª -y^-âÜâx^a ª -{x^a\over üx^2} $$
 This transformation is called an ``inversion".

\x IA6.4  Show that a finite conformal boost can be obtained by
performing a translation sandwiched between two inversions.

\x IA6.5  The conformal group for Euclidean space (or any spacetime
signature) can be obtained by the same construction.  Consider the
special case of D=2 for these SO(D+1,1) transformations.  (This is a
subgroup of the 2D superconformal group:  See exercise IA6.1.)  Use
complex coordinates for the two ``physical" dimensions:
$$ z = \f1{å2}(x^1+ix^2) $$
ªa Show that the inversion is
$$ z ª -{1\over z*} $$
ªb Show that the conformal boost is (using a complex number also for the
boost vector)
$$ z £ {z\over 1+v*z} $$

\x IA6.6  Any parity transformation (reflection in a spatial axis) can be
obtained from any other by a rotation of the spatial coordinates. 
Similarly, when there is more than one time dimension, any time reversal
can be obtained from another (but time reversal can't be rotated into
parity, since a timelike vector can't be rotated into a spacelike one).  Thus,
the complete orthogonal group O(m,n) can be obtained from those
transformations that are continuous from the identity by combining them
with 1 parity transformation and 1 time reversal transformation (for
mn$±$0).  
 ªa For the conformal group, find the rotation (in terms of an angle)
that rotates between the two time directions, and express its action on
$x^a$.  
 ªb Show that for angle $¹$ it produces a transformation that is the
product of time reversal and inversion.  
 ªc Use this to show that inversion is
related to time reversal by finding the continuum of conformal
transformations that connect them.

\refs

£1 M. Planck, ÓS.-B. Preuss. Akad. Wiss.Õ (1899) 440, ÓAnn. der Phys.Õ É1 (1900) 69, 	ÓPhysikal\-ische Abhandlungen und Vortr¬ageÕ, 
	Bd. I (Vieweg, Braunschweig, 1958) 560, 614:\\
	natural units.
£2 I.M. Mills, P. J. Mohr, T. J. Quinn, B.N. Taylor, and E.R. Williams,
	ÓMetrologiaÕ É42 (2005) 71:\\
	some arguments for fixing the value of $\h$ by definition, 
	so the kilogram is a derived unit.
£3 F.A. Berezin, ÓThe method of second quantizationÕ (Academic, 1966):\\
	calculus with anticommuting numbers.
 £4 P.A.M. Dirac, ÓProc. Roy. Soc.Õ ÉA126 (1930) 360:\\
	antiparticles.
 £5 E.C.G. St¬uckelberg, ÓHelv. Phys. ActaÕ É14 (1941) 588, É15 (1942) 23;\\
	J.A. Wheeler, 1940, unpublished:\\
	the relation of antiparticles to proper time.
 £6 S. Mandelstam, ÓPhys. Rev.Õ É112 (1958) 1344.
 £7 H.W. Brinkmann, ÓProc. Nat. Acad. Sci. (USA)Õ É9 (1923) 1:\\
	projective lightcone as conformal to flat space.
 £8 P.A.M. Dirac, ÓAnn. Math.Õ É37 (1936) 429;\\
	H.A. Kastrup, ÓPhys. Rev.Õ É150 (1966) 1186;\\
	G. Mack and A. Salam, ÓAnn. Phys.Õ É53 (1969) 174;\\
	S. Adler, \PRD 6 (1972) 3445;\\
	R. Marnelius and B. Nilsson, \PRD 22 (1980) 830:\\
	conformal symmetry.
 £9 S. Coleman and J. Mandula, ÓPhys. Rev.Õ É159 (1967) 1251:\\
	conformal symmetry as the largest (bosonic) symmetry of the
	S-matrix.
 £10 W. Siegel, ÓInt. J. Mod. Phys. AÕ É4 (1989) 2015:\\
	equivalence between conformal invariance and self-duality in all
	dimensions.

\unrefs

Û5 B. INDICES

In the previous section we saw various spacetime groups (Galilean,
Poincar«e, conformal) in terms of how they acted on coordinates.  This not
only gave them a simple physical interpretation, but also allowed a direct
relation between classical and quantum theories.  However, as we know
from studying rotations in quantum theory in terms of spin, we will often
need to study symmetries of quantum theories for which the classical
analog is not so useful or perhaps even nonexistent.

We therefore now consider some general results of group theory, mostly
for continuous groups.  We use tensor methods, rather than the slightly
more powerful but greatly less convenient Cartan-Weyl-Dynkin methods. 
Much of this section should be review, but is included here for
completeness; it is not intended as a substitute for a group theory course,
but as a summary of those results commonly useful in field theory.

Ü1. Matrices

Matrices are defined by the way they act on some vector space; an n$ð$n
matrix takes one n-component vector to another.  Given some group, and
its multiplication table (which defines the group completely), there is more
than one way to represent it by matrices.  Any set of matrices we find
that has the same multiplication table as the group elements is called a
``representation" of that group, and the vector space on which those
matrices act is called the ``representation space."  The representation of
the algebra or group in terms of explicit matrices is given by choosing a
basis for the vector space.  If we include infinite-dimensional
representations, then a representation of a group is simply a way to
write its transformations that is linear:  $Æ'=MÆ$ is linear in $Æ$.  More
generally, we can also have a ``realization" of a group, where the
transformations can be nonlinear.  These tend to be more cumbersome, so
we usually try to make redefinitions of the variables that make the
realization linear.  A precise definition of ``manifest symmetry" is that all
the realizations used are linear.  (One possible exception is ``affine" or
``inhomogeneous" transformations $Æ'=M_1 Æ+M_2$, such as the usual
coordinate representation of Poincar«e transformations, since these
transformations are still very simple, because they are really still linear,
though not homogeneous.)

\x IB1.1  Consider a general real affine transformation $Æ'=MÆ+V$
on an $n$-component vector $Æ$ for arbitrary real $nðn$ matrices $M$ 
and real $n$-vectors $V$.  A general group element is thus $(M,V)$.
 ªa Perform 2 such transformations consecutively, 
and give the resulting ``group multiplication" rule for
$(M_1,V_1)$ ``$ð$" $(M_2,V_2)$ $=$ $(M_3,V_3)$.
 ªb Find the infinitesimal form of this transformation.  Define 
the $n^2+n$ generators as operators on $Æ$, in terms of $Æ^a$ and
$»/»Æ^a$.
 ªc Find the commutation relations of these generators.
 ªd Compare all the above with (nonrelativistic) rotations and translations.

\x IB1.2  Let's consider some properties of matrix inverses:
 ªa Show $(AB)^{-1}=B^{-1}A^{-1}$ for matrices $A$ and $B$ that
have inverses but don't necessarily commute with each other.
 ªb Show that
$$ {1\over A+B} = {1\over A} - {1\over A}B{1\over A} 
	+{1\over A}B{1\over A}B{1\over A} - ... $$
 (There may be other assumptions; ignore convergence questions.
 Hint: Multiply both sides by $A+B$.)

For convenience, we write matrices with a Hilbert-space-like notation,
but unlike Hilbert space we don't necessarily associate bras directly with
kets by Hermitian conjugation, or even transposition.  In general, the two
spaces can even be different sizes, to describe matrices that are not
square; however, for group theory we are interested only in matrices
that take us from some vector space into itself, so they are square.
Bras have an inner product with kets, but neither necessarily has a norm
(inner product with itself):  In general, if we start with some vector
space, written as kets, we can always define the ``dual" space, written
as bras, by defining such an inner product.  In our case, we may start
with some representation of a group, in terms of some vector space, and
that will give us directly the dual representation.  (If the representation
is in terms of unitary matrices, we have a Hilbert space, and the dual
representation is just the complex conjugate.)

So, we define column vectors $|ÆÔ$ with a basis $|{}^IÔ$, and row vectors
$ÒÆ|$ with a basis $Ò{}_I|$, where $I=1,...,$n to describe n$ð$n matrices. 
The two bases have a relative normalization defined so that the inner
product gives the usual component sum:
$$ |ÆÔ = |{}^IÔÆ_I,¼Ò| = ^IÒ{}_I|;âÒ{}_I|{}^JÔ = ¶_I^JâÜâ
	ҍ|ÆÔ = ^I Æ_I;âÒ{}_I|ÆÔ = Æ_I,¼Ò|{}^IÔ = ^I $$
 These bases then define not only the components of vectors, but also
matrices:
$$ M = |{}^IÔM_I{}^JÒ{}_J|,âÒ{}_I|M|{}^JÔ = M_I{}^J $$
 where as usual the $I$ on the component (matrix element) $M_I{}^J$
labels the row of the matrix $M$, and $J$ the column.  This implies the
usual matrix multiplication rules, inserting the identity in terms of the
basis,
$$ I = |{}^KÔÒ{}_K|âÜâ(MN)_I{}^J = Ò{}_I|M|{}^KÔÒ{}_K|N|{}^JÔ =
	M_I{}^K N_K{}^J $$
 Closely related is the definition of the trace,
$$ tr¼M = Ò{}_I|M|{}^IÔ = M_I{}^IâÜâtr(MN) = tr(NM) $$
 (We'll discuss the determinant later.)  

The bra-ket notation is really just
matrix notation written in a way to clearly distinguish column vectors,
row vectors, and matrices.
We can, of course, also use the usual pictorial notation
$$ |ÆÔ = \pmatrix{ Æ_1\cr Æ_2 \cr \vdots\cr },ââҍ| = ( ^1¼^2¼\ldots ) $$
$$ M = \bordermatrix{ & 1 & 2 & \ldots & J & \ldots \cr 
		1 & M_1{}^1 & M_1{}^2 & \ldots & M_1{}^J & \ldots \cr
		2 & M_2{}^1 & M_2{}^2 & \ldots & M_2{}^J & \ldots \cr 
		\vdots & \vdots & \vdots & \ddots & \vdots & \ddots \cr
		I & M_I{}^1 & M_I{}^2 & \ldots & M_I{}^J & \ldots \cr
		\vdots & \vdots & \vdots & \ddots & \vdots & \ddots \cr } $$
 This is useful only when listing individual components.

We can easily translate transformation laws from
matrix notation into index notation just by using a basis for the
representation space.  We now write $g$ and $G$ to refer to 
either matrix representations of the group and algebra elements, 
or to the abstract elements: i.e., either to a specific representation,
or the most general one.  Again writing $g = e^{iG}$,
$$ g|{}^IÔ = |{}^JÔg_J{}^I,âG|{}^IÔ = |{}^JÔG_J{}^I $$
$$ G = Œ^i G_i,â¶|ÆÔ = iG|ÆÔ = |{}^IÔiŒ^i (G_i)_I{}^J Æ_J
	âÜâ¶Æ_I = iŒ^i (G_i)_I{}^J Æ_J $$
The dual space isn't needed for this purpose.  However, for any
representation of a group, the transpose
$$ (M^T)^I{}_J = M_J{}^I $$
 of the inverse of those matrices also gives a representation of the group,
since
$$ g_1 g_2 = g_3âÜâ(g_1)^{T-1}(g_2)^{T-1} = (g_3)^{T-1} $$
$$ [G_1,G_2] = G_3âÜâ[-G_1^T,-G_2^T] = -G_3^T $$
 This is the dual representation, which follows from defining the above
inner product to be invariant under the group:
$$ ¶ÒÆ|Ô = 0âÜâ¶Æ^I = -iÆ^J Œ^i (G_i)_J{}^I $$

The complex conjugate of a complex representation is also a
representation, since
$$ g_1 g_2 = g_3âÜâg_1*g_2* = g_3* $$
$$ [G_1,G_2] = G_3âÜâ[G_1*,G_2*] = G_3* $$
 From any given representation, we can thus find three others from
taking the dual and the conjugate:  In matrix and index notation,
$$ \li{ Æ' = gÆ:ââ& Æ'_I = g_I{}^J Æ_J \cr
	Æ' = (g^{-1})^TÆ:ââ& Æ'^I = g^{-1}{}_J{}^I Æ^J \cr
	Æ' = g*Æ:ââ& Æ'_{ÀI} = g*_{ÀI}{}^{ÀJ} Æ_{ÀJ} \cr
	Æ' = (g^{-1})ÿÆ:ââ& Æ'^{ÀI} = g*^{-1}{}_{ÀJ}{}^{ÀI} Æ^{ÀJ} \cr} $$
 since $(g^{-1})^T$, $g*$, and $(g^{-1})ÿ$ (but not $g^T$, etc.)¼satisfy the
same multiplication algebra as $g$, including ordering.  We use up/down
and dotted/undotted indices to denote the transformation law of each
type of index; contracting undotted up indices with undotted down
indices preserves the transformation law as indicated by the remaining
indices, and similarly for dotted indices.  These four representations are
not necessarily independent:  Imposing relations among them is how the
classical groups are defined (see subsections IB4-5 below).

Ü2. Representations

For example, we always have the ``adjoint" representation of a Lie
group/algebra, which is how the algebra acts on its own generators:
$$ \hbox{(1) adjoint as operator:} â G = Œ^i G_i,âA = º^i G_i
	âÜâ¶A = i[G,A] = º^j Œ^i f_{ij}{}^k G_k $$
$$ Üⶺ^i = -iº^k Œ^j (G_j)_k{}^i,â(G_i)_j{}^k = if_{ij}{}^k $$
 This gives us two ways to represent the adjoint representation space: as
either the usual vector space, or in terms of the generators.  Thus, we
either use the matrix $A=º^iG_i$ (for arbitrary representation of the
matrices $G_i$, or treating $G_i$ as just abstract generators), 
or we can write $A$ as a row vector:
$$\hbox{(2) adjoint as vector:}â ÒA| = º^i Ò{}_i|âÜâ¶ÒA| = -iÒA|G $$
$$ Üⶺ^i Ò{}_i| = -iº^k Œ^j (G_j)_k{}^i Ò{}_i| $$

The adjoint representation also provides a convenient way to define a
(symmetric) group metric invariant under the group, the ``Cartan metric":
$$ ú_{ij} = tr_A(G_i G_j) = -f_{ik}{}^l f_{jl}{}^k $$
 ($tr_A$ refers to the trace taken with respect to the representation $A$; 
equivalently, we could take the $G$'s inside the trace to be in the $A$ representation.)
For ``Abelian" groups the structure constants vanish, and thus so does
this metric.  ``Semisimple" groups are those where the metric is
invertible (no vanishing eigenvalues).  A ``simple" group has no nontrivial
subgroup that transforms into itself under the rest of the group: 
Semisimple groups can be written as ``products" of simple groups. 
``Compact" groups are those where it is positive definite (all eigenvalues
positive); they are also those for which the invariant volume of the group
space is finite.  For simple, compact groups it's convenient to choose a
basis where
$$ ú_{ij} = c_A ¶_{ij} $$
 for some constant $c_A$ (the ``Dynkin index" for the adjoint
representation).  For some general irreducible representation $R$ of such
a group the normalization of the trace is
$$ tr_R(G_i G_j) = c_R ¶_{ij} = {c_R\over c_A}ú_{ij} $$
 Now the proportionality constant $c_R/c_A$ is fixed by the choice of
$R$ (only), since we have already fixed the normalization of our basis.  

\x IB2.1  What is $c_R$ for an Abelian group?  (Hint: not just 1.)

In general, the cyclicity property of the trace implies, for any
representation, that
$$ 0 = tr([G_i,G_j]) = -if_{ij}{}^k tr(G_k) $$
 so $tr(G_i)=0$ for semisimple groups.  Similarly, we find
$$ f_{ijk} ­ f_{ij}{}^l ú_{lk} = i¼tr_A ([G_i,G_j]G_k) $$
 is totally antisymmetric:  For semisimple groups, this implies the total
antisymmetry of the structure constants $f_{ij}{}^k$, up to factors (which
are absent for compact groups in a basis where $ú_{ij}¾¶_{ij}$).  This also
means the adjoint representation is its own dual.  (For example, for the
compact group SO(3), we have $ú_{ij}=-·_{ikl}·_{jlk}=2¶_{ij}$.)  Thus, we
can write $A$ in a third way, as a column vector
$$ \hbox{(3) adjoint as dual vector:} â |AÔ = |{}^iÔ º_i ­ |{}^iÔ º^j ú_{ji}
	âÜâ¶|AÔ = iG|AÔ  $$
 We can also do this for Abelian groups, by defining an invertible metric
unrelated to the Cartan metric:  This is trivial for Abelian groups, since
the generators themselves are invariant, and thus so is ÓanyÕ metric on
them.

An identity related to the trace one is the normalization of the value
$k_R$ of the ``Casimir operator" for any particular representation,
$$ ú^{ij}G_i G_j = k_R I $$
 Its proportionality to the identity follows from the fact that it commutes
with each generator:
$$ [ú^{jk}G_j G_k,G_i] = -if^j{}_i{}^kÓG_j,G_kÕ = 0 $$
 using the antisymmetry of the structure constants.  (Thus it takes the
same value on any component of an irreducible representation, since they
are all related by group transformations.)  By tracing this identity, and
contracting the trace identity,
$$ {c_R\over c_A}d_A = tr_R(ú^{ij}G_i G_j) = k_R d_R $$
$$ Üâk_R = {c_R d_A\over c_A d_R} $$
 where $d_R­tr_R(I)$ is the dimension of that representation.

Although quantum mechanics is defined on Hilbert space, which is a kind
of complex vector space, more generally we want to consider real
objects, like spacetime vectors.  This restricts the form of linear
transformations:  Specifically, if we absorb $i$'s as $g=e^G$, then in such
representations $G$ itself must be real.  These representations are then
called ``real representations", while a ``complex representation" is one
whose representation isn't real in any basis.  A complex representation
space can have a real representation, but a real representation space
can't have a complex representation.  In particular, coordinate
transformations (of real coordinates) have only real representations,
which is why absorbing the $i$'s into the generators is a useful
convention there.  For semisimple unitary groups, hermiticity of the
generators of the adjoint representation implies (using total
antisymmetry of the structure constants and reality of the Cartan metric)
that the structure constants are real, and thus the adjoint representation
is a real representation.  More generally, any real unitary representation
will have antisymmetric generators ($G=G*=-GÿÜG=-G^T$).  If the complex
conjugate representation is the same as the original (same matrices up to
a similarity transformation $g*=MgM^{-1}$), but the representation is not
real, then it is called ``pseudoreal".  (An example is the spinor of SU(2), to
be described in section IC.)

For any representation $g$ of the group, a transformation
$g£g_0 g g_0^{-1}$ on every group element $g$ for some particular group
element $g_0$ clearly maps the algebra to itself, and preserves the
multiplication rules.  (Similar remarks apply to applying the
transformation to the generators.)  However, the same is true for
complex conjugation, $g£g*$:  Not only are the multiplication rules
preserved, but for any element $g$ of that representation of the group,
$g*$ is also an element.  (This can be shown, e.g., by defining
representations in terms of the values of all the Casimir operators,
contructed from various powers of the generators.)  In quantum
mechanics (where the representations are unitary), the latter is called an
``antiunitary transformation".  Although this is a symmetry of the group,
it cannot be reproduced by a unitary transformation, except when the
representation is (pseudo)real.

\x IB2.2  Show how this works for the Abelian group U(1).
Explain this antiunitary transformation in terms of
two-dimensional rotations O(2).  (U(1)=SO(2), the ``proper
rotations" obtained continuously from the identity.)

A very simple way to build a representation from others is by ``direct
sum".  If we have two representations of a group, on two different
spaces, then we can take their direct sum by just putting one column
vector on top of the other, creating a bigger vector whose size
(``dimension") is the sum of that of the original two.  Explicitly, if we start
with the basis $|{}^îÔ$ for the first representation and $|{}^{î'}Ô$ for
the second, then the union $(|{}^îÔ,|{}^{î'}Ô)$ is the basis for the direct
sum.  (We can also write $|{}^IÔ=(|{}^îÔ,|{}^{î'}Ô)$, where
$î=1,...,m$; $î' =1,...,n$; $I=1,...,m,m+1,...,m+n$.)  The group then
acts on each part of the new vector in the obvious way:
$$  Æ = |{}^îÔÆ_î,⍠= |{}^{î'}ԍ_{î'};ââ
	g|{}^îÔ = |{}^ûÔg_û{}^î,âg|{}^{î'}Ô = |{}^{û'}Ôg_{û'}{}^{î'} $$
$$ Üâ|ïÔ = |{}^îÔÆ_î ¢ |{}^{î'}ԍ_{î'} = |ÆÔ¢|Ô
	âorâ(ï) = {Æ\choose } $$
$$ g|ïÔ = |{}^ûÔg_û{}^î Æ_î ¢ |{}^{û'}Ôg_{û'}{}^{î'}_{î'}
	âorâ(g) = \pmatrix{ g_î{}^û & 0 \cr 0 & g_{î'}{}^{û'} \cr} $$
 (We can replace the $¢$ with an ordinary $+$ if we understand the basis
vectors to be now in a bigger space, where the elements of the first basis
have zeros for the new components on the bottom while those of the
second have zeros for the new components on top.)  The important point
is that no group element mixes the two spaces:  The group representation
is block diagonal.   Any representation that can be written as a direct sum
(after an appropriate choice of basis) is called ``reducible".  For example,
we can build a reducible real representation from an irreducible complex
one by just taking the direct sum of this complex representation with the
complex conjugate representation.  Similarly, we can take direct sums of
more than two representations.

A more useful way to build representations is by ``direct product".  The
idea there is to take a colummn vector and a row vector and use them to
construct a matrix, where the group element acts simultaneously on rows
according to one representation and columns according to the other. 
If the two original bases are again $|{}^îÔ$ and $|{}^{î'}Ô$, the new
basis can also be written as $|{}^IÔ=|{}^{îî'}Ô$ ($I=1,...,mn$).  Explicitly,
$$ |ÆÔ = |{}^îÔ°|{}^{î'}ÔÆ_{îî'},â
	g(|{}^îÔ°|{}^{î'}Ô) = |{}^ûÔ°|{}^{û'}Ôg_û{}^î g_{û'}{}^{î'}
	âÜâg_{îî'}{}^{ûû'} = g_î{}^û g_{î'}{}^{û'} $$
 or in terms of the algebra
 $$ G_{îî'}{}^{ûû'} = 
	G_î{}^û ¶_{î'}{}^{û'} + ¶_î{}^û G_{î'}{}^{û'} $$
 A familar example from quantum mechanics is rotations (or Lorentz
transformations), where the first space is position space (so $î$ is the
continuous index $x$), acted on by the orbital part of the generators, while
the second space is finite-dimensional, and is acted on by the spin part of
the generators.  Direct product representations are usually reducible: 
They then can be written also as direct sums, in a way that depends on
the particulars of the group and the representations.

Consider a representation constructed by direct product:  In matrix
notation
$$ öG_i = G_i°I' +I°G_i' $$
 Using $tr(A°B)=tr(A)tr(B)$, and assuming $tr(G_i)=tr(G_i')=0$, we have
$$ tr(öG_i öG_j) = tr(I')tr(G_i G_j) +tr(I)tr(G_i' G_j') $$
 For example, for SU(N) (see subsection IB4 below) we can construct the
adjoint representation from the direct product of the N-dimensional,
``defining" representation and its complex conjugate.  (We also get a
singlet, but it will not affect the result for the adjoint.)  In that case we
find
$$ tr_A (G_i G_j) = 2NÊtr_D (G_i G_j)âÜâ{c_D\over c_A} = {1\over 2N} $$
 For most purposes, we use $tr_D(G_i G_j)=¶_{ij}$ ($c_D=1$) for SU(N), so
$c_A=2N$. 

Ü3. Determinants

We now ``review" some properties of determinants that will prove useful
for the group analysis of the following subsections.  Determinants can be
defined in terms of the Levi-Civita tensor $·$.  As a consequence of its
antisymmetry,
$$ ·¼totally¼antisymmetric,â·_{12...n} = ·^{12...n} = 1âÜâ
	·_{J_1...J_n}·^{I_1...I_n} = ¶_{[J_1}^{I_1}ò¶_{J_n]}^{I_n} $$
 since each possible numerical index value appears once in each $·$, so
they can be matched up with $¶$'s.  By similar reasoning,
$$ \f1{m!}·_{K_1...K_m J_1...J_{n-m}}·^{K_1...K_m I_1...I_{n-m}} =
	¶_{[J_1}^{I_1}ò¶_{J_{n-m}]}^{I_{n-m}} $$
 where the normalization compensates for the number of terms in the
summation.

\x IB3.1  Apply these identities to rotations in three dimensions:
 ªa Given only the commutation relations 
$[J_{ij},J^{kl}] = i¶^{[k}_{[i}J_{j]}{}^{l]}$
and the definition $G_i­ü·_{ijk}J_{jk}$, 
derive $f_{ij}{}^k=·_{ijk}$.
 ªb Show the Jacobi identity $·_{[ij}{}^l ·_{k]l}{}^m=0$
by explicit evaluation.
 ªc Find the Cartan metric, and thus the value of $c_A$.

This tensor is used to define the determinant:
$$ det¼M_I{}^J = \f1{n!}·_{J_1...J_n}·^{I_1...I_n}M_{I_1}{}^{J_1}ò
	M_{I_n}{}^{J_n} 
	âÜâ·_{J_1...J_n}M_{I_1}{}^{J_1}òM_{I_n}{}^{J_n} 
	= ·_{I_1...I_n}det¼M $$
 since anything totally antisymmetric in $n$ indices must be proportional
to the $·$ tensor.  This yields an explicit expression for the inverse:
$$ (M^{-1})_{J_1}{}^{I_1} = \f1{(n-1)!}·_{J_1...J_n}·^{I_1...I_n}
	M_{I_2}{}^{J_2}òM_{I_n}{}^{J_n} (det¼M)^{-1}$$
 From this follows a useful expression for the variation of the
determinant:
$$ {»\over »M_I{}^J}Êdet¼M = (M^{-1})_J{}^IÊdet¼M $$
 which is equivalent to
$$ ¶¼ln¼det¼M = tr(M^{-1}¶M) $$
 Replacing $M$ with $e^M$ gives the often-used identity
$$ ¶¼ln¼det¼e^M = tr(e^{-M}¶e^M) = tr¼¶MâÜâdet¼e^M = e^{trÊM} $$
 where we have used the boundary condition for $M=0$.  Finally, replacing
$M$ in the last identity with $ln(1+L)$ and expanding both sides to order
$L^n$ gives general expressions for determinants of $nðn$ matrices in
terms of traces:
$$ det(1+L)=e^{trÊln(1+L)}âÜâdet¼L=\f1{n!}(tr¼L)^n
	-\f1{2(n-2)!}(tr¼L^2)(tr¼L)^{n-2}+ò $$

\x IB3.2  Use the definition of the determinant (and not its relation to
the trace) to show
$$ det(AB) = det(A) det(B) $$

These identities can also be derived by defining the determinant in terms
of a Gaussian integral.  We first collect some general properties of
(indefinite) Gaussian integrals.  The simplest such integral is
$$ Ç{d^2 x\over 2¹}¼e^{-x^2/2} 
	= Ç_0^{2¹}{dÏ\over 2¹}Ç_0^¥ dr¼re^{-r^2/2}
	= Ç_0^¥ du¼e^{-u} = 1 $$
$$ ÜâÇ{d^D x\over (2¹)^{D/2}}¼e^{-x^2/2} 
	= \left(Ç{dx\over å{2¹}}¼e^{-x^2/2}\right)^D =
	\left(Ç{d^2 x\over 2¹}¼e^{-x^2/2}\right)^{D/2} = 1 $$
 The complex form of this integral is
$$ Ç{d^D z*¼d^D z\over (2¹i)^D}¼e^{-|z|^2} = 1 $$
 by reducing to real parameters as $z=(x+iy)/å2$.  These generalize to
integrals involving a real, symmetric matrix $S$ or a Hermitian matrix $H$
as
$$ Ç{d^D x\over (2¹)^{D/2}}¼e^{-x^T Sx/2} = (det¼S)^{-1/2},ââ
	Ç{d^D z*¼d^D z\over (2¹i)^D}¼e^{-zÿHz} = (det¼H)^{-1} $$
 by diagonalizing the matrices, making appropriate redefinitions of the
integration variables, and identifying the determinant of a diagonal
matrix.  Alternatively, we can use these integrals to ÓdefineÕ the
determinant, and derive the previous definition.  The relation for the
symmetric matrix follows from that for the Hermitian one by separating
$z$ into its real and imaginary parts for the special case $H=S$.  If we
treat $z$ and $z*$ as independent variables, the determinant can also be
understood as the Jacobian for the (dummy) variable change $z£H^{-1}z$,
$z*£z*$.  More generally, if we define the integral by an appropriate
limiting procedure or analytic continuation (for convergence), we can
choose $z$ and $z*$ to be unrelated (or even separate real variables), and
$S$ and $H$ to be complex.

\x IB3.3  Other properties of determinants can also be derived directly
from the integral definition:
ªa Find an integral expression for the inverse of a (complex) matrix
$M$ by using the identity
$$ 0 = Ç{»\over »z_I}(z_JÊe^{-zÿMz}) $$
ªb Derive the identity $¶¼ln¼det¼M=tr(M^{-1}¶M)$ by varying the
Gaussian definition of the (complex) determinant with respect to $M$.

An even better definition of the determinant is in terms of an
ÓanticommutingÕ integral (see subsection IA2), since anticommutativity
automatically gives the antisymmetry of the Levi-Civita tensor, and we
don't have to worry about convergence.  We then have, for ÓanyÕ matrix
$M$,
$$ Çd^D ½ÿ¼d^D ½¼e^{-½ÿM½} = det¼M $$
 where $½ÿ$ can be chosen as the Hermitian conjugate of $½$ or as an
independent variable, whichever is convenient.  From the definition of
anticommuting integration, the only terms in the Taylor expansion of the
exponential that contribute are those with the product of one of each
anticommuting variable.  Total antisymmetry in $½$ and in $½ÿ$ then yields
the determinant; we define ``$d^D ½ÿ¼d^D ½$" to give the correct
normalization.  (The normalization is ambiguous anyway because of the
signs in ordering the $d½$'s.)  This determinant can also be considered a
Jacobian, but the inverse of the commuting result follows from the fact
that the integrals are now really derivatives.

\x IB3.4  Divide up the range of a square matrix into two (not necessarily
equal) parts:  In block form,
$$ M = \pmatrix{A&B\cr C&D\cr} $$
 and do the same for the (commuting or anticommuting) variables used in
defining its determinant.  Show that
$$ det\pmatrix{A&B\cr C&D\cr} = 
	det¼D É det(A -BD^{-1}C) = det¼A É det(D -CA^{-1}B) $$
 ªa by integrating over one part of the variables first (this requires
off-diagonal changes of variables of the form $y£y+\O x$, which have
unit Jacobian), or
 ªb by first proving the identity
$$ \pmatrix{A&B\cr C&D\cr} = \pmatrix{I & BD^{-1} \cr 0&I\cr}
	\pmatrix{A -BD^{-1}C & 0\cr 0&D\cr}\pmatrix{I&0\cr D^{-1}C &I\cr} $$

We then have, for any ÓantiÕsymmetric (even-dimensional) matrix $A$,
$$ Çd^{2D}że^{-Å^T AÅ/2} = \hbox{\it Pf}¼A,ââ
	(\hbox{\it Pf}¼A)^2 = det¼A $$
 by the same method as the commuting case (again with appropriate
definition of the normalization of $d^{2D}Å$; the determinant of an
odd-dimensional antisymmetric matrix vanishes, since $det¼M=det¼M^T$). 
However, there is now an important difference:  The ``Pfaffian" is not
merely the square root of the determinant, but itself a polynomial, since
we can evaluate it also by Taylor expansion:
$$ \hbox{\it Pf}¼A_{IJ} = \f1{D!2^D}
	·^{I_1...I_{2D}}A_{I_1 I_2}òA_{I_{2D-1}I_{2D}} $$
 which can be used as an alternate definition.  (Normalization can be
checked by examining a special case; the overall sign is part of the
normalization convention.)

Ü4. Classical groups

The rotation group in three dimensions can be expressed most simply in
terms of 2$ð$2 matrices.  This description is the most convenient for not
only spin 1/2, but all spins.  This result can be extended to orthogonal
groups (such as the rotation, Lorentz, and conformal groups) in other low
dimensions, including all those relevant to spacetime symmetries in four
dimensions.

There are an infinite number of Lie groups.  Of the compact ones, all but a
finite number are among the ``classical" Lie groups.  These classical
groups can be defined easily in terms of (real or complex) matrices
satisfying a few simple constraints.  (The remaining ``exceptional"
compact groups can be defined in a similar way with a little extra effort,
but they are of rather specialized interest, so we won't cover them
here.)  These matrices are thus called the ``defining" representation of
the group.  (Sometimes this representation is also called the
``fundamental" representation; however, this term has been used in
slightly different ways in the literature, so we will avoid it.) These
constraints are a subset of:

\vskip-.1in
$$ \matrix{ \hbox{volume:} \vrule height0pt depth6pt width0pt
		\vphantom{(g}
	\hfill \hbox{{\bf S}pecial:}¼Ê& det(g) = 1 & \cr
	\hbox{metric:} \vrule height0pt depth26pt width0pt
			\leftÓ \matrix{
		\hbox{hermitian:}\vphantom{(gçÿ}
			\hfill \hbox{{\bf U}nitary:}ÊÊ\cr
		\hbox{(anti)symmetric:} \vrule height22pt depth0pt width0pt
			\leftÓ \matrix{ \vphantom{(g^T}\hbox{{\bf O}rthogonal:}\cr
				\vphantom{¯^T}\hbox{{\bf S}ym{\bf p}lectic:}\cr}
			\right. \cr} \right. &
		\matrix{ gÿçg = ç \cr \vrule height26pt depth0pt width0pt
			\matrix{ g^Túg = ú \cr g^T¯g = ¯ \cr} \cr} &
		\matrix{ (çÿ = ç) \cr \vrule height22pt depth0pt width0pt
			\matrix{ (ú^T = ú) \cr (¯^T = -¯) \cr} \cr} \cr
	\hbox{reality:} \hfill \leftÓ \hfill \matrix{ 
			\hfill\hbox{{\bf R}eal:}\cr \hbox{pseudoreal ($*$):} \cr}\right.
		¼Ê& \matrix{ g* = úgú^{-1} \cr g* = ¯g¯^{-1} \cr} & \cr} $$
\vskip.1in

\noindent where $g$ is any matrix in the defining representation of
the group, while $ç,ú,¯$ are group ``metrics", defining inner products
(while the determinant defines the volume, as in the Jacobian).  For the
compact cases $ç$ and $ú$ can be chosen to be the identity, but we will
also consider some noncompact cases.  (There are also some
uninteresting variations of ``Special" for complex matrices, setting the
determinant to be real or its magnitude to be 1.)

\x IB4.1  Write all the defining constraints of the classical groups (S, U, O,
Sp, R, pseudoreal) in terms of the algebra rather than the group.

Note the modified definition of unitarity, etc.  Such things are also
encountered in quantum mechanics with ghosts, since the resulting
Hilbert space can have an indefinite metric.  For example, if we have a
finite-dimensional Hilbert space where the inner product is represented in
terms of matrices as
$$ ÒÆ|Ô = Æÿç $$
 then ``observables" satisfy a ``pseudohermiticity" condition
$$ ÒÆ|HÔ = ÒHÆ|ÔâÜâçH = Hÿç $$
 and unitarity generalizes to
$$ ÒUÆ|UÔ = ÒÆ|ÔâÜâUÿçU = ç $$
 Similar remarks apply when replacing the Hilbert-space ``sesquilinear"
(vector times complex conjugate of vector) inner product with a
symmetric (orthogonal) or antisymmetric (symplectic) bilinear inner
product.  An important example is when the wave function carries a
Lorentz vector index, as expected for a relativistic description of spin 1;
then clearly the time component is unphysical.

The groups of matrices that can be constructed from these conditions are
then:

\vskip-.1in
$$ \vtop{\hbox{GL(n,C) [SL(n,C)]}}ââ
\vtop{
\halign{#\hfil\cr
U:  [S]U(n${}_+$,n${}_-$)\cr
O:  [S]O(n,C)\cr
Sp:  Sp(2n,C)\cr
R:  GL(n) [SL(n)]\cr
*:  [S]U*(2n)\cr}}ââ
\vtop{\vskip-8pt\offinterlineskip
\hrule
\halign{ &\vrule#&\hfil\strut¼#¼\hfil\cr
height2pt&\omit&\hskip2pt\vrule&\omit&&\omit&\cr
& U &\hskip2pt\vrule& R && * &\cr 
height2pt&\omit&\hskip2pt\vrule&\omit&&\omit&\cr
\noalign{\hrule}
height2pt&\omit&\hskip2pt\vrule&\omit&&\omit&\cr
\noalign{\hrule}
height2pt&\omit&\hskip2pt\vrule&\omit&&\omit&\cr
& O &\hskip2pt\vrule& [S]O(n${}_+$,n${}_-$) && SO*(2n) &\cr 
height2pt&\omit&\hskip2pt\vrule&\omit&&\omit&\cr
\noalign{\hrule}
height2pt&\omit&\hskip2pt\vrule&\omit&&\omit&\cr
& Sp &\hskip2pt\vrule& Sp(2n) && USp(2n${}_+$,2n${}_-$) &\cr
height2pt&\omit&\hskip2pt\vrule&\omit&&\omit&\cr
}\hrule} $$
\vskip.1in

\noindent Of the non-determinant constraints, in the first column we
applied none (``GL" means ``general linear", and ``C" refers to the
complex numbers; the real numbers ``R" are implicit); in the second
column we applied one; in the third column we applied three, since two of
the three types (unitarity, symmetry, reality) imply the third.  (The
corresponding groups with unit determinant, when distinct, are given in
brackets.)  These square matrices are of size n, n${}_+$+n${}_-$, 2n, or
2n${}_+$+2n${}_-$, as indicated.  n${}_+$ and n${}_-$ refer to the number
of positive and negative eigenvalues of the metric $ç$ or $ú$.  O(n)
differs from SO(n) by including ``parity"-type transformations, which
can't be obtained continuously from the identity.  (SSp(2n) is the same as
Sp(2n).) For this reason, and also for studying ``topological" properties,
for finite transformations it is sometimes more useful to work directly
with the group elements $g$, rather than parametrizing them in terms of
algebra elements as $g=e^{iG}$.  U(n) differs from SU(n) (and similarly for
GL(n) vs.¼SL(n)) only by including a U(1) group that commutes with the
SU(n):  Although U(1) is noncompact (it consists of just phase
transformations), a compact form of it can be used by requiring that all
``charges" are integers (i.e., all representations transform as
$Æ'=e^{iqÏ}Æ$ for group parameter $Ï$, where $q$ is an integer defining
the representation).

Of these groups, the compact ones are just SU(n), SO(n) (and O(n)), and
USp(2n) (all with n${}_-$=0).  The compact groups have an interesting
interpretation in terms of various number systems:  SO(n) is the unitary
group of n$ð$n matrices over the real numbers, SU(n) is the same for the
complex numbers, and USp(2n) is the same for the quaternions. 
(Similar interpretations can be made for some of the noncompact
groups.)  The remaining compact Lie groups that we didn't discuss, the
``exceptional" groups, can be interpreted as unitary groups over the
octonions.  (Unlike the classical groups, which form infinite series, there
are only five exceptional compact groups, because of the restrictions
following from the nonassociativity of octonions.)

Ü5. Tensor notation

Usually nonrelativistic physics is written in matrix or Gibbs' notation.  This is insufficient even for 19th century physics:  We can write a column or row vector $p$ for momentum, and a matrix $T$ for moment of inertia, but how do we write in that notation more general objects?  These are different representations of the rotation group:  We can write how each transforms under rotations:
$$ p' =pA,ââT' = A^T T A $$
The problem is to write ÓallÕ representations.

One alternative is used frequently in quantum mechanics:  A scalar is ``spin 0", a vector is ``spin 1", etc.  Spin $s$ has $2s+1$ components, so we can write a column ``vector" with that many components.  For example, moment of inertia is a symmetric 3$ð$3 matrix, and so has 6 components.  It can be separated into its trace $S$ and traceless pieces $R$, which don't mix under rotations:
$$ T = R +\f13 SI,âtr(T) = S,âtr(R) = 0 $$
$$ Üâtr(T') = tr(A^T T A) = tr(AA^T T) = tr(T)âÜâtr(R')=0,âS' = S $$
using the cyclicity of the trace.  Thus the ``irreducible" parts of $T$ are the scalar $S$ and the spin-2 (5 components) $R$.  But if we were to write $R$ as a 5-vector, it would be a mess to relate the 5$ð$5 matrix that rotates it to the 3$ð$3 matrix $A$, and even worse to write a scalar like $pRp^T$ in terms of 2 3-vectors and 1 5-vector.  (In quantum mechanics, this is done with ``Clebsch-Gordan-Wigner coefficients".)

The simplest solution is to use indices.  Then it's easy to write an object of arbitrary integer spin $s$ as a generalization of what we just did for spins 0,1,2:  It has $s$ 3-vector indices, in which it is totally (for any 2 of its indices) symmetric and traceless:
$$ T^{i_1...i_s}:ââT^{...i...j...} = T^{...j...i...},âT^{...i...j...}¶_{ij} = 0 $$
and it transforms as the product of vectors:
$$ T'^{i_1...i_s} = T^{j_1...j_s}A_{j_1}{}^{i_1}...A_{j_s}{}^{i_s} $$

Similar remarks apply to group theory in general:
Although historically group representations have usually been taught in
the notation where an m-component representation of a group defined by
n$ð$n matrices is represented by an m-component vector, carrying a
single index with values 1 to m, a much more convenient and transparent
method is ``tensor notation", where a general representation carries
many indices ranging from 1 to n, with certain symmetries (and perhaps
tracelessness) imposed on them.  (Tensor notation for a covering group is
generally known as ``spinor notation" for the corresponding orthogonal
group:  See subsection IC5.)  This notation takes advantage of the
property described above for expressing arbitrary representations in
terms of direct products of vectors.  In terms of transformation laws, it
means we need to know only the defining representation, since the
transformation of this representation is applied to each index.  

There are
at most four vector representations, by taking the dual and complex
conjugate; we use the corresponding index notation.  Then the group
constraints simply state the invariance of the group metrics (and their
complex conjugates and inverses), which thus can be used to raise, lower,
and contract indices:

\vskip-.1in
$$ \matrix{ \hbox{volume:} \vrule height0pt depth8pt width0pt
		\vphantom{·^{I_1...I_n}}
	\hfill \hbox{{\bf S}pecial:}¼Ê& ·^{I_1...I_n} \cr
	\hbox{metric:} \vrule height0pt depth26pt width0pt
			\leftÓ \matrix{
		\hbox{hermitian:}\vphantom{ç^{ÀIJ}}
			\hfill \hbox{{\bf U}nitary:}ÊÊ\cr
		\hbox{(anti)symmetric:} \vrule height20pt depth0pt width0pt
			\leftÓ \matrix{\vphantom{ú^{IJ}}\hbox{{\bf O}rthogonal:}\cr
				\vphantom{¯^{IJ}}\hbox{{\bf S}ym{\bf p}lectic:}\cr}
			\right. \cr} \right. &
		\matrix{ ç^{ÀIJ} \cr \vrule height20pt depth0pt width0pt
			\matrix{ ú^{IJ} \cr ¯^{IJ} \cr} \cr} \cr
	\hbox{reality:} \hfill \leftÓ \hfill \matrix{ \hfill\hbox{{\bf R}eal:}
			\vphantom{ú_{ÀI}{}^J}\cr 
			\hbox{pseudoreal ($*$):} \cr}\right.
		¼Ê& \matrix{ ú_{ÀI}{}^J \cr ¯_{ÀI}{}^J \cr} \cr} $$
\vskip.1in

\noindent
 As a result, we have relations such as
$$ Ò{}^I|{}^JÔ = ú^{IJ}âorâ¯^{IJ},ââÒ{}^{ÀI}|{}^JÔ = ç^{ÀIJ} $$
 We also define inverse metrics satisfying
$$ ú^{KI}ú_{KJ} = ¯^{KI}¯_{KJ} = ç^{ÀKI}ç_{ÀKJ} = ¶_J^I $$
 (and similarly for contracting the second index of each pair).  Therefore,
with unitarity/(pseudo)reality we can ignore complex conjugate
representations (and dotted indices), converting them into unconjugated
ones with the metric, while for orthogonality/symplecticity we can do the
same with respect to raising/lowering indices:
$$ \matrix{ \hfill\hbox{{\bf U}nitary:} & Æ^{ÀI} = ç^{ÀIJ}Æ_J \cr
	\hfill\hbox{{\bf O}rthogonal:} & Æ^{I} = ú^{IJ}Æ_J \cr
	\hfill\hbox{{\bf S}ym{\bf p}lectic:} & Æ^{I} = ¯^{IJ}Æ_J \cr
	\hfill\hbox{{\bf R}eal:} & Æ_{ÀI} = ú_{ÀI}{}^J Æ_J \cr
	\hfill\hbox{pseudoreal ($*$):} & Æ_{ÀI} = ¯_{ÀI}{}^J Æ_J \cr} $$

\noindent For the real groups there is also the constraint of reality on the
defining representation:
$$ ÐÆ_{ÀI} ­ (Æ_I)* = Æ_{ÀI} ­ ú_{ÀI}{}^J Æ_J $$

\x IB5.1  As an example of the advantages of index notation, show that
SSp is the same as Sp.  (Hint:  Write one $·$ in the definition of the
determinant in terms of $¯$'s by total antisymmetrization, which then can
be dropped because it is enforced by the other $·$.  One can ignore
normalization by just showing $det¼M = det¼I$.)

For SO(n${}_+$,n${}_-$), there is a slight modification of a sign convention: 
Since then indices can be raised and lowered with the metric, $·^{I...}$ is
usually defined to be the result of raising indices on $·_{I...}$, which
means
$$ ·_{12...n} = 1âÜâ·^{12...n} = det¼ú = (-1)^{n_-} $$
 Then $·^{I...}$ should be replaced with $(-1)^{n_-}·^{I...}$ in the
equations of subsection IB3:  For example,
$$ ·_{J_1...J_n}·^{I_1...I_n} = (-1)^{n_-}¶_{[J_1}^{I_1}ò¶_{J_n]}^{I_n} $$

We now give the simplest explicit forms for the defining representations
of the classical groups.  The most convenient notation is to label the
generators by a pair of fundamental indices, since the adjoint
representation is obtained from the direct product of the fundamental
representation and its dual (i.e., as a matrix labeled by row and column). 
The simplest example is GL(n), since the generators are arbitrary
matrices.  We therefore choose as a basis matrices with a 1 as one entry
and 0's everywhere else, and label that generator by the row and column
where the 1 appears.  Explicitly,
$$ GL(n):â(G_I{}^J)_K{}^L = ¶_I^L ¶_K^JâÜâG_I{}^J = |{}^JÔÒ{}_I| $$
 This basis applies for GL(n,C) as well, the only difference being that
the coefficients $Œ$ in $G=Œ_I{}^J G_J{}^I$ are complex instead of
real.  The next simplest case is U(n):  We can again use this basis,
although the matrices $G_I{}^J$ are not all hermitian, by requiring that
$Œ_I{}^J$ be a hermitian matrix.  This turns out to be more convenient in
practice than using a hermitian basis for the generators.  A well known
example is SU(2), where the two generators with the 1 as an off-diagonal
element (and 0's elsewhere) are known as the ``raising and lowering
operators" $J_à$, and are more convenient than their hermitian parts for
purposes of contructing representations.  (This generalizes to other
unitary groups, where all the generators on one side of the diagonal are
raising, all those on the other side are lowering, and those along the
diagonal give the maximal Abelian subalgebra, or ``Cartan subalgebra".)

Representations for the other classical groups follow from applying their
definitions to the GL(n) basis.  We thus find
$$ \li{ SL(n):â(G_I{}^J)_K{}^L = ¶_I^L ¶_K^J -\f1n ¶_I^J ¶_K^Lâ
		&ÜâG_I{}^J = |{}^JÔÒ{}_I| -\f1n ¶_I^J|{}^KÔÒ{}_K| \cr
	SO(n):â(G_{IJ})^{KL} = ¶_{[I}^K ¶_{J]}^Lâ
		&ÜâG_{IJ} = |{}_{[I}ÔÒ{}_{J]}| \cr
	Sp(n):â(G_{IJ})^{KL} = ¶_{(I}^K ¶_{J)}^Lâ
		&ÜâG_{IJ} = |{}_{(I}ÔÒ{}_{J)}| \cr} $$
 As before, SL(n,C) and SU(n) use the same basis as SL(n), etc.  For SO(n)
and Sp(n) we have raised and lowered indices with the appropriate metric
(so SO(n) includes SO(n${}_+$,n${}_-$)).  For some purposes (especially for
SL(n)), it's more convenient to impose tracelessness or (anti)symmetry on
the matrix $Œ$, and use the simpler GL(n) basis.

\x IB5.2  Our normalization for the generators of the classical groups
is the simplest, and independent of n (except for subtracting out traces):
 ªa  Find the commutation relations of the generators (structure
constants) for the defining representation of GL(n) as given in the text. 
Note that the values of all the structure constants are 0, $ài$.  Show that
$$ c_D = 1 $$
 (see subsection IB2).
 ªb  Consider the GL(m) subgroup of GL(n) (m<n) found by restricting the
range of the index of the above defining representation.  Show the
structure constants are the same as those given by starting with the
above representation of GL(m).
 ªc  Find the structure constants for SO(n) and Sp(n).
 ªd  Directly evaluate $k_D c_A¼(=¶^{ij}G_i G_j)$ for SL(n), SO(n), and Sp(n),
and compare with $c_D d_A/d_D$.

\x IB5.3  A tensor that pops up in various contexts is
$$ d_{ijk} = tr(G_iÓG_j,G_kÕ) $$
 It takes a very simple form in terms of defining indices:
 ªa  Show that for SU(n) this tensor is determined to be, up to an overall
normalization (that depends on the representation),
$$ tr\left( G_{I_1}{}^{J_1}\leftÓ G_{I_2}{}^{J_2},G_{I_3}{}^{J_3} \rightÕ
	\right) ¾
	[ (231) +(312) ] -{2\over n}[ (132) +(213) +(321) ] +{4\over n^2}(123) $$
$$ (abc) ­ ¶_{I_a}^{J_1}¶_{I_b}^{J_2}¶_{I_c}^{J_3} $$
 (where $a,b,c$ are some permutation of $1,2,3$)
from just the total symmetry of $d_{ijk}$ (and $G_I{}^I=0$), since the
only invariant tensor available is $¶_I^J$.  (If $·_{IJ...}$ were used,
$·^{IJ...}$ would also be required, to balance the number of subscripts
and superscripts; but their product can be expressed in terms of just
$¶$'s also.)
 ªb  Check this result by using the explicit $G$'s for the defining
representation, and determine the proportionality constant for that
representation.

With the exception of the ``spinor" representations of SO(n) (to be
discussed in subsection IC5, section IIA, and subsection XC1), general
representations can be obtained by reducing direct products of the
defining representations.  This means they can be described by objects
with multiple indices (up/down, dotted/undotted), where each index is
that of a defining representation, and satisfying various (anti)symmetry
and tracelessness conditions on the indices.

\x IB5.4  Consider the representations of SU(n) obtained from the
symmetric and antisymmetric part of the direct product of two
defining representations.  For simplicity, one can work with the U(n)
generators, since the U(1) pieces will appear in a simple way.
 ªa  Using tensor notation for the generators $(G_I{}^J)_{KL}{}^{MN}$, find
their explicit representation for these two representations.
 ªb  By evaluating the trace, show that the Dynkin index for the two
cases is
$$ c_a = n-2,ââc_s = n+2  $$
 ªc  Show the sum of these two is consistent with the argument at the
end of subsection IB2.  Show each case is consistent with n=2, and the
antisymmetric case with n=3, by relating those cases to the singlet,
defining, and adjoint representations.

\refs

£1 H. Georgi, ÓLie algebras in particle physics: from isospin to unified theoriesÕ, 2nd ed.\\
	(Perseus, 1999):\\
	best book on Lie groups; unlike other texts, covers not only more
	powerful Cartan-Weyl methods, but also more useful tensor methods;
	also has useful applications to nonrelativistic quark model.
 £2 S. Helgason, ÓDifferential geometry and symmetric spacesÕ
	(Academic, 1962);\\
	R. Gilmore, ÓLie groups, Lie algebras, and some of their applicationsÕ
	(Wiley, 1974):\\
	noncompact classical Lie groups.

\unrefs

Û5 C. REPRESENTATIONS

We now consider some of the more useful representations, as explicit
examples of the results of the previous section.  In particular, we
consider symmetries of the quark model.

Ü1. More coordinates

We began our ``review" of group theory by looking at how symmetries
were represented on coordinates.  We now return to coordinates as a
special case (particular representation) of the general results of the
previous section.  The idea is that the coordinates themselves are already
a representation of the group, and the wave functions are functions of
these coordinates.  For example, for ordinary rotations we use wave
functions that depend on position or momentum, which transforms as a
vector.  (This is not always the case:  For example, in our description of
the conformal group the usual space and time coordinates transformed
nonlinearly, and not just by multiplication by constant matrices unless
the extra two coordinates were introduced.)  This is the basic distinction
between classical mechanics and classical field theory:  Mechanics uses
the coordinates themselves as the basic variables, while field theory uses
functions of the coordinates.  (Similarly, in quantum mechanics the wave
functions are functions of the coordinates, while in quantum field theory
the wave functions are ``functionals" of functions of the coordinates.)  

In general, the construction of such a ``coordinate representation" starts
with a given matrix representation (usually finite dimensional)
$(G_i)_I{}^J$ and then defines a new representation
$$ öG_i = q^I (G_i)_I{}^J p_J;ââ[p_I,q^JÕ = ¶_I^J,â[q,qÕ=[p,pÕ=0 $$
 for some objects $q$ and $p$, which are interpreted as either coordinates
and their conjugate momenta (up to a factor of $i$), or as creation and
annihilation operators:  The latter nomenclature is used when the
boundary conditions allow the existence of a state $|0Ô$ called the
``vacuum", satisfying $p|0Ô=0$, so we can define the other states as
functions of $q$ acting on $|0Ô$.  (If the coordinates are fermionic, the
distinction is moot, since by the usual Taylor expansion the Hilbert space
is finite dimensional.  See exercise IA2.3.)  It is easy to check that $öG_i$
satisfy the same commutation relations as $G_i$.  In particular, if the
matrices are in the adjoint representation, $q^i$ can be interpreted as
the group coordinates themselves:  This follows from considering the
action of an infinitesimal transformation on the group element
$g(q)=e^{iq^iG_i}$ (or just the Lie algebra element $G(q)=q^i G_i$).  

If we write these results in bra/ket notation, since
$$ öG_i q^I = q^J (G_i)_J{}^I,âöG_i p_I = -(G_i)_I{}^J p_J $$
 it is more natural to look at the action on bras:
$$ Òq| = q^IÒ{}_I|,â|pÔ = |{}^IÔp_IâÜâöG_iÒq| = Òq|G_i,âöG_i|pÔ = -G_i|pÔ $$
 Note that this vector space is coordinate space itself, not the space of
functions of the coordinates; it is the same space on which $G_i$ is
defined.  (Of course, $öG_i$ is defined on arbitrary functions of the
coordinates; it has a reducible representation bigger than $(G_i)_I{}^J$. 
Effectively,
$(G_i)_I{}^J$ is represented on the space of functions linear in the
coordinates.)  Then, for example
$$ öG_1 öG_2 Òq| = öG_1 Òq| G_2 = Òq| G_1 G_2 $$
 is obviously equivalent, while (ignoring any extra signs for fermions)
$$ öG_1 öG_2 |pÔ = -öG_1 G_2 |pÔ = -G_2 öG_1 |pÔ = G_2 G_1 |pÔ $$
 at least gives an equivalent result for the commutator algebra
$[öG_1,öG_2]$.  This is the expected result for the dual representation
$G_i£-G_i^T$.

Interesting examples are given by using the defining representation for
$G$.  For example, the commonly used oscillator representation for U(n) is
$$ U(n):âöG_I{}^J = aÿ^J a_I,â[a_I,aÿ^JÕ = ¶_I^J $$
 where the oscillators can be bosonic or fermionic.  For the SO and Sp
cases, because we can raise and lower indices, and because of the
(anti)symmetry on the indices, the interesting possibility arises to
identify the coordinates with their momenta, with the statistics
appropriate to the symmetry:
$$ Sp(n):âöG_{IJ} = üz_{(I}z_{J)},â[z_I,z_J] = ¯_{IJ} $$
$$ SO(n):âöG_{IJ} = ü©_{[I}©_{J]},âÓ©_I,©_JÕ = ú_{IJ} $$
 For SO(n) the representation is finite dimensional because of the
Fermi-Dirac statistics, and is called a ``Dirac spinor" (and $©$ the ``Dirac
matrices").  If the opposite statistics are chosen, the coordinates and
momenta can't be identified:  For example, bosonic coordinates for SO(n)
give the usual spatial rotation generators $öG_{IJ}=x_{[I}»_{J]}$.

\x IC1.1  Use this bosonic oscillator representation for U(2)=SU(2)$°$U(1),
and use the SU(2) subgroup to describe spin.  
 ªa Show that the spin $s$ (the
integer or half-integer number that defines the representation) itself has
a very simple expression in terms of the U(1) generator.  Show this holds
in the quantum mechanical case (by interpreting the bracket as the
quantum commutator), giving the usual $s(s+1)$ for the sum of the
squares of the generators (with appropriate normalization).  
 ªb Use this
result to show that these oscillators, acting on the vacuum state, can be
used to construct the usual states of arbitrary spin $s$.

\x IC1.2  Considering SO(2n), divide up $©_I$ into pairs of canonical (and
complex) conjugates $a_1=(©_1+i©_2)/å2$, etc., so $Óa,aÿÕ=1$.  
 ªa Write the
SO(2n) generators in terms of $aa$, $aÿaÿ$, and $aÿa$.  Show that the
$aÿa$'s by themselves generate a U(n) subgroup.  
 ªb Decompose the Dirac
spinor into U(n) representations.  Show that the product of all the $©$'s is
related to the U(1) generator, and commutes with all the SO(2n)
generators.  Show that the states created by even or odd numbers of
$aÿ$'s on the vacuum don't mix with each other under SO(2n), so the Dirac
spinor is reducible into two ``Weyl spinors".

Ü2. Coordinate tensors

We have just seen how groups can be represented on coordinates.  
Depending on the choice
of coordinates, the coordinates may transform nonlinearly (i.e., as a
realization, not a representation), as for the D-dimensional conformal
group in terms of D (not D+2) coordinates.  However, given the nonlinear
transformation of the coordinates, there are always representations
other than the defining one (scalar field) that we can immediately write
down (such as the adjoint).  We now consider such representations: 
These are useful not only for the spacetime symmetries we have already
considered, but also for general relativity, where the symmetry group
consists of arbitrary coordinate transformations.  Furthermore, these
considerations are useful for describing coordinate transformations that
are not symmetries, such as the change from Cartesian to polar
coordinates in nonrelativistic theories.

When applied to quantum mechanics, we write the action of a symmetry
on a state as $¶Æ=iGÆ$ (or $Æ'=e^{iG}Æ$), but on an operator as
$¶A=i[G,A]$ (or $A'=e^{iG}A e^{-iG}$).  In classical mechanics, we always
write $¶A=i[G,A]$ (since classical objects are identified with quantum
operators, not states).  However, if $G=Â^m »_m$ is a coordinate
transformation (e.g., a rotation) and $Ä$ is a scalar field, then in quantum
notation we can write 
$$ ¶Ä(x) = [G,Ä] = GÄ = Â^m »_m Äââ(Ä' = e^GÄe^{-G} = e^GÄ) $$
 since the derivatives in $G$ just differentiate $Ä$.  (For this discussion of
coordinate transformations we switch to absorbing the $i$'s into the
generators.)  The coordinate transformation $G$ has the usual properties
of a derivative:
$$ [G,f(x)] = GfâÜâGf_1 f_2 = [G,f_1 f_2] = (Gf_1)f_2 +f_1 Gf_2 $$
$$ e^G f_1 f_2 = e^G f_1 f_2 e^{-G} = (e^G f_1 e^{-G})(e^G f_2 e^{-G}) =
	(e^G f_1)(e^G f_2) $$
 and similarly for products of more functions.

The adjoint representation of coordinate transformations is a ``vector
field" (in the sense of a spatial vector), a function that has general
dependence on the coordinates (like a scalar field) but is also linear in the
momenta (as are the Poincar«e generators):
$$ G = Â^m(x)»_m,âV = V^m(x)»_mâÜâ
	¶V = [G,V] = (Â^m »_m V^n - V^m »_m Â^n)»_n $$
	$$ Üâ¶V^m = Â^n »_n V^m - V^n »_n Â^m $$
 The same result follows if we use the Poisson bracket instead of the
quantum mechanical commutator, replacing $»_m$ with $ip_m$ in both
$G$ and $V$.  

Finite transformations can also be expressed in terms of transformed
coordinates themselves, instead of the transformation parameter:
$$ Ä(x) = e^{-Â^m »_m}Ä'(x) = Ä'(e^{-Â^m »_m}x) $$
 as seen, for example, from a Taylor expansion of $Ä'$, using
$e^{-G}Ä'=e^{-G}Ä'e^G$.  We then define
$$ Ä'(x') = Ä(x)âÜâx' = e^{-Â^m »_m}x $$
 This is essentially the statement that the active and passive
transformations cancel.  However, in general this method of defining
coordinate transformations is not convenient for applications:  When we
make a coordinate transformation, we want to know $Ä'(x)$.  Working
with the ``inverse" transformation on the coordinates, i.e., our original
$e^{+G}$,
$$ ÷x ­ e^{+Â^m »_m}xâÜâÄ'(x) = e^G Ä(x) = Ä(÷x(x)) $$
 So, for finite transformations, we work directly in terms of $÷x(x)$, and
simply plug this into $Ä$ in place of $x$ ($x£÷x(x)$) to find $Ä'$ as a
function of $x$.

Similar remarks apply for the vector, and for derivatives in general.  We
then use
$$ x' = e^{-G}xâÜâ»' = e^{-G}»e^G $$
 where $»'=»/»x'$, since $»'x'=»x=¶$.  This tells us
$$ V^m(x)»_m = e^{-G}V'^m(x)»_m e^G = V'^m(x')»'_m $$
 or $V'(x')=V(x)$.  Acting with both sides on $x'^m$,
$$ V'^m(x') = V^n(x){»x'^m\over »x^n} $$
 On the other hand, working in terms of $÷x$ is again more convenient: 
Changing the transformation for the vector operator in the same way
as the scalar
$$ V'(x) = V(÷x) $$
$$ ÜâV'^m(x)»_m = V^m(÷x)÷»_m $$
$$ ÜâV'^m(x) = V^n(÷x){»x^m\over »÷x^n} $$
where  $÷»=»/»÷x$ and as usual
$$ (÷»_m x^n)(»_n ÷x^p) = ¶_m^pâÜâ{»x^n\over »÷x^m} = 
	\left.\left[\left({»÷x(x)\over »x}\right)^{-1}\right]_m\right.^n $$
We can also use
$$ V'(x) = e^G V(x) e^{-G} $$
$$ ÜâV'^m(x)»_m = (e^G V^m(x) e^{-G})(e^G »_m e^{-G}) = V^m (÷x)÷»_m $$

A ``differential form" is defined as an infinitesimal $W=dx^m W_m(x)$. 
Its transformation law under coordinate transformations, like that of
scalar and vector fields, is defined by $W'(x')=W(x)$.  For any vector field
$V=V^m(x)»_m$, $V^m W_m$ transforms as a scalar, as follows from the
``chain rule" $d=dx'^m»'_m=dx^m »_m$.  Explicitly,
$$ W'_m(x') = W_n(x){»x^n\over »x'^m} $$
 or in infinitesimal form
$$ ¶W_m = Â^n »_n W_m +W_n »_m Â^n $$
 Thus a differential form is dual to a vector, at least as far as the matrix
part of coordinate transformations is concerned.  They transform the
same way under rotations, because rotations are orthogonal; however,
more generally they transform differently, and in the absence of a metric
there is not even a way to relate the two by raising or lowering indices.

Higher-rank differential forms can be defined by antisymmetric products
of the above ``one-forms".  These are useful for integration:  Just as the
line integral $ÇW=Çdx^m W_m$ is invariant under coordinate
transformations by definition (as long as we choose the curve along which
the integral is performed in a coordinate-independent way), so is a
totally antisymmetric $N$th-rank tensor (``$N$-form") $W_{m_1òm_N}$
integrated on an $N$-dimensional subspace as
$$ Çdx^{m_1}òdx^{m_N}W_{m_1òm_N} :ââ
	W'_{m_1òm_N}(x') = W_{p_1òp_N}(x)
		{»x^{p_1}\over »x'^{m_1}}ò{»x^{p_N}\over »x'^{m_N}} $$
 where the surface element $dx^{m_1}òdx^{m_N}$ is interpreted as
antisymmetric.  (The signs come from switching initial and final limits of
integration, as prescribed by the ``orientation" of the hypersurface.)  This
is clear if we rewrite the integral more explicitly in terms of coordinates
$§^i$ for the subspace:  Then
$$ Çdx^{m_1}òdx^{m_N}W_{m_1òm_N}(x) =
	Çd§^{i_1}òd§^{i_N}ßW_{i_1òi_N}(§) =
	Çd^N §¼·^{i_1òi_N}ßW_{i_1òi_N}(§) $$
 where
$$ ßW_{i_1òi_N}(§) = {»x^{m_1}\over »§^{i_1}}ò{»x^{m_N}\over »§^{i_N}}
		W_{m_1òm_N}(x) $$
 is the result of a coordinate transformation that converts $N$ of the
$x$'s to $§$'s, an interpretation of the functions $x(§)$ that define the
surface.  Then any coordinate transformation on $x£x'$ (not on $§$) will
leave $ßW(§)$ invariant.  In particular, if the subspace is the full space, so
we can look directly at $Çd^N x¼·^{m_1òm_N}W_{m_1òm_N}$, we see
that a coordinate transformation generates from $W$ an $N$-dimensional
determinant exactly canceling the Jacobian resulting from changing the
integration measure $d^N x$.

\x IC2.1  For all of the following, use the exponential form of the finite
coordinate transformation:  
 ªa Show that any (local) function of a scalar
field (without explicit $x$ dependence additional to that in the field) is
also a scalar field (i.e., satisfies the same coordinate transformation law). 
 ªb Show that the transformation law of a vector field or differential form
remains the same when multiplied by a scalar field (at the same $x$). 
 ªc Show that $VÄ=V^m »_m Ä$ is a scalar field for any scalar field $Ä$ and
vector field $V$.  
 ªd Show that $[V,W]$ is a vector field for any vector fields
$V$ and $W$.

\x IC2.2  Examine finite coordinate transformations for integrals of
differential forms in terms of $÷x$ rather than $x'$.  Find the explicit
expression for $W'(x)$ in terms of $W(÷x(x))$, etc., and use  this to show
invariance:
$$ Çdx^{m_1}òdx^{m_N}W_{m_1òm_N}'(x) =
	Çd÷x^{m_1}òd÷x^{m_N}W_{m_1òm_N}(÷x) $$
$$ = Çdx^{m_1}òdx^{m_N}W_{m_1òm_N}(x) $$
 where in the last step we have simply substituted $÷x£x$ as a change of
integration variables.  Note that, using the $÷x$ form of the
transformation rather than $x'$, the transformation generates the
needed Jacobian, rather than canceling one.

From the above transformation law, we see that the curl of a differential
form is also a differential form:
$$ »'_{[m_1}W'_{m_2òm_N]}(x') = 
	»'_{[m_1}(»'_{m_2}x^{p_2})ò(»'_{m_N]}x^{p_N})W_{p_2òp_N}(x) $$
$$ = [»_{[p_1}W_{p_2òp_N]}(x)](»'_{m_1}x^{p_1})ò(»'_{m_N}x^{p_N}) $$
 because the curl kills $»'»'x$ terms that would appear if there were no
antisymmetrization.  Objects that transform ``covariantly" under
coordinate transformations, without such higher derivatives of $x$ (or $Â$
in the other notation), like scalars, vectors, differential forms and their
products, are called (coordinate) ``tensors".  Getting derivatives of
tensors to come out covariant in general requires special fields, and will
be discussed in chapter IX.  An important application of the covariance of
the curl of differential forms is the generalized Stokes' theorem (which
includes the usual Stokes' theorem and Gauss' law as special cases):
$$ Çdx^{m_1}òdx^{m_{N+1}}\f1{(N+1)!}»_{[m_1}W_{m_2òm_{N+1}]} =
	Èdx^{m_1}òdx^{m_N}W_{m_1òm_N} $$
 where the second integral is over the boundary of the space over which
the first is integrated.  (We use the symbol ``$ÈÊ$" to refer to boundary
integrals, including those over contours, which are closed boundaries of 2D
surfaces.)  It is basically just the fundamental theorem of (integral)
calculus ($Ç_a^b dx¼f'(x)=f(b)-f(a)$), as is clear from choosing a 
coordinate system where the boundary is at a fixed value of just one
coordinate (at least in patches).  (A standard example is a pair of
infinite constant-time surfaces, neglecting the boundaries that connect 
them at spatial infinity.)

Ü3. Young tableaux

We now return to our discussion of finite-dimensional representations. 
In the previous section we gave the machinery for describing them using
index notation, but examined only the defining representation in detail. 
Now we analyze general irreducible representations.

All the irreducible finite-dimensional representations of the groups SU(N)
can be described by tensors with lower N-valued indices with various
(anti)symmetrizations.  (An upper index can be replaced with N$-$1 lower
indices by using the Levi-Civita tensor.)  Although detailed calculations
require explicit use of these indices, three properties can be more
conveniently discussed pictorially: 
\item{(1)} the (anti)symmetries of the
indices, 
\item{(2)} the dimension (number of independent components) of the
representation, and 
\item{(3)} the reduction of the direct product of two
representations (which irreducible representations result, and how many
of each).

A ``Young tableau" is a picture representing an irreducible representation
in terms of boxes arranged in a regular grid into rows and columns, such
that the columns are aligned at the top, and their depths are
nonincreasing to the right: for example,
$$ \upõ4{\õ4\õ3\õ2\õ2\õ1} $$
 Each box represents an index, with antisymmetry among indices in any
column, and symmetry among indices in any row.  More precisely, since
one can't simultaneously have these symmetries and antisymmetries, it
corresponds to the result of taking any arbitrary tensor with that many
indices, first symmetrizing the indices in each row, and then
antisymmetrizing the indices in each column (or vice versa; symmetrizing
and then antisymmetrizing and then symmetrizing again gives the same
result as skipping the first symmetrization, etc.).  This gives a simple way
to classify and symbolize each representation.  (We can denote the
singlet representation, which has no boxes, by a dot.)  Note that the
deepest column should have no more than N$-$1 boxes for SU(N) because
of the antisymmetry.

To calculate the dimension of the representation for a given tableau, we
use the ``factors over hooks" rule:  
\item{(1)} Write an ``N" in the box
in the upper-left corner, and fill the rest of the boxes with numbers that
decrease by 1 for each step down and increase by 1 for each step to the
right.  
\item{(2)} Draw (or picture in your mind) a ``hook" for each box --- a
``$ýÊ$" with its corner in the box and lines extending right and down out
of the tableau.  
\item{(3)} The dimension is then given by the formula

$$ \hbox{dimension} = Þ_{\hbox{each box}}{\hbox{integer written there}
	\over\hbox{\# boxes intersected by its hook}} $$
 For the previous example, we find (listing boxes first down and then to
the right)
$$ {N\over 8}É{N-1\over 6}É{N-2\over 3}É{N-3\over 1}É
	{N+1\over 6}É{N\over 4}É{N-1\over 1}É{N+2\over 4}É{N+1\over 2}É
	{N+3\over 3}É{N+2\over 1}É{N+4\over 1} $$

The direct product of two Young tableaux A$°$B is analyzed by the
following rules:  First, label all the boxes in B by putting an ``a" in each
box in the top row, ``b" in the second row, etc.  Then, take the following
steps in all possible ways to find the Young tableaux resulting from the
direct product:  
\item{(1)} Add all the ``a" boxes from B to the right side and
bottom of A, then ``b" to the right and bottom of that, etc., to make a new
Young tableaux.  Any two tableaux constructed in this way with the same
arrangement of boxes but different assignment of letters are considered
distinct, i.e., multiple occurences of the same representation in the direct
product.  
\item{(2)} No more than 1 ``a" can be in any column, and similarly for
the other letters.  
\item{(3)} Reading from right to left, and then from top to
bottom (i.e., like Hebrew/Arabic), the number of a's read should always be
$³$ the number of b's, b's $³$ c's, etc.  

\noindent For example,
 \def\inny#1{õ \hbox to 0pt{\kern-.6em\raise-.6ex\hbox{$^{#1}$}}}
 \def\dumbõ#1{\mkern-3.5mu \inny{#1}}
 \def\dumberõ#1{\llap{\raise-1.5ex\hbox{$\inny{#1}$}}}
 \def\dumbestõ#1{\llap{\raise-3ex\hbox{$\inny{#1}$}}}
 {\hbadness=10000
$$ \upõ1{\õ1} ° \upõ2{\dumbõa\dumberõb\dumbõa}
	= \upõ2{\dumbõ{}\dumberõb\dumbõa\dumbõa} ¢
	\upõ2{\dumbõ{}\dumberõa\dumbõa\dumberõb} ¢ 
	\upõ3{\dumbõ{}\dumberõa\dumbestõb\dumbõa} $$
 }
 Note that A$°$B always gives the same result as B$°$A, but one way
may be simpler than the other.  For a given value of N, a column of N
boxes is equivalent to none (again by antisymmetry), while more than N
boxes in a column gives a vanishing tableau.

\x IC3.1  Calculate
$$ \upõ2{\õ2\õ1} ° \upõ2{\õ2\õ1} $$
 Check the result by finding the dimensions of all the representations and
adding them up.

These SU(N) tableaux also apply to SL(N):  Only the reality properties are
different.  Similar methods can be applied to USp(2N) (or Sp(2N)), but
tracelessness (with respect to the symplectic metric) must be imposed in
antisymmetrized indices, so these trace pieces must be separated out
when considering the above rules.  (I.e., consider USp(2N)$¤$SU(2N).) 
Similar remarks apply to SO(N), which has a symmetric metric, but there
are also ``spinor" representations (see below).  The additional irreducible
representations then can be constructed from taking direct products of
the above with the smallest spinors, and removing the ``gamma-matrix"
traces.  Furthermore, using the Levi-Civita tensor, 
all columns can be reduced to no more than N/2 in height.

Ü4. Color and flavor

We now consider the application of these methods to ``internal
symmetries" (those that don't act on the coordinates) in particle physics. 
The symmetries with experimental confirmation involve only the unitary
groups (U and SU) of small dimension.  However, we will find later that
larger unitary groups can be useful for approximation schemes.  (Also,
larger unitary and other groups continue to be investigated for
unification and other purposes, which we consider in later chapters.)

The ``Standard Model" describes all of particle physics that is well
confirmed experimentally (except gravity, which is not understood at the
quantum level).  It includes as its ``fundamental" particles:  
\item{(1)} the
spin-1/2 quarks that make up the observed strongly interacting particles,
but do not exist as asymptotic states, 
\item{(2)} the weakly interacting spin-1/2
leptons, 
\item{(3)} the spin-1 gluons that bind the quarks together, which couple
to the charges associated with SU(3) ``color" symmetry, but also are not
asymptotic,  
\item{(4)} the spin-1 particles that mediate the weak and
electromagnetic interactions, which couple to SU(2)$°$U(1) ``flavor", and
 \item{(5)} the yet unobserved spin-0 Higgs particles that are responsible for all
the masses of these weakly interacting particles.  

\noindent (However, quarks and gluons are temporarily free at high energy,
eventually recombining to give rise to ``jets", clusters of resulting hadrons.)
These particles, along
with their masses (in GeV) and (electromagnetic) charges ($Q=ÐQ+ëQ$), are:

$$ \vbox{\offinterlineskip
\hrule
\halign{ &\vrule#&\strut¼#\hfilÊ\cr
height2pt&\omit&\omit&\omit&\omit&\omit&\cr
&& \omit & \hfil $s=ü$ & \omit &&\cr 
height2pt&\omit&\omit&\omit&\omit&\omit&\cr
\noalign{\hrule}
height2pt&\omit&\hskip2pt\vrule&\omit&&\omit&\cr
\noalign{\hrule}
height2pt&\omit&\hskip2pt\vrule&\omit&&\omit&\cr
&\hfill color: $£$ &\hskip2pt\vrule& \hfil quark (3) && \hfil lepton (1)&\cr 
& flavor ($ëQ$) &\hskip2pt\vrule& \hfil ($ÐQ=\f16$) && \hfil ($ÐQ=-ü$) &\cr  
height2pt&\omit&\hskip2pt\vrule&\omit&&\omit&\cr
\noalign{\hrule}
height2pt&\omit&\hskip2pt\vrule&\omit&&\omit&\cr
\noalign{\hrule}
height2pt&\omit&\hskip2pt\vrule&\omit&&\omit&\cr
& \hfil $-ü$ &\hskip2pt\vrule& $d$ (.006) && $e$ (.00051099892) &\cr
& \hfil $+ü$ &\hskip2pt\vrule& $u$ (.003) && $Ã_e$ ($<3É10^{-9}$) &\cr
height2pt&\omit&\hskip2pt\vrule&\omit&&\omit&\cr
\noalign{\hrule}
height2pt&\omit&\hskip2pt\vrule&\omit&&\omit&\cr
& \hfil $-ü$ &\hskip2pt\vrule& $s$ (.10) && $µ$ (.105658369) &\cr
& \hfil $+ü$ &\hskip2pt\vrule& $c$ (1.2) && $Ã_µ$ (<.00019) &\cr
height2pt&\omit&\hskip2pt\vrule&\omit&&\omit&\cr
\noalign{\hrule}
height2pt&\omit&\hskip2pt\vrule&\omit&&\omit&\cr
& \hfil $-ü$ &\hskip2pt\vrule& $b$ (4.2) && $ $ (1.7770) &\cr
& \hfil $+ü$ &\hskip2pt\vrule& $t$ (178) && $Ã_ $ (<.0182) &\cr
height2pt&\omit&\hskip2pt\vrule&\omit&&\omit&\cr}
\hrule}
â
\vbox{\offinterlineskip
\hrule
\halign{ &\vrule#&\strut¼#\hfilÊ\cr
height2pt&\omit&\omit&\omit&\omit&\omit&\cr
&& \omit & \hfil $s=1$ & \omit &&\cr 
height2pt&\omit&\omit&\omit&\omit&\omit&\cr
\noalign{\hrule}
height2pt&\omit&\hskip2pt\vrule&\omit&&\omit&\cr
\noalign{\hrule}
height2pt&\omit&\hskip2pt\vrule&\omit&&\omit&\cr
& \hfill color: $£$ &\hskip2pt\vrule& \hfil gluon && \hfil electroweak&\cr
& flavor ($Q$) &\hskip2pt\vrule& \hfil (8) && \hfil (1) &\cr
height2pt&\omit&\hskip2pt\vrule&\omit&&\omit&\cr
\noalign{\hrule}
height2pt&\omit&\hskip2pt\vrule&\omit&&\omit&\cr
\noalign{\hrule}
height2pt&\omit&\hskip2pt\vrule&\omit&&\omit&\cr
& \hfil $0$ &\hskip2pt\vrule& $g$ (0) && $©$ ($<6É10^{-26}$) &\cr
& \hfil $0$ &\hskip2pt\vrule&&&$Z$ (91.188) &\cr
& \hfil $à1$ &\hskip2pt\vrule&&& $W$ (80.42) &\cr
height2pt&\omit&\hskip2pt\vrule&\omit&&\omit&\cr
\noalign{\hrule}
height12.5pt depth0pt width0pt
	&\omit&\omit&\omit&\omit&\omit&\omit\cr
\noalign{\hrule}
height2pt&\omit&\omit&\omit&\omit&\omit&\cr
&& \omit & \hfil $s=0$ & \omit &&\cr 
height2pt&\omit&\omit&\omit&\omit&\omit&\cr
\noalign{\hrule}
height2pt&\omit&\omit&\omit&\omit&\omit&\cr
\noalign{\hrule}
height2pt&\omit&\omit&\omit&\omit&\omit&\cr
& \hfill ($Q=0$) & \omit & \hfil $H$ & \omit & (>114.4) &\cr 
height2pt&\omit&\omit&\omit&\omit&\omit&\cr}
\hrule} $$
 The quark masses we have listed are the ``current quark masses", the
effective masses when the quarks are relativistic with respect to their
hadron (at least for the lighter quarks), and act as almost free.  
But since they are not free, their masses are ambiguous and 
energy dependent, and defined by some convenient conventions.
Nonrelativistic quark models use instead
the ``constituent quark masses", which include potential energy from the
gluons.  This extra potential energy is about .30 GeV per quark in the
lightest mesons, .35 GeV in the lightest baryons; there is also a
contribution to the binding energy from spin-spin interaction.  Unlike
electrodynamics, where the potential energy is negative because the
electrons are free at large distances, where the potential levels off (the
top of the ``well"), in chromodynamics the potential energy is positive
because the quarks are free at high energies (short distances, the bottom
of the well), and the potential is infinitely rising.  Masslessness of the
gluons is implied by the fact that no colorful asymptotic states have ever
been observed.  We have divided the spin-1/2 particles into 3 ``families"
with the same quantum numbers (but different masses).  Within each
family, the quarks are similar to the leptons, except that:  
\item{(1)} the masses
and average charges ($ÐQ$) are different,  
\item{(2)} the quarks come in 3 colors,
while the leptons are colorless, and  
\item{(3)} the neutrinos, to within
experimental error, are massless, so they have half as many components
as the massive fermions (1 helicity state each, instead of 2 spin states
each).  

\noindent This means that each lepton family has 1 SU(2) doublet and 1 SU(2)
singlet.  For symmetry (and better, quantum mechanical, reasons to be
explained later), we also assume the quarks have 1 SU(2) doublet, but
therefore 2 SU(2) singlets.  (Some experiments have indicated small
masses for neutrinos:  This would require generalization of the Standard
Model, such as models with parity broken by interactions.  Some
examples of such theories will be discussed in subsection IVB4.)

We first look at the color group theory of the physical states, which are
color singlets.   The fundamental unobserved particles are the spin-1
``gluons", described by the Yang-Mills gauge fields, and the spin-1/2
quarks.  Suppressing all but color indices, we denote the quark states by
$q^i$, and the antiquarks by $qÿ_i$, where the indices are those of the
defining representation of SU(n), and its complex conjugate.  The quarks
also carry a representation of a ``flavor" group, unlike the gluons.  The
simplest flavorful states are those made up of only (anti)quarks, with
indices completely contracted by one factor of an SU(n) group metric: 
From the ``U" of SU(n), we can contract defining indices with their
complex conjugates, giving the ``mesons", described by $qÿ_i q^i$
(quark-antiquark), which are their own antiparticles.  From the ``S" of
SU(n), we have the ``baryons", described by $·_{i_1...i_n}q^{i_1}...q^{i_n}$
(n-quark), and the antibaryons, described by the complex conjugate
fields.  All other colorless states made of just (anti)quarks can be written
as products of these fields, and therefore considered as describing
composites of them.  Thus, we can approximate the ground states of the
mesons by 
$$ qÿ_i(x)q^i(x) $$
 which describe spins 0 and 1 because of the
various combinations of spins (from $ü°ü=0¢1$).  The first excited level
will then be described by 
$$ qÿ\onª»q ­ qÿ»q-(»qÿ)q $$
The antisymmetric derivative picks out the relative momentum of the two quarks, rather than the total, and thus introduces orbital angular
momentum 1 (and similarly for more such derivatives).  This level thus includes
spins 0, 1, and 2. (Similar remarks apply to baryons.)  We can also have
flavorless states made from just gluons, called ``glueballs":  The ground
states can be described by $F_i{}^j F_j{}^i$, where each $F$ is a gluon
state (in the adjoint representation of SU(n)), and includes spins 0 and 2
(from the symmetric part of $1°1$).  Because of their flavor multiplets
and (electroweak) interactions, many mesons and baryons corresponding
to such ground and excited states have been experimentally identified,
while the glueballs' existence is still uncertain.  Actually, quarks and
gluons can almost be observed independently at high energies, where the
``strong" interaction is weak:  The energetic particle appears as a ``jet"
--- a particle of high energy accompanied by particles of much lower
energy (perhaps too small to detect) in color-singlet combinations. 
(Depending on the available decay modes, the jet might not be observed
until after decaying, but still within a small angle of spread.)

Just as all physical states are singlets of the local color SU(3), they are also singlets of the local SU(2) of electroweak interactions.  As will be explained later (subsection IVB2), there are four Higgs fields, which transform simultaneously as a doublet of this local SU(2) and a doublet of a broken, global isospin SU(2).  (The determinant of this 2$ð$2 matrix gives the observable singlet Higgs.)  For example, the proton and neutron, which have close but different mass, are a doublet of this global SU(2).  Unlike the confinement responsible for SU(3) singlets, which is nonperturbative, the Higgs mechanism responsible for SU(2) singlets is perturbative, since the scalar Higgs fields can be expanded about their ``vacuum values", which are just numbers:  SU(2) singlets can be found from multiplying general fields by Higgs scalars, which trade the local SU(2) for the global one, while there are no scalars that transform under SU(3), and giving a vacuum value to a field with spin would violate Lorentz invariance.
Ironically, while the Higgs is easy to describe theoretically, but hasn't been found yet experimentally, confinement is the opposite. However, they look similar: Both have (lowest mass) composite scalars of the form $Æ Æ$ and vectors of the form $Æ iá_aÆ$ that are singlets under their (nonabelian) gauge group, where $Æ$ is a scalar field for Higgs and a spinor (fermion) field for confinement. Classically they seem quite different, but the quantum relation is still unclear. Supersymmetry might provide some relation.

We now look at the flavor group theory of the physical hadronic states. 
In contrast to the previous paragraph, we now suppress all but the flavor
indices.  Mesons $M_i{}^j=qÿ_iq^j$ are thus in the adjoint representation
of flavor U(m) ($m°Ðm$, where $m$ is the defining representation and $Ðm$
its complex conjugate), for both the spin-0 and the spin-1 ground states. 
The baryons are more complicated:  For simplicity we consider SU(3) color,
which accurately describes physics at observed energies.  Then the color
structure described above results in total symmetry in combined flavor
and Lorentz indices (from the antisymmetry in the color indices, and the
overall antisymmetry for Fermi-Dirac statistics).  Thus, for the 3-quark
baryons, the Young tableaux
$$ \upõ3{\õ3} ¢ \upõ2{\õ2\õ1} ¢ \upõ1{\õ1\õ1\õ1} $$
 for SU(m) flavor are accompanied by the same Young tableaux for spin
indices:  In nonrelativistic notation, the first tableau, being totally
antisymmetric in flavor indices, is also totally antisymmetric in the three
two-valued spinor indices, and thus vanishes.  Similarly, the last tableau
describes spin 3/2 (total symmetry in both types of indices), while the
middle one describes spin 1/2.  Since only 3 flavors of quarks have small
masses compared to the hadronic mass scale, hadrons can be most
conveniently grouped into flavor multiplets for SU(3) flavor:  The ground
states are then, in terms of SU(3) flavor multiplets, 8$¢$1 for the
pseudoscalars, 8$¢$1 for the vectors, 8 for spin 1/2, and 10 for spin 3/2.

\x IC4.1  What SU(flavor) Young tableaux, corresponding to what spins,
would we have for mesons and baryons if there were 
 ªa 2 colors?  
 ªb 4 colors?

However, the differing masses of the different flavors of quarks break
the SU(3) flavor symmetry (as does the weak interaction).  In particular,
the mass eigenstates tend to be pure states of the various combinations
of the different flavors of quarks, rather than the linear combinations
expected from the flavor symmetry.  Specifically, the linear combinations
predicted by an 8$¢$1 separation for mesons (trace and traceless pieces
of a 3$°$3 matrix) are replaced with particles that are more accurately
described by a particular flavor of quark bound to a particular flavor of
antiquark.  (This is known as ``ideal mixing".)  The one exception is the
lighest mesons (pseudoscalars), which are more accurately described by
the 8$¢$1 split, for this restriction to the 3 lighter flavors of quarks, but
the mass of the singlet differs from that naively expected from group
theory or nonrelativistic quark models.  (This is known as the ``U(1)
problem".)  The solution is probably that the singlet mixes strongly with
the lightest psuedoscalar glueball (described by
$tr¼·^{abcd}F_{ab}F_{cd}$); the mass eigenstates are linear combinations
of these two fields with the same quantum numbers.  In any case, the
most convenient notation for labeling the entries of the matrix $M_i{}^j$
representing the various meson states for any particular spin and angular
momentum of the quark-antiquark combination is that corresponding to
the choice we gave earlier for the generators of U(n):  Label each entry by
a separate name, where the complex conjugate appears reflected across
the diagonal.  These directly correspond to the combination of a
particular quark with a particular antiquark, and to the mass eigenstates,
with the possible exception of the entries along the diagonal for the 3
lightest flavors, where the mass eigenstates are various linear
combinations.  (However, the SU(2) of the 2 lightest flavors is only slightly
broken by the quark masses, so in that case the combinations are very
close to the 3$¢$1 split of SU(2).)

For example, for the lightest multiplet of mesons (spin 0, and relative
angular momentum 0 for the quark and antiquark, but not all of which
have yet been observed), we can write the U(6) matrix (for the 6 flavors
of the 3 known families)

$$ M_i{}^j = \pmatrix{Ðdd & Ðdu & Ðds & Ðdc & Ðdb & Ðdt \cr
					Ðud & Ðuu & Ðus & Ðuc & Ðub & Ðut \cr
					Ðsd & Ðsu & Ðss & Ðsc & Ðsb & Ðst \cr
					Ðcd & Ðcu & Ðcs & Ðcc & Ðcb & Ðct \cr
					Ðbd & Ðbu & Ðbs & Ðbc & Ðbb & Ðbt \cr
					Ðtd & Ðtu & Ðts & Ðtc & Ðtb & Ðtt \cr} $$
$$ = \bordermatrix{ & d\hfill & u\hfill & s\hfill 
		& c\hfill & b\hfill & t\hfill \cr
	Ðd¼ & ú_d \hfill& ¹^+ (.1395702) \hfill& ÐK^0 (.49765) \hfill& 
		D^+ (1.8693) \hfill& ÐB^0 (5.2794) \hfill& T^+ \hfill\cr
	Ðu¼ & ¹^- (\dit) \hfill& ú_u \hfill& K^- (.49368) \hfill& 
		D^0 (1.8645) \hfill& B^- (5.2790) \hfill& T^0 \hfill\cr
	Ðs¼ & K^0 (\dit) \hfill& K^+ (\dit) \hfill& ú_s \hfill&
		D_s^+ (1.9682) \hfill& ÐB_s^0 (5.370) \hfill& T_s^+ \hfill\cr
	Ðc¼ & D^- (\dit) \hfill& ÐD^0 (\dit) \hfill& D_s^- (\dit) \hfill&
		ú_c (2.980) \hfill& B_c^- (6.4) \hfill& T_c^0 \hfill\cr
	Ðb¼ & B^0 (\dit) \hfill& B^+ (\dit) \hfill& B_s^0 (\dit) \hfill&
		B_c^+ (\dit) \hfill& ú_b \hfill& T_b^+ \hfill\cr
	Ðt¼ & T^- \hfill& ÐT^0 \hfill& T_s^- \hfill&
		 ÐT_c^0 \hfill& T_b^- \hfill& ú_t \hfill\cr} $$
 where (approximately)
$$ ú_u = \f1{å2}¹^0 (.1349766) +ü[ú'(.9578)+ú(.5478)] $$
$$ ú_d = -\f1{å2}¹^0 +ü(ú'+ú),âú_s = \f1{å2}(ú'-ú) $$
 in terms of the mass eigenstates (observed particles), with masses
again in GeV, and ditto marks refer to the transposed entry.  (We have
neglected the important contribution from the glueball.)  For the
corresponding spin-1 multiplet,
$$ ~M_i{}^j = \hss \bordermatrix{ & d\hfill & u\hfill & s\hfill 
		& c\hfill & b\hfill & t\hfill \cr
	Ðd¼ & ¿_d \hfill& ¨^+ (.7755) \hfill& ÐK*^0 (.8961) \hfill& 
		D*^+ (2.0100) \hfill& ÐB*^0 (5.3250) \hfill& T*^+ \hfill\cr
	Ðu¼ & ¨^- (\dit) \hfill& ¿_u \hfill& K*^- (.8917) \hfill& 
		D*^0 (2.0067) \hfill& B*^- (5.3250) \hfill& T*^0 \hfill\cr
	Ðs¼ & K*^0 (\dit) \hfill& K*^+ (\dit) \hfill& Ä (1.01946) \hfill& 
		D*_s^+ (2.1120) \hfill& ÐB*_s^0 (5.417) \hfill& T*_s^+ \hfill\cr
	Ðc¼ & D*^- (\dit) \hfill& ÐD*^0 (\dit) \hfill& D*_s^- (\dit) \hfill& 
		J/Æ (3.09692) \hfill& B*_c^- \hfill& T*_c^0 \hfill\cr
	Ðb¼ & B*^0 (\dit) \hfill& B*^+ (\dit) \hfill& B*_s^0 (\dit) \hfill& 
		B*_c^+ \hfill& ç (9.4603) \hfill& T*_b^+ \hfill\cr
	Ðt¼ & T*^- \hfill& ÐT*^0 \hfill& T*_s^- \hfill& 
		ÐT*_c^0 \hfill& T*_b^- \hfill& Ï \hfill\cr} $$
 where
$$ ¿_u = \f1{å2}[¿ (.7826) +¨^0 (.7758)],â¿_d =  \f1{å2}(¿-¨^0) $$
 (with $Ðss=Ä$, ideal mixing, also approximate).

\x IC4.2 Check the consistency of the masses in the second mass matrix above by assuming the meson masses are just the sum of the ``constituent" quark masses:  See how close a fit you can get.  (Potential energies have just been lumped into the quark masses, assuming they are the same throughout the multiplet.  Note that masses on the diagonal will come out a bit low from annihilation effects.  The first multiplet was not used because of complications from mixing with glueballs.  Similar mass relations can be obtained from group theory arguments, but the underlying physics is explained by the quark model.)

Ü5. Covering groups

The orthogonal groups O(n${}_+$,n${}_-$) are of obvious interest for
describing Lorentz symmetry in spacetimes with n${}_+$ space and n${}_-$
time dimensions, or conformal symmetry in spacetimes with n${}_+-$1
space and n${}_--$1 time dimensions.  This means we should be
interested in O(n) for n$²$6, and their ``Wick rotations": transformations
that put in extra factors of $i$ to change some signs on the metric. 
Coincidentally, these are just the cases where the Lie algebras of the
orthogonal groups are equivalent to those of some algebras for smaller
matrices.  The smaller representation then can be identified as the
``spinor" representation of that orthogonal group.  Since the ``vector", or
defining representation space of the orthogonal group, itself is
represented as a matrix with respect to the other group (i.e., the state
carries two spinor indices), the other group may include certain phase
transformations (such as $-1$) that cancel in the transformation of the
vector.  The other group is then called the ``covering" group for that
orthogonal group, since it includes those missing transformations in its
defining representation.  (As a result, its group space also has a more
interesting topology, which we won't discuss here.)

One way to discover these covering groups is to first count generators,
then try to construct explicitly the orthogonal metric on matrices.  SO(n)
has n(n$-1$)/2 generators (antisymmetric matrices), Sp(n) has n(n+1)/2
(symmetric), and SU(n) has n${}^2-$1 (traceless).  (These are hermitian
generators, since we applied reality or hermiticity.)  So, for some group
SO(n), we look for another group that has the same number of
generators.  Then, if the new group is defined on m$ð$m matrices, we
look for conditions to impose on an m$ð$m matrix (not necessarily the
adjoint) to get an n-component representation.  This is easy to do by
inspection for small n; for large n it's easy to see that it can't work, since
m will be of the order of n, and the simple constraints will give of the
order of n${}^2$ components instead of n.  We then construct the norm of
this matrix $M$ as $tr(MÿM)$, which is just the sum of the absolute value
squared of the components, for SO(n), and the other orthogonal groups by
Wick rotation.  (Wick rotation affects mainly the reality conditions on
$M$.)

The identifications for the Lie algebras are then:

\nobreak\noindent
SO(2) = U(1),âSO(1,1) = GL(1)\\
SO(3) = SU(2) = SU*(2) = USp(2),âSO(2,1) = SU(1,1) = SL(2) = Sp(2)\\
SO(4) = SU(2)$°$SU(2),âSO(3,1) = SL(2,C) = Sp(2,C),â
SO(2,2) = SL(2)$°$SL(2)\\ 
SO(5) = USp(4),âSO(4,1) = USp(2,2),âSO(3,2) = Sp(4)\\
SO(6) = SU(4),âSO(5,1) = SU*(4),âSO(4,2) = SU(2,2),âSO(3,3) = SL(4)

\noindent Note that the Euclidean cases are all unitary, while the ones
with (almost) equal numbers of space and time dimensions are all real. 
There are also some similar relations for the pseudoreal orthogonal
groups:

\nobreak\noindent\centerline{SO*(2) = U(1),â
	SO*(4) = SU(2)$°$SL(2),âSO*(6) = SU(3,1),âSO*(8) = SO(6,2)}

The norm and conditions for an m-spinor of SO(n${}_+$,n${}_-$) are:
$$ \vbox{\offinterlineskip
\hrule
\halign{ &\vrule#&\strut¼\hfil$#$\hfil¼\cr
height2pt&\omit&\omit&\omit&\omit&\omit&\omit
	&\omit&\hskip2pt\vrule&\omit&&\omit&&\omit&&\omit&\cr
& & \omit & & \omit & & \omit & ââââân_- Ü &
	\hskip2pt\vrule & 0 && 1 && 2 && 3 &\cr
height2pt&\omit&\omit&\omit&\omit&\omit&\omit
	&\omit&\hskip2pt\vrule&\omit&&\omit&&\omit&&\omit&\cr
\noalign{\hrule}
height2pt&\omit&&\omit&\hskip2pt\vrule&\omit&&\omit&
	\hskip2pt\vrule&\omit&\omit&\omit&\omit&\omit&\omit&\omit&\cr
& m && n & \hskip2pt\vrule & norm && symmetry: z^T= &
	\hskip2pt\vrule && \omit & â¼reality: & \omit & z*=â¼ & \omit
	&&\cr  
height2pt&\omit&&\omit&\hskip2pt\vrule&\omit&&\omit&
	\hskip2pt\vrule&\omit&\omit&\omit&\omit&\omit&\omit&\omit&\cr
\noalign{\hrule}
height2pt&\omit&&\omit&\hskip2pt\vrule&\omit&
	&\omit&\hskip2pt\vrule&\omit&&\omit&&\omit&&\omit&\cr
\noalign{\hrule}
height2pt&\omit&&\omit&\hskip2pt\vrule&\omit&
	&\omit&\hskip2pt\vrule&\omit&&\omit&&\omit&&\omit&\cr
& 1 && 2 & \hskip2pt\vrule & z'z &&& \hskip2pt\vrule &
	z' && z¼(z'*=z') && && &\cr
height2pt&\omit&&\omit&\hskip2pt\vrule&\omit&
	&\omit&\hskip2pt\vrule&\omit&&\omit&&\omit&&\omit&\cr
\noalign{\hrule}
height2pt&\omit&&\omit&\hskip2pt\vrule&\omit&
	&\omit&\hskip2pt\vrule&\omit&&\omit&&\omit&&\omit&\cr
& 2 && 3 & \hskip2pt\vrule & z^{Œº}z^{©¶}·_{©Œ}·_{¶º} &
	& z &\hskip2pt\vrule & -·z· && z && && &\cr
&&& 4 & \hskip2pt\vrule & z^{Œº'}z^{©¶'}·_{©Œ}·_{¶'º'} & 
	&& \hskip2pt\vrule & -·z· && z^T && z && &\cr
height2pt&\omit&&\omit&\hskip2pt\vrule&\omit&
	&\omit&\hskip2pt\vrule&\omit&&\omit&&\omit&&\omit&\cr
\noalign{\hrule}
height2pt&\omit&&\omit&\hskip2pt\vrule&\omit&
	&\omit&\hskip2pt\vrule&\omit&&\omit&&\omit&&\omit&\cr
& 4 && 5 & \hskip2pt\vrule & z^{Œº}z^{©¶}·_{¶©ºŒ} &&
	-z¼(z^{Œº}¯_{ºŒ}=0) & \hskip2pt\vrule & ü·z && ü·(çzç) && z 
	&&&\cr 
&&& 6 & \hskip2pt\vrule & z^{Œº}z^{©¶}·_{¶©ºŒ} && -z &
	\hskip2pt\vrule & ü·z && -¯z¯ && ü·(çzç) && z &\cr 
height2pt&\omit&&\omit&\hskip2pt\vrule&\omit&
	&\omit&\hskip2pt\vrule&\omit&&\omit&&\omit&&\omit&\cr}
\hrule} $$
 Note that in all but the 2D cases the norms are associated with
determinants:  For D=3 and 4 the norm is given by the determinant, while
for D=5 and 6 we use the fact that the determinant of an antisymmetric
matrix is the square of the Pfaffian.

\x IC5.1 Show that for D=5 $zz·$ and $zz¯¯$ give the same norm.  (Hint: 
Consider $¯_{[Œº}¯_{©¶]}$.)

Unfortunately, for SO(n) for larger n, the spinor is as least as large as, and
usually larger than, the vector.  In general, the spinor is like the ``square
root" of the vector, in that the vector can be found by taking the direct
product of two spinors.  It is impossible to find the spinor representation
by taking direct products of vectors.  This situation occurs only for
orthogonal groups:  In all other classical groups, all
(finite-dimensional) representations are among those obtained from
multiple direct products of vectors.  Furthermore, in those cases the
``irreducible" representations (those that can't be divided into smaller
representations) can be picked out by (anti)symmetrization, and by
separating trace and traceless pieces (where traces are taken with the
group metrics).  Fortunately, for the above cases of orthogonal groups,
we can perform the same construction starting with the spinor
representations, since those are the ``vectors" of non-orthogonal groups.

\refs

£1 J. Schwinger, On angular momentum, ÓQuantum theory of angular
	momentum: a collection of reprints and original papersÕ, eds. L.C.
	Biedenharn and H. Van Dam (Academic, 1965) p. 229:\\
	spin using spinor oscillators.
 £2 Georgi, Óloc. cit.Õ (IB).
 £3 P.A.M. Dirac, ÓProc. Roy. Soc.Õ ÉA117 (1928) 610.
 £4 H. Weyl, ÓZ. Phys.Õ É56 (1929) 330.
 £5 M. Hamermesh, ÓGroup theory and its application to physical problemsÕ
	(Dover, 1962):\\
	detailed discussion of Young tableaux.
 £6 M. Gell-Mann, \PL 8 (1964) 214;\\
	G. Zweig, preprints CERN-TH-401 and 412 (1964); Fractionally
	charged particles and SU${}_6$, in ÓSymmetries and elementary
	particle physicsÕ, proc. Int. School of Physics ``Ettore Majorana",
	Erice, Italy, Aug.-Sept., 1964, ed. A. Zichichi (Academic, 1965) p. 192:\\
	quarks.
 £7 Particle Data Group (in 2004, 
	S. Eidelman, K.G. Hayes, K.A. Olive, M. Aguilar-Benitez, C. Amsler, D. Asner,
	K.S. Babu, R.M. Barnett, J. Beringer, P.R. Burchat
% ...
	\pdflink{http://pdg.lbl.gov/pdg.html}
	(published biannually, including \PL 592B (2004) 1):\\
	Review of Particle Physics, a.k.a. Review of Particle Properties, a.k.a.
	Rosenfeld tables; tables of masses, decay rates, etc., of all known
	particles, plus useful brief reviews on particle physics and cosmology.

\unrefs

ÚII. SPIN

Special relativity is simply the statement that the laws of nature are
symmetric under the Poincar«e group.  Free relativistic quantum mechanics
or field theory is then equivalent to a study of the representations of the
Poincar«e group.  Since the conformal group is a classical group, while its
subgroup the Poincar«e group is not, it is easier to first study
the conformal group, which is sufficient for finding the massless
representations of the Poincar«e group.  The massive ones then can be
found by dimensional reduction, which gives them in the same form as
occurs in interacting field theories.  In four spacetime dimensions we
use the covering group of the conformal group, which is the easiest way
to include spinors.  These methods extend straightforwardly to
supersymmetry, a symmetry between fermions and bosons that includes
the Poincar«e group.

Û7 A. TWO COMPONENTS

Although we have already specialized to spacetime symmetries, we have
considered arbitrary spacetime dimensions.  We have also noted that
many of the lower-dimensional Lie groups have special properties,
especially with regard to covering groups.  In this section we will take
advantage of those features; specifically, we examine the physical case
D=4, where the rotation group is SO(3)=SU(2), the Lorentz group is
SO(3,1)=SL(2,C), and the conformal group is SO(4,2)=SU(2,2).

Ü1. 3-vectors

The most important nontrivial Lie group in physics is the rotations in
three dimensions.  It is also the simplest nontrivial example of a Lie
group.  This makes it the ideal example to illustrate the properties
discussed in the previous chapter, as well as lay the groundwork for later
discussions.  We have already mentioned the orbital part of rotations,
i.e., the representation of rotations on spatial coordinates.  In this
chapter we discuss the spin part; this is really the same as finding all
(finite dimensional, unitary) representations.

Since the earliest days of quantum mechanics, we know that half-integer spins exist, in nature as well as group theory, e.g., the electron and proton.  This might be expected to complicate matters, but actually simplifies them, due to the well-known inequality
$$ ü < 1 $$
This means that a ``spinor", describing spin 1/2, has only 2 components, compared to the 3 components of a vector, and its matrices (e.g., for rotations) are thus only 2$ð$2 instead of 3$ð$3.

We first consider spinors in matrix notation, then generalize to ``spinor notation" (spinor indices).  The simplest way to understand why rotations can be represented as 2$ð$2 instead of 3$ð$3 is to see why 3-vectors themselves can be understood as 2$ð$2 matrices, which for some purposes is simpler.  (This is equivalent to Hamilton's ``quaternions", which predated Gibbs' vector notation, and were used by Maxwell for his equations.  This way also generalizes in a very simple way to
relativity, in three space and one time dimensions.)  Consider such
matrices to be hermitian, which is natural from the quantum mechanical
point of view.  Then they have four real components, one too many for a
three-vector (but just right for a relativistic four-vector), so we restrict
them to also be traceless:
$$ V = Vÿ,âtr¼V = 0 $$
 The simplest way to get a single number out of a matrix, besides taking
the trace, is to take the determinant.  By expanding a general
matrix identity to quadratic order we find an identity for 2$ð$2 matrices
$$ det(I+M) = e^{trÊln(I+M)}âÜâ-2¼det¼M = tr(M^2) -(tr¼M)^2 $$
 It is then clear that in our case $-det¼V$ is positive definite, as well as
quadratic, so we can define the norm of this 3-vector as
$$ |V|^2 = -2¼det¼V = tr(V^2) $$
 This can be compared easily with conventional notation by picking a
basis:
$$ V = \f1{å2}\pmatrix{V^1 & V^2 -iV^3 \cr V^2 +iV^3 & -V^1 \cr} 
	= \vec VÉ\vec §âÜâdet¼V = -ü(V^i)^2 $$
 where $\vec §$ are the Pauli $§$ matrices, up to normalization.  As usual,
the inner product follows from the norm:
$$ |V+W|^2 = |V|^2 +|W|^2 +2VÉW $$
$$ ÜâVÉW = det¼V +det¼W -det(V+W) = tr(VW) $$

Applying our previous identities for determinants to 2$ð$2 matrices, we
have
$$ MCM^TC = I¼det¼M,ââM^{-1} = CM^T C (det¼M)^{-1} $$
 where we now use the imaginary, hermitian matrix
$$ C = \tat0i{-i}0 $$
 If we make the replacement $M£e^M$ and expand to linear order in $M$,
we find
$$ M + CM^T C = I¼tr¼M $$
 This implies
$$ tr¼V = 0âÛâ(VC)^T = VC $$
 i.e., the tracelessness of $V$ is equivalent to the symmetry of $VC$. 
Furthermore, the combination of the trace and determinant identities tell
us
$$ M^2 = M¼tr¼M - I¼det¼MâÜâV^2 = - I¼det¼V = Iü|V|^2 $$
 Here by ``$V^2$" we mean the square of the matrix, while
``$|V|^2$"$=(V^i)^2$ is the square of the norm (neither of which should be
confused with the component $V^2=V^i ¶_i^2$.)

Again expressing the inner product in terms of the norm, we then find
$$ ÓV,WÕ = (VÉW)I $$
 Also, since the commutator of two finite matrices is
traceless, and picks up a minus sign under hermitian conjugation, we can
define an outer product (vector$ð$vector = vector) by
$$ [V,W] = å2iVðW $$
 Combining these two results,
$$ VW = ü(VÉW)I +\f1{å2}iVðW $$
 In other words, the product of two traceless hermitian 2$ð$2 matrices
gives a real trace piece, symmetric in the two matrices, plus an
antihermitian traceless piece, antisymmetric in the two.  Thus, we have a
simple relation between the matrix product, the inner (``dot") product
and the outer (``cross") product.  Therefore, the cross product is a
special case of the Lie bracket, or commutator.

\x IIA1.1 Check this result in two ways:
ªa Show the normalization agrees with the usual outer product.  Using
only the above definition of $VðW$, along with $ÓV,WÕ=(VÉW)I$, show
$$ -I|VðW|^2 = ([V,W])^2 = -I[|V|^2|W|^2 -(VÉW)^2] $$
ªb Use components, with the above basis.

\x IIA1.2  Write an arbitrary two-dimensional vector in terms of a
complex number as $V=\f1{å2}(v_x-iv_y)$.  
ªa  Show that the phase (U(1)) transformation $V'=Ve^{iÏ}$ generates the
usual rotation.  Show that for any two vectors $V_1$ and $V_2$, $V_1*V_2$
is invariant, and identify its real and imaginary parts in terms of well
known vector products.  What kind of transformation is $V£V*$, and how
does it affect these products?
ªb  Consider two-dimensional functions in terms of $z=\f1{å2}(x+iy)$ and
$z*=\f1{å2}(x-iy)$.  Show by the chain rule that $»_z=\f1{å2}(»_x-i»_y)$. 
Write the real and imaginary parts of the equation $»_{z*}V=0$ in terms
of the divergence and curl.  (Then $V$ is a function of just $z$.)
ªc  Consider the complex integral
$$ È{dz\over 2¹i}¼V $$
 where ``$ÈÊ$" is a ``contour integral": an integral over a closed
path in the complex plane defined by parametrizing $dz=du(dz/du)$ in
terms of some real parameter $u$.  This is useful if $V$ can be Laurent
expanded as $V(z)=Ý_{n=-¥}^¥ c_n (z-z_0)^n$ inside the contour about
a point $z_0$ there, since by considering circles $z=z_0+re^{iÏ}$ we find
only the $1/(z-z_0)$ term contributes.  Show that this integral contains as
its real and imaginary parts the usual line integral and ``surface"
integral.  (In two dimensions a surface element differs from a line
element only by its direction.)  Use this fact to solve Gauss' law in two
dimensions for a unit point charge as $E=1/4¹z$.

\x IIA1.3  Consider electromagnetism in 2$ð$2 matrix notation:  Define the
field strength as a ÓcomplexÕ vector $F=å2(E+iB)$.  Write partial
derivatives as the sum of a (rotational) scalar plus a (3-)vector as
$»=\f1{å2}I»_t+á$, where $»_t=»/»t$ is the time derivative and $á$ is the
partial space derivatives written as a traceless matrix.  Do the same for
the charge density $¨$ and (3-)current $j$ as $J=-\f1{å2}I¨+j$.  Using the
definition of dot and cross products in terms of matrix multiplication as
discussed in this section, show that the simple matrix equation $»F=-J$,
when separated into its trace and traceless pieces, and its hermitian and
antihermitian pieces, gives the usual Maxwell equations
$$ áÉB = 0,âáÉE = ¨,âáðE+»_t B = 0,âáðB-»_t E = j $$
 (Note:  Avoid the Pauli $§$-matrices and explicit components.)

Ü2. Rotations

One convenience of representing three-vectors as 2$ð$2 instead of 3$ð$1
is that rotations are easier to write.  Since vectors are hermitian, we
expect their transformations to be unitary:
$$ V' = UVUÿ,ââUÿ = U^{-1} $$
 It is easily checked that this preserves the properties of these matrices:
$$ (V')ÿ = (UVUÿ)ÿ = V',ââtr(V') = tr(UVU^{-1}) = tr(U^{-1}UV) = tr(V) = 0 $$
 Furthermore, it also preserves the norm (and thus the inner product):
$$ det (V') = det (UVU^{-1}) = det(U) det(V) (det¼U)^{-1} = det¼V $$
 Unitary 2$ð$2 matrices have 4 parameters; however, we can elimimate
one by the condition
$$ det¼U = 1 $$
 This eliminates only the phase factor in $U$, which cancels out in the
transformation law anyway.  Taking the product of two rotations now
involves multiplying only 2$ð$2 matrices, and not 3$ð$3 matrices. 

We can also write $U$ in exponential notation, which is useful for going to
the infinitesimal limit:
$$ U = e^{iG}âÜâGÿ = G,âtr¼G = 0 $$
   This means that $G$ itself can be considered a vector.  Rotations can be
parametrized by a vector whose direction is the axis of rotation, and
whose magnitude is ($1/å2ð$) the angle of rotation:
$$ V' = e^{iG}Ve^{-iG}âÜâ¶V = i[G,V] = -å2GðV $$
 We also now see that the Lie bracket we previously identified as the
cross product is the bracket for the rotation group.

\x IIA2.1 Evaluate the elements of the matrix $e^{iG}$ in closed form for a
diagonal generator $G$.  Generalize this result to arbitrary $G$.  (Hint:  Use
rotational invariance.)

The hermiticity condition on $V$ can also be expressed as a reality
condition:
$$ V = Vÿâandâtr¼V = 0âÜâV* = -CVC,â(VC)* = C(VC)C $$
 where ``$¼*¼$" is the usual complex conjugate.  A similar condition for $U$
is
$$ Uÿ = U^{-1}âandâdet¼U =1âÜâU* = CUC $$
 which is also a consequence of the fact that we can write $U$ in terms of
a vector as $U=e^{iV}$.  As a result, the transformation law for the
vector can be written in terms of $VC$ in a simple way, which manifestly
preserves its symmetry:
$$ (VC)' = UVU^{-1}C = U(VC)U^T $$

\x IIA2.2  We return to our example of D=2:
 ªa Write an arbitrary rotation in two dimensions in terms of the
ÓslopeÕ ($dy/dx$) of the rotation (the slope to which the x-axis is rotated)
rather than the angle.  (This is actually more convenient to measure if you
happen to have a ruler, which you need to measure lengths anyway, but
not a protractor.)  This avoids trigonometry, but introduces ugly square
roots:  Compare Lorentz transformations.  
Also note that this square root form covers only half of the available angles.
 ªb Show that the square roots can be eliminated by using the slope of
ÓhalfÕ the angle of transformation as the variable.  Show the relation to
the variables used in writing 3D rotations in terms of 2$ð$2 matrices,
i.e., the use of complex variables, as in exercise IIA1.2a. 
(Hint:  Consider $U$ and $VC$ diagonal.)

Ü3. Spinors

Note that the mapping of SU(2) to SO(3) is two-to-one:  This follows from
the fact $V'=V$ when $U$ is a phase factor.  We eliminated continuous
phase factors from $U$ by the condition $det¼U=1$, which restricts U(2) to
SU(2).  However,
$$ det(Ie^{iÏ}) = e^{2iÏ} = 1âÜâe^{iÏ} = à1 $$
 for 2$ð$2 matrices.  More generally, for any SU(2) element $U$, $-U$ is
also an element of SU(2), but acts the same way on a vector; i.e., these
two SU(2) transformations give the same SO(3) transformation.  Thus SU(2)
is called a ``double covering" of SO(3).  However, this second
transformation is not redundant, because it acts differently on
half-integral spins, which we discuss in the following subsections.

The other convenience of using 2$ð$2 matrices is that it makes obvious
how to introduce spinors --- Since a vector already transforms with two
factors of $U$, we define a ``square root" of a vector that transforms with
just one $U$:
$$ Æ' = UÆâÜâÆÿ' = ÆÿU^{-1} $$
 where $Æ$ is a two-component ``vector", i.e., a 2$ð$1 matrix.  The
complex conjugate of a spinor then transforms in essentially the same
way:
$$ (CÆ*)' = CU*Æ* = U(CÆ*) $$
 Note that the antisymmetry of $C$ implies that $Æ$ must be complex:  We
might think that, since $CÆ*$ transforms in the same way as $Æ$, we can
identify the two consistently with the transformation law.  But then we
would have
$$ Æ = CÆ* = C(CÆ*)* = CC*Æ = -Æ $$
 Thus the representation is pseudoreal.  The fact that $CÆ*$ transforms
the same way under rotations as $Æ$ leads us to consider the
transformation
$$ Æ' = CÆ* $$
 Since a vector transforms the same way under rotations as $ÆÆÿ$, under
this transformation we have
$$ V' = CV*C = -V $$
 which identifies it as a reflection.

Another useful way to write rotations on $Æ$ (like looking at $VC$ instead
of $V$) is
$$ (Æ^T C)' =  (Æ^T C)U^{-1} $$
 This tells us how to take an invariant inner product of spinors:
$$ Æ' = UÆ,â' = UâÜâ(Æ^T C)' = (Æ^T C) $$
 In other words, $C$ is the ``metric" in the space of spinors.  An important
difference of this inner product from the familiar one for three-vectors is
that it is antisymmetric.  Thus, if $Æ$ and $$ are anticommuting spinors,
$$ Æ^T C = - ^T C^T Æ = ^T C Æ $$
 where one minus sign comes from anticommutativity and another is
from the antisymmetry of $C$.  Thus, it makes sense to take the norm of
an anticommuting spinor as $Æ^T CÆ$, which would vanish if $Æ$ were
commuting.  Of course, since rotations are unitary, we also have the
usual $ÆÿÆ$ as an invariant, positive definite, inner product.

\x IIA3.1  Consider a hermitian but ÓnotÕ traceless 2$ð$2 matrix $M$
($M=Mÿ$, $tr¼M±0$).
 ªa Show
$$ det¼M = 0âÜâM = àÆ°Æÿ $$
 for some ÓcommutingÕ spinor (column vector) $Æ$ (and some sign $à$).
 ªb Define a vector by
$$ V = å2 (M -üI¼tr¼M) $$
 Show $|V|$ (ÓnotÕ $|V|^2$) is simply $ÆÿÆ$.

Ü4. Indices

The best way to discuss general spins is to use index notation, rather
than matrix notation.  Then a spinor rotates as
$$ Æ'_Œ = U_Œ{}^º Æ_º $$
 with two-valued indices 
$$ \boxeq{Œ = ¢, \¢} $$ 
The inner product is ÓdefinedÕ by
$$ \boxeq{Æ^Œ _Œ = Æ^Œ C_{ºŒ} ^º = -Æ_Œ ^Œ} $$
 where we have ÓdefinedÕ raising and lowering of indices by
$$ \boxeq{Æ_Œ = Æ^º C_{ºŒ},âÆ^Œ = C^{Œº}Æ_º} $$
$$ \boxeq{C_{Œº} = -C_{ºŒ} = -C^{Œº} = C^{ºŒ} = \tat0i{-i}0} $$
 paying careful attention to signs.  (In general, we fix signs by using a
convention of contracting indices from upper-left to lower-right.)  Then
objects with many indices transform as the product of spinors:
$$ A'_{Œº...©} = U_Œ{}^¶ U_º{}^· ... U_©{}^½ A_{¶·...½}  $$
 An infinitesimal transformation is then a sum:
$$ -i¶A_{Œº...©} = G_Œ{}^¶ A_{¶º...©} +G_º{}^¶ A_{Œ¶...©} +...+
	G_©{}^¶ A_{Œº...¶} $$
 This is also true for $C_{μ}$, even though it is an invariant constant:
$$ C'_{Œº} = U_Œ{}^© U_º{}^¶ C_{©¶} = C_{Œº}¼det¼U = C_{Œº} $$
 A more interesting case is the vector:  The transformation law is
$$ V'_{Œº} = U_Œ{}^© U_º{}^¶ V_{©¶} $$
 where $V_{μ}$ is the symmetric $VC$ considered earlier (in contrast to
the antisymmetric $C$).

There is basically only one identity in index notation, namely
$$ \boxeq{0 = üC_{[Œº}C_{©]¶} = C_{Œº}C_{©¶} +C_{º©}C_{Œ¶} +C_{©Œ}C_{º¶}} $$
 The expression vanishes because it is antisymmetric in those indices, and
thus the indices must all have different values, but there are three
two-valued indices.  Another way to write this identity is to use the
definition of $C^{μ}$ as the inverse of $C_{μ}$:
$$ \boxeq{C_{Œ©}C^{º©} = ¶_Œ^ºâÜâC_{Œº}C^{©¶} 
	= ¶_{[Œ}^© ¶_{º]}^¶ ­ ¶_Œ^© ¶_º^¶ - ¶_º^© ¶_Œ^¶} $$
 This tells us that antisymmetrizing in any pair of indices automatically
contracts (sums over) them:  Contracting this identity with an arbitrary
tensor $A_{©¶}$,
$$ \boxeq{A_{[Œº]} = C_{Œº}C^{©¶}A_{©¶} = -C_{Œº}A^©{}_©} $$
 That means that we need to consider only objects that are totally
symmetric in their free indices.  This gives all spins:  Such a field with 2s
indices describes spin s; we have already seen spins 0, 1/2, and 1.

We have defined the transformation law of all fields with lower indices
by considering the direct product of spinors.  Transformations for upper
indices follow from multiplication with $C^{μ}$:  They all follow from
$$ Æ'^Œ = Æ^º (U^{-1})_º{}^Œ $$
 Since the vertical position of the index indicates the form of the
transformation law, we define
$$ ÐÆ_Œ ­ (Æ^Œ)* $$
 where the ``$Ñ{\phantom M}$" indicates complex conjugation.  Thus, a
hermitian matrix is written as
$$ M_Œ{}^º = (Mÿ)_Œ{}^º ­ (M_º{}^Œ)* ­ ÑM^º{}_ŒâÜâM_{Œº} = ÑM_{ºŒ} $$
 So, for a vector we have
$$ V_{Œº} = ÑV_{Œº} = V_{ºŒ} $$

Spin s is usually formulated in terms of a (2s+1)-component ``vector". 
Then one needs to calculate Clebsch-Gordan-Wigner coefficients to
construct Hamiltonians relating different spins.  For example, to couple
two spin-1/2 objects to a spin-1 object, one might write something like
$\vec VÉÆÿ\vec § $.  The matrix elements of the Pauli matrices $\vec §$
are the CGW coefficients for the spin-1 piece of $ü°ü=1¢0$.  This
method gets progressively messier for higher spins.  On the other hand,
in spinor notation such a term would be simply $V^{Œº}ÐÆ_Œ _º$; no
special coefficients are necessary, only contraction of indices.  Similarly
the decomposition of products of spins involves only the picking out of
the various symmetric and antisymmetric pieces:  For example, for
$ü°ü$,
$$ Æ_Œ _º = ü(Æ_{(Œ}_{º)} +Æ_{[Œ}_{º]}) =
	üÆ_{(Œ}_{º)} -C_{Œº}Æ^© _© = V_{Œº} +C_{Œº}S $$
 where $(μ)$ means to symmetrize in those indices, by adding all
permutations with plus signs.  We have thus explicitly separated out the
spin-1 and spin-0 parts $V$ and $S$ of the product.  The square roots of
various integers that appear in the CGW coefficients come from
permutation factors that appear in the normalizations of the various
fields/wave functions that appear in the products:  For example,
$$ A^{Œº©}ÐA_{Œº©} = |A^{¢¢¢}|^2 +3|A^{¢¢\¢}|^2 
	+3|A^{¢\¢\¢}|^2 +|A^{\¢\¢\¢}|^2 $$
 In the spinor index method, the square roots never appear explicitly,
only their squares appear in normalizations:  For example, in calculating a
probability for $A°B£C$, we evaluate
$$ {ÒA°B|CÔÒC|A°BÔ\over ÒA|AÔÒB|BÔÒC|CÔ} $$
 where $A$, $B$, and $C$ each have 2s indices for spin s, and $Ò|Ô$ means
contracting all indices (with the usual complex conjugation).
(Normalizing states to other than 1 is often convenient and sometimes
necessary:  For example, plane waves are normalized with $¶$ functions.)

\x IIA4.1  Redo exercise IIA3.1 in index notation:  For $Æ_Œ _º$ above
(both now bosonic), show $V^{μ}V_{μ}=-2S^2$.

Ü5. Lorentz

Consider now a 2$ð$2 matrix, whose elements we label as
$$ (V)^{ŒÀº} = \pmatrix{V^{¢\rdt ¢} & V^{¢\rdt\¢} \cr V^{\¢\rdt ¢} & V^{\¢\rdt\¢} \cr}
	= \pmatrix{V^+ & V^t* \cr V^t & V^- \cr} $$
$$ = \f1{å2}\pmatrix{V^0+V^1&V^2+iV^3\cr V^2-iV^3&V^0-V^1\cr}
	=  V^a (§_a)^{ŒÀº} $$
 which we choose to be hermitian,
$$ V = VÿâÜâV^{ŒÀº} = (Vÿ)^{ŒÀº} ­ (V^{ºÀŒ})* $$
 where we distinguish the right spinor index by a dot because it will be
chosen to transform differently from the left one:  According to our
discussion of subsection IB5, this is the general labeling consistent 
with hermiticity, i.e., $V'=gVgÿ$ (but without the extra restriction
of group unitarity of the previous subsections).  For comparison,
lowering both spinor indices with the matrix $C$ as for SU(2), and the
vector indices with the Minkowski metric (in either the orthonormal
or null basis, as appropriate --- see subsection IA4), we find another
hermitian matrix
$$ (V)_{ŒÀº} = \pmatrix{V_+ & V_t* \cr V_t & V_- \cr}
	= \f1{å2}\pmatrix{V_0+V_1&V_2-iV_3\cr V_2+iV_3&V_0-V_1\cr}
	=  V_a (§^a)_{ŒÀº} $$
 In the orthonormal basis, $§_a$ are the Pauli matrices and the identity,
up to normalization.  They are also the Clebsch-Gordan-Wigner
coefficients for spinor$°$spinor = vector.  In the null basis, they are
completely trivial: 1 for one element, 0 for the rest, the usual basis for
matrices.  In other words, they are simply an arbitrary way (according to
choice of basis) to translate a 2$ð$2 (hermitian) matrix into a
4-component vector.  We will sometimes treat a vector index ``$a$" as
an abbreviation for a spinor index pair ``$ŒÀŒ$":
$$ V^a = V^{ŒÀŒ},âa = ŒÀŒ = (¢\rdt ¢,¢\rdt\¢,\¢\rdt ¢,\¢\rdt\¢) = (+,Ðt,t,-) $$
 where $Œ$ and $ÀŒ$ are understood to be independent indices ($¢±\rdt ¢$, etc.).

Examining the determinant of (either version of) $V$, we find the correct
Minkowski norms:
$$ -2¼det¼V = -2V^+V^- +2V^tV^t*
	= -(V^{0})^2+(V^{1})^2+(V^{2})^2+(V^{3})^2 = V^2 $$
 Thus Lorentz transformations will be those that preserve the hermiticity
of this matrix and leave its determinant invariant:
$$ V' = gVgÿ,âdet¼g = 1 $$
 ($det¼g$ could also have a phase, but that would cancel in the
transformation.)  Thus $g$ is an element of SL(2,C).  In terms of the
representation of the Lie algebra,
$$ g = e^G,âtr¼G = 0 $$
 Thus the group space is 6-dimensional ($G$ has three independent
complex components), the same as SO(3,1) (where
$gúg^T=úÜ(Gú)^T=-Gú$).

\x IIA5.1 SL(2,C) also can be seen (less conveniently) from vector
notation:  
ªa Consider the generators
$$ J^{(à)}_{ab} = ü(J_{ab} àiü·_{abcd}J^{cd}) $$
 of SO(3,1).  Find their commutation relations, and in particular show\\
$[J^{(+)},J^{(-)}]=0$.  Express $J^{(à)}_{0i}$ in terms of $J^{(à)}_{ij}$.  Show
$J^{(à)}_{ij}$ have the same commutation relations as $J_{ij}$.  Finally,
take a general infinitesimal Lorentz transformation in terms of $J_{ab}$
and rewrite it in terms of $J^{(à)}_{ij}$, paying special attention to the
reality properties of the coefficients.  This demonstrates that the algebra
of SO(3,1) is the same as that of SU(2)$°$SU(2), but Wick rotated to
SL(2,C).
ªb Apply the same procedure to SO(4) and SO(2,2) to derive their covering
groups.

\x IIA5.2  Consider relativity in two dimensions (one space, one time):
 ªa  Show that SO(1,1) is represented in lightcone coordinates by
$$ x'^+ = ñx^+,ââx'^- = ñ^{-1}x^- $$
 for some (nonvanishing) real number $ñ$, and therefore SO(1,1) = GL(1). 
Write this one Lorentz transformation, in analogy to exercise IIA1.2a on
rotations in two space dimensions, in terms of an analog of the angle
(``rapidity") for those transformations that can be obtained continuously
from the identity.  Do the relativistic analog of exercise IIA2.2.
 ªb  Still using lightcone coordinates, find the parity and time reversal
transformations.   Note that writing $ñ$ as an exponential, so it can be
obtained continuously from the identity, restricts it to be positive,
yielding a subgroup of GL(1).  Explicitly, what are the transformations
of O(1,1) missing from this subgroup?  Which of P, (C)T, and (C)PT are
missing from these transformations, and which are missing from GL(1)
itself?

In index notation, we write for this vector
$$ V'_{ŒÀº} = g_Œ{}^© g*_{Àº}{}^{À¶}V_{©À¶} $$
 while for a (``Weyl") spinor we have
$$ Æ'_Œ = g_Œ{}^º Æ_º $$
 The metric of the group SL(2,C) is the two-index antisymmetric symbol,
which is also the metric for Sp(2,C):  In our conventions,
$$ \boxeq{C_{Œº} = -C_{ºŒ} = -C^{Œº} = C_{ÀŒÀº} = \tat0i{-i}0} $$
 We also have the identities
$$ det¼L_Œ{}^º = üC^{Œº}C_{©¶}L_Œ{}^©L_º{}^¶ = ü(tr¼L)^2 -ütr(L^2),â
	(L^{-1})_¶{}^º = C^{Œº}C_{©¶}L_Œ{}^© (det¼L)^{-1} $$
$$ \boxeq{A_{[Œº]}=C_{Œº}C^{©¶}A_{©¶},âA_{[Œº©]} = 0}  $$
 discussed earlier in this section.  As there, we use the metric to raise,
lower, and contract indices:
$$ \boxeq{ \eqalign{ Æ_Œ = Æ^º C_{ºŒ},&âÆ_{ÀŒ} = Æ^{Àº}C_{ÀºÀŒ} \cr
	VÉW =& V^{ŒÀº}W_{ŒÀº} \cr} } $$

These results for SO(3,1) = SL(2,C) generalize to SO(4) = SU(2)$°$SU(2)
(relevant to the Standard Model: see subsection IVB2) and
SO(2,2) = SL(2)$°$SL(2).  As described earlier, the reality conditions
change, so now
$$ SO(4):â(V^{Œº'})* = V^{©¶'}C_{©Œ}C_{¶'º'},ââ
	SO(2,2):â(V^{Œº'})* = V^{©¶'} $$
 consistent with the (pseudo)reality properties of spinors for SU(2) and
SL(2), where we now use unprimed and primed indices for the two
independent group factors ($V£gVg'$).

\x IIA5.3 Take the explicit 2$ð$2 representation for a vector given above,
change the factors of $i$ to satisfy the new reality conditions for SO(4)
and SO(2,2), and show the determinant gives the right signatures for the
metrics.

A common example of index manipulation is to use antisymmetry
whenever possible to give vector products.  For example, from the fact
that $V^{ŒÀº}V^{©À¶}C_{À¶Àº}$ is antisymmetric in $Œ©$ we have that
$$ V^{ŒÀº}V_{©Àº} = ü¶^Œ_© V^2 $$
 where the normalization follows from tracing both sides.  Similarly,
$$ V^{ŒÀº}W_{©Àº} +W^{ŒÀº}V_{©Àº} = ¶^Œ_© VÉW $$
 It then follows that
$$ V^{ŒÀº}W_{©Àº}V^{©À¶} = (¶^Œ_© VÉW -W^{ŒÀº}V_{©Àº})V^{©À¶}
	= VÉW V^{ŒÀ¶} -üV^2 W^{ŒÀ¶} $$

Antisymmetry in vector indices also implies some antisymmetry in spinor
indices.  For example, the antisymmetric Maxwell field strength
$F_{ab}=-F_{ba}$, after translating vector indices into spinor, can be
separated into its parts symmetric and antisymmetric in undotted indices;
antisymmetry in vector indices (now spinor index pairs) then implies the
opposite symmetry in dotted indices:
$$ F_{ŒÀ©,ºÀ¶} = -F_{ºÀ¶,ŒÀ©} = \f14 (F_{(Œº)[À©À¶]} +F_{[Œº](À©À¶)}) =
	ÐC_{À©À¶}f_{Œº} +C_{Œº}Ðf_{À©À¶},ââf_{Œº} = üF_{ŒÀ©,º}{}^{À©} $$
 Thus, an antisymmetric tensor also can be written in terms of a
(complex) 2$ð$2 matrix.  (However, our normalization of tensor 
vs.\ symmetric spinor matrix will vary according to application.)

We also need to define complex (hermitian) conjugates carefully because
$C$ is imaginary, and uses indices consistent with transformation
properties:
$$ \boxeq{ ÐÆ^{ÀŒ} ­ (Æ^Œ)*âÜâÐÆ_{ÀŒ} = -(Æ_Œ)*,â(Æ^Œ Æ_Œ)ÿ = ÐÆ^{ÀŒ}ÐÆ_{ÀŒ} } $$
$$ ÑV^{ŒÀº} ­ (Vÿ)^{ŒÀº} ­ (V^{ºÀŒ})*âÜâ Ðx^{ŒÀº} = x^{ŒÀº} $$
 where we assume the spinor is fermionic (when re-ordering for
hermitian conjugation), and
we have used the spacetime coordinates as an example of a real
vector (hermitian 2$ð$2 matrix).  (Sometimes we will drop the
``$Ð{\phantom n}$" on $ÐÆ^{ÀŒ}$, since it is redundant to the 
``$À{\phantom n}$".  Note that, unlike SU(2), $ÐÆ_Œ±(Æ^Œ)*$.)
In general, hermitian conjugation
properties for any Lorentz representation are defined by the
corresponding product of spinors:  For example,
$$ (Æ^{(Œ}^{º)})ÿ = Ѝ^{(ÀŒ}ÐÆ^{Àº)} = -ÐÆ^{(ÀŒ}Ѝ^{Àº)}âÜâ
	Ðf^{ÀŒÀº} ­ -(f^{Œº})*  $$
 More generally, we find
$$ (T^{(Œ_1...Œ_j)(Àº_1...Àº_k)})ÿ ­ 
	(-1)^{j(j-1)/2 +k(k-1)/2}ÐT^{(º_1...º_k)(ÀŒ_1...ÀŒ_j)} $$

As we'll see later, most spinor algebra involves, besides spinors, just
vectors and antisymmetric tensors, which carry only two spinor indices,
so matrix algebra is often useful.  When using bra-ket notation for
2-component spinors, it is often convenient to distinguish undotted and
dotted spinors.  Furthermore, since spinor indices can be raised and
lowered, we can always choose the bras to carry upper indices and the
kets lower, consistent with our index-contraction conventions, to avoid
extra signs and factors of $C$.  We therefore define (see subsection IB1)
$$ \boxeq{ \eqalign{ ÒÆ| = Æ^ŒÒ{}_Œ|,ââ|ÆÔ = |{}^ŒÔÆ_Œ;ââ
	& [Æ| = Æ^{ÀŒ}[{}_{ÀŒ}|,ââ|Æ] = |{}^{ÀŒ}]Æ_{ÀŒ} \cr
	V = |{}^ŒÔV_Œ{}^{Àº}[{}_{Àº}|âÜâV* = -|{}^{ÀŒ}]V^º{}_{ÀŒ}Ò{}_º|;â
	& f = |{}^ŒÔf_Œ{}^ºÒ{}_º|âÜâf* = |{}^{ÀŒ}]f_{ÀŒ}{}^{Àº}[{}_{Àº}| \cr}} $$
where we now use ``angle brackets" to denote the undotted spinor basis and ``square brackets" to denote the dotted.  As a result, we also have
$$ \boxeq{ \eqalign{ ÒÆÔ = ҍÆÔ = Æ^Œ _Œ,â
	[ƍ] = & Æ^{ÀŒ}_{ÀŒ};ââÒƍÔÿ = [ƍ] \cr
	ÒÆ|V|] = Æ^Œ V_Œ{}^{Àº} _{Àº},ââ & ÒÆ|f|Ô = Æ^Œ f_Œ{}^º _º \cr
	VW* +WV* = & (VÉW)I \cr}} $$
 where we have used the anticommutativity of the spinor fields.  From
now on, we use this notation for the matrix representing a vector $V$
($V_Œ{}^{Àº}$), rather than the one with which we started ($V^{ŒÀº}$).

\x IIA5.4  Consider the generators
$$ G_Œ{}^º = x^{ºÀ©}»_{ŒÀ©} +|{}^ºÔÒ{}_Œ| $$
 and their Hermitian conjugates, where $»_{ŒÀº}=»/»x^{ŒÀº}$.  Show their
algebra closes.  What group do they generate?  Find a subset of these
generators that can be identified with (a representation of) the Lorentz
group.

Since we have exhausted all possible linear transformations on spinors
(except for scale, which relates to conformal transformations), the only
way to represent discrete Lorentz transformations is as antilinear ones:
$$ Æ'_Œ = å2n_Œ{}^{Àº}ÐÆ_{Àº}ââ(Æ' = -å2nÆ*) $$
 From its index structure we see that $n$ is a vector, representing the
direction of the reflection.  The product of two identical reflections is
then, in matrix notation
$$ Æ'' = 2n(nÆ*)* = n^2 ÆâÜân^2 = à1 $$
 where we have required closure on an SL(2,C) transformation ($à$1). 
Thus $n$ is a unit vector, either spacelike or timelike.  Applying the same
transformation to a vector, where $V^{ŒÀº}$ transforms like $Æ^Œ ^{Àº}$,
we write in matrix notation
$$ V' = -2nV*n = n^2 V - 2(nÉV)n $$
 (The overall sign is ambiguous, and depends on whether it is a polar or
axial vector.)  This transformation thus describes parity (actually CP,
because of the complex conjugation).  In particular, to describe purely CP
without any additional rotation (i.e., exactly reflection of the 3 spatial
axes), in our basis we must choose a unit vector in the time direction,
$$ å2n^{ŒÀº} = ¶^{ŒÀº}âÜâÆ'^Œ = ÐÆ_{ÀŒ}ââ(ÐÆ'^{ÀŒ} = -Æ_Œ) $$
$$ ÜâV'^{ŒÀº} = -V_{ºÀŒ} $$
 which corresponds to the usual in vector notation, since in our basis
$$ §_a^{ŒÀº} = §^a_{ºÀŒ} $$
 To describe time reversal, we need a transformation that does not
preserve the complex conjugation properties of spinors:  For example,
CPT is
$$ Æ'^Œ = Æ^Œ,âÐÆ'^{Àº} = -ÐÆ^{Àº}âÜâV' = -V $$
 (The overall sign on $V$ is unambiguous.)

In principle, whenever we work on a problem with both spinors and
vectors we could use a mixed vector-spinor notation, converting between
the usual basis for vectors and the spinor-index basis with identities such
as
$$ §^a_{ŒÀº}§_b^{ŒÀº} = ¶_a^b,ââ§^a_{ŒÀº}§_a^{©À¶} = ¶_Œ^© ¶_{Àº}^{À¶} $$
 However, in practice it's much simpler to use spinor indices exclusively,
since then one needs no $§$-matrix identities at all, but only the trivial
identities for the matrix $C$ that follow from its antisymmetry.  For
example, converting the vector index on the $§$ matrices themselves into
spinor indices ($a£ŒÀº$), they become trivial:
$$ (§^{ŒÀº})_{©À¶} = ¶^Œ_© ¶^{Àº}_{À¶} $$
 (This is the same as saying an orthonormal basis of vectors has the
components $(V^a)_b = ¶^a_b$ when the components are defined with
respect to the same basis.)

Thus, the most general irreducible (finite-dimensional) representation of
SL(2,C) (and thus SO(3,1)) has an arbitrary number of dotted and undotted
indices, and is totally symmetric in each: $A_{(Œ_1...Œ_m)(Àº_1...Àº_n)}$.
Treating a vector index directly as a dotted-undotted pair of indices (e.g.,
$a=ŒÀŒ$, which is just a funny way of labeling a 4-valued index), we can
translate into spinor notation the two constant tensors of SO(3,1):  Since
the only constant tensor of SL(2,C) is the antisymmetric symbol, they can
be expressed in terms of it:
$$ \boxeq{ ú_{ŒÀŒ,ºÀº} = C_{Œº}C_{ÀŒÀº},ââ·_{ŒÀŒ,ºÀº,©À©,¶À¶} 
	= i(C_{Œº}C_{©¶}C_{ÀŒÀ¶}C_{ÀºÀ©} -C_{Œ¶}C_{º©}C_{ÀŒÀº}C_{À©À¶}) } $$
 When we work with just vectors, these can be expressed in matrix
language:
$$ \boxeq{ \eqalign{ VÉW = &¼ tr(VW*) \cr
	·_{abcd}V^a W^b X^c Y^d ­ ·(V,W,X,Y) 
		& = i¼tr(VW*XY* -Y*XW*V) \cr }} $$
 (We have assumed real vectors; for complex vectors we should really write
$VÉW*=...$, etc.)

\x IIA5.5  Prove this expression for the $·$ tensor (in either index or
matrix version) agrees with that defined in subsection IB3 (as modified in
subsection IB5) by (1) showing total antisymmetry, (2) explicitly
evaluating a nonvanishing component.

Ü6. Dirac

The Dirac spinor we encountered in subsection IC1 is a 4-component reducible
representation in D=4:  in terms of two (``left" and ``right")
two-component spinors,
$$ ï = \pmatrix{ Æ_{LŒ} \cr ÐÆ_{RÀŒ} \cr} $$
 The Hermitian metric $ç$ that defines the (Lorentz-invariant) Dirac
spinor inner product 
$$ Ðïï = ïÿçï = Æ_L^Œ Æ_{RŒ} +h.c.,ââÐï ­ ïÿç = (Æ_R^Œ¼ÐÆ_L^{ÀŒ}) $$
 takes the simple form
$$ ç = \pmatrix{ 0 & ÐC^{ÀŒÀº} \cr C^{Œº} & 0 \cr} = å2©_0 $$
 The Dirac matrices are given by
$$ ÖV ­ ©ÉV = \pmatrix{ 0 & V_Œ{}^{Àº} \cr V^º{}_{ÀŒ} & 0 \cr}
	=\pmatrix{ 0 & V \cr -V* & 0 \cr} $$
 where the indices have been chosen to insure that the $©$ matrices
always take a Dirac spinor to the same type of spinor.  Since
$ÓÖV,ÖWÕ=-VÉW$, the $©$ matrices satisfy 
$$ Ó©^a,©^bÕ = -ú^{ab} $$
  The extra sign is the result of normalizing the $©$'s to be
pseudohermitian with respect to the metric: $ç©ÿç^{-1}=+©$.  This Dirac
spinor can be made irreducible by imposing a reality condition that
relates $Æ_L$ and $Æ_R$:  The resulting ``Majorana spinor" is then
$$ ï = \f1{å2}\pmatrix{ Æ_Œ \cr ÐÆ_{ÀŒ} \cr} $$

The product of all the $©$'s is a pseudoscalar, and an additional $©$-matrix:
$$ ©_{-1} = \f{2å2}{4!}·_{abcd}©^a ©^b ©^c ©^d =
	\f1{å2}\pmatrix{ -i¶_Œ^º & 0 \cr 0 & i¶_{ÀŒ}^{Àº} \cr} $$
$$ Üâ Ó©_{-1},©_aÕ = 0,â Ó©_{-1},©_{-1}Õ = -1 $$
 (This is usually called ``$©_5$" in the literature for $D=4$, or ``$©_D$" for
$D±4$.  We have renamed it for consistency with dimensional reduction.)
It can be used to project a Dirac or Majorana spinor onto its two
two-component spinors:
$$ ¸_à = ü(Iàå2i©_{-1}) = \tat{I}000,¼\tat000{I} $$

Various identities for these matrices can be derived directly from the
anticommutation relations:  For example,
$$ ©^a ©_a = -2,ââ©^a Öa ©_a = Öa,ââ©^a ÖaÖb ©_a = aÉb,ââ
	©^a ÖaÖbÖc ©_a = ÖcÖbÖa $$
$$ tr (I) = 4,ââtr (ÖaÖb) = -2aÉb,ââtr (ÖaÖbÖcÖd) = aÉb¼cÉd +aÉd¼bÉc -aÉc¼bÉd $$
 The trace identities follow from the fact that the only way to get a
nonvanishing trace out of a product of $©$ matrices is when there are
terms proportional to the identity; since $Ó©^a,©^bÕ=-ú^{ab}$, this only
happens when the indices are pairwise identical.  The above results then
follow from examination of relevant special cases.  (Traces of odd
numbers of $©$ matrices vanish.  An exception is $©_{-1}$, until it is
rewritten in terms of its definition as the product of the other
$©$-matrices.)

Although use of the anticommutation relations is convenient for
generalization of such identities to arbitrary dimensions, 2-spinor
bra-ket notation is easier for deriving 4D identities.  Since a Dirac spinor
is the direct sum of a Weyl spinor and its complex conjugate, we write
$$ ï = |{}^ŒÔÆ_{LŒ} +|{}^{ÀŒ}]ÐÆ_{RÀŒ},ââ
	Ðï = Æ_R^ŒÒ{}_Œ| +ÐÆ_L^{ÀŒ}[{}_{ÀŒ}| $$
 In this notation, there is no need to use a spinor metric $ç$, just as in
Minkowski 4-vector bra-ket notation there is no need for an explicit
matrix to represent the Minkowski metric:  It is included implicitly in the
definition of the inner product for the basis elements
($Ò{}_a|{}_bÔ=ú_{ab}$ or $Ò{}_Œ|{}_ºÔ=C_{Œº}$).  Thus hermitian
conjugation is automatically pseudohermitian conjugation, etc.:  $Ðï$ ÓisÕ
$ïÿ$, from the effect of hermitian conjugating the basis vectors along
with the components of the spinors they multiply.  (See subsections
IB4-5.)  We then have simply
$$ ©_{ŒÀº} = -|{}_ŒÔ[{}_{Àº}| -|{}_{Àº}]Ò{}_Œ|;ââ
	¸_+ = |{}^ŒÔÒ{}_Œ|,ââ¸_- = |{}^{ÀŒ}][{}_{ÀŒ}| $$
 where we have replaced the vector index $a£ŒÀº$ on $©_a$.

\x IIA6.1 Use this representation for the $©$ matrices and projection
operators $¸_à$ for all of the following:
ªa Derive
$$ ©^{ŒÀº}Öa_1òÖa_{2n+1}©_{ŒÀº} = Öa_{2n+1}òÖa_1 $$
$$ ©^{ŒÀº}Öa_1òÖa_{2n}©_{ŒÀº} = 
	-üI¼tr(Öa_1òÖa_{2n}) -©_{-1}¼tr(©_{-1}Öa_1òÖa_{2n}) $$
ªb Rederive the trace identities above.  (Hint:  For the last identity, use
the identity $C_{[Œº}C_{©]¶}=0$ repeatedly.) 
ªc Show that
$$ tr[(¸_+ -¸_-)©_{ŒÀŒ}©_{ºÀº}©_{©À©}©_{¶À¶}] = -i·_{ŒÀŒ,ºÀº,©À©,¶À¶} $$
 by comparison with the expression of the previous subsection for $·$.

\x IIA6.2 Again using this representation:
ªa Show that $©_{[a}©_{b]}$ (up to a proportionality constant)
generates the usual Lorentz transformations of SL(2,C) on the 2 2-component
spinors in the Dirac spinor.
ªb Relate this representation of the $©$ matrices to the defining
representation of Sp(4) as given in subsection IB5, noting that Sp(4)
is the covering group of SO(3,2) (subsection IC5).

Ü7. Chirality/duality

$¸_à$ are often called ``chiral projectors"; 2-component spinors (not
paired into Dirac spinors) are often called ``chiral spinors", and appear in
``chiral theories"; the two 2-component spinors of a Dirac spinor are often
labeled as having left and right ``chirality"; etc.  When these two halves
decouple, a theory can have a ``chiral symmetry"
$$ Æ'_Œ = e^{iÏ}Æ_Œ $$
 Since chirality is closely related to parity (chiral spinors can represent
CP, but need to be doubled to allow C, and thus P), Dirac spinors are often
used to describe theories where parity is preserved, or softly broken, or
to analyze parity violation specifically, using $©_{-1}$ to identify it.

A similar feature appears in electrodynamics.  We first translate the
theory into spinor notation:  The Maxwell field strength $F_{ab}$ is
expressed in terms of the vector potential (``gauge field") $A_a$, with a
``gauge invariance" in terms of a ``gauge parameter" $Â$ with spacetime
dependence.  The gauge transformation $¶A_a = -»_a Â$ becomes
$$ A'_{ŒÀº} = A_{ŒÀº} - »_{ŒÀº} $$
 where $»_{ŒÀº}=»/»x^{ŒÀº}$.  It leaves invariant the field strength
$F_{ab} = »_{[a}A_{b]}$:
$$ F_{ŒÀ©,ºÀ¶} = »_{ŒÀ©}A_{ºÀ¶}-»_{ºÀ¶}A_{ŒÀ©}
	= \f14 (F_{(μ)[˩˦]} +F_{[μ](˩˦)}) $$
$$ = ÐC_{À©À¶}f_{Œº} +C_{Œº}Ðf_{À©À¶},âf_{Œº} = ü»_{(ŒÀ©}A_{º)}{}^{À©} $$
 Maxwell's equations are
$$ »^º{}_{À©}f_{ºŒ} ¾ J_{ŒÀ©} $$
 They include both the field equations (the hermitian part) and the
``Bianchi identities" (the antihermitian part).

\x IIA7.1  We already saw $VW*+WV*$ gave the dot product; show how
$VW*-WV*$ is related to the ``cross product" $V_{[a}W_{b]}$.

\x IIA7.2  Write Maxwell's equations, and the expression for the field
strength in terms of the gauge vector, in 2$ð$2 matrix notation, without
using
$C$'s.  Combine them to derive the wave equation for $A$.

Maxwell's equations now can be easily generalized to include magnetic
charge by allowing the current $J$ to be complex.  (However, the
expression for $F$ in terms of $A$ is no longer valid.)  This is because the
``duality transformation" that switches electric and magnetic fields is
much simpler in spinor notation:  Using the expression given above for the
4D Levi-Civita tensor using spinor indices,
$$ F_{ab}' = ü·_{abcd}F^{cd}âÜâf_{Œº}' = -if_{Œº} $$
 More generally, Maxwell's equations in free space (but not the
expression for $F$ in terms of $A$) are invariant under the continuous
duality transformation 
$$ f_{Œº}' = e^{iÏ}f_{Œº} $$
 (and $J_{ŒÀº}'=e^{iÏ}J_{ŒÀº}$ in the presence of both electric and magnetic
charges).

\x IIA7.3  Prove the relation between duality in vector and spinor
notation.  Show that $F_{ab}+iü·_{abcd}F^{cd}$ contains only $f_{Œº}$
and not $f_{ÀŒÀº}$.

\x IIA7.4  How does complexifying $J_{ŒÀº}$ modify Maxwell's equations in
ÓvectorÕ notation?

In even time dimensions, Wick rotation kills the $i$ (or $-i$) in the
spinor-index expression for $·_{Œº',©¶',·½',úÏ'}$.  Since the (discrete and
continuous) duality transformation now contains no $i$, we can impose
self-duality or anti-self-duality; i.e., that $f_{Œº}$ or $f_{Œ'º'}$ vanishes,
since they are now independent and real instead of complex conjugates.

These continuous chirality and duality symmetries on the field strengths
generalize to the free field equations for arbitrary massless fields in four
dimensions.  For reasons to be explained in the following section, they
distinguish the two polarizations of the waves described by such fields. 
They are closely related to conformal invariance:  In higher dimensions,
where not all free, massless theories are conformal (even on the mass
shell), these symmetries exist exactly for those that are conformal.

\refs

£1 L.D. Landau and E.M. Lifshitz, ÓQuantum mechanics, non-relativistic
	theoryÕ, 2nd ed. (Pergamon, 1965) ch. VIII:\\
	review of SU(2) spinor notation.
 £2 B. van der Waerden, ÓG¬ottinger NachrichtenÕ (1929) 100:\\
	SL(2,C) spinor notation.
 £3 E. Majorana, ÓNuo. Cim.Õ É14 (1937) 171.
 £4 A. Salam, ÓNuo. Cim.Õ É5 (1957) 299;\\
	L. Landau, ÓJETPÕ É32 (1957) 407, ÓNucl. Phys.Õ É3 (1957) 27;\\
	T.D. Lee and C.N. Yang, ÓPhys. Rev.Õ É105 (1957) 1671:\\
	identification of neutrino with Weyl spinor.

\unrefs

\sectskip\bookmark7{B. POINCAR\noexpandÉ}\secty{B. POINCAR«E}

The general procedure for finding arbitrary representations of the
Poincar«e group relevant to physics is to:   
\item{(1)} Describe spin 0.  As we have
seen, this means starting with the coordinate representation, which is
reducible, and apply the constraint $p^2+m^2=0$ to get an irreducible
one.   
\item{(2)} Find arbitrary, finite-dimensional, irreducible representations of
the Lorentz\break group.  This we have done in the previous section.   
\item{(3)} Take
the direct product of these two representations of the Poincar«e group,
which give the orbital and spin parts of the generators.  (The spin part of
translations vanishes.)  We then need a further constraint to pick out an
irreducible unitary piece of this product, which is the subject of this section.

Ü1. Field equations

We have already constrained the momentum:  The equation
$$ \boxeq{p^2+m^2=0} $$
 as an operator equation acting on a field or wave function
is the ``Klein-Gordon (or relativistic Schr¬odinger) equation".  States or
fields that satisfy their field equations are called
``on-(mass-)shell", while those that don't (or for which the equations
haven't been imposed) are ``off-shell".

The next step is to constrain the ``spin" (actually, its Lorentz generalization).
The basic idea of the extra constraint is very simple:  The Lorentz group 
introduces states of negative probability, since the Minkowski space
metric is indefinite.  For example, if we write the naive Lorentz invariant
Hilbert-space norm for a vector wave function, the time component
will have negative probability.  
(Similar remarks apply to spinors, e.g., for the metric $義^0$ for the
Dirac spinor.)
The solution to this problem, in
first-quantized operator language, is to constrain the spin
to eliminate the negative-metric
component, in analogy to the way we have already constrained the
momentum by the Klein-Gordon equation.  We thus impose
$$ \boxeq{S_a{}^b p_b +wp_a = 0} $$
 to kill the part of the Lorentz generators in the direction of the
momentum,
 where ``$w$" is a constant to be determined.  (Its term can be
attributed to ordering ambiguities.)  This
equation is the general field equation for all spins (acting on the field
strength), in addition to the Klein-Gordon equation (which is redundant
except for spin 0).

We will see that this constraint is appropriate for massless particles.
Massive particles then follow from dimensional
reduction: adding a further spatial dimension and fixing its component
of momentum to a constant, the mass, so $p^2£p^2+m^2$.  

Before examining this constraint, we first give an alternative ``derivation"
based on the conformal group.  
The earlier derivation of the massless particle from the
conformal particle for spin 0 can be generalized to all ``spins", i.e., all
representations of the Poincar«e group in arbitrary dimensions.  There is a
way to do this in terms of classical mechanics for all representations of
the conformal group, by generalizing the description of the classical
spinning particle.  However, by analyzing the conformal particle quantum
mechanically instead, applying a set of constraints, it will be clear how to
generalize from conformal particles to general massless particles by
weakening the constraints.  The general idea is that the symmetry group
for massive particles is the Poincar«e group, while that for massless
particles includes also scale transformations, and finally conformal
particles have also conformal boosts.  So, starting with the conformal
group and dropping anything to do with conformal boosts will give
massless particles.  

We begin with a general representation of the conformal group SO(D,2) in
terms of generators $G_{AB}$, where $A,B$ are D+2-component
vector indices.  We then impose constraints that are the conformally
covariant form of $p^2=0$:  Identifying 
$$ (G^{+a},G^{ab},G^{+-},G^{-a}) = (P^a,J^{ab},ë,K^a) $$
 (where $A=(à,a)$) as the generators for translations, Lorentz
transformations, dilatations, and conformal boosts, we see that
$$ \G^{AB} = üG^{C(A}G_C{}^{B)} 
	-\f1{D+2}ú^{AB}G^{CD}G_{CD} = 0 $$
 is an irreducible piece of the product $GG$ (symmetric and traceless) and
includes:
$$ (\G^{++},\G^{+a},\G^{ab},\G^{+-},\G^{-a},
	\G^{--}) = (P^2,üÓJ^{ab},P_bÕ+üÓë,P^aÕ,...) $$
 where ``$...$" all have terms containing $K^a$.

\x IIB1.1 Work out all the $\G$'s in terms of $P$, $J$, $ë$, and $K$.

In general theories, even massless ones, it is not always possible to have
invariance under conformal boosts.  (We'll see examples of this
insubsection IXA7.)  However, all massless theories are scale invariant, at
least at the free level.  (In D=4, free massless theories can always be
made conformal on shell.  However, the fact that even these theories can
have actions that are not invariant under conformal boosts proves that it
is sufficient to add just dilatations to the Poincar«e group.  Furthermore,
the fact that conformal boosts are not always an invariance in D>4 means
that dropping them will give results in a dimension-independent form.) 
Therefore, only $\G^{++}$ and $\G^{+a}$ can be defined in general
massless theories, but we'll see that these are sufficient to define the
kinematics.  The former is just the masslessness condition, which we
used to pick the constraints in the first place.  

As we saw earlier, $ë$ just scales $x^a$:  We can therefore write the
relevant generators as
$$ P^a = »^a,ââJ^{ab} = x^{[a}»^{b]} +S^{ab},ââ
	ë = üÓx^a,»_aÕ +w-1 = x^a »_a +w +\f{D-2}2 $$
 (We have used the antihermitian form of the generators.)  The ``scale
weight" $w+\f{D-2}2$ is the real ``spin" part of $ë$, just as $S_{ab}$ is
the spin part of the angular momentum $J_{ab}$.  To preserve the algebra
it must commute with everything, and thus we can set it equal to a
constant on an irreducible representation.  We'll see shortly that its value
is actually determined by the spin $S_{ab}$.  It is the engineering
dimension of the corresponding field.  It has been normalized for later
convenience; the value of $w$ depends on the representation of $S^{ab}$,
but is independent of $D$.  The dilatation generator $ë$ is not exactly
antihermitian because the integration measure $d^D x$ isn't invariant
under scaling.  This is another reason $w$ is determined, by the free
action.  The form we have given preserves reality of fields.  The
commutation relations for the spin parts, and the total generators, are
the same as those for the orbital parts; e.g.,
$$ [S_{ab},S^{cd}] = -¶_{[a}^{[c}S_{b]}{}^{d]} $$
 (A convenient mnemonic for evaluating this commutator in general is to
use $S^{ab}£x^{[a}»^{b]}$ instead.)

\x IIB1.2 We can also use this method to find the stronger conditions
for the fully conformal case:
 ªa Find an expression for $K^a$ in terms of $x$, $»$, $S$, and $w$
that preserves the commutation relations.  
 ªb Evaluate all the constraints
$\G$, and express the independent ones in terms of just $»$, $S$, and $w$
(no $x$).

Substituting the explicit representation of the generators into the
constraint $\G^{+a}$, and using the former constraint $P^2=0$ (when
acting on wave functions on the right), we find that all $x$ dependence
drops out, leaving for $\G^{+a}$ the condition
$$ S_a{}^b »_b +w»_a = 0 $$
 (paying careful attention to quantum mechanical ordering).

\x IIB1.3  Define spin for the conformal group by starting in D+2
dimensions:  In terms of the (D+2)-dimensional coordinates $y^A$ and
their derivatives $»_A$,
$$ G_{AB} = y_{[A}»_{B]} +S_{AB} $$
 Besides the previous conditions
$$ y^2 = »^2 = Óy^A,»_AÕ = 0 $$
 impose the constraints, in analogy to the D-dimensional field equations,
and taking into account the symmetry between $y$ and $»$,
$$ S_A{}^B y_B +wy_A = S_A{}^B »_B +w»_A = 0 $$
 ªa Show that the algebra of constraints closes, if we include the additional
constraint
$$ üS_{(A}{}^C S_{B)C} +w(w+\f{D}2)ú_{AB} = 0 $$
 ªb Solve all the constraints with explicit $y$'s for everything with an upper
``$-$" index, reducing the manifest symmetry to SO(D$-$1,1), in analogy to
the way $y^2=0$ was solved to find $y^-$.  
 ªc Write all the conformal
generators in terms of $x_a$, $»_a$, $S_{ab}$, and $w$.

Ü2. Examples

We now examine the constraints $S_a{}^b »_b+w»_a=0$ in more detail. 
We begin by looking at some simple (but useful) examples.  The simplest
case is spin¼0:
$$ S_{ab} = 0âÜâw = 0 $$
 The next simplest case (for arbitrary dimension) is the Dirac spinor
(see subsections IC1 and IIA6):
$$ S_{ab} = -ü[©_a,©_b]âÜâ
	S_a{}^b »_b +w»_a = -©_a ©^b »_b +(w-ü)»_a $$
$$ Üâ©^a »_a = 0,âw = ü $$
 where we have separated out the pieces of the constraint that are
irreducible with respect to the Lorentz group (e.g., by multiplying on the
left with $©^a$).  This gives the (massless) ``Dirac equation" $Ö»ï=0$.  The
next case is the vector:  In terms of the basis $|VÔ=V^a|{}_aÔ$, the spin is
(see subsection IB5)
$$ S_{ab} = |{}_{[a}ÔÒ{}_{b]}| $$
 However, the vector yields just another description of the scalar:

\x IIB2.1  Apply the field equations for general field strengths to the case
of a vector field strength. 
ªa  Find the independent field equations (assuming the field strength is
not just a constant)
$$ »_{[a}F_{b]} = 0,ââ»^a F_a = 0,ââw = 1 $$
 Note that solving the first equation determines the vector in terms of a
scalar, while the second then gives the Klein-Gordon equation for that
scalar, and the third fixes the weight of the scalar to be the same as that
found by starting with a scalar field strength. 
ªb  Lorentz covariantly solve the second equation first 
to find a gauge field that is not a scalar.

All other representations can be built up from the spinor
and vector.  As our final example, we consider the case where the field is
a 2nd-rank antisymmetric tensor:  Using the direct product
representation (applied as in subsection IB2 given the vector
representation)
$$ F = F^{ab}|{}_aÔ°|{}_bÔ,â
	S_{ab}(|{}_cÔ°|{}_dÔ) = (S_{ab}|{}_cÔ)°|{}_dÔ + |{}_cÔ°S_{ab}|{}_dÔ $$
$$ Üâ(S_{ab}F)^{cd} = ¶_{[a}^{[c}F_{b]}{}^{d]} $$
 we find the equations
$$ (S_a{}^b »_b +w»_a)F_{cd} =
	ü»_{[a}F_{cd]} -ú_{a[c}»^b F_{d]b} +(w-1)»_a F_{cd} $$
$$ Üâ»_{[a}F_{bc]} = »^b F_{ab} = 0,ââw = 1 $$
 which are Maxwell's equations, again separating out irreducible pieces
(e.g., by tracing and antisymmetrizing).  

\x IIB2.2  Verify the representation of Lorentz spin given above for
$F_{ab}$ by finding the commutation relations implied by this
representation.

\x IIB2.3  Use the definition of the action of the Lorentz generators on a vector
in vector and spinor notations,
$$ S_{ab} = |{}_{[a}ÔÒ{}_{b]}|,â|{}_aÔ = |{}_ŒÔ°|{}_{ÀŒ}Ô $$
$$ S_{Œº} = |{}_{(Œ}ÔÒ{}_{º)}|,âS_{ÀŒÀº} = |{}_{(ÀŒ}ÔÒ{}_{Àº)}|,â $$
 to derive
$$ S_{ŒÀŒ,ºÀº} = -ü(C_{Œº}ÐS_{ÀŒÀº} +ÐC_{ÀŒÀº}S_{Œº}) $$

\x IIB2.4  Consider the field equations in 4D spinor notation for a general
field strength, totally symmetric in its $m$ undotted indices and $n$
dotted indices,
$$ S_Œ{}^º »_{ºÀ©} -m»_{ŒÀ©} = ÐS_{ÀŒ}{}^{Àº}»_{©Àº} -n»_{©ÀŒ} = 0,ââ
	w = ü(m+n) $$
 ªa  Show this implies
$$ »^{ŒÀ©}Æ_{Œ...Àº...} = »^{©Àº}Æ_{Œ...Àº...} = 0 $$
 ªb  Translate the field equations into vector notation (in terms of
$S_{ab}$), finding $S_a{}^b »_b +w»_a=0$ and an axial vector
equation.
 ªc  Show that the two equations are equivalent by deriving the
equations of part {\bf a} from $S_a{}^b »_b +w»_a=0$ alone, and from the
axial equation alone (except that the axial equation doesn't work for
the cases $m=n$).

In each case, choosing the wrong scale weight $w$ would imply the field
was constant.  Note that we chose the field ÓstrengthÕ $F_{ab}$ to
describe electromagnetism:  The arguments we used to derive field
equations were based on physical degrees of freedom, and did not take
gauge invariance into account.  In chapter XII we use more powerful
methods to find the gauge covariant field equations for the gauge fields,
and their actions.

Ü3. Solution

Free field equations can be solved easily in momentum space.  Then the
simplest way to do the algebra is in the ``lightcone frame".  This is a
reference frame, obtained by a Lorentz transformation, where a massless
momentum takes the simple form 
$$ p^a = ¶^a_+ p^+ $$
 (using only rotations), or the even simpler form $p^a=à¶^a_+$ (using also
a Lorentz boost), where again $à$ is the sign of the energy.  In that frame
the general field equation $S_a{}^b »_b+w»_a=0$ reduces to
$$ S^{-i} = 0,ââw =  S^{+-} $$
 The constraint $S^{-i}=0$ determines $S^{+-}$ to take its maximum
possible value within that irreducible representation, since $S^{-i}$ is the
raising operators for $S^{+-}$:  For any eigenstate of $S^{+-}$,
$$ S^{+-}|hÔ = h|hÔâÜâS^{+-}(S^{-i}|hÔ) = 
	(S^{-i}S^{+-} +[S^{+-},S^{-i}])|hÔ = (h+1)(S^{-i}|hÔ) $$
 The remaining constraint then determines $w$:  It is the maximum value
of $S^{+-}$ for that representation.  By parity ($+ª-$), $-w$ is the
minimum, so
$$ w ³ 0;ââw = 0âÛâS^{ab} = 0 $$
 since if $S^{+-}=0$ for all states then $S^{ab}=0$ by Lorentz
transformation.  As we have seen by other methods (but can easily be
derived by this method), $w=ü$ for the Dirac spinor and $w=1$ for the
vector; since general representations can be built from reducing direct
products of these, we see that $w$ is an integer for bosons and
half-integer for fermions.  If we describe a general irreducible
representation by a Young tableau for SO(D$-$1,1) (with tracelessness
imposed), or a Young tableau times a spinor (with also $©$-tracelessness
$©^a Æ_{a...b}=0$), then it is easy to see from the results for the spinor
and vector, and antisymmetry in rows, that $w$ is simply the number of
columns of the tableau (its ``width"), counting a spinor index as half a
column:  $S^{+-}$ just counts the maximum number of ``$-$" indices that
can be stuck in the boxes describing the basis elements.  (In fact, Dirac
spinor $°$ Dirac spinor gives just all possible 1-column representations.)

This leaves undetermined only $S^{ij}$ and $S^{+i}$.  However, $S^{+i}$
(``creation operator") is canonically conjugate to $S^{-i}$ (``annihilation
operator"), so its action has also been fixed:
$$ [S^{-i},S^{+j}] = ¶^{ij}S^{+-} +S^{ij} $$
 ($S^{ij}$ vanishes for $i=j$, so $S^{+i}$ and $S^{-i}$ are conjugate, though
not ``orthonormal".  The constant $S^{+-}$ was fixed above to be
nonvanishing, except for the trivial case of spin 0.)  Equivalently,
$S^{ij}$ preserves $S^{-i}=0$, while $S^{+i}$ doesn't:  $S^{ij}$ are the
only nontrivial spin operators acting within the subspace satisfying
the constraint.

Thus only the ``little group" SO(D$-$2) spin $S^{ij}$ remains nontrivial: 
The original irreducible representation of SO(D$-$1,1) Lorentz spin
$S^{ab}$ was a reducible representation of SO(D$-$2) spin $S^{ij}$; the
irreducible SO(D$-$2) representation with the highest value of $S^{+-}$ is
picked out of this SO(D$-$1,1) representation.  This solution also gives the
field strength in terms of the gauge field:  Working with just the
highest-$S^{+-}$-weight states is equivalent to working with the gauge
field, up to factors of $»^+$.  

As an explicit example, for spin 1/2 we have simply $©^- ï=0$, which kills
half the components, leaving the half given by $©^+ ï$.  For spin 1, we
find
$$ \left. \matrix{ p^b F_{ab} = 0âÜâF^{-a} = 0 \cr
	p^{[a}F^{bc]} = 0âÜâonly¼F^{+a} ± 0 \cr}â\rightÕâ
	Üâonly¼F^{+i} ± 0 $$
 In the ``lightcone gauge" $A^+=0$, we have $F^{+i}=»^+ A^i$, so the
highest-weight part of $F^{ab}$ is the transverse part of the gauge field. 
The general pattern, in terms of field strengths, is then to keep only
pieces with as many as possible upper + indices and no upper $-$ indices
(and thus highest $S^{+-}$ weight).  In terms of the vector potential, we
have
$$ F^{ab} ¾ p^{[a}A^{b]}âÜâonly¼A^i ± 0 $$
 The general rule for the gauge field is to drop $à$ indices, so the field
becomes an irreducible representation of SO(D$-$2).  All + indices on the
field strength are picked up by the momenta, which also account for the
scale weight of the field strength:  All gauge fields have $w=0$ for bosons
and $w=ü$ for fermions.

\x IIB3.1 Using only the anticommutation relations $Ó©^a,©^bÕ=-ú^{ab}$,
construct projection operators from $©^à$:  These are operators $¸_I$
that satisfy
$$ ¸_I ¸_J = ¶_{IJ}¸_I¼\left(no¼Ý\right),âݸ_I = 1 $$
 Because of time reversal symmetry $©^+ª-©^-$ (or parity $©^+ª©^-$),
these project onto two subspaces equal in size.

A method equivalent to using the lightcone frame is to perform a
unitary transformation $U$ on the spin that is the inverse of the
transformation on the coordinates/momentum that would take us to the
lightcone frame:  We want a Lorentz transformation $ñ_a{}^b$ on the
field equations, which are of the form
$$ \O_a{}^b p_b = 0,ââ\O_a{}^b = S_a{}^b +w¶_a^b $$
 that has the effect
$$ U \O_a{}^b U^{-1} = ñ_a{}^c \O_c{}^d ñ^b{}_d,ââ
	ñ^b{}_a p_b = p'_a,ââp'^a = ¶^a_+ p^+âÜ $$
$$ 0 = U \O_a{}^b p_b U^{-1} =
	ñ_a{}^c \O_c{}^d ñ^b{}_d p_b = ñ_a{}^c \O_c{}^d p'_d
	âÜâ\O_a{}^b p'_b = 0 $$
  If $|ÆÔ$ satisfies the original constraint, then $U|ÆÔ$ will satisfy the new
one.  If we like, we can always transform back at the end.  This is
equivalent to a gauge transformation in the field theory.

It is easy to check that the appropriate operator is
$$ U = e^{S^{+i}p^i/p^+} $$
 Any operator $V^a$ that transforms as a vector under $S^{ab}$,
$$ [S^{ab},V^c] = V^{[a}ú^{b]c} $$
 but commutes with $p$, is transformed by $U$ into $UVU^{-1}=V'$ as
$$ V'^+ = V^+,ââV'^i = V^i +V^+{p^i\over p^+},ââ
	V'^- = V^- +V^i{p^i\over p^+} +V^+{(p^i)^2\over 2(p^+)^2} $$
 as follows from explicit Taylor expansion, which terminates because
$S^{+i}$ act as lowering operators (as for conformal boosts in subsection
IA6).  This yields the desired result
$$ V'^a p_a = V^a p'_a +{V^+\over 2p^+}p^2 $$
 when we impose the field equation $p^2=0$.

\x IIB3.2  Check this result by performing the transformation
explicitly on the constraint.  Before the transformation, the lightcone
decomposition of the constraint is
$$ (-S^{+-}+w)p^+ +S^{+i}p^i = 0 $$
$$ -S^{i-}p^+ +S^{ij}p^j +wp^i -S^{i+}p^- = 0 $$
$$ S^{-i}p^i +(-S^{-+}+w)p^- = 0 $$
 Show that after this transformation, the constraint becomes
$$ (-S^{+-}+w)p^+ = 0 $$
$$ -S^{i-}p^+ +(-S^{+-}+w)p^i -üS^{+i}{p^2\over p^+} = 0 $$
$$ S^{-i}p^i +(-S^{+-}+w)p^- -S^{+-}{p^2\over p^+}
	-üS^{+i}p^i{p^2\over p^{+2}} = 0 $$
 Clearly these imply
$$ w = S^{+-},ââS^{-i} = 0 $$
 with $p^2=0$.

On the other hand, if instead of using the lightcone identification of $x^+$
as ``time", we choose to use the usual $x^0$ for purposes of finding the
evolution of the system, then we want to consider transformations that
do not involve $p^0$, instead of not involving the ``energy" $p^-$.  Thus,
by $p^0$-independent rotations alone, the best we can do is to choose
$$ p^i = 0,ââp^1 = ¿ $$
 i.e., we can fix the value of the spatial momentum, but not in a way that
relates to the sign of the energy.  The result is then
$$ \li{ p^0 > 0: &âp^a = ¶^a_+ p^+ \cr
	p^0 < 0: &âp^a = ¶^a_- p^- \cr } $$
 The result is similar to before, but now the positive and negative energy
solutions are separated:  In this frame the field equations reduce to
$$ \li{ p^0 > 0: &âS^{-i} = 0,âS^{+-} = w \cr
	p^0 < 0: &âS^{+i} = 0,âS^{+-} = -w \cr } $$
 Thus, while $w$ takes the same value as before, now the
positive-energy states are associated with the highest weight of
$S^{+-}$, while the negative-energy ones go with the lowest weight (and
nothing between).  The unitary transformation that achieves this result is
a spin rotation that rotates $S^{ab}$ in the field equations with the same
effect as an orbital transformation that would rotate $(p^1,p^i)£(¿,0)$. 
By looking at the special case $D=3$ (where there is only one rotation
generator), we easily find the explicit transformation
$$ U = exp\left[ tan^{-1}\left({|p^i|\over p^1}\right)
	S^{1i}{p^i\over |p^i|}\right] $$

\x IIB3.3 Perform this transformation:
ªa Find the action of the above transformation on an arbitrary vector
$V^a$.  (Hint:  Look at $D=3$ to get the transformation on the
``longitudinal" part of the vector.)  In particular, show that
$$ V'^a p_a = V^a p'_a,ââp'^a = ¶^a_0 p^0 +¶^a_1 ¿ $$
ªb Show the field equations are transformed as
$$ S^{0a}p_a +wp^0âÜâ
	-¿S^{10} +wp^0 = p^0(w-\f{p^0}¿ S^{10}) -\f1¿ S^{10}p^2 $$
$$ S^{1a}p_a +wp^1âÜâ
	\f1¿ p^i(¿S^{1i}-p^0 S^{0i}) + p^1(w-\f{p^0}¿ S^{10}) $$
$$ S^{ia}p_a +wp^iâÜâ
	-[¶^{ij} -\f1{¿(¿+p^1)}p^i p^j](¿ S^{1j}-p^0 S^{0j})
	+ p^i(w-\f{p^0}¿ S^{10}) $$
 Note that the first equation gives the time-dependent Schr¬odinger
equation, with Hamiltonian
$$ H = \f1w(S^{10}p^1 -S^{0i}p^i)¼£¼\f1w S^{10}¿ $$
 This diagonalizes the Hamiltonian $H$ (in a representation where
$S^{10}$ is diagonal).  Thus the only independent equations are
$$ p^2 = 0,ââS^{10} = ·(p^0)w,ââS^{1i} - ·(p^0)S^{0i} = 0 $$
 leading to the advertised result. 
ªc Find the transformation that rotates to the $p^i$ direction instead
of the 1 direction, so
$$ H £ -\f1w S^{0i}{p_i\over |p^j|}¿ $$

Ü4. Mass

So far we have considered only massless theories.  We now introduce
masses by ``dimensional reduction", identifying mass with the component
of momentum in an extra dimension.  As with the extra dimensions used
for describing conformal symmetry, this extra dimension is just a
mathematical construct used to give a simple derivation.  (Theories have
been postulated with extra, unseen dimensions that are hidden by
``compactification":  Space curls up in those directions to a size too small
to detect with present experiments. However, no compelling reason has
been given for why the extra dimensions should want to compactify.)

The method is to: 
\item{(1)} extend the range of vector indices by one additional
spatial direction, which we call ``$-1$";  
\item{(2)} set the corresponding
component of momentum to equal the mass,
$$ p_{-1} = m $$
 and  
\item{(3)} introduce extra factors of $i$ to restore reality, since
$»_{-1}=ip_{-1}=im$, by a unitary transformation.  

\noindent Since all
representations can be constructed by direct products of the vector and
spinor, it's sufficient to define this last step on them.  For the scalar this
method is trivial, since then simply $p^2£p^2+m^2$.  
Except for the last step, the other constraint becomes
$$ \boxeq{ S_a{}^b »_b +S_{a,-1}im +w»_a = 0,â
	S_{-1}{}^a »_a +wim = 0 } $$

For the spinor, since any transformation on the spinor index can be
written in terms of the gamma matrices, and the transformation must
affect only the $-1$ direction, we can use only $©_{-1}$.  (For even
dimensions, we can identify the $©_{-1}$ of dimensional reduction with
the one coming from the product of all the other $©$'s, since in odd
dimensions the product of all the $©$'s is proportional to the identity.) 
We find
$$ U = exp(-¹©_{-1}/2å2):ââ©_{-1} £ ©_{-1},â©_a £ -å2©_{-1}©_a $$
 We perform this transformation directly on the spin operators appearing
in the constraints, or the inverse transformation on the states. 
Dimensional reduction, followed by this transformation, then modifies the
massless equation of motion as
$$ iÖ» £ iÖ»-m©_{-1} £ -å2©_{-1}(iÖ»+\f{m}{å2}) $$
 so $iÖ»ï=0£(iÖ»+\f{m}{å2})ï=0$.

The prescription for the vector is
$$ U = exp(üi¹|{}^{-1}ÔÒ{}^{-1}|):ââ
	|{}^{-1}Ô £ i|{}^{-1}Ô,âÒ{}^{-1}| £ -iÒ{}^{-1}|ââ(Ò{}^{-1}|{}^{-1}Ô = 1) $$
 with the other basis states unchanged.  This has the effect of giving each
field a $-i$ for each ($-1$)-index.  For example, for Maxwell's equations
$$ »_{[a}F_{bc]} £ \cases{ »_{[a}F_{bc]} &\cr »_{[a}F_{b]-1}+imF_{ab} &\cr}
	£ \cases{ »_{[a}F_{bc]}ââ\hbox{(redundant)} &\cr 
	-i(»_{[a}F_{b]-1}-mF_{ab}) &\cr} $$
$$ »^b F_{ab} £ \cases{ »^b F_{ab}+imF_{a-1} &\cr »^a F_{-1a} &\cr} £
	\cases{ »^b F_{ab}+mF_{a-1} &\cr 
	-i»^a F_{-1a}ââ\hbox{(redundant)} &\cr} $$
 Note that only the mass-independent equations are redundant.  Also,
$F_{a-1}$ appears explicitly as the potential for $F_{ab}$, but without
gauge invariance.  Alternatively, we can keep the gauge potential:
$$ F_{ab} = »_{[a}A_{b]} £ \cases{ F_{ab} = »_{[a}A_{b]} &\cr
	F_{a-1} = »_a A_{-1}-imA_a &\cr} £
	\cases{ F_{ab} = »_{[a}A_{b]} &\cr 
	-iF_{a-1} = -i(»_a A_{-1}+mA_a) &\cr} $$
 This is known as the ``St¬uckelberg formalism" for a massive vector,
which maintains gauge invariance by having a scalar $A_{-1}$ in addition
to the vector:  The gauge transformations are now
$$ ¶A_a = -»_a  £ \cases{ ¶A_a = -»_a  &\cr ¶A_{-1} = -im &\cr} £
	\cases{ ¶A_a = -»_a  &\cr -i¶A_{-1} = -im &\cr} $$

\x IIB4.1 Consider the general massive field equations that follow from
the general massless ones by dimensional reduction.  One of these is
$$ S_{-1}{}^a »_a +wim = 0 $$
 (before restoring reality).  This scalar equation alone gives the complete
field equations for $w$=1/2 and 1 (antisymmetric tensors), 0 being trivial.
ªa Show that for $w$=1/2 it gives the (massive) Dirac equation.
ªb Expanding the state over explicit fields, find the covariant field
equations it implies for $w$=1.  Show these are sufficient to describe
spins 0 (vector field strength: see exercise IIB2.1) and 1 ($F_{ab}$ and
$F_{a-1}$).  Note that $S^{-1a}$ act as generalized $©$ matrices (the Dirac
matrices for spin 1/2, the ``Duffin-Kemmer matrices" for $w$=1), where 
$$ S^{ab} = -[S^{-1a},S^{-1b}] $$
ªc Show that these covariant field equations imply the Klein-Gordon
equation for arbitrary antisymmetric tensors.  Show that in D=4 all
antisymmetric tensors (coming from 0-5 indices in D=5) are equivalent to
either spin 0 or spin 1, or trivial.  (Hint:  Use $·_{abcd}$.)
 ªd Consider the ÓreducibleÕ representation coming from the direct
product of two Dirac spinors, and represent the wave function itself as a
matrix:
$$ S^{ij}ï = ÷S^{ij}ï +ï÷S^{ij} $$
 where $i=(-1,a)$ and $÷S^{ij}$ is the usual Dirac-spinor representation. 
Using the fact that any 4$ð$4 (in D=4) matrix can be written as a linear
combination of products of $©$-matrices (antisymmetric products, since
symmetrization yields anticommutators), find the irreducible
representations of SO(4,1) in $ï$, and relate to part {\bf c}.

\x IIB4.2 Solve the field equations for massive spins 1/2 and 1 in
momentum space by going to the rest frame.

The solution to the general massive field equations can also be found by
going to the rest frame ($p^0=m$):  The combination of that and dimensional
reduction is, in terms of the massive analog of lightcone components,
$$ p^+ = \f1{å2}(p^0+p^{-1}) = å2m,ââp^- = \f1{å2}(p^0-p^{-1}) = 0,ââ
	p^i = 0 $$
 where $p^i$ are now the other D$-$1 (spatial) components.  This fixing of
the momentum is the same as the lightcone frame except that $p^1$ has
been replaced by $p^{-1}$, and thus $p^i$ now has D$-$1 components
instead of D$-$2.  The solution to the constraints is thus also the same,
except that we are left with an irreducible representation of the
``little group" SO(D$-$1) as found in the rest frame for the massive
particle, vs.¼one of SO(D$-$2) found in the lightcone frame for the
massless case.

Ü5. Foldy-Wouthuysen

The other frame we used for the massless analysis, which involved only
energy-independent rotations, can also be applied to the massive case by
dimensional reduction.  The result is known as the
``Foldy-Wouthuysen transformation", and is useful for analyzing
ÓinteractingÕ massive field equations in the nonrelativistic limit. 
Replacing $p^1£p^{-1}=m$ in our previous result, we have for the free
case
$$ U = exp\left[ tan^{-1}\left({|\vec pÊ|\over m}\right)
	S^{-1 i}{p^i\over |\vec pÊ|}\right],ââUHU^{-1} = \f1w S^{-1 0}¿ $$
 For purposes of generalization to interactions, it was important that in
the free transformation (1) we used only the spin part of a rotation, since
the orbital part could introduce explicit $x$ dependence, and (2) we used
only rotations, since a Lorentz boost would introduce $p^0$ dependence
in the ``parameters" of the transformation, which could generate
additional $p^0$ (time derivative) terms in the field equation. 

\x IIB5.1 Perform this transformation for the Dirac spinor, and then apply
the reality-restoring transformation to obtain
$$ H¼£¼å2©_0 ¿ $$
 We then can use the diagonal representation $©_0=\tbt{I}00{-I}/å2$. 
(We can define this representation, up to phases, by switching $©_0$ and
$©_{-1}$ of the usual representation.)  In general the reality-restoring
transformation will be unnecessary for any spin, since applying the field
equation $S^{-10}=àw$ picks out a representation of the ``little group"
SO(D$-$1).

In the interacting case the result generally can't be obtained in closed
form, so it is derived perturbatively in $1/m$.  The goal is again a
Hamiltonian diagonal with respect to $S^{-10}$, to preserve the
separation of positive and negative energies; we then can set $S^{-10}=w$
to describe just positive energies.  We thus choose the transformation to
cancel any terms in $H$ that are off-diagonal, which come from odd total
numbers of ``$-1$" and ``$0$" indices from the spin factors in any term:
i.e., odd numbers of $S^{0i}$ and $S^{-1i}$ (e.g., the $S^{0i}p_i$ term in the
original $H$).  For example, for coupling to an electromagnetic field, the
exponent of $U$ is generalized by covariantizing derivatives (minimal
coupling $»£á=»+iA$), but also requires field-strength ($E$ and $B$) terms
to cancel certain ones of those generated from commutators of these
derivatives in the transformation:
$$ á^a = »^a +iA^aâÜâ[á^a,á^b] = iF^{ab} $$

Before performing this transformation explicitly for the first few orders,
we consider some general properties that will allow us to collect similar
terms in advance.  (Few duplicate terms would appear to the order we
consider, but they breed like rabbits at higher orders.)  We start with a
field equation $\F$ that can be separated into ``even" terms $\E$ and
``odd" ones $\O$, each of which can be expanded in powers of $1/m$:
$$ \F = \E +\O:ââ\E = Ý_{n=-1}^¥ m^{-n}\E_n,ââ
	\O = Ý_{n=0}^¥ m^{-n}\O_n $$
 Note that the leading ($m^{+1}$) term is even; thus we choose only odd
generators to transform away the odd terms in $\F$, perturbatively from
this leading term:
$$ \F' = e^G \F e^{-G},ââG = Ý_{n=1}^¥ m^{-n}G_n $$
 Since $\F'$ is even while $G$ is odd, we can separate this equation into its
even and odd parts as
$$ \li{ \F' ={} & cosh(\L_G) \E +sinh(\L_G) \O \cr
		0 ={} & sinh(\L_G) \E +cosh(\L_G) \O \cr} $$
 (with $\L_G=[G,¼]$ as in subsection IA3).  Since we can perturbatively
invert any Taylor-expandable function of $\L_G$ that begins with 1, we
can use the second equation to give a recursion relation for $G_n$: 
Separating the leading term of $\F$,
$$ \E = m\E_{-1} +ë\E,ââ-m[G,\E_{-1}] = [G,ë\E] +\L_GÊcoth(\L_G)\O $$
 which we can expand in $1/m$ [after Taylor expanding $\L_GÊcoth(\L_G)$]
to give an expression for $[G_n,\E_{-1}]$ to solve for $G_n$.  We can also
use the implicit solution for $[G,\E]$ directly to simplify the expression for
$\F'$:
$$ \F' = \E +tanh(ü\L_G)\O $$
 For example, to order $1/m^2$ we have for $\F'$
$$ \F'_{-1} = \E_{-1},ââ\F'_0 = \E_0,ââ\F'_1 = \E_1 +ü[G_1,\O_0] $$
$$ \F'_2 = \E_2 +ü[G_2,\O_0] +ü[G_1,\O_1] $$
 To this order we therefore need to solve
$$ -[G_1,\E_{-1}] = \O_0,ââ-[G_2,\E_{-1}] = \O_1 +[G_1,\E_0] $$

For our applications we will always have
$$ \E_{-1} = -\f1w S^{-10} $$
 unchanged by interactions.  We have oversimplified things a bit in the
above derivation:  For general spin we need to consider more than just
even and odd terms; we need to consider all eigenvalues of $S^{-10}$:
$$ [S^{-10},\F_s] = s\F_s $$
 and find the transformation that makes $\F'$ commute with it ($s=0$). 
The procedure is to first divide into even and odd values of $s$, as above,
then to divide the remaining even terms in $\F'$ into twice even values of
$s$ (multiples of 4) as the new $\E'$ and twice odd as the new $\O'$,
which are transformed away with the new twice odd $G'$, and so on.  This
very rapidly removes the lower nonzero values of $|s|$ ($1£2£4£...$),
which has a maximum value of $2w$ (from the operators that mix the
maximum value $S^{-10}=w$ with the minimum $S^{-10}=-w$).  For
example, for the case of most interest, the Dirac spinor, the only
eigenvalues (for operators) are 0 and $à$1, so the original even part does
commute with $S^{-10}$, and the procedure need be applied only once. 
Furthermore, terms in $\F$ of eigenvalue $s$ can be generated only at
order $m^{1-s}$ or higher; so at any given order the procedure rapidly
removes all undesired terms for any spin.

Since the terms we want to cancel are exactly the ones with nonvanishing
eigenvalues of $S^{-10}$, they can always be written as $[G,S^{-10}]$ for
some $G$, so we can always find a transformation to eliminate them:
$$ [S^{-10},G_{sn}] = sG_{sn}âÜâ
	G_{sn} = -\f{w}sÓ[G,ë\E] +\L_GÊcoth(\L_G)\OÕ_{sn} $$
  (This is just diagonalization of a Hermitian matrix in operator language.) 
In particular for the Dirac spinor, since $\E_{-1}$ has only $à1$
eigenvalues, it's easy to see that not only do all even operators commute
with it, but all odd operators anticommute with it.  (Consider the
diagonal representation of $\E_{-1}$:  $Ó\tbt100{-1},\tbt0ab0Õ=0$.)  We
then have simply
$$ w = üâÜâ(\E_{-1})^2 = 1âÜâ
	[\E_{-1},ë\E] = Ó\E_{-1},\OÕ = Ó\E_{-1},GÕ = 0 $$
$$ ÜâmG = -üÓ[G,ë\E] +\L_GÊcoth(\L_G)\OÕ\E_{-1} $$

As a final step, we can apply the usual transformation
$$ U_0 = e^{imtS^{-10}/w} $$
 which commutes with all but the $p^0$ term in $\E_0$ to have the sole
effect of canceling $\E_{-1}$, eliminating the rest-mass term from the
nonrelativistic-style expression for the energy.

For the minimal electromagnetic coupling described above, we have
besides $\E_{-1}$
$$ \E_0 = ¹^0,ââ\O_0 = \f1w S^{0i}¹^i $$
 where we have written $¹^a=p^a+A^a$ (instead of $¹^a=-iá^a$, to
save some $i$'s).  There are no additional terms in $\F$ for minimal
coupling for spin 1/2, but later we'll need to include nonminimal effective
couplings coming from quantum (field theoretic) effects.  There are also
extra terms for spins 0 and 1 because the field strength is not the same
as the fundamental field, so we'll treat only spin 1/2 here, but we'll
continue to use the general notation to illustrate the procedure.  Using
the above results, we find to order $1/m^2$ for $\F'$
$$ G_1 = S^{-1i}¹^i,ââG_2 = wS^{0i}iF^{0i} $$
 in agreement with with the free case up to field strength terms.  The
diagonalized Schr¬odinger equation is then to this order, including the
effect of $U_0$,
$$ \F'_{-1} = 0,ââ\F'_0 = ¹^0,ââ
	\F'_1 = -\f1{2w}[üÓS^{-1i},S^{0j}ÕiF^{ij} +S^{-10}(¹^i)^2] $$
$$ \F'_2 = -\f14[ÓS^{0i},S^{0j}Õ(»^i F^{0j}) -S^{ij}ÓiF^{0i},¹^jÕ] $$
 For spin 1/2 we are done, but for other spins we would need a further
transformation (before $U_0$) to pick out the part of $\F'_2$ that
commutes with $S^{-10}$ (by eliminating the twice odd part); the final
result is 
$$ \F'_2 = \f14[ü(ÓS^{-1i},S^{-1j}Õ-ÓS^{0i},S^{0j}Õ)(»^i F^{0j})
	+S^{ij}ÓiF^{0i},¹^jÕ] $$
 It can also be convenient to translate into $à$ notation (as for the
massless case, but with index $1£-1$):  We then write
$$ \F'_{-1} = 0,ââ\F'_0 = ¹^0,ââ
	\F'_1 = -\f1{2w}[üÓS^{+i},S^{-j}ÕiF^{ij} +S^{+-}(¹^i)^2] $$
$$ \F'_2 = -\f14[üÓS^{+(i},S^{-j)}Õ(»^i F^{0j}) -S^{ij}ÓiF^{0i},¹^jÕ] $$
 In this notation the eigenvalue of $S^{+-}=S^{-10}$ for any combination
of spin operators can be simply read off as the number of $-$ indices
minus the number of +.

\x IIB5.2 Find the Hamiltonian for spin 1/2 in background
electromagnetism, expanded nonrelativistically to this order, by
substituting the appropriate expressions for the spin operators in terms
of $©$ matrices, and applying
$S^{+-}=àw$ on the right for positive/negative energy.  (Ignore the
reality-restoring transformation.)  $©$-matrix algebra can be performed
directly with the spin operators:  For the Dirac spinor we have the
identities
$$ S_{(a}{}^{(b}S_{c)}{}^{d)} = ü¶_{(a}^b ¶_{c)}^d -ú_{ac}ú^{bd}âÜâ
	ÓS^{+i},S^{-j}Õ = ü¶^{ij} -2S^{ij}S^{+-} $$ 

Ü6. Twistors

Besides describing spin 1/2, spinors provide a convenient way to solve
the condition $p^2=0$ covariantly:  Any hermitian matrix with
vanishing determinant must have a zero eigenvalue (consider the
diagonalized matrix), and so such a 2$ð$2 matrix can be simply expressed
in terms of its other eigenvector.  Absorbing all but the sign of the
nontrivial eigenvalue into the normalization of the eigenvector, we have
$$ p^2 = 0âÜâp^{ŒÀº} = àp^Œ p^{Àº} $$
 for some spinor $p^Œ$ (where $p^{ÀŒ}­(p^Œ)*$).  
Since $p^0$ is the (canonical) energy, the $à$ is
the sign of the energy.  This explains why time reversal (actually CT in the
usual terminology) is not a linear transformation.  Note that $p^Œ$ is a
commuting object, while most spinors are fermionic, and thus
anticommuting (at least in quantum theory).  Such commuting spinors are
called ``twistors".

\x IIB6.1  Show that, in terms of its energy $E$ and the angular direction
$(Ï,Ä)$ (with respect to the ``1" axis)
of its velocity, a massless particle is described by the twistor
$$ p^Œ = 2^{1/4}å{|E|}( cos \fÏ2 e^{iÄ/2} , sin \fÏ2 e^{-iÄ/2} ) $$

One useful way to think of twistors is in terms of the lightcone frame.  In
spinor notation, the momentum is
$$ p^{ŒÀº} = à\tat1000 $$
 If we write an arbitrary massless momentum as a Lorentz
transformation from this lightcone frame, then the twistor is just the
part of the SL(2,C) matrix that contributes:
$$ p'^{ŒÀº} = p^{©À¶}g_©{}^Œ Ðg_{À¶}{}^{Àº}
	= à¶^©_¢ ¶^{À¶}_{\rdt ¢}g_©{}^Œ Ðg_{À¶}{}^{Àº} = àg_¢{}^Œ Ðg_{\rdt ¢}{}^{Àº}
	= àp^Œ p^{Àº} $$
 For this reason, the twistor formalism can be understood as a Lorentz
covariant form of the lightcone formalism.

The twistor construction thus gives a covariant way of constructing
wave functions satisfying the mass-shell condition (Klein-Gordon
equation) for the massless case, $õÆ=0$, where $õ=»^2=-p^2$.  We
simply Fourier transform, and use the twistor expression for the
momentum, writing the momentum-space wave function in terms of
twistor variables (``Penrose transform"):
$$ Æ(x) = Çd^2 p_Œ d^2 Ðp_{ÀŒ}[exp(ix^{ŒÀº}p_Œ Ðp_{Àº})_+(p_Œ,p_{ÀŒ})
	+exp(-ix^{ŒÀº}p_Œ Ðp_{Àº})_-(p_Œ,Ðp_{ÀŒ})] $$
 where $_à$ describe the positive- and negative-energy states,
respectively.  (The integral over $Ðp_{ÀŒ}$ can be performed also,
effectively taking the Fourier transform with respect to that variable
only, treating $àx^{ŒÀº}p_Œ$ as the conjugate.)

We can extend the matrix notation of subsection IIA5-6 to twistors:
$$ \boxeq{ \eqalign{ Òp| = p^ŒÒ{}_Œ|,â|pÔ = |{}^ŒÔp_Œ; 
	& â[p| = p^{ÀŒ}[{}_{ÀŒ}|,â|p] = |{}^{ÀŒ}]p_{ÀŒ} \cr
	P = |pÔ[p|,â & â-P* = |p]Òp| \cr }} $$
 As a result, we also have for twistors
$$ \boxeq{ \eqalign{ ÒpqÔ = & -ÒqpÔ,â[pq]=-[qp];ââÒpqÔ* = [qp] \cr
	& ÒpqÔÒrsÔ +ÒqrÔÒpsÔ +ÒrpÔÒqsÔ = 0 \cr }} $$
 These properties do not apply to physical, anticommuting spinors, where
$ÒƍÔ=+ҍÆÔ$, and $ÒÆÆÔ±0$.

Another natural way to understand twistors is through the conformal
group.  We have already seen that the conformal group in D dimensions is
SO(D,2).  Since this group in four dimensions is the same as SU(2,2), it's
simpler to describe its general representations (and in particular spinors)
in SU(2,2) spinor notation.  Then the simplest way to generate
representations of this group is to use spinor coordinates:  We therefore
write the generators as (see subsection IC1)
$$ G_\A{}^\B = ½^\B н_\A -\f14 ¶_\A^\B ½^\C н_\C $$
 where we have subtracted out the trace piece to reduce U(2,2) to SU(2,2)
and, consistently with the group transformation properties under complex
conjugation, we have chosen the complex conjugate of the spinor to also
be the canonical conjugate:  The Poisson bracket is defined by
$$ [н_\A,½^\B] = ¶_\A^\B $$
 To compare with four-dimensional notation, we reduce this
four-component spinor by recognizing it as a particular use of the Dirac
spinor.  Using the same representation as in subsection IIA6, we write
$$ ½^\A = (p^Œ,п^{ÀŒ}),ââн_\A = (¿_Œ,Ðp_{ÀŒ});ââ
	ç^{À\A\B} = \pmatrix{ 0 & ÐC^{ÀŒÀº} \cr C^{Œº} & 0 \cr } $$
 Now the Poisson brackets are
$$ [¿_Œ,p^º] = ¶_Œ^º,â[п_{ÀŒ},Ðp^{Àº}] = ¶_{ÀŒ}^{Àº} $$
 The group generators themselves reduce to
$$ p_Œ Ðp_{Àº},â¿_Œ п_{Àº},âp_{(Œ}¿_{º)},âÐp_{(ÀŒ}п_{Àº)},â
	p^Œ ¿_Œ +Ðp^{ÀŒ}п_{ÀŒ} -2 $$
 (for $E>0$, with an overall $-$ for $E<0$),
which are translations, conformal boosts, SL(2,C) generators and their
complex conjugates, and dilatations.

Another kind of twistor, related to position space instead of
momentum space, follows from this (D+2)-coordinate description of
conformal symmetry for D=4 (see subsection IA6).  In practice, it's more
convenient to work with invariances than constraints.  In this case, we
can solve the lightcone constraint on Wick-rotated D=3+3 or 5+1 space,
replacing 6-component conformal vector indices with 4-component
conformal spinor indices, with a position-space twistor:
$$ y^2 = \f14 ·_{\A\B\C\D}y^{\A\B}y^{\C\D} = 0âÜâ
	y^{\A\B} = z^{\A Œ}z^\B{}_Œ $$
 where $\A$ is an SL(4) (or SU*(4)) index and $Œ$ is an SL(2) (or SU(2))
index, and $z^{\A Œ}$ is real (with either two real or two pseudoreal
indices).  (Here the SL groups apply to 3+3 dimensions, the SU groups to
5+1.)  Whereas $y$ had $6-1=5$ components due to the constraint, $z$ has
$4É2-3=5$ components due to the SL(2) (SU(2)) gauge invariance of the
above relation to $y$.  These coordinates reduce to the usual by an SL(2)
transformation:
$$ \A = (µ,Àµ),âz^\A{}_Œ = Â_Œ{}^Ã(¶_Ã^µ,x_Ã{}^{Àµ})
	âÜâ\hbox{\it SL(2) gauge}¼Â_Œ{}^à = ¶_Œ^à $$
 where $e=Â^2$.

\x IIB6.2  Substitute this spinor-notation $z(Â,x)$ into $y¾z^2$ and
compare with the vector-notation $y(e,x)$ of subsection IA6.

Ü7. Helicity

A sometimes-useful way to treat the transverse spin operators $S^{ij}$ is
in terms of
$$ W_{abc} = üP_{[a}J_{bc]} = üP_{[a}S_{bc]} $$
 which (like the field equations)
can be written in terms of just the Poincar«e generators.  
This is the part
of $S_{ab}$ whose commutator with the field equations is proportional to
the field equations (i.e., it preserves the constraints).  For the massive case, it reduces to $S^{ij}$ for SO(D$-$1) in the rest frame; for the massless case in the lightcone frame, using the field equations it again reduces to $S^{ij}$, but now for just SO(D$-$2).
In D=4 this is the
``Pauli-Luba«nski (axial) vector"
$$ W_a = \f16 W^{bcd}·_{bcda} $$
 We can choose our states to be eigenstates of a component of it:  For
example, for massless states $W^0/P^0$ is called the ``helicity".  For
massive states the helicity is defined as $W^0/|\vec P|$, but is less
useful, especially since it is undefined (0/0) in the rest frame.  In that
case one instead chooses a component in terms of a
(momentum-dependent axial) vector $s^a$ as $s^a W_a$, where $s^a
P_a=0$ and $s^2=1/m^2$.

\x IIB7.1  Show in both the massless and massive cases that $W_{abc}$
reduces to the little group generators on shell by going to the appropriate
reference frame.

The twistor representation of the conformal group does not give the
most general representation, but it does give all the (free) massless
ones.  The reason it gives massless ones is that this representation
satisfies the constraint (see subsection IIB1)
$$ \G_{[\A\B]}{}^{[\C\D]} = G_{[\A}{}^{[\C} G_{\B]}{}^{\D]} - traces = 0 $$
 which includes $p^2=0$ as well as all the equations that follow from
$p^2=0$ by conformal transformations.  As a consequence, this
representation also satisfies
$$ G_\A{}^\C G_\C{}^\B - trace = h G_\A{}^\B $$
 where $h$ is the helicity.  This equation may be more recognizable in
SO(4,2) notation, as
$$ \f18 ·^{ABCDEF}G_{CD}G_{EF} = ihG^{AB} $$
 This equation includes, as its lowest mass-dimension part (as defined by
dilatations), the Pauli-Luba«nski vector
$$ W^a = ü·^{bcda}P_b J_{cd} = ihP^a $$
 (The ``$i$" appears in the last two equations only when we use the
antihermitian form of the generators $G_{AB}$ and $J_{ab}$.)  Although
any massless representation of the conformal group satisfies the above
conditions (see exercise IIB2.3), the twistor representation satisfies the
unusual property that helicity is realized as a linear transformation on
the coordinates:  For the twistors the implicit definition of helicity can be
solved explicitly to give
$$ h = \f14Óн^\A,½_\AÕ = üн^\A ½_\A +1 = ü(p^Œ ¿_Œ -Ðp^{ÀŒ}п_{ÀŒ}) $$
 (also for $E>0$), 
which is exactly the U(1) transformation of U(2,2)=SU(2,2)$°$U(1). 
(This is similar to SU(2) in terms of ``twistors":  See exercise IC1.1.)  On
functions of $p_Œ$ and $Ðp_{ÀŒ}$, it effectively just counts half the number
of $p_Œ$'s minus $Ðp_{ÀŒ}$'s.

\x IIB7.2  These results are pretty clear from symmetry, but we should do
some algebra to check coefficients:
Express $J_{ab}$ and $P_a$ in terms of the twistors
$p_Œ,Ðp_{ÀŒ},¿_Œ,п_{ÀŒ}$ (see also exercise IIB2.3 for normalization), 
and plug into $·PJ=ihP$ to derive the
above expression of $h$ in terms of twistors.

The simple form of the helicity in the twistor formalism is another
consequence of it being a covariantized lightcone formalism.  In the
lightcone frame, there is still a residual Lorentz invariance; in particular,
a rotation about the spatial direction in which the momentum points
leaves the momentum invariant.  This is another definition of the helicity,
as the part of the angular momentum performing that rotation.  (Only
spin contributes, since by definition the momentum is not rotated.)  Since
the product of two Lorentz transformations is another one, this rotation
can be interpreted as a transformation acting on the Lorentz
transformation to the lightcone frame, i.e., on the twistor, such that the
momentum is invariant.  This is simply a phase transformation:
$$ g'_Œ{}^º = \pmatrix{e^{iÏ} & 0 \cr 0 & e^{-iÏ} \cr}_Œ^{¼©} g_©{}^º
	âÜâp'^Œ = e^{iÏ}p^Œ $$

We can generalize the Penrose transform in a simple way to wave
functions carrying indices to describe spin:
$$ \li{ Æ_{Œ_1...Œ_m Àº_1...Àº_n}(x) = 
	& Çd^2 p_Œ d^2 Ðp_{ÀŒ}¼p_{Œ_1}òp_{Œ_m}Ðp_{Àº_1}òÐp_{Àº_n} \cr
	& ð[exp(ix^{ŒÀº}p_Œ Ðp_{Àº})_+(p_Œ,p_{ÀŒ})
	 +exp(-ix^{ŒÀº}p_Œ Ðp_{Àº})_-(p_Œ,p_{ÀŒ})] \cr} $$
 For the integral to give a nonvanishing result, the integrand must be
invariant under the U(1) transformation generated by the helicity
operator $h$:  In other words, $_à$ must have a transformation under
$h$, i.e., a certain helicity, that is exactly the opposite that of the
explicit $p$ factors that carry the external indices to give a contribution
to the integral, since otherwise integrating over the phase of $p_Œ$ would
average it to zero.  (Explicitly, if we derive the helicity by acting on the
Penrose transform, this minus sign comes from integration by parts.)  This
means that $Æ(x)$ automatically has a certain helicity, half the number of
dotted minus undotted indices:
$$ h = ü(n-m)ââ[ w = ü(m+n) ] $$
 (also for $E>0$),
as given by the above twistor operator expression acting on $_à$. 
(Alternatively, comparing the $x$-space form of the Pauli-Luba«nsky
vector, its action plus that of the twistor-space one must vanish on
$|{}^ŒÔp_Œ$, so the helicity is again minus the twistor-space helicity
operator acting on the prefactor.)

If we work in momentum space, then we use implicitly the relation between momentum and twistors.  Then we can use the abbreviated form of the above relation,
$$ ÷Æ_{Œ_1...Œ_m Àº_1...Àº_n}(p) = p_{Œ_1}òp_{Œ_m}Ðp_{Àº_1}òÐp_{Àº_n}(p_Œ,p_{ÀŒ}) $$
 using $_+$ or $_-$ as appropriate to the sign of energy.

The above transform is just for field ÓstrengthsÕ:  The generalization to on-shell gauge fields is straightforward, though not as simple, since gauge fields contain more than just 2 physical helicities, but also unphysical degrees of freedom.  For example, for the 4-vector potential of electromagnetism, we have
$$ \li{ A_{©À¶}(x) = & Çd^2 p_Œ d^2 Ðp_{ÀŒ}¼ÓÐp_{À¶}
	[exp(ix^{ŒÀº}p_Œ Ðp_{Àº})A_{+©}(p_Œ,p_{ÀŒ}) \cr
	& +exp(-ix^{ŒÀº}p_Œ Ðp_{Àº})A_{-©}(p_Œ,p_{ÀŒ})] +h.c.Õ \cr} $$

\x IIB7.3  Look at the Maxwell field strength in spinor notation $f_{Œº}$ (and its complex conjugate) defined in subsection IIA7, in terms of the above gauge field.  Show it reduces to a special case of the previous general expression, and express $_à$ in terms of $A_{àŒ}$ and $ÐA_{àÀŒ}$.

Since, after restricting to the appropriate helicity, the integral over this
phase is trivial, we can also eliminate it by replacing the ``volume"
integral over the twistor or its complex conjugate (but not both) with a
``surface" (boundary) integral:
$$ Çd^2 p_Œ £ Èp^Œ dp_Œ $$
 (Alternatively, we can insert a $¶$-function in the helicity.)  The result is
equivalent to the usual integral over the three independent components
of the momentum.

This generalization of the Penrose transform implies that $Æ(x)$ satisfies
some equations of motion besides $p^2=0$, namely
$$ p^{ŒÀŒ}Æ_{Œ...ºÀ©...À¶} = p^{©À©}Æ_{Œ...ºÀ©...À¶} = 0 $$
 which are also implied by $S_a{}^b »_b +w»_a=0$ (see exercise IIB2.3).
Besides Poincar«e invariance, these equations are invariant under the
phase transformation
$$ Æ'_{Œ...ºÀ©...À¶} = e^{i2hÏ}Æ_{Œ...ºÀ©...À¶} $$
 that generalizes duality and chiral transformations.  We also see that
(anti-)self-duality and chirality are related to helicity.  Another way to
understand the twistor result is to remember its interpretation as a
Lorentz transformation from the light cone:  In the light cone frame,
where $p^{+À+}$ is the only nonvanishing component of $p^{ŒÀŒ}$, the
above equations of motion imply the only nonvanishing component of
$Æ^{Œ_1...Œ_m Àº_1...Àº_n}$ is $Æ^{+...+À+...À+}$, which can be identified with
$_+$ (for $p^{+À+}>0$) or $_-$ (for $p^{+À+}<0$).

\refs

£1 E. Wigner, ÓAnn. Math.Õ É40 (1939) 149;\\
	V. Bargmann and E.P. Wigner, ÓProc. Nat. Acad. Sci. USÕ É34 (1946)
	211:\\
	little group; more general discussion of Poincar«e representations and
	relativistic wave equations for D=4.
 £2 A.J. Bracken, ÓLett. Nuo. Cim.Õ É2 (1971) 574;\\
	A.J. Bracken and B. Jessup, ÓJ. Math. Phys.Õ É23 (1982) 1925;\\
	W. Siegel, \NP 263 (1986) 93:\\
	conformal constraints.
 £3 E.C.G. St¬uckelberg, ÓHelv. Phys. ActaÕ É11 (1938) 299.
 £4 P.A.M. Dirac, ÓRev. Mod. Phys.Õ É21 (1949) 392:\\
	lightcone gauge.
 £5 G. Nordstr¬om, ÓPhys. Z.Õ É15 (1914) 504:\\
	dimensional reduction.
 £6 O. Klein, ÓZ. Phys.Õ É37 (1926) 895;\\
	V. Fock, ÓZ. Phys.Õ É39 (1927) 226:\\
	mass from dimensional reduction.
 £7 A.S. Eddington, ÓProc. Roy. Soc. AÕ É121 (1928) 524:\\
	$©_{-1}$ (``$©_5$").
 £8 R.J. Duffin, ÓPhys. Rev.Õ É54 (1938) 1114;\\
	N. Kemmer, ÓProc. Roy. Soc.Õ ÉA173 (1939) 91.
 £9 M.H.L. Pryce, ÓProc. Roy. Soc.Õ ÉA195 (1948) 62;\\
	S. Tani, ÓSoryushiron KenkyuÕ É1 (1949) 15 (in Japanese):\\
	free version of Foldy-Wouthuysen.
 £10 L.L. Foldy and S.A. Wouthuysen, ÓPhys. Rev.Õ É78 (1950) 29.
 £11 M. Cini and B. Touschek, ÓNuo. Cim.Õ É7 (1958) 422;\\
	S.K. Bose, A. Gamba, and E.C.G. Sudarshan, ÓPhys. Rev.Õ É113 (1959)
	1661:\\
	``ultrarelativistic" version of Foldy-Wouthuysen transformation.
 £12 R. Penrose, ÓJ. Math. Phys.Õ É8 (1967) 345,
	ÓInt. J. Theor. Phys.Õ É1 (1968) 61;\\
	M.A.H. MacCallum and R. Penrose, ÓPhys. Rep.Õ É6 (1973) 241:\\
	twistors.
 £13 W. Pauli, unpublished;\\
	J.K. Luba«nski, ÓPhysicaÕ ÉIX (1942) 310, 325.

\unrefs

Û5 C. SUPERSYMMETRY

Supersymmetry is a symmetry that relates fermions to bosons.
It includes the Poincar«e group as a subgroup.
We'll see later that quantum field theory requires particles with integer
spin to be bosons, and those with half-integer spin to be fermions.  This
means that any symmetry that relates bosonic wave functions/fields to
fermionic ones must be generated by operators with half-integer spin. 
The simplest (but also the most general, at least of those that preserve
the vacuum) is spin 1/2.  In this section we look
at representations, generalizing the results of the previous sections for
Poincar«e symmetry.

Although supersymmetry has not been experimentally verified yet,
it is a major ingredient in the most
promising generalizations of the Standard Model:
\item{(1)}  The fact that it enlarges the symmetry of nature means
that it further restricts the allowed models, and thus makes stronger
predictions.
\item{(2)}  The greater symmetry also simplifies quantum calculations
in many ways, especially through the use of the concept of ``superspace".
The results of these calculations are also often simplified.
\item{(3)}  Because supersymmetric calculations are simpler, they can be
used to simplify nonsupersymmetric calculations, at both the classical
and quantum levels.
\item{(4)}  This simplification in quantum rules results in improved 
high-energy behavior.  In some cases it even results in the absence of
the infinities in momentum integration that occur in all nonsupersymmetric
theories.  Although these infinities can be removed in perturbation theory,
their effects reappear upon summation of the expansion.
An analogy can be drawn with the elusive Higgs boson:  It also has not been observed, but is needed to remove certain infinities.
\item{(5)}  This improvement at high energies also improves the
experimental agreement of Grand Unified Theories of the strong,
electromagnetic, and weak interactions.

Ü1. Algebra

From quantum mechanics we know that for any operator $A$
$$ ÒÆ|ÓA,AÿÕ|ÆÔ = Ý_n (ÒÆ|A|nÔÒn|Aÿ|ÆÔ +ÒÆ|Aÿ|nÔÒn|A|ÆÔ) $$
$$ = Ý_n (|Òn|Aÿ|ÆÔ|^2 +|Òn|A|ÆÔ|^2) ³ 0 $$
 from inserting a complete set of states.  In particular,
$$ ÓA,AÿÕ = 0âÜâA = 0 $$
 from examining the matrix element for all states $|ÆÔ$.  This means the
anticommutation relations of the supersymmetry generators must be
nontrivial.  

We are then led to anticommutation relations of the form, in Dirac
(Majorana) notation,
$$ Óq,ÐqÕ = ÖpââorââÓq,qÿÕ = p^a ©_a å2©_0 $$
 (We use translations instead of internal symmetry or Lorentz generators
because of dimensional analysis:  Bosonic fields differ in dimension from
fermionic ones by half integers.)  Note that this implies the positivity of
the energy:
$$ trÓq,qÿÕ = å2p^aÊtr(©_a ©_0) 
	= å2p^aÊtr(üÓ©_a, ©_0Õ) = p^0\f1{å2}Êtr¼I $$

Similar arguments imply that the supersymmetry generators are
constrained, just as the momentum is constrained by the mass-shell
condition.  For example, in the massless case,
$$ ÓÖpq,ÐqÖpÕ = ÖpÖpÖp = -üp^2Öp = 0âÜâÖpq = 0 $$

In four dimensions the commutation relations can be written in terms of
irreducible spinors as
$$ Óq^Œ,Ðq^{Àº}Õ = p^{ŒÀº},ââÓq,qÕ = ÓÐq,ÐqÕ = 0 $$
 This generalizes straightforwardly to more than one spinor, carrying a
U(N) index:
$$ Óq_i{}^Œ,Ðq^{jÀº}Õ = ¶_i^j p^{ŒÀº} $$

\x IIC1.1  Show positivity of energy in 2-component spinor notation
for 4D U(N) supersymmetry.

Ü2. Supercoordinates

Since the momentum is usually represented as coordinate derivatives, we
naturally look for a similar representation for supersymmetry.  We
therefore introduce an anticommuting spinor coordinate $Ï^Œ$.  Because
of the anticommutation relations $q$ can't be simply $»/»Ï$, but the
modification is obvious: 
$$ q_Œ = -i{»\over »Ï^Œ} +üÐÏ^{Àº}{»\over »x^{ŒÀº}},ââ
	Ðq_{ÀŒ} = -i{»\over »ÐÏ^{ÀŒ}} +üÏ^º{»\over »x^{ºÀŒ}} $$
 We can also express supersymmetry in terms of its action on the
``supercoordinates":  Using the hermitian infinitesimal generator
$·^Œ q_Œ +з^{ÀŒ}q_{ÀŒ}$,
$$ ¶Ï^Œ =·^Œ,ââ¶ÐÏ^{ÀŒ} = з^{ÀŒ},ââ¶x^{ŒÀº} = üi(·^Œ ÐÏ^{Àº} +з^{Àº}Ï^Œ) $$
 Note that $(q^Œ)ÿ=Ðq^{ÀŒ}$, $(q_Œ)ÿ=-Ðq_{ÀŒ}$.

We can also define ``covariant derivatives": derivatives that
(anti)commute with (are invariant under) supersymmetry.  These are
easily found to be
$$ d_Œ = {»\over »Ï^Œ} +üÐÏ^{Àº}p_{ŒÀº},ââ
	Ðd_{ÀŒ} = {»\over »ÐÏ^{ÀŒ}} +üÏ^º p_{ºÀŒ} $$
 Besides overall normalization factors of $i$, leading to the opposite
hermiticity condition $(d^Œ)ÿ=-Ðd^{ÀŒ}$, these differ from the $q$'s by the
relative sign of the two terms.  These changes combine to preserve
$$ Ód^Œ,Ðd^{Àº}Õ = p^{ŒÀº},ââÓd,dÕ = ÓÐd,ÐdÕ = 0 $$
 as a result of which $p$ is also a covariant derivative as well as being
a symmetry generator (as for the Poincar«e group), but now
$(d_Œ)ÿ=+Ðd_{ÀŒ}$.

In classical mechanics, the fact that $»/»x$ commutes with translations
is ``dual" to the fact that the infinitesimal change $dx$, or the finite
change $x-x'$, is also invariant under translations.  Furthermore, the
d'Alembertian $õ=(»/»x)^2$ being Poincar«e invariant is dual to the line
element $ds^2=-(dx)^2$ being invariant.  This allows the construction of
the action from $Àx{}^2$.  In the supersymmetric case the
infinitesimal invariants under the $q$'s (and therefore $p$) are
$$ dÏ^Œ,ââdÐÏ^{ÀŒ},ââdx^{ŒÀº} +üi(dÏ^Œ)ÐÏ^{Àº} +üi(dÐÏ^{Àº})Ï^Œ $$
 and the corresponding finite ones (by integration) are
$$ Ï^Œ - Ï'^Œ,ââÐÏ^{ÀŒ} -ÐÏ'^{ÀŒ},ââ
	x^{ŒÀº} -x'^{ŒÀº} +üiÏ^Œ ÐÏ'^{Àº} +üiÐÏ^{Àº}Ï'^Œ $$
 Although these can be used to construct classical mechanics actions,
their quantization is rather complicated.  Just as for particles of one
particular spin, direct treatment of the quantum mechanics has proven
much simpler than deriving it by quantization of a classical system.

\x IIC2.1  Check explicitly the invariance of the above infinitesimal and
finite differences under supersymmetry.

Now that we have a (super)coordinate representation of the
supersymmetry generators, we can examine the wave functions/fields
that carry this representation.  Such ``superfields" can be Taylor
expanded in the $Ï$'s with a finite number of terms, with ordinary fields
as the coefficients.  For example, if we expand a real (hermitian) scalar
superfield
$$ ì(x,Ï,ÐÏ) = Ä(x) +Ï^Œ Æ_Œ(x) +ÐÏ^{ÀŒ}ÐÆ_{ÀŒ}(x) +... $$
 and also expand its supersymmetry transformation
$$ ¶ì = ·^Œ Æ_Œ +з^{ÀŒ}ÐÆ_{ÀŒ}
	 +üi·^Œ ÐÏ^{Àº}»_{ŒÀº}Ä +üi·^{ÀŒ}Ï^º »_{ºÀŒ}Ä +... $$
 we find the component field transformations
$$ ¶Ä = ·^Œ Æ_Œ +з^{ÀŒ}ÐÆ_{ÀŒ},ââ¶Æ_Œ = -üiз^{Àº}»_{ŒÀº}Ä +...,ââ
	¶ÐÆ_{ÀŒ} = -üi·^º »_{ºÀŒ}Ä +...,ââ... $$
 which mix the different spins.

An alternative, and more convenient, way to define the $Ï$ expansion is
by use of the covariant derivatives.  Using ``|" to mean ``$|_{Ï=0}$",
we can define
$$ Ä = ì|,ââÆ_Œ = (d_Œ ì)|,ââÐÆ_{ÀŒ} = (Ðd_{ÀŒ}ì)|,ââ... $$
 There is some ambiguity at higher orders in $Ï$ because the $d$'s don't
anticommute, and this can be resolved according to whatever is
convenient for the particular problem, avoiding field redefinitions in
terms of fields appearing at lower order in $Ï$:  Since the field equations
must be covariant under supersymmetry (otherwise there is no advantage
to using superfields), they must be written with the covariant derivatives. 
Then one defines the component expansions by choosing the same
ordering of $d$'s as appear in the field equations (where relevant), which
gives the component expansion of the field equations the simplest form. 
It also gives a convenient method for deriving supersymmetry
transformations, since the $d$'s anticommute with the $q$'s:
$$ ¶[(d...dì)|] = [d...d(¶ì)]| = [d...d(i·qì)]| = [(i·q)d...dì]| = [(·d)d...dì]| $$
 where we have used the fact that $q=-id+Ï$-stuff, where the $Ï$-stuff
is killed by evaluating at $Ï=0$, once it has been pulled in front of all the
$Ï$-derivatives.  Covariant derivatives can also be used for integration, 
since $ÇdÏ=»/»Ï=d$ up to an $x$-derivative, which can be dropped when
also integrating $Çdx$.

Ü3. Supergroups

We saw certain relations between the lower-dimensional classical groups
that turned out to be useful for just the cases of physical interest of
rotational (SO(D$-$1)), Lorentz (SO(D$-$1,1)), and conformal (SO(D,2))
groups.  In particular, the Poincar«e group, though not a classical group, is
a certain limit (``contraction") of the groups SO(D,1) and SO(D$-$1,2), and
a subgroup of the conformal group.  Similar remarks apply to
supersymmetry, but because of its relation to spinors, these classical
``supergroups" (or ``graded" classical groups) exist only for certain lower
dimensions, the same as those where covering groups for the orthogonal
groups exist.  In higher dimensions the supergroups do not correspond to
supersymmetry, at least not in any way that can be represented on
physical states.

We'll consider only the graded generalization of the classical groups that
appear in the bosonic case.  The basic idea is then to take the group
metrics and combine them in ways that take into account the difference
in symmetry between bosons and fermions:
$$ \li{ \hbox{{\bf U}nitary:}ââ& ç^{ÀAB} \cr
	\hbox{{\bf O}rtho{\bf S}ym{\bf p}lectic:}ââ& M^{AB} \cr
	\hbox{{\bf R}eal:}ââ& ú_{ÀA}{}^B \cr
	\hbox{pseudoreal ($*$):}ââ& ¯_{ÀA}{}^B \cr} $$
 where $ú$ is symmetric and $¯$ antisymmetric, as before, while $M$ is
graded symmetric:  For $A=(a,Œ)$ with bosonic indices $a$ and fermionic
ones $Œ$,
$$ M^{[AB)} = 0:âM^{ab}-M^{ba} = M^{aº}-M^{ºa} = M^{Œº}+M^{ºŒ} = 0 $$
 Again we have inverse metrics, e.g.,
$$ M^{KI}M_{KJ} = ¶_J^I $$
 With respect to the usual index-contraction convention (no extra grading
signs when superscript is contracted with subscript immediately
following), we should take the ordering of indices on $¶$ as $¶_J{}^I$.

There is no analog of the $·$ tensor, at least for finite-dimensional
groups, since it would have an infinite number of indices when totally
symmetric.  However,  ``special" supergroups can still be defined by
generalizing the definition of trace and determinant to supermatrices. 
One convenient way to do this is by using Gaussian integrals, since this is
a common way that such expressions will arise.  As a generalization of
the bosonic and fermionic identities we therefore define the
``superdeterminant"
$$ (sdet¼M)^{-1} = NÇdzÿ¼dz¼e^{-zÿMz} $$
 where ``$NÊ$" is a normalization factor defined so $sdet¼I=1$.  By
explicitly evaluating the integral, separating out the commuting and
anticommuting parts, we find (see exercise IB3.4)
$$ sdet\pmatrix{A & B \cr C & D\cr} =
	{det¼A\over det(D-CA^{-1}B)} = {det(A-BD^{-1}C)\over det¼D} $$
where here $A$ and $D$ contain only bosonic elements, while
$B$ and $C$ are completely fermionic.

\x IIC3.1  Generalize exercise IB3.4 to superdeterminants:  
Divide up the range of a square matrix into two (not necessarily
equal) parts, where each of the two parts may include indices
of both fermionic and bosonic grading, so the four resulting blocks
in the matrix may each include both commuting and anticommuting elements.
Show that
$$ sdet\pmatrix{A&B\cr C&D\cr} = 
	sdet¼D É sdet(A -BD^{-1}C) = sdet¼A É sdet(D -CA^{-1}B) $$
by integration.

The ``supertrace" (see also exercise IA2.3c) then can be defined by
generalizing the bosonic identity $det(e^M)=e^{trÊM}$:
$$ sdet(e^M) = e^{strÊM} $$
$$ str(M_A{}^B) = (-1)^A M_A{}^A = M_a{}^a -M_Œ{}^Œ = tr¼A -tr¼D $$
 follows, as in the bosonic case, from $¶¼ln¼sdet¼M=str(M^{-1}¶M)$, which
is derived by varying the Gaussian definition.

\x IIC3.2  Show that for graded matrices we need to use $str$
(and not $tr$) for the identity
$$ str(MN) = str(NM) $$

A useful identity for superdeterminants can be derived by starting with
the following identity for the inverse of a matrix for which the range of
the indices has been divided into two pieces:
$$ \pmatrix{a & b \cr c & d \cr}^{-1} =
	\pmatrix{(a-bd^{-1}c)^{-1} & (c-db^{-1}a)^{-1} \cr
			(b-ac^{-1}d)^{-1} & (d-ca^{-1}b)^{-1} \cr} $$
 We have assumed all the submatrices are square and invertible;
equivalent expressions, which are more useful in other cases, can be
derived easily by multiplying and dividing by the submatrices:  For
example,
$$ \pmatrix{a & b \cr c & d \cr}^{-1} =
	\pmatrix{(a-bd^{-1}c)^{-1} & -a^{-1}b(d-ca^{-1}b)^{-1} \cr
			-d^{-1}c(a-bd^{-1}c)^{-1} & (d-ca^{-1}b)^{-1} \cr} $$
 From either of these we immediately see
$$ \pmatrix{A & B \cr C & D \cr}^{-1} = \pmatrix{÷A & ÷B \cr ÷C & ÷D \cr}âÜâ
    sdet\pmatrix{A & B \cr C & D \cr} = det¼A¼det¼÷D = {1\over det¼D¼det¼÷A} $$

The graded generalizations of the classical groups are then

\vskip.1in

\narrower\noindent
GL(m|n,C) [SL(m|n,C),SSL(n|n,C)]

\vskip.1in

\noindent
U:  [S]U(m${}_+$,m${}_-$|n) [SSU(n${}_+$,n${}_-$|n${}_+$+n${}_-$)]\\
OSp: OSp(m|2n,C)\\
R:  GL(m|n) [SL(m|n),SSL(n|n)]\\
*:  [S]U*(2m|2n) [SSU*(2n|2n)]

\vskip.1in

\noindent\vtop{\offinterlineskip
\halign{#\hfil\cr
\strut U \& OSp \cr
\vrule height2pt width0pt depth0pt\cr
\noalign{\hrule}
\vrule height2pt width0pt depth0pt\cr
\strut R: OSp(m${}_+$,m${}_-$|2n) \cr
\strut *: OSp*(2m|2n) \cr}}

\vskip.1in

\unnarrower

\noindent where ``$(m|n)$" refers to $m$ bosonic and $n$ fermionic
indices, or vice versa.  In the matrices of the defining representation, the
elements with one bosonic index and one fermionic are anticommuting
numbers, while those with both indices of the same kind are commuting. 
In particular, the commuting parts give the bosonic subgroups:
$$ \li{ \hbox{GL(m|n,C)}â\supsetâ& \hbox{GL(m,C)$°$GL(n,C)} \cr
	\hbox{SL(m|n,C)}â\supsetâ& \hbox{GL(m,C)$°$SL(n,C)} \cr
	\hbox{SSL(n|n,C)}â\supsetâ& \hbox{SL(n,C)$°$SL(n,C)} \cr
	\hbox{U(m${}_+$,m${}_-$|n)}â\supsetâ&
		\hbox{U(m${}_+$,m${}_-$)$°$U(n)} \cr
	\hbox{SU(m${}_+$,m${}_-$|n)}â\supsetâ&
		\hbox{U(m${}_+$,m${}_-$)$°$SU(n)} \cr
	\hbox{SSU(n${}_+$,n${}_-$|n${}_+$+n${}_-$)}â\supsetâ&
		\hbox{SU(n${}_+$,n${}_-$)$°$SU(n${}_+$+n${}_-$)} \cr
	\hbox{OSp(m|2n,C)}â\supsetâ& \hbox{SO(m,C)$°$Sp(2n,C)} \cr
	\hbox{GL(m|n)}â\supsetâ& \hbox{GL(m)$°$GL(n)} \cr
	\hbox{SL(m|n)}â\supsetâ& \hbox{GL(m)$°$SL(n)} \cr
	\hbox{SSL(n|n)}â\supsetâ& \hbox{SL(n)$°$SL(n)} \cr
	\hbox{U*(2m|2n)}â\supsetâ& \hbox{U*(2m)$°$U*(2n)} \cr
	\hbox{SU*(2m|2n)}â\supsetâ& \hbox{U*(2m)$°$SU*(2n)} \cr
	\hbox{SSU*(2n|2n)}â\supsetâ& \hbox{SU*(2n)$°$SU*(2n)} \cr
	\hbox{OSp(m${}_+$,m${}_-$|2n)}â\supsetâ&
		\hbox{SO(m${}_+$,m${}_-$)$°$Sp(2n)} \cr
	\hbox{OSp*(2m|2n)}â\supsetâ&
		\hbox{SO*(2m)$°$USp(2n)} \cr} $$
 When the commuting and anticommuting dimensions are equal, we can
impose tracelessness conditions on both bosonic parts of the generators
separately (``SS", also called ``PS":  $tr¼A=tr¼D=0$).  This is related to the fact $str(I)=0$ in
such cases.

Ü4. Superconformal

Since the conformal group is a classical group, its supersymmetric
generalization should be a classical supergroup.  Because the fermionic
generators must include the supersymmetry generators, which are
spinors, the representation of the conformal group that appears in the
defining representation of the supergroup must be the spinor
representation.  However, we have seen that only for n$²$6 (where
covering groups exist) and n=8 (where the spinor of SO(8) is another of
its defining representations) can the spinor representation of SO(n) be
defined by classical group restrictions.  This implies that the
superconformal group exists only in D$²$4 and D=6.

The relevant supergroups can be identified easily by looking at the
bosonic subgroups:
$$ \li{ D = 3: &â\hbox{OSp(N|4)} \cr
		4: &â\hbox{SU(2,2|N) (or SSU(2,2|4))} \cr
		6: &â\hbox{OSp*(8|2N)} \cr} $$
 (We consider only D>2, since the conformal group is infinite-dimensional
in D$²$2.)  These three cases of D=3,4,6 are special for a number of
reasons:  In particular, these three supergroups can be related to SU(N|4)
over the division algebras: the real numbers, complex numbers, and
quaternions, respectively.  (Similar remarks apply to their important
classical bosonic subgroups: the conformal, Lorentz, and rotation groups. 
Attempts have been made to extend these results to the octonions for
D=10, but with less success, and there seems to be no superconformal
group for that case.)  However, just as in the case of the Hilbert space of
quantum mechanics, the complex numbers seems to be the best of these
``division algebras", having the analytic properties the real numbers lack,
while avoiding the noncommutativity of the quaternions.  We'll see later
that nontrivial interacting (local, classical) conformal field theories exist
only in D$²$4.

For example, for D=4 we find that the bosonic generators are the
conformal group and the internal symmetry group U(N) (or SU(4) for N=4),
while the fermionic generators include supersymmetry (N spinors) and its
fraternal twin, ``S-supersymmetry".  As supersymmetry is the ``square
root" of translations, so S-supersymmetry is the square-root of
conformal boosts.

\x IIC4.1  For D=4, write the (graded) commutation relations of the
superconformal generators.  Decompose them into representations of the
Lorentz group, and find their commutation relations.

Ü5. Supertwistors

We saw that a simple way to find representations of SO(4,2) was to use
the coordinate representation for SU(2,2):  The resulting twistors gave all
massless representations ($p^2=0$ for all helicities).  This method
generalizes straightforwardly to the superconformal groups:  The
generators are
$$ G_A{}^B = ½^B н_A $$
 (For the SU case we should also subtract out the trace, but that generator
commutes with the rest anyway.)  The coordinates and their conjugate
momenta satisfy 
$$ [н_A,½^BÕ = ¶_A^B $$
 $½^A$ is then in the defining representation of the supergroup, while the
wave function, which is a function of $½$, contains more general
representations.

For D=3, the reality condition sets $½=н$, so the $½$'s are the graded
generalization of Dirac $©$ matrices.  In fact, the anticommuting $½$'s
are the $©$ matrices of the SO(N) subgroup of the OSp(N|4).  On the other
hand, the commuting $½$'s carry the index of the defining representation
of Sp(4), so they are a spinor of SO(3,2), the 3D conformal group:  They
are the bosonic twistor, and can be used in a similar way to the 4D
twistors discussed earlier.

For D=4, there is a U(1) symmetry acting on $½$ under which $G_A{}^B$ is
invariant, generated by $(-1)^A G_A{}^A$, as in the bosonic case:  This is
the ``superhelicity".

For D=6, $½$ is pseudoreal.  In general, for pseudoreal representations of
groups it is often convenient to introduce a new SU(2) under which the
pseudoreal representation $½^A$ and its equivalent complex conjugate
representation $н^{ÀB}¯_{ÀB}{}^A$ transform as a doublet (SU(2) spinor). 
 This is also obvious from construction, since half of the
components are related to the complex conjugate of the other half.  We
then can write
$$ ½^{Ai} = (½^A,iн^{ÀB}¯_{ÀB}{}^A) = н^{ÀBÀk}¯_{ÀBÀk}{}^{Ai};ââ
	¯_{ÀBÀk}{}^{Ai} = ¯_{ÀB}{}^A C^{ki},âM^{Ai,Bk} = M^{AB}C^{ik} $$
$$ G_A{}^B = ½^{Bi}½_{Ai} $$
 (Thus, OSp*(2m|2n)$¤$OSp(4n|4m), and SO*(2m)$¤$Sp(4m),
USp(2n)$¤$SO(4n).)  This means there is now an SU(2) symmetry on $½$,
generated by
$$ G_{ij} = ½^A{}_{(i}½_{Aj)} $$
 under which $G_A{}^B$ is invariant.  This is the 6D version of superhelicity.
In the D=6 light cone, the manifest part of Lorentz invariance is
SO(D$-$2)=SO(4)=SU(2)$°$SU(2).  This is one of those SU(2)'s.

We now concentrate on D=4 (although our methods generalize
straightforwardly to D=3 and 6).  The simplest way to find (massless)
representations of 4D supersymmetry is to generalize the Penrose
transform.  Just as twistors automatically satisfy the massless field
equations in D=4, supertwistors automatically satisfy their
supersymmetric generalization.  The supertwistor is the defining
representation of SU(2,2|N).  The SU(2,2) part is the usual twistor, while
the SU(N) part is the usual fermionic creation and annihilation operators
for SU(N).  Thus, to relate superspace to supertwistors, we write
$$ p_{ŒÀº} = -i»_{ŒÀº} £ àp_Œ Ðp_{Àº} $$
$$ q_{iŒ} = -i»_{iŒ} +üÐÏ_i^{Àº}»_{ŒÀº} £ àa_i p_Œ,ââ
	Ðq^i_{ÀŒ} = -iл^i_{ÀŒ} +üÏ^{iº}»_{ºÀŒ} £ àaÿ^i Ðp_{ÀŒ} $$
 This determines the Penrose transform from superspace to
supertwistors:
$$ Ä_{Œ... Àº...}(x,Ï,ÐÏ) = Çd^2 p_Œ d^2 Ðp_{ÀŒ} d^N a_i¼p_ŒòÐp_{Àº}ò
	[e^{i\Ä}_+(p_Œ,Ðp_{ÀŒ},a_i) +e^{-i\Ä}_-(p_Œ,Ðp_{ÀŒ},a_i)] $$
$$ \Ä = (x^{ŒÀº} -iüÏ^{iŒ}ÐÏ_i^{Àº})p_Œ Ðp_{Àº} +Ï^{iŒ}a_ip_Œ $$
 where we have used ``chiral superfields" (trivial dependence on $ÐÏ$,
via the constraint $Ðd^i_{ÀŒ}Ä=0$)
without loss of generality.  (Instead of treating $a_i$ as coordinates to be
integrated, we can also treat them as operators; we then make the $$'s
functions of $aÿ$, and replace the integration with vacuum evaluation
$Ò0|¼|0Ô$.)  As for ordinary twistors, this result can be related to the
lightcone:  For given momentum, we can choose the lightcone frame
$p^Œ=¶^Œ_¢$; then $q_i^¢=àa_i$, while $q_i^\¢=0$ is a result of the
supertwistor formalism automatically incorporating $Öpq=0$.

\x IIC5.1  Find the Penrose transform for D=3.  (Warning:  The
anticommuting part of the twistor is now like Dirac matrices rather than
creation/annihilation operators.)

Taylor expanding in $a_i$ (and thus $Ï^{iŒ}$, producing terms
antisymmetric in $i...j$ and symmetric in $Œ...º$), the states then carry
the index structure $Ä,Ä_i,Ä_{ij},...,÷Ä^i,÷Ä$, totally antisymmetric, and
terminating with another singlet, where
$$ ÷Ä = \f1{N!}·^{i_1òi_N}Ä_{i_1òi_N},ââ
	÷Ä^{i_1} = \f1{(N-1)!}·^{i_1òi_N}Ä_{i_2òi_N},ââ... $$
 From our discussion of helicity in subsection IIB7, we see that the states
also decrease in helicity by 1/2 for each $a$ (i.e., ignoring $ÐÏ$, each
$Ï^{iŒ}$ comes with a $p_Œ$, simply because it adds an undotted index). 
Taking the direct product with any helicity (coming from the explicit
$p_Œ$'s and
$Ðp_{ÀŒ}$'s carrying the external Lorentz indices), we see that the states
have helicity $h,h-1/2,h-1,...,h-N/2$, with multiplicty $N\choose n$ for
helicity $h-n/2$:

$$ \vbox{\offinterlineskip
\hrule
\halign{ &\vrule#&\hbox{\vrule height13pt depth5pt width0pt}¼
	\hfil#\hfil¼\cr
height2pt&\omit&\hskip2pt\vrule&\omit&&\omit&\cr
& state &\hskip2pt\vrule& helicity (Poincar«e) && multiplicity [SU(N)] &\cr
height2pt&\omit&\hskip2pt\vrule&\omit&&\omit&\cr
\noalign{\hrule}
height2pt&\omit&\hskip2pt\vrule&\omit&&\omit&\cr
& $Ä$ &\hskip2pt\vrule& $h$ && 1 &\cr
& $Ä_i$ &\hskip2pt\vrule& $h-ü$ && $N$ &\cr
& $Ä_{ij}$ &\hskip2pt\vrule& $h-1$ && $\f{N(N-1)}2$ &\cr
& $\vdots$ &\hskip2pt\vrule& $\vdots$ && $\vdots$ &\cr
& $Ä_{i_1òi_n}$ &\hskip2pt\vrule& $h-\f{n}2$ && $N\choose n$ &\cr
& $\vdots$ &\hskip2pt\vrule& $\vdots$ && $\vdots$ &\cr
& $÷Ä^i$ &\hskip2pt\vrule& $h-\f{N}2+ü$ && $N$ &\cr
& $ր$ &\hskip2pt\vrule& $h-\f{N}2$ && 1 &\cr
height2pt&\omit&\hskip2pt\vrule&\omit&&\omit&\cr}
\hrule} $$

This multiplet structure is carried separately by $_+$ and by $_-$,
which are related by charge (complex) conjugation, one describing the
antiparticles of the other, as for ordinary twistors.  (The existence of
both multiplets also follows from CPT invariance, which is required for
local actions, to be discussed in subsection IVB1.  Here we generalized
from the Penrose transform, which contained both terms as a
consequence of being the most general solution to $S^{ab}p_b+wp^a=0$,
which is CPT invariant.)  Because of the values of the helicities, we can
impose a reality condition, identifying all states with helicity $j$ as the
complex conjugates of those with $-j$, only for $-h=h-N/2$ $£$ $h=N/4$,
when $N$ is a multiple of 4.  We can also get larger representations by
taking the direct product of these smallest representations of
supersymmetry with representations of U(N), in which case the fields will
carry those additional SU(N) indices.

\x IIC5.2  Let's examine these 4D multiplets in detail:
 ªa List the SU(N) representations for each allowed value of N
for each of the cases where the helicity $|h|²$ 
1/2 (``scalar multiplets"), 1 (``vector multiplets"), 
3/2 (``gravitino multiplets"), or 2 (``graviton multiplets"),
assuming the maximum-helicity state is a singlet.
 ªb Show that ``supergravity" (graviton multiplets)
can exist only for N$²$8.  Show
that the relevant representation for N=8, if real, is the same as the one
(complex) for N=7.  
 ªc Find the analogous statements for ``super
Yang-Mills" (vector multiplets).

The explicit form of the reality condition is somewhat complicated in
terms of the chiral superfields, because they are really field strengths of
real gauge fields.  (Consider, for example, expressing reality of $A_{ŒÀº}$
in terms of $f_{μ}$ in the case of electromagnetism.)  However, in terms
of the twistor variables, charge conjugation can be expressed as
$$ C:ââ_à £ _¦ *,ââa_i £ (a_i)ÿ $$
 where the transformation on $a_i$ is required because it carries the
SU(N) ``charge".  Since this violates ``chirality" in these variables
(dependence on $a$ and not $aÿ$), it is accomplished by Fourier
transformation:
$$ C:ââ_à(a_i) £ C Çd÷aÿ^i¼e^{÷aÿ^i a_i} [_¦(÷a_i)]* C^{-1} $$
 for some ``charge conjugation matrix" $C$ (in case the field carries an
additional index).

\refs

£1 P. Ramond, \PRD 3 (1971) 86;\\
	A. Neveu and J.H. Schwarz, \NP 31 (1971) 86, \PRD 4 (1971) 1109:\\
	2D supersymmetry in strings.
 £2 Yu.A. Gol'fand and E.P. Likhtman, ÓJETP Lett.Õ É13 (1971) 323;\\
	D.V. Volkov and V.P. Akulov, \PL 46B (1973) 109;\\
	J. Wess and B. Zumino, \NP 70 (1974) 39:\\
	4D supersymmetry.
 £3 A. Salam and J. Strathdee, \NP 76 (1974) 477;\\
	S. Ferrara, B. Zumino, and J. Wess, \PL 51B (1974) 239:\\
	superspace.
 £4 Gates, Grisaru, Ro×cek, and Siegel, Óloc. cit.Õ;\\
	J. Wess and J. Bagger, ÓSupersymmetry and supergravityÕ, 2nd ed.
	(Princeton University, 1992);\\
	P. West, ÓIntroduction to supersymmetry and supergravityÕ, 2nd ed.
	(World-Scientific, 1990);\\
	I.L. Buchbinder and S.M. Kuzenko, ÓIdeas and methods of
	supersymmetry and supergravity, or a walk through superspaceÕ
	(Institute of Physics, 1995).
 £5 Berezin, Óloc. cit.Õ (IA):\\
	superdeterminant.
 £6 P. Ramond, ÓPhysicaÕ É15D (1985) 25:\\ 
	supergroups and their relation to supersymmetry.
 £7 J. Wess and B. Zumino, \NP 70 (1974) 39:\\
	superconformal group in D=4.
 £8 R. Haag, J.T. \Sll opusza«nski, and M. Sohnius, \NP 88 (1975) 257:\\
	superconformal symmetry as the largest possible
	symmetry of the S-matrix, most general supersymmetries in D=4.
 £9 A. Ferber, \NP 132 (1978) 55:\\
	supertwistors.
 £10 T. Kugo and P. Townsend, \NP 221 (1983) 357;\\
	A. Sudbery, ÓJ. Phys.Õ ÉA17 (1984) 939;\\
	K.-W. Chung and A. Sudbery, \PL 198B (1987) 161:\\
	spacetime symmetries and division algebras.

\unrefs

ÚIII. LOCAL

In the previous chapters we considered symmetries acting on coordinates
or wave functions.  For the most part, the transformations we considered
had constant parameters:  They were ``global" transformations.  In this
chapter we will consider mostly field theory.  Since fields are functions of
spacetime, it will be natural to consider transformations whose
parameters are also functions of spacetime, especially those that are
localized in some small region.  Such ``local" or ``gauge" transformations
are fundamental in defining the theories that describe the fundamental
interactions.

Û5 A. ACTIONS

A fundamental concept in physics, of as great importance as symmetry, is
the action principle.  In quantum physics the dynamics is necessarily
formulated in terms of an action (in the path-integral approach), or an
equivalent Hamiltonian (in the Heisenberg and Schr¬odinger approaches). 
Action principles are also convenient and powerful for classical physics,
allowing all field equations to be derived from a single function, and
making symmetries simpler to check.

Ü1. General

We begin with some general properties of actions.  (For this subsection
we'll restrict ourselves to bosonic variables; however, in the following
subsection we'll find that the only modification for fermions is a more
careful treatment of signs.)  Generally, equations of motion are derived
from actions by setting their variation with respect to their arguments to
vanish:
$$ ¶S[Ä] ­ S[Ä+¶Ä] - S[Ä] = 0 $$
 The solutions to this equation (find $Ä$, given $S$) are
``extrema" of the action; generally we want them to be minima,
corresponding to minima of the energy, so that they will be stable
under small perturbations.

\x IIIA1.1  Often continuous coordinates are replaced with 
discrete ones, for calculational or conceptual purposes.  
Consider
$$ S = -Ý_{n=-¥}^¥ ü(q_{n+1}-q_n)^2 $$
 The integer $n$ is interpreted as a discrete time, in terms of some
``small" unit.
 ªa Show that
$$ ¶S = 0âÜâq_{n+1} -2q_n +q_{n-1} = 0 $$
 ªb Examine the continuum limit of the action and equations of motion:
Introduce appropriate factors of $·$, with $t=n·$, and take the limit
$·£0$.

Now we take the variables $Ä$ to be functions of time; thus, $S$ is a
function of functions, a ``functional".  It just means that $S$ is a 
function of an infinite set of variables.  We can generalize properties of
ordinary functions (derivatives, etc.) as usual by considering discrete
time and taking a continuum limit:
$$ i=1,2,...â£âtã[-¥,¥] $$
$$ Ä_iâ£âÄ(t) $$
$$ Ý_iâ£âÇdt $$
$$ ¶_{ij}â£â¶(t-t') $$
$$ {»\over »Ä_i}â£â{¶\over ¶Ä(t)} $$
$$ ÇdÄ_iâ£âÇDÄ(t) $$
 (the last, a ``functional integral", will appear in quantum theory)
where $¶_{ij}$ is the usual Kronecker delta function, while $¶(t-t')$ is
the ``Dirac delta function".  It's not really a function, since it takes only
the values 0 or $¥$, but a ``distribution", meaning it's defined only by
integration:
$$ Çdt'¼f(t')¶(t-t') ­ f(t) $$
Of course, the variable $Ä(t)$ can also carry an index (or indices).
In field theory, it will also be a function of more coordinates, those
of space.

For example, making these substitutions into the definition of a
(partial) derivative to get a ``functional derivative",
$$ {»f(Ä_i)\over »Ä_j} = \lim_{·£0}{f(Ä_i +·¶_{ij}) -f(Ä_i)\over ·}
¼¼Ü¼¼{¶f[Ä(t)]\over ¶Ä(t')} = \lim_{·£0}{f[Ä(t) +·¶(t-t')] -f[Ä(t)]\over ·} $$
 Sometimes the functional derivative is defined in terms of that of the
variable itself:
$$ {¶Ä(t)\over ¶Ä(t')} = ¶(t-t') $$
 If we apply this definition of the Dirac $¶$ to $¶Ä/¶Ä$, we obtain the
previous definition of the functional derivative.  (Consider, e.g., varying
$S=Çdt¼f(t)Ä(t)$ for a fixed function $f$.)
 However, in practice we never need to use these definitions of the
functional derivative:  
The only thing for which we need a functional derivative is the action,
whose functional derivative is defined by its variation,
$$ ¶S[Ä] ­ S[Ä+¶Ä] - S[Ä] ­ Çdt¼¶Ä(t){¶S\over ¶Ä(t)} $$
 (The fact that the variation can always be written in this form is just
the statement that it is linear in $¶Ä$, since $¶Ä$ is ``infinitesimal".)

A general principle of mechanics is
``locality", that events at one time directly affect only those events an
infinitesimal time away.  (In field theory these events can be also
only an infinitesimal distance away in space.)  This means that the action
can be expressed in terms of a Lagrangian:
$$ S[Ä] = Çdt¼L[Ä(t)] $$
 where $L$ at time $t$ is a function of only $Ä(t)$ and a finite number of
its derivatives.  For more subtle reasons, this number of time derivatives
is restricted to be no more than two for any term in $L$; after integration
by parts, each derivative acts on a different factor of $Ä$.  The general
form of the action is then
$$ L(Ä) = -üÀÄ{}^m ÀÄ{}^n g_{mn}(Ä) +ÀÄ{}^m A_m(Ä) +U(Ä) $$
 where ``$À{\phantom m}$" means $»/»t$, and the ``metric" $g$, 
``vector potential"
$A$, and ``scalar potential" $U$ are not to be varied independently when
deriving the equations of motion.  (Specifically, $¶U=(¶Ä^m)(»U/»Ä^m)$,
etc.  Note that our definition of the Lagrangian differs in sign from the
usual:  Thus, for a particle with kinetic energy $T$ in a potential $U$ with energy $H=T+U$ we have $L=-T+U$.)  The equations of motion following from varying an action that can
be written in terms of a Lagrangian are
$$ 0 = ¶S ­ Çdt¼¶Ä^m{¶S\over ¶Ä^m}âÜâ{¶S\over ¶Ä^m} = 0 $$
 where we have eliminated $¶ÀÄ{}^m$ terms by integration by parts
(assuming $¶Ä=0$ at the boundaries in $t$), and used the fact that $¶Ä(t)$ is
arbitrary at each value of $t$.  For example,
$$ S = -Çdt¼üÀq{}^2âÜâ0 = ¶S = -Çdt¼Àq¶Àq = Çdt¼(¶q)¬q
	âÜâ{¶S\over ¶q} = ¬q = 0 $$

\x IIIA1.2 Find the equations of motion for $Ä^m$ from the above
general action in terms of the external fields $g$, $A$, and $U$ (and their
partial derivatives with respect to $Ä$).

Locality applies only to the classical action; in quantum field theory
we will also find ``effective actions" that include 
nonlocal contributions from  quantum effects. 
Similar effects can appear in classical theories; 
for example, electrodynamics in the Coulomb gauge
includes a (spatially) nonlocal Coulomb interaction term.
The interpretation is always that some quantity has been
eliminated, which would return locality (e.g., the
``longitudinal photon" in the Coulomb gauge).
Such actions can still be varied by the same methods as above.
However, one should always avoid the rule
$»L/»Ä=»_t(»L/»_t Ä)$, since (1) it applies only to actions
that can be expressed in terms of just $Ä$ and $ÀÄ$ (and not
higher derivatives nor nonlocalities), and (2) it arbitrarily
separates terms into two sets.

Such actions can be reduced to ones that are only linear in time
derivatives by introducing additional variables.  First, separate out the
subspace where $g$ is invertible, with coordinates $q$ ($Ä^m=(q^i,r^µ)$);
the Lagrangian is then written as
$$ L(q,r) =- üÀq{}^i Àq{}^j g_{ij}(q,r) +Àq{}^i A_i(q,r) +Àr{}^µ A_µ(q,r) 
	+U(q,r) $$
 This Lagrangian gives equivalent equations of motion to
$$ L'(q,p,r) = [-Àq{}^i p_i +Àr{}^µ A_µ] +[üg^{ij}(p_i +A_i)(p_j +A_j) +U] $$
 where $g^{ij}$ is the inverse of $g_{ij}$.  (Many other forms are possible
by redefinitions of $p$.)  Eliminating the new variables $p$ by their
equations of motion gives back $L(q,r)$.  Note that this works only
because $p$'s equations of motion are algebraic:  For example, eliminating
$x$ from the Lagrangian $-Àxp+üp^2$ by the equation of motion $Àx=p$ is
illegal (it would give the trivial action $S=-Çdt¼üp^2$), since it would
require solving for the time dependence of $x$.  On the other hand, $p$ is
given explicitly in terms of the other variables by its equations of motion
without inverting time derivatives, so eliminating it does not lose any of
the dynamics.  (It is an ``auxiliary variable".)

The result is a Hamiltonian form of the Lagrangian:
$$ L_H(ì) = iÀì{}^M \A_M(ì) +H(ì) $$
 in terms of the Hamiltonian $H$, where $ì=(q,p,r)$.  It has the ``gauge
invariance"
$$ ¶\A_M = »_M ñ(ì) $$
 (where $»_M=»/»ì^M$), since that adds only a total derivative term
$iÀñ$.  Clearly $\A$ will introduce a modification of the Poisson bracket if
it is not linear in $ì$ (e.g., as when we make independent nonlinear
redefinitions of coordinates and momenta on the usual form of the
Lagrangian).  To determine this modification we compare the equation of
motion as defined by a Poisson bracket,
$$ Àì{}^M = -i[ì^M,H] = -i[ì^M,ì^N]»_N H $$
 with that following from varying the action,
$$ -iÀì{}^N \F_{NM} +»_M H = 0,ââ\F_{MN} = »_{[M}\A_{N]} $$
 to find
$$ [ì^M,ì^N] = (\F^{-1})^{NM} $$
 where ``$\F^{-1}$" is the inverse on the maximal subspace where $\F$ is
invertible.    The variables in the directions where $\F$ vanishes are
``auxiliary", since they appear without time derivatives:  Their equations
of motion are not described by the Poisson bracket.  In particular, if they
appear linearly in $H$ they are ``Lagrange multipliers", whose variation
imposes algebraic constraints on the rest of $ì$. 

Finally, we can make redefinitions of the part of $ì$ describing the
invertible subspace so that $\A$ is linear:
$$ \A_M = üì^N ¯_{NM}âÜâL_H(ì) = üiÀì{}^M ì^N ¯_{NM} +H(ì) $$
 where $¯$ is a constant, hermitian, antisymmetric (and thus imaginary)
matrix.  For some purposes it is more convenient to assume this
Hamiltonian form of the action as a starting point.  We now have the
canonical commutation relations as 
$$ [ì^M,ì^N] = ¯^{MN} $$
 where $¯^{MN}$ is the inverse of $¯_{NM}$ on the maximal subspace:
$$ ¯^{MN}¯_{PN} = ¸_P{}^M $$
 for the projection operator $¸$ for that subspace.

\x IIIA1.3  For electromagnetism, define $\vec Æ = \vec E +i\vec B$.  
 ªa Show
that Maxwell's equations (in empty space) can be written as two
equations in terms of $\vec Æ$.  
 ªb Interpret the equation involving the time
derivative as a Schr¬odinger equation for the wave function $\vec Æ$, and
find the Hamiltonian operator.  
 ªc Define the obvious inner product $Çd^3
x¼\vec Æ*É\vec Æ$:  What physical conserved quantity does this
represent?  (Note that, unlike electrons, the number of photons is not
conserved.)

Note that the requirement of the existence of a Hamiltonian formulation
determines that the kinetic term for a particle in the Lagrangian
formulation go as $Àx{}^2$ and not $x¬x$.  Although such terms give the
same equations of motion, they are not equivalent quantum mechanically,
where boundary terms (dropped when using integration by parts for
deriving the equations of motion) contribute.  Furthermore, the
Hamiltonian form of the action
$$ S = Çdt¼H -dx^i p_i $$
 shows that the energy $H$ relates to the time in the same way the
momentum relates to the coordinates, except for an interesting minus
sign that is explained only by special relativity.

Ü2. Fermions

In nonrelativistic quantum mechanics, spin is usually treated as a
quantum effect, rather than being derived from classical mechanics. 
Although it is possible to derive spin from classical mechanics, in general
it is rather cumbersome, and involves first introducing a large number of
spins and then constraining away all the undesired ones, whereas in the
quantum mechanics one can just directly introduce some particular
representation of the spin angular momentum operators.  The one
nontrivial exception is spin 1/2.

We know from quantum mechanics that the spin variables for spin 1/2
are described by the Pauli $§$ matrices.  Since they satisfy
anticommutation relations, and are represented by finite-dimensional
matrices, they are interpreted as fermionic.  We have already seen that
classical fermions are described by anticommuting numbers, so we begin
by considering general quantization of such objects. 

We can now consider actions that depend on both commuting and
anticommuting classical variables, $ì^M = (Ä^m,Æ^µ)$, where now $Ä$
refers to the bosonic variables and $Æ$ to the fermionic ones.  The
Hamiltonian form of the Lagrangian can again be written as
$$ L_H(ì) = üiÀì{}^M ì^N ¯_{NM} +H(ì) $$
 When $¯$ is invertible, the ÓgradedÕ bracket is defined by (see
subsection IA2)
$$ [ì^M,ì^NÕ = \h ¯^{MN},ââ¯^{MN}¯_{PN} = ¶_P^M $$

To describe spin 1/2, we therefore look for particle actions of the form
$$ S_H = Çdt [-Àx{}^i p_i +üiÀÆ{}^i Æ_i +H(x,p,Æ)] $$
 This corresponds to using
$$ ì^M = (Ä^m;Æ^µ) = (Ä^{iŒ};Æ^i) = (x^i,p^i;Æ^i) $$
$$ ¯_{mn} = ¯_{iŒ,jº} = ¶_{ij}C_{Œº},ââ¯_{µÃ} = ¯_{ij} = ¶_{ij},ââ
	¯_{mÃ} = ¯_{µn} = 0 $$
 The fundamental commutation relations are then
$$ [x^i,p_j] = i\h ¶_j^i,ââÓÆ^i,Æ^jÕ = \h ¶^{ij}ââ
	([x,x] = [p,p] = [x,Æ] = [p,Æ] = 0) $$
 We recognize $Æ^i$ as the Pauli $§$ matrices (the Dirac matrices of
subsection IC1 for the special case of SO(3), normalized as in subsection
IIA1), $Æ^i=å\h§^i$.  The free
Hamiltonian is just 
$$ H = {p^2\over 2m} $$
 as for spin 0:  Spin does not affect the motion of free particles.  

A more interesting case is coupling to electromagnetism:  Quantum
mechanically, the Hamiltonian can be written in the simple form
$$ H = {ÓÆ^i[p_i +qA_i(x)]Õ^2\over m\h} -qA^0(x) $$
 in terms of the vector and scalar potentials $A^i$ and $A^0$.  The
classical expression is not as simple, because the commutation relations
must be used to cancel the $1/\h$ before taking the classical limit. 
This is an example of ``minimal coupling",
$$ H(p_i) £ H(p_i+qA_i) -qA^0 $$
 However, this prescription works only if $H$ for spin 1/2 is written in the
above form:  Using the commutation relations before or after minimal
coupling gives different results.  The form we have used is justified only
by considering the nonrelativistic limit of the relativistic theory.

\x IIIA2.1  Use the multipication rules of the $§$ matrices to show that
the quantum mechanical Hamiltonian for spin 1/2 in an electromagnetic
field can be written as a spin-independent piece, identical to the spin-0
Hamiltonian, plus a term coupling the spin to the magnetic field.

Ü3. Fields

Actions for field theories are just a special case (ÓnotÕ a generalization)
of the actions we have just considered:  We just treat spatial coordinates
$\vec x$ as part of the indices carried by the variables appearing in the
action.  In the notation used above,
$$ \li{ M & £ (i,\vec x) \cr
	ì^M(t) & £ ì^i (t,\vec x) \cr} $$
 Then spatial derivatives are just certain matrices with respect to
the $M$ index, $Çd\vec x$ comes from summation over $M$, etc.

The field equations for all field theories (e.g., electromagnetism) are
wave equations.  Wave equations also follow from mechanics upon
quantization.  Although classical field theory and quantum mechanics are
not equivalent in their physical interpretation, they are mathematically
equivalent in that they have identical wave equations.  This is true not
only for the free theories, but also for particles in external fields, and
without direct self-interactions.  This is no accident:  Classical field
theory and classical mechanics are two different limits of quantum field
theory.  They are both called classical limits, and written as $\h£0$,
but since $\h$ is really 1, this limit depends on how one inserts
$\h$'s into the quantum field theory action.

The wave equation in quantum mechanics is the Schr¬odinger equation. 
The corresponding field theory action is then simply the one that gives
this wave equation as the equation of motion, where the wave function
is replaced with the field:
$$ S_{ft} = Çd^4 x¼Æ*(-i»_t +H)Æ $$
 As usual (cf.¼electromagnetism), the field is a function of space and time;
thus, we integrate $d^4 x=dt¼d^3 x$ over the three space and one time
dimensions.  The Hamiltonian is some function of coordinates and
momenta, with the replacement $p_i£-i»_i$, where $»_i=»/»x^i$ are the
space derivatives and $»_t=»/»t$ is the time derivative.

The Hamiltonian can contain coupling to other fields.  For a general
Hamiltonian quadratic in momenta, in a notation implied by the
corresponding Lagrangian quadratic in time derivatives,
$$ H = üg^{ij}(-i»_i +A_i)(-i»_j +A_j) +U $$
 where $g^{ij}$, $A_i$, and $U$ are now interpreted as fields, and thus
depend on both $x^i$ and $t$, as does $Æ$.  In the case $g^{ij}=¶^{ij}$, we
can identify $A_i$ and $U$ as the three-vector and scalar potentials of
electromagnetism, and we can add the usual action for electromagnetism
to the action for $Æ$.  The action then can be varied also with respect to
$A$ and $U$ to obtain Maxwell's equations with a current in terms of $Æ$
and $Æ*$.  We can also treat $g^{ij}$ as a field, in which case it and parts
of $A$ and $U$ are the components of the gravitational field.

Field theory actions can be quantized in the same ways as mechanics
ones.  In this case, we recognize the $Æ*ÀÆ$ term as a special case of the
$Àìì¯$ term in the generic Hamiltonian form of the action discussed
earlier.  Thus, $Æ(x^i)$ and $Æ*(x_i)$ have replaced $x^i$ and $p_i$ as
the variables; $x^i$ is now just an index (label) on $Æ$ and $Æ*$, just as
$i$ was an index on $x^i$ and $p^i$.  The field-theory Hamiltonian is then
identified as
$$ H_{ft}[Æ,Æ*] = Çd^3 x¼\H,ââ\H = Æ*HÆ $$
 In field theory the Hamiltonian will always be a space integral of a
``Hamiltonian density" $\H$.

The classical limit of a quantum theory defined by a classical action $S$
can be defined as follows:  Introduce $\h$ into the theory by replacing
$$ S £ \h^{-1}S $$
This has no effect on the classical equations of motion, but it
introduces $\h$ into the Poisson bracket:  
$$ -Àqp +Hâ£â-{1\over\h}Àqp +{1\over\h}H $$
$$ Üâ{1\over\h}[q,p] = i,â{d\over dt} = {»\over »t} +i{1\over\h}[H,¼] $$
 We can then recognize
the limit $\h£0$ as the classical limit.  
In the quantum theory, it is equivalent to replacing
$$ p £ -i\h »_q,ââi»_t -H £ i\h »_t -H $$
 i.e., all derivatives get a factor of $\h$.
(More details will be possible
when we consider quantization in subsection VA2.)
However, a quantum theory can often be described by more than one
classical action:  This is known as ``duality" (between any two such
actions).

In particular, any free quantum field theory, and many interacting ones,
can be described by both a classical mechanics action and by a
classical field theory action:  This is the well-known
``wave-particle duality".  We have just seen the standard
nonrelativistic example.  Furthermore, since we know the direct
relation between the two actions in terms of the mechanics
Hamiltonian $H$, we can describe both classical limits directly
in terms of just the field theory action.
The classical field theory limit is defined by inserting $\h$ ÓonlyÕ as
$$ S_{ft} £ \h^{-1}S_{ft} $$
 On the other hand, if we put in $\h$'s ÓonlyÕ as
$$ »_i £ \h »_i,ââ»_t £ \h »_t $$
 which gives the usual $\h$ dependence associated with the
Schr¬odinger equation, then the classical limit $\h£0$ gives classical
mechanics.  This defines classical mechanics as the macroscopic limit, the
limit of large distances and times.  

A convenient way to implement this limit is to introduce the mechanics
action $S=Çdt(-Àx{}^i p_i +H)$ into the field theory, and then take the limit
$\h£0$ after the replacement
$$ S £ \h^{-1}S $$
 on the mechanics action instead of on the derivatives.  The mechanics
action can be introduced when solving the field equations:  The solution
to the wave equation can be expressed in terms of the propagator,
which in turn can be written in terms of the mechanics action or
Hamiltonian.

Usually $\h$ is introduced through dimensional analysis; 
it has dimensions of action, whose dimension is defined by conservation laws 
vs.Êsymmetries (see subsection IIB4) as energy $ð$ time or momentum $ð$ length. 
But the various parameters in the action may be assigned different dimensions in mechanics and field theory; for example, in classical mechanics one has $p^2+m^2$, while in classical field theory one has $-õ+m^2$, and these $m$'s differ by a factor of $\h$ in quantum field theory, but are unrelated classically, where one has dimensions of mass and the other of inverse length. (Consider, e.g., coupling a classical, massive particle to a classical, massive field.) 
So it's only how you scale these constants (mass, charge, etc.) as $\h£0$ that gives different classical limits.  Since field theory actions can have fields rescaled to put the inverse coupling in front of the action, the classical field theory limit is generally the limit of weak coupling.  On the other hand, the classical mechanics limit, being macroscopic, is the limit of large mass; it's also a strong coupling limit, as the action coming from a force between particles is proportional to the coupling, and gets a $1/\h$.  (For example, the familiar fine structure constant $e^2/\h$ of quantum mechanics is $e^2\h$ in quantum field theory.)

More generally, we can define actions that are not restricted to be
quadratic in any field.  The Hamiltonian density $\H(t,x^i)$ or Lagrangian
density $\L(t,x^i)$,
$$ S[Ä] = Çdt¼d^3 x¼\L[Ä(t,x^i)] $$
 should be a function of fields at that point, with only a finite number
(usually no more than two) spacetime derivatives.  This is the definition
of locality used for general quantum systems in subsection IIIA1, but
extended from derivatives in time to also those in space.  Although this
condition is not always used in nonrelativistic field theory (for example,
when long-range interactions, such as Coulomb or gravitational, are
described without attributing them to fields), it is crucial in relativistic
field theory.  For example, global symmetries lead by locality to local
(current) conservation laws.  Locality is also the reason that spacetime
coordinates are so important:  Translation invariance says that the
position of the origin is an unphysical, redundant variable; however,
locality is most easily used with this redundancy. 

Field equations are derived by the straightforward generalization of the
variation of actions defined in subsection IIIA1:  As follows from treating
the spatial coordinates in the same way as discrete indices,
$$ ¶S ­ Çdt¼d^3 x¼¶Ä^m(t,x^i){¶S\over ¶Ä^m(t,x^i)} $$
 For example,
$$ S = -Çdt¼d^3 x¼üÀÄ{}^2âÜâ{¶S\over ¶Ä} = ¬Ä $$

\x IIIA3.1  Consider the action
$$ S[Ä] = Çdt¼d^{D-1}x¼[-üÀÄ{}^2 +V(Ä)] $$
 for potential $V(Ä)$ (a function, not a functional).
 ªa Find the field equations.
 ªb Assume $V(Ä)=ÂÄ^n$ for some positive integer $n$ and constant,
ÓdimensionlessÕ $Â$, in units $\h=c=1$.  Use dimensional
analysis to relate $n$ and $D$ (of course, also a positive integer),
and list all paired possibilities of $(n,D)$.

Ü4. Relativity

Generalization to relativistic theories is straightforward, except for the
fact that the Klein-Gordon equation is second-order in time derivatives;
however, we are familiar with such actions from nonrelativistic quantum
mechanics.  As usual, we need to check the sign of the terms in the
action:  Checking the positivity of the Hamiltonian (i.e., the energy), we
see from the general relation between the Lagrangian and Hamiltonian
(subsection IIIA1) that the terms without time derivatives must be
positive; the time-derivative terms are then determined by Lorentz
covariance.

At this point we introduce some normalizations and conventions that will
prove convenient for Fourier transformation and other reasons to be
explained later.  Whenever D-dimensional integrations are involved (as
should be clear from context), we use
$$ Çdx ­ Ç{d^D x\over (2¹)^{D/2}},ââÇdp ­ Ç{d^D p\over (2¹)^{D/2}} $$
$$ ¶(x-x') ­ (2¹)^{D/2}¶^D (x-x'),ââ¶(p-p') ­ (2¹)^{D/2}¶^D (p-p') $$
 In particular, this normalization will be used in Green functions and
actions.  For example, these implicit $2¹$'s appear in functional
variations:
$$ ¶S ­ Çdx¼¶ÄÊ{¶S\over ¶Ä}âÜâ{¶\over ¶Ä(x)}Ä(x') = ¶(x-x') $$

The action for a real scalar is then
$$ S = Çdx¼L,ââL = \f14 (»Ä)^2 +V(Ä) $$
 where $V(Ä)³0$, and we now write $L$ for the Lagrange density.  In
particular, $V=\f14 m^2 Ä^2$ for the free theory.  The free field equation
is then $p^2+m^2=-õ+m^2=0$, replacing the nonrelativistic $-i»_t+H=0$. 
For a complex scalar, we replace $üÄģЍ$ in both terms.

We know from previous considerations (subsection IIB2) that the field
equation for a free, massless, Dirac spinor is $©É»ï=0$.  The
generalization to the massive case (subsection IIB4) is obvious from
various considerations, e.g., dimensional analysis; the action is
$$ S = Çdx¼Ðï(iÖ»+\f{m}{å2})ï $$
 in arbitrary dimensions, again using the notation $Ö»=©É»$.  In four
dimensions, we can decompose the Dirac spinor into its two Weyl spinors
(see subsection IIA6):
$$ L = Ðï(iÖ»+\f{m}{å2})ï = (ÐÆ_L^{ÀŒ}i»^Œ{}_{ÀŒ}Æ_{LŒ}
	+ÐÆ_R^{ÀŒ}i»^Œ{}_{ÀŒ}Æ_{RŒ})
	+\f{m}{å2}(Æ_L^Œ Æ_{RŒ} +ÐÆ_L^{ÀŒ}ÐÆ_{RÀŒ}) $$
 For the case of the Majorana spinor, the 4D action reduces to that for a
single Weyl spinor,
$$ S = Çdx¼[-iÐÆ^{Àº}»_{ŒÀº}Æ^Œ +\f{m}{å2}ü(Æ^ŒÆ_Œ+ÐÆ^{ÀŒ}ÐÆ_{ÀŒ})] $$
 Note that in our conventions $§^0_{ŒÀº}=\f1{å2}¶_{Œº}$ (and similarly for
the opposite indices, since $§^a_{ŒÀº}=§_a^{ºÀŒ}$), so that the time
derivative term is always proportional to $Æÿ(-i»_0)Æ$, as
nonrelativistically (previous subsection).

A scalar field must be complex to be charged (i.e., a representation of
U(1)):  From the gauge transformation
$$ ' = e^{iÂ} $$
 we find the minimal coupling (for $q=1$)
$$ S_ = Çdx¼[ü|(»+iA)|^2 +üm^2||^2] $$
 The electromagnetic current is then defined by varying the matter action with respect to the gauge field:
$$ J ­ {¶S\over ¶A} = Ѝ(-iü\onª»+A) $$
 This action is also invariant under charge conjugation
$$ C:  £ *,ââA £ -A $$
 which changes the sign of the charge, since $*'=e^{-iÂ}*$.  

\x IIIA4.1  Let's consider the semiclassical interpretation of a charged
particle as described by a complex scalar field $Æ$, with Lagrangian
$$ L = ü(|áÆ|^2 +m^2|Æ|^2) $$
 ªa  Use the semiclassical expansion in $\h$ defined by
$$ á £ \h » +iqA,ââÆ £ å¨e^{-i\S / \h} $$
 Find the Lagrangian in terms of $¨$ and $\S$ (and the background field
$A$), order-by-order in $\h$ (in this case, just $\h^0$ and $\h^2$).
 ªb  Take the semiclassical limit by dropping the $\h^2$ term in $L$, to
find
$$ L £ ¨ü[(-»\S +qA)^2 +m^2] $$
 Vary with respect to $\S$ and $¨$ to find the equations of motion. 
Defining
$$ p ­ -»\S $$
 show that these field equations can be interpreted as the mass-shell
condition and current conservation.  Show that $A$ couples to this current
by varying $L$ with respect to $A$.

The spinor field also needs doubling for charge.  (Actually, the doubling
can be avoided in the massless case; however, problems show up at the
quantum level, related to the fact that there is no charge conjugation
transformation without doubling.)  The gauge transformations are similar
to the scalar case, and the action again follows from minimal coupling, to
an action that has the global invariance ($Â$ = constant in the absence of
$A$):
$$ Æ_L'^Œ = e^{iÂ}Æ_L^Œ,âÆ_R'^Œ = e^{-iÂ}Æ_R^Œ $$
$$ S_e = Çdx¼[ÐÆ_L^{Àº}(-i»_{ŒÀº}+A_{ŒÀº})Æ_L^Œ
	+ ÐÆ_R^{Àº}(-i»_{ŒÀº}-A_{ŒÀº})Æ_R^Œ
	+\f{m}{å2}(Æ_L^ŒÆ_{RŒ}+ÐÆ_L^{ÀŒ}ÐÆ_{RÀŒ})] $$
 The current is found from varying with respect to $A$:
$$ J^{ŒÀº} = ÐÆ_L^{Àº}Æ_L^Œ - ÐÆ_R^{Àº}Æ_R^Œ $$
 Charge conjugation
$$ C: Æ_L^Œ ª Æ_R^Œ,ââA £ -A $$
 (which commutes with Poincar«e transformations) changes the sign of the
charge and current.

\x IIIA4.2  Show that this action can be rewritten in Dirac notation as
$$ S_e = Çdx¼Ðï(iÖ»-ÖA+\f{m}{å2})ï $$
 and find the action of the gauge transformation and charge conjugation
on the Dirac spinor.

As a last example, we consider the action for electromagnetism itself.  As
before, we have the gauge invariance and field strength
$$ A'_{ŒÀº} = A_{ŒÀº} - »_{ŒÀº} $$
$$ F_{ŒÀ©,ºÀ¶} = »_{ŒÀ©}A_{ºÀ¶}-»_{ºÀ¶}A_{ŒÀ©}
	= C_{Œº}Ðf_{À©À¶}+ÐC_{À©À¶}f_{Œº},âf_{Œº} = ü»_{(ŒÀ©}A_{º)}{}^{À©} $$
 We can write the action for pure electromagnetism as
$$ S_A = Çdx¼\f1{2e^2} f^{Œº}f_{Œº} = Çdx¼\f1{2e^2} Ðf^{ÀŒÀº}Ðf_{ÀŒÀº}
	= Çdx¼\f1{8e^2}F^{ab}F_{ab} $$
 dropping boundary terms, with the overall sign again determined by
positivity of the Hamiltonian, where $e$ is the electromagnetic coupling
constant, i.e., the charge of the proton.  (Other normalizations can be used
by rescaling $A_{ŒÀº}$.)  Maxwell's equations follow from varying the
action with a source term added:
$$ S = S_A +Çdx¼A^{ŒÀº}J_{ŒÀº}âÜâ\f1{e^2}»^º{}_{À©}f_{ºŒ} = J_{ŒÀ©} $$

\x IIIA4.3  By plugging in the appropriate expressions in terms of $A_a$
(and repeatedly integrating by parts), show that all of the above
expressions for the electromagnetism action can be written as
$$ S_A = -Çdx¼\f1{4e^2}[AÉõA +(»ÉA)^2] $$

\x IIIA4.4  Find all the field equations for all the fields, found from adding
to $S_A$ all the minimally coupled matter actions above.

Having seen many of the standard examples of relativistic field theory
actions, we now introduce one of the most important principles in
field theory; unfortunately, it can be justified only at the quantum level
(see chapter VII):

\medskip\Boxit{\noindent\it
Good ultraviolet behavior:  All quantum field theories should have
only couplings with nonnegative mass (engineering) dimension.}

\noindent (Here ``couplings" means the coefficients of arbitrary terms, when the fields have been defined so that the massless parts of the kinetic terms have no coupling dependence.)

\x IIIA4.5 Show in D=4 using dimensional analysis that this 
restriction on bosons $Ä$ and fermions $Æ$ restricts terms in the 
action to be of the form
$$ Ä, Ä^2, Ä^3, Ä^4, Ä»Ä, Ä^2 »Ä, Ä»»Ä; Æ^2, Æ»Æ; ÄÆ^2 $$
 and find the dimensions of all the corresponding coupling constants.

The energy-momentum tensor for electromagnetism is much simpler in
this spinor notation, and follows (up to normalization) from gauge
invariance, dimensional analysis, Lorentz invariance, and the vanishing of
its trace.  It has a form similar to that of the current in electrodynamics:
 $$ T_{ŒºÀ©À¶} = -\f1{e^2}f_{Œº}Ðf_{À©À¶} $$
 Note that it is invariant under the duality transformations of subsection
IIA7 (as is the electrodynamic current under chirality).

We have used conventions where $e$ appears multiplying only the action
$S_A$, and not in the ``covariant derivative"
$$ á = »+iqA $$
 where $q$ is the charge in units of $e$: e.g., $q=1$ for the proton, $q=-1$
for the electron.  Alternatively, we can scale $A$, as a field redefinition,
to produce the opposite situation:
$$ A £ eA:ââS_A £ Çdx¼\f18 F^2,ââá £ »+iqeA $$
 The former form, which we use unless noted otherwise, has the
advantage that the coupling appears only in the one term $S_A$, while the
latter has the advantage that the kinetic (free) term for $A$ is normalized
the same way as for scalars.  The former form has the further advantage
that $e$ appears in the gauge transformations of none of the fields,
making it clear that the group theory does not depend on the value of
$e$.  (This will be more important when generalizing to nonabelian groups
in section IIIC.)

Note that the massless parts of the kinetic (free) terms in these actions
are scale invariant (in arbitrary dimensions, when the
dimension-independent forms are used), when the fields are assigned the
scale weights found from conformal arguments in subsection IIB2.

\x IIIA4.6 Using vector notation, minimal coupling, and dimensional
analysis, find the mass dimensions of the electric charge $e$ in arbitrary
spacetime dimensions, and show it is dimensionless only in $D=4$.

An interesting distinction between gravity and electromagnetism is that
static bodies always attract gravitationally, whereas electrically they
repel if they are like and attract if they are opposite.  This is a direct
consequence of the fact that the graviton has spin 2 while the photon has
spin 1:  The Lagrangian for a field of integer spin $s$ coupled to a current,
in an appropriate gauge and the weak-field approximation, is
$$ L = -\f1{4s!}Ä^{a_1...a_s}õÄ_{a_1...a_s} + 
	\f1{s!}gÄ_{a_1...a_s}J^{a_1...a_s} $$
 for some coupling $g$, where the sign of the first term is fixed by unitarity in quantum field
theory, or by positivity of the energy in classical field theory:
$$ L_0 = -\f1{4s!}Ä^{a_1...a_s}õÄ_{a_1...a_s}âÛâ
H_0 = \f1{4s!}[ÀÄ{}^{a_1...a_s}ÀÄ{}_{a_1...a_s} +(»_i Ä)^{a_1...a_s}(»_i Ä)_{a_1...a_s}] $$
(Time components of $Ä$ are unphysical, arising from gauge fixing, and so should be ignored as far as arguments of unitarity or positivity of energy are concerned.)
From a scalar field in the semiclassical approximation (see
exercise IIIA4.1 above), starting with
$$ J^{a_1...a_s} = Æ*(-üi\onª»{}^{a_1})ò(-üi\onª»{}^{a_s})Æ $$
 where 
$$ A\onª» B ­ A»B-(»A)B $$
 we see that the current will be of the form
$$ J^{a_1...a_s} = ¨p^{a_1}òp^{a_s} $$
 for a scalar particle, with ``density" $¨$.  (The same follows from comparing the
expressions for currents and energy-momentum tensors for particles as
in subsection IIIB4 below.  The only way to get vector indices out of a
scalar particle, to couple to the vector indices for the spin of the force
field, is from momentum.)  In the static approximation, only time
components contribute:  We then can write this Lagrangian as, taking into
account $ú_{00}=-1$,
$$ L = -(-1)^s \f1{4s!}Ä_{0...0}õÄ_{0...0} + \f1{s!}gÄ_{0...0}¨(p^0)^s $$
 where $E=p^0>0$ for a particle and $<0$ for an antiparticle.  
Solving for $Ä$ by its field equation and plugging back in, we have
$$ L = (-1)^s \f1{s!} g^2 ¨E^s {1\over õ} ¨E^s $$
Since we're looking at the static case, $õ$ can be replaced with the Laplacian $ë$, and the Lagrangian (density) is the same as the Hamiltonian (density), so the ``potential energy" $V$ produced by this interaction (we have neglected the ``kinetic energy", or pure $Æ$ terms in the action) is, in D=4,
$$ V =  -(-1)^s \f1{s!} (2¹g^2)üÇ{d^3 x\over (2¹)^2}{d^3 x'\over (2¹)^2}
(¨E^s)(x){1\over |x-x'|}(¨E^s)(x') $$
where we have used
$$ {1\over ë}¶^3(x-x') = -{1\over 4¹|x-x'|} $$
in terms of the 3D distance $|x-x'|$.  Thus the
spin-dependence of the potential/force between two particles goes as
$-(-E_1 E_2)^s$.  It then follows that all particles attract by forces
mediated by even-spin particles, and a particle and its antiparticle
attract under all forces, while repulsion will occur for odd-spin forces
between two identical particles.  (We can substitute ``particles of the
same-sign charge" for ``identical particles", and ``particles of opposite-sign charge" for ``particle and its antiparticle", where the charge is the
coupling constant appropriate for that force.)

\x IIIA4.7  Show that the above current is conserved,
$$ »_{a_1}J^{a_1òa_s} = 0 $$
 (and the same for the other indices, by symmetry) if $Æ$ satisfies the
free Klein-Gordon equation (massless or massive).

Ü5. Constrained systems

Constraints not only frequently appear in nonrelativistic physics, but are a
general feature of relativistic particles, so we now give a brief
description of how they are incorporated into actions.  Consider a general
action, with constraints, in Hamiltonian form:
$$ S = Çdt (-Àq{}^m p_m +H),âH = H_{gi}(q,p) +Â^i G_i(q,p) $$
 (For simplicity, we consider all physical variables to be bosonic for this
subsection, but the method generalizes straightforwardly paying careful
attention to signs.)  This action is a functional of $q^m,p_m,Â^i$, which
are in turn functions of $t$, where $m$ and $i$ run over any number of
values.  We can think of this as describing a nonrelativistic particle with
coordinates $q$ and momenta $p$ in terms of time $t$, but the form is
general enough to apply to relativistic theories.  The $Àqp$ term tells us
$p$ is canonically conjugate to $q$; the rest of the action gives the
Hamiltonian, usually quadratic in momenta.  The variables $Â^i$ are
``Lagrange multipliers", whose variation in the action implies the
constraints $G_i=0$.  We then can interpret $H_{gi}$ as the usual (``gauge
invariant") Hamiltonian.  We also require that the transformations
generated by the constraints close, and that the Hamiltonian be invariant:
$$ [G_i,G_j] = -if_{ij}{}^k G_k,â[G_i,H_{gi}] = 0 $$
 (More generally, we can allow $[G_i,H_{gi}] = -if_i{}^j G_j$.)  This says
that the constraints don't imply any new constraints that we might have
missed, and that the ``energy" represented by $H_{gi}$ is invariant under
these transformations.  
In general, not all constraints commute with the Hamiltonian, and thus those constraints are not time independent; we are considering here just the ones that do. The ones that don't, including their Lagrange multipliers, are implicitly included in the gauge invariant Hamiltonian.  (Thus, the time-dependent constraints must commute with the time-independent ones.)

We then find that the action is invariant under the canonical
transformations
$$ ¶(q,p) = i[½^i G_i,(q,p)]âÜâ
	¶q^m = ½^i{»G_i\over »p_m},â¶p_m = -½^i{»G_i\over »q^m} $$
$$ 0 = ¶\left({d\over dt}\right) = ¶\left({»\over »t} +iH\right) 
	= i(¶Â^i)G_i -iÀ½{}^i G_i +[Â^j G_j,½^i G_i] $$
$$ Üâ¶Â^i = À½{}^i +½^j Â^k f_{kj}{}^i $$
 (with $¶(d/dt)$ defined as in subsection IA1), where $»/»t$ acts on the
``explicit" $t$ dependence (that in everything except $q$ and $p$):  For
general expressions, the total time derivative and total variation are
given by commutators as
$$ {d\over dt}A = {»\over »t}A +i[H,A],â¶A = ¶_0 A + i[½^i G_i,A] $$
 where $¶_0$ acts on everything except $q$ and $p$.  The action then
varies under these transformations as the integral of a total derivative,
which vanishes under appropriate boundary conditions:
$$ ¶S_H = Çdt{d\over dt}[-(¶q^m)p_m +½^i G_i] = 0 $$

The simplest example is the case with one constraint, which is linear in
the variables:  If the constraint is $p$, the gauge transformation is
$¶q=½$, so we gauge $q=0$ and use the constraint $p=0$.  In general, this
means that for every degree of freedom we can gauge away, the
conjugate variable can be fixed by the constraint.  Thus, for each
constraint we eliminate 3 variables: the variable fixed by the constraint,
its conjugate, and the Lagrange multiplier that enforced the constraint,
which has no conjugate.  (In the Lagrangian form of the action the
conjugate may not appear explicitly, so only 2 variables are eliminated.)
As an example of a constraint that does not generate a gauge invariance, consider a nonrelativistic particle constrained to a sphere by
$G=(x^i)^2-1$:  We can change to spherical coordinates, apply the
constraint to eliminate the radial coordinate, and then eliminate the radial component of the momentum as an auxiliary variable (not appearing with time derivatives), leaving
an unconstrained theory in terms of angles and their conjugates.  In most
cases in field theory a similar procedure can be applied, eliminating both gauge and auxiliary variables:  The result is
called a ``unitary gauge".

\x IIIA5.1  Let's look closer at this example:
 ªa Perform quantization of a nonrelativistic particle on a sphere
($G=(x^i)^2-1$ for $i=1,2,3$), reducing to an action in terms of just
the angles $Ï$ and $Ä$, and their conjugates.
 ªb Repeat this procedure using instead the gauge invariant Hamiltonian
$H_{gi}=(x^i)^2(p^j)^2/2m$ and the time-independent constraint $G'=x^i p_i$,
and compare.  Show the relation of what's left of the Hamiltonian to the angular momentum.

The standard example of a relativistic constrained system is in field
theory --- electromagnetism.  Its action can be written in ``first-order (in
derivatives) formalism" by introducing an auxiliary field $G_{ab}$:
$$ F^2 £ F^2 -G^2 £ F^2 -(G-F)^2 = 2GF -G^2 $$
 where in the first step we added a trivial term for $G$ and in the second
step made a trivial redefinition of $G$, so elimination of $G$ by its
algebraic equation of motion returns the original Lagrangian.  The
Hamiltonian form comes from eliminating only $G_{ij}$ by its field
equation, since only $F_{0i}$ contains time derivatives:
$$ 2GF -G^2 £ (F_{ij})^2 -4G_{0i}F_{0i} +2(G_{0i})^2 $$
$$ = -4ÀA_i G_{0i} +[2(G_{0i})^2 +(F_{ij})^2] -4A_0 »_i G_{0i} $$
 which we recognize as the three generic terms for the action in
Hamiltonian form, with $G_{0i}$ as the canonical momenta for $A_i$, and
$A_0$ as the Lagrange multiplier.  The constraint is Gauss' law, and it
generates the usual gauge transformations.

Thus $Â^i$ are also gauge fields for the gauge (time-dependent)
transformations $½^i(t)$.  They allow construction of the gauge-covariant
time derivative
$$ á = »_t +iÂ^i G_i,ââ{d\over dt} = á +iH_{gi}âÜâ¶_0 á = i[½^i G_i,á] $$
 It is convenient to transform the gauge fields away using these gauge
transformations, so $H=H_{gi}$.  However, with the usual boundary
conditions $Ç_{-¥}^¥ dt¼Â^i$ is gauge invariant under the linearized
transformations, so the most we could expect is to gauge $Â^i$ to
constants.  More precisely, the group element
$$ \T\left[ exp \left( -iÇ_{-¥}^¥ dt¼Â^i(t) G_i \right) \right] $$
 is gauge invariant, where ``$Ê\TÊ$" is time ordering, meaning we
write the exponential of the integral as the product of exponentials of
infinitesimal integrals, and order them with respect to time, later time
intervals going to the left of earlier ones.  (We treat $G_i$ quantum
mechanically or use Poisson brackets when combining the exponentials.) 
This is the quantum mechanical version of the time development resulting
from the corresponding term in the classical action.  It is also the phase
factor coming from the infinite limit of the covariant time translation
$$ e^{-ká(t)} = 
	\T\left[ exp \left( -iÇ_{t-k}^t dt'¼Â^i(t') G_i \right) \right]
	e^{-k»_t} $$
 as seen from reordering the time derivatives when writing $e^{-ká(t)}$
as the product of exponentials of infinitesimal exponents.  This allows us
to write the explicit gauge transformation
$$ e^{-iñ(t)} = 
	\T\left[ exp \left( -iÇ_{t_0}^t dt'¼Â^i(t') G_i \right) \right] 
	= e^{-(ët)á(t)}e^{(ët)»_t},ââët = t-t_0 $$
$$ Üâá'(t) = e^{iñ(t)}á(t)e^{-iñ(t)} = e^{-(ët)»_t}á(t)e^{(ët)»_t}
	=  á(t_0) = »_t +iÂ^i(t_0) G_i $$
 (where we define $»_t$ to vary $t$ while keeping $t-t_0$ fixed).  Thus,
we can gauge $Â$ to its value at a fixed time $t_0$.  Another way to see
this is that varying $Â^i$ in the action at a fixed time gives $G_i=0$ at
that time, but the remaining field equations imply $ÀG_i=0$, so $G_i=0$
always, and $Â^i$ is redundant at other times.  This means that if we
carelessly impose $Â^i=0$ at all times, we must also impose $G_i=0$ at
some fixed time.

Note that this special gauge transformation itself has a very simple
gauge transformation:  Transforming the $Â$ in $ñ$ by an arbitrary finite
transformation $½^i(t)$,
$$ e^{-iñ'(t)} = e^{-i½^i(t)G_i}e^{-iñ(t)}e^{i½^i(t_0)G_i} $$
 consistent with the transformation law of $á'(t)$ above.  Thus, applying
the transformation $ñ$ to any gauge-dependent quantity $Ä$ gives a
gauge-independent quantity $Ä'(Ä,Â)$, which is invariant under the local
transformations $½(t)$ and transforms only under the ``global"
transformations $½(t_0)$.  Thus, fixing the gauge $Â(t)=0$ is equivalent to
working with gauge-invariant quantities.

Fixing an invariance of the action is not unique to gauge invariances: 
Global invariances also need to be fixed, although the procedure is so
trivial we seldom discuss it.  For example, even in nonrelativistic systems
Galilean invariance needs to be fixed:  When analyzing a specific problem,
we often choose some object to be at rest (velocity transformations),
choose another to be oriented or moving in a specific direction
(rotations), and choose a specific event to happen at the origin of space
and time (translations).  Alternatively, we can work with Galilean
invariants, just as in gauge theories we can work with gauge invariants;
however, in practice, for explicit calculations (as opposed to discussing
general properties), it is more convenient to fix the invariance, as this
allows simplification of the equations.

\refs

£1 M.B. Halpern and W. Siegel, \PRD 16 (1977) 2486:\\
	classical mechanics as a limit of quantum field theory.
 £2 Fock, Óloc. cit.Õ (IIB):\\
	gauge transformations (for both electromagnetic potentials and wave functions).
 £3 J.D. Jackson and L.B. Okun, \xxxlink{hep-ph/0012061}, 
	ÓRev.Mod.Phys.Õ É73 (2001) 663:\\
	history of gauge invariance.
 £4 P.A.M. Dirac, ÓProc. Roy. Soc.Õ ÉA246 (1958) 326;\\
	L.D. Faddeev, ÓTheo. Math. Phys.Õ É1 (1969) 1:\\
	Hamiltonian formalism for constrained systems.

\unrefs

Û5 B. PARTICLES

The simplest relativistic actions are those for the ÓmechanicsÕ (as opposed
to field theory) of particles.  These also give the simplest examples of
gauge invariance in relativistic theories.  Later we will find that various
properties of the quantum mechanics of these actions help to explain
some features of quantum field theory.

Ü1. Free

For nonrelativistic mechanics, the fact that the energy is expressed as a
function of the three-momentum is conjugate to the fact that the spatial
coordinates are expressed as functions of the time coordinate.  In the
relativistic generalization, all the spacetime coordinates are expressed as
functions of a parameter $ $:  All the points that a particle occupies in
spacetime form a curve, or ``worldline", and we can parametrize this
curve in an arbitrary way.  Such parameters generally can be useful to
describe curves:  A circle is better described by $x(Ï),y(Ï)$ than $y(x)$
(avoiding ambiguities in square roots), and a cycloid can be described
explicitly only this way.  

The action for a free, spinless particle then can be written in relativistic
Hamiltonian form as
$$ S_H = Çd [-Àx{}^m p_m +vü(p^2 +m^2)] $$
 where $v$ is a Lagrange multiplier enforcing the constraint $p^2+m^2=0$.
This action is very similar to nonrelativistic ones, but instead of
$x^i(t),p_i(t)$ we now have $x^m( ),p_m( ),v( )$ (where ``$À{}¼$'' now
means $d/d $).  The gauge invariance generated by $p^2+m^2$ is
$$ ¶x = ½p,â¶p = 0,â¶v = À½ $$

\x IIIB1.1  Consider the action
$$ S = Çd ¼Ó(ÀtE -Àx{}^i p_i) -vü[·(x)E^2 -(p^i)^2 -m^2]Õ $$
 describing propagation of a particle in a medium with a ``dielectric constant" $·(x)$. Using its equations of motion,
 ªa  Show that the ``group velocity" $dE/dp^i$ is just the usual velocity $dx^i/dt$. (This agrees with the usual interpretation of group velocity as the velocity of information.)
 ªb  Show that the components of the ``wave velocity" $p^i/E$ are conserved, for time-independent $·$, in directions in which $·$ doesn't change. If $·$ is a function of only one spatial dimension (as in the usual light refraction problems), these conservation laws, together with the energy-momentum relation, allow all components of the wave velocity (and thus the group velocity) to be determined from initial values.
 ªc  Show that for a massive particle neither of these is the same as the ``phase velocity" $v^i = ¶x^i/¶t$, defined by 
$$ 0 = ¶(phase) = (¶x^m)p_m ¾ v^i p_i - E $$
 even in empty space. (Since this is only 1 equation for 3 unknowns, it is really more of a ``phase speed".) Examine phase velocity in the rest frame.

A more recognizable form of this invariance can be obtained by noting
that any action
$S(Ä^A)$ has invariances of the form
$$ ¶Ä^A = ·^{AB}{¶S\over ¶Ä^B},â·^{AB} = - ·^{BA} $$
 which have no physical significance, since they vanish by the equations
of motion.  In this case we can add
$$ ¶x = ·(Àx-vp),â¶p = ·Àp,â¶v = 0 $$
 and set $½=v·$ to get
$$ ¶x = ·Àx,â¶p = ·Àp,â¶v = (À{·\hbox{\strut}v}) $$
 We then can recognize this as a (infinitesimal) coordinate transformation
for $ $:
$$ x'( ') = x( ),âp'( ') = p( ),âd 'v'( ') = d ¼v( );ââ ' =   -·( ) $$
 The transformation laws for $x$ and $p$ identify them as ``scalars" with
respect to these ``one-dimensional" (worldline) coordinate
transformations (but they are vectors with respect to D-dimensional
spacetime).  On the other hand, $v$ transforms as a ``density":  The
``volume element" $d ¼v$ of the world line transforms as a scalar.  This
gives us a way to measure length on the worldline in a way independent
of the choice of $ $ parametrization.  Because of this geometric
interpretation, we are led to constrain
$$ v > 0 $$
 so that any segment of the worldline will have positive length.  Because
of this restriction, $v$ is not a Lagrange multiplier in the usual sense.
This has significant physical consequences:  $p^2+m^2$ is treated
neither as a constraint nor as the Hamiltonian.  While in nonrelativistic
theories the Schr¬odinger equation is $(E -H)Æ=0$ and $G_i Æ=0$ is
imposed on the initial states, in relativistic theories $(p^2+m^2)Æ=0$ is
the Schr¬odinger equation:  This is more like $HÆ=0$, since $p^2$ already
contains the necessary $E$ dependence.

The Lagrangian form of the free particle action follows from eliminating
$p$ by its equation of motion $vp=Àx$:
$$ S_L = Çd ¼ü(vm^2 -v^{-1}Àx{}^2) $$
 For $m±0$, we can also eliminate $v$ by its equation of motion
$v^{-2}Àx{}^2+m^2=0$:
$$ S = mÇd å{-Àx{}^2} = mÇå{-dx^2} = mÇds = ms $$
 The action then has the purely geometrical interpretation as the proper
time; however, this last form of the action is awkward to use because of
the square root, and doesn't apply to the massless case.  Note that the
$v$ equation implies $ds=m(d ¼v)$, relating the ``intrinsic" length
of the worldline (as measured with the worldline volume element) to its
``extrinsic" length (as measured by the spacetime metric).  As a
consequence, in the massive case we also have the usual relation
between momentum and ``velocity"
$$ p^m = m{dx^m\over ds} $$
 (Note that $p^0$ is the energy, not $p_0$.)

\x IIIB1.2  Take the nonrelativistic limit of the Poincar«e algebra:
 ªa  Insert the speed of light $c$ in appropriate places for the structure
constants of the Poincar«e group (guided by dimensional analysis) and take
the limit $c£0$ to find the algebra of the Galilean group. 
 ªb  Do the same for the representation of the Poincar«e group generators
in terms of coordinates and momenta.  In particular, take the limit of the
Lorentz boosts to find the Galilean boosts.
 ªc  Take the nonrelativistic limit of the spinless particle action, in the
form $ms$.  (Note that, while the relativistic action is positive, the
nonrelativistic one is negative.)

\x IIIB1.3 Consider the following action for a particle with additional
fermionic variables $©$ and additional fermionic constraint $©Ép$:
$$ S_H = Çd  (-Àx{}^m p_m -üiÀ©{}^m ©_m +üvp^2 +i©Ép) $$
 where $Â$ is also anticommuting so that each term in the action is
bosonic.  
 ªa Find the algebra of the constraints, and the transformations
they generate on the variables appearing in the action.  
 ªb Show that the
``Dirac equation" $©Ép|ïÔ=0$ implies $p^2|ïÔ=0$.  
 ªc Find the Lagrangian
form of the action as usual by eliminating $p$ by its equation of motion. 
(Note $Â^2=0$.)

\x IIIB1.4 Consider a ``supercoordinate" $X^m$ that is a function of both a
fermionic variable $½$ and the usual $ $:
$$ X^m( ,½) = x^m( ) +i½©^m( ) $$
 where the Taylor expansion in $½$ terminates because $½^2=0$.  Identify
$x$ with the usual $x$, and $©$ with its fermionic partner introduced in
the previous problem.  In analogy to the way $©Ép$ was the square root
of the $ $-translation generator $üp^2$, we can define a square root of
$»/» $ by the ``covariant fermionic derivative"
$$ D = {»\over »½} +i½{»\over » }âÜâD^2 = i{»\over » } $$
 We also want to generalize $v$ in the same way as $x$, to make the
action independent of coordinate choice for both $ $ and $½$.  This
suggests defining
$$ E = v^{-1} +i½Â $$
 and the gauge invariant action
$$ S_L = Çd d½¼üE(D^2 X^m) DX_m $$
 Integrate this action over $½$, and show this agrees with the action of
the previous problem after suitable redefinitions (including the
normalization of $Çd½$).

The (D+2)-dimensional (conformal) representation of the massless particle
(subsection IA6) can be derived from the action
$$ S = Çd ¼ü(-Ày{}^2 +Ây^2) $$
 where $Â$ is a Lagrange multiplier.  This action is gauge invariant under
$$ ¶y = ·Ày -üÀ·y,â¶Â = ·À +2À·Â +ü\tdt · $$
 If we vary $Â$ to eliminate it and $y^-$ as in subsection IA6, the action
becomes
$$ S = -Çd ¼üe^2 Àx{}^2 $$
 which agrees with the previous result, identifying $v=e^{-2}$, which also
guarantees $v>0$.

\x IIIB1.5  Find the Hamiltonian form of the action for $y$:  The
constraints are now $y^2$, $r^2$, and $yÉr$, in terms of the conjugate $r$
to $y$ (see exercise IA6.2).  Find the gauge transformations in the
standard way (see subsection IIIA5).  Show how the above Lagrangian
form can be obtained from it, including the gauge transformations.

Using instead the corresponding twistor (subsection IIB6) to satisfy
$y^2=0$, the massless, spinless particle now has a single term for its
mechanics action:
$$ S = Çd ¼\f14 ·_{\A\B\C\D}Àz{}^{\A Œ}Àz{}^\B{}_Œ z^{\C º}z^\D{}_º $$
 Unlike all other relativistic mechanics actions, all variables have been
unified into just $z$, without the introduction of square roots.  

\x IIIB1.6  Expressing $z$ in terms of $Â_Œ{}^µ$ and $x^{µÀµ}$ as in
subsection IIB6, show this action reduces to the previous one.

Ü2. Gauges

Rather than use the equation of motion to eliminate $v$ it's more
convenient to use a gauge choice:  The gauge $v=1$ is called ``affine
parametrization" of the worldline.  Note that the gauge transformation of
$v$, $¶v=À½$, has no dependence on the coordinates $x$ and momenta $p$,
so that choosing the gauge $v=1$ avoids any extraneous $x$ or $p$
dependence that could arise from the gauge fixing.  (The appearance of
such dependence will be discussed in later chapters.)  Since $T=Çd ¼v$, the
intrinsic length, is gauge invariant, that part of $v$ still remains when the
length is finite, but it can be incorporated into the limits of integration: 
The gauge $v=1$ is maintained by $À½=0$, and this constant $½$ can be
used to gauge one limit of integration to zero, completely fixing the
gauge (i.e., the choice of $ $).  We then integrate $Ç_0^T$, where $T³0$
(since originally $v>0$), and $T$ is a variable to vary in the action.  The
gauge-fixed action is then
$$ S_{H,AP} = Ç_0^T d [-Àx{}^m p_m +ü(p^2 +m^2)] $$
 In the massive case, we can instead choose the gauge $v=1/m$; then the
equations of motion imply that $ $ is the proper time.  The Hamiltonian
$p^2/2m$ + constant then resembles the nonrelativistic one.

Another useful gauge is the ``lightcone gauge"
$$   = {x^+\over p^+} $$
 which, unlike the Poincar«e covariant gauge $v=1$, fixes $ $ completely;
since the gauge variation $¶(x^+/p^+)=½$, we must set $½=0$ to maintain
the gauge.  Also, the gauge transformation is again $x$ and $p$
independent.  In lightcone gauges we always assume $p^+±0$, since we
often divide by it.  This is usually not too dangerous an assumption, since
we can treat $p^+=0$ as a limiting case (in D>2).

We saw from our study of constrained systems that, for every degree of
freedom we can gauge away, the conjugate variable can be fixed by the
constraint that generates that gauge invariance:  In the case where the
constraint is $p$, the gauge transformation is $¶q=Â$, so we gauge
$q=0$ and use the constraint $p=0$.  In lightcone gauges the constraints
are almost linear:  The gauge condition is $x^+=p^+ $ and the constraint
is $p^-=...$, so the Lagrange multiplier $v$ is varied to determine $p^-$. 
On the other hand, varying $p^-$ gives
$$ ¶p^-âÜâv = 1 $$
 so this gauge is a special case of the gauge $v=1$.   An important point is
that we used only ``auxiliary" equations of motion: those not involving
time derivatives.  (A slight trick involves the factor of $p^+$:  This is a
constant by the equations of motion, so we can ignore $Àp{}^+$ terms. 
However, technically we should not use that equation of motion; instead,
we can redefine $x^-£x^-+...$, which will generate terms to cancel any
$Àp{}^+$ terms.)  The net result of gauge fixing and the auxiliary equation
on the action is
$$ S_{H,LC} = Ç_{-¥}^¥ d  [Àx{}^- p^+ -Àx{}^i p^i +ü(p^{i2} +m^2)] $$
 where $x^a=(x^+,x^-,x^i)$, etc.  In particular, since we have fixed one
more gauge degree of freedom (corresponding to constant $½$), we have
also eliminated one more constraint variable ($T$, the constant part of
$v$).  This is one of the main advantages of lightcone gauges:  They are
``unitary", eliminating all unphysical degrees of freedom.

\x IIIB2.1  Another obvious gauge is $ =x^0$, which works as well as the
lightcone gauge as far as eliminating worldline coordinate invariance is
concerned.  (The same is true for $ =nÉx$ for any constant vector $n$.)
 ªa  Consider the auxiliary equations of motion: 
Apply this gauge condition; then $p^0$ appears without time derivatives,
so eliminate it and $v$ by their equations of motion.  Show this
gauge is consistent only for $p^0>0$.  
 ªb  The resulting square root is awkward except in the nonrelativistic 
limit:  Take it, and compare with the usual nonrelativistic mechanics.
 ªc  A better type of gauge is
$ =nÉx/nÉp$, what we actually used for the lightcone.
Compare the value of $v$ that results from the field equations
in this case to that of the case $ =nÉx$.  Discuss the consistency
of this case in terms of the allowed signs of $nÉx$ and $nÉp$
vs.¼those of $ $ and $v$.

We have seen that the lightcone gauge is a special case of the covariant (affine) gauge, where more components are eliminated (a unitary gauge). In other textbooks, gauge fixing to a unitary gauge is always performed in two steps, by first going to a covariant gauge, and then using the ``residual" gauge invariance to completely fix the gauge.  (This has been done for particles, strings, gauge theories, and even general relativity.)  When this procedure is explicitly performed, the result can be seen to be a lightcone gauge. Clearly it is easier to perform all the gauge fixing in one step. 

Ü3. Coupling

One way to introduce external fields into the mechanics action is by
considering the most general Lagrangian quadratic in $ $ derivatives:
$$ S_L = Çd  [-üv^{-1}g_{mn}(x)Àx{}^m Àx{}^n +A_m(x)Àx{}^m +vÄ(x)] $$
 In the free case we have constant fields $g_{mn}=ú_{mn}$, $A_m=0$, and
$Ä=üm^2$.  The $v$ dependence has been assigned consistent with
worldline coordinate invariance.  The curved-space metric tensor
$g_{mn}$ describes gravity, the D-vector potential $A_m$ describes
electromagnetism, and $Ä$ is a scalar field that can be used to introduce
mass by interaction.

\x IIIB3.1  Use the method of the problem IIIB1.4 to write the
ÓnonrelativisticÕ action for a spinning particle in terms of a 3-vector (or
(D$-$1)-vector) $X^i( ,½)$ and the fermionic derivative $D$.  Find the
coupling to a magnetic field, in terms of the 3-vector potential $A_i(X)$. 
Integrate the Lagrangian over $½$.  Show that the quantum mechanical
square of $Æ^i[p_i+A_i(x)]$ is proportional to the Hamiltonian.

\x IIIB3.2  Derive the relativistic Lorentz force law
$$ »_ (v^{-1}Àx_m) +F_{mn}Àx{}^n = 0 $$
 by varying the Lagrangian form of
the action for the relativistic particle, in an external electromagnetic field
(but flat metric and $Ä=üm^2$), with respect to $x$.

This action also has very simple transformation properties under
D-dimensional gauge transformations on the external fields:
$$ ¶g_{mn} = ·^p »_p g_{mn} +g_{p(m}»_{n)}·^p,â
	¶A_m = ·^p »_p A_m +A_p »_m ·^p -»_m Â,â¶Ä = ·^p »_p Ä $$
$$ ÜâS_L[x] +¶S_L[x] = S_L[x+·] -Â(x_f) +Â(x_i) $$
 where we have integrated the action $Ç_{ _i}^{ _f}d $ and set
$x( _i)=x_i$, $x( _f)=x_f$.  These transformations have a very natural
interpretation in the quantum theory, where
$$ ÇDx¼e^{-iS} = Òx_f|x_iÔ $$
 Then the $Â$ transformation of $A$ is canceled by the U(1) (phase)
transformation
$$ Æ'(x) = e^{iÂ(x)}Æ(x) $$
 in the inner product
$$ ÒÆ_f|Æ_iÔ = Çdx_f dx_i¼ÒÆ_f|x_fÔÒx_f|x_iÔÒx_i|Æ_iÔ =
	Çdx_f dx_i¼Æ_f*(x_f)Òx_f|x_iÔÆ_i(x_i) $$
 while the $·$ transformation associated with $g_{mn}$ is canceled by the
D-dimensional coordinate transformation
$$ Æ'(x) = Æ(x+·) $$

Ü4. Conservation

There are two types of conservation laws generally found in physics:  In
mechanics we usually have global conservation laws, of the form $ÀQ=0$,
associated with a symmetry of the Hamiltonian $H$ generated by a
conserved quantity $Q$:
$$ 0 = ¶H = i[Q,H] = -ÀQ $$
 On the other hand, in field theory we have local conservation laws, since
the action for a field is written as an integral $Çd^D x$ of a Lagrangian
density that depends only on fields at $x$, and a finite number of their
derivatives.  The local conservation law implies a global one, since
$$ »_m J^m = 0âÜâ
	0 = Ç{d^D x\over (2¹)^{D/2}}¼»_m J^m 
	¾ {d\over dt}Ç{d^{D-1}x\over (2¹)^{D/2}}¼J^0 = ÀQ = 0 $$
 where we have integrated over a volume whose boundaries in space are
at infinity (where $J$ vanishes), and whose boundaries in time are
infinitesimally separated.  Equivalently, the global symmetry is a special
case of the local one.

A simple way to derive the local conservation laws is by coupling gauge
fields:  We couple the electromagnetic field $A_m$ to arbitrary charged
matter fields $Ä$ and demand gauge invariance of the matter part of the
action, the matter-free part of the action being separately invariant.  We
then have
$$ 0 = ¶S_M = Çdx \left[ (¶A_m){¶S_M\over ¶A_m} +
	(¶Ä){¶S_M\over ¶Ä} \right]  $$
 using just the definition of the functional derivative $¶/¶$.  Applying the
matter field equations $¶S_M/¶Ä=0$, integration by parts, and the gauge
transformation $¶A_m=-»_m Â$, we find
$$ 0 = Çdx¼Â\left(»_m {¶S_M\over ¶A_m}\right)âÜâ
	J^m = {¶S_M\over ¶A_m},â»_m J^m = 0 $$
 Similar remarks apply to gravity, but only if we evaluate the ``current",
in this case the energy-momentum tensor, in flat space
$g_{mn}=ú_{mn}$, since gravity is self-interacting.  We then find
$$ T^{mn} = -2\left.{¶S_M\over ¶g_{mn}}\right|_{g_{mn}=ú_{mn}},â
	»_m T^{mn} = 0 $$
 where the normalization factor of $-$2 will be found later for consistency
with the particle.  In this case the corresponding ``charge" is the
D-momentum:
$$ P^m = Ç{d^{D-1}x\over (2¹)^{D/2}}¼T^{0m} $$
 In particular we see that the condition for the energy in any region of
space to be nonnegative is
$$ T^{00} ³ 0 $$

\x IIIB4.1 Show that the local conservation of the energy-momentum tensor
allows definition of a conserved angular momentum
$$ J^{mn} = Ç{d^{D-1}x\over (2¹)^{D/2}}¼x^{[m}T^{n]0} $$
 Note this result (local conservation of energy-momentum 
inplies local conservation of angular momentum) is the same as that of
exercise IA4.3.

To apply this to the action for the particle in external fields, we must
first distinguish the particle coordinates $X( )$ from coordinates $x$ for
all of spacetime:  The particle exists only at $x=X( )$ for some $ $, but
the fields exist at all $x$.  In this notation we can write the mechanics
action as
$$ \li{ S_L = Çdx \left[ \vphantom{Ç}\right. &-g_{mn}(x)
	Çd ¼¶(x-X)üv^{-1}ÀX{}^m ÀX{}^n \cr
	& \left.+A_m(x)Çd ¼¶(x-X)ÀX{}^m +Ä(x)Çd ¼¶(x-X)v\right] \cr } $$
using $Çdx¼¶(x-X( ))=1$.  We then have
$$ J^m(x) = Çd ¼¶(x-X)ÀX{}^m $$
$$ T^{mn} = Çd ¼¶(x-X)v^{-1}ÀX{}^m ÀX{}^n $$
 Note that $T^{00}³0$ (since $v>0$).  Integrating to find the charge and
momentum:
$$ Q = Çd ¼¶(x^0-X^0)ÀX{}^0 = ÇdX^0¼·(ÀX{}^0)¶(x^0-X^0) = ·(p^0) $$
$$ P^m = Çd ¼¶(x^0-X^0)v^{-1}ÀX{}^0 ÀX{}^m 
	= ÇdX^0¼·(ÀX{}^0)¶(x^0-X^0)v^{-1}ÀX{}^m = ·(p^0)p^m $$
 where we have used $p=v^{-1}ÀX$ (for the free particle), where $p$ is the
momentum conjugate to $X$, not to be confused with $P$.  The factor of
$·(p^0)$ ($·(u)=u/|u|$ is the sign of $u$) comes from the Jacobian from
changing integration variables from $ $ to $X^0$.

The result is that our naive expectations for the momentum and charge of
the particle can differ from the correct result by a sign.  In particular
$p^0$, which semiclassically is identified with the angular frequency of
the corresponding wave, can be either positive or negative, while the
true energy $P^0=|p^0|$ is always positive, as physically required. 
(Otherwise all states could decay into lower-energy ones:  There would
be no lowest-energy state, the ``vacuum".)  When $p^0$ is negative, the
charge $Q$ and $dX^0/d $ are also negative.  In the massive case, we
also have $dX^0/ds$ negative.  This means that as  the proper time $s$
increases, $X^0$ decreases.  Since the proper time is the time as
measured in the rest frame of the particle, this means that the particle is
traveling backward in time:  Its clock changes in the direction opposite to
that of the coordinate system $x^m$.  Particles traveling backward in
time are called ``antiparticles", and have charges opposite to their
corresponding particles.  They have positive true energy, but the
``energy" $p^0$ conjugate to the time is negative.

\x IIIB4.2  Compare these expressions for the current and
energy-momentum tensor to those from the semiclassical expansion in
exercise IIIA4.1.  (Include the ÓinverseÕ metric to define the square of 
$-»_m\S+qA_m$ there.)

Ü5. Pair creation

Free particles travel in straight lines.  Nonrelativistically, external fields
can alter the motion of a particle to the extent of changing the signs of
spatial components of the momentum.  Relativistically, we might then
expect that interactions could also change the sign of the energy, or at
least the canonical energy $p^0$.  As an extreme case, consider a
worldline that is a closed loop:  We can pick $ $ as an angular coordinate
around the loop.  As $ $ increases, $X{}^0$ will either increase or
decrease.  For example, a circle in the $x^0$-$x^1$ plane will be viewed
by the particle as repeating its history after some finite $ $, moving
forward with respect to time $x^0$ until reaching a latest time $t_f$,
and then backward until some earliest time $t_i$.  On the other hand, from
the point of view of an observer at rest with respect to the $x^m$
coordinate system, there are no particles until $x^0=t_i$, at which time
both a particle and an antiparticle appear at the same position in space,
move away from each other, and then come back together and
disappear.  This process is known as ``pair creation and annihilation".

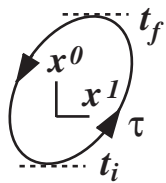
$$ \fig{pair} $$

Whether such a process can actually occur is determined by solving the
equations of motion.  A simple example is a particle in the presence of
only a static electric field, produced by the time component $A^0$ of the
potential.  We consider the case of a piecewise constant potential,
vanishing outside a certain region and constant inside.  Then the
electric field vanishes except at the boundaries, so the particle travels in
straight lines except at the boundaries.  For simplicity we reduce the
problem to two dimensions:
$$ A^0 = -V¼for¼0²x^1²L,â0¼otherwise $$
 for some constant $V$.  The action is, in Hamiltonian form,
$$ S_H = Çd ¼Ó-Àx{}^m p_m +vü[(p+A)^2+m^2]Õ $$
 and the equations of motion are
$$ Àp_m = -v(p+A)^n »_m A_nâÜâp^0 = E $$
$$ (p+A)^2 = -m^2âÜâp^1 = àå{(E+A^0)^2-m^2} $$
$$ v^{-1}Àx = p + AâÜâv^{-1}Àx{}^1 = p^1,âv^{-1}Àx{}^0 = E+A^0 $$
 where $E$ is a constant (the canonical energy at $x^1=¥$) and the
equation $Àp{}^1=...$ is redundant because of gauge invariance.  We assume
$E>0$, so initially we have a particle and not an antiparticle.

$$ \fig{pot} $$

We look only at the cases where the worldline begins at $x^0=x^1=-¥$
(lower left) and continues toward the right till it reaches $x^0=x^1=+¥$
(upper right), so that $p^1=v^{-1}Àx{}^1>0$ everywhere (no reflection). 
However, the worldline might bend backward in time ($Àx{}^0<0$) inside the
potential:  To the outside viewer, this looks like pair creation at the right
edge before the first particle reaches the left edge; the antiparticle then
annihilates the original particle when it reaches the left edge, while the
new particle continues on to the right.  From the particle's point of view,
it has simply traveled backward in time so that it exits the right of the
potential before it enters the left, but it is the same particle that travels
out the right as came in the left.  The velocity of the particle outside and
inside the potential is
$$ {dx^1\over dx^0} = \cases{
	{\displaystyle {å{E^2-m^2}\over E}} & outside \cr
	\noalign{\vskip8pt}
	{\displaystyle {å{(E-V)^2-m^2}\over E-V}} & inside \cr} $$
 From the sign of the velocity we then see that we have normal
transmission (no antiparticles) for $E>m+V$ and $E>m$, and pair
creation/annihilation when
$$ V-m > E > mâÜâV > 2m $$
 The true ``kinetic" energy of the antiparticle (which appears only inside
the potential) is then $-(E-V)>m$.

\x IIIB5.1  This solution might seem to violate causality.  However, in
mechanics as well as field theory, causality is related to boundary
conditions at infinite times.  Describe another solution to the equations of
motion that would be interpreted by an outside observer as pair creation
Ówithout any initial particlesÕ:  What happens ultimately to the particle
and antiparticle?  What are the allowed values of their ÓkineticÕ energies
(maximum and minimum)?  Since many such pairs can be created by the
potential alone, it can be accidental (and not acausal) that an external
particle meets up with such an antiparticle.  Note that the generator of
the potential, to maintain its value, continuously loses energy (and
charge) by emitting these particles.

\refs

£1 A. Barducci, R. Casalbuoni, and L. Lusanna, ÓNuo. Cim.Õ É35A (1976)
	377;\\
	L. Brink, S. Deser, B. Zumino, P. DiVecchia, and P. Howe, \PL
	64B (1976) 435;\\
	P.A. Collins and R.W. Tucker, \NP 121 (1977) 307:\\
	worldline metric; classical mechanics for relativistic spinors.
 £2 R. Marnelius, \PRD 20 (1979) 2091;\\
	W. Siegel, ÓInt. J. Mod. Phys. AÕ É3 (1988) 2713:\\
	conformal invariance in classical mechanics actions.
 £3 W. Siegel, \xxxlink{hep-th/9412011}, \PRD 52 (1995) 1042:\\
	twistor formulation of classical mechanics action.
 £4 R.P. Feynman, ÓPhys. Rev.Õ É74 (1948) 939:\\
	classical pair creation.

\unrefs

Û8 C. YANG-MILLS

The concept of a ``covariant derivative" allows the straightforward
generalization of electromagnetism to a self-interacting theory, once
U(1) has been generalized to a nonabelian group.  Yang-Mills theory is an
essential part of the Standard Model.

Ü1. Nonabelian

The group U(1) of electromagnetism is Abelian:  Group elements commute,
which makes group multiplication equivalent to multiplication of real
numbers, or addition if we write $U=e^{iG}$.  The linearity of this addition
is directly related to the linearity of the field equations for
electromagnetism without matter.  On the other hand, the nonlinearity of
nonabelian groups causes the corresponding particles to interact with
themselves:  Photons are neutral, but ``gluons" have charge and
``gravitons" have weight.

In coupling electromagnetism to the particle, the relation of the canonical
momentum to the velocity is modified:  Classically, the covariant
momentum is $dx/d =p+qA$ for a particle of charge $q$ (e.g., $q=1$ for
the proton).  Quantum mechanically, the net effect is that the wave
equation is modified by the replacement
$$ » £ á = » +iqA $$
 which accounts for all dependence on $A$ (``minimal coupling").  This
``covariant derivative" has a fundamental role in the formulation of
gauge theories, including gravity.  Its main purpose is to preserve gauge
invariance of the action that gives the wave equation, which would
otherwise be spoiled by derivatives acting on the coordinate-dependent
gauge parameters:  In electromagnetism,
$$ Æ' = e^{iqÂ}Æ,âA' = A -»ÂâÜâ(áÆ)' = e^{iqÂ}(áÆ) $$
 or more simply
$$ á' = e^{iqÂ}áe^{-iqÂ} $$
 (More generally, $q$ is some Hermitian matrix when $Æ$ is a reducible
representation of U(1).)

Yang-Mills theory then can be obtained as a straightforward
generalization of electromagnetism, the only difference being that the
gauge transformation, and therefore the covariant derivative, now
depends on the generators of some nonabelian group.  We begin with the
hermitian generators
$$ [G_i,G_j] = -if_{ij}{}^k G_k,ââG_iÿ = G_i $$
 and exponentiate linear combinations of them to obtain the unitary group
elements
$$ {\bf g} = e^{iÂ},â = Â^i G_i;ââÂ^i* = Â^iâÜâ{\bf g}ÿ = {\bf g}^{-1} $$
 We then can define representations of the group (see subsection IB1)
$$ Æ' = e^{iÂ}Æ,âÆÿ' = Æÿ e^{-iÂ};ââ(G_i Æ)_A = (G_i)_A{}^B Æ_B $$

For compact groups charge is quantized:  For example, for SU(2) the spin
(or, for internal symmetry, ``isospin") is integral or half-integral. 
On the other hand, with Abelian groups the charge can take continuous
values:  For example, in principle the proton might decay into a particle of
charge $¹$ and another of charge $1-¹$.  The experimental fact that
charge is quantized suggests already semiclassically that all interactions
should be descibed by (semi)simple groups.

If $Â$ is coordinate dependent (a local, or ``gauge" transformation), the
ordinary partial derivative spoils gauge covariance, so we introduce the
covariant derivative
$$ á_a = »_a +iA_a,ââA_a = A_a{}^i G_i $$
 Thus, the covariant derivative acts on matter in a way similar to the
infinitesimal gauge transformation,
$$ ¶Æ_A = iÂ^i G_{iA}{}^B Æ_B,ââ
	á_a Æ_A = »_a Æ_A +iA_a{}^i G_{iA}{}^B Æ_B $$
 Gauge covariance is preserved by demanding it have a covariant
transformation law
$$ á' = e^{iÂ}á e^{-iÂ}âÜâ¶A = -[á,Â] = -»Â -i[A,Â] $$
 The gauge covariance of the field strength follows from defining it in a
manifestly covariant way:
$$ [á_a,á_b] = iF_{ab}âÜâF' = e^{iÂ}F e^{-iÂ},â
	F_{ab} = F_{ab}{}^i G_i = »_{[a}A_{b]} +i[A_a,A_b] $$
$$ ÜâF_{ab}{}^i = »_{[a}A_{b]}{}^i +A_a{}^j A_b{}^k f_{jk}{}^i $$
 The Jacobi identity for the covariant derivative is the Bianchi identity for
the field strength:
$$ 0 = [á_{[a},[á_b,á_{c]}]] = i[á_{[a},F_{bc]}] $$
 (If we choose instead to use antihermitian generators, all the explicit
$i$'s go away; however, with hermitian generators the $i$'s will cancel
with those from the derivatives when we Fourier transform for purposes
of quantization.)  Since the adjoint representation can be treated as
either matrices or vectors (see subsection IB2), the covariant derivative
on it can be written as either a commutator or multiplication:  For
example, we may write either $[á,F]$ or $áF$, depending on the
context.

Actions then can be constructed in a manifestly covariant way:  For
matter, we take a Lagrangian $L_{M,0}(»,Æ)$ that is invariant under
global (constant) group transformations, and couple to Yang-Mills as
$$ L_{M,0}(»,Æ) £ L_{M,A} =  L_{M,0}(á,Æ) $$
 (This is the analog of minimal coupling in electrodynamics.)  The
representation we use for $G_i$ in  $á_a=»_a+iA_a^i G_i$ is determined by
how $Æ$ represents the group.  (For an Abelian group factor U(1), $G$ is
just the charge $q$, in multiples of the $g$ for that factor.)  For example,
the Lagrangian for a massless scalar is simply
$$ L_0 = ü(á^a Ä)ÿ(á_a Ä) $$
 (normalized for a complex representation).

For the part of the action describing Yang-Mills itself we take (in analogy
to the U(1) case)
$$ L_A(A^i_a) = \f1{8g_A^2}F^{iab}F^j_{ab}ú_{ij} $$
 where $ú_{ij}$ is the Cartan metric (see subsection IB2).  This way of
writing the action is independent of our choice of normalization of the
structure constants, and so gives one unambiguous definition for the
normalization of the coupling constant $g$.  (It is invariant under any
simultaneous redefinition of the fields and the generators that leaves the
covariant derivative invariant.)  Generally, for simple groups we can
choose to (ortho)normalize the generators $G_i$ with the condition
(see subsection IB2)
$$ ú_{ij} = c_A ¶_{ij} $$
 for some constant $c_A$; for groups that are products of simple groups
(semisimple), we might choose different normalization factors (but, of
course, also different $g$'s) for each simple group.  For Abelian groups
(U(1) factors) $ú_{ij}=0$, but then the gauge field has no self-interactions,
so the normalization of the coupling constant is defined only by matter
terms in the action, and we can replace $ú_{ij}$ with $¶_{ij}$ in the
above.  

Usually it will prove more convenient to use matrix notation:  Choosing
some convenient representation $R$ of $G_i$ (not necessarily the adjoint),
we write
$$ L_A(A^i_a) = \f1{8g_R^2}tr_R¼F^{ab}F_{ab} $$
 The normalization of the trace is determined by $R$, and thus so is the
normalization convention for the coupling constant; a change in the
representation used in the action can also be absorbed by a redefinition of
the coupling.  For example, comparing the defining and adjoint
representations of SU(N) (see subsection IB2),
$$ L_A = \f1{8g_D^2}tr_D¼F^{ab}F_{ab} = 
	\f1{8g_A^2}tr_A¼F^{ab}F_{ab}âÜâg_A^2 = 2Ng_D^2 $$
 In general, we specify our normalization of the structure constants by
fixing $c_R$ for some $R$, and our normalization of the coupling constant
by specifying the choice of representation used in the trace (or use
explicit adjoint indices).  As a rule, we find the most convenient choices
of normalization are
$$ c_D = 1,ââg = g_D $$
 (see subsection IB5).

\x IIIC1.1 Write the action for SU(N) Yang-Mills coupled to a massless
(2-component) spinor in the defining representation.  Make all (internal
and Lorentz) indices explicit (no ``$tr$", etc.), and use defining
(N-component) indices on the Yang-Mills field.

We have chosen a normalization where the Yang-Mills coupling constant
$g$ appears only as an overall factor multiplying the $F^2$ term (and
similarly for the electromagnetic coupling, as discussed in previous
chapters).  An alternative is to rescale $A£gA$ and $F£gF$ everywhere;
then $á=»+igA$ and $F=»A+ig[A,A]$, and the $F^2$ term has no extra
factor.  This allows the Yang-Mills coupling to be treated similarly to
other couplings, which are usually not written multiplying kinetic terms
(unless analogies to Yang-Mills are being drawn), since (almost) only for
Yang-Mills is there a nonlinear symmetry relating kinetic and interaction
terms.

Current conservation works a bit differently in the nonabelian case: 
Applying the same argument as in subsection IIIB4, but taking into
account the modified (infinitesimal) gauge transformation law, we find
$$ J^m = {¶S_M\over ¶A_m},ââá_m J^m = 0 $$
 Since $»_m J^m±0$, there is no corresponding covariant conserved
charge.

\x IIIC1.2  Let's look at the field equations:
 ªa  Using properties of the trace, show the entire covariant derivative
can be integrated by parts as
$$ Çdx¼tr (\A [á,\B ]) = -Çdx¼tr ([á,\A] \B ),ââÇdx¼Æÿ፠= -Çdx¼(áÆ)ÿ $$
 for matrices $\A,\B$ and column vectors $Æ,$.
 ªb  Show
$$ ¶F_{ab} = á_{[a}¶A_{b]} $$
 ªc  Using the definition of the current as for electromagnetism
(subsection IIIB4), derive the field equations with arbitrary matter,
$$ \f1{g^2}üá^b F_{ba} = J_a $$
 ªd  Show that gauge invariance of the action $S_A$ implies
$$ á^a(á^b F_{ba}) = 0 $$
 Also show this is true directly, using the Jacobi identity, but not the field
equations.  (Hint:  Write the covariant derivatives as commutators.)

\x IIIC1.3  Expand the left-hand side of the field equation (given in
excercise IIIC1.2c) in the field, as
$$ \f1{g^2}üá^b F_{ba} = \f1{g^2}ü»^b »_{[b}A_{a]} - j_a $$
 where $j$ contains the quadratic and higher-order terms.  Show the
ÓnoncovariantÕ current
$$ \J_a = J_a +j_a $$
 is conserved.  The $j$ term can be considered the gluon contribution to
the current:  Unlike photons, gluons are charged.  Although the current is
gauge dependent, and thus physically meaningless, the corresponding
charge can be gauge independent under situations where the boundary
conditions are suitable.

Ü2. Lightcone

Since gauge parameters are always of the same form as the gauge field,
but with one less vector index, an obvious type of gauge choice (at least
from the point of view of counting components) is to require the gauge
field to vanish when one vector index is fixed to a certain value. 
Explicitly, in terms of the covariant derivative we set
$$ nÉá = nÉ»âÜânÉA = 0 $$
 for some constant vector $n^a$.  We then can distinguish three types of
``axial gauges":  
\item{(1)} ``Arnowitt-Fickler", or spacelike ($n^2>0$),  
\item{(2)} ``lightcone", or lightlike ($n^2=0$), and  
\item{(3)} ``temporal", or timelike
($n^2<0$).  

\noindent By appropriate choice of reference frame, and with the usual
notation, we can write these gauge conditions as $á^1=»^1$, $á^+=»^+$,
and $á^0=»^0$.

One way to apply this gauge in the action is to keep the same set of
fields, but have explicit $n$ dependence.  A much simpler choice is to use
a gauge choice such as $A_0=0$ simply to eliminate $A_0$ explicitly from
the action.  For example, for Yang-Mills we find
$$ A_0 = 0âÜâF_{0i} = ÀA_iâÜâ
	\f18 (F_{ab})^2 = -\f14 (ÀA_i)^2 +\f18 (F_{ij})^2 $$
 where ``$À{\phantom M}$" here refers to the time derivative.  Canonical
quantization is simple in this gauge, because we have the canonical
time-derivative term.  However, the gauge condition can't be imposed
everywhere, as seen for the corresponding gauge for the
one-dimensional metric in subsection IIIB2, and in our general discussion
in subsection IIIA5:  Here we can generalize the time-ordered integral
for the temporal gauge to an integral Ópath-orderedÕ with respect to a
straight-line path in the $n$ direction:
$$ e^{-knÉá(x)} = e^{-iñ(x,x-kn)}e^{-knÉ»},ââe^{-iñ(x,x-kn)} = 
	\P\left[ exp \left( -iÇ_{x-kn}^x dx'ÉA(x') \right) \right]  $$
 Applying this gauge transformation to $nÉá$, as in subsection IIIA5, fixes
$nÉA$ to a constant with respect to $nÉ»$; the effect on all of $á$ is:
$$ á'(x) = e^{iñ(x,x-kn)}á(x)e^{-iñ(x,x-kn)}âÛâ
	á'(x+kn) = e^{knÉá(x)}á(x)e^{-knÉá(x)} $$
 For example, for the temporal gauge, if we choose ``$x$" to be on the
initial hypersurface $x^0=t_0$, then we can choose $k=t-t_0$ so that $á'$
is evaluated at arbitrary time $t$:
$$ á'_a(x) = [(e^{ká_0(x)}á_a(x)e^{-ká_0(x)})|_{x^0=t_0}]|_{k=x^0} $$
 By Taylor expanding in $k$, this gives an explicit expression for $A_a$ at
all times in terms of $A_a$, and $F_{ab}$ and its covariant
time-derivatives, evaluated at some initial time, but with simply
$A_0(t,x^i)=A_0(t_0,x^i)$.  Thus, we still need to impose the $A_0$ field
equation $[á_i,ÀA_i]=0$ as a constraint at some initial time.

\x IIIC2.1  First set $A_0=0$, then derive the field equations for $A_i$ 
from the Yang-Mills action.  Compare the results of exercise IIIC1.2c
for $J=0$.  Show explicitly that these field equations 
imply the Ótime derivativeÕ of the constraint
$[á_i,ÀA_i]=0$.  (Hint:  Write everything in terms of $F$'s and
$á$'s till the end.)

In the case of the lightcone gauge we can carry this analysis one step
further.  In subsection IIB3 we saw that lightcone formalisms are
described by massless fields with (D$-$2)-dimensional (``transverse")
indices.  In the present analysis, gauge fixing alone gives us, again for the
example of pure Yang-Mills,
$$ A^+ = 0âÜâF^{+i} = »^+ A^i,âF^{+-} = »^+ A^-,â
	F^{-i} = »^- A^i -[á^i, A^-] $$
$$ Üâ\f18 (F^{ab})^2 = 
	-\f14 (»^+ A^-)^2 -ü(»^+ A^i)(»^- A^i -[á^i, A^-]) +\f18 (F^{ij})^2 $$
 In the lightcone formalism $»^-$ ($-»_+$) is to be treated as a time
derivative, while $»^+$ can be freely inverted (i.e., modes propagate to
infinity in the $x^+$ direction, but boundary conditions set them to vanish
in the $x^-$ direction).  Thus, we can treat $A^-$ as an auxiliary field. 
The solution to its field equation is
$$ A^- = {1\over »^{+2}}[á^i,»^+ A^i] $$
 which can be substituted directly into the action:
$$ \li{ \f18 (F^{ab})^2 =¼& üA^i »^+ »^- A^i  +\f18 (F^{ij})^2 
		-\f14 [á^i,»^+ A^i]{1\over »^{+2}}[á^j,»^+ A^j] \cr
	= & -\f14 A^i õA^i +iü[A^i,A^j]»^i A^j 
		+iü(»^i A^i){1\over »^+}[A^j,»^+ A^j] \cr
	& -\f18 [A^i,A^j]^2 +\f14 [A^i,»^+ A^i]{1\over »^{+2}}[A^j,»^+ A^j]
		\cr} $$

We can save a couple of steps in this derivation by noting that elimination
of any auxiliary field, appearing quadratically (as in going from
Hamiltonian to Lagrangian formalisms), has the effect
$$ L = üax^2 +bx +c £ -üax^2|_{»L/»x=0} +L|_{x=0} $$
 In this case, the quadratic term is $(F^{-+})^2$, and we have
$$ \f18 (F^{ab})^2 = \f18 (F^{ij})^2 -üF^{+i}F^{-i} -\f14 (F^{+-})^2
	£ \f18 (F^{ij})^2 -ü(»^+ A^i)(»^- A^i) +\f14 (F^{+-})^2 $$
 where the last term is evaluated at
$$ 0 = [á_a,F^{+a}] = -»^+ F^{+-} +[á^i,F^{+i}]âÜâ
	F^{+-} = {1\over »^+}[á^i,F^{+i}] $$
$$ ÜâL = \f18 (F^{ij})^2 +üA^i »^+ »^- A^i 
	-\f14[á^i,»^+ A^i]{1\over »^{+2}}[á^j,»^+ A^j] $$
 as above.

In this case, canonical quantization is even simpler, since interpreting
$»^-$ as the time derivative makes the action look like that for a
nonrelativistic field theory, with a kinetic term linear in time derivatives
(as well as interactions without them).  The free part of the field equation
is also simpler, since the kinetic operator is now just $õ$.  (This is true in
general in lightcone formalisms from the analysis of free theories in
chapter XII.)  In general, lightcone gauges are the simplest for analyzing
physical degrees of freedom (within perturbation theory), since the
maximum number of degrees of freedom is eliminated, and thus kinetic
operators look like those of scalars.  On the other hand, interaction terms
are more complicated because of the nonlocal Coulomb-like terms
involving $1/»^+$:  The inverse of a derivative is an integral.  (However, in
practice we often work in momentum space, where $1/p^+$ is local, but
Fourier transformation itself introduces multiple integrals.)  This makes
lightcone gauges useful for discussing unitarity (they are ``unitary
gauges"), but inconvenient for explicit calculations.  However, in
subsection VIB6 we'll find a slight modification of the lightcone that
makes it the most convenient method for certain calculations.  (In the
literature, ``lightcone gauge" is sometimes used to refer to an axial
gauge where $A^+$ is set to vanish but $A^-$ is not eliminated, and
$D$-vector notation is still used, so unitarity is not manifest.  Here we
always eliminate both components and explicitly use
$(D-2)$-vectors, which has distinct technical advantages.)

Although spin 1/2 has no gauge invariance, the second step of the
lightcone formalism, eliminating auxiliary fields, can also be applied
there:  For example, for a massless spinor in D=4, identifying
$»^{\¢\rdt\¢}=»^-$ as the lightcone ``time" derivative, we vary $ÐÆ^{\rdt\¢}$ (or
$Æ^\¢$) as the auxiliary field:
$$ -iL = ÐÆ^{\rdt\¢}»^{¢\rdt ¢}Æ^\¢ +ÐÆ^{\rdt ¢}»^{\¢\rdt\¢}Æ^¢ 
	-ÐÆ^{\rdt\¢}»^{\¢\rdt ¢}Æ^¢ -ÐÆ^{\rdt ¢}»^{¢\rdt\¢}Æ^\¢ $$
$$ ÜâÆ^\¢ = {1\over »^{¢\rdt ¢}}»^{\¢\rdt ¢}Æ^¢ $$
$$ ÜâL = ÐÆ^{\rdt ¢}{üõ\over i»^{¢\rdt ¢}}Æ^¢ $$
 This tells us that a 4D massless spinor, like a 4D massless vector (or a
ÓcomplexÕ scalar) has only 1 complex (2 real) degree of freedom,
describing a particle of helicity +1/2 and its antiparticle of helicity $-$1/2
($à1$ for the vector, 0 for the scalar), in agreement with our general
discussion of helicity in subsection IIB7.  On the other hand, in the
massive case we can always go to a rest frame, so the analysis is in
terms of spin (SU(2) for D=4) rather than helicity.  For a massive Weyl
spinor we can perform the same analysis as above, with the modifications
$$ L £ L +{im\over å2}(Æ^¢ Æ^\¢ +ÐÆ^{\rdt ¢}ÐÆ^{\rdt\¢})
	âÜâL = ÐÆ^{\rdt ¢}{ü(õ-m^2)\over i»^{¢\rdt ¢}}Æ^¢ $$
 where we have dropped some terms that vanish upon using integration
by parts and the antisymmetry of the fermions.  So now we have the two
states of an SU(2) spinor, but these are identified with their antiparticles. 
This differs from the vector:  While for the spinor we have 2 states of a
given energy for both the massless and massive cases, for a vector we
have 2 for the massless but 3 for the massive, since for SU(2) spin s has
2s+1 states:
$$ \hbox{4D states of given $E$:}ââ
\vcenter{\offinterlineskip
\hrule
\halign{ &\vrule#&\strut¼$#$¼\cr
height2pt&\omit&&\omit&\omit&\omit&\omit&\omit&\omit&\omit&\omit&\omit&\cr
& spin && 0 &\omit& ü &\omit& 1 &\omit& \f32 &\omit& \ldots &\cr 
height2pt&\omit&&\omit&\omit&\omit&\omit&\omit&\omit&\omit&\omit&\omit&\cr
\noalign{\hrule}
height2pt&\omit&&\omit&\omit&\omit&\omit&\omit&\omit&\omit&\omit&\omit&\cr
& m=0 && 1 &\omit& 2 &\omit& 2 &\omit& 2 &\omit& \ldots &\cr
& m>0 && 1 &\omit& 2 &\omit& 3 &\omit& 4 &\omit& \ldots &\cr
}\hrule} $$

\x IIIC2.2  Show that integration by parts for $1/»$ gives just a sign
change, just as for $»$.

In general dimensions, massless particles are representations of the
``little group" SO(D$-$2) (the helicity SO(2) in D=4), as described in
subsection IIB3.  Massive particles represent the little group SO(D$-$1),
corresponding to dimensional reduction from an extra dimension, as
described in subsection IIB4.

Ü3. Plane waves

The simplest nontrivial solutions to nonabelian field equations are the
generalizations of the plane wave solutions of the free theory.  We begin
with general, free, massless theories, as analyzed in subsection IIB3.  In
the lightcone frame only $p^+$ is nonvanishing.  In position space this
means the field strength depends only on $x^-$.  This describes a wave
traveling at the speed of light in the positive $x^1$ direction, with no
other spatial dependence (i.e., a plane wave).  We allow arbitrary
dependence on $x^-$, corresponding to a superposition of waves with
parallel momenta (but different values of $p^+$).  While its dependence
on only $x^-$ solves the Klein-Gordon equation, Maxwell's equations are
solved by giving the field strength as many upper + indices as possible,
and no upper $-$'s.

Generalizing to interactions, we notice that the Yang-Mills field equations
and Bianchi identities differ from Maxwell's equations only by the
covariantization of the derivatives (at least for pure Yang-Mills). 
Because Maxwell's equations were satisfied by just restricting the index
structure, we can do the same for the covariant derivatives by assuming
that only $á^+$ is novanishing on the field strengths.  In other words, we
can solve the field equations and Bianchi identities by choosing the only
nontrivial components of the gauge fields to be those in $á^+$.

The final step is to solve the relation between covariant derivative and
field strength.  This is simple because the index structure we found
implies the only nontrvial commutators are
$$ [»^i,á^+] = iF^{i+},ââ[»^-,á^+] = 0 $$
 In particular, this implies that the gauge fields have no $x^+$
dependence, and only a very simple dependence on $x^i$.  We find directly
$$ A^+ = x^i F^{i+}(x^-) $$
 where $F^{i+}(x^-)$ is unrestricted (other than the explicit index
structure and coordinate dependence).  Of course, this result can also
be used in the free theory, although it differs from the usual lightcone
gauge.

\x IIIC3.1  Gauge transform this solution to the lightcone gauge
$A^+=0$ in the Abelian case.

\x IIIC3.2  Translate the above results into spinor notation in D=4.

Ü4. Self-duality

The simplest and most important solutions to the field equations are
those that are invariant under the ``duality" symmetry that relates
electric and magnetic charge:
$$ [á_a,á_b] = àü·_{abcd}[á^c,á^d] $$
 Applying the self-duality condition twice, we find
$$ ü·_{abef}·^{efcd} = +¶_{[a}^c ¶_{b]}^d $$
 which requires an even number of time dimensions.  For example, since
the action is usually Wick rotated anyway for perturbative purposes, we
might assume that we should do the same for classical solutions that are
not considered as ``small" fluctations about the usual vacuum.  (Such a
Euclidean definition of field theory has been considered for a
mathematically rigorous formalism, called ``constructive quantum field
theory", since the Gaussian path integrals for scalars and vectors are
then well-defined and convergent.  However, other spins, such as for
fermions or gravity, are a problem in this approach.)  Alternatively, we
can replace $·_{abcd}$ with $i·_{abcd}$ and complexify our fields.  The
self-duality condition, when combined with the Bianchi identities, implies
the field equations:  For Yang-Mills,
$$ á_{[a}F_{bc]} = 0âÜâ0 = àü·^{abcd}á_a F_{bc}
	= \f14 ·^{abcd}á_a ·_{bcef}F^{ef} = á_a F^{ad} $$
 Since the self-duality condition is only first-order in derivatives, it's
easier to solve than the usual field equations.

Plane wave solutions provide a simple example of self-duality, since the
field strengths can easily be written as the sum of self-dual and
anti-self-dual parts:  In Minkowski space we define the self-dual part as
helicity +1 ($f_{ÀŒÀº}$), and anti-self-dual as $-$1 ($f_{Œº}$).  For example,
for a wave traveling in the ``1" direction, the $F^{+2}¦iF^{+3}$
components give the two self-dualities for Yang-Mills, describing
helicities $à$1 (the two circular polarizations).

\x IIIC4.1  Generalize the results of the previous subsection to 
more general waves, with an
$A^+$ which is a general function of $x^-$ and $x^i$ (with the other
components of $A$ still vanishing).
 ªa  Find the field strength, and show it satisfies the interacting
field equations if $A^+$ satisfies the free Laplace equation
$$ (»^i)^2 A^+ = 0 $$
 ªb In D=4, the solution to this equation is
$$ A^+ = f(x^-,x^t) +Ðf(x^-,Ðx^t) $$
 Show this decomposition describes the two separate helicities.

Before further analyzing solutions to the self-duality condition, we
consider actions that use self-dual fields directly.  This will allow us to
describe not only theories whose only solutions are self-dual, but also
more standard theories as perturbations about self-duality, and even
massive theories.  The most unusual feature of this approach is that
complex fields are used without their complex conjugates, since this is
implied in D=3+1 by self-duality.  (Alternatively, we can Wick rotate to 2+2
dimensions, where all Lorentz representations are real.)  There are two
stages to this approach:  (1) Use a first-order formalism where the
auxiliary field is self-dual.  The usual first-order actions for spin 1/2
(Weyl or Dirac) already can be interpreted in this way, where
``self-duality" means ``chirality".  (2) For the massive theory, eliminate
the non-self-dual field (as an auxiliary field, as allowed by the mass
term), so that the dynamics is described by the self-dual field, which was
formerly considered as auxiliary.  The massless theory then can be
treated as a limiting case.

The simplest (and perhaps most useful) example is massive spin 1/2
coupled in a real representation to Yang-Mills fields:
$$ L = Æ^{TŒ} iá_Œ{}^{ÀŒ}ÐÆ_{ÀŒ} +\f1{2å2}m(Æ^{TŒ} Æ_Œ +ÐÆ^{TÀŒ}ÐÆ_{ÀŒ}) $$
 where the transposition (``$Ê{}^TÊ$") refers to the Yang-Mills group index
(with respect to which the spinors are column vectors).  Note that $Æ$
must be a real representation of this group ($A^T=-A$) for the mass term
to be gauge invariant (unless the mass term includes scalars: see the
following chapter).  Even though $Æ$ and $ÐÆ$ are complex conjugates,
they can be treated independently as far as field equations are
concerned, since they are just different linear combinations of their real
and imaginary parts.  (Complex conjugation can be treated as just a
symmetry, related to unitarity.)  Noticing that the quadratic term for $ÐÆ$
has no derivatives, we can treat it as an auxiliary field, and integrate it
out (i.e., eliminate it by its equation of motion, which gives an explicit
local expression for it):
$$ L £ -\f{å2}m [\f14 Æ^{TŒ}(õ-m^2)Æ_Œ +üÆ^{TŒ} if_Œ{}^º Æ_º] $$
 where we have used the identity
$$ á_Œ{}^{À©}á^º{}_{À©} = üÓá_Œ{}^{À©},á^º{}_{À©}Õ +ü[á_Œ{}^{À©},á^º{}_{À©}]
	= -ü¶_Œ^º õ -if_Œ{}^º $$
 whose simplicity followed from $Æ$ being a real representation of the
Yang-Mills group.  (Of course, we could have eliminated $Æ$ instead, but
not both.)  For convenience we also scale $Æ$ by a constant
$$ Æ £ 2^{-1/4}åmÆ $$
 to find the final result
$$ L £ -\f14 Æ^{TŒ}(õ-m^2)Æ_Œ -üÆ^{TŒ} if_Œ{}^º Æ_º $$
 Now the massless limit can be taken easily.  This action resembles that
of a scalar, plus a ``magnetic-moment coupling", which couples the
``(anti-)self-dual" (chiral) spinor $Æ_Œ$ to only the (anti-)self-dual part
$f_{μ}$ of the Yang-Mills field strength.

For the same reason, the kinetic operator can be written in terms of just
the self-dual part $S_{μ}$ of the spin operator:
$$ L = -\f14 Æ^T(õ -m^2 -if^{Œº}S_{ºŒ})Æ $$
 This operator is of the same form found by squaring the Dirac operator:  
$$ -2Öá^2 = -2(©Éá)^2 = -(Ó©^a,©^bÕ +[©^a,©^b])á_a á_b 
	= õ -iF^{ab}S_{ba} $$
 except for the self-duality.  The simple form of this result again depends
on the reality (parity invariance) of the Yang-Mills representation;
although this squaring trick can be applied for complex representations
(parity violating), the coupling does not simplify.  This is related to the
fact that real representations are required for our derivation of the
self-dual form.

In the special case where the real representation is the direct sum of a
complex one $Æ_{+Œ}$ with its complex conjugate $Æ_{-Œ}$ (as for
quarks in the Standard Model, or electrons in electrodynamics), we can
rewrite the Lagrangian as
$$ L_c = -üÆ_+^{TŒ}(õ-m^2)Æ_{-Œ} -Æ_+^{TŒ} if_Œ{}^º Æ_{-º} $$
  The method can also be generalized to the case of scalar couplings, but
the action becomes nonpolynomial.

For spin 1, we start with the massless case.  We can write the Lagrangian
for Yang-Mills as
$$ L = tr(G^{Œº}f_{Œº} -üg^2 G_{Œº}^2) $$
 where $G_{μ}$ is a (anti-)self-dual auxiliary field.  Although this action
is complex, eliminating $G$ by its algebraic field equation gives the usual
Yang-Mills action up to a total derivative term ($·^{abcd}F_{ab}F_{cd}$),
which can be dropped for purposes of perturbation theory.  For $g=0$, this
is an action where $G$ acts as a Lagrange multiplier, enforcing the
self-duality of the Yang-Mills field strength.

If we simply add a mass term
$$ L_m = \f14 (\f{m}g)^2 A^2 $$
 then $A$ can be eliminated by its field equation, giving a nonpolynomial
action of the form
$$ L +L_m £ -ü(»G)[(\f{m}g)^2 +G]^{-1}(»G) -üg^2 G^2 $$
 Just as the spin-1/2 action contained only a 2-component spinor
describing the 2 polarizations of spin 1/2, this action contains only the
3-component $G_{μ}$, describing the 3 polarizations of (massive) spin 1.

\x IIIC4.2 Find the Abelian part of this action.  Show the free field
equation is\\
$(õ-m^2)G_{Œº}=0$ (without gauge fixing).

Ü5. Twistors

In four dimensions with an even number of time dimensions, the
``Lorentz" group factorizes (into SU(2)${}^2$ for D=4+0 and SL(2)${}^2$ for
D=2+2).  This makes self-duality especially simple in spinor notation:  For
Yang-Mills (cf.¼electromagnetism in subsection IIA7),
$$ [á^{Œº'},á^{©¶'}] = iC^{Œ©}f^{º'¶'}ââ(f^{Œº} = 0) $$
 where we have written primes instead of dots to emphasize that the
two kinds of indices transform independently (instead of as complex
conjugates, as in D=3+1).  For purposes of analyzing self-duality within
perturbation theory, we can use a lightcone method that breaks only
one of the two SL(2)'s (or SU(2)'s), by separating out its indices into the
$¢$ and $\¢$ components:
$$ [á^{¢Œ'},á^{¢º'}] = 0âÜâá^{¢Œ'} = »^{¢Œ'} $$
 where we have chosen a lightcone gauge:  The vanishing of all field
strengths for the covariant derivative $á^{¢Œ'}$ says that it is pure gauge
(as seen by ignoring all but the $x^{\¢Œ'}$ coordinates). We now solve
$$ [á^{¢[Œ'},á^{\¢º']}] = 0âÜâá^{\¢Œ'} = »^{\¢Œ'} +i»^{¢Œ'}Ä $$
 i.e., $á^{\¢Œ'}-»^{\¢Œ'}$ has vanishing curl, and is therefore a gradient. 
We therefore have
$$ A^{¢Œ'} = 0,âA^{\¢Œ'} = »^{¢Œ'}Ä;ââf^{Œ'º'} = -i»^{¢Œ'}»^{¢º'}Ä $$
 These can also be written in terms of an arbitrary constant twistor $·^Œ$
($=¶^Œ_\¢$ above) as
$$ A^{Œº'} = »^{©º'}(-i·^Œ ·_© Ä),ââ
	f^{Œ'º'} = »^{©Œ'}»^{¶º'}(i·_© ·_¶ Ä) $$
 The final self-duality condition $[á^{\¢Œ'},á^{\¢º'}]=0$ then gives the
equation of motion
$$ üõÄ +(»^{¢Œ'}Ä)(»^¢{}_{Œ'}Ä) = 0 $$

\x IIIC5.1  Show that the sign convention for Wick rotation of the
Levi-Civita tensor consistent with the above equations is
$$ F_{ab} = ü·_{abcd}F^{cd},ââF_{ŒŒ'ºº'} = C_{Œº}f_{Œ'º'}âÜ $$
$$ ·_{ŒŒ'ºº'©©'¶¶'} = C_{Œº}C_{©¶}C_{Œ'¶'}C_{º'©'}
	-C_{Œ¶}C_{º©}C_{Œ'º'}C_{©'¶'} $$

\x IIIC5.2  Look at the action $G^{μ}f_{μ}$ for self-dual Yang-Mills in
the lightcone gauge, using the results above.  Show that this action is
equivalent to the lightcone action for ordinary Yang-Mills (subsection
IIIC2), with some terms in the interaction dropped.

At least for 4D Yang-Mills, advantage can be taken of the conformal
invariance of the classical interacting theory by using a formalism where
this invariance is manifest.  We saw in subsection IA6 that classical
mechanics could be made manifestly conformal by use of extra
coordinates.  Covariant derivatives can be defined in terms of projective
lightcone coordinates, but the twistor coordinates $z^{\AŒ}$ (see
subsections IIB6 and IIIB1) are more useful.  The self-dual covariant
derivatives then satisfy
$$ [á_{\A Œ},á_{\B º}] = iC_{Œº}f_{\A\B} $$
 in direct analogy to 4D spinor notation.  This equation also can be solved
by the lightcone method used above, but now this method breaks only
the internal SL(2) symmetry, leaving SL(4) conformal symmetry manifest.
More general self-dual field strengths in this twistor space are also of
the form $f_{\A...\B}$, totally symmetric in the indices.  We also need to
impose the constraint on the field strength
$$ z^{\A Œ}f_{\A\B} = 0 $$
 (and similarly for the more general case) to restrict the range of indices
to the usual 4D spinor indices (in which the field strengths are totally
symmetric).  Self-duality implies the Bianchi identity
$$ á_{[\A Œ}f_{\B]\C} = 0 $$
 which also generalizes to the other field strengths, and is the
equivalent of the usual first-order differential equations (Dirac, Maxwell,
etc.)¼satisfied by 4D field strengths.  As usual, it in turn implies the
interacting Klein-Gordon equation, which in the Yang-Mills case is
$$ üá_{[\A}{}^Œ á_{\B]Œ}f_{\C\D} = -i[f_{\C[\A},f_{\B]\D}] $$
 The Bianchi together with the $z$ index constraint imply the constraint on
coordinate dependence
$$ (z^{\A Œ}á_{\A º} +¶_º^Œ)f_{\B\C} = 0 $$
 which eliminates dependence on all but the usual 4D coordinates.  These
four equations are generically satisfied by self-dual field strengths.  The
self-duality itself of the field strengths is a consequence of their total
symmetry in their indices, and the fact that they are all lower (SL(4))
indices.  (The $z$ index constraint then reduces them to SL(2) Weyl indices
all of the same chirality.)

\x IIIC5.3 Derive the last three equations from the previous two
(self-duality and $zf$=0).

\x IIIC5.4 Show that non-self-dual Yang-Mills is conformally invariant in
D=4 by extending the (4+2)-dimensional formalism of subsections IA6 and
IIIB1 (especially exercise IIIB1.5):  Show the field strength
$$ F_{ABC} = -iüy_{[A}[á_B,á_{C]}] $$
 satisfies the gauge covariances
$$ ¶A_A = -[á_A,Â] -y_A öÂ $$
 and Bianchi identities
$$ y_{[A}F_{BCD]} = á_{[A}F_{BCD]} = 0 $$
 The duality transformation
$$ F_{ABC} £ \f16 ·_{ABCDEF}F^{DEF} $$
 then suggests the field equations
$$ y^A F_{ABC} = á^A F_{ABC} = 0 $$
 in addition to the usual constraint 
$$ y^2 F_{ABC} = 0 $$
 By reducing to D=4 coordinates with the aid of the above $yF$
conditions, show $F$ reduces to the usual field strength, and the
remaining equations reduce to the usual gauge transformation, Bianchi
identity, duality, and field equation.

Ü6. Instantons

Another interesting class of self-dual solutions to Yang-Mills theory are
``instantons", so called because the field strength is maximum at points
in spacetime, unlike the plane waves, whose wavefronts propagate from
and toward timelike infinity.  A particular subset of these can be
expressed in a very simple form by the 't Hooft ansatz in terms of a
scalar field:  In twistor notation, choosing the Yang-Mills gauge group
GL(2) (in 2+2 dimensions, or SU(2)$°$GL(1) for 4+0),
$$ iA_{\A Œî}{}^û = -¶_Œ^û »_{\A î}ln¼ÄâÜâ
	iA_{\A Œî}{}^î = -»_{\A Œ}ln¼Ä $$
 so the GL(1) piece is pure gauge, and has been included just for
convenience.  Note that this ansatz ties the SL(2) twistor index with the
SL(2) gauge group indices ($î,û$), but in this notation the index that
carries the spacetime (conformal) symmetry is free.  Imposing the
self-duality condition on the field strength, and separating out the terms
symmetric and antisymmetric in
$\A\B$, we find
$$ if_{\A\B î}{}^û = -üÄ »_{(\A}{}^û »_{\B)î}Ä^{-1} $$
$$ Ä^{-1}»_\A{}^Œ »_{\B Œ}Ä = 0 $$

The ``field equation" for $Ä$ is just the twistor version of the (free)
Klein-Gordon equation, and its solution is the projective lightcone version
of 4D point sources (see subsection IA6):  Since for any two 6D lightlike
vectors $y$ and $y'$
$$ y = e(x,1, üx^2)âÜâyÉy' = -üee'(x-x')^2 $$
 we have the solution
$$ Ä = Ý_{i=1}^{k+1}(yÉy_i)^{-1},ây_i^2 = 0 $$
 with $y$ given in terms of $z$ as before, and $y_i$ are constant null
vectors.  ``$k$" is the number of instantons.  (The one term for $k=0$
is pure gauge.)  The usual singularities in the Klein-Gordon equation at
$y=y_i$ are killed by the extra factor of $Ä^{-1}$ in the field equation.

\x IIIC6.1  Let's check the Klein-Gordon equation for $y±y_i$ directly in
twistor space.  We will need the identity
$$ y_i^2 = 0âÜây_{i\A[\B}y_{i\C\D]} = 0ââ\left(no¼Ý_i\right) $$
 in the product
$$ yÉy_i = üy^{\A\B}y_{i\A\B} $$
 Prove this identity in two ways:
 ªa  Show it follows from the definition
$$ y_i^2 = \f14 ·^{\A\B\C\D}y_{i\A\B}y_{i\C\D} $$
 ªb  Show it follows from plugging in the solution to the lightlike
condition,
$$ y_{i\A\B} = z_{i\A}{}^Œ z_{i\B Œ} $$
 ªc  Now use the identity to show the above solution satisfies its field
equation by evaluating the $z$ derivatives.

We can rewrite this in the usual 4D coordinates by transforming from
$z^{\A Œ}$ to $Â^{Œµ}$ and $x^{µÃ'}$ as 
$z^\A{}_Œ=Â_Œ{}^Ã(¶_Ã^µ,x_Ã{}^{µ'})$ (see subsection IIB6):
$$ dz^{\A Œ}A_{\A Œî}{}^û = 
	dx^{µÃ'}A_{µÃ'î}{}^û +[(dÂ^Œ{}_Ã)Â^{-1Ã}{}_º]z^{\A º}A_{\A Œî}{}^û
	= dx^{µÃ'}A_{µÃ'î}{}^û -iÂ^{-1Ã}{}_î dÂ^û{}_à $$
 where in the first step we have used the expression for $z$ in terms of
$Â$ and $x$, and in the second we used the result that
$$ (z^{\A Œ}»_{\A º}+¶_º^Œ)Ä = 0 $$

We now recognize that the gauge transformation that gets rid of all
but the ``$x$ components" of $A$ (whose existence is guaranteed by the
condition $z^{\A Œ}f_{\A\B}=0$) uses $Â$ itself as the gauge parameter:
$$ dz^{\A Œ}A_{\A Œî}{}^û = -iÂ^{-1Ã}{}_î dÂ^û{}_Ã
	+Â^{-1Ã}{}_î (dz^{\A Œ}A'_{\A ŒÃ}{}^µ) Â^û{}_µ $$
 The net result is that $A$ can be reduced to an ordinary 4-dimensional
expression by just setting $Â=¶$ in the original expression.  Then
$$ iA_{µµ'î}{}^û = -¶_µ^û »_{îµ'}ln¼Ä,ââÄ = Ý_i {1\over e_i(x-x_i)^2} $$
 with $y$ in terms of $x$ and a scale factor (worldline metric) $e$ as in
subsections IA6 and IIIB1 (and dropping an overall factor that doesn't
contribute to $A$).  Note that, unlike the expression in twistor space,
where conformal invariance is manifest, here Lorentz invariance is tied to
the Yang-Mills symmetry.

\x IIIC6.2 Show in 4D coordinates that the gauge-invariant quantity
$tr(f^2)$ is finite at the points $x=x_i$, where $A$ is singular.  (This
means that the gauge choice is singular, not physical quantities.)

Another important property of instantons is that they give finite
contributions to the action.  In vector notation, we have
$$ F^{ab} = ü·^{abcd}F_{cd}âÜâ
	S = \f1{8g^2}trÇ{d^4 x\over (2¹)^2}¼F^{ab}F_{ab}
	= \f1{16g^2}trÇ{d^4 x\over (2¹)^2}¼·^{abcd}F_{ab}F_{cd} $$
 The last expression can be reduced to a boundary term, since
$$ \f18 tr¼F_{[ab}F_{cd]} = \f16 »_{[a}B_{bcd]} $$
 in terms of the ``Chern-Simons form"
$$ B_{abc} = tr(üA_{[a}»_b A_{c]} +i\f13 A_{[a}A_b A_{c]}) $$

\x IIIC6.3  Although the Chern-Simons form is not manifestly invariant, its
variation is, up to a total derivative:
 ªa  Show that its general variation is
$$ ¶B_{abc} = ü¼tr[(¶A_{[a})F_{bc]} -»_{[a}A_b ¶A_{c]}] $$
 ªb  Show the gauge transformation of $B$ is
$$ ¶B_{abc} = -ü»_{[a}Â_{bc]},ââÂ_{ab} = ü tr(»_{[a}A_{b]}) $$

If we assume boundary conditions such that $F$ drops off rapidly at
infinity, then $A$ must drop off to pure gauge at infinity:
$$ iA_m £ {\bf g}^{-1}»_m {\bf g} $$
 Since instantons always deal with an SU(2) subgroup of the gauge group,
we'll assume now for simplicity that the whole group is itself SU(2).  Then
the action can be given a group theory interpretation directly, since the
integral over the surface at infinity is an integral over the 3-sphere,
which covers the group space of SO(3), and thus half the group space of
SU(2).  Explicitly,
$$ S = \f1{8¹^2 g^2}Èd^3 §_m¼\f16 ·^{mnpq} B_{npq} $$
$$ £ \f1{8¹^2 g^2}Èd^3 §_m¼\f16 ·^{mnpq}tr({\bf g}^{-1}»_n {\bf g})
	({\bf g}^{-1}»_p {\bf g})({\bf g}^{-1}»_q {\bf g}) $$
$$ = \f1{8¹^2 g^2}Èd^3 x¼\f16 ·^{ijk}tr({\bf g}^{-1}»_i {\bf g})
	({\bf g}^{-1}»_j {\bf g})({\bf g}^{-1}»_k {\bf g}) $$
 where in the last step we have switched to coordinates for the 3-sphere,
using the fact $Çd^4 x¼·^{mnpq}f_{mnpq}$ is independent of coordinate
choice.  In fact, in the case where ${\bf g}$ is a one-to-one map between
the 3-sphere and the group SO(3), this last expression is just the
definition of the invariant volume of the SO(3) group space.  In that case,
the integral gives just the volume of the 3-sphere (2$¹^2$).  In general,
the map ${\bf g}$ will cover the SU(2) group space an integer number $q$
of times, and thus cover the SO(3) group space $2q$ times, so the result
will be
$$ S = {|q|\over 2g^2} $$
 where we have used the fact that self-dual solutions have $q>0$ while
anti-self-dual have $q<0$.  

(Anti-)self-dual solutions give relative minima of the action with respect
to more general field configurations:
$$ 0 ² trÇü(F_{ab} àü·_{abcd}F^{cd})^2 
	= trÇ(F^2 àü·^{abcd}F_{ab}F_{cd}) $$
$$ ÜâS ³ {|q|\over 2g^2} $$
 $q$ is an integer, and thus can't be changed by continuous variations:  It
is a topological property of finite-action configurations.  Thus the
self-dual solutions give absolute minima for a given topology. (All these
solutions will be given implicitly by twistor construction in the following
subsection.  Note that our normalization for the structure constants of
SU(2) differs from the usual, since we use effectively $tr(G_i G_j)=¶_{ij}$
instead of the more common $tr(G_i G_j)=ü¶_{ij}$, which would normalize
the structure constants as in SO(3): $f_{ijk}=·_{ijk}$.  The net effect is
that our $g^2$ contains a relative extra factor of 1/2, in addition to the
effective extra factors coming from our different normalization of the
action.)

\x IIIC6.4  Explicitly evaluate the integral for the instanton number $q$
for the solutions of the 't Hooft ansatz.  Show that the asymptotic form
can be expressed in terms of $({\bf g})_Â{}^û = x^{ûÂ'}$.  ($det¼{\bf g}±1$
because of the GL(1) piece.)  Note that there are boundary contributions
not only at $x=¥$ but also around the singular points $x=x_i$, which are
of the same form but opposite sign.  (Since the singular parts of $A$ are
pure gauge, they cancel in $F$.)

At the quantum level, instantons are important mostly because they are
an example of fields that don't fall off rapidly at infinity, and thus
contribute to $Ç·FF$.  However, once the restriction on boundary
conditions is relaxed, there can be many such field configurations.  The
instantons are then distinguished by the fact that they are the minimal
action solutions for a given topology; this makes them important for
describing low-energy behavior.

Ü7. ADHM

Much more general solutions of this form can be constructed using
twistor methods.  (In fact, they can be shown to be the most general
self-dual solutions that fall off fast enough at infinity in all directions.) 
The first step of the Atiyah-Drinfel'd-Hitchin-Manin (ADHM) construction is
to introduce a scalar square matrix in a larger group space
$$ U_I{}^{I'} = (u_I{}^î, v_{IiŒ})ââ(I' = (î,iŒ)) $$
 The index $Œ$ is the usual two-valued twistor index, for SU(2) in
Euclidean space or SL(2) in 2+2 dimensions.  The other indices are
$$ \vbox{\offinterlineskip
\hrule
\halign{ &\vrule#&\strut¼\hfil#\hfil¼\cr
height2pt&\omit&&\omit&&\omit&\cr
& $î$ (H) && $I$ (G) && $i$ &\cr 
height2pt&\omit&&\omit&&\omit&\cr
\noalign{\hrule}
height2pt&\omit&&\omit&&\omit&\cr
& SO(N) && SO(N+4k) && GL(2k) &\cr
& SU(N) (SL(N)) && SU(N+2k) (SL(N+2k)) && GL(k,C) &\cr
& USp(2N) (Sp(2N)) && USp(2N+2k) (Sp(2N+2k)) && GL(k) &\cr  
height2pt&\omit&&\omit&&\omit&\cr}
\hrule} $$
 The index $î$ is for the defining representation of the Yang-Mills group
H, which is any of the compact classical groups for Euclidean space, but is
its real Wick rotation for 2+2 dimensions.  The index $I$ is for the defining
representation of the group G, a larger version of H, where k is the
instanton number.  Finally, the index $i$ is for a general linear group.  We
also have the matrix
$$ U^I{}_{I'} = (u^I{}_î, v^I{}_{i'Œ}) $$
 For the SO and (U)Sp cases both $U$ matrices are real, for the SU case
they are complex conjugates of each other, and for the SL case they are
real and independent.  We next relate the two $U$'s by
$$ u^I{}_î u_I{}^û = ¶_î^û,âu^I{}_î v_{IiŒ} = v^I{}_{i'Œ}u_I{}^î = 0,â
	v^I{}_{i'Œ}v_{Iiº} = C_{ºŒ}g_{ii'} $$
 so they are almost inverses of each other, except that the ``metric" $g$
is not constrained to be a Kronecker $¶$.  We then write the gauge field
as a generalization of pure gauge:
$$ iA_{\A Œî}{}^û = u^I{}_î »_{\A Œ} u_I{}^û $$
 (This is similar to the method used for nonlinear $§$ models of coset
spaces G/H as discussed in subsection IVA3 below, except for $g$.)

Self-duality then follows from requiring a certain coordinate dependence
of the $U$'s:  This is fixed by giving the explicit dependence of the $v$'s as
$$ v_{IiŒ} = b_{Ii\A}z^\A{}_Œ,âv^I{}_{i'Œ} = b^I{}_{i'\A}z^\A{}_Œ $$
 where the $b$'s are constants.  The orthonormality conditions on the
$U$'s then implies the constraint on the $b$'s
$$ b^I{}_{i'(\A}b_{Ii\B)} = 0 $$
 as well as determining the $u$'s in terms of the $b$'s (with much messier
dependence than the $v$'s), and thus $A$.  Note that the $z$ dependence
of $u$ can be written in terms of just $x$, as follows from rewriting the
$uv$ orthogonality as (after multiplying by $z$)
$$ u^I{}_î b_{Ii\A}y^{\A\B} = u_I{}^î b^I{}_{i'\A}y^{\A\B} = 0 $$
 and noting scale invariance.  Then the $x$ components of $A$ can also be
written in terms of just $x$.  We then can check the self-duality condition
by calculating $f$:  The orthonormality condition on the $U$'s can be
written as
$$ ¶_I^J = u_I{}^î u^J{}_î +v_{Ii}{}^Œ g^{ii'}v^J{}_{i'Œ} $$
 where $g^{ii'}$ is the inverse of $g_{ii'}$.  Then schematically we have
$$ \li{ iF &= »iA +iAiA \cr
		&= (»Ðu)(»u) -(»Ðu)uÐu(»u) \cr
		&= (»Ðu)vgÐv(»u) \cr
		&= Ðu(»v)g(»Ðv)u \cr
		&= ÐubgÐbu \cr } $$
 or more explicitly
$$ if_{\A\B î}{}^û = -(u^I{}_î b_{Ii(\A})g^{ii'}(u_J{}^û b^J{}_{i'\B)}) $$
 where self-duality is $F_{\A Œ,\B º}=C_{Œº}f_{\A\B}$.  We can also
directly show $z^{\A Œ}f_{\A\B}=0$.

\x IIIC7.1  Solve the $bb$ constraint for k=1 and H=SU(2), and compare
to the 1-instanton solution of subsection IIIC6.

Ü8. Monopoles

Instantons are essentially 0-dimensional objects, localized near a point in
4-dimensional spacetime (or many points for multi-instanton solutions). 
Another type of solution is 1-dimensional; this represents a particle (with
a 1D worldline).  Unlike the plane-wave solutions, which represent the
massless particles already described explicitly by fields in the action, we
now look for time-independent solutions, which describe massive (since
they have a rest frame), bound-state particles.

Looking at time-independent solutions is similar to the dimensional
reduction that we considered in subsection IIB4 to introduce masses into
free theories, only (1)¼this mass vanishes, and (2) we reduce the time
dimension, not a spatial one.  In our case, the dimensional reduction of a
4-vector (the Yang-Mills potential) gives a 3-vector and a scalar, both in
the adjoint representation of the group.  Let's consider the reduction in
Euclidean space, so the scalar kinetic term comes out with the right sign. 
Then the 4D Yang-Mills action reduces as
$$ \f18 F_{ab}^2 £ \f18 F_{ij}^2 +\f14 [á_i,Ä]^2 $$
 where we have labeled the scalar $A_0=Ä$ and by dimensional reduction
$»_0£0$.  Note that this is the same action that would have been
obtained by starting out with Yang-Mills coupled to an adjoint scalar in
four dimensions, either Minkowski or Euclidean, and choosing the gauge
$A_0=0$.  Thus, time-independent solutions to Euclidean Yang-Mills
theory are also time-independent solutions to Minkowskian Yang-Mills
coupled to an adjoint scalar (although not the most general, since the
gauge $A_0=0$ is not generally possible globally, especially when we
assume time independence of even gauge-dependent quantities).  In
particular, this means that time-independent solutions to self-dual
Yang-Mills are also solutions of Minkowskian Yang-Mills coupled to an
adjoint scalar.  This allows us to use the first-order differential equations
and topological properties of self-dual Yang-Mills theory to find physical
bound-state particles in this vector-scalar theory.

Dimensionally reducing the (Euclidean) self-duality condition, we have
$$ -[á_i,Ä] = ü·_{ijk}F_{jk} $$
 As for instantons, the simplest solutions are for SU(2).  As for the 't Hooft
ansatz, we look for a solution that is covariant under the combined SU(2)
of the gauge group and 3D rotations:  In SO(3) vector notation for both
kinds of indices (using the SO(3) normalization of the structure constants
$[iG_i,iG_j]=·_{ijk}iG_k$),
$$ Ä_i = x_i \Ä(r),ââ(A_i)_j = ·_{ijk}x_k \A(r) $$
 (We know to use an $·$ tensor in $A$ because of covariance under
parity.)  The self-duality equation then reduces to two nonlinear
first-order differential equations (the coefficients of $¶_{ij}$ and 
$x_i x_j/r^2$):
$$ -\Ä -r^2 \A\Ä = 2\A +r\A',â-r\Ä' +r^2\A\Ä = -r\A' +r^2\A^2 $$
 After some massaging, we find the change of variables
$$ ÷\Ä = \f1r +r\Ä,ââ÷\A = \f1r +r\A $$
 leads to the simplification
$$ ÷\Ä{}' = -÷\A{}^2,ââ÷\A{}' = -÷\A÷\Ä $$
 $÷\Ä$ then can be eliminated, giving an equation for $÷\A$.  Making a final
change of variables,
$$ Æ = ÷\A{}^{-1}âÜâÆÆ'' - (Æ')^2 = -1 $$
 we can guess the solution (with regularity at $r=0$)
$$ Æ = k^{-1}sinh(kr)âÜâ
	\A = {1\over r^2}\left({kr\over sinh(kr)}-1\right),â
	\Ä = {1\over r^2}[kr¼coth(kr) -1] $$

\x IIIC8.1  Repeat this calculation in spinor notation:
 ªa  In Euclidean space we can choose $§^0_{Œº'}¾C_{Œº'}$.  Show that
we can then write the 4-vector potential for the monopole as
$$ i(A_{Œº})^{©¶} = ¶_Œ^© x_º{}^¶ \A_+ (r) +¶_º^¶ x_Œ{}^© \A_- (r) $$
 which is symmetric in neither $Œº$ nor $©¶$.  (Compare the 't Hooft
ansatz in subsection IIIC6.)  However, $x^{μ}$ is now symmetric from
dropping $x^0$.
 ªb  Impose self-duality,  where
$$ »_{Œº}x^{©¶} = ü¶_{(Œ}^© ¶_{º)}^¶ = ¶_Œ^© ¶_º^¶ -üC_{Œº}C^{©¶} $$
 from subtracting out the $»_0 x^0$ piece.  Derive the resulting equations
for $\A_à$, and show they agree with the above for
$$ \A_à = -ü(\A à\Ä) $$

In general, the Lagrangian of a Euclidean theory is the Hamiltonian of the
Minkowskian theory (with the sign conventions we introduced in
subsection IIIA1), since Wick rotation changes the sign of the kinetic
energy and not the potential energy.  In our case, this means the
Minkowskian energy of the Yang-Mills + adjoint scalar theory can be
evaluated in terms of the same topological expression we used for
instantons:  From the previous subsection, using $S=Çdt¼E$ and $»_0=0$,
we evaluate in Euclidean space
$$ E = \f1{16¹^2 g^2}Èd^2 §_i¼·^{i0jk}B_{0jk},â
	·^{i0jk}B_{0jk} £ -·_{ijk}tr(ÄF_{jk}) = 2¼tr(Ä[á_i,Ä]) = »_i¼tr(Ä^2) $$
 where we have used an integration by parts to simplify $B$.  (Compare
exercise\break IIIC6.3a.)  Since at spatial infinity
$$ Ä_i £ x_i\left({|k|\over r}-{1\over r^2}\right),ââ
	A_{ij} £ -·_{ijk}x_k{1\over r^2} $$
 and effectively $Èd^2 §_i£4¹rx_i$, we find
$$ E = {|k|\over 2¹g^2} $$
 Also by similar arguments to those used for instantons, we see that any
solutions with boundary conditions $A£0$, $|Ä|£|k|$ as $r£¥$ have
energy at least as great as this.  There is also a topological interpretation
to this energy:  Writing it as
$$ E = -\f1{16¹^2 g^2}Èd^2 §_i¼·_{ijk}tr(ÒÄÔF_{jk}) $$
 we see that the energy is proportional to the magnetic flux, i.e., the
``magnetic charge" of the monopole.  (The asymptotic value $ÒÄÔ$ of
$Ä$ picks out a direction in isospace, reducing SU(2) to U(1).)  As in
electromagnetism, magnetic charge is quantized in terms of electric
charge.  However, for compact gauge groups, electric charge is also
quantized.  (For the usual U(1), charges are arbitrary, but for SU(2), any
component of the isospin is quantized.)  The energy is thus quantized in
terms of $k$:  It is a multiple of the energy we found for the single
monopole above.

\x IIIC8.2  Perform a singular gauge transformation that makes $ÒÄÔ$
point in a constant (rather than radial) direction in isospin (SU(2)) space. 
Show that the isospin component of the asymptotic form of $A$ describes
a U(1) magnetic monopole: magnetic flux radiating outward from the
origin.

\refs

£1. O. Klein, On the theory of charged fields, in ÓNew theories in physicsÕ,
	proc. Warsaw, May 30 - June 3, 1938 (International Institute of
	Intellectual Co-operation, Paris, 1939) p. 77;\\
	W. Pauli, unpublished (1953):\\
	early work on ``Yang-Mills" theory.
 £2 C.N. Yang and R.L. Mills, ÓPhys. Rev.Õ É96 (1954) 191;\\
	R. Shaw, ÓThe problem of particle types and other
	contributions to the theory of elementary particlesÕ, Cambridge
	University Ph.D. thesis (1955);\\
	R. Utiyama, ÓPhys. Rev.Õ É101 (1956) 1597:\\
	complete formulation of the (classical) theory.
 £3 L. O'Raifeartaigh, ÓThe dawning of gauge theoryÕ (Princeton 
	University, 1997):\\
	early history of Yang-Mills theory, including reprints and previously
	unpublished material.
 £4 R.L. Arnowitt and S.I. Fickler, ÓPhys. Rev.Õ É127 (1962) 1821;\\
	W. Kummer, ÓActa Phys. AustriacaÕ É14 (1961) 149.
 £5 S. Weinberg, ÓPhys. Rev.Õ É150 (1966) 1313;\\
	J.B. Kogut and D.E. Soper, \PRD 1 (1970) 2901:\\
	lightcone field theory.
 £6 S. Coleman, \PL 70B (1977) 59:\\
	Yang-Mills plane waves.
 £7 R. Feynman and M. Gell-Mann, ÓPhys. Rev.Õ É109 (1958) 193;\\
	L.M. Brown, ÓPhys. Rev.Õ É111 (1958) 957;\\
	M. Tonin, ÓNuo. Cim.Õ É14 (1959) 1108;\\
	G. Chalmers and W. Siegel, \xxxlink{hep-ph/9708251},
	\PRD 59 (1999) 045012;\\
	M. Veltman, \xxxlink{hep-th/9712216},
	ÓActa Phys. Polon. BÕ É29 (1998) 783:\\
	description of spin 1/2 with only undotted spinors.
 £8 A. Ashtekar, \PR 57 (1986) 2244, \PRD 36 (1987) 1587;\\
	T. Jacobson and L. Smolin, \PL 196B (1987) 39, 
	ÓClass. Quant. Grav.Õ É5 (1988) 583;\\
	J. Samuel, ÓPramanaÕ É28 (1987) L429:\\
	first-order actions with self-dual auxiliary fields.
 £9 C.N. Yang, \PR 38 (1977) 1377:\\
	reduction of self-dual Yang-Mills field to single component.
 £10 A.N. Leznov, ÓTheor. Math. Phys.Õ É73 (1988) 1233,\\
	A.N. Leznov and M.A. Mukhtarov, ÓJ. Math. Phys.Õ É28 (1987) 2574;\\
	A. Parkes, \xxxlink{hep-th/9203074}, \PL 286B (1992) 265:\\
	lightcone gauge for self-dual Yang-Mills.
 £11 A.A Belavin, A.M. Polyakov, A.S. Shvarts, and Yu.S. Tyupkin,
	\PL 59B (1975) 85:\\
	instantons.
 £12 G. 't Hooft, unpublished;\\
	R. Jackiw, C. Nohl, and C. Rebbi, \PRD 15 (1977) 1642:\\
	't Hooft ansatz for multi-instanton solutions.
 £13 S.-S. Chern and J. Simons, ÓAnn. Math.Õ É99 (1974) 48.
 £14 M.F. Atiyah and R.S. Ward, ÓComm. Math. Phys.Õ É55 (1977) 117;\\
	M.F. Atiyah, V.G. Drinfel'd, N.J. Hitchin, and Yu.I. Manin, \PL 65A (1978)
	185;\\
	E. Corrigan, D. Fairlie, P. Goddard, and S. Templeton, \NP 140 (1978)
	31;\\
	N.H. Christ, E.J. Weinberg, and N.K. Stanton, \PRD 18 (1978) 2013;\\
	M.F. Atiyah, ÓGeometry of Yang-Mills fieldsÕ (Scuola Normale
	Superiore, Pisa, 1979);\\
	V.E. Korepin and S.L. Shatashvili, ÓMath. USSR IzvestiyaÕ É24 (1985)
	307:\\
	general multi-instanton construction, using twistors. 
 £15 G. 't Hooft, \NP 79 (1974) 276;\\
	A.M. Polyakov, ÓJETP Lett.Õ É20 (1974) 194:\\
	monopoles in nonabelian theories.
 £16 E.B. Bogomol'nyi, ÓSov. J. Nucl. Phys.Õ É24 (1976) 449;\\
	M.K. Prasad and C.M. Sommerfeld, \PR 35 (1975) 760:\\
	exact solution for monopole.
 £17 W. Nahm, \PL 90B (1980) 413:\\
	general monopole construction based on ADHM instanton construction.

\unrefs

ÚIV. MIXED

In this chapter we consider ways in which gauge symmetry combines
with global symmetries for new effects.  The interplay
between global internal symmetries of scalar and spinor theories and local
symmetries of Yang-Mills is important for understanding mass
generation for all spins, and is fundamental for the Standard Model. 

Û7 A. HIDDEN SYMMETRY

Symmetries, especially local ones, are clearly very important in the
formulation of interactions.  However, symmetries are not always
apparent in nature:  For example, while most symmetries prefer
massless particles, of all the observed particles the only massless ones
are the graviton, photon (probably), and (some) neutrinos.  Furthermore, of the massive
ones, none with different properties have the same mass, although some
are close (e.g., the proton and neutron).  There are three solutions to this
problem:   
\item{(1)}  The symmetry is not a property of nature, but only an
approximate symmetry.  Some terms in the action are invariant under the
symmetry, but other terms violate it.  We can treat such ``explicit
symmetry breaking" by first studying the symmetry for the invariant
terms, and then treating the breaking terms as a perturbation. 
\item{(2)} Although the laws of physics are symmetric, nature is an asymmetric
solution to them.  In particular, such a solution is the ``vacuum", or state
of lowest energy, with respect to which all other states are defined. 
Since the vacuum is not invariant under the symmetry, the symmetry transformations take the vacuum to other states of the same energy.
This case is called ``spontaneous symmetry breaking".   
For example, in electrodynamics an infinite charge distribution of constant density is translationally and rotationally invariant, but by Gauss' law we know there must be an electric field, whose direction breaks rotational invariance.
\item{(3)} The particles in
terms of which these laws are formulated are not those observed in
nature.  For example, the hydrogen atom is most conveniently described
in terms of a proton and an electron, but in its low-energy physics only
the atom itself is observed as a separate entity:  The U(1) symmetry
related to charge is not seen from the neutral atoms.  The more extreme
case where such particles always appear in bound states is known as
``confinement".

Generally, such broken symmetries are at least partially restored at high
energies.  For example, if the symmetry breaking introduces masses, or
mass differences between related particles, then the symmetry may
become apparent at energies large with respect to those masses. 
Similarly, a hydrogen atom excited to an energy much larger than its
lower energy levels will ionize to reveal its constituent particles.

It often is possible to change to a set of variables that are invariant
under a local symmetry.  
(We saw the analog for the global case when considering translation invariance in subsection IA1.)
For example, if we can define everywhere a
variable that transforms as $¶Ä(x)=Â(x)$, then it can be used to
everywhere undo the invariance.  We can choose the ``gauge" $Â=-Ä$,
transforming $Ä$ to 0 everywhere, leaving no residual invariance, or we
can work with composite, invariant variables:  E.g., $Æ'=Æe^{iÂ}$ is
replaced (invertibly) with $öÆ=Æe^{-iÄ}$, so $öÆ'=öÆ$.

Ü1. Spontaneous breakdown

We first consider symmetry breaking by the vacuum, known as
``spontaneous breakdown".  The action is invariant under the symmetry,
but the vacuum state is not:  Thus, the symmetry acting on the vacuum
produces other zero-energy solutions to the field equations, but this
symmetry is not apparent when considering perturbation about the
vacuum.  In this case, although the symmetry is broken, there are obvious
residual effects, particularly if the breaking can be considered as ``small"
with respect to some other effects.

The ``Goldstone theorem" is an important statement about the effect of
symmetry breakdown:  If a continuous global symmetry is spontaneously
broken, then there is a corresponding massless scalar.  The proof is
simple:  Consider a (relative) minimum of the potential, as the vacuum.  By
definition, we have spontaneous symmetry breaking if this minimum is
not invariant under the continuous symmetry: i.e., applying infinitesimal
symmetry transformations gives a curve of nearby states, which have the
same energy, because the transformations are a symmetry of the
theory.  But the mass of a scalar, by definition, is given by the quadratic
term in its potential, i.e., the second derivative of the potential evaluated
at the vacuum value.  (The first derivative vanishes because the vacuum
is a minimum.)  So, if we look at the scalar defined to parametrize this
curve of constant energy in field space, its mass vanishes.  (This field
may be a function of the given fields, such as an angle in field space.)

We can also formulate this more mathematically, for purposes of
calculation:  Consider a theory with potential $V(Ä^i)$.  (The Lagrangian is
$V$ plus derivative terms.  For simplicity we consider just scalars.)  The
masses of the scalars are defined by the quadratic term in the potential,
expanding about a minimum, the vacuum.  The statement of symmetry of
the potential means that
$$ symmetryâ¶Ä^i = ½^i(Ä)âÜâ0 = ¶V = ½^i »_i Vâfor¼all¼Ä^i $$
 where we allow nonlinear symmetries, and $»_i=»/»Ä^i$.  Differentiating,
and then evaluating at this minimum,
$$ Ò»_i VÔ = 0âat¼minimum¼Ä=ÒÄÔ $$ 
$$ Üâ0 = Ò»_j(½^i »_i V)Ô = Ò(»_j ½^i)(»_i V)Ô +Ò½^i »_i »_j VÔ
	 = Ò½^iÔÒ»_i »_j VÔ $$
 where here the vacuum value $Ò¼Ô$ classically means to just evaluate at
$Ä=ÒÄÔ$.  (So classically $ÒABÔ=ÒAÔÒBÔ$.)  Spontaneous symmetry breaking
means the vacuum breaks the symmetry:  If this symmetry is broken,
then $Ò½^iÔ±0$, so it is a nontrivial eigenvector of $Ò»_i »_j VÔ$ (the mass
matrix) with vanishing eigenvalue.  So, we can write
$$ Ä^i = ÒÄ^iÔ +Ò½^iÔ +... $$
 where $$ is a massless field.

The simplest example is a single free, massless field, $V=0$.  Then $½$ is
simply a constant.  The simplest choice of vacuum is just $ÒÄÔ=0$, which
breaks the symmetry:
$$ L = \f14 (»Ä)^2,ââ¶Ä = constant,ââÒÄÔ = 0 $$
 Then $Ä$ is a ``Goldstone boson".  

The simplest nontrivial example, and a useful one, is a complex scalar
with the potential
$$ V(Ä) = \f14 Â^2(|Ä|^2 -üm^2)^2 $$
 This is invariant under phase transformations $¶Ä=i½Ä$.  There is a
continuous set of minima at $|Ä|=m/å2$.  We choose $ÒÄÔ=m/å2$; then the
Goldstone theorem tells us that the imaginary part of $Ä$ is the
Goldstone field.  Explicitly, separating the field into its real and imaginary
parts,
$$ Ä = \f1{å2}(m +Æ +i)âÜâV = \f14 Â^2 m^2 Æ^2 
	+\f14 Â^2 mÆ(Æ^2 +^2) +\f1{16}Â^2(Æ^2 +^2)^2 $$
 where $ÒÆÔ=ҍÔ=0$.  We could also use the nonlinear separation of the
field into magnitude and phase, $Ä=(m+¨)e^{iÏ}/å2$:  Then $Ï$ drops out
of the potential, and its transformation ($¨$ is invariant) is the same as
that of the free massless scalar.  If $Ä$ had been real, then only the
discete symmetry $Ī-Ä$ would have been broken, and there would be no
Goldstone boson.

\x IVA1.1 Write the complete action in terms of $¨$ and $Ï$.

Note that this model would naively seem to have a tachyon (state with
negative (mass)${}^2$) if we had expanded about $ÒÄÔ=0$.  However, since
the vacuum is defined always as a minimum in the potential (or the
energy), the true states always have nonnegative (mass)${}^2$.  This is
the case for positive spins for similar reasons:  We saw in subsection IIB4
that free massive theories follow from massless ones by dimensional
reduction from one extra spatial dimension.  If we had used an extra time
dimension instead, as required for the ``wrong sign" for the mass term in
$p^2+m^2$, there would also be wrong signs for Lorentz indices, resulting
in kinetic terms with arbitrarily negative energy.

Spontaneous symmetry breaking will also affect the actions for fields
other than those getting vacuum values, that couple to them.
For example, terms of the form $Æ^2 f(Ä)$ will tend to generate
a mass for $Æ$ if $ÒÄÔ±0$ (actually $f(ÒÄÔ)±0$).  Such couplings
exist for $Æ$ of spins $0,ü,1$.  Since masslessness is generally
associated with symmetry (chiral symmetry for spin $ü$ and gauge
symmetry for spin 1), this type of mass generation implies symmetries
other than just those of the scalars are broken by this mechanism
(see subsections IVA4-6).

Ü2. Sigma models

The Goldstone mechanism thus produces massive particles as well as
massless ones, at least for polynomial potentials, to which we are
restricted by quantum considerations, to be discussed later.  We now
look for approximations to polynomial scalar actions that eliminate the
massive fields, but still take them into account through their equations of
motion, in the limit where their masses tend to infinity.  For example, in
the above simple model, we can take the limit $壴$, which takes the
$Æ$ mass ($Âm$) to infinity.  In this limit, the potential energy can
remain finite only if it vanishes:  $|Ä|^2=üm^2$.  (In quantum language,
the potential's contribution to the path integral is just $¶(ÇV)$ in that
limit.  Alternatively, we can neglect the kinetic energy for $|Ä|$ in
comparison to the mass or potential, and then eliminate $|Ä|$ through its
equation of motion in this approximation.)  We can also enforce this limit
directly by using a Lagrange multiplier field $ñ$:
$$ L = ü|»Ä|^2 +ñ(|Ä|^2 -üm^2) $$
 The solution to the constraint is $Ä=\f{m}{å2}e^{iÏ}$, and the action then
describes just a free, real scalar $Ï$.

A less trivial example is a nonabelian generalization of this example: 
Consider $Ä$ as a vector of an internal SO(n) symmetry.  (The previous
example was the case SO(2).)  The Lagrangian is then
$$ L = \f14(»Ä)^2 +üñ(Ä^2 -m^2) $$
 The usual way to solve quadratic constraints without introducing square
roots is to use the identity
$$ |(1+ix)^2|^2 = (|1+ix|^2)^2âÜâ(2x)^2 +(1-x^2)^2 = (1+x^2)^2 $$
 This is often used for trigonometric substitutions or simplifying integrals.
For example, when an integrand has a $å{1-x^2}$, substituting $x=sin¼Ï$
eliminates the square root at the price of requiring trigonometric
identities, which in turn are usually solved by making a second variable
change to $y=tan(Ï/2)$.  On the other hand, the above identity suggests
making instead the variable change $x=2y/(1+y^2)$, which actually gives
the same result, more directly, as the previous two-step method.  (This
identity can also be used for finding integer solutions to the Pythagorean
theorem:  A right triangle with two shorter sides of integer lengths $2mn$
and $m^2-n^2$ has the hypotenuse $m^2+n^2$, where $m,n$ are
integers.)

We then can solve the constraint $Ä^2=m^2$ with the coordinates for the
sphere in terms of an SO(n$-$1) vector $$,
$$ Ä = m \left( {\over 1+\f14 ^2}, 
	{1-\f14 ^2\over 1+\f14 ^2} \right) $$
 Then the kinetic term (now the whole action) becomes
$$ \f14 (»Ä)^2 = \f14 m^2{(»)^2 \over (1+\f14 ^2)^2} $$

\x IVA2.1  For SO(3), express $$ in terms of the usual spherical polar
angular coordinates $Ï$ and $\Ä$, along with the inverse expressions
($Ï$ and $\Ä$ in terms of $$).

Another way to obtain this result is to use the solution of subsection IA6
to the constraint 
$$ 0 = y^2 = (y^a)^2 -2y^+ y^-âÜâ y = e(x^a,1,üx^2) $$
 (but now $(y^a)^2$ is positive definite).
 Then the desired constraint
$$ (y^a)^2+(y^1)^2 = 1 $$
 follows from further constraining 
$$ 1 = y^0 = e\f1{å2}(1+üx^2)âÜâe = {å2\over 1+üx^2} $$
$$ Üâdy^2 = e^2 dx^2 = {2dx^2\over (1+üx^2)^2} $$
 yielding the above result for $x=/å2$.  We thus have a nonpolynomial
action, each term having derivatives.  The original SO(n) symmetry is
nonlinearly realized on the ``angle" variables $$, and the vacuum
($ҍÔ=0$) spontaneously breaks the symmetry to SO(n$-$1).  The constant
$m$ acts as a dimensionful coupling, as seen by scaling $£/m$ to give
the kinetic term the standard normalization.

A complex generalization of this model is described by the Lagrangian
$$ L = ü|áÄ|^2 +ñ(|Ä|^2-m^2) $$
 where $Ä$ is now a complex n-component vector, $á$ is a U(1)-covariant
derivative ($áÄ=(»+iA)Ä$), and $ñ$ is a Lagrange multiplier enforcing
that $Ä$ has magnitude $m$.  This model thus has a U(n) symmetry.  Since
$A$ has no kinetic term ($F^2$), we can eliminate it by its algebraic field
equation:
$$ L £ ü|»Ä|^2 +\f1{8m^2}(Äÿ\onª» Ä)^2 +ñ(|Ä|^2-m^2) $$
 where we have applied the constraint $|Ä|^2=m^2$ (or shifted $ñ$ to
cancel terms proportional to $|Ä|^2-m^2$).  Since the U(1) gauge was not
fixed yet, we still have local U(1) invariance even without an explicit
gauge field.  We can use this invariance to fix the phase of one
component of $Ä$, and use the constraint from $ñ$ to fix its magnitude. 
In terms of the remaining (n$-$1)-component complex vector $$,
$$ Ä = m \left( {\over 1+\f14 ||^2}, 
	{1-\f14 ||^2\over 1+\f14 ||^2} \right) $$
$$ ÜâL = üm^2{|»|^2 +\f14 (ÿ\onª» )^2 \over (1+\f14 ||^2)^2} $$
 (Alternatively, we can solve the constraint and fix the gauge first, then
eliminate $A$ by its field equation.)  This model is known as the CP(n$-$1)
model (``complex projective").

Another example that will prove more relevant to physics is to generalize
$Ä$ to an n$°$n matrix:  We then consider the Lagrangian
$$ L = tr[ü(»Ä)ÿÉ(»Ä) +\f14 Â^2(ÄÿÄ -üm^2 I)^2] $$
 (where $I$ is the identity matrix).  Since $ÄÿÄ$ is hermitian and positive
definite, the minimum of the potential is at $ÄÿÄ=üm^2 I$, and we can
choose
$$ ÒÄÔ = \f{m}{å2}I $$
 using the SU(n)$°$SU(n)$°$U(1) invariance
$$ Ä' = U_L Ä U_R{}^{-1} $$
 (We can include the U(1) in either $U_L$ or $U_R$.)  The vacuum then
spontaneously breaks this invariance to SU(n):
$$ ÒÄ'Ô = ÒÄÔâÜâU_L = U_R $$
 In the large-mass limit, we get the constraint
$$ ÄÿÄ = üm^2 IâÜâÄ = \f{m}{å2}U,âUÿU = I,âÒUÔ = I,â
	L £ \f14 m^2 tr[(»U)ÿ(»U)] $$
 so the field $U$ itself is now unitary.

Ü3. Coset space

The appearance of the scalar fields (Goldstone bosons) as group elements
can be generalized directly in terms of the effective theory, without
reference to massive fields.  Such a theory should be considered as a
low-energy approximation to some unknown theory.  Although the
unknown theory may be better behaved at high energies quantum
mechanically (see later), the low-energy effective theory can be
determined from just (broken) symmetry.  We therefore assume a
symmetry group G that is broken down to a subgroup H by the vacuum. 
(I.e., the vacuum is invariant under the subgroup H, but not the full group
G.)  We are interested in only the Goldstone bosons, associated with all
the generators of the group G less those of H.  These fields are thus
coordinates for the ``coset space" G/H:  They correspond to elements of
the group G, but elements related by the subgroup H are identified. 

Explicitly, we first write the field $g$ as an element of the group G, either
by choosing a matrix representation of the group (as in the U(N) example
above), or explicitly expanding over the group generators $G_I$:
$$ g = e^{iÄ},ââÄ = Ä^I(x) G_I $$
 We then ``factor" out the subgroup H by introducing a gauge invariance
for that subgroup:
$$ g' = gh,ââh = e^{ih^î(x) H_î} $$
 in terms of the H generators $H_î$, which are a subset of $G_I$:
$$ G_I = (H_î,T_i) $$
 where $T_i$ are the remaining generators, corresponding to G/H.  In
particular, we can choose
$$ gauge¼Ä^î = 0âÜâÄ = Ä^i T_i $$
 However, G should still be a global invariance of the theory, though not of
the vacuum.  We therefore assume the global transformation
$$ g' = g_0 g $$
 where $g_0$ is an element of the full group G, but is constant in $x$.  The
vacuum 
$$ ÒgÔ = I $$
 is then invariant under the global subgroup $g_0=h^{-1}$,
where thus $h$ is constant and $g_0ã$H (i.e., G is spontaneously 
broken to H).

$g$ can be used to convert any representation of the global group G
into one (but usually reducible) of the smaller local group H:
$$ Æ' = g_0 ÆâÜâ÷Æ ­ g^{-1}Æ,â÷Æ' = h^{-1}÷Æ $$
 We can apply a similar procedure to find a field strength for $g$,
invariant under the global
group, as an element of the Lie algebra of G:
$$ g^{-1}»_a g = »_a +iA_a^î H_î +iF_a^i T_i = á_a +iF_a^i T_i $$
 This can be evaluated in the $Ä$ parametrization as multiple
commutators, as usual:  $A$ and $F$ are both nonpolynomial functions of
$Ä$, but with only one derivative.  We have absorbed $A$ into a covariant
derivative $á$ because of the remaining transformation law under the
local group H:
$$ (á+iF)' = h^{-1}(á+iF)hâÜâá' = h^{-1}áh,âF' = h^{-1}Fh $$
 where we have assumed $[H_î,T_i]¾T_j$.  (In particular, this is true for
compact groups, where the structure constants are totally
antisymmetric:  Then $f_{îûi}=0Üf_{îiû}=0$.)  Then the action invariant
under global and local transformations can be chosen as
$$ L = \f14 m^2 tr(F^2) $$
 For example, the real vector model we gave in the previous subsection
describes the coset space SO(n)/SO(n$-$1), the complex vector describes
SU(n)/U(n$-$1), and the matrix model describes U(n)$°$U(n)/U(n).

\x IVA3.1  Use the coset-space construction to derive the specific $§$
models explicitly given in the previous subsection, as just identified.
 ªa Find the real and complex vectors by dividing up the adjoint
representation into appropriate blocks.
 ªb For the case of U(n)$°$U(n)/U(n), the direct product means we
use separate group-element fields for the two global groups, with
$$ g_L' = g_{L0}g_L h,ââg_R' = g_{R0}g_R h $$
 for the same $h$.  Find an expression for the field $U$ of the
previous subsection without breaking any global or local symmetries.

Note that the field redefinition between the G-representation matter
field $Æ$ and the H-representation matter field $÷Æ$ modifies the form
of the couplings.  For example, the kinetic term for $Æ$ will have
ordinary partial derivatives $»$, while that for $÷Æ$ will have covariant
ones $á$.  (One or the other will also have $F$ terms.)  
On the other hand, a mass term for $֮$ may turn into a 
potential/Yukawa term
for $Æ$, since the larger group $G$ might not allow mass terms
permitted by the smaller group $H$.  The result is that what appears
as a nonderivative coupling in terms of $Æ$ may appear as a
derivative coupling in terms of $֮$.

We can formulate general spontaneous breakdown in this language:
\item{(1)} Start with a polynomial action with symmetry $G$, 
including scalars $Ä$ that
will suffer the breakdown through expectation, and other fields $Æ$.
\item{(2)} Introduce an appropriate $g$ and define the new scalar fields 
$÷Ä­g^{-1}Ä$, as well as the new matter fields $÷Æ­g^{-1}Æ$.
Thus $S[Ä,Æ]£S[÷Ä,÷Æ,g]$.
In terms of these new fields, the action has a local symmetry $H$,
and $G$ now acts ÓonlyÕ on $g$.
\item{(3)} (H-covariantly) constrain $ր$ in such a way 
that $g$ effectively replaces the missing parts. 
Then $g$ describes all the Goldstone bosons, while the
reduced $÷Ä$ describes the other scalars in the original $Ä$
(in the previous examples, the massive ones, which decoupled at
low energies).

\noindent For example, for
the SO(n) model, $Ä$ is an n-vector, while $g$ parametrizes
the coset
SO(n)/SO(n$-$1), and thus has n(n$-$1)/2 $-$ (n$-$1)(n$-$2)/2 =
n$-$1 non-gauge components --- it is an n$-$1-vector under H.
Thus for $÷Ä$, which is n$£$n$-$1$¢$1 under H, we just constrain the
n$-$1 part to vanish.

We have described how nonpolynomial actions quadratic in derivatives
can arise as a low-energy approximation to polynomial theories.  Further
nonpolynomial terms quartic in derivatives (but no more than quadratic in
time derivatives) can be useful for certain applications, but these arise
from polynomial actions quadratic in derivatives (which are preferred
for quantum reasons) only by quantum effects.
One use is in models which describe (pseudo)scalar mesons by
fundamental fields (i.e., solutions to the free field equations,
which yield interacting solutions through perturbation theory), 
but baryons by nonperturbative solutions to the field equations
of these scalars.  Such an interpretation is suggested by an expansion
in 1/N, where N is the number of colors, since a baryon is made of
N quarks (whereas a meson contains just one quark and one antiquark).
Such models are useful for describing static properties of baryons
(masses, quantum numbers), but the complexity of such solutions
to the field equations prevents their use for interactions of baryons
(especially with other baryons).

Ü4. Chiral symmetry

Later we'll examine a description of the strongly interacting particles
(``hadrons") in which they are all considered as composites (bound
states) of fermionic ``quarks".  However, this theory is extremely difficult
to solve, so we first consider treating the hadrons as fundamental
instead.  Since there are probably an infinite number of kinds of hadrons
(or at least some integer power of $10^{40}$, considering the (Planck
mass)${}^2$), this would require a formulation in terms of a ``string" that
treated all ``mesons" (bosonic hadrons) as a single entity.  That
possibility also will be considered later; for now, we look at the simpler
possibility of studying just the low-energy physics of hadrons by using
fields for just the lightest particles.

So far, the only observed scalar particles have been strongly interacting
ones.  Some of the scalar mesons, especially the ``pions", are not only the
lightest hadrons, but can be considered close to massless on the hadronic
scale.  We therefore look for a description of pions (and some close
relatives) in the massless approximation; then mass-generating
corrections can be considered.

Normally, quantum corrections can affect masses.  The only way to
guarantee masslessness at the quantum level is through some symmetry;
we then can study this symmetry already at the classical level.  We have
seen that (unbroken) gauge invariance can require masslessness for all
fields except the scalar and spinor.  Masslessness for a spinor can be
enforced by ``chiral symmetry":   If there is a U(1) symmetry for all
irreducible spinors $Æ_Œ$, then no mass terms (bilinears $Æ_1^Œ Æ_{2Œ}$)
can be constructed.  (Generally, each spinor can have different U(1)
charges, as long as no two charges add to zero.  Of course, this U(1) can
be a subgroup of a larger chiral symmetry group.)  The only way a scalar
can be guaranteed masslessness is if it is a Goldstone boson.  We
therefore look for a description of pions as Goldstone bosons of some
spontaneously broken symmetry.  (Supersymmetry is another possibility
to enforce massless scalars, but only if there are also massless fermions,
which is not the case for hadrons.)  Furthermore, pions and the other
lightest scalars are actually ÓpseudoÕscalars:  This suggests that they are
the Goldstone bosons of broken chiral symmetry, which simultaneously
generates masses for the fermions.

For simplicity, we consider the coupling of scalar mesons to quarks.  We
could instead couple mesons to ``baryons" (fermionic hadrons), thus
treating only hadrons, but the principles would be the same, only the
indices would be messier.  Combining C invariance with chiral symmetry,
and including a meson potential for spontaneous symmetry breaking, we
can write the action for just the quarks and scalar mesons as
$$ S = Çdx¼tr¼L $$
$$ L = [qÿ_L^{ÀŒ}i»^º{}_{ÀŒ}q_{Lº} +q_R^{Tº}i»_º{}^{ÀŒ}q*_{RÀŒ}]
	+[ü(»Ä)ÿÉ(»Ä) +\f14 Â^2(ÄÿÄ -üm^2 I)^2] $$
$$ +ñ[q_L^Œ Ä q^T_{RŒ} +q*_R^{ÀŒ}Äÿ qÿ_{LÀŒ}] $$
 where $Ä$ is an m$°$m matrix (m ``flavors"), $q_L$ and $q_R$ are
n$°$m matrices (n ``colors"), and $ñ$ is the ``Yukawa coupling". 
Sometimes it will be convenient to drop Lorentz indices to emphasize
internal symmetries:
$$ L = (qÿ_L i» q_{Lº} +q_R^{T} i» q*_R)
	+[ü(»Ä)ÿÉ(»Ä) +\f14 Â^2(ÄÿÄ -üm^2 I)^2] 
	+ñ(q_L Ä q^T_R +q_R* Äÿ qÿ_L) $$
 Besides
color symmetry (local if we had bothered to write in the Yang-Mills fields
for the ``gluons", by $»£á$ on the quarks), we have the (global)
U(m)${}_L°$U(m)${}_R$ chiral (flavor) symmetry
$$ q_L' = q_L U_L,âq_R' = q_R U_R*,âÄ' = U_L^{-1} Ä U_R $$
 including the (global) U(1) ``baryon number" symmetry
$$ U_L = U_R = e^{iÏ}âÜâq_L' = e^{iÏ}q_L,âq_R' = e^{-iÏ}q_R,âÄ'=Ä $$

If we think of baryon number as an SO(2) symmetry, then charge
conjugation is just the reflection that completes this to an O(2)
symmetry (see exercise IIA1.2):
$$ C:ââq_L ª q_R,âÄ £ Ä^T $$
 From this, and the usual CP
$$ CP:ââq_L £ q_L*,âq_R £ q_R*,âÄ £ Ä* $$
 we find the parity symmetry
$$ P:ââq_L ª q_R*,âÄ £ Äÿ $$
 (where for CP and P we also transform the coordinates as usual).

As before, the vacuum $ÒÄÔ=\f{m}{å2}I$ breaks the flavor symmetry to the
diagonal subgroup $U_L=U_R$, which commutes with parity (and is
therefore no longer ``chiral").  It also gives masses to the quarks (since
chiral symmetry is broken); this is a general feature of spinors coupled to
scalars under spontaneous breakdown.  In the limit $壴$ (where the
mass of all bosons but the Goldstones becomes infinite, but the quark
mass $M=ñm$ is fixed), the Goldstone bosons are described by the
unitary matrix $U$, which transforms as $U'=U_L^{-1}UU_R$.

\x IVA4.1  Rewrite this action according to the analysis of 
exercise IVA3.1b:
 ªa Separate the Goldstone bosons from the massive scalars.
 ªb Replace the G-representation quarks with the
H-representation quarks.

An interesting special case is m=1 (one flavor).  The Goldstone boson of
axial U(1) can be identified with the $¹^0$.  In the limit $£¥$, the
Lagrangian becomes (with a $tr$ no longer needed)
$$ L = (qÿ_L^{ÀŒ}i»^º{}_{ÀŒ}q_{Lº} +q_R^{Tº}i»_º{}^{ÀŒ}q*_{RÀŒ})
	+\f14 m^2(»¹)^2
	+\f{M}{å2}(e^{i¹} q^{TŒ}_R q_{LŒ}  +e^{-i¹} qÿ_L^{ÀŒ}q*_{RÀŒ}) $$
 writing $¹$ for the neutral pion field.  $ñ=M/m$ is still the coupling of
the pion to the quarks, as can be seen by rescaling $¹£¹/m$ to give the
kinetic term the usual normalization.  (The coupling $m$ is known as the
``pion decay constant", and is usually denoted $f_¹$.  If we include
leptons with the quarks, then this coupling also describes the decay of
the pion into two leptonic fermions.)  

In this case, the (broken) axial U(1) transformations are
$$ q_L' = e^{iÏ}q_L,âq_R' = e^{iÏ}q_R,â¹' = ¹ -2Ï $$
 The corresponding axial current (determined, e.g., by coupling a gauge
vector) is
$$ J_A^{ŒÀº} = (qÿ_L^{Àº}q_L^Œ -q_R^{TŒ}q*_R^{Àº}) -m^2 »^{ŒÀº}¹ $$
 This current is still conserved, since the field equations aren't changed by
the properties of the vacuum.  The linear term is characteristic of
expanding the Goldstone field about the spontaneously broken vacuum; it
corresponds to the fact that that field has an inhomogeneous
transformation under the broken symmetry.  

However, in reality the pion is not exactly massless, so we should add to
the previous action a mass term for the pion, which explicitly violates the
symmetry.  (It is then a ``pseudogoldstone boson".)  In the general chiral
symmetry model, where the Goldstone bosons are described by a unitary
matrix, a simple term that gives them masses while preserving the polar
(parity-preserving) diagonal symmetry $U_L=U_R$ of the vacuum is, for
some constant $Å$,
$$ L_m = -żtr (Ä+Äÿ-å2mI) $$
 Since this explicitly breaks the axial U(m) symmetries, the corresponding
currents are no longer conserved.  In the U(1) case, we can also add just a
mass term 
$$ L_m = \f14 ½¹^2,ââ½ = m^2 m_¹^2 $$
 (for some constant $½$), which is the leading contribution from the
general term above.  The change in the field equation for $¹$ now
violates the conservation law as
$$ »ÉJ_A = -½¹ $$
 This explicitly broken conservation law is known as ``Partially
Conserved Axial Current" (PCAC).

\subsectskip\bookmark0{5. St\noexpandückelberg}\subsecty{5. St¬uckelberg}

By definition, only gauge-invariant variables are observable.  Although in
general a change of variables to gauge-invariant ones can be complicated
and impractical, there are certain theories where such a procedure can
be implemented very simply as part of the normal gauge-fixing.  Not
surprisingly, the only nonlinearity in these redefinitions involves scalars.

The simplest cases of such redefinitions are free theories, and are thus
contained in our earlier discussion of general free, massive gauge
theories.  The simplest of these is the massive vector.   As described in
subsection IIB4, the Lagrangian and gauge invariance are
$$ L = \f18 F^2 +\f14 (mA+Ȁ)^2 $$
$$ ¶A = -»Â,ââ¶Ä = m $$
 where $F_{ab}$ is the Abelian field strength.  Note that the scalar is pure
gauge:  It is called a ``compensator" for this gauge invariance.  Since it
has a nonderivative gauge transformation, it can easily be gauged to zero
at each point, by just choosing $Â=-Ä/m$.  This means that without loss of
generality we can consider the theory in terms of just the
gauge-invariant field
$$ A' = A +\f1m Ȁ $$
 This ``composite" field can also be considered as a field redefintion or
gauge transformation on $A$.  The lagrangian simplifies to
$$ L =\f18 F'^2 +\f14 m^2 A'^2 $$
 Later we'll see that it is often more useful to keep $Ä$ as an
independent field.

\x IVA5.1 Choose the gauge $A^0=Ä$.  Show that $Ä$ then can be
eliminated by its equation of motion, leaving only the transverse 3-vector
$A^i$, with Lagrangian \hbox{$-\f14 A^i(õ-m^2)A^i$}.  Show the relation
to the lightcone gauge of subsection IIIC2, using the dimensional
reduction langauge of subsection IIB4.  (Hint: You might want to use the fact that 
$f^2 ª f(m-»_0)f/m$ using integration by parts.)

The original Lagrangian can also be considered an unusual coupling of a
massless vector to a massless scalar:  Remember that the massless
scalar is the simplest example of a Goldstone boson, with the
spontaneously broken global symmetry
$$ ¶Ä = ·TÄ,ââTÄ = 1 $$
 where we have defined the symmetry generator $T$ to act
inhomogeneously on $Ä$.  We then couple the ``photon" to this charge: 
After a trivial rescaling of the gauge field,
$$ L = \f1{8m^2}F^2 +\f14 (áÄ)^2,ââá = » +AT $$
 where $m$ is the ``charge" with which $A$ couples to $Ä$, which in this
case happens to have dimensions of mass.  The electromagnetic current
in this case is simply $J=üáÄ$, whose conservation is the scalar field
equation $õÄ=0$ (with gauge-covariantized $õ$).

Because the spontaneously broken symmetry of the corresponding
Goldstone model is now gauged, expanding about $ÒÄÔ=0$ is no longer a
physical statement about the vacuum, since $Ä$ is no longer gauge
invariant.  (As we saw, we can even choose $Ä=0$ as a gauge condition.) 
Therefore, from now on, when we make a statement such as ``$ÒÄÔ=0$"
in such a case, it will be understood to refer to choosing $Ä=0$
as the value about which to perform perturbation expansions (e.g., for
separating actions into kinetic terms and interactions).

Note that the St¬uckelberg action can be generated starting from the
action with just $A'$, and performing a gauge transformation that is not
an invariance:
$$ A' £ A' +\f1m »Ä $$
 Dropping the prime from $A$, this transformation is just the inverse of
the one we used to eliminate the scalar.  If we start from an action that
has also a coupling of $A'$ to matter, we see that conserved currents
decouple from $Ä$:
$$ ÇA'ÉJ £ ÇAÉJ -\f1m ÇÄ »ÉJ $$
 More precisely, if the ÓonlyÕ term in the action for vector + matter that is
not gauge invariant is the vector mass term ($\f14m^2A'^2$), then the
above gauge transformation affects only that term.

Ü6. Higgs

We have seen that spontaneous symmetry breakdown can generate
masses for spinors.  We also saw how a massless vector could become
massive by ``eating" a would-be Goldstone scalar, in the simplest case
of a scalar without self-interactions.  We'll now examine more
interesting models:  Yang-Mills theories, which describe self-interacting
vectors, must couple to self-interacting scalars to become massive.

We can expect, by considering the linearization of any Yang-Mills theory
coupled to scalars, that we will need more scalars than massive vectors,
since each vector needs to eat a scalar to become massive, and some
scalars will become massive and thus uneaten.  (Only would-be Goldstone
bosons can be eaten, as seen by linearization to the St¬uckelberg model.)
For the simple (and most useful) example of U(n) for the gauge group, an
obvious choice for the scalar ``Higgs" field is an n$°$n matrix.  (SU(n)
can be treated as a slight modification.)  The simplest such model is the
one studied in subsection IVA2:  We now consider one of the SU(n)
symmetries (together with the U(1)) as the local ``color" symmetry to
which the Yang-Mills fields couple, and the other SU(n) as the global
``flavor" symmetry (where we use the names ``color" and ``flavor" to
distinguish local and global symmetries, not necessarily related to
chromodynamics).  

The Lagrangian for this ``Gervais-Neveu model" is then
$$ L = tr[\f1{8g^2}F^2 +ü(áÄ)ÿÉ(áÄ) +\f14 Â^2(ÄÿÄ -üm^2 I)^2] $$
 where $á=»+iA$, and $A_a$ and $Ä$ are n$°$n matrices (but $A_a$ are
hermitian).  Now $ÄÿÄ$ is gauge invariant (although not invariant under
the flavor group), so we still have
$$ ÒÄÿÄÔ = üm^2 I $$
 as a gauge-invariant statement (but $ÒÄÔ=\f{m}{å2}I$, or $ÒÄÄÿÔ=üm^2 I$,
still makes sense only for purposes of gauge-dependent perturbation
expansions).  

Since any complex matrix can be written as $Ä=UH/å2$, where $U$ is
unitary and $H$ is hermitian, we can choose the ``unitary gauge" $U=I$
(i.e., $Ä=Äÿ$).  As for the St¬uckelberg case, this is equivalent to working
in terms of the gauge-invariant fields (defined by using this $U$ as a
gauge transformation)
$$ A' = U^{-1}(-i»+A)U,âÄ' = \f1{å2}H = U^{-1}Ä $$
 where $U$ can be defined by
$$ \f1{å2}H = å{ÄÿÄ},âU = Äå2H^{-1} $$
 This is well-defined as long as $H$ is invertible, which is true for small
perturbations about its vacuum value
$$ ÒHÔ = mI $$
 If the perturbation is so large that $H$ has vanishing eigenvalues, then
this is equivalent to looking at states so far away from the vacuum that
some of the broken symmetry is restored.  Expanding about the vacuum
($H£mI+H$), the Lagrangian is now
$$ L = tr[\f1{8g^2}F'^2 +\f14 m^2 A'^2 +\f14 (»H)^2 +\f14 Â^2 m^2 H^2 $$
$$ +A'É\f14 (Hi\onª» H) +ümA'^2 H +\f14 Â^2 mH^3 +\f14 A'^2 H^2
	+\f1{16}Â^2 H^4] $$
 Thus all particles are now massive.  As for the Goldstone case, we can
take the limit $壴$ to get rid of all the massive scalars, which in this
case leaves just the massive vectors, adding only the mass term to the
original Yang-Mills action.  This was clear from the nonlinear $§$ model
that resulted from that limit, by coupling that field ($U$) to Yang-Mills
directly.

\x IVA6.1  Find the chiral action for this model of the type described in
subsection IIIC4, where the massive vectors are described by self-dual
tensors instead of vectors.

\x IVA6.2  Consider again this model, for the case n=2.  We modify this
example by dropping the U(1) gauge field, so we have just SU(2).  Since
SU(2) is pseudoreal, we can further restrict the Higgs field to satisfy the
reality condition $Ä*=CÄC$.  Thus, both color and flavor groups are SU(2),
and $Ä$ is the usual matrix representation of the 4-vector of
SO(4)=SU(2)$°$SU(2) (see subsection IIA5).  Repeat the analysis given
above.

\x IVA6.3 Consider again the gauge group SU(2), but now take the Higgs
field in the ÓadjointÕ representation, with no flavor group (i.e., a real
3-vector).  Show that only 2 of the 3 vectors get mass, leaving a residual
U(1) gauge invariance.  Explain this in terms of the gauge transformations
of the 3-vector.  (Hint: think 3D rotations.)

Ü7. Dilaton cosmology

Some of the ideas in general relativity can be introduced by a simple model that involves introducing only a scalar field.  Although this model does not correctly describe gravitational forces within our solar system, it does give an accurate description of cosmology.  The basic idea is to introduce a dynamical length scale in terms of a real scalar field $Ä(x)$ called the ``dilaton" by redefining lengths as
$$ -ds^2 = dx^m dx^n Ä^2(x) ú_{mn} $$
(Squaring $Ä$ preserves the sign of $ds^2$; we assume $Ä$ vanishes nowhere.)  As explained in our discussion of conformal symmetry, this field changes only how we measure lengths, not angles (which is why it is insufficient to describe gravity):  At any point in spacetime, it changes the length scale by the same amount in all directions.  In fact, it allows us to introduce conformal invariance as a symmetry:  We have already seen that under a conformal transformation the usual proper time of special relativity changes as
$$ dx'^m dx'^n ú_{mn} = Å(x)dx^m dx^n ú_{mn} $$
Thus, by transforming $Ä$ as
$$ Ä'(x') = [Å(x)]^{-1/2}Ä(x) $$
we have
$$ ds'^2 = ds^2 $$
for our new definition above of proper time.  This transformation law for for the dilaton allows any Poincar«e invariant action to be made conformally invariant.  This definition of length is a special case of the general relativistic definition,
$$ -ds^2 = dx^m dx^n g_{mn}(x)âÜâg_{mn} = Ä^2 ú_{mn} $$

The action for a particle is easily modified:  For example,
$$ S_L = Çd ¼ü(vm^2 -v^{-1}Àx{}^2)â£âÇd ¼ü[vm^2 -v^{-1}Ä^2(x)Àx{}^2] $$
since $Àx{}^2=dx^2/d ^2$ (or by using our previous coupling to the metric tensor $g_{mn}$).  It is convenient to rewrite this action by redefining
$$ v( ) £ v( )Ä^2(x( )) $$
The resulting form of the action
$$ S_Lâ£âÇd ¼ü(vm^2Ä^2(x) -v^{-1}Àx{}^2) $$
makes it clear that there is no change in the case $m=0$:  A massless (spinless) particle is automatically conformally invariant.  We have seen this action before:  It is the coupling of a massless particle to an external scalar field $üm^2 Ä^2$.  (What we call the scalar field is irrelevant until we write the terms in the action for that field itself.)

\x IVA7.1  Let's examine these actions in more detail:
ªa Find the equations of motion following from both forms of the particle action with background dilaton $Ä(x)$.  
ªb Find the action that results from eliminating $v$ by its equation of motion from both actions for $m±0$, and show they are the same.  
ªc By a different redefinition of $v$, find a form of the action that is completely linear in $Ä$.

The corresponding change in field theory is obvious if we look at the Hamiltonian form of the particle action
$$ S_Hâ£âÇd [-Àx{}^m p_m +vü(p^2 +m^2Ä^2)] $$
Using the correspondence principle, we see that the Klein-Gordon equation for a scalar field $Æ$ has changed to
$$ (õ -m^2 Ä^2)Æ = 0 $$
The corresponding modification to the field theory action is
$$ Sâ£âÇd^D x¼\f14 [(»Æ)^2 +m^2 Ä^2 Æ^2] $$
Since conformal invariance includes scale invariance, it is now natural to associate dimensions of mass with $Ä$ (or inverse length, if we do classical field theory) instead of $m$, since in scale invariant theories all constants in the field equations (or action) must be dimensionless (otherwise they would set the scale).  
Similarly, this makes $ds^2$ dimensionless, reflecting the fact that it is now scale invariant.

Since this is supposed to describe gravity, at least in some crude approximation that applies to cosmology, where is the (Newton's) gravitational constant?  Since $Ä$ must be nonvanishing, ``empty space" must be described by $Ä$ taking some constant value:  We therefore write
$$ ÒÄÔ = {å3\over û},ââû^2 = {G\over ¹} $$
where ``$Ò{\phantom m}Ô$" means vacuum value, or asymptotic value, or weak-field limit (the value $Ä$ takes far away from matter).  (We have chosen an extra factor here of $¹$ in the definition of Newton's constant $G$ for later convenience, so we effectively use units $G=¹$.  Its normalization can't be determined without introducing true gravity.  Similarly for the $å3$, which simplifies things for cosmology, but differs from our later conventions.)  Thus, the usual mass in the Klein-Gordon equation arises in this way as $må3/û$.  The dilaton $Ä$ is thus defined as the field that spontaneously breaks scale invariance, and also as its Goldstone boson.  Unfortunately, things are more complicated in cosmology, since then $Ä$ is time dependent, even though it's not space dependent.  But physical quantities are scale invariant, just as they are rotationally and translationally; thus only $(dÄ/d )/Ä$ (the ``Hubble constant": see below) and its $ $ derivatives are measurable.

In natural (``Planck") units $û=1$ (i.e., $G=¹$; or some other convenient value):  Fixing $c=\hbar=û=1$ completely determines the units of length, time, and mass.  These units are the convenient ones for quantum gravity; they are also the most obvious universal ones, since special relativity, quantum theory, and gravity apply to everything.  
However, they are presently impractical in general, since
the gravitational constant is not so easy to measure:  Its
presently accepted value is
$$ G=6.6742(10)ð10^{-11}m^3 kg^{-1}s^{-2} $$
(where the numbers in parentheses refer to errors in the last digits),
which is accurate to only a few parts per 10,000, compared to the
standard atomic and nuclear constants, which are known to a few parts per 100,000,000.  
On the other hand, cosmological measurements are even less accurate, so we can use them there:  The orders of magnitude seem inappropriate, but interesting.

In relation to standard units, the Planck units (adjusted to our units $G=¹$) are
$$ å{G\hbar\over ¹c^3} = 9.11867(69)É10^{-36}m $$
$$ å{G\hbar\over ¹c^5} = 3.04166(23)É10^{-44}s $$
$$ å{\hbar c¹\over G} = 3.85786(29)É10^{-8}kg $$

\x IVA7.2  There is another Planck unit, for temperature.  Evaluate it in standard units (Kelvins) by setting to 1 the Boltzmann constant $k$.

We have yet to determine the action for $Ä$ itself:  We write the usual action for a massless scalar in D=4 (for other D we need to replace $Ä$ with a power by dimensional analysis), up to normalization,
$$ S_Ä = -Çd^4 x¼ü (»Ä)^2 $$
but we have written it with the ``wrong" sign, for reasons we cannot justify without recourse to the complete theory of gravity.  However, without this sign change we would not be able to get cosmological solutions with positive energy density for source particles (matter and radiation without self-interaction).

To a good approximation the universe can be described by a spacetime
which is (spatially) rotationally invariant (``isotropic") with respect to a
preferred time direction. Furthermore, it should be (spatially)
translationally invariant (``homogeneous"), so the dilaton should depend
only on that time coordinate.  We therefore look for solutions of the equations of motion which depend only on time.  Thus the proper time is given by
$$ -ds^2 = Ä^2(t)[-dt^2 +(dx^i)^2] $$
By a simple redefinition of the time coordinate, this can be put in a form
$$ -ds^2 = -d ^2 + Ä^2( )(dx^i)^2 $$
 where by ``$Ä( )$" we really mean ``$Ä(t( ))$", and the two time
coordinates are related by
$$ d  = dt¼ÄâÜâ  = Çdt¼Ä(t)âorât = Çd  {1\over Ä(t( ))} $$
In this latter form of $ds$ we can recognize $ $ as the usual time, as measured by a clock at rest with respect to this preferred time frame.  It will prove convenient to calculate with time $t$, so we will work with that coordinate from now on, unless otherwise stated; in the end we will transform to $ $ for comparison to quantities measured by experiment.

To a good approximation the matter in the universe can be approximated
as a ``dust", a collection of noninteracting particles.  It should also be
rotationally invariant with respect to the preferred time direction, so the
momenta of the particles should be aligned in that time direction.  (Really
it is this matter that defines the time direction, since it generates the
solution for $Ä$.)  Furthermore, the dust should be translationally
invariant, so all the momenta should be the same (assuming they all have the same mass), and the distribution should be independent of time.  
(Here, unlike general relativity, we do not think of spacetime itself as changing:  We treat spacetime as ordinary Minkowski space, and $Ä$ as another field on it.)
Varying the Hamiltonian form of the action for a single particle with respect to $Ä$, we find
$$ {¶S_M\over ¶Ä(x)} = m^2Çd ¼v Ä ¶^4(x-X) $$
(We briefly use $ $ again for the worldline parameter, not to be confused with the physical time $ $ just introduced.)
Using the equations of motion following from that action, we also have
$$ vmÄ = å{-Àx{}^2} = \left|{dt\over d }\right| $$
where we have used $dx^i=0$ for this dust, and the fact that $v,m,Ä$ are all positive by definition.  We thus have
$$ {¶S_M\over ¶Ä(x)} = m¶^3(x-X) $$
We can compare this to the energy density, derived as in subsection IIIB4 (since the $ÀX{}^2$ term in the action, which would contain the metric, is unmodified): for matter
$$ T_M^{00} = Çd ¼¶^4(x-X)v^{-1}Àt{}^2 = Çd ¼¶^4(x-X)vm^2 Ä^2 = mĶ^3(x-X) $$
as we could guess from dimensional analysis.  The relation between these 2 quantities is no accident:  Our original introduction of $Ä$ was as $g_{mn}=Ä^2 ú_{mn}$.  If we introduce both $Ä$ and metric independently, so as to calculate both of the above quantities, in the combination $Ä^2 g_{mn}$, then we automatically have
$$ Ä{¶S_M\over ¶Ä} = 2g_{mn}{¶S_M\over ¶g_{mn}} = -T_M{}^m{}_m $$
which is $T^{00}$ in this case (since the other components vanish).

Of course, the dust consists of more than one particle:  It is a collection of particles, each at fixed $x^i$.  That means we should replace $¶^3(x-X)$ with some constant, independent of both $x^i$ (because of homogeneity) and $t$.  (Because of isotropy, the particles don't move.  In this interpretation of the expanding universe, we thus have ``static" particles whose separation increases:  Although $x^i$ is constant for them, distance is measured with an extra factor of $Ä$.)  Actually, we need to average over particles of different masses:  The result is then
$$ {¶S_M\over ¶Ä(x)} = a,ââT_M^{00} = aÄ $$
for some constant $a$.  The equations of motion for $Ä$ are now very simple; since $»_i Ä=0$, we now have simply
$$ € = a $$
where the dots now refer to $t$ derivatives.  If we take this equation and multiply both sides by $ÀÄ$, we get an obvious total derivative.  Integrating this equation, we get
$$ üÀÄ{}^2 = aÄ +üb $$
for some constant $üb$.  This equation has a simple interpretation:  Recognizing $aÄ$ as the energy density $T_M^{00}$ of the dust, and $-üÀÄ{}^2$ as the energy density of $Ä$ (from our earlier discussion of Hamiltonian densities), we see it implies that the total energy density of the Universe is a constant.  

We can also identify the source of this constant energy:  We evaluated the energy density of dust and its coupling to $Ä$.  However, there can also be radiation: massless particles.  As we saw, massless particles do not couple to $Ä$.  Also, we have neglected any interaction of particles with each other.  Thus massless particles in this approximation are totally free; their energy consists totally of kinetic energy, and thus is constant.  (They also move at the speed of light, so components of $T^{ab}$ other than $T^{00}$ are nonvanishing.  However, we average over massless particles moving in all directions to preserve isotropy.)  Therefore we can identify the energy density for radiation,
$$ T_R^{00} = üb $$

\x IVA7.3  Consider general forms of the energy-momentum tensor that have the right symmetry:
ªa Show that the most general form that has spatial isotropy and homogeneity is
$$ T^{mn} = ¨(t)u^m u^n +P(t)(ú^{mn} +u^m u^n),ââu^m ­ ¶_0^m $$
(or the equivalent).  $¨$ is the energy density, while $P$ is the pressure.  This general form is called a ``perfect fluid" (e.g., an ideal gas).
ªb Show that the equation of motion for $Ä$ and energy conservation are now
$$ Ä¬Ä = ¨-3P,ââüÀÄ{}^2 = ¨ $$
Relate pressure to energy density for radiation by using the fact that it doesn't couple to $Ä$.
ªc Derive from these the ``covariant conservation law"
$$ ÄÀ¨ + ÀÄ(3P-¨) = 0 $$

These equations are easily solved.  There is an unavoidable ``singularity" (the ``Big Bang") $Ä=0$ (all lengths vanish) at some time:  Imposing  the initial condition $Ä(0)=0$ (i.e., we set it to be $t=0$) and $ÀÄ(0)>0$ (so $ij0$),
$$ Ä = aüt^2 +åbt $$
The ``physical" time coordinate is then 
$$   = Ç_0 dt¼Ä = a\f16 t^3 +åbüt^2 $$ 
Since $Ä$ can't be expressed simply in terms of $ $, we use the expressions
for both in terms of $t$.  Simple expressions can be found for $a=0$ ($ľå $) and $b=0$ ($ľ ^{2/3}$).

For the case of pure matter ($b=0$), the energy conservation equation
written in terms of the $ $ coordinate becomes, using $d =ļdt$,
$$ ü\left({dÄ\over d }\right)^2 - {a\over Ä} = 0 $$
This is the same as the Newtonian equation for the radial motion of a
particle under the influence of a fixed point mass (or the relative
motion of 2 point particles), with total energy zero.

Since $Ä$ increases with time, distances (as measured by $ds$) between slowly moving objects (such as the dust particles, but also the stars and galaxies to which they are an approximation) also increase.  This is true in spite of the fact that such objects are not moving with respect to the natural rest frame.
The most obvious effect of this cosmological expansion is the
cosmological ``red shift".  The expansion of the universe causes photons
to lose energy, including those of the black-body radiation of the
universe as well as those emitted long ago from distant sources.  

Since the Universe is approximately translation invariant in the spatial directions, spatial momentum $p^i$ is conserved.  (For example, vary the particle action with respect to $x^i$.)  This tells us nothing for the dust, but for the radiation we still have
$$ 0 = p^2 = -E^2 + (p^i)^2  $$
and thus $E$ also is conserved.  But this is $E$ as defined with respect to $t$, not $ $.  (For example, it appeared in the action as $Àt E$.  Also, the above equation is for $p^m=v^{-1}dx^m/d $ with $dx^2=0$.)
However, the time measured by clocks at rest is $ $, and thus the energy $öE$ that is measured is with respect to $ $.  In terms of canonical conjugates as defined in a Lagrangian or Hamiltonian, we see this as
$$ ÀtE = À öEâÜâöE = Ä^{-1}E $$
using $d =ļdt$.  In particular, for the dust particles we have $öE=m$.

Actually, this is true for all components of the (4-)momentum:  At any fixed point $\on\circ x{}^m$, we always choose coordinates near that point such that the proper time looks like the usual one, i.e., $Ä(\on\circ x)=1$.  This can always be accomplished by a scale transformation:  Since we have conformal invariance, we are allowed to choose a reference frame by not only choosing an origin (translation) and orientation of the axes (Lorentz transformation), but also the scale (and even acceleration, via conformal boost).  Rather than make this scale transformation explicitly, we simply note that the measured momentum is actually
$$ öp^m = Ä^{-1}p^m $$
For example, for massive particles we then have $öp{}^2+m^2=0$.

Since $E$ is conserved but $öE$ is measured, we thus have $öE¾Ä^{-1}$.  Therefore, observers measure the photon's energy, frequency, and corresponding
black-body radiation (whose distribution depends only on
energy/temperature) as having time dependence $¾Ä^{-1}$ (and wavelength as $Ä$).  The spectrum of radiation emitted by a distant object is then shifted by this
energy loss, so the amount of shift determines how long ago it was
emitted, and thus the distance of the emitter.

Similar remarks apply to observed energy densities:  When using variations with respect to external fields, we used $¶^4(x-X)$'s:  For the observer's coordinates, this will be multiplied by $Ä^{-4}$ (since $dx$ is multiplied by $Ä$).  Thus the observed energy density is
$$ ö¨ ­ö T{}^{00} = T^{00}Ä^{-4} = aÄ^{-3} +übÄ^{-4} $$

Astronomers use (at least) 3 parameters which are more directly observable.
The ``size" of the Universe $Ä$ is coordinate dependent, but we can
measure the change in time of this scale through red shifts:  
Comparing lengths at different times, we measure $Ä( _2)/Ä( _1)$,
more conveniently represented in terms of the difference of the $ln$:
In terms of the derivative, we have
$$ ln\left({Ä( _2)\over Ä( _1)}\right) ­ Ç_{ _1}^{ _2}d ¼H( ) $$
or
$$ H ­ {dÄ/d \over Ä} $$
The ``Hubble constant" $H$ (constant in space, not time) measures the expansion rate, and
gives an inverse length (time) scale.  Thus it is not predicted, but determined from observations.  As for all cosmological quantities, it is difficult to measure, its value is based on various astrophysical assumptions, and its quoted value has changed often and by large amounts over the years.  A recent estimate for
its present value is
$$ H^{-1} = 13.8(7) É 10^9¼yrs. $$
In ``natural (Planck) units," $c=G/¹=\h=1$,
$H^{-1}=1.43(7)ð10^{61}$.

We can also define a dimensionless  ``(energy) density parameter" $¯$ by using $H^{-1}$ as a length scale:  However, in the simplified model we have used, it is already fixed
$$ ¯ ­ {2ö¨\over 3H^2} = 1 $$
(Sometimes the parameter $§­¯/2$ is used instead.)
Note that in our conventions spatial
integrals are weighted as $Çd^{D-1}x/(2¹)^{D/2}$; thus the relation of our
density to the more standard one is 
$$ ö¨ = (2¹)^2 ¨_{usual}âÜ⯠= {\f{8¹}3 G¨_{usual}\over H^2} $$
 where $G=¹$ in our conventions (but sometimes $G=1$ is useful,
especially for solutions describing stars and planets). 
In the more general (relativity) case, this parameter measures energy density with respect to the amount needed to ``close" the universe; in this case, it takes the ``critical" value, bordering between open and closed.  However, this value agrees with observations to within experimental error.  This alone shows that the dilaton is sufficient to give an accurate cosmological model (although ingredients other than those discussed so far may be needed).

The rate of change of the Hubble constant can be
defined in terms of a dimensionless quantity by comparing its
inverse with the true time:
$$ q ­ {d(H^{-1})\over d } -1 $$
or
$$ q ­ -{ļd^2 Ä/d ^2\over (dÄ/d )^2} $$
The ``deceleration parameter" $q$ tells how fast the 
expansion rate is slowing down.
In the case of pure dust $q=ü$, while for pure radiation $q=1$; otherwise, it's somewhere in between.

\x IVA7.4  Calculate $H$, $q$ and $¯$ in terms of $a$, $b$, and $t$.

Recent supernova observations (together with the assumption of the supernova as a ``standard candle") indicate that $q$ is negative:  The expansion is accelerating.  
Although this ``experimental" value is highly unreliable, and its estimate varies widely from year to year based on methods of measurement and choice of assumptions (as well as author), the existence of measurements indicating $q<ü$ suggests the above model of energy coming from just dust and radiation may be too simple.  In fact, other observations indicate the vast majority of energy in the Universe (about 95\%!) is not in any known form.  While some forms of proposed missing matter (``dark matter") seem to fit into the above types (but are simply not observed by non-gravitational methods), others (``dark energy") do not, and seem to form the majority of the missing energy.  One simple remedy is to introduce a ``cosmological constant" term (or its equivalent) into the action:  In the language of the dilaton, it takes the form
$$ S_ñ = ñÇd^4 x¼Ä^4 $$
where $ñ$ is the cosmological constant.  This term preserves conformal invariance.  (Its scale invariance is obvious by dimensional analysis.)  Unfortunately, it makes the dilaton field equation nonlinear, so we no longer have a simple closed solution as before.  (Numerical methods are required.)  Furthermore, the observed value of this constant corresponds to a length scale of the order of the size of the observed Universe.  While this can be explained for the Hubble constant, since it varies with time, there is no ``natural" way to explain why a true constant should just happen to set a scale comparable to the present value of the Hubble constant (i.e., there is an unexplained $10^{60}$ floating around).  One possibility is that it is dynamically generated as a vacuum value of another scalar field, and thus might vary with time.

\x IVA7.5  Show explicitly that the cosmological term is invariant under a conformal boost.

Various early features of the universe are not well explained by the model presented so far, in particular, why this model works so well, i.e., why the universe is conformally flat.  Furthermore, the observed isotropy of the universe suggests an early period of the universe where all parts of the (now-observed) universe were causally connected so that they could interact in a way to produce this homogeneity.  (The universe as described above would expand too quickly for this to happen, at least for the observable part of the universe.)  The details of this earliest era are not well understood, primarily because they involve physics at the Planck scale.  There are also many models available:  The most popular class of models is ``inflation", the theory that the universe expanded more rapidly initially; another class considers the period before the Big Bang (which may be modified to be less or not singular).  On the technical level, the necessary properties required for such conditions can be described most easily by introducing an extra scalar field (``inflaton") whose changing vacuum value has the effect of a time-dependent cosmological constant.  This field might be either fundamental or composite, or even represent modified dynamics of spacetime itself (by eliminating the inflaton by its equation of motion to modify the action of the dilaton: see exercise IXB5.4).  Unlike the ``dark energy" problem, which would effectively modify gravity at the cosmological scale, this problem would modify gravity at the Planck scale.

\refs

£1 Y. Nambu, \PR 4 (1960) 380;\\
	Y. Nambu and G. Jona-Lasinio, ÓPhys. Rev.Õ É122 (1961) 345,
	É124 (1961) 246:\\
	introduced into relativistic physics massless bosons associated with a
	broken symmetry.
 £2 J. Goldstone, ÓNuo. Cim.Õ É19 (1961) 154;\\
	J. Goldstone, A. Salam, and S. Weinberg, ÓPhys. Rev.Õ É127 (1962) 965:\\
	found the theorem relating the two.
 £3 M. Gell-Mann and M. L«evy, ÓNuo. Cim.Õ É16 (1960) 705:\\
	$§$-models.
 £4 S. Weinberg, \PR 18 (1967) 188:\\
	nonlinear $§$-models.
 £5 S. Coleman, J. Wess, and B. Zumino, ÓPhys. Rev.Õ É177 (1969) 2239,\\
	C.G. Callan, S. Coleman, J. Wess, and B. Zumino, 
		ÓPhys. Rev.Õ É177 (1969) 2247;\\
	A. D'Adda, M. LŸscher, and P. Di Vecchia, ÓPhys. Rep.Õ É49C (1979) 239;\\
	E. Cremmer and B. Julia, 
		\PL 80B (1978) 48, \NP 159 (1979) 141;\\
	A.P. Balachandran, A. Stern, and G. Trahern, \PRD 19 (1979) 2416:\\
	sigma models on coset spaces.
 £6 H. Eichenherr, \NP 146 (1978) 215;\\
	V.L. Golo and A.M. Perelomov, ÓLett. Math. Phys.Õ É2 (1978) 477,
	\PL 79B (1978) 112;\\
	E. Cremmer and J. Scherk, \PL 74B (1978) 341:\\
	classical CP(n).
 £7 T.H.R. Skyrme, ÓProc. Roy. Soc.Õ ÉA260 (1961) 227;\\
	J. Wess and B. Zumino, \PL 37B (1971) 95;\\
	E. Witten, \NP 223 (1983) 422, 433:\\
	higher-derivative terms.
 £8 H. Yukawa, ÓProc. Phys.-Math. Soc. JapanÕ É17 (1935) 48.
 £9 R. Marshak and E.C.G. Sudarshan, ÓProc. Padua-Venice conference on
	mesons and recently discovered particlesÕ, September, 1957;
	ÓPhys. Rev.Õ É109 (1958) 1860;\\
	Feynman and Gell-Mann, Óloc. cit.Õ (IIIC):\\
	chiral symmetry in weak interactions (for coupling to vector
	currents).
 £10 Gell-Mann and Levy, Óloc. cit.Õ;\\
	Nambu, Óloc. cit.Õ;\\
	K.-C. Chou, ÓSov. Phys. JETPÕ É12 (1961) 492:\\
	PCAC.
 £11 St¬uckelberg, Óloc. cit.Õ (IIB).
 £12 P.W. Anderson, ÓPhys. Rev.Õ É112 (1958) 1900, É130 (1963) 439:\\
	nonrelativistic ``Higgs" effect in condensed matter theory.
 £13 F. Englert and R. Brout, \PR 13 (1964) 321;\\
	P.W. Higgs, \PL 12 (1964) 132;\\
	G.S. Guralnik, C.R. Hagen, and T.W.B. Kibble, \PR 13 (1964) 585;\\
	A.A. Migdal and A.M. Polyakov, ÓSov. Phys. JETPÕ É24 (1967) 91:\\
	``Higgs" effect.
 £14 J.L. Gervais and A. Neveu, \NP 46 (1972) 381.
 £15 M. Trodden and S.M. Carroll, \xxxlink{astro-ph/0401547}:\\
	introductory review of cosmology.
 £16 A.G. Riess, A.V. Filippenko, P. Challis, A. Clocchiattia, A. Diercks, P.M. Garnavich, 
	R.L. Gilliland, C.J. Hogan, S. Jha, R.P. Kirshner, B. Leibundgut, M. M. Phillips, 
	D. Reiss, B.P. Schmidt, R.A. Schommer, R.C. Smith, J. Spyromilio, C. Stubbs, 
	N.B. Suntzeff, and J. Tonry, \xxxlink{astro-ph/9805201}, 
	ÓAstron. J.Õ É116 (1998) 1009;\\
	S. Perlmutter, G. Aldering, G. Goldhaber, R.A. Knop, P. Nugent, P.G. Castro, 
	S. Deustua, S. Fabbro, A. Goobar, D.E. Groom, I. M. Hook, A.G. Kim, M.Y. Kim, 
	J.C. Lee, N.J. Nunes, R. Pain, C.R. Pennypacker, R. Quimby, C. Lidman, R.S. Ellis, 
	M. Irwin, R.G. McMahon, P. Ruiz-Lapuente, N. Walton, B. Schaefer, B.J. Boyle, 
	A.V. Filippenko, T. Matheson, A.S. Fruchter, N. Panagia, H.J.M. Newberg, W.J. Couch, 
	\xxxlink{astro-ph/9812133}, ÓAstrophys. J.Õ É517 (1999) 565:\\
	determination of cosmological parameters from observations of supernovae.
 £17 T.R. Choudhury and T. Padmanabhan, \xxxlink{astro-ph/0311622}, 
	ÓAstron. Astrophys.Õ É429 (2005) 807:\\
	recent review of supernovae for cosmology.
 £18 D. N. Spergel, L. Verde, H. V. Peiris, E. Komatsu, M. R. Nolta, C. L. Bennett, 
	M. Halpern, G. Hinshaw, N. Jarosik, A. Kogut, M. Limon, S. S. Meyer, L. Page, 
	G. S. Tucker, J. L. Weiland, E. Wollack, and E. L. Wright, 
	\xxxlink{astro-ph/0302209}, ÓAstrophys. J. Suppl.Õ É148 (2003) 175:\\
	determination of cosmological parameters from observations with the
	Wilkinson Microwave Anisotropy Probe.
 £19 A. Guth, \PRD 23 (1981) 347:\\
	inflation.
 £20 N. Turok, \pdfklink{ÓClass. Quantum Grav.Õ É19 (2002) 3449}%
	{stacks.iop.org/cq/19/3449}:\\
	critical review of inflation.
 £21 M. Gasperini and G. Veneziano, \xxxlink{hep-th/9309023},
	ÓMod. Phys. Lett. AÕ É8 (1993) 3701:\\
	close relation of inflation to ``deflation".

\unrefs

Û4 B. STANDARD MODEL

In this section we discuss the ``Standard Model", the minimal theory that
describes all the observed particles and forces (except gravity). 
We also consider some features of ``Grand Unified Theories" (GUTs), an
extension of the Standard Model that uses fewer multiplets.

Ü1. Chromodynamics

One way in which particles naively described by the action can be hidden
from observation is if the force binding them is too strong to allow them
to exist freely.  Such a condition is often called ``infrared slavery" since
this alleged property of the force is a long-range feature, preventing the
constituent particles from escaping to infinity. This ``confinement" is not
a classical phenomenon, and its occurence even at the quantum level has
not yet been proven.  Therefore, in this section we'll simply assume
confinement, and describe the resultant symmetry properties, leaving the
dynamical properties for later chapters.

The assumption of ``color" confinement is that the color forces are so
strong that they bind any objects of color to other such objects; thus, only
``colorless" states, those that are singlets under the color gauge group,
can exist freely.  Composite fields that are invariant under the local group
transformations can be obtained by multiplying matter fields or
Yang-Mills field strengths, perhaps using also covariant derivatives, and
contracting all color indices.  The color gauge group is generally assumed
to be SU(n): usually SU(3), but sometimes larger n for purposes of
perturbation in 1/n.  Larger n is also used for unification, but in that case
the Higgs mechanism is used to reduce the group of the massless vectors
to SU(3) (times Abelian factors).

Another feature of these confined states, to be considered later, is their
geometrical structure.  The observed spectrum and scattering amplitudes
of the ``hadrons" (strongly interacting particles) indicates a stringlike
identification of at least the excited states.  (The ground states may
behave more like ``bags".)  This picture also fits in with confinement,
since the spatial separation of the quarks and antiquarks in excited
states would force the gluons that convey their interactions (and 
self-interact) to confine themselves as much as possible by collapsing
into ``strings" connecting the quarks.  Thus, we describe a meson with an
``open string", with a quark at one end and an antiquark at the other. 
Similarly, an excited glueball would no longer be a ball, but rather a
``closed string", forming a closed loop.

We will need to reconsider also the discrete symmetries, C, P, and T, and
their combinations.  First of all, we note the ``CPT theorem":  All local,
hermitian, Poincar«e invariant actions are CPT invariant.  This is easy to
see from the fact that CPT only changes the overall sign of the
coordinates, which is effectively the same as changing the sign of each
derivative, as well as giving a $-1$ for each vector index on a field.  Since
CPT also gives signs to dotted spinors and not undotted ones, we
also get $-1$'s for vector combinations of indices on spinors ($Æ^Œ
ÐÆ^{ÀŒ}$; signs cancel when contracting spinor indices on pairs of dotted or
undotted spinors).  Thus, all these signs cancel because Poincar«e
invariance requires an even number of vector indices (in even numbers of
dimensions, from contracting with $ú_{ab}$ and $·_{abcd}$). 
Alternatively, and even more simply, in D=4 we can attribute it to having
even numbers both of undotted spinor indices and of dotted spinor indices,
since we can define CPT by associating a sign with each dotted index
(including those that appear as part of a vector index).  Consequently,
from now on we ignore T and consider only C, P, and CP.

Although we have considered C (and thus CP) in the context of
electromagnetism, nonabelian gauge fields require some (simple)
generalization, since they carry charge themselves.  We start with the
general coupling of massless fermions to nonabelian gauge fields:
$$ L = Æÿ^{Àº}(-i»_{ŒÀº} +A_{ŒÀº})Æ^Œ $$
 where $Æ$ is a column vector with respect to the gauge group, and $A$ a
hermitian matrix.  The CP transformation of the fermions then determines
that of the vectors, needed for invariance:
$$ CP:ââÆ'^Œ = Æ*_{ÀŒ},ââÆ'*^{ÀŒ} = -Æ_Œ,ââ»'_{ŒÀº} = -»^{ºÀŒ},ââ
	A'_{ŒÀº} = A^{TºÀŒ} $$
 (remember $(Æ^Œ)*­ÐÆ^{ÀŒ}$, but $(Æ_Œ)*=-ÐÆ_{ÀŒ}$ because of the factor
of $C_{μ}$), where we have chosen to represent parity on the
coordinates as acting on the explicit $»$ rather than on the arguments of
the fields.  The transformation on the vector is thus parity on the vector
index, combined with charge conjugation $A_a'=-A_a^T=-A_a*$:  The
minus sign can be associated with change in sign of the coupling (as for
the Abelian case), while the complex conjugation takes into account the
charge of the vector fields themselves.  (As discussed in subsection IB2,
$G£-G*$ is an invariance of the algebra, where $g£g*$ and $g=e^{iG}$.) 
Although these terms, as well as the $F^2$ term for the vectors, are CP
invariant, this invariance can be broken by coupling to scalars:  The
Yukawa coupling
$$ L_Y = Æ^{TŒ}Ä Æ_Œ +h.c. $$
 for some matrix $Ä$ of scalars, would require the CP transformation
$$ Ä' = Ä* $$
 (up to perhaps some unitary transformation), but unlike the vectors
there is no guarantee that under complex conjugation the matrix
$Ä=Ä^i M_i$ for real scalars $Ä^i$ and constant matrix (Yukawa
couplings) $M_i$ will preserve this form, i.e., satisfy
$$ Ä'^i M_i = Ä^i M_i* $$
 since the matrices $M_i$ can be complex.

The basic assumption of ``chromodynamics", or in the quantized version
``quantum chromodynamics (QCD)", is that we have a nonabelian gauge
theory without fundamental scalars that couple directly (but scalars will
show up when we introduce electroweak interactions).  Namely, we
assume only Yang-Mills for the ``color" gauge group SU(n), specifically
n=3, with the usual action, minimally coupled to spin-1/2 ``quarks" in the
defining representation of the color group, which may have masses. 
(These masses are actually generated by weakly interacting Higgs
bosons, whose coupling we consider in subsection IVB2; for now we
include just the resulting mass terms.)  Such an action is automatically
invariant under CP and T.  We furthermore assume invariance under
charge conjugation:  Just as an irreducible real scalar describes particles
that are their own antiparticles, and needs doubling (or complexification)
to define charge, an irreducible (massive) spinor cannot describe
distinguishable particle and antiparticle.  But the quarks come in the 
defining representation of SU(n), which is complex, and thus requires
doubling to define mass terms.
Therefore, for every quark field
$q_{LŒ}$ (``L" for ``left") we have an ``antiquark" field $q_{RŒ}$ (``R" for
``right"), and they transform into each other under charge conjugation,
just as a scalar transforms into its complex conjugate.  (A spinor can't
transform into its complex conjugate under C, since C commutes with
spacetime symmetries, like Lorentz transformations.)  Besides this
doubling, and the n colors of the quarks, we also assume a further
multiplicity of m different ``flavors" of quarks.  Gauge invariance
requires the quark masses be independent of color, and C invariance
requires the mass terms couple $q_L$ to $q_R$, but these terms violate
an otherwise global U(m)$°$U(m) flavor symmetry.  

The action is then of the form
$$ tr[\f18 F^2 +(qÿ_L iáq_L +qÿ_R iáq_R) +(\f{M}{å2}q{}^T_R q_L +h.c.)] $$
 where we have written $q_L$ and $q_R$ as matrices with respect to
SU(n) color ($U_c$) and U(m)$°$U(m) flavor ($U_{fL}$ and $U_{fR}$) such
that they transform as
$$ q_L' = U_c q_L U_{fL},ââq_R' = U_c* q_R U_{fR}* $$
 and thus the covariant derivatives can be written as
$$ á_a q_L = (»_a+iA_a)q_L,ââá_a q_R = (»_a -iA_a*)q_R $$
 where $A_a$ are hermitian, traceless, n$ð$n matrices.  (By definition,
charge conjugation takes a representation of an internal symmetry into
the complex-conjugate one.)  

While the color symmetry is a local symmetry, the flavor symmetry is
broken, inducing the transformation on the mass matrix
$$ M' = U_{fL}MU_{fR}^{-1} $$
 This transformation allows the mass matrix $M$ to be chosen real and
diagonal:  Any matrix can be written as a hermitian one times a unitary
one.  A $U_{fR}$ transformation, as a field redefinition, then can be made
to cancel the unitary factor in $M$; then a unitary transformation
$U_{fL}=U_{fR}$ can be made to diagonalize $M$ (while keeping it
hermitian).  These diagonal elements are then simply the masses of the m
different quark flavors.  The most symmetric case is $M=0$, which leaves
the entire U(m)$°$U(m) chiral symmetry unbroken.  (See subsection IVA4.) 
The least symmetric case is where all the masses are nonzero and
unequal, leaving as unbroken only the subgroup U(1)${}^m$, with
$U_{fL}=U_{fR}$.  (In general, $U_{fL}=U_{fR}$ if all masses are
nonvanishing.)

\x IVB1.1 Show the most general case is the product of U(N)'s for various
subspaces, with 2 U(N)'s for the massless subspace.

Since the above transformation allows $M$ to be diagonalized,
in particular it can be made symmetric, which is sufficient
to define charge conjugation:
$$ C:ââq_{LŒ} ª q_{RŒ},ââA_a £ -A_a* $$
 Furthermore, since $M$ can be chosen not only symmetric
but real, in particular it can be made hermitian, which
is enough to define parity:
$$ P:ââq_{L,R}^Œ £ Ðq_{R,LÀŒ},ââA_a £ -A^a $$

The minimal form of this action, besides making CP and T automatic, also
automatically extends the discrete symmetry C to an O(2) symmetry,
whose ``parity" transformation is C and whose continuous SO(2)=U(1)
symmetry  is the U(1) part of the U(m) flavor symmetry, which is not
broken by the mass term.  It corresponds to a charge called ``baryon
number":  Up to an overall normalization factor, it simply counts the
number of quarks $q_{LŒ},Ðq_{RÀŒ}$ (which form a Dirac spinor) minus the
number of ``antiquarks" $q_{RŒ},Ðq_{LÀŒ}$.  However, such an O(2)
symmetry can be defined separately for each flavor, since (after $M$ has
been diagonalized) the action can be written as a sum of independent
terms for each flavor.  In particular, each flavor has its own separately
conserved quark number.  These flavor conservation laws, at the classical
level, are broken only by the weak interactions, which we have not
included in the above action.  (Gravity and electromagnetism do not
violate them.)

Since confinement is a quantum effect, the details of hadronic scattering
cannot be discussed within classical field theory.  However, we saw that
low-energy properties of mesons (and similarly for baryons) could be
described by effective Lagrangians.  The fact that hadrons are made of
quarks can be used to obtain a bit more information even at the classical
level, especially if the relevant quarks are heavy.  (Heavy with respect to
what is unfortunately also a question that can be answered only at the
quantum level.)  For example, in a nonrelativistic approximation,
low-energy properties of hadrons can be found from just the quantum
numbers, spin-spin interactions, and masses of the quarks, while their
velocities are ignored, and the gluons are neglected altogether.  In such
an approximation, reasonably accurate predictions are made for the
masses and magnetic moments of the ground-state hadrons.

Actually, the claim that color nonsinglet states can never be observed
needs a bit of stipulation:  There may be a ``quark-gluon plasma" phase
of hadronic matter that can exist only at extremely high temperatures or
pressures.  Thus, a hypothetical observer during the first moments of the
universe might observe ``free" quarks and gluons.  Similarly, a small
enough observer, living inside an individual hadron, might see individual
quarks and gluons, since the size of his equipment would be much smaller
than what we consider ``asymptotic" distances.  Conversely, we could
consider the possibility of a new chromodynamic force, other than the
one responsible for the hadrons of which we are composed, that has a
confinement scale that is astronomical (extremely low energy), so that
earthly laboratories would fit inside the new ``hadrons".  Thus, any
statement about the observability of color must be a dynamical one, and
does not follow as an automatic consequence of the appearance of a
nonabelian group:  Just as for the Higgs effect, confinement can be
repealed under appropriate circumstances, and the observability of color
depends on the details of the dynamics, and in particular on the values of
the various parameters (momenta and couplings).

Ü2. Electroweak

The weak and electromagnetic interactions are mediated by observed
spin-1 particles, some of which have charge and mass.  Specifically (see
subsection IC4), the massive vectors form a triplet ($W^+$, $W^-$, $Z$),
while there is only one massless vector (the photon).  This suggests a
gauge group of SU(2)$°$U(1), with a Higgs effect that leaves only U(1)
unbroken.  From the table of known fundamental fermions, we can see
that they fall into doublets and singlets of this SU(2), with the U(1) charge
being that of electromagnetism.  (This SU(2)$°$U(1) unification of the
weak and electromagnetic interactions is called the
``Glashow-Salam-Weinberg" model.)

We saw in subsection IVA4 a very simple model of spontaneously
broken chiral U(m)$°$U(m) symmetry where masses were generated for
quarks. In subsection IVA6 we saw how the same scalars could generate
masses for vectors, by coupling to one of the U(m)'s.  We now combine
those two models, specializing to the case m=2, but with two slight
modifications:  (1) Since the defining representation of SU(2) is
pseudoreal, we can impose a reality condition on the Higgs field, which is
in the ($ü$,$ü$) representation of SU(2)$°$SU(2):
$$ Ä* = CÄC $$
 This makes it a vector of SO(4)=SU(2)$°$SU(2)  (See exercises IIA5.3
and IVA6.2.)  It's
also the reality condition satisfied by an element of (the defining
representation of) SU(2).  (See subsection IIA2.)  This is not surprising,
since the group product $U'=U_L U U_R$ allows the interpretation of a
group element itself as a representation of chiral symmetry.  This is the
situation described in subsection IVA2 ($Ä£U$ in the large-mass limit), but
in this case $ÄÿÄ$ is automatically proportional to the identity (it gives
the square of the 4-vector), so in general an SO(4) 4-vector can be
written as the product of a scalar with an SU(2) element. This reality
condition breaks the chiral U(1)$°$U(1) to the diagonal U(1) that leaves
the Higgs invariant.  

(2) The gauged SU(2) is still one of the two chiral SU(2)'s, but the gauged
U(1) must now be a ÓsubgroupÕ of the other SU(2), since the Higgs is now
invariant under the usual U(1)'s.  Thus, the ungauged SU(2) is explictly
broken, and this accounts for the mass splittings in the doublets of known
fundamental fermions.  Remember that observables are singlets of
gauged nonabelian groups (except perhaps for Abelian subgroups), so any
observed internal SU(2) must be a global symmetry, even when it's
broken.  As described in subsection IVA6, these singlets can be
constructed as composite fields resulting from the gauge transformation
obtained from the SU(2) part of $Ä$.

Using the electromagnetic charges of the various particles, we thus
determine their SU(2)$°$U(1) representations:  For spin 1, we have
$W$=(1,0) and $V$=(0,0), where the first entry is the ``isospin" and the second
is the U(1) charge.  For spin 0, we have $Ä$=($ü$,$àü$), choosing the U(1)
generator as the diagonal one from $U_R$.  Finally, for spin 1/2, we have
for the leptons $l_L$=($ü$,$-ü$), which combines with $Ä$ to produce
(0,0)$¢$(0,$-$1), and $l_R$=(0,1).  Similarly, for the quarks we have
$q_L$=($ü,\f16$), and $q_R$=(0,$-\f16àü$).  (We use 
for undotted spinors the convention
``$L$" = fermion, ``$RÊ$" = antifermion.)
The Lagrangian is then
$$ L = L_1 +L_0 +L_{1/2} $$
$$ L_1 = \f1{8g'^2}F^2(V) +\f1{8g^2}tr¼F^2(W) $$
$$ L_0 = tr[\f14(áÄ)ÿ(áÄ) +\f14 Â^2(ÄÿÄ -üm^2)^2] $$
$$ L_{1/2} = tr(ÆÿiáÆ) +tr\left[\tbt{ñ_+}00{ñ_-}q_R^T q_L Ä 
	+ñl_R^T l_L Ä {1\choose 0} +h.c. \right] $$
 where the fermions $Æ=(q_L,q_R,l_L,l_R)$, and the SU(2)$°$U(1)
covariant derivative acts as
$$ áÄ = »Ä +iWÄ -iüVÄ\tat100{-1} $$
$$ áq_L = »q_L -iq_L W +i\f16 V q_L $$
$$ áq_R = »q_R +iüV q_R\left[-\f13 I +\tat100{-1}\right] $$
$$ ál_L = »l_L -il_L W -iüV l_L $$
$$ ál_R = »l_R +iV l_R $$
 (The infinitesimal gauge transformations have the same form, 
dropping the derivative term and
replacing the gauge field with the corresponding gauge parameter.)
 For simplicity we have ignored the indices for color (and its gauge
fields, treated in the previous section), families (treated in the following
subsection), and spin.  We have also used matrix notation with respect to
the local SU(2) (gauged by $W$) and the global SU(2) (explicitly broken in
$L_{1/2}$ by the gauging of a U(1) subgroup, the Yukawa couplings, and
the chirality of the massless neutrinos):  Thus $W$ is a traceless
hermitian 2$ð$2 matrix, $Ä$ is also 2$ð$2 but satisfying the ``reality"
condition given above (traceless antihermitian plus real trace), $q_L$,
$q_R$, and $l_L$ are 2-component rows, and $l_R$ is 
a single component.  (By definition, the diagonal parts of $W$ and $Ä$ are
electromagnetically neutral.)  The quark Yukawa coupling is diagonal in
the broken SU(2) to preserve the local U(1) symmetry.  (The $tr$ here is
trivial for the lepton Yukawa term, but we have left it for generalization
to more than one family.)  Explicitly, we can write
$$ W = \pmatrix{ \f1{å2}W^0 & W^+ \cr W^- & -\f1{å2}W^0 \cr} $$
 and for the lightest family (see subsection IC4)
$$ q_L = ( d_L ¼ u_L ),ââq_R = ( d_R ¼ u_R ),ââ
	l_L = ( e_L ¼ Ã ),ââl_R = e_R $$

In the unitary gauge for the local SU(2),
$$ Ä = \f1{å2}\Ä I,ââÒ\ÄÔ = m $$
 where $\Ä$ is a single real scalar, the simplifications to the Lagrangian
are
$$ L_0 £ \f14 (»\Ä)^2 +\f18 \Ä^2ÊtrÓ[ W -üV\tat100{-1}]^2Õ
	+\f18 Â^2(\Ä^2 -m^2)^2 $$
$$ L_{1/2} £ tr(ÆÿiáÆ) +\f1{å2}\ļtr\left[\tbt{ñ_+}00{ñ_-}q_R^T q_L 
	+ñl_R^T l_L {1\choose 0} +h.c. \right] $$
 We then can expand $\Ä$ about its vacuum value $m$:  The lowest order
terms give masses for most of the vectors and fermions:  The
massless fermions are the neutrinos, while the massless vector gauging
the unbroken U(1) (a combination of the original U(1) with a U(1) subgroup
of the SU(2)) is the photon (of electromagnetic fame).  The mass of the
remaining vectors accounts for the weakness and short range of the
``weak" interactions.

\x IVB2.1 Diagonalize this Lagrangian with respect to the mass
eigenstates.  For convenience, normalize
$$ g = \f1{å2}g_0Êcos¼Ï_W,âg' = g_0Êsin¼Ï_W $$
 where $Ï_W$ is the ``weak mixing (Weinberg) angle".
ªa Find explicitly the masses for all the particles in the Standard Model
(first family for fermions) in terms of the couplings
$m,Â,g_0,Ï_W,ñ_à,ñ$.  Show from the experimental values for the vector
masses given in subsection IC4 that $sin^2 Ï_W ® .223$.
ªb Find all the other couplings of the mass eigenstates.  Show that, with
the conventional electric charge assignments,
$$ {1\over e^2} = {1\over 2g^2} +{1\over g'^2} $$
 (Hint:  Rather than rescaling the vectors, note that the generated mass
term, and the given couplings of $V$ and $W$, suggest defining 
$$ W' = W -üV\tat100{-1}âÜâV = © +k_1 Z,ââå2W^0 = © +k_2 Z $$
 in the conventions of subsection IIA1, for the new fields $Z$ and photon
$©$, for appropriate constants $k_i$.)

Note that, unlike the strong (chromodynamic) or purely electromagnetic
(or even gravitational) interactions, the weak interactions break every
discrete spacetime symmetry possible.  (The others break none.  CP
violation will be discussed in the following subsection.  Of course, CPT is
always preserved.)  Sometimes this is attributed to the presence of a
chiral symmetry, used to reduce 4-component spinors to 2-component;
however, we have already seen that in general chiral and parity
symmetries are unrelated.  (You can have either without the other.  This
fact will be further discussed in subsections IVB4 and VIIIB3.)  A better
explanation is to attribute P and C to doubling, which converts spinors
from 2-component to 4-component:  2-component spinors are the
simplest description of helicity/spin $ü$; 4-component spinors are useful
only to manifest a larger symmetry, when it exists.  The weak
interactions violate parity because the neutrino is not doubled, and
because the fermions that are doubled no longer have a symmetry
relating their two halves.

Ü3. Families

In the Standard Model (and its simpler generalizations) there is no
explanation for the existence of more than one family of fermions. 
However, the existence of 3 families does have interesting
consequences.  Most of these follow from the form of the Yukawa
couplings, and thus the fermion masses.  In subsection IVB1 we considered
redefinitions of the fermion fields as unitary flavor transformations. 
These allowed us to obtain the simplest form of the mass matrices, since
they were not flavor singlets, and thus transformed.  We now perform
similar transformations, but only on the family indices, since
transformations that don't commute with the gauge symmetries would
complicate the other terms in the action.  Now ignoring spin, color, and
local flavor indices, and using matrix notation for the family indices, the
fermions transform as
$$ q_L' = q_L U_{qL},âq_{Rà}' = q_{Rà} U_{qRà}*,â
	l_L' = l_L U_{lL},âl_R' = l_R U_{lR}* $$
 where $q_L$, $q_{Rà}$, $l_L$, and $l_R$ have m components for the m
families.  $q_{Rà}$ are the 2 components of the (explicitly broken) ÓglobalÕ
flavor doublet $q_R$.  We thus have 5 U(m) symmetries, all broken by the
Yukawa couplings:  These field redefinitions induce transformations
on them,
$$ ñ_à' = U_{qL} ñ_à U_{qRà}^{-1},ââñ' = U_{lL}ñU_{lR}^{-1} $$

As in subsection IVB1, $U_{qRà}$ and $U_{lR}$ can be used to make $ñ_à$
and $ñ$ hermitian.  Then $U_{lL}$ can be used to make $ñ$ diagonal, also
as in subsection IVB1, leaving a U(1)${}^m$ symmetry $U_{lL}=U_{lR}$,
corresponding to separate conservation laws for electron number
(including its neutrino), muon number, and tauon number for the 3 known
flavors.  However, the quark sector works a bit differently:  We can use
$U_{qL}$ to diagonalize $ñ_+$ or $ñ_-$, but not both.  This leaves another
U(1)${}^m$ symmetry $U_{qL}=U_{qR+}=U_{qR-}$.  If $ñ_+$ has been
diagonalized, then 1 of the m U(1)'s, corresponding to total quark number
(baryon number) conservation, leaves $ñ_-$ invariant, while the
remaining m$-$1 U(1)'s can be used to eliminate some of the phases of the
complex off-diagonal components of $ñ_-$.

The remaining global flavor symmetries are thus m lepton U(1)'s and 1
quark U(1).  The remaining Yukawa couplings are the real, diagonal $ñ$,
describing the m masses of the massive leptons (the neutrinos remain
massless), the real, diagonal $ñ_+$, giving the m masses of half of the
quarks, and the hermitian $ñ_-$, consisting of m diagonal components,
describing the masses of the other quarks, m(m$-$1)/2 magnitudes of the
off-diagonal components, and (m$-$1)(m$-$2)/2 phases of the
off-diagonal components.  These phases violate CP invariance:  CP,
besides its affect on the coordinates, switches each spinor field with its
complex conjugate.  Since the complex conjugate term in the action uses
the complex conjugates of the $ñ$'s, this symmetry is violated whenever
any of the components have imaginary parts (after taking into account all
possible symmetries that could compensate for this, as we have just
done).  Note that CP is violated only for 3 families or more.  (C and P are
separately violated for any number of families by the SU(2)$°$U(1)
coupling:  As discussed in subsection IVB1, C invariance of the strong
interactions is the symmetry $q_Lªq_R$.)  Since we can choose to
transform away the phases in the subsector of the 2 lighter quark
families, the large masses of the heavier quarks suppress this effect,
accounting for the smallness of CP violation.

Since observed particles are mass eigenstates, it's convenient to perform
a further unitary transformation (the ``Cabibbo-Kobayashi-Maskawa
matrix") that diagonalizes the mass matrix.  Although this is clearly
possible by the arguments of subsection IVB1, it is not part of the unitary
transformations considered in this subsection because it does not
commute with the SU(2) gauge symmetry:  After such a transformation,
we find that the components of each SU(2) quark multiplet are linear
superpositions of different families.

\x IVB3.1 Perform this diagonalization explicitly for the case m=2 (two
families), using the two lightest families of quarks and leptons as listed in
subsection IC4.  Which particles mix?  Parametrize this mixing by an angle
$Ï_c$ (the ``Cabibbo angle").

An important experimental result with which the Standard Model is
consistent is the suppression of ``flavor-changing neutral currents
(FCNC)".  The two electrically neutral gauge fields in this model, the $Z$
and the photon $©$, couple to currents that are neutral with respect to
the U(1) symmetries associated with each of the quark (flavor) numbers. 
This is true by construction before the unitary CKM transformation, but
this transformation also leaves these two currents invariant (the
``Glashow-Iliopoulos-Maiani mechanism").  Thus, at the classical level we
do not see effects such as the decay $K^0£Z£µ^+µ^-$, which would
violate this ``conservation law".  Furthermore, the quantum corrections
are suppressed (though nonvanishing) for similar reasons:  For example,
the lowest-order nonvanishing quantum correction comes from replacing
the $Z$ with a $W^+W^-$ pair.  Without the CKM matrix, this contribution
would vanish; treating CKM, and its resulting contribution to quark
masses, as a perturbtation, the resulting contribution is suppressed by a
factor of $m_q^2/m_W^2$.  The absence of FCNC is an important
constraint on generalizations of the Standard Model.

Ü4. Grand Unified Theories

The Standard Model gives a description of the weak and electromagnetic
interactions that describes the spin-1 particles in terms of gauge fields,
and accounts for all masses by the Higgs effect.  However, it does not
give any unification, in the sense that we still have 3 groups (SU(3), SU(2),
and U(1)) for 3 interactions (strong, weak, and electromagnetic), and a
large variety of spin-1/2 fields that are unrelated except by color and
broken SU(2) flavor.  Grand Unified Theories unify this symmetry by
forcing all 3 gauge groups to be subgroups of a simple group, which is
broken to SU(3)$°$SU(2)$°$U(1) by Higgs (and then broken to SU(3)$°$U(1)
by more Higgs).  This means introducing new spin-1 particles that are
unobserved so far because of their very large masses.  On the other hand,
the known fermions are then grouped together in a small number of
multiplets without introducing new fermions (except perhaps partners for
the neutrinos to allow them to have small masses).  Unfortunately, this
requires a more complicated (and ambiguous) Higgs sector, with separate
spin-0 multiplets and couplings for first breaking to SU(3)$°$SU(2)$°$U(1)
and then breaking to SU(3)$°$U(1); we won't discuss those Higgs fields
here.

The simplest such model uses the group SU(5).  Recall the
SU(3)$°$SU(2)$°$U(1) representations of each family of fermions:
$$ q_L = (3,ü,\f16),âq_{R+} = (Ð3,0,\f13),âq_{R-} = (Ð3,0,-\f23),â
	l_L = (1,ü,-ü),âl_R = (1,0,1) $$
 where the first argument is the dimension of the SU(3) representation
($Ð3$ being the complex conjugate of the $3$), the second is the SU(2)
isospin, and the third is the U(1) charge.  An SU(3)$°$SU(2)$°$U(1)
subgroup of SU(5) can be found easily by taking the 5-component defining
representation and picking 3 components as the defining representation
of SU(3) and the other 2 for that of SU(2):  I.e., consider a traceless
hermitian 5$ð$5 matrix as an element of the SU(5) Lie algebra, and take
$$ \pmatrix{ SU(5) \cr } 
	£ \pmatrix{ SU(3) -\f13 IðU(1) & 0 \cr 0 & SU(2) +üIðU(1) \cr } $$
 or in other words
$$ 5 £ (3,0,-\f13) ¢ (1,ü,ü) $$
 From this we recognize the fermions as falling into a $Ð5¢10$, where the
$10$ is the antisymmetric product of two 5's, which consists of the
antisymmetric product of the two 3's (a $Ð3$), the antisymmetric product
of the two SU(2) doublets, and the product of one of each:
$$ Ð5 £ (Ð3,0,\f13) ¢ (1,ü,-ü) = q_{R+} ¢ l_L $$
$$ 10 £ (Ð3,0,-\f23) ¢ (1,0,1) ¢ (3,ü,\f16) = q_{R-} ¢ l_R ¢ q_L $$

\x IVB4.1  Find the symmetric product of 2 5's, and its decomposition
into representations of SU(3)$°$SU(2)$°$U(1).

A more unifying model is based on SO(10).  A U(5) subgroup can be found
from the spinor representation by dividing up the set of 10 Dirac $©$
matrices into two halves, and taking complex combinations to get 5 sets
of anticommuting creation and annihilation operators.  (See exercise
IC1.2.)  The Dirac spinor is then
$$ (1,-\f52) ¢ (5,-\f32) ¢ (10,-ü) ¢ (Ñ{10},ü) ¢ (Ð5,\f32) ¢ (1,\f52) $$
 in terms of the SU(5) representation and the U(1) charge.  This Dirac
spinor is reducible into Weyl spinors $16¢Ñ{16}$; in fact, $iå2©_{-1}$ is just
$(-1)^{Y+1/2}$ in terms of the U(1) charge $Y$.  (The SO(10) generators
are even in oscillators, and thus do not mix even levels with odd.)  We
then have
$$ 16 £ (1,-\f52) ¢ (10,-ü) ¢ (Ð5,\f32) $$
 Ignoring the U(1) charge, these are the multiplets found for each family
in the SU(5) GUT, plus an extra singlet.  

A simple way to understand this extra singlet is to look at a different
path of breaking to SU(3)$°$SU(2)$°$U(1):  Looking at the vector
(defining) representation of SO(10), we can break it up as 6+4 (in the
same way we broke up the 5 of SU(5) as 3+2) to get the subgroup
SO(6)$°$SO(4)=SU(4)$°$SU(2)$°$SU(2).  We can also see that a Dirac spinor
of SO(10) (16$¢Ñ{16}$) will be a Dirac spinor of SO(6) (4$¢Ð4$) ÓtimesÕ (not
plus) a Dirac spinor of SO(4), while the Dirac spinor of SO(4) is a defining
representation of one SU(2) (($ü$,0)) ÓplusÕ a defining representation of
the other SU(2) ((0,$ü$)).  Thus,
$$ 16 £ (4,ü,0) ¢ (Ð4,0,ü) $$
 where we have used the fact that $©_{-1}$ (used for projection to Weyl
spinors) of SO(10) is proportional to the product of all the $©$-matrices,
and thus the product of $©_{-1}$'s for SO(6) and SO(4).  

Looking at this model (``Pati-Salam model") as an alternative to SU(5)
(but with a semisimple, rather than simple, group, so it unifies only spin
1/2, not spin 1), we now look at breaking SU(4)$£$U(3)= SU(3)$°$U(1)
(using 4=3+1, as we did 5=3+2 for SU(5)), and breaking one SU(2)$£$ U(1). 
We then find
$$ (4,ü,0) £ (3,-\f13,ü,0) ¢ (1,1,ü,0) = q_L ¢ l_L $$
$$ (Ð4,0,ü) £ (Ð3,\f13,0,ü) ¢ (Ð3,\f13,0,-ü) ¢ (1,-1,0,ü)¢ (1,-1,0,-ü) $$
$$ = q_{R+} ¢ q_{R-} ¢ l_R ¢ l_{R-} $$
 where the arguments are the SU(3) representation, the U(1) charge from
SU(4), the SU(2) isospin, and the U(1) charge from the broken SU(2).  If we
choose the U(1) charge of SU(3)$°$SU(2)$°$U(1) as $-$1/2 times the
former of these two U(1) charges plus $1$ times the latter, this agrees
with the result obtained by way of SU(5).  However, we now see that all
the left-handed fermions are contained within one SU(4)$°$SU(2)$°$SU(2)
multiplet, and the right-handed within another, but with a partner for the
neutrino.  Also, one of the SU(2)'s is that of SU(3)$°$SU(2)$°$U(1), while
the other is the other SU(2) of the Standard Model, which was broken
explicitly there to U(1), whereas here it is broken spontaneously. Thus,
there is a local chiral SU(2)$°$SU(2) flavor symmetry.  

$$ \vcenter{\baselineskip=1.1\normalbaselineskip\halign{\strut#&&#\cr
	& SO(10) & \cr
	\hfil $\swarrow$ & & $\searrow$ \hfil \cr
	SU(5) & & SU(4)$°$SU(2)${}_L°$SU(2)${}_R$ \cr
	\hfil $\searrow$ & & $\swarrow$â$®10^{16}$ GeV? \hfil \cr
	\multispan3 SU(3)$°$SU(2)${}_L°$U(1)${}_R$ \hfil \cr
	& \hfil $\downarrow$ \hfil & $®$ 100 GeV \hfil \cr
	& \multispan2 SU(3)$°$U(1) \hfil \cr }} $$

Furthermore, the SU(4)$°$SU(2)$°$SU(2) model is invariant under C:  In
general, C is just a permutation symmetry.  In this case, it simply
switches the two multiplets of each family,
$$ C:ââ(4,ü,0) ª (Ð4,0,ü) $$
 Combining with the usual CP, this model is thus also invariant under P:
$$ P:ââ(4,ü,0) ª (Ð4,0,ü)*ââi.e.,ââÆ^Œ(4,ü,0) ª ÐÆ_{ÀŒ}(4,0,ü) $$
 But both C and P are broken spontaneously on reduction to the Standard
Model.  However, SO(10) lacks C and P invariance (contrary to some
statements in the literature), since there is only a single complex
representation for each family of fermions (and thus no nontrivial C; of
course, there is still CP, at least for the vector-spinor coupling, as
always).  In fact, the C of SU(4)$°$SU(2)$°$SU(2) is just an SO(10)
transformation:  Although SO(10) is not O(10) (which is why it lacks a C), it
still includes reflections in an ÓevenÕ number of ``axes", since reflection in
any pair of axes is a $¹$ rotation (just as for SO(2)).  Thus, breaking
$10£6+4$ includes not only SO(6)$°$SO(4), but also the reflection of an
odd number of the ``6" axes together with an odd number of the ``4"
axes --- a combined ``parity" of both SO(6) and SO(4).  (They are all the
same up to continuous SO(6)$°$SO(4) transformations.)  This parity of the
internal space is the C given above.  (We saw a similar situation for O(2) in
subsection IVB1.)

The identification of C is somewhat semantic in a nonabelian 
gauge theory (except for unbroken U(1) subgroups), since it is 
defined by changes in sign of unobserved charges: The C 
appearing above at an intermediate stage of breaking of the 
SO(10) GUT originates as a global symmetry of only the 
Higgs sector, leaving all ``fundamental" particles with 
spin invariant. After breaking to SU(4)$°$SU(2)$°$SU(2), the 
vectors and the spinors are composites of the original 
ones and the Higgs responsible for the breaking, so they 
pick up this symmetry. (In the same way, the spinning 
particles of the Standard Model pick up the broken global 
SU(2) of its Higgs.)

Since GUTs unify quarks and leptons, they allow decay of the proton. 
However, since this requires simultaneous decay of all 3 quarks into 3
leptons, it is an extremely unlikely (i.e, slow) decay, but barely within
limits of experiment, depending on the model.  Proton decay is still
unobserved:  This eliminates the simplest version of the SU(5) GUT.

\refs

£1 ÓThe rise of the standard model: Particle physics in the 1960s and
	1970sÕ, eds. L. Hoddeson, L. Brown, M. Riordan, and M. Dresden
	(Cambridge University, 1997):\\
	interesting accounts of the development of the Standard Model from
	many of the people responsible.
 £2 J.J.J. Kokkedee, ÓThe quark modelÕ (Benjamin, 1969);\\
	J.L. Rosner, Plenary report on recursive spectroscopy (theory), in
	ÓProc. of the XVII Int. Conf. on High Energy TheoryÕ, London, July,
	1974, ed. J.R. Smith (Science Research Council, 1974) p. II-171;\\
	A. De R«ujula, H. Georgi, and S. Glashow, \PRD 12 (1975) 147:\\
	nonrelativistic quark model.
 £3 A.N. Tavkhelidze, \pdfklink{Color, colored quarks, quantum chromodynamics}%
	{http://www-lib.kek.jp/cgi-bin/kiss_prepri?KN=199501245&OF=4.}, in
	ÓQuarks '94Õ, Vladimir, Russia, May 11-18, 1994,
	eds. D.Yu. Grigoriev, V.A. Matveev, V.A. Rubakov, D.T. Son, and
	A.N. Tavkhelidze (World Scientific, 1995) p. 3:\\
	early history of QCD.
 £3 H. Fritzsch and M. Gell-Mann, ÓProc. XVI International Conference on
	High Energy PhysicsÕ, eds. J.D. Jackson and A. Roberts, Chicago and
	Batavia, Sep. 6-13, 1972 (National Accelerator Laboratory, 1972) v. 2,
	p. 135, \xxxlink{hep-ph/0208010};\\
	H. Fritzsch, M. Gell-Mann, and H. Leutwyler, \PL 47B (1973) 365;\\
	S. Weinberg, \PR 31 (1973) 494, \PRD 8 (1973) 4482:\\
	QCD, with confinement.
 £4 W. Pauli, ÓNiels Bohr and the development of physicsÕ (McGraw-Hill,
	1955) p. 30;\\
	G. L¬uders, ÓAnn. Phys.Õ É2 (1957) 1:\\
	CPT theorem.
 £5 E. Fermi, ÓRic. ScientificaÕ É4 (1933) 491; ÓNuo. Cim.Õ ÉII (1934) 1;
	ÓZ. Phys.Õ É88 (1934) 161:\\
	weak interactions, without vector bosons.
 £6 T.D. Lee and C.N. Yang, ÓPhys. Rev.Õ É104 (1956) 254:\\
	parity violation in weak interactions.
 £7 S.L. Glashow, ÓNuc. Phys.Õ É22 (1961) 579;\\
	A. Salam and J.C. Ward, \PL 13 (1964) 168;\\
	S. Weinberg, \PR 19 (1967) 1264:\\
	unification of weak and electromagnetic interactions.
 £8 N. Cabibbo, \PR 10 (1963) 531;\\
	M. Kobayashi and K. Maskawa, ÓProg. Theor. Phys.Õ É49 (1972) 282.
 £9 S.L. Glashow, J. Iliopoulos, and L. Maiani, \PRD 2 (1970) 1285.
 £10 J. Pati and A. Salam, \PR 31 (1973) 275:\\
	earliest GUT, but semisimple.
 £11 H. Georgi and S. Glashow, \PR 32 (1974) 438:\\
	SU(5) GUT; earliest with simple group.
 £12 H. Georgi, in ÓParticles and Fields --- 1974Õ, proc. AIP conference, 
	Division of Particles and Fields, Sep. 5-7, 1974, Williamsburg, ed.
	E. Carlson (American Institute of Physics, 1975) p. 575;\\
	H. Fritzsch and P. Minkowski, ÓAnn. Phys.Õ É93 (1975) 193:\\
	SO(10) GUT.
 £13 G.G. Ross, ÓGrand unified theoriesÕ (Benjamin/Cummings, 1984).

\unrefs

Û7 C. SUPERSYMMETRY

In section IIC we studied some general properties of supersymmetry in
arbitrary dimensions, and its representations in D=4.  We now consider 4D
interactions, by introducing gauge fields defined on superspace, and their
actions.  A complete discussion of supersymmetry would require (at least)
a semester; but here we give more than just an overview, and include
the basic tools with examples, which is enough for many applications. 
Quantum aspects of supersymmetry will be discussed in chapters VI and
VIII, supergravity in chapter X, and some aspects of superstrings in
chapter XI.

Ü1. Chiral

We first consider some field equations that appear in all free, massless,
supersymmetric theories.  Of course, since the theory is massless it
satisfies the massless Klein-Gordon equation by definition:  $õì=0$. 
From our earlier discussion of general properties of supersymmetry, we
also know that $p^{ŒÀº}Ðq_{Àº}ì=p^{ŒÀº}q_Œì=0$.  These don't look
covariant, but noticing that $pq$ differs from $pd$ only by $Ïõ$ terms
(because of the index contraction), which already vanishes, we have the
field equations
$$ p^{ŒÀº}Ðd_{Àº}ì = p^{ŒÀº}d_Œ ì = 0 $$
 These equations imply the Klein-Gordon equation, as seen by hitting
them with another $d$ and using the anticommutation relations
$Ód_Œ,Ðd_{Àº}Õ=p_{ŒÀº}$.  They imply stronger equations:  By evaluating at
$Ï=0$, $d_Œ ì$ yields a spinor component field $Æ_Œ$, and we find
$$ p^{ŒÀº}ÐÆ_{Àº} = p^{ŒÀº}Æ_Œ = 0 $$
 the usual for massless spin 1/2.

Another equation that can be imposed is the ``chirality" condition
$$ Ðd_{ÀŒ}Ä = 0 $$
where $Ä$ now refers to such a ``chiral superfield" (and thus
$ÐÄ$ to an ``antichiral" one, $d_Œ ÐÄ=0$).
This requires that $Ä$ be complex, otherwise we would also have 
$d_Œ Ä=0$ and thus $pÄ=0$ by the anticommutation relations.  The
component expansion is given completely by just the $d$'s and not the
$Ðd$'s:
$$ Ä| = A,ââ(d_Œ Ä)| = Æ_Œ,ââ(d^2 Ä)| = B $$
 where $A$ and $B$ are complex scalars, and we use the normalization
$$ d^2 = üd^Œ d_Œ $$
 All other components are $x$-derivatives of these, since the $Ðd$'s can be
pushed past the $d$'s (producing $p$'s) until they annihilate $Ä$.  Another
way to state this is to use the fact
$$ d_Œ = e^{-U/2}»_Œ e^{U/2},ââÐd_{ÀŒ} = e^{U/2}л_{ÀŒ}e^{-U/2};ââ
	U = Ï^Œ ÐÏ^{Àº}p_{ŒÀº} $$
 to solve the chirality constraint as
$$ Ä(x,Ï,ÐÏ) = e^{U/2}öÄ(x,Ï) $$
 where $öÄ$ is independent of $ÐÏ$:  It is defined on ``chiral superspace". 
(In this equation $U$ generates a ÓcomplexÕ coordinate transformation.)
Another way to solve the chirality constraint is to use the covariant
derivatives:   Since $d_Œ d_º d_©=0$ by anticommutativity (and similarly
for $Ðd$'s),
$$ Ðd_{ÀŒ}Ä = 0âÜâÄ = Ðd^2 Æ $$
 where $Æ$ is a ``general" (unconstrained) complex superfield.  It is the
``prepotential" for the field $Ä$.

\x IVC1.1  Let's analyze the supersymmetry generators $q_Œ,Ðq_{ÀŒ}$ 
in this case.
 ªa Find similar expressions for $q,Ðq$ in terms of $e^{U/2}$.
 ªb Find $q,Ðq$ on $öÄ$ in terms of just $Ï$ and $»/»Ï$ (no $ÐÏ$ nor
$»/»ÐÏ$).

\x IVC1.2  Show that the prepotential has a gauge invariance,
under which $Ä$ is invariant.
(Hint:  Use the same identity that led to the prepotential.)

From the anticommutation relations we find
$$ [Ðd^{ÀŒ},d^2] = p^{ºÀŒ}d_º $$
 Since this must vanish on $Ä$, we find
$$ d_Œ d^2 Ä = Ðd_{ÀŒ}d^2 Ä = 0âÜâp_{ŒÀº}d^2 Ä = 0âÜâ
	d^2 Ä = constant $$
 (We can safely ignore this constant, at least when considering the free
theory:  It corresponds to a term in the action linear in the fields.)  This
field equation, together with the chirality constraint, is sufficient to
determine the theory:  $A$ is the usual free (complex) scalar, $Æ_Œ$ is
the usual free spinor, and $B$ is a constant.

To describe interactions of this (``scalar") multiplet, we keep the
chirality condition, since that greatly simplifies the field content of the
superfield.  In fact, this is clearly the simplest off-shell superfield we can
define, since it already has the smallest number of fermions (as do the
coordinates of chiral superspace).  (``Off shell" means all components
less gauge degrees of freedom.)  This means that the equation $d^2 Ä=0$
will be generalized, since it implies the Klein-Gordon equation.  The
simplest way to do this is by constructing an explicit action, our next
topic.

Ü2. Actions

The construction of actions in superspace is different from ordinary
theories because the geometrically simple objects, the potentials, are
constrained, while the unconstrained objects, the prepotentials, can be
awkward to work with directly.  (This problem is magnified with
extended supersymmetry, whose actions we don't consider here.)

We start with the simplest supermultiplet, the chiral superfield.  Since
chiral superfields are defined on chiral superspace, a natural
generalization of a potential (nonderivative) term in the action to
superspace is
$$ S_1 = Çdx¼d^2 ϼf(Ä) +h.c. $$
 in terms of some function (not functional) $f$ of chiral superfields $Ä$
(the ``superpotential").  We can ignore any $ÐÏ$ dependence because it
contributes only total derivatives:
$$ Ä = e^{U/2}öÄ(x,Ï)âÜâf(Ä) = e^{U/2}f(öÄ) $$
 Integration over $Ï$ is defined as in subsection IA2; however, now we
can replace partial derivatives with covariant ones, since the
modification is again only by total derivatives:
$$ Çdx¼d^2 Ï = Çdx¼d^2 $$
 with an appropriate normalization.  This turns out to be the most
convenient one, since it allows covariant manipulations of the action, and
the $Ï$ integration can be performed covariantly:  Since we know that
the result of $Ï$ integration gives a Lagrangian that depends only on $x$,
up to total derivative terms, we can evaluate it as
$$ Çdx¼d^2 ϼf(Ä) = Çdx¼[d^2 f(Ä)]| $$
$$ = Çdx¼[f'(Ä|)(d^2 Ä)| +f''(Ä|)ü(d^Œ Ä)|(d_Œ Ä)|] $$
 (suppressing indices on multiple $Ä$'s).  This gives the result directly in
terms of component fields, using the covariant method of defining the
component expansion:  In the conventions of the previous subsection,
this part of the action becomes
$$ Çdx¼[f'(A)B +f''(A)üÆ^Œ Æ_Œ] $$

We now consider integration over the full superspace.  As a
generalization of the above, we can write
$$ Çdx¼d^4 ϼK(Ä,ÐÄ) = Çdx¼(d^2 Ðd^2 K)| $$
 Supersymmetric versions of nonlinear $§$ models can be written in this
way; here we consider just the case where $K$ is quadratic, which is the
one interesting for quantum theory.  Since a function of just $Ä$ (or
just $ÐÄ$) will give zero in the $d^4 Ï$ integral, we choose 
$$ K = -ÐÄÄâÜâS_0 = -Çdx¼d^4 ϼÐÄÄ $$ 
 Explicitly,
$$ L_0 = d^2 Ðd^2(- ÐÄÄ) = -d^2 (Ðd^2 ÐÄ)Ä 
	= -(üõÐÄ)Ä +(i»^{ŒÀº}Ðd_{Àº}ÐÄ)d_Œ Ä -(Ðd^2 ÐÄ)(d^2 Ä) $$
$$ £ -AüõÐA +Æ^Œ i»_Œ{}^{Àº}ÐÆ_{Àº} -BÐB $$
 where we have used the commutation relations of the covariant
derivatives to push all $d$'s past $Ðd$'s to hit $ÐÄ$.  Clearly, this term by
itself reproduces the results derived in the previous subsection based on
kinematics, so it is the desired massless kinetic term.    

We can now see the influence of adding the superpotential term to the
action:  The result of combining the two terms, and then eliminating the
auxiliary field $B$ by its equation of motion, is
$$ S_0 + S_1 £
	L = -AüõÐA + Æ^Œ i»_Œ{}^{Àº}ÐÆ_{Àº} +|f'(A)|^2 +[f''(A)üÆ^Œ Æ_Œ +h.c.] $$
 For example, a quadratic $f$ gives mass to the physical scalar and
spinor.  This action is invariant under modified supersymmetry
transformations, where the auxiliary fields are replaced by their
equations of motion there also; those transformations then become
nonlinear in the presence of interactions.  Note that the scalar potential
is positive definite; this is a consequence of supersymmetry, since it
implies that the energy is always positive.

\x IVC2.1  These results generalize straightforwardly:
 ªa Find the explicit form of the component-field action for
arbitrary $K(Ä^i,ÐÄ_i)$ and $f(Ä^i)$ for an arbitrary number of chiral
superfields $Ä^i$, including all indices.  
 ªb Eliminate the auxiliary fields from
the action, and find the modified supersymmetry transformations.  
 ªc Show by direct evaluation that the action is still invariant.

As a notational convenience, we can drop the ``|" after expanding a
superspace action in components:  For example, we can write simply
$$ Æ_Œ = d_Œ Ä,ââB = d^2 Ä $$
 After performing the $Ï$-integration as above by using derivatives $d$
and $Ðd$, and then ``evaluating" these derivatives on $Ä$ by writing $Æ$
and $B$, the component action is expressed completely in terms of such
superfields and only ÓspacetimeÕ derivatives $»_{ŒÀº}$.  This component
action is independent of $Ï$ (the Lagrangian is independent up to total
spacetime derivatives):  This is the statement of supersymmetry
invariance.  Thus, we can choose to evaluate at $Ï=0$, or $Ï=·$, or
whatever; it is irrelevant.  It is then understood that the relation to the
usual component actions is simply to treat the superfield as a component
field, since the $Ï$-derivatives (in $d$ and $ÇdÏ$) have been eliminated. 
From now on we will generally drop the |'s.

The above results can also be derived from the superfield 
equations of motion by varying the
action.  Since $Ä$ is constrained, it can't be varied arbitrarily; 
we vary instead the prepotential $Æ$ ($Ä=Ðd^2 Æ$).
For example, we find $d^2 Ä=0$ (and the
complex conjugate) from the free action.
Effectively, 
since chiral superfields are essentially independent of $ÐÏ$, not only
integration is modified, but also (functional) variation.  Since a chiral
superfield is (up to a transformation) an arbitrary function on chiral
superspace, we define
$$ ¶S[Ä] = Çdx¼d^2 ϼ(¶Ä){¶S\over ¶Ä} $$
 for an arbitrary variation of a chiral superfield $Ä$, and similarly for
varying $ÐÄ$.  In evaluating such variations, we make use of the identities
$$ Çdx¼d^4 ϼL = Çdx¼d^2 ϼÐd^2 L $$
$$ Ðd^2 d^2Ä = üõÄââ( Ðd^2 d^2 Ðd^2 = üõÐd^2 ) $$
 Thus, to vary a general action, it is convenient to first integrate over $ÐÏ$,
and then vary in the naive way:  For example,
$$ S = -Çdx¼d^4 ϼÐÄÄ +\left[ Çdx¼d^2 ϼf(Ä) +h.c. \right] $$
$$ Üâ0 = {¶S\over ¶Ä} = -Ðd^2 ÐÄ +f'(Ä) $$

\x IVC2.2  Check for this action that the component expansion of the
superfield equations of motion agree with the variation of the
corresponding component action.

Ü3. Covariant derivatives

The supersymmetric generalization of nonabelian gauge theories can be
derived by similar methods.  We first write the supersymmetry covariant
derivatives collectively as
$$ d_A = (d_Œ, Ðd_{ÀŒ}, »_{ŒÀŒ}) = E_A{}^M »_M $$
$$ »_M = (»_µ,»_{Àµ},»_m) = »/»z^M,ââz^M = (Ï^µ,ÐÏ^{Àµ},x^m) $$
 Unlike the nonsupersymmetric case, the ``vielbein" $E_A{}^M$ has $Ï$
dependence even in ``flat" superspace, and thus the ``torsion" $T$ is
nonvanishing:
$$ [d_A,d_BÕ = T_{AB}{}^C d_C $$
$$ T_{ŒÀº}{}^{©À©} = T_{ÀºŒ}{}^{©À©} = -i¶_Œ^© ¶_{Àº}^{À©},ârest = 0 $$

We now gauge-covariantize all the supersymmetry-covariant derivatives:
$$ á_A = d_A +iA_A $$
 The covariant field strengths are then defined as
$$ [á_A,á_BÕ = T_{AB}{}^C á_C +iF_{AB} $$
 From our analysis of general representations of supersymmetry in D=4 in
subsection IIC5, we know that the simplest supersymmetrization of
Yang-Mills is to include a spinor with the vector, in terms of physical
degrees of freedom.  (The spinor and vector each have two physical
degrees of freedom, one for each sign of the helicity.)  Off shell, Fermi
and Bose components must still balance, so there must also be an auxiliary
scalar.  From dimensional analysis, the field strengths must therefore
satisfy
$$ F_{Œº} = F_{ÀŒÀº} = F_{ŒÀº} = 0;ââ
	F_{Œ,ºÀº} = -iC_{Œº}ÑW_{Àº},âF_{ÀŒ,ºÀº} = -iC_{ÀŒÀº}W_º $$
 where $W_Œ|$ is the physical spinor.

The constant piece of the torsion implies stronger relations among the
field strengths than in nonsupersymmetric theories.  For super Yang-Mills
we find from the Jacobi identity for the covariant derivatives the Bianchi
identity for the field strengths
$$ á_{[A}F_{BC)} = T_{[AB|}{}^D F_{D|C)} $$
 Specifically, the dimension-1 constraints above on the field strengths
imply the dimen\-sion-3/2 algebraic constraint that defines $W_Œ$, as
well as
$$ F_{ŒÀŒ,ºÀº} = C_{Œº}üá_{(ÀŒ}ÑW_{Àº)} + C_{ÀŒÀº}üá_{(Œ}W_{º)} $$
 They also imply that $W_Œ$ is covariantly chiral and satisfies a
``reality" condition,
$$ Ñá_{ÀŒ}W_º = 0,ââá^Œ W_Œ + á^{ÀŒ}W_{ÀŒ} = 0 $$

The most straightforward way to derive these results is to just evaluate
the Jacobi identities directly.  We begin with a weaker set of conditions,
both of dimension 1, that will be found (in the following subsection) to be
necessary and sufficient for solving explicitly.  One directly determines
the vector derivative in terms of the spinor ones:
$$ F_{ŒÀº} = 0âÜâ-iá_{ŒÀŒ} = Óá_Œ,Ñá_{ÀŒ}Õ $$
 Since one could always define the vector covariant derivative this way,
imposing this condition simply eliminates redundant degrees of freedom.  

The remaining constraint (including its complex conjugate) allows
coupling of super Yang-Mills to the chiral superfield:
$$ Ñá_{ÀŒ}Ä = 0âÜâ0 = ÓÑá_{ÀŒ},Ñá_{Àº}ÕÄ = iF_{ÀŒÀº}Ä $$
 It also implies the maintenance of certain free identities, such as
$$ á_Œ á_º = ü[á_Œ,á_º] +üÓá_Œ,á_ºÕ = C_{ºŒ}á^2 $$
 (Such constraints appear also for first quantization, e.g., in superstring
theory, whenever a supersymmetric system is put in a background of a
supersymmetric gauge field of higher superspin.  This should not be
confused with background Ófield equationsÕ imposed by any gauge system
put in a background of the ÓsameÕ type: see subsection VIB8.)

Thus, our minimal set of constraints can be written directly in terms of
the field strengths as
$$ F_{Œº} = F_{ÀŒÀº} = F_{ŒÀº} = 0 $$
 but for our purposes it will prove more convenient to write them directly
as (anti)com\-mutators:
$$ Óá_Œ,á_ºÕ = ÓÑá_{ÀŒ},Ñá_{Àº}Õ = 0,ââÓá_Œ,Ñá_{Àº}Õ = -iá_{ŒÀº} $$

 The solution to the dimension-3/2 Jacobis are then
$$ [á_{(Œ},Óá_{º},á_{©)}Õ] = 0âÜâtrivial $$
$$ [á_{(Œ},Óá_{º)},Ñá_{À©}Õ] +[Ñá_{À©},Óá_Œ,á_ºÕ] = 0âÜâ
	[á_Œ,á_{ºÀ©}] = C_{Œº}ÑW_{À©} $$
 for some field $W$, simply applying the constraints to drop $Óá_Œ,á_ºÕ$
and replace $Óá_Œ,Ñá_{Àº}Õ$ with $á_{ŒÀº}$.  Similarly, we find from the
dimension-2 Jacobis
$$ Óá_{(Œ},[á_{º)},á_{©À©}]Õ +[á_{©À©},Óá_Œ,á_ºÕ] = 0âÜâá_Œ ÑW_{À©} = 0 $$
$$ [á_{ŒÀŒ},Óá_º,Ñá_{À©}Õ]+Óá_º,[Ñá_{À©},á_{ŒÀŒ}]Õ+ÓÑá_{À©},[á_º,á_{ŒÀŒ}]Õ=0$$
$$ Üâ[á_{ŒÀŒ},á_{ºÀº}] = i(C_{Œº}Ðf_{ÀŒÀº}+C_{ÀŒÀº}f_{Œº}),ââ
	f_{Œº} = üá_{(Œ}W_{º)},ââá^Œ W_Œ +Ñá{}^{ÀŒ}ÑW_{ÀŒ} = 0 $$
 where we separated the last equation into its (Lorentz) irreducible
pieces.  (The dimension-5/2 and 3 identities are redundant.)

\x IVC3.1  Explicitly evaluate all the remaining Jacobi identities, and show
that they imply no further conditions on $W_Œ$.

Ü4. Prepotential

We saw in the previous subsection that coupling super Yang-Mills to
matter gave directly one of the minimal constraints on the super
Yang-Mills fields themselves.  Hence, as for ordinary Yang-Mills, the
definition of the gauge theory follows from considering the
transformation of matter, and generalizing it to a local symmetry.  As for
self-dual Yang-Mills (see subsection IIIC5), the vanishing of some field
strengths implies that part of the covariant derivative is pure gauge:
$$ Óá_Œ,á_ºÕ = 0âÜâá_Œ = e^{-¯}d_Œ e^¯ $$
 However, since $Óá_Œ,Ñá_{Àº}Õ±0$, this gauge transformation $¯$
(``prepotential") is complex.  We therefore have the covariantly chiral
superfield
$$ Ñá_{ÀŒ}Ä = 0,âÑá_{ÀŒ} = e^{Я}d_{ÀŒ}e^{-Я}âÜâ
	Ä = e^{Я}öÄ,âd_{ÀŒ}öÄ = 0 $$
 Alternatively, we could combine this exponential with that already
contained in the free spinor derivative:
$$ á_Œ = e^{-U/2-¯}»_Œ e^{U/2+¯},âÄ = e^{U/2+Я}öÄ,â»_{ÀŒ}öÄ = 0 $$
 $U+2¯$ is the analog of the covariant derivative for the Yang-Mills
prepotential.  This is a hint at supergravity:  $U$ is just the flat piece of
the supergravity prepotential.  We thus see that supersymmetry
automatically gives gravity the interpretation of the gauge theory of
translations.

Component expansions are now defined with Yang-Mills-covariant
derivatives:
$$ á_Œ Ä = Æ_Œ,ââá^2 Ä = B $$
$$ á_Œ W_º = f_{Œº} +iC_{Œº}D,ââ
	á^2 W_Œ = -iá_Œ{}^{Àº}ÑW_{Àº} $$
 where we have used the Bianchi identities for $W$, and $f_{μ}$ (not to
be confused with $F_{μ}$) is the usual Yang-Mills field strength (in
spinor notation).  The ``vector multiplet" thus consists of the component
fields $A_a$ (the gauge field whose strength is $f$), $W_Œ$, and $D$
(auxiliary).  (As explained earlier, we drop all |'s.)

Note that the gauge parameter is real, while the matter multiplet is
(covariantly) chiral.  The resolution of this apparent inconsistency is that
solving the constraints introduces a new gauge invariance,
just as solving the source-free half of Maxwell's equations 
(really just constraints, not field equations) introduces the potential
and its gauge invariance:
$$ á_A' = e^{iK}á_A e^{-iK},âá_Œ = e^{-¯}d_Œ e^¯âÜâ
	e^{¯'} = e^{iÐñ}e^¯ e^{-iK},âd_Œ Ðñ = 0 $$
$$ Ä' = e^{iK}Ä,âÄ = e^{Я}öÄâÜâöÄ' = e^{iñ}öÄ $$
 This suggests the definition of a new (``chiral") representation, where we
use the obvious field $öÄ$ and the chiral gauge parameter $ñ$ replaces
the real one $K$:  Making a nonunitary similarity transformation,
$$ ßá_A = e^{-Я}á_Ae^{Я}âÜâß{Ñá}_{ÀŒ} = d_{ÀŒ},âßá_Œ = e^{-V}d_Œ e^V,â
	e^V = e^¯ e^{Я} $$
$$ öÄ = e^{-Я}Ä,âö{ÐÄ} = ÐÄe^{Я}âÜâd_{ÀŒ}öÄ = 0,âö{ÐÄ} = (öÄ)ÿe^V $$
$$ ßá_A' = e^{iñ}ßá_A e^{-iñ},âe^{V'} = e^{iÐñ}e^V e^{-iñ} $$
 Alternatively, we can also include $U$ in the transformation as above;
then $U$ and $V$ appear only in the combination $U+V$.

\x IVC4.1  Show that the explicit expression for the field strength $W_Œ$
in terms of the prepotential $V$ in the chiral representation is
$$ W_Œ = -iÐd^2(e^{-V}d_Œ e^V) $$
 Show this expression is chiral.

\x IVC4.2  In the Abelian case, give an explicit component expansion of the
prepotential $V$, such that the vector potential $A_a$, the physical spinor
$W_Œ$, and the auxiliary field $D$ appear as independent components. 
Note that the other components do not appear explicitly in component
expansions when gauge-covariant expansion ($á...|$) is used.  The
component (nonsupersymmetric) gauge where these components are set
to vanish is the ``Wess-Zumino gauge", and is the $Ï$ part of the radial
gauge of subsection VIB1 below.

\x IVC4.3  For some purposes (like quantization) we need the explicit
form of an infinitesimal gauge transformation of $V$.  Show this can be
written as
$$ ¶V = -i\L_{V/2}[(ñ+Ðñ) +coth(\L_{V/2})(ñ-Ðñ)] $$
 (Hint:  Consider $e^{-V}¶e^V$, and think of $¶$ as an operator, as for the
expansion of $á_Œ=e^{-V}d_Œ e^V$.  $\L_A$ was defined in subsection
IA3.)

Ü5. Gauge actions

Generalization of actions to super Yang-Mills theory is straightforward. 
Matter coupling is achieved simply by replacing the chiral superfields of
the matter multiplets with Yang-Mills-covariantly chiral superfields.  The
coupling can be seen explicitly in the chiral representation:  In the kinetic
term,
$$ ÐÄÄ = (öÄ)ÿe^V öÄ $$
 while in the $Çd^2 Ï$ term all $V$-dependence drops out because of
gauge invariance.  (The superpotential is a gauge invariant function of the
$Ä$'s, and the transformation to the chiral representation is a complex
gauge transformation.  The fact that the gauge transformation is complex
is irrelevant, since the superpotential depends only on $Ä$ and not $ÐÄ$.) 
Component expansion can be performed covariantly by replacing $d$'s
with $á$'s in the definition of $Ï$ integration:  Since the Lagrangian is a
gauge singlet, this is the same acting on it, although individual terms in
the expansion differ because the fields are not singlets.  Similarly, $Ðd^2$
can be replaced with $Ñá^2$ also when performing $ÐÏ$ integration for
purposes of varying an action with respect to a chiral superfield.  This is
equivalent to gauge covariantizing the functional derivative (e.g., by
transforming from a chiral representation) as
$$ {¶Ä(x,Ï)\over ¶Ä(x',Ï')} = Ñá^2 ¶(x-x')¶^4(Ï-Ï') $$
 Usually we will drop the ``$ß{\phantom m}$"'s on $Ä$ and $ÐÄ$, when the
representation is clear from the context by the use of explicit $V$'s.

The action for super Yang-Mills itself follows from dimensional analysis: 
Since each $Ï$ integral is really a $Ï$ derivative, $d^2 Ï$ integration has
mass dimension +1, the same as a spacetime derivative.  Since the
Lagrangian for a physical spinor, in this case $W^Œ$, has a single such
derivative, dimensional analysis says the action must be
$$ S_{sYM} =  -\f1{g^2}tr Çdx¼d^2 ϼüW^Œ W_Œ $$
 where the (covariant) chirality of $W^Œ$ allows integration over chiral
superspace.  (Similar analysis applies to the matter multiplet, where
$Çd^4 Ï$ takes the place of a $õ$ for the scalar $Ä$.)  Replacing 
$Çd^2 Ï£á^2$, we evaluate the component expansion as
$$ S_{sYM} = \f1{g^2}tr ÇdxÊ(üf^{Œº}f_{Œº} +W^Œ iá_Œ{}^{Àº}ÑW_{Àº} -D^2) $$
  Another term we
can write, for superelectromagnetism (supersymmetrization of an Abelian
gauge theory) is the ``Fayet-Iliopoulos term"
$$ S_{FI} = ½Çdx¼d^4 ϼV = ½Çdx¼D $$
 which involves only the auxiliary field $D$.  (The analog for the chiral
scalar superfield is $Çdx¼d^2 ϼÄ$.)

\x IVC5.1  Derive the supersymmetric analog of the St¬uckelberg model of
subsection IVA5, by coupling an Abelian vector multiplet to a massless
chiral scalar multiplet using the symmetry generator $T$ defined there. 
($G£-iT$ in transformation laws, covariant derivatives, etc., on $Ä$,
where $TÄ=1ÜT^2 Ä=0$.) 
 ªa  To couple the gauge field it is necessary to start, as usual, with a
(quadratic) matter action that is ÓgloballyÕ invariant under this symmetry:
$$ S_0 = Çdx¼d^4 ϼü(Ä-ÐÄ)^2 $$
 (At this point this is the usual, since only the cross-term survives, but
this will not be the case for the covariantly chiral superfields.)  Find the
supersymmetric gauge coupling, and express the resulting action in terms
of $V$ and $öÄ$.
 ªb  Use this result to find the mass term for $V$ in the gauge $öÄ=0$.

Another interesting form of the action uses a generalization of the
Chern-Simons form defined in the discussion of instantons in subsection
IIIC6.  In superspace, the calculation of the field strength with curved
indices is modified to
$$ á_M = E_M{}^A á_A = »_M +iA_M,â
	-i[á_M,á_NÕ = F_{MN} = E_M{}^A E_N{}^B F_{AB} $$
 where we have left sign factors from index reordering in the last
equation implicit.  Although the curved-index expressions are not as
useful (for example, for seeing which components vanish by constraints),
we can see easily that some arguments used in nonsupersymmetric
theories carry over to superspace.  Thus, we can define the super
Chern-Simons form by
$$ \f18 tr¼F_{[MN}F_{PQ)} = \f16 »_{[M}B_{NPQ)} $$
$$ B_{MNP} = tr(üA_{[M}»_N A_{P)} +i\f13 A_{[M}A_N A_{P)}) $$
 Converting to flat superspace (again with some implicit sign factors),
$$ B_{ABC} = E_A{}^M E_B{}^N E_C{}^P B_{MNP} =
	tr(üA_{[A}d_B A_{C)} -\f14 A_{[A}T_{BC)}{}^D A_D 
		+i\f13 A_{[A}A_B A_{C)}) $$
 In terms of this expression, the super Yang-Mills action can be written
simply in terms of the spinor-spinor-vector part $B_{ŒÀºc}$ of $B_{ABC}$ as
$$ S_{sYM,1} = üi\f1{g^2}tr Çdx¼d^4 ϼB_{Œ,ÀŒ}{}^{ŒÀŒ} $$

Note that the fact that the curl of $B$ is gauge invariant implies that $B$
transforms under a gauge transformation as the curl of something, and
thus the integral of any part of $B$ is gauge invariant (up to possible
torsion terms:  see the exercise below).  Furthermore, we can drop the
$F_{ŒÀº}=0$ constraint on the $A$ in this action; it follows from variation
with respect to $A_{ŒÀº}$.  One simple way to check this action is to use
the chiral representation $A_{ÀŒ}=0$:  Then only the
$A^{ŒÀº}\onªd_{Àº}A_Œ$ and $(A_{ŒÀº})^2$ terms contribute, and
$A_{ŒÀº}=iÐd_{Àº}A_Œ$, while  $W_Œ=Ðd^2 A_Œ$, so $Çd^2 ÐÏ$ integration gives
$-Çdx¼d^2 ϼW^2$.

\x IVC5.2  Derive the expression for $B_{ABC}$ directly using only flat
indices:  
ªa  Start with $F_{[AB}F_{CD)}$ expressed in terms of $T$ and $A$,
and write it as a total derivative plus torsion terms.
ªb  Do the same for the gauge transformation of $B$.  Show that the
torsion terms do not contribute to $¶B^{ŒÀŒ,}{}_{Œ,ÀŒ}$.

The multiplets and couplings we have considered are sufficient to write a
supersymmetric generalization of the Standard Model.  Unfortunately,
supersymmetry provides no unification.  To get the right symmetry
breaking, it turns out to be necessary to provide a supersymmetry
multiplet for each particle of the Standard Model:  The spin-1 gauge
bosons are accompanied by spin-1/2 ``gauginos" (``gluinos", ``photino",
``Wino", ``Zino"), the spin-1/2 leptons by spin-0 ``sleptons", the quarks
by ``squarks", and the spin-0 Higgs' by spin-1/2 ``Higgsinos". 
Furthermore, since a reality condition can't be imposed on chiral scalar
multiplets, the Higgs scalars are themselves doubled.  Ultimately, the
success of supersymmetry depends on the experimental detection of
these particles.

Ü6. Breaking

The methods of section IVA can be generalized straightforwardly to
supersymmetric theories:  Goldstone bosons and Higgs fields become
supermultiplets, etc.  However, to obtain realistic models supersymmetry
itself must be broken, since fermions and bosons with similar mass and
other properties are not observed in nature.  More specifically, since
gravity is observed, any supersymmetric theory of the world must include
supergravity, and thus the breaking must be spontaneous.  (Explicit
breaking would violate gauge invariance.)  Then the gravitino, which
gauges supersymmetry, will become massive by a superhiggs mechanism,
by eating a Goldstone fermion.  (See subsections XB6-7.  If the graviton
and gravitino are treated as composites, then this fermion could also be a
composite.)

We saw in subsection IIC1 that energy is always nonnegative in
supersymmetric theories.  In particular, from the same arguments
used there we see that a state can be invariant under supersymmetry
($q|ÆÔ=qÿ|ÆÔ=0$) if and only if it has zero energy.  Any such state can be
identified as the vacuum, since no state has lower energy.  This means
that the only way to guarantee spontaneous supersymmetry breaking is
to choose a theory which has no zero-energy state.  (Note that energy is
uniquely defined by the supersymmetry algebra; there is no possibility of
adding a constant as in nonsupersymmetric theories.)  In theories with
extended supersymmetry, the relation between supersymmetry and
energy applies for each supersymmetry; thus supersymmetry is either
completely broken spontaneously or completely unbroken.  (An exception
is central charges, which modify the supersymmetry algebra:  See the
following subsection.)

Furthermore, physical scalars appear at $Ï=0$ in matter multiplets, while
auxiliary fields appear at higher order.  Since supersymmetry breaking
requires $Ï$ dependence in a vacuum value of a superfield, this means an
auxiliary field must get a vacuum value.

A simple example of spontaneous supersymmetry breaking is the
O'Raifeartaigh model; it has the Lagrangian
$$ L_{O\mathchar`'\kern-1pt R} = -Çd^4 ϼÝ_{i=1}^3 Ðì_i ì_i 
	+ \left[ Çd^2 ϼÂ(½ì_1 +mì_2 ì_3 +ì_1 ì_2^2) +h.c. \right] $$
 To study symmetry breaking we ignore derivative terms, since vacuum
values are constants.  Then the scalar field equations are:
$$ {¶\over ¶B_i} £ -ÐB_i +»_i f = 0:â
	-ÐB_1 +½ +A_2^2 = -ÐB_2 +mA_3 +2A_1 A_2 = -ÐB_3 +mA_2 = 0 $$
$$ {¶\over ¶A_i} £ B_j »_i »_j f = 0:â
	2A_2 B_2 = mB_3 +2A_2 B_1 +2A_1 B_2 = mB_2 = 0 $$
 (where $»_i=»/»A_i$ on the superpotential $f(A)$).  Since there is no
solution for $B_i=0$, supersymmetry breaking is required.  In general, for
superpotential $f(ì)$, the field equations for $B=0$ are $f'(A)=0$, so a
linear term is always needed for supersymmetry breaking.

With Abelian vector multiplets, a Fayet-Iliopoulos term $Çd^4 ϼV$ can also
generate such breaking, since it also is a linear term of an auxiliary field.

\x IVC6.1  Evaluate the Lagrangian $-Çd^4 ϼÐÄÄ$ for covariantly chiral $Ä$
by using covariant $Ï$-integration, $Çd^4 Ï=á^2 Ñá^2$.  For the case of
U(1) gauge theory, add the action for the gauge superfield with a
Fayet-Iliopoulos term, and find the potential for the physical scalars by
eliminating the auxiliary field $D$ by its field equation.

For simplicity (as in this chapter), we may want to ignore supergravity;
however, we still need to take account of its contribution to breaking
global supersymmetry via the superhiggs effect.  The net low-energy
contribution from the supergravity fields (assuming no cosmological
constant is generated) is to introduce effective explicit supersymmetry
breaking:  Although the original theory is locally supersymmetric, we
neglect the supergravity fields but not their vacuum values (in particular,
those of the auxiliary fields).  In particular, if the supergravity fields are
bound states, then this procedure is essentially the classical introduction
of nonperturbative quantum effects.

Thus we consider adding terms to the classical action that break
supersymmetry explicitly.  The easiest way to do this is to introduce
constant superfields (``spurions"); this allows us to continue to take
advantage of the superspace formalism (at both the classical and
quantum levels).  Since we are neglecting (super)gravity, and in particular
its nonrenormalizability (see chapter VII), we consider only terms that
will preserve the quantum properties of the unbroken theories.  This will
clearly be the case if we consider only the usual terms, with some fields
replaced by spurions:  This is equivalent to using background (fixed)
fields, in addition to (but in the same way as) the usual field variables,
performing all (classical/quantum) calculations as usual, and then setting
the background fields (specifically, the auxiliary fields, which are
responsible for breaking supersymmetry) to constants.

Thus, introducing constant (in $x$) chiral and real spurion fields
$$ \Ä = Ï^2 c,ââ\V = Ï^2 ÐÏ^2 r $$
 in terms of complex and real parameters $c$ and $r$, in addition to the
true fields $Ä$ and $V$, we have terms of the form
$$ Çd^2 ϼ[\Ä Ä, \Ä Ä^2, \Ä Ä^3, \Ä W^2, (Ðd^2 d^Œ\V)ÄW_Œ ],
	ââÇd^2 ϼd^2 Ðϼ\V ÐÄe^V Ä $$ 
 (and complex conjugates).  These terms can preserve the usual gauge
invariances, and can be shown to also preserve the desirable quantum
properties of supersymmetry:  The condition is that replacing the spurion
field by 1 (instead of its above value) gives either 0 or a conventional
term (one with coupling constant of nonnegative mass dimension). 
Another way to introduce these spurions (except perhaps for the $ÄV$
crossterm, which is less useful) is as Ócoupling constantsÕ, rather than as
fields:  Instead of introducing new terms to the action, we generalize
the old ones, so the constant part of each coupling is the usual coupling,
while its $Ï$-dependent terms produce the breaking.

\x IVC6.2  Find the component expansions of the above explicit breaking
terms.  What are the mass dimensions of the constants $c$ and $r$ in the
various cases?

\x IVC6.3  Expand the Lagrangian
$$ L = -Çd^4 ϼÐÄÄ +\left[ Çd^2 ϼ(\f16 Ä^3 +\Ä Ä) +h.c.\right] $$
  in components.  Find the masses.

Ü7. Extended

The supersymmetry we discussed earlier in this chapter, with a single
spinor coordinate, is called ``simple (N=1) supersymmetry"; the
generalization to many spinors is called ``extended (N>1)
supersymmetry" (for N spinor coordinates).  N=1 supersymmetric
theories, at least for spins$²$1, are most conveniently described by
superspace methods.  (There are also some definite advantages for N=1
supergravity at the quantum level.)  On the other hand, the technical
difficulties of extended superspace often outweigh the advantages.  (The
main advantage of extended superspace is proving certain properties of
the quantum theories.  Of course, extended supersymmetric theories are
complicated in any case.)  Alternative formulations of extended
supersymmetry are either  
\item{(1)} on shell,  
\item{(2)} in terms of components
(ordinary spacetime, not superspace), or  
\item{(3)} in simple superspace
(manifesting only one of the supersymmetries).

By going half way, using N=1 superfields to describe extended
supersymmetry, some of the advantages of the superspace approach can
be retained.  In this subsection we will list some of the extended
supersymmetric actions for lower spins in N=1 superspace form.  These
actions can be obtained by: (1) using extended superspace to derive the
component field equations (usually using dimensional reduction: see
subsections XC5-6), and combining components into N=1 superfields, or (2)
writing the extra supersymmetries in N=1 superspace form, and using
them to determine the action.

The simplest example is N=2 supersymmetry.  As for any extended
supersymmetry, its algebra can be modified by including Abelian
generators $Z$ (with dimensions of mass), called ``central charges":
$$ Óq_{iŒ},Ðq^j_{Àº}Õ = ¶_i^j p_{ŒÀº},âÓq_{iŒ},q_{jº}Õ = C_{Œº}C_{ij}Z,â
	ÓÐq^i_{ÀŒ},Ðq^j_{Àº}Õ = C_{ÀŒÀº}C^{ij}Z;â[Z,q] = [Z,Ðq] = 0 $$
 (where $i=1,2$).  In terms of dimensional reduction (for N=2, from D=5 or
6; see subsections XC5-6), the origin of these generators can be
understood as the higher-dimensional components of the momentum.  N=2
supersymmetry is sometimes called ``hypersymmetry", and N=2
supermultiplets, ``hypermultiplets".

Our first example is the free, massive N=2 scalar multiplet: Since we
already know the field content (see subsection IIC5), it's easy to write
the free Lagrangian
$$ L_{sm,N=2} = -Çd^4 ϼÐÄ^{i'}Ä_{i'} 
	+ü\left(Çd^2 ϼm^{i'j'}Ä_{i'}Ä_{j'} +h.c.\right) $$
 where the index ``$i'$" is for an extra SU(2) (not the one acting on the
supersymmetry generators), broken by the mass term, and the mass
matrix $m^{i'j'}$ is symmetric while $m_{i'}{}^{j'}=C_{k'i'}m^{k'j'}$ is
hermitian.  In other words, it represents a 3-vector of this SU(2), and thus
a generator of the preserved U(1) subgroup, which we have used to define
the central charge:
$$ ZÄ_{i'} = m_{i'}{}^{j'}Ä_{j'} $$

The other N=2 multiplet of low spin is the vector multiplet.  It also has a
simple Lagrangian,
$$ L_{sYM,N=2} = -\f1{g^2}tr\left( Çd^2 ϼW^2 +Çd^4 ϼÐÄÄ \right) $$
 where $Ä$ is covariantly chiral and in the adjoint representation of the
Yang-Mills gauge group.  In the Abelian case, we can also add an N=2
Fayet-Iliopoulos term,
$$ L_{FI,N=2} = Çd^4 ϼ½_0 V +\left( Çd^2 ϼ½_+ Ä +h.c. \right) $$
 where $(½_0,½_+,½_-)$ ($½_-=½_+*$) is a constant 3-vector of the SU(2) of
the N=2 supersymmetry:  The 3 scalar auxiliary fields of this N=2
multiplet form a 3-vector of the SU(2).  Unlike the previous example, this
multiplet has all the auxiliary fields needed for an off-shell N=2
superspace formulation:  Not only do the physical components balance
between bosons and fermions (4 of each), but also the auxiliary ones
(also 4 of each).

These 2 N=2 multiplets can be coupled:  The scalar multiplet action is
modified to
$$ L_{sm,N=2} = -Çd^4 ϼÐÄ^{i'}Ä_{i'} 
	+ü\left[Çd^2 ϼ ^{i'j'}Ä_{i'}(Ä+M)Ä_{j'} +h.c.\right] $$
 where now $Ä_{i'}$ is also a representation of the Yang-Mills group (not
necessarily adjoint), with respect to which it is covariantly chiral. 
However, the same SU(2) matrix $ $ that appears in the mass matrix
$m_{i'j'}=M _{i'j'}$ now also appears with the N=2 super Yang-Mills fields,
$$ á_A Ä_{i'} = d_A Ä_{i'} +iA_A^n G_n  _{i'}{}^{j'}Ä_{j'},ââ
	Ä = Ä^n G_n $$
 where $G_n$ are the usual Yang-Mills group generators.  (Without loss of
generality, we can choose $ _{i'}{}^{j'}=\tbt100{-1}$; then $Ä_{+'}$ is some
arbitrary representation of the Yang-Mills group, while $Ä_{-'}$ is the
complex conjugate.)  Note that the mass term appears in exactly the
same way as an Abelian N=2 vector multiplet that has been replaced by a
vacuum value for its physical scalars.  This can also be seen from the
commutation relations for the N=2 super Yang-Mills covariant derivatives
(see below), since the scalars appear in exactly the same way as the
central charge.

By our earlier helicity arguments, the only N=3 supersymmetric theory
with spins $²1$ is N=3 super Yang-Mills.  The analogous statement also
holds for N=4, while no such theories exist for N>4.  Since theories with
N supersymmetries are a subset of those with only N$-$1
supersymmetries, N=3 and N=4 super Yang-Mills must be the same:
Counting states of supersymmetry representations, we see that this
theory is the same as N=2 super Yang-Mills coupled to one N=2 scalar
multiplet in the adjoint representation (in direct analogy to N=2 super
Yang-Mills in terms of N=1 multiplets).  In terms of N=1 multiplets, this is
super Yang-Mills plus 3 adjoint scalar multiplets.  The action then follows
from the above results (without central charges and Fayet-Iliopoulos
terms):
$$ L_{sYM,N=4} = \f1{g^2}tr \left[ -Çd^2 ϼW^2 -Çd^4 ϼÐÄ^I Ä_I 
	+\left( Çd^2 ϼ\f16 ·^{IJK}Ä_I[Ä_J,Ä_K] +h.c. \right) \right] $$
 where ``$IÊ$" is a U(3) index.  (The U(1) part of the U(3) symmetry involves
also a phase transformation of the $Ï$'s.)

For comparison, here are the general (UV well-behaved) actions for
all numbers of supersymmetries (in D=4):
$$ L_{N=1} = -\f1{g^2}tr Çd^2 ϼüW^Œ W_Œ +½Çd^4 ϼV
	-Çd^4 ϼÐÄe^V Ä + \left[ Çd^2 ϼf(Ä) +h.c.\right] $$
$$ L_{N=2} = -\f1{g^2}tr\left( Çd^2 ϼW^2 +Çd^4 ϼe^{-V}ÐÄe^V Ä \right)
	+Çd^4 ϼ½_0 V +\left( Çd^2 ϼ½_+ Ä +h.c. \right) $$
$$ -Çd^4 ϼÐÄ^{i'}(e^{V })_{i'}{}^{j'}Ä_{j'} +ü\left[Çd^2 ϼ ^{i'j'}Ä_{i'}(Ä+M)Ä_{j'} +h.c.\right] $$
$$ L_{N=4} = \f1{g^2}tr \left[ -Çd^2 ϼW^2 -Çd^4 ϼe^{-V}ÐÄ^I e^V Ä_I 
	+\left( Çd^2 ϼ\f16 ·^{IJK}Ä_I[Ä_J,Ä_K] +h.c. \right) \right] $$
where we now use ordinary chiral superfields, making dependence on
$V$ explicit.

As for off-shell N=1 supersymmetry, much information on extended
supersymmetric gauge theories can be gained by examining the
properties of the covariant derivatives and their field strengths.  In fact,
this is more true in the extended case, where the ``obvious" constraints
often imply field equations (which is more than one would want for an
off-shell formulation).  The empty-space covariant derivatives are the
direct generalization of N=1:  Introducing N $Ï$'s as $Ï^{iŒ}$ (and
complex conjugate $ÐÏ_i{}^{ÀŒ}$), where ``$i$" is an N-valued index with as
much as a U(N) symmetry,
$$ d_A = (d_{iŒ},Ðd{}^i{}_{ÀŒ},»_{ŒÀŒ});ââ
	d_{iŒ} = »_{iŒ} -iüÐÏ_i{}^{ÀŒ}»_{ŒÀŒ},â
	Ðd{}^i{}_{ÀŒ} = л{}^i{}_{ÀŒ} -iüÏ^{iŒ}»_{ŒÀŒ} $$
$$ T_{iŒ,}{}^j{}_{Àº}{}^{©À©} = T{}^j{}_{Àº}{}_{,iŒ}{}^{©À©} 
	= -i¶_i^j ¶_Œ^© ¶_{Àº}^{À©},âârest = 0 $$

\x IVC7.1  Find the superspace representation of the extended
supersymmetry generators (which anticommute with these covariant
derivatives).  For N=2, include the central charge.

By definition, extended super Yang-Mills has only spins 1 and less. 
Dimensional analysis then gives the unique result, including physical fields
only,
$$ Óá_{iŒ},Ñá{}^j{}_{Àº}Õ = -¶_i^j iá_{ŒÀº} $$
$$ Óá_{iŒ},á_{jº}Õ = C_{ºŒ}iÐÄ_{ij} $$
$$ [Ñá{}^i{}_{ÀŒ},-iá_{ºÀº}] = C_{ÀºÀŒ}iW^i{}_º $$
$$ [á_{ŒÀŒ},á_{ºÀº}] = C_{Œº}iÐf_{ÀŒÀº} +C_{ÀŒÀº}if_{Œº} $$
 (and complex conjugates of some of these equations).  This corresponds
directly to our discussion in subsection IIC5, where we saw that a general
representation looked like antisymmetric tensors $Ä,Ä^i,Ä^{ij},...$ of
U(N), corresponding to helicities h, h$-$1/2, h$-$1,... .  In this case, h=1,
and these helicities come from the surviving on-shell components of
$f_{Œº},W^i{}_Œ,Ä^{ij},$... .  For N=4 we have self-duality with respect to
charge conjugation (see also subsection IIC5),
$$ Ä^{ij} = ü·^{ijkl}ÐÄ_{kl} $$

\x IVC7.2   Analyze the Bianchi identities of these covariant derivatives:
 ªa Show that for N>2 they imply the field equations.  
 ªb Find a component action that yields these field equations for N=4.

An interesting simplification of extended superspace occurs for
self-duality:  Constraining
$$ f^{Œº} = W^{iŒ} = Ä^{ij} = 0 $$
 and dropping the self-duality condition for N=4 (so $ÐÄ_{ij}±0$), we find
all commutators involving $Ñá^i{}_{ÀŒ}$ are trivial:
$$ Óá_{iŒ},Ñá{}^j{}_{Àº}Õ = -¶_i^j iá_{ŒÀº},ââ
	ÓÑá^i{}_{ÀŒ},Ñá^j{}_{Àº}Õ = [Ñá{}^i{}_{ÀŒ},á_{ºÀº}] = 0 $$
 while all the remaining commutators have a similar form:
$$ Óá_{iŒ},á_{jº}Õ = C_{ºŒ}iÐÄ_{ij},ââ[á_{iŒ},-iá_{ºÀº}] = C_{ºŒ}iÑW_{iÀº},ââ
	[á_{ŒÀŒ},á_{ºÀº}] = C_{Œº}iÐf_{ÀŒÀº} $$
 The latter result suggests we combine the internal and dotted spinor
indices as
$$ \A = (ÀŒ,i) $$
 so that we can combine the nontrivial equations as
$$ [á_{\A Œ},á_{\B º}Õ = iC_{Œº}f_{\A\B} $$
 The former equations then allow us to interpret the remaining
covariant derivatives $Ñá^i{}_{ÀŒ}$ as a subset of the SL(2|N) generators
that rotate the $\A$ index, which form a subgroup of the superconformal
group (S)SL(4|N).  We therefore restrict ourselves to the chiral superspace
described by the coordinates
$$ z^{\A Œ} = (x^{ŒÀŒ},Ï^{iŒ}) $$
 The net result is that we have a superspace with no torsion, with
coordinates that represent half of the supersymmetries as
translations and the other half as rotations.

By comparison with our treatment of the self-dual bosonic theory in
subsections IIIC5-7, we see that we can extend trivially all our results for
the bosonic case to the (extended) supersymmetric case by simply
extending the range of the indices.  In particular, we also have a chiral
twistor superspace:  Extending the range on the twistor coordinates
$z^{\A Œ}$ used there so $\A$ is now an SL(4|N) index, the superconformal
group is now manifest, and all the methods and results there (e.g., the
ADHM construction) apply automatically to the supersymmetric case.

\refs

£1 Wess and Zumino, Óloc. cit.Õ (IIC, ref. 2):\\
	free scalar and vector multiplets, Wess-Zumino gauge.
 £2 J. Wess and B. Zumino, \PL 49B (1974) 52:\\
	interacting chiral scalar multiplets.
 £3 J. Wess and B. Zumino, \NP 78 (1974) 1:\\
	coupling Abelian vector multiplet to scalar multiplets.
 £4 S. Ferrara and B. Zumino, \NP 79 (1974) 413;\\
	A. Salam and J. Strathdee, \PL 51B (1974) 102:\\
	nonabelian vector multiplets.
 £5 J. Wess and B. Zumino, \PL 66B (1977) 361;\\
	J. Wess, Supersymmetry-supergravity, in ÓTopics in quantum field
	theory and gauge theoriesÕ, VIII GIFT Int. Seminar on Theoretical
	Physics, Salamanca, Spain, Jun 13-19, 1977 (Springer-Verlag, 1978)
	p. 81:\\
	covariant derivatives for super Yang-Mills.
 £6 P. Fayet and J. Iliopoulos, \PL 51B (1974) 461.
 £7 S.J. Gates, Jr. and H. Nishino, ÓInt. J. Mod. Phys. AÕ É8 (1993) 3371:\\
	Chern-Simons superform.
 £8 L. O'Raifeartaigh, \NP 96 (1975) 331.
 £9 L. Girardello and M.T. Grisaru, \NP 194 (1982) 65:\\
	general soft explicit supersymmetry breaking, with spurions.
 £10 L.A. Avdeev, D.I. Kazakov, and I.N. Kondrashuk, 
	\xxxlink{hep-ph/9709397}, \NP 510 (1998) 289;\\
	N. Arkani-Hamed, G.F. Giudice, M.A. Luty, and R. Rattazzi,
	\xxxlink{hep-ph/9803290}, \PRD 58 (1998) 115005:\\
	spurions in couplings.
 £11 H.P. Nilles, ÓPhys. Rep.Õ É110 (1984) 1:\\
	review of supersymmetry phenomenology.
 £12 P. Fayet, \NP 113 (1976) 135:\\
	extended supersymmetry.
 £13 P. Fayet, \NP 149 (1979) 137:\\
	central charges.
 £14 F. Del Aguila, M. Dugan, B. Grinstein, L. Hall, G. Ross, and P. West,
	\NP 250 (1985) 225;\\
	L. Girardello, M. Porrati, and A. Zaffaroni, \xxxlink{hep-th/9704163},
	\NP 505 (1997) 272:\\
	semi-realistic N=2 models.
 £15 F. Gliozzi, J. Scherk, and D.I. Olive, \NP 122 (1977) 253;\\
	L. Brink, J.H. Schwarz, and J. Scherk, \NP 121 (1977) 77:\\
	N=4 super Yang-Mills.
 £16 R. Grimm, M. Sohnius, and J. Wess, \NP 133 (1978) 275;\\
	M.F. Sohnius, \NP 136 (1978) 461;\\
	E. Witten, \PL 77B (1978) 394:\\
	covariant super Yang-Mills derivatives for extended supersymmetry.
 £17 W. Siegel, \xxxlink{hep-th/9205075}, \PRD 46 (1992) R3235;
	\xxxlink{hep-th/9207043}, \PRD 47 (1993) 2504; 
	Óloc. cit.Õ (IIIB, ref. 3):\\
	conformal chiral superspace.
 £18 Gates, Grisaru, Ro×cek, and Siegel, Óloc. cit.Õ

\unrefs

\volume PART TWO: QUANTA

Many important new features show up in field theory at the quantum
level.  Probably the most important is ``renormalizability", which states
that all the parameters (masses and couplings) that appear as
coefficients of terms in the action must have nonnegative mass
dimension (when the massless part of the kinetic term has no
dimensionful coefficient).  Since the action is dimensionless, $Çd^4 x$
has dimension $-4$, and the fields have positive dimension, this allows
only a small number of terms for any given set of fields.  This one
condition gives relativistic quantum field theory more predictive power
than any known alternative.

There are many perturbation expansions that can be applied to quantum
field theory.  One is the mechanical JWKB expansion, which is an
expansion in derivatives.  Of the inherently field theoretical expansions,
the simplest is to expand directly in fields, or equivalently, in the coupling
constants.  This expansion is the basis of perturbative quantum field
theory.  However, this expansion does not preserve gauge invariance
term by term.  On the other hand, the terms in this expansion can be
collected into small subsets that do preserve gauge invariance.  There
are three such regroupings, discussed in the four following chapters,
and they are based on perturbation expansions:  
\item{(1)} the field theoretic JWKB (``loop") expansion,  
\item{(2)} expansions in spin or helicity, and  
\item{(3)} expansions in internal symmetry (color or flavor).

ÚV. QUANTIZATION

For the most part, integrals are hard to evaluate, in particular the path
integrals of exponentials that appear in quantum theory.  The only
exponentials that are generally easy to integrate are Gaussians, and the
products of them times polynomials, which can in turn be evaluated as
derivatives of Gaussians.  Such integrals are the basis of perturbation
theory:  We keep the quadratic part of the action, but Taylor expand the
exponential of higher-order terms.  Effectively, this means that we not
only expand in orders of $\h$ to perturb about the classical theory,
but also expand in orders of the coupling constants to perturb about the
free theory.  This makes particularly useful our analysis of relativistic
quantum mechanics (as free field theory).  The JWKB expansion for the
wave function (or S-matrix) expands the exponent in powers of $\h$,
dividing it onto three qualitatively different parts:  
\item{(1)} negative powers of
$\h$ (generally $1/\h$ only), which describe the classical theory (they
dominate the classical limit $\h£0$), whose physical implications have
been considered in previous chapters;  
\item{(2)} $\h$-independent, where
almost all of the important (perturbative) quantum features appear
(including topological ones, and quantum breaking of classical
symmetries); and  
\item{(3)} positive powers, which give more quantum
corrections, but little new physics, except when summed to all orders. 

\noindent
These are generally known as ``trees", ``one-loop", and ``multiloop",
because of their graphical interpretation.

Û5 A. GENERAL

In the Schr¬odinger approach to quantum mechanics one solves a
differential equation.  The Feynman approach is complementary:  There
one performs an integral.  Integrals are solutions to differential equations
(e.g., $f'=gÜf=Çg$), but usually differential equations are easier to solve
than integral equations.  However, there is an important exception: 
Gaussian integrals are easy, and so are their boundary conditions.  In field
theory the most important approximation is one where the integrand is
approximated as a Gaussian, and the exact integral is evaluated as a
perturbation about that Gaussian.  Of course, solving the
corresponding differential equation is also easy, but in that case the
integral is easier because it corresponds to working with the action,
while the differential equation corresponds to working with the field
equations.

A major advantage of Feynman's approach is that it allows space and
time to be treated on an equal footing.  For example, as in classical
electrodynamics, we can solve the wave equation inside a spacetime
volume in terms of conditions on the boundary of that volume:  It is not
necessary to choose the spatial boundary at infinity so that it can be
ignored, and divide the temporal boundary into its ``future" and ``past"
halves so that all conditions are ``initial" ones imposed at the past
boundary.  It is not even necessary to distinguish between preparation
(``if") and measurement (``then") when describing probabilities:  We can
instead ask the probability of a given wave function describing the whole
boundary.  This is a particular advantage for relativistic quantum field
theory, where space and time are more closely related than in
nonrelativistic theories.  We now ``review" Feynman's approach for
general quantum systems, and quantum mechanics in particular, so that it
can be applied without further explanation when we come to quantum
field theory.

Ü1. Path integrals

Before discussing the path integral approach to quantum mechanics,
we first review some features of quantum mechanics.  We can separate
the fundamentals of quantum mechanics into ``kinematics" and
``dynamics":  The kinematics are everything at a fixed time ---
Hilbert space, preparation/measurement, probability, observables.
The dynamics are the time development.  There are several ways to
describe time dependence of matrix elements; we will start with a 
general framework, then specialize.

Time dependence may be associated with either the states
(Schr¬odinger picture) or operators (Heisenberg picture).
We will be more explicit at first, taking all the time dependence
out of the states and operators and putting it into a
``time development operator" $U(t,t')$ that transforms the
Hilbert space from time $t'$ (earlier) to time $t$ (later).
For example, if we want to relate an earlier state to a later one
we evaluate $Òf|U(t,t')|iÔ$; more generally, we can look at things like
 $$ Òf|...\O_2 U(t_2,t_1) \O_1 U(t_1,t_i) |iÔ $$
 which means to prepare an initial state $|iÔ$ at time $t_i$, then
act with an operator $\O_1$ at time $t_1$, operator $\O_2$ at time
$t_2$, etc., and eventually measure the amplitude for a final state
$Òf|$.  

Now the dynamics can be described entirely through the properties of
$U$.  The general physical properties it must satisfy are
$$ \li{ causality¼(locality):â&âU(t_3,t_2)U(t_2,t_1) = U(t_3,t_1) \cr
	unitarity:â&âU(t_2,t_1)ÿU(t_2,t_1) = I \cr } $$
 Causality tells us that things happen
in chronological order:  Each event is determined by those
immediately preceding.
It is a kind of group property; in particular, from considering
$t_3=t_2$ we find that 
$$ U(t,t) = I $$
 We can then write
$$ U(t+·,t) ® I -i·H(t) $$
 by expanding in $·$, for some operator $H(t)$
that we call the Hamiltonian.
Again applying causality, we find
$$ \li{ »_t U(t,t') & = \lim_{·£0}{U(t+·,t') -U(t,t')\over ·} \cr
	& = \left( \lim_{·£0}{U(t+·,t) -I\over ·} \right) U(t,t') 
	= -iH(t) U(t,t') \cr} $$
 which is the Schr¬odinger equation for $U$.  Again applying causality
to build up the finite $U$ from products of the 
infinitesimal ones,
$$ U(t_f,t_i) = e^{-i·H(t_f-·)} ò e^{-i·H(t_i+·)}e^{-i·H(t_i)}
	­ \T \leftÓ exp \left[-iÇ_{t_i}^{t_f}dt¼H(t)\right]\rightÕ $$
 which defines the ``time-ordered product" $\T$.  Finally,
unitarity, another group property, tells us that probability
is conserved; in particular, from applying to $U(t+·,t)$,
$$ H(t)ÿ = H(t) $$
 The expression of $U$ in terms of a hermitian Hamiltonian
guarantees causality and unitarity.  (It ``solves" those
conditions.)  If $H$ is time independent and we have
a (orthonormal) basis of eigenstates of $H$, we can write
$$ H|IÔ = E_I|IÔâÜâU(t,t') = Ý_I |IÔÒI|e^{-i(t-t')E_I} $$

In Feynman's path integral approach to quantum mechanics (based on an
analogy of Dirac), the action is the starting point for quantization.  The
basic idea is to begin with the basic quantity in quantum mechanics, the
transition amplitude, and write it as an integral of the action
$$ Òf|iÔ = ÇDļe^{-iS[Ä]} $$
 where $ÇDÄ$ is a ``functional integral":  Integrate over $Ä(t)$ for each
$t$ (with some appropriate normalization).  The boundary conditions in
$t$ are defined by the choice of initial and final states.  In this subsection
we will define this integral in a more explicit way by breaking up the time
interval into discrete points and taking the continuum limit; in
subsection VA2 we will study ways to evaluate it using its general properties.

The path integral can be derived from the usual Hamiltonian operator
formalism.  Considering for simplicity a single coordinate $q$, the wave
function is given in coordinate space by
$$ Æ(q) = Òq|ÆÔ,ââ|ÆÔ = Ç{dq\over å{2¹}}ÊÆ(q)|qÔ $$
 where we use the convenient normalizations
$$ Ç{dq\over å{2¹}}|qÔÒq| = 1 = Ç{dp\over å{2¹}}|pÔÒp|â
	\left[ Òq|q'Ô = å{2¹}¶(q-q'),¼Òp|p'Ô = å{2¹}¶(p-p') \right] $$
 for coordinate and momentum space.  To describe time development, we
work in the Heisenberg picture, where time dependence is in the
operators (and thus their eigenstates):
$$ Æ(q,t) = Òq,t|ÆÔ $$
 Time development is then given
completely by the ``propagator" or ``Green function"
$$ G(q_f,t_f;q_i,t_i) ­ Òq_f,t_f|q_i,t_iÔâ
	ÜâÆ(q_f,t_f) = Ç{dq_i\over å{2¹}}ÊG(q_f,t_f;q_i,t_i)Æ(q_i,t_i) $$

\x VA1.1  Let's review the relationship between time development in the
Heisenberg and Schr¬odinger pictures.  Using the usual relation
$$ ÒÆ|Q(t)|Ô ­ ÒÆ(t)|Q|(t)Ô $$
 between the time-independent states $|ÆÔ$ and time-dependent
operators $Q(t)$ of the Heisenberg picture and the time-dependent states
$|Æ(t)Ô$ and time-independent operators $Q$ of the Schr¬odinger picture,
define time-dependent eigenstates in two ways:
$$ Q|qÔ = q|qÔâÜâ\leftÓ\matrix{ Òq(t)|Æ(t)Ô ­ Òq|ÆÔ \cr 
	Æ(q,t) ­ Òq|Æ(t)Ô ­ Òq,t|ÆÔ \cr}\right. $$
 Given the time development of a state
$$ |Æ(t)Ô = U(t)|ÆÔ $$
 ($U(t)­U(t,0)$),
find the development of $Q(t)$, $|q(t)Ô$, and $|q,tÔ$, and show in
particular that $|q(t)Ô±|q,tÔ$.  Which is the eigenstate of $Q(t)$?

In general, even for time-dependent Hamiltonians, we can find the
infinitesimal time development explicitly from the definition of the time
derivative and the time-depen\-dent Schr¬odinger equation:
$$ [i»_t -H(-i»_q,q,t)]Òq,t| = 0 $$
$$ ÜâÒq,t+·| = Òq,t|Ó1 -i·H[P(t),Q(t),t]Õ ® Òq,t|e^{-i·H[P(t),Q(t),t]} $$
 and similarly for $Òp,t+·|$ (where $P$ and $Q$ are the Hilbert-space
operators).  To derive the path-integral formalism, we then iterate this
result to obtain finite time development by inserting unity infinitely many
times, alternating between coordinate and momentum,
$$ Òq_f,t_f|q_i,t_iÔ =
	Ç{dp_0\over å{2¹}}{dq_1\over å{2¹}}{dp_1\over å{2¹}}...Òq_f,t_f|... $$
$$ ...|p_1,t_i+3·ÔÒp_1,t_i+3·|q_1,t_i+2·ÔÒq_1,t_i+2·|p_0,t_i+·Ô
	Òp_0,t_i+·|q_i,t_iÔ $$
  to obtain successive infinitesimal exponentials,
$$ Ç{dp_0\over å{2¹}}{dq_1\over å{2¹}}{dp_1\over å{2¹}}...
	Òq_f|e^{-i·H}...e^{-i·H}|p_1ÔÒp_1|e^{-i·H}|q_1ÔÒq_1|e^{-i·H}|p_0Ô
	Òp_0|e^{-i·H}|q_iÔ $$
 where the time dependence follows from the previous equation. 
However, note that all the implicit time dependence of the Heisenberg
picture drops out, because we extracted the $e^{-i·H}$'s, putting all the
factors of each matrix element at the same time:  Although each matrix
element is evaluated at time $·$ earlier than the one to its immediate left,
each is of the form
$$ Òa,t+·|b,tÔ = Òa,t|e^{-i·H[P(t),Q(t),t]}|b,tÔ = Òa|e^{-i·H[P,Q,t]}|bÔ $$
 (where $|bÔ­|b,t_iÔ$, etc.), leaving only any explicit time dependence that
may appear in the Hamiltonian, effectively translating the other $t$'s
$£t_i$.  Then we only need to know
$$ Òq|pÔ = e^{ipq},ââÒp|qÔ = e^{-ipq} $$
 to evaluate the matrix elements in the path integral as
$$ Ç{dp_0\over å{2¹}}{dq_1\over å{2¹}}{dp_1\over å{2¹}}... $$
$$ expÓ-i[q_i p_0 +·H(p_0,q_i,t_i) -q_1 p_0 +·H(p_0,q_1,t_i+·)
	+q_1 p_1 +·H(p_1,q_1,t_i+2·) +...]Õ $$
 More explicitly, this result is
$$ Òq_f,t_f|q_i,t_iÔ = ÇDp¼Dq¼e^{-iS},ââDp¼Dq = 
	Þ_{n=0}^{N-1}{dp_n\over å{2¹}}Þ_{n=1}^{N-1}{dq_n\over å{2¹}} $$
$$ S = Ý_{n=0}^{N-1}Ó-(q_{n+1}-q_n)p_n +·[H(p_n,q_n,t_i+2n·) 
	+H(p_n,q_{n+1},t_i+(2n+1)·)]Õ $$
$$ q_0 = q_i,âq_N = q_f;âât_f -t_i = 2N· $$
 Note that by adding (or subtracting) a step or two we could just as well
evaluate $Òq_f,t_f|p_i,t_iÔ$ or $Òp_f,t_f|q_i,t_iÔ$ or $Òp_f,t_f|p_i,t_iÔ$.

The classical picture is a segmented path, with the particle traveling
along a straight line segment from point $q_n$ to point $q_{n+1}$ with
momentum $p_n$:  Each $q$ is associated with a point, while each $p$ is
associated with the line segment connecting two consecutive points.  In
the ``continuum" limit $·£0$, $N£¥$, $t_f-t_i$ fixed,
$$ S = Ç_{t_i}^{t_f} dt [-Àqp +H(p,q,t)] $$
 (We have dropped some terms in $Òq|H|pÔ$ and $Òp|H|qÔ$ from reordering
the operators $Q$ and $P$ in $H(P,Q)$ to apply $P|pÔ=p|pÔ$ and
$Q|qÔ=q|qÔ$.  These commutator terms alternate in sign, combining to give
terms of order $·^2$, and can be dropped in the continuum limit.)

More generally, we can evaluate an arbitrary transition amplitude as
$$ \A = Òf|iÔ = 
	Ç{dq_f\over å{2¹}}{dq_i\over å{2¹}}Æ_f*(q_f)Òq_f,t_f|q_i,t_iÔÆ_i(q_i)
	= ÇDp¼Dq¼Æ_f*(q_f) e^{-iS} Æ_i(q_i) $$
 where now
$$ Dp¼Dq = Þ_{n=0}^{N-1}{dp_n\over å{2¹}}Þ_{n=0}^N{dq_n\over å{2¹}} $$
 Note that we can combine the initial and final wave function, as
$$ ï(q_i,q_f) ­ Æ_f*(q_f)Æ_i(q_i)âÜâ\A = ÇDp¼Dq¼ï(q_i,q_f) e^{-iS} $$
 The complex conjugation of $Æ_f$ vs.¼$Æ_i$ is due to the complex
conjugation involved in time reversal (as seen, e.g., when comparing an
eigenstate of $p$ at the inital time to the same eigenstate at the final
time).  In field theory, where the ``$p$'s and $q$'s" are functions of space
as well as time, if we choose the boundary in space also to be finite, so
that the space and time boundaries form a single connected and closed
boundary, then $ï$ is simply a function of the $q$'s over all that
boundary.

We now see the relationship of the path integral approach to the
time development operator:  From the
above derivation of the path integral, by integrating back out the
insertions of unity immediately after extracting the infinitesimal
exponentials and translating the time of each matrix element to zero, 
we find
$$ Òq_f,t_f|q_i,t_iÔ = Òq_f|U(t_f,t_i)|q_iÔ $$
$$ U(t_f,t_i) = e^{-i·H(t_f-·)} ò e^{-i·H(t_i+·)}e^{-i·H(t_i)}
	= \T \leftÓ exp \left[-iÇ_{t_i}^{t_f}dt¼H(t)\right]\rightÕ $$
 as previously.  This is effectively a
Schr¬odinger-picture expression (all the $P$'s and $Q$'s are at the initial
time), and can also be derived in that picture by solving for the time
dependence of any state $|Æ(t)Ô$.

Ü2. Semiclassical expansion

The path integral formulation is especially suited for semiclassical
approximations:  The Bohr-Sommerfeld quantization rule follows from the
fact that the functional integral is invariant under $S£S+2¹n$, since
$S$ appears only as $e^{-iS}$; in that sense the action is more like an angle
than a single-valued function.  The JWKB expansion follows from
$S£S/\h$ and expanding in $\h$.  This expansion can be interpreted as an
expansion in (space and time) derivatives, since it leads in the usual way
to the identification $p=-i\h»/»x$ and $E=i\h»/»t$.  

\x VA2.1  For comparison, we review the Schr¬odinger equation approach.
Consider the nonrelativistic JWKB expansion for the propagator
(for an arbitrary Hamiltonian $H$) to the first two orders in $\h$,
writing it as 
$$ G ® å¨e^{-iS/\h} $$
 ªa  Show the corresponding orders in the time-dependent Schr¬odinger
equation at $t>0$ can be written as the classical equation of motion for
the action $S$ and the (probability) current conservation law for the
(probability) density $¨$ (``Hamilton-Jacobi equations"),
$$ H = ÀS,ââ{»\over »q^i}\left( ¨{»H\over »p_i} \right) +À¨ = 0 $$
 when the argument $p$ of $H$ is evaluated at
$$ p_i = -{»S\over »q^i} $$
 (Assume a symmetric ordering of $p$'s and $q$'s in the quantum $H$.) 
Compare the ÓrelativisticÕ case examined in exercise IIIA4.1.
 ªb  The propagator is expressed in terms of $q$ and $q_0$, where
$G(q,q_0,t)¾¶(q-q_0)$ at $t=0$, so the first order in $\h$ is found by
using the solution to the Hamilton-Jacobi equations to write the classical
action in terms of the ``final" position $q$ and initial position $q_0$.  (In
principle; in general even the classical equations may be too difficult to
solve analytically.)  However, the Hamiltonian is given as a function of $p$
and $q$.  Show that the change in variables from $q,p$ to $q,q_0$ gives
$$ {»H\over »p_i} = -(M^{-1})^{ij} {»^2 S\over »q_0^j »t},ââ
	(M)_{ij} = {»^2 S\over »q_0^i »q^j} $$
 Show that 
$$ ¨ = det¼(-i\f1\h M) $$
 (the ``van Vleck determinant") solves the current conservation law,
using the explicit expression for $(det¼M)M^{-1}$ given in subsection IB3. 
Check the normalization, using the initial condition for propagators (or
comparing to the free case).

One way to apply the path integral is as follows:
(1) Find a classical solution to the equations of motion.  This gives the
leading contribution in $\h$ (``stationary phase approximation"),
$$ ÇDļe^{-iS/\h} ® e^{-iS_0/\h} $$
 (The validity of such an approximation with an imaginary exponent
will be discussed in subsection VA5.)

(2) Expand about the classical solution as
$$ Ä = Ä_{cl}+å\h ëÄ $$
Expanding in $ëÄ$ (or $\h$), we have schematically
$$ \h^{-1}S = \h^{-1}S_0 + \h^{-1/2}S'_0 ëÄ 
	+üS''_0 (ëÄ)^2 +\h^{1/2}\f16 S'''_0 (ëÄ)^3 + ... $$
 where ``$_0$" means to evaluate at $Ä=Ä_{cl}$ and the derivatives are
really functional derivatives (so there is also an integral for each derivative).
The first term in the action gives the classical
contribution, while the linear term vanishes by the equations of motion. 
The quadratic term gives an $\h$-independent contribution to the
exponential, so the next order approximation 
to the functional integral comes from integrating just that:
Integrating Gaussians as in subsection IB3,
$$ ÇDļe^{-iS/\h} ® e^{-iS_0/\h}(det¼S''_0)^{-1/2} $$
 where the determinant is now a functional one, which can be defined by
performing the functional integral as in the previous section, as a series
of ordinary Gaussian integrals.  The
boundary conditions are $ëÄ=0$ at $t_i$ and $t_f$
(since $Ä_{cl}=Ä$ there).  Normalization constants
can be determined by comparing the free case, or considering
the limit where the initial and final times converge.   

(3)  We then expand the exponential in the cubic and higher terms (positive
powers of $\h$):  The resulting functional integral is that of an
$\h$-independent Gaussian times a polynomial with positive powers
of $\h$.  Since odd orders in $ëÄ$ vanish by symmetry ($ëÄ£-ëÄ$),
only integer powers of $\h$ appear:
$$ ÇDļe^{-iS/\h} = e^{-iS_0/\h}ÇD(ëÄ)e^{-iS''_0(ëÄ)^2/2}
	\left(1 +Ý_{n=1}^¥ \h^n f_n[ëÄ]\right) $$
 Polynomials times Gaussians are also straightforward to integrate:
The easiest way is to first evaluate integrals of Gaussians with linear terms:
$$ Ç{d^D x\over (2¹)^{D/2}}e^{-x^T Sx/2 +j^T x} 
	= (det¼S)^{-1/2}e^{j^T S^{-1}j/2} $$
$$ Ç{d^D z*¼d^D z\over (2¹i)^D}e^{-zÿHz+zÿj+jÿz} 
	= (det¼H)^{-1}e^{jÿH^{-1}j} $$
 from shifting the integration variables ($x£x+S^{-1}j$, etc.)¼to eliminate
the linear terms, then using the previous results.  In functions of $x$
multiplying the Gaussian, $x$ can be replaced with $»/»j$ (and similarly
for $z$) and then pulled outside the integral.  (If a linear term is not
included, it can be introduced, and the result can be evaluated at $j=0$.) 
The final result then takes the form
$$ Òq_f,t_f|q_i,t_iÔ = e^{-iS_0/\h}(A +\h B +...)
	= exp\left( -i\f1\h Ý_{n=0}^¥ \h^n S_n \right) $$

\x VA2.2  Generalize the above results for integration of Gaussians with
linear terms to the cases with fermionic and mixed (subsection IIC3)
integration variables. 

\x VA2.3  Evaluate
$$ Ç{d^D x\over (2¹)^{D/2}}e^{-x^T Sx/2} x^i x^j x^k x^l $$
 by taking $(»/»j_i)(»/»j_j)(»/»j_k)(»/»j_l)$ on the above result.

As an example, consider the free nonrelativistic particle.  
The separability of the action
translates into factorization of the functional integral, so the result can
be found from the one-dimensional case.  As usual,
$$ L = -ümÀx{}^2âÜâx_{cl}(t) = x_i +{x_f -x_i\over t_f -t_i}(t-t_i) $$
 where we have written the classical solution in terms of the variables
appropriate to the initial and final states, namely $x_i$ for an initial state
localized there at time $t_i$, and $x_f,t_f$ for the final state.  Since the
classical action is itself quadratic, so is its expansion:
$$ S = S_0 + ëS,ââS_0 = -üm{(x_f -x_i)^2\over t_f -t_i},ââ
	ëS = -Çdt¼üm(À{ëx})^2 $$
 In general, a determinant from the $ëS$ integral must be evaluated 
(but see exercise VA2.1).  In this simple case, 
time translation invariance, dimensional analysis, and 
independence from $x_f,x_i$ are enough to determine the result of that 
functional integral up to a constant, fixed by the short-time limit
$t_f£t_i$.  The final one-dimensional result is then 
$$ Òx_f,t_f|x_i,t_iÔ = å{-im\over t_f -t_i}Êe^{im(x_f-x_i)^2/2(t_f-t_i)} $$
 where we have used
$$ å{2¹}¶(x) = \lim_{·£0}\f1{å·}e^{-x^2/2·} $$
 (one way of defining a Dirac $¶$ function) to normalize
$$ Òx_f,t|x_i,tÔ =  å{2¹}¶(x_f -x_i) $$

The Gaussian integral for the free particle can also be performed
explicitly, by using the discretized Hamiltonian path integral of the
previous subsection.  

\x VA2.4  The path integral for the free, nonrelativistic particle can be
evaluated much more easily using the Hamiltonian form of the action. 
First consider the Gaussian integral
$$ Ç_{-¥}^¥ dx¼e^{ipx-·x^2/2} $$
 as a special case of the Gaussians already evaluated, and use it to derive
the identity
$$ Ç_{-¥}^¥ dx¼e^{ipx} = 2¹¶(p) $$
 (The $·$ thus acts as a regulator to make the integral well defined.)  Then
use the discretized expression of subsection VA1, and evaluate the $x$
integrals first.  All but one of the $p$ integrals then can be trivially
evaluated, the last giving a Fourier transform.

\x VA2.5  Consider the one-dimensional harmonic oscillator.
(The multi-dimensional case is again separable.)
 ªa Explicitly evaluate the discretized path integral 
to find the result
$$ Òx_f,t_i+t|x_i,t_iÔ = å{-im¿\over sin¼¿t}¼exp\leftÓ
	{im¿[ü(x_f^2+x_i^2)cos¼¿t -x_f x_i]\over sin¼¿t}\rightÕ $$
 ªb Rederive the result using the result of exercise VA2.1.  
(Hint:  First solve the
classical equations of motion for $x(t)$, then rewrite it in terms of
$x_i=q_0$ and $x_f=q$; plug into $S_0=S$ and apply the above.)

Note that we have been sloppy about the definition of the ``integration
measure":  In going from the Hamiltonian form of the action to the
Lagrangian form, we ignored some $m$ dependence.  Specifically, if we
start with the Hamiltonian form, as derived in the previous subsection,
and derive the Lagrangian form by integrating out $p$, we find the
$1/m$ in $H=p^2/2m$ leads to
$$ Þ_{n=0}^{N-1}{dp_n\over å{2¹}}Þ_{n=1}^{N-1}{dx_n\over å{2¹}}â£â
	m^{N/2}Þ_{n=1}^{N-1}{dx_n\over å{2¹}} $$
 The $m^{(N-1)/2}$ then cancels similar factors from the $N-1$
$x$-integrals, while the remaining $åm$ is that found in the final result
above.

If we had considered a more general Hamiltonian, as in subsection IIIA1,
where $p^2$ appeared as $üg^{ij}(x)p_i p_j$, then we would have
obtained a measure of the form (for $i=1,...,D$)
$$ [det¼g(x_0)det¼g(x_N)]^{-1/4}
	Þ_{n=1}^{N-1}{d^D x_n\over (2¹)^{D/2}å{det¼g(x_n)}} $$
 (We have averaged $g$ as $g(x)p^2£å{g(x_n)g(x_{n+1})}p_n^2$, since
$x_n$ is associated with the point $n$ while $p_n$ is associated with the
link from $n$ to $n+1$.)  Such measure factors are easy to recognize,
since they are always local, without any derivatives:  
If we included it in the action, it would be a
term proportional to
$$ ln¼Þ_n det¼g(x_n) = \f1· Ý_n ·¼ln¼det¼g(x_n) ¾ ¶(0)Çdt¼ln¼det¼g(x(t)) $$
 (The factors at $x_0$ and $x_N$ are for standard normalization of the
wave functions, which we can absorb by a redefinition.)
In practice we just drop all such factors throughout the calculation, and
fix the normalization at the end of the calculation.  Since the Lagrangian
form follows from the Hamiltonian form, which was properly normalized,
we know such factors will cancel anyway.  Auxiliary fields can require
similar factors for proper normalization; then such factors are simply the
Jacobians from the field redefinitions from a form where they appeared
with trivial quadratic terms.

Ü3. Propagators

The amplitude we defined by path integration in subsection VA1 
is the ``propagator" or ``Green
function" for the Schr¬odinger equation.  Explicitly, we define
$$ G(q,t;q',t') ­ Ï(t-t')Òq,t|q',t'Ô $$
 where we have included the ``step function" 
$Ï(t-t')$ (1 for $t>t'$, 0 otherwise) to
enforce that the final time is later than the initial time (retarded
propagator).  This satisfies the free case of the general defining equation
of the propagator
$$ [»_t +iH(-i»_q,q,t)]G(q,t;q',t') =  [-»_{t'} +iH(i»_{q'},q',t')]G(q,t;q',t') $$
$$ = å{2¹}¶(q-q')¶(t-t') $$
 where we have used
$$ »_t Ï(t-t') = ¶(t-t') $$
 and the facts that $G$ without the $Ï$ factor is a homogeneous solution
of the Schr¬odinger equation (no $¶$'s) and becomes a $¶$ in $x$ for small
times.  The propagator then gives a general solution of the Schr¬odinger
equation as
$$ Òq,t| = Ç{dq'\over å{2¹}}ÊÒq,t|q',t'ÔÒq',t'|
	âÜâÆ(q,t) = Ç{dq'\over å{2¹}}G(q,t;q',t')Æ(q',t') $$
 In particular, for $Æ(q,t')=å{2¹}¶(q-q')$ at some time $t'$ for some point
$q'$, $Æ(q,t)=G(q,t;q',t')$ at all later times.
These equations are matrix elements of the corresponding
operator equations; e.g.,
$$ G(q,t;q',t') ­ Òq|U(t,t')|q'Ô $$
$$ [»_t +iH(t)]U(t,t') =  U(t,t')[-\onÁ»{}_{t'} +iH(t')] = ¶(t-t')I $$
 where we now include a step function in the definition of the
time development operator $U$:
$$ U(t,t') ­ Ï(t-t')\T \leftÓ exp \left[-iÇ_{t'}^t dt¼H(t)\right]\rightÕ $$

This solution for the propagator is not unique; as usual, a first-order
differential equation needs one boundary condition.  Another way to say
it is that the inhomogeneous differential equation is arbitrary up to a
solution of the homogeneous equation.  We have eliminated the
ambiguity by requiring that the propagator be retarded, as incorporated
in the factor $Ï(t-t')$; using instead $-Ï(t'-t)$ would give the advanced
propagator.  

This has an interesting translation in terms of the Fourier
transform with respect to the time,
which replaces the so-called ``time-dependent" Schr¬odinger equation
with the ``time-independent" one.  
Fourier transforms are a useful way to solve differential equations
when performed with respect to variables with translational invariance,
since this implies conservation of the conjugate variable:
The result is elimination of the corresponding derivatives.
In this case, it means the time-independent Schr¬odinger equation
needs a time-independent Hamiltonian.
For example, defining
$$ ÷U(E,E') ­ Ç{dt\over å{2¹}}{dt'\over å{2¹}}e^{-i(E't'-Et)}U(t,t') $$
$$ Üâ-i(E-H)÷U(E,E') = ¶(E-E')I $$
$$ Üâ÷U(E,E') = {i\over E-H}¶(E-E') $$

Now inverse Fourier transforming,
$$ U(t,t') = Ç{dE\over å{2¹}}{dE'\over å{2¹}}e^{i(E't'-Et)}÷U(E,E') $$
$$ = Ç{dE\over 2¹}e^{-iE(t-t')}{i\over E-H} $$
 we have an ambiguity in integrating
$E$ past the pole at $E=H$.  We therefore shift the pole slightly off
the real axis, so we can integrate exactly on the real axis.  Closing the
contour by adding to the real axis a semicircle of infinite radius in either
the complex upper- or lower-half-plane, wherever convergent
($\lim_{|t|£¥}e^{-|Et|}=0$, but $\lim_{|t|£¥}e^{+|Et|}=¥$), we find
$$ Ç{dE\over 2¹}e^{-iEt}{i\over E-Hài·} = àÏ(àt)e^{-iHt} $$
 which gives either the retarded or advanced propagator depending
on the choice of sign for the infinitesimal constant $·$ (retarded for
$E-H+i·$).  
Remember from exercise IIA1.2 that complex integration is
essentially just Gauss' law, with poles acting as charges:
The general integral result we used is
$$ È{dz'\over 2¹i}f(z'){1\over z'-z} = f(z) $$
 where the counterclockwise contour of integration encloses 
the pole at $z$ but no singularity in $f$, so we can evaluate 
the integral by Taylor expanding $f$ about $z$.

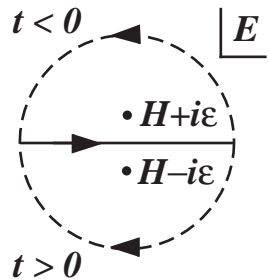
$$ \fig{contour} $$

To perform the inverse Fourier transform, we note that the
exponent needs an infinitesimal negative part to make the integral
convergent:
$$ Çdt¼e^{iEt}(à)Ï(àt)e^{-iHt¦·t} = {i\over E-Hài·} $$

\x VA3.1 Show that
$$ {i\over x+i·} -{i\over x-i·} = 2¹¶(x) $$
 by three methods:
ªa Use the above result for the Fourier transform.
ªb Show that this is the contour integral definition of the $¶$ function,
which is actually a distribution, by integration, multiplying by an arbitrary
(nonsingular) function and integrating along the real axis.  (Hint: Push the
poles onto the real axis, shifting the contours along with them, to find the
integral of a single function along the difference of two contours.)
ªc Prove the identity (checking the normalization)
$$ \lim_{·£0} {2·\over x^2 +·^2} = 2¹¶(x) $$

For the example of the free particle in one dimension we found
by various methods
$$ G(x,t;x',t') = Ï(t-t')å{-im\over t -t'}Êe^{im(x-x')^2/2(t-t')} $$
 However, we could have saved the trouble if we just started in
momentum space,
$$ öG(p,t;p',t') ­ Òp|U(t,t')|p'Ô = Òp|Ï(t-t')e^{-i(t-t')H}|p'Ô = 
	Ï(t-t')e^{-i(t-t')p^2/2m}Òp|p'Ô $$
$$ = Ï(t-t')å{2¹}¶(p-p')e^{-i(t-t')p^2/2m} $$
 in the retarded case.  If we Fourier transform $p$ to $x$
(the same as a change of basis from $|pÔ$ to $|xÔ$), 
the integrals are then simple Gaussians.
Again, the result is simpler in $p$-space because $p$ is conserved.
In the relativistic case we will want to treat energy and momentum
equally; doing the same here for later comparison, we define
$$ ÷Æ(p,E) = Ç{dq\over å{2¹}}{dt\over å{2¹}}e^{-i(pq-Et)}Æ(q,t) $$
 and similarly for $÷G$, and we have
$$ ÷G(p,E;p',E') = {i\over E-p^2/2m+i·}å{2¹}¶(p-p')¶(E-E') $$

Ü4. S-matrices

``Scattering" is defined as a process that starts with a free state
and ends with a free state, with interaction (self- or with external
forces) at intermediate times, e.g., particles coming in from and
going out to spatial infinity and
scattering from a potential of finite spatial extent.
Thus, if the interaction is nonvanishing somewhere between
times $t_1$ and $t_2$, where $t_f>t_2>t_1>t_i$, we can write
$$ U(t_f,t_i) = U(t_f,t_2)U(t_2,t_1)U(t_1,t_i)
	= e^{-i(t_f-t_2)H_0}U(t_2,t_1)e^{-i(t_1-t_i)H_0} $$
 in terms of the ``free term" $H_0$ of the Hamiltonian
$H=H_0+V$, where $V$ is the ``interaction term".
($V$ may be time dependent, but not $H_0$.)
It is more convenient to work with a quantity that is independent
of initial and final times (as long as they are outside of the
interaction region $t_1$ to $t_2$).  We therefore define
the ``S(cattering)-matrix" operator $\S$ as
$$ \S ­ \lim_{t_i£-¥\atop t_f£+¥}e^{it_f H_0}U(t_f,t_i)e^{-it_i H_0} $$
 where we have thrown in the limit because in the real world
interaction doesn't just start and stop, but fades in and out.
However, in our simple example above we find
$$ \S = e^{it_2 H_0}U(t_2,t_1)e^{-it_1 H_0} $$
In the special case of a free theory ($V=0$), we have simply $\S=I$.

In the interacting case, the amplitude we get from the path integral is
the interacting propagator.  To be able to take the limit
describing time development between infinite initial and final times, we
need to choose boundary conditions such that the initial and final basis
states have the time dependence of free particles, described by $H_0$,
assuming that the particle behaves freely at such asymptotically large
times.   This is called the ``interaction picture", to distinguish from the
Heisenberg picture, where the states have no time dependence, and the
Schr¬odinger picture, where the states have the complete interacting time
dependence.  We thus evaluate the limiting amplitude
$$ \A = \lim_{t_i£-¥\atop t_f£+¥}ÒÆ_f(t_f)|Æ_i(t_i)Ô = 
	\lim_{t_i£-¥\atop t_f£+¥}Ç{dq_f\over å{2¹}}{dq_i\over å{2¹}}
	Æ_f*(q_f,t_f)Òq_f,t_f|q_i,t_iÔÆ_i(q_i,t_i) $$
 for the interaction-picture states $|Æ(t)Ô$, relating the
interaction-picture coordinate basis ${}_0Òq,t|$ to the Heisenberg-picture
basis $Òq,t|$ (with initial conditions ${}_0Òq,0|=Òq,0|­Òq|$):
$$ {}_0Òq,t| = Òq|e^{-itH_0}âÜâÒq_f,t_f|q_i,t_iÔ ={}_0Òq_f,t_f| e^{it_f H_0}
	U(t_f,t_i)e^{-it_i H_0}|q_i,t_iÔ_0 $$
$$ Æ(q,t) = {}_0Òq,t|ÆÔâÜâ\A = ÒÆ_f|\S|Æ_iÔ $$
 with $\S$ as defined above.

The fact that time development conserves probability ($H=Hÿ$) is reflected
in the corresponding ÓunitarityÕ condition for the S-matrix:
$$ \Sÿ\S = 1 $$
  A more complicated condition is ÓcausalityÕ:  The basic idea is that
interactions take place in chronological order.  (A stronger statement of
causality will be found in the relativistic case: that any interaction should
take place at a spacetime point, rather than just at a single time.  It
follows from this weaker one in relativistic theories, since event B is
later than event A in every Lorentz frame only when B is in A's lightcone.) 
Causality is the condition that the Hamiltonian at any time involves only
variables evaluated at that time.  ($H(t)$ is a function of only $Ä(t)$, all at
the same time $t$, where $Ä=p,q$ are the quantum variables appearing in
the Hamiltonian.)  A nice way to describe the interactions is by
introducing a classical background as we did for the semiclassical
expansion of path integrals, such as by $Ä(t)£Ä(t)+(t)$, where $$ is just
some function.  The important point is that we have shifted $Ä(t)$ by
$(t)$ at the same $t$, so as not to disturb causality.  We then consider
the effect on the S-matrix of modifying the background $$ by a function
$¶$ localized (nonvanishing) at some particular time $t$, and a function
$¶'$ localized at $t'$, such that $t>t'$.  Picking out the $¶$ pieces in the
time-ordered product, we can therefore write
$$ \S[+¶+¶'] = U(f,t)\V(t)U(t,t')\V(t')U(t',i) $$
$$ \S[+¶] = U(f,t)\V(t)U(t,t')U(t',i) $$
$$ \S[+¶'] = U(f,t)U(t,t')\V(t')U(t',i) $$
$$ \S[] = U(f,t)U(t,t')U(t',i) $$
 where $U(t',i)$ is the time-development operator from time $t_i$ to time
$t'$ (including the canceling factor with $H_0$), $\V(t')$ is the extra factor in the
time development at time $t'$ resulting from the function $¶'$ localized
there, etc.  Thus we replace a $\V$ with the identity if the corresponding
$¶$ is absent.  Then we easily find
$$ \S[+¶+¶'] = \S[+¶]\S^{-1}[]\S[+¶'] $$
$$ Üâ(\S^{-1}[+¶]\S[+¶+¶'] -I) -(\S^{-1}[]\S[+¶'] -I) = 0 $$
$$ Üâ{¶\over ¶(t)}\left( \S[]ÿ {¶\over ¶(t')} \S[] \right) = 0
	âfor¼t > t' $$
 using the infinitesimal functions $¶$ and $¶'$ to define 
functional derivatives (as in subsection IIIA1 for the action).

In general, it is not possible to solve the Schr¬odinger equation for the
propagator or the S-matrix exactly.  One approximation scheme is the
perturbation expansion in orders of the interaction:
$$ \li{ H = H_0 +Vâ& Üâ\T(e^{-iÇdt¼H}) = e^{-i(t_f-t_i)H_0}
	+Ç_{t_i}^{t_f}dt¼e^{-i(t_f-t)H_0}[-iV(t)]e^{-i(t-t_i)H_0} \cr
	& +Ç_{t_i}^{t_f}dt Ç_{t_i}^t dt'¼
	e^{-i(t_f-t)H_0}[-iV(t)]e^{-i(t-t')H_0}[-iV(t')]e^{-i(t'-t_i)H_0} +... \cr} $$
$$ \li{ Üâ\S_{fi} ­ Òf|\S|iÔ =¼& Òf|iÔ +Ç_{-¥}^¥ dt¼Òf,t|[-iV(t)]|i,tÔ \cr
	& +Ç_{-¥}^¥ dt Ç_{-¥}^t dt'¼
	Òf,t|[-iV(t)]e^{-i(t-t')H_0}[-iV(t')]|i,t'Ô +... \cr} $$
 (To get this result, look at the definition of the time-ordered product in terms of infinitesimal integrals.)  The first term in $\S$ is just the identity (i.e., the free piece).  All the
other terms consist of a string of interactions ($-iV$) connected by free
propagators ($e^{-itH_0}$, where $t$ is the time between the
interactions), with each interaction integrated over all time (subject to
time-ordering of the interactions), and the initial/final state (wave
function) evaluated at the initial/final interaction time.

\x VA4.1  Assume the initial and final states are eigenstates of the free
Hamiltonian:
$$ H_0|iÔ = E_i|iÔ,ââH_0|fÔ = E_f|fÔ $$
 Assuming $V$ has no explicit time dependence, explicitly evaluate
the time integrals in the S-matrix, effectively Fourier transforming from
time to energy, to find
$$ \S_{fi} = Òf|iÔ-2¹i¶(E_f-E_i)Òf|(E-H_0){1\over E-H+i·}(E-H_0)|iÔ|_{E=E_i} $$
 (Hints:  Redefine the integration variables to be the times between
interactions.  Taylor expand $1/(E-H+i·)$ in $V$ for comparison.)

In field theory we want to express any state in terms of a basis of
products of 1-particle states, so we can calculate the behavior of these
specified particles.  We try to do this by using field variables (the ``$q$'s"
of field theory):  Each field operator should produce a single particle. 
Unfortunately, this is not the case:  An asymptotic state of given
3-momentum created by such a field operator is not necessarily an
eigenstate of the energy, because such a state can be either 1-particle or
n-particle, due to interactions.  The propagator for the field is then of the
form
$$ öG(p,t;p',t') ¾ ¶(p-p')Ý_I Æ*_I(p)Æ_I(p)e^{-i(t-t')E_I(p)} $$
$$ E_I(p) = Ý_{i=1}^{n_I}E_{I,i}(p_i),ââÝ_{i=1}^{n_I}p_i = p $$
 where ``$E_{I,i}(p_i)$" is the energy of a 1-particle state (the $Ý_I$ will
include an integral in general).  However, as long as all particles have
masses, such an asymptotic 1-particle state is distinguishable as that of
lowest energy $E_0$:  The higher-energy states are n-particle states to
which this particle can couple.  (If some of the n-particle states were
lower energy, the 1-particle state could decay into them, and thus the
1-particle state would be unstable, and not asymptotic.  With massless
particles things are more complicated:  Then 1-particle states are more
difficult to define and to measure.)  In principle, we could define the
1-particle states by constructing the corresponding operator, consisting
of the field plus terms higher order in the fields; in practice, this is rather
complicated.  (Note:  For the above analysis, it might be convenient to use the center-of-mass frame.)

A simpler way to make the asymptotic states unambiguous is by
modifying the definition of the S-matrix:
$$ \S = \lim_{t_i£-¥(1+i·)\atop t_f£+¥(1+i·)}
	e^{it_f H_0}Ê\T\left(e^{-iÇ_{t_i}^{t_f}dt¼H}\right)e^{-it_i H_0} $$
 introducing factors of $1+i·$ for some positive $·$, which may be chosen
small for convenience.  (Actually, we can generally replace $1+i·$ with
just $i$ if it is not too confusing:  The result is the same.)  The effect is
seen by considering a matrix element of particular fields that may be a
superposition of different energies $E$ in the initial state
and $E'$ in the final state, but evaluated between 
an initial state of energy $E_i$ and a final state of energy $E_f$
(which might not be equal for a time-dependent interaction,
e.g., if the number of particles changes).
Since $E³E_i$ initially and $E'³E_f$ finally, the time dependence 
of any such matrix element is proportional to
$$ \S_{fi} ¾ \lim_{t_f£+¥(1+i·)} e^{it_f(E'-E_f)}
	\lim_{t_i£-¥(1+i·)} e^{-it_i(E-E_i)}
	= \leftÓ\matrix{ 1 & for¼E = E_i, E' = E_f \cr 0 & otherwise\hfill\cr}\right. $$
 Alternatively, we can simply impose $E=E_i,E'=E_f$
directly in the definition:
$$ \S = \lim_{t_i£-¥\atop t_f£+¥}e^{it_f H_0}¶_{H(t_f),H_0}Ê
	\T\left(e^{-iÇ_{t_i}^{t_f}dt¼H(t)}\right)¶_{H(t_i),H_0}e^{-it_i H_0} $$
 where the free Schr¬odinger equation $H_0=E_i$ or $E_f$ defines $E_i$ for the
initial state and $E_f$ for the final state, and 
$¶_{H,H_0}$ is evaluated by examining the
asymptotic time-dependence of the time-development operator with
respect to $t_i$ and $t_f$:  Normally field theory is calculated in
energy-momentum space, working with the spacetime Fourier transform
of the above, where this amounts to simply comparing energies
$E=E_i,E'=E_f$.

If we know some details of the interaction, this modification may be
irrelevant:  In particular, in local quantum field theory interactions
happen at a point in space and time.  For example, consider the inner
product between a 1-particle state in its rest frame and a related
n-particle state, which appears in the same propagator.  Because of
locality, the wave function for the n-particle state, when evaluated in
position space (which is where the theory is local) is simply the
product of n 1-particle wave functions evaluated at the same point.  But
we know that for small relative momenta (where a nonrelativistic
approximation holds) that the individual wave functions propagate as 
$$ |Æ| ¾ |t-t'|^{-(D-1)/2} $$
 from the form of the free 1-particle propagator.  (Or, we can use
dimensional analysis, and consider the spread of a particle of
restricted range of momenta from a confined region:  Then $|Æ|^2¾1/V$
and the volume $V¾|t-t'|^{D-1}$.)  This implies that the n-particle wave
function will fall off as the nth power of that, so in the limit of large
times the 1-particle state will dominate.  In a relativistic theory the
length scale associated with this fall-off will be associated with the
masses involved, and thus at a subatomic scale.

Ü5. Wick rotation

In the previous subsection we ensured convergence in the definition of
the S-matrix by effectively making the ``coordinate change"
$$ t £ (1-i·)t = e^{-i·}t $$
 in the definition of the limit
$$ (1-i·)t £ ¥âÜât £ (1+i·)¥ $$
 This affected the time-development operator as
$$ e^{-iHt} £ e^{-iHt-·t} $$
 for $H>0$ to pick out the ground state $H=0$.  The same effective
substitution was made in subsection VA3 in defining the contour integral
for the propagator:
$$ Ç{dE\over 2¹}Êe^{-iE(1-i·)t}{i\over E-H} =
	Ç{dE\over 2¹}Êe^{-iEt}{i\over (1+i·)E-H} $$
$$ = Ç{dE\over 2¹}Êe^{-iEt}{i\over E-(1-i·)H} =
	Ç{dE\over 2¹}Êe^{-iEt}{i\over E-H+i·} $$
 which is the same as the substitution
$$ E £ (1+i·)E = e^{i·}E $$
 (when working with the time-independent Schr¬odinger equation)
since essentially $E=i»/»t$.

In general, having to do contour integrals and keep track of $i·$'s in
propagators is inconvenient.  Fortunately, there is a simple way in
practical calculations to get rid of not only the $i·$'s but (almost) all the
other $i$'s as well.  The method is known as ``Wick rotation".  The basic
idea is to extend the above complex rotation from angle $·$ to angle
$¹/2$:
$$ t £ -it = e^{-i¹/2}t,ââE £ iE $$
 pushing the contour even farther away from the singularities.  Thus, the
Schr¬odinger equation is changed to a ``diffusion equation" (to describe,
e.g., Brownian motion):
$$ (i»_t -H)Æ = 0âÜâ(»_t +H)Æ = 0 $$
 For example, for the free particle the resulting equation has no $i$'s.  The
time-independent Schr¬odinger equation then becomes
$$ (E -H)Æ = 0âÜâ(iE-H)Æ = 0 $$
 The result for the propagator is then
$$ Ç_{-¥}^¥ {dE\over 2¹}¼e^{-iEt}{1\over H-iE} = Ï(t)e^{-Ht}  $$
 Now no $i·$ prescription is needed, since the pole was moved away from
the real axis.  Similar remarks apply to the inverse Fourier transform
$$ Ç_{-¥}^¥ dt¼e^{iEt}Ï(t)e^{-Ht} = {1\over H-iE} $$

\x VA5.1  Find the Wick-rotated retarded propagator $G(x',t';x,t)$ for the
free (1D) particle, satisfying
$$ (»_t +H)G = (-»_{t'} +H')G = å{2¹}¶(x-x')¶(t-t') $$

 Furthermore, if we define the S-matrix directly in this Wick-rotated space
$$ \S = \lim_{t_i£-¥\atop t_f£+¥}
	e^{t_f H_0}Ê\T\left(e^{-Ç_{t_i}^{t_f}dt¼H}\right)e^{-t_i H_0} $$
 then the limiting procedure is unambiguous even in field theory, since
$$ \lim_{t_i£-¥} e^{t_i(E-E_i)}
	= \leftÓ\matrix{ 1 & for¼E = E_i \cr 0 & for¼E > E_i \cr}\right. $$
$$ \lim_{t_f£+¥} e^{-t_f(E-E_f)}
	= \leftÓ\matrix{ 1 & for¼E = E_f \cr 0 & for¼E > E_f \cr}\right. $$

Another important effect is on actions.  For example, in the mechanics
path integral for a particle with kinetic term $T=ümÀx{}^2$ in a potential
$U(x)$, we integrated
$$ e^{-iS}¼:âS = Çdt (U-T) $$
 Upon Wick rotation, this becomes
$$ e^{-S}¼:âS = Çdt (U+T) $$
 The major change on the exponent $-S$ is that it is now not only real, but
negative definite.  (For physical purposes, we assume the potential has a
lower bound, which can be defined to be nonnegative without loss of
generality.)   Thus, the semiclassical approximation we made earlier,
called the ``stationary phase" approximation, has now become the
``steepest descent" approximation, namely fitting $e^{-S/\h}$ to a
Gaussian, which is approximating the integral by the places where the
integrand is largest.  We thus write
$$ S(x) = S(x_0) +ü(x-x_0)^2 S''(x_0) +...,ââS'(x_0) = 0,âS''(x_0) > 0 $$
 for one variable, with the obvious generalization to many variables. 
Explictly, we then have
$$ Ç{dx\over å{2¹\h}}e^{-S(x)/\h} ® 
	\left.{1\over å{S''(x)}}e^{-S(x)/\h}\right|_{S'(x)=0} $$
 plus higher orders in $\h$, expressed in terms of higher derivatives of
$S$.  In the case of many variables, $S''$ is replaced with a determinant,
as for the Gaussian integrals of subsection IB3, and for functional
integrals, with a functional determinant.  (But sometimes the functional
determinant can be replaced with an ordinary determinant:  See exercise
VA2.4.)

So now we can first calculate everything in Wick-rotated spacetime,
where everything is real (more precisely, classical reality properties are
preserved quantum mechanically), and then Wick rotate back to find the
correct result in physical spacetime.  In particular, the appropriate
$·$'s, still needed to correctly position the singularities in physical
spacetime, can be restored by rotating back through an angle $ü¹-·$:
$$ inverse¼Wick:ââ
	t £ (i+·)t = e^{i(¹/2-·)}t,âE £ (-i+·)E = e^{-i(¹/2-·)}E $$

\refs

£1 N. Wiener, ÓJ. Math. and Phys. Sci.Õ É2 (1923) 132:\\
	Euclidean path integrals for Brownian motion.
 £2 P.A.M. Dirac, ÓPhys. Z. der Sowj.Õ É3 (1933) 64:\\
	proposed path integrals for quantum mechanics.
 £3 R.P. Feynman, ÓRev. Mod. Phys.Õ É20 (1948) 367:\\
	formulated path integral approach to quantum mechanics.
 £4 R.P. Feynman, ÓPhys. Rev.Õ É84 (1951) 108:\\
	path integrals in phase space.
 £5 R.P. Feynman and A.R. Hibbs, ÓQuantum mechanics and path integralsÕ
	(McGraw-Hill, 1965):\\
	review of path integrals; interesting side stuff, like D=2.
 £6 R. Shankar, ÓPrinciples of quantum mechanicsÕ, 2nd ed. (Plenum,
	1994):\\
	good quantum mechanics text that includes path integrals.
 £7 J.H. van Vleck, ÓProc. Natl. Acad. Sci. USAÕ É14 (1928) 178.
 £8 P.A.M. Dirac, ÓProc. Roy. Soc.Õ ÉA136 (1932) 453,
	P.A.M. Dirac, V.A. Fock, and B. Podolosky, ÓPhys. Z. Sowj.Õ É2 (1932)
	468:\\
	interaction picture.
 £9 J.A. Wheeler, ÓPhys. Rev.Õ É52 (1937) 1107;\\
	W. Heisenberg, ÓZ. Phys.Õ É120 (1943) 513, 673:\\
	S-matrix.
 £10 E.C.G. St¬uckelberg, ÓHelv. Phys. ActaÕ É19 (1946) 242;\\
	E.C.G. St¬uckelberg and D. Rivier, ÓHelv. Phys. ActaÕ É24 (1949) 215;\\
	E.C.G. St¬uckelberg and T. Green, ÓHelv. Phys. ActaÕ É24 (1951) 153:\\
	causality in quantum field theory.
 £11 N.N. Bogoliubov, ÓDoklady Akad. Nauk USSRÕ É82 (1952) 217,
	É99 (1954) 225:\\
	explicit condition of causality on S-matrix, using functionals.
 £12 L.S. Brown, ÓQuantum field theoryÕ (Cambridge University, 1992) 
	p. 293:\\
	decoupling of asymptotic multiparticle states in field-theory
	propagators.
 £13 M. Kac, On some connections between probability theory and
	differential and integral equations, in ÓProc. 2nd Berkeley Symp.
	Math. Stat. ProbabilityÕ, ed. J. Neyman, 1950 (University of California,
	1951) p. 189;\\
	E. Nelson, ÓJ. Math. Phys.Õ É5 (1964) 332:\\
	Wick rotation of path integral in quantum mechanics.
 £14 G.C. Wick, ÓPhys. Rev.Õ É96 (1954) 1124:\\
	Wick rotation in quantum field theory.

\unrefs

Û4 B. PROPAGATORS

Classically we distinguish between particles and waves.  This can
be consistent with a classical limit of a quantum theory if there is a
conserved charge associated with the classical particles, with respect to
which the classical waves are neutral.  Such a situation is described 
by a field theory Lagrangian (density) of the form
$$ L = Æÿ\O (Ä) Æ +L_Ä (Ä) $$
 where $Æ$ is the field of the charged particle, and $Ä$ the field 
of the waves that carry
the interaction.  $\O$ includes the kinetic operator;
a nonrelativistic example was given in subsection IIIA3.
Thus $\O (Ä) Æ=0$, the field equation for $Æ$, is also
a Schr¬odinger equation, which we can derive from a
classical mechanics action.
(A zero-range interaction, as in billiard-ball collisions,
is described by an $L_Ä$ without derivatives.)
Then we can have continuous
worldlines for the particles:  The statement that the worldlines do not end
or split is associated with charge conservation.  The interaction between
the particles and waves is described by $Ä$ dependence in the particle (mechanics)
action obtained from $\O$ (and not the term $L_Ä$ for the 
wave fields).  If we look at just the
mechanics action, the modification is the same as considering external
fields (like external potentials in nonrelativistic mechanics), since we are
ignoring $L_Ä$, which is needed for the field equations of $Ä$.

$L_Ä$ then can be added separately.  Coupling to such
external fields is a simple way to study properties of particles without
applying field theory.  For example, in nonrelativistic mechanics it helps
to explain charge and spin, which don't appear explicitly in the free
Schr¬odinger equation.

Ü1. Particles

All the information in quantum mechanics is contained in the propagator,
which gives the general solution to the Schr¬odinger equation, and can be
obtained by the Feynman path integral.  Here we discuss the free
propagator for the spinless particle (whose classical description was
given in section IIIB), which is the starting point for relativistic
perturbation theory.

We consider quantization first in the Lorentz covariant gauge $v=1$. 
From subsection IIIB2 we have
$$ S_{H,AP} = Ç_0^T d [-Àx{}^m p_m +ü(p^2 +m^2)] $$
Except for the $T$ integration in the functional integral
(in addition to the functional integration over $x$ and $p$), 
the same methods can be applied as in the
nonrelativistic case, where we had
$$ S_{H,NR} = Ç_{t_i}^{t_f} dt(-Àx{}^i p_i +\f1{2m}p^2) $$
  The simplest expression (and ultimately the most
useful one) is obtained by Fourier transforming with respect to $x$:  In
comparison to the multidimensional nonrelativistic result
$$ öG_{NR}(p^i,t;p'^i,t') = ¶(p-p')Ï(t-t')e^{-i(t-t')p^2/2m} $$
 (where here $¶(p-p')=(2¹)^{(D-1)/2}¶^{D-1}(p^i-p'^i)$ for $D-1$ spatial
dimensions), the relativistic result is
$$ ÷G(p,p') = ÇdT¼¶(p-p')Ï(T)e^{-iT(p^2+m^2)/2} $$
 (where now $¶(p-p')=(2¹)^{D/2}¶^D(p^a-p'^a)$ for $D$ spacetime
dimensions).

There are several simple yet important differences from the
nonrelativistic case:   
\item{(1)} The dependence on the mass $m$ is different.  In
particular, we can set $m=0$ only in the relativistic case.   
\item{(2)} There is an
additional integration $ÇdT$, because the variable $T$, which is the
remaining part of $v$, survives the gauge $v=1$.  
(It is all that remains of a would-be functional integral over $v$.)
This is analogous to the
time integral in the nonrelativistic case for $÷G(p^i,E;p'^i,E')$, if we set the
energy to zero.  This is as expected, since the relativistic classical
mechanics differs from the nonrelativistic one mainly by constraining the
``Hamiltonian" $ü(p^2+m^2)$ to vanish.  This interpretation also leads to
the ``zero-energy" version of the inhomogeneous
(proper-)time-independent Schr¬odinger equation for this case,
$$ -iü(õ-m^2)G(x,x') = ¶(x-x') $$
 \item{(3)} The propagator is automatically ``retarded" in the ``proper time" $T$,
as a consequence of the positivity condition $v>0$, which was motivated
by the geometrical interpretation of $v$ as the worldline metric.  

\noindent When
used in this manner to write the propagator in terms of a Gaussian, $T$ is
known as a ``Schwinger parameter".

Generally, it is convenient to remove the momentum $¶$-function (which
resulted from translational invariance) as 
$$ G(x,x') = ë(x-x')âÜâ÷G(p,p') = ¶(p-p')ë(p) $$
$$ ë(p) = ÇdT¼Ï(T)e^{-iT(p^2+m^2)/2} $$
 where we have simply written $ë(p)$ for the Fourier transform of $ë(x)$
(dropping the tilde).  Performing the $T$ integral, using the same methods
as for the $t$ integral in the nonrelativistic case, we have the final result
$$ ë(p) = {-i\over ü(p^2+m^2-i·)} $$
 Actually, this result is almost obvious from solving the relativistic wave
equation.  The only part that is not obvious is the ``$i·$ prescription":
how to perform the contour integration upon Fourier transformation.  In
the nonrelativistic case, we saw two obvious choices, corresponding to
retarded or advanced propagators; the classical action did not distinguish
between the two, although the retarded propagator has the obvious
convenience of determining later events from earlier ones.  On the other
hand, in the relativistic case the choice of propagator was fixed from
classical considerations.  $T$ is restricted to be positive, and the $i·$ is
needed to make the $T$ integral converge.

\x VB1.1  Take the nonrelativistic limit of the relativistic propagator, and
compare with the propagator of nonrelativistic quantum mechanics. 
Explain the difference in terms of the nonrelativistic limit of the classical
mechanics action.

\x VB1.2  Perform the analysis of exercise VA2.4 for the relativistic
particle.  First replace the integration over $T$ by a sum:  Instead of
dividing up the time into $2N$ intervals of length $·$ and taking the limit
$N£¥,·£0$, with $2N·$ fixed, sum $2·Ý_{N=0}^¥$, and then take the
limit $·£0$.  ($2N·$ is now $T$ instead of $t_f-t_i$, and we integrate over
it instead of keeping it fixed.)  Perform all $x$ integrals and then all but
the last $p$ integral before summing over $N$.  Again, the entire
calculation is much easier than using the Lagrangian (second-order) form
of the path integral.

To understand this point better, we examine the Fourier transformation
with respect to time.  In contrast to the nonrelativistic case, there are
now two poles, at
$$ p^0 = à¿,â¿ = å{(p^i)^2+m^2}:ââë = 
	{i\over ¿}\left({1\over p^0 -(¿-i·)} -{1\over p^0 +(¿-i·)}\right) $$
 where now $p^a=(p^0,p^i)$.  These are also the two classical values of
the canonical energy (as opposed to the true energy, which is the
absolute value), which we saw previously corresponded to particles and
antiparticles.  With our prescription for integrating around the poles,
using the same methods as in the nonrelativistic case, we then find
$$ öG(p^i,t;p'^i,t') = (2¹)^{D/2}¶^{D-1}(p^i-p'^i){1\over ¿}e^{-i¿|t-t'|} $$
$$ = (2¹)^{D/2}¶^{D-1}(p^i-p'^i){1\over ¿}
	[Ï(t-t')e^{-i¿(t-t')} +Ï(t'-t)e^{i¿(t-t')}] $$
 We now see that the particles ($p^0=¿$) have a retarded propagator,
while the antiparticles ($p^0=-¿$) have an advanced propagator.  This is
the quantum version of the classical result we saw earlier, that particles
travel forward in time, while antiparticles travel backward.
The interpretation is simple:  When evaluating matrix elements of the
form $Òf|\O |iÔ$, the resulting propagator ensures that the initial
wave function contains only positive energies, while the final
Ócomplex conjugateÕ wave function contains only negative energies
(i.e., the final wave function itself contains positive energies).

We next compare quantization in the lightcone gauge. 
Again from subsection IIIB2,
$$ S_{H,LC} = Ç_{ _i}^{ _f} d  [Àx{}^- p^+ -Àx{}^i p^i +ü(p^{i2} +m^2)] $$
 Whereas in the
covariant gauge the analog to the nonrelativistic time $t$ was the
``proper time" $T$, the analog is now the lightcone ``time" $ $.  Since
$ =x^+/p^+$, we have $E=p^- p^+$ ($E=i»/» $ vs.¼$p^-=i»/»x^+$), and thus
$$ ë(p) = {i\over E-ü(p^{i2}+m^2)+i·} = {-i\over ü(p^2+m^2-i·)} $$
 as before.  Note that this derivation was almost identical to the
nonrelativistic one:  Unlike the covariant gauge, we did not have to add in
$T$ as a separate variable of integration (but not path integration). 
However, this Schwinger parameter is useful for evaluating momentum
integrals and analyzing momentum dependence.  This is a typical
characteristic of unitary gauges:  They are more useful for keeping track
of degrees of freedom.

Ü2. Properties

As in electrodynamics, the free scalar satisfies a differential equation
second-order in time, so the propagator is used differently from
nonrelativistic quantum mechanics to give a general solution to the wave
equation.  We begin by considering a free ``action" between two different
scalar fields, written in a way where all derivatives act on just one
field or just the other, i.e., where the field equation is explicit.
The two forms are related by integration by parts, but now we
keep boundary terms:
$$ Çd^D x¼[A(õ-m^2)B -B(õ-m^2)A]
	= Çd^D x¼»É(A\onª» B) 
	= Èd^{D-1}§^m¼A\onª»{}_m B $$
 where in the last step we have used the (generalized) Stokes' theorem
(see subsection IC2);
``$Èd^{D-1}§^m$" is the integral over the closed surface bounding
the volume integrated over in $Çd^D x$.  In practice we take the volume to
encompass all spacetime in the limit, neglect the part of the boundary at
spacelike infinity, and choose the parts of the boundary at timelike
infinity to be surfaces at constant time, so the boundary integrals are
over just space:
$$ Èd^{D-1}§^m¼A\onª»{}_m B = Çd^{D-1}x¼A\onª»{}_t B |_{-¥}^¥ =
	Çd^{D-1}x¼A\onª»{}_t B |_¥ -Çd^{D-1}x¼A\onª»{}_t B|_{-¥} $$
 We then have the solution for the wave function inside the volume in
terms of that on the boundary:
$$ (õ-m^2)Æ = 0,â-iü(õ-m^2)G(x,x') = -iü(õ'-m^2)G(x,x') = ¶(x-x') $$
$$ ÜâÈ{d^{D-1}§'^m\over (2¹)^{D/2}}¼G(x,x')üi\onª»{}'_m Æ(x') = Æ(x) $$
 where the wave equation for $Æ$ is the Klein-Gordon equation.

\x VB2.1  For a free nonrelativistic particle, solve $x(t)$'s "1D wave equation" for a Green function that vanishes at $t_i$ and $t_f$ (``Dirichlet" boundary conditions). Use it to find the solution for $x(t)$ in terms of $t_i$, $t_f$, $x_i$, and $x_f$ as given in subsection VA2. (Don't forget the sign from the orientation of the ``boundary", i.e., $t = t_i$ or $t_f$.)

Similarly, this defines a conserved current from any two wave functions
$$ »É(Æ_1*\onª» Æ_2) =  Æ_1*(õ-m^2)Æ_2 -Æ_2(õ-m^2)Æ_1* = 0 $$
 or, evaluating the integral over a volume infinite in space but
infinitesimal in time, the conserved charge
$$ {d\over dt}Çd^{D-1}x¼Æ_1*\onª»{}_t Æ_2 = 0 $$
 This leads to the covariant inner product $Ò¼||¼Ô$ on a hypersurface (as opposed to the naive inner product $Ò¼|¼Ô$ for the full space)
$$ Ò1||2Ô = ·(p^0)Ç{d^{D-1}x\over (2¹)^{D/2}}¼Æ_1*üi\onª»{}_t Æ_2 $$
 where the $·(p^0)$ appears because the contour integral gives a + at
later times (positive energy) and a $-$ at earlier times (negative energy).
Explicitly, we find for the inner product of plane waves
$$ Æ_p(x) = Òx|pÔ = e^{ipÉx} $$
$$ ÜâÒp||p'Ô = (2¹)^{D/2-1}¶^{D-1}(p^i-p'^i)·(p^0)ü(p^0+p'^0) $$
 We have used $p^2+m^2=p'^2+m^2=0$, which also implies that
$|p^0|=|p'^0|$:  Thus, the inner product vanishes if the waves have
opposite-sign energy, while for the same sign
$·(p^0)ü(p^0+p'^0)=|p^0|$.  The result then can be written manifestly
covariantly as
$$ Òp||p'Ô = {¶(p-p')\over 2¹¶[ü(p^2+m^2)]} =
	Ï(p^0p'^0)¿(2¹)^{D/2-1}¶^{D-1}(p^i-p'^i) $$
 Similarly, the solution for the wave function in terms of the Green
function gives only positive-energy contributions from the part of the
surface at earlier times, and only negative-energy contributions from the
part of the surface at later times.
 More general on-shell wave functions, since they depend on only $D-1$
spatial momenta and the sign of the energy, can be written as a
restricted Fourier transform
$$ Æ(x) = Çdp¼2¹¶[ü(p^2+m^2)]e^{ipÉx}÷Æ(p) $$
$$ ÜâÒ1||2Ô = Çdp¼2¹¶[ü(p^2+m^2)]÷Æ_1(p)*÷Æ_2(p) $$
 (Here $֮(p)*$ means to complex conjugate after Fourier transforming to
$p$-space; otherwise, we need to change the sign of the argument.)  In
particular, for a plane wave we have
$$ ÷Æ_{p'}(p) = {¶(p-p')\over 2¹¶[ü(p^2+m^2)]} $$

It will prove useful later to have a collection of solutions to the
homogeneous and inhomogeneous Klein-Gordon equations, and compare
them in 4-momentum space and time-3-momentum space.  Using the
previous nonrelativistic and relativistic results, we find
$$ \vbox{\halign{$ë#$:\hfil&â$#$\hfilâ$Ü$â&$#$\hfil&$#$
	\hfil&â$=â#$\hfil \cr
	& -i/(p^2+m^2-i·) & Ï(t)e^{-i¿t} & +Ï(-t)e^{i¿t} & e^{-i¿|t|} \cr
	* & i/(p^2+m^2+i·) & Ï(t)e^{i¿t} & +Ï(-t)e^{-i¿t} & e^{i¿|t|} \cr
	_R & -i/(p^2+m^2-i·p^0) & Ï(t)e^{-i¿t} & -Ï(t)e^{i¿t} & 
		-2iÏ(t)sin(¿t) \cr
	_A & i/(p^2+m^2+i·p^0) & Ï(-t)e^{i¿t} & -Ï(-t)e^{-i¿t} & 
		2iÏ(-t)sin(¿t) \cr
	_+ & Ï(p^0)2¹¶(p^2+m^2) & Ï(t)e^{-i¿t}&+Ï(-t)e^{-i¿t} & e^{-i¿t} \cr
	_- & Ï(-p^0)2¹¶(p^2+m^2) & Ï(t)e^{i¿t}&+Ï(-t)e^{i¿t}&e^{i¿t}\cr}}$$
 where we have omitted certain common factors (see subsection VB1).  $ë_à$
satisfy the homogeneous equation, while the rest satisfy the
inhomogeneous one.  (These are easily checked in the mixed space,
where the Klein-Gordon operator is $-(»_t^2+¿^2)$.)  This table makes
explicit which sign of the energy propagates in which time direction, as
well as the linear relations between the momentum-space expressions. 
In particular, we see that $ë_+$ propagates just the positive-energy
states, while $ë_-$ propagates just the negative-energy ones.

\x VB2.2  The relativistic propagator uses a particular choice for
integrating around the two poles in the complex energy plane, as encoded
in the $i·$ prescription.  If we ignored the classical determination of that
prescription, there would be four simple choices, integrating either above
or below the two poles.  
 ªa Show these four choices can be enforced by
replacing $i·$ in $p^2+m^2-i·$ with
$$ i·,â-i·,âi·p^0,â-i·p^0 $$
 and derive the results of the table above.  
 ªb Give explicit expressions for
the four propagators in position space in four dimensions for the
massless case.

We can check the propagator's behavior by explicit evaluation, using
plane waves:
$$ ·(p^0)Ç{d^{D-1}x'\over (2¹)^{D/2}}¼ë(x-x')üi\onª»_t{}'Æ_p(x')
	= ·(p^0)ü(p^0 +i»_t){1\over ¿}e^{-i¿|t-t'|}
	e^{i\vec pÉ\vec x-ip^0 t'} $$
$$ = Ï[p^0(t-t')]Æ_p(x) $$
 where we have used the previous result for $öG(\vec p,t;\vec p',t')$ (and
thus $ë(\vec p,t)$).  Again we see that the propagator propagates
positive-energy solutions forward in time and negative-energy backward.

This propagator also applies to relativistic field theory.  (See subsection
IIIA3 for nonrelativistic field theory.)  In comparison to the nonrelativistic
case, the propagator is now $-i/ü(p^2+m^2)$ instead of $-i/(\f1{2m}\vec
pÊ{}^2-E)$, and this determines the kinetic term in the field theory action:
$$ S_0 = -Çdx¼üÄü(õ -m^2)Ä = Çdx¼\f14 [(»Ä)^2 +m^2 Ä^2] $$
 To make the functional integral of $e^{-iS_0}$ converge, we replace
$m^2£m^2-i·$, which is the same $i·$ prescription found in
first-quantization.  Note that we have used a real field $Ä*=Ä$.  (A
complex field can be used by doubling $Æ=(Ä_1+iÄ_2)/å2$.)  This is
possible only in the relativistic case because we have both
positive-energy solutions $e^{-iEt}$ as well as negative ones $e^{+iEt}$. 
(In other words, the relativistic Schr¬odinger equation is a second-order
differential equation, so we get two $i$'s to make the kinetic operator
real.)  Reality simply means identifying particles with antiparticles.  (E.g.,
there is no ``antiphoton" distinct from the photon.)

Ü3. Generalizations

More generally, we will find propagators of the form (in momentum space)
$$ ë = -{i\over K},âK = Kÿ $$
 corresponding to free actions
$$ S_0 = Çdx¼üÄKÄ $$
 where $K=-ü(õ-m^2)$ in the case just considered.  
(We have neglected the $i·$ in $ë$, which destroys its naive antihermiticity.)
Then the inner
product is defined as above in terms of the Green function by 
again using integration by parts,
$$ iÇd^D x¼[(KA)ÿB -AÿKB]
 = ·(p^0)Èd^{D-1}§^m¼AÿM_m B $$
 to define the operator $M_m$,
which was $·(p^0)üi\onª»{}_m$ in the previous case.  
(For the usual equal-time
hypersurfaces, we use $M_0=-M^0$.  There may be additional implicit
matrix factors in the Lorentz-invariant inner product $AÿB$.)
This in turn defines the inner product
$$ Ò1||2Ô = Ç{d^{D-1}§^m\over (2¹)^{D/2}}¼Æ_1ÿ M_m Æ_2 $$
 and thus
$$ Æ(x) = ·(p^0)È{d^{D-1}§'^m\over (2¹)^{D/2}}¼G(x,x')M'_m Æ(x') $$
 This inner
product gives a nonnegative norm on physical bosonic states, but on
physical fermionic states it is negative for negative energy, because
ordering the initial state to the left of the final state (the wrong ordering
for quantum mechanics) produces a minus sign from the
anticommutativity of the fermions.  (From the explicit integral, this
appears because $K$ is generally second-order in derivatives for bosons,
but first-order for fermions, so $M_m$ has one factor of $p^0$ for bosons
and none for fermions.) 

For physical fields, the (free) field equation will always imply the
Klein-Gordon equation (after gauge fixing for gauge fields).  Thus, the
propagator can always be written as
$$ ë = -{i\over K} = -i{N(p)\over ü(p^2 +m^2 -i·)} $$
 in terms of some matrix kinematic factor $N(p)$.  Using this expression
for the propagator in our above position-space inner product, this
implies (e.g., using $öG$ as in the previous subsection
for the denominator by using a Fourier transform)
$$ Æ_p(x) = öÆ(p)e^{ipÉx}âÜâN(p)M_0(p)öÆ(p) = ¿ öÆ(p) $$
 from integrating over a hypersurface at constant time.
If we generalize to other flat hypersurfaces with timelike normals $n^m$
(e.g., by just Lorentz transforming), we have
$$ N(p)nÉM(p)öÆ(p) = -·(p^0)nÉp öÆ(p) $$
 and finally, by taking linear combinations for different $n$'s,
$$ N(p)M_m(p)öÆ(p) = -·(p^0)p_m öÆ(p) $$

If we choose a basis that is orthonormalized with respect to all
quantum numbers other than momenta 
(i.e., with repsect to spin/helicity and internal
symmetry), we have
$$ Òp,i||p',jÔ = ¶_{ij}{¶(p-p')\over 2¹¶[ü(p^2+m^2)]} $$
 If we ignore coordinate/momentum dependence and focus on just these
other quantum numbers, then it is clear that 
$$ N = [·(p^0)]^{2s}Ý_i|iÔÒi| $$
 where we have included an extra sign factor for negative energy and
half-integer spin from the reordering of states, as explained above. 
In an arbitrary basis, we can generalize to
$$ N(p) = [·(p^0)]^{2s} Ý_i öÆÿ_i(p)öÆ_i(p) $$
 The positive-energy
propagator is then given by a sum over all positive-energy states:
$$ ë_+ = N(p)Ï(p^0)2¹¶[ü(p^2 +m^2)] $$
 The
fact that $K$ is not simply the Klein-Gordon operator is a consequence of
unphysical (gauge/auxiliary) degrees of freedom appearing in the action: 
Then $N$ is a projection operator that projects out the auxiliary degrees
of freedom on shell, and the gauge degrees of freedom on and off (in
unitary gauges), as represented above by a sum over physical states. 
However, more general $N$'s are sometimes used that include unphysical
degrees of freedom; these must be canceled by ``ghosts", similar
unphysical degrees of freedom of the opposite statistics.

\x VB3.1 Demonstrate all these properties for spin (helicity) $ü$ (see
subsections IIA6, IIB6-7, and IIIA4):  Find $M_m$ from integration by parts.
Find $N$ both from inverting to get the propagator and
from summing over physical states.
Show the $NMöÆ$ identity is satisfied.
 ªa For the massless case, use twistors for the solution to the field
equation to find
$$ (M_{ŒÀº})^{©À¶} = -¶_Œ^© ¶_{Àº}^{À¶}·(p^0),ââ
	(N)_{ŒÀº}(p) = ·(p^0)p_Œ Ðp_{Àº} = p_{ŒÀº} $$
 ªb Do the same for the massive Dirac spinor, to find
$$ M_m = ©_m ·(p^0),ââN = Öp +\f{m}{å2} $$
 (Hint:  Consider the rest frames for $p^0>0$ and $<0$.)

\x VB3.2  Use the construction of exercise VB1.2 to define the path
integral for the spinning particle of exercise IIIB1.3.  Show that in the
covariant gauge $v=1$, $Â$ = constant, the propagator can be written as
$$ ë(p) ¾ ÇdżdT¼Ï(T)e^{-Tüp^2-iÅ©Ép} 
	¾ {-i©Ép\over üp^2} = {i\over ©Ép} $$
 up to some arbitrary normalization factor, where $Å=Çd ¼Â$ is the only
gauge invariant part of $Â$ (as $T=Çd ¼v$ is for $v$).

We will find that quantum corrections modify the form of the propagator. 
In particular, it may modify the position $m^2$ of the pole in $-p^2$ and
its residue, as well as adding terms that are analytic near that pole.  For
example, consider a scalar propagator of the form
$$ ë(p) = -i{N\over ü(p^2+m^2-i·)} +R $$
 where $N$ is a constant and $R$ is analytic in $p$.  By the procedure of
``renormalization", $N$ can be set to 1, and $m^2$ can be set to its
original value (see chapter VII).  Alternatively, we can cancel $N$ in the
normalization of external states, and redefine the masses of these states
to coincide with what appears in the propagator.

\x VB3.3  Use this propagator to define the inner product between two
plane waves, and evaluate it explicitly.  Show that $R$ gives no
contribution, and the plane waves need factors of $åN$ to maintain
their normalization.  (Hint:  What is the wave equation corresponding to
$ë$, and how is it related to $M_m$?  You can also consider the relation
of $N$ to $ë_+$.)

Away from the pole, at higher values of $-p^2$ than $m^2$, there will
also be cuts corresponding to mutiparticle states.  Although these
higher-energy intermediate states in the propagator will contribute to
the time development even for on-shell states (those satisfying
$p^2+m^2=0$ asymptotically), in S-matrix elements we can ignore such
contributions on external lines, using our modified definition for
evaluating the asymptotic limit for the S-matrix (see subsection VA4).

\x VB3.4 Consider the general scalar propagator
$$ ë(p) = -iÇ_0^¥ dµ¼{¨(µ)\over ü(p^2+µ^2-i·)},ââ
	¨(µ) = ¶(µ-m) +Ï(µ-2m)§(µ) $$
 which contains a pole at mass $m$ and contributions from multiparticle
states at mass $2m$ and higher.  Fourier transform from energy to time. 
Use this propagator to define the time development of a momentum
eigenstate satisfying the free wave equation asymptotically, using the
$1+i·$ prescription of subsection VA4 to define the asymptotic limit:
$$ Æ(t) ­ \lim_{t_i£-¥(1+i·)}ÇG(t,t_i)üi\onª»_{t_i}Æ_0(t_i) $$
 and show that $§(µ)$ does not contribute:
$$ (õ-m^2)Æ_0 = 0âÜâÆ(t) = Æ_0(t) $$


Ü4. Wick rotation

As in the nonrelativistic case, the $i·$ prescription can also be fixed by
the infinitesimal Wick rotation (see subsection VA5)
$$ t £ (1-i·)t,âE £ (1+i·)EâÜâ
	{1\over p^2 +m^2} £ {1\over p^2 +m^2 -i·} $$
 However, in the relativistic case, a finite Wick rotation gets rid of not
only $i$'s but also the annoying minus signs associated with the
Minkowski metric.  We now replace all timelike coordinates, including
proper time, with spacelike coordinates:
$$ t £ -it,ââ  £ -i  $$
 In addition, for every vector field $V^a$ we replace
$$ V^0 £ -iV^0ââ(V^i £ V^i) $$
 and similarly for tensor fields.  (Here we have defined Wick rotation in
the first-quantized sense: on all explicit coordinates and momenta, as
well as on explicit Lorentz indices.  For example, in the field theory action
we rotate the explicit derivatives and the integration measure, rather
than the arguments of the fields.)

$$ \fig{wick} $$

Note that $E$ as defined in the nonrelativistic case was $-p_0$, which is
the same as $p^0$ only in Minkowski space:
$$ p^0 £ -ip^0,ââp_0 £ +ip_0 $$
 Furthermore, there is some apparent ambiguity in how to change the
integration measure, corresponding to how the integration contours are
rotated (i.e., changes in the limits of integration).  In particular, we see
from subsection VA5 that the contour rotation for $E$ is actually in the
opposite direction of that for $t$, consistent with Fourier transformation. 
(Effectively, we keep the extra $i$ for $Çdp$ from rotating $p_0$, while
dropping the $-1$ from $p_0ªp^0$ because of the usual absolute value in
the Jacobian in real changes of variables.) The net result is the naive
change for $Çdx$, while that for $Çdp$ preserves the inverse Fourier
transform:
$$ Çdx £ -iÇdx,ââÇdp £ iÇdp;ââ¶(x) £ i¶(x),ââ¶(p) £ -i¶(p) $$

When manipulating explicit expressions, the factors of $i$ on
coordinates/momenta and fields can be transferred to the constant
tensors contracting their indices:  The net effect is that the
Wick rotation is equivalent to changing just $ $, the integration measures,
the flat-space metric, and the Levi-Civita tensor:
$$   £ -i ,ââ{»\over » } £ i{»\over » } $$
$$ Çd  £ -iÇd ,ââÇdx £ -iÇdx,ââÇdp £ iÇdp $$
$$ ú_{mn} £ ¶_{mn},ââ·_{abcd} £ -i·_{abcd} $$
 So now the inner product is positive definite:  We have gone from
Minkowski space to Euclidean space.

For example, for the relativistic particle in the gauge $v=1$, the
propagator is now
$$ ë(p) = ÇdT¼Ï(T)e^{-T(p^2+m^2)/2} = {1\over ü(p^2+m^2)} $$
 The integral is automatically convergent because $p^2+m^2$ is now
positive definite.  If we examine our transformation of timelike
components in terms of the complex $p^0$ plane, we see that we have
just rotated the contour from the real axis to the imaginary axis, through
an angle $¹/2$, which avoids the poles at $p^0=à(¿-i·)$.  (There is
actually a slight cheat in the massless case, since the two poles converge
near vanishing 3-momentum, where $¿=0$, but this problem can be
avoided by an appropriate limiting procedure.)  Note that in the
relativistic case the Euclidean propagator is completely real also in
momentum space; even the overall $-i$ has been killed.  (Compare the
nonrelativistic case in subsection VA5, where $-i/(H-E)£1/(H-iE)$.)  This
follows from first Wick rotating in position space, then performing the
Fourier transform as usual (avoiding an extra $-i$ from rotating the $Çdt$
in the Fourier transform).

\x VB4.1 Find the propagator in time-3-momentum space, $öG(p^i,t; p'^i,t')$,
after Wick rotation.  (I.e., Wick rotate $ë(p)$ first, then Fourier
transform.)

\x VB4.2 Find the ÓmasslessÕ propagator in 4D Minkowski coordinate
space, including the $i·$, by
 ªa Fourier transforming the Schwinger-parametrized
momentum-space propagator in Minkowski space (including the
``Minkowski" $ $),
 ªb doing the same entirely in Euclidean space, and then Wick rotating the
time back to Minkowski space, and
 ªc Fourier transforming both of the above cases without using the
Schwinger parameter, first doing the energy integrals as in the previous
section.  (Hint:  Use rotational invariance to point $\vec x$ in a particular
direction to simplify the angular integration.)

\x VB4.3 Although the propagator in momentum space is most useful for
scattering of plane waves, its position-space dependence is more useful
for bound states/scattering of localized sources:
 ªa  Evaluate the Wick-rotated propagator in arbitrary dimensions D for
large $x=å{x^2}$ by
\itemutem{(1)} Fourier transforming the Schwinger-prametrized form,
\itemutem{(2)} performing the (Gaussian) $p$ integration before the $T$, and
\itemutem{(3)} using the steepest descent approximation on the exponential to
approximate the $T$ integral (but don't bother sticking the power of $T$
multiplying the exponential in it as a log).
 ª{} Why does this approximation
correspond to large $x$?  (It may be useful to make the redefinition
$T£(x/m) $.)  You should find the result
$$ ë(x) ® å{2¹}m^{(D-3)/2}x^{-(D-1)/2}e^{-mx} $$
 After Wick rotating back, the exponential becomes a phase inside the
lightcone, while outside it gives exponential damping, as in quantum
mechanical barrier penetration.  (In the massless case, there is damping
away from the lightcone both inside and outside, but only by powers,
determined by dimensional analysis.)
 ªb  Show the above result is exact in D=1 and 3 by 
\itemutem{(1)} Fourier transforming ÓwithoutÕ the Schwinger parameter, 
\itemutem{(2)} performing a
contour integral over the magnitude of $p$ by closing it appropriately and
picking up the contributions at the poles at $p=àim$, and 
\itemutem{(3)} for D=3,
fixing $x^m$ in a particular direction and doing the angular part of the $p$
integration.  
 ª{} (This method can also be used to obtain the exact
corrections to the above in terms of elementary functions for higher odd
D.)  For the physical case of the potential produced by a static point
source in 3 spatial dimensions, we find $ë(x)=å{2¹}e^{-mx}/x$, so at short
range the Coulomb potential is unmodified, while it is exponentially
damped at range $1/m$.
 ªc  Check the results for {\bf b} by taking the massless limit, and
comparing to the analogous result using the method of {\bf a}, but doing
the $T$ integral exactly for those cases.  (Warning:  The result is infinite
in D=1, and some type of ``regularization" must be used to subtract an
infinite constant, leaving a finite $x$-dependent remainder.) 

The corresponding effect on the action, where we path-integrated
$$ e^{-iS}¼:âS = Çd ¼ü(vm^2 -v^{-1}Àx{}^m Àx{}^n ú_{mn}) $$
 is to integrate the Wick rotated expression
$$ e^{-S}¼:âS = Çd ¼ü(vm^2 +v^{-1}Àx{}^m Àx{}^n ¶_{mn}) $$
 This method also applies to relativistic field theory.  Wick rotation
$t£-it$ of the kinetic term
$$ e^{-iS_0}¼:âS_0 = Çdx¼\f14[ú^{mn}(»_m Ä)(»_n Ä) +m^2 Ä^2] $$
  now gives
$$ e^{-S_0}¼:âS_0 = Çdx¼\f14[¶^{mn}(»_m Ä)(»_n Ä) +m^2 Ä^2] $$

This has an interesting consequence in the complete action (including
interactions). Positivity of the energy implied the non-time-derivative
terms in the action had to be positive, but the time-derivative terms are
now the same sign in the Wick-rotated action, effectively the
same as adding an extra spatial dimension:  The energy $T+U$ is the same
as the Wick-rotated Lagrangian ($-T+U£T+U$).  So we can replace the
condition of positivity of the energy with positivity of the Wick-rotated
action, which we need anyway for path-integral quantization.  (Note that
for the particle this again requires $v³0$.)

Although Wick rotation is thus very useful for application to
``intermediate" states, it can never be applied to physical
(i.e., initial and final) states:  For example, $p^2+m^2=0$ has no
solution in Euclidean space, since each term is strictly positive.
Typically, this means that one performs quantum calculations
first in Euclidean space, then Wick rotates back to Minkowski
space before applying physical state conditions.

Unfortunately, the simple results for Wick rotation obtained here for spin
0, although they generalize to spin 1, do not work so simply for other
spins.  (Consider, e.g., trying it in spinor notation.)  However, the method
can still be applied to the coordinates, and simplifies momentum integrals,
since one can avoid contours, $i$'s, and Minkowski minus signs.

\refs

£1 V.A. Fock, ÓPhys. Z. Sowj.Õ É12 (1937) 404;\\
	E.C.G. St¬uckelberg, ÓHelv. Phys. ActaÕ É14 (1941) 322, Óloc. cit.Õ (IA);\\
	Y. Nambu, ÓProg. Theo. Phys.Õ É5 (1950) 82,\\
	R.P. Feynman, ÓPhys. Rev.Õ É80 (1950) 440:\\
	quantization of relativistic mechanics in terms of proper time.
 £2  J. Schwinger, ÓPhys. Rev.Õ É75 (1949) 651, É82 (1951) 664:\\
	his parameters.
 £3 J. Polchinski, ÓCommun. Math. Phys.Õ É104 (1986) 37:\\
	gives a more rigorous discussion of path-integral quantization of the
	relativistic particle.
 £4 St¬uckelberg, Óloc. cit.Õ (IA);\\
	R.P. Feynman, ÓPhys. Rev.Õ É76 (1949) 749, 769:\\
	relativistic propagator.

\unrefs

Û9 C. S-MATRIX

A Órelativisitic quantum field theoryÕ is defined by three properties:  
\item{(1)} ÓPoincar«e invarianceÕ (``relativistic") is the basic result of special
relativity.  Its consequences have been accurately observed both
macroscopically and (sub)micro\-scopically, and no violations are known. 
\item{(2)} ÓUnitarityÕ (``quantum") is the main mathematical result of quantum
mechanics:  Any quantum theory can be considered as the corrections to
the classical theory implied by unitarity.  This is one way to define
perturbation theory, and is equivalent to the usual (JWKB) expansion in
$\h$.  (The other major axioms of quantum mechanics are concerned
with the physical interpretation of the quantities calculated, such as the
preparation and measurement of states.)  Quantum mechanics also has
been accurately verified, with no observed violations.    
\item{(3)} ÓCausalityÕ
(``field theory") appears in many areas of physics, formulated in many
ways.  The strongest way to state causality, in a way independent of
special relativity and quantum mechanics, is as locality:  All interactions
happen at a point; there is no action at a distance.  This means that any
force applied by an object at one point in spacetime on another
elsewhere(/when) must be mediated by yet another object that carries
the effect of that force between the two.  The most accurate
verifications of this principle have been through the predictions of
relativistic quantum field theory.  Note that ``locality" is what ÓdefinesÕ spacetime: For example, in quantum mechanics $x$ and $p$ are treated on an equal footing. Usually $p$ is defined as the generator of a symmetry, but this definition can be obscured in a translationally noninvariant potential or in curved spaces (like a particle on a sphere, or in general relativity). But the fact that interactions are local in $x$ (and are no more than quadratic in $p$) tells us that events occur at a point that can sensibly be interpreted as a ``position".

We now define the perturbation expansion of the S-matrix, and give its
general properties.

Ü1. Path integrals

Path integrals for relativistic quantum field theory in four dimensions
are better in every way than canonical quantization.  They are
\item{(1)} easier to learn and apply: just Gaussian integrals.
\item{(2)} more heuristic: no ``Dirac sea" or harmonic oscillators
(where are the springs?).
\item{(3)} less mathematical: no operators, so no time-ordered
products (much less ``T*" ones), Wick contraction, normal ordering, etc.
\item{(4)} more efficient: Functionals make combinatoric factors automatic.
\item{(5)} more rigorous: Constructive quantum field theory has proven
the existence of certain relativistic quantum field theories in less than
four dimensions (working in Euclidean space).
\item{(6)} more covariant: The action and Feynman rules are, so the
middle steps should be.

As we have seen in simpler examples (subsection VA2), perturbation for
path integrals is based on Taylor expansion of the exponential of
the ``interaction" (part of the exponent higher than quadratic in
the functional integration variables).
Since general transition amplitudes involve also wave functions, we also
Taylor expand them:  In field theory, this is an expansion in the number of
particles.  We therefore write
$$ ï[Ä] = Ý_{N=0}^¥ \f1{N!}Ç{d§_1^m...d§_N^n\over (2¹)^{ND/2}}¼
	Ä(x_1)...Ä(x_N)M_{1m}...M_{Nn}Æ_N(x_1,...,x_N) $$
 where we have used the covariant inner product of subsection VB3. 
(The surfaces of integration are at $t=à¥$.)  We have also used the ÓfreeÕ
N-particle wave function $Æ_N$,
$$ K_1 Æ_N = ... = K_N Æ_N = 0 $$
 which is sufficient to describe particles at $t=à¥$.  Amplitudes for such
asymptotic states are elements of the S-matrix, as defined in the
interaction picture (see subsection VA4).  In practice we choose a
particular value of $N$, and use a basis element for $Æ_N$, namely the
product of N 1-particle wave functions:
$$ Æ_N(x_1,...,x_N) = Þ_{i=1}^N Æ_{Ni}(x_i)¼+¼permutationsâÜâ
	 ï[Ä] = Þ_{i=1}^N ÒÄ||Æ_{Ni}Ô $$
 In principle, if there are bound states in the theory, we can consider
similar wave functions, but besides $Ä$ we expand in the composite field
describing the bound state.  It should be possible to discover such states
by looking at the properties of the amplitudes of the $Ä$ states.  (For
example, a two-particle bound state would show up in the amplitude
describing the scattering of two particles.)  

Note that the fields $Ä$ are real (or we sum over both $Ä$ and $Ä*$), 
while the wave functions ($ï$ or $Æ$) are complex:  As usual in
quantum mechanics, we work in a complex Hilbert space, but often 
expand over a real basis.  For example,
$Æ(q)=Æ(0)+Æ'(0)q+üÆ''(0)q^2+...$ or 
$({Æ_1\atop Æ_2})=Æ_1\left({1\atop 0}\right)+Æ_2\left({0\atop 1}\right)$.
 Since in a covariant approach
we treat particles and antiparticles on an equal footing, $ï[Ä]$ should
include both the initial and final wave function:  Thus, the factors in $ï$
at $t=-¥$ can be interpreted as the usual positive-energy multiparticle
wave functions, while the factors at $t=+¥$ can be interpreted as the
complex conjugate of the usual positive-energy multiparticle
wave functions.  Since they each have one sign of the energy, they
are necessarily complex.  However, for a real
field $Ä$ we can have only $ÒÄ|ÆÔ$, while for a complex field $Ä$
we distinguish $ÒÄ|ÆÔ$ as representing a particle at $t=-¥$ or an
antiparticle at $t=+¥$, and $ÒÄ*|ÆÔ$ as an antiparticle at $t=-¥$ or a
particle at $t=+¥$.  (As usual, which we choose to call particle and
which antiparticle is relative, because of CPT invariance.)

We therefore want to evaluate the path integral
$$ \A = ÇDļe^{-iS[Ä]}ï[Ä] $$
 (Because the wave functions have free time dependence, the transition amplitude $\A$ is in this case an S-matrix amplitude $\S$.)
Separating out the free and interacting pieces of the (gauge-fixed)
action,
$$ S = S_0 +S_I = Çdx¼üÄKÄ +S_I $$
and using the integration identity
$$ \boxeq{ Ç{du\over å{2¹}}¼e^{-uMu/2}f(u+v) = 
	Ç{du\over å{2¹}}¼e^{-uMu/2}e^{u»_v}f(v) ¾
	e^{»_v M^{-1}»_v/2}f(v) } $$
 at $v=0$, we can evaluate the path integral as
$$ \A = \left. exp \left( -iÇdx¼ü{¶\over ¶Ä}{1\over K}{¶\over ¶Ä}\right)
	e^{-iS_I[Ä]}ï[Ä] \right|_{Ä=0} $$
$$ -iÇdx¼ü{¶\over ¶Ä}{1\over K}{¶\over ¶Ä} = 
	Çdx¼dx'¼ü{¶\over ¶Ä(x)}ë(x-x'){¶\over ¶Ä(x')} $$
 We have dropped the determinant factor, since in our case it will be
only an overall constant coming from the kinetic operator $K$.
We then absorbed this proportionality constant into the definition of
$DÄ$, as we did for the free particle in subsection VA2.  In this
case, this normalization is fixed by the ``free" part of the S-matrix.

It will prove convenient to distinguish the ends of propagators that
attach to $S_I$ from those that attach to $ï$, so using the identity
$$ f(»_x)g(x,x+y) = f(»_x' +»_y')g(x',y')ââ(x'=x,ây'=y+x) $$
 evaluated at $x=y=0$, we rewrite this expression as
$$ \A = \left. exp \left( -iÇ
	ü{¶\over ¶Ä}{1\over K}{¶\over ¶Ä}
	+{¶\over ¶Ä}{1\over K}{¶\over ¶\Ä}
	+ü{¶\over ¶\Ä}{1\over K}{¶\over ¶\Ä} \right)
	e^{-iS_I[Ä]}ï[\Ä] \right|_{Ä=\Ä=0} $$
 We then evaluate the derivatives that act on only one or only the other:
$$ \eqalign{
	\A = \left. exp \left( -iÇ{¶\over ¶Ä}{1\over K}{¶\over ¶\Ä}\right)
		Z[Ä]÷ï[\Ä] \right|_{Ä=\Ä=0} \cr
	÷ï[Ä] = \leftÓ exp \left( -iÇ
		ü{¶\over ¶Ä}{1\over K}{¶\over ¶Ä}\right) ï[Ä] \rightÕ \cr
	Z[Ä] = \leftÓ exp \left( -iÇ
		ü{¶\over ¶Ä}{1\over K}{¶\over ¶Ä}\right) e^{-iS_I[Ä]}\rightÕ\cr}
	ââ\vcenter{\hbox{\fig{amplitude}}} $$

Then we move the differential operators into the wave functional:
$$ \A = \left. öï\left[{¶\over ¶Ä}\right] Z[Ä] \right|_{Ä=0} $$
$$ öï\left[{¶\over ¶Ä}\right] = \left. exp \left( -iÇ
	{¶\over ¶Ä}{1\over K}{¶\over ¶\Ä}\right) ÷ï[\Ä]\right|_{\Ä=0} $$

$Z[Ä]$ (the ``generating functional" for the S-matrix) contains all
propagators with $S_I[Ä]$'s attached at both ends, and forms the basis of
the perturbation expansion.  From the integration identity above, we can
also write it as
$$ \boxeq{ Z[\Ä]  = ÇDļe^{-i(S_0[Ä] +S_I[Ä+\Ä])} } $$
 Effectively, we have just taken the functional integral $ÇDļe^{-iS[Ä]}$
and separated the field into a ``quantum field" $Ä$ (the integration
variable) and a ``background field" $\Ä$, where $\Ä$ includes the
asymptotic states, which propagate to infinity, while $Ä$ vanishes at
infinity (or at least goes to a constant) fast enough to allow the usual
integration by parts in performing the functional integral.  Thus $\Ä$
gives the boundary value of the field.  This is essentially the same as the
general prescription for path integrals given in subsection VA2, except
that we take $\Ä$ to be arbitrary for convenience of functional
differentiation, and we drop free $\Ä$ terms, which were incorporated
into $֕$ (i.e., we expand just $S_I$).

$÷ï[Ä]$ is the result of contracting some pairs of the one-particle wave
functions with propagators.  This gives the usual inner product of those
one-particle states:  Since for any one-particle wave function satisfying
the free field equations we have the propagator identity
$$ \boxeq{ Ç{d§^m\over (2¹)^{D/2}}Æ(x)M_m ë(x-x') = Æ(x') } $$
 such contractions give
$$ Ç{d§^m d§'^n\over (2¹)^{D}}Æ(x)M_m ë(x-x')M'_n Æ'(x') $$
$$ = Ç{d§^m\over (2¹)^{D/2}}Æ(x)M_m Æ'(x) = ÒÆ*||Æ'Ô = ÒÆ'*||ÆÔ $$
 where the inner product vanishes unless $Æ$ and $Æ'$ have opposite
energies (i.e., one is incoming and one is outgoing).  This is the boring part
of the S-matrix element:  It represents the corresponding particles not
interacting at all.  (For example, in the free case $S_I=0$ we have $Z=1$,
and $\A=÷ï|_{Ä=0}$ consists of only such inner products.)  For most
purposes we just factor out such free inner products, and consider only
processes where all particles interact.  

Finally, the conversion from $÷ï$ to $öï$ replaces all the on-shell inner
products with integrals over all spacetime:  Using the propagator identity
from above,
$$ öï\left[{¶\over ¶Ä}\right] = 
	Ý_{N=0}^¥ \f1{N!}Ç{d^{ND}x\over (2¹)^{ND/2}}
	÷Æ_N(x_1,...,x_N){¶\over ¶Ä(x_1)}...{¶\over ¶Ä(x_N)} $$
 where $÷Æ_N$ appears in $÷ï$ exactly as $Æ_N$ in $ï$: 
$$ ÷ï[Ä] = Ý_{N=0}^¥ \f1{N!}Ç{d§_1^m...d§_N^n\over (2¹)^{ND/2}}¼
	÷Æ_N(x_1,...,x_N)M_{1m}...M_{Nn}Ä(x_1)...Ä(x_N) $$
 Usually we represent $Æ_N$ as the product of single-particle wave
functions,
$$ öï_N\left[{¶\over ¶Ä}\right] = 
	Þ_{i=1}^N \left( ÇÆ_{Ni}{¶\over ¶Ä}\right)âÜâ\A_N = 
	\left. Þ_{i=1}^N \left( ÇÆ_{Ni}{¶\over ¶Ä}\right) Z[Ä]\right|_{Ä=0} $$
 in which case $öï$ simply replaces each field $Ä$ in $Z$ with one of these
wave functions.  

$$ ë £ \leftÓ 
	\eqalign{ ë & : S_I(¼)S_I \cr 1 & : Æ(¼)S_I \cr Ò¼||¼Ô & : Æ(¼)Æ \cr}
	\right.ââ\vcenter{\hbox{\fig{amplified}}} $$

Thus, of the three types of propagators, only the ones that connected
two factors of $S_I$ remain, all inside $Z$; the ones that connect the
wave functions to $Z$ have been replaced with just spacetime integrals,
while those connecting the wave functions to each other have become
the usual spatial integrals for the Hilbert-space inner product.

\x VC1.1 Show that the amplitude can also be written as
$$ \A = \left. ×ï\left[{¶\over ¶Ä}\right]Z[Ä]ï_0[Ä]\right|_{Ä=0} $$
$$ ×ï\left[{¶\over ¶Ä}\right] = 
	Ý_{N=0}^¥ \f1{N!}Ç{d^{ND}x\over (2¹)^{ND/2}}
	Æ_N(x_1,...,x_N){¶\over ¶Ä(x_1)}...{¶\over ¶Ä(x_N)} $$
$$ ï_0 = e^{ÒÄ||ÄÔ/2} $$
 making all inner products of non-scattering particles explicit.  
(Hint:  Write both this and the earlier form of the amplitude
completely explicitly, in terms of operators on $Z[Ä]ï[\Ä]$.)
We can
interpret $ï_0$ as the free ``vacuum wave functional" and $Zï_0$ as the
interacting vacuum wave functional.

Ü2. Graphs

Before giving applications of these rules, we consider a few general
properties.  A convenient way to describe the terms in the expansion of
the two exponentials in $Z$ is pictorially, by ``Feynman
diagrams/graphs".  Each factor of $S_I$ is a ``vertex" of the graph,
represented by a dot in the diagram; each factor of the propagator $1/K$
is a ``link" in the graph, represented by a line connecting the two dots
representing the two factors of $S_I$ on which each $¶/¶\Ä$ acts.  (Both
derivatives can also act on the same factor of $S_I$, giving a loop.)  So
any term in the expansion of $Z$ is represented by a diagram consisting
of a bunch of dots (interaction vertices) connected by lines
(propagators).  When we want to ``draw" the amplitude $\A$ itself, we
also draw additional lines, each with one end attached to a vertex and
one end unattached.  These ``external lines" represent the one-particle
wave functions coming from $ï_N$, and not propagators (``internal
lines").  While in the diagram for $Z$ each vertex can have $\Ä$
dependence, in the diagram for $\A$ there is none, and the number of
lines (internal and external) coming from any vertex explicitly indicates
the order in $\Ä$ of the corresponding term in
$S_I$.  

The physical interpretation of these diagrams is simple:  The lines
represent the paths of the particles, where they act free, while the
vertices represent their collisions, where they interact.  These diagrams
are generally evaluated in momentum space:  We then can associate a
particular momentum with each line (propagator), and momentum is
conserved at each vertex.  An arrow is drawn on each line to indicate the
direction of ``flow" of the momentum.  (Otherwise there is a sign
ambiguity, since complex conjugation in position space changes the
sign of the momentum.)  Then the sum of all momenta flowing into (or all
out of) a vertex vanishes.  The momentum associated with a line is then
interpreted as the momentum of that particle, with the arrow indicating
the direction of flow of the proper time $ $.  ($p$ changes sign with $ $.) 
When evaluating an S-matrix element, the fact that the external-line
wave functions satisfy the free wave equation means the external
momenta are on-(mass-)shell ($p^2+m^2=0$); on the other hand, this is
not true of the momenta on the internal lines, even though those
particles are treated as free.

Some graphs in the S-matrix are disconnected:  They can be divided into
separate parts, each with a subset of the particles that interact with
each other but not with the other subsets.  For convenience, we consider
only the connected graphs:  If we write
$$ Z[\Ä] = e^{-iW[\Ä]},ââ
	\A_c = öï\left[{¶\over ¶\Ä}\right](-i)W[\Ä]|_{\Ä=0} $$
 then $W$ is the generating functional for the connected S-matrix $\A_c$. 
To prove this relation between $Z$ and $W$, we first note that it is just
the combinatorics of the graphs, and has nothing to do with spacetime. 
Therefore, it is sufficient to consider the simple (unphysical) case where
the action has no derivatives.  Since the propagator is then local,
connectedness is equivalent in this case to locality.  We then observe that
the lack of derivatives allows the functional integral to be factorized
explicitly into ordinary integrals at each point in spacetime:
$$ Z[\Ä] = ÇDļe^{-iÇdx¼L(Ä(x),\Ä(x))} = Þ_xÇdÄ(x)¼e^{-iL(Ä(x),\Ä(x))} $$
$$ = Þ_x \Z(\Ä(x)) = e^{-iÇdx¼\W(\Ä(x))} = e^{-iW[\Ä]} $$
 Thus, this $W$ is local, and therefore connected; this implies $W$ is
connected in the general case.  The simplest kind of connected graph is a
``tree" graph, which is a graph that has no closed paths; the rest are
called ``loop" graphs.

``One-particle-irreducible" (1PI) graphs are defined to be those
connected graphs that can't be disconnected by severing a single
propagator.  It then follows that any connected graph can be represented
as a generalized tree graph, whose ``vertices" (ÓincludingÕ two-point
vertices) are actually 1PI graphs.  We then define the ``effective action"
$ý[Ä]$ to be the classical action plus all 1PI loop graphs.  Note that the
vertices of the original action are the 1PI tree graphs; thus $ý$ is also the
classical kinetic term plus all (tree and loop) 1PI graphs.  Actually, since
the 1PI tree graphs are $-iS_I$, we define the classical part of $ý$ to be
$S$, but the quantum part to be the quantum 1PI part of $W$.  Of course,
the effective action is nonlocal.  However, the tree graphs that follow
from this action are exactly all the connected graphs of the original
action:  This is clear for all but the 2-point vertices from the definition. 
For the propagator and its relation to the 2-point 1PI loop graphs, we
simply compute the expression following from $ý$:  Denoting the 2-point
1PI loop operator as $A$, the kinetic operator of $ý$ is $K+A$.  The
propagator following from $ý$ is then
$$ {1\over K+A} = {1\over K} -{1\over K}A{1\over K}
	+{1\over K}A{1\over K}A{1\over K} - ... $$
 But this is exactly the result of the complete propagtor (including loop
graphs) following from the original action.

$$ \fig{connected} $$

This quantum modification of the propagator leads us to reanalyze our
prescription for evaluating the S-matrix:  For example, even in the
simplest case, where this $A$ is just a constant, the full propagator
differs from the free propagator by a change in the mass.  This means the
mass of asymptotic states should also be changed, which invalidates part
of our evaluation of the S-matrix in the previous subsection.  Similar
problems occur when $A$ is proportional to $K$, which changes the
normalization of asymptotic states.  There are two ways to fix these
problems:  (1) We compensate by modifying the kinetic term in the
classical action, replacing $K$ with $K$ minus such local contributions
from $A$.  Treating these new terms as part of the interaction in our
derivation of the S-matrix, so our normalization and mass in the free
propagator are unchanged, these ``interaction" terms cancel the
unwanted terms in the quantum propagator, so it then has the same
residue and pole as the free one.  This procedure is known as
``renormalization", and will be discussed further in chapter VII, primarily
for the purpose of eliminating infinities.  (2) Alternatively, we modify our
derivation of the S-matrix to take the full propagator into account.  The
easiest way to see this change is to remember that by definition the
S-matrix follows from treating the effective action as a classical action
(except for its nonlocality and nonhermiticity), but keeping only the
``tree" graphs.  Then clearly (a) the quadratic part $ý_0$ of $ý$ is used to
define the asymptotic states, and (b) instead of eliminating all free
propagators except those connecting factors of $S_I$, we eliminate all
full propagators (found from $ý_0$) except those connecting factors of
$ý_I$, the nonquadratic part of $ý$.  In other words, we modify our
earlier definition of the S-matrix by dropping all graphs that have ÓanyÕ
quantum correction to external lines.  Thus, this procedure can also be
applied in the case of renormalization; in fact, it should be applied in
general, simply because it allows us to immediately ignore many graphs. 
It also allows us to avoid confusion resulting from attaching wave
functions of the wrong mass to propagator corrections:  E.g., in
momentum space, we would have to interpret ambiguous factors such as
$¶(K)A(1/K)...$, where the factor of $¶(K)$ comes from a plane-wave
wave function.

$$ \fig{amputate} $$

This analysis of the quadratic part of $ý$ also leads us to examine the
terms of lower order: constant and linear.  The constant term is just a
normalization, and should be dropped.  (This is not true in the case of
gravity, where a constant term in the Lagrangian is not gauge invariant
by itself.)  The linear term describes the decay of a particle into the
vacuum:  It implies we have the wrong vacuum.  A linear term necessarily
has no derivatives (otherwise it is a boundary term, which vanishes by
our boundary conditions); it is part of the ``effective potential" (a
generalization of the potential energy, whose contribution to a classical
mechanics Lagrangian contains no time derivatives; see subsection
VIIB2).  The existence of a linear term means that the minimum of the
effective potential, i.e., the true vacuum, is not described by vanishing
fields.  To correct this situation we therefore apply the same procedure
as for the classical action (chapter IV):  (1) Shift the appropriate fields
by constants, to put us at the minimum of the potential, and (2) use the
new quadratic terms in the potential to determine the true masses of
states defined by perturbation about this new vacuum.  Again eliminating
any constant terms, the resulting $ý$ has only quadratic and higher-order
terms.

To summarize, the general procedure for calculating Feynman graphs is:
\item{(1)} Calculate the effective action, i.e., the 1PI graphs.    
\item{(2)} Shift the
scalars to put them at the minimum of the effective potential, dropping
the resultant constant, to reveal the true masses of all particles.    
\item{(3)}
Calculate the S-matrix from diagrams without external-line corrections,
with external wave functions whose normalization and masses are
determined by the zeroes of the kinetic operators in the shifted $ý$.

Another use of the effective action, besides organizing the calculation of
the S-matrix, is for studying low-energy behavior:  This means we apply
an expansion in derivatives, as in first-quantized JWKB (see subsection
VA2).  Of most interest is the lowest order in the approximation, where all
fields are effectively constant:  This gives the effective potential.  (In
practice, ``all fields" means just the scalars, since constant spinor fields
are not generally useful, while higher spins are described by gauge fields,
whose constant pieces can be set to vanish in an appropriate
formulation:  E.g., the constant piece of the metric tensor can be
attributed to a scalar --- see subsection IXA7.)  However, the definition
of ``1PI" graphs is ambiguous, depending on how we define ``particle": 
For example, if we include auxiliary fields in the effective action (as in
supersymmetry, but also for bound-state problems: see subsections VIIB3
and 6), the result at fixed order in any expansion parameter ($\h$,
coupling, etc.) is different, since the auxiliaries get contributions at each
order, so elminating them by their effective-action field equations mixes
orders.  (E.g.,  $B^2+\h Bf(A)£\h^2 f^2$.)  This is crucial when the
composite fields defined by these auxiliaries, and thus the auxiliaries
themselves, obtain vacuum values.  Therefore, the effective action is
most useful for these purposes when, for appropriate choice of fields and
definition of $\h$, a useful first-quantized semiclassical expansion can be
found.  Another important use of the effective action is that it is gauge
invariant (even in the nonabelian case, when using the background-field
gauge; see subsection VIB8):  Sometimes simplifications due to gauge
invariance are thus easier to see in the effective action than in the
S-matrix.

We now consider an interesting topological property of graphs:  
\item{(1)} For any
graph, if we draw an extra propagator from a vertex to itself or another,
that gives an extra ``loop" (closed circuit) and no extra vertices.  
\item{(2)} Adding a
2-point vertex to the middle of a propagator or external line
gives an extra propagator and no extra loops.  
\item{(3)} Adding an external line to a vertex changes nothing
else.  

\noindent Since any nontrivial (not a lone propagator) connected 
graph can be built up this way from a lone vertex, we find
$$ P - V = L - 1 $$
 for $P$ propagators, $V$ vertices, and $L$ loops (and $E$ external lines). 
The same result follows from counting momentum integrals:  In
momentum space there is an internal momentum, and corresponding
integral, for each propagator, and a momentum conservation condition,
and corresponding $¶$ function, for each vertex.  The only independent
momenta are the external ones (associated with each $\Ä$ in $Z[\Ä]$) and
one momentum vector for each loop.  Thus, after integrating out all the
delta functions, except for an overall momentum conservation $¶$
function for each connected graph, we are left with integrations over
only the loop momenta.  So, we are again led to the above result.

\x VC2.1  For the figure at the beginning of this subsection,
check this identity for each of the 3 connected graphs.
Apply the above construction to produce each of them from
a single vertex.

Ü3. Semiclassical expansion

We can define perturbations by inserting $\h$'s in various ways, as
discussed in subsection IIIA3.  The $\h$ that defines classical mechanics
yields an expansion in derivatives on ``matter" fields (those that describe
classical particles in the limit $\h£0$, as opposed to the ``wave" fields). 
This expansion is covariant as long as the $\h$ multiplies covariant
derivatives.  However, it can't be applied to Yang-Mills fields, and it
doesn't correspond to a diagrammatic expansion.  On the other hand, the
$\h$ that defines classical field theory is an expansion in the number of
``loops", and allows us to group graphs in gauge-invariant sets, since
gauge transformations are not $\h$-dependent:  As in quantum mechanics,
we can perform a JWKB expansion by appropriately inserting $\h$:
$$ \li{ Z[\Ä]  & = ÇDļexp
	\leftÓ-{i\over\h}\left(S_0[Ä] +S_I[Ä+\Ä]\right)\rightÕ \cr & =
	exp\leftÓ-i\hÇdx¼ü{¶\over ¶\Ä}{1\over K}{¶\over ¶\Ä}\rightÕ
	e^{-iS_I[\Ä]/\h} \cr} $$
 The order in $\h$ has a simple graphical interpretation.  We see that
there is a factor of $\h$ for each propagator and a factor of $1/\h$ for
each vertex.  Thus, by the above topological identity, for each connected
graph the power in $\h$ is one less than the number of loops.  We
therefore write
$$ Z[\Ä] = e^{-iW[\Ä]/\h},ââW = Ý_{L=0}^¥ \h^L W_L $$
 where $W_0$ generates the connected ``tree" graphs, which have no
loops.

We know that the leading term in the JWKB expansion is associated with
the classical theory.  We can make this more explicit in the field theory
case by finding the general classical (perturbative) solution to the field
equations from the tree graphs.  Graphically the solution is very simple: 
We replace one $\Ä$ on each tree graph with a propagator, and associate
the end of the propagator with the position of the classical field $ì(x)$. 
If we then act on this $ì$, which is a sum over all tree graphs, with $K$, it
cancels the propagator, leaving a bunch of $ì$'s (also sums over all tree
graphs) connected at $x$, with the appropriate vertex factor.  In other
words, we find $Kì=-¶S_I[ì]/¶ì$, the classical field equations.  

$$ \fig{SD} $$

To prove this, it's convenient to again use functionals, to automatically
keep track of all combinatorics.  The quantum field equations can be
derived from the general identity
$$ ÇDļ{¶\over ¶Ä}f[Ä] = 0 $$
 since we only integrate functionals $f$ that are assumed to fall off fast
enough as $Ä£¥$ to kill all boundary terms.  (This follows from the
perturbative definition of the functional integral.)  In particular, for any
action $÷S$,
$$ 0 = ÇDļi\h{¶\over ¶Ä}e^{-i÷S/\h} = ÇDļ{¶÷S\over ¶Ä}e^{-i÷S/\h} $$
 For our present purposes, we choose
$$ ÷S = S_0[Ä] +S_I[Ä+\Ä] $$
$$ Üâ0 =
	ÇDļ\left({¶S_0[Ä]\over ¶Ä}+{¶S_I[Ä+\Ä]\over ¶Ä}\right)e^{-i÷S/\h}
	= ÇDļ\left(KÄ +i\h{¶\over ¶\Ä}\right)e^{-i÷S/\h} $$
$$ ÜâKÒÄÔ_{\Ä} +{¶W[\Ä]\over ¶\Ä} = 0 $$
 where ``$ÒÄÔ_{\Ä}$" is the expectation value of the field in a background:
$$ ÒÄÔ_{\Ä} = {ÇDļÄe^{-i÷S/\h}\over ÇDļe^{-i÷S/\h}} $$
 (and, of course, $ÇDļe^{-i÷S/\h}=e^{-iW/\h}$).

We now examine the classical limit $\h£0$ of this result by noting that if
we impose the free field equation on the background
$$ K\Ä = 0âÜâ{¶÷S\over ¶Ä} = {¶S[Ä+\Ä]\over ¶Ä} $$
 where $S[Ä]=S_0[Ä]+S_I[Ä]$ is the usual action:  In other words, the field
equations following from $÷S$ are just the usual field equations for the
complete field 
$$ ÷Ä = Ä+\Ä,ââ\Ä = \lim_{x£¥}÷Ä $$
 since we chose our boundary conditions so $Ä£0$ as $x£¥$ (including
$|t|£¥$, whereas $\Ä£0$ only at spatial infinity).  We then apply the
stationary-phase approximation (or, after Wick rotation, the
steepest-descent approximation)
$$ \boxeq{ \lim_{\h£0} ÇDļf[Ä]e^{-i÷S/\h} = 
	\left.\left(f[Ä]e^{-i÷S/\h}\right)\right|_{¶÷S/¶Ä=0}} $$
 for $f=Ä$ and 1 to find
$$ Kì +{¶W_0[\Ä]\over ¶\Ä} = 0âÜâ
	ì = \Ä -{1\over K}{¶W_0[\Ä]\over ¶\Ä} $$
$$ ì ­  \lim_{\h£0}Ò÷ÄÔ_{\Ä} = ÷Ä|_{¶S[÷Ä]/¶÷Ä=0} $$
 Thus, $ì$ is the solution to the classical field equations with boundary
condition $ì£\Ä$, and can be found directly from the classical (0-loop)
part of $W$ by replacing one field with a propagator.

A similar result holds for $W_0$ itself, by taking the classical limit as
above for $f[Ä]=1$:
$$ W_0[\Ä] = S_0[Ä] +S_I[Ä+\Ä] $$
 when evaluated at the result of varying the above with respect to either
argument:
$$ {¶\over ¶Ä}âÜâ{¶S[Ä+\Ä]\over ¶Ä} = K\Ä = 0 $$
$$ {¶\over ¶\Ä}-{¶\over ¶Ä}âÜâ{¶W_0[\Ä]\over ¶\Ä} = -KÄ $$
 The former follows directly from the limiting procedure; the latter we
have just proven equivalent by evaluating the limit for $Ò÷ÄÔ_{\Ä}$.  Since
by definition the effective action $ý$ is related to $W$ in exactly the same
way that $S$ is related to $W_0$ (the trees from $ý$ give the full $W$),
we also have
$$ W[\Ä] = ý_0[Ä] +ý_I[Ä+\Ä] $$
$$ {¶ý[Ä+\Ä]\over ¶Ä} = ÷K\Ä = 0âÛâ{¶W[\Ä]\over ¶\Ä} = -÷KÄ $$
 where $÷K$ is the kinetic operator appearing in $ý_0$, the quadratic part
of $ý$.  (Some care must be taken for the fact that the poles and residues
of $÷K$ in $p^2$ may differ from those of $K$, as discussed in the previous
subsection.)

In practice, if one wants to make use of the classical field equations
perturbatively, one looks at tree graphs with a specific number of
external lines:  For example, in a scalar theory with $Ä^3$ interaction
(assuming $ÒìÔ=0$),
$$ Kì +üì^2 = 0,âì = Ý_{n=1}^¥ ì_nâÜâ
	Kì_n +üÝ_{m=1}^{n-1} ì_m ì_{n-m} = 0 $$
 gives a recursion relation for the term $ì_n$ that is $n$th order in $\Ä$.

\x VC3.1  Consider the relativistic Schr¬odinger (Klein-Gordon) equation for
a scalar wave function $Æ$ in an external scalar potential $Ä$:
$$ (K+Ä)Æ = 0 $$
 (If you find it less confusing, you can consider the nonrelativistic case
$K=\vec pÊ{}^2/2m -E$, where $H=\vec pÊ{}^2/2m+Ä$.)  Find the
perturbative solution for the quantum mechanical (one-particle) S-matrix
for $Æ$ (see exercise VA4.1).  Show that this agrees with the contribution
to the field-theoretic S-matrix for the Lagrangian
$$ L(Æ,Ä) = Æ*(K+Ä)Æ +L_Ä(Ä) $$
 coming from tree graphs with an external $Æ$ line, an external $Æ*$ line,
and an arbitrary number of external $Ä$ lines.

\x VC3.2  Consider, instead of the background field $\Ä$, a ``current"
source $J$ that attaches propagators to external lines.  The current is
effectively just a one-point interaction (it caps loose ends of
propagators), so it can be introduced into the generating functional by
the modification
$$ S_I[Ä] £ S_I[Ä] + Çdx¼JÄ $$
 We now have
$$ e^{-iW[J]/\h} = ÇDļexp
	\leftÓ-{i\over\h}\left(S_0[Ä]+S_I[Ä]+Çdx¼JÄ\right)\rightÕ $$
 $Z[J]$ is thus the Fourier transform of $e^{-iS[Ä]/\h}$ with respect to the
conjugate variables $Ä$ and $J$.
 ªa Derive the ``Schwinger-Dyson equations"
$$ \left( \left.{¶S[Ä]\over ¶Ä(x)}\right|_{Ä=i\h ¶/¶J} +J(x)\right)
	e^{-iW[J]/\h} = 0 $$
 ªb Find the classical limit
$$ W_0[J] = S[Ä] +ÇJÄâatâ
	Ä = {¶W_0[J]\over ¶J}âÛâJ = -{¶S[Ä]\over ¶Ä} $$
 (i.e., $W_0[J]$ is the ``Legendre transform" of $S[Ä]$).  Find the
corresponding relation for $ý[Ä]$ and $W[J]$.
 ªc Show that the free part of $W[J]$ is given by
$$ W_{free} = -üÇdx¼J{1\over K}JâÜâ\Ä ­ Ä_{free} = -{1\over K}J $$
 Note that $J$ can be replaced with $\Ä$ in $W[J]$ everywhere ÓexceptÕ
the free term, since in all other terms $J$ appears only in the combination
$(1/K)J$.  Show how this can be done in such a way as to reproduce the
results above for the solution of the classical field equations in terms of
$W[\Ä]$.  Warning:  Before this substitution we use $K\Ä+J=0$, but
afterwards we apply $K\Ä=0$; also beware of integration by parts, since
$\Ä$ does not vanish at $¥$, so the naive substitution $Ä£Ä+\Ä$ is not
very helpful.  (Historically, $Z$ was introduced as a functional of $J$. 
However, the only two applications of Feynman diagrams, the S-matrix
$\A$ and the effective action $ý$, both required that the external-line
propagators resulting from the Feynman rules for $Z[J]$ be ``amputated". 
Therefore, we use background fields exclusively.  The resulting
derivations, generalities, and applications are at least as simple as and
often a little simpler than the corresponding ones with current sources.)

There is a major flaw in this relation between classical field theory
and tree graphs, the ``Klein paradox". The difference is that 
classical field theory uses fields everywhere, not wave functions. 
Thus the propagator must be real (or pure imaginary, 
depending on conventions) to preserve the reality 
(or complex conjugation) properties of the fields; 
usually one uses the retarded propagator. On the other hand, 
in quantum field theory negative-energy states must travel 
backward in time to preserve positivity of the true energy, 
so the complex St¬uckelberg-Feynman propagator must be used. 
Furthermore, in classical field theory the external line factors 
are the fields, which contain both positive and negative energies, 
the same on each line. In quantum field theory, each external line 
carries a different one-particle wave function, 
positive energy in the asymptotic past or negative energy in the 
asymptotic future.

Ü4. Feynman rules

It is usually most convenient to calculate Feynman diagrams in
Wick-rotated (to eliminate $i$'s) momentum space (where massive
propagators are simpler).  The ``Feynman rules" are then read off of the
action as
$$ \boxeq{ S = Çdx¼üÄKÄ +S_I[Ä]âÜâZ[Ä] = e^{-W[Ä]} = 
	exp \left(Çü{¶\over ¶Ä}{1\over K}{¶\over ¶Ä}\right)e^{-S_I[Ä]} } $$
 where in $Z[Ä]$ we simply replace each field $Ä$ with a single-particle
wave function in all possible permutations, since for the case of an
N-particle amplitude we usually write the wave function as the product
of N single-particle wave functions (although more generally it is a linear
combination of these):
$$ \boxeq{ \A_N = öï_N\left[{¶\over ¶Ä}\right] Z[Ä]¼
	\left( \A_{N,c} = -öï_N\left[{¶\over ¶Ä}\right] W[Ä] \right)
	,âöï_N\left[{¶\over ¶Ä}\right]
	= Þ_{i=1}^N \left( ÇÆ_{Ni}{¶\over ¶Ä}\right) } $$

We Fourier transform as
$$ Ä(x) = Çdp¼e^{ipÉx}Ä(p),ââÄ(p) = Çdx¼e^{-ipÉx}Ä(x) $$
 (Of course, $Ä(x)$ and $Ä(p)$ are different functions, but the distinction
should be clear from context.)  In practice we choose the single-particle
wave functions to be eigenstates of the momentum, so
$$ Æ_i(x) = e^{ip_iÉx}öÆ_i(p_i),ââÆ_i(p) = ¶(p-p_i)öÆ_i(p_i)ââ
	(p_i^2 +m^2 =0) $$
 where $öÆ$ is some simple factor (e.g., 1 for a scalar).  Then the external
line factor terms become
$$ Ç dx¼Æ_i(x){¶\over ¶Ä(x)} = öÆ_i(p_i){¶\over ¶Ä(p_i)} $$
 while for propagator terms
$$ Çdx¼ü{¶\over ¶Ä(x)}{1\over K(-i»)}{¶\over ¶Ä(x)} =
	Çdp¼ü{¶\over ¶Ä(p)}{1\over K(p)}{¶\over ¶Ä(-p)} $$
 and for vertex terms
$$ Çdx¼Ä_1(x)...Ä_n(x) = 
	Çdp_1...dp_n¼Ä_1(p_1)...Ä_n(p_n)¶(p_1+...+p_n) $$
 where each of the $Ä$'s in the vertex may represent a field with
derivatives; then we replace $-i»$ on $Ä(x)$ with $p$ on $Ä(p)$.  Thus,
e.g., we have
$$ \A_N = \left[ Þ öÆ_{Ni}(p_i) {¶\over ¶Ä(p_i)}\right] Z[Ä] $$

\x VC4.1  Use the definition
$$ {¶\over ¶÷Ä(p)}÷Ä(p') = ¶(p-p') $$
 to show that
$$ ~{\left({¶\over ¶Ä}\right)}(p) = {¶\over ¶÷Ä(-p)} $$
 where we now use tildes to indicate Fourier transformation.

Note that there is some ambiguity in the normalization of external line
factors, associated with the numerator factor in the propagator
$$ ë = {1\over K} = {N(p)\over ü(p^2 +m^2)} $$
 For nonzero spin (or internal symmetry), we have already discussed the
normalization analogous to that for scalars, namely
$$ ÝöÆÿ(p)öÆ(p) = àN(p) $$
 (with $-$ for negative energy and half-integer spin).  However, there is
already some freedom with respect to coupling constants:  Even for
scalars, if a coupling appears in the kinetic term as a factor of $1/g^2$,
then effectively the kinetic operator is $K=K_0/g^2$, where $K_0$ is the
usual (coupling-independent) one.  Thus $N=g^2 N_0$, so $öÆ=göÆ_0$,
meaning coupling dependence in external lines.  Alternatively, this
external-line factor of the coupling can be included in the definition of
probabilities in terms of amplitudes, which already includes nontrivial
factors because of the use of (non-normalizable) plane waves. 
Furthermore, quantum effects modify the form of the propagator:  Such
effects can be absorbed near the pole $p^2=-m^2$ by a field redefinition,
but often it is more convenient to leave them.  Then $N$ will again have
a constant, (but more complicated) coupling-dependent factor, which
must be canceled in either the external-line factors or probabilities.
(However, note that these questions do not arise in calculations of the
functionals $W[Ä]$ or $ý[Ä]$.)

In tree graphs all momentum integrals are trivial, with the momentum
conservation $¶$ functions at each vertex, and the $¶$ functions of the
external lines, determining internal momenta in terms of external
momenta.  In loop graphs there is a momentum integral left for each loop,
over the momentum of that loop.  The amplitude will always have an
overall $¶$ function for momentum conservation for each connected piece
of the graph.  Since we are always interested in just the connected
graphs, we pull this conservation factor off to define the ``T-matrix": 
Including the factor of ``$i$" from Wick rotating back to Minkowski space,
$$ \S_{connected} = i¶\left(Ýp\right)T $$

In general there will be combinatoric factors associated with a graph. 
These follow automatically from the functional expressions, but can also
be seen from the symmetries of the graph.  Here ``symmetries" means
ways in which the graph can be twisted, with external lines fixed, such
that the graph looks the same, including the types of particles
propagating along the lines.  For example, a graph with 2 vertices that are
connected by n identical propagators would get a factor of 1/n! for that
symmetry.  There are also sign factors from fermions:  Permutation of
external fermion lines gives minus signs, because it involves permutation
of anticommuting fields in the functionals.  Each fermion loop gets a
minus sign for the same reason.  (This is related to the fact that fermionic
integration gives determinants instead of inverse determinants.) 
Explicitly, it comes from evaluating expressions of the form (ignoring
momentum dependence and external fields)
$$ \left( {¶\over ¶ÐÆ}{¶\over ¶Æ} \right) ò 
	\left( {¶\over ¶ÐÆ}{¶\over ¶Æ} \right) (ÐÆÆ) ò (ÐÆÆ) $$
$$ = \left( {¶\over ¶ÐÆ}{¶\over ¶Æ} \right) ÐÆ
	\left[ \left( {¶\over ¶Æ}Æ \right) \left( {¶\over ¶ÐÆ}ÐÆ \right) \right] ò
	\left[\left({¶\over ¶Æ}Æ\right)\left({¶\over ¶ÐÆ}ÐÆ\right)\right]Æ $$
 where the propagator derivatives $(¶/¶ÐÆ)(¶/¶Æ)$ give no signs
connecting up successive vertex factors $ÐÆÆ$, but the last one does in
closing the loop.

The general rules for contributions to the (unrenormalized) effective
action $ý[Ä]$ are then:

\Boxit{\rm\noindent
 (A1) 1PI graphs only, plus $S_0$ ($ÇüÄKÄ$).\\
 (A2) Momenta: label consistently with conservation, with $Çdp$ for each
	loop.\\
 (A3) Propagators: $1/ü(p^2+m^2)$, or $1/K$, for each internal line.\\
 (A4) Vertices: read off of $-S_I$.\\
 (A5) External lines: attach the appropriate (off-shell) fields and $Çdp$,
	 with $¶(Ýp)$.\\
 (A6) Statistics: 1/n! for n-fold symmetry of internal/external lines;\\ 
	\phantom{(7) }$-$1 for fermionic loop; overall $-$1.}\\
 (If we want to calculate $W[Ä]$ instead, then simply replace step 1 with
``Connected graphs only".)  The next step is to analyze the vacuum:

\Boxit{\rm\noindent
 (B1) Find the minimum of the effective potential (for scalars).\\
 (B2) Shift (scalar) fields to perturb about minimum; drop constant in
	potential.\\
 (B3) Find resulting masses; find wave function normalizations.}\\
 Renormalization is performed either before or after this step, depending
on the scheme.  Finally, the trees from $ý$ are identified with the
complete amplitudes from $S$.  Thus, T-matrix elements are given by:

\Boxit{\rm\noindent
 (C1) Connected ``trees" of (shifted, renormalized) $ý$:
	(A2-4) for L=0 with $S£ý$.\\
 (C2) ``Amputate" external $ý_0$-propagators.\\
 (C3) External lines: 1, or appropriate to $ý_0$ wave equation $÷KÆ=0$
	($ÝöÆÿöÆ=àN$).\\
 (C4) External-line statistics: No symmetry factors; $-$1 for fermion
	permutation.}

Note that $ý$ is usually simpler than $T$ with respect to treatment of
external lines:  In $T$ we often have contributions from graphs which are
identical except for permutation of external lines from identical fields.  In
$ý$ only one such graph need be considered, since the statistics of the
attached external fields automatically takes care of this symmetry.  (We
then also drop the 1/n!, or at least reduce it.)

We label the lines of a graph with arrows to indicate
the direction of ``flow" of momenta:  
Momentum conservation means the total momentum flowing
into any vertex
is equal to that flowing out, which we use to eliminate
dependent momenta (integrating out $¶$ functions).
(In tree graphs all internal momenta are determined by 
external ones, which are constrained by conservation
of total external momentum.)
This ``sign" of direction of the arrow is independent of
the sign of the energy; we must combine the two to determine
whether an external state is initial or final:
An incoming external line with positive energy is initial,
negative energy is final; an outgoing line is the opposite
(i.e., for positive energy the arrow indicates the direction of time,
but negative energy means travel backward in time).
The choice of direction of arrows is arbitrary, and the
convenience of any choice depends on the graph and
the theory.  There is no correspondence between this choice
and signs of energy, since generally one wants to apply the same
graph for cases with each external line with either sign,
whereas internal lines in trees may have either sign
depending on the external kinematics, and those in loops
must be summed over both signs.
Often the direction of the arrow is chosen to indicate the
direction of flow of positive charge, when such a quantum
number (U(1) symmetry) exists.

The simplest nontrivial tree graphs are 4-point amplitudes.
We now label all momenta as incoming, which is convenient for
symmetry, and corresponds naturally to using incoming (initial) states
with positive energy and outgoing (final) states with negative energy
(as from the complex-conjugate final wave functions).
These momenta are conveniently expressed in terms of the 
Mandelstam variables (see subsection IA4): with these signs, 
$$ s = -(p_1+p_2)^2,âât = -(p_1+p_3)^2,ââu = -(p_1+p_4)^2 $$
 We also use the convention that $s$ is defined in terms of the
momenta of the two initial particles (and we also use this same definition
when there are more than two final particles); $t$ and $u$ are then more
or less interchangable, but if initial and final particles are pairwise
related we choose $t$ in terms of the momenta of such a pair.

$$ \fig{stu} $$

For example, the simplest nontrivial theory is $Ä^3$ theory, with
$$ K = ü(-õ+m^2) £ ü(p^2+m^2) $$
$$ S_I = Çdx¼\f16 gÄ^3 
	= Çdp_1¼dp_2¼dp_3¼\f16 gÄ(p_1)Ä(p_2)Ä(p_3)¶\left(Ýp\right) $$
 The four-point S-matrix amplitude at tree level (order $g^2$) then comes
from using the following contributions to the factors in $\A$ (calculated in
Euclidean space):
$$ öï_4 = {¶\over ¶Ä(p_1)}{¶\over ¶Ä(p_2)}
	{¶\over ¶Ä(p_3)}{¶\over ¶Ä(p_4)} $$
$$ e^{Ç(¶/¶Ä)(1/K)(¶/¶Ä)/2} = ... 
	+ Çdk¼ü{¶\over ¶Ä(k)}{1\over ü(k^2+m^2)}{¶\over ¶Ä(-k)} + ... $$
$$ e^{-S_I} = ... +üS_I^2 + ... $$
 Using $(¶/¶Ä(p))Ä(k)=¶(p-k)$, keeping only the connected part, and
integrating out the $¶$ functions (except $¶(Ýp_{external})$), we are left
with
$$ \S_c = ¶(p_1 +p_2 +p_3 +p_4) g^2 \left( {1\over ü(m^2 -s)}
	+{1\over ü(m^2 -t)} +{1\over ü(m^2 -u)} \right) $$
  (There would be an extra $i$ for the $¶$ in Minkowski space.)  Note the
symmetry factor $1/3!$ for the $Ä^3$ coupling, which is canceled upon
taking 3 functional derivatives for the vertex factor.  The T-matrix then
comes from just factoring out the $¶$:
$$ T = g^2 \left( {1\over ü(m^2 -s)}
	+{1\over ü(m^2 -t)} +{1\over ü(m^2 -u)} \right) $$
 But this contribution to $W$ is given by a single term:
$$ W = -g^2 Çdp_1 dp_2 dp_3¼
	ü[üÄ(p_1)Ä(p_2)]{1\over ü(m^2-s)}[üÄ(p_3)Ä(-p_1-p_2-p_3)] $$
 or, in position space,
$$ W = -g^2 Çdx¼ü(üÄ^2){1\over ü(m^2-õ)}(üÄ^2) $$
 (The $ü$'s correspond to the various symmetries: switching a pair
connecting to the same vertex, or switching the two pairs.)

\x VC4.2  Find the 5-point tree amplitude for $Ä^3$ theory.  What order in
$g$ is the n-point tree?

The momentum integrals are real in Euclidean space:  There are no
singularities in the integrand, since $p^2+m^2$ is always positive
(although there are some subtleties in the massless case).  Thus,
all these integrals are most conveniently performed in Euclidean
space.  However, eventually the result must be analytically continued back
to Minkowski space: $x^0£ix^0$, which means $p^0£ip^0$ (but
$p_0£-ip_0$, being also careful to distinguish $¶_m^n£¶_m^n$ and
$¶_{mn}£ú_{mn}$ for indices on fields) via a $90^\circ$ rotation.  This
returns some of the $i$ dependence.  There are also $i$'s associated with
integration measures:  Since all the momentum integrals for the S-matrix
elements have already been performed, all that remains is a factor of $i$
to go with $¶(Ýp)$ for each connected graph.  There can also be $i$
dependence in external line factors.

Remember that negative $p^0$ indicates a particle traveling backward in
time; the true motion of such a particle is opposite to that of the arrow
indicating momentum flow.  Thus, external lines with arrows pointing into
the diagram and positive $p^0$, or out of the diagram and negative $p^0$,
both indicate initial states, arriving from $t=-¥$.  Conversely, external
lines with arrows pointing into the diagram and negative $p^0$, or out of
the diagram and positive $p^0$, both indicate final states, departing to
$t=+¥$.

A related issue is particles vs.¼antiparticles.  If a particle is described by a
real field, it is identified as its own antiparticle; but if it is described by a
complex field, then it is identified as a particle if it has a certain charge,
and as an antiparticle if it has the opposite charge.  (For example, a
proton is positively charged while an antiproton is negatively.)  Of course,
this is convention, since a complex field can always be replaced by two
real fields, and we can always relabel which is the field and which the
complex conjugate; generally there should be a useful conservation law
(symmetry) associated with these complex combinations (usually electric
charge), and the one called ``particle" is the one more common to the
observer.  For example, suppose we have a complex scalar external
field/wave function $Ä(p)$.  For $p^0>0$ this describes a particle
propagating to $x^0=+¥$.  Similarly, $Ä*(p)$ for $p^0>0$ describes an
ÓantiÕparticle propagating to $x^0=+¥$.  On the other hand, $Ä*$ for
$p^0<0$ describes a particle propagating from $x^0=-¥$, while $Ä$ for
$p^0<0$ describes an antiparticle propagating from $x^0=-¥$.

Ü5. Semiclassical unitarity

As in nonrelativistic quantum mechanics, the only conditions for unitarity
are that: (1) the metric (inner product) on the Hilbert space is positive
definite (so all probabilities are nonnegative), and (2) the Hamiltonian is
hermitian (so probabilities are conserved).  Both of these conditions are
statements about the classical action.  The second is simply that the
action is hermitian, which is easy to check.  The first is that the kinetic
(quadratic) terms in the action, which define the (free) propagators, have
the right sign.  This can be more subtle, since there are gauge and
auxiliary degrees of freedom.  Therefore, the simplest way to check is by
using the lightcone formalism.

We see from the analysis of subsection IIB3 for field equations, or for
actions in subsection IIIC2 (for spins $²1$, and more generally below in
chapter XII) that after lightcone gauge fixing and elimination of auxiliary
degrees of freedom the kinetic terms for physical theories always reduce
to $-\f14 ÄõÄ$ for bosons and $\f14 Æ(õ/i»^+)Æ$ for fermions, where
there is a sum over all bosons and fermions, and each term in the sum has
a field with a single hermitian component.  Complex fields can multiply
their complex conjugates, but these can always be separated into real
and imaginary parts.  There are never crossterms like $AõB$, since after
field redefinition, i.e., diagonalization of the kinetic operator, this gives
$A'õA'-B'õB'$, so one term has the wrong sign.  Similar remarks apply to
massive fields, but with $õ$ replaced by $õ-m^2$ (as seen, e.g., by
dimensional reduction), or we can treat the mass term as part of the
interactions.

\x VC5.1  What's wrong with $AõA+BõB+m^2 AB$?  (Hint: something,
but none of the above.  Diagonalize.)

Now we only need to check that the single-component propagators of
these two cases define positive-definite inner products.  Since
multiparticle inner products are products of uniparticle inner products,
it's sufficient to look at one-particle states.  We therefore examine the
S-matrix defined in  subsection VC1 for the special case of 1 particle at
$t=-¥$ going to 1 particle at $t=+¥$, using the free, massless lightcone
Lagrangians given above.  For the boson we found that this matrix
element was simply the inner product between the two states, appearing
in the form $Çd§¼d§'¼ÆëÆ'$.  This worked only because the propagator
had the right sign.  (It is essentially $+e^{-i¿|t|}$.)  Thus the sign we use
is required for unitarity.

A simple way to treat the fermion is to use supersymmetry:  Since the
boson and fermion kinetic terms are spin independent in the lightcone
formalism, we can look at any supersymmetric theory, and check the
boson and fermion kinetic terms there.  If the boson term agrees with the
one we just checked, then the fermion term is OK by supersymmetry
(which preserves unitarity by the lightcone-like supertwistor analysis of
subsection IIC5).  Alternatively, we can use the same method applied to
the boson:  The propagator now has an extra factor of $1/(-i»^+)$, or
$1/p^+$ in momentum space.  Since $p^+$ is always positive for positive
energy, these states also appear with the correct-sign norm.  To analyze
negative energy (antiparticles), we note that in the derivation of the path
integral final states always appear to the left and initial states to the
right.  In the fermionic case this is important because it will introduce an
extra sign:  Since
$$ Æ_1{õ\over »^+}Æ_2 = +Æ_2{õ\over »^+}Æ_1 $$
 (with two signs canceling from reordering fermions and integration by
parts), the right sign will always be produced with correct ordering of
initial vs.¼final states (i.e., positive vs.¼negative energy), independent of
the helicity (or whether electron vs.¼positron, etc.)

For similar reasons, it is clear that integral spin is always described by
commuting (bosonic) fields, while half-integral spin is always described by
anticommuting (fermionic) fields:  The number of undotted minus
dotted indices on a field is always odd for half-integer spin, even for
integer, and a derivative carries one dotted and one undotted index, so
contraction of all indices means an even number of derivatives for
integer spin and odd for half-integer.  Without loss of generality, we then
can separate each field into its real and imaginary parts.  Then for each
real field integration by parts gives
$$ Ä(-i»)^n Ä = (-1)^n [(-i»)^n Ä]Ä = (-1)^{n+Ä}Ä(-i»)^n Ä
	âÜâ(-1)^{n+Ä}=1 $$
 where $(-1)^Ä$ is the statistical factor for $Ä$ (1 for bosons, $-$1 for
fermions), and we have included the appropriate $i$'s for hermiticity of
the action.  Thus, integer spin is associated with bosons
($(-1)^n=1=(-1)^Ä$), and half-integer with fermions ($(-1)^n=-1=(-1)^Ä$).
This is the ``spin-statistics theorem".  By using real fields with a diagonal
kinetic term, we have implicitly assumed kinetic terms appear only with
the correct sign:  For example, for a complex bosonic field
$$ Ä = A+iBâÜâÄÿ»Ä = Ai\onª»B = -Bi\onª»A $$
 and thus has indefinite sign.  Thus, spin and statistics follows from
Poincar«e invariance, locality, and unitarity.  If we drop unitarity, we get
``ghosts":  We'll see examples of such wrong-statistics fields when
quantizing gauge theory.

Note that demanding unitarity (the right sign of the kinetic term) is the
same as demanding positivity of the true energy, as least as far as the
kinetic term is concerned:  The energy is given by the Hamiltonian of the
field theory; if the kinetic term changes sign, the corresponding
contribution to the Hamiltonian does also.  (Compare the discussion of
Wick rotation of the action in subsection VB4.)

Using anticommuting fields to describe fermions is more than a formality. 
In general, the significance of describing states by quantizing classical
fields that commute or anticommute has two purposes: (1) to avoid
multiple counting for indistinguishable particles, and (2) to insure that
two identical fermions do not occupy the same state.  Thus, when
describing two particles in different states, the phase associated with
(anti)commutation is irrelevant:  A ``Klein transformation" can be made
that makes anticommuting quantities ÓcommuteÕ for different states, and
anticommute (i.e., square to zero) only for the same state.  However,
such transformations are nonlocal, and locality is crucial in relativistic
field theory.  (See exercise IA2.3e.)

Ü6. Cutting rules

For some purposes it is useful to translate the three defining properties
of relativistic quantum field theory into graphical language.  Poincar«e
invariance is trivial, since the propagators and vertices are manifestly
Poincar«e covariant in covariant gauges.  Unitarity and causality can also
be written in a simple way in functional notation.  We first note that the
inner product for free multiparticle wave functions can be written very
simply in momentum space as
$$ ÒÆ|Ô = \left.\left(Æÿ[Ä]e^{D_+}[Ä]\right)\right|_{Ä=0},ââ
	{D_+} = Çdp¼{\onÁ¶\over ¶Ä(-p)}ë_+(p){¶\over ¶Ä(p)} $$
$$ ë(p) = {-iN(p)\over ü(p^2 +m^2 -i·)}âÜâ
	ë_+(p) = Ï(p^0)2¹¶[ü(p^2+m^2)]N(p) $$
 where $Æ$ and $$ are products of positive-energy single-particle
states, and $ë_+$ (the ``cut" propagator)
projects onto the positive-energy mass shell.  (The
exponential takes care of the usual combinatoric factors.)  We have
written a generic propagator, with numerator factor $N$ (=1 for scalars). 
(Without loss of generality, we have assumed a real basis for the fields,
so $N$ can be taken as real.)  The S-matrix amplitude then can be written
in operator language as
$$ ï[Ä] = Æÿ[Ä][Ä]âÜâ\A = ÇDļï[Ä] e^{-iS[Ä]} 
	= öï\left[{¶\over ¶Ä}\right] Z[Ä] = ÒÆ|\S|Ô $$
 We have used the fact that positive-energy states propagate forward in
time and negative backwards to write the usual Hilbert-space inner
product in terms of initial and final states of positive energy.  The
S-matrix operator $\S$ appears because $$ and $Æ$ satisfy the free
equations of motion, and $\S$ performs time translation from $t=-¥$ for
$$ to $t=+¥$ for $Æ$ to include interactions.

The unitarity condition is then (see subsection VA4)
$$ \Sÿ\S = 1âÜâZ[Ä]ÿ e^{D_+} Z[Ä] = 1 $$
 while causality is
$$ {¶\over ¶Ä(x)}\left( \S[Ä]ÿ {¶\over ¶Ä(y)} \S[Ä] \right) = 0âÜâ
	{¶\over ¶Ä(x)}\left( Z[Ä]ÿ e^{D_+} {¶\over ¶Ä(y)} Z[Ä] \right) = 0â
	for¼x^0 > y^0 $$
 This causality relation, which already holds in nonrelativistic field theory,
can be strengthened by using Lorentz invariance:  If $(x-y)^2>0$
(spacelike separation), then $x^0<y^0$ can be Lorentz transformed to
$x^0>y^0$.  Thus, the above expression vanishes everywhere except on
or inside the backward lightcone with respect to $x-y$.  These functional
forms of unitarity and causality (and Poincar«e invariance) can also be
used as a basis for the derivation of the functional integral form of $Z[Ä]$
in terms of the action, rather than relying on its relation to the
Hamiltonian formalism.

The fact that these conditions are satisfied by Feynman diagrams follows
easily from inspection.  We examine them using the explicit expression
for $Z$ following from the functional integral:
$$ Z[Ä] = e^{D}e^{-iS_I},ââZ[Ä]ÿ = e^{D*}e^{iS_I} $$
$$ D = üÇdp¼{¶\over ¶Ä(p)}ë(p){¶\over ¶Ä(-p)},ââ
	ë(p) = {-iN\over ü(p^2+m^2-i·)} $$
$$ D* = üÇdp¼{¶\over ¶Ä(p)}ë(p)*{¶\over ¶Ä(-p)},ââ
	ë(p)* ={iN\over ü(p^2+m^2+i·)} $$
 These expressions can be translated straightforwardly into position
space as
$$ Çdp¼{¶\over ¶Ä(p)}ë(p){¶\over ¶Ä(-p)} = 
	Çdx¼dx'¼{¶\over ¶Ä(x)}ë(x-x'){¶\over ¶Ä(x')}$$
 etc.

From the results at the end of subsection VB2, we see that the
propagators satisfy the relations
$$ ë_+(x) = ë(x) -ë_A(x) = ë*(x) +ë_R(x) $$
$$ ë_+(x) = ë_-(-x),âë(x) = ë(-x),âë_R(x) = ë_A(-x) $$
 and, of course, $ë_R(x)=0$ for $x^0<0$.  We will now see that the
cancellations in the unitarity and causality relations occur graph by
graph:  There are contributions consisting of a sum of terms represented
by exactly the same diagram, with each term differing only by whether
each vertex comes from $Z$ or $Zÿ$.  First, this affects the sign of the
term, since each vertex from $Zÿ$ gets an extra sign from the
$e^{iS_I}$ in $Zÿ$, as compared to the $e^{-iS_I}$ in $Z$.  Second, this
affects which propagators appear:
$$ \li{ (x,y)¼in:¼(Z,Z) & £ ë(x-y) \cr
			(Zÿ,Z) & £ ë_+(x-y) \cr
			(Z,Zÿ) & £ ë_-(x-y) \cr
			(Zÿ,Zÿ) & £ ë*(x-y) \cr } 
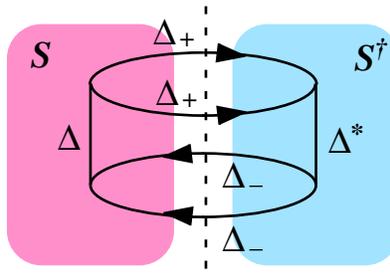
$$

$$ \fig{cross} $$

 Now if we sum over two terms differing only by whether 
the position $x$ of one particular vertex appears
in $Z$ or $Zÿ$, the result before integration is proportional to
$$ Þë(x-y_i)Þë_-(x-z_j) -Þë_+(x-y_i)Þë*(x-z_j) $$
 where $y_i$ are from $Z$ and $z_j$ are from $Zÿ$.  However,
$$ ë(x-y) -ë_+(x-y) = ë_-(x-y) -ë*(x-y) = ë_R(y-x) = 0âforâx^0 > y^0 $$
 Writing $ë=ë_++ë_R$ and $ë_-=ë*+ë_R$ in the difference of the two
products, each surviving term in the difference contains a $ë_R$,
and therefore the two products cancel if $x^0$ is the latest of all the
vertices.  We thus take any sum of the same diagram over different
distributions of the vertices between $Z$ and $Zÿ$ occuring in
the unitarity or causality relation before integration over the 
coordinates of the vertices, separate the sum into pairs which
are identical except for whether the latest vertex is in $Z$ or $Zÿ$,
and apply the above relation to show this difference vanishes.
Thus, the vanishing of a sum of graphs indicated by unitarity
or causality is actually satisfied by cancellation between each pair
of terms before integration over coordinates.
(Which pair is determined by the values of the coordinates,
since we need to find the latest one; after integration, the
cancellation is between the whole set of terms for the same graph.)
In the unitarity relation we sum over whether a vertex occurs
in $Z$ or $Zÿ$ for each vertex, including the latest one, so that condition
is easily satisfied.  (The only diagram that survives is the one with no
particles, which gives 1.)  In the causality relation we perform this sum
for each vertex except $y$ (since $¶/¶Ä(y)$ acts only on $Z$), but since
$y^0<x^0$, $y^0$ is not the latest vertex, so again the latest one is
summed over.

$$ \fig{cuts} $$

\x VC6.1 Consider an arbitrary 1-loop graph.  Why would replacing all the
propagators $ë$ with advanced propagators $ë_A$ (or all with retarded
propagators $ë_R$) all the way around the loop in the same direction give
zero?  Use this result, and the relation between the various propagators,
to show that any one-loop diagram (with normal propagators) can be
expressed as a sum of products of tree graphs, with some summations of
external states (``Feynman tree theorem").  How does this differ from
the cutting rule for unitarity?  (Hint:  Look at the signs of the energies of
external states.)

There is one fine point in this construction:  We may use Feynman rules
from a complex action, such as those used for massive theories in
subsection IIIC4, or when using complex gauges (see section VIB).  In that
case, since the S-matrix $\S$ is gauge-independent, and the original action
$S$ was real (before eliminating complex fields or choosing complex
gauges), it is legal to use the ÓHermitian conjugateÕ action $Sÿ$ to define
the Feynman rules for $\Sÿ$ (and $Zÿ$):  When multiplying $\Sÿ\S$, we use
the usual rules to find the second factor $\S$, and the conjugate rules to
find the $\S$ used in the first factor $\Sÿ$:
$$ \Sÿ\S = [\S(Sÿ)]ÿ\S(S) $$
 The result of conjugating the S-matrix then will be to complex conjugate
twice, and return rules identical to those used for $\S$, except for the
differences noted above for real actions.  That means that the above
proof of the cutting rules goes through unmodified, where we use the
same complex rules in the entire diagram, regardless of whether they are
associated with $Z$ or $Zÿ$.  In particular, this means that vertices from
the two parts of the graph will differ only by sign (conjugating just the
$i$ in $e^{iS_I}$, not the $S_I$), and propagators will differ only by their
(momentum-space) denominators, not their numerators.  This is
particularly important for the complex fields of subsection IIIC4, since
otherwise even the types of indices carried by the fields would differ.

Ü7. Cross sections

In quantum physics, the only measurables are probabilities, the squares
of absolute values of amplitudes.  Since we calculate amplitudes in
momentum space, probabilities are expressed in terms of scattering of
plane waves.  They are more naturally normalized as probabilities per
unit 4-volume (or D-volume in arbitrary dimension), since plane waves
are uniformly distributed throughout space.  This can be seen explcitly
from the amplitudes:  Because of the total momentum conservation
$¶$-function that appears with each connected S-matrix element
$\S_{fi}$, we have for the probability $P$
$$ \S_{fi} = i¶\left(Ýp\right)T_{fi} $$
$$ ÜâP = |\S_{fi}|^2 = |T_{fi}|^2 ¶\left(Ýp\right)¶(0) 
	= |T_{fi}|^2 ¶\left(Ýp\right){V_D\over (2¹)^{D/2}} $$
 where we have found the coordinate D-volume by the Fourier-transform
definition of the $¶$-function:
$$ ¶(0) = Çdx¼1 = Ç{d^D x\over (2¹)^{D/2}}¼1 = {V_D\over (2¹)^{D/2}} $$

A ``cross section" is defined as a probability for the scattering of two
incoming particles into some number of outgoing particles.  The
scattering is ``elastic" if the two final particles are the same as the two
initial particles (they exchange only 4-momentum), ``inelastic"
otherwise.  Generally one particle is in a beam directed at a target (at
rest in the lab frame) containing the other ``incoming" particle, but in
some experiments two beams are directed at each other.  In either case
the cross section is defined by the rate at which one particle interacts
divided by the flux of the other particle, where
$$ flux = {rate¼of¼arrival\over area} = (density) ð (relative¼velocity) $$
 and thus the ``differential cross section" (yet to be integrated/summed
over final states) is
$$ d§ = {P \over V_D}ð{1\over ¨_1 ¨_2 v_{12}} $$
The ``total cross section" $§$ is then given by summation over all
types of final states and integration over all their remaining momentum
dependence.

The spatial density $¨$ is the integrand of the spatial integral that defines
the inner product:  From subsection VB2, for bosonic plane waves 
($Æ_p(x)=e^{ipÉx}$) we have
$$ ÒÆ|ÆÔ = ·(p^0)Ç{d^{D-1}x\over (2¹)^{D/2}}¼Æ* üi\onª»_t Æ
	= Ç{d^{D-1}x\over (2¹)^{D/2}}¼¿ $$
$$ Ü⨠= {¿\over (2¹)^{D/2}} $$
 where $¿=|p^0|$.  (The same result can be obtained for fermions, when
their external line factors are appropriately normalized.)

The expression for $d§$ is actually independent of the frame, as long as
the 3-momenta of the two particles are parallel.  This is the case for the
most frequently used reference frames, the center-of-mass frame
and the ``lab frame" for either particle (where that particle is at rest, as
is the lab if that particle is part of a target).  Then
$$ (¿_1 ¿_2 v_{12})^2 = ¿_1^2 ¿_2^2 
	\left|{\vec p_1\over ¿_1} -{\vec p_2\over ¿_2}\right|^2
	= |¿_2 \vec p_1 -¿_1 \vec p_2|^2 = -ü(p_{1[a}p_{2b]})^2 
	= Â_{12}^2 $$
$$ \boxeq{ Â_{12}^2 ­ \f14 [s-(m_1+m_2)^2][s-(m_1-m_2)^2] } $$
 using again the Mandelstam variables and $Â_{ij}$ introduced in
subsection IA4. Finally, we include the ``phase space" for the final states
to obtain
$$ d§ = |T_{fi}|^2 {(2¹)^D ¶^D\left( Ýp \right) \over Â_{12}}
	Þ_f {d^{D-1} p\over (2¹)^{D/2-1}¿} Þ_{f: n¼ident}{1\over n!} $$
 where the first product is over all final one-particle states, and the
second is over each set of $n$ identical final particles.  The normalization
again follows from the inner-product for plane waves:  By Fourier
transformation, $d^{D-1}x¼¨£d^{D-1}p/(2¹)^{D-1}¨$.  It also appears in
the ``cut propagator" $ë_+$ used in unitarity, as in the previous
subsection:
$$ Ç{d^D p\over (2¹)^{D/2}}¼Ï(p^0)2¹¶[ü(p^2+m^2)] = 
	Ç{d^{D-1} p\over (2¹)^{D/2-1}¿} $$

The simplest and most important case is where two particles scatter to
two particles.  (This includes elastic scattering.)  The ``differential cross
section" $d§/d¯$, where $d¯$ is the angular integration element for
$p_3$, is found by integrating $d§$ over $d^{D-1} p_4$ and 
$d|\vec p_3|$.  The former integration is trivial, using the $¶$ function for
$(D-1)$-momentum conservation.  The latter integration is almost as trivial,
integrating the remaining $¶$ function for energy conservation:
$$ d^{D-1} p_3¼¶ \left( Ýp^0 \right) = d¯¼d|\vec p_3|¼(\vec p_3)^{D-2} 
	\left|{»Ýp^0\over »|\vec p_3|}\right|^{-1}¼
	¶ (|\vec p_3|-|\vec p_3|_0) $$
$$ Ýp^0 = ¿_1+¿_2  -å{(\vec p_3)^2+m_3^2}
	-å{(\vec p_1 +\vec p_2 -\vec p_3)^2+m_4^2} $$
$$ Üâ{»Ýp^0\over »|\vec p_3|} = 
	{-ü(s-m_3^2-m_4^2)¿_3 +m_3^2 ¿_4 \over |\vec p_3|¿_3 ¿_4} $$
 where $|\vec p_3|_0$ is $|\vec p_3|$ evaluated as a function of the
remaining variables at $Ýp^0=0$.  We then find
$$ {d§\over d¯} = (2¹)^2|T_{fi}|^2
	{|\vec p_3|^{D-1} \over Â_{12}[ü(s-m_3^2-m_4^2)¿_3 -m_3^2 ¿_4]} $$

The center-of-mass frame (see subsection IA4) is the simplest for
computations.  In that frame the differential cross section simplifies to
$$ {d§\over d¯} = (2¹)^2|T_{fi}|^2{Â_{34}^{D-3}\over Â_{12}s^{D/2-1}} $$
 and in particular in D=4:
$$ {d§\over d¯} = (2¹)^2|T_{fi}|^2{Â_{34}\over Â_{12}s} $$
 Another convenient form for the differential cross section is $d§/dt$,
trading $Ï$ for $t$ and integrating out the trivial dependence $ÇdÄ=2¹$
(for D=4).  In the center-of-mass frame we have
$$ d¯ = 2¹d(cos¼Ï) = ¹{s\over Â_{12}Â_{34}} dt $$
 Since $t$ is Lorentz invariant, we therefore have in ÓallÕ frames (that
conform to our earlier requirement for the $Â_{12}$ factor in $d§$)
$$ \boxeq{ {d§\over dt} = ü(2¹)^3|T_{fi}|^2{1\over Â_{12}^2} } $$

For example, for the 4-point scalar example considered in subsection VC4,
we have
$$ {d§\over dt} = {(4¹)^3 g^4\over s(s-4m^2)}\left(
	{1\over s-m^2} +{1\over t-m^2} +{1\over u-m^2} \right)^2 $$
	
\x VC7.1  A simpler example than the cross section is the decay rate,
$$ {dP\over dt} = Ý_f {P\over ¨V_D} $$
 for initial density $¨$ (where $t$ is now time).  
 ªa For the case of decay of a particle of mass $M$
into 2 particles of masses $m_1,m_2$ in D=4, 
show that in the rest frame
$$ {dP\over dt} = 4¹{Â_{12}\over M^3}|T_{fi}|^2 $$
 (with a factor of $ü$ for final identical particles).
What happens when $M=m_1+m_2$?
 ªb For the case of coupling $S_I=Çdx¼gÄ_1 Ä_2 Ä_3$,
evaluate the classical (tree) contribution in terms of $g,M,m_1,m_2$.
Give the dimensional analysis of the result.
Consider also the case where the final particles are massless.

\x VC7.2  Consider the cross section for elastic scattering of 
two particles to lowest order (tree graphs), in four dimensions.  
For the following, consider the nonrelativistic limit (small velocities).
ªa Show from the definition
that classically the ``total cross section", as indicated by the
name, is just the cross-sectional area with respect to the beam
(assuming each ``arrival" results in an ``interaction").
ªb A $Ä^4$ interaction correpsonds to a $¶$-function potential
in classical mechanics, since it has zero range.  In classical
mechanics, such ``billiard ball" scattering is purely geometrical,
depending only on the ``size" of the balls.  Find the effective ``radius"
of these classical billiard balls in terms of their mass and coupling.
ªc Replace the $Ä^4$ interaction with a $Ä^2 $ interaction,
where $$ is an intermediate particle with a different mass
(nonzero, and also much larger than the $Ä$ kinetic energies).
(We still consider scattering of $Ä$ particles.)
What is the effective radius of the $Ä$ particles?
ªd In the limit where the $$ mass becomes infinite, but also its coupling,
the cubic interaction is effectively replaced with a $Ä^4$ interaction
(finite when the limit is taken appropriately).  What is this coupling
constant in terms of the cubic coupling and $$ mass?

$$ \fig{cross} $$

For some purposes (such as considerations of unitarity and causality,
as in the previous subsection) it is useful to draw the Feynman
diagrams for the cross section itself (or actually $|T|^2$).  In such a
diagram we draw one of the diagrams from $\S$ and one from $\Sÿ$,
separating the two by a line (dashed, zig-zag, or shaded on one side,
according to your preference), and connecting ÓallÕ the external lines
(initial and final) on one side to the corresponding ones on the other.  The
result is a bubble diagram with a ``cut":  The ``cut propagators" are
$ë_+=2¹Ï(p^0)N¶[ü(p^2+m^2)]$ (or $ë_-$, depending on how we label the
momenta), corresponding to the propagator $ë=N/ü(p^2+m^2)$, if we
sum over all polarizations; otherwise $N$ is replaced by a term in the
sum $N=ÝöÆÿöÆ$.  The momenta of the cut propagators may not be
integrated over, depending on whether they represent final states whose
momenta are summed over (i.e., not measured; in practice the momenta
of initial particles are always measured).  The only other difference in the
Feynman rules from the S-matrix is that in Wick rotating back $\S$ gets
the usual $m^2£m^2-i·$ while $\Sÿ$ gets $m^2£m^2+i·$, and each
connected graph in $\S$ gets an $i¶(Ýp)$ while each in $\Sÿ$ gets a
$-i¶(Ýp)$.  The algebra for the cross section is thus identical to that of a
vacuum bubble (although the momentum integration is not, and the cut
propagators lack the usual denominators).  In particular, instead of
summing over just physical polarizations in a cut vector propagator,
which corresponds to using a unitary gauge, we can include ghosts in the
external states, and use any gauge:  This follows from the cutting rules
derived for unitarity.

Ü8. Singularities

We know from free theories that any propagator has a pole at the
classical value of the square of the mass.  This statement can be
extended to the interacting theory:  The (``Landau") singularities in any
Feynman diagram are exactly at classically allowed (on shell) values of
the momenta.

The simplest way to see this is to write the propagators in a way
reminiscent of the classical theory, where the appearance of the
worldline metric in the action results in the (Wick-unrotated) Schwinger
parametrization of the propagator,
$$ {-i\over ü(p^2+m^2)} = Ç_0^¥ d ¼e^{-i (p^2+m^2)/2} $$
 For simplicity we consider a scalar field theory with nonderivative
self-interactions.  The corresponding form of a Feynman diagram, written
in momentum space by Fourier transformation, is then
$$ Çdx_i' dp_{ij} d _{ij}¼
	e^{-iÝ_{ÒijÔ}[ _{ij}(p_{ij}^2+m^2)/2 -(x_i-x_j)Ép_{ij}]} $$
 where $i,j$ label vertices (and external endpoints), $ÒijÔ$ labels links
(propagators), and $dx'$ indicates integration over just vertices not
attached to external lines.

Integration over these $x'$'s produces $¶$ functions for momentum
conservation:
$$ Ý_j p_{ij} = 0 $$
 i.e., the sum of all momenta flowing into any vertex vanishes.  (Note
$p_{ij}=-p_{ji}$.)  This constraint can be solved by replacing momenta
associated with each propagator with momenta associated with each
loop (and keeping momenta associated with external lines).  For example,
for planar diagrams, we can write
$$ p_{ij} = p_{IJ} = k_I - k_J $$
 where $I,J$ label loops:  $p_{IJ}$ labels the propagator by the two loops
on either side, rather than the vertices at the ends (and similarly for
$ _{ij}= _{IJ}$).  Similar remarks apply to nonplanar diagrams, but the
parametrization in terms of loop momenta $k_I$ is more complicated
because of the way external momenta appear.  (In the planar case, the
above parametrization can also be used for external momenta, and
automatically enforces overall momentum conservation.)  The Feynman
integral then becomes
$$ Çdk_I' d _{IJ}¼e^{-iÝ_{ÒIJÔ} _{IJ}[p_{IJ}(k)^2+m^2]/2} $$
 where $k$ are loop ($k'$) and external momenta after solving the
conservation conditions.

We can now treat the exponent of the Feynman diagram in the same way
as a classical mechanics action, and find the corresponding classical
equations of motion by varying it.  By the stationary phase approximation
(or steepest descent, after Wick rotation of $ $), the classical solutions
give the most important contribution to the integrals (at least for weak
coupling).  This approximation is related to long-distance (i.e., infrared)
behavior (see exercise VB4.3a).  Taking account of the fact that the $ $'s
are constrained to be positive (by treating $ =0$ separately or making a
temporary change of variables $ =º^2$ or $e^º$ to an unconstrained
variable), we find
$$  _{IJ}(p_{IJ}^2+m^2) = 0,ââÝ_J  _{IJ}p_{IJ} = 0 $$
 or, in terms of the original variables,
$$  _{ij}(p_{ij}^2+m^2) = 0,ââp_{ij} = {x_i-x_j\over  _{ij}},ââ
	Ý_j p_{ij} = 0 $$
 These are known as the ``Landau equations".  Their correspondence with
classical configurations of particles follows from treating $ $ as the
proper time, as seen from $p=ëx/ë $.  (Actually, it is the generalization
discussed in section IIIB of proper time to include the massless case; in
the massive case $s=m $.)  The equation $ (p^2+m^2)=0$ says that either
the particle for the line is on-shell or there is no such line (it has vanishing
proper length), while the equation $Ý_J  p=0$ says that the sum of
$ p$ around a loop vanishes, another statement that $pë =ëx$.

\x VC8.1  Consider the ``one-loop propagator correction": the graph 
where a single particle of mass $M$ splits into two of masses $m_1$
and $m_2$, which rejoin.  What is the physical condition relating these
3 masses (with all particles on shell to satisfy the above)?  What happens
when $M=m_1+m_2$?  What if $m_2=0$?

Note that these physical singularities are all in physical Minkowski space: 
In Euclidean space, $p^2+m^2$ is positive definite, so it never vanishes
(except for constant, massless fields $p=m=0$).  Furthermore, in Euclidean
space, one can always rotate any momentum to any direction, whereas in
Minkowski space one can never Lorentz transform to or through either
the forward or backward lightcone.  Thus, calculating in Euclidean space
makes it clear that S-matrix elements for positive-energy states are
given by the same expressions as those for negative-energy states:  To
compare two amplitudes that are the same except for some final particles
being replaced with initial antiparticles, or vice versa, we just change the
sign of the energy.  This is called ``crossing symmetry".  In the case
where all particles are reversed, it is CPT invariance (``CPT theorem").

Ü9. Group theory

Although the manipulation of spin indices in Feynman diagrams is closely
tied to momentum dependence, the group theoretic structure is
completely independent, and can be handled separately.  Therefore, it is
sufficient to consider the simple example of scalars with a global
symmetry.  (The more physical case of chromodynamics will be the same
with respect to group theory, but will differ in dependence on momentum
and spin.)  The simplest case is the U(N) family of groups.  We choose an
action of the form
$$ L = \f1{g^2}tr[\f14 (»Ä)^2 +V(Ä)] $$
 where $Ä$ is a hermitian N$ð$N matrix.  This action is invariant under the
global U(N) symmetry
$$ Ä' = UÄU^{-1} $$
 A simplification we have chosen is that the interaction has only a single
trace; this is the case analogous to pure Yang-Mills theory.  (We also
included the coupling constant as in Yang-Mills.)

$$ \fig{color} $$

When we draw a Feynman diagram for this field theory, instead of a
single line for each propagator, we draw a double (parallel) line, each line
corresponding to one of the two indices on the matrix field.  Because of
the trace in the vertex, the propagator lines connect up there in such a
way that effectively we have continuous lines that travel on through the
vertices, although the two lines paired in a propagator go their separate
ways at the vertex.  These lines never split or join, and begin or end only
on external fields.  We can also draw arrows on the lines, pointing in the
same direction everywhere along a single line, but pointing in opposite
directions on the two lines in any propagator pair:  This keeps track of
the fact that $Ä$ appears in the trace always multipled as $ÄÄ$ and
never as $ÄÄ^T$.   A physical picture we can associate with this is to think
of the scalar as a bound-state of a quark-antiquark pair, with one line
associated with the quark and another with the antiquark; the arrows are
then oriented in the direction of time of the quark (which is the opposite
of the direction of time for the antiquark).  The quark is thus in the
defining representation of U(N) (and the antiquark in the complex
conjugate representation).  Group theory factors are trivial to follow in
these diagrams:  The same color quark continues along the extent of a
``quark line"; thus, there is a Kronecker $¶$ for the two indices appearing
at the ends of the quark line (at external fields); each quark is conserved.

\x VC9.1 Using the quark-line notation, where the lines now represent
flavor, draw all 4-point tree graphs, with 3-point vertices, representing
scattering of $K^+ K^-£¹^+ ¹^-$ (see subsection IC4) via exchange of
other (pseudo)scalar mesons in that U(6) multiplet.  What are the intermediate states
(names of mesons) in each channel?  Note that the ``flavor flow" of both
diagrams can be represented by a single diagram, by separating all pairs
of intermediate lines to leave a square gap in the middle:  This ``duality
diagram" represents the mesons as strings; the gap between quarks
represents the ``worldsheet".

\x VC9.2  Consider quark-line notation for doing group theory in general:
(Calculate using only graphs, with numerical factors --- no explicit indices
or Kronecker $¶$'s, except for translating definitions.)
 ªa  Write the structure constants for U(N) as the difference of two
diagrams, by considering a vertex of the form $tr(A[B,C])$.
 ªb  Find pictorially the resulting expression for the Cartan metric (see
subsection IB2):  Show that it is the identity times 0 for the U(1) subgroup
and 2N for the SU(N) subgroup (as found previously in subsection IIIC1).
 ªc  Also use these diagrams to prove the Jacobi identity (i.e., find the
resulting 6 diagrams and show they cancel pairwise).
 ªd  Derive diagrammatically the value of $d_{ijk}$ of exercise IB5.3b for
the defining representation.

\x VC9.3 Consider the group theory factors for the above scalar theory,
with only a cubic interaction.  Draw all the 1PI 1-loop diagrams with 4
external double-lines, and rewrite the corresponding factors in terms of
traces and products of 4 fields, including factors of N.  (Be careful to
include all permutations of connecting propagators to vertices.)

In general, we find that connected Feynman diagrams may include
diagrams that are disconnected with respect to the above group-theory
diagrams, where we consider the two group-theory lines on an external
line to be connected.  Such diagrams correspond to multiple
traces:  There is a factor of the form $tr(G_i G_j... G_k)$ corresponding to
each connected group-theory graph, where $i,j,...,k$ are the
group-theory indices of the external lines (actually double indices in the
previous notation).  However, all connected trees are group-theory
connected.  Furthermore, they are all ``planar":  Any connected U(N) tree
can be drawn with none of the quark lines crossing, and all the external
lines on the outside of the diagram, if the external lines are
``color-ordered" appropriately.  Of course, one must sum over
permutations of external lines in the T-matrix because of Bose symmetry.
However, in calculating $W[Ä]$ one need consider only one such planar
graph, with the external lines color-ordered; the Bose symmetry of the
fields in $W$ automatically incorporates the permutations.  (Similar
remarks apply to loop and nonplanar graphs.)

As an example, consider the U(N) generalization of the 4-point tree
example of subsection VC4.  The Lagrangian is now (scaling the coupling
back into just the vertex)
$$ L = trÓ\f14 [(»Ä)^2 +m^2 Ä^2] +\f13 gÄ^3Õ $$
 where the interaction term now has a combinatoric factor of $\f13$
instead of $\f16$ because it is symmetric only under cyclic permutations. 
The result for $W$ is now modified to
$$ W = -g^2 tr Çdx¼üÄ^2{1\over ü(m^2-õ)}Ä^2 $$

This analysis for U(N) can be generalized to SU(N) by including extra
diagrams with lines inside propagators short-circuited, representing
subtraction of traces.  It can also be generalized to SO(N) and USp(2N):  In
those cases, antisymmetry or symmetry of the matrices means the lines
no longer have arrows, and we include diagrams where the lines inside
the propagators have been ``twisted", with signs appropriate to
symmetrization or antisymmetrization.  Generalization for all the above
groups to include defining representations is also straightforward:  Such
fields, like the true quarks in QCD, carry only a single group-theory line. 
For example, sticking with our simpler scalar model, we can generalize to
a scalar theory with the same group theory as chromodynamics, with
``free" quark fields, appearing as scalar fields in the defining
representation of the group:
$$ L £ L +tr[ü(»Æ)ÿÉ(»Æ) +Æÿf(Ä)Æ],ââÆ' = UÆU_f^{-1} $$
 where $Æ$ is an N$ð$M matrix with M flavors, and $U_f$ is the flavor
symmetry.  (This color+flavor symmetry was treated in subsections IC4
and IVA4.)  This field has a propagator with a single color line (with an
arrow); however, we can also use another double-line notation, where
$Æ$ propagators carry one line for color and another for flavor.  This
method can be generalized to arbitrary representations obtained by
direct products of defining representations, (anti)symmetrizations, and
subtractions of traces, by giving the propagators the corresponding
number of lines (though usually two lines are sufficient for the
interesting cases).

\refs

£1 P.A.M. Dirac, ÓProc. Roy. Soc.Õ ÉA114 (1927) 243;\\
	W. Heisenberg and W. Pauli, ÓZ. Phys.Õ É56 (1929) 1, É59 (1930) 168;\\
	E. Fermi, ÓRev. Mod. Phys.Õ É4 (1932) 87:\\
	birth of (interacting) quantum field theory (QED).
 £2 S. Tomonaga, ÓProg. Theor. Phys.Õ É1 (1946) 27;\\
	J. Schwinger, ÓPhys. Rev.Õ É74 (1948) 1439, Óloc. cit.Õ (VB,
	ref. 2, first ref.), É76 (1949) 790:\\
	gave relativistic procedure for treating quantum field theory.
 £3 Feynman, Óloc. cit.Õ (VB, ref. 4):\\
	same as above, with diagrams.
 £4 F.J. Dyson, ÓPhys. Rev.Õ É75 (1949) 486, 1736:\\
	related Tomonaga-Schwinger and Feynman approaches; gave
	procedure to all loops.
 £5 J. Schwinger, ÓProc. Natl. Acad. Sci. USAÕ É37 (1951) 452, 455:\\
	generating functionals for field theory.
 £6 Feynman, Óloc. cit.Õ (VA):\\
	path-integrals for field theory.
 £7 J. Schwinger, ÓQuantum electrodynamicsÕ (Dover, 1958):\\
	reprints of most of the important quantum field theory papers of the
	40's and 50's.
 £8 S.S. Schweber, ÓQED and the men who made it: Dyson, Feynman,
	Schwinger, and TomonagaÕ (Princeton University, 1994).
 £9 J. Goldstone, A. Salam, and S. Weinberg, ÓPhys. Rev.Õ É127 (1962) 965;\\
	G. Jona-Lasinio, ÓNuo. Cim.Õ É34 (1964) 1790:\\
	effective action.
 £10 Y. Nambu, \PL 26B (1968) 626:\\
	tree graphs are classical field theory.
 £11 D.G. Boulware and L.S. Brown, ÓPhys. Rev.Õ É172 (1968) 1628:\\
	expression for classical field in terms of tree graphs.
 £12 S. Coleman, ÓSecret symmetry: An introduction to spontaneous
	symmetry breakdown and gauge fieldsÕ, in ÓLaws of hadronic matterÕ,
	proc. 1973 International School of Subnuclear Physics, Erice, July
	8-26, 1973, ed. A. Zichichi (Academic, 1975) part A, p. 139:\\
	relation of Legendre transform to 1PI graphs via path integrals.
 £13 Dyson, Óloc. cit.Õ (ref. 4);\\
	Schwinger, Óloc. cit.Õ (ref. 5).
 £14 W. Pauli, ÓPhys. Rev.Õ É58 (1940) 716:\\
	spin-statistics theorem.
 £15 R.E. Cutkosky, ÓJ. Math. Phys.Õ É1 (1960) 429:\\
	cutting rules for unitarity.
 £16 T. Veltman, ÓPhysicaÕ É29 (1963) 186:\\
	  cutting rules for causality.
 £17 R.P. Feynman, ÓActa Phys. Polon.Õ É24 (1963) 697:\\
	tree theorem.
 £18 L.D. Landau, ÓNucl. Phys.Õ É13 (1959) 181.
 £19 S. Coleman and R.E. Norton, ÓNuo. Cim.Õ É38 (1965) 438:\\
	classical interpretation of Landau singularities.
 £20 W. Siegel, \xxxlink{hep-th/9601002}, ÓInt. J. Mod. Phys. AÕ É13 (1998)
	381:\\
	Landau rules from classical mechanics action.
 £21 R.J. Eden, P.V. Landshoff, D.I. Olive, and J.C. Polkinghorne, ÓThe 
	analytic S-matrixÕ (Cambridge University, 1966):\\
	singularities of Feynman diagrams.
 £22 M. Gell-Mann and M.L. Goldberger, proc. ÓFourth International
	Conference on High Energy PhysicsÕ (University of Rochester, 1954);\\
	J.M. Jauch and F. Rohrlich, ÓThe theory of photons and electronsÕ,
	1st ed. (Springer-Verlag, 1955) p. 161:\\
	crossing symmetry.
 £23 H. Harari, \PR 22 (1969) 562;\\
	J. Rosner, \PR 22 (1969) 689;\\
	T. Matsuoka, K. Ninomiya, and S. Sawada, ÓProg. Theor. Phys.Õ É42
	(1969) 56:\\
	quark lines for group theory factors.
 £24 J.E. Paton and Chan H.-M., \NP 10 (1969) 516:\\
	traces for group theory in planar graphs.
 £25 G. 't Hooft, \NP 72 (1974) 461:\\
	double-line notation in field theory.

\unrefs

ÚVI. QUANTUM GAUGE THEORY

We now consider special features of spin and gauge invariance, and
introduce some special methods for dealing with them.  In quantum
theory, gauge fixing is necessary for functional integration:  Gauge
invariance says that the action is independent of some variable;
integration over that variable would thus give infinity when evaluating
amplitudes for gauge-invariant states.  Eliminating that variable from the
action (a ``unitary gauge") solves the problem, but not always in the most
convenient way.  (If it were, we wouldn't have introduced such a
redundant description in the first place.)  Note that such infinities already
appear for global symmetries:  For example, the functional integral with
wave function 1 (vacuum-to-vacuum amplitude) is infinite by translation
invariance.  This infinity is easier to understand for a nontrivial amplitude
in momentum space, as a factor of a momentum-conservation
$¶$-function (which is either $¥$ or 0, but necessarily $¥$ for the
vacuum amplitude, which has vanishing momentum because the vacuum is
translationally invariant).

Û4 A. BECCHI-ROUET-STORA-TYUTIN

We have seen the relationship of gauge invariances to constraints in
subsection IIIA5.  In this section we consider the quantization of
constrained systems, and its application to gauge theories.  The
Becchi-Rouet-Stora-Tyutin (BRST) method is not only the most powerful,
but also the easiest way to gauge fix:  It replaces the gauge symmetry
with an unphysical, fermionic, global symmetry that acts only on
unphysical degrees of freedom.

Ü1. Hamiltonian

Physical observables commute with the constraints.  Thus, time
development is described by the gauge-invariant Hamiltonian $H_{gi}$, or
we can set the gauge fields $Â^i$ equal to some arbitrary functions of time $f^i(t)$
as a ``gauge choice":
$$ Â^i = f^iâÜâH = H_{gi} +f^i G_i $$
 In quantum mechanics, physical states should be annihilated by the
constraints.  However, more generally we can require only that these
states satisfy the constraints through expectation values:
$$ ÒÆ|G_i|Ô = 0 $$
 This condition is satisfied by dividing up the constraints into:   
\item{(1)} a subalgebra $G_0$ that annihilates all physical states,   
\item{(2)} complex
``lowering operators" $G_-$ that also annihilate these states, and are a
representation of the subgroup generated by $G_0$, and   
\item{(3)} their
hermitian conjugate ``raising operators" $G_+=G_-ÿ$.  

\noindent (We treat $+$, $-$,
and $0$ here as multivalued indices.)  Thus,
$$ (G_0-const.)|ÆÔ = G_-|ÆÔ = ÒÆ|(G_0-const.) = ÒÆ|G_+ = 0 $$
 where we have allowed for ``normal-ordering" constants to be
included as eigenvalues of (an Abelian subeset of) the $G_0$ constraints.
 ÓIn the Abelian case,Õ it then follows that, although $G_+$ do not
annihilate these states, they generate gauge invariances:
$$ ¶|ÆÔ = G_+|½^+Ô,ââ(G_0-const.')|½^+Ô = G_-|½^+Ô = 0 $$
 preserves the inner product of such states as well as the constraints on
$|ÆÔ$.  This may include ``residual gauge invariances" that survive in a solution to the constraints.

Unfortunately, things get more complicated in the nonabelian case.  For
example, the gauge invariance and constraints above are no longer
compatible in general:
$$ 0 = G_- ¶|ÆÔ = G_- G_+|½^+Ô = [G_-,G_+]|½^+Ô = -if_{-+}{}^+ G_+|½^+Ô ±0 $$

However, one example that doesn't have this problem is the simple
case where there are only 3 constraints, forming an SU(2) algebra
(so $f_{-+}{}^+=0$):
If we choose $G_-$ to be the lowering operator and $G_+$ to be the
raising operator, then the constant appearing for $G_0$ is simply
the lowest eigenvalue in some irreducible representation in the Hilbert
space $|ÆÔ$, and the constraints pick out the corresponding state
(or states, if there is more than one representation with that ``spin").

A convenient way to deal with this problem is to replace the
nonabelian algebra $G_i$ with a single operator, which is therefore
Abelian.  We define a BRST operator $Q$ that imposes all constraints $G_i$
by adding a classical anticommuting ``ghost" variable $c^i$, and its
canonical conjugate $b_i$,
$$ Ób_i,c^jÕ = ¶_i^j $$
 for each constraint:
$$ Q = c^i G_i -iüc^i c^j f_{ji}{}^k b_k $$
 The second term has been added to insure the Poisson bracket or
commutator
$$ ÓQ,QÕ = 0 $$
 so that its crossterm cancels the square of the first term, while its own
square vanishes by the Jacobi identity $f_{[ij}{}^l f_{k]l}{}^m=0$.  
Quantum mechanically, the BRST operator is nilpotent:
$$ quantum¼mechanicallyââÓQ,QÕ ­ 2Q^2 = 0 $$
  We can also describe the ghost dependence by the ``ghost number"
$$ J = c^i b_iâÜâ[J,Q] = Q $$
 (Quantum mechanically, we need to normal order these expressions for
$Q$ and $J$.)  
Each anticommuting ghost and its conjugate will serve to ``cancel"
each commuting constraint and its conjugate gauge degree of freedom.
(Similarly, bosonic ghosts are introduced for fermionic constraints,
so each term in $Q$ is fermionic.)

The BRST operator provides a convenient method to treat more general
gauges than $Â=f$, such as ones where the gauge fields become
dynamical, which will prove useful particularly in relativistic theories. 
Now the original physical observables $A$ will satisfy
$$ [G_i,A] = [b_i,A] = [c^i,A] = 0âÜâ[Q,A] = [J,A] = 0 $$
 and similarly the physical quantum mechanical states $|ÆÔ$ will satisfy
$$ (G_0,G_-;b_0,b_-;c^+)|ÆÔ = 0âÜâQ|ÆÔ = J|ÆÔ = 0 $$
 where we have used the fact the only nonvanishing structure constants
are $f_{00}{}^0$, $f_{0+}{}^+$, $f_{0-}{}^-$, $f_{++}{}^+$, $f_{--}{}^-$, and
$f_{+-}{}^k$.  (In the quantum case there are also some subtleties due to
normal ordering.)  We also have the gauge invariances
$$ ¶A = ÓQ,ñÕ,â¶|ÆÔ = Q|ÂÔ $$
 for arbitrary operators $ñ$ and (unrelated) states $|ÂÔ$, since the
$Q$-terms won't contribute when evaluating matrix elements with states
annihilated by $Q$:
$$ (ÒÆ_1|+ÒÂ_1|Q)(A+ÓQ,ñÕ)(|Æ_2Ô+Q|Â_2Ô) = ÒÆ_1|A|Æ_2Ô $$
 The gauge invariances are consistent with the constraints because of
the nilpotence of the BRST operator.  (So $Q(|ÆÔ+Q|ÂÔ)$ still vanishes, etc.) 
States satisfying
$$ Q|ÆÔ = 0,â¶|ÆÔ = Q|ÂÔ $$
 (i.e., we identify states that differ by $Q$ on something) are said to be in
the ``cohomology" of $Q$ (``BRST cohomology"), and operators satisfying
$$ [Q,A] = 0,â¶A = ÓQ,ñÕ $$
 are said to be in its ``operator cohomology".  (The latter cohomology also
has a classical analog.)

\x VIA1.1  Assume that each physical state can be represented as a
physical observable (Hermitian operator) acting on a ground state, which
is itself physical:
$$ |ÆÔ = A|0Ô,ââA = Aÿ,ââQ|0Ô = 0 $$
 Show how this relates the gauge parameters and cohomologies of $|ÆÔ$
and $Q$.

The BRST operator incorporates the ghosts that are necessary to
generalize treatment of the constraints to the nonabelian case:  For
example, to reproduce the gauge transformations of the Abelian case, we
choose
$$ |ÂÔ = b_+|½^+Ô,ââQ|½^+Ô = 0âÜâ¶|ÆÔ = Q|ÂÔ = öG_+|½^+Ô $$
 where 
$$ öG_i = ÓQ,b_iÕ = G_i -i c^j f_{ji}{}^k b_k $$
 are the ``gauge-fixed" constraints, which include an extra term to
transform the ghosts as the adjoint representation. They reduce to just
$G_i$ in the Abelian case, but add ghost terms to the gauge
transformation law otherwise.

In particular, the Hamiltonian is a physical operator describing the energy
and the time development, so we can write
$$ H = H_{gi} +ÓQ,ñÕ,â[Q,H_{gi}] = 0âÜâ
	\T\left(e^{-iÇdt¼H}\right) = \T\left(e^{-iÇdt¼H_{gi}}\right) +ÓQ,ûÕ $$
 for some $û$.  This includes gauge fixing for the gauge $Â^i=f^i$
discussed above, using:
$$ ñ = f^i b_iâÜâH = H_{gi} +f^i öG_i,âL = -(Àq{}^m p_m -iÀc{}^i b_i) +H $$
 The ghost terms in $öG_i$ only affect the time development of unphysical
states in this gauge.

For example, when calculating S-matrix elements, the result is
independent of the gauge choice $ñ$, as long as both the
gauge-invariant Hamiltonian $H_{gi}$ and the states are BRST invariant. 
($H_{gi}$ commutes with $Q$, the initial and final states are annihiliated by
it.)  It is also independent of the gauge choice $|ÂÔ$ for $|ÆÔ£|ÆÔ+Q|ÂÔ$. 
(Such a ``residual gauge invariance" persists even though the asymptotic
states satisfy the free field equations.)

In the cases of interest in relativistic physics, the constraints always
consist of a linear term depending only on the canonical momenta $p$
(conjugate to the fundamental variables $q$), at least after some
redefinitions, plus higher-order terms, which can be treated
perturbatively.  Therefore, as the simplest nontrivial example, we
consider a model with a single variable $q$, with
$$ H_{gi} = 0,âG = pâÜâQ = cp $$
 If we assume boundary conditions on the wave functions such that they
can be Taylor expanded in $q$ (they can always be expanded in $c$), we
can write
$$ Æ(q) = Òq|ÆÔ = Ý_{n=0}^¥ (Œ_n +cº_n)\f1{n!}q^n $$
 and similarly for $Â(q)$.  We then examine $¶|ÆÔ=Q|ÂÔ$, 
comparing terms with the same power of $c$ and $q$ on both sides
of the equation, to find $¶Œ_n$ and $¶º_n$.
We then see that we can easily gauge $º_n=0$ for all $n$
by choosing certain coefficients in $|ÂÔ$ to be $-º_n$
(so $º_n'=º_n+¶º_n=0$).  
Looking at $Q|ÆÔ=0$, we then find that $Œ_n=0$
for all $n$ except $n=0$, so only the constant piece of $Æ$ survives.  In
other words, the cohomology is given by
$$ Q|ÆÔ = 0,â¶|ÆÔ = Q|ÂÔâÜâp|ÆÔ = b|ÆÔ = 0 $$
 So, solving for the cohomology of $Q=cp$ is the same as solving the
constraint $pÆ=0$ without ghosts.

\x VIA1.2  Consider creation and annihilator operators satisfying
$$ [a,dÿ] = [d,aÿ] = Óc,bÿÕ = Ób,cÿÕ = 1 $$
 (the other commutators vanishing):
 ªa Find the cohomology of the BRST operator
$$ Q = caÿ +cÿa $$
 by expanding in creation operators
$aÿ$, $bÿ$, $cÿ$, $dÿ$ about a vacuum state destroyed by the annihilation
operators $a$, $b$, $c$, $d$.  (This is the common alternative to the
boundary conditions used for $Q=cp$ above.)
 ªb Compare the method of constraints used in the Abelian case:
Ignore the fermions, and identify which bosons are constraints and how they are applied.

Ü2. Lagrangian

To obtain more interesting gauges we need some extra bosonic variables,
such as the gauge fields $Â^i$ that we lost along the way, and their
canonical conjugates (``Nakanishi-Lautrup fields") $B_i$,
$$ [B_i,Â^j] = -i¶_i^j $$
 as well as their corresponding ghosts (``antighosts") $÷c_i$.  We can do
this in a trivial way by including constraints that set $B$ to zero:
$$ Q = c^i G_i -iüc^i c^j f_{ji}{}^k b_k + ÷b^i B_i,ââJ = c^i b_i -÷c_i ÷b^i $$
 where $÷c_i$ is conjugate to $÷b^i$,
$$ Ó÷c_i,÷b^jÕ = ¶_i^j $$
 As a simple example, consider
$$ ñ = Â^i b_iâÜâÓQ,ñÕ = Â^i öG_i -i÷b^i b_i $$
 The action now includes the gauge fields and all the ghosts as dynamical
variables:
$$ L = -(Àq{}^m p_m +ÀÂ{}^i B_i -iÀc{}^i b_i -iÀ{÷b}{}^i ÷c_i) +H_{gi} +ÓQ,ñÕ $$
 For this gauge we can eliminate $b$ and $÷b$ by their equations of
motion; assuming $G_i$ is only linear in $p$, we then can eliminate $p$ to
return completely to a Lagrangian formalism:
$$ L = L_{gi}(q,Â) -ÀÂ{}^i B_i -i(á_t c^i)À{÷c}_i $$
 where $L_{gi}$ represents the original gauge-invariant action (which
depended on both $q$ and $Â$, including time derivatives), and $á_t$ is
the covariant (time) derivative:
$$ á_t c^i = Àc{}^i +c^j Â^k f_{kj}{}^i $$
 The gauge condition (from varying $B$) is now $ÀÂ=0$, generalizing the
non-derivative gauges found without the antighosts and
Nakanishi-Lautrup fields.  Correspondingly, the ghost term is now second
order in derivatives.

\x VIA2.1 Consider the general gauge choice
$$ ñ = Â^i b_i +[F^i(q,p) +E^i(B)]÷c_i  $$
$$ ÜâÓQ,ñÕ = Â^i öG_i -i÷b^i b_i +(F^i +E^i)B_i +c^i[G_i,F^j]÷c_j $$
 where $F^i$ are some arbitrary functions of the original variables, and
$E^i$ are functions that effectively average over the types of gauges
produced by $F^i$.  Find the gauge-fixed Hamiltonian and Lagrangian.  In
the case where $E$ is linear in $B$, eliminate $B$, $b$, and $÷b$ from the
Lagrangian by their algebraic equations of motion.

Now that we understand the principles, all these manipulations can be
performed directly in the Lagrangian formalism.  This will have the
advantage that in field theory the Lagrangian is manifestly Lorentz
covariant, while the Hamiltonian (or the Lagrangian in the Hamiltonian
form $-Àqp+H$) is not, because of the way it singles out time derivatives
(and not spatial ones).  (Consider, e.g., electromagnetism.)  Similarly, neither the unitary gauge $G_i=0$ nor the temporal gauge $?_i=0$ is usually Lorentz covariant. 
We can work with just the
original variables $q,Â$ plus the new variables $B,c,÷c$, and define
$Q$ by the transformation it induces (as derived from the Hamiltonian
formalism):
$$ Qq^m = c^i ¶_i q^m,âQÂ^i = -i(Àc{}^i +c^j Â^k f_{kj}{}^i),â
	Qc^i = -iüc^j c^k f_{kj}{}^i,âQ÷c_i = B_i,âQ B_i = 0 $$
 where $¶_i$ is the gauge transformation induced by $G_i$ ($[G_i,¼]$ in the
Hamiltonian formalism).  
In deriving $Q$, we have used the equations of motion of $p,b,÷b$ (which were eliminated).
Note that the BRST transformations of the
original variables are exactly the same as the gauge transformations,
with the gauge parameters replaced with the corresponding ghosts.  We
can also consider the Nakanishi-Lautrup fields $B$ as original variables,
with the fact that they don't occur explicitly implying they have
constraints $B=0$.  Alternatively, we can treat the antighosts $÷c$ as pure
gauge degrees of freedom, with their own nonderivative gauge
transformation $¶÷c=÷Â$ that allows them to be completely gauged away.

The Lagrangian can be gauge-fixed directly as
$$ L = L_{gi} +Qñ_L $$
 where in the case just considered
$$ ñ_L = -ÀÂ{}^i ÷c_i  $$
 gives the same $L$ as above for the $ÀÂ=0$ gauge.  In the simpler case
described earlier (the gauge $Â$ = function of $t$ only)
$$ ñ_L = (Â^i -f^i)÷c_i $$
 This gives the result, for the simplest choice $f=0$,
$$ L = L_{gi}(q,Â) +Â^i B_i -i(Àc{}^i +c^j Â^k f_{kj}{}^i)÷c_iâ
	£âL_{gi}(q,0) -iÀc{}^i ÷c_i $$
 after eliminating the Lagrange multipliers $B$ and $Â$ by their algebraic
equations of motion.  Note that $ñ_L=Â^i ÷c_i$ corresponds to the
Hamiltonian formalism's $ñ=0$.  Thus, in the Hamiltonian formalism we
never quantize with $H=H_{gi}+ÂG$, but can use just $H_{gi}$ and $ñ=0$,
which is equivalent to using $H_{gi}+ÓQ,ñÕ$ for any $ñ$, while in the
Lagrangian formalism we can never quantize with just $L_{gi}(q,Â)$, or
even $L_{gi}(q,0)$, and $ñ_L$ is never zero, but must be chosen so as to
break the gauge invariance. However, the extra term for $L_{gi}(q,0)$ is
just the $Àcb$ term found from converting $H_{gi}$ to the Lagrangian
formalism.

\x VIA2.2 Repeat exercise VIA2.1 directly in the Lagrangian formalism. 
(Find $ñ_L$, etc.)

All our results for quantization apply equally well in the path-integral
formalism, which can be applied to either the Hamiltonian or Lagrangian. 
(Of course, for field theory we will be interested in applying BRST to path
integrals for Lagrangians.)  We then evaluate matrix elements as
$$ \A = ÇDļï[Ä]e^{-iS[Ä]};ââ
	S = S_{gi} +Qñ,âï = ï_{gi} +Qñ_ï;ââQS_{gi} = Qï_{gi} = 0 $$
 $S_{gi}$ and $ï_{gi}$ depend on just the physical fields (no ghosts), so
they are gauge invariant as well as BRST invariant.  For S-matrices, since
$ï$ is an asymptotic state, the BRST operator used for its constraint and
gauge invariance can be reduced to its free part:  $Q$ then acts on only
the gauge fields.  The statement of gauge invariance of $ï_{gi}$ is then
equivalent to the requirement that gauge fields appear in it only as their
Abelian field strengths.  For example, the usual gauge vector $A$
describing electromagnetism appears in single-particle factors in the
wave functional  ($ï[Ä]=Þï_1[Ä]$ as in subsection VC1) only as:
$$ ï_1[A] = ÒA^a||Æ_aÔ,ââ¶A_a = -»_a Â,ââõÆ_a = õ = 0 $$
$$ Üâ0 = ¶ï_1 = Ò¶A^a||Æ_aÔ = ÒÂ||»^a Æ_aÔ $$
 using $»_0ÒÂ||ÆÔ=0$ (where the relativistic inner product $Ò¼||¼Ô$ was
defined in subsection VB2).  The transversality of $Æ_a$ is equivalent to
coupling to the Abelian field strength, since
$$ »^a Æ_a = 0âÜâÆ_a = »^b Æ_{ab}âÜâ
	ÒA^a||Æ_aÔ = üÒF^{ab}||Æ_{ab}Ô $$
 in terms of an antisymmetric-tensor external-line factor $Æ_{ab}$.

Ü3. Particles

We have seen that the relativistic particle (with or without spin) is a
simple example of a contrained system.  For the simplest case, spin 0, the
BRST operator follows simply from the single constraint:
$$ Q = cü(õ-m^2) $$
 Unlike the nonrelativistic case, the relativistic ``Hamiltonian" is
identified with this constraint.  Since we know constraints are treated by
the BRST operator, we can consider writing the field theory action in
terms of it:
$$ S_0 = -Çdx¼dc¼üìQì $$
 Using the explicit $c$ dependence of the field $ì=Ä-icÆ$, we find the
usual scalar kinetic term.  $Ä$ is thus the usual field, while $Æ$ is an
``antifield", which has opposite statistics to $Ä$ (fermion instead of
boson).  We'll see in chapter XII that $Q$ can be constructed
straightforwardly for arbitrary spin, and has a simple expression in term
of generalized spin operators.  (As in nonrelativistic theories, spin is
easier to treat directly in quantum mechanics rather than by
first-quantization of a classical system.)  The kinetic term then generally
can be written as a slight modification of the above.  Then the antifields
will be found to play a nontrivial function, rather than just automatically
dropping out as in this case.

From the constraints and their algebra for spin 1/2 (see also exercise
IIIB1.2) we find the BRST and ghost-number operators:
$$ Q = cü(õ-m^2) -Å(©É»-i\f{m}{å2}) -Å^2 b +÷ŵ,ââJ = cb +Ž +÷Å÷½ $$
 where $Å$ and its conjugate $½$ are bosonic ghosts, and we have added a
nonminimal term with boson $÷Å$ (conjugate $÷½$) and fermion $µ$
(conjugate $û$) to allow gauges general enough for first-quantization:
$$ [½,Å] = [÷½,÷Å] = Óû,µÕ = 1 $$
 For convenience, we also have chosen $Å$ (and $½$) to ÓantiÕcommute
with $©$,
$$ ÓÅ,©^aÕ = Ó½,©^aÕ = 0 $$
 to avoid having to replace $-i\f{m}{å2}$ with $©_{-1}im$; this has the
natural interpretation of treating $Å$ and $½$ as bosonic (ghost)
components of the $©$ matrices (see subsections XIIA4-5,B5).

\x VIA3.1  Find $Q$ and $J$ for spin 1 as constructed from the
direct product of 2 spin 1/2's (see exercise IIB4.1d).

Note that $[Q,Å]=0$, but $ÓQ,AÕ±Å$ for any $A$, so $Å$ is in the operator
cohomology of $Q$.  Normally, this would imply infinite copies of the
physical states in the cohomology, since applying a ``translation" with the
ghost variable $Å$ gives a new state in the cohomology from any given
one.  The nonminimal variables allow us to avoid this problem by
combining $Å$ with $÷Å$ to produce harmonic oscillator creation and
annihilation operators:
$$ Å = \f1{å2}(a+aÿ),â÷Å = \f1{å2}i(aÿ-a),â
	½ = \f1{å2}(÷a-÷a{}ÿ),â÷½ = -\f1{å2}i(÷a{}ÿ+÷a) $$
$$ [a,÷aÿ] = [÷a,aÿ] = 1,ârest = 0 $$
 This allows us to define a ground state
$$ a|0Ô = ÷a|0Ô = 0 $$
 which breaks the translation symmetry of $Å$.  In chapter XII we'll show
in a more general framework how the $ìQì$ type of action then
reproduces the Dirac action.

Ü4. Fields

As described in subsection VIA2, we can perform gauge fixing through
BRST, including the introduction of ghosts, directly on the Lagrangian at
the classical level.  Also, the BRST transformations on the physical fields
are just the gauge transformations with the gauge parameters replaced
by ghosts, and the BRST transformation on the ghosts is quadratic in
ghosts times the structure constants, while on the antighosts it gives the
Nakanishi-Lautrup fields, and it annihilates the NL fields.  In the case of
Yang-Mills we then have
$$ \boxeq{ QA_a = -[á_a,C],âQC = iC^2,âQ÷C = -iB,âQB = 0 } $$
 while for matter transforming as $¶Ä=iÂÄ$ we have
$$ \boxeq{ QÄ = iCÄ } $$
 where we have used matrix notation for the group algebra, as usual. 
There are two minor differences from the transformation rules we used
in our general discussion previously:  (1) We have included an extra ``$i$"
in our definition of the relativistic $Q$, for a convenience that will become
apparent only when we relate relativistic first- and second-quantization
(see chapter XII).  (2) There is a relative sign difference for $QC$ because
now $Q$ is second-quantized while $G_i$ is still first-quantized (i.e.,
matrices).  More explicitly, we have, e.g.,
$$ C^2=üÓC,CÕâÜâQC^i = -üC^j C^k f_{kj}{}^i;ââQÄ = iC^i G_i Ä $$

The gauge-fixed action is then the gauge-invariant action plus the BRST
transformation of some function $ñ$:
$$ S_{gf} = S_{gi} -iQñ $$
 For example, consider Yang-Mills in the most common type of gauge,
where some function of $A$ is fixed:
$$ ñ = trÇü÷C[f(A) +üŒB]âÜ $$
$$ \boxeq{ L_{gf} = L_{gi} -üB[f(A) +üŒB] -üi÷C{»f\over »A}É[á,C] } $$
 for some constant $Œ$.  For $Œ=0$, $B$ is a Lagrange
multiplier, enforcing the gauge $f(A)=0$, while for $α0$, we
can eliminate $B$ by its auxiliary field equation:
$$ -üB[f(A) +üŒB] £ \f1{4Œ}f^2 $$
 Examples will be given in the following section.

In field theory gauge-fixing functions always have linear terms, as do
gauge transformations.  Furthermore, there always exist ``unitary
gauges", where no ghosts are required.  The ghost terms in general
gauges serve simply to provide the appropriate Jacobian factor for the
field redefinition that transforms from the general gauge to the unitary
gauge, which appears at the quantum level from functionally integrating
out the ghosts.  The simplest example is the trivial gauge invariance that
occurs in the St¬uckelberg model (subsection IVA5):
$$ QA = -»C,ââQÄ = mC,ââQ÷C = -iB,ââQC = QB = 0 $$
 which we can fix with
$$ -iQ(÷C\O Ä) = -B\O Ä +im÷C\O C $$
 for some field-independent operator $\O$.  Functionally integrating out
$B$ still sets $Ä=0$, but produces an inverse functional determinant of
$\O$ (from redefinition of $Ä$, or from $¶(\O Ä)$), canceled by that from
integrating out the ghosts.  The advantage of BRST is that all this can be
treated at the classical level, in terms of the classical action, without
regard to functional integration, while directly giving a solution that can
be expressed immediately in terms of Feynman rules.

\x VIA4.1 Show that the gauge fixing
$$ -iQ(÷C\O Ä +÷C\A B +C\B Ä +C\C B) $$
 where $\O$, $\A$, $\B$, and $\C$ are field-independent operators, gives a
result equivalent to the previous, by considering functional determinants
or field redefinitions.

\x VIA4.2 Show that the Lagrangian
$$ A\A B +C\B D £ A\A\B D $$
 by the field redefinition
$$ D £ D +\B^{-1}B $$
 for bosons $A$, $B$, $C$, $D$ and operators $\A$, $\B$.  This is the classical
equivalent of $det(\A\B)=det(\A)det(\B)$.

These methods apply straightforwardly to supersymmetric theories in
superspace:  From the gauge transformations of subsection IVC4,
$$ Qe^V = iÐCe^V -ie^V C;ââQC = iC^2,âQÐC = iÐC^2 $$
$$ Q÷C = -iB,âQÑ{÷C} = -iÐB;ââQB = QÐB = 0 $$
 where $C$, $÷C$, and $B$ are chiral superfields, and $ÐC$, $Ñ{÷C}$, and $ÐB$ their
hermitian conjugates.

In practice, this BRST approach is sufficient for gauge fixing.  In
particular, this is true for the fundamental fields used in the standard
model (including gravity), which have spin$²$2.  Therefore, we'll
use mostly this approach in the rest of this text.  However, some
observed hadrons have much higher spin.  The first-quantized approach
of chapter XII gives a natural and direct way of understanding ghosts and
BRST for the fields describing such particles, and translates directly into
the treatment of Zinn-Justin, Batalin, and Vilkovisky (ZJBV) for field
theory.

\refs

 £1 Feynman, Óloc. cit.Õ (VC, ref. 17);\\
	B. DeWitt, ÓPhys. Rev.Õ É162 (1967) 1195, 1239;\\
	L.D. Faddeev and V.N. Popov, \PL 25B (1967) 29;\\
	S. Mandelstam, ÓPhys. Rev.Õ É175 (1968) 1580:\\
	ghosts.
 £2 N. Nakanishi, ÓProg. Theor. Phys.Õ É35 (1966) 1111;\\
	B. Lautrup, ÓK. Dan. Vidensk. Selsk. Mat. Fys. Medd.Õ É34 (1967) 
	No. 11, 1.
 £3 C. Becchi, A. Rouet, and R. Stora, \PL 52B (1974) 344, 
	ÓAnn. Phys.Õ É98 (1976) 287;\\
	I.V. Tyutin, Gauge invariance in field theory and in statistical 
	physics in the operator formulation, Lebedev preprint FIAN No. 39 
	(1975), in Russian, unpublished;\\
	M.Z. Iofa and I.V. Tyutin, ÓTheor. Math. Phys.Õ É27 (1976) 316;\\
	T. Kugo and I. Ojima, \PL 73B (1978) 459;\\
	L. Baulieu, ÓPhys. Rep.Õ É129 (1985) 1:\\
	BRST.
 £4 E.S. Fradkin and G.A. Vilkovisky, \PL 55B (1975) 224;\\
	I.A. Batalin and G.A. Vilkovisky, \PL 69B (1977) 309;\\
	E.S. Fradkin and T.E. Fradkina, \PL 72B (1978) 343;\\
	M. Henneaux, ÓPhys. Rep.Õ É126 (1985) 1:\\
	Hamiltonian BRST.

\unrefs

Û10 B. GAUGES

There are two important properties of gauges we have examined:  (1)
Gauges which eliminate some degrees of freedom, such as lightcone or
unitary gauges, are simpler classically, which makes them easier to
understand physically.  (2) Gauges that manifest as many global
invariances as possible, such as the Fermi-Feynman gauge, will be found
later to simplify quantum calcuations, because the explicit momentum
dependence of the propagator or vertices is simpler, and keeping a
symmetry manifest makes it unnecessary to check.  In this section we'll
examine these gauges in greater detail, especially as they relate to
intereacting theories.

We'll study also some special gauges, with nontrivial interaction terms,
that have both of these properties to some extent.  In particular, they
are manifestly Lorentz covariant, but avoid many of the complications
associated with ghosts.

Ü1. Radial

We know from nonrelativistic classical and quantum mechanics that the
equations of motion can be solved exactly only for certain simple
external field configurations.  One particular case we have already
emphasized is that of an action quadratic in the dynamical variables, i.e.,
the harmonic oscillator and its generalizations.  Higher-order terms are
then treated as perturbations about the exact solution.  Such an
expansion in the coordinates $x$ is the particle version of the JWKB
expansion in $\h$:  Calling the ``classical" part of $x$ ``$y$", we
substitute $x£y+å\h x$ and Taylor expand in $x$.  (From now on we'll drop
the $\h$'s, and just remember to perturb about the quadratic terms.)  For
the scalar field we write
$$ Ä £ Ä +xÉ»Ä +üx^m x^n »_m »_n Ä +... $$
 where $»...»Ä$ is implicitly evaluated at $y$.

For the gauge fields we would like to be a bit more clever:  For example,
for the electromagnetic potential $A_m$ we know we can always add a
constant, so $A_m(y)$ is irrelevant, while for $»A$ only
$F_{mn}=»_{[m}A_{n]}$ is gauge invariant.  This means we want to choose
a gauge best suited to this calculation: a gauge that both eliminates as
many as possible of the lower-order terms, and expresses $A(y+x)$ in
terms of only $F(y)$ and its derivatives.  Similarly, we should have a
Taylor expansion for charged fields in terms of ÓcovariantÕ derivatives.
The appropriate gauge, which easily can be found explicitly, is the ``radial
gauge"
$$ xÉA(y+x) = 0 $$
 (Note that, unlike $F$, $A$ depends on $x$ independently of $y$, not just
as $y+x$, since the gauge condition itself is $x$-dependent.  We write
$A(y+x)$ only to indicate that $A$ is evaluated at position $y+x$.)  One
way to solve this condition is to use the identity
$$ x^n F_{nm} = (xÉ»+1)A_m -[á_m,xÉA]ââ(» = »/»x) $$
 which follows from the definition of $F$.  Using the gauge condition, we
then can write
$$ A_m = {1\over xÉ»+1}x^n F_{nm} $$
 Alternatively, we can replace $x$ everywhere (including the argument
$y+x$) by $ x$, and then identify $xÉ»= »/» $ to find
$$   x^n F_{nm}(y+ x) = »_    A_m (y+ x) $$
 Integrating both sides over $ $ from 0 to 1, we find
$$ A_m(y+x) = Ç_0^1 d ¼ x^n F_{nm}(y+ x) $$
 Note in particular that $A(y)=0$.

Another way to define this gauge is to consider Ógauge covariantÕ
translation from $y$ to $y+x$ to produce a gauge transformation from an
arbitrary gauge to the radial gauge.  Writing the covariant derivative at
$y$ as
$$ \D = D +iA(y),ââD = »/»y $$
 we know from subsection IIIC2 that
$$ Æ'(y+x) ­ e^{xÉ\D}Æ(y) = e^{iñ}e^{xÉD}Æ(y) = e^{iñ}Æ(y+x) $$
 so that covariant translation produces a $Æ'(y+x)$ that is the same as
$Æ(y+x)$ up to a gauge transformation.  Thus the gauge-transformed $Æ$
can be written as a covariant Taylor expansion (for purposes of
perturbation) about $y$:
$$ Æ'(x+y) = 
	Ý_{n=0}^¥ \f1{n!}x^{a_1}òx^{a_n}(\D_{a_1}ò\D_{a_n}Æ')(y) $$
 In particular, $Æ'(y)=Æ(y)$.

However, we want to define a covariant derivative with respect to $x$
(not $y$), so that
$$ á = » +iA'(x+y),ââáÆ'(y+x) = (\D Æ)'(y+x) = e^{xÉ\D}(\D Æ)(y) $$
 Using
$$ Ȯ(y) = 0 $$
 we find the solution
$$ á = e^{xÉ\D}\D e^{-xÉ\D}¼mod¼e^{xÉ\D}» e^{-xÉ\D} $$
 where the latter term vanishes on $Æ'(y+x)$, so the right amount of it
can be added to the former expression to cancel any $D$ terms.  The
result is
$$ á = e^{xÉ\D}(»+\D)e^{-xÉ\D} $$
 This implies $xÉA'(y+x)=0$ directly:  Contracting both sides with $x$, the
Taylor expansion of the right-hand side terminates after the first couple
of terms.  Taylor expanding the uncontracted expression, we have
$$ A'_a(y+x) = Ý_{n=0}^¥ \f1{n!}\f1{n+2}(xÉ\D)^n x^b F_{ba}(y) $$
 We can also write
$$ » +iA'(x+y) = á = e^{xÉ\D}e^{-xÉD}[» +iA(x+y)]e^{xÉD}e^{-xÉ\D}
	= e^{iñ}[» +iA(x+y)]e^{-iñ} $$
 and
$$ á = e^{xÉ\D}e^{-yÉ»}\D e^{yÉ»}e^{-xÉ\D} $$

\x VIB1.1  Show this Taylor expansion is equivalent to that obtained from
the first method used in this section to solve the gauge condition.  (Hint: 
Look out for hidden $x$ and $y$ dependence --- How does $xÉ»$ on $Æ'$ or
$F'$ relate to $xÉ\D$?  Also beware of notation:  In the first construction
we did not use a gauge transformation, so no primes were used.)

\x VIB1.2  Generalize this construction to superspace (see subsection IVC3):
 ªa First give an expression for the gauge potential $A_A$ in terms of covariant derivatives of field strengths $F_{AB}$.
 ªb Then look at the expansion in just $Ï$.  Give the explicit result for the expansions of $A_Œ$ and $A_a$ about $Ï=0$, applying the constraints:  This is the ``Wess-Zumino gauge" (see exercise IVC4.2 for the Abelian case).

Thus, to just quadratic order in $x$, the mechanics action for a relativistic
particle in external fields (subsection IIIB3) becomes
$$ S_L ® Çd  Ó-üv^{-1}ú_{mn}Àx{}^m Àx{}^n +üÀx{}^m x^n F_{nm}(y) $$
$$ +v[Ä(y) +x^m(»_m Ä)(y) +üx^m x^n(»_m »_n Ä)(y)]Õ $$
 To this approximation the classical equations of motion are linear and
can be solved exactly.  It can also be used to find exact solutions for
constant electromagnetic fields.

Ü2. Lorenz

For purposes of explicit calculations in perturbation theory, it's more
convenient to use gauges where Lorentz covariance is manifest. 
``Lorenz gauges" are a class of gauges using 
$$ f = ȃA $$
 (and similarly for other gauge fields) as the gauge-fixing function.  From
the discussion of subsection VIA4, we have from the usual BRST as applied
to Yang-Mills
$$ L_{gf} = \f18 F_{ab}^2 -iQ¼ü[÷C(»ÉA +üŒB)] $$
$$ = -\f14 AÉõA -\f14(»ÉA)^2 -ü[A^a,A^b](-i»_a A_b +üA_a A_b) 
	-üB»ÉA -\f14ŒB^2 -üi÷C»É[á,C] $$
 After eliminating $B$ by its field equation, the kinetic terms are
$$ -\f14 AÉõA -\f14(»ÉA)^2 +\f1{4Œ}(»ÉA)^2 -üi÷CõC $$
 In particular, for $Œ=1$ we have the ``Fermi-Feynman" gauge, which
gives the nicest propagators.  (It is also the gauge that follows
automatically from a first-quantized BRST construction, which will be
described in chapter XII.)  More generally, we find the propagator from
inverting the kinetic operator:  For the ghosts this is always $2/p^2$, but
for $A_a$,
$$ 2[ú^{ab}p^2 +(\f1Œ-1)p^a p^b]^{-1} = 
	2\left[ {ú_{ab}\over p^2} +(Œ-1){p_a p_b\over (p^2)^2}\right] $$
 For $Œ=0$ this is the ``Landau gauge", which has the advantage that the
propagator is proportional to the transverse projection operator.  (It kills
terms proportional to $p_a$.)  However, $Œ=1$ is clearly the simplest, and
the $1/p^4$ term can cause problems in perturbation theory.

\x VIB2.1 In the Abelian case, consider making a gauge transformation on
the Ógauge-fixedÕ action (including matter), with $¾õ^{-1}B$.  Show that
the only effect is to change the value of the coefficient $Œ$ of the $B^2$
term.  Find a similar transformation for the form of the action where $B$
has been eliminated.  This shows explicitly the decoupling of the
longitudinal mode of the photon.

\x VIB2.2 Show that for general $\A$ and $\B$
$$ (ú^{ab}\A +p^a p^b\B )^{-1} = 
	{1\over \A}\left( ú_{ab} -p_a p_b {\B \over \A +p^2 \B} \right) $$ 

Note that in the Abelian case the lightcone gauge is a special case of the
Landau gauge.  (An analogous situation occurs in the classical
mechanics of the particle for the gauges of the worldline metric, as
discussed in subsection IIIB2.)  Here we have
$$ 0 = n^a (»^b F_{ab}) = nÉ» (»ÉA) -õ (nÉA) $$
 In the lightcone formalism, this is the field equation that comes from
varying the auxiliary field.  In the lightcone gauge $nÉA=0$, it implies
$»ÉA=0$ (and thus also $õA_a=0$), since $nÉ»$ is invertible.  

This is particularly useful in D=4, where we can generalize from the
lightcone to a Lorentz-covariant form by using twistors:  From subsection
IIB6,
$$ p^2 = 0âÜâp^{ŒÀŒ} = ·(p^0)p^Œ Ðp^{ÀŒ},ââ
	n^2 = 0âÜân^{ŒÀŒ} = ·(n^0)n^Œ Ðn^{ÀŒ} $$
 Massless spinors are described on shell in momentum space by (see subsection IIB7)
$$ Æ^Œ = p^Œ Ä,ââÐÆ^{ÀŒ} = Ðp^{ÀŒ} ÐÄ $$
 where external-line factors for Feynman diagrams are given by setting
$Ä=1$.  For massless vectors, we have $pÉA=nÉA =0$ (but $nÉp±0$), so
depending on whether the helicity is +1 (self-dual field strength) or $-$1
(anti-self-dual field strength), we find, respectively,
$$ Ðf^{ÀŒÀº} ¾ Ðp^{ÀŒ}Ðp^{Àº},ââf^{Œº} = 0ââÜââ
	A^{ŒÀº} =  {n^Œ Ðp^{Àº} \over n^© p_©}ÐÄ $$
$$ f^{Œº} ¾ p^Œ p^º,ââÐf^{ÀŒÀº} = 0ââÜââ
	A^{ŒÀº} =  {p^Œ Ðn^{Àº} \over Ðn^{À©}Ðp_{À©}}Ä $$
 The normalization of $A$ has been chosen compatible with $|Ä|=1$ and
$A^a A*_a=1$ for evaluating cross sections. In a general Landau gauge the
arbitrary gauge-dependent polarization spinors $n^Œ$, $Ðn^{ÀŒ}$ can be
chosen independently for each external line, since gauge invariance
means independent gauge parameters for different momenta.  (This
method is known as ``spinor helicity".)  However, in a lightcone gauge the
polarization spinors are constant.

The lightcone gauge condition is thus again a stronger gauge condition
than Lorenz gauges, as expected from the fact that it has fewer
derivatives.  This difference shows itself in various ways:    
\item{(1)} In perturbation theory on shell, in the lightcone frame the
Landau gauge condition $0=pÉA=-p^+ A^-$ kills $A^-$ but says nothing
about $A^+$, which can be eliminated by the residual gauge invariance
$¶A^+=p^+ Â$ to obtain the lightcone gauge.    
\item{(2)} In perturbation theory off
shell, more derivatives in the gauge transformation imply more
derivatives in the ghost kinetic operator.  Thus, more ghost degrees of
freedom are introduced to cancel the extra unphysical degrees of
freedom in the gauge field.    
\item{(3)} Lorenz gauges also have a
nonperturbative ambiguity (the ``Gribov ambiguity") that axial gauges
avoid:  Nonperturbative solutions to the gauge condition can be found
that differ from the perturbative one, in the nonabelian case. 
Specifically, it is possible to find a nontrivial gauge transformation
${\bf g}$ ($á'={\bf g}^{-1}á{\bf g}$) such that 
$$ 0 = »ÉA' = -i»É{\bf g}^{-1}(á{\bf g})âforâ»ÉA = 0 $$
 even when ${\bf g}$ is required to satisfy boundary conditions that it
approach the identity at infinity (except in the Abelian case, where
${\bf g}=e^{iÂ}ÜõÂ=0ÜÂ=0$).  This is not the case for axial gauges, where
$$ 0 = nÉA' = nÉ{\bf g}^{-1}(á{\bf g})âforânÉA = 0âÜâ
	{\bf g}^{-1}(nÉ»{\bf g}) = 0âÜâ{\bf g} = I $$
 by simply integrating from infinity.

Ü3. Massive

In subsection IIB4 we described the introduction of mass for the vector
by dimensional reduction, giving the St¬uckelberg formalism for a
massive (Abelian) gauge field.  The gauge-invariant action (subsection
IVA5) and BRST transformation laws (subsection VIA4) followed from
adding an extra dimension and setting the corresponding component of
the momentum equal to the mass:
$$ L_{gi} = \f18 F_{ab}^2 +\f14(mA_a +»_a Ä)^2 $$
$$ QA_a = -»_a C,âQÄ = mC,âQ÷C = -iB,âQB = 0 $$
 where the scalar is the extra component of the vector.  

There are two obvious covariant gauges for such a vector:  (1) The
``unitary gauge"
$$ f = Ä $$
 simply gauges away the scalar.  Since the scalar has a
nonderivative gauge transformation, the ghosts do not propagate:  The
gauge-fixing term
$$ -iQ(÷CÄ) = -BÄ +im÷CC $$
 simply eliminates the scalar and ghosts as auxiliary fields.  The net result
is that we could have simply chosen
$$ gaugeâÄ = 0 $$
 and ignored ghosts because of $Ä$'s nonderivative transformation law. 
Thus the gauge-fixed Lagrangian is just the result of adding a mass term
to the massless Lagrangian:
$$ L_{gf} = \f18 F_{ab}^2 +\f14 m^2 A^2 $$
 But the propagator is
$$ 2[ú^{ab}(p^2 +m^2) -p^a p^b]^{-1} = 
	2\left[{ú_{ab}\over p^2+m^2} 
		+{p_a p_b\over m^2(p^2+m^2)}\right] $$
 Notice that the second term is higher in derivatives than the first; this
can cause some technical problems in perturbation theory.  

(2) The Fermi-Feynman gauge works similarly to the massless case.  We
then modify the gauge-fixing function to 
$$ f = »ÉA+mÄ $$
 so
$$ -iQ[ü÷C(»ÉA +mÄ +üB)] = -üB(»ÉA +mÄ +üB) -üi÷C(õ-m^2)C $$
$$ \li{ ÜâL_{gf} & = 
	\f18 F^2 +\f14 (mA+»Ä)^2 +\f14(»ÉA+mÄ)^2 -üi÷C(õ-m^2)C \cr
	& = -\f14 AÉ(õ-m^2)A -\f14 Ä(õ-m^2)Ä -üi÷C(õ-m^2)C \cr} $$
 The propagators are again simpler.  The vector has D propagating
components instead of just the D$-$1 physical ones; the 2 ghosts cancel
$Ä$ and the extra component in $A$.

\x VIB3.1 Generalize the Fermi-Feynman gauge for the St¬uckelberg
formalism to the ``re\-normalizable gauges" with gauge-fixing function
$$ f = {m \over µ}»ÉA+µÄ $$
 ªa Find the gauge-fixed action.
 ªb Show that the ghosts and $Ä$ have mass $µ$, while the vector
propagator has the form
$$ 2\left[ \left( ú_{ab} +{p_a p_b\over m^2} \right) {1\over p^2 +m^2}
	-{p_a p_b\over m^2}{1\over p^2 +µ^2} \right] $$
 This shows explicitly the second unphysical bosonic mode of mass $µ$ to
cancel the 2 ghosts, as well as the 3 transverse physical modes of mass
$m$.  
 ªc Look at the cases 
$$ µ = \cases{ 0 & (Landau gauge) \cr 
	m & (Fermi-Feynman gauge) \cr
	¥ & (unitary gauge) \cr} $$

These two choices of gauge also exist for Yang-Mills theories exhibiting
the Higgs mechanism, since those models give the St¬uckelberg model
when linearized about the vacuum values of the fields.  The advantages
are the same:  The unitary gauge eliminates as many unphysical degrees
of freedom as possible (see subsection IVA6 for an example), while the
Fermi-Feynman gauge gives the simplest propagators.

\x VIB3.2 Work out the Fermi-Feynman gauge for an arbitrary Higgs model,
generalizing the analysis for the St¬uckelberg case.

Ü4. Gervais-Neveu

We next consider pure Yang-Mills theory for the gauge group U(N), but
use a ÓcomplexÕ gauge-fixing function
$$ f_0 = ȃA +iA^2 $$
 where $A_a$ is a vector of hermitian N$ð$N matrices, and $A^2­A^a A_a$. 
(The hermitian conjugate, $i£-i$, gives similar results.)  The gauge-fixed
Lagrangian (in the action $S=g^{-2}trÇL$) is then
$$ L_A = \f18F^2 +\f14 f_0^2 = 
	-\f14 AÉõA -iA^a A^b »_b A_a -\f14 A^a A^b A_a A_b $$
 (where $õ$ is the free D'Alembertian) while the ghost action can be
written as
$$ L_C = -üi÷Cá^2 C -ü÷CCf_0 $$
 where $á$ acts on $C$ as if it were in the defining representation
(i.e., $áC=»C+iAC$, not $[A,C]$).  This ``Gervais-Neveu gauge"
already has the simplification that some of the terms in the Yang-Mills
self-interaction have been canceled.

\x VIB4.1  Consider the ``anti-Gervais-Neveu gauge", where the same
gauge-fixing term is added with opposite overall sign.
 ªa  Show the resulting Lagrangian can be written as
$$ L_A ¾ tr[(Ö»ÖA +iÖA^2)^2] $$
 where the trace is with respect to both (N$ð$N) internal and (4$ð$4) Dirac
matrices.  Thus, spin can be treated in a manner closely analogous to
internal symmetry.
 ªb  Show the propagator can be written in the form of the product of 2
(massless) Dirac-spinor propagators.
 ªc  Starting with the complex first-order formulation of Yang-Mills of
subsection IIIC4, show that the action can be written in a way that
replaces the above 4$ð$4 matrices with 2$ð$2 matrices, as
$$ L_A ¾ tr[öG^2 +öG(»A* +iAA*)] $$
 in first-order form, where now $öG$ is neither traceless nor symmetric
in spinor indices (its trace is the Nakanishi-Lautrup field), or in
second-order form as
$$ L_A ¾ tr[(»A* +iAA*)^2] $$
 (Note that this differs from the above Dirac form, as expanded in 2$ð$2
matrices, because it includes the Chern-Simons term.)

Next, consider a model where the Yang-Mills fields couple to scalars that
are also represented by N$ð$N matrices, but that are in the defining
(N-component) representation of the gauge (``color") U(N), while also
being in the defining representation of a second, global (``flavor") U(N). 
(See subsection IVA6.)  This complex field thus has 2N${}^2$ real
components compared to the N${}^2$ gauge vectors, and the 2N${}^2$
ghosts.  We also choose a Higgs potential such that the masses of the
scalar and vector come out the same (but we can also specialize to the
massless case).  The scalar Lagrangian is then (again with $g^{-2}tr$ in
the action)
$$ L_Ä = -üÄÿá^2 Ä +\f14 R^2,ââR = ÄÿÄ -üm^2 $$
 Finally, we modify the gauge-fixing function to
$$ f = f_0 +iR $$ With this choice, the ghost terms are unmodified ($R$ is
gauge invariant), but the scalar self-interaction is completely canceled
(including the mass term).  The total Lagrangian is then
$$ L = (L_A +\f14 m^2 A^2) 
	+( -üÄÿá^2 Ä +iüÄÿÄf_0) +i(-ü÷Cá^2 C +iü÷CCf_0) $$

Since the scalar Lagrangian is identical in form to that of the ghosts, and
neither has self-interactions, functional integration over them will
produce canceling functional determinants, because they have opposite
statistics.  This is a reflection of the fact that both sets of fields now
describe unphysical polarizations, since both describe massless states in
a theory where all physical states are massive (as seen, e.g., in a unitary
gauge).  This has the great advantage that, for this particular model, both
the scalar fields and the ghosts can be dropped altogether, while the
Lagrangian 
$$ L £ L_A +\f14 m^2 A^2 $$
 completely describes the physical massive vector and scalar states.  This
was possible only because of the use of a complex gauge condition:  The
longitudinal component of the vector is now imaginary, which fixes the
wrong sign associated with the Minkowski metric.  A related
characteristic of this gauge is that we nowhere needed to change the
vacuum value of any field, unlike other gauges for actions where there is
a Higgs effect.

We now note that this result for the massive case (and its massless limit)
actually can be obtained more easily than the result for the pure
Yang-Mills case:  Since the final result has no ghosts, it is in a unitary
gauge, where the vector not only ``eats" the usual compensating scalar,
but ``overeats" by absorbing the physical scalar.  The appropriate gauge
condition is still complex and involves the scalars, but is now ÓlinearÕ:
$$ gaugeâÄ = ÒÄÔ = \f1{å2}mI $$
 where $Äÿ$, treated as independent, is unfixed.  (As for the usual unitary
gauge $Im¼Ä=0$, i.e., $Ä=Äÿ$, there are no propagating ghosts, since the
gauge transformation of $Ä$ has no derivatives.)  In this gauge the action
becomes quadratic in $Äÿ$:
$$ L_Ä £ -üÄÿi(»ÉA+iA^2)\f1{å2}m +\f14 (Äÿ\f1{å2}m-üm^2)^2 $$
 In fact, $Äÿ$ appears as an auxiliary field (taking the place of the
Nakanishi-Lautrup field), so we can eliminate it by its equation of motion:
$$ {¶\over ¶Äÿ}âÜâÄÿ = \f{m}{å2} +\f{å2}mif_0âÜâ
	L = L_A +\f14 m^2 A^2 $$
 This procedure is analogous to that used for the lightcone gauge, where
one component of the gauge field is fixed and one is eliminated as an
auxiliary field:  A closer analogy will be found in subsection VIB6.

Of course, such gauges generalize to other Higgs models, but results will
not be as simple when the vector and scalar masses differ:

\x VIB4.2  Make the coefficient of the $R^2$ term in $L_Ä$ arbitrary, so
the masses of the vector and scalar are unequal, but choose the same
gauge $Ä = ÒÄÔ$.  Find the propagator, and compare with that of exercise
VIB3.1.  

Ü5. Super Gervais-Neveu

Nonhermitian gauges are also useful in supersymmetric theories:  Here
we consider the supersymmetric analog of the massive model of the
previous section.  Although we work in N=1 superspace, the model turns
out to automatically have an N=2 supersymmetry.  Just as the bosonic
model ended with only a vector field describing only physical
polarizations, we now want a real scalar superfield to have only physical
polarizations.  Since such a superfield has 8 bosonic components and 8
fermionic, while massless N=1 multiplets have 2+2 physical polarizations,
we need 1 vector multiplet plus 3 scalar multiplets.  Since the bosonic
model had a complex scalar representation, 2 of these scalar multiplets
must form the analogous defining $°$ defining representation of local
$°$ global groups, so the last must be a real (adjoint) representation of
the local group.  The model is then given by (where again
$S=g^{-2}ÊtrÇdx¼L$)
$$ L_{gi} = -Çd^2 ϼW^2 
	-Çd^4 ϼ(e^{-V}ÐÄ_0 e^V Ä_0 +ÐÄ_+ e^V Ä_+ +Ä_- e^{-V}ÐÄ_-) $$
$$ -\left[ Çd^2 ϼ(Ä_+ Ä_- -\f14 m^2)Ä_0 +h.c. \right] $$
 where we have included the only possible scale-invariant potential term,
and introduced a Higgs mechanism by an N=2 Fayet-Iliopoulos term, which
we chose to write in terms of the chiral scalar.  (See subsection IVC7.)

The BRST transformations (which also imply the gauge transformations)
for this action are (see subsection VIA4)
$$ Qe^V = iÐCe^V -ie^V C,ââ
	Qe^{-V} = -e^{-V}(Qe^V)e^{-V} = iCe^{-V} -ie^{-V}ÐC $$
$$ QC = iC^2,âQÐC = iÐC^2;ââ
	Q÷C = -iB,âQ÷{ÐC} = -iÐB;ââQB = QÐB = 0 $$
$$ QÄ_+ = iCÄ_+,ââQÄ_0 = i[C,Ä_0],ââQÄ_- = -iÄ_- C $$
$$ QÐÄ_+ = -iÐÄ_+ ÐC,ââQÐÄ_0 = i[ÐC,ÐÄ_0],ââQÐÄ_- = iÐCÐÄ_- $$

Our nonhermitian choice for the BRST gauge-fixing function is
$$ ñ = -Çd^2 ϼ÷C(Ðd^2 e^{-V} +Ä_0) -Çd^2 Ðϼ÷{ÐC}(d^2 e^V +ÐÄ_0) $$
 Note that $e^V$ is an element of the algebra as well as a ``nonunitary
element" of the group, only because we chose the group U(N) (as was the
case for $A^2$ in the bosonic version).  The gauge-fixing and ghost terms
are then
$$ \li{ -iQñ ={}& Çd^2 ϼB(Ðd^2 e^{-V} +Ä_0) +Çd^2 ÐϼÐB(d^2 e^V +ÐÄ_0) \cr
	 & -Çd^4 ϼ(÷Ce^{-V}ÐC +÷{ÐC}e^V C) +Çd^2 ϼC÷CÄ_0 +Çd^2 ÐϼÐC÷{ÐC}ÐÄ_0 \cr} $$
 where we have used the field equation enforced by the Lagrange
multipliers $B$ and $ÐB$ (or, equivalently, made field redefinitions of the
Lagrange multipliers to generate terms proportional to their constraints).

\x VIB5.1  Make a component analysis of this theory:
 ªa  Expand the gauge-invariant action in components.
 ªb  Do the same for the gauge-fixing terms.
 ªc  Compare the bosonic part of both the gauge-invariant and
gauge-fixed actions to those of the previous subsection, after elimination
of auxiliary fields.

We now see that the ghost terms are identical in form to those for $Ä_à$,
under the identification
$$ (Ä_+,Ä_-,ÐÄ_+,ÐÄ_-) ª (C,÷C,÷{ÐC},ÐC) $$
 (but beware signs from ordering of ghosts).  So again the ghosts cancel
the (N=2) matter fields, leaving only the N=2 vector multiplet.  But we can
also eliminate the N=1 matter half of this N=2 multiplet using the
Nakanishi-Lautrup Lagrange multiplers:  The final simple result for the
gauge-fixed action is thus
$$ L = -Çd^2 ϼW^2 
	-Çd^4 ϼ[e^{-V}(d^2 e^V)e^V Ðd^2 e^{-V} +\f14 m^2(e^V +e^{-V})] $$

A further simplification results from the redefinition (again possible only
for U(N))
$$ e^V £ 1+V $$
 This also simplifies the BRST (and gauge) transformation for $V$:
$$ QV = i(ÐC-C) +i(ÐCV-VC) $$
 whose linear form resembles the bosonic case.  Using the expression (see
exercise IVC4.1)
$$ W_Œ = -iÐd^2 e^{-V}d_Œ e^V £ -iÐd^2{1\over 1+V}d_Œ V $$
 for the field strength, the Lagrangian becomes
$$ L = -Çd^4 ϼ\left[ -ü{1\over 1+V}(d^Œ V)Ðd^2{1\over 1+V}d_Œ V
	+{1\over 1+V}(d^2 V)(1+V)Ðd^2{1\over 1+V} \right. $$
$$ \left. +\f14 m^2\left( V +{1\over 1+V} \right) \right] $$

Although the nonabelian vector multiplet has nonpolynomial
self-interactions in any gauge, this gauge simplifies the lower-point
interactions, which are the ones more frequently used for a fixed number
of external lines.  Expanding this action to cubic order, we use the identity
$$ d^Œ Ðd^2 d_Œ = Ðd^{ÀŒ}d^2 Ðd_{ÀŒ} = -üõ +Ód^2,Ðd^2Õ $$
 for the kinetic term, and
$$ Ðd_{ÀŒ}d^2 = d^2Ðd_{ÀŒ} +i»_{ŒÀŒ}d^Œ $$
 for the gauge-fixing part of the cubic term, with integration by parts. 
(For the gauge-invariant term, some work can be saved by using the
equivalent $ÑW^2$ form.)  The result is
$$ L = Çd^4 Ï¼Ó \f14 V(õ-m^2)V +[\f14 m^2 V^3 +(Ðd^{ÀŒ}V)Vi»_{ŒÀŒ}d^Œ V] 
	+\O(V^4) Õ $$
 Not only are there fewer terms than with linear gauge conditions, but
these terms have fewer spinor derivatives, which yields fewer
nonvanishing contributions in loops (see subsection VIC5).  As for the
bosonic model of the previous subsection, this analysis also applies for
the unbroken case $m=0$.

\x VIB5.2  Find the corresponding form of the kinetic and cubic terms
without the redefinition $e^V£1+V$.

\x VIB5.3  Gauge fix by using the unitary gauge
$$ Ä_+ = ÐÄ_- = \f{m}{å2} $$
 to obtain the same result.

\x VIB5.4  Look at the super ÓantiÕ-Gervais-Neveu gauge, or super
ÓantiÕ-Fermi-Feynman gauge, changing the sign of the gauge-fixing term
for the vector multiplet (see exercise VIB4.1).
 ªa  Show that in the massless case the kinetic operator becomes,
instead of $õ$,
$$ K ¾ d^4 ­ \f1{4!}·^{Œº©¶}d_Œ d_º d_© d_¶ $$
 where we now use 4-component spinor indices.
 ªb  Show that the resulting propagator is of the form, in supercoordinate
space,
$$ ë(x,Ï;x',Ï') ¾ {d^4\over õ^2}¶^4(Ï-Ï')¶^4(x-x') ¾ ln[(x-x'-iüÏ©Ï')^2] $$
 where ``$x-x'-iüÏ©Ï'$" (see subsection IIC2) is the supersymmetry
invariant.  (Hint:  Use $d^4=Çd^4½¼e^{½^Œ d_Œ}$.  Warning:  If derived by Fourier
transformation, the integral is infrared divergent, and requires dropping
an infinite constant.)

Ü6. Spacecone

We have just seen that gauge independence allows complex gauge
conditions, which make the action complex.  (In subsection IIIC4, we also
used complex auxiliary fields, with a similar effect.)  In this subsection we
introduce a complex analog of the lightcone, the ``spacecone", which will
greatly simplify Feynman diagram calculations with massless fields.  The
spacecone gauge condition is
$$ A^2-iA^3 = 0 $$
 or more generally
$$ nÉA = 0,âân^2 = 0,âânÉn* > 0 $$
 (but only $n^a$, not $n*^a$, appears in the action).  While this gauge is
spacelike (in the sense that only spatial components of the gauge field
are fixed), it is also null, by virtue of being complex.  Thus, although
algebraically like the lightcone, it allows canonical quantization with the
usual time coordinate.  In fact, it is just a Wick rotation of the lightcone. 
We then eliminate $A^2+iA^3$ as an auxiliary field.

The spacecone is a new gauge to add to our list of axial gauges $nÉA=0$
from subsection IIIC2, and the related gauges for scalars from
subsections IVA5-6, VIB3-4: 
$$ \vbox{\offinterlineskip
\hrule
\halign{ &\vrule#&\strut¼\hfil$#$\hfil¼\cr
height2pt&\omit&\hskip2pt\vrule&\omit&&\omit&\cr
& axial\,\, gauges &\hskip2pt\vrule& non\hbox{-}null &
	& null\,\, (+\,\, auxiliary\,\, field\,\, eq.) &\cr 
height2pt&\omit&\hskip2pt\vrule&\omit&&\omit&\cr
\noalign{\hrule}
height2pt&\omit&\hskip2pt\vrule&\omit&&\omit&\cr
\noalign{\hrule}
height2pt&\omit&\hskip2pt\vrule&\omit&&\omit&\cr
& (partly)\ temporal &\hskip2pt\vrule& \hfill timelike:¼A^0 = 0 &
	& \hfill lightcone:¼A^+ = 0,¼¶/¶A^- &\cr 
& spacelike &\hskip2pt\vrule& \hfill Arnowitt\hbox{-}Fickler:¼A^1 = 0 &
	& \hfill spacecone:¼A^t = 0,¼¼¶/¶ÐA^tÊ &\cr
& scalar &\hskip2pt\vrule& \hfill unitary:¼Ä = Äÿ &
	& \hfill Gervais\hbox{-}Neveu:¼Ä = ÒÄÔ,¼¶/¶ÄÿÊ &\cr
height2pt&\omit&\hskip2pt\vrule&\omit&&\omit&\cr
}\hrule} $$
 In fact, at least for the free theories, the gauges for the scalars can be
considered as dimensional reductions (from 1 or 2 extra dimensions) of
those for the vector, as used for deriving the St¬uckelberg formalism in
subsection IIB4, where the spacelike components of the vector
associated with gauge fixing become scalars: Arnowitt-Fickler $£$
unitary, spacecone $£$ Gervais-Neveu.

The main advantages of the spacecone over the lightcone are special to
D=4, so we now review the lightcone in a way specialized to physical
spacetime.  We first repeat the results of subsection IIIC2, relabeling
the indices appropriately.
Starting with the gauge condition (see subsections IA4 and
IIA3 for notation)
$$ A^t = 0 $$
 and eliminating $ÐA^t$ by its field equation, the Lagrangian for pure
Yang-Mills becomes
$$ L = A^+ »^t л^t A^- -\f14(F^{+-})^2 +\f14(F^{tÐt})^2 $$
$$ \li{ F^{+-} & = »^+ A^- -»^- A^+ +i[A^+,A^-] \cr
	F^{tÐt} & = »^- A^+ +»^+ A^- 
	+i{1\over »^t}([A^+,»^t A^-] +[A^-,»^t A^+]) \cr} $$
 We simplify the Lagrangian by using the self-dual and anti-self-dual
combinations:  Dropping also the $t$ superscripts on $»$ for simplicity,
$$ \li{ \F^à = &¼ü(F^{tÐt}àF^{+-}) = »^à A^¦ +i{1\over »}[A^à,»A^¦] \cr
	L = &¼A^+ »Ð»A^- +\F^+ \F^-  \cr
	= &¼A^+üõA^- -i\left({»^-\over »}A^+\right)[A^+,»A^-]
		-i\left({»^+\over »}A^-\right)[A^-,»A^+] \cr
	&ââ+[A^+,»A^-]{1\over »^2}[A^-,»A^+] \cr} $$

\x VIB6.1 Label all the fields and derivatives in the above forms of the
Lagrangian, $\F$ in particular, in spinor notation.

\x VIB6.2 Show that the field redefinitions $A^à£(»)^{à1}Ä^à$, when
applied to just the first two terms of the above Lagrangian, produce a
local action describing the self-dual field equations of the lightcone
formalism of subsection IIIC5 (taking into account the difference
between the lightcone and spacecone). Compare the results of exercise
IIIC5.2.  Thus, by treating the latter two terms separately from the
former two, Yang-Mills can be treated as a perturbation about self-dual
Yang-Mills.

Another simplification for massless D=4, and closely related to the use of
helicity, is twistors.  For our Feynman diagram calculations for spins $²1$,
almost all spinor algebra involves objects carrying at most two spinor
indices (spinors, vectors, self-dual tensors), so we use the twistor matrix
notation of subsection IIB6.  In particular, in a general class of gauges the
external line factors for Yang-Mills fields in this notation (see subsection
VIB2) read
$$ A = ·_+ = {|·Ô[p|\over Ò·pÔ}âÜâf* = i|p][p|,âf = 0 $$
 for + helicity or
$$ A = ·_- = {|pÔ[·|\over [·p]}âÜâf = i|pÔÒp|,âf* = 0 $$
 for $-$, where $(·_à)^a$ are the polarization 4-vectors for helicity $à1$
in terms of a twistor $·^Œ,з^{ÀŒ}$, which can vary from line to line, and
whose choice defines the gauge, as a special case of the Landau gauge. 
(Positive helicity is the same as self-duality, negative is anti-self-dual. 
The Landau gauge condition is generally applied in arbitrary Lorenz
gauges to external lines, to eliminate the redundant longitudinal degrees
of freedom.)  

One special case is the lightcone gauge
$$ nÉA = 0,ân = |·Ô[·| $$
 in terms of an arbitrary constant lightlike vector $n$.  A more convenient
gauge is the spacecone gauge, which can be written in terms of two
twistors:
$$ n = |·_+Ô[·_-| $$
 These two twistors are sufficient to define a complete reference frame: 
We can convert all spinor indices into this basis, as
$$ Æ^Œ = Æ^à ·_à{}^Œ $$
 etc.  This corresponds to using two lightlike vectors to define the
spacecone gauge, $n^à=|·_àÔ[·_à|$.  For simplicity, we write $|·_àÔ=|àÔ$;
then a vector in this basis can be written as
$$ p = p^+ |+Ô[+| +p^- |-Ô[-| +p^t|-Ô[+| +Ðp^t|+Ô[-| $$
 if we use the normalization
$$ Ò+-Ô = [-+] = 1 $$
 E.g., for massless momentum $p=|pÔ[p|$,
$$ p^+ = Òp-Ô[-p],âp^- = Ò+pÔ[p+],âp^t = Ò+pÔ[-p],âÐp^t = Òp-Ô[p+] $$
 We will also drop the superscript $t$ in contexts where there is no
ambiguity.  This basis is related to our previous spinor basis up to phase
factors, $|àÔ¾|{}_àÔ$, $|à]¾|{}_{Àà}]$, and we assume them to be commuting
(rather than anticommuting); these changes are more convenient for
dealing with twistors (commuting spinors).

The advantage of the spacecone is that we can Lorentz covariantize the
Feynman rules by identifying these two lightlike vectors with physical
on-shell massless momenta.  We need two such ``reference" vectors
because we are not allowed to have $n=p$ on any line.  Since only $|+Ô$
appears in the external line factors for helicity +1, and only $|-]$ in those
for $-$1, the simplest choice is to pick the momentum of one external line
with helicity +1 to define $|-]$ for all lines with helicity $-$1, and pick the
momentum of one line with helicity $-$1 to define $|+Ô$ for lines with
helicity +1.  (In the presence of massless external spinors, we can also
choose a helicity +1/2 line to define $|-]$, etc.)  Our above normalization
means that we have chosen the phase $Ò+-Ô/[-+]=1$ as allowed by the
usual ambiguity of twistor phases, while our choice of the magnitude
$Ò+-Ô[-+]=-Ò+|[+|É|-Ô|-]=1$ is a choice of (mass) units.  In explicit
calculations, we restore generality (in particular, to allow momentum
integration) by inserting appropriate powers of $Ò+-Ô$ and $[-+]$ at the
end of the calculations, as determined by simple dimensional and helicity
analysis.  (This avoids a clutter of normalization factors $å{Ò+-Ô[-+]}$ at
intermediate stages.)  For example, looking at the form of the usual
spinor helicity external line factors, and counting momenta in the usual
Feynman rules, we see that any tree amplitude (or individual graph) in
pure Yang-Mills must go as
$$ Ò¼Ô^{2-E_+}[¼]^{2-E_-} $$
 where $E_à$ is the number of external lines with helicity $à$.

We now return to external line factors.  The naive factors for the above
Lagrangian are 1, since the kinetic term resembles that of a scalar. 
However, this would lead to unusual normalization factors in
probabilities, which are not obvious in this complex gauge.  Therefore, we
determine external line factors from the earlier spinor helicity
expressions for external 4-vectors.  In Lorenz gauges $(·_à)^a$ would
be the polarization for helicity $à1$ for the complete 4-vector, but in the
spacecone formalism only $A^à$ appear.  Furthermore, in the spacecone
we find
$$ (·_+)^- = -|+Ô[+|É{|+Ô[p|\over Ò+pÔ} = 0 $$
$$ (·_-)^+ = -|-Ô[-|É{|pÔ[-|\over [-p]} = 0 $$
 since by antisymmetry $Ò++Ô=[--]=0$, so that $A^+$ carries only helicity
$+1$ and $A^-$ only $-1$.  (This statement has literal meaning only on
shell, but we can make this convenient identification more general by
using it as a definition of helicity off shell.) The appropriate external line
factors for these fields are thus
$$ (·_+)^+ = -|-Ô[-|É{|+Ô[p|\over Ò+pÔ} = {[-p]\over Ò+pÔ} $$
$$ (·_-)^- = -|+Ô[+|É{|pÔ[-|\over [-p]} = {Ò+pÔ\over [-p]} $$
 Note that these factors are inverses of each other, consistent with
leaving invariant (the inner product defined by) the kinetic term.

An exception is the external line factors for the reference momenta
themselves, where $|pÔ=|¦Ô$ for helicity $à$ gives vanishing results. 
However, examination of the Lagrangian shows this zero can be canceled
by a $1/»$ in a vertex, since $p=Ðp=0$ for the reference momenta by
definition.  (Such cancellations occur automatically from field
redefinitions in the lightcone formulation of the self-dual theory.)  The
actual expressions we want to evaluate, before choosing the reference
lines, are then
$$ {p^-\over p}(·_+)^+ = {Ò+pÔ[p+]\over Ò+pÔ[-p]}¼{[-p]\over Ò+pÔ} = 
	{[p+]\over Ò+pÔ} $$
$$ {p^+\over p}(·_-)^- = {Òp-Ô[-p]\over Ò+pÔ[-p]}¼{Ò+pÔ\over [-p]} =
	{Òp-Ô\over [-p]} $$
 Evaluating the former at $|pÔ=|-Ô$ and the latter at $|pÔ=|+Ô$, we get 1 in
both cases.  In summary, for reference lines: (1) use only the 3-point
vertex of the corresponding self-duality ($àà¦$ for helicity $à$), and use
only the term associating the singular factor with the reference line (the
other term and the other vertices give vanishing contributions); (2)
including the momentum factors on that line from the vertex, the
external line factor is 1.

Ü7. Superspacecone

To generalize these results to high-energy (massless) QCD, we consider
supersymmetric QCD, i.e., Yang-Mills coupled to massless fermions in the
adjoint representation.  For tree graphs, this is equivalent to ordinary
massless QCD except for group theory, which can be evaluated
separately.  We first apply the spacecone approach to the component
action for supersymmetric QCD:  The modification of this action for
ordinary massless QCD is trivial (replacing the adjoint quark current with
defining).  From this we derive the ``superspacecone" formalism,
rewriting the action more simply in terms of spacecone superfields.  (This
form can also be derived from the usual superspace, but we will not
consider that here.)

We now combine the spacecone approach to pure Yang-Mills of
subsection VIB6 with the spacecone version of the lightcone treatment of
the massless spinor in subsection IIIC2.  We modify the lightcone to the
spacecone for the quarks by instead eliminating $Æ^¢$ and $ÐÆ^{\rdt\¢}$ as
auxiliary.  For later convenience, we also write the remaining fermionic
fields as
$$ ÐÆ^{\rdt ¢} £ Æ^+,ââÆ^\¢ £ -Æ^- $$
 Then we directly find the terms in the Lagrangian
$$ L = A^+ »Ð» A^- +\F^+\F^- +iÆ^+(л -á^-\f1» á^+)Æ^- $$
$$ \F^à = »^à A^¦ -\f1»([A^à,-i»A^¦] +ÓÆ^+,Æ^-Õ) $$
 where the quark term in $\F^à$ comes from the quark coupling to $A^{Ðt}$
when using its equation of motion to solve for $F^{tÐt}$.  Collecting terms,
we have
$$ L = L_2 +L_3 +L_4 $$
$$ L_2 = A^+ üõA^- +Æ^+{üõ\over -i»}Æ^- $$
$$ L_3 = \left({»^¦\over »}A^à\right)([A^à,-i»A^¦] +ÓÆ^+,Æ^-Õ)
	+\left({»^¦\over »}Æ^à\right)[A^à,Æ^¦] $$
$$ L_4 = -([A^+,-i»A^-] +ÓÆ^+,Æ^-Õ){1\over »^2}([A^-,-i»A^+] +ÓÆ^+,Æ^-Õ)
	-[A^+,Æ^-]{1\over -i»}[A^-,Æ^+] $$
 where terms with $à$ have only a single sum over it.  Although this
Lagrangian is much messier than the original covariant one, one again
saves work by expanding terms once in the action rather than repeatedly
for each Feynman diagram.

External-line factors for the spinors follow from the covariant ones of
subsection VIB2 as they did for the spacecone vectors of subsection
VIB6.  We thus have
$$ Æ^+ = [-p],ââÆ^- = Ò+pÔ $$
 Compared with those for $A^à$, we see $Æ^+ Æ^-$ has an extra factor of 
$p=Ò+pÔ[-p]$ as compared with $A^+ A^-$, as expected from the extra
factor of $1/(-i»)$ in the kinetic operator.  Similarly, if we choose to use
external quark lines as reference lines, we use
$$ {p^-\over p}Æ^+ = [p+],ââ{p^+\over p}Æ^- = Òp-Ô $$
 which also reduce to 1 for the appropriate reference momenta.

Noting that the bosonic and fermionic terms are the same except for
factors of $-i»$, we can combine them into chiral superfields that depend
on only two anticommuting coordinates, $Ï^+$ and $Ï^-$ (really $Ï^¢$ and
$ÐÏ^{\rdt\¢}$):
$$ S = \f1{g^2}¼tr Çdx¼dÏ^+ dÏ^-¼\L,ââÇdÏ^+ dÏ^- = d_+d_-¼or¼-d_-d_+,
	ââÓd_+,d_-Õ = -i» $$
$$ d_à Ä^¦ = 0;ââÄ^à| = A^à,ââd_à Ä^à| = Æ^à $$
 (no sum on $à$).  These spinor derivatives (and their corresponding
supersymmetry generators) describe only spatial supersymmetry, since
they contain no time derivatives.  Then, using the identity
$$ d_à d_¦[Ä^à,Ä^¦]| = [A^à,-i»A^¦] +ÓÆ^+,Æ^-Õ $$
 we easily combine the terms in the Lagrangian $L$ into the
superspacecone Lagrangian $\L$:
$$ \L = Ä^+{üõ\over -i»}Ä^- +\left({»^¦\over »}Ä^à\right)[Ä^à,Ä^¦]
	+[Ä^+,d_- Ä^-]{1\over »^2}[Ä^-,d_+ Ä^+] $$
 The last term can also be written as
$$ -[Ä^+,Ä^-]{d_+ d_-\over »^2}[Ä^+,Ä^-] $$

\x VIB7.1  Introduce another pair of chiral superfields as auxiliary.  Show
the above $\L$ then can be rewritten in local form, with no spinor
derivatives, where the kinetic term resembles the covariant one for a
massless spinor, while the interaction term contains no derivatives and is
only cubic.  (Hint:  $d_à d_¦/(-i»)$ are projection operators for chiral
superfields.)  Thus this Lagrangian resembles the Chern-Simons one that
appears on the 3D boundary for the topological term in Yang-Mills (see
subsection IIIC6).  Expand the action in components, and separate out the
pure Yang-Mills part.

\x VIB7.2  Repeat exercise VIB6.2 to obtain the superspacecone action
for selfdual super Yang-Mills, quadratic in $Ä^+$ and linear in $Ä^-$.  Use
the field redefinition
$$ Ä^- £ d_+ \Ä^-,ââd_- \Ä^- = 0 $$
 and integrate the action over just $Ï^-$ (by acting with $d_-$) to obtain
a ``chiral" action, with no spinor derivatives, and superfields that are
functions of just $Ï^+$ integrated over just $Ï^+$.  After further
redefinitions as in VIB6.2, obtain an action ÓidenticalÕ to the
nonsupersymmetric one obtained there, ÓexceptÕ for the $ÇdÏ^+$.  Expand
in components, and relate to the nonsupersymmetric case.

Ü8. Background-field

A more general type of gauge choice is the background field gauge.  As we
saw in subsection VC1, the generating functional can be written in a form
where the quantum field is expanded about a background field in the
interaction part of the classical action.  The basic steps of the background
field gauge method are:   
\item{(1)} choose gauge fixing that is gauge invariant in
the background gauge field,   
\item{(2)} show that the quantum/background
splitting of the entire gauge-invariant action is also gauge invariant in
the background gauge field, and   
\item{(3)} show that the effect of splitting the
kinetic term in the gauge-invariant action can be neglected.  (Only the
interaction terms should have been split.)  

\noindent The result is then that the
effective action $ý$, which depends only on the background fields, is
gauge invariant in them.  This gauge invariance is a strong condition
which not only simplifies the effective action but allows a ``background
gauge" to be chosen for it that is independent of the ``quantum gauge"
applied to the path integral:  The background fields and quantum fields
can be in different gauges.  For example, for a relativistic treatment of
the quantum corrections to bound states whose constitutents are
nonrelativistic (such as the hydrogen atom), it is convenient to use a
Fermi-Feynman gauge (convenient for relativistic matter coupling to
electromagnetism or chromodynamics) for the quantum fields and a
Coulomb gauge (convenient for static or nonrelativistic matter) for the
background fields.

A simple way to formulate the background expansion is in terms of the
covariant derivative:
$$ A £ ÷A = \A +AâÜâá £ \D +iA,ââ\D = » +i\A $$
 where $A$ is the quantum field (the variable of path integration) and
$\D$ is the ``background covariant derivative" in terms of the
background field $\A$.  We then find for the field strength
$$ F_{ab} £ -i[\D_a +iA_a,\D_b +iA_b] 
	= \F_{ab} +\D_{[a}A_{b]} +i[A_a,A_b] $$
 and similarly for the action.  Matter fields are split as usual,
$$ Ä £ ÷Ä = \Ä +Ä $$

We now have two gauge invariances, corresponding to the two gauge
fields.  Both transformations are defined to have the same, usual form on
$á=\D+iA$ (and on $÷Ä=\Ä+Ä$), and thus both leave the action inert, but (1)
the ``background gauge invariance" is defined to transform the
background fields covariantly
$$ background:ââ\D' = e^{iÂ}\D e^{-iÂ}â(\Ä' = e^{iÂ}\Ä),ââ
	á' = e^{iÂ}áe^{-iÂ}ââ(÷Ä' = e^{iÂ}÷Ä) $$
$$ ÜâA' = e^{iÂ}A e^{-iÂ}ââ(Ä' =e^{iÂ}Ä) $$
 and thus the quantum field transforms as a matter (non-gauge) field,
while (2) the ``quantum gauge invariance" is defined to leave the
background fields inert
$$ quantum:ââ\D' = \Dâ(\Ä'=\Ä),ââá' = e^{iÂ}áe^{-iÂ}ââ
	(÷Ä' = e^{iÂ}÷Ä) $$
$$ ÜâA' = e^{iÂ}[(-i\D +A) e^{-iÂ}]ââ[Ä' = e^{iÂ}(\Ä+Ä) -\Ä] $$
 The latter then determines the new BRST transformations
$$ QA_a = -[\D_a+iA_a,C],âQC = iC^2,âQ÷C = -iB,âQB = 0ââ
	[QÄ = iC(\Ä+Ä)] $$

The key to the background field gauge is to break the quantum
invariance, so a propagator can be defined, but preserve the background
invariance, so the path integral is gauge invariant.  Since $Q$, the BRST
operator for the ÓquantumÕ gauge invariance, is now background gauge
invariant, we need only choose a gauge-fixing function $ñ$ that is also
background gauge invariant.  Many gauges are possible:  The basic rule is
to modify any normal gauge condition simply by replacing any partial
derivatives $»$ with background covariant derivatives $\D$.  For example,
for a Lorentz covariant gauge
$$ »ÉA £ \DÉA $$
 We then gauge fix in the usual way, and now the gauge-fixing terms and
the ghost terms are background gauge invariant, as long as we define the
ghosts to transform covariantly:
$$ background:ââC' = e^{iÂ}C e^{-iÂ},ââ÷C' = e^{iÂ}÷C e^{-iÂ},ââ
	B' = e^{iÂ}B e^{-iÂ} $$
 For example, for Lorenz gauges the ghost term is modified, by the
modification of the gauge condition and the quantum BRST
transformation, as
$$ ÷C»ÉáC £ ÷C\D^2 C +÷C\DÉi[A,C] $$
 Furthermore, even axial gauges are modified:  For example, even though
the gauge condition $A_0=0$ allows elimination of a component of the
quantum field, it doesn't affect the background field, which now appears
in the ghost Lagrangian
$$ ÷C»_0 C £ ÷C\D_0 C $$

Since the S-matrix is gauge-independent (when BRST is used to perform
gauge fixing, as we have), we can use the background field gauge version
of the generating functional (now using $Ä$ to represent all quantum
fields and $\Ä$ all background),
$$ Z[\Ä] = ÇDļe^{-i÷S},ââ
	÷S = S_0[Ä] +S_I[Ä+\Ä] -iQñ[Ä,\Ä] = öS -(S_0[Ä+\Ä] -S_0[Ä]) $$
$$ öS = S[Ä+\Ä] -iQñ,ââS[Ä] = S_0[Ä] +S_I[Ä] $$
 where $S[Ä]$ is the original gauge-invariant action, $Qñ$ is the
gauge-fixing as described above, and $öS$ is the sum of this gauge
fixing and the background-expanded gauge-invariant action.  We have
thus separated the total action $÷S$ appearing in the
background-gauge-fixed generating functional into the
background-gauge-invariant part $öS$ minus the noninvariant terms
$S_0[Ä+\Ä] -S_0[Ä]$.

As usual, the classical part of the effective action $ý[\Ä]$ is given by
adding the kinetic term $S_0[\Ä]$ of the (gauge-invariant) classical action
to the 1PI tree graphs, which are just the vertices for the background
fields.  
(The $Q$ term doesn't contribute because it has no pure background piece.)
Thus,
$$ ý_{class}[\Ä] = ÷S|_{Ä=0} +S_0[\Ä] = öS|_{Ä=0} = S[\Ä] $$

We now note that, as far as calculating just the effective action is
concerned, we can drop all terms in the gauge-fixed action independent
of or linear in $Ä$:  Any independent term contributes only classically; any
linear term will generate one-particle ÓreducibleÕ graphs (``tadpoles"). 
This means we can drop the noninvariant terms $S_0[Ä+\Ä] -S_0[Ä]$ from
$÷S$.  Thus, the Feynman rules for calculating $ý$ are:  (1)¼Use the classical
gauge-invariant action $S[\Ä]$ for the classical contribution to $ý$; and
(2) for the quantum contribution, use all the 1PI loop graphs coming from
$öS$.  The result is background gauge invariant, since $öS$ is manifestly so.  

Another important feature of the quantum-gauge-fixed background field
action is that it is background-gauge-invariant order-by-order in the
quantum fields.  In fact, every term in the corresponding ordinary gauge
action has been replaced by one (or more, if there are $\F$ terms)
background-gauge-invariant term.

\x VIB8.1 Consider the Fermi-Feynman background-field gauge for the
quantum field of pure Yang-Mills theory.  Write all terms (both
gauge-invariant and gauge-fixed) quadratic in the quantum field.  Show
that these combine as
$$ -\f14 AÉõA -iüA^a[\F_{ab},A^b] $$
 where $õ=(\D)^2$ (and ``$\D A$" means ``$[\D,A]$", etc.).

Since all external lines are associated with background fields, if we draw
graphs in such a way as to exhibit only the quantum fields, they will all
look like vacuum graphs: graphs with no external lines.  However, any
particular such vacuum graph will represent many of the original graphs,
since the background lines can be attached in many ways.  Furthermore, in
background field gauges any such vacuum graph, considered as a
contribution to the effective action, will be gauge invariant with respect
to the background gauge transformations, since it results from the
non-background gauge true vacuum graph by the replacement of the
ordinary derivative with the background covariant derivative
$ȣ\D$ (plus perhaps some noniminimal $\F$ terms), including in the
propagator.  In particular, the complete one-loop contribution to $ý$ is
given by the vacuum graph with no quantum interactions:  It can be
obtained from just the part of the $öS$ that is quadratic in the quantum
fields.

$$ \fig{back} 
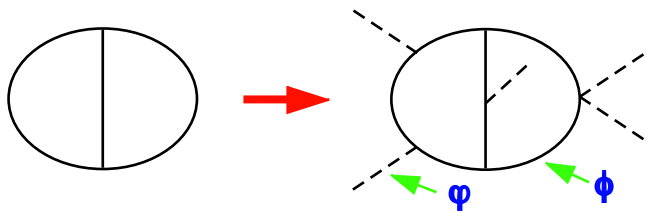
$$

\x VIB8.2 Consider an arbitrary gauge-invariant Yang-Mills action $S[÷A]$
with $÷A=\A+A$ in terms of the background field $\A$ and quantum field
$A$.  Taylor expand the action in $A$ as
$$ S[÷A] = S[\A] + A{¶S[\A]\over ¶\A} + ... $$
 The infinitesimal ÓquantumÕ gauge transformation mixes different-order
terms in the expansion.  Show that the term quadratic in $A$ is invariant
under an ÓAbelianÕ quantum gauge transformation only if the background
satisfies the field equations, $¶S[\A]/¶\A=0$.  Similar remarks apply to
BRST transformations and the gauge-fixed action.  (Since quadratic
actions, even in background fields, yield only a propagator, they can be
described by first-quantization:  Thus gauge invariance implying
background field equations occurs whenever a gauge field appears as
both a quantum mechanical state and a background field, for example in
string theory.  See subsection XIIB7 for a simpler example.)

The S-matrix is then given in the usual way from $ý[\Ä]$, after adding
another gauge-fixing term for the ÓbackgroundÕ gauge invariance.  Since
the total S-matrix is given by just the trees following from treating $ý$
as a classical action, we need only a gauge-fixing term for the physical
fields, and we can ignore background ghosts.  (Of course, quantum ghosts
were already used to calculate $ý$.)  This background gauge fixing is
independent of the quantum gauge fixing.  In particular, we can choose
different quantum and background gauges:  For example, when treating
spontaneously broken gauge theories, it's often more convenient to
choose a Fermi-Feynman quantum gauge and a unitary background gauge;
i.e., we expand the background fields about the physical vacuum to make
the physical states obvious, but leave the quantum fields unexpanded to
avoid complicating the Feynman rules.  This also avoids the complication
of having to expand about the vacuum twice, since vacuum values get
quantum corrections to those appearing in the classical action.

The gauge invariance of the effective action in the background-field
formalism is a big advantage over other quantum gauges, where the
effective action is only BRST invariant, since gauge invariance is a much
stronger constraint than BRST invariance:  Gauge symmetry is local, while
BRST is only global.  Thus, the background-field gauge produces a much
simpler effective action.  In other words, the background-field gauge
produces an effective action without ghosts:  Although we can drop ghost
terms from the effective action in general, because there are no physical
external ghost states (since we calculate only the ``tree" graphs of the
effective action), the result is not normally BRST invariant; but in the
background-field gauge it is still BRST invariant, since it is gauge
invariant.  This means that the background-field gauge yields not only
simpler results, but fewer calculations:  Many terms can be determined
by ``gauge covariantization".

\x VIB8.3  Consider an effective action for Yang-Mills plus matter in a
background-field gauge.  Its gauge invariance can be used to derive
``Ward-Takahashi identities".  (These were originally expressed as
properties of the S-matrix, but are much simpler to understand in terms
of the effective action.)
 ªa Show that the part of the effective action ÓquadraticÕ in the
Yang-Mills fields, and independent of the matter fields, is invariant under
the ÓAbelianÕ gauge transformations.  (Hint:  Taylor expand.)  Thus, in such
gauges the quantum correction to the gluon propagator is transverse.
 ªb By the same method, find a relation between any quantum 3-point
vertex coupling matter to Yang-Mills and the corresponding matter
propagator correction.  (Note a simpler case:  Since the renormalization
counterterms are local, gauge invariance just says that the coefficients
of the two corresponding counterterms are the same, i.e., they occur in
the combination $¾ÐïÖáï$.)

However, this does not mean we can completely ignore BRST and ghosts
by using background-field gauges:  Although the ÓeffectiveÕ action is
gauge invariant and ghost free, ghosts and BRST still appear in the
(quantum-gauge-fixed) ÓclassicalÕ action.  In practice this means, as
far as calculating the Feynman graphs that contribute to the effective
action, that in the background-field gauge calculations are about Óone
loop simplerÕ than in other gauges.  For example, for one-loop graphs we
effectively calculate free one-loop vacuum bubbles (including ghosts)
covariantly coupled to background fields:  There are fewer of the
complications of nonabelian theories, since the quantum fields appear
only as non-gauge fields with covariant couplings and no
self-interactions.  However, already at two loops we have
self-interactions of the quantum fields, which include the same kinds of
terms that would have appeared had we not used a background-field
formalism.

Another complication is that BRST invariance is not as restrictive as gauge
invariance:  It can be shown that in general gauges at the quantum level
BRST invariance is preserved only up to ``wave-function
renormalizations" (rescalings) of the quantum fields.  However, in the
background-field gauge wave-function renormalizations of the quantum
fields can be ignored, since the quantum field is a dummy variable:  There
are no external quantum fields, so all such factors cancel.  (Actually, we
can also ignore wave function renormalization counterterms in
non-background-field gauges, since when calculating S-matrix elements
such divergences will be canceled by corresponding divergences in the
external-line factors.  In general, external-line normalization factors
may be nontrivial even when wave-function renormalization is
performed, depending on the renormalization scheme.)

An exception is Abelian gauge theories, such as QED:  Because the
gauge-invariant action for just the gauge fields is free, background field
gauges are identical to ordinary gauges.  Also, the ghosts decouple (for
linear gauge conditions).

The main point of the background-field gauge is that two gauge choices can be made.  This method can be further generalized so that there are ÓthreeÕ independent gauge choices: (1) First we choose the quantum gauge as before, to obtain an effective action that is gauge invariant with respect to background gauge transformations.  In terms of S-matrix diagrams, this is a choice of gauge for propagators inside loops.\\  (2) Then we choose the background gauge as before, to obtain S-matrix elements.  This is a choice of gauge for propagators external to 1PI subgraphs.  (3) Finally, there is still a gauge invariance of the external fields:  These fields describe asymptotic states, and hence have a linearized gauge invariance.  This means that the generating functional $Z(A)$, or $W(A)$, can always be written in a form invariant under Abelian gauge transformations.  (In fact, this will be the case in general, without a background-field gauge, since the S-matrix is gauge independent.)  As a consequence, $Z$ and $W$ can always be rewritten in terms of Abelian field strengths, making this invariance manifest.  (However, the Feynman rules generally will not give them directly in this form.)  Writing them in this form has the same advantages as manifesting background gauge invariance in the effective action:  There are fewer possible terms one can write, Lorentz covariance is manifest, comparison is easier, and more gauge choices are available.  (In practice, we usually choose some unitary gauge for the external fields, to isolate the physical polarizations.)  Furthermore, since asymptotic states are on shell, these external Abelian field strengths satisfy the free, Abelian, gauge-covariant (Maxwell) field equations, giving further restrictions on the number of independent ways they can appear (with respect to derivatives acting on them).

Ü9. Nielsen-Kallosh

So far we have considered only gauges where the gauge-fixing term is
the square of the gauge-fixing function.  More generally, we'll need
gauge-fixing terms of the form $f\O f$ for some operator $\O$. 
Straightforwardly, we can write
$$ iQü÷C[f(A)-üŒ\O^{-1}B] = üB(f-üŒ\O^{-1}B) +üi÷C{»f\over »A}É[á,C] $$
 However, $B$ is no longer auxiliary, so we can't eliminate it by its field
equation.  But we can diagonalize the Lagrangian by the
corresponding redefinition,
$$ B £ B +\f1Œ \O f $$
 (The Jacobian of such redefinitions is unity, the determinant of a
triangular matrix of the form $\tat10x1$.)  The gauge-fixing terms are
then
$$ -\f14 ŒB\O^{-1}B +\f1{4Œ}f\O f $$
 The inverse operator is inconvenient for Feynman rules.  We know
that integrating out $B$ gives a functional determinant, so $\O^{-1}$ can
be replaced by an $\O$ if we change the statistics of the
Nakanishi-Lautrup field.  However, this is a bit formal, since technically
$\O$ must be symmetric between the two $B$'s, while it should be
antisymmetric between two fermions.

\x VIB9.1  Instead of $f(A)-üŒ\O^{-1}B$, use $\O [f(A)-üŒB]$, and again diagonalize. The extra factor in the ghost kinetic term can then be put in a separate term by (the inverse of) the method of exercise VIA4.2. This method avoids any symmetry problems with $\O$.

A useful example is gauge fixing for super Yang-Mills in superspace. 
Gauge fixing for massless Yang-Mills is actually more difficult than
for the massive (Higgs) case, considered in subsection VIB5.  We'll look at
the Abelian theory, to determine what kind of gauge fixing we need to
define the propagator.  (With slight generalization, this is also sufficient
for the background-field gauge:  See the following subsection.)  In that
case the BRST transformations are
$$ QV = i(ÐC-C),ââQ÷C = -iB,ââQ÷{ÐC} = -iÐB,ââQC = QB = QÐC = QÐB = 0 $$
 where $C$, $÷C$, and $B$ are chiral.  Using the result (the Abelian case of
exercise IVC4.1)
$$ W_Œ = -iÐd^2 d_Œ V $$
 the gauge-invariant kinetic term is (rearranging derivatives and using
integration by parts; see subsection VIB5)
$$ L_0 =  -Çd^2 ϼüW^Œ W_Œ = -Çd^4 ϼüVd^Œ Ðd^2 d_Œ V
	= Çd^4 ϼV(\f14 õ -Ðd^2 d^2) V $$
 To gauge-fix to the Fermi-Feynman gauge we choose
$$ \li{ L_1 & =  -iQÇd^4 ϼ[(÷C+÷{ÐC})V +÷C(üõ)^{-1}ÐB] \cr
	& = Çd^4 ϼ[(÷{ÐC}C-÷CÐC) -(B+ÐB)V -B(üõ)^{-1}ÐB] \cr} $$
 (dropping $d^4 Ï$ integrals of totally chiral or totally antichiral terms,
which vanish).  If we were to simply redefine $B$ by
$$ B £ B -Ðd^2 d^2 V,ââÐB £ ÐB -d^2 Ðd^2 V $$
 the gauge-fixing terms would diagonalize as (using $Ðd^2d^2Ðd^2=üõÐd^2$)
$$ -(B+ÐB)V -B(üõ)^{-1}ÐB £ VÐd^2 d^2 V -B(üõ)^{-1}ÐB $$
 giving the desired result for $V$:  At this stage the total result is
$$ L = L_0 +L_1 £ Çd^4 ϼ[\f14 VõV +÷{ÐC}C -÷CÐC -B(üõ)^{-1}ÐB] $$

Because $B$ is complex, the replacement of $B$ with a fermionic
superfield can be performed classically, just like the rest of the
gauge-fixing procedure.  We thus introduce ghosts for a trivial gauge
invariance as described in subsection VIA4:
$$ QD = E,ââQÐE = -iÐD,ââQE = QÐD = 0 $$
 We have treated the ghosts and their hermitian conjugates
independently; alternatively, we can consider $ÐD$ and $ÐE$ as not being the
conjugates of $D$ and $E$.  The gauge fixing is simply
$$ L_2 = -iQ(-ÐED) = ÐDD -iÐEE $$
 We next make the redefinition
$$ D £ D +(üõ)^{-1}Ðd^2 ÐB,ââÐD £ ÐD +(üõ)^{-1}d^2 B $$
 which has the effect
$$ Çd^4 ϼ[ÐDD -B(üõ)^{-1}ÐB] £ 
	Çd^4 ϼÐDD +Çd^2 ϼDB +Çd^2 ÐϼÐDÐB $$
 which vanishes, after using the now-algebraic field equations from
varying $B$ and $ÐB$.  Alternatively, we can make this field redefinition
ÓinsteadÕ of the previous field redefinition:  We then have the terms
$$ Çd^4 ϼ[ÐDD -(B+ÐB)V] +Çd^2 ϼDB +Çd^2 ÐϼÐDÐB
	£ Çd^4 ϼVÐd^2 d^2 V $$
 after using the still-algebraic $B$ equations.

The net result
$$ L = L_0 +L_1 +L_2 £ Çd^4 ϼ(\f14 VõV +÷{ÐC}C -÷CÐC -iÐEE) $$
 is that the original nonlocal $B$ term has been replaced classically with
the local $ÐEE$ term, which yields the same determinant upon
quantization, but gives simple Feynman rules more directly.  (The
determinant is nontrivial in the background-field gauge.  A similar
procedure can be applied to gauge fixing for spin 3/2.)

\x VIB9.2  Apply the method of exercise VIB9.1 to super Yang-Mills, where $\O$ is now $d^2$ or $Ðd^2$ (as implied by the form of $L_1$ above). Thus, the expression in $L_1$ on which $Q$ acts will have terms $Vd^2÷C$ and $÷C÷B$, and their h.c. Show the result, instead of 3 fermionic ghost pairs, is 4 fermionic and 1 bosonic ghost pairs.

\x VIB9.3  Perform the analogous quantization for the nonabelian case of
pure super Yang-Mills (no matter), using the super Gervais-Neveu gauge. 
Compare with the limit $m£0$ of the model considered in subsection VIB5,
and show the $V$ part of the action agrees.

\x VIB9.4  Use this method to produce a gauge-fixing term $Œ(nÉA)õ(nÉA)$
for a gauge vector $A$ in terms of a parameter $Œ$ and constant vector
$n$.  Find all propagators.  Look for simplifying special cases of $Œ$ and
$n$.

Ü10. Super background-field

Although in principle the background-field formalism is the same for
supersymmetric theories as nonsupersymmetric, there are some technical
differences because of the nonlinearity in the prepotentials.  (Similar
remarks apply to nonlinear $§$ models.)  The basic idea is that we want
to expand the full covariant derivative in quantum fields about
background-covariant derivatives:  As for the nonsupersymmetric case,
$á£\D+iA$, but now $A$ is further expressed in terms of $\D$ and the
prepotential because of the constraints.  The generalization in this case
(and for nonlinear $§$ models) is easy because the solution to the
constraints makes the prepotentials appear as (complex) group
elements:  Because of the closure of group multiplication, we can write
$$ {\bf g} £ {\bf g}_B {\bf g}_Q $$
 in terms of quantum (${\bf g}_Q$) and background (${\bf g}_B$) group
elements (fields).  More explicitly, for our case we write (see subsection
IVC4)
$$ e^¯ £ e^{¯_B}e^{¯_Q} $$
 and thus for the covariant derivatives
$$ á_Œ £ e^{-¯_Q}\D_Œ e^{¯_Q} $$
 absorbing the background prepotential completely into the background
covariant derivative
$$ \D_Œ = e^{-¯_B}d_Œ e^{¯_B} $$
 In other words, as the name suggests, the full covariant derivative $á$
has been expanded about an arbitrary background, described by $¯_B$. 
(This is even clearer in the supergravity case, where we simply replace
the flat-space $d_Œ$ with the curved-space $\D_Œ$, since $d_Œ$ is more
than a partial derivative, and already contains the flat-space part of the
metric tensor.)  For purposes of quantization, it is most convenient to go
to a chiral representation for the quantum field.  For the background field
we need not be so specific, since it is hidden in the background covariant
derivatives.  The result is then
$$ á_Œ £ e^{-V}\D_Œ e^V,ââÑá_{ÀŒ} £ Ð{\D}_{ÀŒ},ââ
	á_{ŒÀŒ} £ iÓÐ{\D}_{ÀŒ},e^{-V}\D_Œ e^VÕ $$
 where $V$ is the quantum field.

\x VIB10.1 Solve the  rest of the commutator algebra to find expressions
for all the field strengths in terms of $V$ and $\D_A$.

The rest of the quantization procedure then follows as for the
nonsupersymmetric case, except for the Nielsen-Kallosh ghost described
in the previous subsection.  In particular, for the terms in the gauge-fixed
classical action quadratic in the quantum field $V$,
$$ W_Œ £ -iü[\D^{ÀŒ},Ó\D_{ÀŒ},e^{-V}\D_Œ e^VÕ]
	= \W_Œ -iÐ{\D}^2\D_Œ V +iüÐ{\D}^2[V,\D_Œ V] +\O(V^3) $$
$$ ÜâS_{2V} = 
	Çdx¼d^4 ϼV(-ü\D^Œ Ð{\D}^2\D_Œ +iü\W^Œ\D_Œ +Ð{\D}^2\D^2)V $$
 Pushing the $\D$ in the first term to the right, we find
$$ -ü\D^Œ Ð{\D}^2\D_Œ = i\f14 (\D^{ŒÀŒ}Ð{\D}_{ÀŒ}\D_Œ 
	+Ð{\D}_{ÀŒ}\D^{ŒÀŒ}\D_Œ) -Ð{\D}^2\D^2 $$
 Using integration by parts on all the derivatives in the second term so
they act to the left, then switching the $V$'s so they again act to the right,
$$ \li{ \D^{ŒÀŒ}Ð{\D}_{ÀŒ}\D_Œ +Ð{\D}_{ÀŒ}\D^{ŒÀŒ}\D_Œ
	& £ \D^{ŒÀŒ}Ð{\D}_{ÀŒ}\D_Œ +\D_Œ\D^{ŒÀŒ}Ð{\D}_{ÀŒ} \cr
	& = \D^{ŒÀŒ}Ð{\D}_{ÀŒ}\D_Œ +\D^{ŒÀŒ}\D_ŒÐ{\D}_{ÀŒ}
		 +[\D_Œ,\D^{ŒÀŒ}]Ð{\D}_{ÀŒ} \cr
	& = -iõ +2Ñ{\W}^{ÀŒ}Ð{\D}_{ÀŒ} \cr} $$
 where $õ=\D^a \D_a$.  The final result is similar to the bosonic case
(exercise VIB8.1):
$$ S_{2V} = Çdx¼d^4 ϼ\f14 V(õ +2i\W^Œ\D_Œ +2iÑ{\W}^{ÀŒ}Ð{\D}_{ÀŒ})V $$
 (This result is invariant under integration by parts because of the Bianchi
identity $\D_Œ\W^Œ+Ð{\D}_{ÀŒ}Ñ{\W}^{ÀŒ}=0$.)

Ghosts and matter are quantized straightforwardly:  For matter we have
$$ Ñá_{ÀŒ}Ä = á_Œ ÐÄ = 0âÜ $$
$$ Ä £ \Ä +Ä,âÐÄ £ e^V(Ð{\Ä} +ÐÄ);ââÐ{\D}_{ÀŒ}\Ä = \D_Œ Ð{\Ä} = 0 $$
 The action thus looks the same as usual ($(Ð{\Ä} +ÐÄ)e^V(\Ä +Ä)$, etc.),
except that all chiral superfields are now background-chiral.  For the
standard ghosts we have for the ghost action 
$S_C=Çdx¼d^4 ϼL_C$ (remembering there are no background ghosts, and
using the full nonlinear transformation law from exercise IVC4.3)
$$ L_C = (÷C +÷{ÐC})\L_{V/2}[coth(\L_{V/2})(C-ÐC) +(C+ÐC)] 
	= (÷C +÷{ÐC})(C-ÐC) +\O(V) $$
$$ £ (÷{ÐC}C -÷CÐC) +\O(V) $$
 the same as in non-background gauges, except again the ghosts are
background-chiral.  Now the Nielsen-Kallosh ghost of the previous
subsection is nontrivial:  We again have
$$ L_{NK} = -iÐEE $$
 but these ghosts also are background-chiral.  This means they contribute
to the effective action ÓonlyÕ at one loop, through ``vacuum bubbles".

\refs

£1 Fock, Óloc. cit.Õ (VB);\\ 
	Schwinger, Óloc. cit.Õ (VB, ref. 2, second ref.):\\ 
	radial gauge.
 £2 L. Lorenz, ÓPhilos.Mag.Õ É34 (1867) 287:
	Lorenz gauge, and independent discovery of Maxwell's equations.
 £3 Fermi, Óloc. cit.Õ (VC);\\
	Feynman, Óloc. cit.Õ (VB, ref. 4).
 £4 A.A. Abrikosov, I.M. Khalatnikov, and L.D. Landau, ÓDoklady Akad. Nauk
	USSRÕ É95 (1954) 773:\\
	Landau gauge.
 £5 P. De Causmaecker, R. Gastmans, W. Troost, and T.T. Wu, 
	\NP 206 (1982) 53;\\
	F.A. Berends, R. Kleiss, P. De Causmaecker, R. Gastmans, W. Troost,
	and T.T. Wu, \NP 206 (1982) 61;\\
	Z. Xu, D.-H. Zhang, and L. Chang, \NP 291 (1987) 392;\\
	J.F. Gunion and Z. Kunszt, \PL 161B (1985) 333;\\
	R. Kleiss and W.J. Sterling, \NP 262 (1985) 235:\\
	spinor helicity.
 £6 V.N. Gribov, \NP 139 (1978) 1.
 £7 G. 't Hooft, \NP 35 (1971) 167;\\
	K. Fujikawa, B.W. Lee, and A.I. Sanda, \PRD 6 (1972) 2923;\\
	Y.-P. Yao, \PRD 7 (1973) 1647:\\
	renormalizable gauges.
 £8 Gervais and Neveu, Óloc. cit.Õ (IVA).
 £9 G.F. Chew and M. Levinson, ÓZ. Phys. CÕ É20 (1983) 19:\\
	proposed Feynman rules similar to anti-Gervais-Neveu gauge in a
	context related to string theory.
 £10 W. Siegel, \xxxlink{hep-th/9502163}, \PRD 52 (1995) 1035:\\
	super Gervais-Neveu.
 £11 M. Mangano and S.J. Parke, ÓPhys. Rep.Õ É200 (1991) 301:\\
	review of modern methods for tree graphs.
 £12 E.T. Newman and R. Penrose, ÓJ. Math. Phys.Õ É3 (1962) 566:\\
	use of ``null tetrad" in gravity, complex basis vectors in terms of
	constant spinors.
 £13 G. Chalmers and W. Siegel, \xxxlink{hep-ph/9801220},
	\PRD 59 (1999) 045013:\\
	spacecone.
 £14 W. Siegel and S.J. Gates, Jr., \NP 189 (1981) 295;\\
	S. Mandelstam, \NP 213 (1983) 149;\\
	L. Brink, O. Lindgren, and B.E.W. Nilsson, \NP 212 (1983) 401:\\
	lightcone superfields.
 £15 Feynman, Óloc. cit.Õ (VC, ref. 17);\\
	DeWitt, Óloc. cit.Õ (VIA);\\
	J. Honerkamp, \NP 48 (1972) 269;\\
	G. 't Hooft, The background field method in gauge field theories, in
	ÓFunctional and probabilistic methods in quantum field theoryÕ, proc.
	12th Winter School of Theoretical Physics, Karpacz, Feb. 17-Mar. 2,
	1975, v. 1, ÓActa Univ. Wratislav.Õ É368 (1976) 345;\\
	L.F. Abbott, \NP 185 (1981) 189:\\
	background-field gauge.
 £16 S. Deser, ÓClass. Quant. Grav.Õ É4 (1987) L99:\\
	quantum-quadratic action implies on-shell background .
 £17 J.C. Ward, \PR 78 (1950) 182;\\
	Y. Takahashi, ÓNuo. Cim.Õ É6 (1957) 370.
 £18 N.K. Nielsen, \NP 140 (1978) 499;\\
	R.E. Kallosh, \NP 141 (1978) 141.
 £19 M.T. Grisaru, W. Siegel, and M. Ro×cek, \NP 159 (1979) 429:\\
	super background-field gauge.

\unrefs

Û5 C. SCATTERING

We have seen how covariant expansions of the S-matrix can be based on
various definitions of $\h$.  Covariant expansions can also be based on
spacetime quantum numbers:  For example, we can perturb in mass; this
is equivalent to adding low-energy corrections to the high-energy
approximation.  Also, the first-quantized version of the $\h$ expansion,
which expands in powers of momenta, is effectively an expansion in
inverse powers of mass (low-energy approximation).

The only other spacetime property of a particle is spin, or helicity for
massless particles in D=4.  It is possible to define expansions in terms of
it by describing the leading order by a complex action.  This violates
semiclassical unitarity at that order; however, the loop expansion
violates unitarity at tree order also, so the expansion is still useful as
long as unitarity returns once the expansion has been summed. 
Furthermore, we have already seen that gauges where unitarity is not
manifest have some advantages over unitary gauges.  In particular, the
Gervais-Neveu gauge uses a complex gauge condition.

Ü1. Yang-Mills

We first consider calculations for massless theories; these are simpler
than massive ones in D=4 because the little group of the Lorentz group is
SO(D$-$2) instead of SO(D$-$1), and is thus Abelian:  We can label the spin
of a state by an integer or half-integer, the helicity, by use of the
spacecone formalism.  To simplify notation, we drop the transverse index
($p^t£p$), and distinguish 4-momentum $P$ from its transverse
component $p$ by using upper- and lower-case.  We also use color
ordering; i.e., we examine only planar diagrams for each permutation of
external lines.  

We begin by summarizing the spacecone rules for pure Yang-Mills found
in subsection VIB6:  The Lagrangian appearing in the action
$S=g^{-2}trÇdx¼L$, writing derivatives as momentum operators for later
convenience, is
$$ L = A^+(-üP^2)A^- +(\f{p^-}p A^+)[A^+,pA^-]
	+(\f{p^+}pA^-)[A^-,pA^+] +[A^+,pA^-]\f1{p^2}[A^-,pA^+] $$
 Twistor notation (see subsection IIB6) is used:
$$ ÒpqÔ = -ÒqpÔ,âÒpqÔÒrsÔ +ÒqrÔÒpsÔ +ÒrpÔÒqsÔ = 0,âÒpqÔ* = [qp] $$
$$ p^+ = Òp-Ô[-p],âp^- = Ò+pÔ[p+],âp = Ò+pÔ[-p],âÐp = Òp-Ô[p+] $$
$$ Ò+-Ô = [-+] = 1 $$
 The propagator and vertices are read from $L$ in the usual way, but in
addition we have further simplification from the choice of external line
factors
$$ ·_+ = {[-p]\over Ò+pÔ},ââ·_- = {Ò+pÔ\over [-p]};ââ
	{p^-\over p}·_¢ = {p^+\over p}·_\¢ = 1 $$
 where $¢$ and $\¢$ are the reference lines, with + and $-$ helicity,
respectively (not to be confused with the earlier notation for spinor
indices $Œ=(¢,\¢)$).  However, the reference ÓmomentaÕ for helicities
$à$ are taken from lines with helicities $¦$:
$$ P_¢ = |-Ô[-|,ââP_\¢ = |+Ô[+| $$
$$ ÜâP_¢^a = ¶^a_-,ââP_\¢^a = ¶^a_+ $$
 The reference external line factors occur only in the above combinations,
because only 1 term of 1 of the 3-point vertices contributes to each.

The simplest examples are classes of diagrams that vanish by virtue of
their ``maximal helicity violation":  By simple counting of +'s and $-$'s, we
see that the tree graphs with the fewest external $-$'s, those with only
self-dual vertices (++$-$), have a single external $-$.  Thus the all +
amplitude vanishes automatically.  Furthermore, the diagrams with a
single external $-$ must have that line chosen as one of the reference
lines.  However, by the above rules that line can carry only the
ÓantiÕ-self-dual vertex ($--$+), so those amplitudes also vanish.

The simplest nonvanishing amplitude is ++$--$.  We consider the case
where the helicities are cyclically ordered as ++$--$; we label them 1234,
and choose 1 and 4 as the reference lines; this amplitude can be denoted
as $¢$+$-\¢$.  ($P_4=|+Ô[+|$, $P_1=|-Ô[-|$:  The positive-helicity reference
line gives the reference momentum for negative helicity, and vice versa.) 
We label all external momenta as flowing inward.  There are only three
diagrams; however, the + reference line uses only the ++$-$ vertex, while
the $-$ reference line uses only the $--$+ vertex, so the 4-point-vertex
diagram vanishes, as does the diagram with both reference lines at the
same vertex.  Thus, we are left with only 1 graph. 

$$ \fig{4gl} $$

Furthermore, we know that the 3-point vertices contribute only 1 term to
the reference line, so this graph has only 1 term.  This means we can
immediately write down the answer (dropping the factors of $-g$ at each
vertex):
$$ ·_{2+}·_{3-}p_2 p_3 {1\over ü(P_3+P_4)^2} = 
	{[-2]\over Ò+2Ô}{Ò+3Ô\over [-3]}Ò+2Ô[-2]Ò+3Ô[-3]
	{1\over Ò34Ô[34]}{1\over Ò+-Ô[-+]} $$
$$ = {[12]^2 Ò34Ô\over [34][41]Ò14Ô} $$
 where we have restored helicity and dimensions
(trees go as $Ò¼Ô^{2-E_+}[¼]^{2-E_-}$), and used $p_1=p_4=0$.
We have omitted the usual group theory factor (see subsection VC9). 
(Note that the propagator is $-1/üP^2$, because of the signature for the
$à$ spacecone components.  This extra sign cancels that coming from the
fact that one vertex has cyclic ordering and one anticyclic with respect
to group theory, i.e., the commutators in the action.)  Using the identities,
following from overall momentum conservation,
$$ (P_1+P_4)^2 = (P_2+P_3)^2âÜâ[41]Ò14Ô = [23]Ò32Ô $$
$$ Ý|pÔ[p| =0âÜâÒ34Ô[14] = -Ò32Ô[12] $$
 this can be put in the standard form
$$ {[12]^4\over [12][23][34][41]} $$

\x VIC1.1  Using similar manipulations, cast it into the form
$$ {Ò34Ô^4\over Ò12ÔÒ23ÔÒ34ÔÒ41Ô} $$

Another simple form can be obtained from the original form by doing a little less cancellation:
$$ {[12]^2 Ò34Ô^2\over Ò34Ô[34][41]Ò14Ô} = 
	-{tr[f*(1)f*(2)]tr[f(3)f(4)]\over (üs)(üt)} $$
using $f=i|pÔÒp|$ and $f*=i|p][p|$ (from subsection VIB6).  Unlike the others, this form is directly in terms of physical quantities, namely momentum invariants and (linearized) field strengths (see subsection VIB8).  Although similar expressions hold in other dimensions, where twistors may not exist, twistors allow for a simpler derivation.

\x VIC1.2  Repeat the calculation for the $+-+-$ (color-ordered) amplitude:
 ªa Find the form in terms of just $Ò¼Ô$'s, or just $[¼]$'s.
 ªb Find the form in terms of momentum invariants and field strengths.

The corresponding differential cross section is very simple:  Using
$$ ÒpqÔ* = -[pq]âÜâ|ÒpqÔ|^2 = |[pq]|^2 = -PÉQ $$
 and momentum conservation, we find (after including the coupling $g$)
$$ |T|^2 = g^4{s^2\over t^2}¼or¼g^4{t^2\over s^2} $$
 depending on the orientation of the diagram with respect to time, for
this color-ordered contribution.  (Depending on the color quantum
numbers of the external states, this can be the only contribution.)  Then
(see subsection VC7)
$$ {d§\over dt} = 2(2¹)^3 g^4 ð 
	\left( {1\over t^2}¼or¼{t^2\over s^4} \right) $$

A more complicated example is the +++$--$ amplitude.  Again taking
color-ordered (planar) amplitudes, we choose the amplitude cyclically
ordered as +++$--$ with lines labeled 12345, picking 1 and 5 as the
reference lines, which we denote as $¢$++$-\¢$.  Again dropping all
graphs with a reference line at a 4-point vertex or 2 references lines at a
3-point, all 5 graphs with a 4-point vertex are killed, and only 3 of the
remaining 5 survive.  (We also need to consider various combinations of +
and $-$ indices, but only 1 survives for each graph because of the
chirality of 3-vertices with reference lines.)

$$ \fig{5gl} $$

Since 3-point vertices with (without) a reference line have 1 (2) terms,
we are left with only 6 terms.  The initial result for the amplitude is then
$$ -·_{2+}·_{3+}·_{4-}\left[ 
	{p_4^3 \left( 
		{p_2^- \over p_2} -{p_3^- \over p_3} \right) \over
		(P_2ÉP_3)(P_4ÉP_5)}
	-{p_2 p_4^2 \left( 
		{p_2^- +1 \over p_2} -{p_3^- \over p_3} \right) \over
		(P_1ÉP_2)(P_4ÉP_5)}
	+{p_2^2 p_4 \left( 
		{p_2^- +1 \over p_2} -{p_3^- \over p_3} \right) \over
		(P_1ÉP_2)(P_3ÉP_4)}
	\right] $$
 where we have used the fact that the reference lines have trivial
momenta: 1 for the component with $à$ index opposite to its helicity, 0
for the remaining components.  The two terms for each diagram simplify
to one, using
$$ {p^-\over p} = {[p+]\over [-p]}âÜâ
	{p_2^- \over p_2} -{p_3^- \over p_3} =
	{[2+][-3] -[3+][-2]\over [-2][-3]} = {[23]\over [-2][-3]} $$
 (applying the cyclic identity) with our normalization.  Using this result, we
find the similar result
$$ {p_2^- +1 \over p_2} -{p_3^- \over p_3} =
	{Ò+-Ô[-3] +Ò+2Ô[23] \over Ò+2Ô[-2][-3]} = {Ò+4Ô[34]\over Ò+2Ô[-2][-3]} $$
 applying momentum conservation.    We next translate the momentum
denominators into twistor notation, and also substitute the spacecone
expressions for the polarizations and numerators.  Canceling identical
factors in numerator and denominator (but no further use of identities),
the amplitude becomes ($+=5$, $-=1$)
$$ {Ò+4Ô^3 \over Ò+2ÔÒ+3Ô}\left( {[-4]^2 \over Ò23Ô[4+]} 
	+{[-4][34]\over Ò2-Ô[4+]} +{Ò+2Ô[-2] \over Ò2-ÔÒ34Ô} \right) $$
$$ = {Ò+4Ô^3 \over Ò+2ÔÒ+3Ô}\left( -{Ò+2Ô[-4]\over Ò2-ÔÒ23Ô} 
	+{Ò+2Ô[-2] \over Ò2-ÔÒ34Ô} \right)
	= -{Ò+4Ô^3 \over Ò2-ÔÒ23ÔÒ34Ô} = -{Ò45Ô^4 \over Ò12ÔÒ23ÔÒ34ÔÒ45ÔÒ51Ô} $$
 applying momentum conservation twice, restoring normalization, and
replacing the numerals for $à$.

\x VIC1.3  Using the spacecone gauge, evaluate all diagrams contributing
to the six-point gluon (Yang-Mills) scattering tree amplitude (T-matrix)
with color-ordered helicities $++++--$, that correspond to the symmetric
diagram with a central 3-point vertex each of whose legs is connected to
another 3-point vertex, each of which carries 2 of the external lines.

These results can be generalized to arbitrary (color-ordered) $n$-point
tree amplitudes with two $-$ helicities, labeled $i$ and $j$, and the rest
$+$ (``Parke-Taylor amplitudes"):  The result is (in an obvious notation),
including now the coupling $(-g)^{n-2}$,
$$ Ò+_1 ò +_{i-1} -_i +_{i+1} ò +_{j-1} -_j +_{j+1} ò +_nÔ = 
	g^{n-2}{ÒijÔ^4\over Ò12ÔÒ23ÔòÒn-1,nÔÒn1Ô} $$

\x VIC1.4 Rewrite this result in terms of field strengths and momenta.
(Hint:  Multiply top and bottom by the complex conjugate of the bottom.
Unlike the simpler $n=4$ case, there will be some momenta contracted with field strengths.)

Ü2. Recursion

A simple way to derive higher-point amplitudes is using the classical field
equations.  (See subsection VC3.  In the literature, the field has often been
mistaken for the current, since $ľ¶/¶J$, $J¾¶/¶Ä$.  As usual, these are
distinguished by the fact the field always has an external propagator,
while the current has it amputated, since $KÄ+...=-J$.)  The steps are:    
\item{(1)}
Calculate the first few terms in the series (enumerated by the number of
external lines).    
\item{(2)} Guess the general result.    
\item{(3)} Prove that it is correct
by induction, using the classical field equations.  

\noindent Of course, the second
part is the hardest in general (at least when one simplifies the third step
by using spacecone methods), and has been possible for just a couple of
cases, only because the results for those cases are so simple.  Since these
results are for off-shell fields, and not S-matrix elements, they are gauge
dependent:  For example, if they are inserted into larger diagrams, the
same choice of reference lines must be used.

The solution to the classical field equations is given by tree graphs with
all external lines but one (the field itself) amputated and put on shell. 
(The usual external-line wave functions describe the asymptotic field,
which is free.)  The two cases with known solutions are those where all
the on-shell lines have the same helicity, or one different.  Note that the
field $A^à$ has a $¦$ associated with the opposite end of its external
propagator.  We then see in the former case, with all +'s on on-shell lines,
that $A^-$ vanishes because there are no fully-amputated diagrams, even
off-shell, with only +'s externally (again counting +'s and $-$'s on
vertices).  Similarly, for the latter case, with only one $-$ on an on-shell
line, we see that $A^-$ has only ++$-$ vertices; but setting that one
on-shell $-$ to be a reference line (which by definition must be on-shell),
it is not allowed such a vertex, so $A^-$ vanishes also in this case.  By
similar reasoning, we see that $A^+$ in the former case consists entirely
of ++$-$ vertices; and in the latter case consists of all ++$-$ except for
one $--$+ (no ++$--$), which must have the $-$ reference line directly
attached.  

The appearance of only the self-dual field ($A^+$) and almost only the
self-dual vertex (++$-$) means that in both cases one is essentially
solving equations in the self-dual theory:  If we take just the kinetic
term and ++$-$ vertex from the action, and make the field redefinitions
(see exercise VIB6.2)
$$ A^+ = pÄ,ââA^- = p^{-1}öÄ $$
 we obtain (after integration by parts and rearrangement inside the trace)
$$ L_{+-,++-} = öÄ(üP^2 Ä +[p^- Ä,pÄ]) $$
 These redefinitions make the ++$-$ vertex local.  $öÄ$ appears only as a
Lagrange multiplier, and its variation gives the self-dual field equation
$$ üõÄ +i(»^{\¢ÀŒ}Ä)(»^\¢{}_{ÀŒ}Ä) = 0 $$
 (which differs from the result of subsection IIIC5 by an $i$ from the use
of $p$ instead of $»$ in the field redefinition, and $¢£\¢$ from the use
of the spacecone instead of the lightcone).

We now consider in more detail the simpler (former) example (the one
which does not directly give a nontrivial scattering amplitude).  As a slight
simplification, we look at the recursion relation for the field $Ä$ as
defined in the self-dual theory.  The recursion relation is now (see
subsection VC3), scaling the coupling out of the kinetic term,
$$ Ä(1,n) = -{g\over üP^2(1,n)}Ý_{i=1}^{n-1}Ä(1,i)Ä(i+1,n)
	[p^-(1,i)p(i+1,n) -p(1,i)p^-(i+1,n)] $$
$$ P(j,k) ­ Ý_{m=j}^k P_m $$
 where we again use color ordering, number the external lines cyclically,
and $Ä(j,k)$ denotes the field with on-shell lines with momenta $P_j$
through $P_k$.  (Thus, on the left-hand side of the equation the field has
$n$ on-shell lines, while on the right-hand side the two fields have $i$ and
$n-i$.)  Plugging in the twistor expressions for the vertex momenta, we
find
$$ p^-(1,i)p(i+1,n) -p(1,i)p^-(i+1,n) = Ý_{j=1}^i Ý_{k=i+1}^n Ò+jÔ[jk]Ò+kÔ $$

If we are clever we can guess the general result from explicit evaluation
of the lower-order graphs; instead we find in the literature, after the
above redefinition,
$$ Ä(i,j) = (-g)^{N-1}{1\over Ò+iÔÒi,i+1ÔòÒj-1,jÔÒ+jÔ} $$
 where $N$ is the number of background momenta ($P_i,...,P_j$) for
$Ä(i,j)$.  For the initial-condition case $N=1$ this is simply the statement
that the external line factor for $Ä$ is now
$$ ·_Ä = {·_+\over p} = {1\over Ò+pÔ^2} $$
 The induction hypothesis is also easy to check:  The product of the two
$Ä$'s from the induction hypothesis gives the desired result by itself up
to a simple factor:
$$ Ä(1,i)Ä(i+1,n) = -\f1g Ä(1,n){Òi,i+1Ô\over Ò+iÔÒ+,i+1Ô} $$
 (The algebra of the color indices works as usual.)  We then perform the
sum over $i$ before that over $j$ and $k$ (the complete sum is over all
$i,j,k$ with $1²j²i<k²n$), making use of the identity
$$ {ÒabÔ\over Ò+aÔÒ+bÔ} +{ÒbcÔ\over Ò+bÔÒ+cÔ} = {ÒacÔ\over Ò+aÔÒ+cÔ}âÜâ
	Ý_{i=j}^{k-1}{Òi,i+1Ô\over Ò+iÔÒ+,i+1Ô} = {ÒjkÔ\over Ò+jÔÒ+kÔ} $$
 Multiplying this by the vertex momentum factor gives a sum over $j<k$ of
$ÒjkÔ[jk]=P_jÉP_k$, canceling the external propagator, yielding the desired
result.

\x VIC2.1  Work out the analog of the above for the anti-selfdual case,
paying careful attention to signs.

This result gives the general perturbative solution to the selfdual field equations as an expansion in free fields.  By similar methods the more complicated case we mentioned can also be solved, yielding the Parke-Taylor amplitudes given above, when the one external line is amputated and put on shell.  (For this simpler case that gives zero, since there is no pole in that line.)  We can see the same characteristic denominator in both expressions.

Ü3. Fermions

We have seen in subsection VIB7 how these methods can be applied to
massless spinors.  Rather than applying the rules directly, in this
subsection we examine the relation of the results in QCD to those in pure
Yang-Mills theory.  We also saw in subsection VIB7 how supersymmetry
could be used to relate different QCD amplitudes.  However, in practice
supersymmetry relations give only a few useful relations, and
only ones that can already be seen directly from the spacecone rules,
which give more results than can be seen by supersymmetry alone.

The simplest relations that follow from supersymmetry are the vanishing
of tree graphs with fewer than two negative helicities, which we saw in
subsection VIC1 follows automatically from the spacecone rules.  The
remaining useful supersymmetry relation for tree graphs is the relation
between Parke-Taylor amplitudes for pure Yang-Mills and those with one
external line each of positive and negative helicity replaced with spinors
or scalars.  The easiest way to see this result is to make use of the
conventions of the selfdual theory, as in the preceeding subsection.  In
Parke-Taylor amplitudes only one vertex is a non-selfdual vertex, which
accounts for the simplicity of these amplitudes.  (Tree amplitudes with
only selfdual vertices vanish.)  Furthermore, after transforming to the
selfdual conventions, all (nonvanishing) selfdual vertices are identical ---
independent of spin.  Finally, the nonselfdual 3-point vertex with
one negative-helicity gluon chosen as a reference line (the only
non-selfdual vertex we'll need for this relation) is independent of the
spins of the remaining two lines.  Consequently, the only difference
between the two amplitudes we are relating comes from the difference
in normalization of external line factors for gluons and quarks (and
scalars).

We will not review the superspace formulation of selfdual
supersymmetric theories here.  The main features will be evident from
the example of supersymmetric QCD that we now examine in more detail. 
The main result follows from treating the selfdual field of the
nonsupersymmetric theory as a spacecone (or lightcone) superfield,
as in subsection VIB7. 
Dimensional analysis then tells us that the field of helicity $h$ has
dimension $1-h$.  The appropriate redefinitions of the spacecone fields
are then
$$ A^+ £ pA^+,âÆ^+ £ pÆ^+,âÄ £Ä,âÆ^- £ Æ^-,â
	A^- £ {1\over p}A^- $$
 for the Yang-Mills fields $A^à$, spinors $Æ^à$, and scalars $Ä$.  The
resulting external line factors are then simply
$$ Ò+pÔ^{-2h} $$
 After these redefinitions, the kinetic terms, selfdual (++$-$) vertices, and
antiselfdual vertices for $-$ gluon reference line (referencing positive
helicity) are
$$ L_2 = A^+ üP^2 A^- +Æ^+ üP^2 Æ^- $$
$$ L_{3,sd} = (p^-A^+)([pA^+,A^-] +ÓpÆ^+,Æ^-Õ) +(p^- Æ^+)[pA^+,Æ^-] $$
$$ L_{3,Ñ{sd},¢} = \left({p^+\over p^2}A^-\right)([pA^+,A^-]+ÓpÆ^+,Æ^-Õ)$$
 for supersymmetric QCD.  (In the $A^3$ term in the last line we have used
integration by parts, and dropped a $(p^+/p)A^-$ term that vanishes for
the reference line:  There $(p^+/p^2)·_-=1$ now, so $(p^+/p)·_-=0$
vanishes for that line since $p£0$.)

\x VIC3.1  Apply these redefinitions to the full action for
supersymmetric QCD given in subsection VIB7:
 ªa Find the action and external line factors (especially for reference lines).
 ªb Evaluate the 4-gluon tree amplitude for 2 positive and 2 negative
helicities with these modified rules.

We now see easily that the terms $L_2$ and $L_{3,sd}$ that define the
selfdual theory are independent of whether boson or fermion is chosen
for the positive helicity fields and the negative helicity one (only the
helicities of the fields must add up to 0 for $L_2$ and 1 for $L_{3,sd}$ for
Lorentz invariance).  Thus, supersymmetry is a much stronger restriction
in a selfdual theory than a nonselfdual one.  Finally, the current that
couples to the reference line $(p^+/p^2)A^-$ is also the same for bosons
and fermions.  We therefore have, for example, the relation
$$ (-,-ü,+ü,+ò+) = {Ò13Ô\over Ò12Ô}(--+ò+) $$
 for the color-ordered tree amplitudes (where we have labeled helicities
$à1$ by $à$).  This follows from choosing line 1 as reference line $\¢$ (for
positive helicity, from a line with negative helicity).  For example, from
our result for the 4-gluon tree, we have the 2-quark, 2-gluon tree
$$ (-,-ü,+ü,+) = {Ò12Ô^2 Ò13Ô\over Ò23ÔÒ34ÔÒ41Ô} $$

\x VIC3.2 Repeat these calculations using scalars in place of the spinors.

In the maximally supersymmetric case (N=4 supersymmetric Yang-Mills), there is a very simple form for the combined result of Maximally Helicity Violating amplitudes, n-point amplitudes whose external helicities sum to n$-$4 (or the opposite; amplitudes of the selfdual theory would have helicities summing to n$-$2, except they vanish).  They can be derived by the methods described above.  In the supertwistor space of subsection IIC5, with coordinates
$$ Òp| = p^Œ ,ââ[p| = (p^{ÀŒ},p^i) $$
(with $p^i=aÿ^i$ in terms of the notation there) we can write the amplitude as
$$ g^{n-2}{¶(Ý|iÔ[i|)\over Ò12ÔÒ23ÔòÒn-1,nÔÒn1Ô} $$
where we have included explicitly the usual momentum conservation $¶$-function (which is nontrivial in twistor space) as part of its supertwistor generalization.  Note that the actual supertwistor space used is more of a (anti)chiral supertwistor space, as is appropriate for describing selfdual theories (in analogy to that described in subsection IVC7 for ADHM twistors).  For example, the $Ò--+ò+Ô$ (Parke-Taylor) amplitude of subsection VIC1 is obtained since the helicity +1 appears at zeroth order in $p^i$ in the external twistor superfields multiplying this amplitude, and helicity $-1$ at highest.

\x VIC3.3 Extract from this amplitude the result for $Ò-,-ü,+ü,+Ô$ given above.

Ü4. Masses

The spacecone formalism yields the simplest method for deriving
S-matrix elements in massless theories (at least for trees; for loops it
may be preferable to use background field gauges, with a Lorenz gauge,
like Gervais-Neveu, for the quantum gauge and spacecone for the
background gauge).  The analogous method for the massive case is to use
actions based on self-dual fields, as described in subsection IIIC4.  The
advantage of these two methods is that they use fields that are
representations of the little group, so in the massive case fields have
2s+1 components and only undotted spinor indices (SO(3)=SU(2)), while in
the massless case they have only 2 components and no indices
(SO(2)=U(1)).  Although the actions used are more complicated, this is just
a reflection of the fact that algebra that is usually done repetitively in
graphs has been performed once and for all in the action.

However, in the massive case the simplification is not as drastic as in the
massless one:  S-matrix elements are just simpler in massless theories,
with many vanishing; the spacecone method takes advantage of this
simplification in the final results by simplifying the intermediate steps. 
The massive examples we will consider in this subsection, taken from
QED, are somewhat simple in any case, so we will stick to the older
methods (although the uses of methods based on self-duality are still
being explored).

The major difference in simplicity between the massless and massive
cases (in any approach) is in the external line factors.  The ambiguity in
the explicit expressions for the external line factors is just the little
group:  In the massless case the solutions to the field equations (one
solution and its complex conjugate) are unique up to a phase factor, which
is why the twistor formalism is so useful.  In the massive case the
solutions (2s+1) are ambiguous up to an SU(2) transformation, which
means they are messy for any choice.  Just as in the massless case the
twistor is part of the Lorentz transformation from an arbitrary frame to
the lightcone frame, in the massive case the solutions are part of the
transformation to the rest frame.  In other words, the external line
factors simply convert Lorentz indices to little-group indices; this makes
indices trivial for the massless case (in D=4), and not as nice for the
massive.

The result is that in practice whenever any of the external particles are
massive their external line factors are left as implicit in S-matrix
elements, and only their squares are explicitly evaluated, in cross
sections.  This was common in older experiments (especially QED), since
recent experiments are mostly at energies so high that masses of
external, stable particles are usually neglected.  This adds to the algebra,
since it means that Lorentz algebra is performed in each of $n^2$ terms
in the cross section rather than $n$ terms in the S-matrix.

Furthermore, the algebra is usually simplified by considering experiments
where polarization is determined in neither the preparation of the initial
states nor the measurement of the final states.  This was also common in
older experiments, when devices for polarization were not well
developed.  The result is that one averages over initial states and sums
over final states, producing the same algebraic factors that appear in the
propagator, as described in subsection VB3:  $ë$ is replaced with $ë_+$.
One then applies the rules for Feynman diagrams for cross sections, as
described in subsection VC7.

The standard S-matrices in QED are the 4-point tree graphs, with 2
3-point vertices and 1 internal propagator.  There are just 2 graphs to
consider, with various labelings of momenta:  (1) The graph with 4
external fermions (electrons/positrons) connected by 1 internal photon
describes both M\o ller (electron-electron) and Bhabha (electron-positron)
scattering, 2 labelings each.  (2) The graph with 2 external photons and 2
external fermions, as a continuous line that includes the 1 internal
fermion, describes Compton (electron-photon) scattering as well as
electron-positron creation/\-annihilation, also 2 labelings each.  In each
case, the 1 S-matrix diagram results in 2 cross section diagrams, each
with 2 momentum labelings (for a total of 2$ð$2=4): 1 diagram from
multiplying similar terms and 1 from cross-terms.

In Dirac spinor notation the Lagrangian for QED is (see subsection IIIA4)
$$ \f18 F^2 +Ðï(iÖ»-eÖA+\f{m}{å2})ï $$
 where we have scaled the ``$e$" into the vertex.  The Feynman rules are
now (Fermi-Feynman gauge):
$$ \vbox{\halign{#:\hfilâ&$#$\hfil\cr
Photon propagator& ú_{ab}/üp^2 \cr
Fermion propagator& (Öp+\f{m}{å2})/ü(p^2+m^2) \cr
Vertex& e©_a \cr}} $$
 where we have used $Öp^2=-üp^2$.
 (We use the Fermi-Feynman-gauge propagator also for defining the cut
propagator; ghosts decouple in QED.)

The cross section diagrams contain closed fermion loops, resulting in
traces of products of $©$ matrices (with a $-$1 for each loop by
Fermi-Dirac statistics).  The algebra is manageable for the present case,
using the 4D $©$-matrix identities from subsection IIA6:
$$ ©^a ©_a = -2,ââ©^a Öa ©_a = Öa,ââ©^a ÖaÖb ©_a = aÉb,ââ
	©^a ÖaÖbÖc ©_a = ÖcÖbÖa $$
$$ tr (I) = 4,ââtr (ÖaÖb) = -2aÉb,ââtr (ÖaÖbÖcÖd) = aÉb¼cÉd +aÉd¼bÉc -aÉc¼bÉd $$
 The traces encountered in the above processes are of the form
$$ N_1 = tr(©^a A ©^b B) tr(©_a C ©_b D) $$
$$ N_2 = tr(©^a A ©_a B ©^b C ©_b D),ââ
	N_3 = -tr(©^a A ©^b B ©_a C ©_b D) $$
 where $A=Öa+\f{m}{å2}$, etc.  Using the above identities, these are
evaluated as
$$ N_1 = 4m^4 +2m^2(aÉb+cÉd) +2(aÉc¼bÉd +aÉd¼bÉc) $$
$$ N_2 = 4m^4 +m^2[2(a+c)É(b+d)-aÉc-4bÉd] +(aÉb¼cÉd+aÉd¼bÉc-aÉc¼bÉd) $$
$$ N_3 = 2m^4 +m^2(aÉb+aÉc+aÉd+bÉc+bÉd+cÉd) +2aÉc¼bÉd $$

\x VIC4.1 Generalize the above identities and expressions for the $N$'s to
arbitrary dimension D.

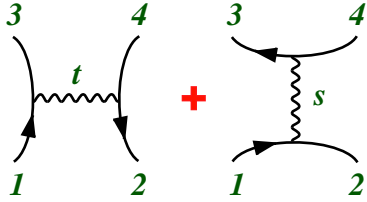
$$ \fig{bha0} $$

Our first example is $e^+ e^- £ e^+ e^-$ (``Bhabha scattering").  We have
aligned all momenta to be that of the electrons (i.e., minus that of the
positrons), so that all numerator factors are $Öp+\f{m}{å2}$ without signs. 
Specifically, we have chosen $p_1$ for the (positive-energy) momentum
for the incoming electron, $-p_2$ for the incoming positron, $p_3$ for the
outgoing electron, and $-p_4$ for the outgoing positron.  With these
conventions,
$$ -s = (p_1-p_2)^2 = (p_3-p_4)^2,ââ-t = (p_1-p_3)^2 = (p_2-p_4)^2 $$
$$ -u = (p_1+p_4)^2 = (p_2+p_3)^2ââ(p_i^2 = -m^2,ââs+t+u = 4m^2) $$
$$ Üâp_1Ép_2 = p_3Ép_4 = üs-m^2,âp_1Ép_3 = p_2Ép_4 = üt-m^2,â
	p_1Ép_4 = p_2Ép_3 = -üu+m^2 $$

For the squared amplitude we have for the average over initial
polarizations and sum over final
$$ \f14 Ý_{pol}|T|^2 = {N_1(1342)\over t^2} +{N_1(1243)\over s^2}
	+{N_3(1243)+N_3(1342)\over st} $$
$$ = ü{f(s)+f(u)\over t^2} +ü{f(t)+f(u)\over s^2} +{f(u)\over st} $$
 not including the overall factor of $e^4$, where
$$ f(x) ­ (x-2m^2)(x-6m^2) $$
 Every other $N$ term is the result of switching $sªt$ ($p_2ªp_3$, or
$p_1ªp_4$) in the previous, since that is the relation of the 2 Feynman
graphs contributing to the S-matrix.  The $N_1$ terms are the squared
diagrams, while the $N_3$'s are the cross terms.  The ``$-$" in $N_3$
comes from Fermi-Dirac statistics, switching two fermion lines.  
(This is the same relative "$-$" as for 1 fermion loop vs.¼2.)

$$ \fig{bha} $$

Finally, from subsection VC7 we have the factors to get the 
differential cross section,
$$ {d§\over dt} = ü(2¹)^3|T|^2Â_{12}^{-2},â
	Â_{12}^2 = \f14[s-(m_1+m_2)^2][s-(m_1-m_2)^2] $$
 so in this case
$$ \left({d§\over dt}\right)_{\hbox{\scriptrm Bhabha}}
	= {(2¹)^3 e^4\over s(s-4m^2)}\left[
	{f(s)+f(u)\over t^2} +{f(t)+f(u)\over s^2} +2{f(u)\over st}\right] $$
 The probabilities $|T|^2$ for $e^-e^-£e^-e^-$ (``M\o ller scattering"), or
$e^+e^+£e^+e^+$, are related by crossing symmetry $sªu$ ($p_1ª-p_3$
or $p_2ª-p_4$):
$$ \left({d§\over dt}\right)_{\hbox{\scriptrm M\o ller}}
	= {(2¹)^3 e^4\over s(s-4m^2)}\left[
	{f(s)+f(u)\over t^2} +{f(s)+f(t)\over u^2} +2{f(s)\over tu}\right] $$

A convenient frame for any of these cross sections is the center-of-mass
frame (subsection IA4).  In these cases all the external masses are
equal, so the Mandelstam variables have simple expressions in terms of
the energy (which is the same for all 4 particles) and the scattering angle:
$$ s = 4E^2,âât = -4(E^2-m^2)sin^2Ê\fÏ2,ââu = -4(E^2-m^2)cos^2Ê\fÏ2 $$

$$ \fig{compton0} $$

Another famous example is $e^-©£e^-©$ (``Compton scattering").  Now
we label $p_1$ for the incoming electron, $p_3$ for the incoming photon,
$p_2$ for the outgoing electron, and $-p_4$ for the outgoing photon, so
the Mandelstam variables are
$$ -s = (p_1+p_3)^2 = (p_2-p_4)^2,ââ-t = (p_1-p_2)^2 = (p_3+p_4)^2 $$
$$ -u = (p_1+p_4)^2 = (p_2-p_3)^2ââ
	(p_1^2 = p_2^2 = -m^2,âp_3^2 = p_4^2 = 0;âs+t+u = 2m^2) $$
$$ Üâp_1Ép_2 = üt-m^2,ââp_1Ép_3 = -ü(s-m^2),ââ
	p_1Ép_4 = -ü(u-m^2) $$
$$ p_2Ép_3 = ü(u-m^2),ââp_2Ép_4 = ü(s-m^2),ââp_3Ép_4 = -üt 
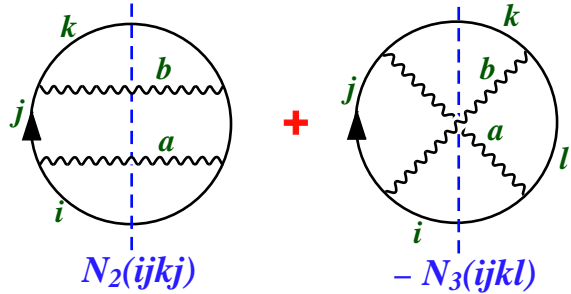
$$

$$ \fig{compton} $$

The probability is
$$ \f14 Ý_{pol}|T|^2 = {N_2(1,1+3,2,1+3)\over (s-m^2)^2} 
	+{N_2(1,1+4,2,1+4)\over (u-m^2)^2} $$
$$ -{N_3(1,1+4,2,1+3)+N_3(1,1+3,2,1+4)\over (s-m^2)(u-m^2)} $$
$$ = ü{m^4+m^2(3s+u)-su\over (s-m^2)^2}
	+ü{m^4+m^2(3u+s)-su\over (u-m^2)^2}
	-{m^2(t-4m^2)\over (s-m^2)(u-m^2)} $$
 where now every other term comes from switching $sªu$ ($p_3ªp_4$),
and the cross section is, after some rearrangement,
$$ \left({d§\over dt}\right)_{\hbox{\scriptrm Compton}}
	= {(2¹)^3 e^4\over (s-m^2)^2}\left[
	4m^4\left({1\over s-m^2}+{1\over u-m^2}\right)^2
	+4m^2\left({1\over s-m^2}+{1\over u-m^2}\right)\right. $$
$$ -\left.\left({u-m^2\over s-m^2}+{s-m^2\over u-m^2}\right)\right] $$

A useful frame is the lab frame (i.e., the rest frame of the electron),
where in terms of the initial and final (positive) energies ($E$ and $E'$)
and scattering angle of the photon we have
$$ s = m^2+2mE,âu = m^2-2mE',ât = 2m(E'-E);ââ
	{1\over E'} -{1\over E} = {2¼sin^2Ê\fÏ2\over m} $$

By crossing symmetry, $sªt$, we get $e^+e^-£2©$ (``pair annihilation")
and $2©£e^+e^-$ (``pair creation"):
$$ \left({d§\over dt}\right)_{\hbox{\scriptrm annihil.}}
	= {(2¹)^3 e^4\over s(s-4m^2)}\left[
	4m^4\left({1\over t-m^2}+{1\over u-m^2}\right)^2
	+4m^2\left({1\over t-m^2}+{1\over u-m^2}\right)\right. $$
$$ -\left.\left({u-m^2\over t-m^2}+{s-m^2\over t-m^2}\right)\right] $$
$$ \left({d§\over dt}\right)_{\hbox{\scriptrm creation}}
	= {(2¹)^3 e^4\over s^2}\left[
	4m^4\left({1\over t-m^2}+{1\over u-m^2}\right)^2
	+4m^2\left({1\over t-m^2}+{1\over u-m^2}\right)\right. $$
$$ -\left.\left({u-m^2\over t-m^2}+{s-m^2\over t-m^2}\right)\right] $$

\x VIC4.2  Calculate all the corresponding massless cross sections using
the spacecone gauge.  Show they agree with the $m=0$ case of the above.

\x VIC4.3  Calculate all the above massive cross sections using scalars
in place of the fermions.

\x VIC4.4  Calculate all the above massive cross sections 
replacing the photons with massless
 ªa scalars
 ªb pseudoscalars (with a $©_{-1}$ at the vertex).

Ü5. Supergraphs

In supersymmetric theories the easiest way to calculate Feynman
diagrams is in superspace.  Supersymmetric cancellations are then
automatic, and new special properties of supersymmetric theories are
revealed.  The derivation of the ``supergraph" rules is similar to that
of subsection VC1, except for some fine points in the treatment of chiral
superfields.  The path integral required the explicit evaluation of only a
Gaussian for perturbation.  Since we dropped proportionality constants,
this was equivalent to substituting the solution to the classical, free field
equations back into a quadratic action.  For real scalar superfields (used
for super Yang-Mills) this is trivial, but chiral scalar superfields (used for
scalar multplets) satisfy the chirality constraint, and have superpotential
terms: integrals over chiral superspace ($Çdx¼d^2 Ï$), not the full
superspace ($Çdx¼d^4 Ï$).

 We want to make use of the identity for evaluating the path integral (see
subsection VC1) 
$$ Ç{du\over å{2¹}}¼e^{-uMu/2}f(u+v) = 
	Ç{du\over å{2¹}}¼e^{-uMu/2}e^{u»_v}f(v) ¾
	e^{»_v M^{-1}»_v/2}f(v) $$
 Then the ``action" we need to integrate is
$$ ÷S = -Çdx¼d^4 ϼÐÄÄ -\left[ Çdx¼d^2 ϼ(\f{m}{å2})üÄ^2 +h.c. \right]
	-\left( Çdx¼d^2 Ï¼Ä {¶\over ¶\Ä} +h.c. \right) $$
 consisting of the (derivative part of the) kinetic term, mass term, and
(minus the) term that acts on $e^{-S_I[\Ä]}$.  Solving the field equations
(see subsection IVC2)
$$ Ðd^2 ÐÄ +\f{m}{å2}Ä +{¶\over ¶\Ä} 
	= d^2 Ä +\f{m}{å2}ÐÄ +{¶\over ¶Ð{\Ä}} = 0 $$
 we find
$$ Ä = {1\over ü(-õ+m^2)}
	\left( Ðd^2{¶\over ¶Ð{\Ä}} -\f{m}{å2}{¶\over ¶\Ä} \right),ââ
	ÐÄ = {1\over ü(-õ+m^2)}
	\left( d^2{¶\over ¶\Ä} -\f{m}{å2}{¶\over ¶Ð{\Ä}} \right) $$
 The propagator exponent $Çü(¶/¶\Ä)(1/K)(¶/¶\Ä)$ thus becomes
(putting back $\Ä£Ä$)
$$ Çdx¼d^4 ϼ{¶\over ¶ÐÄ}\left( -{1\over ü(-õ+m^2)}\right){¶\over ¶Ä}
	+\left[ Çdx¼d^2 ϼü{¶\over ¶Ä}
	\left( \f{m}{å2}{1\overü( -õ+m^2)}\right){¶\over ¶Ä} +h.c. \right] $$

Before writing the Feynman rules, we first note that functional
differentiation with respect to a chiral superfield, as follows from the
above variation, gives
$$ {¶\over ¶Ä(x,Ï)}Ä(x',Ï') = Ðd^2 ¶^4(Ï-Ï')¶(x-x') $$
 This means that there will be an extra $Ðd^2$ at the $Ä$ end of any chiral
propagator and an extra $d^2$ at the $ÐÄ$ end.  We could associate these
directly with the propagator, but we will use one factor of $Ðd^2$ to
convert a $Çd^2 Ï$ into $Çd^4 Ï$ at any superpotential vertex, and
similarly for the complex conjugate.  Therefore, we include such factors
explicitly as a separate Feynman rule for the ends of chiral propagators.
However, this means the $ÄÄ$ propagator (and similarly for $ÐÄÐÄ$) gets
an extra factor of $d^2/üõ$ to compensate for the fact that we include
two $Ðd^2$ factors, whereas it really had only one because its integral was
only $d^2 Ï$.  Furthermore, we Fourier transform $x$ as usual, but not $Ï$,
basically because there is no translation invariance in $Ï$, but also for a
better reason to be explained soon.  The Feynman rules of subsection VC4
are then modified as:

\vskip.1in

\Boxit{\rm\noindent
(A2$ü$) Theta's:  one for each vertex, with an $Çd^4 Ï$.\\
 (A3${}'$)  Propagators:  
$$ \li{ VV:ââ& {1\over ü(p^2+m^2)}¶^4(Ï-Ï') \cr
	ÐÄÄ:ââ& -{1\over ü(p^2+m^2)}¶^4(Ï-Ï') \cr
	ÄÄ:ââ& \f{m}{å2}\left({d^2\over -üp^2}\right)
		{1\over ü(p^2+m^2)}¶^4(Ï-Ï') \cr
       ÐÄÐÄ:ââ&\f{m}{å2}\left({Ðd^2\over -üp^2}\right)
		{1\over ü(p^2+m^2)}¶^4(Ï-Ï')\cr}$$
 (A4$ü$)  Chiral vertex factors:  $Ðd^2$ on the $Ä$ end(s) of every chiral
propagator,\\
\phantom{(A4$ü$)} $d^2$ on the $ÐÄ$ end(s), but drop any one such factor
at superpotential vertex.}

We next explain how $Ï$ integrations are performed on any connected
graph.  Consider any two vertices directly connected by a propagator.  All
the spinor derivatives acting on its $¶^4(Ï-Ï')$ can be removed from that
propagator by integration by parts.  We then can use that $¶$ function to
trivially integrate over $Ï'$, removing the $Çd^4 Ï'$ and that $¶^4(Ï-Ï')$,
and replacing $Ï'$ everywhere with $Ï$.  Effectively, those two vertices
have been contracted to the same point in $Ï$ space, eliminating that
propagator as far as $Ï$ dependence is concerned.  We can repeat this
procedure until ÓallÕ vertices are contracted to a single point.  However,
we are then left with a ``tadpole" for each loop:  Contracting propagators
this way sequentially around a loop identifies all the vertices of
that loop, and leaves the loop as a single propagator with both ends at
that point.  To evaluate this tadpole, we note that
$$ [Ðd^2 d^2 ¶^4(Ï-Ï')]|_{Ï'=Ï} = 1 $$
 ($Ï'$ derivatives can be converted into minus $Ï$ derivatives when
acting directly on the $¶$; this is basically integration by parts.)  Fewer
derivatives give 0; more can be reduced to terms of 4 or less.  This
completes all the evaluation in $Ï$ space, leaving an expression in terms
of fields (some with $d$'s acting on them) with different momenta, times
the usual momentum factors, with the usual momentum integrals, but all
at the same $Ï$, with a single $Çd^4 Ï$.  This means that the
generating functionals $W$ and $ý$ are completely local in $Ï$.

There is a further consequence of this evaluation.  We have obtained
terms with $Çd^4 Ï$, but none with $Çd^2 Ï$.  However, to do it we had to
introduce the factors $d^2/(-üp^2)$ into the $ÄÄ$ propagators.  On the
other hand, such a factor can easily be killed by a $Ðd^2$ from a vertex: 
We sandwich the $d^2$ between a $Ðd^2$ from each vertex, using the
identity
$Ðd^2 d^2 Ðd^2=-üp^2 Ðd^2$, and return the $Ðd^2$ to one vertex.  The only
time we can't do that everywhere is if ÓeveryÕ vertex is a superpotential
(so every propagator in the graph is $ÄÄ$ and every external field is $Ä$),
since otherwise we can inductively borrow $Ðd^2$'s from some non-$Çd^2
Ï$ vertex.  Any such 1PI graph always vanishes, because there are
exactly enough $d$'s left to make the tadpoles nonvanishing, leaving an
$Çd^4 Ï$ of a product of $Ä$'s with no $d$'s, which vanishes.  On the other
hand, for a tree graph there is exactly one $d^2/p^2$ left, which converts
the $Çd^4 Ï$ to an $Çd^2 Ï$.  

The net result is that not only are $W$ and $ý$ local in $Ï$, but only their
classical parts can have $Çd^2 Ï$ terms, and the spurious $d^2/p^2$
factors (which should not appear in massive theories) are always
canceled.  In particular, this implies that all UV divergences are $Çd^4 Ï$
terms:  All terms in the superpotential are unrenormalized (no loop
corrections) to all orders in perturbation theory.

\x VIC5.1 Calculate all the contributions to $W[Ä,ÐÄ]$ from 4-point trees in
massive super-$Ä^3$ theory, and write the result in both $p$- and
$x$-space (in analogy to the nonsupersymmetric example at the end of
subsection VC4).

Improvements again result from using background-field gauges.  We have
already seen in subsection VIB10 the modification to the quantization for
supersymmetric background-field gauges. The background-field
expansion can be defined by solving the constraints on the full covariant
derivatives in terms of quantum prepotentials but background
ÓpotentialsÕ ($A_A$, not $V$), essentially by covariantizing
$d_A$ to the background $á_A$.  Then $á_Œ$ can be manipulated
(integration by parts, etc.)¼in the graphs in the same way as $d_Œ$ was,
leaving only $A_a$ (not $A_Œ$) and its derivatives ($W_Œ$, etc.)¼as
background fields.  This leads to improved power counting, and can be
used to prove ``superrenormalizability" (finiteness beyond one loop) for
N=2 extended supersymmetric theories, and finiteness for N=4.

As for other gauges (background-)chiral superfields need special
treatment, now to get the most out of background gauge invariance. 
Variation can be defined in the obvious way, but now we also need the
covariantized identity
$$ Ð{\D}^2 \D^2 Ä = ü(õ +i[\W^Œ,\D_Œ])Ä $$
 from pushing the $Ð\D$'s to the right and using the commutation
relations.  The functional integral over the quantum background-chiral
superfields can also be performed in the same way as for other gauges,
the only modifications being background covariantization (including the
above ``nonminimal" term for $õ$), and the fact that we can no longer
neglect the ``vacuum" contribution (one-loop diagrams with only
background fields externally).  Specifically, if we look at the general
derivation of the Feynman rules in subsection VC1, we see it gave rules
for all graphs except the one-loop vacuum bubble, since this graph has no
(quantum) vertices.  These rules, as adapted to superspace earlier in this
subsection, are now modified only by the covariantization just discussed,
which only adds background potentials (not prepotentials) and field
strengths to propagators and vertices.  The background-covariantized
propagators then can be further expanded about the free $õ$.  The net
result is that in ÓallÕ diagrams ÓexceptÕ (perhaps) these chiral one-loop
vacuum bubbles the background fields appear only in the form of
potentials and field strengths.  These vacuum bubbles then can be
evaluated by the usual methods, since the formerly neglected Gaussian
path integral for these ``quantum-free" fields is just the usual
one-loop path integral for a chiral superfield with external Yang-Mills
superfields, only the external fields are now identified as background
instead of quantum.  In some cases, this last calculation can be further
simplified to again yield an expression directly in terms of potentials
without explicit prepotentials (see subsection VIIIA6 below).

\refs

£1 M.T. Grisaru, H.N. Pendleton, and P. van Nieuwenhuizen, \PR 15
	(1977) 996; \\ 
	 M.T. Grisaru and H.N. Pendleton, \NP 124 (1977) 333:\\
	supersymmetry identities for trees.
 £2 M.T. Grisaru and W. Siegel, \PL 110B (1982) 49:\\
	generalization to one loop.
 £3 S.J. Parke and T. Taylor, \NP 269 (1986) 410, \PR 56 (1986) 2459;\\
	F.A. Berends and W.T. Giele, \NP 306 (1988) 759:\\
	Parke-Taylor amplitudes.
 £4 W.A. Bardeen, ÓProg. Theor. Phys. Suppl.Õ É123 (1996) 1;\\
	D. Cangemi, \xxxlink{hep-th/9605208}, \NP 484 (1997) 521;\\
	G. Chalmers and W. Siegel, \xxxlink{hep-th/9606061}, \PRD 54 (1996)
	7628:\\
	relation of Parke-Taylor amplitudes to self-dual Yang-Mills.
 £5 R. Brooks (May, 1993), unpublished;\\
	N. Berkovits and W. Siegel, \xxxlink{hep-th/9703154}, \NP 505 (1997)
	139:\\
	use of first-order actions with self-dual auxiliary fields for
	perturbation about self-duality.
 £6 V.P. Nair, \PL 214B (1988) 215:\\
	supertwistor form of N=4 MHV amplitudes.
 £7 Feynman, Óloc. cit.Õ (VB, ref. 4):\\
	Feynman diagrams for QED, from the horse's mouth.  The original
	derivation, not counting his later-published first-quantized
	path-integral approach.  Basically a mechanics point of view.
 £8 H.J. Bhabha, ÓProc. Roy. Soc.Õ ÉA154 (1936) 195.
 £9 C. M\o ller, ÓAnn. Phys.Õ É14 (1932) 531.
 £10 O. Klein and Y. Nishina, ÓZ. Phys.Õ É52 (1929) 853:\\
	Compton scattering.
 £11 A. Salam and J. Strathdee, \PRD 11 (1975) 1521;\\
	K. Fujikawa and W. Lang, \NP 88 (1975) 61;\\
	J. Honerkamp, M. Schlindwein, F. Krause, and M. Scheunert, 
	\NP 95 (1975) 397;\\
	S. Ferrara and O. Piguet, \NP 93 (1975) 261;\\
	D.M. Capper, ÓNuo. Cim.Õ É25A (1975) 259;\\
	R. Delbourgo, ÓNuo. Cim.Õ É25A (1975) 646, ÓJ. Phys. GÕ É1 (1975) 800;\\
	R. Delbourgo and M. Ram«on Medrano, \NP 110 (1976) 473;\\
	W. Siegel, \PL 84B (1979) 193, 197:\\
	early supergraphs.
 £12 Wess and Zumino, Óloc. cit.Õ (IVC, ref. 2);\\
	J. Iliopoulos and B. Zumino, \NP 76 (1974) 310;\\
	Capper, Óloc. cit.Õ;\\
	Delbourgo, Óloc.cit.Õ (first ref. above);\\
	B. Zumino, \NP 89 (1975) 535;\\
	P.C. West, \NP 106 (1976) 219;\\
	D.M. Capper and M. Ram«on Medrano, ÓJ. Phys. GÕ É2 (1976) 269;\\
	S. Weinberg, \PL 62B (1976) 111:\\
	nonrenormalization theorems for chiral superfields, from components
	or old-fashioned supergraphs.
 £13 Grisaru, Siegel, and Ro×cek, Óloc. cit.Õ (VB, ref. 17);\\
	M.T. Grisaru and W. Siegel, \NP 201 (1982) 292;\\
	Gates, Grisaru, Ro×cek, and Siegel, Óloc. cit.Õ;\\
	M.T. Grisaru and D. Zanon, \PL 142B (1984) 359, \NP 252 (1985) 578,
	591:\\
	supergraphs as done today; more general nonrenormalization
	theorems.

\unrefs

ÚVII. LOOPS

Although our analysis so far is sufficient to evaluate the lowest-order
term in the $\h$ expansion (``trees"), certain new features arise at
higher orders.

Û6 A. GENERAL

When infinities were first found in perturbative quantum field theory,
they were thought to be a serious problem.  A prescription can be given to
remove these infinities, called ``regularization".  It was later shown that
this regularization can be defined in such a way as to preserve all the
desirable physical properties of the theory, called ``renormalization". 
Unfortunately, it seems that the original infinities, exiled by
renormalization from any finite order of perturbation theory, return to
plague field theory when all orders of perturbation theory are summed. 
Therefore, renormalization should not be considered a cure to the disease
of infinities, but only a treatment that allows divergent theories to be
more useful.

Ü1. Dimensional renormalization

``Perturbative renormalization" is defined to preserve the three
properties that define relativistic quantum field theory (Poincar«e
invariance, unitarity, causality).  The most general prescription is to start
with a classical theory that is causal, Poincar«e invariant, and satisfies the
semiclassical part of unitarity (as described in subsection VC5).  This
gives the tree graphs of the theory.  One then applies unitarity to define a
perturbation expansion, determining the higher orders (loop diagrams)
from the lowest (trees).  Although the usual loop diagrams are divergent,
there is enough ambiguity in the unitarity condition to allow removal of
the divergences.

The only practical way to implement this procedure is to slightly modify
the divergent graphs one obtains from the naive use of the Feynman
graph rules (obtained, e.g., from path integral quantization of a classical
action).  The steps are:  (1)¼``Regularize" each divergent graph by
modifying the momentum integrals, introducing a parameter(s), the
``regulator(s)", giving a finite result that reproduces the original
divergent integral in a certain limit.  (2) ``Renormalize" each regularized
graph by subtracting out the ``divergent part" of the regularized graph,
keeping only the ``finite part".  Once the graph has been rendered finite,
the regulator can be dropped.

One then has to check that the method of removing divergences,
order-by-order in the perturbation expansion, preserves the three
properties of relativistic quantum field theory.  The easiest way to do this
is to use both a regularization scheme and a subtraction scheme that
preserve these properties manifestly.  The standard way for the
subtraction scheme to do this is to change the coefficients of terms in the
ÓclassicalÕ action by (real) constants that depend on both the regulator
and $\h$.  Since the classical action already satisfies Poincar«e
invariance, causality, and the semiclassical part of unitarity, this will
automatically preserve all the desired properties.  In this manner of
renormalization, there thus remains only two steps to prove that a
theory can be renormalized: (1) the existence of a regularization that
manifestly preserves the three properties, and (2) that the modification
of the action by making the coefficients regulator- and
$\h$-dependent is sufficient to cancel all divergences that might
reappear in the limit as the regulator is removed.  

The latter step, discussed in subsection VIIA5 below, can be further
divided into substeps, proving: (a) The ultraviolet divergence in any graph
corresponding to scaling all internal (integration) momenta to infinity (the
``superficial" divergence) comes from a term in that graph polynomial in
external momenta, and can therefore be canceled by a local term from
the action; and (b) recursively in the number $L$ of loops, if the
renormalization procedure has already been successfully applied through
$L-$1 loops, the resulting modification of the action is sufficient to cancel
all divergences appearing at $L$ loops ÓexceptÕ the superficial ones.  The
former substep is the one that detemines whether the theory can be
renormalized.

The former step is satisfied by dimensional regularization, the standard
method of regularization in relativistic quantum field theory (and for
practical purposes beyond one loop, the only one).  It is defined by
writing the theory under consideration in arbitrary dimensions D, and
treating integrals over loop momenta as analytic functions of D.  These
integrals are then analytically continued from lower D, where they are
(ultraviolet) convergent.  The resulting expressions have pole
singularities in D at integer D, so these poles can be subtracted out as the
divergent parts.

There are two main reasons why dimensional regularization is so useful: 
(1) Most classical actions can be written as easily in arbitrary dimensions
as in D=4.  (The important exception is those that in some way involve the
Levi-Civita tensor $·_{ab...c}$.  The difficulty with such theories is not a
drawback of dimensional regularization, but a general property of
quantum field theory, and is related to the quantum breakdown of
classical symmetries, to be discussed later.)  In particular, this means it
manifestly preserves gauge invariance (which is a part of unitarity), the
property of relativistic quantum field theory most difficult to preserve. It
is also the only workable scheme of regularization to do so.

(2) It requires only one regulator, the number of dimensions D itself. 
(Most other regularization schemes require at least one regulator for each
loop.) This is the main reason why this scheme is the only practical
method of regularization at higher loops.  (An enormous number of
regularization schemes have been proposed, almost all of which work
well at one loop, but all of them are more difficult than dimensional
regularization already at two loops, and even worse at three loops and
higher.)  

The renormalization scheme based on this regularization is also very
simple:  Defining $D=D_0-2·$ (where $D_0$ is the physical dimension,
usually 4), we can Laurent expand any $L$-loop amplitude in $·$, starting
at $1/·^L$.  These $1/·^n$ terms arise from $n$ or more divergent
$D$-momentum integrals.  If we cancel all the negative powers of $·$, we
can take the limit $D£D_0$ ($·£0$) by just dropping all the positive
powers of $·$, i.e., evaluating the remainder at $·=0$.  The procedure is
to  modify the coefficients of the terms in the ``classical" action
(couplings, masses, and field normalizations) by making them $·$- and
$\h$-dependent, giving them $\h^{L-1}/·^n$ (``counter")terms.  Such
terms can cancel any local divergence at $L$ loops.  One then has to show
that they also cancel all nonlocal divergences at higher loops resulting
from this divergence appearing in an $L$-loop subgraph.  Thus, the
procedure is recursive: (1) apply the counterterms obtained from
calculations at less than $L$ loops to cancel subdivergences; (2) cancel
the remaining, local, superficial divergences by introducing new $L$-loop
counterterms.  The form of the superficial divergence can be determined
by evaluating the divergence coming from the region of momentum space
where all loop momenta go to infinity at the same rate.  The superficial
divergence is determined by $1/·$ terms of this loop and of subloop
divergences; however, if the $1/·$ piece of a prospective counterterm
vanishes at a certain loop order, so do all higher powers at that loop
order.  Thus, simple power counting (as well as global and local
symmetries) is sufficient to determine what counterterms can appear at
any particular loop order.

These rules are sufficient for evaluating momentum integrals to the point
where renormalization can be applied.  However, further simplifications
are possible where spin is involved:  Techniques specific to D=4 are useful
to simplify algebra in general, and required to preserve manifest
supersymmetry in particular.  These methods treat spin indices as 4D, in
contrast to the vector indices on momenta (and coordinates), which are
analytically continued away from D=4 by the definition of dimensional
regularization.  This is natural in 4D N=1 supersymmetric theories
formulated in superspace (or 4D N>1 in N=1 superspace), since there spins
$²1$ are described by scalar prepotentials:  There the simple prescription
is to continue in the dimension of the commuting coordinates ($x$), while
fixing the dimension of the anticommuting coordinates ($Ï$).  These rules,
for either supersymmetric or nonsupersymmetric theories, have a simple
physical interpretation for integer D<4:  dimensional reduction.  The
reduced theories differ from those produced by simple dimensional
regularization:  Vectors get (4$-$D) extra scalars; spinors may become
multiple spinors.  For supersymmetric theories this is again natural, since
vector and scalar multiplets remain irreducible after dimensional
reduction.  Unfortunately, vectors in nonsupersymmetric theories reduce
to vectors plus scalars that are not related by any symmetry, so their
renormalization is independent.  However, the complications produced by
the extra renormalizations are usually smaller than the algebraic
simplifications resulting from the restriction to 4D spin algebra, especially
for lower loops.  Another complication is that Levi-Civita ($·$) tensors
can't be treated consistently in the dimensional reduction scheme. 
Although serious in principle, in practice this is not a problem as long as
axial anomalies cancel, which is required anyway for unitarity.  (See
subsection VIIIB3.  Also, axial anomalies are easier to calculate using
Pauli-Villars regularization than with any form of dimensional
regularization.)

\x VIIA1.1  Consider the identity
$$ ö¶^a_{[b}p_c q_d r_e s_{f]} = 0 $$
 which holds in $D<5$, where $ö¶$ is a $D$-dimensional Kronecker
$¶$, as appears from momentum integrals (since momenta
themselves are $D$-dimensional by definition), and $p,q,r,s$ are
momenta.  Show by index contraction that an inconsistency arises
when trying to analytically continue away from $D=4$.  This
difficulty in defining totally antisymmetrized $D$-dimensional
objects is why Levi-Civita tensors don't exist in dimensional
regularization. 

Ü2. Momentum integration

The first step in performing momentum integration is to make all
integrals Gaussian by exponentiating propagators using Schwinger
parameters
$$ {1\over ü(p^2+m^2)} = Ç_0^¥ d ¼e^{- (p^2+m^2)/2} $$
 The general momentum integral in an arbitrary Feynman diagram is then
$$ \A = N(p,-i»_x)\left.Ç{d^{LD}k\over (2¹)^{LD/2}}Çd^P  ¼
	e^{-k^TÉA( )k/2 -k^TÉB( ,p) +ik^TÉx -C( ,p,m)}\right|_{x=0} $$
 where $A,B,C$ are first-order in $ $; $B$ is first-order in $p$ while $C$ is
second-order in $p,m$; and $L$ is the number of loops and $P$ the number
of propagators.  Also, we have used matrix notation with respect to the
$L$-dimensional loop-space, with respect to which $A$ is a matrix,
$B$ is a vector, and $C$ is a scalar.  Finally, $N$ represents all
``numerator" factors (propagator numerators and vertices; everything
but propagator denominators) and has been brought outside the integral
by Fourier transformation (i.e., introducing $x$ to produce a generating
functional for all numerators).  The momentum integrals are now
Gaussian, and can be evaluated by the methods of subsection IB3 (and
VA2):
$$ \A = N(p,-i»_x)\left.Çd^P  ¼(det¼A)^{-D/2}
	e^{(B-ix)^T A^{-1}(B-ix)/2 -C}\right|_{x=0} $$

Some of the $ $ integrations can be performed by various scalings of
subsets of the $ $'s.  For example, to see the superficial divergence of
the graph we scale all of the $ $'s and insert the identity as
$$ 1 = Ç_0^¥ d¼¶\left(Â-Ý_i  _i\right),ââ _i = Œ_i $$
 where $Œ$ are ``Feynman parameters".  The amplitude then becomes
$$ \A = N(p,-i»_x)Çd¼Â^{P-1-LD/2}¼d^P Œ¼
	¶\left(1-݌\right)[det¼A(Œ)]^{-D/2} $$
$$ ðe^{Â[B(Œ)^T A(Œ)^{-1}B(Œ)/2 -C(Œ)]
	-iB(Œ)^TÉA(Œ)^{-1}x -x^TÉA(Œ)^{-1}x/2Â}|_{x=0} $$
 (This method of introducing Feynman parameters is equivalent to directly
changing variables from $ $'s to $Â$ and one less $Œ$, and finding the
Jacobian.)

The $x$ derivatives in $N$ must now be taken.  For a contribution from
these derivatives of order $Â^{-n}$, we have $Â$ integrals of the form
$$ Çd¼Â^{P-1-LD/2-n}e^{Â[B^T A^{-1}B/2 -C]}
	= ý(P-üLD-n)[C-üB^T A^{-1}B]^{-P+LD/2+n} $$
 where we have used the definition of the $ý$ function
$$ ý(z) = Ç_0^¥ d¼Â^{z-1}e^{-Â} $$
 This integral converges only for $Re¼z>0$, but we can extend it to
(almost) all $z$ by analytic continuation:  Using integration by parts,
$$ zý(z) = Ç_0^¥ d¼e^{-Â}{d\over dÂ}Â^z
	= (e^{-Â}Â^z)|_0^¥ +Ç_0^¥ d¼Â^z e^{-Â} = ý(z+1) $$
 in the convergent region.  Analytic continuation then says to evaluate the
integral for $ý(z)$ in the region $0³Re¼z>-1$ as $ý(z+1)/z$ in terms of the
integral for $ý(z+1)$, and so on:
$$ Ç_0^¥ d¼Â^{z-1}e^{-Â} = 
	\left(Þ_{k=0}^n{1\over z+k}\right)Ç_0^¥ d¼Â^{z+n}e^{-Â} $$
 Thus, $ý(z)$ has simple poles in $z$ at all the nonpositive integers.

\x VIIA2.1 Using the identity $ý(z+1)=zý(z)$, derive the following special
cases for nonnegative integer $n$:
$$ ý(n+1) = n!,ââ
	ý(n+ü) = (n-ü)(n-\f32)...üå¹ = {(2n)!\over n!2^{2n}}å¹ $$

 Feynman parameter integrations give more complicated functions.  A
simple but common example is the Beta function
$$ B(x,y) = Ç_0^1 dŒ¼Œ^{x-1}(1-Œ)^{y-1} = {ý(x)ý(y)\over ý(x+y)} $$
 The latter expression for the Beta function can itself be derived by
similar methods:

\x VIIA2.2  Derive the following $B$ function identities:
ªa Use the integral representation of the $ý$ function to write an
expression for $ý(a)ý(b)$ as an integral over two Schwinger parameters,
and introduce a scaling parameter (as with general two-propagator
Feynman graphs, except here there is no momentum).  Show the result is
$ý(a+b)B(a,b)$, where $B$ is given by the integral definition, thus proving
the Beta function can be expressed in terms of Gamma functions.
ªb  Use the integral definition of $B$ to prove 
$$ B(x,x) = 2^{1-2x}B(ü,x)âÜâ
	{ý(x)\over ý(2x)} = {2^{1-2x}å¹\over ý(x+ü)} $$
 Use this result to find the expression in exercise VIIA2.1 for $ý(n+ü)$.
ªc Derive the identity
$$ Ç_0^¥ d ¼ ^{a-1}(1+ )^{-a-b} = B(a,b) $$
 from the substitution $ =z/(1-z)$.  Use this to show 
$$ ý(z)ý(1-z) = ¹¼csc(¹z) $$
 by changing variables $ =e^u$, and closing the contour in the complex
plane to pick up the contributions from the poles.  We thus have
$$ ý(z+n)ý(1-z-n) = (-1)^n ý(z)ý(1-z) $$
 (which is also seen inductively from $ý(z+1)=ý(z)$).
(Hint: To drop the contour at $¥$ in various directions, it might help to work in a particular region of the complex $z$ (not $u$) plane, and analytically continue.)

In general UV divergences will come from powers of $1/·$ in a $ý$
resulting from integration over some scaling parameter.  We thus need an
expression for the Laurent expansion of $ý(z)$.  This can be obtained
directly from the integral expression:  Using the definition of $e$ as a
limit,
$$ ý(z) = \lim_{n£¥}Ç_0^n d¼Â^{z-1}(1-\fÂn)^n 
	= \lim_{n£¥}n^z B(z,n+1) 
	= \f1z \lim_{n£¥}n^z Þ_{m=1}^n {1\over 1+{z\over m}} $$
 from the change of variables $Â=nŒ$, and the above identities.  Defining
the ``Euler-Mascheroni constant" $©$ by
$$ © = \lim_{n£¥}\left( -ln¼n +Ý_{m=1}^n \f1m \right) 
	= 0.5772156649... $$
$$ Üâe^{-©z} = \lim_{n£¥}n^z Þ_{m=1}^n e^{-z/m} $$
 we then can write
$$ ý(z) = \f1z e^{-©z} Þ_{n=1}^¥ {e^{z/n}\over 1+{z\over n}} $$
 which is an alternate definition of $ý$.  We then have
$$ ln¼ý(1-z) = ©z +Ý_{n=1}^¥ [-ln(1-\f{z}n) -\f{z}n]âÜ $$
$$ ln¼ý(1-z)  = ©z +Ý_{n=2}^¥ {½(n)\over n}z^n,ââ
	½(y) = Ý_{m=1}^¥{1\over m^y}$$
 by Taylor expansion of the $ln$, where $½$ is the ``Riemann zeta
function".

\x VIIA2.3   Derive the following $ý$ identities from the previous:
ªa Find the first two terms in the Laurent expansion of $ý(z)$:
$$ \lim_{z£0}ý(z) = \lim_{z£0}\f1z ý(z+1) = \f1z -© + \O(z),ââ
	© = -Ç_0^¥ d¼(ln¼Â)e^{-Â}  $$
ªb Do the same for expansions about other integers:
$$ ý(n+1+·) = 
	n!\left[ 1+·\left( -© +Ý_{m=1}^n {1\over m} \right) +\O(·^2)\right] $$
$$ ý(-n +·) = (-1)^n {1\over n!}{1\over ·}
	\left[ 1+·\left( -© +Ý_{m=1}^n {1\over m} \right) +\O(·^2) \right] $$
ªc Using the $csc$ relation in exercise VIIA2.2c and the above expansion of\\ 
$ln¼ý(1-z)$, show that
$$ cot¼z = {1\over z} \left( 1 -2 Ý_{n=1}^¥ {½(2n)\over ¹^{2n}}z^{2n}
	\right) $$
 and thus $½(2n)$ can be written as a rational number times $¹^{2n}$.

Ü3. Modified subtractions

The convenient normalization for the quadratic part of the
gauge-invariant action for an arbitrary field theory we use is
$$ S_0 = {(üµ^2)^{(D-4)/2}\over g^2}Ç{d^D x\over (2¹)^{D/2}}¼üÄKÄ $$
 for a real field $Ä$ and some coupling constant $g$, where for bosons
$$ K = ü(-õ+m^2) +... $$
 The explicit factor of 1/2 cancels the factor of 2 obtained when varying
the action with respect to $Ä$.  (Similar permutation factors are used for
interaction terms.)  Equivalently, it gives the natural normalization for
Gaussian functional integration over $Ä$ in the field theoretic path
integral.  For complex fields we instead have $ÐÄKÄ$ without the 1/2,
since then $Ä$ and $ÐÄ$ can be varied (or integrated over) independently. 

The ``renormalization mass scale" $µ$ has been introduced to preserve
the mass dimension of $g$ in arbitrary spacetime dimension; it appears
naturally with the normalization $üµ^2$ because the kinetic operator
contains $üp^2$ and $üm^2$.  Generally there will be more than one
coupling, but only one $µ$.  For some purposes it may be convenient to
scale the fields so that the coupling and $µ$ dependence appear only in
the interaction terms.  We will usually suppress the $µ$ dependence,
since it is determined by dimensional analysis, and is relevant only for
quantum corrections, where $D±4$ becomes important.

Note that our normalization differs from that normally chosen in the
literature.  It has been chosen to give the normalization appropriate for
Gaussian integrals, which appear in both the first- and second-quantized
theories.  The first difference is the factor of $(2¹)^{-D/2}$ for coordinate
and momentum integration (rather than the usual 1 for coordinate and 
$(2¹)^{-D}$ for momentum); the second is the factor of 1/2 multiplying
the kinetic operator $p^2+m^2$ (contained in $K$), rather than 1.  Our
normalization is more natural not only for Gaussian integration and
Fourier transformation, but also slightly simplifies perturbative field
theory calculations, allowing one to ignore spurious factors of 2 and
especially 2$¹$.  The net effect is only to change the normalization of
coupling constants, since such factors can be absorbed into the $1/g^2$
sitting in front.  

For example, the most accurately experimentally verified prediction of
quantum theory is the anomalous magnetic moment of the electron, to be
discussed later.  The result for the total magnetic moment, in various
normalizations, to second order in perturbation theory, is
$$ µ_{mag} = 1+e^2 = 1+{e_m^2\over 2¹} = 1+{e_{ft}^2\over 8¹^2} $$
 where ``$e$" is for our normalization (obviously the simplest), ``$e_m$"
is the normalization you first learned in classical mechanics (the one that
gives $e_m^2/r^2$ as the electrostratic force between two electrons,
assuming you used cgs units),
and ``$e_{ft}$" is the one you would see in other quantum
electrodynamics courses
(or in classical mechanics in mks/SI units, if you ignore the $·_0$).  
To complicate matters, you may have also seen
the definition $Œ=e_m^2=e_{ft}^2/4¹$, of which the only merit is
supposed to be that $1/Œ$ is very close to the integer 137, which is silly
since $1/e^2$ is even closer to the integer 861.  (Actually, $Œ$ is the
natural expansion parameter for nonrelativistic quantum mechanics,
which is basically 3D, but our $e^2$ is more natural for the loop
expansion, which is inherently 4D.)  We have also used units $\h=1$;
restoring it introduces the further complication 
$Œ=e_m^2/\h=e_{ft}^2\h/4¹$ because of the difference in the
semiclassical expansions for quantum mechanics and quantum field
theory.  

Furthermore, for nonabelian groups we have an extra factor of $ü$
compared to the standard normalization because we normalize 
$tr_D(G_i G_j)=¶_{ij}$ instead of $tr_D(G_i G_j)=ü¶_{ij}$:  The latter
originated from the case of SU(2), where it cancels the $å2$ in the
diagonal generator (and in the others, if one uses hermitian ones rather
than raising and lowering).  Unfortunately, for SU(N) with $N³3$ this
historical normalization ÓintroducesÕ a $å2$, while not canceling factors
like $åN$ (and making $tr_D(G_i G_j)$ N-dependent would wreak havoc
when considering subgroups, as well as for raising and lowering
operators).  Thus, in the above notation,
$$ nonabelian:ââg^2 = {g_{ft}^2\over 16¹^2} $$

Although we have chosen a normalization for the coupling constants that
is natural for Gaussian momentum integration, and for symmetry with
respect to Fourier transformation, in divergent integrals it has the
disadvantage of having $©$'s (Euler-Mascheroni constant) in the finite
parts.  Another natural normalization that gets rid of this irrational
number from all graphs is to divide out the angular part of the
integrals:  The volume of the unit D$-$1-dimensional sphere (the surface
of unit radius in D-dimensional Euclidean space), is easily evaluated with
Gaussians:
$$ 1 = Ç{d^D k\over (2¹)^{D/2}}¼e^{-k^2/2} =
	(2¹)^{-D/2}\left(Çd^{D-1}¯\right)Ç_0^¥ dk¼k^{D-1}e^{-k^2/2} $$
$$ = üý(\f{D}2)¹^{-D/2}Çd^{D-1}¯ $$
 We might therefore choose our normalization to cancel this factor in
momentum integrals, along with the $(2¹)^{-D}$ from Fourier
transformation.  Then the action, e.g., for a massless scalar, might be
normalized as (with conventional kinetic term)
$$ S_0 = {µ^{D-4}\over g^2}Çd^D x¼{(Çd^{D-1}¯)\over (2¹)^D}ü(»Ä)^2
	= {µ^{D-4}\over g^2}Ç{d^D x\over (4¹)^{D/2}ý(\f{D}2)}(»Ä)^2 $$
 This differs from our previous normalization by a factor of $1/ý(\f{D}2)$,
which is 1 in exactly D=4, but differs infinitesimally far away.  The result
is that the two schemes will differ effectively by finite renormalizations: 
For example, in divergent one-loop graphs the $1/·$ divergences will
have the same coefficient in the two schemes, but the finite remainders
will differ by constants, since effectively the coupling has been redefined
by a factor of $1+\O(·)$.  Thus, the same result can be achieved by
modifiying the counterterm to be proportional to $1/·+constant$.  Hence,
the earlier version of dimensional regularization is called ``minimal
subtraction (MS)", while the modification inspired by the volume of the
sphere is called ``modified minimal subtraction ($Ñ{\rm MS}$)".

We now examine explicitly the difference between the two schemes. As
we can see from our previous example, momentum integrals from scaling
various subsets of Schwinger parameters in a multiloop diagram will
produce $ý$ function factors of the form, with this new normalization,
$$ Þ_i [ý(\f{D}2)]^{L_i}ý(n_i-L_i\f{D}2) $$
 for some integers $n_i$, where $Ý_i L_i=L$.  In even dimensions $D_0$
(especially 4), we can use $ý(z+1)=zý(z)$ to write each $L_i$-loop factor
as a rational function of $·=(D_0-D)/2$ times
$$ [ý(1-·)]^{L_i}ý(1+L_i·) ¾ (e^{©·})^{L_i}e^{-L_i ©·} = 1 $$
 where we have written only the $©$ dependence.  Thus all $©$'s cancel. 
This modification allows some further simplification by eliminating extra
terms arising at 2 loops involving $½(2)$ (but this is only $¹^2/6$; see
exercise VIIA2.3).  Similar results can be obtained by using $ý(1-·)$
instead of $ý(\f{D}2)$.

Another subtraction scheme, the ``G scheme", is defined by normalizing
momentum integration so that the coefficient of the 1-loop massless
propagator correction in $Ä^3$ theory in 4 dimensions (see subsection
VIIB4 below) is ÓexactlyÕ $1/·$ (without extra finite terms) times a power
of $üp^2$, or that up to a sign in higher even dimensions.  (The
normalization factor must be positive, and also finite and nonvanishing as
$·£0$.)  As for $Ñ{\rm MS}$, we can also pull out rational factors to get just
the $ý(1+n·)$'s.  The net effect of these two schemes, as compared to MS,
is to modify $\h$, which appears as $Çdx/\h$ in the classical action or
$\hÇdp$ for loop integrals, as
$$ Ñ{\rm MS}:ââ\h £ ý(\f{D}2)\hâor¼ý(1-·)\h $$
$$ {\rm G}:ââ\h £ {(-1)^{D_0/2}\over ·ý(2-\f{D}2)B(\f{D}2-1,\f{D}2-1)}\h
	âor¼{ý(1-2·)\over ý(1+·)[ý(1-·)]^2}\h $$
 (If we want to be picky, we can also normalize the former forms
appropriately for $D=D_0$, by including an extra factor of $1/ý(\f{D_0}2)$
for $Ñ{\rm MS}$ and $ý(\f{D_0}2-1)/ý(D_0-2)$ for G.)

\x VIIA3.1  Explictly evaluate the difference between the MS, $Ñ{\rm MS}$,
and G schemes to order $·$, including the picky $D_0$ factors.

This particular fix for eliminating irrational numbers works only for those
arising at one or two loops:  In general, because subdivergences produce
expressions of the form $ý(1+·)/·^L$ at $L$ loops, we encounter finite
terms involving $½(L)$ at $L$ loops, which is irrational (worse than just
$¹$'s, $e$'s, $å2$'s, etc.)¼for odd $L$.  So, for example, $½(3)$ appears at 3
loops, and it can be shown that $½(3)$ (and higher $½(n)$) can't always be
canceled.

The ``momentum subtraction scheme (MOM)", rather than simplifying
numerically, is designed to give a more physical interpretation of the
coupling constants appearing in the action:  It is defined so that they take
their on-shell values.  Thus, it is particularly suited to low-energy
calculations, which involve an expansion about the mass shell.  For
example, consider the quantum kinetic operator $K+ëK$, appearing in the
quadratic part of the effective action.  It depends on the momentum
through one variable: $p^2$ or $Öp$, etc.  We then can consider Taylor
expanding it in this variable about its classical on-shell value.  (For
reasons to be explained later, this can be dangerous for massless fields,
when it requires infrared regularization.)  This is equivalent to expanding
in powers of the classical kinetic operator $K$ itself.  The MOM
prescription is then to use subtraction terms $-¶K$ to cancel the terms in
the quantum correction $ëK$ to the kinetic operator that are linear in $K$:
$$ ëK = a +bK +\O (K^2)âÜâ¶K = -a -bK $$
$$ ÜâK_{ren} ­ K +ëK +¶K = K +\O (K^2)âÜâ
	{1\over K_{ren}} = {1\over K} +\O (K^0) $$
 The two renormalizations are related directly to the ``wave function"
and mass renormalizations, one being proportional to the entire kinetic
term, the other to the constant (mass) term.  The result is that the
renormalized propagator has the same pole and residue as the classical
one.

Note that the MOM scheme, unlike the others, does not introduce an
independent mass scale $µ$:  Only physical masses set the scale.  This is
a consequence of the fact that the MOM scheme is designed for studying
low-energy (near-mass-shell) behavior, while the others are more suited
for studying high-energy behavior.  This will be important for our explicit
calculations later, when we see that MOM is more useful for QED, which is
better defined (and thus more useful), in terms of perturbation theory, at
low energies, while QCD is better defined at high energies.  More
precisely, the on-shell values of QED masses and couplings are observed
experimentally, whereas those of QCD are almost meaningless, since the
corresponding particles are not observed as asymptotic states.  On the
other hand, in QCD the introduction of the arbitrary scale $µ$ allows the
definition of a more physical mass scale, and its arbitrariness can actually
improve the accuracy of perturbative calculations.

Ü4. Optical theorem

Writing the S-matrix as $\S=1+i\T$, unitarity can be written as
$$ 1 = \Sÿ\S = (1-i\Tÿ)(1+i\T) = 1 +i(\T-\Tÿ) +\Tÿ\TâÜâ\Tÿ\T = i(\Tÿ-\T) $$
 (Actually, the more useful statement is in terms of $\S=e^{i\T}$, since
then $\T$ represents the connected graphs, but the result of the argument
is the same.)  Summation of a probability over final states then yields the
``optical theorem":
$$ Ý_f |\T_{fi}|^2 = Ý_f Òi|\Tÿ|fÔÒf|\T|iÔ = Òi|\Tÿ\T|iÔ = 2¼Im¼\T_{ii} $$
 Applying unitarity in terms of the cutting rules (subsection VC6), we see
that this condition can be applied to $\T_{ii}$ diagram by diagram, using
any combination of parts of $\T_{fi}$ and $\Tÿ_{fi}$ that fit together to
form the graph considered for $\T_{ii}$.  (For example, the probability
coming from a tree graph with two final states is the imaginary part of a
one-loop graph with two intermediate states.)  Separating out the
momentum-conservation $¶$-function for a connected S-matrix element,
we have finally
$$ \T = ¶\left(Ýp\right)TâÜâÝ_f ¶\left(Ýp\right)|T_{fi}|^2 = 2¼Im¼T_{ii}$$

The simplest example of an experimental measurement of an interaction
is a decay rate.  (The only particle properties contained in the free
Lagrangian are mass and spin.)  
For example, at the tree (classical) level, decay into 2 particles is described by just the 3-pt.¼vertex, while for decay into 3 particles, it can be described by a 4-pt.¼vertex or a tree graph with 2 3-pt.¼vertices.
More generally, by the optical theorem the decay
probability is given by the imaginary part of the propagator correction,
evaluated on shell.  

We then find the total decay probability per unit time by dividing the
probability by the spatial density $¨$ times the spatial volume times the
time duration, and summing over final states (see exercise VC7.1):
$$ {dP\over dt} = Ý_f{P\over ¨ V_D} = {2¼Im¼T_{ii}\over ¿} $$
 using the expressions for $¨$, and $P$ in terms of $|T_{fi}|^2$, given in
subsection VC7.  The optical theorem can be applied similarly for the total
cross section:
$$ § =2(Im¼T_{ii}) {(2¹)^{D/2}\over Â_{12}} $$

The decay rate for a particle is frame dependent, but we
usually pick the rest frame for massive particles, where $¿=m$. 
Alternatively, we can define the total decay probability per unit
ÓproperÕ time:
$$ ¿ = m{dt\over ds}âÜâ{dP\over ds} = {2¼Im¼T_{ii}\over m} $$
 (where $s$ and $t$ should not be confused with the Mandelstam
variables).  For the massless case, we can use instead the parameter $ $,
as it appears in classical mechanics in the gauge $v=1$, or as the classical
value of the Schwinger parameter from the Landau equations:
$$ p^a = {dx^a\over d }âÜâ{dP\over d } = 2¼Im¼T_{ii} $$

\x VIIA4.1  Compare this result for $dP/ds$ with that of exercise VC7.1a
to obtain an explicit expression of $Im¼T_{ii}$ in terms of $|T_{fi}|^2$
for the decay of a particle of mass $M$ into two particles of masses
$m_1,m_2$.  What happens to $Im¼T_{ii}$ for $M=m_1+m_2$, and for
$M<m_1+m_2$?

Now the decay rate of a particle can also be associated with the
imaginary part of the mass, since
$$ M = m -irâÜâ|Æ|^2 ¾ |e^{-iMt}|^2 = e^{-2rt} $$
 so the wave function for the particle at rest automatically includes a
decay factor. 
 The probability of decay, normalized by dividing by $|Æ|^2$, is thus
$$ {dP\over dt} = {d(1-|Æ|^2)\over |Æ|^2 dt} = 2r $$
 The analogous statement in momentum space is found by
Fourier transforming the propagator/wave function from time to energy:
$$ e^{-iMt} £ {1\over E-M} = {1\over E-m} +{1\over E-m}(-ir){1\over E-m}
	+ ... $$
 This expansion is in terms of the free propagator $1/(E-m)$ and the
connected graphs $-ir$.

We now check that this result agrees with that obtained from the
effective action.  The quantum propagator has a pole at $p^2=-M^2$ for
some ÓcomplexÕ constant $M$:
$$ \lim_{p^2£-M^2}ë = {1\over ü(p^2 +M^2)} $$
 We have normalized the propagator at the pole by rescaling the field; we
also keep the real part of the mass the same as the classical value
through renormalization of the mass term.  (Only the real part can be
renormalized consistently with unitarity.)  The on-shell condition is then
at physical (real) momentum $p^2=-m^2$, so the kinetic operator on-shell
is
$$ ü(p^2+M^2) = -üm^2 +ü(m -ir)^2 = -ür^2 -imr $$
 Remembering that ``interaction" terms from $ý$ contribute with a minus
sign to amplitudes, we then have
$$ {dP\over dt} = {2¼Im¼T_{ii}\over m} = {2mr\over m} = 2r $$

Ü5. Power counting

We now consider why subtracting out divergences, as poles in D, can be
implemented by giving singular D-dependence to the coupling constants. 
This is based on dimensional analysis, which tells us how divergent a
graph is at large momenta: the ``ultraviolet" (UV) divergence.  (There are
also infrared divergences, which occur for physical reasons, and do not
require renormalization.  They occur only for massless particles, and will
be considered later.)  Consider first any 1-loop 1PI graph.  In momentum
space, it has an integral over the loop momentum, $Çd^D p$.  It will
diverge if the integrand (before introducing Schwinger parameters) goes
as $p^{-D}$ or slower to infinite momentum (UV limit).  In this limit we can
ignore masses.  If we differentiate this graph with respect to any
of the external momenta, it will become more convergent, since the
power of momenta in the integrand decreases.  (The numerator of the
integrand is a polynomial, while each factor in the denominator
depends on the loop momentum.)  With enough derivatives, it becomes
convergent.  This means that the divergent part of the graph is a
polynomial in the external momenta.  Similar remarks apply to any
1PI graph, if we consider the divergence coming from letting all loop
momenta go to infinity, known as the ``superficial divergence".  Of
course, the superficial divergence is also polynomial in the coupling
constants, as is the graph as a whole; but the superficial divergence is
also polynomial in the masses, since differentiation with respect to them
has the same effect as with respect to external momenta.

We can determine several more properties of this local, but divergent,
contribution to the effective action.  First of all, it is Poincar«e invariant
and invariant under all global symmetries of the classical action, since the
effective action is invariant for all values of D, so poles in D are also. 
(Consider, e.g., contour integration in D to pick out the pole.)   If we use
the background field gauge (or consider only Abelian gauge fields), then it
is also gauge invariant.  (However, possible exceptions are conformal
invariance and invariances involving $©_{-1}$, since those are not
invariances of the classical action for all D.)  The other property we need
is that the coefficient of the divergence is real.  This follows from the fact that the S-matrix satisfies
unitarity, as preserved manifestly by dimensional regularization.  As
discussed in the previous subsection, we have unitarity
$$ \Sÿ\S = I,â\S = I +i\TâÜâi(\Tÿ-\T) = \Tÿ\T $$
 As we saw in subsection VC6, this identity actually can be applied to a
single graph, where the element of $\T$ on the left-hand side of the
equation is that graph, while on the right-hand side the summation over
intermediate states gives a sum where each term divides the graph into
two parts, one for $\T$ and one for $\Tÿ$.  We then see that at any loop
the imaginary part of a 1PI graph in $\T$ is given by ``sewing" together
diagrams from lower loops.  This means that any new divergence at any
number of loops must be real, since sewing doesn't introduce new (UV)
divergences:  Sewed lines are on shell, and phase space for on-shell
states is always finite.  (The 3-momentum of each state is bounded by
the energy, and each positive energy of an outgoing state is bounded by
the total energy of the system.)

Since Poincar«e and gauge invariances, locality, and semiclassical unitarity
were used as properties to determine the classical action, this suggests
that the divergent terms in the effective action might all be of the form
of terms already in the classical action.  Such a property is called
(perturbative) ``renormalizability".  When it holds, all infinities can be
absorbed by a redefinition of the coupling constants (and masses)
appearing in the classical action.  This is physically important because all
infinities are defined only up to finite pieces:  For example, in dimensional
regularization, we saw in subsection VIIA3 that the D-dependent
normalization of the classical action is ambiguous, resulting in an
ambiguity in the finite pieces left over after subtraction of $1/·$
terms.  Since we now know that superficial divergences are local, we see
that renormalization can produce arbitrary finite, local terms in the
effective action, corresponding to the divergent terms.  But if the
divergent terms are all of the same form as in the classical action, all
such finite terms can be absorbed by a redefintion of the coupling
constants.  On the other hand, if such finite terms did not already appear
in the classical action, we would be forced to introduce them, to make the
renormalization procedure unambiguous.  (Of course, we could give an
unambiguous prescription by definition, but from the point of view of
another prescription this would be the same as including the extra terms
in the classical action, and using the first prescription to arbitrarily fix the
nonzero values of the couplings.)  Thus, the condition of renormalizability
is necessary to prevent the appearance of an infinite number of coupling
constants, which would result in the loss of predictability.  (If
divergences require only a finite number of such couplings to be added,
we simply include those, to obtain a renormalizable theory with a number
of couplings that is finite, although larger than that with which we
started.)

Since dimensional analysis determines the form of the divergent terms of
any momentum integral, it also determines which theories are
renormalizable.  By appropriate rescaling of fields by constants, write the
classical action in a form where the derivative parts of the kinetic terms
have no dependence on any couplings.  Then define the couplings to be the
coefficients of the interaction terms.  It is easy to see that the
renormalizable theories are the ones that include all terms which satisfy
all the properties required of the classical action (including preservation
of all appropriate symmetries that are manifestly preserved by
dimensional regularization), where all couplings have engineering
dimensions that are nonnegative powers of mass:  Consider first the case
where all couplings are dimensionless (and there are no masses, or at
least we ignore them at high energies for purposes of considering UV
divergences).  Then the theory is renormalizable simply because there are
no dimensionful parameters around, so any local term must be of the
form of those originally in the classical action.  If we now introduce
couplings with positive mass dimension, then perturbatively they can
occur only to nonnegative power in any diagram, so any divergence thus
produced again has a coefficient with nonnegative mass dimension.  Since
the fields themselves have positive mass dimension, there are only a
finite number of such terms possible.  On the other hand, if we were to
allow couplings with negative dimension, then terms with arbitrarily high
powers of such couplings would also allow arbitrarily high powers of
fields, and thus lead to nonrenormalizability.  (By similar arguments,
theories with only couplings of positive mass dimension, called
``superrenormalizable", can have divergences only to a certain finite
number of loops.)

In particular, in D=4 the derivative part of the kinetic term for bosons is
of the form $Çd^4 x¼Ä»»Ä$, and for fermions $Çd^4 x¼Æ»Æ$, so bosonic
fields have dimension 1 and fermions 3/2.  That means that bosons can
appear only quartically and fermions only quadratically.  More
specifically, renormalizable theories can only have terms of the form
$$ Ä, Ä^2, Ä^3, Ä^4, Ä»Ä, Ä^2 »Ä, Ä»»Ä; Æ^2, Æ»Æ; ÄÆ^2 $$
 (where each $Ä$ can be any boson with any spin, and each $Æ$ any
fermion).  There can also be constant terms (field-independent), which
we always drop, since they don't contribute to perturbative amplitudes
(after appropriate normalization).  The terms, and their relations, are
restricted by Lorentz, gauge, and internal symmetries.  The potential for
scalar fields must also be bounded from below, to allow the existence of
a vacuum (state with lowest energy); otherwise nothing would be stable,
continually decaying into states of lower energy (i.e, the energy of the
scalars converting to other particles):  Thus, $Ä^3$ terms for scalars
requires also $Ä^4$ terms.

Spin 1 can't couple minimally to spins >1.  (One way to show this is to
covariantize the general field equation of IIB1 to $S_a{}^b á_b+ká_a$, and
show the commutator algebra of this constraint, and $õ+...$, doesn't close
unless the spin $²$1 or the external field strength vanishes.)  Furthermore,
gauge invariance for spins >1 prevents them from having renormalizable
gauge couplings in D=4:  For example, we saw in subsection IIIA4 that
spin-2 (gravity) couplings include terms of the form $ĻĻÄ$. 
Renormalizability therefore restricts us to spins 0, 1/2, and 1.  Using
Poincar«e and gauge invariance, the most general action is then of the form
$$ L = tr \leftÓ \f1{8g^2}F^2 +Æ^Œ iá_Œ{}^{ÀŒ}ÐÆ_{ÀŒ} +\f14 (áÄ)^2
	+V(Ä) +[üÆ^Œ (Ä +\f{m}{å2}) Æ_Œ +h.c.] \rightÕ $$
 where all group matrices are implicit:  They may appear in all fields, in
$m$, and even in $g$, which has independent values for the different
factors of the Yang-Mills gauge group.  Also, the matrices may differ for
the same field in different terms:  $A$ (in $á$) has different Yang-Mills
representations on different scalar and spinor fields, and $Ä$ appears
with some matrix in its Yukawa coupling $ÆÄÆ$.  (Of course, all matrices
must be chosen consistently with gauge and global invariances.)  The
potential $V(Ä)$ is no higher than quartic.  Note that the Higgs mechanism
is required to give nonabelian gauge fields mass:  $A^2$ is not gauge
invariant, and $(áÄ)^2$ is the only way to dress it up with scalars in a
renormalizable way.  (For the Abelian case we can use St¬uckelberg
fields, with $á=»+mAT$ as in subsection IVA5.  In the nonabelian case,
introducing scalars by a gauge transformation, as for St¬uckelberg,
results in a nonrenormalizable $(e^{-iÄ}áe^{iÄ})^2$ term.)  If we ignore
gauge invariance, the $A^2$ term produces unitary-gauge propagators
with bad high-energy behavior (see subsection VIB3), which leads to the
same nonrenormalizable behavior in the absence of a Higgs mechanism.

\x VIIA5.1  Show by power counting that interacting renormalizable
theories with potentials that are bounded from below exist only in D$²$4. 
Show that for D$²$2 there are an infinite number of possible
renormalizable terms in the action.  What are the kinds of renormalizable
terms possible in D=3?  

\x VIIA5.2 Superrenormalizable theories aren't realistic, but they give
oversimplified examples of many quantum features of field theory.
ªa What theories are superrenormalizable in D=3?  Show that the only
superrenormalizable interaction in D=4 is (scalar) $Ä^3$.
ªb Let's do power counting (dimensional analysis) for 4D $Ä^3$ theory. 
Write the action in the form
$$ S = \f1{g^2}Çdx¼[-\f14 Ä(õ-m^2)Ä +Ä^3] $$
 so $g^2$ counts loops.  ($Ä£gÄ$ gives the form where $g$ counts
vertices.)  What are the dimensionless terms 
$$ ëS = (g^2)^{L-1}Çdx¼Ä^n $$
 for all $n$ (including the vacuum bubbles $n=0$; of course, $L³0$)?  Since
superficial divergences are polynomial in everything (fields, momenta,
couplings, masses), this gives the maximum number of loops $L(n)$ for a
superficial divergence to appear in an $n$-point 1PI amplitude.  Make a
similar analysis for 3D $Ä^4$ theory.
ªc Find all the divergent 1PI diagrams in 4D $Ä^3$ theory.  (Hint:  There
are 5, excluding the meaningless 1-loop graph with no vertices.)
Which of these are local, and thus can be completely renormalized away?
What kind of counterterms are required for the remaining graphs?

Such global and local symmetry requirements can also be applied to the
effective action.  In background-field gauges $ý$ is gauge invariant (see
subsection VIB8), which restricts the form of the effective potential,
and even nonlocal terms.  In QED charge conjugation, in addition to
switching 2-spinors of opposite charge, changes the sign of the
electromagnetic potential:  Consequently, any pure-$A$ term in $ý$ must
be even in $A$'s (``Furry's theorem").  Such classical symmetries can be
applied at the quantum level only in the absence of ``anomalies",
quantum violations (discussed in chapter VIII below).  However, even the
anomalies themselves are restricted by symmetries:  Anomalies occur
only in symmetries that can't be manifestly preserved by regularization,
which means only conformal or axial symmetries.  Thus, when the
couplings of gauge vectors are parity invariant, the axial anomaly (which
violates parity by definition) is irrelevant.

Ü6. Infrared divergences

Although ultraviolet (UV) divergences represent a serious problem, in the
sense that they strongly restrict which theories can be useful, and require
renormalization, infrared (IR) divergences are merely a consequence of
poor semantics:  The definition of the S-matrix assumes the existence of
well-defined one-particle asymptotic states.  Unfortunately, these do not
exist when massless particles are present, even in classical mechanics: 
(1) Any particle can be accompanied by an arbitrary number of massless
particles with vanishing 4-momentum, and such a collection of particles
can be indistinguishable from the lone particle if the (measured) quantum
numbers are the same.  (These are physical states, since $p^a=0£p^2=0$.) 
(2) Any massless particle can be indistinguishable from an arbitrary
number of massless particles, with the same total 4-momentum, and each
with the same sign energy, if their 4-momenta are all proportional, since
then they are all traveling in the same direction at the same speed.  (This
situation is not important for QED, since the photon can't decay
directly into two photons.)

Experimentally, because detectors have finite accuracy, in the first case
there can be such ``soft" particles with total energy below some small
upper limit, and in the second case there can be such ``colinear" particles
within some small angle of resolution.  In principle this means we should
change our definition of asymptotic states accordingly; in practice this is
too complicated, but to any particular order in perturbation theory only a
finite number of such additional massless particles will couple.  The
procedure is then to   
\item{(1)} infrared regularize the S-matrix amplitudes (by
dimensional regularization, or introducing masses for all particles, or
keeping massless particles off-shell);   
\item{(2)} calculate probabilities/cross
sections, including contributions from soft and colinear particles, as a
function of some upper limit on their energy/angle (representing the
experimental accuracy); and   
\item{(3)} remove the regularization.  

\noindent No infrared
renormalization is necessary.  (Examples will be given in later sections.
Of course, for total cross sections, all energies and angles are integrated
over anyway.)  In general, such a procedure must be applied to both
initial and final states (the``Kinoshita-Lee-Nauenberg theorem"), but in
QED (as opposed to QCD) it is sufficient to treat only the final ones in
cases of physical interest.

The reason why infinities appear in cross sections if we ignore this
careful prescription, and in S-matrix elements in any case, is the
long range of forces mediated by massless particles.  A cross section,
although it represents a probability, is normalized in such a way that it is
an area, representing the effective cross-sectional area of a particle
being targeted by another particle.  The range of the interaction sets
the scale of this area; this is related to the mass of the particle mediating
the interaction.  Since massless particles produce infinite-range forces,
the result is infinite cross sections.  One might expect that these
infinities would appear only in total cross sections, where the momenta
of particles in the final state are integrated over.  However, by the
optical theorem, this total cross section is given by the imaginary part of
an S-matrix amplitude, which thus must also have this infinity.

The fact that these infrared divergences are physical also follows from
the fact that these kinematic situations occur in classical mechanics:  In
subsection VC8 we saw that physical singularities occur in S-matrices for
momenta that are allowed classically.

\x VIIA6.1 Consider 2$£$2 scalar scattering 
at the tree level (as in the example of subsection VC4):
 ªa Evaluate the total cross section with all
particles massless, and show it has an infrared divergence.
Relate this divergence to a classical situation.
 ªb Do the same with external masses $M$ and internal mass $m$, where there is no divergence. Find first the explicit general result, then look at the limit $m^2 ø s-4M^2$.

\refs

£1 P.A.M. Dirac, ÓProc. Cambridge Phil. Soc.Õ É30 (1934) 150;\\
	W. Heisenberg, ÓZ. Phys.Õ É92 (1934) 692;\\
	V. Weisskopf, ÓKon. Dan. Vid. Sel. Mat.-fys. Medd.Õ ÉXIV (1936) 1:\\
	charge renormalization.
 £2 W. Pauli and M. Fierz, ÓNuo. Cim.Õ É15 (1938) 167;\\
	H. Kramers, ÓNuo. Cim.Õ É15 (1938) 134:\\
	mass renormalization.
 £3 Dyson, Óloc. cit.Õ (VC):\\
	renormalization to all loops.
 £4 A. Salam, ÓPhys. Rev.Õ É82 (1951) 217, É84 (1951) 426:\\
	completed Dyson's proof.
 £5 Bogoliubov and Shirkov, Óloc. cit.Õ:\\
	general approach to renormalization.
 £6 G. 't Hooft and M. Veltman, \NP 44 (1972) 189;\\
	C.G. Bollini and J.J. Giambiagi, ÓNuo. Cim.Õ É12B (1972) 20;\\
	J.F. Ashmore, ÓLett. Nuo. CimentoÕ É4 (1972) 289;\\
	G.M. Cicuta and E. Montaldi, ÓLett. Nuo. Cim.Õ É4 (1972) 329:\\
	dimensional regularization.
 £7 G. 't Hooft and M. Veltman, Diagrammar, in ÓParticle interactions at
	very high energiesÕ, proc. 2nd Summer Institute on Particle
	Interactions at Very High Energies, Louvain, Belgium, Aug 12-25, 1973,
	eds. D. Speiser, F. Halzen, J. Weyers (Plenum, 1974) part B, p. 177; 
	and in ÓUnder the spell of the gauge principleÕ, ed. G. 't Hooft
	(World Scientific, 1994) p. 28:\\
	dimensional renormalization.
 £8 Siegel, Óloc. cit.Õ (VIC, ref. 11, next to last),
	W. Siegel, \PL 94B (1980) 37;\\
	Gates, Grisaru, Ro×cek, and Siegel, Óloc. cit.Õ:\\
	regularization by dimensional reduction.
 £9 J. Polchinski, \NP 231 (1984) 269:\\
	simplest proof of renormalization; uses renormalization group,
	doesn't use geometry of graphs (``forests" or ``skeletons", etc.)¼or
	``Weinberg's theorem"; not yet simply generalized to gauge theories.
 £10 Feynman, Óloc. cit.Õ (VB, ref. 4):\\
	Feynman parameters.
 £11 E.T. Whittaker and G.N. Watson, ÓA course of modern analysisÕ, 4th ed.
	(Cambridge University, 1927) p. 235:\\
	detailed discussion of $ý$ functions.
 £12 A.A. Vladimirov, ÓTheor. Math. Phys.Õ É36 (1978) 732;\\
	W.A. Bardeen, A.J. Buras, D.W. Duke, and T. Muta, \PRD 18 (1978)
	3998:\\	
	$Ñ{\rm MS}$ scheme.
 £13 K.G. Chetyrkin, A.L. Kataev, and F.V. Tkachov, \NP 174 (1980) 345:\\
	G scheme.
 £14 E. Feenberg, ÓPhys. Rev.Õ É40 (1932) 40;\\
	N. Bohr, R.E. Peierls, and G. Placzek, 1947 manuscript, in ÓNiels Bohr
	collected worksÕ, v.9, ed. S.R. Peierls (North-Holland, 1986) p. 487:\\
	optical theorem.
 £15 W.H. Furry, ÓPhys. Rev.Õ É51 (1937) 125.
 £16 F. Bloch and A. Nordsieck, ÓPhys. Rev.Õ É52 (1937) 54;\\
	D.R. Yennie, S.C. Frautschi, and H. Suura, ÓAnn. Phys.Õ É13 (1961) 379:\\
	infrared divergences.
 £17 T. Kinoshita, ÓJ. Math. Phys.Õ É3 (1962) 650;\\
	T.D. Lee and M. Nauenberg, ÓPhys. Rev.Õ É133B (1964) 1549.
 £18 R. Gastmans and R. Meuldermans, \NP 63 (1973) 277;\\
	W.J. Marciano and A. Sirlin, \NP 88 (1975) 86:\\
	earliest applications of dimensional regularization to infrared
	divergences.

\unrefs

Û8 B. EXAMPLES

We now give some explicit examples of the evaluation of S-matrices and
contributions to the effective action --- momentum integration,
regularization, and renormalization --- and some examples of their
application.

Ü1. Tadpoles

$$ \fig{tad} $$

The simplest examples of dimensional regularization are one-loop
``tadpoles", graphs with only one external line.  By the Schwinger
parameter method described in subsection VIIA2, we find
$$ \A_1(x,m^2) = Çdk¼e^{ikÉx}{1\over ü(k^2 +m^2)} =
	Ç_0^¥ d ¼ ^{-D/2}e^{-( m^2 +x^2/ )/2} $$
 Further evaluation requires Taylor expansion in $x$ (which we'll need
anyway to evaluate a specific integral of $k...k/(k^2+m^2)$):
$$ \A_1 = Ý_{n=0}^¥\f1{n!}(-üx^2)^n
	{ý(1-\f{D}2-n)\over (üm^2)^{1-D/2-n}} $$
 The mass dependence, as well as the argument of the $ý$ function, are
as expected by dimensional analysis:  $Çd^D k¼k^{2n}/k^2$ is ultraviolet
divergent (large $k$) for $D³2(1-n)$, and infrared divergent (small $k$) in
the limit $m£$0 for $D²2(1-n)$.  The ultraviolet divergence is reflected in
$ý(z)$, which has poles at the nonpositive integers.  

To analyze the massless case, we evaluate the $ $ integral for $D<2(1-n)$
and $m>0$, where it is finite and well-defined, analytically continue to
the region $Re¼D>2(1-n)$ (but not exactly at the points where $D$ is an
even integer), take the limit $m£0$ there, and finally analytically
continue this vanishing result to all $D$.  Therefore, all massless tadpoles
can be taken to vanish in dimensional regularization:
$$ Çdk¼{k_a ... k_b\over ük^2} = 0 $$
 or more generally
$$ Çdk¼{k_a ... k_b\over (ük^2)^a} = 0 $$
 This includes negative $a$, particularly integrals of polynomials of
momenta.  Such an integral can result from ``measure factors", as
discussed in subsections VA2 and VC1:  For example, if an auxiliary field
appears in the action with its quadratic term multiplied by a function of
other fields, then functionally integrating it out of the action results in a
functional determinant (in addition to replacing it in the classical action
by the solution to its field equation).  This is represented in terms of
Feynman graphs as one-loop diagrams whose propagators are all those of
the auxiliary field, namely 1.  The result is then regularized as
$$ Çdx¼1 ¾ ¶(0) £ Çdk¼1 = 0 $$
 consistent with the fact that such factors would cancel corresponding
factors we should include in the functional integration measure.  (In other
words, since we can always arrange to have all $¶(0)$ factors cancel, we
ignore them.)

On the other hand, massive tadpoles contribute both divergent and finite
pieces under minimal subtraction:  For example, for $D=4-2·$,
$$ \A_1(0,m^2) = Çdk¼{1\over ü(k^2 +m^2)} =
	ý(1-\f{D}2)(üm^2)^{D/2-1} $$
$$ = -üm^2Ó\f1· +[-© +1 -ln(üm^2)]Õ $$
 (see exercise VIIA2.3b), using $A^{-·}=e^{-·¼ln(A)}$.  The $©$ can be
killed by using an $Ñ{\hbox{MS}}$ or G scheme (see subsection VIIA3):  At
1-loop order any version of those schemes has the effect of just
canceling the $©$ (but differences appear at 2 loops: see subsection VIIB8
below).  To include the $µ$ dependence of the coupling, we just replace
everywhere (see also subsection VIIA3)
$$ ln(üm^2) £ ln\left({m^2\over µ^2}\right) $$
 (and similarly for any momentum factors such as $ln(üp^2)$ that might
appear more generally); effectively we are using units $üµ^2=1$.  Note
that we are not allowed to Taylor expand in $m$:  Doing so before
integration would give an incorrect result; after integration it's
impossible.  Similar remarks apply to the exponential $e^{ikÉx}$ in $\A_1$
if we interpret it as the definition by Fourier transformation of the
propagator in position space.

\x VIIB1.1  Find the 2D massless propagator in position space by Fourier
transformation.  (But Ódon'tÕ Taylor expand in $x$.)  Note that this Fourier
transform is infinite, and requires ``renormalization" (of a constant of
integration).  Compare this with the result obtained by solving the
integral form of the Klein-Gordon (Laplace) equation (i.e., Gauss' law in
D=2).

$$ \fig{tad2} $$

Two-loop tadpole integrals are not much more difficult if one line is
massive, or two are massive with the same mass.  (Again, tadpoles with
only massless lines can be taken to vanish in dimensional regularization.) 
If two propagators are massless, then they can be treated first as a
one-loop propagator graph:  By dimensional analysis, the result of that
one-loop subintegral must be a power of the momentum squared.  (The
explicit result will be calculated in subsection VIIB4.)  We
therefore consider more general one-loop tadpole integrals with more
complicated propagators that may result from subintegrations in a
higher-loop graph.  For example, we consider
$$ ö\A_1(a,x,m^2) =
	Çdk¼e^{ikÉx}{ý(a)\over [ü(k^2+m^2)]^a} $$
 Using the definition of the $ý$ function, we can write
$$ {ý(a)\over [ü(k^2+m^2)]^a} = Ç_0^¥ d ¼ ^{a-1}e^{- (k^2+m^2)/2} $$
 Performing the resultant Gaussian momentum integration and Taylor
expanding in $x$, we easily find
$$ ö\A_1(a,x,m^2) =
	Ý_{n=0}^¥\f1{n!}(-üx^2)^n{ý(a-\f{D}2-n)\over (üm^2)^{a-D/2-n}} $$

A more complicated example is
$$ ×\A_1(a,b,m^2) = Çdk¼{ý(a)\over (ük^2)^a}
	{ý(b)\over [ü(k^2+m^2)]^b} $$
$$ = Ç_0^¥ d _1 d _2¼ _1^{a-1} _2^{b-1}Çdk¼
	 e^{-[ _1 k^2 + _2(k^2+m^2)]/2} $$
$$ = Ç_0^¥ d _1 d _2¼ _1^{a-1} _2^{b-1}
	( _1+ _2)^{-D/2}e^{- _2 m^2/2} $$
 We then introduce a scaling parameter $Â$ (also described in VIIA2),
scaling $ _i=Œ_i$ in the insertion
$$ 1 = Ç_0^¥ d¼¶(Â- _1- _2) = Ç_0^¥ d¼Â^{-1}¶(1-Œ_1-Œ_2) $$
 and integrating the $¶$ over $Œ_2$ to get $Œ_2=1-Œ_1$.  This gives (with
$Œ_1=Œ$)
$$ ×\A_1(a,b,m^2) = Ç_0^1 dŒ¼Œ^{a-1}(1-Œ)^{b-1}
	Ç_0^¥ d¼Â^{a+b-D/2-1}e^{-Â(1-Œ)m^2/2} $$
$$ = {ý(a+b-\f{D}2)\over (üm^2)^{a+b-D/2}}B(a,\f{D}2-a) $$

When two of the propagators in the two-loop tadpole graph have the
same nonvanishing mass, we consider directly the two-loop integral
$$ \A_{1,2}(a,b,c,m^2) = 
	Çdk_1 dk_2¼{ý(a)\over [ü(k_1+k_2)^2]^a}
	{ý(b)\over [ü(k_1^2+m^2)]^b}{ý(c)\over [ü(k_2^2+m^2)]^c} $$
 (This integral also represents the physically less interesting 2-loop
``vacuum bubble": no external lines, and thus field independent.) 
Introducing the Schwinger parameters and performing the momentum
integration, we find
$$ Ç_0^¥ d^3  ¼ _1^{a-1} _2^{b-1} _3^{c-1}
	[ _2  _3 + _1( _2+ _3)]^{-D/2} e^{-( _2+ _3)m^2/2} $$
 where we have included the power of $det¼A$ for
$$ A = \pmatrix{  _1+ _2 &  _1 \cr  _1 &  _1+ _3 \cr } $$
 Since $ _1$ does not appear in the exponential we integrate over it first
directly, using the second integral form for the Beta function, from
exercise VIIA2.2c.  Then $ _2$ and $ _3$ can be handled by introducing a
scaling parameter for them only, leading to the previous types of
integrals.  The result is then
$$ \A_{1,2}(a,b,c,m^2) = {ý(a+b+c-D)\over (üm^2)^{a+b+c-D}}
	B(a+b-\f{D}2,a+c-\f{D}2)B(a,\f{D}2-a) $$

Ü2. Effective potential

A propagator in an external field represents a certain class of Feynman
tree diagrams.  Thus, some tree graphs can be described by quantum
mechanics.  (In principle this means we can start from classical
mechanics and first-quantize, by either operator or path-integral
methods.  However, as we'll see in chapter XII, in practice we save some
effort if we start directly with the quantum mechanics.)  If we take the
ends of such a propagator and sew them together, we can describe
arbitrary 1PI 1-loop graphs by the background field method.  While tree
graphs describe classical field theory, one-loop graphs contain many of
the important quantum properties, partly because they are the
lowest-order quantum correction, and partly because they are associated
with the functional determinant part of the (second-quantized) path
integral.  (In terms of the exponent, classical is the only negative power
in $\h$, 1-loop is $\h$-independent, and higher loops are positive powers.)

In quantum mechanics, the expansion in $\h$ is an expansion in
derivatives (since it appears only as $p_a=-i\h »_a$).  In terms of the
contribution of one-loop graphs to the effective action, this means an
expansion in the number of derivatives acting on the fields.  This
definition can be applied in general in quantum field theory, without
reference to quantum mechanics.  However, the simplest one-loop
calculations of this expansion are most easily expressed in quantum
mechanical terms.  In practice, this means expanding the external fields
in $x$ about some fixed point, expanding the exponentiated (by a
Schwinger parameter) propagator about the part Gaussian in $p$ and $x$,
and using any of the usual methods to exactly evaluate the matrix
element of a polynomial times a Gaussian.

Since we generally want arbitrary orders in a field and some of its lower
derivatives for this method to have any advantage over the usual
diagrammatic methods, in this approach one generally cuts off the
expansion at the approximation that gives just the Gaussian.  This means
we can keep up to two derivatives of an external scalar, but only a
constant field strength for an external gauge vector.  (See subsection
VIB1.)  The simplest, and most useful, example is a constant scalar field. 
The part of the effective action that consists of only scalars without
derivatives is called the ``effective potential", since it generalizes the
potential term of the classical action.  This potential determines the
quantum corrections to spontaneous symmetry breaking and the Higgs
effect, and this is important for describing mass generation for all spins.

Consider a complex scalar running around a loop, under the influence of
an external real scalar.  The Lagrangian is
$$ L = Æ*[ü(-õ+m^2)+Ä]Æ +L_Ä $$
 where the form of $L_Ä$ won't be important for calculating the $Æ$ loop.
A constant external scalar field is effectively the same as a mass term,
modifying $m^2£m^2+2Ä$.  Thus the effective potential in this case can
be evaluated by summing tadpoles:
$$ V = -Ý_{n=1}^¥ \f1n (-1)^n Ä^n Çdp¼[ü(p^2 +m^2)]^{-n} $$
 for our complex scalar; for a real scalar running around the loop there
would be an extra factor of 1/2.  We can integrate before summing:
$$ V = -Ý_{n=1}^¥ (-1)^n Ä^n {ý(n-\f{D}2)\over n!}(üm^2)^{-n+D/2} $$
 Using the identities (from Taylor expansion in $a/b$, and $ý(z+1)=zý(z)$)
$$ (a+b)^x = Ý_{n=0}^¥ {x\choose n}a^n b^{x-n},â{x\choose n} = 
	{ý(x+1)\over n!ý(x+1-n)} = (-1)^n {ý(n-x)\over n!ý(-x)} $$
 we have
$$ V = -ý(-\f{D}2)\left[(üm^2 +Ä)^{D/2} -(üm^2)^{D/2}\right] $$
 We can also integrate after summing:  Using the identities
$$ ln(a+b) -ln¼b = Ç_0^a {du\over u+b} = 
	Ç_0^a du¼{1\over b}Ý_{n=0}^¥ (-1)^n \left({u\over b}\right)^n =
	-Ý_{n=1}^¥ \f1n (-1)^n a^n b^{-n} $$
$$ Ç_0^a {du\over u+b} = Ç_0^a du Ç_0^¥ d ¼e^{- (u+b)} =
	-Ç_0^¥ {d \over  }¼\left( e^{- (a+b)} -e^{- b} \right) $$
 we have
$$ V = -Ç_0^¥ {d \over  }Çdp¼
	\left( e^{- [Ä+(p^2+m^2)/2]} -e^{- (p^2+m^2)/2} \right) $$
 which gives the same result.

For D=4, we find (after subtracting divergent counterterms, and some
corresponding finite pieces, corresponding to a MOM type of subtraction)
$$ V = ü(üm^2 +Ä)^2 ln \left( 1 +{2Ä\over m^2} \right) $$
 Since this modifies the classical potential, it demonstrates that quantum
effects can generate spontaneous symmetry breaking where there was
none classically, or vice versa (the ``Coleman-Weinberg mechanism").

\x VIIB2.1  Generalize this renormalized result to arbitrary even
dimensions.

For more complicated cases we need a more general procedure:  The basic
idea is that any Gaussian integral gives a (inverse) determinant, of which
we must take (minus) the logarithm for the effective action, and we
use $ln¼det=tr¼ln$.  (The trace includes integration over $x$ or $p$.)  After
subtracting out the field-independent part (vacuum bubble), this gives an
expression as above:  For a general kinetic operator $H=H_0+...$
(generally $H_0=ü(p^2+m^2)$), we want
$$ ý = - \left[ tr¼ln(H^{-1}) -tr¼ln(H_0^{-1}) \right] = -Ç_0^¥ {d \over  }
	Çdx¼Òx| e^{- H} -e^{- H_0} |xÔ $$
 $H$ (and $H_0$) is now treated as an operator, in terms of the coordinate
operator $X$ and momentum operator $P$, and $X|xÔ=x|xÔ$.  External fields
depend on $X$, but are Taylor expanded about $x$: e.g.,
$$ Ä(X) = Ä(x) +(X-x)É»Ä(x) +... $$
 We then can use translation invariance to write
$$ Òx|e^{- H[P,X-x,Ä(x)]}|xÔ = Ò0|e^{- H[P,X,Ä(x)]}|0Ô $$
 When $H$ is quadratic in $P$ and $X$, we can use (see exercise VA2.5)
$$ Òx|e^{- H}|yÔ = å{det¼{»^2(-S)\over »x»y}}¼e^{-S} $$
 where $S$ is the classical ``action" corresponding to the ``Hamiltonian"
$H$.  (Further examples will be given in subsection VIIIB1.)

Note that the (one-loop) vacuum bubble, with no background fields 
of any kind, must always be dropped, as it is totally meaningless 
(although how it is subtracted may be regularization dependent): 
In terms of the graphs summed here, which have equal numbers 
$n$ of propagators and vertices ($P-V=L-1$ by the usual 
$\h$ counting), it is the term $n$=0.
 Thus, in a coordinate space calculation, where there are also 
$n$ integrations $d^D x$, this term would have no propagators, 
no vertices, and no integrals (contrary to some statements 
in the literature, where this graph is misidentified as a 
one-propagator graph with one integration). 
All that remains is the permutation factor, 
1/$n$, but in this case that is an undefined 1/0.

In actual applications, closer examination reveals the 
used graph to be the cut 1-loop tadpole ($P=V=L=1$). 
Since the cut propagator gives a sum over states, 
the result is to evaluate the trace of the operator 
inserted at the vertex; in particular, a trivial vertex yields 
$str(I)$, i.e., the number of states, bosons minus fermions. 
Similar use can be made of the cut propagator correction 
($P=V=2$) for (super)traces of operator products or mass sum rules.

Ü3. Dimensional transmutation

The 2D version of the CP(n) model described in subsection IVA2 is an
interesting model in that it demonstrates generation of bound states at
the one-loop level.  Its Lagrangian is:
$$ L = ü|áÄ|^2 +ñ(|Ä|^2 -\f1{g^2}) $$
 where $g$ is now dimensionless.  For the effective potential for the
Lagrange multiplier $ñ$ from a $Ä$ loop, we find (modifying the
calculation of the previous subsection for D=2)
$$ V_1 = -ñ \left[ ln \left( {ñ\over üµ^2} \right) -1 \right] $$
 after including the renormalization mass scale $µ$ to make the argument
of the logarithm dimensionless, and the coupling dimensionless
in all dimensions.
Note that we have effectively added a mass as an infrared regulator, then taken it to 0 at the end. Equivalently, we treat $ñ$ itself as a mass, as mentioned in the previous section. We would miss this with dimensional regularization, which is OK for ultraviolet divergences, but has some difficulty with infrared divergences, because of their nonlocality.

Now the coupling can be absorbed into the definition of this scale: 
Adding to the classical term $V_0=-ñ/g^2$, the total effective potential
for $ñ$ up to one loop is
$$ V = -ñ \left[ {1\over g^2} +ln \left( {ñ\over üµ^2} \right) -1 \right]
	= -ñ \left[ ln \left( {ñ\over üM^2} \right) -1 \right] $$
 where $M$ is the ``renormalization group invariant mass scale":
$$ M^2 = µ^2 e^{-1/g^2} $$
 Since this was the only place the coupling $g$ appeared in the action, the
mass scale $M$ has now replaced it completely.  This replacement of a
dimensionless coupling ($g$) with a dimensionful one ($M$) is called
``dimensional transmutation".  It is also a common feature of quantum
high-energy behavior (see below); its importance at low energies depends
on whether the classical theory already has dimensionful parameters (like
masses).

Varying the effective potential to find the minimum, which we identify
as the (quantum) vacuum value of the field $ñ$,
$$ ln \left( {ÒñÔ\over üM^2} \right) = 0âÜâÒñÔ = üM^2 $$
 Because $ñ$ has a vacuum value, $Ä$ now has a mass (as seen by
expanding $ñ$ about its vacuum value).  Furthermore, since $ñ$ now has
more than just linear terms in the effective action, it is no longer a
Lagrange multiplier.  In fact, by calculating a massive $Ä$ loop with two
external $ñ$'s, we see that $ñ$ is now a massive physical scalar also. 
Without a Lagrange multiplier, $Ä$ is now unconstrained, so it has an
additional physical degree of freedom.  This leads to a restoration
of the spontaneously broken U(N) symmetry.  This is related to $Ä$
gaining mass, since we no longer have Goldstone bosons associated
with the symmetry breaking.  Finally, if we calculate a massive
$Ä$ loop with two external gauge vectors, we see that at low energies
there is an $F^2$ term, so $A$ is now a physical, massive vector instead
of an auxiliary field.

\x VIIB3.1  Expand $V$ about $ÒñÔ$ to show that $ñ$ gets a mass term.
Expand $ñ(x)$ to quadratic order in $x$ according to the prescription
of the previous subsection to calculate the effective action in terms
of $ñ$, $»ñ$, and $»»ñ$ (using the harmonic oscillator result of 
exercise VA2.5) to show that a $ñõñ$ term is also generated, so
$ñ$ becomes propagating.

Ü4. Massless propagators

$$ \fig{prop} $$

For the massless one-loop propagator corrections, we also introduce a
scaling parameter to convert to Feynman parameters (see the examples
of subsection VIIB1, or the general method in subsection VIIA2), with
the result
$$ \li{\A_2(x,p^2) =
	&Çdk¼e^{ikÉx}{1\over ü(k+üp)^2 ü(k-üp)^2}\cr
	=& Ç_0^¥ d¼Â^{1-D/2}Ç_0^1 dŒ_1 dŒ_2¼¶(1-Œ_1-Œ_2) ð \cr
	& ð expÓ-Â\f18 p^2 [1-(Œ_1-Œ_2)^2] -iü(Œ_1-Œ_2)pÉx
		-Â^{-1}üx^2Õ \cr} $$
 Making the change of variables
$$ Œ_1 = ü(1+º),ââŒ_2 = ü(1-º) $$
 the amplitude takes the form
$$ \A_2 = Ç_0^1 dº¼ü(e^{iºpÉx/2} +e^{-iºpÉx/2})
	Ç_0^¥ d¼Â^{1-D/2} exp[-Â\f18 (1-º^2)p^2 -Â^{-1}üx^2] $$

The integrals can be simplified if we make use of gauge invariance:  For
example, the electromagnetic current for a complex scalar is of the form
$Ä*\onª»Ä$, so the gauge field couples to the difference of the momenta
of the two scalar lines, which is $2k$ for the above as applied to the
scalar-loop correction to the photon propagator.  On the other hand gauge
invariance, or equivalently current conservation, says that such a vertex
factor should give a vanishing contribution when contracted with the
external momentum, which is $p$ in that case.  Checking this explicitly,
we do in fact find
$$ Çdk¼{kÉp\over ü(k+üp)^2 ü(k-üp)^2} = 
	Çdk¼\left[{1\over ü(k-üp)^2} -{1\over ü(k+üp)^2}\right] = 0 $$
 (even with an arbitrary additional polynomial factor in the numerator),
using the facts that the integral of the sum is the sum of the integrals
when regularized, and that massless tadpoles vanish.  (This also tells us
that replacing the numerator $kÉp$ with $k^2+\f14 p^2$ gives 0. 
Furthermore, without an extra numerator factor the integral vanishes by
antisymmetry under $k£-k$.)  Thus, if $x$ is proportional to $p$ in $A_2$,
the only contribution is from the $x=0$ term in the Taylor expansion.  
Since then $pÉ(»/»x)\A_2 = 0$, it depends on only the ``transverse" part of $x$. 
This implies that the dependence on $x$ is only through the combination
$$ u = (pÉx)^2 -p^2 x^2 $$
 so we can evaluate the integral by either of the substitutions
$$ x^2 £ 0,âpÉx £ åuââorââpÉx £ 0,âx^2 £ -u/p^2 $$

We'll consider now the latter choice.  (The former gives the same result: 
See the exercise below.)  Again, since we need to Taylor expand in $x$
anyway to find the result for a particular numerator, we expand and
perform the $Â$ integration:
$$ \A_2 = Ý_{n=0}^¥ \f1{n!}\left({u\over 2p^2}\right)^n 
	(\f18 p^2)^{n+D/2-2} ý(2-\f{D}2-n) Ç_0^1 dº¼(1-º^2)^{n+D/2-2} $$
 Performing the change of variables $º^2=©$ to convert the remaining
integral to a Beta function, and using the identities
$$ ý(ü) = å¹,ââý(z)ý(1-z) = ¹¼csc(¹z) $$
 (see the exercises in subsection VIIA2), the final result is
$$ \A_2 = -ü ¹^{3/2}csc(D\f¹2)(\f18 p^2)^{D/2-2}
	Ý_{n=0}^¥{1\over n!ý(n+\f{D}2-ü)}Ó\f1{16}[p^2 x^2-(pÉx)^2]Õ^n $$
 From the $csc$ factor we see the integral is divergent for all even $D$: 
These are ultraviolet divergences for $D³4$ and infrared ones for $D²4$;
dimensional regularization does not carefully distinguish between the
two, although the difference can usually be told by examining momentum
dependence (here from the exponent $D/2-2$).  Also notice that the two
can be mixed up by the conversion to Feynman parameters.

\x VIIB4.1  Evaluate the general massless one-loop propagator correction
using $x^2£0$, $pÉx£åu$.
ªa  Show it gives the same result as $pÉx£0$, $x^2£-u/p^2$ by using the
$ý$ and $B$ identities in subsection VIIA2.
ªb  Show it can also be written as (for convenience of expansion about
D=4)
$$ \A_2 = (üp^2)^{D/2-2}ý(\f{D}2-1)ý(2-\f{D}2) Ý_{n=0}^¥
	{ý(n+\f{D}2-1) \over n!ý(2n+D-2)}Ó\f14[p^2 x^2-(pÉx)^2]Õ^n $$

As discussed in subsection VIIB1, sometimes certain subdiagrams of
higher-loop diagrams can be evaluated explicitly, particularly propagator
corrections that themselves involve only massless propagators. 
Furthermore, such a formula might be used recursively in appropriate
diagrams.  For example, a higher-loop diagram that is itself a propagator
correction might reduce, as a final integration, to something of the form
$$ Çdk¼{ý(a)\over (ük^2)^a}{ý(b)\over [ü(k+p)^2]^b}
	= {ý(a+b-\f{D}2)\over (üp^2)^{a+b-D/2}}B(\f{D}2-a,\f{D}2-b) $$
 again using the above methods, finding similar integrals to the previous.

\x VIIB4.2 Let's examine this integral more carefully.
ªa Evaluate it in two different ways: first, by the method used above;
second, by Fourier transforming each factor using
$$ Çdk¼e^{ikÉx}{ý(a)\over (ük^2)^a} = {ý(\f{D}2-a)\over (üx^2)^{D/2-a}} $$
 (derive this also) and its inverse, simply multiplying the resulting factors
in $x$ space, and inverse transforming.
ªb Show that the $Ñ{\rm MS}$ scheme cancels $©$'s and $½(2)$'s in
iterated massless propagator corrections to all orders in $·$ by examining
$$ ý(\f{D}2)Çdk¼{1\over (ük^2)^{n_1+L_1 ·}[ü(k+p)^2]^{n_2+L_2 ·}} $$
 where $L_i$ are the numbers of loops in the propagator subgraphs
(show this by dimensional analysis) and $n_i$ are other integers.  Show
the G scheme does the same.

\x VIIB4.3 Calculate the ``phase space" for $n$ massless particles
$$ V_P = Ç\left[Þ^n{d^{D-1}p_i\over (2¹)^{D/2-1}¿_i}\right]
	(2¹)^{D/2}¶^D\left(p-Ý^n p_i\right) $$
 where $p$ is the total momentum of the $n$ particles, by using the
optical theorem:
 ªa  Consider the scalar graph with $n$ massless propagators connecting
2 vertices.  Show, both by induction in the number ($n-1$) of loops, and by
Fourier transformation (as in the previous problem), that this graph (for
distinguishable particles) gives
$$ {[ý(\f{D}2-1)]^n\over ý[n(\f{D}2-1)]}
	{ý[n-(n-1)\f{D}2]\over (üp^2)^{n-(n-1)D/2}} $$
 ªb  Wick rotate back to Minkowski space ($p^2<0$) and take (twice) the
imaginary part to obtain the result for continuous real $D>2$
$$ V_P = 2¹{[ý(\f{D}2-1)]^n\over ý[n(\f{D}2-1)]ý[(n-1)(\f{D}2-1)]}
	(-üp^2)^{-n+(n-1)D/2} $$
 which simplifies in D=4 to
$$ V_P = 2¹{1\over (n-1)!(n-2)!}(-üp^2)^{n-2} $$
 (Hint:  $(üp^2 -i·)^r = (-üp^2)^r e^{-i¹r}$.)

Ü5. Bosonization

A common method in field theory is to consider simpler models where
calculations are easier, and see if they are analogous enough to give
some insight.  In particular, two-dimensional models sometimes have
perturbative features that are expected only nonperturbatively in four
dimensions:  For example, we saw in subsection VIIB3 the generation of
bound states at one loop in the 2D CP(n) model.  Of course, some of the
features may be misleadingly simple, and may have no analog in D=4. 
Two-dimensional theories, especially free, massless ones, are also useful
to describe the quantum mechanics of the worldsheet in string theory (see
chapter XI).  In this subsection we consider free, massless 2D theories: 
Essentially, this means just the scalar and the spinor, since there are no
transverse dimensions to give gauge fields nontrivial components.

Spinor notation is very simple in D=2, since the Lorentz group is
SO(1,1)=GL(1).  For that purpose it's convenient to use lightcone notation. 
2D $©$ matrices can be chosen as
$$ ©_+ = \tat0i00,â©_- = \tat00{-i}0,â©_{-1} = \f1{å2}\tat{-i}00i;â
	ç = \tat0i{-i}0 = å2©_0 $$
$$ ï = \pmatrix{ Æ_¢ \cr Æ_{\¢} \cr },â
	Ðï = \pmatrix{ iÐÆ_{\¢} & -iÐÆ_¢ \cr } $$
 In general, even-D $©$ matrices can be constructed as direct products of
D/2 sets of 2D $©$ matrices, so $tr(I)=2^{D/2}$.  (For details, see
subsection XC1.)  

The Lagrangian for a massless, complex spinor can be written this way as
$$ L = ÐïiÖ»ï = ÐÆ_¢(-i»_{\¢\¢})Æ_¢ +ÐÆ_{\¢}(-i»_{¢¢})Æ_{\¢} $$
 (This also follows from truncation of 4D spinor notation.)  Note that
$Æ_¢$ and $Æ_{\¢}$ transform independently under proper Lorentz
transformations, as do their real and imaginary parts.  Thus, we can not
only impose a reality condition, but also a chirality condition, dropping
$Æ_¢$ or $Æ_{\¢}$:  A single real component is enough to not only define
a spinor Lorentz representation, but also construct an action.

Upon Wick rotation to Euclidean space, the lightcone coordinates become complex conjugates of each other.  These complex coordinates are convenient because they are still null coordinates, and their derivatives occur separately in massless fermion kinetic operators (and $õ$ is just their product).  For later application to string theory it will prove convenient to avoid some $å2$'s, and define
$$ z = x^0+ix^1,ââÐz = x^0-ix^1 $$
$$ Üââüõ = 2»Ð»,ââ{d^2 §\over 2¹} = ü{dz¼dÐz\over 2¹i},ââ
	¶^2(§) = 2i¶(z)¶(Ðz) $$
where $»­»/»z$ and $л­»/»Ðz$.
(The sign for $dz¼dÐz$ depends as usual on order of integration.)

The action for a real scalar is then
$$ S = Ç{dz¼dÐz\over 2¹i}L,ââL = üÄ(-»Ð»)Ä $$
For a chiral spinor we then use either of
$$ L = ÐÆлÆââ orââÐÆ»Æ $$
where the $i$ from the usual energy operator $-i»_0$ has been absorbed into the normalization of the fermions for later convenience.  
(Reality is funny anyway in Euclidean space: $»$ vs.¼$л$; see section XIB.)

In Euclidean position space, the propagator for a massless scalar is (see exercise
VIIB1.1)
$$ {1\over -üõ}2¹¶^2(x-x') = -ln[(x-x')^2] = -ln[(z-z')(Ðz-Ðz')] $$
 up to a real, dimensionful constant:  We use units $µ=1$.
The propagators for massless spinors are then
$$ {1\over л}2¹i¶(z-z')¶(Ðz-Ðz') = »¼ln[(z-z')(Ðz-Ðz')] = {1\over z-z'} $$
$$ {1\over »}2¹i¶(z-z')¶(Ðz-Ðz') = л¼ln[(z-z')(Ðz-Ðz')] = {1\over Ðz-Ðz'} $$
We then find
$$ »{1\over Ðz} = л{1\over z} = 2¹i¶(z)¶(Ðz) $$

The apparent inconsistency of this result is resolved by noting the $·$ prescription for the Euclidean spinor propagator:  If we regularize
$$ ln(zÐz)¼£¼ln(zÐz+·) $$
for any $·$ that is not 0 nor negative (i.e., is positive or complex), we find
$$ {1\over z}¼£¼{1\over z+·/Ðz} $$
and the wave equation for either propagator is satisfied, where the place of the $¶$ function is taken by
$$ 2¹¶^2(§)¼£¼{2·\over (§^2+·)^2} $$
whose normalization is easily checked.

\x VIIB5.1  Show (e.g., by an infinitesimal Wick rotation) that the correct
$i·$ prescription for the spinor propagator in Minkowski space is
$$ {-i\over (x-x')^à -i· ·(t-t')} = 
	Ï(t-t'){-i\over (x-x')^à -i·} +Ï(t'-t){-i\over (x-x')^à +i·} $$
 and that it satisfies the wave equation.  ($t-t'$ can be replaced with
$(x-x')^¦$ in the above.)

As a simple example we consider the 2D phenomenon of ``bosonization/fermioni\-za\-tion", that fermions and bosons can be converted into each other, even when they are free.  
First we examine bosonic currents created from fermions:
Taking the product of 2 such currents inside the functional integral (as usual, time ordering is implicit),
$$ J ­ iÐÆƼܼJ(z)J(z') ® -\left({1\over z-z'}\right)^2
	= »»'[-ln(|z-z'|^2)] ® ÷J(z)÷J(z'),â÷J ­ »Ä $$
where ``$®$" means we look only at the most singular terms as $z'£z$, from using all these fields to generate propagators.  From this we see that the fermionic current $J=iÐÆÆ$ (the $i$ is from Wick rotation) has the same ``propagator" as the bosonic current $÷J=»Ä$ (actually just the complete, chiral part of a boson).  Thus, by integration we can define a ``chiral boson" in terms of fermions.  

By being a little more tricky we can do the reverse, define the fermion in terms of a boson.  Unlike the previous procedure, this would seem to violate statistics, and has no classical analog.  We start by separating the scalar on shell into its
``chiral" and ``antichiral" parts (which were left- and right-propagating in Minkowski space):   
$$ Ä(z,Ðz) = Ä(z) +ÐÄ(Ðz) $$
 since $»Ð»Ä(z,Ðz)=0$.  (This can be accomplished
by differentiating with respect to $z$ or $Ðz$ and then integrating back.
We use ``$Ä$" to represent either the full boson or its chiral part, which should be unambiguous by context.)
The chiral boson has propagator (the chiral half of an ordinary boson's) $-ln(z-z')$.

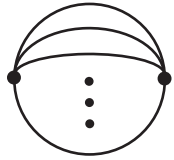
$$ \fig{bos} $$

The inverse relation is (quantum mechanically, not classically)
$$ ÐÆ = e^{-iÄ},ââÆ = e^{iÄ} $$
You might think $ÆÐÆ=1$, but this product is singular, as we would expect if we are to get the correct propagator.  First, we find
$$ Ä(z) e^{iÄ(z')} = ¼:[Ä(z) -i¼ln(z-z')]e^{iÄ(z')}: $$
where this expression is exact, and we have used an explicit normal-ordering symbol ``:¼:" to indicate we have already evaluated all propagator terms, even though we still have fields at different points.  Similarly,
$$ [Ä(z)]^n e^{iÄ(z')} = ¼:[Ä(z) -i¼ln(z-z')]^n e^{iÄ(z')}: $$
$$ Üâe^{-iÄ(z)} e^{iÄ(z')} = ¼:e^{-iÄ(z) -ln(z-z')} e^{iÄ(z')}:¼ 
	= {1\over z-z'}:e^{-iÄ(z) +iÄ(z')}: $$

Taylor expanding both $Ä(z')$ about $z$, and the exponential, we find in the short distance limit
$$ \lim_{z'£z}e^{-iÄ(z)}e^{iÄ(z')} = {1\over z-z'} -i»Ä(z) $$
The nonsingular, i.e., $Æ$-normal-ordered (not the same as $Ä$-normal-ordered), term agrees with the current defined above.  Note that, in terms of 2D quantum field theory, this calculation is a sum of diagrams to an arbitrary number of loops!  Thus $Æ$ and $Ä$ are unrelated classically, and their quantum relation is another example of duality:  If we stick in the $\hbar$ from the $Ä$ action as a factor in the $Ä$ propagator, the $Æ$ propagator looks like $(z-z')^{-\hbar_Ä}$, whose expansion reproduces the powers of $ln$, and which becomes 1 in the $Ä$-classical limit.

Although this gives the appearance of a scalar being the bound state of
spinors, and vice versa, even in the free theory, there is a simpler
interpretation, even classically:  Massless particles in D=2 travel at the
speed of light in one of two possible directions.  Thus, a collection of free
``left-(or right-)handed" massless particles travels along together, not
separating, and thus acting like a bound state.  (As shown in subsection
VC8, singularities in perturbative quantum field theory directly
correspond to configurations in classical mechanics.)

A related calculation is for the energy-momentum tensor of the fermions, which we'll apply in section XIB to conformal transformations:  By taking again the boso\-nized form of the point-split operator product above, and taking derivatives before the short-distance limit, we find
$$ \lim_{z'£z} ü [ ÐÆ(z)»'Æ(z') - »ÐÆ(z)Æ(z') ] = {1\over (z-z')^2} +ü(»Ä)^2 $$
relating the tensors for bosons and fermions (after $Æ$-normal-ordering away the singular term).

Bosonization extends to massive fermions:  The ``massive Thirring model"
$$ L = ÐÆ_¢(-i»_{\¢\¢})Æ_¢ +ÐÆ_{\¢}(-i»_{¢¢})Æ_{\¢} 
	+\f{m}{å2}(ÐÆ_¢Æ_{\¢} +ÐÆ_{\¢}Æ_¢) +gÐÆ_¢ÐÆ_{\¢}Æ_¢Æ_{\¢} $$
 (in Minkowski space) is equivalent to the ``sine-Gordon model"
$$ L = \f1{º^2}[\f14(»Ä)^2 +üµ^2(1-cos¼Ä )] $$
 with the above relation between the spinor and scalar fields, and
$$ {1\over º^2} = 1 +2g,ââ{µ^2\over º^2} ¾ m $$
 (Note in particular the free massive fermion for $º=1$.)  In this case the
bound states are dynamical.  Note that the relation is between strong
coupling in one theory and weak in the other (``duality").

Ü6. Massive propagators

Another way to distinguish infrared divergences is by introducing masses
(being careful not to break any invariances, or restoring them in the
massless limit).  For example, we again evaluate the one-loop propagator
correction, without numerator factors, but with different masses on the
internal propagators.  By the same steps as before, the Feynman
parameter integral is
$$ ö\A_2(p^2,m_1^2,m_2^2) = Çdk¼
	{1\over ü[(k+üp)^2 +m_1^2]ü[(k-üp)^2 +m_2^2]} $$
$$ = ý(2-\f{D}2) üÇ_{-1}^1 dº¼\B^{D/2-2},ââ
	\B = \f18 p^2(1-º^2) +\f14(m_1^2 +m_2^2) +\f14 º(m_1^2 -m_2^2) $$
 Now the $º$ integral is harder for all $D$, but the masses eliminate the
IR divergences (and the UV divergences are already explicit in the $ý$), so
we immediately expand about $D=4-2·$:
$$ ö\A_2 ®  ý(·) üÇ_{-1}^1 dº¼(1 -·¼ln¼\B) $$
 We then use integration by parts
$$ Ç_{-1}^1 dº¼ln¼\B = (º¼ln¼\B)|_{-1}^1-Ç_{-1}^1 dº¼º{d\over dº}ln¼\B $$
$$ \B = aº^2 +bº +c = a(º-º_+)(º-º_-),âº_à = 
	{-b àå{b^2 -4ac}\over 2a} = {m_1^2 -m_2^2 à2Â_{12}\over p^2} $$
$$ Üâº{d\over dº}ln¼\B = {º\over º-º_+} +{º\over º-º_-}
	= 2 +{º_+\over º-º_+} +{º_-\over º-º_-} $$
 in terms of $Â_{12}(s)$ of subsection IA4 for $s=-p^2$.  Note that in
Euclidean space
$$ 2Â_{12} = å{(p^2+m_1^2-m_2^2)^2+4m_2^2 p^2}
	= å{(p^2+m_2^2-m_1^2)^2+4m_1^2 p^2} ³ p^2 +|m_1^2 -m_2^2| $$
$$ Üâàº_à ³ 1 $$
 where the strict inequality holds for both masses nonvanishing.  The
integrals then take the simple form
$$ Ç_{-1}^1 dº¼\left( {º_+\over º-º_+} +{º_-\over º-º_-} \right)
	= º_+ ln \left( {º_+ -1\over º_+ +1} \right)
	+º_- ln \left( {º_- -1\over º_- +1} \right) $$
 Putting it all together,
$$ ö\A_2 = ý(1+·)\left[ {1\over ·} -ln(üm_1 m_2) +2
	+üº_+ ln \left( {º_+ -1\over º_+ +1} \right)
	+üº_- ln \left( {º_- -1\over º_- +1} \right)  \right] $$
 (We can cancel the $ý(1+·)$ by nonminimal subtraction.)  By analytic
continuation from Euclidean space, taking $p^2$ from positive to negative
along the real axis, we see there is no ambiguity at $p^2=0$ or
$-(m_1-m_2)^2$, and $ö\A_2$ remains real until we reach
$p^2=-(m_1+m_2)^2$, where it gets an imaginary part (whose sign is
determined by $(m_1+m_2)^2£(m_1+m_2)^2-i·$), corresponding to the
possibility of real 2-particle intermediate states.

\x VIIB6.1 Let's consider some special cases:
ªa Show for equal masses $m_1=m_2=m$ that this result simplifies to
$$ ö\A_2(p^2,m^2,m^2) = ý(1+·)\left[ {1\over ·} -ln(üm^2) +2
	+º¼ln \left( {º -1\over º +1} \right)  \right] $$
$$ º = å{p^2 +4m^2\over p^2} $$
ªb Consider the case with one internal particle massless, $m_1=m$,
$m_2=0$, and find
$$ ö\A_2(p^2,m^2,0) = ý(1+·)\left[ {1\over ·} -ln(üm^2) +2
	-{p^2+m^2\over p^2}ln \left( {p^2 +m^2\over m^2} \right)\right] $$
ªc Show both these results agree with the previously obtained massless
result in the limit $m£0$.  However, note that both these cases, unlike the
massless case, are IR convergent at $p^2=-(m_1+m_2)^2$.

\x VIIB6.2 Find the phase space for 2 massive particles, again using
the optical theorem (as in exercise VIIB4.3):
 ªa  The calculation is easier if
one takes the imaginary part before performing the Feynman parameter
integration:  Show the result is then 
$$ V_P = ¹ {1\over ý(\f{D}2 -1)}Ç_{º_+}^{º_-} dº¼(-\B )^{D/2-2} $$
 In particular, show from the explicit parameter integral expression for
the propagator that the only cut is at $-p^2³(m_1+m_2)^2$, as expected
from the optical theorem.  
 ªb  Make the change of variables
$$ Œ = {º-º_+\over º_--º_+} $$
 to find the result
$$ V_P = 2¹{ý(\f{D}2-1)\over ý(D-2)}(-üp^2)^{1-D/2}Â_{12}^{D-3} $$
 which simplifies in D=4 to
$$ V_P = 2¹{Â_{12}\over -üp^2} $$
 ªc  Show this result (in all D) agrees with the result of
the explicit phase space integral of subsection VC7, using the
expression for $Çd^{D-2}¯$ from subsection VIIA3.
(Hint:  Use the identity from exercise VIIA2.2b.)
 ªd Show the massless case agrees with exercise VIIB4.3.

In subsection VIIA3 we considered the application of the MOM
subtraction scheme to propagator corrections.  We assumed the
propagator corrections were Taylor expandable in the classical kinetic
operator.  From the above explicit expression for the 1-loop correction in
scalar theories, we see this is possible except near the branch point at
$p^2=-(m_1+m_2)^2$, i.e., when the external particle (whose propagator
we're correcting) has a mass equal to the sum of the internal ones.  To
analyze this more carefully, let's recalculate the propagator correction,
performing the Taylor expansion before evaluating the integrals.  We
consider the case with one vanishing mass, $m_1=m$, $m_2=0$, to
generate an IR divergence.  Assuming the external mass is also
$m$, we expand around the branch point in $p^2+m^2$.  The Feynman
parameter integral is then, to linear order in $p^2+m^2$, in terms of
$Œ=ü(1+º)$,
$$ ö\A_2(p^2,m^2,0) = ý(·)Ç_0^1 dŒ \left[ üm^2 Œ^2 
	\left( 1 +{1-Œ\over Œ}Ê{p^2+m^2\over m^2} \right)\right]^{-·} $$
$$ ® ý(1+·)(üm^2)^{-·}Ç_0^1 dŒ 
	\left[{1\over ·}Œ^{-2·} -(1-Œ)Œ^{-1-2·}{p^2+m^2\over m^2}\right] $$
$$ = ý(1+·)(üm^2)^{-·}\left[ {1\over ·}{1\over 1-2·}
	-\left({1\over -2·}-{1\over 1-2·}\right){p^2+m^2\over m^2}\right] $$
$$ ® ý(1+·)(üm^2)^{-·}\left[\left({1\over ·_{UV}} +2\right)
	+ü\left({1\over ·_{IR}} +2\right){p^2+m^2\over m^2}\right] $$
 where we have distinguished the UV divergence (in the $Â$ integral for
$·³0$) from the IR one (in the $Œ$ integral for $·²0$).  After including the
$(üµ^2)^·$ in the coupling, the $(üm^2)^{-·}$ converts each $1/·$ into a
$1/· -ln(m^2/µ^2)$.  (Of course, we can choose $µ=m$ for convenience.)
Note that this infrared divergence was a consequence of trying to Taylor
expand about a branch point due to a massless particle.

\x VIIB6.3 Do MOM subtraction for external mass $M=m_1+m_2$, with
ÓneitherÕ internal mass vanishing, and show there is no divergence other
than the UV divergence of the minimal scheme.

Later, we will encounter propagator corrections in gauge theories with
massive internal lines, and with various numerators.  Here, we examine
these purely from the point of view of the integrals.  First, consider
$$ \A_a = Çdk¼{k_a\over ü[(k-üp)^2+m_1^2]ü[(k+üp)^2+m_2^2]} $$
 Since $p$ is the only external momentum for a propagator, by Lorentz
invariance we have
$$ \A_a = p_a {1\over p^2} pÉ\A $$
 so it is sufficient to evaluate the integral of $pÉ\A$.  In analogy with the
earlier massless expression, we look at
$$ {1\over ü[(k-üp)^2+m_1^2]} -{1\over ü[(k+üp)^2+m_2^2]} =
   {kÉp +ü(m_2^2-m_1^2)\over ü[(k-üp)^2+m_1^2]ü[(k+üp)^2+m_2^2]} $$
 from which we find
$$ \A_a = p_a {m_1^2 -m_2^2\over 2p^2}
	[ö\A_2(p^2,m_1^2,m_2^2) -ö\A_2(0,m_1^2,m_2^2)] $$
 in terms of our result $ö\A_2$ above for the integral without numerator.

As a more complicated (but important) example, we examine
$$ \A_{ab} = Çdk¼{k_a k_b\over ü[(k+üp)^2+m^2]ü[(k-üp)^2+m^2]} $$
 Following our procedure of the previous example, we note
$$ Çdk¼{(pÉk)k\over ü[(k+üp)^2+m^2]ü[(k-üp)^2+m^2]} $$
$$ = Çdk¼{k\over ü[(k-üp)^2+m^2]} -{k\over ü[(k+üp)^2+m^2]} $$
$$ = Çdk¼{k+üp\over ü(k^2+m^2)} -{k-üp\over ü(k^2+m^2)}
	= p Çdk¼{1\over ü(k^2+m^2)} $$
 Thus transversality again determines the amplitude in terms of a scalar:
$$ \li{ ö\A_{ab} & = Çdk¼{k_a k_b\over ü[(k+üp)^2+m^2]ü[(k-üp)^2+m^2]}
	-{ú_{ab}\over ü(k^2+m^2)} \cr
	\noalign{\vskip.1in}
	& = (ú_{ab}p^2 -p_a p_b) ×\A(p^2,m^2) \cr} $$
 (This amplitude actually will be more useful than $\A_{ab}$.)  We also
have the identity
$$ Çdk¼{ü(k^2+m^2)+\f18 p^2\over ü[(k+üp)^2+m^2]ü[(k-üp)^2+m^2]} $$
$$ = üÇdk¼{1\over ü[(k-üp)^2+m^2]} +{1\over ü[(k+üp)^2+m^2]}
	= Çdk¼{1\over ü(k^2+m^2)} $$
 Taking the trace of the previous expression,
$$ (D-1)p^2 ×\A(p^2) = Çdk¼{k^2\over ü[(k+üp)^2+m^2]ü[(k-üp)^2+m^2]}
	-{D\over ü(k^2+m^2)} $$
$$ = -(\f14 p^2+m^2)Çdk¼{1\over ü[(k+üp)^2+m^2]ü[(k-üp)^2+m^2]}
	-(D-2) Çdk¼{1\over ü(k^2+m^2)} $$
$$ = -(\f14 p^2+m^2)ö\A_2(p^2,m^2,m^2) -(D-2)\A_1(0,m^2) $$
 in terms of the $Ä^3$ propagator and tadpole graphs evaluated earlier. 
This result can be reorganized if we make use of the $p=0$ case:
$$ 0 = -m^2ö\A_2(0,m^2,m^2) -(D-2)\A_1(0,m^2) $$
 (which also follows easily from the earlier explicit expression for
$ö\A_1(a,0,m^2)$).  We then find
$$ ×\A = -\f1{4(D-1)}ö\A_2(p^2,m^2,m^2)
	-\f1{D-1}m^2{ö\A_2(p^2,m^2,m^2) -ö\A_2(0,m^2,m^2)\over p^2} $$

\x VIIB6.4  Check that these results are consistent in the massless limit
with the expressions obtained in the previous subsection, by relating
the first two terms in $\A_2(x,p^2)$ for arbitrary D.

\x VIIB6.5  Calculate the one-loop propagator corrections for $ñ$ and $A$
in the 2D CP(n) model.

Ü7. Renormalization group

An interesting, useful, and simple application of the propagator correction
is to study the high-energy behavior of coupling constants.  For example,
we have seen that, by a change in normalization of gauge fields $A£A/g$,
gauge couplings can be moved from the covariant derivative to the kinetic
term:  $á=»+igA£»+iA$, $L_0=\f18 (»A+igAA)^2£\f1{8g^2}(»A+iAA)^2$. 
Thus, quantum corrections to gauge couplings can be found from just the
propagator (kinetic-operator) correction.  A simpler example is a scalar
field; a $Ä^4$ self-interaction has a dimensionless coupling in D=4, like
Yang-Mills.  However, unlike Yang-Mills, this model has no cubic coupling,
and thus no 1-loop propagator correction.  Furthermore, in Yang-Mills the
one-loop propagator correction contribution to the effective action gives
a multiloop contribution to the propagator itself, from the expansion of
$1/(K+A)$.  This corresponds to the graph consisting of a long string of
these corrections connected by free propagators.  There is a 1-loop
4-point correction in $Ä^4$ theory, and this graph resembles a propagator
correction, but with two external lines at each vertex instead of one. 
Such corrections can also be strung together, resembling the Yang-Mills
string, but with no free propagators inserted.  Since all the intermediate
states in this graph are 2-particle, it is 1PI, so the effect of this string is
not contained in just the 1-loop effective action, even though it is an
iteration of a 1-loop effect.  

$$ \fig{auxiliary} $$

This difficulty can be avoided by introducing the $Ä^4$ interaction
through an auxiliary field, just as it appears in supersymmetric theories
(see subsection IVC2):
$$ L = üÄ(-üõ +)Ä -\f1{2g}^2 $$
 where we have neglected the mass term since we will be concentrating
on the high-energy behavior.  Here the coupling is introduced through the
Óauxiliary-fieldÕ ``kinetic" term.  The diagrams just discussed now appear
through the 1-loop correction to the auxiliary-field propagator:  Since its
free propagator is just a constant, it can be contracted to a point in these
multiloop diagrams.  The definition of 1PI graphs has now changed, since
we can now cut auxiliary-field propagators, which would not exist in the
usual $Ä^4$ form of the action.  This modification of the effective
action simplifies the analysis of quantum corrections to the coupling, as
well as making it more analogous to gauge theories.  Note in particular
the change in interpretation already at the tree level:  We have used the
conventional normalization of 1/n!¼for factors of $Ä^n$ in the potential,
since canceling factors of n!¼arise upon functional differentiation. 
However, the result of eliminating $$ from the classical action produces
$\f18 Ä^4$ instead of $\f1{24}Ä^4$.  The reason is that in the diagrams
with $$ there are 3 graphs contributing to the 4-$Ä$-point tree,
corresponding to $$ propagators in the $s$, $t$, and $u$ ``channels". 
(See subsection VC4.)  Although this is a trivial distinction for the trees,
this is not the case for the loops, where the propagator string consists of
pairs of $Ä$ particles running in one of these three channels.  

The contribution to the 1-loop effective action for $$ is then given by the
above calculations, after including the factors of 1/2 for symmetries of
the internal and external lines, and the usual $-$1 for the effective action:
$$ L' = -\f14  \left[ ý(·)B(1-·,1-·)(-üõ)^{-·} \right]  $$
 where as usual $·=2-D/2$.  Expressing the Beta function in terms of the
Gamma function, and expanding as in previous subsections,
$$ ý(·)B(1-·,1-·)(-üõ)^{-·} ® \f1· +[ -© +2 -ln(-üõ)] $$
 Renormalizing away the constant pieces, we find for the classical action
plus this part of the 1-loop effective action
$$ L +L' = üÄ(-üõ +)Ä
	-ü \left[ \f1g -üln \left( -\fõ{µ^2} \right) \right] 
	= üÄ(-üõ +)Ä +\f14 ¼ln \left( -\fõ{M^2} \right)  $$
 where the renormalization group invariant mass scale $M$ is given by
$$ M^2 = µ^2 e^{2/g} $$
 Thus, the constant coupling $1/g$ has been replaced by an effective
``running coupling" $-üln(p^2/M^2)$, with energy dependence set by the
scale $M$.  (This is sometimes called the ``renormalization group", the
group being related to scale invariance, which is broken by the
introduction of the mass scale $M$.)

We saw the same dimensional transmutation occuring in the effective
potential in massless theories in subsection VIIB3.  The form is similar
because both are related to the appearance of the renormalization mass
scale $µ$ from the breaking of scale invariance by quantum corrections,
at either low or high energy:  In both cases dimensional transmutation
comes from a finite $ln¼µ^2$ term arising from the infinite
renormalization.  The difference is that in the effective potential case we
ignore higher derivatives, so the $µ^2$ must appear in a ratio to scalar
fields, while in the high energy case we look at just the propagator
correction, so it appears in the combination $µ^2/p^2$.  (More
complicated combinations will appear in more general amplitudes.)

\x VIIB7.1  Generalize this model to include internal symmetry:  Write an
analog to the scalar analog to QCD discussed in subsection VC9, where
the ``quark" $Ä$ now carries color and flavor indices, while the ``gluon"
$$ (classically auxiliary) carries just color.  Find $M$, especially its
dependence on the numbers $n$ of colors and $m$ of flavors.  Write the
same model with the gluons replaced by ``mesons" carrying just flavor
indices (so that classical elimination of the auxiliary fields yields the
same action), and repeat the calculation.  What are the different
approximation schemes relevant to the two approaches? 

Ü8. Overlapping divergences

$$ \fig{over} $$

We now perform some 2-loop renormalizations.  Our first example is part
of the propagator correction in $Ä^4$ theory.  By restricting ourselves to
the mass renormalization (coefficient of the mass term), we need
evaluate the graph only at vanishing external momentum.  (It is then
equivalent to a vacuum bubble in $Ä^3$ theory, or a tadpole graph in the
mixed theory.)  Furthermore, we consider the case where some of the
fields are massless.  In such a theory, we encounter (a special case of)
the 2-loop graph of subsection VIIB1, where 1 propagator is massless and
2 are massive.  Expanding in $·$, and keeping only the divergent terms
($1/·^2$ and $1/·$), we find (including a symmetry factor of 1/2 for the 2
massive scalar lines for real scalars)
$$ \T_2 = ü{ý(3-D)\over (üm^2)^{3-D}} B(2-\f{D}2,2-\f{D}2)B(1,\f{D}2-1) =
	-{(üm^2)^{1-2·}\over 2(1-·)(1-2·)}[ý(·)]^2 $$
$$ ® \f14 m^2[ý(·)]^2 [-1-3·+2·¼ln(üm^2)] $$

To this we need to add the counterterm graph, coming from inserting into
the 1-loop massive tadpole $\T_1$ (with 2 external lines) the counterterm
$ë_4$ (for renormalizing the $Ä^4$ term) from the 1-loop divergence in
the 4-point graph with 1 massive and 1 massless propagator.  (Since the
massless tadpole vanishes in dimensional regularization, we need not
consider the counterterm from the 4-point graph with 2 massive
propagators.)  From section VIIB6, we use the corresponding integral for
a 1-loop propagator correction $\A$, which is
$$ \A = ý(·) + finiteâÜâë_4 = - ý(·) $$
 We use a ``modified minimal subtraction", using the $ý(·)$ as the
subtraction instead of just the $1/·$ part of $ý(·)®1/·-©$.

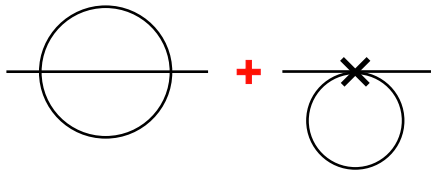
$$ \fig{overthecounter} $$

The 1-loop massive tadpole without coupling is
$$ \T_1 = {ý(1-\f{D}2)\over (üm^2)^{1-\f{D}2}} =
	-{(üm^2)^{1-·}\over 1-·}ý(·) ® -üm^2 ý(·) [1+·-·¼ln(üm^2)] $$

Combining these results, the divergent part of the 2-loop propagator
correction, with 1-loop coupling counterterm contributions included, is
$$ \T_2 +ë_4\T_1 = 
	\f14 m^2[ý(·)]^2 [-1-3·+2·¼ln(üm^2)+2+2·-2·¼ln(üm^2)] $$
$$ = [ý(·)]^2 (1-·) \f14 m^2 $$
 Thus, the $ln¼m^2$ divergences cancel, as expected.  (Divergences must
be polynomial in masses as well as couplings.)  The
surviving divergence is the superficial divergence, to be canceled by
the 2-loop mass counterterm.

\x VIIB8.1 Calculate the $p^2$ part of the 2-loop kinetic counterterm by
writing the above 2-loop propagator graph with nonvanishing external
momentum, introducing the Schwinger parameters, doing the
loop-momentum integration, taking the derivative with respect to $p^2$,
and then evaluating at $p=0$.  Why is there no subdivergence ($1/·^2$)?

\x VIIB8.2 Calculate the ÓcompleteÕ (all graphs, infinite and finite parts of
the) 2-loop propagator correction for ÓmasslessÕ $Ä^4$.  (See exercise
VIIB4.3a.)

For our next example we consider massless $Ä^3$ theory, and work in 6
dimensions, where the theory is renormalizable (instead of
superrenormalizable, as in 4 dimensions).  For the 2-loop propagator
correction, there are only two graphs (plus 1-loop graphs with 1-loop
counterterm insertions), one of which is simply a 1-loop propagator graph
inserted into another.  

$$ \fig{over2} 
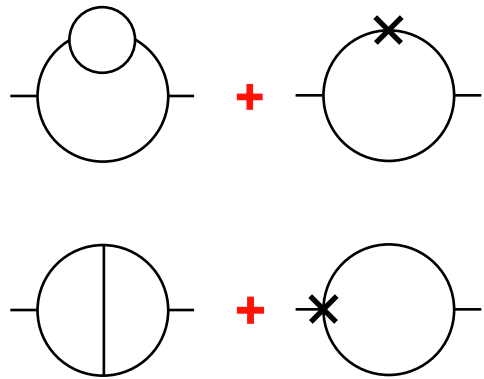
$$

$$ \fig{over3} $$

The other graph is
$$ \P =
  Çdk¼dq¼{1\over ü(k+q)^2ü(k+üp)^2ü(k-üp)^2ü(q+üp)^2ü(q-üp)^2} $$
 (with a symmetry factor of $ü$ for real scalars).  This graph can be
rewritten as iterated propagator corrections by use of integration by
parts in momentum space.  This is legalized by dimensional
regularization, since boundary terms vanish in low enough dimensions. 
All invariants can be expressed as linear combinations of the propagator
denominators (there are 5 of each, not counting the square of the
external momentum $p^2$), so any product of momentum times derivative
acting on the integrand will give terms killing one denominator and
squaring another, except for $p^2$ terms, which can be canceled by
appropriate choice of the momentum multiplying the derivative:
$$ Çdk¼dq¼{»\over »k}É
	{k+q\over ü(k+q)^2ü(k+üp)^2ü(k-üp)^2ü(q+üp)^2ü(q-üp)^2} =0 $$
 This operation effectively gives the factor
$$ {»\over »k}É(k+q) £ (D-4) 
	+{(q-üp)^2\over (k+üp)^2} +{(q+üp)^2\over (k-üp)^2}
	-{(k+q)^2\over (k+üp)^2} -{(k+q)^2\over (k-üp)^2} $$
 We thus have
$$ (\f{D}2-2)\P = \P_1 -\P_2 $$
$$ \P_1 = Çdk¼{1\over [ü(k+üp)^2]^2ü(k-üp)^2}
	Çdq¼{1\over ü(q+üp)^2ü(q-üp)^2} $$
$$ \P_2 = Çdk¼{1\over [ü(k+üp)^2]^2ü(k-üp)^2}
	Çdq¼{1\over ü(k+q)^2ü(q+üp)^2} $$
 The former term is the product of two 1-loop propagator graphs, the
latter is the insertion of one 1-loop propagator graph into another.

Both graphs can be evaluated by repeated application of the generalized
massless one-loop propagator correction (with arbitrary powers of free
propagators) given at the end of subsection VIIB4.  The result can be
expressed as
$$ \P_1 = -(D-3){(\P_0)^2\over üp^2} $$
$$ \P_2 = c\P_1,ââ
	c = {[ý(D-3)]^2 ý(5-D)\over [ý(3-\f{D}2)]^2 ý(\f{3D}2-5) ý(\f{D}2-1)} $$
 in terms of the 1-loop propagator correction $\P_0$.  We therefore
modify our minimal subtraction so that $\P_0$ has the simplest form (G
scheme): 
$$ \P_0 = -\f16 \f1· (üp^2)^{1-·} $$
 where $D=6-2·$, and we calculated the coefficient of the $1/·$ term and
threw in a normalization factor that canceled the rest:
$$ \h £ \N\h $$
$$ \N = {1\over 3(D-6)ý(2-\f{D}2)B(\f{D}2-1,\f{D}2-1)} =
	(1-\f23 ·)(1-2·){ý(1-2·)\over ý(1+·)[ý(1-·)]^2} $$
 Further evaluating $c$, we find
$$ c = -\f13 · {1-2·\over (1-\f32 ·)(1-3·)}
	{[ý(1-2·)]^2 ý(1+2·)\over [ý(1+·)]^2 ý(1-3·) ý(1-·)} $$
 Using the expansion of $ln¼ý(1-z)$ in terms of $©$ and $½(n)$, it is easily
checked that this combination of $ý$'s is $1+\O(·^3)$, so we can just drop
them.  Collecting our results, we have
$$ \P = {1-c\over \f{D}2-2}\P_1 = -\f1{36}\f1{·^2}
	{3-2·\over 1-·}\left[ 1 +\f13 · {1-2·\over (1-\f32 ·)(1-3·)} \right]
	(üp^2)^{1-2·} $$

\x VIIB8.3 Calculate the same graph in four dimensions.  It's finite there,
so no counterterms are necessary.  However, in this case integration by
parts gives a factor of $1/·$, and each of the two resulting graphs has an
additional factor of $1/·^2$.  The result then has a factor of 1 minus the
previously obtained combination of $ý$'s, which we already saw was of
order $·^3$.  The final result is thus obvious except for a factor of a
rational number:
$$ 6½(3) {1\over üp^2} $$
 (The on-shell infrared divergence is as expected from power counting.)

We next calculate the counterterm graphs.  These are the ones that
cancel the subdivergences coming from the 1-loop 3-point subgraphs.  We
therefore need the divergent part of this subgraph.  This is easy to
evaluate by our previous methods:  The result of Schwinger
parametrization, scaling, etc., doing all integration exactly except over
the Feynman parameters is
$$ Çdq¼{1\over üq^2 ü(q+k+üp)^2 ü(q+k-üp)^2} $$
$$ = Ç_0^1 d^3 Œ¼¶(1-݌) ý(3-\f{D}2)
	[üŒ_+(1-Œ_+)(k+üp)^2 +üŒ_-(1-Œ_-)(k-üp)^2]^{D/2-3} $$
$$ = ü \f1· + finite $$
 by simply replacing the factor in brackets by 1 (since it is raised to the
$-·$ power), where we have used
$$ Ç_0^1 d^3 Œ¼¶(1-݌) = Ç_0^1 dŒ_+ Ç_0^{1-Œ_+} dŒ_-  = ü $$
 Since we know the divergence is momentum-independent, we can obtain
the same result from a (infrared regularized) tadpole graph with its
propagator raised to the third power:  In the notation of subsection VIIB1,
$$ {1\over ý(3)}ö\A_1(3,0,m^2) = {ý(3-\f{D}2)\over ý(3)}(üm^2)^{D/2-3}
	= ü \f1· + finite $$
 The contribution of the 2 counterterm graphs (or one for the effective
action if we drop the symmetry factor) is thus
$$ 2ë_3\P_0 = 2(-ü \f1·)\P_0 $$

Collecting terms, we have
$$ \P +2ë_3\P_0 = -\f1{12}\f1{·^2}{1-\f23 ·\over 1-·}
	\left[ 1 +\f13 ·{1-2·\over (1-\f32 ·)(1-3·)} \right] (üp^2)^{1-2·}
	+\f16\f1{·^2} (üp^2)^{1-·} $$
 After a little algebra, dropping terms that vanish as $·£0$, we find
$$ \P +2ë_3\P_0 = (üp^2)[\f1{12}\f1{·^2} -\f1{18}\f1·
	-\f1{12}(ln¼üp^2)^2 +\f19 ln¼üp^2 -\f{23}{216}] $$

Note that modifying minimal subtraction is equivalent to redefining
$üµ^2$, which we have set to 1, but which appears only in the $ln$'s, as
$ln(üp^2)£ln(p^2/µ^2)$.  Thus, modifying $\N$, which appears only in the
combination $\N(üp^2)^{-·}$, is the same as shifting $ln¼üp^2$:
$$ \N £ \N e^{·a}âÜâln¼üp^2 £ ln¼üp^2 -a $$
 For example, choosing $a=-\f23$,
$$ \P +2ë_3\P_0 £ (üp^2)[\f1{12}\f1{·^2} -\f1{18}\f1·
	-\f1{12}(ln¼üp^2)^2 -\f5{72}] $$
 Only the $\O(·)$ part of the normalization factor affects the final result: 
More generally,
$$ (üp^2)^· £ \N (üp^2)^{-·}âÜâln¼üp^2 £ ln¼üp^2 -\f1· ln¼\N $$
 Since after adding counterterms, which cancel nonlocal divergences
arising from subdivergences, $ln$'s appear only in finite terms, only the
$\O(·)$ part of $ln¼\N$ will contribute.  Thus, we can approximate any
normalization factor as 
$$ \N ® e^{·a},ââa = © + rational $$
 as far as the renormalized results are concerned.  (The $©$ identifies this
normalization as modified minimal subtraction, as the $Ñ{\hbox{MS}}$ or G
schemes.)  This is just the statement of the renormalization group, that
the final result in minimal subtraction schemes depends only on the choice
of scale:  The complete normalization factor is really 
$$ \N_{total} = \N (üµ^2)^·âÜâln¼üµ^2 £ ln¼üµ^2 +\f1· ln¼\N $$

However, the higher-order terms can be convenient for intermediate
stages of the calculation.  In this particular case, the nonlocal
divergences appearing before cancellation are of the form $(1/·)ln¼p^2$,
so the $\O(·^2)$ part of $\N$ contributes at intermediate stages.  For
example, replacing the original $\N$ with
$$ \N' = (1-\f23 ·)(1-2·)ý(1-·) $$
 would have given the same result even before cancellation, since the
change is by another combination of $ý$'s that give $1+\O(·^3)$. 

\x VIIB8.4 Complete the 6D calculation of the exact 2-loop propagator
correction in $Ä^3$ theory, including the missing graph and counterterms,
to find the total renormalized 2-loop propagator and its 2-loop
counterterms.

\refs

£1 Goldstone, Salam, and Weinberg; Jona-Lasinio; Óloc. cit.Õ (VC):\\
	effective potential.
 £2 S. Coleman and E. Weinberg, \PRD 7 (1973) 1888.
 £3 A. D'Adda, M. L¬uscher, and P. Di Vecchia, \NP 146 (1978) 63;\\
	E. Witten, \NP 149 (1979) 285:\\
	quantum CP(n).
 £4 L.M. Brown and R.P. Feynman, ÓPhys. Rev.Õ É85 (1952) 231;\\
	G. Passarino and M. Veltman, \NP 160 (1979) 151;\\
	W.L. van Neerven and J.A.M. Vermaseren, \PL 137B (1984) 241:\\
	reduction of 1-loop integrals with numerators to scalar 1-loop
	integrals.
 £5 A.C.T. Wu, ÓMat. Fys. Medd. Dan. Vid. Selsk.Õ É33 (1961) 72:\\
	scalar box (1-loop 4-point) graph.
 £6 R.P. Feynman, unpublished;\\
	L.M. Brown, ÓNuo. Cim.Õ É22 (1961) 178;\\
	F.R. Halpern, \PR 10 (1963) 310;\\
	B. Petersson, ÓJ. Math. Phys.Õ É6 (1965) 1955;\\
	D.B. Melrose, ÓNuo. Cim. AÕ É40 (1965) 181:\\
	reduction of scalar 1-loop graphs to $²$D-point graphs.
 £7 G. 't Hooft and M. Veltman, \NP 153 (1979) 365;\\
	G.J. van Oldenborgh and J.A.M. Vermaseren, ÓZ. Phys. CÕ É46 (1990)
	425;\\
	A. Denner, U. Nierste, and R. Scharf, \NP 367 (1991) 637:\\
	further evaluation and simplification of 1-loop graphs.
 £8 K.G. Chetyrkin and F.V. Tkachov, \NP 192 (1981) 159:\\
	integration by parts in momentum space for exact evaluation of
	higher-loop massless propagators.
 £9 D.I. Kazakov, \PL 133B (1983) 406, ÓTheor. Math. Phys.Õ É58 (1984) 223,
	É62 (1985) 84;
	 D.I. Kazakov and A.V. Kotikov, ÓTheor. Math. Phys.Õ É73 (1988) 1264:\\
	more techniques for exact evaluation of such graphs.
 £10 Dyson, Óloc. cit.Õ (VC):\\
	treatment of lowest-order overlapping divergences.
 £11 Salam, Óloc. cit.Õ (VIIA):\\
	overlapping divergences to all loops.

\unrefs

Û4 C. RESUMMATION

So far we have only assumed that confinement arises nonperturbatively in
4D QCD.  However, to connect with known, successful results of
perturbation theory, we need to understand how the same methods used
to give these perturbative results can be generalized to include the
nonperturbative ones.  The simplest method would be to take the
perturbation expansion as is, and find a good method for evaluating (or
perhaps redefining) its sum, with the hope that summation to all orders
by itself would reveal features invisible at finite orders.

Besides the technical difficulties associated with such an approach, the
main problem is that the summation of the perturbation expansion does
not converge.  Parts of this problem can be solved by appropriate
redefinitions, but other parts indicate a serious problem with
perturbation theory, caused by the very renormalization that was
supposed to solve the main problem of finite-order perturbation theory
(infinities).

Ü1. Improved perturbation

We saw in the previous section that dimensional transmutation replaced
the dimensionless coupling constant with a mass scale.  In principle, we
would like to explicitly make this replacement as the basis of our
perturbation expansion, not only to make the perturbative parameter
physical, but also to take into account the running of the original
coupling.  Unfortunately, this is not possible in practice; however, we can
choose the arbitrary (unphysical) renormalization scale $µ$ to be in the
range of energies in the problem at hand, so that the $ln(p^2/µ^2)$
corrections are small.  A change in scale from one value of $µ$ to another
is related to a resummation of graphs:  Although the one-loop term in the
effective action containing $ln(p^2/µ^2)$ comes from a single
1PI amplitude, it contributes an infinite number of terms at different loop
orders to the propagator when inserted into any higher-loop 1PI graph, as
$1/(K+A)=1/K-(1/K)A(1/K)+...$ .  Although $K+A$ depends only on $M$, $K$
depends only on $g$ and $A$ depends only on $µ$.  Thus, any redefinition
of $µ$ that leaves the physical quantity $M$ unchanged requires a
corresponding redefinition of $g$:
$$ M^2 = µ^2 e^{-1/g^2}âÜâ
	g^2(µ^2) = {1\over ln\left({µ^2\over M^2}\right)} $$
 and thus changing $µ$ redistributes the contributions to $1/(K+A)$ (and
therefore to the summation of graphs in any amplitude) over the
different loop orders.  For example, if the amplitude is most sensitive to
the momentum in a particular propagator (independent of loop momenta),
and we choose $µ^2®p^2$, then although we can't use the resummed
perturbation expansion directly, we can at least push most of it into the
lower orders.

Things get more complicated at higher loops:  It becomes difficult to
associate the running of the coupling with the resummation of a
particular subset of all the graphs.  However, we already know that
this effect can be derived from the breaking of scale invariance by
renormalization.  For example, let's consider Yang-Mills theory, since
gauge invariance restricts it to have only a single coupling parameter. 
(This makes it the simplest case conceptually, although not
computationally.  Here we use only the fact that it has a single coupling;
its explicit renormalization constants won't be considered until 
chapter VIII.  As an alternative, we can consider the scalar QCD analog of
subsections VC9 and VIIB7, a $Ä^4$ theory with an auxiliary field, if we
ignore mass renormalization, or arbitrarily renormalize the mass to
zero.)  For convenience of dimensional analysis, we use only coupling
constants that are dimensionless in all dimensions, by scaling with an
appropriate power of $µ$.  (In general, we can do this even for masses.) 
The classical Yang-Mills action, before and after the addition of
counterterms, is then
$$ S_{class} = {1\over g^2(üµ^2)^·}Çdx¼tr¼\f18 F^2,ââ
	S_{class} +ëS = {1\over ög{}^2}Çdx¼tr¼\f18 F^2 $$
 where
$$ {1\over ög{}^2} = {1\over (üµ^2)^·}\left[ {1\over g^2} 
	+Ý_{n=1}^¥ {1\over ·^n}c_n(g^2) \right],ââ
	c_n(g^2) = Ý_{L=n}^¥ (g^2)^{L-1}c_{nL} $$
 for some numerical constants $c_{nL}$.  (We can also include $\h$'s as
$g^2£g^2\h$.)  We use $üµ^2$ to produce the combination
$(üp^2/üµ^2)^{-·}$ in graphs.  (In practice, one uses units $üµ^2=1$
until the end of the calculation, and restores units.)  The $µ$ dependence
is then given by varying $µ$ for fixed $ög$:
$$ µ^2{»\over »µ^2}g^2 ­ -·g^2 -º(g^2),ââµ^2{»\over »µ^2}ög^2 ­ 0 $$
 where the $·g^2$ term is the classical contribution, and $ög$'s
independence from $µ$ is the statement that the physics is independent
of the choice of $µ$ (i.e., $ög$ depends on only $M$ and $·$).  
$º$ is independent of $·$ (except indirectly through $g^2$):
By definition, $g$ is finite for all $D$ and $µ$, so $º$ has no 1/$·$ divergences; but also $º$ can have no positive powers of $·$, since that would create such contributions in the derivative of $ög$ that could not be canceled at any finite order in the loop expansion.
We then find
$$ 0 = µ^{2·}µ^2{»\over »µ^2}{1\over ög{}^2}âÜâ
	º = g^4{»\over »g^2}(g^2 c_1),ââ
	{»\over »g^2}(g^2 c_{n+1}) = -º{»\over »g^2}c_n $$
 Thus, the coefficients of the $1/·$ terms determine those of both the
higher order terms and $º$.

This gives us an expression for $º$,
$$ º = Ý_{L=1}^¥ (g^2)^{L+1}º_L,ââº_L = Lc_{1L} $$
 Since $g^2$ is itself unphysical, the information we can get from
analyzing the running of this coupling is arbitrary up to redefinitions.  For
example, assume that all $º_L$ are nonvanishing, and write the definition
of $º$ as (in $D=4£·=0$)
$$ µ^2{»\over »µ^2}{1\over g^2} = f = {1\over g^4}º 
	= Ý_{L=0}^¥ (g^2)^L º_{L+1} $$
 Then under a redefinition $g^2£g^2(g'^2)$ we have
$$ µ^2{»\over »µ^2}{1\over g'^2} = 
	\left({»(1/g^2)\over »(1/g'^2)}\right)^{-1}f(g^2(g'^2)) ­ f'(g'^2) $$
 Now we consider a ``perturbative" type of redefinition, as results from
changing renormalization prescriptions, so $g^2$ gets only ``$\O(\h)$"
corrections: Taylor expanding
$$ g^2 = g'^2 +k_1 g'^4 + +k_2 g'^6 +\O(g'^8) $$
$$ Üâ{1\over g^2} = {1\over g'^2} + constant + \O(g'^2) $$
 we find
$$ {»(1/g^2)\over »(1/g'^2)} = 1 +\O(g'^4),ââ
	f(g^2(g'^2)) = f(g'^2)  +\O(g'^4) $$
$$ Üâf'(g'^2) = f(g'^2)  +\O(g'^4) $$
 Thus, the first two coefficients of $º$ ($º_1$ and $º_2$) are unaffected,
while terms found at 3 loops and beyond can be modified arbitrarily, and
even be set to vanish.  In the more general case of more than 1 coupling,
it is sometimes possible to eliminate also some of the 2-loop
contributions.

Therefore, to consider the general behavior of the coupling as a function
of energy ($µ^2$), it is sufficient to solve the equation
$$ µ^2{»\over »µ^2}g^2 = -º_1 g^4 -º_2 g^6 $$
 (using, e.g., the change of variables $t=ln¼µ^2$ and $u=1/º_1 g^2$) as
$$ {µ^2\over M^2} = e^{1/º_1 g^2}
	 \left({1\over g^2} +{º_2\over º_1}\right)^{-º_2/º_1^2} $$
 with $M^2$ as the constant of integration.  Using an allowed type of
redefinition for $g^2$, and also redefining the arbitrary constant of
integration $M^2$, we can simplify this to
$$ {1\over g^2} £ {1\over g^2} -{º_2\over º_1},ââ
	M^2 £ M^2 e^{º_2/º_1^2} $$
$$ Ü⺠= {º_1 g^4\over 1 -{º_2\over º_1}g^2},ââ
	{µ^2\over M^2} = e^{1/º_1 g^2}(g^2)^{º_2/º_1^2} $$
 (This redefinition changes the range of what $g^2$ is called negative and
what positive.  However, $g^2$ is just a parameter, not a physical
coupling:  As far as the unitarity of the kinetic term is concerned, only the
residues near the poles of the propagator are relevant.  Also, our allowed
class of redefinitions does not affect behavior for small $g^2$, and thus
perturbation theory.)

\x VIIC1.1  Let's analyze this solution in more detail:
ªa Graph the function $y(x)=e^{ax}x^b$ (or graph $ln¼y$ to make it
simpler) for $a$ and $b$ positive, negative, and vanishing, to study the
behavior of the function $µ^2(g^2)$.  The analysis can be simpliifed (and
the behavior for different values of $a$ and $b$ related) by considering
$g^2$ positive and negative, and the symmetries
$$ a £ -a,ââx £ -x,âây £ (-1)^b y $$
$$ b £ -b,ââx £ -x,âây £ (-1)^b{1\over y} $$
 Note that $g^2$ can be nonpositive for some values of $µ^2$:  For
example, even for $º_2=0$, we have $g^2=1/º_1 ln(µ^2/M^2)$, which is
negative for $µ<M$ or for $µ>M$.  What happens for $º_2±0$?
ªb After applying the above redefinition, apply the second redefinition
$$ {1\over g} £ {1\over g} +{º_2 \over º_1}g $$
 Find the new $º$ and $µ^2(g^2)$.  Compare to the behavior of $µ^2(g^2)$
before this redefinition, for the cases $º_2/º_1<0$, noting the ``duality"
symmetry $gª(-º_1/º_2)/g$.

\x VIIC1.2  Consider some theory with a single dimensionless coupling
$g^2$, but now also a single mass $m$.  By the above methods we find
$$ {µ^2 »\over »µ^2}m^2 = m^2 [-1 -º_m(g^2)] $$
 (The $m$ dependence follows from dimensional analysis.)  Solve for
$m^2$ as a function of $g^2$, as an integral over $g^2$ in terms of $º$
and $º_m$.  Show that after an appropriate redefinition
$$ {µ^2 »\over »µ^2}m^2 = m^2 (-1 -º_{m1}g^2) $$
 for some constant $º_{m1}$.  Solve for $m^2$ explicitly in terms of
$g^2$, when we have also redefined $º$ to $º_1 g^4 +º_2 g^6$.  Then
make the final redefinition $1/g^2£1/g^2-º_2/º_1$ used to simplify
$M^2$.

For purposes of perturbation theory, it is useful to invert this:  For small
$g^2$, we have approximately
$$ {1\over g^2} ® º_1 ln {µ^2\over M^2}
	+{º_2\over º_1}ln\left( º_1 ln {µ^2\over M^2}\right) $$
 This implies that the terms in the effective action that carry the $M$
dependence are given by
$$ ý_M ® trÇdx¼\f18 F \left[ º_1 ln {-õ\over M^2}
	 +{º_2\over º_1}ln\left( º_1 ln {-õ\over M^2}\right)\right] F $$
 (We can also replace $-õ£õ$ in this limit, ignoring $i¹$'s in comparison to
$ln$'s.)

The general class of coupling redefinitions we considered are allowed by
perturbation theory:  If we knew the exact solution to a field theory, we
would be more restrictive, requiring invertibility.  However, in
perturbation theory, given two renormalization prescriptions related by
some such coupling redefinition, we might know this redefinition only
perturbatively, and perhaps only to a few orders.  Even if we knew it
exactly, and knew it to be noninvertible, it still might not be clear which
of the two prescriptions were the correct one, if either.  Therefore, the
renormalization group alone is sufficient to draw conclusions about the
behavior of a theory only at ``small" ($ø1$) coupling.

Similar remarks apply to propagators, S-matrix elements, etc.  Consider
any renormalized function $G_n$ appearing as the coefficient of $n$ fields in a term in
the effective action.  
The renormalization of the unrenormalized $öG_n$ is taken care of by the combination of the use of $ög$ for the coupling and wave-function renormalization factors $\Z$:
$$ G_n(g^2,µ^2) = \Z^{-n}(ög^2(üµ^2)^{-·},·)öG_n(ög^2,·),ââ
	{µ^2 d\over dµ^2}öG = 0 $$
$$ Üâ\left( {µ^2 »\over »µ^2} +º{»\over »g^2} +n© \right) G = 0,ââ
	© = {µ^2 »\over »µ^2}ln¼\Z $$
 ($\Z$ is not required for
pure Yang-Mills in the background gauge; or we can examine ratios of
such quantities where the $\Z$'s cancel, which are more physical, such as
S-matrix elements.)  

Unfortunately, this behavior of the theory at high energy leads to problems upon resummation of the perturbation expansion.  The basic idea of dimensional transmutation is that the effective action will depend on $g$ and $µ$ only through $M$.  For example, in an asymptotically free theory at high energies, propagators (which are where this effect shows up) will depend on momentum as a function of only $p^2/M^2$.  More specifically, if we restrict ourselves to just the 1-loop contribution to the $º$ function for simplicity, which will appear in the effective action as a logarithmic correction, propagators will depend on
$$ º_1 ln\left({p^2\over M^2}\right) = 
	{1\over g^2} + º_1 ln\left({p^2\over µ^2}\right) $$
(In fact, at one loop in pure Yang-Mills, this is the exact modification of the kinetic term.)  We now consider analytic continuation of the propagators in this argument:  If we continue in $p^2$, we know we must find the usual cuts from multiparticle intermediate states, at negative $p^2$, extending to $p^2=-¥$.  But since the propagators depend on $g^2$ and $p^2$ only through this combination, we must find the same result if we keep $p^2$ fixed and analytically continue instead in $g^2$:  The cuts in $g^2$ are thus located at
$$ {1\over g^2} = real +º_1(2n+1)¹i $$
for arbitrary integer $n$, where ``$real$" means starting at some real value and running to $+¥$.  If we plot these in the complex $g^2$ plane, we can recognize this equation as describing circular arcs running through the origin, with centers on the imaginary axis:
$$ |z-ir|^2 = r^2âÛâ{1\over z} - {1\over z*} = {1\over ir} $$
These arcs approach the origin from the positive side (but from either the upper or lower plane), with radii $1/º_1(2n+1)2¹$.  Normally one would like a small region of analyticity about the origin for the perturbation expansion to converge (the nearest singularity giving the radius of convergence).  Barring that, a wedge of small angle about the real axis will do (for a ``Borel sum": see subsection VIIC3).  In this case, any of these arcs prevent even that.  In the following subsections we will examine and interpret the causes and effects of this behavior.

Ü2. Renormalons

The perturbation expansion in general can't be resummed in the naive
way because the number of diagrams increases as $n!(constant)^n$ at
$n$ loops.  The simplest example of this is a self-interacting scalar in D=0:
$$ Z = Ç_{-¥}^¥ {dÄ\over å{2¹}}e^{-üÄ^2-\f14 g^2 Ä^4} = 
	Ý_{n=0}^¥ g^{2n} Z_n $$
 Note that $Z=1$ for $g^2=0$, and $Z<1$ for $g^2>0$, but $Z=¥$ for $g^2<0$. That's why the perturbation expansion doesn't converge. It also suggests that for any fixed $g^2$ the expansion will start to diverge when $n$ is of the order of $1/g^2$.
 
Since there is no momentum integration, each diagram is just 1 (times
some permutation factors), so $Z_n$ just counts the number of diagrams
at $n$ loops.  We use $g^2$ so the coupling is similar to that in
Yang-Mills:  As usual, we can rescale $Ä£Ä/g$ to recognize $g^2$ as
$\h$:
$$ Ä' = gÄâÜâüÄ^2 +\f14 g^2 Ä^4 = \f1{g^2}(üÄ'^2 +\f14 Ä'^4) $$
 (Of course, we can be more explicit by writing $\h g^2$ in place of just
$g^2$ or $\h$, but the effect is identical, since they both appear only in
that combination.)  This integral can be evaluated exactly at any order of
perturbation theory:
$$ Z_n  = Ç_{-¥}^¥ {dÄ\over å{2¹}}\f1{n!}(-\f14 Ä^4)^n e^{-üÄ^2}
	=  \f1{n!}(-1)^n \f1{å¹}ý(2n+ü) ® {1\over å2¹}(n-1)!(-4)^n $$
 where we have used the Stirling approximation for $ý(z)$ at large $z$.

\x VIIC2.1  Find the following properties of the $ý$ function for large
argument:
ªa Derive the Stirling approximation 
$$ \mathop{lim}_{z£¥}ý(z) ® å{2¹\over z}\left({z\over e}\right)^z $$
 by applying the method of steepest descent to the integral definition of\\
$ý(z+1)$.  (See subsections VA2 and VA5.)  
ªb Use this approximation, and $\mathop{lim}_{z£¥}(1+\f1z)^z=e$, to
show 
$$ \mathop{lim}_{z£¥}ý(az+b) ® å{2¹}(az)^{az+b-1/2}e^{-az} $$

Thus we might as well apply the steepest descent approximation directly
to the original integral:  Using also an integral for $\h$ ($=g^2$ in this
case),
$$ Z_n = È{d\h\over 2¹i \h^{n+1}}
	ÇD \left( {Ä\over å\h} \right) e^{-S/\h} $$
 we first apply steepest descent to the $Ä$ integral, yielding the usual
first two terms in the JWKB expansion.  Then the $\h$ integral can be
approximated as $ý(n)$ by keeping only the part of the contour on the
positive real axis:
$$ \left. È{d\h\over 2¹i \h^{n+1}} e^{-S/\h}\right|_{¶S/¶Ä=0} ® 
	\left. {1\over 2¹i}ý(n) \left( {1\over S} \right)^n \right|_{¶S/¶Ä=0} $$
$$ ÜâZ_n ® Ý_{\astop{¶S/¶Ä=0}{S±0}}{1\over 2¹i} 
	\left[ det \left( {¶^2 S\over ¶Ä^2} 
	\right) \right]^{-1/2} (n-1)! \left( {1\over S} \right)^n $$
 ($S=0$ solutions contribute only to $Z_0$.  A similar result can be
obtained by simultaneously using steepest descent for the $\h$ integral,
yielding a ``classical value" of $\h$ in terms of $S$.)  
This approximation is the same poor approximation as that used for the perturbation expansion: ignoring negative $g^2$. In fact $e^{-S/\h}$ is not Taylor expandable, and the coefficients in its Taylor expansion (from doing the integral exactly) all vanish!
In the present case,
the nontrivial classical solutions are
$$ S = üÄ^2 +\f14 Ä^4âÜâÄ = ài $$
 which gives the same $Z_n$ as previously (being careful to sum the two
terms for the two solutions).  Thus, we see that in general we have to
sum $Ý_{n=0}^¥ n!(\h/S)^n$, which does not converge.  Furthermore, this
divergence is associated with finite-action (``instanton") solutions to the
classical equations of motion.

The simplest example of a resummation problem is the one-loop
propagator correction.  We have seen that the classical and one-loop
kinetic terms can be combined to give a kinetic operator of the
form $º_1 K(p^2) ln(-p^2/M^2)$ in massless theories, or at high energy in
massive theories, where $K$ is the classical kinetic operator.  The free (or
asymptotic) theory has solutions where this kinetic operator has a zero
(the propagator blows up).  Besides the classical solution at
$K(p^2)=0$, there is another at $p^2=-M^2$:
$$ {1\over º_1 K(p^2)¼ln\left(-{p^2\over M^2}\right)} 
	= {1\over º_1 K(p^2)¼ln\left(1-{p^2+M^2\over M^2}\right)} 
	® -{M^2\over º_1 K(-M^2)}ð{1\over p^2+M^2} $$
 This might be expected to be a bound state, called a ``renormalon"
because of its relation to the renormalization group.  However, the
residue of this pole in the propagator can have the wrong sign, indicating
the appearance of a ghost (``Landau ghost"), and thus a violation of
unitarity. 

\x VIIC2.2 The Landau ghost itself is not necessarily a problem in quantum
field theory, although it indicates the possibility of such problems. 
Examine the behavior of this ghost after taking into account the 2-loop
correction ($º_2$), before and after the simplifying redefinition of the
previous subsection, for all the various signs of $º_1$ and $º_2$.  Since
the expression for $µ^2(g^2)$ can't be inverted, use the fact that the
propagator follows from the coupling $g^2(µ^2)$ as
$$ ë ¾ {g^2(p^2)\over p^2} $$
 (Field redefinitions can't remove the momentum dependence of
couplings.)  Then new poles (or other singularities) in the propagator
correspond to the limit $g^2£¥$, so find $p^2(g^2)$ there.

$$ \fig{renormala} $$

This causes problems similar to those from instantons when the quantum
propagator is inserted into another graph.  We set external momenta to
vanish, as an approximation for high energy for the loop momenta, or to
evaluate low-energy quantities such as anomalous magnetic moments. 
In any one-loop 1PI graph with $n$ 1-loop propagator insertions and $l$
external lines, we get an integral at high energy of the form (e.g., in QCD
or the scalar analog with auxiliaries of subsections VC9 and VIIB7)
$$ Çd^4 k¼(k^2)^{-l} \left[ -º_1¼ln \left( {k^2\over µ^2} \right) \right]^n
	¾ (-º_1)^n Ç_0^¥ du¼e^{-(l-2)u}u^n 
	¾ n! \left( -{º_1\over l-2} \right)^n $$
 where we changed variables to $u=ln(k^2/µ^2)$ (remembering
$ý(n+1)=n!$).  We have used effectively an infrared cutoff by
approximating the $u$-integral from 0 to $¥$ instead of $-¥$ to $¥$.  If
we look instead at the low-energy (of the loop momentum) behavior, now
taking $l-1$ massive classical propagators with 1 massless propagator
(to insure IR convergence) with $n$ insertions, we find
$$ Çd^4 k¼(k^2+m^2)^{-(l-1)} (k^2)^{-1} 
	\left[ -º_1¼ln \left( {k^2\over µ^2} \right) \right]^n
	¾ º_1^nÇ_{-¥}^0 du¼e^u (-u)^n = n!º_1^n $$
 Since the former comes from UV behavior it's called a ``UV renormalon",
while the latter coming from IR behavior is called an ``IR renormalon". 
The essential difference is the relative factor of $(-1)^n$.  In fact, the
former expression is also the high-energy limit of the latter (neglecting
masses then), so the complete integral ($u$ from $-¥$ to $¥$, so $k^2$
from 0 to $¥$) can be approximated as the sum of the UV renormalon and
IR renormalon contributions.

Ü3. Borel

Since renormalons and instantons cause the perturbation expansion to
diverge by a factor of $n!$, we look for a method to formally sum such
series, by relating them to series that do converge.  In general, we
consider the series
$$ A(\h) = Ý_{n=0}^¥ \h^n a_n $$
 and define the ``Borel transform" as:
$$ ÷A(z) = Ç_{r-i¥}^{r+i¥} {d(1/\h)\over 2¹i} e^{z/\h}A(\h) $$
 (for some real number $r$ to the right of all singularities of $A$) in
anticipation of instanton-like contributions.  The inverse is
$$ A(\h) = Ç_0^¥ dz¼e^{-z/\h} ÷A(z) $$
 The inverse Borel transform is related to the Laplace transform (with the
variable change $x=1/\h$) and the Mellin transform ($x=1/\h$ and
$y=e^z$).  Evaluating explicitly for the above series,
$$ ÷A(z) = ¶(z)a_0 +Ý_{n=0}^¥ z^n \f1{n!} a_{n+1} $$
 So the Borel-transformed sum converges faster by a factor of $n!$,
which is just what we need for perturbation theory.  The idea for
resumming the perturbation expansion is to first do the Borel sum, then
inverse Borel transform the resulting function.  Of course, this procedure
does not necessarily fix the original problem, which might merely be
translated into problems of convergence or ambiguity for integration of
the inverse transform.  In particular, we need $÷A(z)$ to be well defined
along the positive real axis.

We saw that generically the sums involved were approximately of the
form
$$ A(\h) ¾ Ý_{n=1}^¥ \h^n (n-1)!(-k)^n $$
 In that case
$$ ÷A(z) ¾ Ý_{n=0}^¥ (-1)^n z^n k^{n+1} = {1\over z +\f1k} $$
 When $k<0$, this leads to a singularity in the integral defining the inverse
Borel transform.  It can be ``regularized" by choosing a contour that goes
around the pole, but the choice of contour is ambiguous, and choosing an
arbitrary linear combination of the two contours introduces a free
parameter.  Explicitly, we have
$$ A(\h) = A_0(\h) +½e^{-1/|k|\h} $$
 where $A_0$ is the result of a particular prescription (e.g., principal
value), and $½$ is the new parameter.  The $½$ term is clearly
nonperturbative, since each term in its Taylor expansion in $\h$ vanishes.
This new parameter can be interpreted as a new (nonperturbative)
coupling constant in the theory, just like ambiguities in renormalization
of new counterterms in perturbatively nonrenormalizable theories.

Now we more carefully analyze the explicit sums we found in the
previous subsection.  The first example is $å\h Z$ for D=0:
$$ ÷A(z) = Ç_{r-i¥}^{r+i¥} {d(1/\h)\over 2¹i} e^{z/\h}
	Ç_{-¥}^¥{dÄ\over å{2¹}}e^{-S/\h} 
	= Ç_{-¥}^¥{dÄ\over å{2¹}}¶(z-S) 
	= {1\over å{2¹}}Ý_{S=z} (S')^{-1} $$
 (The contribution from $S=0$ is artificial, coming from our using
$A=å{\h}Z$ instead of $Z$.)  So, this integral can be explicitly evaluated. 
(For example, for the action we used in the previous subsection, 
we can explicitly solve for
$Ä$ at $S=z$.)  However, there is then a problem in inverting the Borel
transform:  Near $z=z_0­S(Ä_0)$ for classical solutions $Ä_0$, we have
$$ S(Ä) ® S(Ä_0) +üS''(Ä_0)(Ä-Ä_0)^2¼Ü¼
	S'(Ä) ® S''(Ä_0)(Ä-Ä_0),¼z -z_0 ® üS''(Ä_0)(Ä-Ä_0)^2 $$
$$ Üâ(S')^{-1} ® [2S''(Ä_0)]^{-1/2}(z-z_0)^{-1/2} $$
 Therefore, there are cuts with branchpoints at classical values of the
action, leading to ambiguities in the result for $A(\h)$.  We thus see that
new coupling constants are introduced for each solution to the classical
field equation with positive action.  (For our D=0 example $S<0$, and there
is no problem, but more realistic examples, like Yang-Mills instantons,
have $S>0$.)

\x VIIC3.1  Consider the D=0 action
$$ S = üÄ^2 -\f14 Ä^4 $$
 which differs from our previous example by the sign of the interaction.
Now we have classical solutions with $S>0$.  (The interaction is the
wrong sign for the integral to be well defined, but the ``kinetic" term
is the right sign for it to be defined perturbatively.)  Explicitly evaluate
$Ä(S)$ (i.e., $Ä(z)$), and show it has the above behavior near
$z=S(Ä_0)$.

In the case of renormalons, we see from the previous subsection that the
large-$n$ behavior gives singularities at $z=N/º_1$ for positive integer
$N$.  This integral also is easier to evaluate after Borel transforming:  We
consider a one-loop graph, but replace one internal line with the ``full"
quantum propagator coming from the 1-loop effective action (the same
as summing a string of 1-loop propagator insertions), while using massive
propagators for the remaining lines.  We thus examine first the transform
of the quantum propagator
$$ Ç{d(1/\h)\over 2¹i} e^{z/\h}{1\over k^2}
		{1 \over \f1\h +º_1¼ln(k^2/µ^2)}
	= {1\over k^2} \left( {k^2 \over µ^2} \right)^{-º_1 z} $$
 by closing the contour on the left.  Then inserting this transformed
propagator into the complete diagram,
$$ Çd^4 k {1\over (k^2 +m^2)^{l-1}}{1\over k^2}
	 \left( {k^2 \over µ^2} \right)^{-º_1 z}
	¾ \left( {m^2\over µ^2} \right)^{-º_1 z} ý(1-º_1 z) ý(l-2+º_1 z) $$
 using the integrals of subsection VIIB1.  This expression is the sum
over $n$ of the UV/IR renormalon example at the end of the previous
subsection, except that we have done the summation over $n$ as the first
step (and used the Borel transform to assist in the evaluation).  The first
$ý$ has poles at $z=N/º_1$ for positive $N$, representing the IR
renormalon, which are relevant for $º_1>0$, but the second $ý$ has poles
at $z=-(N+l-3)/º_1$ for positive $N$ (and $l³3$ for the original diagram to
be UV convergent), representing the UV renormalon, which are relevant
for $º_1<0$.  To the one-loop approximation for the $º$-function we have
used, the singularities are just poles, but if the two-loop propagator
insertions are used, these singularities become the branchpoints for cuts.

The new coupling constants that appear nonperturbatively can be given a
physical interpretation in terms of vacuum values of polynomials of the
fields.  The basic idea is analogous to perturbative tadpoles:  In that case
corrections to S-matrices due to vacuum expectation values of scalar
fields can be expressed by propagators that end at a ``one-point vertex",
whose coefficient is the vacuum value of the field:
$$ ÒÄ(x)Ô = ÇDļe^{-iS}Ä(x) = c $$
 in position space for some constant $c$, or in momentum space as
$$ ÇDļe^{-iS}Ä(p) = c¶(p) $$
 Similarly, we could expect graphs to have two propagators that end at a
two-point vertex representing the vacuum value of the product of the
two fields associated with the ends of the two propagators, and so on for
higher-point vertices.  For example, for a $Ä^2$ vertex in a scalar theory,
it would correspond to a contribution of the form
$$ ÇDļe^{-iS}Ä(x)Ä(y) = (c^2 +c') + ... $$
 in position space, or in momentum space
$$ ÇDļe^{-iS}Ä(p)Ä(q) = (c^2 +c')¶(p)¶(q) + ... $$
 where $c^2$ is the contribution from $ÒÄÔ^2$, so $c'$ represents
$(ëÄ)^2=ÒÄ^2Ô-ÒÄÔ^2$.  Such vacuum values do not appear in perturbation
theory for higher than one-point; we get only one $¶(p)$ for each
connected part of any graph.  However, such contributions would be
expected to give similar contributions to those we have found for
renormalons:  By dimensional transmutation, a contribution to an
amplitude of the form $½e^{-n/º_1\h}$ must appear in the combination 
$$ ½e^{-n/º_1\h} £ ½e^{-n/º_1\h} \left( {µ^2 \over p^2} \right)^n 
	= ½ \left( {M^2 \over p^2} \right)^n $$
 This is the type of contribution expected from a propagator with tadpole
insertions, or in the same way from any other type of vacuum value.  In
particular, in QCD there are no fundamental scalar fields, but only scalar
fields can get vacuum values, by Lorentz invariance.  Thus, the vacuum
values come from composite scalars, like $tr(F^2)$, $Ðqq$, etc.

Note that renormalons are a feature of renormalizable theories:  They do
not appear in superrenormalizable or finite theories.  In particular, the
path-integral methods of ``constructive quantum field theory" have been
used to show that certain interacting field theories in lower dimensions
can be proven rigorously to exist --- superrenormalizable theories with
unique vacuua.

Ü4. 1/N expansion

Perturbation theory is insufficient to evaluate all quantities in quantum
physics, since   
\item{(1)} such expansions don't always converge;   
\item{(2)} if they do
converge, they might not converge to the complete result; and   
\item{(3)} even if
they do give the complete answer, their summation might not be
practical.

There are many perturbative expansions in quantum field theory.  When
we say ``perturbation theory" in this context, we generally mean an
expansion in the number of fields (or, in diagrammatic terms, number of
vertices), since in the path integral we kept the exact quadratic part of
the action but expanded in powers of the interaction terms (cubic and
higher).  (This is usually also an expansion in the coupling constants,
depending on how we define the fields, which can be redefined by factors
of the couplings.)  One disadvantage of this expansion is that it violates
manifest gauge invariance:  Nonabelian gauge transformations are
nonlinear in the fields, and thus mix diagrams with different numbers of
fields.  (These are the internal fields; external fields are asymptotic, and
approximated as free.)  Graphs that are related by gauge transformations
must be added together to obtain gauge-invariant, and thus physically
meaningful, expressions.  Also, in practice individual graphs contain
``gauge artifacts" that complicate them in certain gauges, but cancel in
gauge-invariant objects, like S-matrix elements.

There can be a large number of graphs contributing to a particular
physical process (given set of external states) at any particular loop
order.  There is another gauge-invariant expansion that can be applied to
Yang-Mills theory to subdivide these sets of graphs, based on the
freedom of choice of the Yang-Mills group itself:  We have seen that the
classical groups are defined in terms of N$ð$N matrices, where N is
arbitrary.  Clearly, S-matrix elements must depend on N, even if the
external states are restricted to be group singlets or representations of
an N-independent subgroup, since the number of internal states increases
as some polynomial in N.  We now examine how this can be used to define
a perturbation expansion in terms of N.  

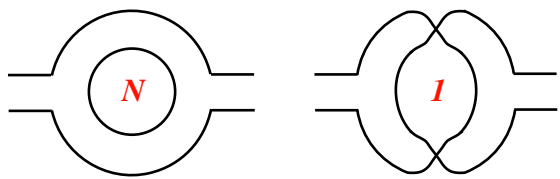
$$ \fig{N} $$

We have already seen in subsection VC9 that the group theory of any
graph can be detached from the momentum and spin (so we considered
there a simple model of scalars $Ä$).  We also saw there that the group
theory of such matrices is most conveniently graphed by a double-line
notation, where each line acts group-theoretically as a bound (anti)quark,
reducing the group theory to trivial Kronecker $¶$'s.  We now notice that
in some loop graphs, depending on how the lines are connected, some of
the quark lines form closed loops.  Again the group theory is trivial:  There
is a factor of N for each such loop, from the sum over the N colors.  We
can also give a physical picture to these numerical factors:  Since we
draw the scalar propagator as quark and antiquark lines with finite
separation, think of the scalar as a (very short) string, with a quark at
one end and antiquark at the other.  This gives a two-dimensional
structure to the diagram, by associating a surface with the area between
the quarks and antiquarks (including the area at the vertices).  We can
extend this picture by associating a surface also with the area inside (i.e.,
on the other side of) each closed quark loop.  In particular, for any
``planar" diagram, i.e., any diagram that can be drawn on a sheet of
paper without crossing any lines, and with all external lines on the
outside of the diagram, the entire diagram forms an open sheet without
holes, and with the topology of a disk (simply connected).  It is also clear
that, for a fixed number of loops and a fixed number of external lines, a
planar diagram has the greatest number of factors of N, since crossing
lines combines quark loops and reduces the power of N.

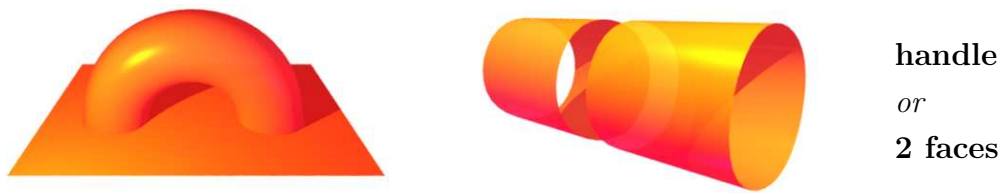
$$ \vcenter{\hbox{\figscale{handle}{2in}}}ââ
	\vcenter{\hbox{\figscale{faces}{2in}}}ââ
	\vcenter{\hbox{\bf handle}\hbox{\it or}\hbox{\bf 2 faces}} $$

We can be more quantitative about this N dependence, and relate it to the
topology of the graph.  In subsection VC2 we saw the number of
propagators, vertices, and loops were related by $P-V=L-1$.  This relation
treats a Feynman diagram as just a graph, points connected by lines.  We
now consider a diagram as a polyhedron, with propagators as the edges,
and closed quark loops as the faces, as defined by our use of matrices for
fields.  We then have as an additional relation for ÓclosedÕ surfaces
``Euler's theorem",
$$ F = P-V -2(H-1) $$
 (in terms of the ``Euler number" $­-2(H-1)=V-P+F$), where $F$ is the
number of faces and $H$ is the number of ``handles": 0 for the sphere, 1
for the torus (doughnut), etc.  This follows from the previous relation: 
First combining them as
$$ L = F +2H -1 $$
 we note that ``cutting" any handle along a loop (without separating the
pieces) produces 2 faces; in other words, introducing two faces (as a
``lens") into a loop that circles a handle changes the surface without
changing the diagram, replacing 1 handle with 2 faces.  The last relation
then follows from the case with no handles, where each face gives a loop,
except that the no-loop case corresponds to 1 face (or start with a less
trivial case, like a cube, if that's easier to picture and count momenta for).

Using the fact that the $g^2$ appears in Yang-Mills the same way as $\h$,
and that each face gets a factor of N, we find the $g$ and N dependence of
any graph is
$$ (g^2)^{L-1}N^{(L-1)-2(H-1)} = (Ng^2)^{L-1}N^{-2(H-1)} $$
 We thus see that effectively $Ng^2$ is the coupling squared suited to
planar graphs, counting the number of loops, while $1/N^2$ is a new
coupling squared, counting the number of handles.  Therefore, we can
sum over both $Ng^2$ and $1/N^2$:  Each Feynman graph is a particular
order in each of these two couplings.  The sum of all graphs at fixed
orders in both couplings gives a gauge-invariant subset of the graphs
contributing to a particular S-matrix element.  (This is sometimes called
``color decomposition".  Note that $g^2$ is the coupling normalized for
matrices of the defining representation, which was required here to
define the 1/N expansion, while $Ng^2=üg_A^2$ is the coupling
normalized for the adjoint:  If we had used matrices for the adjoint
representation, a factor $1/g_A^2$ would appear in front of the action,
because of the difference in normalization of the trace of the matrices.)

\x VIIC4.1 Consider $Ä^4$ theory in D=4, where $Ä$ is now an N$ð$N
hermitian matrix.  Generalize the auxiliary-field propagator correction
calculation of subsection VIIB7 to leading order in 1/N, showing the
N-dependence at all steps.  Show that now, to this leading order, both the
N- and $g$-dependence of the effective action can be absorbed into $M$.

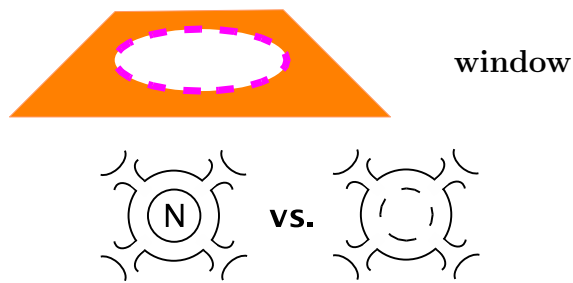
$$ \vcenter{\hbox{\figscale{window}{2in}}}ââ\hbox{\bf window} $$

$$ \figscale{down}{2in} $$

We can also consider more complicated models, such as chromodynamics,
with fields appearing in the defining representation of the group, such as
quarks.  When a quark field makes a closed loop, it looks like a planar
loop of a gluon, except that the closed quark line is missing, along with a
corresponding factor of N.  Thus, there is effectively a ``hole" in the
surface.  Since only one factor of N is missing, a hole counts as half a
handle.  We can also draw a flavor-quark line for the quark propagator
alongside the color-quark line.  Since this line closes in quark-field loops,
we also get a factor of M (for M flavors) for each quark loop.

The fact that the 1/N expansion is topological (the power of 1/N is the
number of holes plus twice the number of handles) closely ties in with the
experimental observation that hadrons (in this case, mesons) act like
strings.  Thus, we can expand in 1/N as well as in loops.  While the leading
order in the loop ($Ng^2$) expansion is classical (particle) field theory, the
leading order in the 1/N expansion is classical open-string theory (planar
graphs).  However, seeing the dynamical string properties requires
summing to all orders in $Ng^2$ for leading order in 1/N.

Thus, 1/N acts as the string coupling constant.  (N appears nowhere else
in the action describing string states, since they are all color singlets.) 
The experimental fact that the hadronic spectrum and scattering
amplitudes follow so closely that of a string (more on this later) indicates
that the perturbative expansion in 1/N is accurate, i.e., that quantum
corrections are ``small" in that sense.  One application of the smallness
of 1/N (largeness of N) is the ``Okubo-Zweig-Iizuka rule":  A planar graph
describes classical scattering of open strings (mesons).  It corresponds
topologically to a disk, which is a sphere with one hole, and is therefore
order 1/N.  Compare this to two planar graphs connected by a handle.  It
describes classical scattering of open strings with one intermediate closed
string (glueball), where the handle is a closed-string propagator
connecting two otherwise-disconnected classical open-string graphs.  It
corresponds to a cylinder, which is a sphere with two holes, and is
therefore order 1/N${}^2$.  In terms of ÓflavorÕ lines, the latter graph
differs from the former in that it has an intermediate state (the glueball)
with no flavor lines.  The OZI rule is that amplitudes containing an
intermediate glueball are always smaller than those with an
intermediate meson.  This rule also has been verified experimentally,
giving a further justification of the 1/N expansion (though not necessarily
of string behavior).

Generalizing to groups SO(N) and USp(2N) gives more varied topologies: 
Since the left and right sides of propagators are no longer distinguishable,
the string surface is no longer orientable (the surface no longer has two
distinguishable sides), so we can also have unorientable surfaces such as
M¬obius strips and Klein bottles.  One can also perform a separate
expansion in the number M of flavors.

The fact that the leading (planar) contributions are of order
$(Ng^2)^{L-1}$ requires a modification of the Borel transform of the
previous subsection:  We now identify
$$ \h = Ng^2 $$
 instead of just $\h=g^2$, so we can use the 1/N expansion in conjunction
with the Borel transform.  In particular, this means removing the factor
of N from $º_1$ and absorbing it into $\h$.  The result is that the position
of the renormalon singularities in the $z$ plane is independent of N. 
However, the same is not true for the instantons:  A one-instanton
solution corresponds to choosing a single component of $Ä$ nonvanishing
in our scalar model, so that the classical solution $Ä_0$ for the action
$S[Ä_0]$ has no N-dependence.  (Choosing $Ä$ proportional to the identity
matrix yields an N-instanton solution.)  The analog in the Yang-Mills case
is using just a (S)U(2) subgroup of the full U(N) to define the instanton. 
(Note that the structure constants for U(N) are N-independent for the
defining representation:  See exercise IB5.2.)  Then $S/g^2=NS/\h$.  The
result for the positions of the singularities in $z$ is then at integer
multiples (positive or negative, depending on considerations given in the
previous subsection) of $z_0$, where
$$ z_0 = \cases{ 1/º_1 & for renormalons \cr 
	NS[Ä_0] & for instantons \cr} $$
 where $º_1$ and the one-instanton action $S$ are N-independent.  

The net result is that instantons are unimportant for large N.  Thus, if we
take the 1/N approach of using a resummation to define a string theory,
the instantons do not take a role in defining the string.  (They might
return in another form when considering classical solutions to the string
theory, or their contribution might be just a small part of the total
nonperturbative contribution.)  On the other hand, approaches that
analyze just the low-energy behavior of a theory can make use of the
instantons:  If the physical value of N is small, or the U(N) theory is
spontaneously broken to give a small effective N at low energies (as in
GUTs), then instantons may be treated as the dominant nonperturbative
contribution to low-energy effects such as chiral symmetry breaking. 
This can be sufficient for studying low-energy bound states, but is
insufficient for studying confinement, whose physical definition is the
existence of bound states of very high energy.

\refs

£1 E.C.G. St¬uckelberg and A. Petermann, ÓHelv. Phys. ActaÕ 
	É26 (1953) 499;\\
	M. Gell-Mann and F.E. Low, ÓPhys. Rev.Õ É95 (1954) 1300;\\
	C.G. Callan, \PRD 2 (1970) 1541;\\
	K. Symanzik, ÓCommun. Math. Phys.Õ É18 (1970) 227:\\
	renormalization group.
 £2 G. 't Hooft, \NP 61 (1973) 455:\\
	renormalization group via dimensional regularization.
 £3 G. 't Hooft, Can we make sense out of ``quantum chromodynamics", in
	ÓThe whys of subnuclear physicsÕ, proc. 1977 Int. School of
	Subnuclear Physics, Erice, ed. A. Zichichi (Plenum, 1979) p. 943:\\
	arbitrariness of all but first two nonvanishing coefficients in $º$,
	convergence of singularities in the complex coupling plane at zero
	coupling.
 £4 L.D. Landau and I. Pomeranchuk, ÓDoklady Akad. Nauk USSRÕ É102 (1955)
	489:\\ 
	 Landau ghost.
 £5 F.J. Dyson, ÓPhys. Rev.Õ É85 (1952) 631:\\
	divergence of perturbation expansion in quantum field theory.
 £6 L.N. Lipatov, ÓSoviet Physics JETPÕ É45 (1977) 216:\\
	D=0 renormalon.
 £7 G. 't Hooft, Óloc. cit.Õ (ref. 3), \PL 109B (1982) 474;\\
	B. Lautrup, \PL 69B (1977) 109;\\
	G. Parisi, \PL 76B (1978) 65, \NP 150 (1979) 163;\\
	Y. Frishman and A. White, \NP 158 (1979) 221;\\
	C. DeCalan and V. Rivasseau, ÓComm. Math. Phys.Õ É82 (1981) 69;\\
	F. David, \NP 234 (1984) 237;\\
	A.H. Mueller, \NP 250 (1985) 327, \PL 308B (1993) 355;\\
	L.S. Brown, L.G. Yaffe, and C.-X. Zhai, \xxxlink{hep-ph/9205213},
	\PRD 46 (1992) 4712:\\
	renormalons.
 £8 A.S. Wightman, Should we believe in quantum field theory, in
	ÓThe whys of subnuclear physicsÕ, Óibid.Õ p. 983:\\
	review of the relation of renormalons to constructive quantum field
	theory.
 £9 't Hooft, Óloc. cit.Õ (VC):\\
	1/N.
 £10 S. Okubo, \PL 5 (1963) 165;\\
	Zweig, Óloc. cit.Õ (IC);\\
	J. Iizuka, ÓProg. Theor. Phys. Suppl.Õ É37-38 (1966) 21;\\
	J. Iizuka, K. Okada, and O. Shito, ÓProg. Theor. Phys.Õ É35 (1966) 1061.
 £11 G. Veneziano, \PL 52B (1974) 220:\\
	expansion in (1/) the number of flavors.

\unrefs

ÚVIII. GAUGE LOOPS

Gauge invariance plays an important role in quantum corrections.  It not
only simplifies their form, but leads to new effects.  In particular, it not
only improves high-energy behavior, but can eliminate divergences
altogether, in the presence of supersymmetry.

In general, the first thing to calculate in quantum field theory is the
effective action.  Once this has been calculated, other properties can be
determined: the vacuum, S-matrix, etc.  In particular, in spontaneously
broken theories, the effective action should be calculated with the
symmetric (unbroken) vacuum, which has simpler Feynman rules; once the
effective action has been calculated, vacuum values of the fields can be
determined, and the S-matrix can be calculated as a perturbation about
this quantum vacuum.  (The alternative of defining Feynman rules for the
classical broken vacuum and then calculating quantum corrections
doubles the work in finding vacuum values.)

Û7 A. PROPAGATORS

We first consider propagator corrections in some specific theories with
spin.  In the following calculations we assume the gauge coupling appears
only as an overall factor in the classical action:  It thus also counts loops,
so our 1-loop graphs are coupling-independent.  All the integrals have
been performed in subsections VIIB4 and VIIB6; all that remains is the numerator
algebra, which follows the examples of subsection VIC4.  As we have
seen, such corrections are important in analyzing high-energy behavior;
as we'll see in the following section, they are also important for low
energy.  (Of course, for massless particles the two are related by
conformal invariance, even when quantum corrections break it.)

Ü1. Fermion

$$ \fig{selfe} $$

Our first calculation is the one-loop correction to the electron kinetic
operator in QED:  The S-matrix element is
$$ \A_{2e} =
	Çdk {©^a(Ök+üÖp+\f{m}{å2})©_a \over ü(k-üp)^2 ü[(k+üp)^2+m^2]} $$
 At this loop level the only difference between using D-dimensional
$©$-matrix algebra (dimensional regularization) and 4-dimensional
(dimensional reduction) is an unphysical finite renormalization, so for
simplicity we'll use the latter method.  Then the numerator is
$$ Ök+üÖp-å2m $$
 The result of the integral is then
$$ \A_{2e} = -Öp{m^2\over 2p^2}[ö\A_2(p^2,0,m^2)-ö\A_2(0,0,m^2)] 
	+(üÖp-å2m)ö\A_2(p^2,0,m^2) $$
 in the notation of subsection VIIB6.  The UV divergent part follows from
$$ ö\A_2(p^2,0,m^2) = \f1· +finite $$
 The contribution to $ý$ is minus the S-matrix element, but the
counterterm has a second minus sign to cancel the divergence:
$$ ëS = \f\h·Çdx¼Ðï(-üiÖ»-å2m)ï $$
 The calculation for the quark self-energy in QCD is the same except for
group-theory factors (see subsection VIIIA5).

\x VIIIA1.1 Repeat the calculation with D-dimensional $©$-matrix algebra. 
What is the difference in the ÓfiniteÕ part, and why doesn't it matter?

In subsection VIIB6 we considered MOM subtraction (see subsection
VIIA3) for scalar propagators.  The analysis in this case is similar, but now
we expand in
$Öp$ instead of $p^2$:
$$ ëK = a +b(\f{m}{å2}-Öp) +\O[(\f{m}{å2}-Öp)^2] $$
 However, since $ëK$ is normally expressed as functions of $p^2$ times 1
and $Öp$, we need to translate:  Using 
$(\f{m}{å2}+Öp)(\f{m}{å2}-Öp)=ü(p^2+m^2)$,
$$ ëK = a +b'(\f{m}{å2}-Öp) +cü(p^2+m^2) 
	+\O[(\f{m}{å2}-Öp)(p^2+m^2), (p^2+m^2)^2] $$
$$ = a +(b'+2\f{m}{å2}c)(\f{m}{å2}-Öp) +\O[(\f{m}{å2}-Öp)^2] $$

We next reevaluate the fermion propagator correction, to linear order in
$\f{m}{å2}-Öp$.  Starting with
$$ ÷\A(x,p^2,m_1^2,m_2^2) = Çdk¼e^{ixÉk}
	{1\over ü[(k+üp)^2+m_1^2]ü[(k-üp)^2+m_2^2]} 
	= Çd^2  ¼Â^{-D/2}e^{-E} $$
$$ E = ü\f1Âx^2 +ixÉüºp +\f18Â(1-º^2)p^2 
	+\f14 Â[(m_1^2 +m_2^2) +º(m_1^2 -m_2^2)] $$
 we keep only linear order in $x$ and $p^2+m^2$, and set $m_1=m$,
$m_2=0$ (switching back to $Œ=ü(1+º)$):
$$ E ® ixÉ(Œ-ü)p +üŒ(1-Œ)(p^2+m^2) +üÂm^2 Œ^2 $$
 To clearly separate UV divergences (from $®0$) and IR divergences
(from $Ψ0$), we scale
$$  £ {Â\over Œ^2}âÜâE ® (Œ-ü)ixÉp +üÂ(\f1Œ-1)(p^2+m^2) +üÂm^2 $$
$$ ÷\A ® Ç_0^¥ d¼Â^{·-1}e^{-Âm^2/2}
	 Ç_0^1 dŒ¼Œ^{-2·}[1-(Œ-ü)ixÉp][1-üÂ(\f1Œ-1)(p^2+m^2)] $$
 The integrals are easily performed in either order:
$$ ÷\A ® ý(1+·)(üm^2)^{-·}\left[ {1\over ·_{UV}}{1\over 1-2·}
	+ü{1\over (1-·)(1-2·)}ixÉp \right. $$
$$ +\left.\left( ü{1\over ·_{IR}}+{1\over 1-2·} \right){p^2+m^2\over m^2} 
	+\left( \f14 {1\over ·_{IR}} +\f32 {1\over 1-2·} -ü{1\over 1-·} \right)
	ixÉpÊ{p^2+m^2\over m^2} \right] $$
 and in the limit $·£0$,
$$ ÷\A ® ý(1+·)(üm^2)^{-·}\left[ {1\over ·_{UV}} +2 +üixÉp
	+\left( ü{1\over ·_{IR}}+1 \right){p^2+m^2\over m^2} \right. $$
$$ +\left.\left(\f14{1\over ·_{IR}} +1\right)
	ixÉpÊ{p^2+m^2\over m^2} \right] $$
 The electron propagator correction to linear order in $\f{m}{å2}-Öp$ is then
$$ \A_{2e} ® ý(1+·)\left({m^2\over µ^2}\right)^{-·}\leftÓ\left[ü +\left(
	\f14 {1\over ·_{IR}} +1\right){p^2+m^2\over m^2}\right]Öp\right. $$
$$ +\left.\left[{1\over ·_{UV}} +2 +\left(ü{1\over ·_{IR}} +1\right)
	{p^2+m^2\over m^2}\right](üÖp -2\f{m}{å2})\rightÕ $$
$$ ® ý(1+·)\left({m^2\over µ^2}\right)^{-·}(-ü)\left[ \f{m}{å2}
	\left( 3{1\over ·_{UV}} +5 \right) +\left( {1\over ·_{UV}}
	+2{1\over ·_{IR}} +5 \right)(\f{m}{å2}-Öp) \right] $$
 The $1/·_{UV}$ terms are the same as the $1/·$ terms obtained above
for minimal subtraction.  In the MOM scheme, this entire contribution
($\O(K^0)$ and $\O(K^1)$) is canceled by counterterms.

\x VIIIA1.2  Repeat the above calculations replacing the fermion
with a scalar.

\x VIIIA1.3  Repeat the above calculations replacing the photon
with a (massless) 
 ªa scalar
 ªb pseudoscalar (with a $©_{-1}$ vertex).

Ü2. Photon

$$ \fig{selfp} $$

We next calculate the spin-1/2 contribution to the photon (or gluon) self
energy:  The S-matrix element is
$$ Çdk {tr[-©_a(Ök-üÖp+\f{m}{å2})©_b(Ök+üÖp+\f{m}{å2})] \over
	ü[(k-üp)^2+m^2]ü[(k+üp)^2+m^2]} $$
 The result of the trace (again using 4-dimensional algebra) is
$$ -2[k_a k_b -ú_{ab}ü(k^2+\f14 p^2 +m^2)] -ü(ú_{ab}p^2 -p_a p_b) $$
 The first part is the expression appearing in $ö\A_{ab}$ in subsection
VIIB6, once we recognize its $ú_{ab}$ terms as the average of the
denominator factors, yielding tadpoles.  The integral thus gives
$$ (ú_{ab}p^2 -p_a p_b)(-2×\A -üö\A_2) ®
	(\f1· -ln¼p^2)(-\f13)(ú_{ab}p^2 -p_a p_b) $$
 for the divergent and Óhigh-energyÕ terms.  Using
$$ A^a(-p)(ú_{ab}p^2 -p_a p_b)A^b(p) = üF^{ab}(-p)F_{ab}(p) $$
 in terms of the linearized field strength $F$, the corresponding
contributions to the unrenormalized one-loop effective action are
(including a factor of $ü$ for identical external lines)
$$ ý_1 ® \h \f23 Çdx¼\f18 F^{ab}(\f1· -ln¼õ)F_{ab} $$
 (neglecting the ``$-1$" part of $ln(-õ)$) and the counterterm is thus
$$ ëS = \f\h·(-\f23)Çdx¼\f18 F^{ab}F_{ab} $$
 in the case of QED.  For QCD, we must include the group-theory factor
$tr(G_i G_j)$ multiplying $F^{iab}F^j_{ab}$.  (Examples will be given
in the following subsections.)

This propagator correction is easier to analyze in the MOM scheme than
the electron propagator, since there are no internal massless particles,
and thus no IR divergence to distinguish from the UV one.  We therefore
just take the explicit expressions for the integrals from subsection VIIB6
and Taylor expand in $p^2$ about 0 (or actually in $1/º$ of VIIB6.1a,
substituting for $p^2$ only at the end).  The Ólow-energyÕ part of the
renormalized effective action for the photon, exhibiting the momentum
dependence of the coupling, is then
$$ ý_{0+1,2©,r} ® Çdx¼
	\f18 F^{ab}\left({1\over e^2} +\f2{15}{õ\over m^2}\right) F_{ab} $$
 where we have applied MOM subtraction by canceling constant (infinite
and finite) constributions to the coupling.

\x VIIIA2.1 Evaluate this contribution to the ÓunrenormalizedÕ effective
action to this order.  Show that the constant contributions to the coupling
(to be canceled by renormalization) are
$$ \f1{e^2} £ \f1{e^2} +\f23 [ \f1· -© -ln (üm^2)] $$

Ü3. Gluon

The most interesting case is the propagator of the Yang-Mills field,
in a theory of Yang-Mills coupled to lower spins.  There is an important
simplification in this calculation in the background field gauge:  Writing
the classical Yang-Mills Lagrangian as $tr¼F^2/g^2$, the covariant
derivative appears as $á=»+iA$ without coupling constant, so the gauge
transformation of $A$ is coupling independent, as in general for the
matter fields.  (In terms of a group element ${\bf g}$, $Ä'={\bf g}Ä$ and
$á'={\bf g}á{\bf g}^{-1}$.)  The effective action is gauge invariant, which
means the only divergent terms involving the Yang-Mills field are the
gauge-covariantized kinetic (less mass) terms of the various fields.  The
divergences for the non-gauge fields are not so interesting, since they
can be absorbed by rescaling those fields (``wave-function
renormalization"), but the divergence of the $tr¼F^2/g^2$ term can be
absorbed only by rescaling the coupling $g$ itself.  (On the other hand, if
we use $á=»+igA$, then renormalization of $g$ requires the opposite
renormalization of $A$ to preserve gauge invariance.)  Thus this
divergence is related to the UV behavior of this coupling (as discussed in
subsection VIIB7, and further later).  The important point is that there is
no wave-function renormalization for the Yang-Mills field (since there is
no corresponding gauge-invariant counterterm), so the coupling-constant
renormalization (like mass renormalizations) can be found from just the
propagator correction, while in other gauges one would need also a much
messier vertex (3-point) correction:  BRST invariance is not enough to
give the result from a single graph.

$$ \fig{selfg} $$

We now consider the contributions of spins 0 (including ghosts) and 1
(including gluon self-interactions), and redo the spin-1/2 contribution in a
way that resembles the bosons.  It is based on the observation that there
is a universal form for the gauge-covariantized Klein-Gordon equation for
spins 0,1/2,1, which can also be shown by supersymmetry.  The kinetic
operator in a background Yang-Mills field is
$$ K = -ü(õ -iF^{ab}S_{ba}) $$
 where now $õ=(á)^2$ is gauge covariantized.  This form is true in
arbitrary dimensions.  For spin 0 it is obvious.  For spin 1/2, we use the
fact that the one-loop contribution to the functional integral is the trace
of the logarithm of the propagator, as follows from Gaussian integration,
$$ ÇDƼDÐƼe^{-ÐÆKÆ} = det¼K = e^{tr¼ln¼K} $$
 where the trace is over all indices, including the coordinates.  Then the
contribution to the effective action from kinetic operator $K$ is 1/2 the
contribution from $K^2$.  (See also exercise VIA4.2.)  We then use (see
subsection IIIC4)
$$ -2Öá^2 = -2(©Éá)^2 
	= -(Ó©^a,©^bÕ +[©^a,©^b])á_a á_b = õ +iS^{ab}F_{ab} $$
 where we have used
$$ S^{(1/2)}_{ab} = -ü©_{[a}©_{b]},ââÓ©_a,©_bÕ = -ú_{ab} $$
 In the case D=4, this is equivalent to the result obtained in subsection
IIIC4 in terms of just the undotted spinor, but there the 1/2 is
automatically included because there are half as many fields, so the
range of the trace is half as big.  

For spin 1, we use the result of the
background-field version of the Fermi-Feynman gauge:  At quadratic
order in the quantum fields, from exercise VIB8.1 we have
$$ \f18 F^2 +\f14(»ÉA)^2 £ 
	\leftÓ\f18 (\D_{[a}A_{b]})^2 +i\f14 \F^{ab}[A_a,A_b]\rightÕ 
	+\f14 (\DÉA)^2 $$
$$ = -\f14 AÉõA -iüA^a[\F_{ab},A^b] = -\f14 AÉ(õ -i\F^{ab}S_{ba})A $$
 where $õ=(\D)^2$ contains only the background gauge field, and in the
last step we have written the quantum field $A$ as a column vector in the
group space and the background fields (like $\F$) as matrices for the
adjoint representation (which replaces commutators with multiplication),
and used the explicit expression
$$ S^{(1)}_{ab} = |{}_{[a}ÔÒ{}_{b]}|,ââÒ{}_a|{}_bÔ = ú_{ab} $$
 To this order in the quantum fields, the kinetic operator for the two
ghosts looks just like that for two physical scalars, but gives a
contribution to the effective action of opposite sign because of statistics.

This method can be used for arbitrary one-loop graphs with external
gluons, and easily generalizes to massive fields; we now specialize to
propagator corrections.  There are two kinds of vertices, the spin-0 kind
and the vertex with the spin operator.  Since $tr¼S_{ab}=0$, we get
only graphs with either 2 spin vertices or none.  There is only one spin
graph, with 2 internal free propagators; the 2 spinless graphs include
such a graph but also a tadpole, which vanishes by dimensional
regularization in the massless case.  Since the spinless graphs give the
complete result for internal spin-0, their sum is separately gauge
invariant; the spin graph is obviously so, since it is expressed directly in
terms of the field strength.  (We refer here to the Abelian part of the
gauge invariance, which is all you can see from just 2-point graphs.)  As
far as Lorentz index algebra is concerned, we need to evaluate only
$tr(S_{ab}S_{cd})$.  For the vector, we have
$$ tr(S^{(1)}_{ab}S^{(1)}_{cd}) = 2ú_{b[c}ú_{d]a} $$
 For spin 1/2, the traces are the same as in D=4 except for overall
normalization; using earlier identities, or using the same methods for
this case directly,
$$ tr(S^{(1/2)}_{ab}S^{(1/2)}_{cd}) = \f14 tr(I)ú_{b[c}ú_{d]a} $$
 where $tr(I)$ is the size of the spinor.

\x VIIIA3.1  Let's look at other ways to interpret the last two identities:
 ªa  Use the double-line notation (subsection VC9) for the defining
representation of the orthogonal group to derive the above expression
for the trace of two $S^{(1)}$'s.
 ªb  Use the fermion action of IIIC4 in terms of just undotted spinors for
D=4. Evaluate
$$ tr(S^{(1/2)}_{Œº}S^{(1/2)}_{©¶}) $$
 using both bra-ket notation and double-line notation for SL(2,C).  Show
the result is the same as from vector notation (by relating $F_{ab}$ and
$f_{μ}$).

All diagrams will also have a group-theory factor of $tr(G_i G_j)¾¶_{ij}$.
We'll be interested mostly in SU(N) for Yang-Mills theory (as appropriate
to describe color in the Standard Model for N=3, or arbitrary N for
applying the 1/N expansion).  Then the most interesting representations
are the adjoint (for the gluons and their ghosts) and the defining (for the
quarks).  As explained in subsection IB2, or as follows from the
double-line notation of subsection VC9, we use the normalization
$$ tr_D(G_i G_j) = ¶_{ij}âÜâtr_A(G_i G_j) = 2N¶_{ij} $$

Finally, there are the momentum-space integrals, which have already
been evaluated in subsection VIIB4 for the massless case (which is
sufficient for determining the high-energy behavior, and thus the UV
divergences) and VIIB6 for the massive case.  The integral for the spin
graph is the same as that for $Ä^3$ theory (using the $S_{ab}$ vertex
from $-üõ-üiF^{ab}S_{ab}$).   As labeled there, the external line has
momentum $p$ and the internal lines $kàüp$.  Then the vertex factors in
the spinless graph with two propagators are both simply $-k$ (from
$-üõ=-ü»^2+üAÉ(-i»)+ü(-i»)ÉA+üA^2$), giving $\A_{ab}$, while the
addition of the tadpole, with vertex factor $-ú$, converts it to $ö\A_{ab}$. 
(By comparison, the tadpole graph that was apparently avoided in the
Dirac-spinor calculation of the previous subsection appeared anyway
after evaluating the trace algebra.)  This contribution also gets an overall
$tr(I)$ factor, simply counting the number of degrees of freedom.  Note
that the scalar factor $×\A$ that appears in $ö\A_{ab}$ is the sum of a
divergent term proportional to the $Ä^3$ graph and a convergent term
that vanishes in the massless case.

We now combine all factors to obtain the contributions to the two-gluon
part of the unrenormalized 1-loop effective action (including the $-$1 for
getting the effective action from the S-matrix, a $-$1 for internal
fermions, either spin $ü$ or ghost, the $ü$ for identical external gluon
lines, the $ü$ for the spinor to compensate for squaring the propagator,
and yet another $ü$ for identical internal lines if the group
representation was real.)  The result is the sum of contributions of the
form
$$ ý_{1,2g} = \h¼tr Çdx¼
	\f18 F^{ab}(üc_R)(-1)^{2s}[\f1{D-1}\B_1(-õ) -4s^2\B_2(-õ)]F_{ab} $$
 where $c_R$ is the group theory factor from the trace, which for the
interesting cases is
$$ c_R = \cases{2 & for $N¢ÐN$ (defining) \cr 2N & for adjoint (real)\cr} $$
 This result applies to spins $s=0,ü,1$, with the understanding that it is
the result for two polarizations, so there is an implicit extra factor of $ü$
for a single scalar, while for massive spin 1 (spontaneously broken gauge
theories) the third polarization in the (background-field) Fermi-Feynman
gauge is carried by a scalar field.  (The result above for $s$=1 is the sum
of the contributions from the vector field and the two fermionic ghosts.) 
The functions $\B_1$ and $\B_2$ are the spinless and spin contributions,
related to the massive $Ä^3$ propagator correction $ö\A_2(p^2,m^2,m^2)$
as
$$ \B_2(p^2) = ö\A_2(p^2,m^2,m^2),â
	\B_1(p^2) = \B_2(p^2) +4m^2{\B_2(p^2)-\B_2(0)\over p^2} $$

Note that 
$$ \B_1 ® \B_2 ® \f1· - ln¼p^2 $$
 as far as divergent (at D=4) or high-energy (i.e., massless) terms are
concerned.  Also note that all contributions ÓexactlyÕ cancel if all spins
are in the adjoint and have the same mass, and appear in the ratio 1:4:6
for spins 1 (including ghosts), $ü$, 0:  For the massless case, this is N=4
super Yang-Mills, which is also the massless sector of the dimensional
reduction of the open superstring from D=10.  The massless sector of the
reduction of the open bosonic string from D=26 yields Yang-Mills plus 22
adjoint scalars, which cancels near D=4 up to a finite, local term ($F^2$), which can be removed by a nonminimal renormalization.

In examining the contribution of this term to the running of the coupling
constant with energy, we see that the vectors contribute with opposite
sign to lower spins.  In particular, in terms of the coefficient $º_1$ (of
subsection VIIC1), only nonabelian vectors make positive contributions
(since Abelian vectors are neutral).  This means that nonabelian vectors
are responsible for any weakening in a coupling at high energies, known
as ``asymptotic freedom", an important experimental feature of the
strong interactions (see section VIIIC).  Note that while the sign of $º_1$
for $Ä^4$ theory, using the method of subsection VC9, is independent of
the coupling (since all 1-loop corrections are coupling-independent when
the coupling appears as an overall factor in the classical action, like $\h$),
changing the sign of the coupling changes its sign relative to $º_1$:  The
result is that this theory can be made asymptotically free only if its
potential has the wrong sign (negative for large $Ä$).  Thus, although
nonabelian vectors are required for asymptotic freedom in physical
theories, ``wrong-sign $Ä^4$" can be used as a toy model for studying
features associated with asymptotic freedom (especially resummation of
the perturbation expansion: see section VIIC).

Note that for (massless) fermions that couple chirally to vectors (as in
electroweak interactions), $c_R$ consists of the contribution from a
complex representation but not its complex conjugate:  Only one of the
two Weyl spinors of the Dirac spinor contributes.  The result is that the
contribution to the vector propagator is half that of the parity-invariant
case.  This fact follows from comparing the calculations of the chiral and
nonchiral cases without the squared-propagator trick:  In Dirac
(4-component) spinor notation, the $©_{-1}$'s drop out of the calculation;
in Weyl (2-component) spinor notation, the left- and right-handed-spinor
diagrams are identical except for (internal) group theory.  (Things are
more complicated for higher-point functions, because the group theory
gives more than just $tr_R(G_i G_j)$:  See subsection VIIIB3.)

\x VIIIA3.2  Find the conditions for exact cancellation if spins $ü$ and 0
include both adjoint and defining representations.  Find the weaker
conditions if only the divergent (and therefore also high-energy) terms
cancel.

\x VIIIA3.3  Use the optical theorem to find the decay rate for a massive
vector (e.g., Z boson) into massive particle-antiparticle pairs of various
spins.

\x VIIIA3.4  Find the propagator correction for internal particles of
ÓdifferentÕ masses on each of the two lines (e.g., for a W boson
propagator).

In the case of QCD, with color gauge group $SU(N_c)$ and $N_f$ flavors of
quarks in the defining representation of color, the divergent and
high-energy contributions to this term in the unrenormalized 1-loop
effective action are
$$ ý_{1,2g,QCD} ® \h¼tr Çdx¼
	\f18 F^{ab}\f13(2N_f-11N_c)(\f1· - ln¼õ)F_{ab} $$

At higher loops the effective action will still be gauge invariant in
background-field gauges (for the quantum fields), so the renormalization
of the Yang-Mills coupling can still be determined from just the gluon
propagator correction.  On the other hand, in other gauges a three-point
vertex must also be calculated:  It can be shown that the gauge-fixed
classical action, including counterterms, is BRST invariant only up to
wave-function renormalizations; i.e, the most general counterterms
needed (with a BRST preserving regularization) are BRST-invariant terms
with additional multiplicative renormalizations of the quantum fields. 
Thus, BRST invariance, unlike gauge invariance, is not strong enough to
relate the gluon coupling and wave-function renormalizations.  Not only
does this mean evaluating many more graphs, but graphs which make the
propagator correction look easy by comparison.  (This is not so difficult
for just the one-loop divergences we have considered, but the difficulty
grows exponentially with the number of loops.)  

However, in the background-field gauge the $L$-loop propagator
correction has ($L-1$)-loop vertex subdivergences, similar to those in
other gauges.  The net result is:  (1) We still have a BRST-invariant
``classical" action, containing the same (counter)terms that appear in
other gauges (including quantum ghosts), but covariantized with respect
to background gauge fields (and including coupling to other background
fields).  However, the coefficients need be calculated only to order $L-1$
for the $L$-loop effective action, one loop less than in other gauges.  (2)
In addition, we have background-field-only terms in the classical action
whose $L$-loop coefficients do need to be calculated, but with a
relatively small amount of additional effort, due to gauge invariance. 
Thus renormalization consists of two steps:  (1) adding BRST-invariant
counterterms for the quantum fields (background covariantized) to cancel
subdivergences, and (2) adding gauge-invariant counterterms for the
background fields (which can be interpreted as vacuum renormalization
for the quantum fields) to cancel superficial divergences.  Consequently,
background-field gauges save about one loop of difficulty as far as
renormalization is concerned.  Furthermore, similar simplifications occur
for calculations of finite parts (e.g., effective potentials), because of
simplifications from gauge invariance.

Ü4. Grand Unified Theories

The best result of GUTs is their prediction that the gauge couplings of the
Standard Model coincide at some high energy, as a consequence of the
running of the couplings with energy.  (Mixed results have been obtained
for masses, arguably because renormalization group arguments are
accurate only for high energies, and thus leptons with large masses.  A
``failed" prediction is proton decay, which has already eliminated the
nonsupersymmetric SU(5) model with minimal Higgs.)  The numerical
details of this prediction are model dependent (and thus easy to fudge,
given enough freedom in choice of nonminimal fields), but the fact that all
three couplings come close together at high energies is already strong
evidence in favor of unification.

Thus we make only the crudest form of this calculation, using only the
one-loop results of the previous subsection.  The main assumption is that
there is a ``desert" between the Standard Model unification scale (around
the masses of the intermediate vector bosons W and Z) and the Grand
Unification scale $M_{GUT}$, with no fundamental particles with masses in
that range (although, of course, a huge number of hadrons appear there). 
This allows us to crudely approximate all fundamental particles below
that region (i.e., those of the Standard Model) as massless, and all above
as infinitely massive.  In particular, in the framework of the minimal SU(5)
GUT, this means ÓallÕ the fermions are treated as massless.

Therefore the calculation is to use the one-loop results to calculate the
running of the couplings in the Standard Model, and use the relation of
the gauge couplings in the SU(5) GUT to identify those of the Standard
Model in terms of that of this GUT.  From the previous subsection, the
running of the couplings is given by
$$ {1\over g^2(-p^2)} ® 
	{1\over g_0^2} -º_1 ln \left( {M_{GUT}^2\over -p^2} \right),ââ
	º_1 = Ý_{R,s} üc_R(-1)^{2s}(4s^2-\f13) $$
 for two helicities of spin $s$ (with an extra factor of $ü$ for only
1 helicity of spin 0), where $g_0­g(M_{GUT}^2)$.  

If we use $g_1,g_2,g_3$ to label the couplings of U(1), SU(2), and SU(3)
that are identified with the single SU(5) gauge coupling at the unification
scale, then their relation to those of the Standard Model (as normalized in
exercise IVB2.1) is
$$ {1\over g_1^2} = \f65 {1\over g'^2} = \f65 {cos^2 Ï_W\over e^2},ââ
	{1\over g_2^2} = {1\over g^2} = 2 {sin^2 Ï_W\over e^2},ââ
	{1\over g_3^2} = {1\over g_s^2} $$
 where $g_s$ and $g$ the usual SU(3) and SU(2) couplings, and the factor
of $\f65$ is because the U(1) generator (see subsection IVB4) satisfies
$tr_D(G^2)=\f56$ in terms of SU(5) matrices.  (We generally normalize to
$tr_D(G^2)=1$ for each generator.  Physical couplings are preserved if
changes in normalization of generators are accompanied by changes in
coupling normalization so as to preserve $g_i G_i$.)

Then the values of the $º_1$'s for the Standard Model are
$$ º_{1,1} = 0 -4 -\f1{10} = -\f{41}{10},ââ
	º_{1,2} = \f{22}3 -4 -\f16 = \f{19}6,ââ
	º_{1,3} = 11 -4 +0 = 7 $$
 where we have listed the contributions from spins 1, $ü$ (for 3 families),
0, respectively. (Note that the spinors contribute the same to each
because they are all effectively massless:  They don't notice the SU(5)
breaking.  Also, we can ignore SU(2)$°$U(1) breaking when calculating
these $º$'s, since we have neglected the corresponding masses.)  

\x VIIIA4.1  Calculate the contribution of the spinors to the $º_1$'s, in
terms of both SU(5) and SU(3)$°$SU(2)$°$U(1) multiplets.  (Note the chiral
couplings for spinors, so for $c_R$ a complex representation and its
complex conjugate might not both contribute.)

The experimental values of the couplings (in the $Ñ{\rm MS}$ prescription)
at $µ=M_Z®$ 91 GeV are
$$ {1\over e^2} ® 804,ââsin^2 Ï_W ® .231,ââ{1\over g_s^2} ®  106 $$
 Unfortunately, taking any two of the equations for $1/g_i^2$ gives widely
varying answers: e.g.,
$$ M_{GUT} ® 10^{15à2} GeV $$
 Alternatively, since we have used only two parameters to fit three
experimental numbers, we can try to predict the value of any one of $e$,
$Ï_W$, or $g_s$ from the rest: e.g., from $e$ and $g_s$ we can find
$$ sin^2 Ï_W ® .207 $$
 which shows the same disagreement (but looks better than the
exponentiated error for $M_{GUT}$).

The result is not very accurate, since we have made many
approximations, which can be improved with some effort:  Two-loop
corrections add $ln¼ln$ terms to the one-loop $ln$ terms; including the
mass dependence of the effective couplings also adds significant
corrections.  But the most important approximation assumption we made
was the desert:  Undiscovered particles, such as new fermions,
nonminimal Higgs, or supersymmetric partners, change even the one-loop
expressions $º_1$.  Specifically, since by definition the unification scale is
where the masses of all unobserved vectors reside, these new particles
will all have spins 0 or $ü$, and thus make the $º$'s more negative.  In
particular, supersymmetrization yields a result consistent with
experiment, with
$$ M_{GUT} ® 2.2 ð 10^{16} GeV $$
 (This has been interpreted as the only experimental verification of
supersymmetry.)

\x VIIIA4.2  Let's examine the effects of supersymmetry:
 ªa  Supersymmetrize the Standard Model contributions to $º_1$ by
adding the supersymmetric partners to each spin: $1£1¢ü$,
$ü£ü¢0¢0$, $0£ü¢0¢0$ (where the Higgs scalars have doubled because
chiral scalar superfields can't satisfy reality conditions) to find the result
$$ º_{1,1} = 0 -6 -\f35 = -\f{33}5,ââ
	º_{1,2} = 6 -6 -1 = -1,ââ
	º_{1,3} = 9 -6 +0 = 3 $$
 ªb  Solve for $1/g_0^2$, $ln(M_{GUT}^2/M_Z^2)$ (and thus $M_{GUT}$),
and $sin^2 Ï_W$ in terms of $1/e^2$ and $1/g_s^2$.  Then plug in to find
the numerical values.
 ªc  Show the consistency condition relating the 3 couplings is
$$ {º_{1,2}-º_{1,3}\over g_1^2} +{º_{1,3}-º_{1,1}\over g_2^2} +
	{º_{1,1}-º_{1,2}\over g_3^2}  = 0 $$
 and that the closest integer values for the couplings from the above data,
$$ {1\over g_1^2} = 742,ââ{1\over g_2^2} = 371,ââ
	{1\over g_3^2} = 106 $$
 satisfy it exactly.  (OK, so this is just a numerical coincidence,
considering experimental inaccuracies and theoretical approximations,
but isn't it still nice?)  Also, note that
$$ {1\over g_1^2} = 2{1\over g_2^2}âÜâsin^2 Ï_W = \f3{13} $$
 ªd  Drop the contributions of the Higgs (and its superpartners) to the
$º$'s in both the supersymmetric and nonsupersymmetric cases, and
reevaluate $sin^2 Ï_W$, showing both give the same (poor) value.  (Thus,
Higgs can make a difference.)

Ü5. Supermatter

Although the problem with infrared renormalons may be only technical,
the appearance of this same problem in several different approaches
(including a nonperturbative one; see subsection VIIIB7) strongly suggests that the
``correct" approach to quantum field theory, in the sense of a practical
method for unambiguously (i.e., with predictive power) calculating
perturbative and nonperturbative effects, might be to consider only
theories that are perturbatively finite.  In this subsection we will analyze
general properties of supersymmetric field theory using superspace, and
in particular improved UV behavior, concentrating on finite theories.

Finite supersymmetric theories must be in particular one-loop finite.  This
turns out to be enough to guarantee finiteness to all loops:  Two-loop
finiteness is automatic, while an appropriate renormalization prescription
is required to guarantee finiteness is preserved order by order in
perturbation theory.  (No constraints on the coupling constants are needed
beyond those found at one loop, but without the renormalization
prescription infinities cancel between different loop orders.)  Of course,
wave-function renormalizations are gauge dependent:  N=1
supersymmetric gauges eliminate some of these unphysical divergences
(and gauges with higher supersymmetry more), as do background-field
gauges even in nonsupersymmetric theories.  So, ``finite theory" in
general gauges refers only to the ``physical" divergences --- those that
affect the high-energy behavior of the theory, namely those that appear
in couplings and masses.

Because of the nonrenormalization of chiral terms in the action (see
subsection VIC5), it might seem that the corresponding couplings and
masses are always unrenormalized.  However, the kinetic terms of chiral
superfields can receive quantum corrections, and the true couplings are
defined by field redefinitions that eliminate these rescalings.  This means
that all such renormalizations are related, and given by the wave
function renormalizations.  The only other couplings are the Yang-Mills
ones, whose renormalization is also given by kinetic terms in
background-field gauges.  Thus, all ``physical" renormalizations in
supersymmetric theories can be found from just propagator corrections. 
In particular, this means that if the effective action is calculated with
background-field supergraphs, then it is completely finite in a finite
theory.

A possible exception to our statement of all physical renormalizations
coming from propagator corrections would seem to be the
Fayet-Iliopoulos tadpole term $Çd^4 ϼV$.   However, massless tadpoles
vanish in dimensional regularization, and massive ones require real
representations, which cannot generate explicit-prepotential terms.  (In
particular, at more than one loop such terms never appear in the
background-field gauge for any representation.)

$$ \fig{supermatter} $$

The simplest one-loop propagator correction is to $ÐÄÄ$.  (The $ÄÄ$
correction vanishes, since $Çd^4 ϼÄ^2=0$:  See subsection VIC5.)  There
are two graphs to consider, one with two internal $ÐÄÄ$ propagators, and
one with internal $ÐÄÄ$ and $VV$ propagators.  The $d$ algebra for the
two graphs is identical:  Both get a $d^2$ and a $Ðd^2$ inside the loop,
exactly enough to give a nonvanishing graph (using 
$[Ðd^2 d^2 ¶^4(Ï-Ï')]|_{Ï'=Ï}=1$).  There is also a symmetry factor of
$ü$ for the two $ÐÄÄ$ propagators, and a $-1$ for the mixed graph
because the two different types of internal propagator have opposite
sign (and, as usual, an overall $-1$ to get $ý$ from the T-matrix).  Thus,
the supersymmetry (spin) part of the algebra is almost trivial in this case.

\x VIIIA5.1  Use component methods to evaluate the first graph with
external fermions: the contribution of the Yukawa interaction to the
fermion propagator.  Show it agrees with the supergraph evaluation.

On the other hand, the internal group theory is slightly messy, so we treat
the general case immediately:  We take vector multiplets $V^i$ for an
arbitrary group (though we will need a semisimple group for finiteness,
since Abelian groups are not even asymptotically free).  Sums $Ý_G$ are
over each simple subgroup (or each Abelian factor), since they can have
independent coupling constants $g_{RG}$ for representation $R$
(especially $g_{AG}$ for the adjoint, which we use for the pure super
Yang-Mills term for definiteness; except for the Abelian factors, where a
nontrivial representation should be substituted).  Similarly, sums $Ý_R$
are over irreducible representations of the group; $¸_{RI}{}^J$ is the
corresponding projection operator.  For the simple (or single-component
Abelian) factors of the group 
$$ ú_{Gij} = c_{AG}¸_{Gij} $$
 is used (but again, with a different normalization for the Abelian
factors).  We also use the group theory identities (normalizations) from
subsection IB2, now generalized to these nonsimple groups and reducible
representations:
$$ G_{iI}{}^K G_{jK}{}^J ¸_{RJ}{}^I = Ý_G c_{RG}¸_{Gij} 
	= Ý_G {c_{RG}\over c_{AG}}ú_{Gij} $$
$$ G_{iI}{}^K G_{jK}{}^J ú_G^{ij} = Ý_R k_{RG}¸_{RI}{}^J
	= Ý_R {c_{RG}d_{AG}\over c_{AG}d_R}¸_{RI}{}^J $$
 Then from the Lagrangian
$$ L = -Çd^4 ϼÐÄ^I(e^V)_I{}^J Ä_J 
	+\left( Çd^2 ϼ\f16 Â^{IJK}Ä_I Ä_J Ä_K +h.c. \right) $$
$$ -Ý_G{1\over g_{AG}^2}Çd^2 ϼüW^{iŒ}W^j_Œ ú_{Gij} $$
 (ignoring mass terms) the result is simply
$$ ý_{1,ÄÐÄ} = \h Çd^4 ϼÐÄ^I M_I{}^J ö\A_2 Ä_J,ââ
	M_I{}^J = Ý_{R,G}g_{AG}^2{c_{RG}d_{AG}\over c_{AG}d_R}¸_{RI}{}^J
	-üÐÂ_{IKL}Â^{JKL} $$
 where again $ö\A_2$ is the operator representing the one-loop
propagator correction (T-matrix) for self-interacting scalars.  (Of course,
this operator may vary depending on the internal masses; here we are
concerned mostly with the divergences and leading high-energy behavior,
which is mass-independent.  As usual, we can rescale the gauge fields by
their couplings in the Lagrangian; this moves these couplings from the
propagators into the vertices, giving the same result for this term in
$ý$, since it has no $V$'s.)  Of course, this is the identical group theory
that appears in the nonsupersymmetric case; we have been more general
here because we want to consider exact cancellation, while in the
nonsupersymmetric case simplicity is usually more important.

Ü6. Supergluon

The supergluon self-energy calculation is similar to the
nonsupersymmetric cases considered in subsections VIIIA2-3.  Examining
the Feynman rules, we see that those for the vector multiplets are similar
to the nonsupersymmetric ones for vectors (as expected), while those for
the scalar multiplets are similar to those for spinors:  $d^2$ and $Ðd^2$ are
analogous to (in 2$ð$2 matrix notation) $»$ and $»*$, etc.  

There are now only two kinds of loops to consider, vector and scalar
multiplets:  As for the nonsupersymmetric case, ghosts in
background-field gauges couple the same as matter, since at one loop the
only coupling is to background fields and thus covariant, even for ghosts. 
For the real scalar superfield describing the quantum vector multiplet,
looking at the terms in the action quadratic in $V$ (from subsection VIB10)
$$ S_{2V} = Çdx¼d^4 ϼ\f14 V(õ +2i\W^Œ\D_Œ +2iÑ{\W}^{ÀŒ}Ð{\D}_{ÀŒ})V $$
 we see that vertices have only 1 spinor derivative at most.  However, we
need at least 4 spinor derivatives (2 $d$'s and 2 $Ðd$'s) per loop (see
subsection VIC5), since the result of reducing any loop to a point in $Ï$
space always leaves the tadpole $Ï$-integral $[d...d¶^4(Ï-Ï')]|_{Ï'=Ï}$,
which vanishes for fewer than 4 derivatives.  Thus, a $V$ loop in a super
Yang-Mills background vanishes for fewer than 4 external lines.  This
means the entire contribution of quantum super Yang-Mills to the
supergluon propagator correction (or 3-point correction from real
representations) in the background-field gauge comes from the 3 ghosts
(including the Nielsen-Kallosh ghost), which couple the same as $-3$
scalar multiplets in the adjoint representation.  Thus, for example, we
see without evaluating a single graph that this correction vanishes for
N=4 super Yang-Mills, which has also 3 physical adjoint scalar multiplets. 
(See subsection IVC7.)

For the scalar multiplets, we can find the analog of the
squared-propagator trick:  The easiest way is by the method of
subsection IIIC4, which automatically takes care of factors of $ü$, and
can be applied classically, without worrying about functional
determinants.  This method requires we consider the massive theory at
intermediate stages of the calculation, although the mass can be dropped
at the end.  The only resulting limitation is that we must restrict to real
representations of the gauge group.  (In other words, the couplings must
preserve parity:  For these terms, CP invariance is automatic, and reality
means C invariance, so P invariance is implied.)  However, this is a
restriction of the usefulness of the squared-propagator trick anyway: 
Otherwise we get expressions like $(Ö»+iÖA)(Ö»-iÖA*)$ which do not yield
useful simplifications.  (They require as much work as without the trick.) 
In such cases we are stuck with doing the calculations the hard way.  This
is not just a technical difficulty, it is a consequence of the final result
being messier in such cases:  For example, for real representations there
is no possibility of anomalies.  However, we can separate the generators
into the real (scalar) and imaginary (pseudoscalar) ones:  Then this trick
simplifies the real (polar vector) couplings but not the imaginary (axial
vector) couplings.  (As for Pauli-Villars in subsection VIIIB2 below, but
also for the physical fields before taking the mass to vanish after the
trick has been applied, the mass term can be chosen to preserve the polar
symmetries and thus violate the axial ones.)

However, by comparison of the propagator correction for complex and
real representations without (the supersymmetric version of) the
squared-propagator trick, we see that the only difference between the
two is in the (Yang-Mills) group theory.  Thus, we can calculate for real
representations first, using the trick, and then for complex
representations by simply replacing the group-theory factor in the result
for the real ones.  

Repeating the procedure of subsection IIIC4 with spinors replaced with
chiral superfields, we begin with the Lagrangian ($S=Çdx¼L$)
$$ L = -Çd^4 ϼÐÄÄ +\f{m}{å2}\left(Çd^2 ϼüÄ^2 +Çd^2 ÐϼüÐÄ^2\right) $$
 where the chiral superfields are ÓcovariantlyÕ chiral (or
background-covariantly chiral)
$$ Ñá_{ÀŒ}Ä = á_Œ ÐÄ = 0 $$
 Treating $ÐÄ$ as auxiliary (the $ÐÄ^2$ term has no Yang-Mills coupling, as
can be seen, e.g., in an ``antichiral" representation), we eliminate it by its
algebraic field equation
$$ ÐÄ = \f{å2}m á^2 Ä $$
 After a trivial rescaling
$$ Ä £ 2^{-1/4}åmÄ $$
 (and using $Çd^4 Ï=Çd^2 ϼÑá^2$) we obtain the action
$$ L_Ä = -Çd^2 ϼüÄ(Ñá^2 á^2 -üm^2)Ä 
	= -Çd^2 ϼ\f14Ä(õ -m^2 +i[W^Œ,á_Œ])Ä $$
 ($õ=á^a á_a$) using an identity from subsection VIC5.

In the chiral vacuum-bubble loop, we no longer have an explicit chiral
superfield to convert $Ñá^2 á^2$ to $õ+...$.  However, using the chiral
representation $Ñá^2=Ðd^2$, we can write the kinetic operator as
$$ Ðd^2 á^2 = Ðd^2 d^2 +Ðd^2(á^2 -d^2) $$
 to separate the truly free part from the background interactions.  Then
quantization can be preformed as usual (see subsection VIC5): 
Essentially, we can now use the free $-õ_0+m^2=p^2+m^2$ as kinetic
operator, since at each vertex there is a $Ðd^2$ to project back to chiral
superfields.  Of course, in general we need only one projector in any
trace over a subspace:  In this case that result is obtained by integrating
the $Ðd^2$'s by parts in the loop back and forth across the free
propagators, since sandwiching any $á^2-d^2$ between them produces
$$ Ðd^2(á^2 -d^2)Ðd^2 = ü(õ-õ_0 +i[W^Œ,á_Œ])Ðd^2 $$
 Repeating the procedure till only one $Ðd^2$ is left, the Feynman rules for
this loop become
$$ \li{ propagator:â&â{1\over ü(p^2+m^2)}¶^4(Ï-Ï') \cr
	one¼vertex:â&âÐd^2(á^2-d^2) \cr
	other¼vertices:â&âü(õ-õ_0 +i[W^Œ,á_Œ]) \cr} $$

We thus see that one vertex has at most 3 derivatives ($Ðd^2 d$) while the
other has at most 1 ($d$):
$$ Ðd^2(á^2-d^2) = Ðd^2[iA^Œ d_Œ +üi(d^Œ A_Œ) -üA^Œ A_Œ] $$
$$ ü(õ-õ_0 +i[W^Œ,á_Œ]) = iW^Œ d_Œ +üi(d^Œ W_Œ) -ü[W^Œ,A_Œ]
	+iA^a »_a +üi(»^a A_a) -üA^a A_a $$
 exactly the minimum needed.  (Thus, there are insufficient derivatives
for a tadpole contribution to the propagator.)  The result for this diagram
is then the same as the corresponding diagram in bosonic $\Ä^3$ theory,
with a group theory factor $tr(G_i G_j)$, and replacing $\Ä(-p)\Ä(p)$ with
$$ Çd^4 ϼd^4 Ï'¼[iW^{iŒ}(-p,Ï')d'_Œ ¶^4(Ï-Ï')]
	[iA^{jº}(p,Ï)Ðd^2 d_º ¶^4(Ï-Ï')] $$
$$ =  Çd^4 ϼüW^{iŒ}(-p,Ï)A^j_Œ(p,Ï) =  Çd^2 ϼüW^{iŒ}(-p,Ï)W^j_Œ(p,Ï)$$
 using $d¶=-d'¶$, integration by parts, $[Ðd^2 d^2 ¶^4(Ï-Ï')]|_{Ï'=Ï}=1$,
and $W_Œ=Ðd^2 A_Œ$ (chiral representation).  Written in the notation of
subsection VIIIA3, the 2-supergluon part of the unrenormalized 1-loop
effective action is then
$$ ý_{1,2sg} = -\h¼trÇ dx¼d^2 ϼüW^Œ (üc_R)ö{\A}_2 W_Œ $$
 for a scalar multiplet, and exactly $-3$ times that for a vector multiplet,
including the massive case.  Thus, cancellations again survive the
introduction of masses.  Also, if the masses of the various scalar
multiplets are equal the entire propagator correction is canceled in such
theories, while for unequal masses only the divergence, and the
corresponding leading (logarithmic) high-energy term, is canceled.

\x VIIIA6.1  Show this result agrees with the restriction to N=1
supersymmetric theories of the component result of subsection VIIIA3. 

\x VIIIA6.2  Take the result of subsection VIIIA3 literally for ÓallÕ spins
$s$ (arbitrarily large).  Using the fact that multiplets with N+1
supersymmetries can be written as 2 multiplets with N supersymmetries,
differing in maximum helicity by 1/2, recursively find the result for
general $s$ (now labeling maximum helicity) for all values of N$³1$, and
show it vanishes for N$³3$.

\x VIIIA6.3  Calculate the chiral scalar contribution to the one-loop
supergluon propagator correction without the squared-propagator trick. 
(Hint:  There are 8 spinor derivatives in the loop.  Integrating them by
parts off one propagator produces 3 terms, since the number of
$d$'s and $Ðd$'s inside must be equal, because what's left is always
spacetime derivatives on $[Ðd^2 d^2 ¶^4(Ï-Ï')]|_{Ï'=Ï}$.)

Generalizing the group theory as in the previous subsection, we have the
total result
$$ ý_{1,VV} = -\h¼Ç dx¼d^2 ϼÝ_G üW^{iŒ}(üM_G) ú_{Gij}ö{\A}_2 W^j_Œ,ââ
	M_G = Ý_R{c_{RG}\over c_{AG}} -3 $$
 (For Abelian factors, irrelevant for finiteness, we should take the
$c_{AG}$ factor out of $ú_{Gij}$ and put it into $M_G$; then $c_{AG}=0$ for
Abelian groups, so $M_G£Ý_R c_{RG}>0$.)  Therefore, combining with the
results of the previous subsection, the conditions for finiteness are
$$ \boxeq{ Ý_R{c_{RG}\over c_{AG}} = 3,ââ
	Ý_{R,G}g_{AG}^2{c_{RG}d_{AG}\over c_{AG}d_R}¸_{RI}{}^J
	= üÐÂ_{IKL}Â^{JKL} } $$
 In particular, for the case of N=4 super Yang-Mills written in terms of
N=1 superfields (see subsection IVC7), we have 3 adjoint chiral scalars
$Ä_I$ with $I=iI'$, where $i$ is the adjoint label and $I'=1,2,3$ (which
appeared as the label $I$ in subsection IVC7, where the adjoint label was
implicit in matrix notation).  Then
$$ Â^{IJK} = g_A f^{ijk}·^{I'J'K'} $$
 and the above two finiteness conditions reduce to
$$ ¶_{I'}^{I'} = 3,ââ¶_{I'}^{J'} = ü·_{I'K'L'}·^{J'K'L'}ââ
	(¶_i^j = f_{ikl}f^{jkl}) $$
 As explained in the previous subsection, in the general case the finiteness
conditions may receive quantum corrections at 3 loops and beyond,
depending on the model and renormalization prescription, but no new
conditions are added.

Presently there is no deep understanding for the finiteness of these
models (at least, not deep enough to always avoid the quantum
corrections to the finiteness conditions).  Note that they are finite for
arbitrary values of the couplings, up to the two above restrictions:  For
example, we can scale all the couplings by a common factor.  Thus, they
are finite order-by-order in perturbation theory (loops). 
Nonsupersymmetric theories can also be finite, but only for specific
numerical values of the coupling, i.e., not for arbitrarily small values of
the coupling, and thus not order-by-order in the loop expansion; they
therefore suffer from the renormalon problem.  (The renormalon-like
behavior of instantons is not a problem in the framework of the 1/$N_c$
expansion.)  The finiteness of theories with extended supersymmetry has
been explained by various arguments (in particular, for N=2 there are no
divergences beyond 1 loop even for theories that are just
renormalizable), but none of these applies to the general case of simple
supersymmetry.

To obtain more realistic models, we may want to consider adding ``soft"
supersymmetry breaking terms (those which have little effect on
high-energy behavior), as introduced in subsection IVC6, to these finite
theories.  Finiteness can be maintained, but the conditions become
considerably more complicated in the general case.  Note that
spontaneous breaking of supersymmetry is not allowed, because the first
condition prohibits U(1) factors (with $c_{AG}=0$; thus no $Çd^4 ϼV$
terms), while the second prohibits gauge-singlet matter (with $c_{RG}=0$;
thus no $Çd^2 ϼÄ$ terms).

Ü7. Schwinger model

The simplest interacting model in D=2 is the ``Schwinger model", massless
QED.  This theory is even simpler than scalar theories because its
interactions occur only through a massless gauge vector, which has no
physical polarizations in two dimensions (D$-$2=0).

The most interesting feature of the Schwinger model is that all amplitudes
with external vectors can be calculated exactly.  In fact, the only
nonvanishing 1PI vector amplitude is the one-loop propagator correction,
which gives just a mass term.  In that sense the theory is trivial, and
describes just a massive vector.  However, the methods of calculation are
instructive.  We first consider some simple methods of calculation of just
the propagator correction, and then show that it is the only 1PI vector
graph.  One method we have already considered is dimensional
regularization; from subsection VIIIA2-3 we have the contribution to
the effective action (correcting for the 2D normalization $tr(I)=2$)
$$ ý_1 = Çdx¼F{1\over õ}F $$
 where we write $F_{ab}=·_{ab}F$ in D=2.   Although this calculation needs
no renormalization, regularization is still necessary to allow naive
manipulation of the integrand:  Using dimensional regularization, we see
from the result of subsection VIIIA3 that we get a factor of
$\f1{D-1}-4s^2¾·$ in $D=2+2·$, canceling the $1/·$ pole from the scalar
integral.

It can also be calculated in position space, using the methods of the
previous subsection.  The Lagrangian in lightcone notation is
$$ L = -\f1{4e^2}F^2 +[ÐÆ_¢(-i»_{\¢\¢}+A_{\¢\¢})Æ_¢
	+ÐÆ_{\¢}(-i»_{¢¢}+A_{¢¢})Æ_{\¢}] $$
$$ F = »_{¢¢}A_{\¢\¢} -»_{\¢\¢}A_{¢¢} $$
 We can calculate separately the contributions of $Æ_¢$ and $Æ_{\¢}$ to
fermion loops.  The ``photon" propagator correction consists of the
product of two fermion propagators, as given in the previous subsection.
We then find for the effective action (including another $-$1 for $T£ý$
and a $ü$ for identical external lines), after including a ÓfiniteÕ
counterterm to restore gauge invariance,
$$ üA_-(»_+)^2{1\over -üõ}A_- +üA_+(»_-)^2{1\over -üõ}A_+ -A_+ A_- 
	= F{1\over õ}F $$
 (after integration by parts).  

This same calculation also gives the ``axial anomaly":  Consider an axial
vector gauge field $B$ that couples to the current $àÐÆ_ŒÆ_Œ$ (not
summed), in addition to $A$'s coupling to $ÐÆ_ŒÆ_Œ$.  (In D=2,
$W_a=·_a{}^b V_bÜW_à=àV_à$.)  The contribution to the 1-loop effective
action with one of each vector externally is, after including a
counterterm to preserve $A$ gauge invariance (and therefore break $B$
gauge invariance),
$$ -B_-(»_+)^2{1\over -üõ}A_- +B_+(»_-)^2{1\over -üõ}A_+ 
	+B_- A_+ -B_+ A_- = -(»ÉB){1\over -üõ}F $$
 The anomaly is the breaking of $B$ gauge invariance,
$$ ¶B = -»ÂâÜâ¶ý = Ç»É{¶ý\over ¶B} = -2ÇÂF $$

An anomaly is by definition a quantum effect:  As we have seen from the
2D axial anomaly, it is related to a divergence that violates naive classical
arguments, since the regulator itself violates the symmetry.  In the axial
case there is no actual divergent term in the effective action, but a finite
term results from a $·/·$ type of cancellation.  Dimensional analysis
immediately reveals that the propagator correction is the only graph in
D=2 that can contribute such a term from the fermion loop.  (Fermion
propagators go as $1/p$, while the vertex is a constant:  The electric
charge has dimension in D$±$4.)

The complete one-loop effective action for the vectors then follows
directly from the complete anomaly for the axial current, and the
vanishing of the anomaly for the polar current:  By separating out the
anomalous term in the effective action,
$$ ý = ÇF{1\over õ}F +ëý;ââJ = {¶ý\over ¶A},âëJ = {¶(ëý)\over ¶A} $$
$$ »ÉJ = 0,¼»ðJ = -2F¼Ü¼»É(ëJ) = »ð(ëJ) = 0¼Ü¼ëJ = 0¼Ü¼ëý = 0 $$
 (up to an irrelevant constant), where $»ðJ=·^{ab}»_a J_b$ is the curl of
the polar current, but also the divergence of the axial current.  (There are
some questions of boundary conditions in solving the divergence- and
curl-free conditions as $ëJ=0$, but these are resolved by working in
Euclidean momentum space.)

Similar remarks apply to external gravity:  From a similar calculation,
replacing the vector current with the energy-momentum tensor, we find
$$ »_m T^{mn} = 0,¼»_m ·^m{}_n T^{np} ¾ ·^{pm}»_m RâÜâ
	T^m{}_m ¾ R,¼ý ¾ R{1\over õ}R $$
 where $R$ is the 2D curvature (which is just a scalar, as the vector field
strength is a pseudoscalar).  While in the vector case the finite local
counterterm was chosen to preserve polar gauge invariance and thus
violate axial, for the tensor case a term is chosen to preserve local
conservation of energy-momentum and thus violate conformal invariance
$T^m{}_m=0$.  (The above expressions are linearized, but the results can
be generalized to fully nonlinear gravity.)

\x VIIIA7.1  Calculate the gravitational anomalies from a massless spinor
loop in D=2, using the classical expressions (as follow from dimensional
and Lorentz analysis)
$$ T_{++} =  ÐÆ_¢ (-üi)\onª»{}_+ Æ_¢,ââ
T_{--} =  ÐÆ_\¢ (-üi)\onª»{}_- Æ_\¢,ââT_{+-} = 0 $$
 (If you work in terms of $ý$ you can define the perturbative field
$h_{ab}$ such that $¶ý/¶h_{ab}=T^{ab}$.)

The simple form of the effective actions in the Schwinger model is a
consequence of bosonization:  Thus, including coupling to
electromagnetism and gravity, the action for the massless spinor is
equivalent to
$$ L = -\f14 ÄõÄ +(F+R)Ä $$
 Integrating out the scalar generates the above effective actions
ÓclassicallyÕ.  

\x VIIIA7.2 The above action is dual to the mass term of the St¬uckelberg
action:
ªa Consider the first-order Lagrangian
$$ L = -G^2 +G^a(mA_a +»_a Ä) $$
 Eliminating the auxiliary field $G_a$ by its field equation yields the usual
mass term for the St¬uckelberg model.  Show that if we vary $Ä$ instead
and solve the resulting constraint on $G$, we obtain (the nongravitational
part of) the previous action.
ªb Generalize this construction to D=4, where the field dual to the
St¬uckelberg scalar is now an antisymmetric tensor gauge field.  (See
exercise IIB2.1.)

\refs

£1 A.A. Slavnov, ÓTheor. Math. Phys.Õ É10 (1972) 99;\\
	J.C. Taylor, \NP 33 (1971) 436:\\
	graphical form of unitarity for nonabelian gauge theories (later
	simplified as BRST).
 £2 G. 't Hooft, \NP 33 (1971) 173, 
	G. 't Hooft and M. Veltman, \NP 44 (1972) 189, É50 (1972) 318;\\
	B.W. Lee and J. Zinn-Justin, \PRD 5 (1972) 3121, 3137, 3135, 
	É7 (1973) 1049:\\
	renormalization proof for nonabelian gauge theories.
 £3 V.S. Vanyashin and M.V. Terent'ev, ÓSov. Phys. JETPÕ É21 (1965) 375:\\
	first 1-loop gluon coupling renormalization, but neglecting ghosts.
 £4 I.B. Khriplovich, ÓSov. J. Nucl. Phys.Õ É10 (1970) 235:\\
	first complete 1-loop gluon coupling renormalization.
 £5 G. 't Hooft, unpublished comment after K. Symanzik's talk at
	ÓColloquium on renormalization of Yang-Mills fields and 
	applications to particle physicsÕ, Marseille, June 19 -23, 1972;\\
	H.D. Politzer, \PR 30 (1973) 1346;\\
	D.J. Gross and F. Wilczek, \PR 30 (1973) 1343:\\
	asymptotic freedom.
 £6 R.J. Hughes, \PL 97B (1980) 246:\\
	analyzed 1-loop gluon propagator corrections in terms of spinless +
	spin contributions.
 £7 T.L. Curtright, \PL 102B (1981) 17:\\
	extended this analysis to higher spins, supersymmetry, and massless
	sectors of reduced strings.
 £8 Z. Bern, G. Chalmers, L. Dixon, and D.A. Kosower,
	\xxxlink{hep-ph/9312333}, \PR 72 (1994) 2134;\\
	G.D. Mahlon, \xxxlink{hep-ph/9312276}, \PRD 49 (1994) 4438:\\
	generalization of Parke-Taylor amplitudes to one loop.
 £9 Z. Bern, L. Dixon, D.A. Kosower, \xxxlink{hep-ph/9602280},
	ÓAnn. Rev. Nucl. Part. Sci.Õ É46 (1996) 109:\\
	review of modern methods for one-loop graphs.
 £10 H. Georgi, H. Quinn, and S. Weinberg, \PR 33 (1974) 451:\\
	test of running couplings in SU(5) GUT.
 £11 S. Dimopoulos, S. Raby, and F. Wilczek, \PRD 24 (1981) 1681:\\
	test of running couplings in supersymmetric SU(5) GUT.
 £12 P. Langacker and N. Polonsky, \PRD 47 (1993) 4028:\\
	a review of accurate tests of supersymmetric and
	nonsupersymmetric GUTs, emphasizing the improvement from
	supersymmetry.
 £13 A. Parkes and P. West, \PL 138B (1984) 99;\\
	D.R.T. Jones and L. Mezincescu, \PL 138B (1984) 293:\\
	one-loop finiteness implies two-loop finiteness.
 £14 D.I. Kazakov, \PL 179B (1986) 352;\\
	A.V. Ermushev, D.I. Kazakov, and O.V. Tarasov, \NP 281 (1987) 72;\\
	O. Piguet and K. Sibold, \PL 177B (1986) 373:\\
	one-loop finiteness implies all-loop finiteness.
 £15 D.R.T. Jones, L. Mezincescu, and Y.-P. Yao, \PL 148B (1984) 317:\\
	one-loop finite theories with soft breaking.
 £16 I. Jack and D.R.T. Jones, \xxxlink{hep-ph/9405233}, 
	\PL 333B (1994) 372:\\
	one-loop finiteness of softly-broken theories implies two-loop
	finiteness.
 £17 I. Jack, D.R.T. Jones and A. Pickering, \xxxlink{hep-ph/9712542}, 
	\PL 426B (1998) 73,
	I. Jack and D.R.T. Jones, \xxxlink{hep-ph/9907255}, 
	\PL 465B (1999) 148:\\
	one-loop finiteness of softly-broken theories implies all-loop
	finiteness.
 £18 S. Hamidi, J. Patera, and J.H. Schwarz, \PL 141B (1984) 349,
	S. Hamidi and J.H. Schwarz, \PL 147B (1984) 301;\\
	D.R.T. Jones and S. Raby, \PL 143B (1984) 137,
	J.E. Bjorkman, D.R.T. Jones, and S. Raby, \NP 259 (1985) 503;\\
	J. Le«on, J. Perez-Mercader, M. Quiros, and J. Ramirez-Mittelbrunn,
	\PL 156B (1985) 66;\\
	X.-D. Jiang and X.-J. Zhou, \PL 197B (1987) 156;\\
	D. Kapetanakis, M. Mondrag«on, and G. Zoupanos, 
	\xxxlink{hep-ph/9210218}, ÓZ. Phys. CÕ É60 (1993) 181;\\
	D.I.Kazakov, M.Yu.Kalmykov, I.N.Kondrashuk, and A.V.Gladyshev,
	\xxxlink{hep-ph/9511419}, \NP 471 (1996) 389:\\
	realistic finite models.
 £19 Mandelstam; Brink, Lindgren, and Nilsson; Óloc. cit.Õ (VIB):\\
	finiteness proof for N=4 using lightcone superfields.
 £20 Grisaru and Siegel, Óloc. cit.Õ (VIC, ref. 13):\\
	nonrenormalization theorems for extended supersymmetry.
 £21 P. Howe, K.S. Stelle and P. West, \PL 124B (1983) 55:\\
	finiteness proof for N=2 (+ matter) and N=4 using nonrenormalization
	theorems.
 £22 V.A. Novikov, M.A. Shifman, A.I. Vainshtein, and V.I. Zakharov,
	\NP 229 (1983) 381:\\
	finiteness proof for N$³$2 based on instantons.
 £23 P. Jordan, ÓZ. Phys.Õ É93 (1935) 464;\\
	M. Born and N.S. Nagendra Nath, ÓProc. Ind. Acad. Sci.Õ É3 (1936) 318;\\
	A. Sokolow, ÓPhys. Z. Sowj.Õ É12 (1937) 148;\\
	S. Tomonaga, ÓProg. Theo. Phys.Õ É5 (1950) 544:\\
	early attempts at bosonization.
 £24 W. Thirring, ÓAnn. Phys.Õ É3 (1958) 91.
 £25 T.H.R. Skyrme, ÓProc. Roy. Soc.Õ ÉA262 (1961) 237;\\
	D. Mattis and E. Lieb, ÓJ. Math. Phys.Õ É6 (1965) 304;\\
	B. Klaiber, The Thirring model,
	in ÓLectures in theoretical physicsÕ, eds. A.O. Barut
	and W.E. Brittin (Gordon and Breach, 1968) v. X-A, p. 141;\\
	R.F. Streater and I.F. Wilde, \NP 24 (1970) 561;\\
	J. Lowenstein and J. Swieca, ÓAnn. Phys.Õ É68 (1971) 172;\\
	K. Bardak'ci and M.B. Halpern, \PRD 3 (1971) 2493;\\
	G. Dell'Antonio, Y. Frishman, and D. Zwanziger, \PRD 6 (1972) 988;\\
	A. Casher, J. Kogut, and L. Susskind, \PR 31 (1973) 792, \PRD 10 (1974)
	732;\\
	A. Luther and I. Peschel, ÓPhys. Rev.Õ ÉB9 (1974) 2911;\\
	A. Luther and V. Emery, \PR 33 (1974) 598;\\
	S. Coleman, \PRD 11 (1975) 2088;\\
	B. Schroer, ÓPhys. Rep.Õ É23 (1976) 314;\\
	S. Mandelstam, \PRD 11 (1975) 3026;\\
	J. Kogut and L. Susskind, \PRD 11 (1975) 3594:\\
	bosonization (mostly for sine-Gordon $ª$ Thirring).
 £26 J. Schwinger, ÓPhys. Rev.Õ É128 (1962) 2425.

\unrefs

Û7 B. LOW ENERGY

In general, the only loop corrections that can be evaluated exactly in
terms of elementary functions are the one-loop propagator corrections. 
However, limiting forms of vertex corrections, for various low- or
high-energy limits, explicitly yield the most important pieces for certain
applications.

Ü1. JWKB

Some low-energy contributions to the effective action can be obtained by
various quantum mechanical JWKB approximations.  This involves an
expansion of the external field about its vacuum value in spacetime
derivatives (momenta).  Such an expansion makes sense if this field is
massless, since then small spatial momentum means also small energy, in
the relativistic sense.  (Otherwise one needs to expand
nonrelativistically, about  $\vec p=0$ but $E=m$.  Such treatments were
considered in subsection IIB5, and will be applied to loops in subsection
VIIIB6.)  It also can be useful when the mass of the external field is small
compared to the mass scale relevant to the interactions, such as for
chiral symmetry breaking in the low-energy description of light mesons
(subsection IVA4).  

On the other hand, the fields we are integrating out must be massive,
with a mass greater than the energy we want to investigate:  Otherwise,
the internal particles would show up as poles (and cuts) in the
amplitudes, where Taylor expansion in momenta would be a poor
approximation.  The basic principle for analyzing the behavior of such a
theory in a certain energy range is thus to first find contributions to the
effective action where: (1) only particles with masses of lower energy
appear on external (background) lines, and (2) only particles with masses
of higher energy appear on internal (quantum) lines.  These contributions
are approximated by Taylor expansion to finite order in external
momenta, yielding a local effective action.  We could then consider
finishing the functional integration by integrating out the lighter particles
on internal lines:  However, in this approximation it would be inaccurate to
consider such particles in loops, since there they would include energies
above the approximation scale.  Thus, the effective action obtained by
integrating out just the heavier fields is useful only when the lighter
fields are treated classically.  We apply the same approximation scheme
to the classical action:  Eliminate the heavier fields by their classical
equations of motion, and Taylor expand their propagators in momenta to
the desired order to get a local result.

In subsection VIIB2 we saw the simplest example, the effective
potential:  In that case the constant background scalar field acted as just
a correction to the mass.  We now consider more complicated cases,
where spin and gauge invariance play roles for the internal or external
fields.  In particular, adding coordinate dependence to the background
fields means we need to consider more general propagators for quadratic
kinetic operators, such as harmonic oscillators.

We saw in subsection VIB1 the most general relativistic particle action for
a scalar in external fields that was quadratic in $x$ and $Àx$.  We now
consider such actions in more detail:  They are the most general ones for
which we can derive one-loop results to all orders in the external fields
(i.e., without performing the JWKB expansion beyond the first quantum
correction, which requires Taylor expanding the exponential in terms that
are beyond quadratic, thus expanding in the number of external fields). 

Without loss of generality, we can consider Lagrangians that are
homogeneous of second order in $x$ and $Àx$:  Terms linear in $Àx$ are
boundary (in $ $) terms, and were already eliminated by a gauge
transformation (radial gauge).  Terms linear in $x$ can be removed by a
translation, in the presence of an $x^2$ term (which is needed to bound
an $x$ term in the potential).  (Both these kinds of terms can be restored
trivially at the end.)  A constant term is also trivial, giving a contribution
to the classical action that is just that times
$T$ (after integration $Ç_0^T d $), and can be treated separately.  (It
doesn't contribute to the equations of motion.)  The remaining
contribution to the mechanics action is then of the form (as usual, in the
gauge $v=1$)
$$ S = Ç_0^T d ¼ü[-Àx{}^2 +x\A Àx +x\B x]âÜâ¬x +\A Àx +\B x = 0 $$
$$ ÜâS = Ç_0^T d ¼ü(-Àx{}^2 -x¬x) = -ü(xÀx)|_0^T $$
 where $\A$ is an antisymmetric matrix and $\B$ symmetric.  The steps to
this contribution to the one-loop effective field action are then:    
\item{(1)} Solve
the equations of motion, which are homogeneous second-order
differential equations.    
\item{(2)} Change variables from the two parameters
used for each $x$ to $x(0)$ and $x(T)$.  (Second-order differential
equations require two initial conditions, or one initial and one final.)    
\item{(3)}
Find $S(x(0),x(T))$, including separately the contribution from the
constant term in the Lagrangian.    
\item{(4)} Find the propagator for ``time"
$T$, including the $e^{-iS}$ and the van Vleck determinant.  (See exercise
VA2.4.)    
\item{(5)} Integrate the propagator over $T$ to find $ý$.  (See subsection
VIIB2.)

For example, consider in QED the contribution to $ý$ from a fermion loop. 
If we are interested in only the properties of photons, then this gives the
entire contribution to the functional integral from integrating out the
fermions:  This contribution, plus the classical (free) Maxwell action,
gives a nonlocal ``classical" action of self-interacting photons, which can
itself be quantized to give the exact QED result for external photons. 
Although this one-loop effective action is too difficult to calculate exactly,
the first-quantized JWKB approximation can give an accurate description
at energies small compared to the electron mass.  Note that we are
simultaneously approximating to the first quantum correction in JWKB
expansions of both the field (second-quantized) type (one-loop) and the
mechanics (first-quantized) type.

The mechanics action for a massive particle in a constant external
electromagnetic field strength (the lowest nontrivial order, but also the
highest that keeps the action quadratic), in the radial gauge for the
background field and affine parametrization of the worldline, is (see
subsection VIB1)
$$ S = Çd ¼ü(-Àx{}^2 +x^a F_{ab}Àx{}^b +M^2) $$
 To include spin, we identify (see subsection VIIIA3)
$$ M^2 = m^2 -iS^{ab}F_{ab} $$
 Since the only appearance of spin operators in the calculation of the
propagator (denominator) is this constant matrix, it commutes with
everything, so we can treat it as a number till the last step.  The equation
of motion
$$ ¬x +FÀx = 0 $$
 is easily solved in matrix notation.  (Hint:  Solve for $Àx$ first.)  Finding
$x_i=x(0)$ and $x_f=x(T)$ in terms of our integration parameters and
inverting, then expressing $Àx(0)$ and $Àx(T)$ in terms of $x_i$ and $x_f$
(and $T$ and $F$), and making use of the antisymmetry of $F$, the result is
$$ S = -\f14 (x_f -x_i)F¼coth(\f{FT}2)(x_f -x_i) -üx_f Fx_i +üM^2 T $$
 The propagator is then given by (see subsections VA2 and VIIB2)
$$ Òx_f|e^{-iTH}|x_iÔ = å{detÊ{»^2(-iS)\over »x_f »x_i}}Êe^{-iS} $$
 Plugging in, and then Wick rotating $T£-iT$, we find for the propagator
with ends tied together
$$ Òx|e^{-TH}|xÔ = å{detÊ{iF\over 1-e^{-iFT}}}Êe^{-M^2 T/2} $$
 Finally, the contribution to the effective action is (see subsection VIIB2)
$$ ý = -cÇdx Ç_0^¥ {dT\over T}¼tr\left(
	å{detÊ{iF\over 1-e^{-iFT}}}Êe^{-M^2 T/2} 
	-å{detÊ{I\over T}}Êe^{-m^2 T/2} \right) $$
$$ = -cÇdxÇ_0^¥ dT¼T^{-D/2-1}e^{-m^2 T/2}
	\left[å{detÊ{iFT\over 1-e^{-iFT}}}¼tr(e^{iSÉFT/2}) -tr(I)\right] $$
 where $c=-ü$ for fermions, for statistics and squaring the propagator.
(The ``$det$" is for the vector indices on $F_a{}^b$, the ``$tr$" is for the
spin indices from powers of $S^{ab}$ in ``$SÉFÊ$".)

\x VIIIB1.1  Explicitly evaluate the determinant and trace for D=2.

\x VIIIB1.2  Expand $ý$ in $F$ and show the resulting $F^2$ terms agree
with those obtained in subsection VIIIA2-3.

\x VIIIB1.3  Consider the quadratic action
$$ S = Çd ¼ü[-Àx{}^2 +x(a-a^T)Àx -xaa^T x] $$
 where the matrix $a$ commutes with its transpose ($[a,a^T]=0$).  Solve
the field equations for $S(x_i,x_f;T)$.  Find $Òx|e^{-TH}|xÔ$.

Ü2. Axial anomaly

The axial anomaly comes from a finite graph, as we have already seen in
subsection VIIIA7 for the case D=2.  However, the naive manipulations
that would show the graph to preserve gauge invariance involve
evaluating the finite difference between divergent graphs, each of which
needs regularization.  Although in some cases the graph can be evaluated
explicitly, and then shown to be anomalous, it is generally easier, and
more instructive, to analyze the anomaly by itself.

The axial anomaly is associated with the use of $·$ tensors.  In
renormalizable theories in D=4, these occur only through $©_{-1}$'s for
spinors.  (In nonrenormalizable theories, or in D=2, $·$ tensors can occur
in scalar theories.  There is also the term $·^{abcd}F_{ab}F_{cd}$, which is
a total divergence, and has no effect in perturbation theory.)  In general
even dimensions, the massless kinetic term for a spinor is invariant under
transformations generated by $©_{-1}$, but the mass term is not.  Chiral
symmetry is thus related to masslessness; this is also true for conformal
invariance, so it's not surprising that quantum corrections can break
both.  (In fact, in supersymmetric theories conformal symmetry is related
to a particular chiral symmetry by supersymmetry, so breaking of one
requires breaking of the other if supersymmetry is to be preserved.)

Dimensional regularization manifestly preserves neither conformal nor
chiral invariance; no regularization does.  The existence of these
anomalies proves the impossibility of such a regularization.  Furthermore,
dimensional reduction has difficulty dealing with $©_{-1}$; it even has
inconsistencies in the presence of axial anomalies.  On the other hand,
Pauli-Villars regularization is especially convenient for dealing with axial
anomalies because it regularizes by introducing masses.  Thus, it breaks
chiral symmetry explicitly but softly, conveniently parametrizing the
breaking by mass parameters.  We therefore will use Pauli-Villars
regularization for the single purpose of evaluating the axial anomaly.

The basic idea of Pauli-Villars regularization is to include massive
``ghost" fields which would cancel graphs from physical fields if they had
the same mass.  But the masses of the ghosts are used as regulators;
after subtracting local divergences, the regulator mass is taken to
infinity.  In our case, as we'll see by explicit evaluation, the anomaly
itself is finite, so no subtraction is necessary.

$$ \fig{anomaly} $$

The graph whose anomaly we want to evaluate is a one-loop 1PI graph
with external vectors and a massless internal spinor.  Of the vectors, all
but one is a ``polar" vector, coupling to $ÐÆ©_a Æ$, while the last is an
``axial" vector, coupling to $ÐÆ©_{-1}©_a Æ$.  These are the currents
associated with the symmetries $Æ'=e^{iÏ}Æ$ and $Æ'=e^{Ï©_{-1}}Æ$
($©_{-1}^2=-ü$).  We add to this graph a similar one, but with a massive
spinor, and give the second graph an overall relative minus sign.  Since
the mass breaks chiral invariance, we have explicitly broken the gauge
invariance of the axial vector, while preserving those of the polar
vectors.  Note that this is a feature of the regularization:  If a
regularization existed that preserved chiral symmetry, then we could
freely move the $©_{-1}$ around the graph from one vertex to the next
using the usual naive anticommutation relations, thus moving also the
anomaly from one vertex to the next (i.e., violating gauge invariance in
any vector we choose).

Gauge invariance is represented by vanishing divergence of the
corresponding current:  At each vertex we have the coupling $ÇAÉJ$, with
gauge invariance $¶A=»Â$, implying $»ÉJ=0$ by integration by parts,
where $J$ may be polar or axial depending on the vertex.  These currents
are conserved classically.  We know they are also conserved quantum
mechanically in the absence of $©_{-1}$'s, since dimensional
regularization and renormalization preserve the gauge invariance of the
effective action.  In graphical terms, taking the divergence at a vertex
kills a propagator (since $ȃJ$ is proportional to the field equations of the
internal field), and this can be shown to lead to vanishing of the graph.

However, with the Pauli-Villars regulator, the classical conservation of
the axial current is explicitly broken.  The result is that the complete
axial anomaly can be found by looking at just the contribution coming
from this explicit classical violation of current conservation (inserted into
the one-loop graph).  (The classically vanishing contributions are actually
nonvanishing because of the anomaly, but they cancel between the
physical and regulator fields, precisely because the regularization allows
the naive manipulations that justify dropping them.)

We therefore want to evaluate the anomaly
$$ »_a \J^a(x) ­ »_a {¶ý \over ¶A_a(x)} $$
 where we start with a term in the classical action $ÇAÉJ$, so classically
$J=¶S/¶A$, and then evaluate its quantum correction by looking at
$\J­¶ý/¶A$ in terms of the one-loop part of the effective action $ý$. 
Classically, we find a contribution from only the regulator,
$$ »ÉJ ­ »É(-å2iÐÆ©_{-1}©Æ) = 2mÐÆ©_{-1}Æ $$
 So, all we need to evaluate is a one-loop diagram with the axial vector
coupling to the regulator replaced with a pseudoscalar coupling
$ÇÄÐÆ©_{-1}Æ$, and look at the graphs with one external pseudoscalar and
the rest polar vectors.  Clearly this is the same as coupling the
pseudoscalar to the propagator of a ÓbosonicÕ spinor regulator in an
external vector field:
$$ »É\J = 2m¼tr \left( ©_{-1} {1\over iÖá+\f{m}{å2}} \right) =
	2m¼tr \left[ ©_{-1} (-iÖá+\f{m}{å2}) {1\over Öá^2+üm^2} \right] $$
$$ = å2m^2¼tr \left( ©_{-1} {1\over Öá^2+üm^2} \right) $$
 where the trace is in the $©$-matrix space.  In the limit $m£¥$, graphs
with more external lines vanish more rapidly.  On the other hand, we
need at least D/2 factors of $S^{ab}$ (D $©$-matrices) to give a
nonvanishing $©$-matrix trace.  Thus, the leading contribution will be,
using $-2Öá^2=õ+iS^{ab}F_{ab}$ from subsection IIIC4,
$$ »É\J = å2m^2¼tr \left[ ©_{-1} {1\over ü(m^2-õ_0)}
	 üiS^{ab}F_{ab}{1\over ü(m^2-õ_0)} ò
	üiS^{ab}F_{ab}{1\over ü(m^2-õ_0)} \right] $$
 with D/2+1 propagators, where $õ_0=(»_a)^2$.

Thus, the only Feynman diagram we actually need to evaluate is the
one-loop 1PI diagram with external and internal scalars.  The limit
internal $m£¥$ is the same as the limit external $p£0$.  (The result does
not depend on the internal momentum, which is integrated over, nor the
external mass, which would appear only in external propagators.)  Thus,
this is just an effective potential calculation.  We therefore have the
integral (see subsection VIIB1)
$$ Çdk {1\over [ü(k^2+m^2)]^{D/2+1}} = {1\over ý(\f{D}2 +1)üm^2} $$
$$ Üâ»É\J = {2å2\over (\f{D}2)!}¼tr[©_{-1}(üiS^{ab}F_{ab})^{D/2}] $$

\x VIIIB2.1  Check this result by using the expression from subsection
VIIIB1 for the propagator in a constant external electromagnetic field
(strength).

To evaluate in arbitrary even D, we note that the normalization of
$©_{-1}$ is such that we can choose
$$ (©_{-1})^2 = -üâÜâ©_{-1} = (-i)^{D/2}2^{(D-1)/2}©^0 ©^1 ò ©^{D-1} $$
$$ tr(I) = 2^{D/2},â·^{01òD-1} = -1âÜâtr[©_{-1}(üiS^{ab}F_{ab})^{D/2}] =
	\f1{å2}(ü)^{D/2}·^{abòcd}F_{ab}...F_{cd} $$
$$ Üâ»É\J = 2{1\over 2^{D/2}(\f{D}2)!}·^{abòcd}F_{ab}...F_{cd} $$
 Thus, for example, for the Schwinger model (D=2) we have
$$ ȃ\J = -2F $$
 in agreement with subsection VIIIA7, while for D=4
$$ »É\J = \f14 ·^{abcd}F_{ab}F_{cd} $$

\eject

Ü3. Anomaly cancellation

When the anomaly occurs in a current that couples to a gauge field,
unitarity is destroyed, since gauge invariance implies current
conservation.  This is a potential problem, since axial vector couplings
occur in the Standard Model.  (Actually, they are ``V$-$A":
(vector)$-$(axial vector).)  The only way to avoid this problem is to have
an anomaly cancellation between the different spinors:  The coefficient of
the anomaly is given purely by group theory, as $tr(AÓB,CÕ)$, where
$A,B,C$ are the matrices representing the couplings of the three vectors
to all spinors, and the anticommutator comes from Bose symmetrization
(from the crossed and uncrossed graph in the S-matrix, or the single
contribution multiplying commuting fields in the effective action).  We
therefore require this trace (which represents the sum over all spinors)
to vanish.  (See exercises IB5.3 and VC9.2d for an example of the
calculation of this trace.)  The representations in the Standard Model
have been chosen so this cancellation occurs in each family.  

We already know in terms of Dirac notation that axial anomalies appear
only in the presence of $©_{-1}$'s.  An absence of $©_{-1}$'s is equivalent
in terms of Weyl notation to the use of a (pseudo)real representation for
undotted Weyl spinors.  For example, consider a real representation that
is reducible to a smaller (by half) representation ``$\R$" and its complex
conjugate ``$Ð\R$":  Then we can complex conjugate the
complex-conjugate representation to produce a dotted Weyl spinor that
is the same representation as the undotted spinor.  The undotted and
dotted spinor can then be combined, as usual, to form a Dirac spinor,
which transforms as the complex representation, without $©_{-1}$'s, and
thus the same goes for the coupling of the gauge vector:
$$ Æ_{\RŒ}¢Æ_{Ð\RŒ} £ Æ_{\RŒ}¢ÐÆ_{\RÀŒ} £ ï_{\R} $$
 So, in Dirac notation we can see that such representations do not
contribute to anomalies because of the absence of $©_{-1}$'s.  Similar
remarks apply to general real or pseudoreal representations:  We can
take an arbitrary (pseudo)real representation and make a ÓMajoranaÕ
spinor, as
$$ Æ_{\RŒ} £ Æ_{\RŒ}¢ÐÆ_{\RÀŒ} £ ï_{\R} $$
 where now $ÐÆ_{\RÀŒ}$ is simply the complex conjugate of $Æ_{\RŒ}$
since $\R=Ð\R$.

This cancellation also can be seen directly in terms of Weyl spinors:  The
(pseudo)\-reality of the representation is charge conjugation invariance
(which is equivalent to parity invariance for spin-1 couplings to spinors,
since such couplings are always CP invariant).  Anomaly cancellation is
then a generalization of Furry's theorem (see subsection VIIA5).  Real and
pseudoreal representations of the generators (including complex +
complex conjugate) are antisymmetric, up to a unitary transformation,
since they are hermitian:
$$ G^T = G* = -UGU^{-1} $$
 (so $¶Æ=iGÆ$ preserves reality or pseudoreality).  Thus
$$ tr(AÓB,CÕ) = tr(-A^TÓ-B^T,-C^TÕ) = -tr(AÓB,CÕ)âÜâtr(AÓB,CÕ) = 0 $$
 In particular, any mass term (without Higgs) $Æ^{TŒ}Æ_Œ$ requires a
real representation (so its variation yields $G+G^T=0$); a pseudoreal
representation won't work because it uses an antisymmetric metric
which, when combined with $C_{Œº}$, makes $Æ^{TŒ}Æ_Œ$ vanish by
symmetry (since $Æ$ is anticommuting).  A related way to see in Weyl (or
Dirac) notation that real representations are nonanomalous is to use the
same squared-propagator trick we used for the propagator correction in
subsection VIIIA3 (or related complex action from subsection IIIC4),
which resulted in simplified Feynman rules only for real representations: 
With those rules, there are no potentially divergent 3-point graphs other
than those that already occur for scalars (as part of the covariantization
of the propagator divergence).

The absence of $©_{-1}$'s is a special case of parity invariance.  However,
even parity invariance is not enough to enforce cancellation of anomalies,
since some parity invariant theories have axial gauge vectors, which
couple to axial currents $Ðï©_{-1}©_a ï$, and the appearance of these
$©_{-1}$'s can be sufficient to introduce anomalies.  In these anomalous
cases, even if there is a C, the charge conjugation argument above does
not apply because the C following from the usual CP and the obvious P
does not simply replace $A£-A^T$, but is some other permutation of
similar representations.  Thus, in general P (and C) invariance is unrelated
to anomaly cancellation:  We can have one without the other.  Having real
representations (i.e., no $©_{-1}$'s) is a special case of both.

\x VIIIB3.1  Consider chiral symmetry (as in subsection IVA4 or IVB1) for a
single flavor --- U(1)${}_L°$U(1)${}_R$.  Now gauge that symmetry:
 ªa  In Weyl spinor notation, write the action for massless Weyl spinors 
$Æ_{LŒ}$, $Æ_{RŒ}$ each coupled to their own gauge vector.  Clearly
there is one anomaly for 3 external $A_{La}$'s, due to $Æ_L$, and another
for $A_{Ra}$, due to $Æ_R$, and no mixing.  Now assume the left and right
coupling constants are equal (so the anomalies are equal).  Write the
resulting symmetry transformations on all fields under CP, C, and P.
 ªb  Rewrite this theory in Dirac notation.  Using P, find the combinations
of $A_L$ and $A_R$ that are (polar) vector and axial vector.  Relate the
anomaly calculations in the two notations.  Show that dropping the axial
vector gives (massless) QED.  Find the theory that results from dropping
the vector instead:  Give the gauge symmetry, and show it is anomalous,
and explain the anomaly (vs.¼the cancellation of the anomaly in QED) in
both Weyl and Dirac language.
 ªc  Generalize all the above results to U(n)${}_L°$U(n)${}_R$.  (Note that
C will now include complex conjugation on the hermitian matrices for the
vectors, so that P won't.)

The simplest way to prove anomalies cancel in the Standard Model is to
use our previous results for GUTs (subsection IVB4):  (1) One way is to
consider the GUT gauge group SU(4)$°$SU(2)$°$SU(2).  First, we note that
$tr(G_i)=0$ because the group is semisimple, so there are no mixed
anomalies.  Then we see that the SU(4) couplings are the usual
``color"-type couplings, without $©_{-1}$'s (i.e., $4¢Ð4$), so it has no
anomalies.  On the other hand, SU(2) has only (pseudo)real
representations, so neither SU(2) has anomalies.  Thus, anomalies cancel
in the  SU(4)$°$SU(2)$°$SU(2) GUT.  Finally, breaking to
SU(3)$°$SU(2)$°$U(1) (which also spontaneously breaks parity) leaves an
extra singlet per family, which decouples, showing the cancellation for the
Standard Model.  

(2) Another way is to start with SO(10), which is anomaly free for any
representation of fermions:
$$ tr(G_{ab},ÓG_{cd},G_{ef}Õ) = 0 $$
 simply because there is no combination of Kronecker $¶$'s with the
appropriate symmetry (and similarly for SO(N), except for N=2 or 6,
where such a term can be produced with the $·$ tensor).  Breaking to the
Standard Model again drops just a singlet (as does breaking to SU(5),
showing its anomaly cancellation; breaking to SU(4)$°$SU(2)$°$SU(2)
drops nothing, again showing its cancellation).  In general, proving
anomaly cancellation requires (a) using such arguments about real
representations, or (b) the absence of anomalies for certain groups
(namely, only SU(N) for N>2, or U(1), can have anomalies), or (c) explicitly
calculating the relevant traces.

\subsectskip\bookmark0{4. Pi\noexpandº -\noexpand> 2 gamma}
	\subsecty{4. {\sectmath\char'031}${}^{\hbox{\bf 0}}$
	{\sectsymb\char'041} 2{\sectmath\char'015}}	

When an anomalous axial symmetry appears only as a global symmetry
classically, unitarity is preserved, since no gauge field couples to that
current.  This can be a useful way to explain approximate global
symmetries.  The fact that the anomaly is always a total derivative
(because of the $·$ tensor and the Bianchi identity for $F$) means that
the global symmetry is not broken perturbatively.  (However, when the
external vectors are nonabelian, there can be contributions from field
configurations like  instantons:  See subsection IIIC6.)  In subsection
IVA4, we saw that the neutral pion ($¹^0$), the lightest hadron, could be
considered as the pseudogoldstone boson of an axial U(1) symmetry.  We
also want to consider the pion as a bound state of a quark and antiquark: 
If we knew the wave function, we could write the coupling, and calculate
directly the decay of the neutral pion into two photons ($¹^0£2©$) via
quark-antiquark annihilation, or at least find the leading
low-quark-energy contribution from the $¶$-function part of the wave
function (in the relative coordinates of the quark and antiquark),
corresponding to the coupling to $ÐÆ©_{-1}Æ$.  (An expansion of the wave
function in derivatives of the $¶$ function would give coupling to currents
containing derivatives.)  

Lacking such detailed information, the best we can do is extend the
nonlinear $§$ model approach, which is to look for the terms in the
phenomenological Lagrangian (expressed in terms of composite meson
fields, not fundamental quark fields) with fewest derivatives (i.e., those
most important at low energy), applying the condition of (approximate)
chiral symmetry.  Specifically, the global axial symmetry $¹'=¹-2Ï$,
where $A'=A$ for the photon field, along with the electromagnetic gauge
invariance for $A$, under which the neutral pion field is invariant, would
suggest couplings of pion to photon involving only $»¹$ and $F$.  

However, the anomaly allows the existence of another term:  Since by
definition (from considering coupling to an unphysical axial gauge field)
the anomaly is given from a ÓlocalÕ axial transformation, while the pion
field transforms in a trivial way under this transformation, we can
attribute the anomaly to the pion coupling as
$$ ¶¹ = -2Ï,â¶ý = -ÇÏ»É\J,âý = ý_0 +ëý,â¶(ëý) = 0 $$
$$ Üâý_0 = Çü¹»É\J = 
	ǹ{1\over 2^{D/2}(\f{D}2)!}·^{abòcd}F_{ab}...F_{cd} $$
 Thus, in four dimensions we find the contribution
$$ ý_0 = ǹ\f18 ·^{abcd}F_{ab}F_{cd} $$
 Using the abelian form of the Chern-Simons form (subsection IIIC6), we
also can write this as
$$ ý_0 = -Ç\f16 ·^{abcd}(»_a ¹)B_{bcd} $$
 (In the nonabelian case, we can neglect the surface term only if the
vacuum value of $¹$ has already been subtracted.)  Adding this term to
those found previously (the $¹$ and $A$ kinetic terms, as well as the
quark terms that define the normalization of the $¹$ field through its
coupling), the decay rate for $¹^0£2©$ can be calculated (including the 2
relevant flavors of quarks, and 3 colors, using the values of their
electromagnetic charges), and is found to agree closely with the
experimental value.

\x VIIIB4.1  What is $ý_0$ in D=2?  What is the interpretation of the pion
field in terms of the fields of the Schwinger model (subsection VIIIA7)?

The global anomaly in the nonperturbative case can be applied to the
strong interactions (QCD), although not as straightforwardly:  Considering
the external vectors to be gluons (so there is an implicit trace above over
the group indices), $ý_0$ gives a coupling of a neutral meson to a
pseudoscalar glueball, as discussed in subsection IC4.  If the vacuum
gives a nontrivial value to $tr(·^{abcd}F_{ab}F_{cd})$ (as for instantons),
this also leads to anomalous CP violation in the strong interactions.

Ü5. Vertex

One-loop triangle graphs can't be evaluated in terms of elementary
functions.  However, in QED the most important effects are at low energy.
We therefore will evaluate the effective action in the quantum
mechanical version of the JWKB expansion, as an expansion in
derivatives.  The resulting approximation to the effective action thus will
be local, but include terms of higher dimension than the classical action,
whose coefficients are therefore finite and unrenormalized:  By
dimensional analysis, this means their coefficients will have powers of
the inverse electron mass, which can be considered as the expansion
parameter.  (See also subsection VIIB8, where a scalar 1-loop vertex
divergence was evaluated.)

$$ \fig{vertex} $$

The propagator corrections have been found already in subsection
VIIIA1; now we calculate the vertex correction.  The integral is
$$ \A_{a,3,QED} = Çdk¼{\N_a\over\D} $$
$$ \N_a = ©^b(Ök +Öp' +\f{m}{å2})©_a(Ök +Öp +\f{m}{å2})©_b,ââ
	\D = ük^2 ü[(k+p')^2+m^2]ü[(k+p)^2+m^2] $$
 Without loss of generality, we can drop terms that vanish by the free
fermion field equations; this corresponds to canceling them by fermion
field redefinitions.  We then evaluate the numerator by applying the
identities
$$ p^2 = p'^2 = -m^2,ââq = p'-pâÜâ(p+p')^2 = -4m^2 -q^2 $$
$$ Öv©Öv = üv^2 © -vÖv $$
 as well as the identities of subsection VIC4 for $©^b...©_b$, and the field
equations $Öp=\f{m}{å2}$ on the far right and $Öp'=\f{m}{å2}$ on the far left,
to obtain
$$ \N = (Ök+Öp)©(Ök+Öp') +\f{m^2}2© +\f{m}{å2}(2k+p+p') $$
$$ (Ök+Öp)©(Ök+Öp') = (Ök+Öp+Öp')©(Ök+Öp+Öp') -Öp©Öp -Öp'©Öp' -Öp'©Öp -Ök©Öp -Öp'©Ök $$
$$ = [ük^2 +kÉ(p+p') -2m^2 -üq^2]© -(k+p+p')(Ök+2\f{m}{å2}) +m^2 ©
	+\f{m}{å2}(p+p') -\f{m^2}2 © +\f{m}{å2}k $$
$$ Üâ\N = (ük^2 © -kÖk) +[kÉ(p+p')© -(p+p')Ök +\f{m}{å2}k] -(m^2 +üq^2)© $$

For the momentum integral we evaluate
$$ \A_3(x,m^2,q^2) = Çdk¼{e^{ikÉx}\over\D} = Çd^3  ¼Â^{-D/2}e^{-E} $$
$$ E = ü\f1 x^2 +ixÉü[(Œ_1+Œ_2)(p+p') +(Œ_1-Œ_2)q] 
	+üÂ[(Œ_1+Œ_2)^2 m^2 +Œ_1 Œ_2 q^2] $$
 again on the fermion mass shell.  We also have
$$ Çd^3   = Ç_0^¥ d¼Â^2 Ç_0^1 d^3 Œ¼¶\left( 1-݌ \right)
	= Ç_0^¥ d¼Â^2 Ç_0^1 dŒ_1 Ç_0^{1-Œ_1} dŒ_2 $$
 using, e.g., the definitions
$$ Ç_a^b dx¼¶(x)f(x) = Ï(-a)Ï(b)f(0),ââ
	Ç_{-¥}^¥ dx¼Ï(x-a)Ï(b-x)f(x) = Ç_a^b dx¼f(x) $$
 As for the fermion propagator, we clearly separate UV and IR divergent
integrals by the changes of variables
$$ Œ = Œ_1 +Œ_2,â⺠= Œ_1 -Œ_2âÜâ
	Çd^3   = Ç_0^¥ d¼Â^2 Ç_0^1 dŒ¼üÇ_{-Œ}^Œ dº $$
 followed by
$$  £ {Â\over Œ^2},â⺠£ ŒºâÜâ
	Çd^3   £ Ç_0^¥ d¼Â^2 Ç_0^1 dŒ¼Œ^{-5}¼üÇ_{-1}^1 dº $$
 which modifies the integral to
$$ \A_3 = Ç_0^¥ d¼Â^· Ç_0^1 dŒ¼Œ^{-1-2·}¼üÇ_{-1}^1 dº¼e^{-E} $$
$$ E = üÓ\f1 Œ^2 x^2 +iŒxÉ(p+p'+ºq) +Â[m^2 +\f14 (1-º^2)q^2]Õ $$

We now expand to $\O(x^2)$ and $\O(q^2)$.  The $º$ integral is then
trivial (the integrand becomes quadratic in $º$), the $Â$ integral gives the
usual, and the $Œ$ integral is similar to the case of the fermion
propagator.  The result is
$$ \A_3 ® -üý(1+·)(üm^2)^{-·}\leftÓ-\f16 {q^2\over m^2}+
	\left(1-\f16 {q^2\over m^2}\right)\left[{1\over ·_{IR}} +ixÉ(p+p')
	\right.\right. $$
$$ \left.\left.\phantom{q^2\over m^2}+\f18 (xÉ(p+p'))^2 +\f14 m^2 x^2
	\right] +\f1{24}(xÉq)^2 +{1\over ·_{UV}}\f14 m^2 x^2\rightÕ $$
 This leads to the expression for the vertex correction
$$ \A_{3,QED} ® ý(1+·)(üm^2)^{-·}\leftÓ\left[\left(1+\f13{q^2\over m^2}
	\right){1\over ·_{IR}} +ü{1\over ·_{UV}} +\f72 +\f1{12}{q^2\over m^2}
	\right] © \right. $$
$$ +\left.\f14\left(1 -\f16{q^2\over m^2}\right){p+p'\over m/å2}\rightÕ $$
 where we have used
$$ ¶_a^a = D $$
 in evaluating the contribution from the $k^2$ term.  (Remember that all
algebra from indices on the fields is done in 4 dimensions, while all
algebra from indices on momenta is done in $D$ dimensions.  Since the
two parts of the calculation are usually done separately, this should
cause no confusion; however, the difference in evaluating $¶_a^a$ is the
main thing to watch.)  Using the on-shell identity
$$ 4\f{m}{å2}© = ÓÖp+Öp',©Õ +[Öq,©] = -(p+p') +[Öq,©] $$
 we can rewrite this as (again keeping only $\O(q^2)$)
$$ \A_{3,QED} ® ý(1+·)(üm^2)^{-·}\leftÓ\left[\left( {1\over ·_{IR}} 
	+ü{1\over ·_{UV}} +\f52\right) +\left(\f13{1\over ·_{IR}}
	+\f14\right){q^2\over m^2}\right] © +\f14{[Öq,©]\over m/å2}\rightÕ $$

The next step is to cancel the UV divergence by adding the counterterm
for electron wave-function renormalization from subsection VIIIA1:
$$ \A_{3,QED,r} = \A_{3,QED} +¶\A_{3,QED} ® 
	\left[\f13\left({1\over ·_{IR}} -lnÊ{m^2\over µ^2}\right)+\f14\right]
	{q^2\over m^2}© +\f14{[Öq,©]\over m/å2} $$
 Equivalently, we can take the $q=0$ piece of $\A_{3,QED}$, and note that
it combines with $\A_{2e}$ of subsection VIIIA1 to gauge-covariantize the
term proportional to $Öp£Öp+ÖA$.  (The unrenormalized effective action is
thus automatically gauge invariant, as is the counterterm.)  At this point
we can see the anomalous magnetic moment:  Combining the tree and
1-loop result (including coupling), and writing as spinless + magnetic
moment contributions, we have
$$ © +e^2\A_{3,QED,r} ® \leftÓ1 +e^2
	\left[\f13\left({1\over ·_{IR}} -lnÊ{m^2\over µ^2}\right)+\f14\right]
	{q^2\over m^2}\rightÕ(-\f14){p+p'\over m/å2}
	+(1+e^2)\f14{[Öq,©]\over m/å2} $$

We can translate these 1-loop corrections into a contribution to the
effective action as (with the usual $-1$ for the effective action)
$$ ý_{1,3,QED,r} = -Ðï(-p')\A_{3,QED,r}ï(p)ÉA(q) $$
 We then note, again using the spinor (free) field equations to imply
$(p+p')Éq=0$, to $\O(q^2)$,
$$ -4\f{m}{å2}q^a ©^b q_{[a}A_{b]} ® q^a (p+p')^b q_{[a}A_{b]} 
	= q^2 (p+p')ÉA $$
 The low-energy part of the renormalized effective action exhibiting up
to order $q^2/m^2$ corrections to the coupling is then, in gauge invariant
form,
$$ ý_{0+1,2e,r} ® Çdx¼Ñï\BiggÓiÖ» -ÖA+\f{m}{å2} 
	-{e^2\over 2å2m}iS^{ab}F_{ab} $$
$$ \left. -{e^2\over m^2}
	\left[\f13\left({1\over ·_{IR}} -lnÊ{m^2\over µ^2}\right)+\f14\right]
	©^a(»^b F_{ab})\rightÕï $$
	
\x VIIIB5.1  Perform the supergraph version of this calculation:
a massless Abelian vector multiplet coupled to a massive
chiral scalar multiplet.

Ü6. Nonrelativistic JWKB

As for other processes, the application of quantum field theory to bound
states has two steps:  (1) Calculate the (gauge-invariant) effective
action; (2) find solutions to the field equations following from the
effective action (``on-shell" states).  For bound states such solutions are
nonperturbative; however, their determination is easier for nonrelativistic
systems, since we can ignore production and annihilation of additional
nonrelativistic (massive) particles in the second step because their effect
already has been included as small corrections to the effective action.

The Lamb shift is the (field theoretic) quantum contribution to the energy
levels of the hydrogen atom, which is described accurately even at one
loop.  The relativistic solution is found by perturbing the relativistic
effective action in derivatives about the nonrelativistic one, whose
solutions are the usual exact ones of the nonrelativistic Schr¬odinger
equation.  For atoms the electron speed $p/m$ is of the order of
$Œ$ ($=2¹e^2$), so the loop and derivative expansions are in the same
small parameter.

The effective action is more conveniently calculated with manifestly
relativistic methods, since the internal (``virtual") particles can be
relativistic (especially those that contribute to the UV divergences).  On
the other hand, the solutions to the field equations are more conveniently
calculated in a representation that takes better advantage of the
nonrelativistic expansion, since the external particles are nonrelativistic. 
Therefore, the second step begins by performing a field redefinition that
converts the manifestly Lorentz invariant effective action to a form
recognizable as nonrelativistic field theory with low-energy relativistic
and loop corrections.  (Originally the Lamb shift was calculated without
this transformation.  Higher-order calculations then required use of the
relativistic Bethe-Salpeter equation, which made collection of terms of a
given order more difficult.)  In subsection IIB5 we considered the
generalized Foldy-Wouthuysen transformation and its application to
minimal coupling; we now apply it to the nonminimal coupling introduced
by loop corrections.  (In the literature this step has been performed on the
Feynman diagrams themselves; however, as usual we can save some
effort by working directly with the effective action.)  Here the
nonminimal correction to the transformation is easy, since the
nonminimal terms are already near the order to which we work.

We first perform some dimensional analysis, using the fact that the
leading behavior is given by the usual nonrelativistic Schr¬odinger
equation.  Then the only parameters in units $\h=1$ (but there is no $c$
in the nonrelativistic theory with just Coulomb interaction) are the mass
$m$ and ÓspeedÕ $e^2$, so (in the notation of subsection IIB5)
$$ ¹^i ¾ me^2,ââ¹^0 ¾ me^4 $$
 (neglecting the rest mass contribution).  It is then convenient to
reorganize the expansion in $1/m$ to relate to the expansion in $e^2$: 
For example, we can identify the two by choice of units
$$ \f1m ¾ e^2âÜâ¹^i ¾ 1,ââ¹^0 ¾ \f1m $$
 along with $c=1$ (since we will include relativistic corrections).

The relativistic form of the Schr¬odinger equation is obtained by
multiplying $2©^0$ in front of the kinetic operator of the electron in a
background electromagnetic field, as obtained from the effective action. 
Approximating the proton as infinitely massive (for which we can
partially correct by using the reduced mass for the electron), we take the
electric field as described by the usual static ``scalar" potential, and drop
the magnetic field along with the ``vector" potential.   

We therefore modify the expansion of subsection IIB5 by   
\item{(1)} reorganizing
the $1/m$ expansion according to our dimensional analysis,   
\item{(2)} using only a static electric background, and   
\item{(3)} working directly in terms of $©$
matrices:  We can either plug in the Dirac case of the spin operators into
the expressions of subsection IIB5, including the reality-restoring
transformation of subsection IIB4,
$$ S^{ab} £ -ü©^{[a}©^{b]},ââS^{-1a} £ -\f1{å2}©^a $$
 or just expand the Dirac operator directly,
$$ 2©^0 (-Ö¹ +\f{m}{å2}) = ¹^0 -2©^0 ©^i ¹^i +å2m©^0 $$
 (and similarly for the loop correction terms).  

\noindent From the reuslts of the
previous subsection, we thus choose to order $1/m^4$
$$ \E_{-1} = å2©^0,ââ\O_0 = 2©^ip^i ©^0,ââ\E_1 = m¹^0 $$
$$ \O_3 = -m^2 e^2\f1{å2}i©^i F^{0i},ââ\E_4 = m^2 e^2
	\left[\f13\left({1\over ·_{IR}} -lnÊ{m^2\over µ^2}\right)+\f14\right]
	»^i F^{0i} $$
 (others vanishing), where we have included explicit $m$ dependence so
that the coefficients $\E_n$ and $\O_n$ are of order $m^0$ according to
our above dimensional analysis (so our expansion in $m$ makes sense). 
Using
$$ tanh¼x ® x -\f13 x^3 $$
 the relevant commutators from IIB5 are then, for the nonvanishing
generators to this order
$$ mG = -üÓ[G,ë\E] +\L_G coth(\L_G)\OÕ\E_{-1} $$
$$ \li{ ÜâG_1 = {}& -ü\O_0 \E_{-1} \cr
	G_3 = {}& -ü([G_1,\E_1] +\f13[G_1,[G_1,\O_0]])\E_{-1} \cr
	G_4 = {}& -ü\O_3 \E_{-1} \cr} $$
 and for the transformed kinetic operator
$$ \F' = \E +tanh(ü\L_G)\O $$
$$ \li{ Üâ\F'_1 = {}& \E_1 +ü[G_1,\O_0] \cr
	\F'_3 = {}& ü[G_3,\O_0] -\f1{24}[G_1,[G_1,[G_1,\O_0]]] \cr
	\F'_4 = {}& \E_4 +ü[G_1,\O_3] +ü[G_4,\O_0] \cr} $$
 Remembering that $\E_{-1}$ commutes with even and anticommutes
with odd, we have identities like
$$ (\L_{G_1})^n \O_0 = (-1)^{n(n-1)/2}(\O_0)^{n+1}(\E_{-1})^n,ââ
	 (\E_{-1})^2 = 1 $$
 Substituting for $G$ into $\F'$:
$$ \li{ \F'_1 = {}& \E_1 +ü(\O_0)^2 \E_{-1} \cr
	\F'_3 = {}& -\f18[\O_0,[\O_0,\E_1]] -\f18 (\O_0)^4 \E_{-1} \cr
	\F'_4 ={}& \E_4 +üÓ\O_0,\O_3Õ\E_{-1} \cr} $$
 The final result is, using 
$$ (\O_0)^2 = (p^i)^2,ââme^2[\O_0,\E_1] = -2\O_3\E_{-1} $$
 and setting $\E_{-1}=-1$ on the right for positive energy,
$$ \li{ \F'_1 = {}& m¹^0 -ü(p^i)^2 \cr
	\F'_3 = {}& -\f14m[ü(»^i F^{0i}) -iS^{ij}ÓF^{0i},p^jÕ] +\f18(p^i)^4 \cr
	\F'_4 ={}& m^2 e^2 \left[
		\f13\left({1\over ·_{IR}} -lnÊ{m^2\over µ^2}\right)»^i F^{0i}
		+iüS^{ij}ÓF^{0i},p^jÕ \right] \cr} $$
 As expected from dimensional analysis, $\F'_1$ is the nonrelativistic
result, $\F'_3$ is the lowest-order relativistic correction, and $\F'_4$ is
the lowest-order part of the one-loop correction.  Putting it all together,
to this order we have
$$ \F' ® ¹^0 -\left[ {(p^i)^2\over 2m} -{(p^i)^4\over 8m^3} \right]
	-\f1{8m^2}\left[1 -\f83 e^2
	\left({1\over ·_{IR}} -lnÊ{m^2\over µ^2}\right)\right]»^i F^{0i}
	+{1+2e^2\over 4m^2}iS^{ij}ÓF^{0i},p^jÕ $$

\x VIIIB6.1 Find the additional terms in $\F'$ to this order when the
electromagnetic field is arbitrary (magnetic field, time derivatives of
background), assuming the same dimensional analyis for the background.

Ü7. Lattice

Integrals are defined as limits of sums.  For some cases it can be
convenient to define quantum theories on discrete spacetimes
(``lattices"), perform all calculations there, and then take the limit of
continuous spacetime.  Two types of such lattices will be considered here:
(1) Physical four-dimensional spacetime can be treated as a regular
hypercubic lattice.  Then the existence and uniqueness of a continuum
limit where Lorentz invariance is restored must be proven.  (2) In
first-quantization of particles or strings, the worldline or worldsheet can
be approximated as a random lattice (see subsection XIA7).  Integration over the metric of the
worldline or worldsheet is then replaced with summation over lattices
with different geometries.  The continuum limit is not required by
physical criteria, but only for purposes of comparison to the theory as
defined in the continuum.

The use of a regular 4D lattice for quantizing QCD has three main
advantages:  
\item{(1)} The lattice acts as a gauge invariant regulator for UV
divergences (and, if the lattice is finite, also IR ones).  
\item{(2)} Gauge fixing is
no longer necessary, since the path integral can be performed without it. 
\item{(3)} Nonperturbative calculations are possible, some analytically and some
numerically (if the lattice is small enough).

Gauge fields are associated with translations through the covariant
derivative.  However, on a lattice, even a regular one, infinitesimal
translations are no longer possible:  For example, scalar fields are defined
only at vertices of the lattice.  We therefore consider covariantizing
finite translations, as in subsections IIIA5 and IIIC2,
$$ e^{-k^m á_m} = \P\left[ exp \left( 
	-iÇ_{x-k}^x dx'ÉA \right) \right] e^{-kÉ»}
	= U_{x,x-k} e^{-kÉ»} $$
 Without loss of generality, we can restrict ourselves to translations
along links, from one vertex straight to an adjacent one (keeping all
coordinates but one constant), and successive combinations of these. 
Then the gauge field is replaced with the group element $U_{x,x-k}$
associated with each link, where $k$ is now any of the 4 orthonormal
basis vectors (in Euclidean space).  The gauge transformation of this
representation of the gauge field follows from either the path-ordered
definition or the covariant-translation definition:
$$ e^{-kÉá(x)'} = {\bf g}(x)e^{-kÉá}{\bf g}^{-1}(x)âÜâ
	U_{x,x-k}' = {\bf g}(x)U_{x,x-k}{\bf g}^{-1}(x-k) $$
 Note that, while the gauge field is a group element associated with a
link, the gauge transformation is a group element associated with a
vertex.  Furthermore, the field strength can be associated with the
product of these group elements of the links bounding a ``plaquet":
$$ U_{x,x-k}U_{x-k,x-k-k'}U_{x-k-k',x-k'}U_{x-k',x}
	= \P\left( e^{-iÈdxÉA} \right) = e^{-kÉá}e^{-k'Éá}e^{kÉá}e^{k'Éá} $$
$$ ® e^{[kÉá,k'Éá]} ® 1 +ik^a k'^b F_{ab} $$
 where we have used
$$ e^B e^C = e^{B+C+ü[B,C]+...} $$
 (In general, there is a geometric prescription associating a scalar with a
point, a vector with a line, a second-rank antisymmetric tensor with a
surface, etc.)

We now define a gauge-invariant action by looking for an expression in
terms of these group elements that approximates the usual Yang-Mills
action to lowest order in the lattice spacing, while involving the least
number of factors of the group elements.  The result is:
$$ S = -\f1{g^2}tr Ý_{plaquets}
	(U_{x,x-k}U_{x-k,x-k-k'}U_{x-k-k',x-k'}U_{x-k',x} -1) $$
$$ ® -\f1{g^2}tr Ý_{plaquets}ü(ik^a k'^b F_{ab})^2
	¾ \f1{g^2}tr Ý_x F^2(x) $$
 (expanding the exponential as above to quadratic order, and noting that
total commutators vanish when traced).  Since our fields are now
represented by group elements, we no longer need to fix the gauge to
make the functional path integral well defined:  In contrast to the
continuum case, where integrating a gauge-invariant action over gauge
transformations would produce an infinite factor, here such an integral at
any one point is just an integral over the group space, which is finite (for
compact groups, which have finite volume).  The functional integration is
now integration over $U$ for each link, where the range of $U$ is the
group space (which is finite, since the group is compact).

Matter can also be introduced:  Scalars are naturally associated with
vertices, just as vectors are with links, and second-rank antisymmetric
tensors with plaquets.  However, fermions do not have such a natural
geometric interpretation.  In particular, it has been proven (the
``Nielsen-Ninomiya theorem") that massless fermions can't be defined in a
useful way on the lattice without ``fermion doubling":  There must be a
multiple of $2^D$ massless fermion fields for D lattice dimensions.  This is
closely related to the existence of axial anomalies:  The absence of an
anomaly is implied by the existence of a regularization that manifestly
preserves a symmetry (in this case, chiral symmetry as a consequence of
the existence of lattice-regularized fermions).  However, massless
fermions can be defined as limits of massive ones (so chiral invariance is
not manifest).  Alternatively, ÓnonlocalÕ spinor kinetic operators can be
found that preserve masslessness and chirality without doubling.  (The
nonlocality can be controlled, but at the cost of a significantly more
complicated action.)

\x XIC1.1  For the lattice action for a spinor in D=1, use
$$ S = üi Ý_n Æ_{n+1} Æ_n $$
 where $Æ$ is a real one-component fermion.
 ªa Show this has the correct continuum limit.
 ªb Find the equations of motion.
 ªc Solve the equations of motion for both the lattice and
continuum cases, and show the lattice has twice as many solutions.
 ªd Repeat all the above for the single-component complex
(Dirac) fermion,
$$ S = i Ý_n (ÐÆ_{n+1} -ÐÆ_n)Æ_n $$
 ªe Make the same analysis for the D=1 scalar, and show it has no
such problems.

\x XIC1.2  In the book by Feynman and Hibbs, exercise 2-6 states
rules for the path integral for a Dirac spinor in D=2.  These rules are
equivalent to the use of a lightlike lattice, where the lightcone
coordinates are discretized.  The rules are to consider all
paths that are piecewise lightlike forward in time, 
with a factor of $im·$ for each
right-angle ``kink" (where $m$ is the mass and $·$ the lattice
spacing).  Show these rules follow from the 2D action for a Dirac spinor
(see subsection VIIB5, and include a mass term), using a term as
in the previous exercise for the derivative term for each of
the two component fields (each of which has a derivative in only
one of the two lightlike coordinates).

In general, fermions are more difficult to integrate over, particularly
when using ``numerical methods" (computers), since fermions are not
numbers.  In principle one can integrate out the fermions analytically to
produce functional determinants in terms of bosonic fields, but
nonlocality makes them hard to evaluate by iterative schemes.  In
practice fermion loops are usually ignored (``quenched approximation"),
which corresponds to leading order in an expansion in the inverse of the
number of flavors, or the approximation of heavy quarks.  The resulting
accuracy of QCD calculations for low-energy parameters (masses of light
hadrons, decay constants, etc.) is of the order of 5-10\%.  (Getting good
numbers in nonperturbative calculations is significantly harder than in
perturbative ones.  The situation is expected to improve somewhat with
the advent of faster computers.)  Finding scattering amplitudes, or other
properties that involve high-mass hadrons, is presently beyond the scope
of lattice methods.  However, lattice QCD is one of the few methods so far to
obtain numbers for comparison with experiment from nonperturbative
calculations with the QCD action.  (Other nonperturbative methods
have also been restricted to low-mass hadrons, and basically study
effects of chiral symmetry breaking, not confinement.)

The spacetime lattice allows a direct nonperturbative analysis of
confinement.  For example, consider the potential between a heavy
quark-antiquark pair.  The heaviness again allows us to ignore pair
creation, and to treat the quarks as static.  For simplicity, consider scalar
quarks, as described by first-quantization.  Since we approximate the
quarks as static, the only relevant term in the quark mechanics action is
the interaction term $Çd ¼ÀxÉA=ÇdxÉA$.  Taking into account the nonabelian
nature of the group, and ignoring the first-quantized path integration
$Dx$ (since $x$ is assumed fixed), the factor $e^{-S}$ for the
quark becomes just the path-ordered expression $\P(e^{-iÇdxÉA})$ we
have been considering, while for the antiquark we get the inverse
expression.  To get a gauge-invariant expression, we connect the paths
at top and bottom, since the fields will be fixed at the boundaries at
$t=à¥$.  (Functional integration over any gauge-field link picks out just
the singlet part of the integrand, since the integral is over the group, and
nonsinglet representations can be rotated to minus themselves by an
appropriate group element, canceling the contribution.)  The result is a
``Wilson loop"
$$ tr¼\P\left( e^{-iÈdxÉA} \right) $$

The strong-coupling expansion is applied by expanding the functional
integrand $e^{-S}$ in powers of $S$, which is an expansion in powers of
$1/g^2$, and which is also an expansion in the number of plaquets. 
Clearly the dominant term in this expansion is the one with the fewest
factors of $S$.  To be nonvanishing, each link variable must appear in a
singlet combination:  The function of that link, when expanded in
irreducible group representations, must include a term that is
proportional to the identity.  For example, for any unitary group, this is
true for the product $U_i{}^j (U^{-1})_k{}^l$, where the two $U$'s are for
the same link, and the indices are the group indices; this has the constant
piece $U_i{}^j (U^{-1})_j{}^i$.  For the case of the Wilson loop, if we
assume the simplest case where the path is a rectangle, then we need at
least a factor of $S$ for each plaquet enclosed by the loop, so there will a
$UU^{-1}$ for each link on the boundary (one factor from the loop, one
from $S$), as well as for each link enclosed by it (both factors from the
contribution to $S$ from either side).  The result for the path integral is
then
$$ \A = \leftÒtr¼\P\left( e^{-iÈdxÉA} \right)\rightÔ
	 ¾ e^{-Vt} ¾ \left( {1\over g^2} \right)^{rt} $$
 where $V$ is the (potential) energy ($S=Çdt(V+T)$ in Euclidean space), $t$
is the time separation between the top and bottom of the rectangle, and
$r$ is the spatial separation between the two sides.  We thus have a linear
quark-antiquark potential
$$ V(r) ¾ (ln¼g) r $$
 so the quark-antiquark pair is confined.

Unfortunately, we can get a similar result from QED, by defining the U(1)
group in terms of a phase factor (so effectively the range of group
integration is $2¹$, defining a ``compact" group).  The reason is that
for this U(1) theory this strong coupling expansion is not accurate.  The
approximation is better for nonabelian theories, but the persistence of
confinement has not been proven in the continuum limit (small coupling). 
In fact, while the transition to deconfinement in Abelian theories has
been found at finite coupling, it has been proven that such a phenomenon
can occur in the nonabelian theory only near zero coupling.  However, the
perturbative properties of the continuum theory show that this is exactly
where one expects the appearance  of ambiguities in the theory (known
in lattice terminology as ``nonuniversality"):  As seen in subsection VIIC1, analytic continuation of the coupling near the positive real axis runs into trouble only near $g$=0.  Although these problems might be resolved in finite theories, such theories require supersymmetry, which is difficult to treat on a lattice because of its problems with fermions, as discussed above.

\refs

£1 W. Pauli and F. Villars, ÓRev. Mod. Phys.Õ É21 (1949) 434.
 £2 J. Steinberger, ÓPhys. Rev.Õ É76 (1949) 1180;\\
	Schwinger, Óloc. cit.Õ (VB, ref. 2, second ref.):\\
	triangle graph, for $¹^0£2©$.
 £3 J.S. Bell and R. Jackiw, ÓNuo. Cim.Õ É60A (1969) 47:\\
	identified axial anomaly.
 £4 S. Adler, ÓPhys. Rev.Õ É177 (1969) 2426:\\
	explained axial anomaly.
 £5 S.L. Adler and W.A. Bardeen, ÓPhys. Rev.Õ É182 (1969) 1517:\\
	showed axial anomaly is only at 1 loop.
 £6 H. Georgi and S.L. Glashow, \PRD 6 (1973) 429:\\
	anomaly cancellation for general GUTs.
 £7 H. Bethe, ÓPhys. Rev.Õ É72 (1947) 339:\\
	calculation of Lamb shift.
 £8 H. Fukuda, Y. Miyamoto, and S. Tomonaga, ÓProg. Theo. Phys.Õ É4 (1949)
	47, 121;\\
	N.M. Kroll and W.E. Lamb, Jr., ÓPhys. Rev.Õ É75 (1949) 388;\\
	Y. Nambu, ÓProg. Theo. Phys.Õ É4 (1949) 82;\\
	J.B. French and V.F. Weisskopf, ÓPhys. Rev.Õ É75 (1949) 1240:\\
	relativistic calculation of Lamb shift.
 £9 W. E. Caswell and G. P. Lepage, \PL 167B (1986) 437:\\
	nonrelativistic effective actions.
 £10 L.S. Brown, Óloc. cit.Õ (VA), p. 511;\\
	A. Pineda and J. Soto, \xxxlink{hep-ph/9711292}, \PL 420B (1998)
	391:\\
	Lamb shift with only dimensional regularization.
 £11 ÓQuantum electrodynamicsÕ, ed. T. Kinoshita (World Scientific,
	1990):\\
	higher-loop QED.
 £12 K.G. Wilson, \PRD 10 (1974) 2445:\\
	QCD lattice.
 £13 H.B. Nielsen and M. Ninomiya, \NP 185 (1981) 20,
	É195 (1982) 541, É193 (1981) 173, \PL 105B (1981) 219.
 £14 P.H. Ginsparg and K.G. Wilson, \PRD 25 (1982) 2649;\\
	P. Hasenfratz, \xxxlink{hep-lat/9709110},
	ÓNucl. Phys. Proc. Suppl.Õ É63A-C (1998) 53,\\
	\xxxlink{hep-lat/9802007}, \NP 525 (1998) 401,
	P. Hasenfratz, V. Laliena, and F. Niedermayer, 
	\xxxlink{hep-lat/9801021}, \PL 427B (1998) 125;\\
	H. Neuberger, \xxxlink{hep-lat/9707022}, \PL 417B (1998) 141, 
	\xxxlink{hep-lat/9801031}, \PL 427B (1998) 353;\\
	M. L¬uscher, \xxxlink{hep-lat/9802011}, \PL 428B (1998) 342:\\
	avoiding Nielsen-Ninomiya, while preserving chiral invariance.
 £15 Feynman and Hibbs, Óloc. cit.Õ (VA).
 £16 C. Destri and H.J. de Vega, \NP 290 (1987) 363, 
	\PL 201B (1988) 261, ÓJ. Phys. AÕ É22 (1987) 1329:\\
	regular (square) 2D lightcone lattice.
 £17 M. Creutz, \PRD 21 (1980) 2308;\\
	E. Marinari, G. Parisi, and C. Rebbi, \PR 47 (1981) 1795:\\
	computer calculations with lattice QCD.
 £18 E.T. Tomboulis, \PRD 25 (1982) 606, \PR 50 (1983) 885:\\
	deconfinement can occur only at zero coupling.
 £19 H.J. Rothe, ÓLattice gauge theories, an introductionÕ (World Scientific,
	1992).

\unrefs

Û5 C. HIGH ENERGY

The sign of the one-loop correction to the gauge coupling is opposite in
QCD to that of QED:  The photon coupling is weak at ``low" energies
(actually, any observable energy, since the coupling runs so slowly), while
the gluon coupling is weak at high energies (with respect to the hadronic
mass scale).  Thus, typically perturbation in loops is used to study
high-energy behavior of QCD, while the low-energy behavior awaits the
discovery of a general nonperturbative approach.  Although such an approach is usually referred to as ``perturbative QCD", it is really a mixed approach, where amplitudes are generally factored into a high-energy piece, which is calculated with the usual Feynman diagrams, and a low-energy piece, which is found only from experiment.  The ``high" and ``low" energy here refers to a parton that is liberated from a hadron, having low energy before and high energy after.  In the processes that are best understood, this liberation is performed by an electroweak boson (photon, W, or Z), so one is actually calculating the electroweak interactions of a strongly interacting particle (quark), and its QCD corrections.

Ü1. Conformal anomaly

Symmetries of the classical action that are violated at the quantum level
are called ``anomalous".  There are two major sources for such
``anomalies" in renormalizable quantum field theory:  (1) There are
anomalies associated with the totally antisymmetric matrix
$·_{a_1...a_D}$, called ``axial" (see subsections VIIIA7 and VIIIB2-4). 
When they occur, they are found in graphs with at least (D+2)/2 external
lines.  They are associated with graphs that have no divergences, yet
require regularization.  (2) The existence of divergences requires the
introduction of a mass scale even in theories that are classically
conformal.  If anywhere, these show up at least in the most divergent
graphs, the propagator corrections.  Normally, both kinds of anomalies
will first appear at one loop.

When anomalies are associated with global symmetries, they provide a
natural way to explain approximate symmetries, in the sense of the
perturbative approximation.  However, when they occur in local
symmetries, they destroy the gauge invariance needed to prove
unitarity.  The latter type of theory therefore must be avoided by applying
the condition of anomaly cancellation in local symmetries.

We have already seen the appearance of the conformal anomaly in our
renormalization of divergent loop graphs:  The introduction of a
renormalization mass scale breaks the scale invariance of a classically
scale-invariant theory.  The simplest example, and generally the most
important, is the one-loop propagator correction.  If we examine only
high-energy behavior, then we can neglect masses from the classical
action. 

Using dimensional regularization, the generic effect on the effective
action of the complete one-loop propagator correction is to modify the
kinetic term of an arbitrary massless theory to
$$ üÄK\left({1\over g^2} +º_1¼ln{õ\over µ^2}\right) Ä $$
 where $Ä$ is an arbitrary-spin field that we have normalized $Ä£Ä/g$
for some appropriate coupling $g$ (like the Yang-Mills coupling if $Ä$ is
the Yang-Mills vector), $K$ is the classical kinetic operator, $µ$ is the
renormalization mass scale, and $º_1$ is a constant determined by the
one-loop calculation.  As long as $º_1$ is nonvanishing (i.e., the theory is
not finite) we can rewrite this as
$$ üÄKº_1¼ln{õ\over M^2}Ä $$
 where
$$ M^2 = µ^2 e^{-1/º_1g^2} $$
 is a renormalization-independent mass scale:  Any physical measurement
will observe $g$ and $µ$ in only this combination.  A choice of different
renormalization mass scale is equivalent to a finite renormalization of
$g^2$, such that $M$ is unchanged.  In the case where $g$ is
dimensionless (the relevant one, since we are studying the conformal
anomaly), the coupling constant has undergone dimensional
transmutation, being replaced with a dimensionful constant.

\x VIIIC1.1  Show this is the case for massless (scalar) $Ä^3$ theory
in D=6 from the explicit one-loop correction.

If there is more than one coupling constant, things are more complicated,
but the same phenomenon occurs:  One dimensionless coupling is replaced
with a mass.  A particularly interesting case is pure Yang-Mills theory: 
Then we can write an important contribution to the effective action as
$$ Fº_1¼ln{õ\over M^2}F $$
 where $F$ is now the complete nonabelian field strength, and $õ$ is the
square of the covariant derivative.  Since this contribution by itself gives
the complete 1-loop conformal anomaly, the rest of the 1-loop effective
action is conformally invariant.  (All its $M$ dependence cancels.)

Note that the (one-loop) anomaly itself is local:  If we perform an
infinitesimal conformal transformation on the one-loop part of the
effective action, this variation gives a local quantity.  This is clear from
the way this anomaly arose in dimensional regularization:  If there were
no infinities, there would be no anomaly, since the naive conformal
invariance of the classical theory would be preserved at each step. 
However, to regularize the divergence we needed to continue the theory
to arbitrary dimensions, and the theory is not conformal away from 4
dimensions.  The scale variation of a 4D conformal action in 4-2$·$
dimensions is proportional to
$·$ times that action, as follows from dimensional analysis; this scaling
can be associated with the nonvanishing (engineering) dimension of the
coupling away from D=4.  (Usually, we write the coupling as $gµ^·$,
where $g$ is dimensionless.  The fields have engineering dimension
independent of D, defined by the value in D=4:  E.g., in $á=»+A$, $A$ has
the same dimension as $»$.)  However, the one-loop effective action is
coupling independent; thus, when dimensionally regularized but
unrenormalized, it's scale invariant.  For example, in the propagator
correction discussed above, we get a regularized term
$-\f1·º_1ÄK(üõ)^{-·}Ä$, which is scale invariant but divergent.  On the
other hand, the counterterm added to make it finite is from the 4D
conformal action, and thus is not scale invariant in D$±$4; so the breaking
of scale invariance can be associated entirely with the counterterm.  (I.e.,
the anomaly coming from the renormalized, nonlocal effective action is
equal to that coming from the infinite, local counterterm.)  Since the
counterterm is local, the anomaly is local.  It's also finite, since it's
proportional to $·$ (from the variation) times 1/$·$ (from the divergent
coefficient of the counterterm).  In our propagator example, we have
$$ -\f1·º_1ÄK(üõ)^{-·}Ä +\f1·(üµ^2)^{-·}º_1ÄKÄ 
	® º_1ÄK¼ln{õ\over µ^2}Ä $$
 A similar situation occurs for the axial anomaly with Pauli-Villars
regularization:  After regularization, the anomaly comes entirely from the
regulator graph, which is not only finite by power counting, but local in
the infinite-mass limit because that is the zero-momentum (effective
potential or JWKB) limit.

There is a physical significance to the sign of the constant $º_1$.  (We saw
some evidence of this already in our analysis of renormalons in section
VIIC.)  Instead of thinking in terms of the renormalization-independent
mass scale $M$, we can treat $g$ as an effective energy-dependent
(``running") coupling,
$$ {1\over g^2(p^2)} = {1\over g^2} +º_1¼ln{p^2\over µ^2} $$
 In the case $º_1>0$ the coupling gets weaker at high energy
(``asymptotic freedom"), while for $º_1<0$ the coupling gets stronger at
high energy (until it reaches the Landau ghost).  (For low energy the
situation is generally more subtle, since we usually have complications
from physical masses.)  For QCD, this weakening of the coupling at high
energy allows the separation of an amplitude into a nonperturbative
low-energy piece (describing the observed particles, the bound-state
hadrons), which is determined experimentally, and a perturbative
high-energy piece (describing the non-asymptotic, fundamental
``partons", gluons and quarks), which can be calculated.  (This sometimes
goes under the somewhat misleading name of ``perturbative QCD".)  This
strongly contrasts with QED, where the weakening of the coupling at low
energy means both fundamental particles (photons, electrons, etc.)¼and
bound states (positronium, atoms, etc.)¼can be treated perturbatively,
and the only experimentally determined quantities are the values of
masses and the electron charge (coupling at low-energy).  Thus, in QED
one in principle can calculate anything, while in QCD one is restricted to
parts of certain amplitudes.  (Various nonperturbative methods also have
been developed for QCD, but so far they have successfully calculated only
a few low-energy constants, as used in $§$ models, i.e., masses and
low-energy couplings.)  Although experimental verification of these
results is sufficient to confirm the QCD description of hadrons, a practical
description of hadronic cross sections at all energies would seem to
require a string model that can incorporate behavior attributed to both
strings and partons.

\subsectskip\bookmark0{2. e+e- -\noexpand> hadrons}
	\subsecty{2. e\kern1pt\raise1ex\hbox{\vrule height1pt width7pt}
	\kern-9.5pt\raise.5ex\hbox{\vrule height7pt width1pt}
	e\kern1pt\raise1ex\hbox{\vrule height1pt width7pt}
	 {\sectsymb\char'041} hadrons}

If quarks and gluons are confined, how can QCD be useful?  QED is useful
because the coupling is small: $e^2®1/861$ is the perturbation parameter
in relativistic (quantum field theory, or 4D) calculations,
$Œ=2¹e^2®1/137$ in nonrelativistic (quantum mechanics, or 3D).  Energy
levels of the hydrogen atom can be calculated quite accurately, without
the question of freely existing electrons and protons coming up.  The
speed of the bound electron is also $Œ$, another way to understand why
pair creation/annihilation and other relativistic or multiparticle effects
are small, and can be treated perturbatively.

Therefore, the real usefulness of a field theory depends not on how
``physical" the choice of fields is, but how accurate the perturbation
expansion is.  ``Nonperturbative" results may give some nice qualitative
features, but they are ultimately useless unless they can be used as the
basis of a new perturbation expansion.  (Attempts at nonperturbative
approaches to 4D quantum field theory continue, but so far the results
are meager compared to perturbative results, or to nonperturbative
results in quantum mechanics or 2D quantum field theory.)

The simplest application of QCD is to the production of hadrons by a
photon created by the annihilation of an electron and a positron.  The
total cross section for such an event is given (according to the optical
theorem) by the imaginary part of quark contributions to the photon
propagator:  Since hadrons are made up of partons (quarks and gluons),
we assume a sum over hadrons can be written as a sum over partons. 
This assumption, that hadrons can be described by a resummation of the
perturbation expansion, should be good at least at high energies, where
the partons' asymptotic freedom takes effect (and perhaps at lower
energies by an appropriate extrapolation).  To lowest order for the
process under consideration this is a 1-loop graph, with a quark in the
loop.  If we compare this to the production of, e.g., muon-antimuon pairs
(but not back to electron-positron pairs, because that includes the
crossed diagram) by the same procedure, and we neglect masses (at high
enough energies), then the only difference should be in the group theory: 
Hadron production should be greater by a factor of the number of colors
times the sum over flavors of the square of the quark's electric charge:
$$ R ­ {P(e^+ e^-£h)\over P(e^+ e^-£µ^+ µ^-)} ® N_c Ý_f q_f^2 $$
 Experimentally this relation is confirmed for $N_c=3$, if the only flavors
included in the sum are those with masses below the photon energy
($(2m_f)^2<s$).

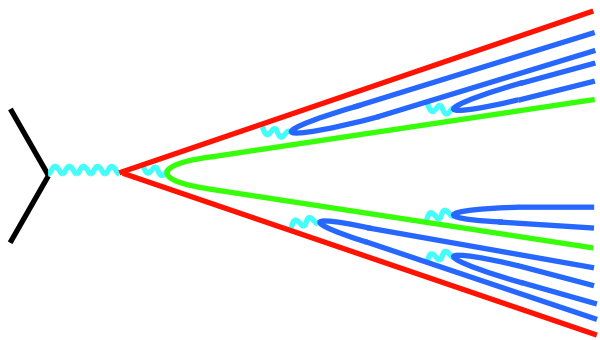
$$ \fig{jets} $$

This result can be extended to the case where the momenta of hadrons
are observed (not summed over):  Although individual partons are not
observed as asymptotic states, the dominant contribution to the cross
section at high energies is given by the conversion of the quarks into
hadrons by the creation from the vacuum of parton pairs with energies,
and angular deviation from the partons created by the photon, smaller
than experimental accuracy.  We treat all partons as approximately
massless, with respect to the energy scale of the photon.  Thus, each
parton created by the photon starts out initially as free, is then
accompanied by parallel partons of small energy to form hadrons, and
then these hadrons may further decay, but with a small angular spread
with respect to the directions of each of the initial partons.  Such
collections of final-state hadrons are called ``jets".  For high-energy
electron-positron annihilation, the dominant hadronic decay mode of this
off-shell photon is thus into two jets.  This experimental result is further
verification of QCD, and in particular a jet is the most direct observation
of a parton.  Of course, even for asymptotic states the directness of
experimental observations varies widely:  For example, compare a photon
or electron to a neutrino.  A closer analogy is unstable particles:  For
example, the neutron is observed as a constituent of the nucleus (as
quarks are constituents of hadrons), but eventually decays outside (as
quarks ``decay" into jets of hadrons).

A similar analysis can be applied to the creation of any electroweak boson
by annihilation of a lepton with an antilepton.

\x VIIIC2.1  Find the corresponding process (particles) for 
positron-neutrino annihilation.  Find the expected numerical value of
both this and the above $R$ in the Standard Model for energies well
above the masses of all the fundamental particles.

Ü3. Parton model

We have already seen that in quantum field theory coupling constants are
usually energy-dependent.  However, the dependence is only logarithmic,
and thus can be treated as perturbative unless the relevant energy scale
is within a few orders of magnitude of the mass scale that appears by
dimensional transmutation.  In QED, the value quoted for the electron
charge is at the scale of the electron mass $m$.  Using the result of
subsection VIIIA2 (or VIIIA3) for the 1-loop propagator correction, we
find (neglecting higher-loop corrections)
$$ {M_{QED}\over m} = e^{3/4e^2} = 2.8380185(62) ð 10^{280}â
	ÜâM_{QED} = 1.4502244(32) ð 10^{277} GeV $$
 (where for fun we have included the 1-standard-deviation uncertainties
for this 1-loop result as the figures in parentheses; the $e$ in the
exponent is the electron charge).  Since the mass of the observable
universe is of the order of $10^{80}$ GeV, and the Planck mass (beyond
which a particle will gravitationally collapse from its Compton radius
falling within its Schwarzschild radius) is of the order of ``only" $10^{20}$
GeV, there is little worry of observing the QED Landau ghost, even if QED
were correct to that scale.

On the other hand, the mass scale for QCD is (in the $Ñ{\rm MS}$ scheme)
$$ M_{QCD} = .217(24)¼GeV $$
 (This result depends on renormalization scheme, and is also an effective
mass in the sense that the usual experimental energy scale is among the
quark masses, so the high-energy approximation of the renormalization
group is inaccurate, and the full propagator correction with quark mass
dependence should be used.  The above number is for above the bottom but below the top threshold.)  This indicates that perturbative QCD is
inadequate to describe properties for which the energy of the quarks is
low, such as hadron masses (although nonrelativistic quark models have
had partial successes).  

\x VIIIC3.1  Take masses into account in the simplest approximation:
Treat particles as massless for energies above (twice) their mass,
infinitely massive for energies below.  Approximate the masses of
Higgs and superpartners of Standard Model particles as about the mass
of the Z boson.  Then graph the strong coupling $1/g^2$ in the 
supersymmetric Standard Model (see subsection VIIIA4) as a function 
of the $ln$ of the energy from the Grand Unification scale down to 
where it vanishes ($g=¥$), $M_{QCD}$.

However, in certain processes a single ``parton" (quark or gluon) in a
hadron is given a high energy with respect to the other partons, usually a
quark by electroweak interaction.  In those cases, the ``strong"
(chromodynamic) interaction of that parton with the others in its original
hadron is negligible:  It has been liberated.  The approach is then to factor
the amplitude into a piece with the electroweak and high-energy
(``hard") chromodynamic interactions of this parton, which can be
calculated perturbatively, and the low-energy (``soft") chromodynamic
part of the remaining partons, which is left as an unknown, to be
experimentally determined.  (Thus, the hard part is the easy part, while
the soft part is the difficult part.)  The predictive power is thus limited to
the dependence of the amplitude on the energy of this parton, and on the
particulars of the electroweak particles involved.

Another possible complication would be the effect of exciting many
partons within a hadron, indirectly through the first parton's interactions
with the rest:  Then one would have several terms to sum in an
amplitude, each with a different unknown soft factor, making the
approach useless.  Originally, it was thought that the high energy alone
was enough to explain the parton acting as free once liberated from the
hadron (based on ``intuitive" arguments), but soon it was realized that
this possibility depended totally on the high-energy behavior of the
theory:  It requires the decrease of the coupling with increasing energy,
asymptotic freedom (or superrenormalizability, or finiteness with
effective asymptotic freedom through the Higgs effect).  Based on this
property, one can show from the usual perturbation expansion that one
soft factor (per each hadron with an excited parton) is sufficient as a
leading approximation, a property known as ``factorization".  This
feature is a consequence of the fact that the dominant contributions to
Feynman graphs in this high-energy limit are those where the values of
the momenta of some of the partons are those corresponding to their
classical mechanics, as described in subsection VC8 and VIIA6.

This new approximation scheme is effectively a perturbation expansion in
the inverse of the energy being channeled into this parton.  One neglects
terms that are smaller by such powers (including those from masses and
renormalons), but incorporates logarithms through the renormalization
group and other loop corrections to the hard factor.  Since available
energy scales are much nearer to $M_{QCD}$ than to $M_{QED}$ in QED,
such an approximation scheme tends to break down around two loops,
where the corrections compete with the neglected terms, ambiguities in
renormalization schemes, and the relative size (convergence) of
successive terms in the expansion.  Although the accuracy of the
predictions of this approach cannot compare numerically with those of
QED, it is the only method to describe such processes that can lay claim to
being a theory, and provides direct experimental evidence of the validity
of QCD, both as a qualitative description of nature and as a valid
perturbation scheme.  (As in the previous subsection, we also have
processes where all the partons appear only in intermediate states, or
effectively so for final states in total cross sections via the optical
theorem, so factorization is unnecessary.)

$$ \fig{dis} $$

The most effective application of factorization is to ``Deep(ly) Inelastic
Scattering (DIS)".  (An equivalent method for this process is the ``operator
product expansion", but unlike factorization there is no useful
generalization of it to general processes.)  In this process a high-energy
photon (or intermediate vector boson) is exchanged between a lepton
(usually an electron) and a quark.  (This is the leading-electroweak-order
interaction of a lepton with a hadron.)  The quark and rest of the hadron
do not interact again:  Color singlets are obtained by the creation of soft
partons from the vacuum, which split from their own singlets and
eventually combine with the separated quark and hadron.  For this
process one calculates only the total cross section, at least as far as all
the strongly interacting particles are concerned (``inclusive
scattering") but again this can be generalized to the observation of jets
(``exclusive scattering").  Applying the optical theorem, and ignoring the
leptons, the leading contribution to this process is given by the tree graph
for scattering of a vector boson off a quark, where the intermediate
quark has a cut propagator.  This is the perturbatively calculated hard
part, which is later attached to the soft factor.  Thus the hard part is the
lepton-parton cross section, while the soft part is the ``parton
distribution", giving the probability of finding a parton in the hadron with
a particular fraction $Å$ ($³0$, $²1$) of its momentum $p$.  To leading
order this fraction is determined by kinematics:  Since the hadron and
scattered parton are treated as on-shell and massless,
$$ p^2 = (q+Åp)^2 = 0âÜâÅ = x ­ -{q^2\over 2qÉp} $$
 so the ``(Bjorken) scaling variable" $x$ is a useful dimensionless
parameter even when (at higher orders) $űx$.  The energy scale is set by
the square of the momentum $q$ of the vector boson.

There are several approximations used in this analysis, all of which can
be treated as the beginnings of distinct perturbation expansions:  
\item{(1)} The
hard part is expanded in the usual (loop/coupling) perturbation expansion
of field theory.  The leading contribution is that of the naive (pre-QCD)
parton model (``leading order"), where the quark that scatters off the
photon is treated as free with respect to the strong interactions. 
One-loop corrections (``next-to-leading-order") introduce the running of
the coupling associated with asymptotic freedom, which justifies the
validity of the parton picture.  This is usually the only
perturbation expansion considered, because such corrections are
logarithmic in the energy of the exchanged parton (rather than powers),
and thus more important and easier to isolate from the data. 
Furthermore, by the usual renormalization group methods such
logarithmic corrections can be reduced by careful choice of
renormalization scale ($µ^2$ close to $q^2$ in $ln(q^2/µ^2)$).    
Two-loop corrections lead to various ambiguities,
and have not proven as useful yet.
In particular, the $º$ function is scheme-dependent past two loops,
making the dependence on the separation between hard and soft 
harder to fix.
\item{(2)} In
calculating the hard part ``light" quarks are approximated as massless. 
One can rectify this by also perturbing in the masses, as a Taylor
expansion in the square of each mass divided by the square of the vector
boson's energy ($m^2/q^2$).    
\item{(3)} In the explicit calculation the
momentum $Åp$ of the excited parton is assumed to be proportional to
the momentum $p$ of the initial hadron.  In the rest frame of the initial
hadron (which is massive in real life), this corresponds to the
nonrelativistic approximation of motionless quarks; one quark is then set
into relativistic motion by the photon, liberating it from the hadron.  
Thus the parton model simultaneously uses a ÓnonrelativisticÕ approximation for a parton ÓbeforeÕ it's scattered, and an ÓultrarelativisticÕ (near-speed-of-light) approximation ÓafterÕ it's scattered.
This nonrelativistic
approximation can be corrected by a JWKB expansion (expressed in
operator language, the operator product expansion), also known as an
expansion in ``twist" (effectively, the power of momentum transverse to
$p$).  However, this means a separate soft part for each term in the
expansion:  Since these are determined experimentally, such an expansion
would lead to a loss of predictability.  Thus generally (with few
exceptions), parton model predictions are restricted to high enough
energies ($q^2$) that such corrections can be neglected.    
In this sense, this approach is very similar to low-energy approaches
to hadronic physics, e.g., nonlinear $§$ models:
Useful results are obtained at lowest order for describing physics in a
certain energy range, but outside that range the increasing loss of
predictability, e.g., nonrenormalizability, makes the approach less
and less applicable.  (Another, related, similarity between this approach
and nonlinear $§$ models is that both were originally described in the
language of the operator product expansion, as applied to currents.
However, this language was later replaced in both cases because of the 
difficulty of evaluating operator products of more than two currents.)
\item{(4)} Expansions in
renormalons (see subsections VIIC2-3) introduce new coupling constants,
effectively nonperturbative corrections to the otherwise perturbative
hard part.  Like all but the first of these expansions, this leads to
correction terms that are down by powers of $1/q^2$.  This type of correction could be absorbed into the previous one, since in
principle the soft parts should contain all nonperturbative corrections by
definition.
However, this would be begging the question, since it would mean more parameters to be determined by experiment.

$$ \fig{dy} $$

The other common application of the parton model is to ``Drell-Yan
scattering":  In this case two hadrons scatter producing, in addition to
hadrons, a photon (or other electroweak boson) that decays into a
lepton-antilepton pair.  To lowest order, the relevant diagram is the same
as for DIS (crossing some of the lines).  Because both of the initial
particles are hadrons, 2 soft parts are required; however, each of these
is the same as that used in DIS (``universality") so they do not need to be
redetermined.  In fact, there is a direct progression from $e^+e^-$ to DIS
to Drell-Yan:  The above diagrams are similar except for the number
($0£1£2$) of soft parts (corresponding to the number of initial hadrons);
the leading contribution comes from the same diagram, rotated to various
positions (crossing).

More generally, we can consider not only hard parts involving identified
quarks in the initial state of the hard part, but also in the final state, by
examining jets.  Thus, for soft parts we have not only the ``parton
distribution functions", which are energy-dependent probabilities to find specific partons in specific hadrons, found from amplitudes for
an initial hadron $£$ parton + anything (summing over anything), we have
``fragmentation functions", which are probabilities from amplitudes for
parton $£$ final hadron + anything.  In principle these are related by
crossing symmetry:  The diagrams are similar to the previous, with the
partons connecting to the hard part, but the external hadron lines may be
either initial or final (and the opposite for the corresponding parton with
respect to the hard subgraph). As for the parton distributions, the
fragmentation function for any particular parton and hadron is measured
in one particular experiment, then used universally.  (The simplest is deep
inelastic scattering for the parton distribution, and $e^+e^-$ annihilation
with one of the two jets $£$ hadron + anything for fragmentation.)  Then
the cross section is generally of the form
$$ d§_{A...B} = Ý_{a...b}ÇdÅ_a ò dÅ_b¼f_{Aa}(Å_a) ò f_{Bb}(Å_b) 
	d§_{a...b} (Å_i) $$
 where $d§_{A...B}$ is the observed (differential) cross section, $a...b$
(not to be confused with vector indices)
label the different partons and $A...B$ their hadrons (we leave off the
labels for non-strongly interacting particles), the sum is over different
kinds (and flavors) of partons (and perhaps over different hard parts, if
corrections down by powers are desired), $Å_a$ is the momentum fraction
for parton $a$ of hadron $A$'s momentum, $f_{Aa}$ is either the parton
distribution function for $A£a+X$ ($X$ = ``anything") or the fragmentation
function for $a£A+X$, and $d§_{a...b}$ is the hard cross section (calculated
perturbatively), which is just the original with all the hadrons replaced
by partons.  For the parton distributions we integrate $Ç_0^1 dÅ$, while
for fragmentation we integrate $Ç_1^¥ dÅ$, or change variables to the
hadron's fraction of the parton's momentum $½=1/Å$ and integrate $Ç_0^1
d½$.

Note that, while physical cross sections are independent of the
renormalization mass scale $µ$, the same is not true of the hard cross
sections calculated perturbatively in the above factorized expressions,
since they are expressed in terms of unphysical quark ``states". 
However, these hard parts satisfy renormalization group equations, as
calculated in the usual perturbative way.  (Of course, nontrivial
contributions require calculating beyond leading order.)  This implies
corresponding renormalization group equations (see subsection VIIC1), the
``evolution" or ``Gribov-Lipatov-Dokshitzer-Altarelli-Parisi (GLDAP)
equations", to be satisfied by the parton distributions, so that $µ$
dependence cancels in the complete cross sections.  This determines the
energy dependence of the parton distributions.  The equations take the
form
$$ µ^2{d\over dµ^2} f_{Aa}(Å,µ^2) = Ý_b Ç_Å^1 {d½\over ½}¼
	f_{Ab}(½,µ^2) P_{ba}\left({Å\over ½},g^2(µ^2)\right) $$
 where $f_{Aa}$ describes $A£a+X$, $f_{Ab}$ describes $A£b+X'$, the
``splitting functions" $P_{ba}$ describe $b£a+X''$ ($X=X'+X''$), and the
sum is over the intermediate parton $b$.  For hadron $A$ with momentum
$p$, the intermediate parton $b$ has momentum $½p$, and parton $a$ has
momentum $Åp$, so $Å/½$ is $a$'s fraction of $b$'s momentum.  The
kinematics are such that $0²x²Å²½²1$ (momentum is lost to $X$'s as
$A£b£a$).  The splitting functions can be calculated perturbatively from the
corresponding renormalization group equation for the hard part, since the
combined $µ$ dependence must cancel in the physical cross section.
Specifically, one considers the same equation with $A$ replaced by another parton:  Since $P_{ba}$ is independent of $A$, it can then be found from a completely perturbative equation.

For similar reasons, the hard cross sections are infrared divergent; the
soft parts of the complete cross sections deal with low energies.  This
leads to complications beyond next-to-leading order, due to the fact that
the renormalization group scale $µ$, which relates to ultraviolet
divergences (high-energy behavior), and the ``factorization scale", which
relates to infrared divergences (it determines the division between hard
and soft energies), are in principle independent scales.  This allows an
ambiguity in factorization prescriptions, in addition to the usual
ambiguity in UV renormalization prescriptions.  (In more general
processes there can be other energy scales than just $q^2$, each with its
own factorization scale, further complicating matters.)

Ü4. Maximal supersymmetry

The results of loop calculations simplify when the amount of supersymmetry is increased; in particular, more things vanish.  We have already seen this with respect to divergences in subsections VIIIA5-6.  Furthermore, in the massless case, vanishing of propagator divergences implies vanishing also of the finite parts of propagator corrections, since the unrenormalized corrections are always proportional to $õ^{-·}/·$ (except in D=2 from anomalies: see subsection VIIIA7).  These simplifications make supersymmetric theories useful models; if supersymmetry is used to eliminate the renormalon problem, these results are also physically relevant.  (We have already seen that supersymmetric methods are useful to derive nonsupersymmetric results for tree graphs, where unwanted particles can decouple.  Similar results can hold in 1-loop graphs, where supersymmetric results can be used to trade particles with spin in the loop for scalars, which are easier to calculate.)  In this subsection we will examine this behavior in 3- and 4-point functions.  We will find cancellations from just algebra, without momentum integration.  In general our analysis will apply to any $D²10$ (since super Yang-Mills doesn't exist in $D>10$).

Specifically, we will calculate amplitudes for external gauge bosons, using the same methods as for propagator corrections in subsection VIIIA3.  There we found a unified kinetic operator in background Yang-Mills, allowing us to separate the coupling into spinless (covariant $õ$) and spin ($F^{ab}S_{ab}$) pieces.  
In general the contributions will differ in form depending on the number of $S$ vertices that contribute, so we will require separate cancellation for each case.  We now consider these contributions order-by-order in $S$, but with an arbitrary number of non-$S$ vertices (i.e., whatever number is needed to give an $n$-pt.¼graph for whatever $n$ we are considering).  Since the no-$S$ terms are by definition spin-independent, they cancel for supersymmetric theories.  (The ghosts, together with the $-1/2$ factor for spinors, guarantee that the trace for spin gives the supertrace for physical degrees of freedom.)  Cancellation of $S^1$ terms is trivial, since $tr¼S=0$ (trace in states, not indices).  As we saw in subsection VIIIA3, tracing the $S^2$ terms gives the usual Casimir of SO(D):
$$ tr(S_{ab}S_{cd}) = ú_{b[c}ú_{d]a} ð \cases{ 0 & for spin 0 \cr 
	\f14 tr(I) & for spin $ü$ \cr 2 & for spin 1 \cr} $$
Its cancellation fixes the number of spinors to be that of maximally supersymmetric Yang-Mills.  The $S^3$ terms (in 3-point and higher graphs) can be separated into $tr(S[S,S])$, which reduces to $S^2$ (already canceled), and $tr(SÓS,SÕ)$, as in 4D anomalies for internal group SO(D).  The latter could give $·$ terms, but only in $D=2$ or 6.  We would miss them because of our (parity) doubling, but such parity invariance occurs anyway in maximally supersymmetric Yang-Mills in $D<10$.  The net result is that the 1-loop graphs completely cancel at less than 4-point, and the 4-point contribution comes completely from $S^4$ terms, but only for maximally supersymmetric Yang-Mills.  (For these calculations we needed only the same field content, but supersymmetry will then kill these lower-point graphs with fields other than Yang-Mills externally.)  The amplitude is then proportional to $F^4$ times a scalar box graph.  This $F^4$ factor turns out to be the same one (including Lorentz index structure) that appears in the tree amplitude (although the tree factor requires much more work to derive, except when one uses methods specific to D=4, even though it is the same as for pure Yang-Mills).

For example, representing group theory by the 't Hooft double-line notation, let's look specifically at the graphs with 2 external fields on one line and 2 on the other.  Then we get $tr(FF)tr(FF)$ for the Óinternal symmetryÕ traces.  By Bose symmetry, the 4-Lorentz-index color singlet $tr(FF)$ can consist only of the symmetric part of the direct product of 2 2-forms, i.e.,  a tensor with the symmetry of the Riemann tensor (Young tableau a 2$ð$2 box, including traces) and a 4-form (single column of 4).  First consider the case where the 2 $F$'s in a trace are adjacent on the loop.  Then the form of $S_{ab}$ for spin 1 requires its spin trace always to give traces for those 2 $F$'s, and thus contributes only to ``graviton" (traceless symmetric tensor) and ``dilaton" (scalar) type couplings to these singlets.  On the other hand, the form of $S$ for spin 1/2 always gives forms (times traces) for these 2 $F$'s, and thus contributes only to dilaton and ``axion" (4-form) type couplings.  
(In D=4 this axion is the usual pseudoscalar; for the relation to the usual string 2-form by duality see subsection XIC6.)
For the case where the 2 $F$'s are not adjacent, we commute the corresponding $S$'s so that they are:  The commutator terms give traces of fewer $S$'s, which cancel from our conditions, so the result is the same.  (In the string case the momentum integral is different, so these graviton, dilaton, and axion couplings actually couple to those states, appearing as poles:  see subsection XIC6.)

\x VIIIC4.1  Work out all the explicit Lorentz and group theory traces for all graphs.

\x VIIIC4.2  For D=4, reproduce these results using the methods of subsection VIIIA6, thus automatically including external fermions.

Similar methods can be applied when coupling gravity externally, essentially by replacing the Yang-Mills generators by a second set of spin operators, and the field strength with the Weyl tensor $W$.  (Without loss of generality, we can drop the Ricci tensor, since it vanishes by the free field equations, allowing us to drop other terms, not of this form, that might appear.  These tensors are discussed in subsection IXA4.)  This is not surprising if we know string theory, where gravity vertex operators are obtained as the product of left and right-handed Yang-Mills vertex operators.  

Explicilty, if we perform the same procedure for external gravity as for external Yang-Mills (background gauge fixing, squaring fermion kinetic operators), we find a universal kinetic operator
$$ õ +\f14 W^{abcd}S_{ab}S_{cd} $$
when coupling to spins 0, 1/2, 1 (or arbitrary forms), 3/2, and 2.  (For spins 3/2 and 2, we need to introduce ``compensators" for local S-supersymmetry and Weyl scale symmetry: see subsections IXA7 and XA3.)  In particular, the $WSS$ term vanishes for spins 0, 1/2, and 1 (and forms) by explicit evaluation.  ($W$ is traceless and has no 4-form piece.)  

We can then write all these spins as linear combinations of direct products of just spins 0, 1/2, and 1:
$$ S_{ab} = S^L_{ab} +S^R_{ab} $$
But when we plug this into the $WSS$ term above, only the cross terms will contribute.  Thus the kinetic operator becomes
$$ õ +ü W^{abcd}S^L_{ab}S^R_{cd} $$
Consequently, all vertex operators for all interesting spins (as in supergravity) can be expressed as direct products of those for spins 0, 1/2, and 1 (as for super Yang-Mills).  (Note that $p^L=p^R$, and the connection term in the covariant derivative becomes $¿_a{}^{bc} (S^L_{bc}+S^R_{bc})$.)

For example, taking the direct product of left and right maximally super Yang-Mills of the same chirality, we find the simple result $W^4$, with no lower-point diagrams, for maximal supergravity (twice the number of supersymmetries of maximally super Yang-Mills).

Ü5. First quantization

In subsection VIIIB1 we found some simple low-energy results for gauge loops applying JWKB methods.  The approach was essentially quantum mechanical, using the Hamiltonian formalism.  Here we use the quantization procedure in a more explicit way:  We work now in the 1D Lagrangian formalism, using 1D propagators and vertices, for calculating complete loops.  This method will be the most useful one when applied in chapter XI to 2D Lagrangians for strings.

$$ \figscale{propagator}{2in} $$

The propagator
$$ {1\over H} = {1\over H_0 -V} = {1\over H_0} +{1\over H_0}V{1\over H_0}
	+{1\over H_0}V{1\over H_0}V{1\over H_0} +... $$
gives the $N$-point graph
$$ A_N = 
	Ò-k_N|V_{N-1}{1\over H_0}V_{N-2}{1\over H_0}...V_3{1\over H_0}V_2|k_1Ô $$
(We can see see this from the usual Feynman diagrams, or by the relativistic generalization, along the lines of section IIIB, of the nonrelativistic quantum mechanics of subsection VA4, especially exercise VA4.1.)
Restricting all states to scalars,
$$ H_0 = ü(p^2 +m^2),ââV_i = g e^{ik_iÉx} $$
The initial and final states can also be defined by the same vertex operators that created the external states:
$$ p|kÔ ­ k|kÔâÜâ|kÔ = e^{ikÉx}|0Ô $$
(We have used $-k_N$ so all states propagate inward: $Ýk=0$.)

We now translate this to the Heisenberg picture by absorbing the free propagators $1/H_0$ into $ $ dependence for the $V$'s:  First we introduce Schwinger parameters,
$$ {1\over H_0} = Ç_0^¥ d ¼e^{- H_0} $$
$$ ÜâA_N = 
	Ç_0^¥ d^{N-3}  Ò-k_N|V_{N-1}e^{- _{N-2}H_0} ... e^{- _2 H_0}V_2|k_1Ô $$
Then we change variables from the ``relative" $ $'s $ _i$ to the ``absolute" $ $'s $÷ _i$,
$$  _i ­ ÷ _{i+1} - ÷ _i,â÷ _{N-1} ­ 0âÛâ÷ _i ­ -Ý_{j=i}^{N-2}  _j $$
and use the $÷ $-dependent $V$'s
$$ V(÷ ) = e^{÷ H_0}V(0)e^{-÷ H_0},ââV(0) = V $$
to write
$$ A_N = Ç_{-¥ ² ÷ _i ² ÷ _{i+1} ² 0}d^{N-3}÷ ¼
	Ò-k_N|V_{N-1}(0)V_{N-2}(÷ _{N-2}) ... V_2(÷ _2)|k_1Ô $$
when the initial and final states are on shell,
$$ H_0|kÔ = 0âÛâk^2 +m^2 = 0 $$
This is the same form that appears in nonrelativistic quantum mechanics (in the Heisenberg picture), simply evaluating operators for the potential at arbitrary times, where one time is not integrated over because of time translation invariance.  (Its integral gives the usual $¶$ function for energy conservation.  See subsection VA4.)

We can also write the initial and final states in terms of the same vertices, for arbitrary $÷ $'s:
$$ V(÷ )|0Ô =  e^{÷ H_0}Ve^{-÷ H_0}|0Ô = e^{÷ H_0}V|0Ôe^{-÷ m^2/2} =
	ge^{÷ H_0}|kÔe^{-÷ m^2/2} = |kÔge^{-÷ m^2/2} $$ 
Then the amplitude is
$$ A_N = g^{-2}\lim_{{÷ _1£-¥\atop ÷ _{N-1}=0}\atop ÷ _N£+¥}
	e^{(÷ _1 -÷ _N)m^2/2}Çd^{N-3}÷ ¼
	Ò0|V_N(÷ _N)V_{N-1}(÷ _{N-1}) ... V_2(÷ _2)V_1(÷ _1)|0Ô $$
This is the form of the amplitude we might have expected from a first-quantized path integral in terms of $X( )$, with an interaction term $-Çd ¼V( )$, except that 3 of the $V$'s are not integrated.

Now we can easily evaluate the S-matrix element by path integration, since $X$ appears everywhere in exponentials.  We saw in subsection VIIB5 that multiplication of 2 exponentials inside a path integral for a free theory yields their ``normal-ordered" product times the exponential of a Green function.  Since all the vertex operators (including those for initial and final states) are exponentials, we can also see this result from completing the square in the functional integral.  Either way, the result is
$$ \leftÒ Þ_i :exp[ik_iÉX(÷ _i)]: \rightÔ = exp \left[ - Ý_{i<j}k_iÉk_j ÒX(÷ _i)X(÷ _j)Ô \right] $$
 (The normalization is clear from Taylor expansion.)
As a consequence of normal ordering each vertex, we drop any
terms coming from connecting a vertex to itself with a Green function.

We normalize the Green function as
$$ ÒX¼XÔ = üG $$
Then the amplitude is simply
$$ \boxeq{ A_N = g^{N-2}\hskip-.06in
	\lim_{{÷ _1£-¥\atop ÷ _{N-1}=0}\atop ÷ _N£+¥}\hskip-.01in
	e^{(÷ _1 -÷ _N)m^2/2}\hskip-.13in
	Ç\limits_{÷ _i ² ÷ _{i+1}}\hskip-.14in
	d^{N-3}÷ ¼exp\left[-üÝ_{i<j}k_iÉk_j G(÷ _i,÷ _j)\right] } $$

We can use the propagator for $X( )$ 
$$ -ü¬G = ¶âÜâG( , ') = -| - '| $$
where we have applied the boundary conditions
$$ G( , ') = G( ', ) ,ââ G( , ) = 0 $$
to avoid ``renormalization" for the $i=j$ terms.  We then get
$$ A_N = g^{N-2}\hskip-.06in
	\lim_{{÷ _1£-¥\atop ÷ _{N-1}=0}\atop ÷ _N£+¥}\hskip-.01in
	e^{(÷ _1 -÷ _N)m^2/2}\hskip-.13in
	Ç\limits_{÷ _i ² ÷ _{i+1}}\hskip-.14in
	d^{N-3}÷ ¼exp\left[üÝ_{i<j}(÷ _j -÷ _i)k_iÉk_j\right] $$
At this point we could convert back to the original $ $'s to make the integrals easier and arrive essentially at the starting point.

Now that we understand the approach for trees, we can analyze loops.
1-loop 1PI graphs are easy to relate to the tree graphs we have considered:  Starting with a tree graph with 1 long line out of which branch external lines (but no trees), we connect the 2 ends of the long line with another propagator.  (There are fancier arguments, but they are less convenient:  Unitarity gives only the imaginary part of the loop, and requires then a dispersion relation.  Feynman's tree theorem gives a cut propagator, plus additional multi-cut graphs that must be argued away.)  This is obvious from the diagrammatic point of view; the reason we start from a tree is that normally quantum mechanics is just matrix mechanics, and is geared toward sandwiching a product of matrices (operators) between 2 vectors (states).  A loop is then a trace of a product of matrices, where the initial and final states have been replaced by a sum over all states (trace), and the initial and final ``times" have been identified, making ``time" periodic.

$$ \figscale{sew}{4in} $$

We thus start with an amplitude of the form
$$ A_N^{(1)} = tr \left( V_N {1\over H_0} ... {1\over H_0} V_1 {1\over H_0} \right);ââ
	H_0 = ü(p^2 +m^2),ââV_i = e^{ik_i Éx} $$
This is the same expression we used earlier for trees, except for the extra propagator $1/H_0$ and the trace, explained above.  Again as for trees, we introduce Schwinger parameters
$$ {1\over H_0} = Ç_0^¥ d ¼e^{- H_0} $$
and change variables from the $N$ ``relative" $ $'s, $ _i$, to the $N-1$ ``absolute" $ $'s, $÷ _i$, and the ``overall" $ $, $T$:
$$  _i ­ ÷ _i - ÷ _{i-1},â÷ _N ­ 0âÛâ÷ _i ­ -Ý_{j>i}^N  _j,âT ­ -÷ _0 = Ý_1^N  _i $$
and use the $÷ $-dependent $V$'s
$$ V(÷ ) = e^{÷ H_0}V(0)e^{-÷ H_0},ââV(0) = V $$
to write
$$ \li{ A_N^{(1)} & = Ç_{-¥ ² -T ² ÷ _i ² ÷ _{i+1} ² 0}dT¼d^{N-1}÷ ¼
	tr[V_N(0)V_{N-1}(÷ _{N-1}) ... V_1(÷ _1)e^{-T H_0}] \cr 
	& = Ç_{-¥ ² -T ² ÷ _i ² ÷ _{i+1} ² 0}dT¼d^{N-1}÷ ¼
	Ý_nÒn,0|V_N(0)V_{N-1}(÷ _{N-1}) ... V_1(÷ _1)|n,-TÔ \cr} $$
where we have written the trace as a sum over all states to show that the effect of the surviving propagator is to guarantee that there is a length of ``time" $T$ between the initial and final times, which are sewn.

The amplitude can now be evaluated by 1D Feynman graphs (by operator or path integral methods) as (compare the tree result above)
$$ \boxeq{ A_N^{(1)} = Ç_{-¥ ² -T ² ÷ _i ² ÷ _{i+1} ² 0}dT¼d^{N-1}÷ ¼\V (T)
	exp\left[-üÝ_{i<j}k_iÉk_j G(÷ _i,÷ _j)\right] } $$
where the Green function $G$ is the $X$-propagator for this worldline (see below).  The ``volume element" $\V(T)$ (called the ``partition function" in statistical mechanics, where $T$ is the ``inverse temperature") comes from the ``vacuum" amplitude found by evaluating the general amplitude for vanishing sources $V=1$.  (In the path integral approach, it comes from the determinant of the Green function.)  It depends only on the parameter $T$ that defines the geometry (i.e., the range of ``time" for $G$).  Comparing our ``result" to the original expression, we see
$$ \V (T) = tr \left( e^{-TH_0} \right) = 
	Ç{d^D p\over (2¹)^{D/2}}¼e^{-T(p^2+m^2)/2} =
	T^{-D/2}e^{-Tm^2/2} $$

A standard change of variables for loops is to factor out the scale $T$ from the times:
$$ ÷ _i = -TŒ_iÜ $$
$$ A_N^{(1)} = Ç_{0²Œ_{i+1}²Œ_i²1}d^{N-1}ŒÇ_0^¥ dT¼T^{N-1}\V (T)
	exp\left[-üÝk_iÉk_j ÷G(Œ_i,Œ_j,T)\right] $$
where $Œ_i$ are (some linear combination of) the Feynman parameters.

At this point we need to note that the Green function can't be defined in the strict sense, since we now have closed lines:  In a closed space, the total ``charge" must vanish, by comparing Gauss' law inside and outside any closed ``surface".  This is related to the existence of ``zero-modes", functions that are killed by the wave operator (d'Alembertian), on which this operator therefore has no inverse.  In this case the zero-mode is a constant, since a constant is the only periodic function that is a homogeneous solution to the wave equation.  This zero-mode corresponds to an invariance (translations), and must drop out anyway.  It appears in general when solving the wave equation
$$ -üõ2X = jâÜâ2X = {1\over -üõ}j,âõ{1\over õ}j = j $$
where the $õ$ kills any zero-modes mistreated by $1/õ$.  
More explicitly, this is satisfied for any space with coordinates $ $ if
$$ -üõG( , ') = ¶( - ') +h( ) $$
$$ Üâ-üõÇd ' G( , ')j( ') = j( ) +h( )Çd ' j( ') $$
for some $h$, since the latter term vanishes by momentum conservation, the aforementioned invariance:  Since the source comes from vertex operators $e^{ikÉX}$,
$$ j = Ýk_i ¶( - _i)âÜâÇj = Ýk = 0 $$
The Green function $G( , ')$ should be symmetric in $ $ and $ '$, since only the symmetric part contributes to $j(1/õ)j$.  For the loop the only choice for $h$ is a constant, representing a constant ``background charge" distribution in addition to the point charge represented by the $¶$ function in $G$'s wave equation.  The value of the constant follows from integrating the Green function's equation over the loop (just $ '$):
$$ h = -{1\over length} $$
(or ``length" $£$ ``volume" for a general space) so the total charge vanishes.  The Green function itself is now determined up to a constant.

We easily modify our earlier tree result for $G( , ') = G( - ')$ to
$$ -ü¬G = ¶( ) -{1\over T}âÜâG = -| | +{ ^2\over T} $$
We have written the result in a form valid for $| |²T$, which is sufficient in terms of $ - '$ for $0² , '²T$.  This function is a repeating parabola, rather than the ``V"-shape of the tree case.  $T$ then scales out of $G$ in a simple way:
$$ G(÷ _i-÷ _j) = -T[|Œ_i-Œ_j| -(Œ_i-Œ_j)^2] $$
Using the result for the volume element above,
the $T$ integration is then of the form
$$ Ç_0^¥ dT¼T^{N-1-D/2}e^{-TF(Œ_i,m^2)} = ý(N-\f D2)F^{D/2-N} $$
for a function $F$ found from the above prescription, where the $ý$ function shows the usual divergence structure for $Çd^D p¼p^{-2N}$ ($N$ propagators with no derivatives at vertices).  All that remains are the usual, messy Feynman parameter integrations.

\x VIIIC5.1  Use this method to explicitly evaluate the propagator correction.

Finally we consider generalization from scalars to vectors.
The basic idea is to make a ``stringy" generalization of the exponent of $e^{ikÉx}$:  In terms of some $§$-dependent parameter $P(§)$,
$$ Ç{d§\over 2¹}P(§)ÉX(§) £ kÉX +ûÉ»X $$
keeping only the first ``excitation" (so we can describe massless vectors),
where the mode expansion is implemented as a Taylor expansion of $X$ about one of the boundaries.  The first term gives just the momentum dependence we have already used for the scalar vertex while the second term, if expanded in the exponential to lowest nontrivial order, gives the vector vertex.  (Ghost dependence can be included by the same method.)  In a more general approach, this corresponds to introducing arbitrary external states with a ``source" $P(§)$ by writing the wave functional $ï[X(§)]$ in terms of its functional Fourier transform $÷ï[P(§)]$.

Although this derivation was motivated by string theory, this result can be applied to particles.  In fact, the usual contribution of an external vector to the particle action is $gÇd ¼ÀXÉA(X)$, which upon Fourier expansion in $X$ gives vertices
$$ V_{i,vector} = gû_iÉÀXe^{ik_iÉX} $$
where $û$ is the polarization vector.  This is the same result obtained by expanding the exponential described above to first order in $û$, zeroth order reproducing the scalar vertex.  (The $§$ derivative gets replaced by a $ $ derivative for $X$ satisfying the 2D wave equation: see chapter XI.)

Evaluation of the amplitude is simpler if we keep the original exponential in both $k$ and $û$,
$$ V_i = ge^{i(kÉX +ûÉÀX)} $$
Then the only change in the amplitude is the replacement
$$ k_iÉk_j G_{ij} £ k_iÉk_j G_{ij} + û_iÉk_j »G_{ij} + k_iÉû_j »'G_{ij} + û_iÉû_j »»'G_{ij} $$
where $»$ is the derivative with respect to $÷ _i$ and $»'$ to $÷ _j$.  Since $G$ depends on them only through the difference $÷ _i-÷ _j$, we can write $»'=-»$.  (For some applications it may be useful to integrate these derivatives by parts in the amplitude.)  

The last term gives a simple expression:
$$ »»'G = -»^2 G = 2¶( ) $$
in terms of $ _{ij}­÷ _i-÷ _j$.  Unlike the other $û$-dependent terms, this piece gives direct contraction of vector indices on polarizations, instead of contraction of them with momenta.  The $¶$ term makes the 2 vertices corresponding to the 2 polarizations coincide:  It correpsonds to the $A^2$ term in the coupling of an external electromagnetic field to this scalar.

A simple example of this method is to keep the full exponential in $û$, but choose
$$ û_i = àk_i $$
in some units.  This can be considered a more ``stringy" model with higher-derivative couplings.  The $à$ will be interpreted as corresponding to the 2 boundaries of the string:  We have written $X£Xà»X$ as an infinitesimal expansion of $X$ at either boundary about the center of an infinitesimal string.  We can enforce this interpretation with group theory by using the 't Hooft double-line notation, so the ``inside" and ``outside" vertices couple to different lines, i.e., with different group theory factors.  The above modification is now
$$ G_{ij} £ G_{ij} +(à_i - à_j)ÀG_{ij} + (à_i à_j) 2\left[ ¶( _{ij}) - {1\over T} \right] $$
where $à_i$ indicates which boundary the vertex is on.  After again scaling $÷ _i=-TŒ_i$, the first term goes as $T$ again, the second is $T$-independent, the last goes as $1/T$.  Ignoring the uninteresting 4-point contribution (which results in terms in the amplitude similar to those for fewer external lines), the most important new contribution for small $T$ is the $Œ_i$-independent $1/T$ term.  It modifies the $T$ integration for small $T$:  Using
$$ Ý_{i<j}(à_i à_j)k_iÉk_j = üÝ_{i,j}(à_i à_j)k_iÉk_j = ü\left(Ý_i à_i k_i\right)^2 $$
$$ = ü\left( Ý_I k_I -Ý_{I'}k_{I'} \right)^2 = 2\left( Ý_I k_I\right)^2 ­ -2s $$
writing $i=(I,I')$ for the two ``sides" of the particle, and choosing the external vector to be massless ($k_i^2=0$), we find
$$ Ç_0^¥ dT¼T^{N-1-D/2}e^{-2s/T} = Ç_0^¥ dT'¼T'^{D/2-N-1}e^{-2T's} 
	= ý(\f{D}2 -N) (2s)^{N-D/2} $$
where $T'=1/T$.  Thus, for the interesting case of the 4-point amplitude in 10 dimensions, we find a massless pole $1/s$ replacing the usual 1-loop divergence, which now appears only for the ``planar" case where all vertices are on the same ``side".

\x VIIIC5.2  To what explicit modification of the field action for a scalar does this stringy vertex correspond?

The above method gives coupling of external vectors to an internal scalar.  Generalizing the internal particle to a spinor or vector is not as simple classically, but we can make the generalization from what we know about such coupling from the field theory methods of subsection VIIIA3:  Using the universal kinetic operator
$$ õ +igF^{ab}S_{ab} $$
with quantum spin operator
$$ S_{ab} = \cases{ 0 & for spin 0 \cr -\f14 ©_{[a}©_{b]} & for spin $ü$ \cr
	|_{[a}ÔÒ_{b]}| & for spin 1 \cr} $$
we expand to linear order in the vector field:
$$ -ü(õ +igF^{ab}S_{ab}) ® -üõ_0 +g[AÉ(-i») +(-i»^a A^b)S_{ab}] $$
where $õ_0$ is the free $õ$, and for convenience we have ignored ordering by using the (background) gauge $»ÉA=0$.  Choosing the external field to have definite momentum,
$$ A_a(x) = û_a e^{ikÉx} $$
In terms of the momentum $p$ of the internal particle and $k$ of the external field, we then have
$$ V = gû^a e^{ikÉx} (p_a -k^b S_{ab}) $$
In Lagrangian language, we simply replace $p£ÀX$:
$$ V = gû^a e^{ikÉx} (ÀX_a -k^b S_{ab}) $$

\refs

£1 R.P. Feynman, \PR 23 (1969) 1415;\\
	J.D. Bjorken and E.A. Paschos, ÓPhys. Rev.Õ É185 (1969) 1975:\\
	parton model.
 £2 J.C. Collins, D.E. Soper, and G. Sterman, Factorization of hard
	processes in QCD, in ÓPerturbative quantum chromodynamicsÕ, ed. A.H.
	Mueller (World-Scientific, 1989) p. 1;\\
	G. Sterman, Introduction to perturbative QCD, in ÓPerspectives in the
	standard modelÕ, proc. TASI '91, Boulder, Colorado, June 2-28, eds. 
	R.K. Ellis, C.T. Hill, and J.D. Lykken (World-Scientific, 1992) p. 475;\\
	Sterman, Óloc. cit.Õ;\\
	G. Sterman, \xxxlink{hep-ph/9606312}, Partons, factorization and
	resummation, in ÓQCD \& beyondÕ, proc. TASI '95, Boulder, Colorado,
	June 4-30, ed. D.E. Soper (World-Scientific, 1996) p. 327:\\
	reviews of factorization.
 £3 G. Sterman and S. Weinberg, \PR 39 (1977) 1436:\\
	justification for jets from perturbative QCD.
 £4 V.N. Gribov and L.N. Lipatov, ÓSov. J. Nucl. Phys.Õ É15 (1972) 438, 675;\\
	Yu.L. Dokshitzer, ÓSov. Phys. JETPÕ É46 (1977) 641;\\
	G. Altarelli and G. Parisi, \NP 126 (1977) 298.
 £5 M.B. Green, J.H. Schwarz, and L. Brink, \NP 198 (1982) 474:\\
	1-loop S-matrices in superfield theory as limits of superstring loops.
 £6 M.T. Grisaru and W. Siegel, \PL 110B (1982) 49:\\
	direct 1-loop S-matrices from supergraphs.
 £7 M.J. Strassler, \xxxlink{hep-ph/9205205}, \NP 385 (1992) 145:\\
	first-quantized gauge loops based on string methods.
 £8 K. Lee and W. Siegel, \xxxlink{hep-th/0303171}, \NP 665 (2003)179:\\
	loops in stringy modification of field theories.

\unrefs

\volume PART THREE: HIGHER SPIN%

Higher-spin (unstable) particles have been observed experimentally. 
Whether they are considered elementary depends on how their theory is
formulated.  In particular, a description of hadrons in terms of strings
would have many advantages, such as unification of all hadrons,
manifestation of duality symmetry, and calculability through an accurate
perturbation scheme.  

Gravity and supergravity also include higher-spin particles.  String theory
might also yield some solutions to some of their problems, especially
renormalizability and unification of all particles.  Such gravitational
strings would differ from hadronic strings in their mass scale and in the
appearance of massless particles, including the graviton itself (in
contrast to the massive ``pomeron", the analog of the graviton in
hadronic strings).  Gravitational strings might require supergravity.

For these and other reasons supergravity and strings are two of the major
areas of research in theoretical high energy physics today (although not
the only ones).  Most of the discussion of this part is introductory, and can
be covered earlier, but it is not essential to the course; however, its
inclusion in a field theory text is essential at least for reference.

Gravity is uniquely defined as the force carried by a massless spin-2 particle:  There are no such particles other than the graviton, and there is no massless spin-0 particle.  Similarly, the photon is the only massless spin-1 particle.  (Gluons do not appear outside of hadrons.)  Thus, gravity and electromagnetism are the only long-range forces.  But there are massive strongly interacting particles of all spins.  Thus, at short distances gravity might not be so clearly defined:  Hadrons couple to sums of various spin-2 fields, weighted by various functions of spin-0 (scalar) fields, and in a way that depends on the type of hadron.  This means that the ``equivalence principle", which basically says to replace the flat-space Minkowski metric with the ``curved" metric of gravity as a type of minimal coupling, holds only at macroscopic distances.  (Similar remarks apply to nonminimal coupling, involving adding to the metric some function of the curvature tensor, which involves derivatives of the metric, and its covariant derivatives, which is possible even for weakly interacting particles.)  For these and similar reasons, the success of general relativity at macroscopic distances should not be taken too seriously when applied to interactions at the submicroscopic scale, as in the earliest stages of cosmology (``inflation") or the latest stages of gravitational collapse of stars (``black holes").

ÚIX. GENERAL RELATIVITY

Before discussing supergravity we need to study ordinary gravity.  Both
can be treated as generalizations of Yang-Mills theory.  We use this
approach rather than the traditional one, based on the metric, which is
insufficient for describing spinors or supersymmetry:  There is no useful
definition of distance in anticommuting directions in curved (super)space.

Gravity is the only observed long-range (massless) force mediated by a
higher-spin (2) field.  It is relevant for astrophysics, cosmology, and
unification, all of which have applications to particles of lower spin.

Û7 A. ACTIONS

We begin with the general principles that define pure gravity as a
nonabelian gauge theory, and use them to derive actions and couple to
matter.

Ü1. Gauge invariance

General relativity can be described by a simple extension of the methods
used to describe Yang-Mills theory.  The first thing to understand is the
gauge group.  We start with coordinate transformations, which are the
local generalization of translations, since gravity is defined to be the
force that couples to energy-momentum in the same way that
electromagnetism couples to charge.  However, these are not enough to
define spinors.  This is easy to see already from the linear part of
coordinate transformations:  Whereas SO(3,1) is the same Lie group as
SL(2,C), GL(4) (a Wick rotation of U(4)) does not have a corresponding
covering group; there is no way to take the square root of a vector under
coordinate transformations.  So we include Lorentz transformations as an
additional local group.  We therefore have a coordinate transformation
group, which includes translations and the orbital part of Lorentz
transformations, and a local Lorentz group, which includes the spin part
of Lorentz transformations.

Clearly the coordinates $x^m$ themselves, and therefore their partial
derivatives $»_m$, are not affected by the (spin) Lorentz generators.  We
indicate this by use of ``curved" vector indices $m,n,...$.  On the other
hand, all spinors should be acted on by the Lorentz generators, so we give
them ``flat" indices $Œ,º,..$, and we also have flat vector indices $a,b,...$
for vectors that appear by squaring spinors.  Flat indices can be treated
the same way as in flat space, with metrics $C_{Œº}$ and $ú_{ab}$ to
raise, lower, and contract them.  

Some gravity texts, particularly the more mathematical ones, emphasize
the use of ``index-free notation".  An example of such notation is matrix
notation:  Matrix notation is useful only for objects with two indices or
fewer, as we saw in our treatment of spinor indices in chapter II.  Such
mathematical texts consider the use of indices as tantamount to
specifying a choice of basis; on the contrary, as we have seen in previous
chapters, indices in covariant equations usually act only (1) as place
holders, indicating where contractions are made and how to associate
tensors on either side of equations, and (2) as mnemonics, reminding us of
representations and transformation properties.  Thus, the full content of
the equation can be seen at a glance. In contrast, many
mathematical-style equations (when indeed equal signs are actually
used) say little more than ``$A=B$Ê", with the real content of the equation
buried in the text of preceding paragraphs.

We therefore define the elements of the group as
$$ g = e^Â,ââ = Â^m »_m +üÂ^{ab}M_{ba} $$
 where $»_m$ acts on all coordinates, including the arguments of the real
gauge parameters $Â^m$ and $Â^{ab}$ and any fields.  $M_{ab}=-M_{ba}$
are the Ósecond-quantizedÕ Lorentz generators:  They act on all flat
indices, including those on $Â^{ab}$ and any fields that carry flat indices. 
As a shorthand notation, sometimes we will also write
$$ üÂ^{ab}M_{ba} = Â^I M_I $$
 (and similarly for other appearances of the antisymmetric index pair
$ab$).  We thus have a combination of the matrix generators of section IB
and the coordinate generators of subsection IC2.

\x IXA1.1  Sometimes it's more convenient to perform explicit finite
coordinate transformations in terms of new coordinates as
functions of old, as in subsection IC2.  As an example for curved
space we consider the sphere in arbitrary dimensions.
Rather than the usual cumbersome angles, which introduce
trigonometric functions into measurements of distances, we use
coordinates which manifest the slightly smaller rotational invariance 
of the corresponding flat space, as we did for scalar fields in
subsection IVA2.
 ªa As in subsection IVA2, we can derive coordinates for the sphere
by constraining flat space in Cartesian coordinates to have unit
radius.  Rather than looking for an explicit solution as in subsection
IVA2, we can enforce the constraint by the replacement
$$ x £ {y\over |y|} $$
 so the flat coordinate ``vector" $x$ automatically has magnitude 
$|x|=1$ at the expense of introducing the scale invariance
$$ y' = Â(y)yââ(x' = x) $$
 Show the infinitesimal distance $ds$ is given by
$$ ds^2 = dx^2 = {dy^2\over y^2} - {(yÉdy)^2\over y^4}
	= {(y^{[a}dy^{b]})^2\over 2y^4} $$
 Check scale invariance of the last form.
 ªb Ultimately we'll need to use the scale invariance to
eliminate one coordinate.  Writing $y^a=(y^0,y^i)$, consider the
coordinate transformation
$$ y^0 = (z^0)^2 - (z^i)^2,ây^i = 2z^0 z^i $$
 (This is just a generalization of the substitution used in
subsection IVA2.)
What is the interpretation in two dimensions ($y^i=y^1$) in terms
of complex coordinates?  (This generalizes to quaternions in four
dimensions.)  Show that this results in
$$ ds^2 = {(2z^{[0}dz^{i]})^2\over [(z^0)^2+(z^i)^2]^2} $$
 Compare the result on $ds^2$ of the scale gauge $y^0=1$
(on the previous form) to that of $z^0=1$ (on this form).

\x IXA1.2  More general scale gauges for the previous problem
come from considering the kind of
projections made in map making, looking at the result of shining a
point light source through a transparent globe onto a plane, where
the ray from the source through the center of the globe exits it at
the point tangent to the plane.  Instead of looking at the geometry
of the rays, we consider expanding this globe of unit radius through
the plane in such a way that the source remains at the same scaled
position inside the globe.  (The center of the sphere moves while
the source and plane remain fixed, at least with respect to each other.)
The globe continues to expand until it intersects a chosen point on
the plane.  Explicitly, in terms of coordinates $y$ of that point with
respect to the origin of the expanded globe, the distance on the 
original (unit-radius) globe is
$$ ds^2 = dx^2 = \left( d {y\over |y|} \right)^2 $$
 while the position of the source is
$$ x_s^0 = -a,âx_s^i = 0âÜây_s = (-a|y|,0) $$
 in terms of the constant $a$ that defines the gauge (projection),
so the condition that it hasn't moved relative to the plane is
$$ y^0 +a|y| = 1+a $$
 ªa Show the solution is
$$ y^0 = {(1+b) -(1-b)å{1+b(y^i)^2}\over 2b},ââb = {1-a\over 1+a} $$
 or
$$ z^0 = å{1+b(z^i)^2} $$
ªb Find $ds^2$ in terms of both $y^i$ and $z^i$ for the special cases
$$ \li{ gnomonic:â& a=0â(b=1) \cr
	stereographic:â& a=1â(b=0) \cr
	orthographic:â& a=¥â(b=-1) \cr} $$

In general, the relation between first- and second-quantized group
generators is the same as the relation between active and passive
transformations, and the relation between a matrix representation and
the corresponding coordinate representation, as discussed in subsection
IC1, where in this case the fields are the coordinates.  In particular,
the second-quantized Lorentz operators $M_{ab}$ have the same action
as the first-quantized Lorentz operators $S_{ab}$ introduced in subsection
IIB1:  For any field $Æ$, $MÒÆ|=ÒÆ|S$, etc.  The action of the Lorentz
generators on vector indices is thus given by
$$ [M_{ab},V_c] = V_{[a}ú_{b]c}âÜâ
	Â^I[M_I,V_a] = üÂ^{bc}[M_{cb},V_a] = Â_a{}^b V_b $$
 This implies the commutation relations
$$ [M_{ab},M^{cd}] = -¶_{[a}^{[c}M_{b]}{}^{d]} $$
 In explicit calculations, only two indices in the commutator will match,
and they reduce to simple expressions such as
$$ \boxeq{[M_{12},V_2] = ú_{22}V_1,ââ[M_{12},M_{23}] = ú_{22}M_{13}}$$
 As for derivatives, when acting on functions instead of operators we can
write the action of the Lorentz generator as simply $M_{ab}V_c$ without
the commutator.

When spinors are involved in four dimensions, it's simpler to convert all
flat indices to spinor indices.  In that case, we can write
$$ Â = Â^m »_m +Â^I M_I,ââ
	Â^I M_I = üÂ^{ab}M_{ba} = üÂ^{Œº}M_{ºŒ} +üÂ^{ÀŒÀº}M_{ÀºÀŒ} $$
$$ [M_{Œº},Æ_©] = Æ_{(Œ}C_{º)©}âÜâ
	Â^I[M_I,Æ_Œ] = üÂ^{º©}[M_{©º},Æ_Œ] = Â_Œ{}^º Æ_º $$
$$ [M_{Œº},M^{©¶}] = ¶_{(Œ}^{(©}M_{º)}{}^{¶)} $$
 in terms of the SL(2,C) generators $M_{Œº}=M_{ºŒ}$.  Note that
$(M_{Œº})ÿ=+ÑM_{ÀŒÀº}$ because $(M_{ab})ÿ=-M_{ab}$.  We have used
conventions consistent with OSp generators
$$ üÂ^{BC}[M_{CB},Æ_AÕ = Â_A{}^B Æ_B,ââ
	ú_{AB} = (ú_{ab},C_{Œº},C_{ÀŒÀº}) $$
 Relating vector to spinor indices as usual as $V_a=V_{ŒÀŒ}$, etc., then
fixes the Lorentz subgroup of the OSp group as (see exercise IIB7.2a)
$$ M_{ŒÀŒºÀº}V_{©À©} = V_{ŒÀŒ}C_{º©}C_{ÀºÀ©} -V_{ºÀº}C_{Œ©}C_{ÀŒÀ©} $$
$$ = -ü(C_{ÀŒÀº}V_{(ŒÀ©}C_{º)©} +C_{Œº}V_{©(ÀŒ}C_{Àº)À©}) =
	-ü(C_{ÀŒÀº}M_{Œº} +C_{Œº}M_{ÀŒÀº})V_{©À©} $$
$$ ÜâM_{ŒÀŒºÀº} = -ü(C_{ÀŒÀº}M_{Œº} +C_{Œº}M_{ÀŒÀº}) $$
$$ ÜâÂ_{ŒÀŒºÀº} = C_{ÀŒÀº}Â_{Œº} +C_{Œº}Â_{ÀŒÀº} $$
  For most of the remaining discussion of gravity, we'll limit ourselves to
bosonic fields in vector notation, which is easy to generalize to arbitrary
dimensions.  For spinors, we must either choose a dimension and use its
corresponding spinor notation (for $D²6$), or work in mixed spinor-vector
notation (which is much messier).

Matter representations of the group work similarly to Yang-Mills.  We
define such fields to have only flat indices.  Then their transformation
law is
$$ Æ' = e^Â Æ $$
 where the transformation of a general Lorentz representation follows
from that for a vector (or spinor, if we include them), as defined above. 
Alternatively, the transformation of a vector could be defined with curved
indices, being the adjoint representation of the coordinate group:
$$ V = V^m »_mâÜâV' = V'^m »_m = e^{Â^m »_m}V e^{-Â^m »_m} $$
 However, as in Yang-Mills theory, it is more convenient to identify only
the gauge field as an operator in the group.  In any case, only the adjoint
representation (and direct products of it) has such a nice operator
interpretation.

As an example of this algebra, we now work out the commutator of two
transformations in gory detail:  We first recall that the coordinate
transformation commutator was already worked out in subsection IC2,
using the usual quantum mechanical relations (see also subsection IA1)
$$ [f,f] = [»,»] = 0,ââ[»,f] = (»f) $$
 for any function $f$.  For the Lorentz algebra we will use the additional
identities
$$ [M_{ab},»_m] = [M_{ab},Â^m] = [M_{ab},Â^{cd}M_{dc}] = 0 $$
 all expressing the fact the Lorentz generators commute with anything
lacking free flat indices (i.e., Lorentz scalars).  The commutator algebra is
then
$$ \li{ [Â_1^m »_m+üÂ_1^{ab}M_{ba},& Â_2^n »_n +üÂ_2^{cd}M_{dc}]\cr
	={}& Â_{[1}^m [»_m,Â_{2]}^n]»_n +Â_{[1}^m [»_m,üÂ_{2]}^{ab}]M_{ba}
	+üÂ_1^{ab}[M_{ba},üÂ_2^{cd}]M_{dc} \cr
	={}& (Â_{[1}^n »_n Â_{2]}^m)»_m 
		+ü(Â_{[1}^m »_m Â_{2]}^{ab}+Â_{[1}^{ac}Â_{2]c}{}^b)M_{ba}\cr}$$

One fine point to worry about:  We may consider spaces with nontrivial
topologies, where it is not possible to choose a single coordinate system
for the entire space.  For example, on a sphere spherical coordinates have
singularities at the two poles, where varying the longitude gives the
same point and not a line.  (However, the sphere can be described by
coordinates with only one singular point.)  We then either treat such
points by a limiting procedure, or choose different sets of nonsingular
coordinates on different regions (``patches") and join them to cover the
space.

Ü2. Covariant derivatives

We can also define covariant derivatives in a manner similar to
Yang-Mills theory; however, since $»_m$ is now one of the generators,
the ``$»$" term can be absorbed into the ``$A$" term of $á=»+A$:
$$ á_a = e_a{}^m »_m +ü¿_a{}^{bc}M_{cb} $$
 in terms of the``vierbein (tetrad)" $e_a{}^m$ and ``Lorentz
connection" $¿_a{}^{bc}$.  Now the action of the covariant derivative on
matter fields looks even more similar to the gauge transformations: e.g.,
$$ ¶Ä = Â^m »_m Ä,ââá_a Ä = e_a{}^m »_m Ä $$
$$ ¶V_a = Â^m »_m V_a +Â_a{}^b V_b,ââ
	á_a V_b = e_a{}^m »_m V_b +¿_{ab}{}^c V_c $$
$$ ¶Æ_Œ = Â^m »_m Æ_Œ +Â_Œ{}^º Æ_º,ââ
	á_a Æ_º = e_a{}^m »_m Æ_º +¿_{aº}{}^© Æ_© $$
 I.e., the covariant derivative $á_a$ is essentially $D$ elements (labeled
by ``$a$") of the gauge algebra.

\x IXA2.1  Write the transformation law and covariant derivative of an
antisymmetric tensor in spinor notation ($f_{μ}$), and compare to vector
notation as above.

Note that the free index on the covariant derivative is flat so that it
transforms nontrivially under
$$ á'_a = e^Â á_a e^{-Â} $$
 Explicitly, for an infinitesimal transformation $¶á=[Â,á]$ we have
$$ ¶e_a{}^m = (Â^n »_n e_a{}^m -e_a{}^n »_n Â^m) +Â_a{}^b e_b{}^m $$
$$ ¶¿_a{}^{bc} = Â^m »_m ¿_a{}^{bc}
	+(-e_a{}^m »_m Â^{bc} +¿_a{}^{d[b}Â_d{}^{c]} +Â_a{}^d ¿_d{}^{bc}) $$
 This commutator is the same as for $[Â_1,Â_2]$ in the previous
subsection, except for the two additional terms coming from the Lorentz
generators acting on the free index on $á_a$.  In particular, the vierbein
$e_a{}^m$ transforms on its flat index as the vector (defining)
representation of the local Lorentz group, and on its curved index (and
argument) as the vector (adjoint) representation of the coordinate
group.  Also, it should be invertible, since originally we had $á=»+A$:  We
want to be able to separate out the flat space part as
$e_a{}^m=¶_a^m+h_a{}^m$ for perturbation theory or weak gravitational
fields.  That means we can use it to convert between curved and flat
indices:
$$ V^m = V^a e_a{}^mâÛâV^a = V^m e_m{}^a $$
 where $e_m{}^a$ is the inverse of $e_a{}^m$.  Furthermore, if we want
to define the covariant derivative of an object with curved indices, we
can simply flatten its indices, take the covariant derivative with $á$, and
then unflatten its indices.

Flat indices are the natural way to describe tensors:  (1) They are the
ÓonlyÕ way to describe half-(odd-)integer spin.  (2) Even for integer spin, they
correspond to the way components are actually measured.  
In fact, the above conversion of vectors from curved to flat indices
is exactly the one you learned in your freshman physics course!
The special cases you saw there were curvilinear coordinates (polar
or spherical) for flat space.  Then $e_a{}^m$ was the usual orthonormal
basis.  Only the notation was different:  Using Gibbs' notation for 
the curved but not the flat indices, $\vec V = V^a \vec e_a$, where, e.g.,
$a=(r,Ï,Ä)$ for spherical coordinates and $\vec e_a=(ör,öÏ,öÄ)$
are the usual orthonormal basis.  Thus, you probably learned 
about the vierbein years before you ever saw a ``metric tensor".
Similarly, when you learned how to integrate over the volume element
of spherical coordinates, you found it from this basis, and only
learned much later (if yet) to express it in terms of the square root
of the determinant of the metric.  (With the orthonormal basis,
there was no square root to take; the determinant came from the cross product.)
You also learned how to do this for curved space:  Considering
again the sphere, vectors in the sphere itself can be expressed in terms
of just $öÏ$ and $öÄ$.  And the area element of the sphere 
(the volume element of this smaller space) you
again found from this basis.

For example,
consider nonrelativistic momentum in two flat spatial dimensions, but
in polar coordinates:  Now using $x^m$ to represent just the
nonrelativistic spatial coordinates,
$$ x^m = (r,Ï),ââp^m = m{dx^m\over dt} = (mÀr, mÀÏ) = p^a e_a{}^m $$
$$ e_a{}^m = \pmatrix{ 1 & 0 \cr 0 & r^{-1} \cr },ââ
	p^a = ( mÀr, mrÀÏ ) $$
 Then the two components of $p^a$ (with the
simplest choice of $e_a{}^m$) are the usual components of momentum in
the radial and angular directions.  On the other hand, one component of
$p^m$ is still the radial component of the momentum, while the other
component of $p^m$ is the ÓangularÕ momentum --- a useful quantity,
but not normally considered as a component along with the radial
momentum, which doesn't even have the same engineering dimensions.
In writing the Hamiltonian, one simply squares $p^a$ in the naive way, 
whereas squaring $p^m$ would require use of the metric.

\x IXA2.2  Show that the above choice of $e_a{}^m$ actually describes
flat space:  Use the fact that $p^a$ transforms as a scalar under the
coordinate transformations that express $r$ and $Ï$ in terms of Cartesian
coordinates $x$ and $y$, and as a vector under local ``Lorentz"
transformations, which are in this case just 2D rotations, to transform it
to the usual Cartesian $p'^a=(mÀx,mÀy)$.

This direct conversion between curved and flat indices also leads directly
to the covariant generalization of length:  In terms of momentum (as
would appear in the action for the classical mechanics of the particle),
$$ p^m = m{dx^m\over ds},â¼-m^2 = p^2 = p^a p^b ú_{ab}âÜâ
	-ds^2 = dx^m dx^n e_m{}^a e_n{}^b ú_{ab} ­ dx^m dx^n g_{mn} $$
 Equivalently, the metric tensor $g_{mn}$ is just the conversion of the
flat-space metric $ú_{ab}$ to curved indices.  Also, in terms of
differential forms,
$$ ¯^a = dx^m e_m{}^aâÜâ-ds^2 = ¯^a ¯^b ú_{ab} $$
These curved generalizations of the energy-momentum relation and definition of proper time imply the corresponding generalization of the definitions of timelike, lightlike, and spacelike.

The field strengths are also defined as in Yang-Mills:
$$ \boxeq{ [á_a,á_b] = T_{ab}{}^c á_c +üR_{ab}{}^{cd}M_{dc} } $$
 where we have expanded the field strengths over $á$ and $M$ rather
than $»$ and $M$ so that the ``torsion" $T$ and ``curvature" $R$ are
manifestly covariant:
$$ M' = e^ M e^{-Â} = MâÜâT' = e^ T e^{-Â},âR' = e^ R e^{-Â} $$
 The commutator can be evaluated as before, with the same change as
for going from $[Â_1,Â_2]$ to $[Â,á_a]$  (i.e., now there are two free
indices on which the Lorentz generators can act), except that now we
rearrange terms to convert $»_m£e_a{}^m»_m£á_a$.  Making the further
definitions
$$ e_a = e_a{}^m »_m,â[e_a,e_b] = c_{ab}{}^c e_câÜâc_{ab}{}^c = 
	(e_{[a}e_{b]}{}^m)e_m{}^c = - e_a{}^m e_b{}^n »_{[m}e_{n]}{}^c $$
 for the ``structure functions" $c_{ab}{}^c$, we find the explicit
expressions
$$ T_{ab}{}^c = c_{ab}{}^c +¿_{[ab]}{}^c
	= -e_a{}^m e_b{}^n(»_{[m}e_{n]}{}^c +e_{[m}{}^d ¿_{n]d}{}^c) $$
$$ R_{ab}{}^{cd} 
	= e_{[a}¿_{b]}{}^{cd} -c_{ab}{}^e ¿_e{}^{cd} +¿_{[a}{}^{ce}¿_{b]e}{}^d
	= e_a{}^m e_b{}^n(»_{[m}¿_{n]}{}^{cd} +¿_{[m}{}^{ce}¿_{n]e}{}^d) $$
 If we ignore the action of $á$ on curved indices (it doesn't act on them,
but alternatively we could flatten them, act, then curve them back), we
can also write
$$ \boxeq{ T_{mn}{}^a = -á_{[m}e_{n]}{}^a,ââ
	[á_m,á_n] = üR_{mn}{}^{ab}M_{ba} } $$
 where
$$ á_m = e_m{}^a á_a = »_m +ü¿_m{}^{ab}M_{ba} $$
 is essentially a covariant derivative for the Lorentz group only.

From this expression for the torsion we find the following expressions for
the curl and divergence of a vector in terms of curved indices:  Defining
$$ {\bf e} ­ det¼e_a{}^m $$
$$ Üâc_{ba}{}^b = (e_{[b}{}^m »_m e_{a]}{}^n)e_n{}^b
	= (e_b{}^m »_m e_a{}^n)e_n{}^b -e_a{}^m[(»_m e_b{}^n)e_n{}^b]
		= »_m e_a{}^m -e_a{}^m »_m ln¼{\bf e} $$
 we have
$$ e_a{}^m e_b{}^n »_{[m}V_{n]} 
	 = e_{[a}(e_{b]}{}^m V_m) -c_{ab}{}^c e_c{}^m V_m
	= á_{[a}V_{b]} -T_{ab}{}^c V_c $$
$$ {\bf e}Ê»_m{\bf e}^{-1}V^m = e_a{}^m »_m V^a +c_{ba}{}^b V^a
	= á_a V^a +T_{ba}{}^b V^a $$

\x IXA2.3  Relate the two above identities by comparing (in D=4)
$á_{[a}V_{bcd]}$ (generalizing $á_{[a}V_{b]}$) and $á_a V^a$ for
$V_{abc}=·_{abcd}V^d$.

In practice, a useful way to evaluate the commutator is to first
evaluate the commutators of the Lorentz generators with the whole
covariant derivative, and then subtract out the double-counted $[M,M]$
term.  This is particularly convenient when considering some explicit
solution to the field equations with a reduced set of components (e.g.,
spherically symmetric), so that explicit indices may be lost except on the
Lorentz generators.  Schematically, we then calculate
\vskip.1in
\Boxeq{ $$ [á_1,á_2] = [e_1+¿_1,e_2+¿_2] $$
$$ = Ó[e_1,e_2]+(e_1 ¿_2)M_2-(e_2 ¿_1)M_1Õ
	+Ó¿_1[M_1,á_2]-¿_2[M_2,á_1]-¿_1 ¿_2[M_1,M_2]Õ $$ }

\noindent This method turns out to be one of the two simplest ways 
to calculate explicit solutions (as opposed to discussing general 
properties).  (For examples, see subsection IXC5 below.)

The covariant derivative satisfies the Bianchi (Jacobi) identities
$$ 0 = [á_{[a},[á_b,á_{c]}]] 
	= [á_{[a}, T_{bc]}{}^d á_d +üR_{bc]}{}^{de}M_{ed}] $$
$$ = (á_{[a}T_{bc]}{}^d)á_d +ü(á_{[a}R_{bc]}{}^{de})M_{ed}
	-T_{[ab|}{}^e (T_{e|c]}{}^f á_f +üR_{e|c]}{}^{fg}M_{gf}) 
	-R_{[abc]}{}^d á_d $$
$$ ÜâR_{[abc]}{}^d = á_{[a}T_{bc]}{}^d -T_{[ab|}{}^e T_{e|c]}{}^d,ââ
	á_{[a}R_{bc]}{}^{de} -T_{[ab|}{}^f R_{f|c]}{}^{de} = 0 $$

To make the transformation laws manifestly covariant we can define
instead
$$  = Â^a á_a  +üÂ^{ab}M_{ba} $$
 which is just a redefinition of the gauge parameters.  The infinitesimal
transformation law of the covariant derivative is then
$$ \li{ ¶á_a & = [(¶e_a{}^m)e_m{}^b]á_b +ü(e_a{}^m ¶¿_m{}^{bc})M_{cb}
	= [Â^b á_b +üÂ^{bc}M_{cb},á_a] \cr
	& = (-á_a Â^b +Â^c T_{ca}{}^b +Â_a{}^b)á_b 
	+ü(-á_a Â^{bc} +Â^d R_{da}{}^{bc})M_{cb} \cr} $$
$$ Üâ(¶e_a{}^m)e_m{}^b = -á_a Â^b +Â^c T_{ca}{}^b +Â_a{}^b,â
	e_a{}^m ¶¿_m{}^{bc} = -á_a Â^{bc} +Â^d R_{da}{}^{bc} $$

\x IXA2.4 Show that a ÓfiniteÕ local Lorentz transformation takes the form
$$ á'_a = ñ_a{}^b [ á_b - ü (á_b ñ^{cf})ñ^d{}_f M_{dc}] $$
 in the case of vanishing torsion by starting with the more general
expression
$$ á'_a = ñ_a{}^b á_b +üë_a{}^{bc}M_{cb} $$
 examining the commutator $[á',á']$, and applying $T'=0$ to determine $ë$.

A ``Killing vector" is a transformation that leaves the covariant
derivative invariant:
$$ [K,á_a] = 0,ââK = K^a á_a  +üK^{ab}M_{ba} $$
 (The term is usually used to refer to just the
general coordinate part $K^m$ of the transformation, but we'll use it in
a generalized sense to refer to the complete $K$.)  It represents a
symmetry; the existence of Killing vectors depends on the particular
space described by the covariant derivative.  It then follows from the
Jacobi identity for $[K_1,[K_2,á]]$ that the Killing vectors form a group,
the symmetry group of that space.  Invariance of the covariant
derivative requires:
$$ -á_a K^b +K^c T_{ca}{}^b +K_a{}^b = 0âÜâ
	á_{(a}K_{b)} = K^c T_{c(ab)},ââ
	K_{ab} = üá_{[a}K_{b]} -üK^c T_{c[ab]} $$
$$ -á_a K^{bc} +K^d R_{da}{}^{bc} = 0âÜâ
	á_a K^{bc} = K^d R_{da}{}^{bc} $$
 These equations are referred to as the ``Killing equations".  (Again,
usually it is just the first equation, on $K_a$, that is called by this name,
but we'll use it to refer also to the equations for $K_{ab}$, which are
needed to describe the symmetry when acting on spinors, etc.)

\x IXA2.5  Express the Hamiltonian of the classical relativistic particle in
terms of the vierbein:
$$ e_a = e_a{}^m p_m,âH = ü(ú^{ab}e_a e_b +m^2) $$
 Doing the same for general coordinate transformations $K = K^a e_a$,
examine the condition for invariance $[K,H] = 0$ using the Poisson
bracket.  Using the commutation relations for the $e_a$'s, show that this
implies the Killing equation $á_{(a}K_{b)}=K^c T_{c(ab)}$.

\x IXA2.6  Solve the Killing equations explicitly in the case of flat space
$á_a=»_a$.  Show this gives the Poincar«e group, including both orbital and
spin pieces.

Ü3. Conditions

There are two kinds of conditions we can impose to eliminate some
degrees of freedom: Ógauge choicesÕ and ÓconstraintsÕ.  Gauge choices
explicitly determine degrees of freedom that drop out of the action
anyway.  If the gauge is not completely fixed, the form of the residual
gauge transformations may change, since using particular gauge
parameters to fix the gauge, rather than eliminating those parameters,
may just determine them in terms of the remaining parameters:  We
require that the residual transformations do not violate the gauge
conditions that have already been applied.  Similar remarks apply to
global symmetries:  If they do not commute with the gauge
transformations for the gauge that was fixed, then they may aquire extra
gauge-transformation terms to preserve the gauge choice.  On the other
hand, constraints are chosen to be covariant under the transformation
laws, and thus do not alter them, while eliminating degrees of freedom
that might otherwise appear in the action (although not in all possible
terms).  Furthermore, the simplest explicit solution to constraints can
itself introduce new gauge invariances.  (An example of this situation is
supersymmetric Yang-Mills: see subsections IVC3-4.)  In this subsection
this analysis will be applied to Lorentz invariance:  We already saw that
global Lorentz transformations are included in coordinate
transformations, and that local Lorentz invariance is unnecessary when
only integer spin (and in particular, pure gravity) is treated.  We now
examine the consequences of eliminating this useful but redundant
invariance and the gauge field associated with it.

Of course, we can eliminate local Lorentz transformations by hiding flat
indices:  For the vierbein itself, we have the local Lorentz invariant
$$ g^{mn} = ú^{ab}e_a{}^m e_b{}^n $$
 which is the inverse ``metric tensor".  However, we have seen that
tensors with flat indices have simpler coordinate transformations, and
there is no way to get rid of flat indices when spinors are involved. 
Furthermore, the metric has the constraint that it have Minkowski
signature:  This constraint is solved by expressing the metric in terms of
the flat-space Minkowsi metric $ú$ and the vierbein.  Thus, solving the
constraint introduces local Lorentz invariance.  (However, in this case the
constraint does not eliminate degrees of freedom, but only limits their
range.)

The Lorentz transformations in $Â^{ab}$ are redundant to those in
$Â^m$.  The extra gauge parameters also can be fixed by an appropriate
gauge choice:  For example, consider the gauge
$$ ¶e_a{}^m = Â_a{}^b e_b{}^mâÜâ
	Lorentz¼gauge¼ú_{m[a}e_{b]}{}^m = 0 $$
 A coordinate transformation takes us to a different Lorenz gauge, since
the Lorenz gauge condition is not a scalar.  This means that any
coordinate transformation $Â^m$ must be accompanied by a Lorentz
transformation $Â^{ab}$ to preserve this gauge, where this $Â^{ab}$ is
completely determined in terms of $Â^m$.  This is easy to see
perturbing $e_a{}^m$ about $¶_a^m$:  To lowest order we have simply
$$ 0 = ¶(ú_{m[a}e_{b]}{}^m) ® -2Â_{ab} +»_{[a}Â_{b]}âÜâ
	Â_{ab} ® ü»_{[a}Â_{b]} $$

\x IXA3.1  Let's further analyze this gauge condition:
 ªa  By looking at the transformation of a vector, identify the specific
terms in the Taylor expansions of $Â^m$ and $Â^{ab}$ whose coefficients
can be identified with global Lorentz transformations, in the
approximation used above.
 ªb  Using the same methods as exercise IVC4.3, and writing in matrix
notation $e_a{}^m = (e^h)_a{}^m$ for some matrix $h$, solve explicitly for
$Â^{ab}$ in terms of $Â^m$ and $e_a{}^m$ to all orders.

Similarly, the Lorentz connection $¿_a{}^{bc}$ that gauges the $Â^{ab}$
transformations is redundant to the vierbein that gauges $Â^m$:  $¿$ can
be completely determined in terms of $e$ by constraining the torsion to
vanish.  To see this, we first notice that in the general case the
expression for the torsion in terms of the structure functions and
connection can be inverted to give the connection in terms of the other
two.  One way to do this is to use the definition and permute the indices
$a£b£c$ (odd permutations are redundant because of the antisymmetry
of the equation in the first two indices):
$$ T_{abc} = c_{abc} + ¿_{abc} - ¿_{bac},ââ
	T_{bca} = c_{bca} + ¿_{bca} - ¿_{cba},ââ
	T_{cab} = c_{cab} + ¿_{cab} - ¿_{acb} $$
 Using the antisymmetry of the connection in its last two indices, we add
the first and last equation and subtract the second to obtain
$$ ¿_{abc} = ü(÷c_{bca}-÷c_{a[bc]}),ââ÷c_{abc} = c_{abc} -T_{abc} $$
 Since the torsion is a covariant tensor, we can freely set it to vanish
without affecting the transformation laws of the remaining objects (it's a
covariant constraint, not a gauge condition):
$$ T_{ab}{}^c = 0âÜâ¿_{abc} = ü(c_{bca}-c_{a[bc]}) $$
 From now on we assume this constraint is satisfied.  This simplifies the
form of curls and divergences, which implies that $á$ can be integrated
by parts in covariant actions (see below).  However, we have already
seen that the torsion is nonvanishing in superspace (subsection IVC3):  In
that case the symmetry on flat indices is constrained, so the connection
has fewer components than the torsion, and can be determined by setting
only part of the torsion to vanish.  (See subsection XA1 below.)

\x IXA3.2  Show explicitly that when the torsion vanishes the
Killing equations from $¶e_a{}^m=0$ imply those from $¶¿_m{}^{ab}=0$:
$$ á_{(a}Â_{b)} = 0,ââÂ_{ab} = üá_{[a}Â_{b]}âÜâ
	á_a Â^{bc} = Â^d R_{da}{}^{bc} $$

\x IXA3.3  Consider using the group GL(D) on the flat indices instead of
SO(D$-$1,1).  (This construction is not useful for fermions.)  Compensate
for the extra gauge invariance by replacing the Minkowski metric
$ú_{ab}$ with a ``flat"-index metric $g_{ab}$ (and its inverse $g^{ab}$)
that is Ócoordinate dependentÕ, but covariantly constant:
$$  = Â^m »_m +Â_a{}^b G_b{}^a;ââÂ_a{}^b[G_b{}^a,V_c] = Â_c{}^a V_a,
	ââÂ_a{}^b[G_b{}^a,V^c] = -V^a Â_a{}^c $$
$$ á_a = e_a +¿_{ab}{}^c G_c{}^b,ââ
	[á_a,á_b] = T_{ab}{}^c á_c +R_{abc}{}^d G_d{}^c $$
$$ g^{mn} = g^{ab}e_a{}^m e_b{}^n,ââá_a g^{bc} = 0 $$
 where now there is no (anti)symmetry associated with the indices on
$G_a{}^b$ (or $Â_a{}^b$, etc.).  As a result, $g_{ab}$ transforms
nontrivially under both coordinate and GL(D) transformations.  Use it (in
place of $ú$) to raise and lower flat indices.
 ªa Find the explicit expressions for the torsion and curvature in
terms of the vierbein and connection.  Solve these, and $ág=0$, for the
connection in terms of the torsion and vierbein as
$$ ¿_{abc} = ü(÷c_{bca}-÷c_{a[bc]}) +ü(e_c g_{ab} -e_{(a}g_{b)c}),ââ
	÷c_{abc} = c_{abc} -T_{abc} $$
 Show that there exists a GL(D) ÓgaugeÕ 
$$ g_{ab} = ú_{ab} $$
 (ÓassumingÕ $g_{ab}$ has the right signature), that gauge has as a
residual flat-index invariance SO(D$-$1,1), and the resulting
covariant derivative is identical to that used earlier in this subsection.
 ªb Show that one can instead choose a GL(D) gauge
$$ e_a{}^m = ¶_a^mâÜâg^{mn} = g^{ab} $$
 and that this completely fixes the GL(D) invariance.  Since the vierbein
has a curved index, the covariant derivatives are no longer covariant: 
Unlike the previous gauge, to maintain this gauge any coordinate
transformation must be accompanied by a GL(D) transformation whose
parameter is determined by the coordinate transformation parameter. 
Find the solution for $Â_a{}^b$ in terms of $Â^m$ in the infinitesimal
case.  Compare with the transformation law for curved indices (see
subsection IC2).  In this gauge the connection is known as the
``Christoffel symbols".

The vanishing of the torsion simplifies the Bianchi identities on the
curvature:
$$ T_{ab}{}^c = 0âÜâR_{[abc]d} = á_{[a}R_{bc]de} = 0
	âÜâR_{abcd} = R_{cdab} $$
 In terms of SU(N)-like Young tableaux, this means the curvature is of the
form $\upõ2{\õ2\õ2}$.  For SO(N) Young tableaux, we subtract out the
trace pieces:
$$ R_{abcd} £ \upõ2{\õ2\õ2} ¢¼ \õ1\õ1 ¢ \bullet $$
 where the first term is the ``Weyl tensor" $W_{abcd}$ (traceless), the
last two terms combine to give the ``Ricci tensor"
$R_{ab}­R_a{}^c{}_{bc}$, and the last (singlet) term is the ``Ricci scalar"
$R­R^a{}_a=R^{ab}{}_{ab}$. They're simpler in spinor notation in $D=4$: 
Since $[ab]£(Œº)$ and $(ÀŒÀº)$,
$$ R_{abcd} £ R_{(Œº)(ÀŒÀº)} ¢
	( R_{(Œº)(©¶)} = W_{(Œº©¶)} +C_{(Œ(©}C_{¶)º)}R ) $$
 in terms of Weyl $W_{(Œº©¶)}$, the traceless part of Ricci $R_{(Œº)(ÀŒÀº)}$,
and the Ricci scalar $R$.  Later we'll see that the Ricci tensor is fixed
exactly by the equations of motion.  That leaves the Weyl tensor as the
on-shell field strength.  As explained in subsection IIB7, it describes
helicity $à$2.

\x IXA3.4  Prove that $R_{abcd}=R_{cdab}$ follows from the Bianchi
identity $R_{[abc]d}=0$ and the antisymmetry of $R_{abcd}$ in both $ab$
and $cd$.

Ü4. Integration

The antihermitian form of the group generators was a convenient choice
because partial derivatives are antihermitian, and the generators of the
Lorentz group (which is real and orthogonal) are antisymmetric in the
vector representation.  Thus, the generators are real.  However, the group
elements are not unitary, since hermitian conjugation reorders $Â^m$
with respect to $»_m$.  The fix comes from noticing that
$$ {\bf e} = det¼e_a{}^mâÜâ
	¶¼ln¼{\bf e} = e_m{}^a ¶e_a{}^m = Â^m »_m ln¼{\bf e} -»_m Â^m $$
$$ Üâ¶({\bf e}^{-1}) = {\bf e}^{-1}\onÁ = {\bf e}^{-1}Â^m\onÁ»{}_m
	âÜâ({\bf e}^{-1})' = {\bf e}^{-1}e^{\onÁÂ} $$
  where the derivatives $\onÁ»$ act on everything to the left, $Â$ now
includes just coordinate transformations, and we have exponentiated by
the same method as for Lie groups in subsection IA3.  (Note that if we 
expand the exponential in a Taylor series such derivatives in all but the
first factor will hit $Â$'s, just as for those in $e^Â$ acting to the right.)
Any function that transforms in this way is known as a ``density" (see
subsection IIIB1 for the 1D case).  We can easily see from the
infinitesimal transformation that a density times any scalar is also a
density.  This allows invariant actions to be constructed as
$$ S = Çdx¼{\bf e}^{-1}L $$
 for any scalar $L$.  For cases without spinors we can also use
$$ g ­ det¼g_{mn} = -{\bf e}^{-2}âÜâ{\bf e}^{-1} = å{-g} $$
 where $g_{mn}$ is the inverse of $g^{mn}$.  
(In spaces of general signature, i.e., arbitrary numbers of
time dimensions, we should write $å{|g|}$ so, e.g., in
Euclidean space we actually use $å{g}$.  If we were even more general,
and used $|det¼ú|±1$, then it would also appear.)
This can also be understood in terms of differential forms, since
$$ ¯^a = dx^m e_m{}^aâÜâ¯^4 = 
	dx^m dx^n dx^p dx^q e_m{}^0 e_n{}^1 e_p{}^2 e_q{}^3
	= d^4 x¼{\bf e}^{-1} $$
$$ ¯'^a(x')=¯^a(x)âÜâ\left(ǯ^4 L\right)' = ǯ^4 L $$
 under coordinate transformations.  

\x IXA4.1  Let's look at some properties of transformations acting
backwards:
 ªa  Show that for any function $f$
$$ Âf = [Â,f] = [f,\onÁÂ]âÜâ
	e^Â f = e^Â f e^{-Â} = e^{-\onÁÂ} f e^{\onÁÂ} $$
 and use it to show that the product of ${\bf e}^{-1}$ with any scalar
transforms the same way as ${\bf e}^{-1}$ (i.e., is a density) under a
finite coordinate transformation.
 ªb  Derive
$$ fe^{\onÁÂ} = \left(1Ée^{\onÁÂ}\right)(e^Â f) $$
 (where the derivatives in each factor of $\onÁÂ$ act on everything to
the left, but vanish on ``1").

\x IXA4.2  We now examine finite transformations in terms of
transformed coordinates (see subsection IC2):
 ªb  Show that
$$ det \left( {»x'\over »x} \right) = 1Ée^{-\onÁÂ} $$
by evaluating
$$ Çdx¼{\bf e}^{-1}(x) = Çdx'¼{\bf e}'^{-1}(x'),ââ
	dx' = dx¼det \left( {»x'\over »x} \right) $$
 ªc  Show that
$$ det \left( {»÷x\over »x} \right) = 1Ée^{\onÁÂ} $$
 by similarly evaluating
$$ Çdx¼{\bf e}^{-1}(x) = Çd÷x¼{\bf e}^{-1}(÷x) = Çdx¼{\bf e}'^{-1}(x) $$

From the results of subsection IXA2, we then have that covariant
derivatives can be integrated by parts in such actions, since
$$ T_{ba}{}^b = 0âÜâ
	Çdx¼{\bf e}^{-1}á_a V^a = Çdx¼»_m {\bf e}^{-1}V^m $$

\x IXA4.3  ÓDeriveÕ the expression for the covariant divergence in terms
of {\bf e} and the partial divergence by assuming integration by parts:
$$ Ç{\bf e}^{-1}Äá_a V^a = -Ç{\bf e}^{-1}V^a á_a Ä $$
 Use this to find a simple form for the covariant d'Alembertian on a
scalar:
$$ õÄ ­ á^2 Ä = {1\over å{-g}}»_m å{-g}g^{mn}»_n Ä $$

Actions for matter are constructed in a similar way to Yang-Mills: 
Starting with the flat-space action, replace ordinary derivatives with
covariant derivatives.  The new ingredient is the extra factor of 
${\bf e}^{-1}$.  This prescription, as for Yang-Mills, is unambiguous up to
only field-strength (curvature) terms, which can usually be eliminated by
symmetry requirements and dimensional analysis.  (At least for low
energies, we want terms of the lowest mass dimension.)  This uniqueness
(at low energies or long distances) is known as the ``equivalence
principle":  Inertial ``mass" (really energy, but also momentum), as
determined by the kinetic term, is the same as gravitational mass, as
determined by the coupling of the gravitational field.

A simple example of matter is a real scalar field:
$$ S = Ç{\bf e}^{-1}\f14[(á)^2 + m^2 ^2 + aR^2] $$
 The constant $a$ can sometimes be fixed by symmetry:  In the massless
case, to preserve the global symmetry $¶=·$, we must have $a=0$.  (For
self-interacting scalars, this generalizes to a global nonabelian
symmetry.)  To preserve conformal symmetry (see subsection IXA7),
also for the massless case, we need  $a=\f14{D-2\over D-1}$.  

This form of actions in terms of scalar Lagrangians also suggests we
modify the definition of functional variation for convenience and
covariance:
$$ ¶S = Çdx¼{\bf e}^{-1}(¶Ä){¶S\over ¶Ä} $$
 or, equivalently, we use the covariant form of the $¶$ function,
$$ {¶Ä(x)\over ¶Ä(x')} = {\bf e}(x)¶(x-x') $$

As in flat space, the action for electromagnetism follows from gauge
invariance:
$$ S = \f1{8e^2}Ç{\bf e}^{-1}F_{ab}^2 =
	\f1{8e^2}Çå{-g}g^{mn}g^{pq}F_{mp}F_{nq} $$
where $F_{mn}=»_{[m}A_{n]}$.  Integration by parts then gives a simple
form for Maxwell's equations.  Such simple covariant equations of
motion that don't require explicit expressions for the Lorentz
connection appear only for antisymmetric tensors (which in practice
means just spin 0 and 1 in 4D).

\x IXA4.4  Methods related to differential forms can be applied to these
special cases:
 ªa  Rewrite the above action for electromagnetism in terms of
$A_a$ and ÓcovariantÕ derivatives.  Find the field equations following
from both forms of the action, and use this to find a simple expression for
the covariant divergence of an antisymmetric tensor with curved indices
using just the metric.  Compare the results of the previous exercise.
 ªb  By converting flat indices on the covariant tensor $·_{abcd}$ to
curved, show that 
$$ å{-g}·_{mnpq}âandâ{1\over å{-g}}·^{mnpq} $$
 are also covariant tensors.  Use these, and the covariance of the curl (see
subsection IC2), to arrive at the same expression for the covariant
divergence of an antisymmetric tensor.

Another example is a Dirac spinor:
$$ S = Ç{\bf e}^{-1} Ðï(©^a iá_a + \f{m}{å2})ï $$
where $©^a$ are the usual constant Dirac matrices, in terms of which the
spin operator appearing in $á$ is the usual
$-M_{ab}£S_{ab}=-ü©_{[a}©_{b]}$.  In 4D, we can rewrite this in spinor
notation by simply replacing $»_{ŒÀº} £ á_{ŒÀº}$ in the flat-space
expressions given in subsection IIIA4, and replacing $M_{ab}£M_{Œº}$ as
described in subsection IXA1, as well as
$Çd^4x £ Çd^4x¼{\bf e}^{-1}$.

Ü5. Gravity

The Einstein-Hilbert action for gravity follows from choosing the only
available scalar second-order in derivatives, the Ricci scalar:
$$ L_G = -\f14 R = -\f14 R_{ab}{}^{ab} $$
This action normally has a coefficient of $1/û^2$ (compare Yang-Mills),
but we'll generally use (natural/Planck) units $û=1$; then $û$ is used only
to parametrize expansion about the vacuum and define the weak-field
limit.  (Actually, Planck units normally use $G=1$, whereas in our
conventions $û=1£G=¹$.)  In any case, the $û$'s can always be absorbed
(unlike Yang-Mills) by a field redefinition of $e_a{}^m$, and then appear
only in the definition of the ``vacuum" (perturbative ground state, or
solution that defines the boundary conditions at infinity):
$$ Òe_a{}^mÔ = û^{2/(D-2)}¶_a{}^m $$
 This makes $e_a{}^m »_m$, and thus $dx^m e_m{}^a$ and $ds^2$,
dimensionless.  In this sense, gravity is a theory with ``spontaneous
breakdown" of conformal invariance:  Coordinate transformations include
conformal transformations, but this invariance is broken by the vacuum,
which introduces a length scale ($û$).

\x IXA5.1  Consider the covariant derivative for nonvanishing torsion.  By
solving for the Lorentz connection in terms of the structure functions and
torsion, express the covariant derivative in terms of the torsion-free
covariant derivative $\ron\circá$ and the torsion.  Thus, any action in
terms of $á$ can be rewritten in terms of $\ron\circá$ and $T$, so any
theory with a nonvanishing torsion is equivalent to a similar one with
vanishing torsion (assuming the action is only second-order in derivatives
of the vierbein, and thus algebraic in the torsion).  Take the commutator
of two $á$'s to find the curvature in terms of the torsion-free curvature
$\on\circ R_{abcd}$.  Write the Einstein-Hilbert action with nonvanishing
torsion in terms of $\on\circ R$, $\ron\circá$, and $T$, to find:
$$ R = \on\circ R  -(T_{ab}{}^b)^2 -üT^{abc}T_{bca} 
	+\f14 T^{abc}T_{abc} -2\ron\circá{}^a T_{ab}{}^b $$
 Since the last term vanishes upon integration, $T$ appears as an auxiliary
field, so $R$ is equivalent to just $\on\circ R$.

\x IXA5.2  For some general applications, where the form of the vierbein is
not specified, it is useful to have a more explicit expression for the action
in terms of the vierbein.  We found in subsection IXA2 that for vanishing
torsion
$$ á_a V^a = {\bf e}Ê»_m ({\bf e}^{-1}e_a{}^m V^a) =
	e_a V^a -c_{ab}{}^b V^a $$
 Use this to show
$$ R =  (c_{ab}{}^b)^2 +üc^{abc}c_{bca} -\f14 c^{abc}c_{abc}
	-2{\bf e}Ê»_m[e^{am}»_n(e_a{}^n{\bf e}^{-1})] $$
 We can drop the last term in the action integral under appropriate
boundary conditions.  (Hint:  Use the result of the previous exercise for
$¿=0$.)

\x IXA5.3  In two dimensions there is a single Lorentz generator,
$$ M_{ab} = ·_{ab}MâÜâá_a = e_a +¿_a M,ââ
	[á_a,á_b] = -ü·_{ab}RM $$
 ªa  Show that the connection and the only surviving part of the curvature
then take the simple forms
$$ ¿_a = -·_{ab}{\bf e}»_m {\bf e}^{-1}e^{bm},ââ
	R = -2{\bf e}»_m[e^{am}»_n({\bf e}^{-1}e_a{}^n)]
		= -2{\bf e}^{-1}\onÁe{}^a \onÁe_a{\bf e} $$
 ªb  Derive, for the sphere in spherical coordinates,
$$ e_a{}^m = \pmatrix{ 1 & 0 \cr 0 & {1\over sin¼Ï} \cr } $$
 (Hint:  First use $ds^2 = dx^m dx^n g_{mn}$ in 3D flat space.)
Then show the covariant derivative is
$$ á_Ï = »_Ï,ââá_Ä = {1\over sin¼Ï}»_Ä +cot¼Ï¼M_{ÏÄ} $$
 ªc  Use these results to calculate $Çdx¼{\bf e}^{-1}R$ for the sphere in
two ways: (1) by showing $R$ is a constant and pulling it out of the
integral, and (2) by converting it into a boundary term, where the
``boundary" consists of infinitesimal circles around the coordinate
singularities at the poles.  (In general, even for spaces without true
boundaries, one has to treat the boundaries of patches as such.)

It's also possible to add a ``cosmological term" to the gravitational action:
$$ S_{cos} = ñ Çdx¼{\bf e}^{-1} $$
 with the ``cosmological constant" $ñ$.  This term has no derivatives, and
is thus analogous to a mass term.  However, it only contributes to the
nonpropagating spin-0 mode of the vierbein (see later), so it doesn't give
a physical mass, but does modify the vacuum.

\x IXA5.4  Show that the action for gravity can be made ÓpolynomialÕ in
$e_a{}^m$ by a field redefinition (rescaling) of the form
$$ e_a{}^m £ {\bf e}^k e_a{}^m $$
 when $k$ takes the values
$$ k = -{n+1\over D-2},âân = 2,3,4,... $$
 and that the resulting action is order $Dn+2$ in the field.  In what cases
(of $n$ and $D$) is the cosmological term also polynomial?

The variation of the curvature can be obtained directly by varying its
definition in terms of $[á,á]$.  We start with the definition
$$ ¶e_a{}^m ­ ½_a{}^b e_b{}^mâÛâ½_a{}^b ­ (¶e_a{}^m) e_m{}^b $$
 and work in terms of the flattened object $½_{ab}$.  Then we drop its
Lorentz piece, choosing $½_{ab}=½_{ba}$.  We find:
$$ ¶á_a = ½_a{}^b á_b + ü½_a{}^{bc} M_{cb} $$
$$ Ü ü(¶R_{ab}{}^{cd})M_{dc} = [á_{[a},¶á_{b]}] =
	(á_{[a}½_{b]}{}^c)á_c - ½_{[a}{}^c üR_{b]c}{}^{de}M_{ed}
	+üá_{[a}½_{b]}{}^{cd}M_{dc} + ½_{[ab]}{}^c á_c $$
$$ Üâá_{[a}½_{b]}{}^c + ½_{[ab]}{}^c = 0,ââ
	¶R_{ab}{}^{cd} = á_{[a}½_{b]}{}^{cd} - ½_{[a}{}^e R_{b]e}{}^{cd} $$
$$ Üâ½_{abc} = á_{[b}½_{c]a} $$
$$ Üâ¶R_{ab}{}^{cd} = üÓá_{[a},á^{[c}Õ½_{b]}{}^{d]}
	-ü(½_{[a}{}^eR_{b]e}{}^{cd} + {}_{ab} ª {}^{cd}) $$
$$ ¶{\bf e}^{-1} = -{\bf e}^{-1}¶¼ln¼{\bf e} = -{\bf e}^{-1}e_m{}^a ¶e_a{}^m
	= -{\bf e}^{-1}½^a{}_a $$
$$ Üâ¶({\bf e}^{-1}R) = 
	2{\bf e}^{-1}[(ú^{ab}õ-á^aá^b) +(R^{ab}-üú^{ab}R)]½_{ab} $$
where $õ­á^a á_a$.  Thus for pure gravity we have the field equations
$$ ¶S_G = 0âÜâR_{ab}-üú_{ab}R = 0âÜâR_{ab} = R = 0 $$
 while with a cosmological constant we have
$$ ¶S_G +¶S_{cos} = 0âÜâR_{ab}-üú_{ab}(R-4ñ) = 0âÜâ
	R_{ab} -\f1D ú_{ab}R = 0,âR = 4\f{D}{D-2}ñ $$

Note that calculating a variation is the same as performing a perturbation
to lowest order:  We will use this result in subsection IXB1.

\x IXA5.5  For gravity, a first-order formalism follows from not imposing
the torsion constraint (see exercise IXA5.1), so either the torsion or the
Lorentz connection can be treated as the auxiliary variable.  
 ªa  Find a first-order action for gravity (in all D) by treating $e_m{}^a$
and $¿_m{}^{ab}$ as the independent variables.  In D=4, using $·^{mnpq}$,
write this action as polynomial in these variables, eliminating the explicit
$\bf e$, to obtain
$$ S_G = Çd^4 x¼\f1{16}·^{mnpq}·_{abcd}e_m{}^a e_n{}^b R_{pq}{}^{cd} $$
 with $R_{pq}{}^{cd}$ in terms of just $¿$.
 ªb  Vary this action with respect to both $e$ and $¿$ (independently) to
find the field equations, expressed in terms of torsion and curvature,
using $¶[á_m,á_n]$ to find the variation of $R_{pq}{}^{cd}$ (see
subsection IXA2).

\x IXA5.6  As discussed in subsection IIIC4 for Yang-Mills, in four
dimensions we can write a complex first-order action for gravity that
yields the usual gravity action up to a surface term.  For Yang-Mills, the
complex action was obtained by starting with a normal first-order
formalism and replacing the auxiliary field with its self-dual part.
 ªa  Starting with the first-order action of the previous problem, find the
analog for gravity by keeping just the part of $¿_m{}^{ab}$ self-dual in
$ab$, using spinor notation.
 ªb  Associate the coupling $û$ with the term quadratic in $¿$
(analogously to the Yang-Mills case).  As for Yang-Mills, associate the
self-dual theory with the limit $û£0$.  Find the equation for $e_m{}^a$
that follows from varying $¿$ in this case, and show that it is equivalent
to setting the self-dual part of $\on\circ ¿_m{}^{ab}$ to zero, where 
$\on\circ ¿$ is the usual torsion-free connection.  Show this is equivalent
to setting the self-dual part of the curvature $R_{mn}{}^{ab}$ to vanish, in
an appropriate gauge.  (Technically, this means we must either
complexify the fields, or Wick rotate to 4+0 or 2+2 space+time
dimensions, where the Lorentz group factorizes.)

Ü6. Energy-momentum

In subsection IIIB4 we saw that in the same way as a current in
electrodynamics or Yang-Mills is defined as the matter contribution to
the gauge field's equation of motion, $¶S_M/¶A_a=J^a$ (in that case
$S_M$ excludes only the pure Yang-Mills action), the ``energy-momentum
tensor" is defined as the matter contribution to the gravitational field
equation (in this case $S_M$ excludes only the pure gravity action):
$$ ¶S_M = Ç{\bf e}^{-1}½^{ab}T_{ab} = üÇå{-g}(¶g^{mn})T_{mn}
	= -üÇå{-g}(¶g_{mn})T^{mn} $$
 The case where $½_{ab}$ represents the invariances of the action implies
restrictions on this tensor:  Using the separate gauge invariance of the
matter action $¶_{gauge}S_M=0$ and the matter field equations
$¶S_M/¶(matter)=0$ (as for the Yang-Mills case), gauge variation of the
gravity fields in $S_M$ implies
$$ ½_{ab} = \cases{ Â_{ab} = -Â_{ba}âÜâT^{[ab]} = 0 : & Lorentz \cr
	-üá_{(a}Â_{b)}âÜâá_a T^{ab} = 0 : & coordinate \cr} $$
 so coordinate invariance of the action implies local conservation of
energy-momentum.  

For example, for a real scalar field:
$$ S = Ç{\bf e}^{-1}\f14[(á)^2 + m^2 ^2 + aR^2] $$
$$ ܼ2T_{ab} = (á_a )(á_b )-üú_{ab}[(á)^2 + m^2 ^2]
	+ a[(ú_{ab}õ-á_a á_b)+(R_{ab}-üú_{ab}R)]^2 $$
 Notice that for $a±0$, the energy-momentum tensor gets extra
total-derivative terms which are separately conserved  in flat space
(since they come from the $R^2$ term, which is separately covariant).

\x IXA6.1  Show that for $a=\f14$ (using the scalar's free field equation) one obtains a result in agreement with that at the end of subsection IIIA4 in flat space. This is the simplest form of the energy-momentum tensor, and the most physical (since it involves only the relative momentum of the two fields, not the total). This choice for $a$ is also favored by string theory, as we'll see later.

\x IXA6.2  Using the action given in subsection IXA4 and the variation of
the covariant derivative from subsection IXA5, find the
energy-momentum tensor for the Dirac spinor, and use its field equations
to show this tensor is conserved.

Note that this is not the same as ordinary conservation $»_mT^{mn}=0$: 
$Çå{-g}T^{0n}$ does not define a conserved total energy-momentum.  This
is in contrast with the conserved current in electrodynamics, since we
then can derive the usual global conservation law
$$ 0 = Çd^Dx¼{\bf e}^{-1}á_a J^a = Çd^Dx¼»_m{\bf e}^{-1}J^m
	¾ {d\over dt}Çd^{D-1}x¼{\bf e}^{-1}J^0 $$
 On the other hand, it's closely related to Yang-Mills, where $¶A_a=-á_aÂ$
leads to $á_a J^a=0$ in terms of the derivative $á$ covariantized with
respect to the Yang-Mills field (as well as gravity, if in curved space), so
$»_m{\bf e}^{-1}J^m=-{\bf e}^{-1}[iA_m,J^m]±0$ (see subsection IIIC1).

However, if there is a Killing vector $K_a$, then the component of
momentum in that direction is conserved:
$$ J^a ­ K_b T^{ba}âÜâá_a J^a = (á_a K_b)T^{ba} + K_b(á_a T^{ba})
	=0 $$
(Remember $á_{(a}K_{b)}=0$.)  Some simple examples of this in flat space
are $(K_a)^b=¶_a^b$ (translational invariance), for which the
corresponding ``charge" is the total momentum, and
$(K_a)^{bc}=¶_a^{[b}x^{c]}$ (Lorentz invariance), for which the charge is
the total angular momentum.

Including the variation of the gravitational action, we get the
gravitational field equations
$$ R_{ab}-üú_{ab}R = 2T_{ab} $$
Coordinate invariance of $S_G$ implies $á_a(R^{ab}-üú^{ab}R)=0$,
which also follows from the Bianchi identities:  In that sense gauge
invariance is said to be ``dual" to Bianchi identities, one implying the
other through variation of the action:  In general, for any gauge field $Ä$
with gauge parameter $Â$
$$ ¶Ä = \O Â,ââ
	0 = ¶S = Çdx¼(\O Â){¶S\over ¶Ä}âÛâ\O^T{¶S\over ¶Ä} = 0 $$
 where the ``transpose" $\O^T$ is defined by integration by parts. 
Positivity of the energy (contained in any infinitesimal volume) is the
condition $T^{00}³0$.  The addition of the cosmological term modifies the
left-hand side of the above equation of motion by adding a term
$2ú_{ab}ñ$.

Although there is no covariant definition of total energy-momentum, in
the case where spacetime is asymptotically flat (the metric falls off to
the flat metric sufficiently fast at infinity), one can define a noncovariant
energy-momentum tensor $t_{ab}$ for gravity itself which is covariant
with respect to coordinate transformations that themselves fall off at
infinity.  (See exercise IIIC1.2 for the analogous Yang-Mills case.)  This
tensor satisfies $»_m(T^{mn}+t^{mn})=0$ (where $T^{mn}$ is the usual
tensor for matter), so the usual conservation laws can be derived for the
total energy-momentum coming from integrating $T+t$.  Many equivalent
expressions exist for $t$.  One way to derive it is to expand the field
equations order-by-order in $h$ as
$$ ü(R_{ab}-üú_{ab}R) ­ L_{ab} -t_{ab} $$
where $L_{ab}$ is the linearized part of the field equations (see
subsection IXB1) and $-t_{ab}$ is the quadratic and higher-order parts.  By
the linearized Bianchi identities, we know
$$ 0 = »_a L^{ab} ­ »_a (üR^{ab}-\f14 ú^{ab}R +t^{ab})
	= »_a (T^{ab} +t^{ab}) $$
where we used the field equations in the last step.  Note that there is a
great deal of ambiguity here:  We could have linearized by expanding
the metric around its flat space value instead of the vierbein, or by
expanding $R_{mn}$ or $R^{mn}$ instead of $R_{ab}$, etc.  Because of
the expression in terms of $L_{ab}¾»»h$, the integral of $T+t$, which gives
the total energy-momentum vector, can be expressed as a surface term,
just as Gauss' law in electrodynamics.  Since space was assumed to be
asymptotically flat, only the quadratic part of $t$ contributes in the
surface integral, which is why there is so much freedom in the
definition of $t$.  Since $t$ is not covariant, the energy-momentum of
the gravitational field is not localized (coordinate transformations
shift it around).  However, since the total energy-momentum is
invariant, one can ask questions about how much energy is radiated to
infinity, etc.

Ü7. Weyl scale

The simplest way to describe conformal transformations in field theory is
as a local scale transformation.  If the theory is not coupled to gravity, we
couple it to gravity as in Yang-Mills theory by replacing a Poincar«e
invariant Lagrangian $L(»,Æ)$ with $L(á,Æ)$ (where all fields $Æ$ have
flat indices), but also including the ${\bf e}^{-1}$ factor in the action.  We
then transform the fields as
$$ e_a{}^m £ ì e_a{}^m,ââÆ £ ì^{w+(D-2)/2} Æ $$
 where $ì$ is the gauge parameter and $w+\f{D-2}2$ is the engineering
dimension (scale weight) of the field $Æ$.  (See subsection IIB1.) 
Effectively, $e_a{}^m$ has dimension 1, since it's the only field with
curved indices, and thus any derivative must appear in the combination
$e_a{}^m »_m$, while the measure appears as $dx¼{\bf e}^{-1}$.  Of
course, the action won't be locally scale invariant unless it is globally
scale invariant, i.e., has only dimensionless coupling constants (and thus
no masses).

If the gravity-coupled theory is invariant under this local scale
transformation, then the theory will be conformally invariant after
decoupling gravity.  This follows from the fact that the most general
combined coordinate and local scale and Lorentz transformation that
preserves the flat-space vierbein $e_a{}^m=¶_a^m$ is exactly a
conformal transformation.  This is equivalent to our previous definition in
terms of the scaling of the flat-space $ds^2$ under conformal
transformations, since $dx'^m dx'^n g'_{mn}(x')=dx^m dx^n g_{mn}(x)$
under coordinate transformations.

\x IXA7.1  Derive the usual conformal transformations by finding the most
general local scale + Lorentz + coordinate transformation that preserves
the flat-space vierbein.

A simple example is Yang-Mills theory.  We look at the Yang-Mills field
with curved index, since its gauge transformation does not depend on
the vierbein.  ($¶A_m=-»_m Â+...$ vs.¼$¶A_a=-e_a{}^m »_m Â+...$.)  To avoid
interference with the Yang-Mills gauge transformation, the Yang-Mills
field with curved index must be scale invariant.  Then the action
$$ S = \f1{8e^2} Ç{\bf e}^{-1}e^{am}e^{bn}e_a{}^p e_b{}^q F_{mn}F_{pq} $$
 transforms with a factor $ì^{4-D}$, and so is invariant in $D=4$ only.

\x IXA7.2  Consider a more general gauge field $A$ and field strength $F$
defined by
$$ ¶A_{m_1òm_N} = -\f1{(N-1)!}»_{[m_1}Â_{m_2òm_N]},ââ
	F_{m_1òm_{N+1}} = \f1{N!}»_{[m_1}A_{m_2òm_{N+1}]} $$
 where $A$ is totally antisymmetric in its $N$ indices.  (Such theories were
encountered in exercise IIB2.1b.)
 ªa  Define an action in terms of $F^2$.  In what dimension $D(N)$ is it
conformally invariant?
 ªb  Show that this theory is related by a ``duality transformation"
(switching Bianchi identities and field equations) to the theory with $N'$
indices on a new $A$, where $N'=D-2-N$, and $D(N')=D(N)$.
 ªc  Examine the cases $N=D,D-1,D-2$.  Note that the scalar obtained by
duality does not have an $RÄ^2$ term in its action, and thus is conformal
only in $D=2$.

Gravity is not scale invariant, but it will prove useful to examine its scale
breaking explicitly.  To preserve gauge covariance and dimensional
analysis, the scale transformation law of the covariant derivative must
take the form
$$ á'_a = ìá_a +k(á^b ì)M_{ab} $$
 where the $ì$ scaling of $e_a{}^m$ was defined above, and the linearity
of $¶¿$ in $ì$ follows from the homogeneity of $á$ in $e$. 
(Alternatively, we could put in something more arbitrary, but it would be
eliminated by the rest of the procedure anyway.)  From the
variation of commutation relations we then find
$$ üR'_{ab}{}^{cd}M_{dc} = [á'_a,á'_b] $$
$$ = ì^2[á_a,á_b]
	+(1-k)ì(á_{[a}ì)á_{b]} +kì(á_{[a}á^c ì)M_{b]c} +k^2(áì)^2 M_{ab}$$
$$ Üâk = 1,ââR'_{ab}{}^{cd} = ì^2 R_{ab}{}^{cd} 
	+ì¶_{[a}^{[c}á_{b]}á^{d]}ì -¶_{[a}^c ¶_{b]}^d (áì)^2 $$
 If we make the redefinition (at least for $ì$ positive)
$$ ì = Ä^{-2/(D-2)} $$
 then we find the very simple scaling law for the integrand of the
Einstein-Hilbert action:
$$ ({\bf e}^{-1}R)' = {\bf e}^{-1}(Ä^2 R -4\f{D-1}{D-2}ÄõÄ) $$

\x IXA7.3 Consider a scale factor that is invariant under a
Killing vector (see subection IXA2).
 ªa Show the Killing vector survives the scale transformation; i.e.,
$$ á'_a = ìá_a +(á^b ì)M_{ab},ââ[K,á_a] = [K,ì] = 0âÜâ[K,á'_a] = 0 $$
 directly using commutators (rather than the Killing equations).
 ªb Although the operator $K$ is the same, the Killing vector is
different:
$$ K = K^a á_a +üK^{ab}M_{ba} = K'^a á'_a +üK'^{ab}M_{ba} $$
 Find $K'^a$ and $K'^{ab}$ in terms of $K^a$ and $K^{ab}$.

\x IXA7.4  Covariant derivatives for flat space in spherical
coordinates can be obtained from those of Cartesian
coordinates by a combination of coordinate and local
Lorentz (rotation) transformations.  However, there are
simpler methods, using a combination of transformations
of a single coordinate and Weyl scale transformations:
 ªa Take the direct product of a sphere with metric $d¯^2$ 
(in arbitrary coordinates) and a line as
$$ d\on\circ s{}^2 = d(ln¼r)^2 +d¯^2 $$
 Then derive flat space in spherical coordinates by making a scale
transformation
$$ ì = \f1r $$
 to yield the metric
$$ ds^2 = dr^2 +r^2 d¯^2 $$
 Show that the resulting covariant derivatives are
$$ á_r = »_r,ââá_i = \f1r ( \ron\circ á_i +M_{ri}) $$
 where $\ron\circ á_i$ are the covariant derivatives on the sphere
corresponding to the metric $d¯^2$.
 ªb Find $á$ in terms of $(r,Ï,Ä)$ using the result of exercise
IXA5.3.  Find $á$ in terms of $r$ and conformally flat coordinates
$x^i$ for the sphere by first deriving
$$ \ron\circ á_i = (1+\f14 x^2)»_i +üx^j M_{ij} $$
 from flat 2D space by another Weyl scaling.

Many special cases of covariant derivatives can be derived
completely by Weyl scalings.  This includes the most commonly used
ones, for cosmology and for static spherical sources.
The general procedure uses the following facts in the following order:
\item{(1)} In a space of one dimension, we can choose 
 $$ D = 1âÜâá = » $$
 (There is no curvature in $D=1$.)
\item{(2)} For a direct product space, i.e., where the metric $ds^2$
can be written as the sum of the metrics of two (or more) spaces, 
the problem for solving for the covariant derivatives is separable.
We can divide up the components into the covariant derivative for
one space and that of the other, each using only its own coordinates
and flat indices (and thus Lorentz generators):
$$ ds^2 = ds_1^2 +ds_2^2âÜâá = (á_1,á_2) $$
 and similarly for the curvature.
\item{(3)} Under a coordinate transformation, each component of
the covariant derivative (and of the curvature)
transforms as a scalar.  We need only
apply the redefinitions of the coordinates, including those that
appear in the partial derivatives:
$$ á_a(x) £ á_a(x') $$
 (This actually applies the alternative $÷x$ definition of coordinate
transformation of subsection IC2.)
\item{(4)} Under a Weyl scale transformation,
$$ ds'^2 = ì^{-2}ds^2âÜâá'_a = ìá_a +(á^b ì)M_{ab}, $$
$$ R'_{ab}{}^{cd} = ì^2 R_{ab}{}^{cd} 
	+ì¶_{[a}^{[c}á_{b]}á^{d]}ì -¶_{[a}^c ¶_{b]}^d (áì)^2 $$
	
\noindent These steps can then be repeated as necessary.
(The first two steps alone lead to Cartesian coordinates for flat space.)

\x IXA7.5  Use this method (as opposed to that of exercise IXA5.3)
to derive the covariant derivatives for the
sphere in the usual spherical coordinates:
 ªa Use steps (1) and (2) to find $á$ for the flat space with metric
$$ ds^2 = du^2 +dÄ^2 $$
 ªb For step (3), apply the transformation
$$ du = {dÏ\over sin¼Ï}â(u = ln¼tan¼\fÏ2) $$
 ªc For step (4), use
$$ ì = {1\over sin¼Ï} $$
 to get the usual metric and covariant derivatives for the (2-)sphere.
We also note that exercise IXA7.4a is just a repetition of these steps,
for a new 1D coordinate $v$ which is redefined as $v=ln¼r$, with
a new $ì=1/r$.

Consider a field theory without gravity that has a conformally invariant
action.  Spontaneous breakdown of scale invariance produces a Goldstone
boson for that symmetry, the ``dilaton" (see subsection IVA7).  Any theory can be made globally
conformally invariant trivially by performing a local scale transformation
and making the parameter the dilaton field.

The dilaton can also act as a Higgs field:  If we couple the dilaton to
conformal gravity (gravity with local Weyl scale invariance), the Higgs
effect reduces conformal gravity to ordinary (Einstein) gravity.  For
example, if we introduce the dilaton into pure gravity by the local scale
transformation above (in analogy to the St¬uckelberg model),
$$ S_G = 4\f{D-1}{D-2}Çdx¼{\bf e}^{-1}\f14 Ä(õ-\f14\f{D-2}{D-1}R)Ä $$
 Up to an (important) overall negative factor, this is the action for a
conformal scalar.  The dilaton field $Ä$ is a compensator for local scale
transformations, and acts as a Higgs field for this gauge symmetry: By
gauging it to its vacuum value $ÒÄÔ={1\over û}$, we regain the usual form
of the gravity action.  (Alternatively, we can set $ÒÄÔ=1$, and introduce
$û$ through the proportionality constant in $Òe_a{}^mÔ¾¶_a{}^m$.)  In this
formalism, where we require the action to be locally scale invariant, the
terms which were conformally invariant before coupling to gravity are
easy to recognize:  They're just the ones which have no $Ä$-dependence. 
(This may require some field redefinition: typically rescaling the matter
fields according to their weight as above.)  The cosmological term
becomes  $S_{cos}=Ç{\bf e}^{-1}ñÄ^{2D/(D-2)}$, which is a conformal
self-interaction term for a scalar.

Because what was the vierbein now appears only in the combination
$e_a{}^m£Ä^{-2/(D-2)}e_a{}^m$, there is now the local scale invariance
$$ e_a{}^m£ìe_a{}^m,âÄ£ì^{(D-2)/2}Ä $$
since this transformation leaves the combination invariant.  Gauge
invariance of the matter action is then (using the infinitesimal
parameter $ì=1+½$):
$$ 0=¶S_M ¾ e_a{}^m{¶S_M\over ¶e_a{}^m}+\f{D-2}2 Ä{¶S_M\over ¶Ä}$$
$$ ÜâT^a{}_a = -\f{D-2}2 Ä{¶S_M\over ¶Ä} $$
Thus, conformal matter has vanishing $T^a{}_a$, since it decouples from
$Ä$.  (Actually, we also need to scale the matter as above to achieve this
decoupling, and there is a corresponding $¶S_M/¶Æ$ term in the above
derivation, so the trace may vanish only after applying the matter field
equations, as in the derivation of $á_a T^{ab}=0$ from coordinate
invariance in the previous subsection.)  In particular, this is easy to check
for the massless point particle, where $T^a{}_a¾ÀX{}^m ÀX{}^n g_{mn}=0$.

An interesting effect is obtained by eliminating the compensator by its
field equation.  (We'll consider just the classical theory here:  In the
quantum case, integrating out this field produces an additional 1-loop
contribution to the effective action.)  Because this manipulation involves
integration by parts, we first expand the compensator about its vacuum
(asymptotic) value:
$$ Ä = 1+üâÜâL = \f14[ (\f{D-1}{D-2}õ -\f14 R) -R -R] $$
 Then eliminating $$ by its field equation,
$$ L £ \f14\left( R{1\over R -4\f{D-1}{D-2}õ}R -R \right) $$
 This action still describes Einstein gravity, but is locally scale invariant
(though not globally, because of the extraction of the vacuum value,
and the way boundary terms were neglected).  Of course, it is nonlocal,
and the nonlocality becomes more complicated if nonconformal matter is
included.  Such terms also appear quantum mechanically:  In two
dimensions, dimensionally regularizing D=2+2$·$, in a Weyl scale
invariant theory we can get a divergent, yet still Weyl scale invariant,
contribution to the effective action proportional to
$$ {1\over ·}\left( R{1\over R -4\f{D-1}{D-2}õ}R -R \right)
	® -{1\over ·}R -üR{1\over õ}R $$
 After renormalizing the divergent term, which is topological and thus
locally scale invariant in exactly D=2, but not in D=2+2$·$, the remaining
finite term contributes a conformal anomaly (see subsections VIIIA7 and
C1).

\x IXA7.6  The statement that the $R$ term is topological in D=2 neglects
boundaries.  In general the topological invariant (the ``Euler number") is
(the ``Gauss-Bonnet theorem")
$$  = Ç{d^2 x\over 2¹}¼ü{\bf e}^{-1}R 
	+È{1\over 2¹}{·_{ab}t^a Dt^b\over ú_{ab}t^a t^b} $$
 where $t^a$ is a tangent vector to the boundary $X^m( )$, as for the
worldline of the particle, and $D$ is the covariant differential (as for the
particle equation of motion and the radial gauge; see subsections IXB2 and
4 below):
$$ t^m = v^{-1}ÀX{}^m,ââDt^a = dX^m e_m{}^b á_b t^a = d ¼vtÉá t^a
	= d (Àt{}^a -vt^b t^c ¿_{bc}{}^a) $$
 (We have used the usual counterclockwise contour, and our convention
$·_{01}=1$, or $·_{xy}=1$ in Euclidean space.)  The additional term in $$
is the angle subtended by the boundary with respect to the surface
($/2¹$), as obtained from the cross product of $t$ and $t+Dt$.  We have
written it in a form that is manifestly invariant under the
reparametrization of $ $, and the $v$'s cancel.  (Of course, it is also
manifestly coordinate invariant.)  
 ªa  Prove that it is also scale invariant by showing that the connection
part of the $D$ exactly cancels the contribution of $R$ to the boundaries,
leaving
$$  = \left.\left( Ç{d^2 x\over 2¹}¼ü{\bf e}^{-1}R
	\right)\right|_{patch¼boundaries}
	+È{1\over 2¹}{·_{ab}t^a dt^b\over ú_{ab}t^a t^b} $$
 where we have turned the $R$ term into a boundary term, and its
remaining contribution is from the fake boundaries at the borders of
patches (or surrounding singularities; $R=»¿$ because the 2D Lorentz
group is Abelian: see exercise IXA5.3).
 ªb  Note that the $dt$ term doesn't contribute if we choose a gauge
where
$$ t^a = ¶^a_1 $$
 (i.e., $t^m=e_1{}^m$).  Demonstrate this by evaluating $$ in polar
coordinates for a disk, and in spherical coordinates for the half-sphere. 
Show the result is half that for a whole sphere (exercise IXA5.3).  Repeat
the calculation for the disk in Cartesian coordinates (so then ÓonlyÕ the
$dt$ term contributes).

\refs

£1 A. Einstein, ÓSitz. Preuss. Akad. Wiss. Berlin, Math.-phys. Kl.Õ (1914)
1030, (1915)
	778, 799, 831, 844, ÓAnn. der Phys.Õ É49 (1916) 769:\\
	general relativity.
 £2 «E. Cartan, ÓLe'cons sur la g«eom«etrie des espaces de RiemannÕ
	(Gauthier-Villars, 1928):\\
	general relativity in terms of vierbein and (GL(D) or Lorentz)
	connection.
 £3 Weyl, Óloc. cit.Õ (IC):\\
	covariant derivatives on spinors.
 £4 L. Infeld and B.L. van der Waerden, ÓSitz. Preuss. Akad. Wiss. Berlin,
	Math.-phys. Kl.Õ (1933) 380:\\
	general relativity in two-component spinor notation.
 £5 H. Weyl, ÓMat. Z.Õ É2 (1918) 384:\\
	Weyl tensor.
 £6 D. Hilbert, ÓNachrichten K¬onigl. Ges. Wiss. G¬ottingen, Math.-phys. Kl.Õ
	(1915) 395:\\
	found the action for Einstein's equations slightly earlier than Einstein.
 £7 A. Palatini, ÓRend. Circ. Mat. PalermoÕ É43 (1919) 203:\\
	first first-order action for gravity (in terms of metric and Christoffel
	symbols).
 £8 H. Weyl, ÓSitz. Preuss. Akad. Wiss. Berlin, Math.-phys. Kl.Õ (1918) 465,\\
	ÓRaum$É$zeit$É$materieÕ (Springer, 1919) p. 246
	[English: ÓSpace-time-matterÕ (Do\-ver, 1952)]:\\
	Weyl scale.
 £9 B. Zumino, Effective Lagrangians and broken symmetries, in ÓLectures
	on elementary particles and quantum field theoryÕ, proc. 1970
	Brandeis University Summer Institute in Theoretical Physics, eds.
	S. Deser, M. Grisaru, and H. Pendleton (MIT, 1970) v. 2, p. 437:\\
	dilaton.
 £10 E.S. Fradkin and V.I. Vilkovisky, \PL 73B (1978) 209:\\
	nonlocal action.
 £11 C.W. Misner, K.S. Thorne, and J.A. Wheeler, ÓGravitationÕ 
	(Freeman, 1970), 1279 pp.:\\
	introductory, long-winded.
 £12 S.W. Hawking and G.F.R. Ellis, ÓThe large-scale structure of spacetimeÕ
	(Cambridge University, 1973), 400 pp.:\\
	mathematical; emphasis on singularity theorems and global
	properties (e.g., Penrose diagrams).
 £13 R.M. Wald, ÓGeneral relativityÕ (University of Chicago, 1984), 492 pp.:\\
	intermediate between the above two.
 £14 S. Weinberg, ÓGravitation and cosmologyÕ (Wiley \& Sons, 1972),
	657 pp.:\\
	old-fashioned, but with more detail on astrophysics and cosmology.

\unrefs

Û5 B. GAUGES

We now consider various gauge choices for coordinate, Lorentz, and scale
transformations.

Ü1. Lorenz

We begin with gauges that preserve global Lorentz invariance, which are
useful for perturbation theory.  Therefore, we look first at perturbation
by finding the kinetic term, which is sufficent for finding linear gauge
conditions.  (It can also be derived from general principles, as will be
shown in subsection XIIA5.)  We expand the vierbein about its flat value,
$$ e_a{}^m = ¶_a{}^m + h_a{}^m $$
 At the linearized level, local Lorentz invariance implies that only the
symmetric part of the field, $üh_{(ab)}$, appears in the curvature and the
action; we will denote this by $h_{ab}$ to simplify notation. (In other
words, the linearized curvature is invariant under the linearized local
Lorentz transformations, which gauge away the antisymmetric part of
the field.  This is equivalent to working directly with the metric.)  We then
can find the linearized curvature, e.g., from the results of subsection
IXA5 for the variation of the curvature, by considering variation about flat
space: i.e., replacing $½_{ab}£h_{ab}$ and $á_a£»_a$.  The result is
$$ R_{ab}{}^{cd} ® »_{[a}»^{[c}h_{b]}{}^{d]} $$
$$ ÜâR_{ab}-üú_{ab}R ® õh_{ab}+»_a »_b h^c{}_c - »_{(a}»^c h_{b)c}
	-ú_{ab}(õh^c{}_c - »^c »^d h_{cd}) $$
Since this comes from varying the action, the quadratic part of the
gauge-invariant action must be
$$ S_G ® -Ç\f14 [h^{ab}õh_{ab} + 2(»^b h_{ab})^2
	- h^a{}_a õ h^b{}_b + 2h^a{}_a »^b »^c h_{bc}] $$
This part of the action, and the linearized curvature, are invariant under
the linearized gauge transformations $¶h_{ab}=-»_{(a}Â_{b)}$. 

\x IXB1.1  Take the Newtonian (weak-field, nonrelativistic) limit of
gravity:  (1) Linearize the action by perturbing about flat space
($e_a{}^m=¶_a^m+h_a{}^m$).  Keep just the part of the pure gravity
action quadratic in the perturbation, the part of the matter coupling
linear in it, and the complete flat-space matter action.  (2)¼Assume small
velocities.  Now consider the problem of a massive point particle in the
field of a much more massive point particle (or spherical body);  in the
above approximations:  
ªa  Show the effect of the gravitational field
generated by the heavier particle on the lighter particle is given by the
action for the lighter particle (in the gauge $x^0­t= $)
$$ S = ms ® Çdt¼(m -ümÀx{}_i^2 +mh_{00}) $$
ªb  Show this field is given by solving Laplace's equation
$$ R_{00} ® ëh_{00} = T_{00} $$
ªc  Show that, with our conventions for normalizing functional
differentiation, a point mass $M$ in D=4 generates
$$ T_{00} = M(2¹)^2 ¶^3(x)âÜâh_{00} = -{M¹\over r} $$
  using the usual solution to Laplace's equation for a
point source.  Combining these results, we see that the potential energy
for the particle is 
$$ U = mh_{00} = -{Mm¹\over r} $$
 which agrees with Newtonian gravity if we identify $G=¹$.  (If we
restore units, this becomes $G=û^2 ¹$.)

The BRST transformations (see subsection VIA4) for gravity again follow
from the gauge transformations:
$$ Qe_a{}^m = C^n »_n e_a{}^m -e_a{}^n »_n C^m +C_a{}^b e_b{}^m $$
$$ QC^m = C^n »_n C^m,âQC_{ab} = C^n »_n C_{ab} +C_a{}^c C_{cb} $$
$$ Q÷C^m = -iB^m,âQ÷C_{ab} = -iB_{ab} $$
 (Other forms follow from different parametrizations of the gauge
transformations, and are equivalent to field redefinitions.  For theories
without spinors, we can work in terms of the metric, and avoid Lorentz
gauge fixing.)

Lorenz gauges for coordinate invariance are similar to Yang-Mills.  For
gravity, the gauge-fixing function is
$$ f_a = »^b h_{ab} -ü»_a h^b{}_b $$
 The BRST procedure works similarly to Yang-Mills.  Looking at just the
graviton kinetic term, the gauge-fixed quadratic Lagrangian for gravity is
then, in the Fermi-Feynman gauge,
$$ L_G¼£¼L_{G,FF} = L_G +ü(»^b h_{ab}-ü»_a h^b{}_b)^2 = 
	-\f14 h^{ab}õh_{ab} +\f18 h^a{}_aõh^b{}_b $$
plus ghost terms.  Note that the trace part of $h$ appears with opposite
sign to the traceless part.  This prevents any redefinition which would
allow rewriting the Lagrangian in the simple form $-\f14 h^{ab}õh_{ab}$. 
However, all derivatives have been absorbed into $õ$'s, which makes the
linearized field equation a simple Klein-Gordon equation.

There are various generalizations of this gauge condition to include
nonlinear terms, such as the ``de Donder (harmonic) gauge", which uses
the gauge-fixing function
$$ f^n = ü»_m(å{-g}g^{mn}) $$
 For example, this allows the field equation for a scalar to be written with
only terms with both partial derivatives acting on the scalar.

Ü2. Geodesics

Consider the field equations for coupling gravity and electromagnetism to
a scalar particle:  From subsection IIIB3, the action for a particle in
external fields, rewritten in Hamiltonian form, is
$$ S_H = Çd  Ó-Àx{}^m e_m{}^a (x)[¹_a -A_a(x)] +vHÕ,ââH = ü¹^2 +Ä(x) $$
 where we have pulled the $v$ out of $H$ for convenience, and use the
``covariant momentum"
$$ ¹_a = e_a{}^m p_m +A_a(x) = e_a{}^m(p_m +A_m) $$
in place of $p_m$ (the canonical conjugate to $x^m$) for covariance.  All
the equations of motion except the Lagrange-multiplier constraint
$$ ü¹^2 + Ä = 0 $$
 follow from the usual Poisson-bracket relation
$$ v^{-1}À\O = i[H,\O] $$
 which can be evaluated by using the canonical commutation relations
(following from the simpler ones for $p_m$)
$$ i[¹_a,x^m] = e_a{}^m,ââi[¹_a,¹_b] = c_{ab}{}^c ¹_c +F_{ab},ââ
	[x,x] = 0 $$
 Thus, $¹_a$ acts effectively like $-ie_a+A_a$, which is the covariant
derivative for gravity and electromagnetism, less the Lorentz term.  The
$Àx$ equation is the obvious
$$ v¹^a = Àx{}^m e_m{}^a $$
 that follows from varying $S_H$ with respect to $¹_a$, while the
equation of motion for $¹$ is
$$ v^{-1}À¹_a = -c_a{}^{bc}¹_b ¹_c -F_a{}^b ¹_b -á_a Ä $$
 Using the relation
$$ 0 = T_{abc} = c_{abc}+¿_{[ab]c}âÜâc_{a(bc)}+¿_{(bc)a} = 0 $$
 we find
$$ v^{-1}À¹{}^a -¹^b ¹^c ¿_{bc}{}^a +F^a{}_b ¹^b +á^a Ä = 0 $$
 This is the coordinate-covariant form of the Lorentz force law (plus
scalar field).  With only the gravitational effects we have the
covariantization of the free particle equation,
$$ Dp^a ­ Àp{}^a -vp^b p^c ¿_{bc}{}^a = 0 $$
 where ``$D$" is understood as a covariantized $ $ derivative (along a
worldline with metric $v$).

It's useful to consider a continuum of particles (``dust") moving under
the influence of these fields, such that any two infinitesimally close
particles have infinitesimally different velocities, and only one particle
passes through any particular point in spacetime (at least within some
small region of spacetime).  We then can treat $¹_a$ (or $p_m$) as a field
defined for all $x$:  Choosing a point $x$ also chooses a curve $X( )$ for
which $x=X( )$ for some $ $, so we can write $¹(x)$ in place of $¹( )$. 
Specifying the field $¹$ also determines this family of curves, since the
tangent to any curve is given by the $X$ equation of motion
$$ ÀX{}^m = v¹^a e_a{}^m $$
 (To determine the $ $ parametrization, we also specify $v$, and the
hypersurface given by the collection of points $X(0)$ from each curve.)
Then we can express the $ $ derivative in terms of $x$ derivatives:
$$ {d\over d } = ÀX{}^m »_m = v¹^a e_a $$
 which gives the manifestly covariant form of the equation of motion
$$ ¹^b á_b ¹_a +F_a{}^b ¹_b +á_a Ä = 0 $$
 For vanishing $F$ (and thus $A$) and constant $Ä$ ($=üm^2$), this
equation
$$ p^b á_b p_a = 0 $$
 describes ``geodesics", which are curves of extremal length,
since the action is 
$$ S = ms,ââ-ds^2 = dx^m dx^n g_{mn} $$
 for massive particles.  These are the analogs of straight lines in flat
space.  (For positive-definite metric, they are shortest lines.  Because of
the indefinite signature of the Minkowski metric, the worldlines of
massive particles are actually longest lines, while massless particles
travel along lines with no length.)

\x IXB2.1  Uniform circular motion in 2D flat Euclidean space,
constant $dÏ/dt$ in polar coordinates, is associated with
acceleration of constant magnitude.  (Or, without time, we can say
that a circle has constant ``extrinsic"
curvature with respect to the 2D space.)  Show that an analogous
situation in 2D Minkowski space can be obtained by Wick rotation:
ªa  Starting with the metric for 2D flat Euclidean space in polar
coordinates, Wick
rotate $Ï$ to make it a time coordinate (``Rindler coordinates").  
Show by a transformation to ``Cartesian" coordinates
that this describes 2D flat Minkowski space.
ªb  Show that any curve described by constant $r$ describes
acceleration of constant magnitude, by evaluating
$(d^2 x(Ï)/ds^2)^2$ in ``Cartesian" coordinates.
Note that the direction of this 2-vector is fixed to be orthogonal
to $dx/ds$ (since $(dx/ds)^2=1$ by definition), so this is just
the acceleration as measured in the rest frame.
ªc  Define the acceleration in arbitrary curved coordinates
(in terms of $pÉá p$) and evaluate it in Wick-rotated polar coordinates,
to obtain the same result as in Cartesian coordinates.
(Use the covariant derivative of exercise IXA7.4a.)

\x IXB2.2  Equations of motion for particles can be derived from
conservation laws.  We know this already nonrelativistically,
for a particle in a potential using energy conservation.
Now consider a dust with $T^{mn}=¨¹^m ¹^n$ and
current $J^m=¨¹^m$.  (Compare subsection IIIB4.
We could use those single-particle expressions here,
but using dust instead avoids integration.
Note that using $¹$ or $p$ allows us to describe also
massless particles.  The existence of a conserved current
corresponds to a complex field with a global U(1) symmetry.)  
ªa In the case with no external fields except gravity, show that
the geodesic equation follows from covariant conservation
of both of these quantities.  (Of course, in flat space this
gives the usual free particle result.)
ªb Generalize to the case of external fields by adding to
$T^{mn}$ that of the external fields themselves.
When taking the divergence of those terms, use
appropriate source terms to the field equations 
in terms of the particle variables $¨$ and $¹$.
(In the case of a single nonrelativistic particle in 
a static electric field, this is the usual derivation of the
force on a particle from the electric field's pressure.)

For some purposes we need a weaker (but equivalent) form of the
geodesic equation:  If for some scalar $f$ and vector $n^a$
$$ (nÉá)n^a = fn^aâÜâp^a = un^a,ââ(pÉá)p^a = 0,ââf = -(nÉá)ln¼u $$
for some scalar $u$, which we can determine by integrating $f$.
In particular, we can identify
$$ u = v^{-1}âÜân^m = ÀX{}^m $$
 Thus, the more general geodesic equation allows arbitrary
parametrization of the geodesics, while the stricter version ($f=0$)
corresponds to affine parametrization ($v=1$) if we still want to identify
$p$ with $ÀX$.  (Remember, as with all constrained systems, the equations
of motion $pÉáp=0$ imply $(d/d )p^2=0$, so any geodesic satisfying the
stricter equation will have some fixed mass along that particular curve.)

\x IXB2.3  Show that in D=2 (one space dimension, one time) ÓanyÕ
lightlike curve is a geodesic, using the weaker form of the geodesic
equation.  (Find $f$.)  This is a consequence of the fact that it is impossible
to change direction in D=2 without slowing down.

The particle (geodesic) version of the conservation of momentum in the
direction of a Killing vector is
$$ pÉá¼p_a = 0âÜâpÉá¼KÉp = 0âÜâ{d\over d }KÉp = 0 $$
 where covariant conservation $pÉá$ has become ordinary conservation
$d/d $ (no connection term) because $KÉp$ is a scalar.  (See also exercise
IXA2.4.)  This is the same as for the conserved current $J^a=K_b T^{ba}$
(subsection IXA6).

Ü3. Axial

The definition of axial gauges in terms of the covariant derivative is
the same as for Yang-Mills ($nÉá=nÉ»$).  In terms of the explicit gravity
fields,
$$ nÉá = nÉ»âÜân^m ­ n^a e_a{}^m = n^a ¶_a^m,ân^a ¿_a{}^{bc} = 0 $$
 In the case of gravity, this implies that lines in the $n^a$ direction are
geodesics (see previous subsection):
$$ »_m n^a = 0âÜâ(nÉá)n = (nÉ»)n = 0 $$

To analyze the consequences of axial gauge conditions for the metric, we
need a further identity:  For any vector field $n^a$, consider the action of
$nÉá$ on $n_m=e_m{}^a n_a$, treating it as a scalar; in this calculation we
ignore any indirect action of $á$ on curved indices.  Then
$$ (nÉ»)e_m{}^a n_a = (nÉá)e_m{}^a n_a = 
	e_m{}^a(nÉá)n_a +n_a(nÉá)e_m{}^a $$
 The last term simplifies for vanishing torsion, since:
$$ \li{ n^n á_n e_m{}^a & = n^n á_{[n}e_{m]}{}^a + n^n á_m e_n{}^a
	= -n^n T_{nm}{}^a + n^n á_m e_n{}^a = á_m n^a -e_n{}^a á_m n^n\cr
	& = á_m n^a -e_n{}^a »_m n^n \cr} $$
 We thus have
$$ (nÉ»)e_m{}^a n_a = e_m{}^a(nÉá)n_a +»_m(ün_a^2)
	-(e_n{}^a n_a)»_m n^n $$

Applying this identity to the axial gauge condition, we find
$$ nÉá = nÉ»,ââ»_m n^a = 0âÜâ(nÉ»)e_m{}^a n_a = 0âÜâ
	n_m ­ e_m{}^a n_a = ¶_m^a n_a $$
 by choosing the appropriate constants of integration.  (This amounts to
fixing a residual gauge invariance.)  
In fact, we can weaken the assumptions in this derivation:
$$ (nÉá)n = 0,âân^a e_a{}^m = n^a ¶_a^mâÜâ(nÉ»)e_m{}^a n_a = 0âÜâ
	n_m ­ e_m{}^a n_a = ¶_m^a n_a $$
We can now determine the form of
the gauge condition on the metric:
$$ n^m = n^a ¶_a^m,âân_m = ¶_m^a n_aâÜâ
	n_m ­ n^n g_{nm} = n^n ú_{nm} $$

Applying these results to perturbation theory, as
$$ e_a{}^m = ¶_a^m +h_a{}^b ¶_b^m $$
 we then have
$$ n^a e_a{}^m = n^a ¶_a^mâÜân_b h^{ba} = 0 $$
$$ e_m{}^a n_a = ¶_m^a n_aâÜâe_a{}^m ¶_m^b n_b = n_a
	âÜân_b h^{ab} = 0 $$
 and thus $n_b h^{(ba)}=0$, so we can again work with just the symmetrized
$h$.

The lightcone gauge is again useful for eliminating unphysical degrees of
freedom.  The lightcone gauge conditions are 
$$ n^a = ¶_-^aâÜâh^{+a} = h^{a+} = h^{(+a)} = 0 $$
 For the rest of this discussion we work with just the symmetrized $h$.
Separating out the trace part as $h_{ij}=h^T_{ij}+¶_{ij}h$, where
$h^T_{ij}$ is traceless, we find for the linearized gauge-fixed action
$$ \li{ L'_G & ® -\f14 h^{ab}õh_{ab} +\f18 h^a{}_a õh^b{}_b -üf^2 \cr
	& = -\f14 h^{Tij}õh^{Tij} -ü(h'^{-i}){}^2 +\f{D-2}2 hh'^{--} \cr} $$
$$ \li{ h'^{-i} & ­ f^i = -»^+ h^{-i} + »^j h^{Tij} -\f{D-4}2 »^i h \cr
	h'^{--} & ­ »^+ f^- +\f{D-4}4 õh =
	-»^{+2}h^{--} +»^+ »^i h^{-i} -\f{D-2}2 »^+ »^- h +\f{D-4}4 õh \cr
	\f{D-2}2 h & = -{1\over »^+}f^+ \cr } $$
 where we have simplified some algebra by writing the gauge-invariant
action as the Lorenz gauge one ÓminusÕ its gauge-fixing term $üf^2$.  (There is
some ambiguity in that we can shift $h'^{-i}$ by a $»^i h$ term, and
absorb the generated terms into $h'^{--}$.)  We see that all but $h^T_{ij}$
are auxiliary fields (we redefined $h^{-i}$ and $h^{--}$ by just shifting
and applying $»^+$), and can be eliminated (but watch out if there are
matter couplings, when eliminating them gives Coulomb-like interactions).
(Again, this procedure is much simpler than quantizing in the de Donder gauge and then applying a further analysis to extract the physical polarizations, as is always done in other texts when analyzing radiation in general relativity.)

The temporal gauge (known also as ``Gaussian normal coordinates") is
used when treating time and space separately:  In this case we have for
the metric
$$ n^m = ¶_0^mâÜâg_{0m} = ú_{0m} $$
 An alternate way of defining the temporal gauge is to start with a spatial
hypersurface, and determine the geodesics normal to this hypersurface
($g_{0i}=0$), where the positions on this hypersurface define $x^i$,
constant along the geodesics, and the proper times along the geodesics
define $x^0$ ($g_{00}=-1$), with $x^0=0$ at the hypersurface.  The fact
that these are geodesics guarantees that the hypersurfaces of fixed, but
nonvanishing, (proper) time are still orthogonal to the geodesics ($g_{0i}$
stays zero):  Consider some constant $V^m$, representing the separation $dx^m$ of 2 ``fixed" nearby points in any hypersurface, $(nÉV)=0$. Then the statement $nÉá(nÉV)=0$ that the separation of those 2 points remains in the hypersurface is just the equation derived above, i.e., $(nÉ»)e_m{}^a n_a=0$.

Equivalently, we can consider a dust of massive particles and choose an
initial hypersurface orthogonal to their (timelike) geodesics to define
$x^0=s=0$.  This coordinate system is thus the ``rest frame" of the dust;
all the information about the geometry of the space is contained in the
time dependence of the spatial separation of the particles ($g_{ij}$). 
There is still the residual coordinate ambiguity of how to assign $x^i$ on
the initial hypersurface. 

Gaussian normal coordinates thus can be useful for studying the dynamics
of particles:  For example, we can study a gravitational field of distant,
unknown (or ignored) origin (i.e., the curvature of spacetime) by watching
the relative motion of two nearby particles of such a dust, neglecting the
gravitational force/curvature effect acting between the two particles
themselves.  If the two particles start out relatively at rest at some
initial time (which is well-defined only if they are close and relatively
slow), then in the temporal gauge the paths of both particles are
described by fixed $x^i$, independent of $x^0$, since their geodesics are
simply lines in the time ($n^a=¶^a_0$) direction, and the proper times of
both particles are the same as the time $x^0$.  Then the distance between
the particles at any given time is given by the magnitude of $dx^m
e_m{}^a$, with $dx^0=0$ and $dx^i$ their infinitesimal separation.  Thus,
since the $x^i$'s, and thus $dx^i$, are fixed, we want to study the change
in $e_m{}^a$ (really just $e_i{}^a$; $e_0{}^a=¶_0^a$) with time.  Using our
evaluation of $(nÉá)e_m{}^a$ from above, we find
$$ (nÉá)^2 e_m{}^a = (nÉá)á_m n^a = [nÉá,á_m]n^a = n^n[á_n,á_m]n^a
	= -n^b n^c R_{bdc}{}^a e_m{}^d $$
 using the axial gauge condition $nÉá=nÉ»$.  For the Gaussian case
$n^a=¶_0^a$, we then have
$$ ¬e_m{}^a = - R_{0b0}{}^a e_m{}^b $$
 (Of course, vanishing curvature implies geodesics that start parallel
remain that way, because the space is then flat.)  By observing different
sets of particles initially at rest with respect to each other, we can
choose different timelike directions $n$, and determine all the curvature
components from their linear combinations.

\x IXB3.1  Let's examine some 2D examples of axial gauges in spaces with
positive-definite metric:
 ªa  Gaussian normal coordinates need not be Cartesian in flat space. 
Show that polar coordinates for the plane define an axial gauge.  What is
the coordinate in the ``$n^a$" direction?  Give the geodesic interpretation.
 ªb  Repeat the above for a curved space --- the (2D) sphere in spherical
coordinates.
 ªc  Apply the above ``equation of motion" ($¬e=-Re$) to the sphere.  (See
exercise IXA5.3.)  Show its solution agrees with the obvious.

Ü4. Radial

Another useful gauge similar to the axial gauge is the radial gauge
(``Riemann normal coordinates"), discussed for Yang-Mills in subsection
VIB1.  In this case we have
$$ n^m = x^mâÜâ(nÉá)n^a = x^m »_m x^n ¶_n^a = n^a $$
 a case of the more general form of the geodesic equation.  Applying the
same identity as for the axial, we again have
$$ nÉá = nÉ»,ââ»_m n^a = ¶_m^aâÜâ
	(nÉ»)e_m{}^a n_a = (nÉ»)¶_m{}^a n_aâÜâe_m{}^a n_a = ¶_m^a n_a $$
 but now the boundary condition is already implied by the gauge
condition near the origin:  For any infinitesimal $x^m=·^m$,
$$ \li{ ·^m e_m{}^a (0) = ·^m ¶_m^aâ& Üâe_m{}^a (0) = ¶_m^a \cr
	·^m ¿_m{}^{ab} (0) = 0â& Üâ¿_m{}^{ab} (0) = 0 \cr} $$
 Thus, there is no residual gauge invariance, unlike axial gauges (where
the coordinates of the initial hypersurface need additional determination).
 Any reference frame satisfying these conditions at the origin is called a
``local inertial frame", and is the most natural for an observer at that
point in spacetime.  (In flat space, this yields Cartesian coordinates.)

\x IXB4.1  We can think of Gaussian normal coordinates as defined by a
dust of particles with affine parametrization $v=1$ and unit mass $m=1$,
with $ =s=x^0$ and $x^i$ constant for any particle ($ÀX=p=n$).  For
Riemann normal coordinates we can think of particles radiating out from
the origin $x^m=0$ in all possible directions in space and time, but then
some must be antiparticles (traveling backward in time), some must be
massless (for the lightlike geodesics), and some must be tachyons, with
$m^2<0$ (for the spacelike geodesics).  However, as for the Gaussian
case, we can still identify
$$ n^m = ÀX{}^m $$
 Using the radial gauge condition, show that these can be chosen as
geodesics with
$$ v = e^ ,ââX( ) = e^  X(0),ââp = X(0) $$
 so all particles start at the origin at $ =-¥$, and their position at $ =0$
is determined by their initial (constant) momentum.  (Thus particles with
proportional momenta travel the same path, but arrive at different points
at $ =0$; however, in this case $ $ is neither the time $x^0$ nor the
proper time $s$, but just an arbitrary parameter.)

As we saw in subsection VIB1, the radial gauge is related to
gauge-covariant translation (in general relativity, ``parallel transport")
as, for any tensor $Æ$,
$$ ÷Æ(÷y) = e^{x^a(y)\D_a}Æ(y) = e^ñ Æ(÷y),ââ
	÷y^m = e^{x^a(y)\E_a{}^m(y) D_m}y^m $$
 where $y$ is the ``origin", $ñ=ñ^I M_I$ is just a Lorentz transformation,
and $\D$ is the covariant derivative acting at $y$:
$$ \D_a = \E_a{}^m(y) D_m +\¿_a{}^I(y) M_I,ââD_m = {»\over »y^m};ââ
	[\D_a,\D_b] = \T_{ab}{}^c\D_c +\R_{ab}{}^I M_I $$
 As in general for coordination transformation parameters $Â^a$, $x^a$
now transforms under local Lorentz transformations.  (In background
field language, this ``quantum field" transforms under the ``background"
Lorentz transformations.)  Thus, $x^a$ is now a function of $y$; it cannot
be made even covariantly constant in general:
$$ \D_a x^b = 0âÜâ0 = [\D_a,\D_b]x^c = -x^d\R_{abd}{}^c $$
 (For more practical reasons, if we defined it to be invariant or constant,
the manipulations that follow would break down.)  At this point we have
only made a Lorentz transformation on $Æ$, since it and $÷Æ$ are
evaluated at the same point $÷y$.  However, as for Yang-Mills in
subsection VIB1, for the next step we want to identify $x^a$ as the new
coordinate:
$$ Æ'(x^a) = ÷Æ(÷y^m(y^m,x^a)) = e^{x^a\D_a}Æ(y) $$
 where $Æ'$ has implicit dependence on $y$, since in radial gauges the
choice of origin is gauge parameters that define the gauge.  (The
coordinates are defined as radial with respect to the origin $y$.)  Thus,
we have made a Lorentz transformation $Æ£÷Æ$ followed by a coordinate
transformation $÷Æ£Æ'$.  

We also want to define a covariant derivative for $x$ by
$$ áÆ' = (\D Æ)' = e^{x^a\D_a}\D_a Æ(y) $$
 where $á$ (as for Yang-Mills) contains only $»_a=»/»x^a$ and not $D_m$:
$$ á_a = e_a{}^b(x)»_b +¿_a{}^I(x)M_I $$
 At this point we no longer distinguish flat and curved indices, since the
Lorenz gauge has been fixed.  We have then transformed
$$ y, Æ, \D £ x, Æ', á $$
 Note that the $y$-coordinate tensors are the $x$-coordinate tensors
evaluated at the origin:
$$ Æ'(0) = Æ(y),ââ(áÆ')(0) = (\D Æ)(y) $$
 We can identify this as the radial gauge when $x(y)$ satisfies the
geodesic condition, since then
$$ (xÉ\D)x = 0âÜâx' ­ e^{xÉ\D}x = x $$
$$ ÜâxÉáÆ' = xÉ(\D Æ)' = (xÉ\D Æ)' = xÉ\D Æ' = xÉ»Æ' $$
 making use of $Æ'(x)=e^{xÉ\D}Æ(y)$. 

Unfortunately, it is somewhat difficult to continue this construction in
terms of the covariant derivative, but simpler in terms of the ``dual"
differential forms.  We therefore define the (Lorentz-covariantized) Lie
derivative as
$$ \L_{xÉ\D}öÆ = xÉ\D öÆ,ââ\L_{xÉ\D}ö\D = [xÉ\D,ö\D] $$
 for any ``tensor" (object carrying only flat indices) $öÆ$ and any
``covariant derivative" (object with a flat vector index free, but
multiplying partial derivatives and Lorentz generators) $ö\D$.  We
generalize to evaluate on not only $Æ$ and $\D$, but to apply the Lie
derivative also as part of the transformation $exp(\L_{xÉ\D})$.  For that
reason, for the remainder of this section we will abbreviate $\L_{xÉ\D}$ as
just $\L$.  We then have
$$ [xÉ\D,\D_a] = -(\D_a x^b)\D_b +x^b(\T_{ba}{}^c\D_c +\R_{ba}{}^I M_I) $$
 Defining Lie derivatives to satisfy the usual Leibniz and distributive
rules like any derivative (since we use them as infinitesimal
transformations), we then find
$$ (\L \E_a{}^m)\E_m{}^b = -\D_a x^b +x^c\T_{ca}{}^b,ââ
	\E_a{}^m \L \¿_m{}^I = x^b\R_{ba}{}^I $$

In terms of differential forms, defined as
$$ \E^a = dy^m\E_m{}^a,ââ\¿^I = dy^m\¿_m{}^I $$
$$ D = \E^a\D_a = d +\¿^I M_IâÜâDx^a = dx^a - x^b\¿_b{}^a $$
 we then have
$$ \L x^a = 0,ââ\L (\T, \R) = xÉ\D (\T, \R) $$
$$ \L\E^a = Dx^a -\E^b x^c\T_{cb}{}^a $$
$$ \L\¿^I = \E^a x^b\R_{ba}{}^IâÜâ
	\L(Dx^a) = -x^b\E^c x^d\R_{dcb}{}^a $$
 which covers all the quantities that appear in evaluating $e^\L$ on $\E^a$
and $\¿^I$.  The geodesic condition prevents higher derivatives of $x^a$
from appearing in the transformation law, and allows us to freely reorder
all the $x$'s to the left at the end of the calculation for identifying the
coefficients of the Taylor expansion, at which point we can forget that
$x$ depends on $y$.  Thus, these few equations for the action of $\L$
allow any transformed quantity to be evaluated straightforwardly by
iteration, Taylor expanding $e^\L$ in powers of $\L$.

The important distinction between the transformation laws for $\D_a$ and
$\E^a$ is that for $\E^a$ the derivatives of $x$ appear only in the
combination $dx$, which makes changing coordinates from $y$ (or $÷y$) to
$x$ easier.  Specifically, by iterating the above Lie derivatives, we find a
solution of the form
$$ \E'^a = e^\L \E^a = \E^b\A_b{}^a +(Dx^b)\B_b{}^a,ââ
	\¿'^I = e^\L \¿^I = \E^a\A_a{}^I +(Dx^a)\B_a{}^I $$
 where $\A_b{}^a$, $\B_b{}^a$, $\A_a{}^I$, $\B_a{}^I$ are functions of $x$
and of tensors evaluated at the ``origin" ($(\Dò\D\T)(y)$,
$(\Dò\D\R)(y)$).  For Riemann normal coordinates, we want to fix $y$
(e.g., $y=0$), so we evaluate the above at $dy=0$.  Furthermore, we can
choose the gauge $¿(0)=0$ (at least for vanishing torsion), so also
$Dx£dx$.  Then the solution is
$$ E^a = dx^b\B_b{}^a,ââ¿^I = dx^a\B_a{}^I $$
 Thus, $\B_b{}^a$ and $\B_a{}^I$ are the inverse vierbein $e_m{}^a$ and
Lorentz connection $¿_m{}^I$ for the new coordinate system,
$$ á_a = (\B^{-1})_a{}^b(»_b +\B_b{}^I M_I) $$
 written explicitly as a Taylor expansion in $x$ by the above method, all
of whose coefficients are tensors (torsions and curvatures and their
derivatives) evaluated at the origin.

However, we can also use these results for first-quantization (where
actions are expressed in terms of, e.g., $\E'^a/d $ for the particle) in
background field gauges, by choosing $y$ as the background and $x$ as
the quantum coordinate (see subsection VIB1); then we keep both the
$dy$ and $Dx$ terms.

\x IXB4.2  Find the first few orders of this expansion.
 ªa  Using the above method, show that for vanishing torsion
$$ E^a = dx^b(¶_b^a -\f16 x^c x^d\R_{cbd}{}^a +...),ââ
	¿^I = üdx^a (x^b\R_{ba}{}^I +...) $$
 ªb  Check the validity of this result by evaluating $[á,á]$ to this order
from the $á$ given by this $E$ and $¿$.
 ªc  Use instead the covariant-derivative method of subsection VIB1.  In
this case, we find
$$ á_a = e^{xÉ÷\D}÷\D_a e^{-xÉ÷\D} +h_a{}^b e^{xÉ÷\D}»_b e^{-xÉ÷\D} $$
 where $h_a{}^b$ is chosen to cancel all $D_m$ terms in $á$, and we have
defined
$$ ÷\D_a = \E_a{}^m(y) D_m +\¿_a{}^I(y) ~M_I $$
 where now $x^a$ is ``constant", so $D_m$ and $~M$ do not act on it. 
(Otherwise, in this approach, we would be stuck with tons of $\Dò\D x$
terms.)  In terms of the previously defined Lorentz generators,
$$ M_{ab} = ~M_{ab} +x_{[a}»_{b]} $$
 Find $h$ to this order, and use it to obtain
$$ á_a = (¶_a^b +\f16 x^c x^d\R_{cad}{}^b +...)»_b 
	+ü(üx^b\R_{ba}{}^{cd} +...)M_{dc} $$
 restoring $~M$ to $M$.

Ü5. Weyl scale

The gauge-fixed kinetic term can be simplified by including the conformal
compensator (see subsection IXA7). The quadratic part of the
gauge-invariant Lagrangian is then ($Ä=1+ü$)
$$ L_0 = -\f14{\bf e}^{-1}Ä(R-4\f{D-1}{D-2}õ)Ä $$
$$  ® -\f14 [h^{ab}õh_{ab} + 2(»^b h_{ab})^2
	- h^a{}_a õ h^b{}_b + 2h^a{}_a »^b »^c h_{bc}]
	-ü(õh^a{}_a - »^a »^b h_{ab}) +\f14\f{D-1}{D-2} õ $$
The nicest (globally Lorentz) covariant gauge comes from choosing the
coordinate and scale gauge-fixing functions
$$ f_a = »^b h_{ab} - ü»_a h^b{}_b + ü»_a ,ââf = -h^a{}_a $$
 We use these to obtain the gauge-fixed Lagrangian (see subsection VIB9)
$$ \li{ L¼&=¼L_0 +ü(»^b h_{ab} - ü»_a h^b{}_b + ü»_a )^2
	-\f18 (-h^a{}_a)õ(-h^b{}_b) \cr
	&= -\f14 h^{ab}õh_{ab} +\f14\f1{D-2}õ \cr } $$
plus ghost terms.  Now the $h$ kinetic term is simpler.  Also, remember
that $$ decouples from conformal matter. These features of gauge
fixing make this formalism closely analogous to the St¬uckelberg
formalism for the massive vector.  We can also define nonlinear versions
of these gauge-fixing functions, such as $»_m(ÄÊ{\bf e}^{-1/2}e_a{}^m)$
or $»_m(Ä^2å{-g}g^{mn})$ for the coordinate gauge, and $ÄÊ{\bf
e}^{-1/2}$ or $Ä^2å{-g}$ for the scale.

\x IXB5.1  Find the ghost terms for linearized gravity in the
Fermi-Feynman gauge, and its simplification with the compensator.

The scale gauge can also be fixed in terms of the vierbein/metric alone: 
For example, we can fix the gauge
$$ {\bf e} = 1 $$
 in which case $Ä$ acts simply as a renaming of ${\bf e}$.  A more unusual
gauge is
$$ R = 0 $$
 This is not a restriction on the geometry, since the physical Ricci scalar is
effectively replaced by its scale transform
$$ R' = Ä^{-(D+2)/(D-2)}(R -4\f{D-1}{D-2}õ)Ä $$
 which is scale invariant.  In the gauge $Ä=1$, $R'=R$, but in the gauge
$R=0$ it is proportional to $õÄ$.

\x IXB5.2 Show that the ghosts for scale transformations propagate in the
gauge $R=0$:  Find their contribution to the action.

More general gauges are possible when matter fields appear.  For
example, consider coupling gravity, with compensator, to a physical
conformal scalar $Æ$.  With appropriate normalization of the
compensator and physical scalar, the kinetic terms for the two fields are
identical except for sign:  There is a manifest O(1,1) symmetry.  We can
take advantage of this by using a ``lightcone" basis for these fields: 
Defining $Ä_à=ÄàÆ$, the full nonlinear (in gravity) Lagrangian $L$ becomes
($S=Çdx¼{\bf e}^{-1}L$)
$$ L = Ä_+(\f{D-1}{D-2}õ -\f14 R)Ä_- $$
 The overall normalization is arbitrary, including sign, since we can
rescale either field by a constant.  Many Weyl scale gauges are possible,
and somewhat more transparent than making field redefinitions on the
corresponding action without compensator.  Effectively, we can redefine
the fields $Ä_à$ arbitrarily as long as we don't fix $Ä_+/Ä_-$ to a
constant, since that combination is scale invariant.  (I.e., $Ä_+/Ä_-$ can
be redefined, but not fixed.)  

Some of the more interesting choices are:
$$ \li{ Ä_à = 1à\ÄâÜâ& L = -\f14 R -\Ä(\f{D-1}{D-2}õ -\f14 R)\Ä \cr
	Ä_à = e^{à\Ä}âÜâ& L = -\f14 R -\f{D-1}{D-2}\Äõ\Ä \cr
	Ä_à = \Ä^{1àa}âÜâ& L = \Ä[(1-a^2)\f{D-1}{D-2}õ -\f14 R]\Ä \cr
	Ä_+ = \Ä,âÄ_- = 1âÜâ& L = -\f14 R\Ä \cr} $$
 We can also have any of these gauge-fixed Lagrangians with opposite
overall sign, simply by changing the choice of either $Ä_+$ or $Ä_-$ by a
sign.  The first two choices are useful because they put the action in
standard form, as the usual gravity action plus a physical scalar kinetic
term.  (Thus, coupling a massless scalar to gravity either conformally or
minimally is equivalent, and the two cases are distinguished only by
interactions.)  In fact, the first choice, or ``temporal gauge"
$Ä_++Ä_-=constant$, just returns us to the form without compensator,
$Ä=1$.  On the other hand, changing the sign of $Ä_-$ yields the ``axial
gauge" $Ä_+-Ä_-=constant$, which is fixing the ÓphysicalÕ scalar as
$Æ=1$.  The overall sign of the action changes because the physical scalar
is traded for the compensator, or the corresponding part of the metric. 
This gauge is closely related to the ``string gauge":  In our third choice
above the gravity action is invisible until the surviving scalar has been
expanded about its vacuum value.  The constant $a$ is arbitrary except
that it must not vanish (so that $Ä_+/Ä_-$ is not a constant).  In
particular, this action appears in string theory, with the choice
$$ a = {1\over å{D-1}}âÜâL = \Ä(õ -\f14 R)\Ä $$
 which eliminates explicit $D$-dependence.  Again the scalar appears with
the wrong-sign kinetic term, but $R$ appears with the right sign (or vice
versa), because of more complicated redefinitions.  The sign of the $õ$
changes back to the usual for $|a|>1$.  However, for $|a|=1$, it disappears
completely.  A similar result occurs for the last choice, or ``lightcone
gauge" $Ä_-=1$.

\x IXB5.3  The property that distinguishes this kinetic term for a scalar
coupled to gravity is the O(1,1) symmetry:
 ªa  Before fixing the Weyl scale gauge, the continuous SO(1,1) subgroup
of this symmetry is just the scaling $Ä_à£ñ^{à1}Ä_à$.  After gauge
fixing, this transformation may change the gauge, and thus may need to
be combined with a constant Weyl scale transformation to preserve the
gauge.  In that case the vierbein will also transform under the resulting
modified SO(1,1) transformation.  Find the SO(1,1) transformations for
$\Ä$ and $e_a{}^m$ in the above 4 gauges.
 ªb  There is also the ``parity" transformation of this O(1,1), $Ä_+ªÄ_-$. 
Find the modified form of this transformation for $\Ä$ and $e_a{}^m$. 

\x IXB5.4  Add to the above action a term proportional to $(Ä_+-Ä_-)^2 Ä_-{}^{4/(D-2)}$. By considering various gauges, show this action is equivalent to (1) the action for gravity plus a scalar conformally coupled to it, with a renormalizable self-interaction, and (2) an $R+R^2$ action with no scalar.

Although all these choices are equivalent in perturbation theory (though
the physical scalar may require a nonvanishing vacuum value), they aren't
necessarily so nonperturbatively, depending on the ranges of the various
scalars.  Unfortunately, nonperturbative gravity is not understood well
enough (even classically) to make such distinctions, even though they
may be important physically.  The above considerations generalize
straightforwardly to the case with many physical scalars, where we may
consider symmetry groups such as O(n,1).  If the physical scalars form a
nonlinear $§$ model, the compensator may join in to make the $§$-model
groups noncompact:  Examples of this appear in supergravity and strings
(see below).

The appearance of a physical scalar can also affect the way scale gauges
are chosen in conjunction with coordinate gauges.  For example, a result
similar to the one found at the beginning of this subsection can be
obtained from the (linearized) action with both compensator and physical
scalar $Æ$ (where $ÒÆÔ=0$),
$$ L ® L_0 -\f14 ÆõÆ $$
 choosing the same $$-dependent coordinate-fixing term $(f_a)^2$, but
imposing the scale gauge
$$ Æ = \f1{å2}( -h^a{}_a) $$
 The result is identical to the one given at the beginning of this subsection,
except that now no scale ghosts appear:  The scalar that appears as
$h^a{}_a$ is now physical, and no longer needs a ghost to cancel it.  This
is the perturbative ``string gauge" for scale invariance, which appears
automatically in covariantly gauge-fixed string theory.

\x IXB5.5  Let's investigate such gauge choices further:
ªa Starting with the Fermi-Feynman-gauge-fixed linearized gravity
action of subsection IXB1, add the physical-scalar kinetic term 
$-\f14 ÆõÆ$.  Separate the traceless and trace pieces of $h_{ab}$.  Show
that the string-gauge action (i.e, the one given at the beginning of
this subsection if we ignore ghosts) follows from simply switching
$$ Æ ª \f1{åD}h^a{}_a $$
 and then identifiying the new $Æ$ with $å2/(D-2)$.
ªb The way the physical scalar of string theory appears in the
gauge-invariant and gauge-fixed action is slightly more clever than as
described above.  (See subsections XIB5-6 below.)  The kinetic term
(already in the string gauge for scale invariance) is
$$ S = Çdx¼ì(õ -\f14 R)ì $$
 where the missing ${\bf e}$ has been absorbed into $ì$ by a field
redefinition.  (Since $ì$ is thus not a scalar, we define $õì$ by 
${\bf e}^{-1/2}õ{\bf e}^{1/2}ì$, since ${\bf e}^{1/2}ì$ is a scalar.) 
Expanding $ì=1+$, the linearized gauge fixing is now simply
$$ L £ L +ü(»^b h_{ab} +»_a )^2 $$
 (or we can use the nonlinear gauge-fixing function $»_m(ìe_a{}^m)$). 
Show the result is the same as above.

\refs

£1 T. de Donder, ÓLa gravifique einsteinienneÕ (Paris, 1921).
 £2  J. Scherk and J.H. Schwarz, ÓGen. Rel. and Grav.Õ É6 (1975) 517;\\
	M. Kaku, \NP 91 (1975) 99;\\
	M. Goroff and J.H. Schwarz, \PL 127B (1983) 61:\\
	lightcone gauge for gravity.
 £3 Cartan, Óloc. cit.Õ (IXA), ch. X;\\
	M. Spivak, ÓA comprehensive introduction to differential geometryÕ
	(Publish or Perish Inc., 1979) v. II, p. 299:\\
	Riemann normal coordinate expansion.
 £4 U. M¬uller, C. Schubert, and A.E.M. van de Ven, A closed formula for the
	Riemann normal coordinate expansion, \xxxlink{gr-qc/9712092}.
 £5 W. Siegel, \PL 149B (1984) 162, É151B (1985) 396, É211B (1988) 55:\\
	string gauge.
 £6 B. Whitt, \PL 145B (1984) 176:\\
	equivalence of $R+R^2$ to gravity + scalar.

\unrefs

Û7 C. CURVED SPACES

There are some important solutions of general relativity that have no
close analog in Yang-Mills.   Here we consider the ones relevant to the
only experimental verifications of this theory:  Solutions outside
approximately spherical matter distributions (like the Sun and Earth), and
those describing the Universe itself.

Ü1. Self-duality

Plane wave solutions can be constructed for gravity in the same way as
for Yang-Mills (see subsection IIIC3):  A little more work (solving the
torsion constraint, or using the result of the free theory) gives
$$ á^+ = »^+ -üx^i x^j R^{+i+j}(x^-)»^- -x^i R^{+i+j}(x^-)M^{-j}ââ
	(á^- = »^-,âá^i = »^i) $$
 where $R^{+i+j}$ is an arbitrary function of $x^-$, but symmetric in $ij$,
and the empty-space field equations imply it is also traceless:
$$ R^{+i+i} = 0 $$
 If we want to couple Yang-Mills to gravity, then we can still write exact
solutions as long as both waves are parallel; then
$$ R^{+i+i} = 2T^{++} = \f1{g^2}tr(F^{+i}F^{+i}) $$
 where here $g^2$ refers to the Yang-Mills coupling.  (Similarly, we can
add in other fields, such as massless, neutral scalars or particles.)

\x IXC1.1  Check that the gravitational plane wave solution satisfies the
field equations and torsion constraint.  Show that we can also find
more-special solutions of this form satisfying
$$ g_{mn} = ú_{mn} +\f1{g^2}tr(A_m A_n),ââ
	R_{mnpq} = \f1{g^2}tr(F_{mn}F_{pq}) $$
 This has the interpretation that the ``graviton" is the bound-state of
two ``gluons".  However, it is only a kinematic effect, since the two
gluons happen to be traveling in the same direction at the same speed. 
(We saw in subsection VIIB5 that a similar effect always occurs in D=2,
since there only two spatial directions exist.)

Self-duality for Yang-Mills was discussed in subsection IIIC4.  Similar
remarks apply to gravity:  We again impose
$$ [á_a,á_b] = àü·_{abcd}[á^c,á^d] $$
 Self-duality again implies the field equations, by dualizing the Bianchi
identities:  For gravity
$$ R_{[abc]d} =  0âÜâ0 = àü·^{abcd}R_{bcde}
	= \f14 ·^{abcd}·_{bcfg}R^{fg}{}_{de} = -R^a{}_e $$
 (Note there is no extra minus from $·^2$ in even time dimensions.)
While it might appear that the self-duality condition is still second-order
because solving the torsion constraint makes the Lorentz connection the
derivative of the vierbein, the self-duality allows the gauge where the
connection is also self-dual, and this condition effectively becomes a
first-order field equation:
$$ R_{abcd} = R_{cdab}âÜâR_{abcd} = àü·_{cdef}R_{ab}{}^{ef}
	âÜâ¿_{abc} = àü·_{bcde}¿_a{}^{de} $$

\x IXC1.2  Apply exercises IIIC3.2 and IIIC4.1 to gravity:
 ªa  Rewrite all the above results of this subsection in spinor notation
for D=4.
 ªb  For arbitrary dimension D, generalize $e_-{}^+$ to an arbitrary
function of $x^-,x^i$, find the covariant derivative and curvature in terms
of it, show the source-free Einstein's equations imply it satisfies
$$ (»^i)^2 e_-{}^+ = 0 $$
 and in D=4 identify the pieces analytic and anti-analytic in $x^t$
with the two polarizations.

In four dimensions (2 space + 2 time), lightcone methods can again be
applied (see subsection IIIC5):  Now
$$ [á^{Œº'},á^{©¶'}] = C^{Œ©}üR^{º'¶'·'½'}M_{·'½'}ââ(¿_{Œº'}{}^{©¶} = 0) $$
 The fact that $[á^{(¢Œ'},á^{\¢)º'}]$ has only an $M_{Œ'º'}$ term poses an
additional constraint; the full solution is then
$$ á^{¢Œ'} = »^{¢Œ'},âá^{\¢Œ'} = »^{\¢Œ'} +(»^{¢Œ'}»^{¢º'}Ä)»^¢{}_{º'}
	+ü(»^{¢Œ'}»^{¢º'}»^{¢©'}Ä)M_{º'©'} $$
$$ R^{Œ'º'©'¶'} = -i»^{¢Œ'}»^{¢º'}»^{¢©'}»^{¢¶'}Ä $$
 (In this case, the existence of covariantly constant spinors is a
consequence of self-duality.)  The equation of motion that follows from
the final condition is now
$$ õÄ -i(»^{¢Œ'}»^{¢º'}Ä)(»^¢{}_{Œ'}»^¢{}_{º'}Ä) = 0 $$

Ü2. De Sitter

The simplest spaces are those where the Ricci scalar is constant, and
the other parts of the curvature vanish:
$$ R_{ab}{}^{cd} = k ¶_{[a}^c ¶_{b]}^d $$
These are special solutions of the field equations without matter, but
with a cosmological term, where there are no physical gravitons (the
Weyl tensor vanishes), and thus represent the vacuum.
Since there are no physical degrees of freedom, we can represent
this space by just the conformal compensator: i.e.¼the vierbein (metric)
is just the flat one up to a local Weyl scale transformation.  We thus
have (from subsection IXA7)
 $$ R_{ab}{}^{cd} £ ì ¶_{[a}^{[c} »_{b]} »^{d]}ì
	- ¶_{[a}^c ¶_{b]}^d(»ì)^2 = k ¶_{[a}^c ¶_{b]}^d $$
where we have written the curvature as a scale transformation of flat
space $R_{ab}{}^{cd}=0,¼á=»$:  The space is ``conformally flat". 
Separating this equation into its irreducible parts with respect to the
Lorentz group, the Weyl tensor part vanishes identically, leaving
$$ 2ìõì-D(»ì)^2=Dk,âD»_a »_b ì=ú_{ab}õì $$
(The latter equation isn't implied in $D=2$, where the global conformal
group is larger, and more general coordinate choices are possible for this
solution.  However, we can still use it consistently.) The latter equation
can be solved easily:  Looking at $a±b$, we see that $ì$ is a sum of
functions of one variable.  Then looking at $a=b$ tells us that these
functions are quadratic and have the same quadratic coefficient, while
the former equation gives $k$: 
$$ ì = A+B^a x_a+Cüx^a x_a,ââk = 2AC-B^2 $$
 We can choose any $A$, $B^a$, and $C$ that give the desired value of $k$:
For example, we can choose the solution $ì=1+\f14 kx^2$ (giving the
usual flat-space coordinates for $k=0$), or $ì=B^a x_a$ (choosing the
direction of $B^a$ as appropriate to $k=-B^2$ --- spacelike, lightlike, or
timelike).

\x IXC2.1  Show that after a Weyl scale transformation the action for
gravity including a cosmological term is (up to a sign) that of a conformal
self-interacting scalar coupled to gravity.  Use this to show that the de
Sitter space solution (in the R=0 gauge) yields an ``instanton" for this
scalar theory, and compare with the Yang-Mills instanton of subsection
IIIC6.  Show that similar solutions exist for massless scalars in arbitrary
dimensions with potentials $¾Ä^n$ for arbitrary $n$ (but then $k=0$).

The geometry of this space can be understood most easily as that of a
D-dimensional hyperboloid embedded in a flat (D+2)-dimensional space,
where we add one space and one time dimension:  Again using the
methods of subsections IA6 and IVA2, we now supplement the constraint
$$ y^2 = 0âÜây^A = ew^A,ââ(w^+,w^-,w^a) = (1,üx^a x_a,x^a) $$
 with the additional constraint 
$$ n^A y_A = 1âÜâ
	e = {1\over n^A w_A} = {1\over -n^- -n^+üx^a x_a +n^a x_a} $$
 where $n^A$ is a (D+2)-vector, yielding the intersection of a cone and
plane.  In particular, for $n^2±0$ we can write the metric on the space
whose coordinates are all but $nÉy$:
$$ y^A = (|n^2|^{-1/2},z^{\A})âÜâz^2 +n^{-2} = 0,ââ-ds^2 = dz^2 $$
 which is the definition of a hyperboloid.  Comparing the metric, we find
the previous result:
$$ -ds^2 = dy^2 = e^2 dx^a dx_a,ââe = ì^{-1}âÜâk = -n^2 $$

Of course, by appropriate choice of the original flat space, we can choose
a space of any signature.  In particular, we see that for a unit sphere
$k$ is normalized to 1.  Thus, with our conventions we have in that case
$$ unit¼sphere:âR_{ab}{}^{cd} = ¶_{[a}^c ¶_{b]}^d $$
(but the constant value of the Ricci scalar will depend on the dimension).

This gives the most general coordinate system for de Sitter space as a
local scale of flat space, since conformal transformations are the most
general coordinate transformations that will just replace this scale factor
with another, and they just rotate $n^A$.  The symmetry group of the
$D$-dimensional subspace that satisfies these two constraints is as big as
the PoincarŽ group, namely SO(D,1), ISO(D$-$1,1), or SO(D$-$1, 2),
depending on whether $n^2$<, =, or >0:  The former constraint preserves
the conformal group, while the latter kills a timelike, lightlike, or
spacelike coordinate.

\x IXC2.2  We can also start instead with a D+1-dimensional space, which
is a natural choice for the symmetry group of de Sitter space:  Consider
the metric and constraint
$$ -k¼ds^2 = dz^2 = k¼dz^a dz^b ú_{ab} +dz_{D+1}^2,ââ
	1 = z^2 = k¼z^a z^b ú_{ab} +z_{D+1}^2 $$
 Both equations have the same global symmetry group, determined by the
sign of $k$; $k=0$, flat space, can be considered as a limiting case of the
others.  
ªa  Solve the constraint $y^2=1£z(x)$ as in subsection IVA2 for
$Ä^2=m^2£Ä()$, and substitute to find the metric in terms of $x$.
ªb  Find the conformal transformation on $x^a$ that relates this
coordinate system to the more general one above.  (Hint:  Use $z$ of the
D+2-dimensional construction.)

Ü3. Cosmology

As discussed in subsection IVA7, the universe is approximately
isotropic (rotationally invariant) and homogeneous
(spatially translationally invariant), so the metric should depend
only on time.  
This means that the 3D subspace at any
fixed time should be 3D spherical, flat, or de Sitter space, up to an overall
time-dependent scale factor:
$$ -ds^2 = -d ^2 + Ä^2( )ç $$
 where $ç$ is the de Sitter metric for the 3 other dimensions for
$k=1,0,-1$ (given, e.g., by the coordinates in the previous subsection.)  By
a simple redefinition of the time coordinate, this can be put in a form
which is conformal to a static space:
$$ ds^2 = Ä^2(t)ß{ds}{}^2,âß{ds}{}^2=-dt^2 + ç $$
 where by ``$Ä(t)$" we really mean ``$Ä( (t))$", and the two time
coordinates are related by
$$ d  = dt¼ÄâÜât = Çd  {1\over Ä( )}âorâ  = Çdt¼Ä( (t)) $$
 Using previous results for 3D de Sitter space, we find $ß{ds}{}^2$ has
curvature
$$ öR_{ij}{}^{kl}=k¶_{[i}^k ¶_{j]}^l,ârest=0 $$ where $k=1,0,-1$.
The case $k=0$ (in good agreement with observations, at least for anything but times shortly after the Big Bang) reduces to the dilaton cosmology of subsection IVA7; here we will generalize the results given there, with a different (and sometimes better) derivation, from general relativity.

Again we begin by considering a universe filled with noninteracting dust, its rest frame defining the preferred time direction, its homogeneity and isotropy the source of those properties of spacetime.  Working directly with the energy-momentum tensor (rather than deriving it from that of the particle, as in subsection IVA7),
we then can write (compare exercises IIIB4.2 and IXB2.2)
$$ T_M^{ab} = ¨_M(t)u^au^b,âu^a = ¶^a_0 $$
 where $¨_M$ is just the spatial density of particles in the ``rest" frame. 
One way to derive the $Ä$ dependence of $¨_M$ that generalizes
straightforwardly to other cases is by using conservation laws:  By
considering particles all of the same mass in units $m=1$, or by
considering $J$ and $T$ for each individual particle (since in this case we
neglect interactions), we have from current conservation
$$ J^a = ¨_M u^aâÜâ0 = á_a J^a = {\bf e}»_m{\bf e}^{-1}J^a e_a{}^m
	= (Ä^{-4})»_0(Ä^4)(¨_M Ä^{-1}) $$
$$ Üâ¨_M = 3aÄ^{-3} $$
 for some nonnegative constant $3a$, and using (covariant)
energy-momentum conservation as a check,
$$ u^a á_a u_b = á_0 u_b = Ä^{-1}»_0 u_b = 0â(geodesic)âÜâ
	á_a T_M^{ab} = u^b á_a J^a +J^a á_a u^b = 0 $$
 where we have used (from the result of subsection IXA7 for scaling
covariant derivatives)
$$ ßá_0 = »_0âÜâá_0 = Ä^{-1}»_0 $$
 (Note, however, that $á_i$ has Lorentz pieces, and $M_{0i}J_i¾J_0±0$
even though $J_i=0$.)

\x IXC3.1  Find completely explicit expressions for the covariant
derivatives in this case (choosing some coordinates for $ç$ for
$k=0,à1$) using just the Weyl transformation method of
subsection IXA7.

For radiation, the momenta of the photons can't be timelike (they're
lightlike, of course), but we can still use rotational and translational
invariance, together with the fact that the trace of the
energy-momentum tensor vanishes (from scale invariance: see
subsection IXA7).  Then
$$ T_R^{ab}=¨_R(t)\f13(4u^a u^b+ ú^{ab}) $$
 There is no conserved current, but energy-momentum conservation alone
determines
$$ 0 = á_a T_R^{ab} = \f43 ¨_M u^a u^b á_a{¨_R\over ¨_M}+á^b\f13 ¨_R
	= Ä^{-1}¶^b_0(\f43 ¨_M »_0{¨_R\over ¨_M}-\f13 »_0¨_R) $$
$$ = Ä^{-1}¶^b_0 ¨_M^{4/3}»_0 ¨_M^{-4/3}¨_R
	âÜâ¨_R = \f32 bÄ^{-4} $$
 for some nonnegative constant $\f32 b$.

Writing the vierbein as a scale-transformation of the constant
curvature space discussed above (de Sitter in spatial directions, flat in
other directions), the gravitational field equations with matter
and radiation become (using results from subsection IXA6 or 7):
$$ 6[(ßá_a Ä)(ßá_b Ä)-üú_{ab}(ßáÄ)^2] + [(ú_{ab}ßõ-ßá_a ßá_b)
	+ (öR_{ab}-üú_{ab}öR)]Ä^2 = 2(T_{Mab}+T_{Rab})Ä^4 $$
The only independent components of this equation are the 00-component
and trace, which are, after multiplying by an appropriate
power of $Ä$:
$$ üÀÄ{}^2+ükÄ^2=aÄ+üb,ââ¬Ä+kÄ=a $$
For $k=1$, these are just energy conservation and the equation of motion
for a harmonic oscillator (centered at $Ä=a$).  The 00 equation gave
energy conservation because $T_{00}$ is the energy density.  The trace
equation gave the field equation for $Ä$,
which is proportional to the time derivative of the 00 equation,
due to the relation of $T^a{}_a$
to $¶S/¶Ä$ given earlier.  These equations are easily solved:  Imposing
the initial condition $Ä(0)=0$ (i.e., we set the ``Big Bang", when
curvatures and energy density were infinite, to be $t=0$) and $ÀÄ(0)>0$ (so
$ij0$),
$$ k=\left.\cases{1\cr 0\cr -1\cr}\rightÕ:â
	Ä=a\left.\cases{1-cos¼t\cr üt^2\cr cosh¼t-1\cr}\rightÕ
	+åb\left.\cases{sin¼t\cr t\cr sinh¼t\cr}\rightÕ $$
The ``physical" time coordinate is then $ =Ç_0 dt¼Ä$.  In general $Ä$
can't be expressed directly in terms of $ $, so we use the expressions
for both in terms of $t$.  For example, for $k=1$ and $b=0$ (just
matter), we get a cycloid, which has only such a parametric expression.
Explicit expressions can be found for $a=0$ (just radiation):  $Ä( )$ is
then a circle, parabola, or hyperbola for $k$=1, 0, $-1$.  Also, for $k=0$
and $b=0$, $ľ ^{2/3}$ (vs.¼$å $ for $a=0$).

\x IXC3.2  Find the modification to the equations of motion when a
cosmological term is included.

Returning to the case of pure matter ($b=0$) as in subsection IVA7, we now find the energy conservation equation for general $k$
$$ ü\left({dÄ\over d }\right)^2 - {a\over Ä} = -{k\over 2} $$
Again comparing to the Newtonian equation for  a
particle, we see that $-ük$ corresponds to the total energy, determining whether expansion is eternal or leads to collapse.

Ü4. Red shift

We now generalize the results of subsection IVA7 on
cosmological red shift by considering Killing vectors.  Since the
cosmological solutions are related to static, isotropic, homogeneous
spaces by a time dependent (but space independent) scale
transformation, the symmetries of this space are just in the spatial
directions, and are basically the same as before the scale
transformation.  Specifically, the Killing vectors that survive the scale
transformation $á_a = ìßá_a + (ßá^b ì)M_{ab}$ satisfy
(see excercise IXA7.3)
$$ ßKÉßáì=0âÜâK = ßKâÜâK_a = ì^{-1}ßK_a $$
We then find for conserved momenta $K^a p_a ¾ ì^{-1} p_a$.  Since the
$K$'s which survive are just the spatial ones, we at first find only the
spatial components of $ì^{-1}p_a$ conserved, but the conservation of
the time component follows from $p^a p_a =0$ for photons.  Thus,
$p_a¾ì¾Ä^{-1}$.  Since $p_a$ is what an observer measures as the
components of momentum (in his ``local inertial frame", a gauge where
at his location the metric is flat and its first derivative vanishes),
observers measure the photon's energy as having time dependence $¾Ä^{-1}$.

\x IXC4.1  Using this result for the $Ä$ dependence of the momenta of
individual particles, we can now rederive the $Ä$ dependence of $¨$'s of
the previous subsection directly from the explicit expressions for $J$ and
$T$ of the point particle.
 ªa  Rederive $J$ and $T$ in curved space as in subsection IIIB4 and show
that
$$ J^m ¶_m^0 = ·(p^0){\bf e}(2¹)^2 ¶^3(x-X),ââ
	T^{mn} = J^m p^n = J^n p^m $$
 ªb  From Killing vectors we just saw that
$$ p^a ¶_a^0 ¾ 
	\leftÓ \matrix{ Ä^0 & (m±0) \cr Ä^{-1} & (m=0) \cr} \right. $$
 where the massive particles are at rest ($p^a=m¶^a_0$).  Combine these
results to find
$$ J^a ¶_a^0 ¾ Ä^{-3},â⨠¾ T^{ab}¶_a^0 ¶_b^0 ¾
	\leftÓ \matrix{ Ä^{-3} & (m±0) \cr Ä^{-4} & (m=0) \cr} \right. $$
 ªc  Find the factors multiplying the $¨$'s in $T^{ab}$ for the two cases of
dust and radiation from the explicit expression for $T^{mn}$.  For the
massive case (dust) all particles can be taken at rest, but for the
massless case the particles travel at the speeed of light, so average over
particles traveling in the three spatial directions and their opposites.  ($¨$ is a continuous
function obtained by summing the $¶$ functions of all the particles. 
However, for the above results it is sufficient to consider each individual
particle for the massive case, and 6 particles at the same point going in
$à$ orthogonal directions for the massless case.)

As discussed in subsection IVA7, astronomers use the parameters $H$, $q$, and $¯$ to measure general features of cosmology.  Here we can generalize the analysis to $k±0$, which we have solved above.
In the case of pure matter, and with
vanishing cosmological constant, $¯=2q$.   Then the ``critical" value is
$q=ü$, for which $k=0$:  For $q>ü$, $k=1$, while for $q<ü$, $k=-1$.  In
this case we also see that for a given value of $H$ the critical value of the
matter density is $¨_c=\f32H^2$.  If the matter of the universe has this
density, we have $k=0$, and spacetime is conformally flat.  If it has
greater density, we have $k>1$, and space is closed.

\x IXC4.2  Solve for $¯$ and $q$ in terms of just $a,b,k,$ and $Ä$ (but no
time derivatives).  In particular, show
$$ \li{ b = 0â& Ü⯠= 2q = (1 -\f{k}{2a} Ä)^{-1} \cr
	a = 0â& Ü⯠= q = (1 -\f{k}b Ä^2)^{-1} \cr} $$

Ü5. Schwarzschild

All gravitational experiments outside of cosmology are based on the
``Schwarz\-schild solution", which describes spherical symmetry outside
the region with matter.  Assuming also time independence, which is a
consequence of spherical symmetry (Birkhoff's theorem), we look for a
metric of the form
$$ -ds^2 = -A^{-2}(r)dt^2 +B^{-2}(r)dr^2 +r^2(dÏ^2+sin^2ϼdÄ^2) $$
(Other coordinate choices are possible,
e.g.¼$-A^{-2}(r)dt^2 +B^{-2}(r)[dr^2 +r^2(dÏ^2+sin^2ϼdÄ^2)]$.)
 The first step is the choice of a vierbein:  The simplest choice following
from this metric is
$$ e_t=A»_t,âe_r=B»_r,âe_Ï={1\over r}»_Ï,â
	e_Ä={1\over r¼sin¼Ï}»_Ä $$
 (This can also be used as a starting point in place of the metric.)  The
next step is to find the commutators of the $e$'s, which tells us what $¿$
terms the $á$'s must have to cancel these $c_{ab}{}^c$'s (vanishing
torsion):
$$ \eqalign{ [e_Ï,e_Ä]&=-r^{-1}cot¼Ï¼e_Ä \cr
	[e_r,e_t]&=B(ln¼A)'e_t \cr
	[e_r,e_Ï]&=-r^{-1}Be_Ï \cr
	[e_r,e_Ä]&=-r^{-1}Be_Ä \cr}
¼Ü¼\eqalign{&á_ļhas¼M_{ÏÄ} \cr
	&á_t¼has¼M_{tr} \cr
	&á_ϼhas¼M_{rÏ} \cr
	&á_ļhas¼M_{rÄ} \cr}
¼Ü¼\eqalign{á_r&=B»_r\cr
		á_t&=A»_t+ŒM_{tr}\cr
		á_Ï&=r^{-1}»_Ï+ºM_{rÏ}\cr
		á_Ä&=(r¼sin¼Ï)^{-1}»_Ä+©M_{rÄ}+¶M_{ÏÄ}\cr} $$
 where $Œ,º,$ and $©$ depend only on $r$, while $¶$ depends also on $Ï$. 
(Their explicit forms are already clear at this point, but we'll collect the
results below.)

We can now determine these Lorentz connections and compute the
curvatures by calculating the $á$ commutators.  Since we now use
explicit functions for the vierbein and connections, we use the method
described in subsection IXA2 for this situation:  Using the identities
$$ [M_{12},V_2] = ú_{22}V_1,ââ[M_{12},M_{23}] = ú_{22}M_{13} $$
$$ [á_1,á_2] = [e_1+¿_1,e_2+¿_2] $$
$$ = Ó[e_1,e_2]+(e_1 ¿_2)M_2-(e_2 ¿_1)M_1Õ
	+Ó¿_1[M_1,á_2]-¿_2[M_2,á_1]-¿_1 ¿_2[M_1,M_2]Õ $$
 we then find:
$$ \li{ [á_t,á_Ï] & = -ŒºM_{tÏ}âÜâR_{tÏtÏ} = -Œº \cr
	[á_t,á_Ä] & = -Œ©M_{tÄ}âÜâR_{tÄtÄ} = -Œ© \cr
	[á_t,á_r] & = -B(ln¼A)'e_t -BŒ'M_{tr} +Œá_t
				â= [Œ-B(ln¼A)']e_t +(Œ^2-BŒ')M_{tr} \cr
				& Ü⌠= B(ln¼A)',ââR_{trtr}=Œ^2-BŒ' \cr
	[á_r,á_Ï] &= -{B\over r}e_Ï +Bº'M_{rÏ} +ºá_Ï
				â= (º-{B\over r})e_Ï +(º^2+Bº')M_{rÏ} \cr
				& Üâº={B\over r},ââR_{rÏrÏ}=-(º^2+Bº') \cr
	[á_r,á_Ä] & = -{B\over r}e_Ä +B©'M_{rÄ} +B¶'M_{ÏÄ} +©á_Ä \cr
				& = (©-{B\over r})e_Ä +(©^2+B©')M_{rÄ}
					+(©¶+B¶')M_{ÏÄ} \cr
				& Üâ©={B\over r},ââR_{rÄrÄ} = -(©^2+B©'),ââ
					R_{rÄÏÄ} = -(©¶+B¶') \cr
	[á_Ï,á_Ä] &= -{cot¼Ï\over r}e_Ä +{1\over r}(»_Ï ¶)M_{ÏÄ} +¶á_Ä
					+º©M_{ÏÄ} -º¶M_{rÄ} \cr
				& = (¶-{cot¼Ï\over r})e_Ä +(©-º)¶M_{rÄ}
					+(¶^2+º©+{1\over r}»_Ï ¶)M_{ÏÄ} \cr
				& Üⶠ= {cot¼Ï\over r},ââ
					R_{ÏÄÏÄ}=-(¶^2+º©+{1\over r}»_Ï ¶),
					ââR_{rÄÏÄ}=0 \cr } $$
Collecting the results:
$$ \eqalign{ á_t & = A»_t +B(ln¼A)'M_{tr} \cr
	á_r & = B»_r \cr
	á_Ï & = {1\over r}»_Ï +{B\over r}M_{rÏ} \cr
	á_Ä & = {1\over r¼sin¼Ï}»_Ä +{cot¼Ï\over r}M_{ÏÄ}
		+{B\over r}M_{rÄ} \cr }ââââ
\eqalign{ R_{trtr} & = BA[B(A^{-1})']' \cr
	R_{tÏtÏ} & = R_{tÄtÄ} = -{B^2\over r}(ln¼A)' \cr
	R_{rÏrÏ} & = R_{rÄrÄ} = -{BB'\over r} \cr
	R_{ÏÄÏÄ} & = {1-B^2\over r^2} \cr } $$

\x IXC5.1  Find the covariant derivative for the 2-sphere in spherical
coordinates
$$ ds^2 = dÏ^2 + sin^2 ϼdÄ^2 $$
 in terms of the single SO(2) generator $M_{ab}=·_{ab}M$ by the above
methods (and not that of exercises IXA5.3 nor IXA7.5).  
Calculate the curvature.  Find
the three Killing vectors.  (Hint:  What is the symmetry of the sphere?)

A simpler method of finding covariant derivatives and curvatures
in this case is the Weyl scale method of subsection IXA7.
(We already applied this method to the much simpler example of
cosmology in subsection IXC3.)
We start with the trivial covariant derivatives for the 2D metric
$$ -ds^2 = -dt^2 +dr^2 $$
which are just partial derivatives (zero curvature).
Then we make the coordinate transformation
$$ dr £ {A(r)\over B(r)} dr $$
 (explicit integration of this expression isn't needed in either
the metric or the covariant derivatives),
which modifies one of the covariant derivatives,
$$  -ds^2 £ -dt^2 +{A^2\over B^2}dr^2;ââ
	á_r £ {B\over A}»_r,ââá_t £ »_t $$
while the curvature still vanishes.  (We have only chosen
non-Cartesian coordinates for flat space.)
Next, we make the scale transformation
$$ ì = rA $$
 to obtain the metric
$$ -ds^2 = -(rA)^{-2}dt^2 +(rB)^{-2}dr^2 $$
Applying the general formula
$$ ds'^2 = ì^{-2}ds^2âÜâá'_a = ìá_a +(á^b ì)M_{ab}, $$
we find
$$ á_r £ rB»_r,ââá_t £ rA»_t +{B\over A}(rA)'M_{tr} $$
Since the space is only 2D, the general equation
$$ R'_{ab}{}^{cd} = ì^2 R_{ab}{}^{cd} 
	+ì¶_{[a}^{[c}á_{b]}á^{d]}ì -¶_{[a}^c ¶_{b]}^d (áì)^2 $$
simplifies to
$$ R'_{ab}{}^{cd} = üR'¶_{[a}^c ¶_{b]}^d,ââ
	üR' = ì^2 (üR +õ¼ln¼ì) $$
so at this stage we have
$$ üR £ (rA)^2{B\over A}\left[{B\over A} (ln¼rA)'\right]' $$
Any 2D space can be expressed as a scale transformation of
flat space locally, essentially because the curvature has only one
component.  (Globally this is not true, since the integral of
the curvature, which is scale invariant in D=2, 
is different for different topologies.
This is related to the fact that for nontrivial topologies
more than one coordinate patch is needed; the
missing part of the integral can be hidden in the boundaries
of the patches: see exercise IXA7.6.)

Now we should repeat this procedure for $Ï$ and $Ä$ to get
the covariant derivatives for the (2-)sphere, but this has already been
done earlier.  Besides, we don't need those expressions explicitly,
since spherical symmetry means they vanish on anything, and
we already know the curvature of a sphere.  (So, we can also avoid
choosing a coordinate system for the sphere.)
Thus we can immediately take the direct product of the sphere 
with the above 2D space, and make the final scale transformation
$$ ì = \f1r $$
The result for the final covariant derivatives is
$$ á_r = B»_r,âá_t = A»_t +{B\over A}A'M_{tr},â
	á_i = \f1r \ron\circ á_i -\f1r B M_{ir} $$
where $\ron\circ á_i$ are the covariant derivatives for the sphere,
in agreement with the previous method.  The curvatures are also
easy to find:  Besides the $á_r ì$ we needed for the covariant
derivatives, the only second-order derivatives we need are
$á_r^2 ì$ and $á_t^2 ì$.  ($á_t ì$ vanishes, but $á_t^2 ì$ is
nonvanishing because of the $M_{tr}$ connection term in $á_t$
converting $á_t ì$ into $á_r ì$.)  Thus we need to evaluate
$$ \li{ R'_{ij}{}^{kl} & = ¶_{[i}^k ¶_{j]}^l [\f1{r^2}É1-(á_r \f1r)^2] \cr
	R'_{ii'}{}^{jj'} & = ¶_i^j [\f1r á_{i'}á^{j'}\f1r -¶_{i'}^{j'}(á_r \f1r)^2] \cr
	R'_{i'j'}{}^{k'l'} & = ¶_{[i'}^{k'} ¶_{j']}^{l'} \f1{r^2} [üR +(á_r^2 -á_t^2)ln¼\f1r] \cr} $$
where $i'=(t,r)$, and the $á$'s and $R$ on the right refer to the 2D $t$-$r$ space
just before or after the direct product.  The result also reproduces the previous.
The final result for $R_{trtr}$ can be obtained even more simply by
noting that it agrees with what we would have obtained
by a single scaling for the 2D space $-ds^2 = -A^{-2}dt^2 +B^{-2}dr^2$, 
because of the triviality of the $Ï$ and $Ä$ derivatives.

Having all the curvatures, we can now calculate the Ricci tensor, which
appears in the field equations.  The nonvanishing components are:
$$ R_{tt}=R_{trtr}+2R_{tÏtÏ},âR_{rr}=-R_{trtr}+2R_{rÏrÏ},â
	R_{ÏÏ}=R_{ÄÄ}=-R_{tÏtÏ}+R_{rÏrÏ}+R_{ÏÄÏÄ} $$
Vanishing of $R_{ab}-üú_{ab}R$ is equivalent to vanishing of $R_{ab}$.
In terms of these curvatures, we see it also implies
$$ -R_{trtr}=2R_{tÏtÏ}=-2R_{rÏrÏ}=R_{ÏÄÏÄ} $$
These are easy to solve:  First,
$$ R_{tÏtÏ}=-R_{rÏrÏ}âÜâ(ln¼A)'=-(ln¼B)'âÜâA=B^{-1} $$
where we have fixed the proportionality constant by requiring $A,B£1$
as $r£¥$ (redefining $t$ by a constant scale transformation).  Also,
$$ -2R_{rÏrÏ}=R_{ÏÄÏÄ}âÜâ(1-B^2)'=-{1\over r}(1-B^2)
	âÜâ1-B^2={k\over r} $$
$$ ÜâB=å{1-{k\over r}} $$
for some constant $k$.  The last field equation is then redundant.  (As
usual, the field equations are related by the Bianchi identity.)  The
constant $k$ can be related to the nonrelativistic result by comparing at
large distances.  (See exercise IXB1.1.)  We then find $k=2GM$, so the
final result is:
$$ \boxeq{ -ds^2 = -\left(1-{2GM\over r}\right)dt^2
	+\left(1-{2GM\over r}\right)^{-1}dr^2 +r^2(dÏ^2+sin^2ϼdÄ^2) } $$

\x IXC5.2  Repeat this Weyl scale derivation of covariant derivatives 
and curvatures for the Schwarzschild metric in dimensions D>4.
(Hint:  Do not use explicit expressions for the covariant derivatives of
the higher-dimensional sphere.)  Solve for $A$ and $B$.

More generally, if we have some spherically symmetric, static matter
distribution, then the only nonvanishing components of the
energy-momentum tensor will be $T_{tt}$, $T_{rr}$, and $T_{ÏÏ}=T_{ÄÄ}$
(representing energy density, radial pressure, and isotropic pressure), all
functions of just $r$.  Repeating the above procedure, we integrate
$$ [r(1-B^2)]' = 2r^2 T_{tt},ââ[ln(AB)]' = -{r\over B^2}(T_{tt}+T_{rr}) $$
 while the remaining equation is redundant.  

\x IXC5.3  Use the local conservation law for energy-momentum to determine $T_{ÏÏ}$ in terms of $T_{tt}$ and $T_{rr}$.

For example, for a spherically symmetric, static electromagnetic field the
only nonvanishing components of the field strength are $F_{tr}$ and
$F_{ÏÄ}$, corresponding to electric and magnetic charges, respectively. 
Then the invariance of $T$ under a duality transformation (see
subsections IIA7, IIIA4) implies
$$ T_{ÏÏ} = T_{ÄÄ}âÜâT_{tt} = -T_{rr}âÜâA = B^{-1} $$
 again, since on this $F_{ab}$ duality effectively replaces $(Ï,Ä)ª(it,r)$,
with the $i$ from Wick rotation.  Local scale invariance (see subsection
IXA7) then tells us
$$ T_a{}^a = 0âÜâT_{tt} = -T_{rr} = T_{ÏÏ} = T_{ÄÄ} $$

\x IXC5.4  Let's rederive these results by brute force:
 ªa  Derive $T_{ab}$ for a general electromagnetic field by varying its
action with respect to $e_a{}^m$ or $g^{mn}$.
 ªb  Find each of $T$'s components explicitly in terms of $F_{tr}$ and
$F_{ÏÄ}$ in the case where those are the only nonvanishing components,
and show they appear only in the combination $F_{tr}^2+F_{ÏÄ}^2$.

As usual, these field strengths can be found easily from the integral form
of Gauss' law by integrating over a sphere:  For example, for the magnetic
field
$$ magnetic¼charge ¾ üÇdx^m dx^n¼F_{mn} = 4¹r^2 F_{ÏÄ} $$
 for the $F_{ÏÄ}$ component of $F_{ab}$ (integrating over $Ï$ and $Ä$),
since the metric (and vierbein) for $Ï$ and $Ä$ is the same as for flat
space.  By duality, the solution for $F_{tr}$ in terms of the electric charge
is the same.  The result is
$$ T_{tt} = {Q\over r^4},ââQ = ¹^2(e^2+g^2) $$
 for electric charge $e$ and magnetic charge $g$.  The $1/r^4$
dependence also follows from scale invariance, since the charges are
dimensionless (and the matter field equations decouple from $A$ and $B$). 
(Again, since the solution does not extend to $r=0$, we normalize by
comparing $F_{ab}$ or $T_{ab}$ at $r=¥$ to the flat-space solution.)  The
net effect on the Schwarzschild metric is
$$ 1-{2GM\over r} £ 1-{2GM\over r} +{2Q\over r^2} $$
 Our solution relates to the usual mechanics normalization of the charges
(see subsection VIIA3), restoring $G$, as
$$ 2Q = G2¹(e^2+g^2) = G(e_m^2+g_m^2) $$

\x IXC5.5  Let's also apply brute force to solving Maxwell's equations
$á_a F^{ab}=á_{[a}F_{bc]}=0$ (outside the matter).
 ªa As a warm-up, using directly the above covariant derivatives, show
that Óin flat spaceÕ
$$ V^a = ¶^a_r V_râÜâá_a J^a = r^{-2}»_r r^2 J_r $$
 Note that the covariant derivative of a vanishing component doesn't
necessarily vanish (just as the ordinary derivative of a function that
vanishes at some point doesn't necessarily vanish at that point): 
Components of $á$ other than $á_r$ contain Lorentz generators that
rotate other components of $V$ to $V_r$.
 ªb  Solve Maxwell's equations in differential form for $F_{ab}$ in
the above case.  Use the empty-space solution to define the
normalization at infinity.  (Actually, in this case, the charge is
well-defined in terms of the flux of the fields, as described above, but
gives the same result here because the space is asymptotically flat.)

\x IXC5.6  Spherically symmetric solutions can also be written in
Eddington-Finkelstein coordinates as
$$ -ds^2 = -dt^2 +dr^2 +r^2(dÏ^2+sin^2ϼdÄ^2) +H(r)(dt+dr)^2 $$
 where $H=1-B^2=G[2M/r-2¹(e^2+g^2)/r^2]$ in terms of the above 
results, and is linear in $G$.  (In these coordinates, gravity looks 
Abelian for this solution.)  Note that this form (and its Abelian
nature) closely resembles the general wave solutions of
exercise IXC1.2b (but the ``Cartesian" coordinate $x^1$ is now
replaced with $r$).
 ªa Obtain this form from the above forms (where $A=B^{-1}$) by
a coordinate transformation.  (Hint:  The angular term didn't change.)
 ªb Find $á$ directly from this form of the metric.
(Note:  It might differ from the previous by not only coordinate but
also local Lorentz transformations.)

\x IXC5.7  Consider the plane wave in the coordinates
$$ -ds^2 = -2dx^+dx^- + L^2(x^-)
	\left(e^{2º(x^-)}dy^2 +e^{-2º(x^-)}dz^2\right) $$
 Calculate the covariant derivatives and curvature tensor by the first method
of this subsection (double-counting and subtracting, not the Weyl scale
method).  Show that the field equations reduce to 
$$ L'' +(º'{}^2)L = 0 $$

\x IXC5.8  Use the first method of this subsection to calculate
the covariant derivative and curvature tensor for the metric
 $$ -ds^2 = - dt^2 + 2 e^x dt¼dy - üe^{2x}dy^2 + dx^2 + dz^2 $$
 Show that this metric satisfies the field equations with a cosmological
term for a dust at rest with respect to this time coordinate; i.e.
 $$ R^{mn} - üg^{mn}(R-4ñ) = ¨¶^m_0 ¶^n_0 $$
 where $ñ$ and $¨$ are both constants.

\x IXC5.9  Use this method to calculate the covariant derivative and
curvature tensor for the cylindrically symmetric metric
 $$ -ds^2 = - A^{-2}(r)dt^2 + B^{-2}(r)dr^2 +r^2 dÏ^2 + dz^2 $$
 Assume the matter in this problem is a ``perfect fluid",
$$ T^{ab} = ¨u^a u^b +P(ú^{ab} +u^a u^b)ââ(u^2 = -1) $$
 Solve the equations of motion for the gravitational field to find $A$ and
$B$, as well as the pressure $P$ and particle density $¨$.  What is the
implied relation between $P$ and $¨$?

\x IXC5.10  Use this method to calculate the covariant derivative and
curvature tensor for the following metric, corresponding to that outside a
planar mass distribution:
 $$ -ds^2 = - A^{-2}(z)dt^2 + B^{-2}(z)(dx^2 +dy^2) + dz^2 $$
 Solve Einstein's equations in empty space to find $A$ and $B$ (up to some
constants of integration).

Ü6. Experiments

When comparing to the real world, it is useful to know some 
astrophysical radii:
\item{(1)}  Earth's orbit (1 AU): 1.5 $ð 10^8$ km
\item{(2)}  Solar radius: 7 $ð 10^5$ km
\item{(3)}  Earth radius: 6000 km
\item{(4)}  Solar gravitational (Schwarzschild) radius ($2GM/c^2$): 3 km
\item{(5)}  Earth gravitational radius: 0.9 cm (1 shoe size).

\noindent To see how these fit in with other physical criteria, consider the following diagram of mass vs.¼radius (in natural/Planck units) for various physical objects:

$$ \figscale{massradius}{6in} $$
Particles have the Compton radius $R=\h/Mc$ according to quantum mechanics, black holes (see subsection IXC7) have the Schwarzschild radius $R=2GM/c^2$.  Condensed matter (solids and liquids) is basically atoms packed together, and has the same density regardless of size, up to an order of magnitude or so.  The size of an atom is about the same as the Compton radius of an electron, up to a factor of the fine structure constant, while stars are more or less condensed matter near their gravitational radii, up to a few orders of magnitude.  So, known objects tend to lie near the lines drawn above, to within a few orders of magnitude (perhaps related to the fine structure constant $Œ®1/137$ or the proton-electron mass ratio $®1836$), which is close compared to the tens of orders of magnitude that set the scale of the diagram.

\x IXC6.1  Consider the following very crude approximations to various types of stars:  
ªa Assume a star has the density of a neutron,
i.e., of a sphere with the mass $M$ of the neutron and radius
equal to the Compton radius $\h /Mc$.  Assume also that the radius of this
(spherical) star is equal to its gravitational radius.  
(This is roughly a ``neutron star".)  Find the mass
and radius, in terms of both physical constants and
conventional units.   Note the appearance of the large dimensionless number, the ratio of the Planck mass to the neutron mass.
ªb Assume a star has the density of a ``compressed" hydrogen atom,
a sphere with the mass of the hydrogen atom (which we can take as roughly equal to the neutron mass) and radius
equal to the Compton radius of the ÓelectronÕ, $\h/mc$ for electron mass $m$.  Assume the mass of this star is equal to that of the neutron star found in the previous example.  (This is roughly a ``white dwarf".)  Find the radius, again in terms of both physical constants and
conventional units.  
ªc Assume the same mass again, but now assume the density of an ordinary hydrogen atom, which has the Bohr radius $\h/mcŒ$.   (This is roughly an ``ordinary" star.)  Compare to the mass and radius of the Sun.

All experiments (excluding cosmology) are based on the Schwarzschild
metric.  The first type of experiment involves gravitational redshift,
but unlike the cosmological case, the relevant reference frames of
observation are not local inertial frames but the static reference frame
in which the Schwarzschild metric is defined.  (There are also
measurements of redshift from airplanes, whose reference frame is
defined with respect to the Schwarzschild one.)  In this reference
frame the relevant Killing vector is the one which expresses the fact
that the space is static, $K^m »_m=»/»t$.  The momentum which is
measured by the observer is $p^a$, not $p^m$ or $p_m$, since the
observer still uses a reference frame for which the metric at his position
is flat (but not its first derivative, since he is not in free fall).  (In fact,
this is one of the purposes for using a vierbein, as a frame of reference.) 
The conserved quantity is then
$$ E = -K_a p^a = å{1-{2GM\over r}}ßE $$
where the energy of a particle $ßE$ is the time component of $p^a$ as
measured in this frame.  Thus, conservation of $E$ for a photon gives the
$r$-dependence of the observed energy $ßE$ (and thus the frequency,
which in turn determines the wavelength, since $p^2=0$).

To compare with nonrelativistic mechanics, we instead evaluate $E$ for a
massive particle in the Newtonian limit:
$$ E ® \left(1 -{GM\over r}\right)(m+K) ® m +K -{GMm\over r} $$
 giving the ``conserved energy" $E$ in terms of the ``particle energy" $ßE$
(rest mass $m$ + kinetic $K$), including the potential energy.

The other type of experiment involves properties of geodesics, so we
need to solve the geodesic equations of motion.  Without loss of
generality, we can choose the angular coordinates such that the initial
position and direction of the particle is in the equatorial plane $Ï=¹/2$,
where it remains because of the symmetry $Ϫ¹-Ï$, as in the
nonrelativistic case.  Also as in the nonrelativistic case, we can find
constants of the motion corresponding to the energy $E$ and
($z$-component of) angular momentum $L$ by using the Killing vectors
$K^m »_m = »/»t$ and $»/»Ä$ to find the conserved quantities 
$K^a p_a=K^m g_{mn}Àx{}^n$ (in the parametrization $v=1$):
$$ E ­ -g_{tm}Àx{}^m = \left(1-{2GM\over r}\right)Àt,â
	L ­ g_{Äm}Àx{}^m = r^2 ÀÄ $$
In the case where the particles come from infinity, these are the initial
kinetic energy and angular momentum.  We have chosen an affine
parametrization, which requires
$$ -m^2 = g_{mn}Àx{}^m Àx{}^n = -\left(1-{2GM\over r}\right)Àt{}^2
	+\left(1-{2GM\over r}\right)^{-1}Àr{}^2 + r^2 ÀÄ{}^2 $$
Solving the previous equations for $Àt$ and $ÀÄ$, this reduces to the radial
equation
$$ 0 = -E^2 + Àr{}^2 +
	\left(1-{2GM\over r}\right)\left({L^2\over r^2}+m^2\right) $$
$$ ÜâüÀr{}^2 + \left(-{GMm^2\over r}+{L^2\over 2r^2}-{GML^2\over r^3}
	\right) = ü(E^2-m^2) $$

This looks like a typical nonrelativistic Hamiltonian for ``energy"
$ü(E^2-m^2)$ with the same terms as in the Newtonian case but with
an extra $r^{-3}$ term.  (To take the nonrelativistic limit for the
massive case, first scale the affine parameter $ £s/m$.)  Since there
are good coordinate systems for a ``black hole" using $r$ as a coordinate
(e.g., see the following subsection: $r$ and $r''+t''$, as seen from the figure
for Kruskal-Szkeres), this equation can even be used to descibe a fall into
a black hole.  (For example, for $L=0$ we get the same cycloid solution
as in cosmology and in Newtonian gravity, reaching the singularity at
$r=0$ in finite proper time.)

Because of the $r^{-3}$ term in the potential, noncircular orbits are no
longer closed.  In particular, let's consider orbits which are close to
circular.  Circular orbits are found by minimizing the potential for the
$r$-equation:
$$ 0 = {dV\over dr} = {GMm^2\over r^2} -{L^2\over r^3}
	+{3GML^2\over r^4} $$
$$ 0 < {d^2 V\over dr^2} = -{2GMm^2\over r^3} +{3L^2\over r^4}
	-{12GML^2\over r^5} $$
The near-circular orbits are described by small (harmonic) oscillations
about this minimum, with angular frequency given by
$$ ¿_r^2 = {d^2 V\over dr^2} = {GMm^2(r-6GM)\over r^3(r-3GM)} $$
from solving for $L^2=GMm^2r^2/(r-3GM)$.  On the other hand, the
frequency of the circular orbit itself in terms of its angular dependence
is just $ÀÄ=L/r^2$, giving
$$ ¿_Ä^2 = {GMm^2\over r^2(r-3GM)} $$
This means that the perihelion (closest approach to the Sun) of an
orbit, which occurs every period $2¹/¿_r$ of the radial motion, results
in the change of angle
$$ 2¹ + ¶Ä = Ç_0^{2¹/¿_r} d {dÄ\over d } = {2¹\over ¿_r}¿_Ä =
	2¹\left(1-6{GM\over r}\right)^{-1/2} $$
$$ Üâ¶Ä ® 6¹{GM\over r} $$
in the weak-field approximation.  This effect contributes to the
measurement of the precession of the perihelion of the (elliptical) orbit
of Mercury, but so do the precession of Earth's axis, the oblateness of
the Sun, and gravitational interaction with other planets.  As a result,
this relativistic effect contributes less than 1\% to the observed
precession.  In particular, the solar oblateness is difficult to measure.

The effects on geodesics of photons are much easier to measure, since
there are no Newtonian effects.  As a result, the weak field
approximation is sufficient.  We first consider bending of light by the
Sun:  A photon comes in from infinity and goes back out to infinity
(actually to the Earth, which we assume is much farther from the Sun
than the photon's closest approach to it), and we measure what angle
its trajectory was bent by.  (For example, we look at the apprarent
change of position in stars when the Sun passes in their direction
during an eclipse.)  Starting with the exact solution for a photon's
geodesic (case $m^2=0$ above), we use the equations for $Àr$ and $ÀÄ$
to find
$$ {dr\over dÄ} = å{{E^2\over L^2}r^4 -r^2 +2GMr} $$
Changing variables,
$$ u ­ {b\over r},âb ­ {L\over E},âa ­ {GM\over b}â
	ÜâdÄ = {du\over å{1 -u^2 +2au^3}} $$
The Óimpact parameterÕ $b­L/E$ would be the closest approach to the
Sun neglecting gravitational effects ($L=rp=bE$).  We now make the weak
field approximation:  For $a$ small,
$$ \li{ dÄ &® {du\over å{1-u^2}}\left(1 -a{u^3\over 1-u^2}\right) \cr
	&= d \left(1 -a{sin^3 \over cos^2 }\right)â(u­sin¼) \cr
	&= d\left[ -a\left(cos¼ +{1\over cos¼}\right)\right] \cr } $$
Defining $Ä=0$ at $r=¥$, the integral is
$$ Ä ®  -a{(1 -cos¼)^2\over cos¼} $$
 The ends of the path ($r=¥$) are at exactly
$$  = 0,¼¹âÜâÄ = 0,¼¹+4a $$
Therefore the deviation of $Ä$ from a straight line is $4GME/L$.
(Mathematical note:  All variable changes were those suggested by the
flat space case $a=0$: E.g., $b/r=sin¼$, where $$ is what $Ä$ would be
in flat space.)

A similar experiment involves measuring the round-trip travel time for
radio waves from Earth to some reflector (on another planet or an
artificial solar satellite), with and without the Sun near the path of the
waves.  Now, instead of $dr/dÄ$ we want, in units $b=1$
$$ {dr\over dt} = \left(1 -{2a\over r}\right)
	å{1 -{1\over r^2} +{2a\over r^3}} $$
$$ \li{ Üâdt &® {rdr\over å{r^2-1}} +2a{dr\over å{r^2-1}} \cr
	&= d\left[ å{r^2-1} +2a¼cosh^{-1}r\right] \cr
	&= d\left[ å{r^2-b^2} +2GM¼cosh^{-1}{r\over b}\right] \cr } $$
putting the $b$'s back.  
The first term (which is actually bigger) is the nongravitational piece (so we examine only the rest); it is the length of the side of a triangle whose other side has length $b$ and whose hypotenuse has length $r$. 
We have neglected the $a$ correction inside the square root in the original, exact expression: It can be estimated by (1) noting the argument of the square root is exactly 0 at $r_{min}$, and (2) looking at $d(rå...)$, and noting its deviation from the exact result goes as $a/r^3$ times the usual, which is less than $a/b^3$, giving a contribution of order $2GMb/r_{max}$, and thus negligible.

For simplicity we assume both orbits are approximately circular, so $r_{Earth}$ and
$r_{reflector}$ are fixed (at least for the duration of the experiment); the change in $b$ then comes from those radii differing from each other, so they revolve around the sun at different rates. 
We then integrate from $r=r_{min}®b$ to
$r=r_{Earth}$, add the integral from $r=r_{min}$ to $r=r_{reflector}$,
multiply by 2 for the round trip, and throw in a factor to convert to the
proper time $s$ of the observer (which turns out to have a negligible
effect to this order in $a$).  This result is then compared to the same
measurement when both observer and reflector have revolved further
about the Sun, so $b$ changes significantly (but not $r_{Earth}$ nor
$r_{reflector}$).  
For $xù1$, $cosh^{-1}x®ln(2x)$, so for
$bør_{Earth}$ and $r_{reflector}$ we find  
 $$ ës ® -8GMë(ln¼b) $$

Ü7. Black holes

For physical massive bodies the Schwarzschild solution applies only
outside the body, where $T_{ab}=0$.  The form of the solution inside
the body depends on the distribution of matter, which is determined
by its dynamics.  Generally the surface of the body is at $rùGM$, but we
can try to find a solution corresponding to a point mass by extending
the coordinates as far as possible, till the curvature components
$R_{ab}{}^{cd}$ blow up.  The Schwarzschild metric is singular at
$r=2GM$.  In fact, $r$ and $t$ switch their roles as space and time
coordinates there.  There is no corresponding singularity there in the
curvatures, which are $¾r^{-3}$.  This unphysical singularity can be
eliminated by first making the coordinate transformation, for $r>2GM$,
 $$ r' = Çdr \left(1-{2GM\over r}\right)^{-1}
	= r +2GM¼ln\left({r\over 2GM}-1\right) $$
and then making a second coordinate transformation by rescaling the
``lightcone" coordinates as
$$ r''àt'' = 4GMe^{(r'àt)/4GM} = 4GMå{{r\over 2GM}-1}¼e^{(ràt)/4GM} $$
The result is the ``Kruskal-Szekeres (`Sack-er-ash') coordinates"
$$ -ds^2 = {2GM\over r}e^{-r/2GM}(-dt''{}^2+dr''{}^2)
	+r^2(dÏ^2+sin^2ϼdÄ^2) $$
where $r(r'',t'')$ is defined by
$$ r''{}^2-t''{}^2 = (4GM)^2\left({r\over 2GM}-1\right)e^{r/2GM} $$
This can now be extended past $r=2GM$ down to the physical
singularity at $r=0$.

The complete space now looks like (plotting just $r''$ and $t''$):

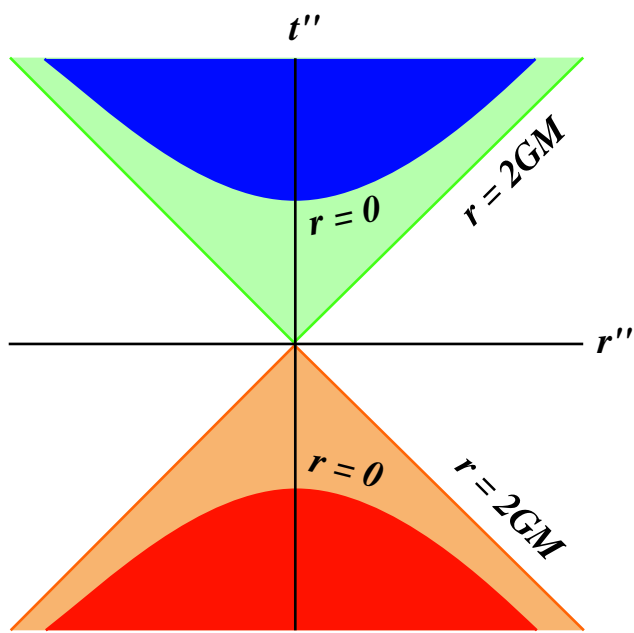
$$ \fig{hole} $$

\noindent In this diagram lines at 45${}^\circ$ to the axes represent radial
lightlike geodesics.  Since nothing travels faster than light, this indicates
the allowed paths of physical objects.  Curves of fixed $r$ are
hyperbolas:  In particular, the physical singularity is the curve
$t''{}^2-r''{}^2=(4GM)^2$ ($r=0$), while $t''{}^2-r''{}^2=0$ ($r=2GM$) is the
``event horizon" which allows things to go only one way (out from the
bottom half or into the top half), and $r=¥$ is both $r''=à¥$.  Nothing can
communicate between the 2 ``outside worlds" of the left and right
90${}^\circ$ wedges.  In particular, a star which collapses (``gravitational
collapse") inside its ``gravitational radius" $2GM$ is crushed to a
singularity, and the spherically symmetric approximation to this collapse
must be represented by part of the Kruskal-Szekeres solution (outside
the star) by Birkhoff's theorem, patched to another solution inside the
star representing the contribution of the matter (energy) there to the
field equations.  This means using just the top and right 90${}^\circ$
wedges, with parts near the left edge of this modified appropriately.  The
top wedge is called a ``black hole".  (If a situation should exist described
by just the bottom and right wedges, the bottom wedge would be called a
``white hole".)  Similarly, stable stars are described by just the right
wedge, patched to some interior solution.  This right wedge represents
the original Schwarzschild solution in the region $r>2GM$ where its
coordinates are nonsingular.  In that region lines of constant $t$ are just
``straight" radial lines in the Kruskal-Szekeres coordinate system
($r''¾t''$).

Besides the fact that nothing can get out, another interesting feature of
the black hole is that an outside observer never sees something falling in
actually reach the event horizon:  Consider an observer at fixed $r>2GM$
using Schwarzschild coordinates, so his proper time $s¾t$.  Then light
radiating radially from an in-falling object is received later and later, up
till $t=¥$, by the observer as the object approaches the event horizon,
although it takes the object a finite amount of proper time to reach the
event horizon and the physical singularity.  

\x IXC7.1  Apply the methods of subsection IXC3 to the equations of
motion in a Schwarzschild metric of subsection IXC6 for a massive
object falling straight into a black hole (angular momentum $L=0$):
Solve for $r, $ in an appropriate parametrization to show that it takes
a finite proper time to reach the event horizon from any finite $r$
outside it.  (Hint: You can also try using the gauge $v=r$ instead of 1.)

There are also more complicated black-hole solutions with spin and
electric charge.

Another interesting effect of the event horizon is the eventual decay of
the black hole (``Hawking radiation"):  Pair creation can result in a similar
way to that in an electrostatic potential of sufficient strength  (see
exercise IIIB5.1).  Particles are emitted near the event horizon (the edge
of the gravitational barrier), carrying energy off to infinity, while their
antiparticles fall into the singularity.

There are two features of the black hole that are less than desirable: the
existence of singularities indicates a breakdown in the field equations,
and the existence of event horizons results in an ``information loss". 
Both these properties might be avoidable quantum mechanically:  For
example, quantum effects can generate curvature-squared terms in the
effective action, which modify the short-distance behavior of the theory. 
One might think that such short-distance effects would have an effect
only at short distances away from regions of high curvature such as the
singularity, and thus remove the singularities but not the event
horizons.  However, it is possible (and examples of such solutions have
been given) that the prevention of the creation of the singularity in
stellar collapse would eventually result in a reversal of the collapse
(``gravitational bounce"):  The would-be black hole solution is patched to
a would-have-been white hole by short-distance modifications, resulting
in an exploding star that initially resembled a black hole but has no true
event horizon.

Although ``compact" bodies have been observed (e.g., at the center of our galaxy) with masses large enough to be black holes (i.e., too large to be neutron stars), their sizes have not been determined to be as small as their event horizons, although our present knowledge of astrophysics does not provide for an alternative explanation.

\refs
	
£1 H.W. Brinkmann, ÓMath. AnnalenÕ É94 (1925) 119;\\
	I. Robinson, unpublished lectures (1956);\\
	J. H«ely, ÓCompt. Rend. Acad. Sci.Õ É249 (1959) 1867;\\
	A. Peres, \xxxlink{hep-th/0205040}, \PR 3 (1959) 571;\\
	J. Ehlers and W. Kundt, Exact solutions of gravitational field equations,
	in ÓGravitation: An introduction to current researchÕ, ed. L. Witten
	(Wiley, 1962) p. 49:\\
	nonlinear gravity waves.
 £2 J.F. Pleba«nski, ÓJ. Math. Phys.Õ É16 (1975) 2395:\\
	reduction of self-dual metric to single component.
 £3 Siegel, Óloc. cit.Õ (IVC, 2nd ref. 17):\\
	lightcone gauge for self-dual supergravity.
 £4 W. de Sitter, ÓProc. Kon. Ned. Akad. Wet.Õ É19 (1917) 1217, É20 (1917)
	229.
 £5 A. Friedmann, ÓZ. Phys.Õ É10 (1922) 377;\\
	H.P. Robertson, ÓAstrophys. J.Õ É82 (1935) 284, É83 (1936) 187, 257;\\
	A.G. Walker, ÓProc. Lond. Math. Soc.Õ É42 (1936) 90:\\
	cosmology in general relativity.
 £6 Particle Data Group, Óloc. cit.Õ (IC):\\
	astrophysical ``constants".
 £7 K. Schwarzschild, ÓSitz. Preuss. Akad. Wiss. Berlin, Math.-phys. Kl.Õ
	(1916) 189.
 £8 A.S. Eddington, ÓNatureÕ É113 (1924) 192;\\
	D. Finkelstein, ÓPhys. Rev.Õ É110 (1958) 965.
 £9 M.D. Kruskal, ÓPhys. Rev.Õ É119 (1960) 1743;\\
	G. Szekeres, ÓPubl. Mat. DebrecenÕ É7 (1960) 285.
 £10 S.W. Hawking, ÓComm. Math. Phys.Õ É43 (1975) 199.
 £11 V.P. Frolov and G.A. Vilkoviskii, \PL 106B (1981) 307:\\
	possibility that quantum corrections to gravity eliminate black holes.
 £12 R. Narayan, Black holes in astrophysics, \xxxlink{gr-qc/0506078}:\\
	review of recent observational status of black holes.

\unrefs

ÚX. SUPERGRAVITY

In the previous chapter we studied the symmetry principles behind
general relativity; now we add supersymmetry to the picture.
Supergravity is a fundamental part of many of the applications of
supersymmetry.  

Û4 A. SUPERSPACE

We first need to understand the ``geometry" associated with local
supersymmetry.

Ü1. Covariant derivatives

In subsection IVC3 we discussed superspace covariant derivatives for
super Yang-Mills.  Similar methods can be applied to supergravity, the
theory of the graviton (spin 2) and gravitino (spin 3/2).  In that case we
want to gauge the complete (unbroken) global symmetry of the theory: 
Besides the obvious Poincar«e and supersymmetry, there is also the axial
U(1) (``R") symmetry that transforms the spin-3/2 field.  (The best we
might have expected is superconformal symmetry, which also has
conformal boosts and scale, but which are broken by the vacuum just as
in ordinary gravity, and S-supersymmetry, which is also broken because
it's the square root of conformal boosts.)  It is introduced in the same
way as local Lorentz invariance in ordinary gravity, and acts on flat spinor
indices (but cancels on vector indices).  We therefore want to gauge the
translations $»_M$ (which have been generalized naturally to superspace
from $»_m$ appearing in ordinary gravity to include supersymmetry), the
Lorentz generators $M_{Œº}$, $M_{ÀŒÀº}$ of ordinary gravity, and the
(second-quantized) ÓhermitianÕ U(1) generator $Y$, defined to act on the
covariant derivatives as
$$ [Y,á_Œ] = -üá_Œ,ââ[Y,Ñá_{ÀŒ}] = üÑá_{ÀŒ},ââ[Y,á_a] = 0 $$
 We now use the ``$Ñ{\phantom M}$" to refer to hermitian conjugation
without reordering, i.e., keeping the partial derivatives and other
generators on the right.  (As in ordinary gravity, transformations are not
truly unitary, and covariant derivatives truly hermitian, because of
ordering.)  

Then the gauge parameter, covariant derivative, and field strengths are
expanded over these generators, as in ordinary gravity:
$$ K = K^M »_M +üK^{Œº}M_{ºŒ} +üK^{ÀŒÀº}M_{ÀºÀŒ} +iK_{-1} Y $$
$$ á_A = E_A{}^M »_M +ü¯_A{}^{º©}M_{©º} +ü¯_A{}^{ÀºÀ©}M_{À©Àº} +iA_A Y $$
$$ [á_A,á_BÕ = T_{AB}{}^C á_C +üR_{AB}{}^{©¶}M_{¶©}
	+üR_{AB}{}^{À©À¶}M_{À¶À©} +iF_{AB}Y $$
 ($E_A{}^M$ is known as the ``supervierbein" or ``vielbein".) 
Alternatively, we can write the $M$ and $Y$ terms collectively as
$üK^{AB}M_{BA}$ (and similarly for the covariant derivative and field
strengths), where $M_{AB}$ are the generators of OSp(3,1|4), by
algebraically constraining $K^{AB}$ to contain just the appropriate pieces
(and relating $K^{ab}$ to $K^{μ}$ in the usual way).  Also, the
shorthand $K^I M_I$ now includes Lorentz and U(1) terms.

\x XA1.1  Use the definition in the above commutation relations to
express the torsion $T_{AB}{}^C$ directly in terms of the structure
functions $C_{AB}{}^C$, Lorentz connection $¯_A{}^{º©}$, and U(1)
connection $A_A$.

The constraints in supergravity are a combination of the kinds used in
ordinary gravity and super Yang-Mills: those that (1) define the vector
derivative in terms of the spinor ones
$$ -iá_{ŒÀº} = Óá_Œ,Ñá_{Àº}Õ $$
 (2) define the spinor (Lorentz and R) connections
$$ T_{Œº}{}^© = T_{Œ,º(Àº}{}^{ºÀ©)} = T_{Œb}{}^b = 0 $$
 and (3) allow the existence of chiral (scalar) superfields
$$ á_Œ Ðì = 0âÜâÓá_Œ,á_ºÕÐì = 0ââ(Yì = 0) $$
 (The first two constraints imply the generalization of this chirality
condition to $Y±0$ and chiral superfields with undotted indices, like the
Yang-Mills field strength.)

\x XA1.2  Rewrite the first and last set of constraints directly in terms of
field strengths.

The explicit solution of all these constraints is a bit messy, but we will
need only a certain subset of them to find the prepotentials and
supergravity action.  The form of the solution is a generalization of super
Yang-Mills in a way similar to how general relativity generalizes ordinary
Yang-Mills.  In particular, just as the vierbein $e_a=e_a{}^m »_m$ is a
generalization of the Yang-Mills vector $A_a$ to describe gauging of the
translations, the generalization of the super Yang-Mills prepotential $V$
to supergravity is $H=H^m(-i)»_m$, which appears in an exponential $e^H$
just as $V$ appears as $e^V$:  The chirality-preserving constraints,
expressed explicitly in terms of the vielbein, is
$$ ÓE_Œ,E_ºÕ = C_{Œº}{}^© E_© $$
 If the commutator vanished like the Yang-Mills case, $E_Œ$ would be
partial derivatives on some complex two-dimensional subspace, the usual
$»_Œ$ up to some complex (super)coordinate transformation, as for
$d_Œ$ in flat superspace.  However, the fact that their algebra still closes
means they still generate translations within such a subspace, and are
thus linear combinations of such partial derivatives:
$$ E_Œ = ÆN_Œ{}^µ öE_µ,ââöE_µ = e^{-¯}»_µ e^¯,â⯠= ¯^M(-i)»_M $$
 where we have separated the matrix coefficient into a local complex
scale (scale $¢$ U(1)) $Æ$ and a local Lorentz transformation $N_Œ{}^µ$. 
For most purposes we will find it convenient to fix all these invariances
by choosing the gauge
$$ ÆN_Œ{}^µ = ¶_Œ^µâÜâE_Œ = öE_Œ $$

As for Yang-Mills, solution of the chirality condition introduces a new,
chiral gauge invariance:
$$ e^{¯'} = e^{iÐñ}e^¯ e^{-iK};ââñ = ñ^M(-i)»_M,âK = K^M(-i)»_M $$
$$ [л_{Àµ},ñ] ¾ л_{ÀÃ}âÜâл_{Àµ}ñ^m = л_{Àµ}ñ^µ = 0 $$
 where $ñ^{Àµ}$ is not chiral, since it generates terms in the
transformation law of $E_{˵}$ that can be canceled by including
$л_{Àµ}ñ^{ÀÃ}$ terms in the transformation law of $ÐÆÐN_{ÀŒ}{}^{Àµ}$.  This
means we can use $ñ^{Àµ}$ and $K^µ$ to gauge
$$ ¯^µ = ¯^{Àµ} = 0âÜ⯠= ¯^m(-i)»_mâÜâöE_µ = »_µ +öE_µ{}^m »_m $$
 where $öE_µ{}^m=-i»_µ ¯^m+...$ from expanding the exponentials as
multiple commutators.  We can again transform to a chiral
representation, and work in terms of
$$ e^U = e^¯ e^{Я},ââe^{U'} = e^{iÐñ}e^U e^{-iñ};ââ
	öE_{Àµ} = »_{Àµ},ââöE_µ = e^{-U}»_µ e^U $$
 where $U$ now generalizes the constant $ÒUÔ=Ï^µ ÐÏ^{Àµ}(-i)»_{µÀµ}$ used
in flat superspace.  Also as for Yang-Mills, the usual local component
transformations (now for coordinate, supersymmetry, scale, U(1), and
S-supersymmetry) reappear in the chiral parameters $ñ^m$ and $ñ^µ$.

For a component analysis, we look at the linearized transformation (see
exercise IVC4.3)
$$ ¶U^m ® i(Ðñ-ñ)^m -iü[ÒUÔ,Ðñ+ñ]^m $$
$$ = i(Ðñ-ñ)^m -üÏ^à ÐÏ^{ÀÃ}»_{ÃÀÃ}(Ðñ+ñ)^m +(Ï^µ Ðñ^{Àµ} -ÐÏ^{Àµ}ñ^µ)
	+iü(ÐÏ^2 Ï^à »_Ã{}^{Àµ} ñ^µ -Ï^2 ÐÏ^{ÀÃ}»^µ{}_{ÀÃ} Ðñ^{Àµ}) $$
 where for $ñ$ we use as independent just the chiral parameters $ñ^m$
and $ñ^µ$; the nonchiral $ñ^{Àµ}$, having already been used to gauge
away $U^µ$, is now fixed in terms of the others as
$$ ñ^{Àµ} = e^{-U}Ðñ^{Àµ}e^U $$
 to maintain $U^µ=0$.  The first term in the transformation tells us that
the surviving component fields are the same as for super Yang-Mills, with
``$m$" as the group index:
$$ U^m = 
	e_a{}^m (ÏÐÏ), Æ_Œ{}^m (ÐÏ^2 Ï), ÐÆ_{ÀŒ}{}^m (Ï^2 ÐÏ), A^m (Ï^2 ÐÏ^2) $$
 The second term in the transformation law gives $¶e_a{}^m®-»_a Â^m$
from $Ðñ^m|=ñ^m|=Â^m$.  Then $ñ^µ$ contains the rest of the gauge
parameters:
$$ \li{ ñ^µ & = ·^µ, a+ib (Ï), Â_Ã{}^µ (Ï), ½^µ (Ï^2) \cr
 & = \hbox{supersymmetry, scale +iU(1), Lorentz, S-supersymmetry}\cr} $$
 The third term in the transformation law then shows scale and Lorentz
gauge away pieces of the vierbein, as usual, while S-supersymmetry
gauges away the trace of $Æ_Œ{}^m$.  It also forces $ñ^m$ to include
$з^{Àµ}$ at order $Ï$ to maintain the gauge; we then see that $Æ_Œ{}^m$
is the gauge field for supersymmetry, with contributions from the second
and fourth terms.  Finally, the fourth term also shows that $A^m$ is the
gauge field for U(1).  The resulting component content is that of
``conformal supergravity", which will be transformed later to ordinary
supergravity through a compensator superfield.

For perturbation theory, or comparison with global supersymmetry, we
should expand about ``flat" superspace (which is nontrivial because of
nonvanishing torsion $T_{ŒÀº}{}^c$ in empty superspace).  We then modify
the chiral representation:
$$ e^¯ e^{Я} = e^U = e^{ÒUÔ/2}e^H e^{ÒUÔ/2}âÜâ
	 öE_{Àµ} = d_{Àµ},ââöE_µ = e^{-H}d_µ e^H $$
 We now expand all derivatives over the covariant derivatives $d_M$ of
global supersymmetry (constructed from $ÒUÔ$ as before)
$$ H = H^M(-i)d_M,ââE_A = E_A{}^M d_M $$
 instead of over partial derivatives $»_M$, which is just a change of
basis.  This also modifies the description of the $ñ$ gauge parameters:
$$ e^{H'} = e^{iÐñ}e^U e^{-iñ};ââñ = ñ^M(-i)d_M,âK = K^M(-i)d_M $$
$$ [Ðd_{Àµ},ñ] ¾ Ðd_{ÀÃ}âÜâ
	ñ^µ d_µ +ñ^m »_m = üÓÐd^{ÀÃ},[Ðd_{ÀÃ},L^µ d_µ]Õ $$
$$ Üâñ^µ = Ðd^2 L^µ,ââñ^{µÀµ} = iÐd^{Àµ}L^µ $$
 in terms of a new parameter $L^µ$.  From this we find the linearized
transformation law
$$ ¶H^{µÀµ} ® d^µ ÐL^{Àµ} -Ðd^{Àµ}L^µ $$

\x XA1.3  Expand this transformation law in components, and compare
with the previous analysis.

Besides chirality, we also need a certain combination of the other
constraints:
$$ 0 = T_{Œb}{}^b -T_{ŒÀº}{}^{Àº} = (-1)^B C_{ŒB}{}^B +C_{Œº}{}^º -iA_Œ $$
 where the $A$ term comes from the contribution $üiA_Œ ¶_{Àº}^{À©}$ to
$T_{ŒÀº}{}^{À©}$.  (See exercise XA1.1.)  Using the gauge $E_Œ=öE_Œ$
without loss of generality, we then find (comparing similar
manipulations in subsection IXA2)
$$ -iA_Œ = E»_M E^{-1}E_Œ{}^M = E^{-1}\onÁE_Œ E $$
 where the backwards arrow on $\onÁE_Œ=E_Œ{}^M\onÁ»_M$ means all
derivatives act on everything to the left (see subsection IA2), and
$$ E ­ sdet¼E_A{}^M $$
 (The superdeterminant was defined in subsection IIC3.)

We now need the general identity, for any function $A$ and first-order
differential operator $B$,
$$ \li{ Ae^{\onÁB} & = (1Ée^{\onÁB}e^{-\onÁB})Ae^{\onÁB}
	= (1Ée^{\onÁB})(e^{-\onÁB}Ae^{\onÁB})
	= (1Ée^{\onÁB})(e^B Ae^{-B}) \cr
	& = (1Ée^{\onÁB})(e^B A) \cr} $$
$$ Üâ1 = (1Ée^{-\onÁB})e^{\onÁB} = (1Ée^{\onÁB})[e^B (1Ée^{-\onÁB})] $$
 The final result in the gauge $Æ=1$ ($N_Œ{}^µ$ is trivially restored) is
then
$$ iA_Œ = E_Œ T,ââe^T = E(1Ée^{-\onÁ¯}) $$

This can be used to solve chirality conditions on matter fields:  In this
gauge, we have
$$ Yì = yìâÜâ0 = Ñá_{ÀŒ}ì = (ÐE_{ÀŒ} +iyÐA_{ÀŒ})ì $$
$$ Üâì = e^{yÐT}e^{Я}Ä,ââл_{Àµ}Ä = 0 $$
 Again as for Yang-Mills, the chiral-representation field $Ä$ transforms
under only the $ñ$ transformations:
$$ Ä' = (1Ée^{i\onÁñ})^y e^{iñ}Ä,ââñ = -i(ñ^m »_m +ñ^µ »_µ) $$
 Thus, scalars in the real representation become densities in the chiral
representation (except for $y=0$).  In particular, we have for the special
case
$$ y = 1âÜâÄ' = Ä e^{i\onÁñ}âÜâ¶Çdx¼d^2 Ï¼Ä = 0 $$
 which will prove useful later for chiral integration.  For now, we note
that such a chiral scalar, with $y±0$, can be seen to compensate from the
$»_µ ñ^µ$ term:  This term allows $U^m$ to eat the complex ``physical"
scalar and spinor, fixing scale, U(1), and S-supersymmetry, while the
complex auxiliary scalar survives, along with coordinate, Lorentz, and
supersymmetry invariance.  We also note that for perturbation about flat
superspace we have
$$ i(1É\onÁñ) = »_m ñ^m -d_µ ñ^µ = -Ðd^2 d_µ L^µ $$

\x XA1.4  Show that preservation of the chirality of $Ä$ implies the
previous chirality conditions on $ñ$.  Thus, as for nonsupersymmetric or
nongravitational theories, the gauge group follows more simply from
starting with matter representations.

We also note that in the gauge $Æ=1$, and also $¯^µ=¯^{Àµ}=0$, the
superdeterminant is simply (see subsection IIC3)
$$ E^{-1} = det(E_m{}^a) $$
 where $E_m{}^a$ is a component of $E_M{}^A$ (not the inverse of
$E_a{}^m$).

Ü2. Field strengths

These constraints can be completely solved for all the field strengths. 
Alternatively, we can impose them, together with the Bianchi identities
(Jacobi identities of the covariant derivatives), to find a smaller set of
algebraically independent field strengths, and the differential equations
that relate them.  The method is analogous to the case of super
Yang-Mills treated in subsection IVC3.  We begin with the constraints
analogous to the Yang-Mills ones:
$$ Óá_Œ,Ñá_{Àº}Õ = -iá_{ŒÀº},ââÓá_Œ,á_ºÕ = R_{Œº}{}^I M_I $$
 (where the latter will simplify from later results).  From just the latter,
we find
$$ [á_{(Œ},Óá_º,á_{©)}Õ] = 0âÜâR_{(Œº©)}{}^¶ -iü¶_{(Œ}{}^¶ F_{º©)}
	= á_{(Œ}R_{º©)}{}^I = 0 $$
 Using both constraints, we also have 
$$ [á_{(Œ},Óá_{º)},Ñá_{À©}Õ] +[Ñá_{À©},Óá_Œ,á_ºÕ] = 0 $$
$$ Üâ[á_{(Œ},á_{º)À©}] = 
	iR_{ŒºÀ©}{}^{À¶}Ñá_{À¶} -üF_{Œº}Ñá_{À©} -i(Ñá_{À©}R_{Œº}{}^I)M_I $$
$$ Üâ[á_Œ,á_{ºÀ©}] = -iC_{Œº}Ñ\W_{À©}
	+üiR_{ŒºÀ©}{}^{À¶}Ñá_{À¶} -\f14 F_{Œº}Ñá_{À©} -üi(Ñá_{À©}R_{Œº}{}^I)M_I $$
 for some operator $Ñ\W_{À©}=Ñ\W_{À©}{}^A á_A+Ñ\W_{À©}{}^I M_I$.  So far the
exercise has been analogous to the super Yang-Mills case (where the
extra ``$i$" in the definition of $Ñ\W$ is due to our use of antihermitian
generators, except for $Y$).  Now we impose the remaining constraints,
which can be combined conveniently as
$$ 0 = T_{Œ,ºÀ©}{}^{ºÀ¶} = -iC_{Œº}Ñ\W_{À©}{}^{ºÀ¶}âÜâ
	Ñ\W_{ÀŒ} = Ñ\W_{ÀŒ}{}^º á_º +Ñ\W_{ÀŒ}{}^{Àº}Ñá_{Àº} +Ñ\W_{ÀŒ}{}^I M_I $$

Following again the steps for Yang-Mills, we analyze the
next-higher-dimension Jacobis, beginning with
$$ 0 = Óá_{(Œ},[á_{º)},á_{©À¶}]Õ +[á_{©À¶},Óá_Œ,á_ºÕ]
	= iC_{©(Œ}Óá_{º)},Ñ\W_{À¶}Õ +ë_{©ŒºÀ¶} $$
$$ \li{ ë_{©ŒºÀ¶} = & üi(á_{(Œ}R_{º)©À¶}{}^{À·})Ñá_{À·} 
		-\f14(á_{(Œ}F_{º)©})Ñá_{À¶} -üi(á_{(Œ}Ñá_{À¶}R_{º)©}{}^I)M_I \cr
	& +üR_{©(ŒÀ¶}{}^{À·}á_{º)À·} +\f14 iF_{©(Œ}á_{º)À¶}
		-üi(Ñá_{À¶}R_{©(Œº)}{}^¶)á_¶ -\f14 (Ñá_{À¶}F_{©(Œ})á_{º)} \cr
	& +(á_{©À¶}R_{Œº}{}^I)M_I -R_{Œº©}{}^¶ á_{¶À©} 
		-R_{ŒºÀ¶}{}^{À·}á_{©À·} \cr} $$
 By inspection, or applying the previous Jacobis, we see
$$ ë_{(©Œº)À¶} = 0âÜâë_{©ŒºÀ¶} = C_{©(Œ}ë_{º)À¶},ââ
	ë_{ŒÀº} = -\f13 ë^©{}_{©ŒÀº} $$
 automatically, so the only new information comes from the trace of this
Jacobi,
$$ Óá_Œ,Ñ\W_{Àº}Õ = ië_{ŒÀº} $$
 Evaluating $Óá,Ñ\WÕ$ in terms of its pieces, we find
$$ R_{Œº}{}^{À©À¶} = F_{Œº} = 0,ââR_{Œº}{}^{©¶} = ¶_{(Œ}^© ¶_{º)}^¶ ÐB,ââ
	Ñ\W_{ÀŒ}{}^{Àº} = -ÐB¶_{ÀŒ}^{Àº} $$
$$ ÑW_{ÀŒ} = Ñá_{ÀŒ}ÐB +á^º Ñ\W_{ÀŒº},ââ
	Ñ\W_{ÀŒ}{}^{º©} = -üá^{(º}Ñ\W_{ÀŒ}{}^{©)} $$
$$ á_Œ ÑW_{Àº} = á_Œ Ñ\W_{Àº}{}^{À©À¶} = 0,ââ
	á_Œ Ñ\W_{Àº}{}^{©¶} = -¶_Œ^{(©}(Ñ\W_{Àº}{}^{¶)} +üiá^{¶)}{}_{Àº})ÐB $$
 where $W_Œ$ is the $Y$ part of $\W_Œ$ ($=W_Œ iY +...$).

\x XA2.1  Show that
$$ Óá_Œ,á_ºÕ = ÐBM_{Œº},ââá_{[Œ}á_º á_{©]} = 0âÜâ
	á_Œ (á^2 +ÐB) = -üÐB á^º M_{ºŒ} $$
 and thus $Ñá^2 +B$ gives a chiral superfield when acting on any superfield
without dotted indices.

For the other Jacobi of this dimension, we have
$$ 0 = [á_{ŒÀŒ},Óá_º,Ñá_{À©}Õ] +Óá_º,[Ñá_{À©},á_{ŒÀŒ}]Õ 
	+ÓÑá_{À©},[á_º,á_{ŒÀŒ}]Õ $$
$$ = -i[á_{ŒÀŒ},á_{ºÀ©}] -iÓá_º,C_{À©ÀŒ}\W_Œ +ü(á_Œ B)ÑM_{À©ÀŒ}Õ
	-iÓÑá_{À©},C_{ºŒ}Ñ\W_{ÀŒ} +ü(Ñá_{ÀŒ}ÐB)M_{ºŒ}Õ $$
$$ = -iC_{ÀŒÀ©}[f_{Œº} -Óá_º,\W_ŒÕ +ü(Ñá^2 ÐB)M_{Œº}] -h.c. $$
$$ Üâf_{Œº} = üÓá_{(Œ},\W_{º)}Õ -ü(Ñá^2 ÐB)M_{Œº},ââ
	Óá^Œ,\W_ŒÕ +ÓÑá^{ÀŒ},Ñ\W_{ÀŒ}Õ = 0 $$
 (Here ``$h.c.$" means ``hermitian conjugate" without the reordering,
which would generate non-operator terms.) 

Evaluating $Óá,\WÕ$ in terms of its pieces, and combining with the results
of the previous Jacobi, we obtain the final result:

 \Boxeq{ $$ ÓÑá_{ÀŒ},Ñá_{Àº}Õ = BÑM_{ÀŒÀº},ââÓá_Œ,Ñá_{Àº}Õ = -iá_{ŒÀº} $$
$$ [Ñá_{ÀŒ}, -iá_{ºÀº}] = C_{ÀºÀŒ}\W_º -ü(á_º B)ÑM_{ÀŒÀº},ââ
	[-iá_{ŒÀŒ},-iá_{ºÀº}] = C_{ÀºÀŒ}f_{Œº} -h.c. $$
$$ \li{ \W_Œ ={}& -Bá_Œ -G_Œ{}^{Àº}Ñá_{Àº} +ü(Ñá{}^{Àº}G_Œ{}^{À©})ÑM_{ÀºÀ©} 
		+üW_Œ{}^{º©}M_{©º} +iW_Œ Y +i\f16 W^º M_{ºŒ} \cr
	f_{Œº} ={}& iüG_{(Œ}{}^{À©}á_{º)À©} -ü(á_{(Œ}B +i\f13 W_{(Œ})á_{º)}
		+W_{Œº}{}^© á_© -ü(á_{(Œ}G_{º)}{}^{À©})Ñá_{À©} \cr
	& -(üÑá{}^2 ÐB +BÐB +\f1{12}iá^© W_©)M_{Œº} 
		-i\f1{16}[(á_{(Œ}{}^{À¶}G_{©)À¶})M_º{}^© +Œªº] \cr
	& +üW_{Œº}{}^{©¶}M_{¶©} 
		+\f14(á_{(Œ}Ñá{}^{À©}G_{º)}{}^{À¶})ÑM_{À¶À©} +iü(á_{(Œ}W_{º)})Y \cr
	W_{Œº©¶} ={}& \f1{4!}á_{(Œ}W_{º©¶)} \cr} $$ }\\
 The ``reduced tensors" $B,G_a,W_Œ,W_{Œº©}$ satisfy the ``reduced
Bianchi identities"

 \Boxeq{ \vskip-.1in
	$$ G_a = ÐG_a,ââÑá_{ÀŒ}B = Ñá_{ÀŒ}W_Œ = Ñá_{ÀŒ}W_{Œº©} = 0,ââ
	Ñá{}^{ÀŒ}G_{ŒÀŒ} = á_Œ B -iW_Œ $$
$$ á^Œ W_{Œº©} -i\f13 á_{(º}W_{©)} = -iüá_{(º}{}^{ÀŒ}G_{©)ÀŒ},ââ
	á^Œ W_Œ +Ñá{}^{ÀŒ}ÑW_{ÀŒ} = 0 $$ }

\noindent  Note that $B$ or $W_Œ$ may vanish in certain gauges, for
reasons to be explained in subsection XA4.

\x XA2.2  In IXA4 we saw that integrals of total covariant derivatives
vanished in curved space by virtue of the identity $T_{ab}{}^b=0$.  Show
that these torsions satisfy the superpace generalization
$$ (-1)^B T_{AB}{}^B = 0 $$

\x XA2.3  Using the expression for $·_{abcd}$ in terms of spinor indices
from subsection IIA5, show
$$ T_{bcd} = G^a ·_{abcd} $$
 Thus $G_a$ is an axial vector.

\x XA2.4  By hermitian conjugation, find the commutators not written
explicitly above, and show the result is essentially the same as switching
dotted and undotted indices (and similarly for bars), except that $G_a$,
$Y$, and $W^Œ$ (and $ÑW^{ÀŒ}$) get extra minus signs.  This illustrates
CP invariance, and the fact that $G_a$ is an axial vector, while $Y$ is a
pseudoscalar (and similarly for $W^Œ$).

\x XA2.5  Derive the Bianchi identities in the absence of constraints, in
terms of the torsions and curvatures (as follow from the Jacobi identity):
$$ á_{[A}T_{BC)}{}^D -T_{[AB|}{}^E T_{E|C)}{}^D = R_{[ABC)}{}^D $$
$$ á_{[A}R_{BC)}{}^I -T_{[AB|}{}^E R_{E|C)}{}^I = 0 $$

\x XA2.6  Show that in Ó4-componentÕ notation we can write
$$ T_{Œº}{}^c = -i©^c_{Œº};ââT_{aº}{}^© = ©_{aº¶}G^{¶©},ââ
	G^{Œº} = -G^{ºŒ},ââá_º G^{ºŒ} = W^Œ $$
 This gives another way to see the result of exercise XA2.2.  Show that
this expression for $G^{Œº}=(G^a,B,ÐB)$ gives it an interpretation as an
SO(3,3) 6-vector in SL(4) notation (see subsection IC5).

Ü3. Compensators

Just as in ordinary gravity (see subsection IXA7), compensators for scale transformations can
be introduced, but for supergravity the compensator should be a
supersymmetric multiplet.  The simplest choice is the chiral scalar
superfield $ì$ considered earlier:  Its complex ``physical" scalar
(scalar +$i$ pseudoscalar) compensates local scale (the real part) and U(1)
(the imaginary part), its spinor compensates local S-supersymmetry, and
its auxiliary complex scalar appears as one of the auxiliary fields of
supergravity.

Compensators are much more important in supergravity than in ordinary
gravity:  Almost any flat space action can be coupled to gravity by the
minimal coupling prescription --- replacing derivatives with covariant
ones, and throwing a factor of ${\bf e}^{-1}$ in for the measure.  In
supergravity this is not the case:  As we'll see in section XB, we
have both integrals over all superspace, which use $E^{-1}$, but also
integrals over chiral superspace (for integrating chiral superfields), which
instead use $ì$ for the measure.  The minimal coupling procedure is then:
\item{(1)} Use $ì$ (and $Ðì$) to make a flat-superspace action superconformally
invariant,   
\item{(2)} replace the flat derivatives $d_A$ with the curved ones $á_A$, and   
\item{(3)} throw in the measure factors appropriate for the
integrals.  

\noindent (The last two steps couple conformal supergravity to a globally
conformally invariant theory.)

Another compensator that is commonly used is the ``tensor multiplet". 
(This is sometimes confused in the literature with the ``(complex) linear
multiplet", another version of the scalar multiplet with no gauge fields
whatsoever.)  Treated as a matter multiplet, it has the same physical
content as the scalar multiplet, but the pseudoscalar is replaced with a
second-rank antisymmetric tensor gauge field,
$$ ¶B_{mn} = »_{[m}Â_{n]} $$
 To make things simpler, let's look at flat space.  We first note that this
tensor is ``dual" to a pseudoscalar in the sense of switching field
equations and constraints of the field strength (see exercises IIB2.1 and
VIIIA7.2):  For the free fields,
$$ F_a = »_a \ÄâÜâ»_{[a}F_{b]} = 0,ââ
	G_a = ü·_{abcd}»^b B^{cd}âÜâ»^a G_a = 0 $$
 with the field equations following from ``self-duality" under $FªG$:
$$ F_a = G_aâÜâ»^a F_a = »_{[a}G_{b]} = 0 $$
 Since the theory of $B_{ab}$ must be described in terms of $G_a$ alone
(because of gauge invariance), no renormalizable self-interactions are
allowed; thus, this field is of little interest in quantum field theory
outside of supergravity.  In terms of the scalar, the fact that only the
field strength $F_a$ appears in the field equations means there is the
global symmetry
$$ ¶\Ä = ½ $$
 for constant parameter $½$.  This generalizes to the nonabelian
symmetries of nonlinear $§$ models, resulting in derivative interactions
(again nonrenormalizable) but no potentials.

\x XA3.1  Consider coupling the tensor field to Yang-Mills:  To preserve
the tensor's own gauge symmetry, this coupling must be nonminimal.  To
produce such a coupling, we start with the scalar and duality transform. 
The coupling we choose is another 4D analog to the 2D model we
considered in exercise VIIIA7.2, replacing the pseudoscalar and total
derivative $ü·^{ab}F_{ab}$ with $tr(\f18 ·^{abcd}F_{ab}F_{cd})$.  (In
general dimensions, the dual to a scalar is a rank-D$-$2 antisymmetric
tensor.)  We start with the Lagrangian
$$ L = -\f14 ÄõÄ +ÂÄ\f1{16}tr(·^{abcd}F_{ab}F_{cd}) $$
 for some coupling constant $Â$.  Making use of the Chern-Simons form
$B_{abc}$ of subsection IIIC6 to write $Ä$ in this action only as $»_a Ä$,
write a first-order form of this action and perform a duality
transformation to obtain
$$ L' = \f1{24}~H{}^2,ââ~H_{abc} = ü»_{[a}B_{bc]} +ÂB_{abc} $$
 Find the Yang-Mills gauge transformation of $B_{ab}$.  (Hint:  $~H$ is
gauge invariant.)

The tensor multiplet is described by a chiral spinor gauge field
$$ ¶Ä_Œ = iÐd^2 d_Œ Kââ(K = ÐK) $$
 Duality is then described in terms of the real scalar superfield strength (in
the free case)
$$ F = Ä+ÐÄâÜâÐd^2 d_Œ F = 0,ââ
	G = ü(d_Œ Ä^Œ +Ðd_{ÀŒ}ÐÄ^{ÀŒ})âÜâÐd^2 G = 0 $$
 with the field equation
$$ F = G $$
 ($F_a$ appears at order $ÏÐÏ$ in $G$, and $B_{ab}$ at order $Ï$ in $Ä_Œ$.) 
Again the pseudoscalar has a global symmetry:  In terms of the superfield,
$$ ¶Ä = i½ $$

Now we return to curved space, covariantizing the above with respect to
conformal supergravity.  We now identify the above global symmetry
with the local axial U(1) (R-)symmetry of supergravity.  Thus, the
superfield $G$ does not compensate for this symmetry; it remains as a
symmetry in actions that use this compensator.  In particular, there are
no $Çd^2 Ï$ terms in such theories, except those that are locally
superscale invariant (so the compensator decouples).  The matter tensor
multiplet also differs from the scalar multiplet in that it has no auxiliary
fields (except for the auxiliary components of the gauge field).

Ü4.  Scale gauges

Since all the covariant derivatives are built up from the spinor part of the
vielbein, we define the local superscale transformations for the covariant
derivatives by first defining
$$ E'_Π= LE_Π$$
 where $L$ is a real, unconstrained superfield.  The constraints then imply
$$ á'_Œ = Lá_Œ + 2(á^º L)M_{ºŒ} +6(á_Œ L)Y,ââ
	Ñá'_{ÀŒ} = LÑá_{ÀŒ} +2(Ñá^{Àº}L)ÑM_{ÀºÀŒ} -6(Ñá_{ÀŒ}L)Y $$
 From the anticommutator we find
$$ \li{ -iá'_{ŒÀŒ} = & L^2(-i)á_{ŒÀŒ} +4L(á_Œ L)Ñá_{ÀŒ} +4L(Ñá_{ÀŒ}L)á_Œ \cr
	& +üL^{-2}(á_Œ Ñá^{Àº}L^4)ÑM_{ÀºÀŒ} +üL^{-2}(Ñá_{ÀŒ}á^º L^4)M_{ºŒ}
		-\f32 L^{-2}([á_Œ,Ñá_{ÀŒ}]L^4)Y \cr} $$
 Using the commutation relations, we then can show
$$ B' = L^6(Ñá{}^2 +B)L^{-4},ââW'_Œ = L^3[W_Œ -12i(Ñá{}^2 +B)á_Œ¼ln¼L] $$
$$ G'_{ŒÀŒ} = (2[á_Œ,Ñá_{ÀŒ}] +G_{ŒÀŒ})L^2,ââW'_{Œº©} = L^3 W_{Œº©} $$
 From the way they appear in the commutators we also have that
$$ YG_a = 0,ââYW_Œ = üW_Œ,ââYW_{Œº©} = üW_{Œº©},ââYB = B $$
 From linearization, we see that $B$ and $W^Œ$ pick out exactly the
two irreducible halves of the real scalar superfield $L$:  The ``vector
multiplet" in $W'_Œ$ and the ``scalar multiplet" in $B'$.  (Compare the
vector multiplet field strength and chiral scalar gauge fixing for the
prepotential $V$ as described in subsections IVC4 and VIB9.)  This means
we can completely fix the superscale gauge by the choice
$$ B = W_Π= 0 $$
 as the generalization of the scale gauge in ordinary gravity that fixes the
Ricci scalar to vanish.

\x XA4.1  Derive the superscale transformations by use of the Bianchi
identities:
 ªa  Use the commutation relations of the covariant derivatives
(and the solution to the Jacobi identities) to find all the transformations
above.  Show they imply $E'=L^4 E$.
 ªb  An easier way is to use the ÓreducedÕ Bianchi identities:  Determine
the transformations of the reduced field strengths, up to constants, using
chirality, dimensional analysis, etc., and then solve for the constants by
plugging into the reduced identities.

We then define the scale (and U(1)) transformations of the compensators:
$$ ì' = L^2 ì,ââYì = \f13 ì;ââÑá_{ÀŒ} ì = 0 $$
$$ G' = L^4 G,ââYG = 0;ââ(Ñá{}^2 +B)G = 0,ââG = ÐG $$
 where the scale weights follow from the U(1) weights (vanishing for $G$
by reality) by consistency with the constraints they satisfy.

\x XA4.2  Show that a superfield can be chiral only if it has no dotted
indices.  Then show the relation that any such superfield has between
scale and U(1) weights.

All these transformations can be derived either by consistency with the
constraints, or by using the solution of the constraints:  In terms of the
unconstrained superfields that solve the constraints, the superscale
transformation is trivial:
$$ Æ' = LÆ,ââN'_Œ{}^µ = N_Œ{}^µ,ââ¯' = ¯ $$
 The net result, as for super Yang-Mills, is that all the superficial
transformations of the constrained covariant derivatives are completely
replaced with the new invariances that appear upon solving the
constraints:  $K_{-1}$ and $L$ eliminate $Æ$, $K^{Œº}$ kills $N_Œ{}^µ$, and
$K^M$ reduces $¯^M$ to its real part $U^M$, which transforms only under
$ñ^M$.

\x XA4.3  Rederive $A_Œ$ as in subsection XA1, but in a general gauge, to
find
$$ iA_Œ = E_Œ T,ââe^T = Æ^2 E(1Ée^{-\onÁ¯}) $$
 Show this result gives a superscale transformation for $A_Œ$ that agrees
with the result above.  Show the explicit solution for $ì$ in terms of $Ä$
and $T$ also gives it a superscale transformation that agrees with the
above.

\x XA4.4  Often it is easier to use the solution to the constraints than the
Jacobi identities:
 ªa  Solve for $F_{AB}$ in terms of $A_Œ$, and use the solution for $A_Œ$
from subsection XA1, to derive
$$ W_Œ = -i(Ñá^2 +B)á_Œ (T+ÐT) $$
 and use this to rederive the superscale transformation above.  (Hint: 
Define and use the chiral representation.)  
 ªb  Find an explicit expression for $B$, and use it to rederive its
superscale transformation.  (Hint:  You will need to find $¯_Œ{}^{º©}$
first.  Since $B$ is a scalar, you can choose the Lorenz gauge
$N_Œ{}^µ=¶_Œ^µ$.)

In subsection XA1 we found that a convenient way to simultaneously fix
Lorentz, U(1), and scale gauges was to choose $E_Œ=öE_Œ$.  (However, the
corresponding component invariances reappeared in the chiral gauge
invariances.)  Compensators allow more freedom for gauge fixing: For
example, we can fix the gauge $B=W_Œ=0$ as described above, or we can
fix to 1 the compensator or a ÓphysicalÕ matter multiplet (string gauge, as
for gravity in subsection IXB5):  The possibility of gauges such as $ì=1$
or $G=1$ depends on the existence in the action of such fields, and not on
the details of how they appear (as long as the gauge choice is consistent
with the allowed vacuum values).  In particular, it does not depend on the
signs of their kinetic terms, which is the only thing that determines what
is physical and what is a compensator.  

Note that either $ì=1$ or $G=1$ completely fixes the superscale gauge, in
spite of the constraints on these superfields.  (E.g, $ì=ì'=1ÜL=1$.)  This is
due to the appearance of the U(1) connection:  For example, before
fixing the scale and U(1) gauges the chirality condition on $ì$, rather than
constraining $ì$, actually determines the spinor U(1) connection $A_Œ$:
$$ á_Œ Ðì = (E_Œ -i\f13 A_Œ)Ðì = 0âÜâA_Œ = -3iE_Œ¼ln¼Ðì $$
 (But the chirality of the ratio of two chiral superfields with the same
weights really fixes it to be chiral; in other words, chirality of scalars
makes all but one truly chiral, since the U(1) connection can be
determined only once.)  As a result, the scale ($ìÐì=1$) and U(1) ($ì/Ðì=1$)
gauge choice $ì=1$ determines $W_Œ$:
$$ ì = 1âÜâA_Œ = 0âÜâW_Œ = 0 $$
 Similarly,
$$ G = 1âÜâ0 = (Ñá{}^2 +B)G = B $$

Conversely, we see that whenever one of the two field strengths $B$
and $W^Œ$ is eliminated by a superscale(/U(1)) gauge choice in terms of
one of the two compensators $ì$ and $G$, the other field strength can be
made superscale invariant:  If we introduce the compensator by a
superscale transformation (as for gravity in subsection IXA7),
substituting either
$$ L^{-4} £ Ðìì¼or¼G $$
 in the above transformation laws, we find
$$ ÷B = (Ðìì)^{-3/2}(Ñá^2 +B)Ðìì = ì^{-1/2}Ðì^{-3/2}(Ñá^2 +B)Ðì $$
$$ ~W_Œ = G^{-3/4}[W_Œ +3i(Ñá^2 +B)á_Œ ln¼G] $$
 as locally superscale invariant, where using $Ðìì$ for $~W_Œ$ or $G$ for
$÷B$ yields zero.  We can therefore interpret gauging away the
compensators as gauging them into the field strengths:  We have a choice
of either
$$ ì = 1âÜâW_Œ = 0,ââ÷B = B $$
$$ G = 1âÜâB = 0,ââ~W_Œ = W_Œ $$
 This is analogous to St¬uckelberg gauges (and their nonlinear
generalizations):  One of these two tensors (gauge fields with respect to
superscale) ``eats" the compensator.  However, it differs from
St¬uckelberg in that a second ``gauge field" is completely gauged away.

In fact, we'll see in section XB that the pure supergravity actions
constructed using either of these compensators gives the corresponding
field strength as its field equation:
$$ {¶\over ¶ì}âÜâ÷B = 0 $$
$$ {¶\over ¶ì_Œ}âÜâ~W_Œ = 0 $$
 Thus, either compensator can be used to eliminate both $B$ and $W_Œ$,
one as a field equation and the other as a gauge choice.  This result is
already clear at this point from dimensional analysis and chirality;
similarly, we must have
$$ {¶\over ¶U^m}âÜâ÷G_a = 0 $$
 where $÷G_a$ is the result of applying a superscale transformation to
$G_a$ with whichever of the two compensators is being used in the
action.  This leaves $W_{Œº©}$ as the on-shell field strength.  The analogy
to ordinary gravity is
$$ (R,R_{ab}-üú_{ab}R,W_{abcd}) ª (B/W_Œ,G_a,W_{Œº©}) $$

Although the Ricci tensor must appear in $G_a$, superscale invariance
allows the choice of gauges where the $Ï=0$ component is arbitrary: 
From the above we find the linearized transformations
$$ ¶G_{ŒÀŒ} ® 4[á_Œ,Ñá_{ÀŒ}]L,ââ¶A_{ŒÀŒ} ® -6[á_Œ,Ñá_{ÀŒ}]L $$
 (For purposes of evaluating at $Ï=0$ we can neglect $A_Œ|$ in $¶A_a$.) 
Thus, this axial vector component field can be moved around as
convenient for component expansions.  In $÷G_a$, they appear only in their
invariant combination, $G_a+\f23 A_a$ in this approximation.  In
Ónonsupersymmetric gaugesÕ, we can even gauge $G_a|=0$.

\x XA4.5  Use the Bianchi identities instead of explicit superscale to track
down the axial vector:
 ªa  Use the relation of $W_{Œº©}$ (which is scale covariant) to $W_Œ$
(the field strength for $A_A$) and $G_a$ to show that it is just this
combination $G_a+\f23 A_a$ that appears in $W_{Œº©}$ (as its curl, 
for U(1) invariance).
 ªb  Show that
$$ G = 1âÜâB = 0âÜâá^a G_a = 0 $$
 Thus, in this gauge the axial vector gauge field $A_a$ has been gauged
out of $G_a|$ (although its field strength may appear at higher order: the
gauge $G=1$ doesn't fix U(1)).  What replaces it?  (Hint:  What's in $G$?)

\refs

£1 D.Z. Freedman, P. van Nieuwenhuizen, and S. Ferrara,
	\PRD 13 (1976) 3214:\\
	supergravity.
 £2 V.P. Akulov, D.V. Volkov, and V.A. Soroka, ÓJETP Lett.Õ É22 (1975) 187:\\
	covariant derivatives for supergravity.
 £3 V. Ogievetsky and E. Sokatchev, \NP 124 (1977) 309;\\
	V.P. Akulov, D.V. Volkov, and V.A. Soroka, ÓTheor. Math. Phys.Õ É31
	(1977) 285;\\
	S. Ferrara and B. Zumino, \NP 134 (1978) 301:\\
	linearized off-shell supergravity (including auxiliary fields).
 £4 P. Breitenlohner, \PL 67B (1977) 49; \NP 124 (1977) 500:\\
	off-shell supergravity in components.
 £5 Wess and Zumino; Wess; Óloc. cit.Õ (IVC, ref. 5):\\
	superspace field equations.
 £6 W. Siegel, \pdfklink{Supergravity superfields without a supermetric}%
	{http://www-lib.kek.jp/cgi-bin/kiss_prepri?KN=197802021&OF=4.},
	Harvard preprint HUTP-77/A068 (November 1977);\\
	\pdfklink{The superfield supergravity action}%
	{http://www-lib.kek.jp/cgi-bin/kiss_prepri?KN=197803065&OF=4.}, 
	Harvard preprint HUTP-77/A080 (December 1977);\\
	\pdfklink{A polynomial action for a massive, self-interacting 
	chiral superfield coupled to supergravity}%
	{http://www-lib.kek.jp/cgi-bin/kiss_prepri?KN=197803054&OF=4.}, 
	Harvard preprint HUTP-77/A077 (December 1977);\\
	\pdfklink{A derivation of the supercurrent superfield}%
	{http://www-lib.kek.jp/cgi-bin/kiss_prepri?KN=197804090&OF=4.}, 
	Harvard preprint HUTP-77/A089 (December 1977):\\
	off-shell supergravity in superspace, compensator superfield.
 £7 M. Kaku, P.K. Townsend, and P. van Nieuwenhuizen, \PRD 17 (1978)
	3179:\\
	conformal supergravity in components.
 £8 M.F. Sohnius and P.C. West, \PL 105B (1981) 353:\\
	R-symmetry generator in superspace covariant derivative.
 £9 V.I. Ogievetsky and I.V. Polubarinov, ÓSov. J. Nucl. Phys.Õ É4 (1967)
	156;\\
	K. Hayashi, \PL 44B (1973) 497;\\
	M. Kalb and P. Ramond, \PRD 9 (1974) 2273;\\
	E. Cremmer and J. Scherk, \NP 72 (1974) 117:\\
	antisymmetric tensor gauge field.
 £10 J. Wess, ÓActa Phys. AustriacaÕ É41 (1975) 409:\\
	N=2 supersymmetric tensor multiplet.
 £11 W. Siegel, \PL 85B (1979) 333:\\
	N=1 supersymmetric tensor multiplet.
 £12 H. Nicolai and P.K. Townsend, \PL 98B (1981) 257;\\
	E. Bergshoeff, M. de Roo, B. de Wit, and P. van Nieuwenhuizen,
	\NP 195 (1982) 97;\\
	G.F. Chapline and N.S. Manton, \PL 120B (1983) 105:\\
	Yang-Mills appearing as Chern-Simons contribution to tensor field
	strength.
 £13 P.S. Howe and R.W. Tucker, \PL 80B (1978) 138:\\
	superscale transformations.
 £14 Gates, Grisaru, Ro×cek, and Siegel, Óloc. cit.Õ:\\
	complete superspace treatment of N=1 supergravity.

\unrefs

Û7 B. ACTIONS

Now that we understand the structure of superfields in curved
superspace, we analyze various supergravity theories through their
actions.  We will use several methods for finding and evaluating
supergravity actions.  These actions are significantly more complicated
than those we have encountered previously, and it is difficult to see all
their features simultaneously, so for any particular application we use
the method which best simplifies the property we most need:    
\item{(1)}
Superspace methods are the best for finding general actions and their
symmetry properties, manifesting supersymmetry, using globally
supersymmetric gauges, and performing quantum calculations.    
\item{(2)}
Component methods are useful for comparing actions and other
properties to nonsupersymmetric theories.  Such approaches sometimes
make some use of superspace, but not superspace integration.    
\item{(3)}
Compensators are useful in conjunction with either of these methods, and
can extract many important features and terms in the action with little
more than the results of global supersymmetry.  They reveal useful
broken symmetries, and are the simplest way to analyze the ``superhiggs
effect" (Higgs for local supersymmetry).

Ü1. Integration

The action for supergravity follows from dimensional analysis:  Since the
usual Einstein-Hilbert Lagrangian has dimension +2, as does $Çd^4 Ï$, the
superspace Lagrangian must be dimensionless.  The only covariant
possibility in terms of the potentials ($E_A{}^M$ and $¯_A{}^I$) is thus
$$ S_{SG} = 3Çdx¼d^4 ϼE^{-1} $$
 including a normalization factor that will prove convenient later. 
Introducing the compensator and local scale invariance must also make
the usual action for supergravity look like the kinetic term for the
compensator multiplet, i.e.,
$$ S_{SG,c} = 3Çdx¼d^4 ϼE^{-1}Ðìì $$
 (For simplicity we will restrict ourselves for the most part to the
simplest compensator, the chiral scalar.)  The previous form then
corresponds to the scale gauge $Ðìì=1$; often the scale + U(1) gauge
$Æ=1$ is more convenient.

There should also be a supersymmetrization of the cosmological term. 
This might seem difficult, requiring explicit prepotentials.  However, we
know from our study of de Sitter space that the cosmological term is
basically a statement about the conformal compensator.  Therefore, the
cosmological term for supergravity, in terms of the superconformal
compensator $ì$, should be the supersymmetrization of the
corresponding term in ordinary gravity, a dimensionless self-interaction
for a scalar.  The solution of the chirality condition can be written as
$$ Yì^3 = ì^3âÜâì^3 = Ä^3 e^{\onÁ{Я}}E $$
 in the gauge $Æ=1$.  The result for the cosmological term is then
$$ S_{scosmo} = Çdx¼d^2 ϼE^{-1}ì^3 +h.c. = Çdx¼d^2 ϼÄ^3 +h.c. $$
 independent of scale or U(1) gauge:  As we saw in subsection XA1, this
expression is invariant under $ñ$ transformations, and the integrand
itself is invariant under $K$ and $L$ transformations.

\x XB1.1  Although this final result for the cosmological term in terms of
$Ä$ is locally superscale invariant, the derivation started in the gauge
$Æ=1$.  Generalize the derivation, and the result in terms of $ì$, to
arbitrary gauges (see exercise XA4.3).  

A chiral expression for $S_{SG}$ can be found by similar methods:
$$ S_{SG} = 3Çdx¼d^2 ϼE^{-1}B = 3Çdx¼d^2 ÐϼE^{-1}ÐB $$
 Thus, as for super Yang-Mills, the action can be expressed as a real,
chiral, or antichiral integral.  With the compensator,
$$ S_{SG} = 3Çdx¼d^2 ϼE^{-1}ì(Ñá^2 +B)Ðì = 3Çdx¼d^2 ÐϼE^{-1}Ðì(á^2 +ÐB)ì $$
 or, more generally,
$$ Çdx¼d^4 ϼE^{-1}{\bf L} = Çdx¼d^2 ϼE^{-1}(Ñá^2 +B){\bf L} = 
	Çdx¼d^2 ÐϼE^{-1}(á^2 +ÐB){\bf L} $$
 which is just the naive covariantization of the flat-space result (at least
in the gauge $Æ=1$: see exercise XB1.1).  Clearly this method generalizes
to coupling to other multiplets, and allows both $Çd^2 Ï$ and $Çd^4 Ï$
integrals to be generalized to curved superspace.  In fact, the analysis of
the compensator is much simpler than that of the conformal supergravity
that couples to it to produce ordinary supergravity:  Just as in ordinary
gravity, where use of just the compensator allowed us to study certain
interesting solutions in gravity, namely de Sitter space and cosomology,
some properties of supergravity can be analyzed in terms of just the
compensator.

A simple expression (as simple as the super Yang-Mills case) can be
written for the supergravity action in terms of unconstrained
superfields.  We first give a first-order action, analogous to the one for
Yang-Mills (subsection IVC5):  In that case the action was in terms of $V$
and $A_a$; here it is in terms of $U^m$ and $E_m{}^a$.  Using just the
constraints solved in subsection XA1, we write the action as
$$ S_{SG,1} = 3Çdx¼d^4 ϼE^{-1}Ðìì(1 -\f14 T_{Œ,ÀŒ}{}^{ŒÀŒ}) $$
 in terms of the torsion $T_{ŒÀº}{}^c$.  (Note the similarity to the
Yang-Mills case, replacing the Chern-Simons form with the same
component of the torsion.)  We already evaluated everything except this
torsion, which is also easily found in the gauge $Æ=1$, $N_Œ{}^µ=¶_Œ^µ$,
$¯^µ=¯^{Àµ}=0$:
$$ S_{SG,1} = 3Çdx¼d^4 ϼdet(E_m{}^a)Ðìì(1 -\f14 öE_a{}^m E_m{}^a) $$
$$ ì = [det(E_m{}^a)]^{-1/3}(1Ée^{\onÁ{Я}})^{1/3}e^{Я}Ä,ââ
	ÓöE_Œ,ß{ÐE}_{ÀŒ}Õ = -iöE_{ŒÀŒ}{}^m »_m $$
 In the chiral representation this simplifies to
$$ S_{SG,1} = 
	3Çdx¼d^4 ϼ[det(E_m{}^a)]^{1/3}(1Ée^{-\onÁU})^{1/3}Ä(e^{-U}ÐÄ)
	(1 -\f14 E_m{}^{ŒÀŒ}iл_{ÀŒ}öE_Œ{}^m) $$

\x XB1.2  Find the algebraic field equation for $E_m{}^a$.  Use this to
eliminate it from the action, yielding expressions for $E{}_m{}^a$, $E$, and
the (second-order) action in terms of $U^m$ only.

The expansion of the superspace action in terms of unconstrained
superfields is needed for supergraphs, the most efficient way to do
quantum calculations.  We will not consider quantization here; the
methods are similar to those described in subsections VIB5, 9-10, and C5
for super Yang-Mills.  In particular, one uses background field methods: 
For example, as for super Yang-Mills,
$$ e^¯ £ e^{¯_B}e^{¯_Q}âÜâöE_Œ £ e^{-H}ö\E_Œ e^H,ââ
	e^H = e^{¯_Q}e^{Я_Q} $$
 The end result is that the expansion is about background-covariant
derivatives $\D_A$, e.g.,
$$ ßá_Œ = e^{-H}\D_Œ e^H,ââá_Œ = e^{-H}(Æ\D_Œ +¯_Œ{}^I M_I)e^H $$
 etc.  However, $\D_A$ satisfy the same constraints as the full covariant
derivatives:  For example, they have nonvanishing torsion
$T_{ŒÀº}{}^c=-i¶_Œ^© ¶_{Àº}^{À©}$.  This differs from the expansion implied
above in $U^m$ about partial derivatives, which anticommute without
torsion:  For a perturbation expansion useful for quantum calculations,
one must expand in $h_{ab}$ about $Òe_a{}^mÔ$, rather than in $e_a{}^m$
itself; thus (at least) the vacuum value $ÒUÔ$ must be separated from $U$.

Ü2. Ectoplasm

Although all supersymmetric theories can be analyzed directly in
superspace (including classical solutions, effective potentials, Feynman
graphs, etc.), for comparison to nonsupersymmetric theories it is
necessary to expand superfields in components.  Since all fundamental
theories are described by actions, it is sufficient to give a prescription for
evaluating any action in terms of component fields, as in subsection IVC2
for global supersymmetry.  In locally supersymmetric theories, the
vielbein needs to be expanded in terms of the prepotentials for
supergraphs.  We can also find component actions by a straightforward
Taylor expansion in $Ï$ of the prepotentials in the superspace action.
However, for component expansions of classical actions, one can get by
more simply by applying Bianchi identities to the covariant derivatives
and differential forms (antisymmetric tensors).  It is unnecessary to
know even the explicit form of the measure in terms of the vielbein or
prepotentials.

The fundamental idea is to think of the Lagrangian not as a scalar times a
measure, but more ``geometrically" as an antisymmetric tensor. 
Although this approach does not work for the usual superspace
Lagrangians because of the peculiarities of fermionic integration, it can be
applied to component Lagrangians integrated over 4D spacetime, treated
as the bosonic subspace of superspace.  We thus write the component
action as
$$ S = \f1{4!}Çdx^m dx^n dx^p dx^q¼L_{mnpq}(x,Ï) $$
 where $L_{MNPQ}$ is a graded antisymmetric superfield.  Of course, the
action should be independent of $Ï$, even though we have integrated
over only $x$.  This is equivalent to requiring that the integral should be
independent of the choice of 4D hypersurface in superspace.  We are
familiar with a similar requirement for conserved charges, which are
defined as integrals over 3D hypersurfaces:  Treating the conserved
current in terms of the 3-form dual to the vector,
$$ J_{mnp} = å{-g}·_{mnpq}J^qâÜâQ =\f1{3!}Çdx^m dx^n dx^p J_{mnp} $$
 the dual to the usual conservation law as vanishing (covariant)
divergence of the vector is vanishing curl of the 3-form:
$$ {d\over dt}Q = 0âÜâ»_{[m}J_{npq]} = 0 $$
 An important point is that neither the definition of the charge nor the
conservation law requires a metric, since integration in general does not. 
We thus require for our supersymmetric action
$$ {»\over »Ï}S = 0âÜâ»_{[M}L_{NPQR)} = 0,â
	¶L_{MNPQ} = \f1{3!}»_{[M}Â_{NPQ)} $$
 where the gauge invariance allows us to drop terms in the Lagrangian
that are total derivatives (surface terms).  Note that both $L$ and $Â$ are
assumed to be local functions of the fields and their (finite-order)
derivatives.  (As a result, this is not the usual ``cohomology", where both
would be allowed to be arbitrary functions of the coordinates.)

Converting the curl-free condition to flat indices (see subsection IVC5 for
the Chern-Simons superform),
$$ \f1{4!}á_{[A}L_{BCDE)} - \f1{2!3!}T_{[AB|}{}^F L_{F|CDE)} = 0,ââ
	¶L_{ABCD} = \f1{3!}á_{[A}Â_{BCD)} -\f1{(2!)^2}T_{[AB|}{}^E Â_{E|CD)} $$
 The plan is then to find $L_{ABCD}$, in terms of which the action can be
written as
$$ S = Çdx¼(-\f1{4!})·^{mnpq}E_q{}^D E_p{}^C E_n{}^B E_m{}^A L_{ABCD} $$
 where $E_m{}^A$ is exactly the nontrivial part of the inverse vielbein
$E_M{}^A$:
$$ E_m{}^A = (e_m{}^a,Æ_m{}^Œ) $$
 namely the inverse vierbein and the gravitino.  (If we also write
$Æ_m{}^Œ=e_m{}^a Æ_a{}^Œ$, we can collect all $e_m{}^a$ factors into a
factor of ${\bf e}^{-1}$ using the $·$ tensor.)

The next step is to explicitly solve the curl-free condition on the 4-form
in terms of the usual scalar superspace Lagrangian.  (An alternative is to
solve the Bianchis for the field strength of the 3-form gauge field, which
is also a 4-form.)  The procedure is the same as that used to solve the
Bianchi identities for covariant derivatives (in subsection XA2):  We start
with the lowest-dimension equations and work up.  The equations that
include the constant (vacuum/flat-space) part of the torsion can be
solved algebraically, the rest give differential constraints.  Of course, we
will need to use the results of subsection XA2 for the torsions and their
constraints.  The result is
$$ L_{Œºcd} = ·_Œ{}^{ÀŒ}{}_{,ºÀŒ,cd}Ð{\L},ââ
	L_{Œbcd} = i·_{ŒÀŒ,bcd}Ñá^{ÀŒ}Ð{\L},ââ
	L_{abcd} = ·_{abcd}[(Ñá^2 +3B)Ð{\L} +h.c.] $$
 and their complex conjugates, the rest vanishing, where $\L$ is the
usual chiral superspace Lagrangian (superpotential):
$$ Ñá_{ÀŒ}\L = 0 $$
 (We use the shorthand notation $a=(ŒÀŒ)$, etc.)  

The result is thus the component expansion of the usual curved
superspace action
$$ S = Çdx¼d^2 ϼE^{-1} \L +h.c. $$
 Using the Bianchi identities of the covariant derivatives one can also
covariantize the usual solution to the chirality condition:
$$ \L = (Ñá^2 +B){\bf L} $$
 which allows us to identify the action as
$$ S = Çdx¼d^4 ϼE^{-1}({\bf L} +Ð{\bf L}) $$
 so ${\bf L}$ can be taken real without loss of generality for general $d^4
Ï$ integrals.  On the other hand, for supergravity we can take
$$ {\bf L}_{SG} = 3,ââÐ{\bf L}_{SG} = 0âÜâ\L_{SG} = 3B,ââÐ\L_{SG} = 0 $$
 or vice versa, and the curvature appears in terms of $R_{μ}{}^{μ}$ and
not $ÐR_{ÀŒÀº}{}^{ÀŒÀº}$ (like the corresponding $f_{Œº}$ without $Ðf_{ÀŒÀº}$ for
Yang-Mills), with half as many terms to collect (for the same final result).
In general, we thus have an expression for $S$ in terms of $E_m{}^A$ and
the components of $L_{ABCD}$, and for the latter in terms of curvatures
and covariant derivatives of ${\bf L}$, which can be evaluated by the same
methods as in flat space (except that the commutation relations of the
covariant derivatives are more complicated).  

Components of a superfield are again defined by evaluating its covariant
derivatives at $Ï=0$.  However, as in the case of global supersymmetry,
the value of $Ï$ is arbitrary, since the result for the action is $Ï$
independent:  We therefore will generally drop the ``Ê|Ê" in component
expansions of actions; any superfield then implicitly refers to the
corresponding component.  This $Ï$ independence also means that it is
not necessary to make any gauge choices:  These methods automatically
express the action in terms of just the component fields that cannot be
completely gauged away.  (For example, $E_Œ{}^m|$ never appears.)

\x XB2.1 Collect all the above results:
ªa Find the complete component expression for the most general chiral
($Çd^2 Ï$) action in terms of covariant derivatives of the superpotential,
and the supergravity fields.
ªb Do the same for a real ($Çd^4 Ï$) action.

\x XB2.2 Evaluate the above actions for massive $Ä^3$ theory (for a
chiral matter field $Ä$) in terms of the components of $Ä$.

\x XB2.3 Do the same for super Yang-Mills:
ªa Solve the Bianchi identities for super Yang-Mills in curved superspace.
ªb Use this result to evaluate the component expansion of its action.

Ü3. Component transformations

We saw in the previous subsection that in component expansions the
supergravity gauge fields naturally appear as $E_m{}^A$, since by
definition we restrict to the bosonic submanifold.  Similar remarks apply
to the component form of their supercoordinate transformations (i.e.,
local supersymmetry), and the related component expansion of their field
strengths:  In subsection IXB4, we saw that coordinate transformations
(as applied to solving the radial gauge condition) were simpler for
$E_M{}^A$ because the derivative term on the parameter was just $»_M
K^A$.

We therefore begin by rewriting the gauge (coordinate) transformations
in terms of $E_m{}^A$.  As for gravity (see subsections IXA2 and IXB4), we
can choose to make the transformation laws more manifestly covariant by
writing the generators in terms of covariant derivatives.  Then, as for
gravity,
$$ K = K^A á_A + K^I M_I,ââ¶á_A = [K,á_A]âÜ $$
$$ ¶E_M{}^A = á_M K^A -E_M{}^B K^C T_{CB}{}^A +K^I M_I E_M{}^A,ââ
	¶¯_M{}^I = -á_M K^I +E_M{}^B K^C R_{CB}{}^I $$
$$ ¶(T_{AB}{}^C,R_{AB}{}^I) = K(T_{AB}{}^C,R_{AB}{}^I) $$
 using $(¶E_A{}^M)E_M{}^B=-E_A{}^M ¶E_M{}^B$.  Here 
$á_M=E_M{}^A á_A=»_M+¯_M{}^I M_I$, so these transformations on
$E_m{}^A$ and $¯_m{}^A$ contain only bosonic derivatives $»_m$, other
than those implicit in the torsions and curvatures.  (Similar remarks apply
to only $E_m{}^A$ and $¯_m{}^I$ appearing on the right.)  Since the
component expansion is an expansion in $Ï$, it is really only
supersymmetry $K^Œ=·^Œ$ which is no longer manifest; specializing to
that, the transformations on the physical gauge fields become, using the
results of subsection XA2 for the torsions,
$$ ¶e_m{}^{ŒÀŒ} = -i(·^Œ ÐÆ_m{}^{ÀŒ} +з^{ÀŒ}Æ_m{}^Œ),ââ
	¶Æ_m{}^Œ = á_m ·^Œ +ie_m{}^{ºÀº}(·_º G^Œ{}_{Àº} -з_{Àº}¶_º^Œ B) $$

We have not yet determined the solution for the Lorentz connection
$¿_m{}^{ab}$ and the transformation laws for the auxiliary fields.  We
will also need the relation between the usual curvature and the
components of the superfield strengths.  All of these can be found by use
of the identities (see subsection IXA2):
$$ \li{ -E_n{}^C E_m{}^B T_{BC}{}^A = -T_{mn}{}^A & =
	»_{[m}E_{n]}{}^A +E_{[m}{}^B ¯_{n]B}{}^A \cr
	E_n{}^D E_m{}^C R_{CD}{}^{ab} = R_{mn}{}^{ab} & =
	»_{[m}¯_{n]}{}^{ab} +¯_{[m}{}^{ac}¯_{n]c}{}^b \cr} $$
 The application of these identities is similar to that for expansion of the
action in the previous section:  The separation of the factors of $E_m{}^A$
into bosonic and fermionic parts yields an expansion in powers of the
gravitino field.  For the torsion case where the index $A=a$ we solve for
$¿_{mab}$ in terms of the torsions (auxiliary fields and constants) and
vielbein; for the torsion case $A=Œ$ we solve for $T_{ab}{}^©$, used for
the transformation law of the auxiliary fields, and for the curvature case
we solve for $R_{ab}{}^{cd}$, used in the component expansion of the
action, in terms of these auxiliaries, the vielbein, and the just-determined
connection.  The U(1) connection $A_m$ needs no solution:  It is pure
(superscale) gauge, and will cancel in actions (after perhaps an
appropriate redefinition of $G_a$).  For these manipulations we use the
relations that $T_{AB}{}^C$ and $R_{AB}{}^{cd}$ have to $B$, $G_a$, $W_Œ$,
$W_{Œº©}$, and their derivatives (as expressed by the solution to the
Bianchi identities given in subsection XA2).  The solution is
$$ ¿_{mbc} = e_m{}^a[\on\circ{¿}_{abc}-ü(öT_{bca}-öT_{a[bc]})],ââ
	öT_{ab}{}^c ­ e_a{}^m e_b{}^n T_{mn}{}^c =
	·_{ab}{}^{cd}G_d +iÆ_{[a}{}^© ÐÆ_{b]}{}^{À©} $$
$$ T_{ab}{}^© = -e_a{}^m e_b{}^n á_{[m}Æ_{n]}{}^©
	+i(Æ_{aº}G^©{}_{Àº} -ÐÆ_{aÀº}¶_º^© B -aªb)
	­ C_{Œº}t_{ÀŒÀº}{}^© +C_{ÀŒÀº}t_{Œº}{}^© $$
$$ ¶B = -\f23 ·^Œ t_{Œº}{}^º,ââ
	¶G_{ŒÀŒ} = -·^º(t_{ºŒÀŒ} +\f13 C_{ºŒ}t_{À©Àº}{}^{Àº}) +h.c. $$
$$ \li{ R_{ab}{}^{©¶} = {}&
	e_a{}^m e_b{}^n R_{mn}{}^{©¶} +ÐBÆ_a{}^{(©}Æ_b{}^{¶)} \cr
	& -2i[Æ_a{}^·(C_{º·}t^{©¶}{}_{Àº} +\f13 ¶_{(º}^© ¶_{·)}^º t_{ÀºÀ·}{}^{À·})
	-ÐÆ_a{}^{Àº}\f16 t_{(º}{}^{©¶)} -aªb] \cr} $$
 where ``$\ron\circ{\phantom m}$" refers to the usual expression for pure
gravity, $Æ_a{}^º­e_a{}^m Æ_m{}^º$, and we have chosen the superscale
gauge $W^Œ=0$ for simplicity.

\x XB3.1  Find the extra terms for $W^α0$.

Ü4. Component approach

We can now take the superspace action of subsection XB1, as expanded in
components by the ectoplasm method of subsection XB2, and substitute
the component expansions of the field strengths found in subsection XB3,
to find the component action
$$ L_{SG} = L_G +L_Æ +{\bf e}^{-1}L_a $$
$$ L_G = -\f14 {\bf e}^{-1}R,ââ
	L_Æ = ·^{mnpq}ÐÆ_{mÀŒ}üÓe_n{}^{ŒÀŒ},á_pÕÆ_{qŒ},ââ
	L_a = -\f38 (G_a)^2 +3ÐBB $$
 (Note the signs are again consistent with $G_a$ and $B,ÐB$ forming a
6-vector of SO(3,3), though not in the same way as in exercise XA2.6.)
Here $á$ and $L_G$ are the usual covariant derivative and Einstein-Hilbert
action of general relativity in terms of $e$ and $¿$, but $¿$ is slightly
different from any of the connections used previously (see exercise
XB4.1 below).  It also differs from the $¿$ given above in that we have
explicitly extracted the $G_a$ piece (which is the sole source of $¿$ in
the ectoplasm approach).  An alternative to ectoplasm to determine $¿$ is
to use a first-order formalism:  Rather than imposing the usual torsion
constraint, we can leave the Lorentz connection as an independent field
in $R$ and in the $á$ in $L_Æ$.  Eliminating the Lorentz connection by its
field equation yields a modified torsion constraint, and produces $Æ^4$
terms in the action.  We have written $L_Æ$ in a form manifestly
symmetric with respect to integration by parts.  (Alternatively, we can
write $ÐÆeáÆ-ÆeáÐÆ$.)

As an alternative to deriving the component action from the
simpler superspace expression, we can postulate the component action
directly.  In the component approach writing the action in components is
more direct than the superspace approach by definition, but proving
supersymmetry invariance is less so.  This is not so true when coupling to
matter, where writing component actions can also be as complicated as
deriving them from superspace, so here we consider the simplest case,
pure supergravity.  We thus begin by postulating $L_{SG}=L_G+L_Æ$; the
first term is obvious, while the second follows from minimal coupling for
the free gravitino action, which can be derived easily by many methods
(see, e.g., subsection XIIA5 below).  We ignore the auxiliary fields, which
are necessary for off-shell closure of the supersymmetry algebra, but not
for supersymmetry invariance of the action.  

We write the action for gravity in a form that more resembles the
gravitino action (see exercise IXA5.5):
$$ L_G = -\f14 {\bf e}^{-1}R =
	\f1{16}·^{mnpq}·_{abcd}e_m{}^a e_n{}^b R_{pq}{}^{cd}
	= \f18 ·^{mnpq}e_m{}^a e_n{}^b ÷R_{pqab} $$
$$ = i\f18 ·^{mnpq}e_m{}^{ŒÀŒ}
	(ÐR_{npÀŒ}{}^{Àº} e_{qŒÀº} -R_{npŒ}{}^º e_{qºÀŒ}) $$
 where we have switched to spinor notation for the curvature (see
subsection IXA1) and used duality in both vector and spinor notation (see
subsection IIA7).  (Alternatively, we can regard this as the definition of
the gravity action.)  We then have for the variation of this part of the
action (after some integration by parts)
$$ ¶L_G = i\f14 ·^{mnpq}e_m{}^{ŒÀŒ}Ó
	[ÐR_{npÀŒ}{}^{Àº} ¶e_{qŒÀº} -R_{npŒ}{}^º ¶e_{qºÀŒ}]
	-[(¶Ð¿_{nÀŒ}{}^{Àº})T_{pqŒÀº} -(¶¿_{nŒ}{}^º)T_{pqºÀŒ}]Õ $$
 where we have used (see exercise IIIC1.2)
$$ ¶R_{mn}{}^I = á_{[m}¶¿_{n]}{}^I,ââÇdx¼á_m V^m = Çdx¼»_m V^m = 0 $$

Next, we pick the obvious transformation law for the gravitino field as
the gauge field of supersymmetry:
$$ ¶Æ_m{}^Œ = á_m ·^Œ $$
 The transformation laws for $e$ and $¿$ will be derived as a by-product
of the invariance proof, as will the explicit expression for $¿$ in terms of
$e$ and $Æ$.  Substituting this expression for $¶Æ$ into $L_Æ$,
$$ \li{ ¶L_Æ = ·^{mnpq}[&·_Œ á_m üÓe_n{}^{ŒÀŒ},á_pÕÐÆ_{qÀŒ}
		-з_{ÀŒ}á_m üÓe_n{}^{ŒÀŒ},á_pÕÆ_{qŒ} \cr
	&+üÐÆ_{mÀŒ}(¶e_n{}^{ŒÀŒ})á_p Æ_{qŒ}
		- üÆ_{mŒ}(¶e_n{}^{ŒÀŒ})á_p ÐÆ_{qÀŒ} \cr
	&+üÐÆ_{mÀŒ}e_n{}^{ŒÀŒ}(¶¿_{pŒ}{}^º)Æ_{qº}
		-üÆ_{mŒ}e_n{}^{ŒÀŒ}(¶Ð¿_{pÀŒ}{}^{Àº})ÐÆ_{qÀº} ] \cr} $$
 where we have integrated by parts to free the supersymmetry
parameters of derivatives.  We then use the antisymmetrization on all
curved indices to collect the resulting terms into torsion and curvature as
$$ áÓe,áÕ = Óe,ááÕ +(áe)á = üÓe,RÕ -üTá $$

The curvature terms then cancel those from $¶L_G$, if we choose for $¶e$
in $¶L_G$ the transformation law
$$ ¶e_m{}^{ŒÀŒ} = -i(·^Œ ÐÆ_m{}^{ÀŒ} +з^{ÀŒ}Æ_m{}^Œ) $$
 Then we also substitute this expression for $¶e$ in $¶L_Æ$, and note that
half those terms immediately drop out, since
$$ Æ_{[m}{}^Œ Æ_{n]Œ} = 0 $$
 by antisymmetry.  The remaining terms from both $L_G$ and $L_Æ$ then
can be collected as
$$ ¶L_{SG} = -\f18 ·^{mnpq}~T_{mn}{}^{ŒÀŒ}ë_{pqŒÀŒ} $$
$$ ~T_{mn}{}^{ŒÀŒ} ­ T_{mn}{}^{ŒÀŒ} -iÐÆ_{[m}{}^{ÀŒ}Æ_{n]}{}^Œ $$
$$ ë_{pqŒÀŒ} ­ 
	i(¶Ð¿_{[pÀŒ}{}^{Àº})e_{q]ÀºŒ} -i(¶¿_{[pŒ}{}^º)e_{q]ºÀŒ}
	+·_Œ á_{[p}ÐÆ_{q]ÀŒ} -з_{ÀŒ}á_{[p}Æ_{q]Œ} $$

We now note that the former factor in $¶L_{SG}$ vanishes by virtue of the
equation of motion from varying the connection:  Rather than vanishing,
the torsion now satisfies
$$ second-order:â~T_{mn}{}^{ŒÀŒ} = 0 $$
 We can regard $~T$ as the ``supersymmetrized torsion"; this is equivalent
on shell to the result we found in the previous subsection from
superspace.  We can therefore quit now, since in a second-order
formalism the torsion (and thus the Lorentz connection) would satisfy
this equation even off shell.  (This approach, using the second-order
formalism but not bothering to substitute the supersymmetry variation of
the connection, is called the ``1.5-order formalism".)  On the other hand,
we can just as easily recognize that in the first-order formalism
cancellation of $¶L_{SG}$ is also guaranteed by allowing vanishing of the
ÓlatterÕ factor to define the supersymmetry variation of the
(independent) connection:
$$ first-order:âë_{pqŒÀŒ} = 0 $$
 Thus, use of the first-order formalism requires no more work than
1.5-order (contrary to remarks in the literature), which is really the same
as second-order, and provides the bonus of yielding the transformation
law for $¿$.  However, it is useful to note that not all quantities should
have their longer forms substituted at the beginning of a calculation (just
as we learned in high-school algebra not to plug in numbers till the end).

\x XB4.1  Let's complete this calculation to the bitter end, finding all the
properties of the connection:
 ªa  Solve the torsion constraint for $¿$ (see subsection IXA3).
 ªb  Find the transformation law for $¿$ that follows from cancellation of
the above terms off shell (i.e., without imposing the torsion constraint).
 ªc  Show the above two results are consistent (modulo terms with $Æ$
field equations, which can be canceled by contributions from auxiliary
fields) by plugging the expressions for $¶e$ and $¶Æ$ into the variation of
the result for part {\bf a} and comparing with the result for part {\bf b}.
 ªd  Compare these results with the connection found in the previous
subsection.  How does the appearance of $G_a$ affect the transformation
law?

Ü5. Duality

Although antisymmetric tensor gauge fields can be avoided in general,
they tend to turn up in string theory, so we now look at them a little
more generally, examining their actions and how they relate to those for
scalars.  In particular, we note that a sensible action for such a tensor
alone cannot be constructed that is conformally invariant:  From the same
analysis as for electromagnetism or Yang-Mills (subsection IXA7), we see
that $(F_a)^2$ does not give a scale-invariant action in four dimensions. 
Thus, such a field is not suitable as a compensator for pure gravity. 
However, in supergravity the tensor multiplet (see subsection XA3) also
has an ordinary scalar, and an appropriate power of it can make the
tensor's action conformal.  Therefore, we now examine general duality
transformations for the supersymmetric case, which is more relevant for
understanding its use in gravity.  We will consider explicitly a flat
superspace background for simplicity, but generalization to curved
superspace by covariantization is straightforward, replacing flat
superspace derivatives with (superconformal) covariant derivatives,
introducing supergravity field strengths where necessary (in this case,
just $Ðd^2£Ñá^2+B$ for chirality), and using the covariant integration
measures.

Duality transformations can be performed directly in the action by use of
first-order formulations.  Starting with the general tensor multiplet action
$$ S_{tm} = Çdx¼d^4 ϼK(G) $$
 where $K$ is some function and $G=d^Œ Ä_Œ+h.c.$, we write this in
first-order form as
$$ S'_{tm} = Çdx¼d^4 ϼ[~K(V) +VG] $$
 where $V$ is an unconstrained real superfield and $~K$ is the Legendre
transform of $K$:  For this action to reduce to the previous upon applying
the algebraic field equation of $V$, we must have
$$ [~K(V) +VG]|_{{»÷K(V)\over »V}=-G} = K(G) $$
 The duality transformation is then performed by varying $Ä_Œ$ instead
of $V$ in $S'_G$:  Remembering that $Ä_Œ$ is chiral, so
$$ ¶Çd^4 x¼d^4 ϼVG = üÇd^4 x¼d^2 ϼ(¶Ä^Œ)Ðd^2 d_Œ V +h.c. $$
 we solve the condition on $V$ as
$$ Ðd^2 d_Œ V = 0âÜâV = Ä+ÐÄ $$
 since thinking of $V$ as the prepotential for a vector multiplet says that
it is pure gauge.  The dualized action is then
$$ S_Ä = Çdx¼d^4 ϼ~K(Ä+ÐÄ) $$
 We can also reverse the procedure through another first-order action
$$ S'_Ä = Çdx¼d^4 ϼ[K(V) -V(Ä+ÐÄ)] $$
 where in this case varying with respect to $Ä$ implies
$$ Ðd^2 V = 0âÜâV = G $$
 while varying with respect to $V$ gives the inverse Legendre transform
$$ [K(V) -V(Ä+ÐÄ)]|_{{»K(V)\over »V}=Ä+ÐÄ} = ~K(Ä+ÐÄ) $$

The simplest case is the Lagrangian $üG^2$:  We then find
$$ K(V) = üV^2âÛâ~K(V) = -üV^2 $$
 so the duality is
$$ L_{tm} = üG^2âÛâL_Ä = -ü(Ä+ÐÄ)^2 $$
 In flat space, this gives the usual free result $L_Ä = -ÐÄÄ$, but in curved
space the $-üÄ^2+h.c.$ part does not vanish because $E^{-1}$ is not
chiral.  Consequently, this action is not the conformal one (as we already
knew from the component argument above).  However, the conformal one
is easy to find by starting with $-ÐÄÄ$:  Making the field redefinition
$Ä£e^Ä$ expresses the action in terms of $Ä+ÐÄ$.  Legendre transforming,
$$ ~K(V) = -e^VâÛâK(V) = V(ln¼V-1) $$
 so the duality is
$$ L_Ä = -e^{Ä+ÐÄ}âÛâL_{tm} = G(ln¼G-1) $$
 These two conformal actions for matter, when coupled to conformal
supergravity, become the two ``minimal" actions for supergravity, when
the overall sign is changed to make the matter fields into compensators: 
The version with $Ä$ as the compensator is called ``old minimal", while
that with $Ä_Œ$ is called ``new minimal".  They differ only off-shell, in
their choice of auxiliary fields.  Note that the field equations for the two
conformal multiplets,
$$ Ðd^2 Ðì = 0ââ(ì ­ e^Ä),ââÐd^2 d_Œ ln¼G = 0 $$
 reproduce the compensator part of the supergravity field equations $÷B=0$
and $~W_Œ=0$ described in subsection XA4.  Again, the full expressions
follow from the usual supergravitational and superscale invariances,
which were used to find $÷B$ and $~W_Œ$; the compensator dependence is
enough to identify them as the appropriate covariantizations.

\x XB5.1  We saw in subsection IVC5 for the Chern-Simons form, or
XB2 for ectoplasmic integrals, that differential forms can be defined in
superspace.  Do the same for the tensor multiplet:
 ªa By generalizing the bosonic case to superspace with curved indices,
and then ``flattening" the indices (as for the Chern-Simons superform),
show that the super 2-form $B_{AB}$ with field strength $H_{ABC}$ and
gauge parameter $Â_A$ is described by
$$ ¶B_{AB} = á_{[A}Â_{B)} -T_{AB}{}^C Â_C,ââ
	H_{ABC} = üá_{[A}B_{BC)} -üT_{[AB|}{}^D B_{D|C)} $$
 (Hint:  Show that replacing $E_A£á_A$ and $C_{AB}{}^C£T_{AB}{}^C$ yields
only canceling connection terms.)
 ªb  Show that the torsions given in subsection XA2 satisfy
$$ Çdx¼d^4 ϼE^{-1}H_{Œ,ÀŒ,}{}^{ŒÀŒ} = 0 $$
 Note that this also implies the gauge invariance of the Chern-Simons
form of the super Yang-Mills action in curved superspace.
 ªc  Show that the constraints
$$ H_{Œº©} = H_{ŒºÀ©} = H_{Œºc} = 0,ââH_{Œ,Àº,©À©} = -iC_{Œ©}C_{ÀºÀ©}G $$
 (and complex conjugates) can be solved by
$$ B_{Œº} = B_{ŒÀº} = 0,ââB_{ÀŒ,ºÀº} = -iC_{ÀŒÀº}Ä_º;ââÑá_{ÀŒ}Ä_º = 0 $$
$$ B_{ab} = C_{ÀŒÀº}b_{Œº} +C_{Œº}Ðb_{ÀŒÀº},ââb_{Œº} = üá_{(Œ}Ä_{º)};ââ
	G = ü(á_Œ Ä^Œ +Ñá_{ÀŒ}ÐÄ^{ÀŒ}) $$
 Relate the results for the Ógauge fieldsÕ $B_{AB}$ to those for the
Yang-Mills Ófield strengthsÕ $F_{AB}$ (subsection IVC3).
 ªd  Supersymmetrize the construction of exercise XA3.1: 
Show one can define a field strength
$$ ~H_{ABC} = H_{ABC} +B_{ABC} $$
 using the Chern-Simons superform $B_{ABC}$.

The supergravity component action with the tensor multiplet compensator
differs from the one of the previous subsection in that $B$ and the
longitudinal part of $G$ have been replaced by the gauge field $B_{mn}$:
$$ L_a £ ü·^{mnpq}G_m »_n B_{pq} $$
 which is the only possibility that preserves the gauge invariances of both
$G$ and $B$ while leaving them both auxiliary.  (Their field equations are
that their field strengths vanish.)

Ü6. Superhiggs

Supergravity affects spontaneous supersymmetry breaking in a simple
way:  From the discussion of the immediately preceding subsections, we
know that supergravity can be described more simply as conformal
supergravity coupled to a compensator.  Simple (N=1) conformal
supergravity contains no scalars:  It consists of only conformal gravity
(the traceless part of the metric), the conformal (traceless) part of the
gravitino field, and an auxiliary gauge vector.  Since symmetry breaking
involves giving vacuum values to only scalars, we can replace
supergravity by just its compensator for these purposes.  

For a general analysis, consider a kinetic term
$$ S_K = Çdx¼d^4 ϼ3ÐÄÄe^{-K(^i,Ѝ_i)/3} $$
 (The exponential form will prove convenient for later component analysis.) 
This is the most general kinetic term with the usual number of spacetime
derivatives:  Any term of the form $f(Ä,ÐÄ,^i,Ѝ_i)$ can be rewritten in
this form after appropriate field redefinitions.  In particular, if we start
with fields with arbitrary Weyl scale weight, then this form follows after
rescaling fields so only $Ä$ carries scale weight, since all terms in the
Lagrangian must have the same scale weight, fixed by (super)conformal
invariance.  $Ä$ is then the only field to carry U(1) weight, which is
proportional to scale weight by superconformal invariance.  Then $Ä$
appears only as $ÐÄÄ$, while $K$ is an arbitrary function of $^i$ and
$Ѝ_i$.  The first step in evaluating this action in components is to simply
ignore conformal supergravity altogether, and evaluate this action as is,
in terms of matter and compensator multiplets, by the methods we have
considered previously for evaluating $Ï$ integration.  The next step is to
add back in some parts of conformal supergravity:   
\item{(1)} the conformal
graviton, which can be put back in easily and uniquely using coordinate
and local scale invariance;   
\item{(2)} the U(1) axial gauge vector, whose coupling
is minimal, and thus follows directly from U(1) covariantizing the
spacetime derivatives; and   
\item{(3)} the conformal gravitino, whose quartic
couplings can be quite complicated, but as a practical matter we are
interested in only the mass term, which is determined from the mass of
the Goldstone fermion it eats, which appears in the compensator (and the
kinetic term, which is the usual one).   

Before considering the general case, we look at the pure supergravity
case under this analysis:  Looking at just the bosons, we find
$$ S_{SG,b} = 
	Çdx¼{\bf e}^{-1}üÓ-3[(á-i\f13 A)ÐÄ]É[(á+i\f13 A)Ä] -üÐÄÄR +6ÐBBÕ $$
 where $á$ is the usual covariant derivative of general relativity, $A$ is
the U(1) gauge vector, and the relative coefficient of the $R$ term was
fixed by local scale invariance (see subsection IXA7).  Note that here $B$
is the usual auxiliary field from $Ä$, and is not associated with conformal
supergravity.  Choosing the component U(1) and scale gauges $Ä|=1$, this
reduces to
$$ S_{SG,b} £ Çdx¼{\bf e}^{-1}(-\f14 R -\f16 A^2 +3ÐBB) $$
 Relating to $G_a$, we recall that if we had included it from conformal
supergravity, for this compensator $÷G_a=G_a+\f23 A_a$, so we can
identify $A_a$ with $\f32 ÷G_a$.  Thus, the compensator method
immediately yields the bosonic action, including auxiliary fields.

Returning to the general case, the part of the action for the ``physical"
scalars ($Ä|$ and $|$) then starts out as
$$ S_{K,ps} = Çdx¼{\bf e}^{-1}e^{-K/3}üÓ-3[(á-i\f13 A)ÐÄ]É[(á+i\f13 A)Ä]
	+ÐÄ[(á+i\f13 A)Ä]É(л{}^i K)áЍ_i $$
$$ +Ä[(á-i\f13 A)ÐÄ]É(»_i K)á^i
	+ÐÄÄ[(»_i л{}^j K) -\f13(»_i K)(л{}^j K)](áЍ_j)É(á^i) -üÐÄÄRÕ $$
 ignoring until the following subsection the auxiliary scalars, which are
irrelevant for the kinetic term.  We use the notation $»_i=»/»^i$,
$л{}^i=»/»Ð_i$.  We then choose the U(1) and scale gauges
$$ Ä| = e^{K(^i|,Ѝ_i|)/6} $$
 where we have explicitly written the |'s to emphasize that this is a
nonsupersymmetric gauge choice for the component $Ä|$.  Finally, we
eliminate $A$ by its algebraic field equation.  We thus obtain
$$ S_{K,ps} £ Çdx¼{\bf e}^{-1}ü[(»_i л{}^j K)(áЍ_j)É(á^i) -üR] $$
 Except for the $R$ term and covariant derivatives, this is what would
follow in flat superspace from the action $-Çdx¼d^4 ϼK$.  For
supersymmetry breaking, we also need the super cosmological term
$$ S_c = Çdx¼d^2 ϼÂÄ^3 +h.c. $$
 for some constant $Â$.  We could consider more general potentials $Ä^3
e^{f(^i)}$ (again the power of $Ä$ is fixed by scale and U(1)), but then
the field redefinition $Ä£Äe^{-f/3}$ would remove it while replacing
$K£K+f+Ðf$.  (This invariance, and the form of the ``metric" on the space
of fields $$ and $Ѝ$ appearing in the action, identify $K$ as a ``K¬ahler
potential".)

The analysis for $S_K$ can also be made by performing a duality
transformation on the compensator.  Following the same steps as
described in the previous subsection for the case without matter
(factoring the overall $-3$ out of the process for convenience), we find
$$ S_K £ -Çdx¼d^4 ϼ[3G¼ln¼G +GK(,Ѝ)] $$
 Since in this form $A$ decouples, the result is obvious from the flat-space
result. 

\x XB6.1  Repeat the above analysis using the compensator $G$:
Evaluate explicitly all the contributions from the bosons in $G$,
couple $A$, find the $R$ term, show the result is the same.

Normally any kind of symmetry breaking will generate a cosmological
term, since a scalar getting a vacuum value implies the potential itself
getting one, giving a term $Çdx¼{\bf e}^{-1} constant$.  This would
require adding a cosmological term to the action by hand to cancel the
generated one, since  the constant generated would
correspond to a subatomic length scale, whereas a realistic cosmological
constant requires a cosmological length scale, which means a constant,
going as 1/length${}^2$, of the order of $10^{-80}$ in subatomic units.
An exception is when the potential is flat in some direction:  In
supersymmetry energy is always positive, and the supersymmetric
vacuum has zero energy, but some potentials allow other, perhaps
nonsupersymmetric, vacuua that also have zero energy, and thus
generate no cosmological constant.  This avoids the ad hoc procedure of
``fine tuning" the cosmological constant of an added term for exact
cancellation (or at least to order $10^{-80}$).

Ü7. No-scale

A useful example of the superhiggs effect with a flat potential is
``no-scale supergravity".  This theory has an explicit super-cosmological
term, but the kinetic term is such that this term does not generate a
component cosmological term, but does spontaneously break
supersymmetry.  The simplest example describes supergravity coupled to
a single chiral scalar multiplet.  The kinetic term has an SU(1,1) symmetry,
and also appears in N=4 supergravity (see subsection XC6 below). 
Written in terms of just the compensator part of supergravity, it is
$$ S_K = Çdx¼d^4 ϼ3(Ðč+ЍÄ) $$
 where $Ä$ is the compensator and $$ is the matter.  We have written it
in a manifestly U(1,1) covariant form, where the U(1,1) metric is
off-diagonal ($\tbt0110$ instead of the usual diagonalized
$\tbt100{-1}$).  For the above component analysis we redefine
$$  £ čâÜâS_K £ Çdx¼d^4 ϼ3ÐÄÄ(Ѝ+)âÜâK = -3¼ln(Ѝ+) $$
 (Many other superfield redefinitions are possible to put this in more
conventional forms, such as $(3ÐÄÄ-Ѝ)$, $ÐÄÄ(3-Ѝ)$, etc.)  The kinetic
term for the physical scalars follows from the same analysis we applied
to the CP(1) model in subsection IVA2.  The only differences here are: (1)
the symmetry is U(1,1), not U(2), and (2) the constraint on the norm of the
complex 2-vector follows not from a Lagrange multipler (or a low-energy
limit), but as a local scale gauge chosen to give the Einstein-Hilbert
curvature term the usual normalization.  Alternatively, we can use the
analysis given in the previous subsection for the general case to find
$$ S_{K,ps} £ Çdx¼{\bf e}^{-1}ü\left[3{|á|^2\over (Ѝ+)^2}-üR\right] $$

However, to study just the supersymmetry breaking, we want to look at
the ``potential" terms: terms that involve the auxiliary scalars instead of
spacetime derivatives.  We thus now need to include the super
cosmological term, which breaks the SU(1,1) invariance.  Again evaluating
at first without conformal supergravity, then putting some (all but the
conformal gravitino) back in, we find the contributions from $S_K$ and
$S_c$
$$ S_{aux} = Çdx¼{\bf e}^{-1}3[ÐBB(Ѝ+) +(ÐBÄb+BÐÄÐb) +Â(BÄ^2+ÐBÐÄ^2)] $$
 where $B=d^2 Ä$ and $b=d^2 $.  We then see that eliminating the
auxiliaries gives nothing, so there is no potential to generate a
cosmological term.  However, there is still a mass term for the gravitino: 
As always, $S_c$ also contains the spinor term
$$ 6Â(Ľ^2+h.c.) $$
 where $½_Œ=d_Œ Ä$ is the trace of the gravitino.  The gravitino in this
model therefore has a mass proportional to $ÂÒЍ+Ô^{-1/2}$.

\x XB6.1  Explicitly evaluate the spinor part of the kinetic term,
and thus determine the exact value of the mass of the spinor,
and thus the gravitino.

SU(1,1) invariant kinetic terms also appear in superstring theory, but
unlike N=4 and no-scale supergravity, the kinetic term is $(Ðč+ЍÄ)^{1/3}$
instead of just $Ðč+ЍÄ$.  (See subsection XIA6.)  When applying no-scale
supergravity to nature, more matter multiplets are added,
$$ S = Çdx¼d^4 ϼ3(Ðč+ÐÄ -Ѝ_i ^i) 
	+\left(Çdx¼d^2 ϼÂÄ^3e^{f(^i/Ä)} +h.c.\right) $$
 generalizing SU(1,1) to SU(n,1) in the first term.  (N=5 supergravity has
such an SU(5,1) symmetry; see below.)  Then $$ acts as the ``hidden"
matter sector that doesn't directly couple to the observed matter $^i$,
but serves only to break supersymmetry.

\refs

£1 W. Rarita and J. Schwinger, ÓPhys. Rev.Õ É60 (1941) 61:\\
	action for spin 3/2.
 £2 Freedman, van Nieuwenhuizen, and Ferrara, Óloc. cit.Õ (XA):\\
	complete component formalism for supergravity, except for auxiliary
	fields.
 £3 S. Deser and B. Zumino, \PL 62B (1976) 335:\\
	first-order formalism for supergravity.
 £4 A.H. Chamseddine and P.C. West, \NP 129 (1977) 39;\\
	P.K. Townsend and P. van Nieuwenhuizen, \PL 67B (1977) 439:\\
	1.5-order formalism.
 £5 P. van Nieuwenhuizen, ÓPhys. Rep.Õ É68 (1981) 189:\\
	general review of supergravity.
 £6 S.J. Gates, Jr., \xxxlink{hep-th/9709104}, Ectoplasm has no topology:
	the prelude, in ÓSupersymmetries and quantum symmetriesÕ, proc.,
	July 22-26, 1997, Dubna (Lecture notes in physics, v. 524), eds. J.
	Wess and E.A. Ivanov (Springer, 1999) p. 46;\\
	S.J. Gates, Jr., M.T. Grisaru, M.E. Knutt-Wehlau, and W. Siegel,
	\xxxlink{hep-th/9711151}, \PL 421B (1998) 203:\\
	ectoplasm.
 £7 Wess and Zumino; Wess; Óloc. cit.Õ (IVC, ref. 5):\\
	component expansions for supergravity.
 £8 W. Siegel, Óloc. cit.Õ (XA);\\
	K.S. Stelle and P.C. West, \PL 74B (1978) 330;\\
	S. Ferrara and P. van Nieuwenhuizen, \PL 74B (1978) 333:\\
	old-minimal supergravity.
 £9 Sohnius and West, Óloc. cit.Õ (XA):\\
	new-minimal supergravity.
 £10 D.V. Volkov and V.A. Soroka, ÓJETP Lett.Õ É18 (1973) 312;\\
	S. Deser and B. Zumino, \PR 38 (1977) 1433:\\
	superhiggs.
 £11 B. Zumino, \PL 87B (1979) 203:\\
	K¬ahler in N=1 nonlinear $§$ models.
 £12 E. Cremmer, S. Ferrara, C. Kounnas, and D.V. Nanopoulos, 
	\PL 133B (1983) 61:\\
	no-scale supergravity.

\unrefs

Û6 C. HIGHER DIMENSIONS

A convenient method for describing extended supersymmetry in D=4 is to
apply dimensional reduction to supersymmetry in D>4, since (1) spinors
are bigger in D>4, so even simple supersymmetry reduces to extended
supersymmetry, and (2) the Lorentz group is bigger in D>4, so some 4D
scalars arise as parts of higher-D vectors, etc., meaning fewer Lorentz
representations in the multiplet in D>4.  

Ü1. Dirac spinors

We saw in subsection IC1 that coordinate representations of orthogonal
groups SO($D$) could be defined in terms of self-conjugate fermions,
$$ G_{ab} = ü[ ©_a, ©_b ],ââÓ ©_a, ©_b Õ = ¶_{ab} $$
 We now will construct explicit matrix representations of the Dirac
matrices for arbitrary $D$, and examine their properties.  This is useful for
understanding: 
\item{(1)} representations of internal symmetries, such as in
Grand Unified Theories; 
\item{(2)} theories in higher dimensions, which give
simpler formulations of certain four-dimensional theories when the extra
dimensions are eliminated, and appear in string theory; and 
\item{(3)} properties
of spinors that are independent of $D$, or their dependence on $D$, which
is useful for comparison and for perturbation in quantum field theory.

An explicit solution can be found easily by first looking at even
dimensions, and breaking up the problem into $D$/2=$n$ two-dimensional
problems.  Furthermore, we can look first at the Euclidean case (SO($D$)),
and solve for the other cases (SO($D_+$,$D_-$)) by Wick rotation.  The
solution for SO(2) is just two of the Pauli $§$ matrices.  The general
solution then comes from the direct product of the two-dimensional
cases, using the third $§$ matrix to introduce appropriate ``Klein factors"
(see exercise IA2.3) to insure that the $©$ matrices from one
two-dimensional subspace anticommute with those from another.  The
resulting $©$ matrices are then:
$$ {1\over å2}(å2§_3 ° ò ° å2§_3 ° å2§_i ° I ° ò ° I),ââ
	{1\over å2}(å2§_3 ° ò ° å2§_3) $$
 where $i=1,2$, there are a total of $n$ factors, and the number of
$å2§_3$ and $I$ factors in the first expression ranges from 0 to $n-$1.
The last matrix can always be included to extend SO(2$n$) to SO(2$n$+1);
in fact, up to normalization, it's simply the product of all the other $©$'s. 
(In other words, the product of all the $©$ matrices is proportional to the
identity.)

\x XC1.1  Apply exercise IC1.2 to this construction:  Show how this
representation relates simply to creation and annihilation operators.
Show that these Klein factors are identical to those of exercise IA2.3.

The next step is to notice that this construction generally gives a
reducible representation.  Reducibility comes from two properties: (1) For
SO(2$n$) we really have SO(2$n$+1); and (2) the representation may be
real.  In fact, most of the interesting cases involve SO(2$n$) (in particular,
SO(3,1) for Lorentz and SO(4,2) for conformal in four dimensions).  In that
case we can call the first (or any other) $©$ matrix ($§_1°I°ò°I$) for
SO(2$n$+1) ``$©_{-1}$", and take the rest as those for SO(2$n$).  Then the
projection operators
$$ ¸_à = ü(1àå2©_{-1})âÜâ¸_à{}^2 = ¸_à,¼¸_+¸_- = ¸_-¸_+ = 0,¼
	¸_+ + ¸_- = 1 $$
 commute with the SO(2$n$) generators $G_{ab}¾©©$, so they can be used
to project the representation of the $©$'s into two representations of
SO(2$n$).  These two halves of a Dirac spinor are known as ``Weyl spinors".
A convenient representation of the $©$ matrices for this purpose
is the one given in subsection IIA6, with the representation of the Pauli
matrices used in our SU(2)/SL(2,C) discussion of subsections IIA1 and 5,
$$ §_1 = \f1{å2}\tat100{-1},â§_2 = \f1{å2}\tat0110,â
	§_3 = \f1{å2}\tat0i{-i}0 $$
 We then can write the spinor, which has $2^n$ components (since it
represents the direct product of $n$ representations of $§$ matrices, each
of which has two components) as two $2^{n-1}$-component spinors
projected by
$$ ¸_à = §_à ° I ° ... ° I,ââ§_+ = \tat1000,â§_- = \tat0001 $$
 The $©$ matrices then take the block-diagonal form
$$ ©_{-1} = \f1{å2}\tat{I}00{-I},ââother¼© = \tat0§{÷§}0 $$
 We will refer to these reduced matrices $§$ (and $÷§$), and the $©$
matrices themselves for SO(2$n$+1), as generalized Pauli ($§$) matrices.

The reality properties of the representation depend on the existence of a
metric $ú_{ÀA}{}^B$ (or $¯_{ÀA}{}^B$ for psuedoreality, which doesn't reduce
the representation), as in our discussion of classical groups of subsection
IB5.  In fact, all the spinor representations of any orthogonal group are
also defining representations of another group:  For less than seven
dimensions, this leads to the identification of covering groups discussed
in subsection IC5; for more than six dimensions, it only identifies the
orthogonal group as a subgroup of this new group.  (An interesting
exception is SO(8), where the spinor representations are also
8-dimensional, and are the two other defining representations of SO(8).) 
In matrix notation, we look for a matrix
$C=ú$ or $¯$ such that we can define the operation of charge conjugation
as
$$ ï £ C^{-1}ï*,¼Gï £ C^{-1}(Gï)*âÜâG = C^{-1}G*C $$
 If we like, we can also choose
$$ C = Cÿ = C^{-1} $$
 without loss of generality.  For a representation to be invariant under
charge conjugation (i.e., real)
$$ ï = C^{-1}ï*âÜâC* = C^{-1} $$

For our $©$ matrix representation, the matrix to look at is
$$ C = ... ° C_2 å2§_3 ° C_2 ° C_2 å2§_3 ° C_2 $$
 where
$$ C_2 = \tat0i{-i}0 $$
 independent of the representation used for the Pauli matrices.  (Our
representation is simplest, since then $C_2 å2§_3=I$, and $C=Cÿ$.  In
other representations, $C$ may also need an $n$-dependent factor of $i$
if we want $C=Cÿ$.)  Using properties of $§$ matrices we found in our
discussion of SO(3) in subsection IIA2, such as $C_2 §* C_2=-§$, we find
$$ C^{-1}©*C = (-1)^n ©,âC* = (-1)^{n(n+1)/2}C^{-1},â
	 C^T = (-1)^{n(n+1)/2}C,âCÿ = C^{-1} $$
 This distinguishes 8 cases, where the irreducible spinors are:
$$ \vbox{\offinterlineskip
\halign{ \strut SO(8$m$#):\hfilââ &#\hfil \cr
	& Weyl and real \cr
+1& real \cr
+2& Weyl \cr
+3& pseudoreal \cr
+4& Weyl and pseudoreal \cr
+5& pseudoreal \cr
+6& Weyl \cr
+7& real \cr
}} $$
 For SO(4$m$+2), charge conjugation does not preserve $©_{-1}$, and thus
$¸_à$.  Therefore, in those cases there is no metric $ú_{ÀA}{}^B$ or
$¯_{ÀA}{}^B$ on the irreducible spinor:  The Dirac spinor consists of two
irreducible spinors that are complex conjugate representations of each
other.  In general, a Dirac spinor has $2^n$ complex components for
SO(2$n$) and SO(2$n$+1); the Weyl condition reduces this a factor of two
for SO(2$n$), as does reality where applicable.  (Pseudoreality does
nothing.)

It is useful to know the other group metrics, if they exist.  For unitarity
properties we look for a metric $ç^{ÀAB}$ such that
$$ G = -ç^{-1}Gÿ çâÜâç = çÿ $$
 (Thus $g=e^G$ satisfies $çg^{-1}=gÿç$.)  We can also choose
$$ ç = ç^{-1} $$
 without loss of generality.  We therefore look for a metric satisfying
$$ ç^{-1}©ÿç = © $$
 so $G¾[©,©]$ is antihermitian with respect to $ç$.  For SO(D), we have
simply
$$ ç = I $$
 since the hermiticity of the $§$ matrices implies that of the $©$
matrices.  We then can also define a metric to raise and lower indices in
terms of these two metrics, by contracting the dotted (or undotted)
indices:  In matrix notation, we then have
$$ (C^T ç)^{-1}©^T(C^T ç) = (-1)^n ©âÜâG = - (C^T ç)^{-1}G^T(C^T ç) $$
 For all cases except SO(4$m$), this also defines the symmetry properties
of the generalized $§$ matrices:  They can be defined as $C^T ç©$ for
SO(2$n$+1), and as its diagonal blocks with respect to $¸_à$ for
SO(4$m$+2); but for SO(4$m$) it's off-diagonal, so the generalized $§$
matrices appearing there carry one each of the two different kinds of
spinor indices, and thus have no symmetry.  Then we rewrite the above
result as
$$ (C^T ç©)^T = (-1)^{n(n-1)/2}(C^T ç©) $$

Ü2. Wick rotation

The indefinite-metric groups SO($D_+$,$D_-$) ($D_+±0±D_-$) can be
treated by Wick rotation: giving $i$'s to $D_-$ of the $©_a$'s, so the
corresponding components of $ú_{ab}$ get minus signs.  Since this affects
$©^T$ in the same way as $©$, the metric $C^T ç$ is unchanged.  In other
words, 
$$ C^T ç = C_E^Tââ(ç_E=I) $$
 in terms of the Euclidean $C$ of the previous subsection.  However, $©*$
and $©ÿ$ are affected in the opposite way to $©$ and $©^T$ ($-i$'s instead
of $i$'s).  $ç$ then becomes (up to normalization) the product of the
timelike $©$'s,
$$ ç ¾ Þ_{ú_{aa}<0} å2©_a $$
 which also determines the modification of $C$.  In the above equations
for $©ÿ$ and $©^T$, we then find a factor of $-$1 for each rotated
dimension, coming from anticommutation with each timelike $©$ in $ç$,
so we redefine all $©$'s by an overall factor of $i^{D_-}$ to preserve their
pseudohermiticity.  This changes the normalization to
$$ G_{ab} = (-1)^{D_-}ü[ ©_a, ©_b ],ââÓ ©_a, ©_b Õ = (-1)^{D_-}ú_{ab} $$
 In odd dimensions the $§$ matrices are the $©$ matrices (up to
multiplication by one of the metrics), while in even dimensions the $©$
matrices consist of two off-diagonal blocks of the $§$ matrices.  To write
actions we also need the ``dual" Dirac spinor, in the sense of a
Hilbert-space inner product,
$$ Ðï = ïÿç $$
 with $ç$ as defined above.  In particular, it is just $å2©_0$ in $D_-=1$.

We then find (e.g., using the explicit representation given above) that
$C$ has the same properties with regard to symmetry and $©_{-1}$ for
SO($D_+$,$D_-$) as for SO($D_+ -1$, $D_- -1$).  Thus, the properties of
these metrics on the ÓirreducibleÕ (as opposed to Dirac) spinors follow
easily from the Euclidean case by using
$$ C^T ç:ââSO(D_+,D_-) Á SO(D_+ +D_-) $$
$$ C:ââSO(D_+,D_-) Á SO(D_+ -D_-) $$
 Then the properties of $ç$ follow from the above two in the cases where
all 3 exist; the few cases where only $ç$ exists, which have $D$ even,
follow from the next higher $D$ (increasing $D_+$ by 1).

\x XC2.1 Find the explicit $©$ matrices for $D=2$ and 4 from the
construction of the previous section, and apply this Wick rotation. 
Compare the results with the conventions of subsections VIIB5 and IIA6.

We can use these properties to determine that the number of real
components $D'$ of an irreducible spinor is
$$ D' = 2^{[D-2+f(D_+ -D_-)]/2},ââ
	\vbox{\offinterlineskip \halign{ &¼\hfil#\hfil¼\cr
	\strut $x¼mod¼8$ & \vrule & 0 & 1 & 2 & 3 & 4 & 5 & 6 & 7 \cr
	& \vrule height2pt &&&&&&&&\cr
	\noalign{\hrule}
	& \vrule height2pt &&&&&&&&\cr
	\strut $f(x)$ & \vrule & 0 & 1 & 2 & 3 & 2 & 3 & 2 & 1 \cr}} $$

The complete results can be summarized by the following table, showing
for each case of SO($D-D_-$,$D_-$), for $D$ mod 8 and $D_-$ mod 4, the
types of irreducible spinors $Æ$, the types of metrics $ú$ (symmetric)
and $¯$ (antisymmetric) for these irreducible spinors, and the type of
generalized $§$ matrices (and its symmetry, where relevant):

$$ \vbox{\offinterlineskip
\hrule
\halign{ &\vrule#&\strut¼\hfil#\hfil¼\cr
height2pt&\omit&\omit&\omit&\hskip2pt\vrule&\omit&&
	\omit&&\omit&&\omit&\cr
&&\omit& $D_-$ &\hskip2pt\vrule& 0 && 1 && 2 && 3 &\cr
& $D$ &\omit&&\hskip2pt\vrule& Euclidean && Lorentz && 
	conformal &&&\cr 
height2pt&\omit&\omit&\omit&\hskip2pt\vrule&\omit&&
	\omit&&\omit&&\omit&\cr
\noalign{\hrule}
height2pt&\omit&\omit&\omit&\hskip2pt\vrule&\omit&&
	\omit&&\omit&&\omit&\cr
\noalign{\hrule}
height2pt&\omit&\omit&\omit&\hskip2pt\vrule&\omit&&
	\omit&&\omit&&\omit&\cr
&&\omit&&\hskip2pt\vrule& $Æ_Œ$ $Æ_{Œ'}$ && $Æ_Œ$ $Æ_{ÀŒ}$ &&
	$Æ_Œ$ $Æ_{Œ'}$ && $Æ_Œ$ $Æ_{ÀŒ}$ &\cr
& 0 &\omit&&\hskip2pt\vrule& $ú^{Œº}$ $ú_{ÀŒ}{}^º$ $ú^{ÀŒº}$ &&
	$ú^{Œº}$ && $ú^{Œº}$ $¯_{ÀŒ}{}^º$ $¯^{ÀŒº}$ && $ú^{Œº}$ &\cr
&&\omit&&\hskip2pt\vrule& $§_{Œº'}$ && $§_{ŒÀº}$ && $§_{Œº'}$ &&
	$§_{ŒÀº}$ &\cr 
height2pt&\omit&\omit&\omit&\hskip2pt\vrule&\omit&&
	\omit&&\omit&&\omit&\cr
\noalign{\hrule}
height2pt&\omit&\omit&\omit&\hskip2pt\vrule&\omit&&
	\omit&&\omit&&\omit&\cr
&&\omit&&\hskip2pt\vrule& $Æ_Œ$ && $Æ_Œ$ && $Æ_Œ$ && $Æ_Œ$ &\cr
& 1 &\omit&&\hskip2pt\vrule& $ú^{Œº}$ $ú_{ÀŒ}{}^º$ $ú^{ÀŒº}$ && 
	$ú^{Œº}$ $ú_{ÀŒ}{}^º$ $ú^{ÀŒº}$ && $ú^{Œº}$ $¯_{ÀŒ}{}^º$ $¯^{ÀŒº}$
	&& $ú^{Œº}$ $¯_{ÀŒ}{}^º$ $¯^{ÀŒº}$ &\cr
&&\omit&&\hskip2pt\vrule& $§_{(Œº )}$ && $§_{(Œº )}$ && $§_{(Œº )}$ 
	&& $§_{(Œº )}$ &\cr 
height2pt&\omit&\omit&\omit&\hskip2pt\vrule&\omit&&
	\omit&&\omit&&\omit&\cr
\noalign{\hrule}
height2pt&\omit&\omit&\omit&\hskip2pt\vrule&\omit&&
	\omit&&\omit&&\omit&\cr
&&\omit&&\hskip2pt\vrule& $Æ_Œ$ $Æ^Œ$ && $Æ_Œ$ $Æ^Œ$ && $Æ_Œ$
	$Æ^Œ$ && $Æ_Œ$ $Æ^Œ$ &\cr
& 2 &\omit&&\hskip2pt\vrule& $ú^{ÀŒº}$ && $ú_{ÀŒ}{}^º$ && 
	$¯^{ÀŒº}$ && $¯_{ÀŒ}{}^º$ &\cr
&&\omit&&\hskip2pt\vrule& $§_{(Œº )}$ $§^{(Œº )}$ && $§_{(Œº )}$
	$§^{(Œº )}$ &&$§_{(Œº )}$ $§^{(Œº )}$ && $§_{(Œº )}$ $§^{(Œº )}$ &\cr 
height2pt&\omit&\omit&\omit&\hskip2pt\vrule&\omit&&
	\omit&&\omit&&\omit&\cr
\noalign{\hrule}
height2pt&\omit&\omit&\omit&\hskip2pt\vrule&\omit&&
	\omit&&\omit&&\omit&\cr
&&\omit&&\hskip2pt\vrule& $Æ_Œ$ && $Æ_Œ$ && $Æ_Œ$ && $Æ_Œ$ &\cr
& 3 &\omit&&\hskip2pt\vrule& $¯^{Œº}$ $¯_{ÀŒ}{}^º$ $ú^{ÀŒº}$ &&
	$¯^{Œº}$ $ú_{ÀŒ}{}^º$ $¯^{ÀŒº}$ && $¯^{Œº}$ $ú_{ÀŒ}{}^º$ $¯^{ÀŒº}$ 
	&& $¯^{Œº}$ $¯_{ÀŒ}{}^º$ $ú^{ÀŒº}$ &\cr
&&\omit&&\hskip2pt\vrule& $§_{(Œº )}$ && $§_{(Œº )}$ && 
	$§_{(Œº )}$ && $§_{(Œº )}$ &\cr 
height2pt&\omit&\omit&\omit&\hskip2pt\vrule&\omit&&
	\omit&&\omit&&\omit&\cr
\noalign{\hrule}
height2pt&\omit&\omit&\omit&\hskip2pt\vrule&\omit&&
	\omit&&\omit&&\omit&\cr
&&\omit&&\hskip2pt\vrule& $Æ_Œ$ $Æ_{Œ'}$ && $Æ_Œ$ $Æ_{ÀŒ}$ &&
	$Æ_Œ$ $Æ_{Œ'}$ && $Æ_Œ$ $Æ_{ÀŒ}$ &\cr
& 4 &\omit&&\hskip2pt\vrule& $¯^{Œº}$ $¯_{ÀŒ}{}^º$ $ú^{ÀŒº}$ &&
	$¯^{Œº}$ && $¯^{Œº}$ $ú_{ÀŒ}{}^º$ $¯^{ÀŒº}$ && $¯^{Œº}$ &\cr
&&\omit&&\hskip2pt\vrule& $§_{Œº'}$ && $§_{ŒÀº}$ && $§_{Œº'}$ &&
	$§_{ŒÀº}$ &\cr 
height2pt&\omit&\omit&\omit&\hskip2pt\vrule&\omit&&
	\omit&&\omit&&\omit&\cr
\noalign{\hrule}
height2pt&\omit&\omit&\omit&\hskip2pt\vrule&\omit&&
	\omit&&\omit&&\omit&\cr
&&\omit&&\hskip2pt\vrule& $Æ_Œ$ && $Æ_Œ$ && $Æ_Œ$ && $Æ_Œ$ &\cr
& 5 &\omit&&\hskip2pt\vrule& $¯^{Œº}$ $¯_{ÀŒ}{}^º$ $ú^{ÀŒº}$ &&
	$¯^{Œº}$ $¯_{ÀŒ}{}^º$ $ú^{ÀŒº}$ && $¯^{Œº}$ $ú_{ÀŒ}{}^º$ $¯^{ÀŒº}$ 
	&& $¯^{Œº}$ $ú_{ÀŒ}{}^º$ $¯^{ÀŒº}$ &\cr
&&\omit&&\hskip2pt\vrule& $§_{[Œº ]}$ && $§_{[Œº ]}$ && $§_{[Œº ]}$ 
	&& $§_{[Œº ]}$ &\cr 
height2pt&\omit&\omit&\omit&\hskip2pt\vrule&\omit&&
	\omit&&\omit&&\omit&\cr
\noalign{\hrule}
height2pt&\omit&\omit&\omit&\hskip2pt\vrule&\omit&&
	\omit&&\omit&&\omit&\cr
&&\omit&&\hskip2pt\vrule& $Æ_Œ$ $Æ^Œ$ && $Æ_Œ$ $Æ^Œ$ && $Æ_Œ$
	$Æ^Œ$ && $Æ_Œ$ $Æ^Œ$ &\cr
& 6 &\omit&&\hskip2pt\vrule& $ú^{ÀŒº}$ && $¯_{ÀŒ}{}^º$ && $¯^{ÀŒº}$ 
	&& $ú_{ÀŒ}{}^º$ &\cr
&&\omit&&\hskip2pt\vrule& $§_{[Œº ]}$ $§^{[Œº ]}$ && $§_{[Œº ]}$
	$§^{[Œº ]}$ &&$§_{[Œº ]}$ $§^{[Œº ]}$ && $§_{[Œº ]}$ $§^{[Œº ]}$ &\cr 
height2pt&\omit&\omit&\omit&\hskip2pt\vrule&\omit&&
	\omit&&\omit&&\omit&\cr
\noalign{\hrule}
height2pt&\omit&\omit&\omit&\hskip2pt\vrule&\omit&&
	\omit&&\omit&&\omit&\cr
&&\omit&&\hskip2pt\vrule& $Æ_Œ$ && $Æ_Œ$ && $Æ_Œ$ && $Æ_Œ$ &\cr
& 7 &\omit&&\hskip2pt\vrule& $ú^{Œº}$ $ú_{ÀŒ}{}^º$ $ú^{ÀŒº}$ && 
	$ú^{Œº}$ $¯_{ÀŒ}{}^º$ $¯^{ÀŒº}$ && $ú^{Œº}$ $¯_{ÀŒ}{}^º$ $¯^{ÀŒº}$ 
	&& $ú^{Œº}$ $ú_{ÀŒ}{}^º$ $ú^{ÀŒº}$ &\cr
&&\omit&&\hskip2pt\vrule& $§_{[Œº ]}$ && $§_{[Œº ]}$ && $§_{[Œº ]}$ 
	&& $§_{[Œº ]}$ &\cr
height2pt&\omit&\omit&\omit&\hskip2pt\vrule&\omit&&
	\omit&&\omit&&\omit&\cr}
\hrule} $$

\x XC2.2  The vectors of SO($D_+$,$D_-$) for 3$²D_++D_-²$6 were
expressed as tensors with two spinor indices, with the appropriate
symmetry and tracelessness conditions, in subsection IC5.
ªa Show that the spinor metrics found from the Dirac analysis are
sufficient to identify each of these covering groups.
ªb Show the equivalence of each orthogonal group to its covering group
by showing that (1) the Lie algebras have the same dimension, and (2) the
determinant (or its square root) of this tensor gives the appropriate
orthogonal metric.  (Hint:  Do the cases $D_++D_-$ = 4 and 6 first, and
specialize to 3 and 5.)

\x XC2.3  Consider the groups SO(n,n) and SO(n+1,n).  Explicitly construct a
real representation of the $©$ matrices by modifying the method of
subsection XC1, demonstrating that all such spinors are real.

Besides the Dirac spinors and Dirac matrices $©$, and the irreducible
spinors and Pauli matrices $§$, it is also useful to introduce irreducible
real (``Majorana") spinors and corresponding matrices $ý$.  When the
irreducible spinors are already real these are the same, but when the
irreducible spinors are complex this real spinor is just the direct sum of
the irreducible spinor and its complex conjugate, a spinor with twice as
many components.  In general, these generalized Majorana spinors and
matrices have many properties that are independent of the number of
dimensions, but depend on the number of time dimensions:

$$ \vbox{\offinterlineskip
\hrule
\halign{ &\vrule#&\strut¼\hfil#\hfil¼\cr
height2pt&\omit&\hskip2pt\vrule&\omit&&
	\omit&&\omit&&\omit&\cr
& $D_-$ &\hskip2pt\vrule& 0 && 1 && 2 && 3 &\cr
&&\hskip2pt\vrule& Euclidean && Lorentz && 
	conformal &&&\cr 
height2pt&\omit&\hskip2pt\vrule&\omit&&
	\omit&&\omit&&\omit&\cr
\noalign{\hrule}
height2pt&\omit&\hskip2pt\vrule&\omit&&
	\omit&&\omit&&\omit&\cr
&&\hskip2pt\vrule& $Æ_Œ$ $Æ_{Œ'}$ && $Æ_Œ$ $Æ^Œ$ &&
	$Æ_Œ$ $Æ_{Œ'}$ && $Æ_Œ$ $Æ^Œ$ &\cr
&&\hskip2pt\vrule& $ú^{Œº}$ &&&& $¯^{Œº}$ &&&\cr
&&\hskip2pt\vrule& $ý_{Œº'}$ && $ý_{(Œº)}$ $ý^{(Œº)}$ && $ý_{Œº'}$ &&
	$ý_{[Œº]}$ $ý^{[Œº]}$ &\cr 
height2pt&\omit&\hskip2pt\vrule&\omit&&
	\omit&&\omit&&\omit&\cr}
\hrule} $$
 For $D$ odd, there is only one irreducible spinor, so there is a metric
$M^{μ}$ or $M^{μ'}$ to relate the two spinors listed.  For $D-2D_-$
twice odd (2 $mod$ 4), the original irreducible spinor was complex, so
there is a metric representing a U(1) generator that rotates the complex
spinor and its complex conjugate oppositely.  (I.e., it's the identity on the
complex spinor and minus the identity for the complex conjugate.)  For
$D-2D_-=3,4,5¼mod¼8$, the original spinor was pseudoreal, and this U(1)
can be extended to an SU(2):  Since the complex spinor and its complex
conjugate transform the same way under the orthogonal group, they can
be paired as a doublet of SU(2).  This doubled representation is a real
representation of the orthogonal group $°$ SU(2), since the direct product
of the two antisymmetric charge-conjugation matrices is symmetric (two
$-$'s under transposition). 

Ü3. Other spins

Before considering supersymmetry in higher dimensions, we first study
representations of the Poincar«e group there.  From the general analysis
of section IIB, we know that general on-shell representations follow
from the massless ones, which can be classified by their representation
of the lightcone little group SO(D$-$2).  Specifically, the bosons can be
described as traceless tensors of a certain symmetry (labeled by a Young
tableau), while the fermions can be labeled as the direct product of such
tensors with an irreducible spinor, with a tracelessness condition imposed
between any vector index and the spinor index using a $©$ or $§$ matrix. 
Similar methods can be used to find the off-shell representations in
terms of representations of SO(D$-$1,1), but without subtracting traces. 
(For full details, see chapter XII.)  The gauge degrees of freedom can be
subtracted from these Lorentz representations by dropping all lower
vector indices with the value ``$-$", by the usual lightcone gauge
condition; this tells us the number of total physical + auxiliary degrees of
freedom.

In practice, the only interesting massless fields in higher dimensions are:
\item{(1)} the metric (graviton), 
\item{(2)} totally antisymmetric tensors (including scalars and vectors), 
\item{(3)} spin-3/2 (gravitino), desribed by
vector$°$spinor, and 
\item{(4)} spinors.  

\noindent By the methods described above, the
counting of physical, auxiliary, and gauge degrees of freedom for these
fields is (where $D'$ is the number of components of an irreducible spinor
of SO(D$-$1,1) --- see the previous subsection):

$$ \vbox{\offinterlineskip
\hrule
\halign{ &\vrule#&\hbox{\vrule height13pt depth5pt width0pt}¼
	\hfil#\hfil¼\cr
height2pt&\omit&\hskip2pt\vrule&\omit&&\omit&&\omit&\cr
& field &\hskip2pt\vrule& physical && auxiliary && gauge &\cr
height2pt&\omit&\hskip2pt\vrule&\omit&&\omit&&\omit&\cr
\noalign{\hrule}
height2pt&\omit&\hskip2pt\vrule&\omit&&\omit&&\omit&\cr
& $h_{(ab)}$ &\hskip2pt\vrule& $üD(D-3)$ && $D$ && $D$ &\cr
& $A_{[a_1...a_n]}$ &\hskip2pt\vrule& $D-2\choose n$ &
	& $D-2\choose n-1$ && $D-1\choose n-1$ &\cr
& $Æ_{aŒ}$ &\hskip2pt\vrule& $üD'(D-3)$ && $üD'(D+1)$ && $D'$ &\cr
& $_Œ$ &\hskip2pt\vrule& $üD'$ && $üD'$ && $0$ &\cr
height2pt&\omit&\hskip2pt\vrule&\omit&&\omit&&\omit&\cr}
\hrule} $$

\x XC3.1  Derive all the entries in the table.  For each type of field, find the
ÓminimumÕ $D$ for which physical degrees exist.

We next consider exactly how many higher dimensions are relevant.  From
the previous subsection, we see that an irreducible spinor (which we use
for the supersymmetry generators) has 1 component in D=2, 2 in D=3, 4 in
D=4, 8 in D= 5 or 6, 16 in D=7, 8, 9, or 10, 32 in D=11, etc.  Since the
maximal Lorentz symmetry can be obtained by looking at the maximum D
for which a certain size spinor exists, we see that the appropriate D for
which an irreducible spinor reduces to N irreducible spinors (for
N-extended supersymmetry) in D=4 is

$$ \vbox{\offinterlineskip
\hrule
\halign{ \vrule # & \vrule # \vrule \cr
height2pt & \hfil \cr
¼N & ¼D \cr
height2pt & \hfil \cr
\noalign{\hrule}
height2pt & \hfil \cr
\strut ¼1 & ¼4 \cr
\strut ¼2 & ¼6 \cr
\strut ¼4 & ¼10 \cr
\strut ¼8 & ¼11 \cr }
\hrule} $$
 etc.  From the discussion of subsection IIC5, we know that supergravity
exists only for N$²$8, and super Yang-Mills only for N$²$4.  This means
that simple supergravity (i.e., any supergravity) exists only for D$²$11,
and simple super Yang-Mills for D$²$10.  Since theories with massless
states of spin>2 are not of physical interest (in fact, no interacting
examples have been constructed), we can restrict ourselves to looking at
just D=4, 6, 10, and 11.  In general, an irreducible multiplet in some D can
become reducible in lower D.  However, since irreducible multiplets of
supersymmetry are constructed as the direct product of the smallest
representation of supersymmetry with an arbitrary representation of the
Poincar«e group, this reducibility corresponds directly to the reducibility of
that Poincar«e representation, which occurs simply because the Lorentz
group gets smaller upon reduction.  In particular, the smallest
representation of supersymmetry is itself irreducible.  For the case of
simple supersymmetry, this is the scalar multiplet (scalars and spinors) in
D=6, the vector multiplet (super Yang-Mills: vectors, spinors, and scalars)
in D=10, and supergravity in D=11.  The statement that it is the smallest
multiplet in that number of dimensions is directly related to the fact that
it does not exist in higher dimensions.

Ü4. Supersymmetry

We first generalize to arbitrary dimensions some definitions used earlier: 
To discuss the properties of supersymmetry that are common to all
dimensions (but one time), it's most convenient to use the Majorana form
$$ Óq_Œ,q_ºÕ = ý^a_{Œº}p_a $$
 which is consistent with the general symmetry of these matrices. The
supersymmetry generators are then
$$ q_Œ = -i{»\over »Ï^Œ} +üý^a_{Œº}Ï^º{»\over »x^a} $$
 and $·^Œ q_Œ$ generates the infinitesimal transformations
$$ ¶Ï^Œ =·^Œ,ââ¶x^a = iüý^a_{Œº}·^Œ Ï^º $$
 where $(q_Œ)ÿ=-q_Œ$.  The covariant derivatives are
$$ d_Œ = {»\over »Ï^Œ} +üý^a_{Œº}Ï^º p_a $$
 and they satisfy the same algebra as supersymmetry
$$ Ód_Œ,d_ºÕ = ý^a_{Œº}p_a $$
 but with the opposite hermiticity condition $(d_Œ)ÿ=+d_Œ$.  The
invariant infinitesimals are
$$ dÏ^Œ,ââdx^a +iü(dÏ^Œ)ý^a_{Œº}Ï^º $$
 Superfields can be expanded as either
$$ ì(x,Ï) = Ä(x) +Ï^Œ Æ_Œ(x) +... $$
 or
$$ Æ_Œ = d_Œ ì,ââ... $$
 giving the transformations
$$ ¶Ä = ·^Œ Æ_Œ,ââ¶Æ_Œ = -i·^{º}üý^a_{Œº}»_a Ä +...,ââ... $$

Representations can be found as for D=4; we don't have twistors in
general, but we can always use a lightcone frame.  We first need to
define $ý^{aŒº}$, which in general is independent of $ý^a_{Œº}$ (only the
latter was needed to define supersymmetry above):  The analog of the
Dirac anticommutation relations (which can be reconstructed if we
combine the two $ý$'s, as generalized $§$'s, to form a generalized $©$) is
$$ ý^{(a}_{Œ©}ý^{b)©º} = ú^{ab}¶_Œ^º $$
 In the lightcone frame the momentum is just $p^a=¶^a_+ p^+$ with
$p^+=à1$ being the sign of the (canonical) energy.  In this frame we have
the constraint $ý^-q=0$.  This projects away half the $q$'s, since
$-ý^àý^¦$ are projection operators: Using the anticommutation relations
of $ý^à$,
$$ ¸_à = -ý^àý^¦âÜâ¸_à{}^2 = ¸_à,¼¸_+¸_- = ¸_-¸_+ = 0,¼
	¸_+ + ¸_- = 1 $$
 The equality of the sizes of the two subspaces follows from parity
symmetry, $ý^àªý^¦$.  We thus need to consider only half of the $q$'s,
namely $ý^+q$.  We therefore switch to a notation where we consider the
truncated spinor $q_µ$ with just that half of the components.  This
``lightcone spinor" is an irreducible spinor of SO(D$-$2).  In a Majorana
basis it satisfies the same commutation relations as Dirac matrices,
$$ Óq_µ,q_ÃÕ = ¶_{µÃ} $$
 Since $q_µ$ has an even number of components ($2^n$, $n>0$) 
in $D>3$, the states that
represent this algebra form a Dirac spinor of SO($2^n$) that is
reducible to two Weyl spinors.  (These spinors should not be confused
with those of SO(D$-$2), such as $q_µ$, which is a vector of this
SO($2^n$).)  Since supersymmetry takes each of these ``spinors" into the
other, one spinor contains all the bosons, while the other contains all the
fermions.  There are an equal number of physical boson and fermion
states because the two Weyl spinors are equal in size.  Since
SO(D$-$2)$¤$SO($2^n$), each Weyl spinor of SO($2^n$) is reducible with
respect to SO(D$-$2).  The only exceptions are (1) D=4, where
SO(D$-$2)=SO($2^n$)=SO(2), and there is one bosonic state and one
fermionic one, and (2) D=10, where SO(D$-$2)=SO($2^n$)=SO(8).

\x XC4.1  Let's look more closely at these exceptions:
 ªa Show that SO(D$-$2)=SO($2^n$) only in D=3,4,6,10.
 ªb Show that in D=6 the bosons form a reducible representation of
the little group SO(D$-$2).  How is this possible, when the group
SO($2^n$) is the same?
 ªc For D=10, what representations of the little group are the bosons
and the fermions?  Compare this to the representations of SO($2^n$)
formed by the bosons, fermions, and $q$ itself, and apply this
``symmetry" to the cases D=4,6.

This ``Dirac spinor" of SO($2^n$) is the smallest representation of
supersymmetry.  It can also be represented in terms of anticommuting
coordinates, by dividing up $q_µ$ into two halves, one of which is complex
coordinates, the other half being both the complex and canonical
conjugate (as for the fermionic harmonic oscillators of exercise IA2.3). 
The most general representation of supersymmetry is then the direct
product of this one with an arbitrary representation of the Poincar«e
group.

All the results of this section can be extended to ``extended
supersymmetry", with supersymmetry generators $q_{iŒ}$ for an
N-valued ``internal" index $i$, as expected from our discussion of
supergroups in subsection IIC4:  For example, in D=4 the supergroup
describing extended conformal supersymmetry, SU(2,2|N), includes
conformal symmetry SU(2,2), internal symmetry U(N), N supersymmetries,
and N S-supersymmetries.  In general, the supersymmetries then satisfy
the algbera
$$ Óq_{iŒ},q_{jº}Õ = ¶_{ij}ý^a_{Œº}p_a $$
 The smallest representation of an extended supersymmetry follows
as before, where now the complete lightcone $q$ acts as Dirac
matrices for SO(N$2^n$).  Other representations are again found by
direct product, now between this smallest supersymmetry
representation and an arbitrary representation of both Poincar«e and the
internal symmetry.  For the more interesting cases, where N itself is a
power of 2, the smallest representation can also be derived by
dimensional reduction from higher dimensions of N=1 (``simple")
supersymmetry, changing the higher-dimensional algebra only by setting
some components of the momentum to vanish, and noting that a spinor of
higher dimensions reduces to many spinors, as clear from our explicit
construction earlier.  (Other representations tend to be reducible, since
the Poincar«e representation in the direct product is reducible upon
dimensional reduction.)  Dimensional reduction can also be defined for an
action (for supersymmetric or nonsupersymmetric theories), by again
setting the derivatives with respect to the ``extra" coordinates to
vanish, and also restricting the integration to the reduced set of
coordinates.  Another interpretation is that we expand the fields over all
momentum modes in the extra coordinates, and then drop all but the zero
(constant) modes.

We also recall from subsection XC2 the index structure of spinors in D=6,
10, and 11, which we need to write supersymmetry covariant
derivatives.  We thus have, for simple supersymmetry,
$$ \li{ D=6: &âÓd_{iŒ},d_{jº}Õ =  -C_{ji}i»_{Œº} \cr
	D=10: &âÓd_Œ,d_ºÕ =  -§^a_{Œº}i»_a \cr
	D=11: &âÓd_Œ,d_ºÕ = -§^a_{Œº}i»_a  \cr} $$
 where in the case of D=6 we have taken advantage of the fact that
SO(5,1)=SU*(4) to eliminate vector indices, and introduced the SU(2) index
$i$ for spinors to make them Majorana.

Ü5. Theories

We first consider the scalar multiplet in D=6.  The constraints and field
equations are given by the statement, in terms of supersymmetry
covariant derivatives, that there are only scalars and spinors on shell,
and by supersymmetry their physical polarizations must be equal in
number.  Since a spinor has 4 polarizations in D=6, we must have 4 real
scalars, and thus
$$ d_{iŒ}Ä_{jk'} = C_{ji}Æ_{k'Œ} $$
 The second SU(2) index $k'$ is introduced again to make a spinor (this
time the field) Majorana, and performs a similar service for the scalars. 
This one equation is sufficient to completely describe this multiplet on
shell in the free case; interactions require derivatives, so we won't
consider them here.  This multiplet reduces to N=2 in D=4 in a very simple
way:  The SU(2) index on $d$ labels the 2 supersymmetries, and the
4-component spinor index reduces in the obvious way to SL(2,C) indices,
$Œ£(Œ,ÀŒ)$, with appropriate 6D spinor conventions.

\x XC5.1  Show the equations given for the 6D scalar multiplet give the
complete field equations for all the components, and that only the scalars
and spinors shown explicitly in that equation survive on shell.

This six-dimensional theory gives a simple example of nontrivial
dimensional reduction:  Assume we have a 5-dimensional theory
with a nontrivial U(1) symmetry.  Then we can dimensionally reduce by
choosing the fields to depend on the fifth coordinate in such a way that
the fifth component of the momentum of each field is equal to a constant
$m$ (with dimensions of mass) times its U(1) charge $Q$:
$$ p_4 = Z = mQ $$
 This is consistent at the interacting level because each term in the action
satsifies conservation of the U(1) charge as well as conservation of
momentum.  This is equivalent to how we introduced masses by
dimensional reduction in subsection IIB4 for free fields, 
since any free field can be
``complexified".  This has an interesting effect on the supersymmetry
algebra:  It introduces a U(1) charge $Z$ (called ``central" because it
commutes with the rest of the algebra).  For example, if we start with
the 6D supersymmetry algebra (like the above algebra for the
supersymmetry covariant derivatives), introduce the central charge in
reducing to 5, and then do an ordinary reduction to 4 (or vice versa), the
supersymmetry algebra becomes (see subsection IVC7)
$$ Óq_{iŒ},Ðq^j_{Àº}Õ = ¶_i^j p_{ŒÀº},âÓq_{iŒ},q_{jº}Õ = C_{Œº}C_{ij}Z,â
	ÓÐq^i_{ÀŒ},Ðq^j_{Àº}Õ = C_{ÀŒÀº}C^{ij}Z $$
 If the higher-dimensional theory was massless, then $p^2+Z^2=0$ for
the 4D theory.  More generally, if the higher-dimensional theory already
had masses before the central charge was introduced, then by
supersymmetry it satisfied $p^2+M_0^2=0$, $M_0^2³0$ (since
supersymmetry always has positive potentials), while afterwards the 4D
theory satisfies
$$ p^2 +Z^2 +M_0^2 = 0âÜâM^2 = M_0^2 +Z^2 ³ Z^2 $$
 where $M$ is the 4D mass, in terms of the higher-D mass $M_0$. 
However, in general, in the absence of central charges, massive
representations of supersymmetry are bigger than massless ones
(because there are twice as many independent supersymmetry
generators on shell, since $q$ is a spinor with 1 helicity for the massless
case, but an SU(2) doublet for the massive).  So, $M^2=Z^2>0$ has the
advantage of allowing smaller massive representations than when
$M^2>Z^2=0$ or when $M^2>Z^2>0$.  Note that when $M^2=Z^2$, so all
masses arise from the central charge, (total) mass is conserved, just as in
nonrelativistic physics, although in the relativistic case the mass $Z$ can
be negative.  (Of course, its square is always positive, as is physical
energy.  The relation between the relativistic and nonrelativistic cases
can be understood through dimensional reduction:  See exercise IA4.5. 
The mass is also a central charge for the Galilean group, but there the
reduction is for a lightlike dimension.)

In the present case, we can choose our U(1) symmetry to be a subgroup
of the extra SU(2) internal symmetry ($k'$ index) of the 6D scalar
multiplet.  Note that the algebra of the $d$'s is modified in the same way
as that of the $q$'s.

Super Yang-Mills is a bit more interesting, because interactions are
easier to introduce.  From the counting arguments given in subsection XC3,
we see that a supersymmetric theory consisting of 1 vector and 1 spinor
can exist in D=3, 4, 6, or 10.  This corresponds directly with our analysis of
the largest dimensions for simple supersymmetries:  Dimensional
reduction of a vector gives also scalars, so the condition of no scalars
gives maximum dimensions.  We now make an analysis similar to that of
the previous subsection:  By dimensional analysis for physical fields, and
using single-Majorana-spinor-index notation,
$$ Óá_Œ,á_ºÕ = -ý^a_{Œº}iá_a $$
$$ [á_Œ,á_a] = ý_{aŒº}W^º $$
$$ [á_a,á_b] = iF_{ab} $$
 Applying the Jacobi (Bianchi) identities, we find
$$ ý_{a(Œº}ý^a_{©)¶} = 0 $$
 This identity can be satisfied only in D=3, 4, 6, or 10.  The Bianchi
identities imply the field equations for D=10.

\x XC5.2  Multiply the identity $ý_{a(Œº}ý^a_{©)¶}=0$ by $ý^{bŒº}$, and
use the $ý$ matrix anticommutation relation
$ý^{(a}_{Œº}ý^{b)º©}=ú^{ab}¶_Œ^©$ to show that D=3, 4, 6, or 10.

Similar methods can be applied to D=11 supergravity.  Our component
counting for general dimensions, and our helicity analysis for general
extended supersymmetric theories in D=4 (applied to the dimensionally
reduced theory), can be satisfied by adding to the metric (44 physical
components) and gravitino (128) a third-rank antisymmetric tensor gauge
field (84) $A_{mnp}$ (with field strength $F_{mnpq}=\f16
»_{[m}A_{npq]}$).  The action for the graviton and gravitino are like those
in 4D N=1, while $A$ has not only the obvious quadratic term but also a
``Chern-Simons term":
$$ L = {\bf e}^{-1}[-\f14 R +ÐÆ_m ©^{mnp} á_n Æ_p +\f1{96}(F_{abcd})^2
	+Æ^2 F +Æ^4] $$
$$  +\f1{4É3!(4!)^2}·^{mnpqrstuvwx}A_{mnp}F_{qrst}F_{uvwx} $$
 (There are also more-complicated fermion interaction terms than in 4D
N=1.)  The necessity of the last term can be shown by finding the
component form of the supersymmetry transformations, or by finding the
field equations implied by the superspace formulation.

Ü6. Reduction to D=4

We now look instead at the component formulation of higher-dimensional
super Yang-Mills.  This formulation is off shell except for the lack of
auxiliary fields.  Since the fields are just a vector and a spinor, the
Lagrangian consists of just that of super Yang-Mills coupled to a spinor in
the adjoint representation of the Yang-Mills group.  Upon dimensional
reduction, the vector produces some scalars.  For example, the D=10
theory has an SO(9,1) symmetry, which reduces in D=4 to the
SO(3,1)$°$SO(6) subgroup.  The SO(6) symmetry of the 6 flattened
dimensions is the SU(4) symmetry of the N=4 supersymmetries.  Under
this reduction, the vector becomes $10£(4,1)¢(1,6)$, namely a 4-vector
and scalars that form a 6 of SU(4), while the spinor becomes $16£(4,4)$, a
4D spinor that is also a 4 of SU(4) (like the supersymmetry generators). 

Although $©$ (or $§$, or $ý$) matrices are necessary in D=10, in D=4 we
can convert to spinor notation for both SO(3,1) (=SL(2,C)) and SO(6)
(=SU(4)).  Thus vectors and the Minkowski metric reduce as
$$ V^a £ (V^{ŒÀŒ},V^{ij});ââ
	(V^{ºÀŒ})* ­ ÐV^{ŒÀº} = V^{ŒÀº},ââ(V^{ij})* ­ ÐV_{ij} = ü·_{ijkl}V^{kl} $$
$$ ú_{ab} £ (C_{Œº}C_{ÀŒÀº},·_{ijkl}):ââ
	VÉW £ V^{ŒÀŒ}W_{ŒÀŒ} +üV^{ij}ÑW_{ij} $$
 while spinors and Pauli matrices reduce as
$$ Æ^Œ £ \f1{å2}\pmatrix{ Æ^{iŒ} \cr ÐÆ_i{}^{ÀŒ} \cr },ââ
	§^a_{Œº}V_a £ \pmatrix{ 
	iC_{ºŒ}ÐV_{ij} & ¶_i^j V_{ŒÀº} \cr ¶^i_j V_{ºÀŒ} & iC_{ÀºÀŒ}V^{ij} \cr} $$
$$ ÆÖV £ V_{ŒÀŒ}ü(Æ^{iŒ}Ѝ_i{}^{ÀŒ} -^{iŒ}ÐÆ_i{}^{ÀŒ})
	+üi(ÐV_{ij}Æ^{iŒ}^j{}_Œ +V^{ij}ÐÆ_i{}^{ÀŒ}Ѝ_{jÀŒ}) $$
 The two terms in the 10D Lagrangian then reduce as
$$ \li{ \f18 F^2 &â£â\f18 F^2 +\f18[á,ÐÄ_{ij}]É[á,Ä^{ij}]
		-\f1{32}[ÐÄ_{ij},ÐÄ_{kl}][Ä^{ij},Ä^{kl}] \cr
	Æ^Œ §^a_{Œº}[-iá_a,Æ^º] &â£âÐÆ_i{}^{ÀŒ}[-iá_{ŒÀŒ},Æ^{iŒ}] 
		+üi(Æ^{iŒ}[ÐÄ_{ij},Æ^j_Œ] +ÐÆ_i{}^{ÀŒ}[Ä^{ij},ÐÆ_{jÀŒ}]) \cr} $$

\x XC6.1  Looking at the SU(3) subgroup of SU(4), decompose the states of
N=4 super Yang-Mills into those of N=3.  (Use the analysis of subsection
IIC5 to count states, in SU(N) representations.)  Do the same to
decompose N=4 into N=2 super Yang-Mills plus scalar multiplet, this time
using the SU(2)$°$SU(2) subgroup for which $4£(ü,0)¢(0,ü)$ (i.e.,
$i£(i,i')$).  This is another way of understanding where the second SU(2)
of the scalar multiplet comes from.

\x XC6.2  Derive the commutation relations of the N=4 Yang-Mills
covariant derivatives of subsection IVC7 by dimensional reduction of
those for 10D N=1 given in the previous subsection.  (Don't forget the
scalars come from the components of the vector covariant derivative in
the extra dimensions.)

Dimensional reduction of (super)gravity is an example of the comparative
simplicity of the vierbein (covariant derivative) formalism vs.¼the metric
or even inverse vierbein (differential form) formalisms.  The reason in
this case is that gravity is treated like Yang-Mills theory, and gauge
vectors result from reducing the graviton.  This is seen most easily from
comparison of the coordinate transformation laws:
$$ ¶e_a{}^m = Â^n »_n e_a{}^m -e_a{}^n »_n Â^m $$
$$ ¶e_m{}^a = Â^n »_n e_m{}^a +e_n{}^a »_m Â^n $$
$$ ¶g_{mn} = Â^p »_p g_{mn} +g_{p(m}»_{n)}Â^p $$
 Fixing the index $m=-1$ on $Â^m$ to get the gauge transformations of
an Abelian vector resulting from reduction from one extra dimension, and
setting $»_{-1}=0$ when acting on any field as the definition of reduction,
we see the identification (in an appropriate gauge for the
SO(D,1)/SO(D$-$1,1) generators $M_{-1a}$)
$$ e_a{}^mâ£â\bordermatrix{ & m & -1 \cr 
	\hfill a & e_a{}^m & A_a \cr -1 & 0 & Æ \cr} $$
 where $A$ transforms in the usual way for a gauge vector, and $Æ$ is an
additional scalar.  A more transparent way to write this is as
$$  ­ Â^m »_m,ââe_a ­ e_a{}^m »_m;ââ¶e_a = [Â,e_a] $$
$$  £  +Â^{-1}»_{-1},ââe_a £ (e_a +A_a »_{-1}, Æ»_{-1}) $$
 which makes it clear that reduction has simply U(1)-covariantized the
gauge parameter, transformation, and field, where $»_{-1}$ is the U(1)
generator.  (Under reduction all fields are U(1) neutral.) 
On the other hand, the reduction of $e_m{}^a$, being the inverse of
$e_a{}^m$, and $g_{mn}$, being the square of that, yields nonlinear
reductions, and the U(1) covariantization is not manifest.  (In
particular, in the metric formalism the metric, and thus the U(1) vector,
does not even appear in the covariant derivative, except in terms with its
derivatives.)

\x XC6.3  Derive the result of exercise IXC1.1 by dimensional reduction.

\x XC6.4  Let's work out the details of this simple example, reduction of
pure gravity from one extra dimension:
 ªa  Find the reduction of $c_{ab}{}^c$ by examining the commutators of
the reduced $e_a$.  ($F_{ab}$ comes out directly.)  Using the expression of
the Lagrangian in terms of the $c$'s from exercise IXA5.2, find the
reduced action, including a cosmological term.  (Drop the $Çdx^{-1}$.  We
can think of this as compactification on a circle, independence from
$x^{-1}$ yielding a constant factor upon integration, which can be
absorbed.)
 ªb  The scalar appears in a funny way, seen previously in subsection
IXB5.  Rather than field redefinitions, it is more convenient to reintroduce
local scale invariance (after the reduction), as in subsection IXB5,
introducing the dilaton $Ä$.  Then make a simple redefinition that
replaces $Æ$ and $Ä$ with the ``canonical" fields $Ä_à$.  (The $F^2$ and
cosmological terms then appear with powers of $Ä_à$.)

Reduction from one extra dimension can give only a single (Abelian) gauge
vector, but two or more dimensions can yield nonabelian gauge groups as
the spacetime symmetries of the compactified dimensions.  For example,
compactifying n extra dimensions into an n-sphere gives SO(n+1).  
(However, compactifying to a box with periodic boundary conditions 
gives an Abelian group again.)  The generalization is then
$$ Ââ£â +Â^I G_I,ââe_a⣼( e_a + A_a{}^I G_I, Æ_i{}^I G_I ) $$
where the only dependence on the extra dimensions is implicit in the
group generators $G_I$.  If we add matter fields (before reduction),
then the fields can be constrained to be independent of the extra dimensions
(i.e., singlets of $G_I$) when their indices are flat.

Another possible modification is to make the action of the generators
on matter fields nontrivial.  If we already have an internal symmetry group, 
with generators $öG_I$, identical to that of the $G_I$, then we can impose
on all matter fields $Ä$
$$ G_I Ä = öG_I Ä $$
 to determine their dependence on the extra coordinates.  The fact that
the original higher-dimensional action was invariant under the $öG_I$
guarantees that the resultant dependence on the extra coordinates
will cancel.  The simplest example was applied to supersymmetry
in the previous subsection:  In the Abelian case we can set
$$ -i»_{-1}Ä = möGÄ $$
 where we are free to scale the Abelian generator by a mass parameter $m$
(unlike the nonabelian case, where it would change the algebra).  

Similar results can be obtained for supergravity, but the results are more
complicated, because the scalars (which appear for N>3) appear in
nonlinear $§$ models.  Furthermore, although these models can be
constructed by the coset method discussed in subsection IVA3, the coset
space G/H is noncompact, because the group G is noncompact, although
the subgroup H is compact.  This is a consequence of the fact that the
``compensating" scalars of the group H=U(N) (or SU(8) for N=8) appear with
the wrong-sign kinetic term (as the dilaton even in ordinary gravity). 
Thus, conformal supergravity is coupled to ``matter" with scalars in the
adjoint representation of the noncompact group G, while gauging away
the compensating scalars leaves the physical scalars of the coset space
G/H.  A simpler analog is N=1 supergravity coupled to a scalar multiplet
(see subsection XB7).  This is the same as conformal supergravity coupled
to the matter action $ÐÄÄ-Ѝ$, which has a symmetry G=U(1,1), while N=1
supergravity has a gauge group U(1).  Including Weyl scale invariance
GL(1), the physical scalars then inhabit the coset space
U(1,1)/U(1)$°$GL(1)=SU(1,1)/GL(1).  

In the case of extended supergravity, the group G can be found by noting
that the physical scalars parametrizing G/H form the representation
$Ä^{ijkl}$ (totally antisymmetric, and complex conjugate) of the group H. 
We then look for the group G whose adjoint representation transforms
under the H subgroup as these scalars + adjoint of H.  We can also
determine G by defining group generators for G as $M_i{}^j$ for H, and
$M_{ijkl}$ (and hermitian conjugate $ÑM^{ijkl}$) for G/H, and write
commutation relations consistent with covariance under H.  For N=8 we
also have $ÑM^{ijkl}=\f1{4!}·^{ijklmnpq}M_{mnpq}$ (and the same for the
corresponding physical scalars).  The result for the coset space G/H is
$$ \li{ N=4: &¼SU(4)°SU(1,1)/U(4) = SU(1,1)/U(1) \cr
		5: &¼SU(5,1)/U(5) \cr
		6: &¼SO*(12)/U(6) \cr
		8: &¼E_{7(+7)}/SU(8) \cr } $$
 where $E_{7(+7)}$ is a noncompact form (Wick rotation) of the
exceptional group $E_7$.

An additional complication is that the vectors represent the full H
symmetry only on shell.  For example, for N=2 we have a single vector, as
in electromagnetism.  Maxwell's equations without sources have a U(1)
symmetry, ``S-duality", that transforms $f_{μ}$ by a phase (and
$Ðf_{ÀŒÀº}$ by the opposite), that mixes the field equations with the Bianchi
identities.  With sources, it mixes electric and magnetic charge, since it
mixes electric and magnetic fields.  So, in general we must introduce
both electric and magnetic potentials for each vector.  Furthermore, for
N=6 the vectors appear as both $f^{ij}_{μ}$ and $f_{ijklmnμ}$ (one
extra vector).  (For N=8, the two are related by an $·$ tensor, just as $Ä$
and $ÐÄ$.)  In this version of extended supergravity, all the vectors are
Abelian.  There is also a version where they gauge SO(N), but that theory
has a cosmological constant.

\refs

£1 Georgi,  Óloc. cit.Õ (IB):\\
	Dirac and irreducible spinors for SO(n).
 £2 Kugo and Townsend, Óloc. cit.Õ (IIC):\\
	Dirac and irreducible spinors for SO(n${}_+$,n${}_-$).
 £3 Siegel,  Óloc. cit.Õ:\\
	spinors and $©,§,ý$ matrices in SO(n${}_+$,n${}_-$).
 £4 Brink, Schwarz, and Scherk, Óloc. cit.Õ (IVC):\\
	D=10 super Yang-Mills.
 £5 E. Cremmer, B. Julia, and J. Scherk, \PL 76B (1978) 409:\\
	D=11 supergravity.
 £6 Haag, \Sll opusza«nski, Sohnius, Óloc. cit.Õ (IIC):\\
	central charges.
 £7 W. Nahm, \NP 135 (1978) 149:\\
	arbitrary free representations of supersymmetry in arbitrary
	dimensions, central charges from dimensional reduction.
 £8 W. Siegel, \PL 80B (1979) 220:\\
	higher-dimensional superspace.
 £9 E. Cremmer and B. Julia, \PL 80B (1978) 48, \NP 159 (1979) 141:\\
	N=8 supergravity.
 £10 L. Brink and P. Howe, \PL 88B (1979) 268;\\
	W. Siegel, \NP 177 (1981) 325:\\
	N=8 superspace formulation of supergravity.
 £11 E. Cremmer and S. Ferrara, \PL 91B (1980) 61;\\
	L. Brink and P. Howe, \PL 91B (1980) 384:\\
	superspace formulation of D=11 supergravity.
 £12 Th. Kaluza, ÓSitz. Preuss. Akad. Wiss. Berlin, Math.-phys. Kl.Õ 
	(1921) 966;\\
	Klein, Óloc. cit.Õ (IIB);\\
	H. Mandel, ÓZ. Phys.Õ É39 (1926) 136:\\
	dimensional reduction/compactification of general relativity,
	Abelian gauge field.
 £13 Klein, Pauli, Óloc. cit.Õ (IIIC):\\
	same with nonabelian gauge fields.
 £14 J. Scherk and J.H. Schwarz, \PL 82B (1979) 60:\\
	identifying symmetries of compactified spaces with existing
	internal symmetries --- masses from dimensional reduction in
	interacting theories.

\unrefs

ÚXI. STRINGS

\def\righthead{\rgbo{1 0 1}{A. GENERALITIES}}

There are three areas of application of QCD, as defined by the region of
momentum space they address:  (1) One is perturbative QCD, which applies
to  large relative, ``transverse" velocity of (some of) the constituents of
the hadrons.  In this approach such an amplitude is divided into a half
consisting of high-energy, asymptotically-free partons, which is
calculated perturbatively in the gauge coupling, and a half consisting of
low-energy, confined partons, which is nonperturbative, and therefore not
calculated.

(2) Another area deals with the low-energy behavior of QCD --- with
properties of the vacuum (e.g., broken chiral symmetry), or the
lowest-mass hadrons, scattering at small relative velocities.  This
approach is nonperturbative with respect to the gauge coupling, and
instead perturbs in derivatives, as in first-quantized JWKB.  The methods
used include instantons, lattice QCD, current algebra, dispersion relations,
nonlinear $§$ models, and duality.  This low-energy behavior really says
nothing about confinement, just as the low-energy states of the
hydrogen atom tell us nothing about ionization.

A closely related problem is that the nonperturbative information about
QCD that comes from (electromagnetic-type) duality considerations,
which relates ``weak" coupling to ``strong" coupling as $gª1/g$, is not
really relating quark-gluon physics to hadronic physics, but is relating
quark-gluon physics to monopole physics; i.e., it relates a description of
weakly coupled ``electric" color charges to a similar looking theory of
weakly coupled ``magnetic" color charges.  Thus, the dual theory, being
formally of the same type as the original, except for a relabeling of what
is called ``electric" and what is called ``magnetic", does not give
anything that looks any more like hadrons, or make it any easier to
calculate.

(3) The one nonperturbative approach that does deal with high (hadron)
energies is string theory:  It incorporates hadrons of arbitrarily high
mass, and studies their scattering at high energies.  It also shows that
stringy (hadron-like) behavior is a characteristic of QCD coupling $g®1$,
while $g®0$ or $¥$ have non-stringy (parton-like) behavior:  String
perturbation expands in $G=ln¼g$, not $g$ nor $1/g$; duality is the
symmetry $Gª-G$.  Furthermore, this $G$ is the coupling that defines the
free string, i.e., how partons bind to form strings.  The coupling that
determines how hadrons couple to each other is $1/N_c$, as described
topologically in subsection VC9.  (However, the relation of duality to the
$1/N_c$ expansion is unclear, since duality has been studied so far only in
relation to theories where the group is spontaneously broken to U(1), so
effectively $N_c=1$, or with respect to instantons, which are always
defined for SU(2) subgroups, so effectively $N_c=2$.   Thus, it has not
been possible to apply such duality arguments simultaneously with a
$1/N_c$ analysis.  Similarly, since both these duality approaches deal only
with low-energy behavior, they are difficult to relate to confinement.)

Most of the research effort on string theory has been directed toward
models with ``critical dimension" D$ù$4 (10, 11, or 26):  
To describe physics in the real world of D=4, it is usually
assumed that the extra dimensions choose to ``compactify" to
submicroscopic dimensions, corresponding to length scales well below the
range of present experiments.  
(The extra dimensions cannot be completely
eliminated without losing renormalizability.)  
Such a solution to the classical field
equations with minimal energy is chosen as the ``vacuum", about which
perturbations are performed, but nothing is known to preclude
contributions to the functional integral from other vacuua, whether
4-dimensional, 10-dimensional, or elsewhere.  
Moreover, although superstrings have been
chosen to describe quantum gravity because of their renormalizability
(finiteness), this advantage is lost after compactification, since the
arbitrariness in choice of compactification is tantamount to the loss of
predictability in nonrenormalizable theories.
Furthermore, D=10 superstring theories are
(compactifications of) D=11 membrane theories in disguise, where the
eleventh dimension shows up only nonperturbatively.  Not only is using a
formalism where not all of the dimensions are manifest a technical
obstacle, but the quantum mechanics of membranes suffers from several
problems, including nonrenormalizability.  This suggests that D=10 superstrings
are nonrenormalizable at the nonperturbative level.
On the other hand, renormalizabilty of theories with a finite number of
fields predicts D=4, since theories in higher dimensions are all
nonrenormalizable (or have unbounded potentials: $Ä^3$ theory). 
Furthermore, both experiments with hadrons and theoretical arguments
in QCD suggest the existence of an inherently 4D string theory.  (For
example, the existence of a continuum limit for confining
spacetime-lattice theories requires asymptotic freedom.)

However, historically the true usefulness of such string theories has been
for the concepts and features of field theory they have revealed:  For
example, supersymmetry (sections IIC and IVC, and chapter X), the
Gervais-Neveu gauge (subsection VIB4), topological (1/N) expansion
(subsection VIIC4), first-quantized BRST approach to gauge theory
(chapter XII), and certain simplifications in one-loop amplitudes were all
discovered through studies of 10- and 26-dimensional string theory, even
though they all are now understood more simply through ordinary field
theory.  This is due to the fact that string theories are so complex and
restrictive that they require the most powerful techniques available. 
Clearly such strings are useful toy models for learning about particle field
theories, and about general properties of string theory that might lead to
generalizations to include realistic 4-dimensional string theories.  (In
fact, the first paper on string theory was written in 1747 by d'Alembert,
and was the first appearance of the wave equation and the
d'Alembertian.  Thus, field theory, quantum mechanics, and special
relativity can trace their origins to string theory.)

In subsections IVB1 and VIIC4 we briefly discussed how hadrons are
expected to arise as strings from QCD.  In this chapter we analyze the
dynamics of this mechanism.  We begin by formulating the theory in terms
of strings directly.  Perturbative calculations are performed using
first-quantized path integrals.  (These methods are based on the corresponding ones for the particle from section IIIB and subsections VB1 and VIIIC5, and massless 2D field theory in subsection VIIB5.)  The only experimental evidence for
strings is as a description of hadrons; to some extent the way that QCD
leads to strings can be understood with similar first-quantized methods,
based on random lattices.  

Û7 A. GENERALITIES

In this section we examine some of the general properties of
string theory, shared by all known models, but expected to apply also to
more realistic strings.  These features can be used for phenomenological
applications of string theory, but also may help point to new
generalizations.

String theories are the only known theories that exhibit (S-matrix)
duality.  Unlike other S-matrix approaches, they provide an explicit
perturbative calculational scheme, like field theory, and string theory can
be formulated as a field theory.  Also like field theory, string theory has
consistency conditions at the classical and quantum levels, related to
gauge invariance and renormalizability.

When string theory is used as a unified theory of gravity and other
forces, the most interesting predictions are those for the ``low-energy"
(with respect to the Planck mass) part of the theory.  Although the
possible low-energy limits of known string theories have not all been
explored, the indications are that there are only a few restrictions
beyond the usual field theoretic ones:  
\item{(1)} In a term in the effective action, the power of the dilaton counts the number of loops, since the string coupling is the vacuum value of the dilaton.  
\item{(2)} The spectrum of the closed string is given by the direct
product of two open strings.  
\item{(3)} String theory has noncompact
symmetries, ``S-duality" and ``T-duality", resulting  from the amplitudes
also having this direct-product structure.  

\noindent All these properties survive the
low-energy limit.  In the supersymmetric case, the last property follows
from the first two.  (However, the D=10 superstring is actually a D=11
supermembrane nonperturbatively.  Since the observed features of
hadrons, as well as qualitative arguments from QCD, indicate stringy but
not membrane-like behavior, we will abstract only the perturbative
features of higher-dimensional strings.)

Ü1. Regge theory

In principle there is no difference between a fundamental state and a
bound state:  We can always write an action with every state
represented by an independent field.  Of course, such an action might not
be renormalizable, but that seems more of a formal distinction.  A more
physical one is based on the qualitative property that bound states have
radial and other excitations with related properties, while fundamental
states are more unique.

Regge theory is an approach to bound states that treats them as
fundamental.  A family of states that are different excitations of the
same ground state is treated as a single entity.  Although basically an
approach based on fundamental properties of the S-matrix, when
combined with perturbation theory it leads directly to string theory.

A quantitative definition of this concept follows from a generalization of
a concept seen in perturbative field theory.  In amplitudes following from
Feynman diagrams the nature of intermediate states can be seen from
the momentum-space behavior:  Single-particle states appear as poles (in
the sense of complex analysis) in some momentum invariants,
$1/(p^2+m^2)$, where this $p$ is the sum of some of the external
momenta, representing the momentum of the internal state.  (Any tree
graph is a simple example.)  Two-particle states appear as cuts in these
invariants, where the branch point represents the state where the two
particles are at rest with respect to one another, and the rest of the cut
corresponds to arbitrary relative velocities.  (For example, a one-loop
propagator correction has a branch point at $-p^2=(m_1+m_2)^2$ for
intermediate particles of masses $m_1$ and $m_2$.)  Similar remarks
apply to other multi-particle states.  ``Analytic S-matrix theory" was an
attempt to formulate particle physics in terms of the S-matrix by
replacing the property of locality of the action with ``maximal"
analyticity of the S-matrix in momentum space.  (Of course, unitarity and
Poincar«e invariance can be described easily in terms of the S-matrix; even
analogs of renormalizability can be formulated in terms of certain
properties of the high-energy behavior.)  Unfortunately, the most general
form of such nonanalytic behavior (poles, cuts, etc.), as discovered from
analyzing Feynman diagrams, proved to be too complicated to provide a
practical method for defining a theory.

$$ \figscale{ladder}{3in} $$
\vskip-.1in

Since Poincar«e invariance means that not only momentum is conserved
but also angular momentum, a natural next step was to consider the
analytic behavior in that variable as well.  This behavior is seen already in
nonrelativistic theories; here we will approach the concept in a language
most relevant to relativistic physics.  The simplest example of ``Regge
behavior" is the 4-point S-matrix; this is the relativistic analog of a
nonrelativistic particle in a potential.  (We can think of an infinitely
massive second particle as producing the potential, or separate
center-of-mass and relative coordinates for two finite-mass
particles.)  Also, the Feynman diagrams that appear in the nonrelativistic
problem are ``ladder diagrams":  The sides of the ladder represent the
two scattering particles, while the rungs represent a perturbation
expansion for the potential.  It can be shown that such diagrams give the
leading behavior of this amplitude at high energies.  Here the appropriate
high energy limit is defined in terms of the Mandelstam variables
(see subsection IA4); by high energy we mean, e.g., 
$$ s £ -¥,ât¼fixed $$
We should really look at $s£+¥$ for a physical amplitude, i.e., total center-of-mass energy $£¥$.  But then we would run into the poles in that channel, from ``annihilation", for intermediate states of positive (mass)${}^2$, so instead we take $s£-¥$, which has a well-defined limit, and later analytically continue $Re(s)£+¥$.  

The high-energy behavior of ladder diagrams can be shown to be of the
form (in  units of an appropriate mass)
$$ \A_4(s,t) = k g^2 ý[-Œ(t)](-s)^{Œ(t)},ââŒ(t) = a +g^2 b(t) $$
 where $a$ is a constant that describes the behavior of the tree graph,
and $b(t)$ is determined by the one-loop graph, but gives the leading contribution of all the higher-loop graphs.  (Both, and the constant
$k$, are independent of $g$.)  
This amplitude takes a simple form under a
modified type of ``Sommerfeld-Watson transform":
$$ \A_4(s,t) = È{dJ\over 2¹i}¼ý(-J)(-s)^J ÷\A_4(J,t) 
	= Ý_{J=0}^¥ \f1{J!}s^J ÷\A_4(J,t) 
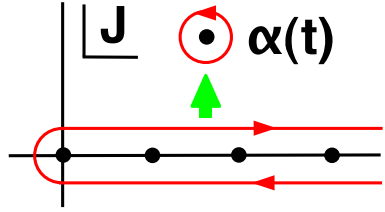
$$
$$ \figscale{Regge}{2in} $$
 The contour integral is taken as clockwise about the positive real
axis to obtain the last form, where it picks up the poles of $ý(-J)$ (see
exercise VIIA2.3b), but can be deformed to surround the singularities of
$÷\A_4$.  In this case, a pole at $J=Œ(t)$ in $÷\A_4$ will reproduce our original
ladder amplitude, while integrating it around the positive real axis gives
$$ ÷\A_4(J,t) = k{g^2\over J-Œ(t)}âÜâ
	 \A_4(s,t) = kÝ_{J=0}^¥ \f1{J!}s^J {g^2\over J-Œ(t)} $$
 which shows that particles of spin $J$ contribute simple poles in $t$ to
the amplitude at $Œ(t)=J$ when $Œ(t)$ can be approximated as linear
near that value.  The spin of the intermediate particle follows from the
$s^J$ factor.  (This is clear from examining a 4-point tree graph where the
external lines are scalars and the internal line carries $J$ indices, and
must contract the momenta of its two ends.  There are also contributions
of lower spins from traces.)  Thus the ``Regge trajectory"
$Œ(t)$ determines not only the high-energy behavior of the amplitude (for
negative $t$), but also the spins and masses of the bound states (for
positive $t$):  Looking at the graph for $J=Œ(t)$, there is a bound
state of spin $J$ and mass $åt$ whenever the curve crosses an integer
value of $J$.  The contribution at $n$ loops in perturbation theory to
$÷\A_4(J,t)$ is a multiple pole $(J-a)^{-(n+1)}$, which contributes to $\A_4(s,t)$ a
term proportional to $(-s)^a[ln(-s)]^n$.  

\x XIA1.1  Calculate the tree scattering amplitude of two spinless particles of equal mass due to the exchange of a particle of spin $J$ with coupling and propagator as given at the end of subsection IIIA4. Show that at the pole in $t$ the leading contribution in $s$ goes as $s^J$.

\x XIA1.2  Consider the amplitude
$$ \A_4 = Ç_0^¥ d ¼e^{ s} [f( )]^{-Œ(t)-1},ââf(0) = 0,âf'(0) ± 0 $$
 where $f$ is Taylor expandable.  By expanding $f$, show that a sum of
Regge amplitudes is obtained, where the ``leading trajectory" is
$Œ(t)$, and there are ``daughter trajectories" $Œ(t)-n$ for positive
integer $n$.

\x XIA1.3  Consider the energy spectrum of the hydrogen atom (nonrelativistic, with spinless constituents).  Show that this corresponds to a leading Regge trajectory of the form
$$ Œ(E) = å{-{E_0\over E}} -1 $$
for some constant $E_0$, with daughter trajectories.

Unfortunately, for field theories with a finite number of fundamental
particles, the trajectories are rather boring, containing only a finite
number of bound states.  In certain cases a trajectory may include one of
the fundamental particles itself (``Reggeization").  Because of the usual
infrared divergences, such calculations can be applied directly to S-matrix
elements only for fundamental massive particles; for fundamental
massless particles, as in confining theories (like QCD), these results
require external lines to be off-shell, and some knowledge of the parton
wave functions is needed.  Regge behavior thus gives a measurable
definition of confinement:  If the scattering amplitudes of color-singlet
states (or color-singlet channels of off-shell amplitudes of
color-nonsinglet states) have linear trajectories, the constitutent
color-nonsinglet particles can be said to be ``confined".  On the other
hand, if the Regge trajectory rises only to finite spin and then falls, as
with the Higgs effect, then there is only ``color screening"; color-singlet
states might not be observable, but we do not see the infinite number of
radial excitations characteristic of confinement.  Another possibility is
that arbitrarily high spin is reached at finite energy:  This is characteristic
of Coloumb binding, and indicates that a new, ``ionized" phase is reached
above that energy.

$$ \figscale{trajectories}{3.5in} $$

Experimentally, hadrons are observed to have Regge behavior with
respect to both high-energy behavior and spectrum.  However, those
Regge trajectories are approximately linear, thus indicating an (near)
infinite number of bound states.  The linearity of the trajectories can be
shown to be related to the relative stability of these unstable particles
(as compared to what is found in ladder approximations).  This suggests a
formulation of the theory of hadrons where the whole Regge trajectory is
treated as fundamental.  It can be shown that in any such ``Regge
theory" based on a perturbation expansion where the ``tree" graphs
have only poles in the angular momentum $J$ (whose accuracy is implied
by the linearity of the observed trajectories), that the theory has a
further property called Dolen-Horn-Schmid ($s$-$t$) ``duality":  This property states that the
amplitude can be expressed as a sum of poles in either the $s$ or $t$
``channel", rather than as a sum over both:
$$ \A_4(s,t) = Ý_n {C_n(t)\over s-s_n} = Ý_n {÷C_n(s)\over t-t_n}ââââ
	\vcenter{\hbox{\fig{dual}}} $$
 This holds even when the sets of particles exchanged in the two channels
are different, due to quantum numbers of the external states.  This
relation has also been experimentally verified (approximately).

Explicit realizations of such ``dual models" of the S-matrix in terms of
first-quantized systems are called ``string theories".  They explain the
linearity of the Regge trajectories by the harmonic-oscillator structure of
the string Hamiltonian, and the duality of the amplitudes by the conformal
invariance (``stretchiness") of the string worldsheet.

Ü2. Topology

The defining concept of the string is that it is a two-dimensional object: 
Just as the particle is defined as a point object whose trajectory through
spacetime is one-dimensional (a worldline), the string has as its
trajectory a two-dimensional surface, the ``worldsheet".  
There are two types of free strings: open (two ends)
and closed (no boundary).  Their worldsheets are a rectangle and a tube
(cylinder).

$$ \fig{string} $$

This leads to a much simpler picture of interactions for strings than for particles.
For particles, one rarely uses first-quantization to describe self-interactions.  Generally, relativistic quantum mechanics is limited to free particles, or particles in a fixed background.  On the other hand, the quantum mechanics of strings is often the best way to describe quantum strings perturbatively in the string coupling, for two reasons:  

(1) For interacting particles, the geometric picture of a worldline becomes a graph, whose geometry is not differentiable at the interaction points, where the curves split.  
For interacting strings, we have instead a differentiable surface (worldsheet) with nontrivial topology: sphere, disk, torus (doughnut), etc.  
The external states are described by boundaries that become disjoint at time $t=à¥$ (not worldsheet parameter $ $).  
For example, a tree graph now looks more like a real
tree, in that the branches now have thickness, and they join smoothly to
the rest of the tree.  

(2) The quantum mechanics of strings is invariant under 2D conformal transformations of the worldsheet.  (But the quantum field theory isn't conformal in spacetime, because there is a discrete mass spectrum.)  As a result, the worldsheet can be ``stretched" to the extent that field theory tree diagrams are described by the same surfaces as propagators.

$$ \fig{tree} $$

The fact that the string worldsheet is described by conformal geometry rather than the usual geometry means that the worldsheet metric is reduced to just a few parameters (called by mathematicians ``moduli"), and topology (which doesn't even require a metric).  These parameters are similar to those that appear in Feynman diagrams for particles (so the 2D metric in some sense has been reduced to a 1D metric), but the topology of surfaces is much different from that of stick graphs.

From this topological point of view, string diagrams are equivalent if they can be ``stretched" into one another.  An explicit way to show this is using Dolen-Horn-Schmid duality.  We have mentioned in the special case of the 4-point amplitude that summing over poles in one channel is equivalent to summing in the other.  This result can be generalized:  We can write any string graph as an ordinary Feynman diagram with just cubic interactions, but with any 4-point subgraph satisfying duality.  (This is ÓnotÕ string field theory, whose graphs are not separately dual.)  So we can use duality to relate graphs of the same 2D topology, and must not double-count by summing graphs that are topologically equivalent.

$$ \vcenter{\hbox{\fig{stretch}}} $$

In particular, in any loop graph any loop can be moved anywhere, into a propagator or an external line, or can be pulled out to form a tadpole (string going into the vacuum).  The result is that any graph is equivalent to
a tree graph with insertions of some one-loop open- or closed-string
tadpoles.  However, this does not mean that any graph constructed with
only open-string propagators and interactions can be expressed as an
ÓopenÕ-string tree graph with tadpole insertions:  The one-loop open-string
graph with two ``half-twists" on the open-string propagators in the loop
is equivalent to a tree graph with a closed-string intermediate state, as
can be seen by stretching the surface, or by tracing the routes of the
boundaries.  (For example, drawing this graph in a psuedo-planar way, as
a flat ring with external states connected to both the inner and outer
edges, pulling the inner edge out of the plane reveals a closed string
connecting the two edges.)  This phenomenon is similar to 2D
bosonization:  A closed string can be represented as the ``bound state" of
two ÓfreeÕ open strings just as a massless scalar in D=2 can be
represented as the bound state of two free massless spinors.

$$ \fig{bound} $$

There are only 3 types of 1-loop insertions to consider (and for ``orientable" strings only 2):  

£1) handle

£2) window (hole)

£3) cross-cap (like a nonorientable hole)

\noindent In string
theory the coupling is topological, in the sense that the power of the
coupling constant is counted by (minus) the ``genus" of the worldsheet, the ``Euler number" $$, given by the integral of the worldsheet
curvature (see exercise IXA7.3).
However, in counting string loops, the last two of the 3 listed above count as open-string loops, while the first counts as a closed-string loop, which is equivalent to 2 open-string loops.  Consequently, the closed-string coupling is the square of the open-string one:  The Euler number is
$$  = 2 -2h -w -c 
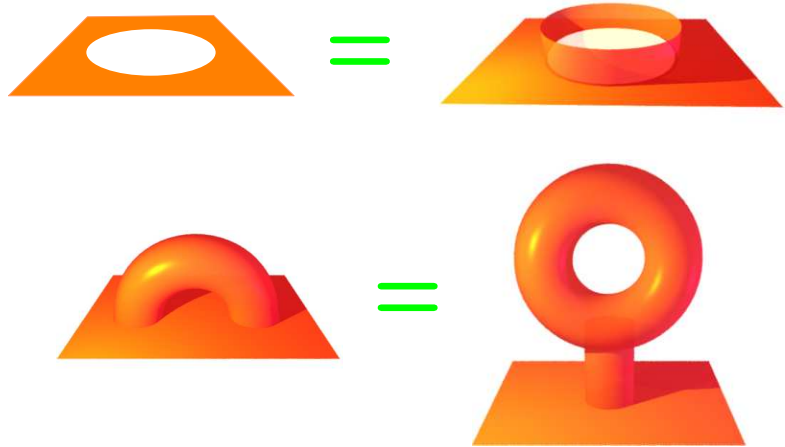
$$
$$ \vcenter{\hbox{\figscale{tadopenf}{1.5in}}}
	\vcenter{\hbox{\rgbo{0 1 0}{\biggest ¼=¼}}} 
	\vcenter{\hbox{\figscale{tadopen}{2in}}} $$

$$ \vcenter{\hbox{\figscale{handle}{1.5in}}}
	\vcenter{\hbox{\rgbo{0 1 0}{\biggest ¼=¼}}} 
	\vcenter{\hbox{\figscale{tadpole}{1.5in}}} $$

The cross-cap is a hole with opposite points identified:  It thus actually does not introduce a boundary, but does introduce nonorientability.  If the number of cross-caps is more than 2, it can be reduced to 1 or 2 by replacing pairs of cross-caps with handles.  Notable examples of surfaces with cross-caps are the projective plane (sphere with 1 cross-cap), Klein bottle (sphere with 2 cross-caps), and M¬obius strip (sphere with 1 cross-cap and 1 window).

$$ \vcenter{\hbox{ââââ\figscale{crosscap}{2in}}}ââ{\bf cross-cap} $$

Analyzing the 1-loop insertions as tadpoles also makes it easy to interpret divergences and how to renormalize them:  Tadpoles contribute to vacuum values of scalars.  In string theory, coupling constants are also vacuum values of scalars (the string coupling $g$ from the vacuum value of the dilaton, the slope $Œ'$ of the string Regge trajectory from the vacuum value of the determinant of the metric tensor).  Thus, string divergences correspond to renormalization of couplings.  However, we know that divergences in quantum gravity can lead to difficulties, so it may be useful to try and cancel them.  The handle is a closed-string tadpole with a closed-string loop (torus) at the end.  Since the propagator connecting a tadpole to the rest of the graph is at zero momentum (by momentum conservation), the divergence of this graph reduces to essentially a counting of states in the loop.  In the superstring, the bosonic and fermionic contributions running around this loop cancel.  On the other hand, the 2 remaining types of tadpoles turn out not to be finite by themselves.  However, their divergences can cancel each other, for either the string or superstring, if the gauge group of the open string is SO(2$^{D/2}$).  For one such insertion into the sphere, this is cancellation between the disk and projective plane.  For 2 such insertions, it is between the annulus (cylinder), Klein bottle, and M¬obius strip.

$$ \figscale{cancel}{2.5in} $$

Loops in open-string graphs can have half-twists in them.  Such graphs are orientable if the number of half-twists in a loop is even.  At 1 loop, such twisting is the same as putting some external lines on the inner boundary of a planar loop and some on the outside.  On a loop with no strings attached to one boundary, that boundary is just a hole, a closed string extending into the vacuum.  (An annulus is topologically the same as a cylinder.)  But a loop with  open strings attached to both boundaries is the same as a tree graph with a closed string attaching the two sets of states.  If one calculates such a graph in open-string theory, no divergences are found, except for the poles of these closed-string states.

\vskip-.25in
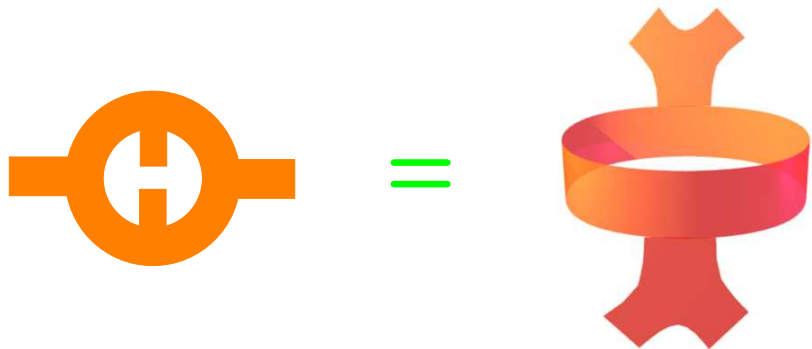
$$ \vcenter{\hbox{\figscale{bound2f}{1.5in}}}
	\vcenter{\hbox{\rgbo{0 1 0}{\biggest â=â}}} 
	\vcenter{\hbox{\figscale{bound2}{1.5in}}} $$
\vskip-.15in

\x XIA2.1  An exercise in pictures for a subsection on pictures:  Draw a 2-loop open string diagram that looks planar when 2 external open-string states are drawn coming from each of the 3 boundaries, when the external states are drawn inward for inner boundaries and outward for outer (as for the 1-loop diagram above).  Show this is equivalent to a tree graph with a 3-closed-string vertex.  Generalize to an arbitrary number of loops.

Ü3. Classical mechanics

We now consider string theory as derived by first quantization.  As for
particles, the first step is to study the classical mechanics, which
determines the appropriate set of variables, the kinetic term of the field
theoretic action, some properties of the interactions, and some
techniques useful for perturbation.  Just as the simplest such action for
the particle produces only the relatively uninteresting case of the scalar,
the most obvious action for the string yields a model that is not only too
simple, but quantum mechanically consistent only in 26 dimensions. 
However, this toy model exhibits many relevant qualitative features, such
as Regge behavior and duality.  Later we'll consider the source of its
problems by relating to four-dimensional particle theories.

The simplest classical mechanics action for the string is a direct
generalization of that for the massless scalar particle: 
For the Lagrangian form of this action we write
$$ S_L = {1\over Œ'}
	Ç{d^2 §\over 2¹}å{-g}g^{mn}ü(»_m X^a)(»_n X^b)ú_{ab} $$
 where $X^a(§^m)$ is the position in spacetime of a point at worldsheet
coordinates $§^m=(§^0,§^1)=( ,§)$, $g^{mn}(§^m)$ is the (inverse)
worldsheet metric, and $Œ'$ is a normalization constant related to the string
tension.  It can also be associated with the flat-space spacetime metric
$ú_{ab}$; if we couple a spacetime metric, then its vacuum value can be
taken as $ú_{ab}/Œ'$, where $Œ'$ is the gravitational coupling, as
discussed in subsection IXA5.  If we vary this action with respect to $X$, we get its 2D wave equation, covariant with respect to the curved worldsheet:
$$ õX^a ­ {1\over å{-g}}»_m å{-g}g^{mn}»_n X^a = 0 $$

A new feature of this action (compared to the particle's) is that it is (2D)
Weyl scale invariant (see subsection IXA7).  This gauge invariance can be used to gauge away
one component of the metric, in addition to the two that can be gauged
away using 2D general coordinate invariance.  The net result is that the
worldsheet metric can be completely gauged away (except for some bits
at boundaries), just as for the particle.  However, this same invariance
prevents the addition of a worldsheet cosmological term:  In the particle
case, such a term was needed to introduce mass.  Here, mass is introduced
through the coefficient $1/Œ'$ of the $(»X)^2$ term:  The same scale
invariance that prevents use of a cosmological term also prevents this
coefficient from being absorbed into the definition of the worldsheet
metric.

Just as for the particle, the metric can be eliminated by its equation of
motion, resulting in a more geometrical, but less useful, form of the
action:  In this case the equation of motion (``Virasoro constraints")
$$ (»_m X)É(»_n X) = üg_{mn}g^{pq}(»_p X)É(»_q X) $$
 after taking the determinant of both sides, gives
$$ S = {1\over Œ'}Ç{d^2 §\over 2¹}å{-÷g},ââ÷g_{mn} = (»_m X)É(»_n X) $$
 This is the area of the string in terms of the ``induced" (intrinsic) metric $÷g_{mn}$,
analogously to the particle case.  The induced metric measures length as
usually measured in spacetime:
$$ d§^m d§^n ÷g_{mn} = (d§^m »_m X)É(d§^n »_n X) = (dX)^2 $$
 Equivalently, this action can be written in terms of the area element
$dX^a\wedge dX^b$:
$$ S = {1\over 2¹Œ'}Çå{-ü(dX^a\wedge dX^b)^2},ââ
	dX^a\wedge dX^b = (d§^0 »_0 X^{[a})(d§^1 »_1 X^{b]}) $$

For purposes of quantization, it's also useful to have the Hamiltonian form
of the action.  This also allows us to see how the Virasoro constraints
generalize the Klein-Gordon equation, and then find the BRST operator. 
By the usual methods of converting from Lagrangian to Hamiltonian, we
find
$$ S_H = Ç{d^2 §\over 2¹}(-ÀXÉP +\H),ââ
	\H = {å{-g}\over g_{11}}ü(Œ'P^2 +Œ'^{-1}X'^2) 
	+{g_{01}\over g_{11}}X'ÉP $$
 where $À{\phantom n}=»_0$ and ${}'=»_1$.  Various combinations of components of
the worldsheet metric now appear explicitly as Lagrange multipliers.  If
we define
$$ öP_{(à)} = \f1{å2}(Œ'^{1/2}PàŒ'^{-1/2}X')âÜâ[öP_{(+)},öP_{(-)}] = 0 $$
 the constraints can be written as two independent sets $öP_{(à)}^2$.

\x XIA3.1  Show that if we call $g_à$ the Lagrange multipliers for
$öP_{(à)}^2$, then in convenient local Lorentz and Weyl scale (but not
coordinate) gauges we can write in a lightcone basis
$$ e_à = e_à{}^m »_m = \f1{å2}(»_0 àg_¦»_1) $$
 while in another scale gauge we can write
$$ dx^m e_m{}^à = \f1{å2}(dx^0 g_à àdx^1) $$

\x XIA3.2  Find the (equal-$ $) commutation relations $[öP_{(à)},öP_{(à)}]$. 
Show that the (semiclassical) commutation relations of the constraints
$öP_{(à)}^2$ close.  (Hint:  Use the identity
$$ f(a)¶'(a-b) = f(b)¶'(a-b) -f'(b)¶(a-b)â.â) $$

Since 2D general coordinate (and even just Lorentz) invariance is no
longer manifest in the Hamiltonian formalism, for some purposes we need to generalize this to a form
that is first-order with respect to both $ $ and
$§$ derivatives:
$$ S_1 = -{1\over Œ'}Ç{d^2 §\over 2¹}
	[(»_m X)ÉP^m +(-g)^{-1/2}g_{mn}üP^m ÉP^n] $$
 obviously reproduces $S_L$ after eliminating $P^m$ by its equation of motion
$$ P^m = -å{-g}g^{mn}»_n X $$  
Eliminating just $P^1$ gives a simpler way of deriving $S_H$ (with $P^0=Œ'P$).

Since open strings have boundaries, the action implies boundary
conditions, originating from integration by parts when deriving the field
equations.  In the last form of the action variation of the first term gives,
in addition to the $Çd^2§$ terms $(¶P)É»X$ and $-(¶X)É»P$ for the field
equations, a boundary term $Èd§^m ·_{mn}(¶X)ÉP^n$, where $d§^m$ is a
line integral along the boundary, and the $·_{mn}$ picks the component
of $P^m$ normal to the boundary.  We thus have
$$ n_m P^m = 0¼at¼boundaries $$
 where $n_m$ is a vector normal to the boundary.  This condition on the derivative of $X$ (``Neumann" boundary condition) causes waves propagating in the string to be reflected at the boundaries. 

A simple interpretation of this boundary condition is to consider an open string as a closed string ``folded over" on itself:  At any fixed $ $, following $X(§)$ for increasing $§$ takes one along the usual open string, but then doubles back at a boundary to backtrack along the same path, and the same at the opposite boundary, becoming periodic as for the closed string.  This periodicity is convenient for $§$-Fourier expanding in exponentials, rather than sines and cosines.  Continuity in $X$ upon reversal at the boundaries implies the Neumann boundary condition, but implemented in the usual way for 2D problems, by the method of images, due to this doubling.

 From the constraint
imposed by varying $g_{mn}$, it then follows that
$$ (t_m P^m)^2 = 0¼at¼boundaries $$
 where $t_m$ is a vector tangent to the boundary (or any vector, for that
matter).  Since by the field equations $P^m¾g^{mn}»_n X$, this means
that the boundary is lightlike in spacetime:  The ends of the string travel
at the speed of light.

Ü4. Types

There are various types of known string theories:  Some are supersymmetric and some are not.  (The supersymmetric ones appear to be equivalent to each other nonperturbatively, and equivalent to membranes, but here we restrict ourselves to perturbation theory.)  

Besides supersymmetry, there are geometric distinctions.  One of these is between open and closed strings:  Closed strings have modes that are either left- or right-handed, i.e., propagating in either $§$-direction.  For open strings these modes are identified, since left-handed modes become right-handed upon reflection at the boundary (and vice versa).  For closed strings they are independent, and may have different supersymmetry properties.  

Since the open strings have ends, we can associate internal symmetry indices (``Chan-Paton factors") with them, as found in subsection VIC4, following from the same in subsection VC9 for ordinary particle field theory.
These indices can also be
associated with worldsheet variables that live only on string boundaries. 
As in the field theory case, these indices are associated with orientation
of the boundaries (arrows) only for U groups, not for SO or USp. 

As we'll see in section XIB, quantization of known open strings always produces Yang-Mills at the massless level (super Yang-Mills for open superstrings).  Since closed strings have effectively two sets (left and right) of open string modes, the closed-string Hilbert space is effectively the direct product of two (perhaps different) open-string Hilbert spaces (with an added restriction implied by $§$-translation invariance, to be discussed later).  In particular, at the massless level this direct product of two vectors can give a graviton (symmetric, traceless tensor), scalar (trace), and axion (antisymmetric).
In the case that the two open strings are the same, it is possible to
restrict this direct product to its symmetric part.  This eliminates the
axion, but not the scalar.  Thus, a massless scalar appears even in the
simplest case.

Another geometric property we discussed topologically was orientability of the worldsheet.  To understand orientability, we examine the discrete
symmetries of the worldsheet.  Wave functions or fields describing the string can be expressed as functionals of $X$ at fixed $ $ (just as for particles).  We also choose $§$ to run from 0 to $¹$ for the open string, and from 0 to $2¹$ (periodic) for the closed.  (This choice as a gauge condition will be discussed in more detail in subsection XIB1.)  As in D=4, local, unitary, Poincar«e
invariant 2D field theories are always CPT invariant.  In particular, CPT
doesn't switch left- and right-handed modes (the ``velocity" $d§/d $ is invariant), which differ in some string theories.  
Thus, we can always impose invariance of the wave
function/field under the worldsheet CPT transformation:
As the generalization of the particle condition $Ä(x)=Äÿ(x)$, we have for the string
$$ worldsheetâCPT:ââì[X(§)] = ìÿ[X(¹-§)] $$
 where parity transforms $§£¹-§$ to preserve $§·[0,¹]$ for the open
string; for the closed string the $¹$ is irrelevant because of periodicity
and invariance under $§$ translation.  (We have written only the $X$ coordinate explicitly for simplicity; similar remarks apply to other coordinates, such as ghosts,
with possible extra signs due to 2D Lorentz indices.)  Hermitian
conjugation for the open string (for the closed string the field is not a
matrix), instead of just complex conjugation (for C), simply switches the
internal symmetry factors associated with the left and right
ends of the open string (matrix transposition), as also required by parity. 
In particular, this implies that the matrices associated with the Yang-Mills
fields are hermitian, so the Yang-Mills group is unitary.

In addition to this reality condition, if the 2D theory is also invariant
under CP and T, it is also possible, though not necessarily required, to
impose such a quantum mechanical invariance under CP, and thus T:
As a generalization of the particle's T condition (see subsection IA5),
$Ä(x)=MÄ*(x)M^{-1}$ $Û$ $Ä(x)=MÄ^T(x)M^{-1}$, for the string
$$ worldsheetâT:ââì[X(§)] = Mì*[X(§)]M^{-1}âÛ $$
$$ worldsheetâCP:ââì[X(§)] = Mì^T[X(¹-§)]M^{-1} $$
As for Yang-Mills in the particle case, for the open
string the matrix ``$M$Ê" is the group metric (we drop the $M$ and the ${}^T$ for the closed string, which
is not a matrix), either symmetric or
antisymmetric depending on whether the group is orthogonal or
symplectic; without imposing T and CP the group is just unitary.  Thus, all the classical groups are allowed (at least in the
classical field theory; exceptional groups cannot be described by associating indices with the ends).  

Since imposing invariance of the states (not just
the action) under CP and T makes it impossible to observe the left/right
handedness of the worldsheet, such strings are ``unoriented", as opposed
to the ``oriented" strings that satisfy just the CPT condition.  Thus,
orientability of the surface is directly related to orientability of its
boundaries (oriented for U, unoriented for SO or USp). 

 Also, as in the
particle field theory, unorientability allows ``twisted" worldsheets that
are prohibited in the oriented case (because we can distinguish the
``front" of the worldsheet from the ``back"):  This allows such exotic
geometries as M¬obius strips and Klein bottles.  Open strings produce
closed ones as bound states (open and closed strings are parts of the
same worldsheet with different boundaries); in theories of open and
closed strings, they must be both oriented or both unoriented.  Since the
worldsheet-CP and -T switch lefty and righty modes, this invariance on
the closed string results in the restriction introduced earlier, keeping only
the symmetric part of the direct product.

\x XIA4.1  What is the difference between an unoriented closed string
(satisfying this worldsheet CP condition)
and the interpretation of subsection XIA3 of the open string as
a closed string folded over on itself?

The known string models all have massless particles.  A string model with
massless particles can be applied to hadrons only if masses are given to
all these states through the Higgs mechanism or some other change in the
vacuum.  An alternative is to use such a model to describe fundamental
massless particles (graviton, photon, gluons, neutrinos), although this
would also require the usual Higgs of the Standard Model for generating
masses for some particles (W, Z, quarks, charged leptons, Higgs).  In
particular, all known string models have a graviton, and there is no
known method whereby this graviton would gain mass, so these models
seem suited only for unified theories of gravity plus matter. For this
purpose, the massive fields have little phenomenological interest.  They
might improve high-energy behavior, but only near the Planck scale,
which is effectively unobservable.  Therefore, it is necessary to analyze
the massless subsector of such string theories to find signs of
fundamental strings in nature.

The massless sector of the open string includes spin 1 and no higher.  This
is true for the known string models, and also is expected to be a general
result, since otherwise the closed string would include massless states
with spin higher than 2, for which no consistent interacting theory is
known.  Spin 1/2 leads to supersymmetry, as described 
below; we first consider bosonic strings.
For convenience (and ultimate utility) we consider 4D states; in presently known strings these are the massless states in perturbation about the compactified vacuum.

The bosonic string contains at least the graviton, a scalar (usually going
by the misnomer of ``dilaton"), and a pseudoscalar (the ``axion",
described by an antisym\-metric-tensor gauge field), and can contain
additional vectors and scalars (if the open strings had scalars in addition
to the vector).  This analysis can be performed covariantly, but it is
simpler to use a helicity or lightcone analysis.  Then the helicities of the
closed-string states are just the sums of those of the open strings:  For
the product of two vectors (the minimal case),
$$ ( +1 ¢ -1 ) ° ( +1 ¢ -1 ) = +2 ¢ 0 ¢ 0 ¢ -2 $$
 giving the graviton, scalar, and axion.  Similarly, additional scalars for
one open string give additional vectors for the closed, while additional
scalars for both open strings give also additional scalars.

\x XIA4.2 Make the same analysis in terms of covariant fields, both for the
fields themselves ÓandÕ their gauge transformations.  Note that the trace
of the gravitational field $h_{ab}$ (determinant of the metric $g_{mn}$) is
missing.  (It's unphysical, and can be found from the ghost sector, as
explained in chapter XII.)

Considering the massless spectrum of superstrings, we now look at the
restrictions imposed by supersymmetry whenever fermions are included.  The open string can also contain
massless spin 1/2, but only if it is related by supersymmetry to its
massless spin 1, since it leads to spin 3/2 in the closed string, and
massless spin 3/2 is known to be inconsistent in an interacting theory
unless related by supersymmetry to the graviton.  
(Spin 3/2 gauges supersymmetry.  But spin 1 can't couple minimally
to spin 3/2: see exercise XIIB7.2b below.  So, spin 3/2 needs spin 2
as its supersymmetric partner.)  Thus there are two
possibilities for the massless sector of each open string: (1) vectors and
scalars for an open bosonic string, or (2) vector multiplets (vectors,
spinors, and scalars, all related by some number of supersymmetries) for
an open superstring.  From our analysis of subsection IIC5, there are
furthermore 3 types of vector multiplets in D=4, corresponding to N=1,2,
or 4 supersymmetries.  (These result from compactification from N=1 in D=10, depending upon how much supersymmetry is broken.)

This leads to four types of closed strings:  
\item{(1)} The bosonic string, from
bosonic $°$ bosonic was discussed above (actually 2 types, if we distinguish oriented and unoriented).  
\item{(2)} The
``heterotic" string comes from bosonic $°$ super.  It thus can have
N=1,2, or 4 supersymmetries.  
\item{(3)} The ``(Type II) superstring" comes from
super $°$ super.  The total number of its supersymmetries is the sum of
those from the open strings:  Depending on the type of supersymmetric
open strings used, the superstring can have N=2,3,4,5,6, or 8 (in other
words, anything greater than 1, since N=7 supersymmetry is equivalent to
N=8, and 8 is the maximum for supergravity).  
\item{(4)} In the super case, if the
left and right open strings are the same, we can impose symmetry as in
the bosonic case (``Type I").  Then we may also include the open strings
in the spectrum:  This symmetrization also identifies the left and
right supersymmetries, so then N=1,2 or 4, the same for open and closed
states (so they can be consistently coupled).

The spectrum again can be analyzed by helicity:  For example, for the N=1
heterotic string, we have
$$ ( 1 ¢ ü ¢ -ü ¢ -1 ) ° ( 1 ¢ -1 ) = 
	( 2 ¢ \f32 ¢ -\f32 ¢ -2 ) ¢ (  ü ¢ 0 ¢ 0 ¢ -ü ) $$
 which is supergravity plus a scalar multiplet.  As for the bosonic string,
all supersymmetric closed strings include the scalar, again coming from
vector $°$ vector.

\x XIA4.3 Make the same analysis for some other supersymmetric cases:
ªa N=2 and 4 heterotic.  Also make a simpler analysis using ``superhelicity", writing any supermultiplet as the lowest-helicity one $°$ some helicity.
ªb N=2 Type II.
ªc N=1 Type I.

Ü5. T-duality

Another symmetry of all known string models is ``T-duality".  It is closely
related to the open $°$ open structure of closed string states, and thus
expected to be a general property of string theory.  We consider the
simple bosonic model as an example.  Including constant background
fields, working with flat worldsheet metric (the ``conformal gauge": see subsection XIB1) for convenience, the Lagrangian is
$$ L = -(»_+ X^m)(»_- X^n)M_{mn},ââM_{mn} = G_{mn}+B_{mn} $$
 where $+,-$ are lightcone worldsheet indices, the curved indices $m,n$ now refer to spacetime, $G_{mn}$ is the
spacetime metric, and $B_{mn}$ is an antisymmetric tensor gauge field
(``axion").  Writing the action in first-order form
$$ L' = -P_{+m} »_-X^m -P_{-m} »_+X^m +P_{+m}P_{-n}M^{mn} $$
 where $M^{mn}$ is the inverse of $M_{mn}$, we vary $X$ instead of $P$ to
solve the field equation
$$ »_+P_{-m}+ »_-P_{+m} = 0âÜâP_{+m} = »_+÷X_m,âP_{-m} = -»_-÷X_m $$
 and substitute to find the ``dual" Lagrangian
$$ L'' = -(»_+ ÷X_m)(»_- ÷X_n)M^{mn} $$

Thus the ``duality transformation" from $X$ to $÷X$ is an invariance of the
theory, as long as we also transform the background:
$$ X^m £ ÷X_m,ââM_{mn} £ M^{mn} $$
 Note that in flat space ($M^{mn}=ú^{mn}$), using the $P$ equation of
motion in $L'$, we have
$$ P_{+m} = ú_{mn}»_+X^n,ââP_{-m} = ú_{mn}»_-X^n $$
 so duality just changes the sign of the right-handed modes
($»_-X=-»_-÷X$) while leaving invariant the left-handed ones
($»_+X=»_+÷X$).  (The treatment of the ``zero-modes", those killed by the derivatives acting on $X$ or $÷X$, is more tricky:  We
have ignored them by taking the background constant.)  We can see this
to lowest order in the background, since $M£M^{-1}$, to lowest order in
perturbation about $ÒMÔ=ú$, changes the sign of the field, corresponding
to the fact that their ``vertex operators" (coefficients of linearized background fields) are linear in both left- and
right-handed modes ($(»_+X)(»_-X)$).  However, in full nonlinearity,
duality mixes the spacetime metric $G_{mn}(X)$ and axion $B_{mn}(X)$.

This invariance can be generalized to a continuous O(D,D) symmetry by
combining it with (global) Lorentz transformations.  The above discrete
symmetry is a kind of ``parity" for this larger group:  There are also
``reflections" from performing the duality on just one component of
$X^m$.  The easiest way to see the full symmetry is in the Hamiltonian
formalism, where it can be made manifest:  We first combine $X'^m$ and
the canonical momentum $P_m$ into an O(D,D) vector:
$$ Z_M = (P_m, X'^m) $$
$$ Üâ[Z_M(1),Z_N(2)] = -2¹i¶'(2-1)ú_{MN},ââ
	ú_{MN} = \pmatrix{0 & ¶_m^n \cr ¶_n^m & 0 \cr} $$
 where the O(D,D) metric $ú_{MN}$ is constant even in curved space.  (We
have abbreviated ``1" for ``$§_1$", etc.)  The Virasoro constraints are
then
$$ üú^{MN}Z_M Z_N = üM^{MN}Z_M Z_N = 0 $$
 where $M$ is not only symmetric but also an element of the O(D,D) group:
$$ M^{MN} = \pmatrix{G^{mn} & G^{mp}B_{pn} \cr
	-B_{mp}G^{pn} & G_{mn} -B_{mp}G^{pq}B_{qn} \cr} 
	= M^{NM} = ú^{MP}(M^{-1})_{PQ}ú^{QN} $$
 If the fields are constant in only d of the D dimensions, than the
symmetry is reduced to O(d,d); thus O(d,d) is a symmetry of the
dimensionally reduced theory with arbitrary fields.

\x XIA5.1  Show that the conditions on $M$ can be solved in a
manifestly O(D,D) covariant way by use of a ``vielbein" $E_A{}^M$:
$$ M = M^T,âMúM = úâ(ú = ú^T = ú^{-1}) $$
$$ ÜâM = E^T ÷úE,âEúE^T = ú,â÷ú = \pmatrix{ ú^{ab} & 0 \cr 0 & ú^{a'b'} \cr} $$
 Show that $M$ is invariant under a local O(D$-$1,1)$°$O(D$-$1,1) 
transformation on $E$, so $E$ is an element of the
coset space O(D,D)/O(D$-$1,1)$°$O(D$-$1,1) (see subsection IVA3).

Ü6. Dilaton

We can extend the spectrum analysis of subsection XIA4 off shell:  The procedure (to be
justified in chapter XII) includes the ghost and antighost (multiplets) for
the vector (multiplet) as a doublet of the ghostly Sp(2) symmetry.  The
direct product of  vector $°$ vector now clearly gives a traceless
symmetric tensor (graviton), the corresponding trace (physical scalar),
and an antisymmetric tensor (axion).  In the direct product of the ghosts,
the Sp(2) singlet gives the trace part of the metric tensor, which is the
true dilaton.  This dilaton (the determinant of the metric tensor in the
nonlinear case) is required in gravity for constructing local actions (see
subsection IXA7), but does not contain a physical degree of freedom.  The
physical polarizations of the graviton are contained in the traceless
(actually $det=-1$) part of the metric, which describes the conformal part
of gravity.  The direct products involving ghosts also give Sp(2)
nonsinglets, which are the ghosts of the massless sector of the closed
string.  BRST transformations (and thus gauge transformations) can also
be obtained by this direct-product procedure.

The natural coupling of background fields in the classical mechanics of
the string reflects this direct-product structure, as seen in
subsection XIA5.  This means that the background metric as we have defined it
has as its determinant not the usual one, but that times a power of the
physical scalar:  It is a physical degree of freedom.  T-duality mixes
physical degrees of freedom with each other.

If we try to construct a low-energy action for the massless fields of the
bosonic string, it is not too difficult to find a scalar invariant under
T-duality to act as the Lagrangian.  However, it is impossible to use the
usual measure $Çdx¼å{-g}$ because $g$ is not invariant under T-duality. 
This problem is solved by including the spacetime dilaton field $ì(X)$:  It
couples to the string as
$$ S_{dil} = -Ç{d^2 §\over 2¹}å{-g}¼ür¼ln¼ì(X) $$
 where we denote the worldsheet curvature by $r(§)$ (only in this
subsection) to distinguish it from the spacetime curvature $R(x)$.  (There
are also boundary contributions: see exercise IXA7.3.)  This term can also
be expressed as a coupling to the worldsheet ghosts (according to the
above arguments), allowing the worldsheet metric to be completely fixed
by gauge transformations, as usual.

Since there is no $X$ dependence of $S_{dil}$ for constant dilaton field (no
$»X$ factors, unlike $G$ and $B$), the constant dilaton is invariant under
T-duality. Furthermore, since it couples to the worldsheet curvature,
which counts the number of loops, the dilaton must appear
homogeneously in the classical action.  The dilaton that appears as above
in the string action transforms as a density under general coordinate
transformations, allowing the construction of actions invariant under
both T-duality and coordinate transformations.  The resulting spacetime
action for the massless fields of the oriented, closed bosonic sting is
$$ S_{massless} = Çdx¼ì(õ -\f14 R +\f1{24}H^{abc}H_{abc} +ñ)ì $$
 where $H_{abc}=üá_{[a}B_{bc]}$ is the field strength for the axion. 
T-duality determines the only arbitrary coefficient, the relative weight of
the $õ$ and $R$ terms.  Note the absence of the factor ${\bf e}^{-1}$,
which has been absorbed into the definition of $ì$:  The covariant
derivative acting on $ì$, since it is a density that transforms as 
${\bf e}^{-1/2}$, acts as 
$$ á_a ì={\bf e}^{-1/2}e_a{\bf e}^{1/2}ì $$
(We have included a cosmological term, allowed by T-duality, but not appearing at tree level.)

\x XIA6.1  Find the field equations following from this action.  ÓThenÕ make
the field redefinition  $ì={\bf e}^{-1/2}e^Ä$, to find the result:
$$ \li{ {¶\over ¶ì}âÜâ&(áÄ)^2 +õÄ -\f14 R +\f1{24}H^2 +ñ = 0 \cr
	{¶\over ¶B_{ab}}âÜâ&á^a H_{abc} +2H_{abc}á^a Ä = 0 \cr
	{¶\over ¶e_a{}^m}âÜâ&R_{ab} = 2á_a á_b Ä +üH_a{}^{cd}H_{bcd} \cr} $$

Both coupling constants in string theory can be associated 
with vacuum values: 
(1)  The string coupling appears as the vacuum value of 
the dilaton, since it counts loops.
(2)  $Œ'$ comes from the vacuum value of the (spacetime) metric, 
as can be seen from the worldsheet action. 
This is the string-gauge equivalent of the fact that 
the gravitational constant naturally arises as the vacuum 
(or asymptotic) value of the metric in ordinary gravity 
(see subsection IXA5).
In string theory, the fact that the gravitational constant is 
a combination of $Œ'$ and the string coupling is equivalent 
to the field redefinition from the string gauge to the 
particular Weyl gauge where the Einstein term in the 
action appears in the usual way.

\x XIA6.2  This action is in the string gauge (see subsection IXB5).  
ªa Make the physical scalar explicit in the action by the field redefinition
(Weyl scaling: see subsection IXA7)
$$ e_a{}^m £ e_a{}^m $$
 leaving $ì$ and $B_{mn}$ unchanged.
ªb The resulting scalar action can be (off-)diagonalized by further
redefinitions:  Noting that the known string theories are defined for
$å{D-1}$ an (odd) integer (5 or 3), write the dimension in general as (for
any $D>1$, $n$ not necessarily integer)
$$ D = n^2 +1 $$
 Restoring the ${\bf e}^{-1}$ to the action, redefine
$$ ì = {\bf e}^{-1/2}Ä_+^{(n-1)/2(n+1)}Ä_-^{(n+1)/2(n-1)},ââ
	 = Ä_+^{1/(n+1)}Ä_-^{-1/(n-1)} $$
 which also gives the scalars $Ä_{à}$ the canonical Weyl scale weights, to
obtain the final result for the Lagrangian $L$ (where 
$S=Çdx¼{\bf e}^{-1}L$)
$$ L = Ä_+(\f{n^2}{n^2-1}õ -\f14 R)Ä_- 
	+\f1{24}Ä_+^{(n+5)/(n+1)}Ä_-^{(n-5)/(n-1)}H^2 $$
$$ +ñÄ_+^{(n-1)/(n+1)}Ä_-^{(n+1)/(n-1)} $$
ªc This redefinition is singular for $D=2$ ($n=1$).  Fix this by making the
additional redefinition
$$ Ä_à £ Ä_à^{(nà1)/2} $$
 and then taking the limit $n£1$.  (We can also use redefinitions equivalent
in the limit, such as $Ä_-£Ä_-^{(D-2)/4}$.)  Show the result is then
$$ L £ \f14 Ä_+(õ¼ln¼Ä_- -R) +\f1{24}Ä_+^3 Ä_-^{-2}H^2 +ñÄ_- $$
 We can no longer choose the gauge $Ä_+=1$, since it is now scale
invariant, but we can still choose $Ä_-=1$.

When applied to the supersymmetric cases (superstring or heterotic
string), the inclusion of ghosts in the direct-product procedure also gives
the auxiliary fields.  (The dilaton itself is an auxiliary field.)  For example,
in the N=1 heterotic case, the direct product of the physical parts of the
vector and vector multiplet give conformal supergravity (the
supersymmetrization of the traceless part of the metric) and a physical
tensor multiplet (the supersymmetrization of the axion and scalar).  On
the other hand, the ghosts of the vector multiplet form a chiral scalar
superfield; its procuct with the scalar ghost of the vector gives another
chiral scalar superfield, the compensator, containing the dilaton.  (See
subsection XA3.)

The two conditions of supersymmetry and that the dilaton must appear
homogeneously (quadratically after an appropriate field redefinition) are
now enough to fix the form of the action (except for the nonminimal
heterotic cases, where the open string's scalars introduce extra vector
multiplets).  For convenience we redefine the chiral scalar compensator as
$Ä£Ä^{2/3}$ so that it appears quadratically in the cosmological term
$Çd^4 x¼d^2 ϼÄ^2$.  Thus, by dimensional analysis $Ä$ now has scale
weight $\f32$.  The axial-vector field strength of the axion appears as
$[á_Œ,Ñá_{ÀŒ}]G$, so $G$ has scale weight 2.  (Gauge fields are Weyl scale
invariant with curved indices for consistency with gauge transformations;
thus $H_{mnp}$ has weight 0 while $H_{abc}$ has weight 3.)  Of course,
these weights also follow from local superscale transformations, the
global part of which transforms fields as $L^{2w}$ (see subsection XA4).
The only action quadratic in the dilaton consistent with global scale and
U(1) (R) invariance is then (with implicit covariantization with respect to
conformal supergravity, which makes these invariances local)
$$ S = Çdx¼d^4 ϼÐÄÄG^{-1/2} +\left( ñÇdx¼d^2 ϼÄ^2 +h.c. \right) $$

\x XIA6.3  Use the methods of subsection XB6 to find all of the
terms in this action involving only bosonic fields.  Compare to the bosonic
string action considered above.

Besides T-duality, string theories also have ``S-duality" symmetries that
are realized only on the field equations, or after performing
electromagnetic-type duality transformations on the fields:  If we
convert $G$ into a second, physical chiral multiplet $$ by such a duality
as described in subsection XB5, the above action is converted to
$$ S = Çdx¼d^4 ϼ(ÐÄÄ)^{2/3}(+Ѝ)^{1/3} 
	+\left( ñÇdx¼d^2 ϼÄ^2 +h.c. \right) $$
 After the redefinitions
$$ Ä^2 £ Ä,ââÄ^2  £  $$
 the first term becomes manifestly SU(1,1) invariant (see subsection XB7):
$$ S = Çdx¼d^4 ϼ(Ðč+ЍÄ)^{1/3} +\left( ñÇdx¼d^2 Ï¼Ä +h.c. \right) $$
 It is now the original T-duality that can be realized only on shell.  Also, in
this form the condition that the dilaton should appear homogeneously is
obscured.  (Such S-dualities were first seen in extended supergravity
theories, especially when obtained by reduction from higher dimensions,
where antisymmetric tensors are often required.)

\x XIA6.4  Apply the results of exercise XB5.1 to include vector multiplets
in the above actions by replacing $G£÷G$ in the first action and performing
duality transformations.  (The super Yang-Mills appears in the spectrum
from the product (vector $¢$ scalars) $°$ vector multiplet in the
heterotic string.)  This substitution is dictated by homogeneity in the
dilaton, which prevents the usual conformal $Çd^2 ϼW^2$ term.  Such
terms occur naturally in higher-dimensional couplings of supergravity to
super Yang-Mills.

Classical and quantum symmetries of mechanics formulations of particle
and string theories in background fields are often used to derive
equations for those backgrounds.  These features are not peculiar to
these theories or their formulations:  They are a general feature of
describing a particle/field of some (super)spin in a gauge background. 
These equations fall into two distinct types:  (1) A ÓsupersymmetricÕ
system in a gauge background of ÓhigherÕ superspin generates
ÓconstraintsÕ on the background, necessary for consistently defining the
coupling (see subsections IVC4 and XA1).\\
(2) Any gauge system in a
background of the ÓsameÕ gauge field generates Ófield equationsÕ for the
background (see exercise VIB8.2).

For example, the classical symmetries of the superparticle always
generate constraints on its background, but give field equations for it
only if the number of supersymmetries is enough to insure its superspin is
as high as that of its background (e.g., 10D N=1 in background super
Yang-Mills or 11D N=1 in background supergravity).  Similarly, the bosonic
string generates field equations for background gravity at the quantum
mechanical level because quantization is required to reveal the massless
graviton excited state contained in the string itself.  On the other hand,
the 10D superstring already generates field equations for background
supergravity classically, since the ground state of the superstring (closed
if boundary conditions are ignored), the only part that is evident
(semi)classically, already contains supergravity.

Ü7. Lattices

In string theory there are two spaces, the two-dimensional space of the
worldsheet, and physical spacetime.  In subsection VIIIB7 we
considered approximating spacetime by a lattice; in this subsection we
instead approximate the worldsheet by lattices.  For the spacetime of QCD
we used a regular lattice, representing the fixed geometry of flat
spacetime.  In string theory we considered worldsheets of arbitrary
geometry, described by a worldsheet metric, so our lattices should be
more arbitrary; in fact, functional integration over the worldsheet metric
must be replaced by summation over different lattices.  We saw that the
topological expansion of QCD in 1/N generated polyhedra analogous to the
worldsheet, with 1/N acting as the string coupling.  We therefore identify
the Feynman diagrams themselves, with faces chosen by the 1/N
expansion, as these lattices, to give a more precise correlation between
the second-quantized path integral of QCD (and other field theories) and
the first-quantized path integral of string theory.

Presently the relation between such field theories and string theory is
not well understood, and has been described only for the bosonic string. 
Since the bosonic string has only the worldsheet metric and spacetime
coordinates as degrees of freedom, it corresponds to a  (N$ð$N-matrix)
scalar field theory.  Since a lattice requires a scale, while conformal
invariance includes scale invariance, we must break the conformal
invariance of the worldsheet.  The simplest coordinate-invariant yet
scale-variant property of a space is its volume, so we add a volume
(area) term to the string action.  Furthermore, to describe interactions we
need to include a term containing the string coupling constant; in string
theory the power of the
coupling constant is counted by the integral of the worldsheet
curvature.  Our worldsheet action thus consists of
the three terms
$$ S = Ç{d^2 §\over 2¹}å{-g}\left[{1\over Œ'}g^{mn}ü(»_m X)É(»_n X)
	+µ +(ln¼û)üR\right] $$
 A lattice version of this action is (with $$ given in subsection VIIC4)
$$ S_1 =  {1\over Œ'}Ý_{ÒjkÔ}ü(X_j-X_k)^2 +µÝ_j 1 
	+(ln¼û)\left( Ý_j 1 -Ý_{ÒjkÔ} 1 +Ý_J 1\right) $$
 where $j$ are vertices of the lattice, $ÒjkÔ$ are the links, and $J$ are the
``plaquets" (faces, loops).

\x XIA7.1  Put the particle on a random lattice ``Minkowski" worldline. 
(See exercise VB1.2.)  Show the propagator for a massless particle,
written in momentum space, before taking the limit lattice spacing $·£0$,
is
$$ ë = {2·\over 1 -e^{-i·p^2}} $$
 Show this has unphysical poles at $p^2=2¹n/·$ for arbitrary integer $n$. 
How do these results differ if the propagator is defined for Wick-rotated
$ $?

The corresponding field theory is easily found, according to our earlier
discussions, by (1) identifying the worldsheet lattice with a
position-space Feynman diagram (the vertices of the lattice being
those of the diagram, the links of the lattice being the Feynman
propagators; see subsection VC8), and (2) using the 1/N expansion to
associate the faces of the worldsheet polyhedra with the U(N) indices of
the scalar field (see subsection VIIC4).  

We then can immediately identify the three terms in the string action
with their counterparts in the scalar field theory:  
\item{(1)} The $X$ term gives the propagators, 
\item{(2)} the area term (which counts the vertices) gives the
vertex factor (coupling constant), and 
\item{(3)} the curvature term (which
classifies the topology) gives the 1/N factors of the topological
expansion.  

\noindent Thus, the three constants in the string action can be
identified with the mass, coupling, and number of colors of the scalar
field theory.  Explicitly, the field theory action is
$$ S_2 =  N¼trÇ{d^D x\over (2¹Œ')^{D/2}} (üÄe^{-Œ'õ/2}Ä -G\f1n Ä^n) $$
 where we have identified
$$ G = e^{-µ},â\f1N = û,âm^2 = \f2{Œ'} $$
 and we have put an overall factor of $N$ (associated with $1/\h$) so that
$G$ (and $m^2$) is fixed (rather than $G$ times some power of $N$) when
the $1/N$ expansion is performed.  (The reverse can be made true by
rescaling $Ä$.)  The unusual kinetic operator
$$ e^{-Œ'õ/2} = \f1{m^2}(m^2-õ +...) $$
 comes from identifying the second-quantized particle propagator as it
appears in the first-quantized path integral for the string:
$$ \A = ÇÞ_j {d^D X_j\over (2¹Œ')^{D/2}}¼e^{-S_1}âÜâ
	ë(x,y) = e^{-(x-y)^2/2Œ'} $$

Unlike the spacetime lattice, the worldsheet lattice preserves spacetime
Poincar«e symmetry, so it's unnecessary to take any limits to define a
physical theory (or at least taking limits won't improve the physical
relevance of this model).  This model thus describes a stiff or lumpy
string.  The usual continuum-worldsheet string then can be identified
with a particular limit of this more general string.  Explicit calculations
have demonstrated that this lattice regularization of the worldsheet
reproduces the results of the continuum approach.  These results have
been limited to spacetime dimension $²1$ because of the inconsistencies
introduced by the tachyon, which is the ground state in higher
dimensions.  Unfortunately, this prevents study of the more interesting
properties, such as scattering amplitudes and the precise form of the
potential (we have left $n$ arbitrary in $S_2$), since it's
superrenormalizable in D$²$2 regardless of its form.  However, these
limitations probably would not appear in a corresponding formulation of
the superstring, which has no tachyons.

An interesting feature of this model is the use of Gaussian propagators to
get rid of the usual perturbative divergences of momentum integration. 
Naively, one might suspect that such field theories were completely
finite.  However, we know in this case that the bosonic string does have
divergences perturbatively in the string coupling, and that there are
further problems unless D=26.  This demonstrates that modifying a theory
to fix problems seen in perturbation theory does not preclude the
reappearance of such difficulties nonperturbatively.

These Gaussian propagators lead to Gaussian behavior of fixed-angle scattering
(as we will see in subsection XIB6), in conflict with hadronic physics, where power-law
behavior is observed for partons with large transverse momenta, and is a
theoretical consequence of asymptotic freedom with the usual
propagators.  (In fact, it is the main empirical verification of QCD.)

Since nonrelativistic first-quantization gives Gaussian propagators
$e^{-x^2/t}$, it is not surprising that the simplest strings should result in
partons with Gaussian propagators $e^{-x^2}$. However, the fact that
first-quantization for particles leads instead to, e.g., $1/x^2$
propagators for massless particles in 4D position space suggests that
an analogous treatment for strings should be possible.  We thus attempt
to follow the derivation above from parton to string,
but starting with realistic parton propagators.  The first step is to
exponentiate the propagator so that the exponent can be identified with
a first-quantized action.  The easiest way, and that most analogous to
the nonrelativistic case, is to use the Schwinger parametrization of the
propagator, which follows from the appearance of the worldline metric in
the action:
$$ {1\over üp^2} = Ç_0^¥ d ¼e^{- p^2/2} $$
 As we saw in subsection VC8, a Feynman diagram in a scalar field theory
with nonderivative self-interactions is then written as
$$ Çdx_i' dp_{ij} d _{ij}¼e^{-Ý_{ÒijÔ}[ _{ij}p_{ij}^2/2 -i(x_i-x_j)Ép_{ij}]} $$
 In the (worldsheet) continuum limit of this expression, $p$ becomes a
worldsheet vector, so $ $ must become a symmetric worldsheet tensor. 
Since on a regular square lattice (``flat" worldsheet) there are two
propagators per vertex (for the two independent directions), $ $ must be
a traceless tensor.  (This also explains why $ $ can't be just a scalar.)
Imposing this tracelessness through a Lagrange multiplier $Â$, we can
write the (Wick rotated) continuum action as
$$ S = Ç{d^2 §\over 2¹}Ó-iP^mÉ»_m X
	+ü _{mn}(P^mÉP^n +Âgg^{mn}) +å{-g}[µ +(ln¼û)üR]Õ $$
 Thus $ $ acts as a kind of second worldsheet metric.  However, since
Schwinger parameters are positive, $ _{mn}$ must be positive definite,
and thus a Euclidean metric.  This also implies that $g_{mn}$ must be
Minkowskian, to be consistent with the tracelessness condition.  Note
that if we set $Â$ equal to a constant, and ignore the positivity condition
on $ $, then eliminating $ $ by the equation of motion from varying
$g_{mn}$ reproduces the usual string action, where we can identify
$Œ'=µ/ÒÂÔ$.  This indicates a possible approximation scheme.

The two components of $ $ that survive this tracelessness condition
correspond to the two lightlike directions defined by $g_{mn}$:  If we use
a ``zweibein", defined as usual by $g_{mn}=-e_{(m}{}^+ e_{n)}{}^-$, to
flatten the indices on $ $, then the Lagrange multiplier constraint can be
solved by simply setting $ _{+-}=0$.  The action is then
$$ S = Ç{d^2 §\over 2¹}å{-g}[iP_àÉe_¦{}^m »_m X
	+ü _{àà}P_¦ÉP_¦ +µ +(ln¼û)üR] $$
 Back on the lattice, this implies that the directions chosen by the
propagators (links) on which $P$ is defined are lightlike.  Thus, the matrix
model defined by this theory should have only 4-point vertices, with the
four propagators coming from any vertex forming the worldsheet
lightcone at that point on the worldsheet.  The field theory action is thus
$$ S_2 =  N¼trÇ{d^D x\over (2¹)^{D/2}} (-\f14ÄõÄ -G\f14 Ä^4) $$
  For $D=4$, this action describes an asymptotically free theory,
``wrong-sign" $Ä^4$ theory.

Unlike conventional strings, this ``QCD string" has critical dimension D=4 for
renormalizability.  (In conventional strings all momentum integrals are
Gaussian and thus converge.)  Another reason for D=4 is T-duality: 
T-duality interchanges the positions of the vertices with the momenta of
the loops.  This is clear from our discussion of the classical mechanics of
Feynman diagrams in subsection VC8, if we note that the procedure we
used there to translate from coordinates to loop momenta is exactly the
random lattice version of the T-duality transformation performed in
subsection XIA5 (with $÷X$ as the loop momenta).  Thus, invariance of a
string theory under T-duality must include invariance of the propagators
of the underlying field theory under Fourier transformation.  This is trivial
for conventional strings, since the Fourier transform of a Gaussian is a
Gaussian.  However, by dimensional analysis (or explicit evaluation: see
exercise VIIB4.2), we see that the Fourier transform of $1/p^2$ is
$1/x^2$ ÓonlyÕ in D=4:  T-duality implies both D=4 and masslessness. 
Furthermore, we can look at interactions by considering the simplest
case:  The flat worldsheet is represented by a regular, flat lattice.  For
$Ä^4$ theory we have the usual square lattice, which is self-dual under
switching vertices with loops (T-duality).  On the other hand, triangular
and hexagonal lattices, corresponding to $Ä^6$ and $Ä^3$ theory, are dual
to each other (i.e., $Ä^n$ is dual to $Ä^{2n/(n-2)}$, as follows from
geometry).  Thus T-duality also implies the $Ä^4$ interaction.

\x XIA7.2  Let's examine T-duality for the random lattice more carefully:
 ªa Repeat the T-duality transformation of subsection XIA5, but for the
QCD string (see subsection VC8), without a background
($M^{mn}=ú^{mn}$).  Show that invariance under $X^m£÷X_m$ requires
that the matrix $ $ also be replaced by its inverse, with some factors of
the 2D $·$ tensor.  ($Â$ also transforms; you can avoid this complication
by using the zweibein form of the action.)
 ªb Write the massless scalar propagator in momentum space of arbitrary
dimension D as an exponential using a Schwinger parameter $ $.  Show
that after T-duality --- Fourier transformation combined with $ £1/ $
(which leaves the exponent invariant) --- a $ $-dependent ``measure"
factor is introduced, ÓexceptÕ for D=4.

\refs

£1 T. Regge, ÓNuo. Cim.Õ É14 (1959) 951, É18 (1960) 947.
 £2 R. Dolen, D. Horn, and C. Schmid, \PR 19 (1967) 402:\\
	duality.
 £3 D.J. Gross, A. Neveu, J. Scherk, and J.H. Schwarz, \PL 31B (1970)
	592;\\
	C. Lovelace, \PL 34B (1971) 500;\\
	E. Cremmer and J. Scherk, \NP 50 (1972) 222:\\
	string loops as 1-loop tadpole insertions.
 £4 Y. Nambu, Quark model and the factorization of the Veneziano
	amplitude, in ÓProc. of International conference on Symmetries and
	quark modelÕ, ed. R. Chaud, Wayne State U., June, 1969
	(Gordon and Breach, 1970) p. 269;\\
	L. Susskind, \PRD 1 (1970) 1182, ÓNuo. Cim.Õ É69A (1970) 457;\\
	H.B. Nielsen, An almost physical interpretation of the integrand of the
	n-point\\ Veneziano model, 15th international conference on high
	energy physics, Kiev, 1970:\\
	dual model as string, through mechanics.	
 £5 Y. Nambu, lectures at Copenhagen Symposium, 1970, unpublished;\\
	O. Hara, ÓProg. Theo. Phys.Õ É46 (1971) 1549;\\
	T. Goto, ÓProg. Theo. Phys.Õ É46 (1971) 1560;\\
	H. Noskowitz, unpublished:\\
	area action for string.
 £6 M. Virasoro, \PRD 1 (1970) 2933;\\
	I.M. Gelfand and D.B. Fuchs, ÓFuncts. Anal. PrilozhenÕ É2 (1968) 92:\\
	Virasoro constraints.
 £7 P.A. Collins and R.W. Tucker, \PL 64B (1976) 207;\\
	L. Brink, P. Di Vecchia, and P. Howe, \PL 65B (1976) 471;\\
	S. Deser and B. Zumino, \PL 65B (1976) 369:\\
	mechanics of string with worldsheet metric.
 £8 Ramond; Neveu and Schwarz; Óloc. cit.Õ (IIC):\\
	almost superstrings (fermions included).
 £9 M.B. Green and J.H. Schwarz, \PL 109B (1982) 444, É149B (1984) 117:\\
	superstrings.
 £10 D.J. Gross, J.A. Harvey, E. Martinec, and R. Rohm, \PR 54 (1985) 
	502, \NP 256 (1985) 253, É267 (1986) 75:\\
	heterotic string.
 £11 W. Siegel, \PL 134B (1984) 318;\\
	T.H. Buscher, \PL 194B (1987) 59, É201B (1988) 466:\\
	T-duality as transformation on worldsheet.
 £12 K. Kikkawa and M. Yamasaki, \PL 149B (1984) 357;\\
	N. Sakai and I. Senda, ÓProg. Theor. Phys.Õ É75 (1986) 692;\\
	V.P. Nair, A. Shapere, A. Strominger, and F. Wilczek, \NP 287 (1987)
	402;\\
	B. Sathiapalan, \PR 58 (1987) 1597;\\
	R. Dijkgraaf, E. Verlinde, and H. Verlinde, ÓComm. Math. Phys.Õ É115
	(1988) 649;\\
	K.S. Narain, M.H. Sarmadi, and E. Witten, \NP 279 (1987) 369;\\
	P. Ginsparg, \PRD 35 (1987) 648;\\
	P. Ginsparg and C. Vafa, \NP 289 (1987) 414;\\
	S. Cecotti, S. Ferrara, and L. Girardello, \NP 308 (1988) 436;\\
	R. Brandenberger and C. Vafa, \NP 316 (1988) 391;\\
	A. Giveon, E. Rabinovici, and G. Veneziano, \NP 322 (1989) 167;\\
	A. Shapere and F. Wilczek, \NP 320 (1989) 669;\\
	M. Dine, P. Huet, and N. Seiberg, \NP 322 (1989) 301;\\
	J. Molera and B. Ovrut, \PRD 40 (1989) 1146;\\
	K.A. Meissner and G. Veneziano, \PL 267B (1991) 33;\\
	A.A. Tseytlin and C. Vafa, \xxxlink{hep-th/9109048},
	\NP 372 (1992) 443;\\
	M. Ro×cek and E. Verlinde, \xxxlink{hep-th/9110053},
	\NP 373 (1992) 630;\\
	J.H. Horne, G.T. Horowitz, and A.R. Steif, \xxxlink{hep-th/9110065},
	\PR 68 (1992) 568;\\
	A. Sen, \PL 271B (1991) 295;\\
	A. Giveon and M. Ro×cek, \xxxlink{hep-th/9112070}, 
	\NP 380 (1992) 128:\\
	T-duality as symmetry of spacetime fields.
 £13 M.J. Duff, \NP 335 (1990) 610;\\
	 J. Maharana and J.H. Schwarz, \xxxlink{hep-th/9207016},
	\NP 390 (1993) 3;\\
	W. Siegel, \xxxlink{hep-th/9302036}, \PRD 47 (1993) 5453,
	\xxxlink{hep-th/9305073}, \PRD 48 (1993) 2826,
 	\xxxlink{hep-th/9308133}, 
	Manifest duality in low-energy superstrings, in 
	ÓProc. of the Conference Strings '93Õ, Berkeley, CA, May 24-29,
	eds. M.B. Halpern, G. Rivlis, and A. Sevrin (World Scientific, 1995) p. 353:\\
	Metric and axion as coset space for T-duality.
 £14 Siegel, Óloc. cit.Õ (IXB, 1st and 2nd papers of ref. 5):\\
	coupling of (unphysical) dilaton to string (through ghosts).
 £15 E.S. Fradkin and A.A. Tseytlin, \PL 158B (1985) 316, \NP 261 (1985) 1:\\
	coupling of dilaton through worldsheet curvature.
 £16 T. Banks, D. Nemeschansky, and A. Sen, \NP 277 (1986) 67:\\
	relation between above two couplings.
 £17 Fradkin and Tseytlin, Óloc. cit.Õ;\\
	C.G. Callan, D. Friedan, E.J. Martinec, and M.J. Perry, \NP 262 (1985)
	593:\\
	low-energy closed-string actions for massless fields.
 £18 S. Cecotti, S. Ferrara, and M. Villasante, ÓInt. J. Mod. Phys. AÕ É2 (1987)
	1839:\\
	superspace action for 4D massless part of heterotic string.
 £19 W. Siegel, \PL 211B (1988) 55:\\
	massless part of heterotic string is old-minimal supergravity coupled
	to tensor multiplet, as follows from direct product of open strings.
 £20 W. Siegel, \xxxlink{hep-th/9510150}, \PRD 53 (1996) 3324:\\
	massless part of string actions from direct product and dilaton
	homogeneity.
 £21 H.B. Nielsen and P. Olesen, \PL 32B (1970) 203;\\
	D.B. Fairlie and H.B. Nielsen, \NP 20 (1970) 637;\\
	B. Sakita and M.A. Virasoro, \PR 24 (1970) 1146:\\
	worldsheet lattice as Feynman diagrams.
 £22 F. David, \NP 257 [FS14] (1985) 543;\\
	V.A. Kazakov, I.K. Kostov, and A.A. Migdal, \PL 157B (1985) 295;\\
	J. Ambj\o rn, B. Durhuus, and J. Fršhlich, \NP 257 (1985) 433:\\
	integration over worldsheet metric as sum over Feynman diagrams.
 £23 M.R. Douglas and S.H. Shenker, \NP 335 (1990) 635;\\
	D.J. Gross and A.A. Migdal, \PR 64 (1990) 127;\\
	E. Br«ezin and V.A. Kazakov, \PL 236B (1990) 144:\\
	continuum limit for worldsheet lattice with metric and 1/N
	expansion.
 £24 G. Veneziano, ÓNuo. Cim.Õ É57A (1968) 190;\\
	V. Alessandrini, D. Amati, and B. Morel, ÓNuo. Cim.Õ É7A (1971) 797;\\
	D.J. Gross and P.F. Mende, \PL 197B (1987) 129, \NP 303 (1988) 407;\\
	D.J. Gross and J.L. Ma÷nes, \NP 326 (1989) 73:\\
	Gaussian behavior of string amplitudes.
 £25 P.A. Collins and R.W. Tucker, \NP 112 (1976) 150;\\
	B. de Wit, M. L¬uscher, and H. Nicolai, \NP 305 (1988) 545:\\
	problems quantizing membranes.
 £26 C.M. Hull and P.K. Townsend, \xxxlink{hep-th/9410167}, \NP 438
	(1995) 109:\\
	the 11th dimension arises nonperturbatively in string theory.
 £27 Siegel, Óloc. cit.Õ (VC):\\
	QCD string with Schwinger parameters as second worldsheet metric.
 £28 A.M. Polyakov, \NP 268 (1986) 406:\\
	two-metric formulation of usual string.
 £29 M.B. Green, J.H. Schwarz, and E. Witten, ÓSuperstring
	theoryÕ, 2 v. (Cambridge University, 1987);\\
	J. Polchinski, ÓString theoryÕ, 2 v. (Cambridge University, 1998):\\
	comprehensive texts on strings.
 £30 P.H. Frampton, ÓDual resonance models and string theoriesÕ (World
	Scientific, 1986);\\
	Siegel, Óloc. cit.Õ;\\
	M. Kaku, ÓIntroduction to superstrings and M-theoryÕ, 2nd ed.
	(Springer-Verlag, 1999):\\
	other string texts.
 £31 S. Fubini; G. Veneziano; V. Alessandrini, D. Amati, M. Le Bellac,
	and D. Olive; J.H. Schwarz; C. Rebbi; S. Mandelstam; ÓDual theoryÕ, ed.
	M. Jacob (North-Holland, 1974);\\
	J. Scherk, ÓRev. Mod. Phys.Õ É47 (1975) 123:\\
	old string reviews.

\unrefs

Û7 B. QUANTIZATION

We saw in subsections VIIB5 and VIIIA7 some unusual features of massless
theories in D=2.  Since the mechanics of the string is mathematically
equivalent to 2D field theory (as the mechanics of the particle is to 1D
field theory), we now examine such field theories in a little more detail. 
In particular, since the string we studied in section XIA possessed local
Weyl scale invariance on the worldsheet, we are directed to 2D conformal
field theories coupled to 2D gravity.

One confusion for beginners in string theory that unfortunately is supported by some of the terminology is the distinction between first- and second-quantization:  The first quantization of the string is often described as ``2D (conformal) field theory", with the justification that they are allegedly the same mathematically.  In the same spirit, one might also say that addition and multiplication are mathematically the same, but no mathematician would ever say that when applied to, e.g., the real numbers, even though they share some properties.  For the same reasons, we must distinguish between first- and second-quantization of string theory.

As we saw in subsection IIIA3, they use different perturbation expansions, corresponding to whether the $\h$ is put in front of the mechanics action of subsection XIA3 or a corresponding ``string field theory" action:  (1) For the string, the expansion about classical mechanics is an expansion in $Œ'$.  This is again a JWKB expansion, an expansion in powers of momenta, since $Œ'$ has dimensions of (mass)${}^{-2}$.  (2) The expansion about classical field theory is as usual an expansion in the (string) coupling constant $g$.
$$ \li{ \hbox{1st-q:} & ââŒ' £ \h Œ' \cr
	\hbox{2nd-q:} & ââg^2 £ \h g^2 \cr} $$

In what follows we will often use the terminology ``conformal field theory" to describe this situation, keeping in mind that as far as application to string theory is concerned a more appropriate term would be ``conformal mechanics".  (True conformal field theory does appear in string theory when applied to the Anti-de Sitter/Conformal Field Theory correspondence, where the relevant 4D conformal field theory is maximally supersymmetric Yang-Mills.)  The mathematical methods of conformal string mechanics are also applied as true 2D conformal field theory in statistical mechanics, for the purpose of studying 2D systems, or as a toy model for better understanding 4D conformal field theory.

Ü1. Gauges

We begin by considering gauge choices for the various forms of the bosonic string action presented in subsection XIA3.
In direct analogy to the particle (subsection IIIB2), the two most useful
gauges are the ``conformal gauge", defined by completely fixing the
worldsheet metric, and the lightcone gauge, which is not manifestly
globally covariant but is a complete fixing of the residual gauge
invariance of the conformal gauge.  In the conformal gauge we set
$$ g_{mn} = ú_{mn} $$
 by using the 2 coordinate invariances and the 1 scale invariance to fix
the 3 components of the symmetric tensor $g_{mn}$.  The coordinate
part of this gauge is essentially the temporal gauge $g_{0m}=ú_{0m}$,
just as for the particle ($-g_{00}=v^2=1$).  Also as for the particle, this
gauge can't be fixed everywhere (see also subsections IIIA5 and IIIC2),
but the equation of motion from the metric is implied everywhere by
imposing it at just the boundaries in $ $.  In this gauge the equations
of motion for $X$ are just the 2D Klein-Gordon equation, which is easy to
solve in 2D lightcone coordinates:
$$ »_+ »_- X = 0âÜâX = X_{(+)}( +§) +X_{(-)}( -§) $$
 (We have used $ à§$ in place of $§^à$ for later convenience.)  The
constraints are then $öP_{(à)}^2¾(X_{(à)}')^2=0$.  This directly relates to
the form of 2D conformal transformations, which are infinite-dimensional
in D=2:
$$ ds^2 = 2d§^+ d§^-âÜâ§'^+ = f_{(+)}(§^+),â§'^- = f_{(-)}(§^-) $$
 The constraints are the generators of these conformal transformations. 
(As described in subsection IIIA5, the constraints generate the gauge
transformations; the global transformations are those that preserve the
temporal gauge.)

For the lightcone gauge, we again fix the (spacetime) +-components of
the variables, and solve the +-components of the equations of motion
(found by varying the $-$-components).  Looking at the equations of
motion first, using the first-order form of the action,
$$ 0 = {¶S\over P^{-m}} ¾ »_m X^+ +(-g)^{-1/2}g_{mn}P^{+n} $$
$$ Üâ(-g)^{-1/2}g_{mn} = (AÉB)^{-1}(·_{mp}A^p ·_{nq}A^q -B_m B_n);ââ
	A^m = P^{+m},âB_m = »_m X^+ $$
 (as seen, e.g., by using $·_{mn}A^n,B_m$ as a basis), and
$$ 0 = {¶S\over X^-} ¾ »_m P^{+m}âÜâ{d\over d }Çd§¼P^{+0} = 0 $$
 which identifies $Çd§¼P^{+0}$ as the conserved momentum $p^+$, up to a
factor of $2¹Œ'$ (since $p$ is really the coefficient of $Àx$ in the action,
where $X(§)=x+...$).  Similarly, $¶S/¶g^{mn}$ determines $P^{-m}$, and
thus $X^-$.

We then choose as our main set of gauge conditions
$$ X^+ = k ,ââP^{+0} = k $$
 for some constant $k$, which explicitly determines $ $, and determines
$§$ up to a function of $ $:  An equivalent way to define the lightcone $§$
in terms of an arbitrary spacelike coordinate $§'$ is
$$ § = k^{-1}Ç_0^§ d§'¼P^{+0}(§') $$
 which identifies $§$ as the amount of momentum $p^+$ between that
value of $§$ and $§=0$ (at fixed $ $).  We thus have that the length of the
string (the range of $§$, not the physical length) is
$$ l = k^{-1}Çd§¼P^{+0} = 2¹Œ'p^+ k^{-1} $$
 We then need to fix the location of $§=0$ as some function $§'( )$:  Since
in this gauge
$$ »_1 P^{+1} = 0 $$
 so $P^{+1}$ is also a function of just $ $, we further fix the gauge for
$§$ by choosing
$$ P^{+1} = 0âÜâ(-g)^{-1/2}g_{mn} = ú_{mn} $$
 Thus the lightcone gauge is a special case of the conformal gauge, after
also fixing scale gauge $g=-1$. For the open string, this almost fixes
$§'( )$ at $§=0$, which we can take as one boundary:  The boundary
condition for $X^+$ is now
$$ 0 = nÉ» X^+ ¾ n_0 $$
 since in this (and any conformal) gauge $»_m X¾ú_{mn}P^n$.  Thus the
normal to the boundary must be in the $§$ direction, so the boundary is at
constant $§$.  This means we have one constant left to fix:
$$ § = 0¼at¼one¼boundary¼(open¼string) $$
 This invariance was left because all our previous gauge conditions
preserved global $§$ translation.  Unfortunately, there is no
corresponding convenient gauge choice for the closed string, so there we
leave just this one invariance.  In summary, our complete set of
lightcone gauge conditions is now:
$$ gauge:ââX^+ = k ,ââP^{+m} = k¶^m_0,ââ
	§ = 0¼at¼one¼boundary¼(open¼string) $$
 The lightcone action is now, in Hamiltonian form,
$$ S_{lc} = Çd \leftÓÀx{}^- p^+ +Ç{d§\over 2¹}
	\left[-ÀX_i P_i +ü(Œ'P_i^2 +Œ'^{-1}X_i'^2)\right]\rightÕ $$

\x XIB1.1  Analyze the classical mechanics of the string by approximating $§$ by a set of discrete points, so $X'(§)£X_{n+1}-X_n$, etc.  Show that the string then acts as a bunch of particles connected by springs, and find all the usual spring properties: tension, speed of wave propagation, etc.  (Note:  You may need some lightcone modifications of nonrelativistic variables.)

The only distinction between open and closed strings is the boundary
condition (since closed strings by definition have no boundary).  For
closed strings we have only periodicity in $§$ (by definition of ``closed"),
while for open strings we have
$$ X'( ,0) = X'( ,l) = 0 $$
 One consequence, as we just saw, is that closed strings have one residual
gauge invariance in the lightcone gauge.  As described in subsection XIA3, these two strings can be made
to resemble each other more closely by extending the open string to
twice its length, defining $X$ for negative $§$ by
$$ X( ,-§) = X( ,§) $$
 This is the known as the ``method of images":  $X( ,-§)$ is identified
with its mirror image in the $ $ axis, $X( ,-§)$.  Then the two strings both
satisfy periodic boundary conditions, while the open string has this one
additional condition.  We also choose
$$ k = 2ûŒ'p^+,ââ
	û = \leftÓ \matrix{ 1 & (open) \cr ü & (closed) \cr} \right.
	âÜâl = {¹\over û} $$
 so the length of the closed string is $2¹$, while the open string has
original length $¹$ that has now been doubled to match the closed string.
Our choice of ``phase" in relating $X$ for positive and negative $§$ for
the open string automatically enforces the boundary condition $X'( ,0)=0$
at one end of the string, while the condition $X'( ,¹)=0$ at the other end
is implied in the same way by the closed string ``boundary condition" of
periodicity, which can be written as $X( ,¹) = X( ,-¹)$.  The picture is
then that the open string is a closed string that has collapsed on itself, so
that for half of the range of $§$ $X$ doubles back over the path it
covered for the other half.  

Because $§$ has a finite range, $X$ can always be expanded in Fourier
modes in that variable; the boundary conditions slightly restrict the form
of this expansion.  We saw that the equations of motion, being
second-order in $ $-derivatives, gave two modes for each initial state: a
left-handed one and a right-handed one.  We need to be a bit more
precise about the ``zero-modes" (modes independent of $§$):  We can separate
them out as
$$ X( ,§) = x +{2¹Œ'\over l}p  
	+å{\f{Œ'}2}[Y_{(+)}( +§) +Y_{(-)}( -§)],ââÇd§¼Y_{(à)} = 0 $$
 where $Y$ contains only nonzero-modes.  (The normalization of $p$,
conjugate to $x$, comes from the $-ÀxÉp$ term in the Lagrangian.)  Then
$x$ represents the ``center of mass" of the string, and $p$ its total
momentum.  Note that this implies $X_{(à)}$ aren't quite periodic:
$$ X_{(à)}(§+2¹) = X_{(à)}(§) +2¹ûŒ'p $$
 Now the periodicity boundary conditions shared by open and closed
strings imply
$$ Y_{(à)}(§+2¹) = Y_{(à)}(§) $$
 while the extra boundary condition for the open string implies
$$ Y_{(+)}(§) = Y_{(-)}(§) = Y(§) $$
 allowing us to drop the subscript in that case.  Thus, the closed string has
twice as many modes as the open, except for the nonperiodic part,
corresponding to the total momentum and average position.  This is
related to the interpretation that the open string is a closed string with
its two halves occupying the same path.  This doubling also shows up in
the constraints: For the closed string we have $öP_{(à)}^2$, while for the
open string we can consider just $öP_{(+)}^2$, since
$öP_{(-)}^2(§)=öP_{(+)}^2(-§)$.  In the lightcone gauge we solve these
constraints for $X^-$, by integrating
$$ 0 = öP_{(à)}^2 ¾ ÀX_{(à)}^2 
	= \left( ÀX_{(à)}^i \right)^2 -kÀX_{(à)}^- ¾ (ÀY_{(à)} +ûå{2Œ'}p)^2 $$

\x XIB1.2  Rederive the solution to the boundary conditions for the open
string without using $X( ,-§)=X( ,§)$ (and periodicity):  The string, as
originally, extends between boundaries at 0 and $¹$.

This separation of zero-modes from nonzero-modes also allows us to find
the spin and mass of the string:  In any conformal gauge,
$$ 0 = p^2 +M^2 = {1\over ûŒ'^2}Ç_0^{¹/û}{d§\over 2¹}¼ü(ÀX{}^2 +X'^2)
	âÜâM^2 = {1\over 2ûŒ'}Ý_àÇ{d§\over 2¹}¼ÀY{}^2_{(à)} $$
$$ J^{ab} = x^{[a}p^{b]} +S^{ab}
	= {1\over Œ'}Ç_0^{¹/û}{d§\over 2¹}¼X^{[a}ÀX{}^{b]}
	âÜâS^{ab} = Ý_àÇ{d§\over 2¹}¼Y{}^a_{(à)}ÀY{}^b_{(à)} $$
 (using the Hermitian form of the Lorentz generators, for classical
purposes), where for the open string we can replace
$$ Ý_àÇ_0^¹{d§\over 2¹} £ Ç_0^{2¹}{d§\over 2¹},ââY_{(à)} £ Y $$
 For the lightcone gauge we then have the gauge condition to
determine $X^+$ and Virasoro constraints to determine $X^-$:
$$ Y^+_{(à)} = 0,ââûŒ'p^- +å{üŒ'}ÊÀY{}^-_{(à)} = 
	{1\over 2ûŒ'p^+}\left(ûŒ'p^i +å{üŒ'}ÊÀY{}^i_{(à)}\right)^2 $$

\x XIB1.3  Consider gauge fixing in the temporal gauge, replacing $X^+$
with $X^0$.  The classical interpretation is now simpler, since $ $ and
$X^0$ can now be identified with the usual time.  Everything is similar
except that the Virasoro constraints can't be solved (e.g., for $X^1$) in
general without square roots. 
ªa Show that some 3D solutions (2 space, 1 time) for the open string are
given by, for $p^1 = p^2 = 0$,
$$ \f1{å2}(ÀY{}^1 -iÀY{}^2)( ) = c e^{-in },ââ $$
 for nonzero integer $n$.  (Without loss of generality, we can choose $c$
real and positive.)  Find the mass (energy) and spin as
$$ M = {c\over å{Œ'}},ââS^{12} = {c^2\over n} = {Œ'\over n}M^2 $$
 Find $X$ explicitly, and show it describes an ``n-fold spinning rod".
ªb Show that the above solution can be generalized to closed strings by
using two such $Y$'s, and fixing the relative magnitude of the two $c$'s. 
Consider the special cases where $n_-=àn_+$.  Find the explicit
masses, spins, and $X$'s, and show that one describes another n-fold
spinning rod, while the other is an ``n-fold oscillating ring".

Ü2. Quantum mechanics

The more interesting features of the string don't appear until
quantization.  In particular, we can already see at the free level the
discrete mass spectrum characteristic of Regge theory, or of bound
states in general.

Canonical quantization is simplest in the lightcone gauge.  As for
particles, canonical quantization is convenient only in mechanics (first
quantization), not field theory (second quantization).  As can be seen
from the lightcone action, the Hamiltonian is part of the constraints:  For
the spinless particle, we had only the constraint $p^2+m^2=0$, which
became $E=H$ in the lightcone gauge $X^+=p^+ $ after identifying the
lightcone ``energy" $E=p^+p^-$ and its Hamiltonian $H=ü(p_i^2+m^2)$. 
(See subsection IIIB2.)  The string Hamiltonian can be rewritten
conveniently in terms of $öP$.  Since the closed string is effectively just a
doubling of the open string, we treat the open string first.  The
Hamiltonian is simply
$$ H = Ç_{-¹}^¹{d§\over 2¹}üöP_i^2 $$
 where $öP=öP_{(+)}$.  Since we have chosen $X^+=2Œ'p^+ $, we have
$E=2Œ'p^+p^-$.

To identify the individual particle states, we Fourier expand the
worldsheet variables in $§$.  As for the particle, we can work at $ =0$,
since all the dynamics is contained in the constraints.  Equivalently,
from the nonrelativistic view of the lightcone formalism, we can work in
the Schr¬odinger picture where the $ $ dependence is in the wave function
instead of the operators.  We expand as
$$ öP(§) = Ý_{n=-¥}^¥ ÷a_n e^{-in§},ââ÷a_0 = å{2Œ'}p,ââ÷a_{-n} = ÷a_nÿ $$
The canonical commutation relations for $P$ and $X$ are
$$ [P_i(§_1),X_j(§_2)] = -2¹i¶(§_2-§_1)¶_{ij} $$
 as the direct generalization of the usual $[p,q]=-i$.
(The $2¹$ is from our normalization $d§/2¹$.)  From the definition of $öP$, we then have
$$ [öP_i(§_1),öP_j(§_2)] = -2¹i¶'(§_2-§_1)¶_{ij} $$

We can then decompose this into modes by multiplying by
$e^{i(m§_1+n§_2)}$ and integrating, where
$$ Ç{d§\over 2¹}Êe^{in§} = ¶_{n0} $$
 We then find
$$ [÷a_{im},÷a_{jn}] = m¶_{m+n,0}¶_{ij} $$
 as well as the usual $[p_i,x_j]=-i¶_{ij}$, and thus can relate the modes to
the usual harmonic oscillator creation and annihilation operators:
$$ ÷a_n = åna_n,ââ÷a_{-n} = åna_nÿâÜâ[a_m,a_nÿ] = ¶_{mn} $$
 for positive $n$.  After normal ordering, we find for the Hamiltonian
$$ H = Œ'p_i^2 +N -Œ_0,ââN = Ý_{n=1}^¥ na_{in}ÿa_{in} $$
$$ ÜâH-E = Œ'(p_a^2 +M^2),ââM^2 = Œ'^{-1}(N -Œ_0) $$
 for some constant $Œ_0$, which we introduce as a ``renormalization constant" for removing an infinity in normal ordering.

From the expression for the mass in terms of the number operator $N$,
we see that the $n$th oscillator $a_{in}ÿ$ raises the mass-squared of the
ground state $|0Ô$ by $n$ (and similarly for multiple applications of these
oscillators).  For any given mass, the highest-spin state is the symmetric,
traceless tensor part of multiple $a_{i1}ÿ$'s acting on $|0Ô$:  This describes
the leading Regge trajectory, with spins 
$$ j = Œ'M^2 +Œ_0 $$
 Let's look first at the first excited level, obtained by acting on the scalar
ground state $|0Ô$ with the lowest-mass oscillators $a_{i1}ÿ$.  Clearly this
describes a (lightcone) transverse vector, with no St¬uckelberg scalar for
describing a massive vector.  (I.e., it has only D$-$2 components, not the
D$-$1 necessary for a massive vector.)  Thus this state describes a
massless vector, so 
$$ Œ_0 = 1 $$
 The ground state is then a scalar tachyon with $M^2=-Œ'^{-1}$.  For any
given level past the first excited level, one can check explicitly that the
states coming from the various oscillators include the necessary
St¬uckelberg fields.  For example, at the second excited level,
$a_{i1}ÿa_{j1}ÿ$ contains a traceless, symmetric tensor and a scalar
(coming from the trace), while $a_{i2}ÿ$ is a vector; they combine to
describe a massive tensor.  The proof that this works to all mass levels
is closure of the Poincar«e algebra quantum mechanically:  The only
nontrivial commutator is $[J^{-i},J^{-j}]=0$, since only $J^{-i}$ is higher
than quadratic (cubic) in oscillators (from the form of $X^-$ and $P^-$
after solving constraints), so normal-ordering ambiguities lead to more
than just constant terms.  For reasons to be explained in chapter XII, the
algebra(ic computations) in calculating this commutator in any
first-quantized theory for the lightcone gauge is the same as the first-quantized BRST algebra for general gauges.  Of course,
the proof of closure is already the same in principle because both
algebras are a consequence of the constraints, the conformal algebra. 
Thus, any anomaly must show up in the conformal algebra itself, which
will be considered in subsection XIB4. 

\x XIB2.1  Check the third excited level massive representations.

The closed string works similarly to the open, but with two sets of
harmonic oscillators, and with 
$$ p_{(+)} = p_{(-)} = üp $$
 In that case we find
$$ M^2 = 2Œ'^{-1}(N_{(+)} +N_{(-)} -2) $$
 where $N_{(+)}$ and $N_{(-)}$ are the number operators for the two
independent sets of oscillators.   In the lightcone gauge the closed string
has the residual gauge invariance generated by $Çd§¼X'ɶ/¶X$; this gives
the residual constraint
$$ N_{(+)} = N_{(-)} $$
 The closed-string states are thus the direct product of two open-string
states of the same mass:  For example, the ground state is a scalar
tachyon with $M^2=-2Œ'^{-1}$, while the first excited states are
massless ones from the product of two vectors --- a scalar, an
antisymmetric tensor, and a symmetric, traceless tensor.  The leading
Regge trajectory consists of states created with equal numbers of
$a_{i1(+)}ÿ$'s and $a_{i1(-)}ÿ$'s, with 
$$ j = üŒ'M^2 +2 $$
 In summary, the leading trajectory for open or closed string is given by
$$ \boxeq{ j = ûŒ'M^2 +{1\over û},ââ
	û = \leftÓ \matrix{ 1 & (open) \cr ü & (closed) \cr} \right. } $$

Covariant quantization of the string can be performed in several ways: 
One is to use the OSp methods of chapter XII, as applied to the Lorentz
generators derived from the lightcone analysis (see subsection XIIB8).  Another is to use the
usual BRST of subsection VIA, as applied to gravity in subsection IXB1,
treating the mechanics of the string as a 2D field theory.  For the case of
the conformal gauge, introducing ghosts $C^m$ corresponding to the
gauge parameters, and antighosts $B_{mn}$ paired with the Lagrange
multipliers of the gauge conditions (Nakanishi-Lautrup fields), we find the
ghost action
$$ å{-g}g^{mn} = ú^{mn},ââ
	¶å{-g}g^{mn} = á^{(m}Â^{n)} -g^{mn}g_{pq}á^p Â^q $$
$$ ÜâL_g = B_{++}á_- C_- +B_{--}á_+ C_+ $$
 where in the last step we have introduced a background ``zweibein" for
applications such as the background field gauge, or geometries that do
not admit the conformal gauge globally, and flattened the indices on the
ghosts so the tracelessness of $B$ (which follows from that of
$¶å{-g}g^{mn}$) can be solved explicitly.  
(The Weyl scale transformation of the gauge-fixing condition for the conformal gauge does not involve derivatives, so the Weyl scale ghosts are just algebraic.)

In covariant gauges, for purposes of calculating more complicated quantities than the spectrum, it will prove useful to work directly in terms of 2D (conformal) field theory in the position space of the worldsheet, rather than Fourier transforming to a mode expansion.  As usual, we Wick rotate to Euclidean space, after which we work in terms of complex coordinates
$$ ¨ =   +i§ $$
Then the usual spin $ü$ fields on shell are directly functions of just $¨$ or $Ш$, while the usual scalars break up into a sum of both (see subsection VIIB5).  In particular, for $X$ we can write
$$ X = å{\f{Œ'}2}[ X_L(¨) +X_R(Ш) ] $$
where we have introduced normalization consistent with earlier parts of the book, since the action we used in for a scalar $Ä$ is $Œ'/2$ times what we used for $X$ in subsection XIA3.

As described above, 
for the open string we always combine the two chiralities, such as $Æ_L(¨)$ and $Æ_R(Ш)$, on the interval $§ã[0,¹]$, into a single chirality on $§ã[-¹,¹]$ as
$$ öÆ(§) = Ï(§)Æ_L(§) +Ï(-§)Æ_R(-§) $$
so the open string then looks like a closed string with one handedness. 
We can then use the same function $öÆ(¨)$, evaluated in different halves of the complex plane, to give $Æ_L$ and $Æ_R$, which are both defined in only one half:
$$ Æ_L(¨) = öÆ(¨),ââÆ_R(Ш) = öÆ(Ш) $$
In particular, the previous separation of $X$ into its chiral halves becomes
$$ X = å{\f{Œ'}2}[ öX(¨) +öX(Ш) ] $$
for the open string, while $X_L$ and $X_R$ remain independent for the closed string.  On the boundary of the open string, where we place vertex operators for external open-string states, we have simply $X=å{2Œ'}öX$.

As in 2D electrostatic problems, it's often convenient to use conformal invariance to transform various surfaces with various topologies and boundary conditions to ones with boundaries whose shapes are simple enough (e.g., straight lines) to use techniques like the method of images to solve for propagators.  
(Otherwise, we are restricted to looking at just short-distance behavior, which is independent of the boundaries.)
For now we consider just the simplest examples, the strip (open string) and cylinder (closed string).  
In general (e.g., interactions in the lightcone gauge) we would need to consider strings of various lengths; for now we simplify matters by assuming the length of the string has been scaled to $¹$ for the open string and $2¹$ for the closed, for reasons explained previously.  
(For lightcone treatment of interacting strings, the string length is proportional to $p^+$, so length is ``conserved" when they split or join at the ends.)  

$$ \fig{transform} $$

We then map the open string to the upper-half plane, or the closed string to the whole plane, via
$$ z = e^¨ $$
Since $¨= +i§$, any closed string at fixed $ $ is mapped to a circle, while any open string is mapped to the upper half of a circle.  The two boundaries of the open string at $§=0$ and $¹$ are then mapped to the positive and negative real axis, while the ends of either string at $ =-¥$ and $+¥$ are mapped to the points $z=0$ and $¥$.  The fields will be singular at $z=0$ and $¥$, so they should really be thought of as singular limits of circles.  ($z=¥$ isn't really much of a point anyway.)  

If we use ``Osterwalder-Schrader reality", determining reality in Euclidean space also by Wick rotation from Minkowski, then $¨$ is pure imaginary ($§$ is real, $ $ is now imaginary), so under Wick roation of Minkowki complex conjugation, we find $z£1/z$.  As usual, this switches $ =à¥$ as $z=0ªz=¥$.  The reality condition on real 2D fields is therefore
$$ *(z) = (\f1z) $$

For the closed string the scalar propagator is then as we found in subsection VIIB5, while for the open string we use an image in the lower-half plane to give the appropriate Neumann boundary conditions (vanishing of normal derivative) on the real axis:
$$ G_{closed}(z,z') = -ln(|z-z'|^2) $$
$$ G_{open}(z,z') = G_{closed}(z,z') + G_{closed}(z,Ðz') = -ln(|z-z'|^2) -ln(|z-Ðz'|^2) $$
where the normalization is
$$ ÒļÄÔ = Gâ orâÒX¼XÔ = üŒ'G $$
for the conformal scalar of subsection VIIB5 and the $X$ of the string.

Note that propagators do not transform simply under conformal transformations, because of conformal-weight factors (see subsection XIB4).  However, the vacuum itself is not invariant under conformal transformations.  The vacuum $|0Ô$ we use is the one natural for identifying the string with the entire complex plane (and the one that comes from the path integral in these coordinates, using ``1" for the vacuum wave functional).  It is with respect to that vacuum that the propagators take the simple form we have used above.

Ü3. Commutators

Since for the most part we will be interested in free fields, quantization
will be described most easily by the path-integral method.  Although 2D
field theory already looks quite different from the 1D field theory of
particle mechanics, free 2D massless fields depend on only one of the two
lightcone coordinates $§^à$ (or are the sum of two such terms), and
hence 2D conformal field theory is similar to 1D massive field theory. 
Consequently some of the features of particle mechanics or nonrelativistic
field theory, such as the commutator, can still be useful and 2D Lorentz
covariant.  In particle mechanics, the (equal-time) commutator is
evaluated by path-integral methods as
$$ Ò[\A,\B](t)Ô ­ \lim_{·£0}Ò\A(t+·)\B(t) -\B(t+·)\A(t)Ô $$
 (and similarly for the anticommutator), since $\A$ and $\B$ are treated
as classical functions when evaluating the path integral
$$ ÒfÔ ­ ÇDļf e^{-iS} $$
 where now ``$Ò¼Ô$" refers not to just the vacuum expectation value, but
incorporates arbitrary initial and final states through the boundary
conditions, or explicit wave functions in the path integral (see
subsection VA1, and XIB6 below).  In general, this definition of $Ò¼Ô$ actually gives
the time-ordered expectation value, as follows from the derivation of
subsection VA1:  The $·$'s were introduced to enforce the appropriate
ordering.  For the rest of this subsection time ordering will be implicit in
expectation values.

In the Hamiltonian formalism for ordinary quantum mechanics, we have the term $-Çdt¼Àqp$ in $S$, which defines $q$ and $p$ as canonically conjugate, and gives the propagator
$$ -i»_t üi·(t-t') = ¶(t-t') $$
As a result,
$$ \lim_{·£0}[q(t+·)p(t) -p(t+·)q(t)] = üi·(·) -üi·(-·) = i $$
Similar results can be obtained in Lagrangian approaches, where for fields satisfying second-order differential equations we use the propagator (see subsection VIIIC5)
$$ »_t^2 ü|t-t'| = ¶(t-t') $$

\x XIB3.1 Find the equal-time commutator of a massive scalar field with its time derivative, in arbitrary dimensions, from the propagator.  (Hint:  Start with the form expressed in terms of time and spatial momentum, and Fourier transform.)

For the analogous result in (Wick-rotated) D=2 we consider again the fermionic $L=ÐÆлÆ$.  The $·$ regulator used in subsection VIIB5 is not needed; now the $·$ we use for time ordering plays that role.  Using the fermionic propagator $1/(z-z')$ (from subsection VIIB5), and the identity (see exercise VA3.1)
$$ {1\over u-i·} -{1\over u+i·} = 2¹i¶(u) $$
we find
$$ \lim_{·£0}[ÐÆ( +·,§)Æ( ,§') -ÐÆ( ,§)Æ( +·,§')] = 
	{1\over ·+i(§-§')} -{1\over -·+i(§-§')} $$
$$ = 2¹¶(§-§') $$
 Although we have taken the commutator at equal $ $, in 2D conformal field theory
the fields' time-dependence is given by their depending on just
$z$ or just $Ðz$, and since even interacting string calculations in
terms of these ``free" (with respect to first-quantization) 2D fields
factorizes into two separate calculations for the left-handed fields and
for the right-handed fields, we generally treat just $z$ or just $Ðz$ as
the only argument.

In arbitrary dimensions, we can similarly evaluate commutators of conserved ``charges" with local operators as (under suitable boundary conditions)
$$ Q = Çd^{D-1}x¼J^0 = Çd^{D-1}ê_m¼J^m $$
$$ Üâ[Q,A(x)Õ = Èdê'_m¼J^m(x')A(x) $$
where the last integral is over a boundary enclosing $x$.  (Conservation implies the result is independent of the boundary.  In the 1D case such a boundary consists of just 2 points.)  The graded commutator $[¼Õ$ (commutator or anticommutator, as appropriate) is automatic in the path integral because of the classical grading of the variables.

$$ \figscale{commutator}{3in} $$

Since such ``surface" integrals in D=2 are basically contour integrals (see exercise
IIA1.2c), working with functions of just $z$ and not $Ðz$, this becomes
$$ Q = È{dz\over 2¹i}JâÜâ[Q,A(z)Õ = È_z {dz'\over 2¹i}J(z')A(z) $$
where the contour that gives the commutator encircles the $z$ where $A$ is evaluated.  Then we can avoid taking a limit, since the contour just picks up the simple pole at $z'=z$.  (For the cases of interest, the current has both divergence and curl vanishing, so the time component of the current is the sum of chiral and antichiral parts.)

In fact, this is the only contour that is relevant:  We need not define the contour for $Q$ as an abstract charge, only a contour for how it acts on an operator.  States themselves are defined by operators:  For example, we can pick an arbitrary point in the complex plane as $t=-¥$ (because of conformal invariance); $z=0$ is conventional.  A state, defined as an operator acting on the vacuum, is then represented as that operator at $z=0$.  (This is true also in path-integral language, where the operator is included in the path integral as the wave function of that state.)  Later times are circles about this origin, so $Q$ is integrated about such a circle, representing its action on that state.  But we see that this is the same as evaluating the commutator of $Q$ with its corresponding operator, which is consistent if $Q$ vanishes on the vacuum.  This picture is natural in Euclidean space, where there is no time in the usual sense, and will prove particularly useful later for string theory, where the worldsheet time is unrelated to physical time.

The contour integral definition of commutators is also convenient to avoid worrying about time ordering and limiting procedures.  For example,
$$ \left[ È{dz'\over 2¹i}½(z')Æ(z'), ÐÆ(z) \rightÕ = È_z{dz'\over 2¹i}½(z'){1\over z'-z}
	= ½(z) $$
$$ ÜâÓÆ(z'),ÐÆ(z)Õ = 2¹i¶(z'-z) $$
directly picking up the contribution from the pole in $z'-z$ (see exercise
VA3.1b),
where $½$ is an arbitrary (classical) function, and the $¶$ function in $z$ is understood at equal times as $2¹i¶(z'-z)£2¹¶(§'-§)$.  More generally, derivatives of $¶$ functions follow from more singular terms:
$$ È_z{dz'\over 2¹i}½(z'){1\over (z'-z)^{n+1}} = 
	{1\over n!}\left({»\over »z}\right)^n ½(z) $$

Ü4. Conformal transformations

Conformal invariance in D=2 is infinite dimensional, and looks like two copies of general coordinate invariance in D=1:
$$ dz'¼dÐz' = h(z,Ðz)dz¼dÐzâÜâz' = f(z),âh(z,Ðz) = (»f)(лÐf) $$
Effectively, we can treat $dz$ and $dÐz$ as independent (except for complex conjugation) 1D line elements.  From the above examples of the scalar and spinor (and we know is also true for the reparametrization ghosts) we see that fields generally depend (on shell) on just $z$ or just $Ðz$ (``chiral" and ``antichiral", or ``holomorphic" and ``antiholomorphic", or ``left-handed" and ``right-handed"), except for the scalar, which can be written as a sum of two such terms.  (There is some ambiguity on what to do with the zero-modes, which we'll have to deal with separately.  But $»Ä$ and $лÄ$ don't suffer from this problem.)

As usual the conformal transformation of a field depends on its scale weight $w$ (which is the same as the 2D engineering dimension):  If we think of a field $$ as a 1D tensor, we can write 
$$ (dz')^w '_{(w)}(z') = (dz)^w _{(w)}(z) $$
for a field with $w$ covariant 1D indices, but since the 1D index takes 1 value, we can trivially generalize to $w$ that is negative and non-integer.  We can also write this as
$$ '_{(w)}(z') = \left({»z'\over »z}\right)^{-w} _{(w)}(z) $$
For example, for the scalar and spinor we have
$$ w(»Ä) = 1,ââw(Æ) = ü $$
Of course, as in general relativity, not everything is a tensor:  For a scalar $Ä$, $Ä$ and $»Ä$ are, but $»»Ä$ isn't.  Unfortunately, after eliminating the 2D metric, we have no 1D metric left to define 1D ``covariant derivatives", so we'll have to live with such noncovariant objects.

Now that we know how conformal transformations act on fields and how to evaluate commutators easily, we can write expressions for conformal generators, i.e., the energy-momentum tensor $T$.  For the infinitesimal form $z'=z-Â(z)$ of the above conformal transformations, working with just one chirality for convenience, we need
$$ -\left[È{dz'\over 2¹i}Â(z')T(z'),_{(w)}(z)\right] = 
	Â(z)(»_{(w)})(z) +w(»Â)(z)_{(w)}(z) $$
We first consider the case where $$ has an action linear in derivatives:
$$ L = Ѝ_{(1-w)}л_{(w)} $$
for either fermionic or bosonic $$.  (In the fermionic case and with $w=ü$ we can identify $$ with $Ѝ$ and include an extra normalization factor of $ü$.)  By conformal invariance, the canonical conjugate $Ѝ$ of $$ has weight $1-w$, so both the left-handed ($z$) weights (from $$ and $Ѝ$) and right-handed ($Ðz$) ones (from $л$) sum to 1 (for invariance under $dz¼dÐz$ integration).  
Using the propagator
$$ (z)Ѝ(z') ® {1\over z-z'} $$
 it's easy to see that to give the above transformation law $T$ must consist of terms with one derivative,
$$ T = Ѝü\onª» +(ü-w)»(Ѝ) $$
 where the coefficient of the second term is obvious from the special cases $w=0,ü,1$.

Originally this Lagrangian came from a 2D coordinate invariant one with a 2D metric (zweibein), and the conformal weight $w$ came from the Lorentz weight (``spin", but in D=2 the Lorentz group is Abelian:
SO(1,1)=GL(1)), the number of ``+" minus ``$-$" indices, as used in the 2D spinor
notation of subsections VIIB5 and VIIIA7.  The Lorentz connection was the derivative of the metric, giving the total derivative  term in $T$, which was the variation of the action with respect to the metric.  (In Minkowski space this chiral part of $T$ is $T_{++}$; the tracelessness condition $T_{+-}=0$ follows from conformal invariance.)

 This includes the
classical string action
$$ P_+ É P_- +P_+ É á_- X +P_- É á_+ X $$
 as well as the conformal-gauge ghost action
$$ B_{++}á_- C_- +B_{--}á_+ C_+ $$
 (The factor of ${\bf e}^{-1}$ can be absorbed into the fields by a local
scale transformation; it's irrelevant for defining $T_{àà}$.)

\x XIB4.1  In general, the commutators we write for conformal transformations come from the singular (pole) parts of products of unintegrated quantities.
ªa  Show that in the special case of an operator with weight $w=1$, we can write
$$ T(z) _{(1)}(z') ® - {1\over (z-z')^2}Ê _{(1)}(z) $$
This is the case where $ȍ$ is conformally invariant.
ªb  Another interesting case is $w=2$:  Show
$$ T(z) _{(2)}(z') ® - {1\over (z-z')^2}[_{(1)}(z) +_{(1)}(z')] $$
ªc  Show that $T$ itself has weight 2, using the Jacobi identity.
ªd  Show directly that $T$ has weight 2 by evaluating $TT$ using the explicit expression in terms of $$'s.  Consider only the semiclassical (1-propagator) terms.  (2-propagator terms will be considered below.)

Closure of the BRST algebra, or lightcone Poincar«e algebra, is nontrivial
because of the infinite summations over oscillators, or equivalently
because of integration over the two-dimensional ``momentum" (which is
quantized as mode number in the $§$ direction because of the finite
extent of $§$).  As usual, BRST invariance is equivalent to gauge
invariance, and we can check for anomalies in the usual way, now applied
to the 2D ``field theory" corresponding to the mechanics of the string. 
The anomaly calculations are similar to those applied to the Schwinger model
in subsection VIIIA7.  (In particular, see exercise VIIIA7.1.)  The gauge
invariances in question are coordinate invariance and local scale
invariance, whose preservation is the vanishing of the divergence and
trace of the energy-momentum tensor.  As seen from our analysis for the
Schwinger model, this implies that the quantum corrections to the
energy-momentum tensor must themselves vanish when the external
fields are restricted to gravity only.  The calculation again involves
one-loop propagator corrections; performed in position space, we get the
product of two propagators between the same two points, with various
numbers of derivatives acting on either end of either propagator.  

$$ \figscale{anomalyT}{1in} $$

We start with the case of fields with first-order field equations, whose $T$ we rewrite as
$$ T = (1-w)Ѝ» -w(»Ð) $$
The 2-propagator terms correspond to 1-loop propagator corrections for the 2D graviton, which couples to $T$.  These should vanish for the 2D metric to consistently remain gauged away, or in other words, for conformal invariance to be preserved at the 1-loop level.  (There are no higher loops for this calculation because the theory is free.)

We get 2 kinds of terms, depending on whether both derivatives from the 2 $T$'s hit the same propagator, or one hits each.  In the former case one gets a term proportional to $z^{-1}»»z^{-1}=2z^{-4}$, in the latter $(»z^{-1})^2=z^{-4}$.  (The $z^{-4}$ means that the contribution to the commutator is proportional to $»^3 ¶(z-z')$.)  The anomaly is thus proportional to
$$ 2[2w(w-1)] +[w^2 +(w-1)^2] = 6(w-ü)^2 -ü $$
with an extra minus sign if the fields were fermionic (from the usual reordering).  For example, a complex fermion with $w=ü$ gives a contribution $+ü$, while a pair of real chiral bosons with weights $w=0,1$ give a contribution of 1.  Thus, as we saw from bosonization, a complex fermion with $w=ü$ gives the same contribution as a single real chiral boson with weight $w=0$.  

In fact, we can use the above expression directly to obtain $T$ for the second-order action for a single real chiral boson:  For the complex case, take either of
$$ \left. \matrix{ w = 0 & Ü & T = Ѝ»; & =Ä, & Ѝ=»ÐÄ \cr
	w = 1 & Ü & T = -(»Ð); & =-»Ä, & Ѝ=ÐÄ \cr} \rightÕ
	âÜâT = (»ÐÄ)(»Ä) $$
where the substitution is justified by the relation between the $Ä$ and $$ propagators.  The anomaly for the complex case is then just twice that for the real one.

For the bosonic string, we have $D$ real scalars, and fermionic ghosts with $w=-1,2$.  The anomaly is then
$$ bosonic:ââüD -13 = 0âÜâD = 26 $$
The superstring can be treated with variables that are the worldsheet supersymmetrization of those of the bosonic string (``Ramond-Neveu-Schwarz formalism").  Thus, there are $D$ real fermions ($w=ü$) as supersymmetry partners of the $D$ $X$'s ($w=0$), as well as bosonic ghosts ($w=-ü,\f32$) as partners of the fermionic ones ($w=-1,2$).  (The weights $-1$ and $-ü$ correspond to those of the gauge parameters of coordinate and supersymmetry transformations, while the weights $2$ and $\f32$ correspond to those of the worldsheet metric and gravitino.)  Thus
$$ RNS:ââüD +\f14 D -13 +\f{11}2 = 0âÜâD = 10 $$

Consequently, we have two conditions on known strings
that make them unsuitable for describing mesons: unphysical intercept
$Œ_0$ for the leading Regge trajectory (massless particles) and
unphysical spacetime dimension $D$.

Ü5. Triality

For the case of a Lagrangian quadratic in derivatives only bosons $Ä$ are interesting.  Then we can consider adding a term to the Lagrangian proportional to $RÄ$, where $R$ is the (2D) curvature, proportional to the second derivative of the metric.  We then find
$$ T = ü(»Ä)^2 -µ»»Ä $$
where $µ$ is the coefficient of the curvature term.  The above transformation law is then modified except for $µ=0$ (where $w=0$) to
$$ -\left[È{dz'\over 2¹i}Â(z')T(z'),Ä(z)\right] = Â(z)(»Ä)(z) +µ(»Â)(z) $$

This inhomogeneity (and familiarity with bosonization: see subsection VIIB5) leads us to consider the fields
$$  = e^{iaÄ},ââЍ = e^{-iaÄ} $$
for some constant $a$.  Since the $µ$ term is linear, it doesn't affect the propagator of $Ä$, so we have
$$ (z)Ѝ(z') ® (z-z')^{-a^2} $$
and we still need $a=1$ to get canonically conjugate fermions (with the usual anticommutation relations).  If we now evaluate the conformal transformation of $$ using the $T$ of $Ä$, we find, using
$$ (»Ä)(z')e^{iaÄ(z)} ® {ia\over z-z'}e^{iaÄ(z)} $$
that it transforms covariantly, with
$$ w() = üa^2 +iµa,ââw(Ѝ) = üa^2 -iµa $$
where the $a^2$ term comes from a 2-propagator term, and is thus a (``anomalous") quantum correction that would not be seen from a Poisson bracket.  
In particular
$$ a = 1âÜâw() = ü+iµ,âw(Ѝ) = ü-iµ $$
The last result should have been expected from the expression of $T$ for $$, since the $w-ü$ term multiplies $»(Ѝ)=-i»»Ä$ from our earlier study of bosonization (where we already found the $w=ü$ term).

\x XIB5.1  The $iµa$ term in $w$ is classical, since it comes from a single
propagator:
 ªa  Derive the Lagrangian for a scalar with $R$ term by starting with the
nonlocal term $R(1/õ)R$ and applying a local Weyl scale transformation,
introducing the scalar as the compensator.
 ªb  Find the classical scale weight of $e^{iaÄ}$ from its local scale
transformation.  (Hint:  In deriving the local scale transformation in part
{\bf a}, an exponential will be needed, so that $e^Ä$ transforms
homogeneously.)

These results generalize easily to general linear exponentials for multiple
scalars
$$ e^{ia_i Ä^i} $$
Including an indefinite metric $ú_{ij}$,
$$ T = üú_{ij}(»Ä^i)(»Ä^j) -µ_i »»Ä^i $$
will modify all the above expressions in an obvious way, including the metric (or its inverse) where needed to contract indices of the same kind (both up or both down).
For example, $a^2£aÉa$ and $µa£µÉa$ in the expression for $w$, and
$$ e^{iaÉÄ(z)}e^{i÷aÉÄ(z')} ® (z-z')^{aÉ÷a}e^{i[aÉÄ(z)+÷aÉÄ(z')]} $$

The most important use of such exponentials, outside of bosonization, is
for external fields (or ``vertex operators"):  The conformal
generators (energy-momentum tensor) have conformal weight 2;
requiring that background fields preserve conformal invariance implies
that such vertex operators must have conformal weight 1 and be local on
the worldsheet (see exercise XIB4.1):
$$ öT(z)öT(z') ® -{1\over (z-z')^2}[öT(z) +öT(z')] $$
$$ öT(z) = T(z) +~W(z)âÜâ~W¼has¼w=1,ââ~W(z)~W(z') ® 0 $$
$$ Üâ~W(z) = 2¹i¶(z-z_0)W(z_0) $$
 where we have assumed the conformal anomaly cancels (or ignored its
contribution), and solved for closure of the algebra perturbatively in the
background.

If we write a background spacetime field $ì(öX(z))$ as a Fourier
transform, then we see that its conformal weight is proportional to
$k^2$, the square of the external momentum.  Hence a vertex operator
consisting of just a scalar field produces the tachyonic ground state ($w=ük^2=1$). 
Excited states are created by products of derivatives of $X$ times fields
(with spacetime Lorentz indices contracted); the derivatives add to the
conformal weight, forcing $k^2$ to decrease in compensation, resulting in
massless and massive ($m^2>0$) states.  For example, $»X$ already has $w=1$, so the vector multiplying it must have $k^2=0$.  

\x XIB5.2  Show this without Fourier transformation:  Evaluate the conformal transformation of an arbitrary function (not functional; this is a particle field, not a string field) $ì(öX(z))$, using $T=ü(»öX)^2$.  Show that $ì$ transforms covariantly, with the number $w$ replaced by $-õ$ (with respect to $öX$) acting on $ì$.

Bosonization can also be applied to representations of groups (Lorentz or
internal).  In particular, to obtain the correct anticommutation relations
for a fermion $e^{iaÉÄ}$ and its conjugate $e^{-iaÉÄ}$ we require $a^2=1$;
to get the usual conformal weights, we require $µ=0$.

The Ramond-Neveu-Schwarz formulation of superstring theory uses fermions that are a representation of SO(D) (especially D=10) by simply carrying a D-valued (vector) index.  The simplest way to obtain these fermions from bosonization for even D is to define a (D/2)-vector $Ä^i$ with $ú_{ij}=¶_{ij}$.
Then we find for our SO(D)-vector fermion
$$ vector:ââa_i = (à1,0,0,...,0),â(0,à1,0,...,0),â... $$
in a (complex) null basis.  This construction follows that
for the Dirac matrices in subsection XC1:  Each scalar corresponds to a two-dimensional subspace of SO(D), and each component of a (D/2)-vector $a_i$ is the corresponding eigenvalue of the two-dimensional spin.  
Klein factors should be included to make fermions using different
scalars anticommute (see subsection IA2).  

From the above we can also find the generators of SO(D):  The raising and lowering operators come as for $©$-matrices from multiplying fermions from different pairs, then integrating.  This gives expressions like the above, but with different $a$'s (sums of 2 different $a$'s of the vector).  For the Cartan subalgebra, taking products from the same pairs and subtracting the divergent constant gives $»Ä^i$.  But then from the product relation for $»Ä$ times $e^{iÄ}$ we see that $a_i$ are the weights of the representation (eigenvalues of the Cartan subalgebra).

Then we can try to make SO(D) spinors the same way:  We try the weights (also obvious from two-dimensional spinors as the square root of two-dimensional vectors, and how spinors come from direct products of two-dimensional spinors)
$$ spinors:ââa_i = (àü,àü,...,àü) $$
where all the $à$'s are independent, except that their product is +1 for one Weyl spinor and $-1$ for the other.  
(The conventions are slightly
different from subsection XC1:  Now we use a representation where
$§_3$ is diagonal, and $©_{-1}$ is chosen as the ÓlastÕ $©$.)

However, these spinors can have the usual commutation relations and conformal weights only for D=8.  This is significant for two reasons:  (1) D=8 is the number of physical (i.e., transverse) fermions for the RNS superstring, and (2) SO(8) is the only simple Lie group with the property of ``triality", a symmetry between the vector and two spinor representations.  In fact, if we start out by defining the basis for one of the spinors with the same $a$ we used above to define the vector, and rewrite the above $a$'s for the vector and other spinor in terms of that new basis, we see that we have just permuted the 3 $a$'s.

This relation between (fermionic) vectors and spinors is important for superstrings because it relates in several ways to supersymmetry.  For example, we know that supersymmetry representations must have equal numbers of (physical) bosons and fermions; in D=10, the vector and (Weyl) spinor both have 8.  Since bosonization allows bosons to be defined from fermions and vice versa, triality allows the SO(8) vector fermion to be defined from either SO(8) spinor fermion, and vice versa.  So, at least in the lightcone gauge, we can translate anything in the RNS formalism to the ``Green-Schwarz" formalism, which uses a spinor fermion.  This allows supersymmetry (or at least the lightcone version) to be manifest, since superspace is defined by adding a spinor fermion to the usual spacetime coordinates.  For RNS there is also ``Gliozzi-Scherk-Olive projection" to get a supersymmetric spectrum:  Keeping only integer, not $ü$-integer (mass)${}^2$ for the bosons of the NS string (see exercise XIIC1.1), and using a chiral ground state for the fermions of the R string.  For GS, this means using states created just by one Weyl spinor field and not the other.

Similar triality constructions apply to lower dimensions by taking into
account supersymmetry, which also relates a vector to a spinor.  In D=6,
simple supersymmetry has an internal SU(2) (R) symmetry:  Thus, there is
a triality relating this SU(2) to the two SU(2)'s of the lightcone's SO(4).  In
terms of these, the vector is the $(ü,ü,0)$ representation, while the
spinors are $(ü,0,ü)$ and $(0,ü,ü)$.  The resulting operators are given by
$$ a_V = (à\f1{å2},à\f1{å2},0),ââa_S = (à\f1{å2},0,à\f1{å2}),ââ
	a_{S'} = (0,à\f1{å2},à\f1{å2}) $$
 To relate to the SO(8) results we use the vector to define the basis,
yielding
$$ a_V = (à1,0,0), (0,à1,0) $$
$$ Üâa_S = (ü,ü,à\f1{å2}), (-ü,-ü,à\f1{å2});â
	a_{S'} = (ü,-ü,à\f1{å2}), (-ü,ü,à\f1{å2}) $$
 This also follows directly from the SO(8) result by dropping the third and
fourth scalars for the vector, and using only ($1/å2ð$) their sum for the
spinors.  (I.e., it represents only internal symmetry.)  For D=4 the
construction is even simpler:  Besides the SO(2)=U(1) of the lightcone,
there is a second U(1) for R symmetry.  In terms of the complex plane
defined by these two quantum numbers, there is an obvious triality for
the three cube roots of 1; thus
$$ a_V = à(1,0),ââa_S = à(-ü,\f{å3}2),ââa_{S'} = à(-ü,-\f{å3}2) $$
 which again also follows from SO(8), now combining its last 3 scalars.

\x XIB5.3  We now extend the analogy to the construction of subsection
XC1:
 ªa  Show for general SO(2n), in analogy to the Dirac $©$'s, that the
(integral of) products of two vector fermions, antisymmetrized in the
vector indices, act in the same way as the group generators, by
examining their commutators with each other and with the vector and
spinor operators.
 ªb  Show for the triality cases that the (anti)commutator of two
representations yields the third (supersymmetry).

These constructions can be generalized from the lightcone to manifest
Lorentz covariance by adding equal numbers of scalars of positive and
negative metric (at least one of each):  Their contributions to the spinors'
operator product (power of $z$) then cancel, preserving the
anticommutation relations.  One of the extra scalars of positive metric
yields the two ``longitudinal" spacetime directions to complete the
SO(D$-$2) vector and spinor representations to SO(D$-$1,1).  The rest of
the scalars come from ``ghosts".  Note that the spacetime metric is
unrelated to the metric for the scalars:  The Minkowski spacetime metric
comes from Wick rotation of the scalars, as applied in subsection XC2 to
the construction of subsection XC1 for Dirac spinors.  For either Euclidean
or Minkowski spacetime the basis is null; the only difference is in reality.

Ü6. Trees

Unlike particles, Feynman diagrams for strings can be treated by
first-quantized methods for arbitrary loops.  The basic idea is that
interacting strings are just strings with nontrivial geometries:  For
example, while an open-string propagator can be described by a
rectangle, an open-string tree graph can be described by a rectangle that
has parallel slits cut from two opposite ends of the rectangle part-way
into the interior; this describes initial strings that join and split at their
ends (interactions).  This is the lightcone picture of interactions, where
conservation of total $p^+$ means conservation of the sum of the lengths
of the strings.  This corresponds to the choice $k=1$ in the language of
subsection XIB1, since the worldsheet coordinates must be chosen
consistently over the whole worldsheet:
$$ X^+ =  âÜâl = 2¹Œ'p^+ $$
 More general conformal gauges are defined by conformal
transformations of this configuration:  For example, the boundary of this
slit rectangle can be transformed to a single straight line by the usual
methods of complex analysis, so the worldsheet becomes simply a
half-plane.  (For the infinite rectangle, relevant for asymptotic states,
the transformation is $¨=Ýp_r^+ln(z-Z_r)$.)  Then even the geometry is
irrelevant; all that matters is the topology, which tells how many loops
the diagram has (see subsection XIA2).

$$ \fig{tree} $$

First-quantized path integrals are then the easiest way to calculate
arbitrary S-matrix elements in string theory.  However, the calculations
still can be quite complicated (as expected from a theory with an infinite
number of one-particle states), so we first consider just the
tree-level scattering of ground states, which is sufficient to illustrate the
qualitative features.  We can start from a gauge where the string is an
infinite strip (a rectangle of infinite length but finite width), with all but
two of the external states associated with points on one side of the
strip, the remaining two states being at the ends at infinity.  This is
equivalent to a picture of a propagator in an external field, with all but
two of the external states associated with the external field.  Similar
calculations are possible for particles, but give only a single graph; for
strings this gives the only graph, since different cyclic orderings are
related by conformal transformations.  (We are restricted to cyclic
orderings by group theory, as for the 1/N expansion for particles.)  This
method can be used in either the lightcone gauge or Lorentz covariant
conformal gauge.  

We begin by adapting the results of subsection VIIIC5 for the particle to the string:  We expand the propagator $1/(H_0-V)$, restricting all external states to tachyons, where now
$$ H_0 = ü(p^2 +m^2),ââV_i = g e^{ik_iÉX(0)} $$
for vertices at one end of the string $§=0$ for convenience.  (All other choices are equivalent by duality.  When acting on the tachyon ground state we'll see this gives the same result as using the usual string zero-mode $x$ after an appropriate limiting procedure.)
The amplitude is again
$$ A_N = g^{-2}\lim_{{÷ _1£-¥\atop ÷ _{N-1}=0}\atop ÷ _N£+¥}
	e^{(÷ _1 -÷ _N)m^2/2}Çd^{N-3}÷ ¼
	Ò0|V_N(÷ _N)V_{N-1}(÷ _{N-1}) ... V_2(÷ _2)V_1(÷ _1)|0Ô $$
We can again evaluate this operator by a path integral.  This time normal ordering removes infinite terms coming from connecting a vertex to itself with a Green function, and we again keep only terms with Green functions connecting different points.  (Normal ordering can also be treated in a more careful way by
taking the vertices to correspond to finite-width strings, as they would in
the lightcone approach, and taking the limit where their widths vanish.) 

Now we normalize the Green function as
$$ ÒX¼XÔ = üŒ'G $$
where we have inserted the $Œ'$ because of the difference in normalization of the action.  We have also used a $ $ for the string normalized to $Œ'=ü$, since the ``time"-development for the string (in normalization where $§$ goes from 0 to $(2)¹$) is really given by $Œ'(p^2+M^2)=(2Œ')ü(p^2+M^2)$.  Thus there is an extra factor of $2Œ'$ for each propagator to restore the original generic $1/ü(p^2+M^2)$.  Then the amplitude is simply
$$ A_N = g^{N-2}(2Œ')^{N-3}\hskip-.06in
	\lim_{{÷ _1£-¥\atop ÷ _{N-1}=0}\atop ÷ _N£+¥}\hskip-.01in
	e^{(÷ _1 -÷ _N)(2Œ')m^2/2}\hskip-.13in
	Ç\limits_{÷ _i ² ÷ _{i+1}}\hskip-.14in
	d^{N-3}÷ ¼exp\left[-üŒ'Ý_{i<j}k_iÉk_j G(÷ _i,÷ _j)\right] $$

We next make the conformal transformation
$$ z = e^{÷ } $$
used earlier.  Vertex operators in general have conformal weight 1 so they can be integrated:  $dzÊV(z)=dz'V'(z')$.  This is true for the tachyon vertex $e^{ikÉX}$ on shell because the tachyon has $m^2=-2$ in units $Œ'=ü$, so the weight $ük^2=1$.  Under this transformation we find $V(÷ )£zV(z)$ for the 3 unintegrated vertices $V_1$, $V_{N-1}$, and $V_N$.  The amplitude is then
$$ A_N = g^{N-2}(2Œ')^{N-3}
	\lim_{{z_1£0\atop z_{N-1}=1}\atop z_N£¥}
	z_N^2Ç_{z_i²z_{i+1}}d^{N-3}z¼
	exp\left[-üŒ'Ý_{i<j}k_iÉk_j G(z_i,z_j)\right] $$
Inserting the propagator
$$ G_{open}(z,z') = -ln(|z-z'|^2) -ln(|z-Ðz'|^2) $$
where all $z$'s are real, we find (again dropping the divergent terms from propagators connecting a vertex to itself)
$$ A_N = g^{N-2}(2Œ')^{N-3}\lim_{{z_1£0\atop z_{N-1}=1}\atop z_N£¥}
	z_N^2Ç_{z_i²z_{i+1}}d^{N-3}z¼Þ_{i<j}(z_j -z_i)^{2Œ'k_iÉk_j} $$
We can now simply evaluate at $z_1=0$ (and $z_{N-1}=1$), and use momentum conservation (and again the ground-state mass-shell condition $k^2=1/Œ'$ for the final state) to cancel the dependence on $z_N$:
$$ A_N = g^{N-2}(2Œ')^{N-3}Ç_{z_i²z_{i+1}}d^{N-3}z¼Þ_{1²i<j²N-1}(z_j -z_i)^{2Œ'k_iÉk_j} $$

The simplest case is the four-point function (it was how string theory began),
$$ A_4 = g^2(2Œ')Ç_0^1 dz¼z^{-Œ(s)-1}(1-z)^{-Œ(t)-1} $$
in terms of the Mandelstam variables
$$ s = -(k_1 +k_2)^2,ât = -(k_1 +k_4)^2 $$
 where the tachyon lies on the Regge trajectory
$$ Œ(s) = Œ's +Œ_0,ââŒ_0 = 1 $$
 which we recognize as the Beta function (see subsection VIIA2)
$$ {1\over 2Œ'g^2}A_4 = B[-Œ(s),-Œ(t)] = {ý[-Œ(s)]ý[-Œ(t)]\over ý[-Œ(s)-Œ(t)]} $$

Similar methods can be used for calculating closed string diagrams:  
There the interactions (vertices) are inside the string, instead of on the boundaries, 
so the integrals are over both $z$ and $Ðz$, and thus not ordered.
Since the 
closed-string Hilbert space is the direct product of two open-string
Hilbert spaces (except for momentum), for left- and right-handed modes,
the closed-string vertices are the product of 2 such open-string vertex operators.
Also, the exponential of the Green function is the product of a function of $z$ times a function of $Ðz$ (which are no longer equal).
Thus the $z$-$Ðz$ integrands are products of two open-string integrands (one for $z$ and one for $Ðz$), but with $p£üp$.  

For example, for the closed-string 4-point tachyon amplitude,
$$ A_{4,closed} ¾ Ç{d^2 z\over 2¹}¼(|z|^2)^{-üŒ(s)-1}(|1-z|^2)^{-üŒ(t)-1} $$
but now
$$ Œ(s) = üŒ's +2 $$
in terms of the same $Œ'$ used for the open string.  The integral is evaluated in the same way as Feynman diagrams (not surprisingly, since those are full of Beta functions, too):  Consider a 2D, massless, 1-loop propagator correction, where each of the internal propagators itself has quantum corrections, and so is some power of $p^2$.  This has the same form as above if we interpret $z$ and $1-z$ as the momenta of these 2 propagators.   We therefore use the usual Schwinger parameterizations
$$ f^{-h} = {1\over ý(h)}Çd ¼ ^{h-1}e^{- f} $$
for $f=|z|^2$ or $|1-z|^2$, introduce a scaling parameter $Â= _1+ _2$, $ _i=Œ_i$ to get Feynman parameters $Œ_i$, etc.  We eventually find
$$ A_{4,closed} ¾ ü{ý[-üŒ(s)]ý[-üŒ(t)]ý[-üŒ(u)]\over 
	ý[-üŒ(s)-üŒ(t)]ý[-üŒ(s)-üŒ(u)]ý[-üŒ(t)-üŒ(u)]} $$
which is dual between all 3 channels, where we have introduced $u$ via the identity
$$ Œ(m^2) = 0âÜâs +t +u = 4m^2 = -{16\over Œ'} $$
to manifest this symmetry.

If we expand the integrand of the Beta function for the open-string amplitude in powers of $z$, we find (see exercise XIB6.1 below)
$$ {1\over 2Œ'g^2}A_4 =  
	Ý_{J=0}^¥{[Œ(t)+J][Œ(t)+J-1]...[Œ(t)+1]\over J!}{1\over J-Œ(s)} $$
which shows Regge behavior:  The leading trajectory comes from the $t^J$ term, while ``daughter" trajectories come from $t^{J-n}/(J-Œ)=t^{J'}/[J'-(Œ-n)]$.  Since the amplitude is symmetric in $s$ and $t$, we can also write this as a sum over $t$-channel poles:  This is duality.

\x XIB6.1  Derive the pole structure of the 4-point string amplitude by
Taylor expanding the integrand of the Beta function in $z$.

If we go back to the Schwinger parameter via $z=e^{- }$ and expand $1-e^{- }$ in $ $, we find the Regge limit (see exercise XIB6.2 below)
$$ {1\over 2Œ'g^2}\lim_{s£-¥\atop t¼fixed}A_4 = ý[-Œ(t)][-Œ(s)]^{Œ(t)} $$
where higher orders in the $ $ expansion give contributions from daughter trajectories.  The same result can be obtained by using the Stirling approximation for the $ý$'s, or by applying the Sommerfeld-Watson transform on the pole expansion above.

\x XIB6.2  Use the Stirling approximation (see exercise VIIC2.1) to derive
the Regge limit of the 4-point string amplitude.  Show that the same
result can be obtained directly from the integral (Beta function)
representation of the amplitude, where the main contribution comes from
$z$ near 1.  (Hint: see exercise XIA1.2.)

\x XIB6.3  Show the Regge limit can be obtained from the
Sommerfeld-Watson transform of subsection XIA1.  (Hint:  The Beta
function is a sum of Regge trajectories.)

Another high-energy expansion is at fixed angle ($t/s$), as used in perturbative QCD (large
``transverse" energies).  Again using the Stirling approximation,
$$ \lim_{s£-¥\atop ϼfixed}A_4 ¾ e^{-f(cos¼Ï)Œ(s)} $$
$$ cos¼Ï ® 1 +2{t\over s},ââ
	f ® {t\over -s}ln\left({-s\over t}\right) +{u\over -s}ln\left({-s\over u}\right) $$
This Gaussian behavior in momenta is not the power behavior expected (at lowest order) from asymptotic freedom.  The interpretation is that the ``partons" that make up these strings have Gaussian propagators instead of the usual $1/(p^2+m^2)$ (see subsection XIA7).  Similar effects show up if we heat up strings past the ``Hagedorn temperature" (see subsection XIC2), where they break up into a parton plasma:  A QCD string would then show quarks and gluons, whereas known strings show almost no degrees of freedom, since Gaussian propagators have no poles.

$$ \figscale{pole}{4in} $$

Similar analyses can be made for the higher-point functions:  For example, we can show the poles are in the same places not only from the derivation (splitting up the path integral) but looking for momentum-space singularities directly in the amplitudes.  Consider some number $n+1$ of consecutive (because of ordering) vertex insertion points approaching each other:  The worldsheet picture is that these external lines are much closer to each other than the rest of the diagram, so relatively they have been stretched away, emphasizing a propagator connecting that bunch to the rest, carrying the sum of their momenta.  The $z$ integration diverges in that region as
$$ A ¾ Ç_0^· dz¼z^{n-1 +Ýk_iÉk_j} $$
where we have ``scaled" all $n$ of the converging $z_j-z_i$'s by the same variable $z$ to treat that region with a single integral, and the sum is over $i<j$ for those $n+1$ $k$'s.  Then using the on-shell condition for the tachyons (in units $2Œ'=1$)
$$ -s_{m,m+n} ­ \left( Ý_{i=m}^{m+n}k_i \right)^2 
	= 2Ý_{m²i<j²m+n}k_iÉk_j +2(n+1) $$
we have
$$ A ¾ Ç_0^· dz¼z^{-s_{m,m+n}/2-2} ¾ {1\over s_{m,m+n} +2} $$
giving the tachyon pole.  Corrections to this result come from mutiplying the integrand by a polynomial in $z$, yielding higher-mass poles. 

There are several ambiguities that might have arisen in calculating the normalization of these amplitudes, but they all would have only 2 effects:  (1) There could have been an extra factor of (constant)${}^N$.  This could come from normalization of the vertex operator, due to normal-ordering prescription (from dropping an infinite constant) or coupling-constant normalization.  It could also come from choosing the 2D-dimensionful constant $µ$ in the propagator $ln(µ|z|)$.  (In the above amplitudes, momentum conservation translates such $k_iÉk_j$ terms into $k_i^2$ terms.)  (2) There could also have been an overall $N$-independent constant.  This ambiguity could arise from normalization of any of the integration measures.  Another source is possible ambiguity in definition of the 3 vertices without $z$ integration as compared to those with.  These 2 effects are identical to those in ordinary field theory:  Consider a Lagrangian
(with normalization appropriate to matrix fields)
$$ L = -Z^2\f14Ä(õ-m^2)Ä -Z^3 g\f13Ä^3 $$
Then the ``wave-function" normalization $Z$ combined with the definition of the coupling $g$ are equivalent to the ambiguities we have described.  The normalization of $g$ is arbitrary, since it must eventually be fixed only by experiment.  The value of $Z$ is also a convention, but must be consistent with our convention for calculating probabilities.  This is not manifest in general conformal field theory methods, since only the amplitudes are defined.  One way to fix it is to use a pole expansion and compare the contribution of that particular particle to the corresponding result of particle field theory.  For example, we can look at the 3-tachyon amplitude $A_3$, which is just a coupling constant:
$$ A_3 = g $$
and at the contribution to the 4-tachyon amplitude $A_4$ from a tachyon in the $t$ channel,
$$ A_4 ® {2Œ'g^2\over -Œ(t)} = {g^2\over ü[(k_1+k_4)^2 +m^2]} $$
Both of these agree with the results obtained from the field theory Lagrangian above by the usual Feynman diagrams, with the same $g$ and with $Z=1$, as a result of our derivation of the string result from assembling vertices and propagators.

There are also generalizations to strings with worldsheet fermions; the
main differences are supersymmetry and D=10 (instead of 26).

Ü7. Ghosts

We did not really prove conformal invariance for the above calculations of scattering amplitudes, because 3 of the vertices were not integrated over.  In particular, if we would have picked different $z$'s for those 3 vertices, the result would have differed (though only by a constant in momenta, a function of those $z$'s).  This is the problem of determining the right integration measure for the $z$'s.  There are several solutions, but the easiest is to use the ghosts.  The full reason for the importance of the ghosts when calculating amplitudes for states that apparently don't involve ghosts will become clearer later when we examine string field theory (see especially subsection XIIB8).  For now we note that it is natural to consider the BRST operator when considering the definition of physical states, since it defines them in a way that is independent of gauge.  For this discussion we'll restrict ourselves to the open bosonic string.

Since the mode expansion was used to define the vacuum, isolate zero-modes, etc., we need to look at how this is affected by the conformal transformation that took us from the strip to the complex plane:
$$ _{(w)}(¨) = Ý_n _n e^{-n¨}âÜâ'_{(w)}(z) = Ý_n _n z^{-n-w} $$
This implies a certain definition of the vacuum:  $ =-¥$ is now $z=0$, so states are created by operators that are nonsingular as $z£0$.  We therefore have
$$ _{n-w}|0Ô = 0âforân>0 $$
which differs from the naive $_n|0Ô=0$ when $w±0$.  For example, for the open, bosonic string we apply it to $»X$, $C$, and $B$, to find
$$ (÷a_n,p,c_{n+1},b_{n-2})|0Ô = 0 $$
In relation to the tachyonic vacuum $|tÔ$, describing an off-shell, zero-momentum tachyon, satisfying
$$ (÷a_n,p,c_n,b_n,b_0)|tÔ = 0 $$
we thus have
$$ |0Ô = b_{-1}|tÔ,ââ|tÔ = c_1|0Ô $$
which describes an on-shell, zero-momentum Yang-Mills ghost (see subsection XIIB8).  

\x XIB7.1  What state is $c_{-1}|tÔ$?  Check the mass level and ghost number, and that it couples to the right state in the propagator $1/H_0$.  (Ignore the mode $c_0$, which gets special treatment, as for the particle.)

One result of this change in vacuum is that the number operator $N-1$ is replaced with $N$ in $H$ because the 1 is canceled by the new normal ordering.  (Now $b_{-1}$ is treated as an annihilation operator, $c_1$ as creation.)  This is clear from the fact that $|0Ô$ corresponds to a massless state, so the $p^2$ term (as well as the normal-ordered oscillators) in $N$ vanishes on it.

We know how to get physical fields from the tachyonic vacuum $|tÔ$, by hitting it with physical oscillators.  But the tachyon itself is created from the vacuum $|0Ô$ by hitting it with $c_1$.  In conformal field theory langauge, since $C$ has conformal weight $w=-1$,
$$ C(z) = Ýc_n z^{-n+1} = ... +c_0 z +c_1 +c_2 z^{-1} + ... $$
$$ Üâ|tÔ = \lim_{z£0}C(z)|0Ô $$
This state is off shell, and so is not in the BRST cohomology.  An on-shell tachyon is described by
$$ |t,kÔ = \lim_{z£0}C(z)e^{ikÉöX(z)}|0Ô,ââk^2 = 2 $$
As we saw earlier, choosing this value of momentum gives the exponential $w=1$. 
We can thus write an arbitrary tachyon state in terms of the tachyon field $Ä$ as
$$ \lim_{z£0}C(z)Ä(öX(z))|0Ô,ââ(õ+2)Ä(x) = 0 $$
(Here $Ä$ is a function of $öX(z)$, not a functional of $öX$.)

Similar remarks apply to excited states:  For example, for the vector, using the generalization of the operator $»X$ used for the particle (see subsection VIIIC5) we have
$$ |û,kÔ = \lim_{z£0}C(z)ûÉ(»X)(z)e^{ikÉöX(z)}|0Ô,ââk^2 = 0 $$
since $»X$ has conformal weight $w=1$.  Near $z=0$ only the first creation operator in $»X$ will contribute when acting on the vacuum:
$$ (»X)(z) = Ý_n Œ_n z^{-n-1}âÜâ\lim_{z£0}(»X)(z)|0Ô = Œ_{-1}|0Ô $$
Thus an arbitrary vector state is
$$ \lim_{z£0}C(z)A(öX(z))É(»X)(z)|0Ô,ââõA_a(x) = 0 $$

\x XIB7.2  What is the local (in $z$) operator that, when acting on $|0Ô$, produces the state of exercise XIB7.1?

More generally, any operator $C(z)W(z)$, which has weight 0 if $W$ has weight 1, will create a physical state from the vacuum if $ÈW$ creates a physical state from the tachyonic vacuum.  (Taking $z£0$ is a conformal transformation, since $CW$ has $w=0$, and corresponds to choosing a gauge.)  
Any operator $W(z)$ with $w=1$ has conformal transformation
$$ -\left[È{dz'\over 2¹i}Â(z')T(z'),W(z)\right] = (»ÂW)(z) $$
so its integral is conformally invariant.  If it is ghost free, it is therefore also BRST invariant,
$$ [Q,{\textstyle È}WÕ = 0 $$
(Gauge-invariant operators without ghosts are BRST invariant.)  Then $CW$ is also  BRST invariant:  Using $ÓQ,CÕ=-C»C$ (see subsection IXB1),
$$ ÓQ,CW(z)Õ = ÓQ,CÕW -C[Q,W] = -C(»C)W +C»(CW) $$
$$ = -C(»C)W +C(»C)W +CC(»W) = 0 $$
Since both $ÈW$ and $CW$ are BRST invariant, and the vacuum is, they can be used to construct a BRST invariant amplitude.

Massless vertices for the closed string are similarly the product of left and right open-string vertices (as are arbitrary vertex operators):
$$ V = û_{mn}(»X^m) (лX^n) e^{ikÉöX} $$
(The ghost structure is a little more complicated because of the zero-mode for 
$T_0-ÐT_0$.)

The fact that $CW$'s are needed at all (and not just $ÈW$'s) is related to the fact that the choice of vacuum does not completely fix conformal invariance:
$$ T_1|0Ô = T_0|0Ô = T_{-1}|0Ô = 0 $$
since $T$ has $w=2$.  (The third is satisfied because $c_0 b_{-1}|0Ô=0$.)  These 3 operators generate Sp(2).  (This unbroken invariance is related to the gauge invariance of the massless vector.)  Fixing this invariance in the path integral requires an extra ghost-dependent factor.

One way to determine this extra dependence is to note that for the scalar particle there is also a ghost, the analog of $c_0$ for the string.  The integration measure is then simply $Çdc_0$, so we have
$$ particle:ââÇdc_0¼c_0 = 1âÜâÒc_0Ô_C = 1 $$
where $Ò¼Ô_C$ means we neglect $X$.  The analog of the scalar for the string is the tachyon, so
$$ 1 = Òt|c_0|tÔ_C = Ò0|c_{-1}c_0 c_1|0Ô_C $$
which are the ghosts of Sp(2).  We then find
$$ C(z) = ... +c_{-1}z^2 +c_0 z +c_1 + ... $$
$$ ÜâÒC(z_1)C(z_2)C(z_3)Ô_C = -(z_1-z_2)(z_2-z_3)(z_3-z_1) $$
We can now solve the problem of conformal invariance using BRST:  $ÈW$ and $CW$ are both BRST invariant, so for general (tree) amplitudes we use 3 $CW$'s and the rest $ÈW$'s.  This eliminates 3 integrals and introduces 3 $C$'s, which produce the above factor.  Later, when we consider string field theory, we'll see that the reason for this counting is that an $n$-point graph has $n-3$ propagators, each of which has a factor of $ÈB$, so all but 3 of the true vertex operators $CW$ are converted into $W$'s.

The $ÒCCCÔ$ factor replaces the factor $z_N^2$ in the open-string tachyon amplitude we produced previously by other arguments, and agrees with it for the previous choices $z_1=0$, $z_{N-1}=1$, $z_N£¥$.  We therefore replace the previous result with
$$ A_N = -(z_1-z_{N-1})(z_{N-1}-z_N)(z_N-z_1)
	Çd^{N-3}z¼Þ_{i<j}(z_j -z_i)^{k_iÉk_j} $$
To show this result agrees with the previous, we use the Sp(2) invariance of the vacuum (and conformal invariance of everything else).  The (``zeroth-quantized") transformation on $z$ from the above 3 Virasoro operators are found by noting that on expressions $Ä$ with conformal weight 0 (no ``spin" piece to the transformation),
$$ T_n = È{dz\over 2¹i}z^{n+1}TâÜâ-[T_n, Ä] = z^{n+1}»Ä $$
The finite transformations generated from the infinitesimal ones ($n=0,à1$) are
$$ z £ {az+b\over cz+d},ââad -bc =1 $$
(e.g., by examining the infinitesimal case and checking the group property).  Then we use the fact that such a transformation can be used to transform any 3 points to fixed values.  In particular,
$$ z £ {z-z_1\over z-z_N}{z_{N-1}-z_N\over z_{N-1}-z_1} $$
transforms $z_1£0$, $z_{N-1}£1$, $z_N£¥$.

If you know some group theory, you might recognize this representation of Sp(2) as the coset space Sp(2)/GL(1), i.e., 2-component vectors with an antisymmetric metric, with scale transformations as a gauge invariance:
$$ ½ = \pmatrix{ ½_1\cr ½_2\cr},ââ½É½' = ½_1 ½'_2 -½_2 ½'_1,âⶽ = ½ $$
After gauging $½_2=1$ (for all $½$'s), we then have simply
$$ ½ = \pmatrix{ z\cr 1\cr}âÜâ½É½' = z-z' $$
But now an Sp(2) (=SL(2)) transformation will change the gauge, so we need to supplement it with a compensating scale transformation:
$$ \pmatrix{ z\cr 1\cr} £ \pmatrix{ a & b \cr c & d \cr} \pmatrix{ z\cr 1\cr}
	= \pmatrix{ az+b\cr cz+d\cr} $$
$$ £ {1\over cz+d}\pmatrix{ az+b\cr cz+d\cr} = \pmatrix{ (az+b)/(cz+d)\cr 1\cr} $$
as above.  From this the result on $z-z'$ is clear:  It is invariant under the original Sp(2), so its transformation comes from just the scalings,
$$ z-z' £ {z-z' \over (cz+d)(cz'+d)} $$
as is easily confirmed by performing the compensated Sp(2) transformation.  For the same reason, $dz$ also has a simple transformation law, so Sp(2) invariance of the amplitude is easy to check explicitly, using momentum conservation and the mass-shell condition $k_i^2=2$.

\refs

 £1 P. Goddard, J. Goldstone, C. Rebbi, and C.B. Thorn, \NP 56 
	(1973) 109:\\
	lightcone quantization for string.
 £2 A.M. Polyakov, \PL 103B (1981) 207:\\
	worldsheet ghosts.
 £3 Virasoro; Gelfand and Fuchs; Óloc. cit.Õ (XIA);\\
	S. Fubini and G. Veneziano, ÓNuo. Com.Õ É67A (1970) 29, 
	ÓAnn. Phys.Õ É63 (1971) 12;\\
	A. Galli, ÓNuo. Cim.Õ É69A (1970) 275;\\
	J.L. Gervais, \NP 21 (1970) 192;\\
	J.-L. Gervais and B. Sakita, \NP 34 (1971) 477;\\
	A. Chodos and C.B. Thorn, \NP 72 (1974) 509:\\
	theory of free conformal fields.
 £4 R. Marnelius, \NP 211 (1983) 14:\\
	effect of $RÄ$ term on Virasoro operators.
 £5 C. Lovelace, \PL 34B (1971) 500:\\
	discovery of D=26 requirement.
 £6 M. Kaku, ÓStrings, conformal fields, and topology: an introductionÕ
	(Springer-Verlag, 1991);\\
	P. Ginsparg and G. Moore, \xxxlink{hep-th/9304011}, Lectures on 2D
	gravity and 2D string theory, in ÓRecent directions in particle theory:
	from superstrings and black holes to the standard modelÕ,
	Proceedings of the Theoretical Advanced Study Institute in
	Elementary Particle Physics (TASI 1992), Boulder, CO,  June 1-26,
	1992, eds. J. Harvey and J. Polchinski (World Scientific, 1993) p. 277;\\
	S.V. Ketov, ÓConformal field theoryÕ (World Scientific, 1995);\\
	P. Di Francesco, P. Mathieu, and D. S«en«echal, ÓConformal field theoryÕ
	(Springer, 1997).
 £7 E. Witten, D=10 superstring theory, in ÓFourth workshop on grand
	unificationÕ, University of Pennsylvania, Philadelphia, April 21-23,
	1983, eds. H.A. Weldon, P. Langacker, and P.J. Steinhardt
	(Birkh¬auser, 1983) p. 395:\\
	vectors and spinors of SO(8) as different exponentials of bosons.
 £8 F. Gliozzi, J. Scherk, and D.I. Olive, \PL 65B (1976) 282, \NP 122 (1977) 253.
 £9 W. Siegel and B. Zwiebach, \NP 263 (1986) 105:\\
	bosonization of ghosts (bosonic string).
 £10 D. Friedan, E. Martinec, and S. Shenker, \PL 160B (1985) 55,
	\NP 271 (1986) 93;\\
	V.G. Knizhnik, \PL 160B (1985) 403:\\
	vectors and spinors of SO(9,1) as different exponentials of bosons.
 £11 Veneziano, Óloc. cit.Õ (XIA);\\
	M. Suzuki, unpublished:\\
	birth of string theory (as dual models).
 £12 K. Bardak'ci and H. Ruegg, ÓPhys. Rev.Õ É181 (1969) 1884;\\
	C.J. Goebel and B. Sakita, \PR 22 (1969) 257;\\
	Chan H.-M. and T.S. Tsun, \PL 28B (1969) 485;\\
	Z. Koba and H.B. Nielsen, \NP 10 (1969) 633, É12 (1969) 517:\\
	generalization of 4-point amplitude.
 £13 M.A. Virasoro, ÓPhys. Rev.Õ É177 (1969) 2309:\\
	first closed string amplitude.
 £14 J.A. Shapiro, \PL 33B (1970) 361:\\
	generalization of closed 4-point amplitude.
 £14 F. Gliozzi, ÓNuovo Cim. Lett.Õ É2 (1969) 846:\\
	Sp(2) in string theory.
 £15 D. Friedan, Introduction to Polyakov's string theory, in ÓRecent
	advances in field theory and statistical mechanicsÕ, eds. J.B. Zuber
	and R. Stora, proc. 1982 Les Houches summer school (Elsevier, 1984)
	p. 839:\\
	covariant vertex operators with ghosts.
 £16 S. Mandelstam, \NP 64 (1973) 205:\\
	lightcone path integrals for interacting strings.
 £17 O. Alvarez, \NP 216 (1983) 125:\\
	covariant path integrals for interacting strings.
	
\unrefs

Û7 C. LOOPS

Note that in the following, although we sometimes use operator or path-integral notation, we never actually calculate by performing explicit oscillator evaluations (using, e.g., coherent states) or the infinite-dimensional integrals of the path integral:  As we did previously for tree amplitudes, we just use general properties of quantum theory.  Specifically, we use spacetime or worldsheet Feynman diagrams, which are just perturbation theory, but can be derived from oscillators, path integrals, or other methods.

We first examine the planar loop, with external tachyons.  There are 3 parts to the calculation:

£1) 2D Green function

£2) volume element (or ``integration measure")

£3) partition function

\noindent The first two we encountered for tree graphs; the last (really a part of the volume element, but a new one) comes from summing over the infinite number of states of the string that circle around the loop.

Ü1. Partition function

As for trees, we generalize the results of subsection VIIIC5 for particles to strings.  That method allows the volume element to be determined unambiguously.  Often symmetry arguments are used to determine the volume element, but that has 4 major drawbacks:

£1) Sometimes symmetry is not enough even to determine functional dependence.

£2) Symmetry will never determine overall constants, since constants are invariant.

£3) In particular, BRST symmetry only guarantees gauge independence of the result.  If BRST is used, a separate evaluation in a unitary gauge is needed.

£4) Anomalies can violate symmetries, so a symmetry-independent evaluation is need\-ed to check for anomalies.  (I have even seen symmetry arguments use to conclude certain asymmetric contributions must have vanishing coefficient, when in fact nonvanishing anomalous contributions were found from direct evauation.)

In string theory, because of duality, 1-loop graphs can always be represented without external trees.  Thus, unlike the particle case, we will automatically find the S-matrix, and not the effective action, just as for trees the single graph we considered automatically gave the complete S-matrix element for strings only.  
We now find the amplitude
$$ A_N^{(1)} = Ç_{-¥ ² -T ² ÷ _i ² ÷ _{i+1} ² 0}dT¼d^{N-1}÷ ¼\V (T)
	exp\left[-üŒ'Ý_{i<j}k_iÉk_j G(÷ _i,÷ _j)\right] $$
(There is also an overall normalization of $(2Œ'g)^N$, in analogy to subsection XIB6, which we'll ignore from now on, and we choose units $Œ'=ü$.  The trace is normalized to agree with that for a single particle, together with a sum over particles, as explained below.  This ``$ÊTÊ$" should not be confused with the 2D energy-momentum tensor.)

The volume element for the string can be factorized into a sum over states (now called ``partition function") and an integral over the momentum of each state as for the particle:
$$ \V(T) = tr(e^{-TH_0}) = T^{-D/2}Ý_{states}e^{-TM^2/2} $$
To evaluate the sum we use $Œ'M^2=N-1$ and evaluate the sum as the product of independent summations over the oscillators of each of the $D-2=24$ transverse directions:  Again making the conformal transformation
$$ z = e^{÷ },ââw = e^{-T} $$
and using units $2Œ'=1$, we have
$$ Ý_{states}e^{-TM^2/2} = w^{-1}\left(Ý_{D=1}w^{N'}\right)^{D-2} $$
where $N'=Ý_n na_nÿa_n$ is the contribution to the number operator of any one dimension.  This sum is itself the product of contributions of any one oscillator to $N'$:  For each oscillator we get a sum of terms, one each from each excitation level of that oscillator.  For the $n$th oscillator,
$$ Ý_{excitations}w^{na_nÿa_n} = Ý_{j=0}^¥ w^{nj} =
	{1\over 1 - w^n} $$
The final result for the volume element is then, 
$$ \V = (-ln¼w)^{-D/2}w^{-1}[f(w)]^{2-D},ââf(w) = Þ_{n=1}^¥ (1-w^n) $$
Putting the pieces together, we have (with $w²z_i²z_{i+1}²1$)
$$ 
\boxeq{
A_N^{(1)} = Ç_0^1 {dw\over w^2}[f(w)]^{2-D}(-ln¼w)^{-D/2} 
	Ç_w^1\left({dz\over z}\right)^{N-1}
	exp\left[-\f14Ý_{i<j}k_iÉk_j G(z_j/z_i,w)\right]
} $$

We have summed over just the transverse oscillators, representing the physical states.  This can be justified by quantizing the string in a lightcone gauge, where only the transverse oscillators (but all components of momenta) appear.  For lower-point functions, this result can be obtained (assuming Lorentz invariance is preserved) by working in a reference frame where all the $k^+=0$, so none of the $x^-$'s, which are quadratic in the transverse oscillators, appear.  (By Wick rotation, we can always assume $p^+$ is a complex combination of any 2 spatial components of momentum, a ``spacecone gauge", so no restriction is imposed on $p^0$ on shell.)  A more general way is to use a Lorenz gauge (like the conformal gauge).  Then the contributions of the ghost oscillators will cancel 2 dimensions of the bosonic ones:  For any fermionic oscillator $d_n$, remembering that a fermionic state gets a minus sign in a loop (so the trace is really a ``supertrace"),
$$ \mathop{{Ý}'}_{excitations} w^{nd_nÿd_n} = 1 -w^n $$
where we have summed over the 2 excitations, and $Ý'$ means the fermionic term gets a minus sign in the sum.  (In functional integral language, fermionic integrals of Gaussians give determinants while bosonic ones give inverse determinants.)  $f(w)$ is thus a partition function for 1 fermion.  We have dealt with the ghost zero-modes by using the $b_0=0$ gauge.  (Similar arguments can be given at 1 loop using just conformal field theory, thus showing as for trees how the ghost contribution preserves conformal invariance, but they become more obscure at higher loops.  A better understanding of this gauge comes from string field theory.)

\x XIIC1.1  All the 2D fields we have explicitly considered effectively have periodic boundary conditions:  They are expanded over $e^{in§}$ for integer $n$.  Consider instead a single fermionic field with ``antiperiodic" boundary conditions (as in, e.g., the NS, or bosonic, sector of the RNS string, which we haven't studied in detail), expanded in $e^{i(n+1/2)§}$.  The masses${}^2$ for the oscillators now go as $n+ü$ instead of $n$, with the ground state mass chosen so that the first excited state is massless.
ªa Find the contribution to the partition function for this field.
ªb For GSO projection (for supersymmetry), one looks at only masses${}^2$ that are integer (dropping, e.g., the tachyonic ground state).  Find the contribution to the partition function for this reduced set of states.

Ü2. Jacobi Theta functions

It will prove useful to consider a more general type of partition function, one for energy and ghost number, for the ghosts in their fermionic and bosonized versions.  
We use the variables
$$ z = e^¨,âw = e^{-T} $$
(We began with $T$ positive, so $0²w²1$, but for the following manipulations $Re¼T³0$, so $|w|²1$, is OK, and later we analytically continue.)
We define the partition function
$$ Z(w,z) = str( e^{-TH+¨J}) = tr[w^H (e^{-i¹}z)^J] $$
where for ``energy" $H$ and ghost number $J$ we use the equivalent bosonic/fermionic expressions
$$ H = üp^2 +Ý_{n=1}^¥ naÿ_n a_n = Ý_{n=1}^¥ n(cÿ_n b_n +bÿ_n c_n) +\f18 $$
$$ J = p = ü[c_0,b_0] +Ý_{n=1}^¥ (cÿ_n b_n -bÿ_n c_n) $$
(Bosonization works the same way for ghosts as for physical fermions, but with $i$'s missing, since $c$ and $b$ are each real, instead of complex conjugates.  The $\f18$ is because the momentum $p$ of the boson is an integer plus $ü$, since total ghost number is, and thus $H$ as defined above for the boson has minimum value $\f18$.)  We have defined the ``supertrace" by including a factor of $e^{-i¹J}$ into the trace:  It gives the usual $-1$ for fermion states as defined for fermion oscillators, which we carry over to the bosonic formulation.

\x XIC2.1  Look at the first excited level (corresponding to massless ghost states of the string):
ªa Express all the states of this level in terms of fermionic oscillators acting on the vacuum.  Translate this into bosonized operators.
ªb Evaluate $H$ and $J$ for these states, and sum their contribution to $Z(w,z)$.

In terms of the variables
$$ Ã = {¨\over 2¹i},â  = -{T\over 2¹i}ââ( z = e^{2¹iÃ},âw = e^{2¹i } ) $$
if we express the result in terms of the Jacobi $Ï$ function, and the partition function of subsection XIC1,
$$ Z(w,z) ­ {Ï_1(Ã| ) \over f(w)},ââf(w) = Þ_{n=1}^¥ (1 -w^n) $$
we find the two equivalent forms for the result

\vskip.05in
\Boxit{\vskip-.18in
$$ \li{ Ï_1(Ã| ) & = iÝ_{n=-¥}^¥ (-1)^n z^{n-1/2}w^{(n-1/2)^2/2} \cr
	& = -iw^{1/8}(z^{1/2}-z^{-1/2})
		Þ_{n=1}^¥ (1 -w^n)(1 -z w^n)(1 -z^{-1}w^n) \cr} $$
\vskip-.24in}
\vskip-.1in

\noindent from the bosonic and fermionic versions, respectively.  (For the bosons, the sum is over eigenvalues of $p$.  The factor of $f$ in the definition of $Ï$ cancels the oscillator contribution in the bosonic case, and is included as a product in the fermion case.)

From either of the above forms we can easily see the Jacobi $Ï$ function satisfies the ``quasiperiodicity" conditions
$$ Ï_1(Ã+1| ) = -Ï_1(Ã| ),ââÏ_1(Ã+ | ) = -e^{-i¹(2Ã+ )}Ï_1(Ã| ) $$
As a function of $Ã$, it vanishes only at $Ã=0$, up to these periods.  It is also odd:
$$ Ï_1(-Ã| ) = -Ï_1(Ã| ) $$
From the second (product) form we also have
$$ Ï'_1(0| ) = 2¹w^{1/8}[f(w)]^3 $$
where the prime means derivative with respect to the first argument.  

We will need transformations of these functions under the subgroup of conformal transformations SL(2,Z)=Sp(2,Z), namely
$$   £ {a +b \over c +d},ââà £ {Ã\over c +d} $$
(or similarly for $T$ and $¨$) in terms of the SL(2,Z) group element
$$ \pmatrix{ a & b \cr c & d \cr} $$
whose elements are integers and determinant $ad-bc$ is constrained to 1.  The simplest is $a=b=d=1$, $c=0$:  From the product form of $Ï_1$,
$$ Ï_1(Ã| +1) = e^{i¹/4}Ï_1(Ã| ) $$

The next simplest one is $b=-c=1$, $a=d=0$:  This can be derived by considering
$$ Ý_{n=-¥}^¥ Ç_{-¥}^¥ du¼e^{2¹[-Au^2/2 +i(B+n)u]} $$
This can be evaluated by doing either the sum or the integral first.  To evaluate the sum first, we use the identity
$$  Ý_{n=-¥}^¥ e^{2¹inu} =  Ý_{n=-¥}^¥ ¶(u-n) $$
(This can be checked by multiplying on either side with a function and integrating, noting that only the periodic part of the function, with period 1, is picked out, so that function can be written as a Fourier sum, and the integral evaluated over 
$[-ü,ü]$.)  Thus we relate the 2 forms
$$ Ýe^{2¹(-An^2/2+iBn)} = 
	A^{-1/2}Ýe^{2¹(-A^{-1}n^2/2-A^{-1}Bn-A^{-1}B^2/2)} $$
In this ``Poisson (re)summation formula" we make the replacements
$$ A = -i ,ââB = Ã -ü  +ü $$
(so the integral is well defined again for $Re¼T³0$).
After straightforward algebra (replacing also $n£n-1$ on the right-hand side), we find
$$ Ï_1\left(-{Ã\over  }\Bigg|-{1\over  }\right) = 
	e^{i¹(Ã^2/ +1/4)} ^{1/2}Ï_1(Ã| ) $$
From applying this to $Ï_1'$ we find, for $w=exp(2¹i )$ and $w'=exp[2¹i(-1/ )]$, the Hardy-Ramanujan formula
$$ \boxeq{w^{1/24}f(w) = å{-ln¼w'\over 2¹}w'^{1/24}f(w'),ââ
w' = e^{(2¹)^2/lnÊw} } $$

The general case can then be found by combining arbitrary multiples of these 2, yielding
$$ \boxeq{Ï_1\left({Ã\over c +d}\Bigg|{a +b\over c +d}\right) =
	½(c +d)^{1/2}e^{i¹cÃ^2/(c +d)}Ï_1(Ã| ) } $$
where $½$ is an eighth root of unity.  (We won't need it, since we'll use only $|Ï_1|^2$ below.)

Hagedorn noticed that the multiplicity of observed hadron states as mass increased (and other features) was characteristic of a thermodynamic system with maximum temperature around the pion mass.  In QCD language this is the temperature of the deconfining phase transition, above which hadronic matter is replaced with a quark-gluon plasma.  Later this behavior was found to follow from strings.  However, in the string case the number of states above this temperature is found to be less than that of an ordinary particle theory.  This can be attributed to the fact that a random lattice worldsheet approach to quantization of known string theories describes a theory whose partons have Gaussian propagators, without poles.

To derive this temperature from the bosonic string, we begin with the Hardy-Ramanujan formula above:  The counting of states at high ``temperature" (mass) is then given by looking at $w=1-·$, $·£0$, which is $w'£0$:
$$ \lim_{·£0} f(w) ® å{2¹\over ·}¼e^{-¹^2/6·} $$
up to terms which are smaller by powers of $·$ (or worse yet, powers of the above exponential factor).  The number of states at the $n$th excited level is then given by
$$ N(n) = È{dw\over 2¹i}{[f(w)]^{2-D}\over w^{n+1}} $$
with the contour a small circle near the origin. 
We can evaluate this integral by the saddle point approximation for the ``action"
$$ S = (D-2)ln¼f +(n+1)ln¼w ® -(D-2)\left[{¹^2\over 6·}+üln¼·\right] -(n+1)· $$
$$ Üâ· ® ¹å{D-2\over 6n} $$
$$ ÜâN(n) ¾ {1\over å{S''}}e^{-S} ¾ n^{-(D+1)/4}e^{2¹å{n(D-2)/6}} $$
To pick up this contribution we have widened the circular contour to run through the saddle point:  Since this point is real (and at $|w|<1$ so $f$ is still well defined), the contour runs in the imaginary direction for the infinitesimal region near the saddle point where it contributes, so this point is a minimum.
Using $n®Œ'm^2$, $dn¼N(n)=dm¼¨(m)$, we have
$$ ¨(m) ¾ m^{-(D-1)/2} e^{m/m_0},ââm_0 = {1\over 2¹}å{6\over (D-2) Œ'} $$
The pion mass is about $2¹$ smaller than $1/å{Œ'}$ (in terms of the hadronic string tension $Œ'$), so this is in the right ballpark.

Ü3. Green function

We begin our determination of the Green function by analogy with the particle in subsection VIIIC5.  Again we have only a constant zero-mode, since a constant is the only periodic function (since we will need to consider only tori) that is a homogeneous solution to the wave equation (as seen by Fourier expansion):
$$ h = -{1\over area} $$
in terms of the area of the worldsheet.

Unlike the tree case, $§=0$ and $§=¹$ correspond to the 2 different boundaries of the orientable loop.  If all the vertices are on 1 boundary (say, $§=0$), we have a planar loop, otherwise nonplanar.  To include both boundaries, we need the complex variable
$$ ¨ = ÷  +i§ $$
in terms of which the torus coming from doubling the open-string surface is a periodic rectangle, with corners at $ài¹$, $Tài¹$.  (For the closed string, this will be distorted to a parallelogram, to take arbitrary twists in $§$ into account.)  Again as for trees, the open-string Green function follows from that of the closed string using image charges for reflection about the boundary:
$$ G_{open}(¨,¨') = G_{closed}(¨,¨') +G_{closed}(¨,Ш') $$
Unfortunately, the Green function for the closed string does not quite separate into holomorphic and antiholomorphic parts, but this is not a complication for low-excitation vertices.  

The closed-string Green function can be found in various ways:  For example, using the method of images, an infinite sum is obtained, which can be recognized as an expression of a Jacobi $Ï$ function.  Alternatively, a Jacobi $Ï$ function can be recognized as the solution to the wave equation with the correct periodicity conditions.  
To see that $Ï_1$ is useful for the Green function, we first note that it is analytic, and for small $¨$, $Ï_1¾¨$, so a $-ln|Ï_1|^2$ term gives the $-ln|¨|^2$ term of the complex plane, yielding the correct $¶$ function term in its wave equation from its nonanalytic behavior at $¨=0$.  Then we see that it has periodicity under $¨£¨+2¹i$ and almost under $¨£¨+T$; the latter is fixed by an extra term, similar to that for the particle, $(Re¼¨)^2/(Re¼T)$ (generalizing to complex $T$ for later application to unoriented strings or twisting of closed strings), which contributes to the wave equation the extra term inversely proportional to the area of the torus $2¹Re¼T$:  With our previous normalization,
$$ »Ð»G = -¹\left[ ¶^2(§) -{1\over 2¹Re¼T} \right] $$
The result for the closed Green function is then
$$ G(¨,T) = -ln\left|Ï_1\left({¨\over 2¹i}\Bigg|{-T\over 2¹i}\right)\right|^2 
	+{(Re¼¨)^2\over Re¼T} +H(T) $$
which appears in the amplitude as $G(¨_i-¨_j,T)$, where we have added a function $H(T)$ that is constant with respect to $z$, but depends on the geometry ($T$).  

In principle constants should not contribute, because of conservation of momentum.  But we sum only over $i±j$, dropping $i=j$ terms by normal ordering.  Normal ordering is not conformally invariant, so we add back the constant so the short-distance behavior, and thus normal-ordering, is the same.  In the tree case the geometry was trivial, so the true constant was fixed as a wave-function/coupling normalization.  But now the function is nontrivial, and our normalization must also be consistent with the tree case.  This ``boundary condition" (actually, there are no boundaries for the closed string) can be imposed by requiring that ``constant" be fixed in the short-distance limit.  In principle, the Green function should always look the same at short distances, and not be affected by boundaries or topology.  For 2 vertices on the same boundary,
$$ \lim_{¨£0}G(¨,T) = -ln(|¨|^2) +\O (¨) $$
where the term constant in $¨$ must be canceled by $H$, or
$$ \lim_{¨£0}e^{-G(¨,T)/2} = |¨| +\O(¨^2) $$
where the coefficient of $|¨|$ must be canceled.  Then we find
$$ H(T) = ln\left|»_¨ Ï_1\left({¨\over 2¹i}\Bigg|{-T\over 2¹i}\right)\right|^2_{¨=0}
	= ln\left|{1\over 2¹} Ï'_1\left(0\Bigg|{-T\over 2¹i}\right)\right|^2 $$
The final result for the closed Green function is thus
$$ 
\boxeq{
G(¨,T) = -ln\left|{2¹ Ï_1({¨\over 2¹i}|{-T\over 2¹i}) \over
	Ï'_1(0|{-T\over 2¹i})}\right|^2 +{(Re¼¨)^2\over Re¼T} 
}$$
This result can also be obtained from the tree result (Green function for the cylinder), which is already periodic in $¨£¨+2¹i$, by using an infinite sum to make it so under $¨£¨+T$:  This results in the infinite product form of $Ï_1$ above.

\x XIC3.1  Do this sum:
ªa First find the Green function for the cylinder from that for the plane by $z=e^¨$.  Then add a homogeneous zero-mode solution to get an expression in terms of just $¨-¨'$.  (The zero-mode contribution needs separate evaluation for the reasons given above.)
ªb Now make it periodic by summing over $¨£¨+nT$ for $n=-¥,...,¥$.  For each $n±0$ you need to fix the short-distance behavior as above.  Finally, add $(Re¼¨)^2/(Re¼T)$ to fix the zero-mode as above.

Using an identity for Jacobi $Ï$ functions from subsection XIC2 we find, writing $G(¨,T)­G(Ã| )$ in terms of the arguments of $Ï_1$,
$$ \boxeq{G\left( {Ã\over c +d} \Bigg| {a +b\over c +d}\right) = 
	G(Ã| ) +ln|c +d|^2 } $$
which states that $e^{-G/2}$ transforms with weight 1 under SL(2,Z) transformations of $ $.  In proving this identity one needs to cancel the phase factor from the $Ï_1$ transformation with the contribution from the non-$Ï$ part of $G$; the following identities are then useful:  First, from the result for SL(2,C),
$$ {az_1+b\over cz_1+d} -{az_2+b\over cz_2+d} = {z_1-z_2\over (cz_1+d)(cz_2+d)}
	âÜâ\boxeq{ Im¼{a +b\over c +d} = {Im¼ \over |c +d|^2} } $$
Then, from the imaginary part of the vector identity
$$ 0 = iÐV_{[i}V_j V_{k]} = 2¼Im(V_i ÐV_j)V_k +cyc. $$
choosing
$$ V_1 = {1\over c +d},ââV_2 = ÐÃ,ââV_3 = {Ã\over c +d} $$
multiplying by $|c +d|^2/Im¼ $, and using the previous result, we find
$$ {(Im¼{Ã\over c +d})^2 \over Im¼{a +b\over c +d}} = {(Im¼Ã)^2 \over Im¼ }
	-c¼Im¼{Ã^2 \over c +d} $$

When performing this transformation we will also need for the volume element, using the same identity, in terms of $w( )$ and $w'( ')$, 
$ =(a '+b)/(c '+d)$,
$$ \boxeq{ w[f(w)]^{24} = (c '+d)^{12}w'[f(w')]^{24} } $$
(The relation for $f$ itself, without the 24th power, has a 24th root of unity, which is conveniently eliminated in this form that appears in the volume element.)  

To analyze singularities in the open string, we'll need to transform variables in the expression given in subsection XIC1 for the amplitude:
$$ A_N^{(1)} = Ç_0^1 {dw\over w^2}[f(w)]^{2-D}(-ln¼w)^{-D/2} 
	Ç_w^1\left({dz\over z}\right)^{N-1}
	exp\left[-\f14Ý_{i<j}k_iÉk_j G_O(z_j/z_i,w)\right] $$
where effectively the open-string Green function $G_O=2G$ in terms of the closed-string one $G$.  We start with the change of variables
$$ Ã(Ã', ') =  {Ã'\over c '+d},ââ (Ã', ') = {a '+b\over c '+d} $$
In the general case, the Green function we start with will not necessarily be $G(Ã| )$, so the SL(2) transformation of $G$ given above will not always be the same as the one just given to change variables:
$$ G\left( {÷Ã\over ÷c÷ +÷d} \Bigg| {÷a÷ +÷b\over ÷c÷ +÷d}\right) = G(÷Ã|÷ ) +ln|÷c÷ +÷d|^2 $$
The second argument of $G$ will also be that of $f$ (as $w=e^{2¹i }$), so this same transformation will be used on $f$.  Performing these procedures on the amplitude, using the above identities for $G$ and $f$, and also the transformations for the measures
$$ {dw\over w} = {dw'\over w'}(c '+d)^{-2},ââ
	{dz\over z} = {dz'\over z'}(c '+d)^{-1} $$
we get for $D=26$ an expression for the amplitude of the same form as the above, but with 
$$ \boxeq{ z,w £ z',w';¼ÊÊG £ G(÷Ã|÷ );¼ÊÊ
	(-ln¼w)^{-13} £ (-2¹i )^{-13}(c '+d)^{-N-1}|÷c÷ +÷d|^{N-12} } $$

Ü4. Open

There are various types of singularities that occur in open string diagrams, all of which are expected from Feynman diagrams, as long as we take topology into account (and satisfy the usual conditions for dimension and string intercept):

£1) poles, from external trees

£2) cuts, from internal 2-particle states

£3) external line divergences, from the loop sitting there

£4) closed-string poles, from recognizing the open-string loop as a cylinder

\noindent (The closed-string pole goes into the vacuum if the diagram is planar, i.e., if all external states sit on the same boundary.)

$$ \figscale{poleloop}{3in} $$

The easiest poles to find are the open-string ones, since they are in the same place as in the string trees (see subsection XIB6):  When some number $n+1$ of consecutive insertion points approach each other, their Green function looks like the one for trees.  The worldsheet picture is that these external lines are stretched away into a tree, emphasizing a propagator connecting that tree to a 1-loop graph.  The $¨$ integration diverges in that region in the same way as for trees, and the calculation is the same, giving the same pole structure. 

The most interesting poles are the closed-string ones, since they appear neither in the trees nor in ordinary Feynman diagrams for particles (though higher-derivative modifications of ordinary field theories can produce them.)  We saw in the particle case the usual UV divergence coming from the $T$ integration near $T=0$, so we now examine $w=e^{-T}®1$ for the string.  The worldsheet picture is that the annulus is very short in the $ $ (periodic) direction, but still $¹$ in the $§$ direction, so it looks like a narrow cylinder, emphasizing the poles propagating along the cylinder: closed-string poles.  For the planar graph, this closed string goes into the vacuum (zero momentum), but for the nonplanar ones there are states connected at each end, so we can see the momentum dependence.

We begin by making some changes of variables.  The first is the same as for the particle case, which we already evaluated:  Separate the Schwinger parameters into a scaling parameter and Feynman parameters
$$ ¨_i = -TŒ_i $$
Next, in the language of the closed-string surface (found from doubling the open-string one), we want to switch the 2 directions of periodicity.  We want to use the same (closed-string) Green function (periods $2¹$ and $T$), but noting that it is only the ratio of the 2 periods that is invariant under a scale transformation (a conformal transformation that doesn't change its shape), the effect of this switch is
$$ {T \over 2¹} £ {2¹\over T}:ââT £ T' = {(2¹)^2\over T} $$
Since now we replace the periodicities
$$ ¨ £ ¨+2¹i,â⨠£ ¨+T $$
with
$$ ¨' £ ¨'+2¹i,ââ¨' £ ¨' +T' $$
we also define
$$ Ã' = ŒâÜâ¨' = 2¹iŒ = -{2¹i¨\over T} $$
These imply
$$   = {2¹i\over T'} = -{1\over  '},âà = {¨'\over T'} = -{Ã'\over  '};ââ
	w = e^{-(2¹)^2/T'} = e^{(2¹)^2/ln¼w'},âz = e^{-2¹iÃ'/ '} $$
In the closed string case this transformation is a symmetry, but in the open-string case it replaces our point of view ($§ª $) from an open-string loop to a closed string propagator.  Thus the divergence at $w=1$ ($T=0$) is at $w'=0$.

In terms of the discussion of the subsection XIC3, this means
$$ ÷Ã,÷  = Ã', ',ââ\pmatrix{ a & b \cr c & d \cr} = \pmatrix{ ÷a & ÷b \cr ÷c & ÷d \cr} = 
	\pmatrix{ 0 & 1 \cr -1 & 0 \cr} $$
$$âÜ (-ln¼w)^{-13} £ [-2¹i (c '+d)]^{-13}\left({c '+d\over |c '+d|}\right)^{12-N} 
	= -i^{N-1}(2¹)^{-13} $$
The $2¹$'s can be attributed to our normalization of momentum integration, while the $i$'s are because, although $w'$ is still real, $z'$ is now a phase ($¨'$ is imaginary), and the $-1$ is because the limits of integration for $w'$ have switched from $w$ ($0ª1$).  The elimination of the $ln¼w$'s is an indication of the replacement of the 2-open-string cuts with closed-string poles.  Looking at just the $w'$ dependence, we find
$$ A ¾ Ç_0^1 {dw'\over w'^2}P(w') $$
for some $P$ that can be Taylor expanded in $w'$.  (The most convenient expansion for this result is the product form of $Ï_1/Ï'_1$.)  There are thus 2 divergences at $w'=0$, coming from the first 2 terms in the expansion of $P$.

The nonplanar case can be obtained by the same method:  The only difference is that when $V_i$ and $V_j$ are on opposite boundaries, $Im(¨_i-¨_j)=ài¹$.  (Also, the $¨$'s on each boundary are ordered separately.)  We thus only need to replace for those $G$'s:
$$ ¨ £ ¨+i¹:âz £ -z,âà £ à +ü,âÃ' £ Ã' -ü ',â¨' £ ¨' +üT',âz' £ z'w'^{-1/2} $$
($¨$, which was real, gets an imaginary part while $¨'$, which was imaginary, gets a real part.)  The only effects on $e^{-G/2}$ for $w'$ near 0 (besides the form of $P(w')$) are from the factors
$$ z'^{1/2}-z'^{-1/2} £ w'^{-1/4},ââe^{-(Re¼¨')^2/2T'} £ w'^{1/8} $$  
The resulting extra contribution, again writing $i=(I,I')$ for the 2 boundaries, comes from the exponent
$$ Ý_{I,I'}k_IÉk_{I'} = -\left( Ý_I k_I \right)^2 ­ s $$
for $w'^{-1/8}$.  The singular integral in $w'$ is thus now
$$ A ¾ Ç_0^1 dw'¼w'^{-s/8-2} ÷P(w') $$
This generates the usual closed-string poles at $üŒ's = 2(n-1) = -2,0,2,...$ .  For $D±26$, there would be extra $ln¼w'$ factors generating cuts as for $w£0$, but now the cuts would be associated with closed strings (color singlets) instead of open, and thus be inconsistent with duality (which implies single-closed-string states).

\x XIC4.1  Compare this analysis of generation of poles to the corresponding ``stringy" higher-derivative particle loop of subsection VIIIC5.

The interpretation of the singularities of the planar graph is now clear:  They represent a special case of the nonplanar one, where $s=0$ because there is only the vacuum at the end of the closed-string propagator, representing scalar fields getting vacuum values.  If we transform back to the usual Schwinger parameters,
$$ A ¾ Ç_0^1 dw'¼w'^{-s/8-2}Ý_{n=0}^¥ c_n w'^n = 
	Ý_{n=0}^¥ c_nÇ_0^¥ dT' e^{-T'(n-1-s/8)} =
	Ý_{n=0}^¥ {c_n\over n-1-s/8} $$
We can always make the integral converge by analytic continuation from $s<8(n-1)$.  This always works in Euclidean space, except for tachyons ($n=0$).  But for the planar case $s=0$:  We can ignore the leading divergence in $A$ ($Çdw'/w'^2$), evaluating it as above by comparison with the nonplanar $s±0$.  But the next-to-leading divergence ($Çdw'/w'$) remains, coming from the dilaton pole.  In both cases these divergences are recognized as due to perturbing about the wrong vacuum.

The nonorientable (``occidental"?) loop (M¬obius strip) is also easy to get from the planar one.  Note the interpretation of the open string as the closed string with reflection about the real axis in the $¨$ plane.  The M¬obius strip is like the planar graph, but after a period of $T$ in the real direction there is a half-twist (flip).  But because of the reflection, a shift by $i¹$ is the same as this twist.  Thus unlike the nonplanar (oriented) case, where we replaced (sometimes) $¨£¨+i¹$, we now replace (always)
$$ T £ T+i¹âÜâ  £  -ü,âw £ -w $$
in the Green function.  (In other words, we use the same expression as before for $G_{open}$ in terms of $G_{closed}$, but with $Im¼T=¹$, so what we write as $T$ below is really $Re¼T$.)  This makes it periodic instead for $¨£¨+Re¼T+i¹$, while for $¨£¨+Re¼T$ we instead get a flip.  We do the same for the partition function $f$:  In operator language, the expression $e^{-(T+i¹)N}$ (in terms of the number operator $N$) performs this flip on the ``initial" states used to define the trace.

To look at the singularity near $w=1$, we again use the transformation
$$ \pmatrix{ a & b \cr c & d \cr} = \pmatrix{ 0 & 1 \cr -1 & 0 \cr} $$
but now this has the effect on $G$
$$ G(Ã| -ü) = G\left( -{Ã'\over  '} \Bigg| -{1\over  '} -ü \right) $$
We therefore use
$$ \pmatrix{ ÷a & ÷b \cr ÷c & ÷d \cr} = \pmatrix{ 1 & 0 \cr -2 & 1 \cr} $$
$$ ÷à = {Ã'\over 2},â÷  = { '\over 4} +üâÛâÃ' = 2÷Ã,â ' = 2(2÷  -1) $$
$$ ÜâG\left( -{Ã'\over  '} \Bigg| -{1\over  '} -ü \right)
	= G\left( {÷Ã\over 1-2÷ } \Bigg| {÷ \over 1-2÷ } \right) 
	= G(÷Ã|÷ ) +ln|1-2÷ |^2 $$
$$ = G\left( {Ã'\over 2} \Bigg| { '\over 4} +ü \right) +ln|ü '|^2 $$
The amplitude is then modified by
$$ (-ln¼w)^{-13} £ [-2¹i (c '+d)]^{-13}\left({c '+d\over |÷c÷ +÷d|}\right)^{12-N}
	= -i^{N-1}(2¹)^{-13}2^{12-N} $$
and the fact that the arguments of $G$ and $f$, compared to the planar case, are
$$ z' £ z'^{1/2},âw' £ -w'^{1/4} $$
We therefore make a second change of variables
$$ z'^{1/2} = z'',âw'^{1/4} = w'' $$
This also generates a factor $2^{N-1}$ from the $dz/z$ measure and a 4 from 
$dw/w$, so now the nonorientable diagram looks the same as the planar one except for $wª-w$ in $G$ and $f$, and an extra factor of $2^{13}$.  However, the planar graph has an extra group-theory factor of N for N ``quarks" from tracing over its second, vertex-free boundary, while the M¬obius strip has an extra group-theory factor, coming from the twist, of
$$ \cases{ 1 & for USp(2N) \cr 0 & for U(N) \cr -1 & for SO(N) \cr} $$
This factor is most easily seen from the massless vector (adjoint) propagator:  To preserve the symmetry of the adjoint representation, the massless vector propagator must be symmetric in its 2 group (fundamental) indices for USp, antisymmetric for SO, and asymmetric (no twist, but orientable) for U.  We thus find the (leading, at least,) divergences of these 2 graphs can cancel only for SO($2^{13}$), at least for a regularization scheme that respects this symmetric choice of integration variables.

Ü5. Closed

For the closed string we'll consider just the orientable loop (torus).  The one-loop amplitudes for the closed string can be obtained by methods similar to those for the open string (except for one cheat, which we'll discuss below).  The main modification is that we have a constraint to impose, namely $ëN­T_0-ÐT_0=0$.  The easiest way to impose this is by including a projection operator in the propagator,
$$ ¸ ­ Ç_{-¹}^¹ {d§\over 2¹}¼e^{-i§ëN} = ¶_{ëN,0} $$
If we combine this with the Schwinger parametrization of the propagator, we have
$$ {1\over H_0}¸ = Ç_0^¥ d  Ç_{-¹}^¹ {d§\over 2¹}¼e^{-( H_0 +i§ëN)} $$
We then rewrite the exponent as
$$  H_0 +i§ëN =  (T_0+ÐT_0) +i§(T_0-ÐT_0) = ¨T_0 +ШÐT_0 $$
Here $T_0$ is the same as for the open string, but with the replacement $p£üp$:
$$ T_0 = Œ'(üp)^2 +N-1 $$
(where this $N$ is just for left-handed modes).
The net effect is thus to double modes, with real $¨_i$ and $T$ becoming complex:
$$ d÷ ¼V(÷ ) £ {d^2 ¨\over 2¹}¼V(¨),ââ
	dT¼e^{-TH_0} £ {d^2 T\over 2¹}¼e^{-(TT_0 +ÐTÐT_0)} $$
So the vertices are now anywhere on the strip instead of just the boundary, while the sum over states in the trace includes averaging over arbitrary twists.  Thus for the partition function we have
$$ w^{-1} £ |w|^{-2},ââ[f(w)]^{2-D} £ [|f(w)|^2]^{2-D} $$
while the contribution to the volume element from the momentum integral is modified by the replacement $p£üp$ (or equivalently, by the closed string having half the slope of the open string):
$$ T^{-D/2} = (-ln¼w)^{-D/2} £ (üRe¼T)^{-D/2} = (-üln|w|)^{-D/2} $$
The amplitude is then
$$ A_N^{(1)} = Ç{d^2 w\over 2¹|w|^4}|f(w)|^{2(2-D)}(-üln|w|)^{-D/2} 
	Ç\left({d^2 z\over 2¹|z|^2}\right)^{N-1}
	exp\left[-\f14Ý_{i<j}k_iÉk_j G_{ij}\right] $$

Unfortunately this procedure has one flaw:  The torus is invariant under the group SL(2,Z), which divides the naive integration region for $T$ into an infinite set of copies.  If we blindly follow the procedure above, the result for the amplitude will be infinite, simply because of the over-counting.  This error is easy to see from 2D geometry, and can be fixed by hand.  What is not clear is the error in the derivation, i.e., the relationship of the loop to the trees, without which the calculation is meaningless.  (If the loops don't follow from the trees, they don't belong to the same theory, regardless of any symmetry arguments.  This is exactly the problem of anomalies.)  The solution should be found from string field theory, and is probably due to our implementing $ëN=0$ in too Abelian a way.

The ``modulus" $T$ is the residual part of the original 2D metric not gauged away by the original invariances.  Similarly, this SL(2,Z) ``modular" invariance is the residual discrete part of the original 2D coordinate plus Weyl scale invariance on the torus, represented as a field transformation on the residual part of the metric $T$.  (There is also continuous translation invariance in the $§$ and $ $ directions.)  We can describe the torus as a parallelogram on the (flat) complex plane, with corners $0$, $z_1$, $z_2$, $z_1+z_2$.  Now consider an arbitrary point $z$ somewhere on the torus:  This point is identified with the points
$$ z £ z +n^i z_i $$
for integers $n^i$ ($i=1,2$).  If we now consider SL(2,C) transformations of $z_i$ (not the conformal SL(2,C), but just transforming $z_i$ linearly)
$$ z_i' = g_i{}^j z_j $$
we see that to preserve the torus, as defined by its periodicity,
$$ z £ z +n^i z'_i = z +n'^i z_iâÜân'^i = n^j g_j{}^i $$
that the matrix $g$ must be integers, i.e., an element of SL(2,Z).  We thus have
$$ g_i{}^j = \pmatrix{ a & b \cr c & d \cr} $$
for integers $a,b,c,d$ with $ad-bc=1$.  Actually, the determinant condition is automatic:  It must be an integer, but the inverse of the group element is also in the group, so its determinant is the inverse of an integer, but also an integer.  Thus the determinant must be 1.  (The other possibility of $-1$ is uninteresting:  It is the result of combining the SL(2,Z) transformations with a switch of $z_1$ with $z_2$.)

The usual conformal transformations include complex scale transformations, under which only the ratio 
$$   ­ {z_1\over z_2} $$ 
which we identify with the modulus, is invariant.  Under a modular transformation
$$  ' = {a  +b\over c  +d} $$
Similarly, for any point $z$, the ratio
$$ Ã ­ {z\over z_2} $$
is also conformally scale invariant.  It transforms under a modular transformation (where $z$ is invariant) as
$$ Ã' = {Ã\over c  +d} $$
Effectively, we have ``gauged" (by conformal transformation) $z_2£1$, and $ $ and $Ã$ are $z_1$ and $z$ in this gauge, so the $c +d$ denominators in the modular transformations are compensating conformal transformations to maintain this gauge.

First we need to check that the closed-string amplitude is invariant under modular transformations.  (We already saw that the open-string amplitudes transformed in a simple way.)  Taking the transformations obtained in subsection XIC3, we need only modify the results by taking some $|\phantom{n}|^2$'s in a few places.  We also need the result (the previous identity for $Im¼ '$)
$$ ln|w| = ln|w'||c '+d|^{-2} $$
for SL(2,Z).  Remembering that for the closed string the tachyon has $k^2=8$, so $Ý_{i<j}k_iÉk_j=-4N$, we find that each piece of the amplitude gives the same multiplied by the following exponent for $|c '+d|$ after replacing $Ã=Ã(Ã', '), = (Ã', ')$:
$$ {1\over 2¹}\left|{dw\over w}\right|^2 £ -4,ââ
	\left( {1\over 2¹}\left|{dz\over z}\right|^2 \right)^{N-1} £ -2(N-1),ââ
	|w|^{-2}|f|^{-48} £ -24 $$
$$ (-üln|w|)^{-13} £ 26,ââe^{-ÝGkÉk/4} £ 2N $$
which cancel, proving invariance.

Next we need to divide up the region of integration for $ $ into ``fundamental regions":  For any such region, any point in the upper-half complex plane can be mapped into it in a unique way by a ``modular transformation" (which is unfortunately also a ``unimodular transformation" due to poor semantics).  
We started with the conditions
$$ -¹ ² Im¼T ²¹,âRe¼T ³0âÜâ-ü ² Re¼  ² ü,âIm¼  ³ 0 $$
The former condition already takes care of the transformation $ £ +1$.  All modular transformations can be obtained from that and $ £-1/ $, which takes the inside of the unit circle to the outside:  Choosing the outside, we get the final conditions
$$ -ü ² Re¼  ² ü,ââIm¼  ³ 0,ââ| | ³ 1 $$
It can then be shown that these choose one fundamental domain.  (For example, by showing that an arbitrary transformation on the unit circle can produce only vertical lines, and circles centered on the real axis whose inverse radius is an integer, where all the lines, or circles of the same radius, are related by $ £ +n$.)
The extra restriction also eliminates the usual UV singularity near $ =0$.

$$ \figscale{fundamental}{2in} 
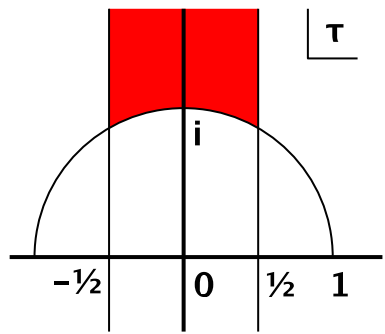
$$

\x XIC5.1 Find several of the other fundamental regions.

Divergences in the closed-string loop are similar to those in the open string, except that now the divergence from the loop integral associated with a closed string going into the vacuum has the torus on the end of that tadpole.  Since restriction to the fundamental region has already eliminated the UV divergence, this divergence now shows up only as an IR divergence from all the $Ã$'s near 0, i.e., factoring the graph as a closed-string tree times the tadpole.   (For the open string we had to perform a modular transformation to change the UV divergence into closed-string IR, since taking $Ã$'s to 0 shows only the usual tree divergences.)  Unlike the closed tree graph, where 3 of the vertices had fixed positions, so no more than $N-2$ vertices could converge (which is symmetric with the fact that no less than 2 can), for the loop only 1 vertex is fixed, so all $N$ can converge (and there is no symmetry $nªN-n$ because the loop itself is associated with vertices that don't converge).  The calculation is the same as for the open string, except now we scale $|Ã|^2$ instead of just $Ã$, and the poles are closed-string instead of open.  The divergence is again from the tadpole propagator, and its interpretation is as for the nonplanar graph, coming from the tachyon and dilaton, but with different coefficients.

Note that the dilaton that appears here as a 1-loop correction (at least for the closed-string case) is really the determinant of the metric, in the usual string gauge for (26D) local Weyl symmetry:  Since the vacuum value of the true dilaton $Ä$, defined as the field that couples to the ghosts (or, at order $Œ'$, to the worldsheet curvature, and thus the Euler number, which counts loops) and not to $X$, generates the string coupling through its classical vacuum value, it appears in the effective action along with $\hbar$, homogeneously as $(Ä^2/\hbar)^{1-L}$, and thus not at all at 1 loop (except through derivatives, as $»Ä/Ä$, etc.).  Thus, the 1-loop term that couples to the determinant (or trace, at linearized order) of the metric is actually the cosmological term.  (Of course, if field redefinitions of the metric are made to get the usual classical $R$ term, this will generate $Ä$ dependence, and $Ä$ will no longer count loops.)

Ü6. Super

The calculation for 1-loop open superstring amplitudes with 4 or less external vectors is very similar to the superparticle case (see subsection VIIIC4), combined with the results for the open bosonic string.  To begin, the vertex operator for the vector takes the same form as for the particle (see subsection VIIIC5):
$$ V = û^a e^{ikÉöX} (ÀX_a -k^b S_{ab}) $$
where the spin operator $S_{ab}$ is now represented by an appropriate current (evaluated at the same $z$ as $X$) expressed in terms of worldsheet fermions (see subsection XIB5).
The simplest way to do these calculations (so far) is in the lightcone.  For 4 or fewer external lines we can choose the $²4$ external momenta ($²3$ independent) and $²4$ external polarizations to all point in just the transverse directions, avoiding the complications of nonlinearities in the longitudinal components.  Then the tranverse part SO(8) of the spin current is simply (in somebody's normalization)
$$ S_{ij} = üÆ_{[i}Æ_{j]} = \f18 \S ©_{[i}©_{j]} \S $$
in terms of the vector fermions $Æ$ of Ramond-Neveu-Schwarz and the spinor ones $\S$ of Green-Schwarz, related by triality and bosonization.  

In the RNS case we still have to consider summing over R and NS strings, so GS is simpler.  Among the various parts of the trace in evaluating the loop, we have in particular the (super)trace over the zero-modes of $\S$.  In fact, this is the same trace considered for the particle in subsection VIIIC4:  It is the trace over all the massless states created by those zero-modes.  (As usual, the trace factorizes into that trace times those for the oscillators that create massive states.)  If we sum over those states individually, we are performing (the lightcone version of) the same analysis we made for the particle.  But we can also make the analysis for all states together, by making the same analysis as for a ÓsuperÕparticle in the lightcone:  For example, note that the $\S$'s form an SO(8) Clifford algebra -- they are the usual $©$ matrices with indices switched by triality.  Then we see that the supertrace is just the usual trace with an extra factor of ``$©_{-1}$" (since $\S$ takes the bosonic vector to the fermionic spinor and vice versa).  So the first nonvanishing supertrace is that of 8 $\S$'s (giving an $·$ tensor in its spinor indices), i.e., 4 $S_{ij}$'s.  The result will then have the same structure as the particle case: 4 $F$'s times the loop integral (less the fermionic zero-modes) for 4 scalar vertices (tachyon-like, but massless).  The explicit form of the $F^4$ factor is the same as for the particle, and also the same as for the superstring tree.  (Again the tree calculation of this factor is more complicated.  For the bosonic string even the tree factor itself is more complicated, containing $Œ'$ corrections.)

\x XIC6.1 Repeat the calculation of exercise VIIIC4.1 of the zero-mode supertraces, again using components (i.e., evaluating the spinor and vector contributions separately), but now using just physical components, corresponding to the SO(8) transverse spin operators.  Compare to the SO(10) calculation reduced to a lightcone gauge.  For SO(8) there is also an $·$ tensor contribution (remember the spinor is Weyl), not appearing in SO(10):  What happens to it?

Now that we have reduced the loop to a ``kinematic" factor for the polarizations, identical to that in the particle case, times a bosonic-string-like expression, we can do the rest of the calculation in analogy to the bosonic case.  The Green function is the same as in the bosonic case, since we now have to deal with just $X$ in the vertex.  The partition function is actually simpler:  For the oscillators, we now have 8 bosons and 8 fermions with the same boundary conditions, so their supertrace is just 1.  (Remember that ghost fermions gave the inverse of $X$ bosons.)  For the remaining, bosonic zero-modes, the only difference is that the ground state is massless, so we don't get the extra $w^{-1}$.  Thus the volume element is simply
$$ \V = (-ln¼w)^{-5} $$
coming from the momentum integration, now in $D=10$.  In summary, the result for the 4-point amplitude is the same as the bosonic tachyon amplitude (see subsection XIC1), except for: 
\item{(1)} the $F^4$ factor, 
\item{(2)} no power of $f(w)$, 
\item{(3)} $w^{-2}£w^{-1}$, 
\item{(4)} $(-ln¼w)$ appears to the power $-5$, and
\item{(5)} external lines are massless.  

\noindent This analysis applies to planar, nonplanar, and nonorientable loops.  

In the net transformation given at the end of subsection XIC3, used for changes of variables in analyzing divergences, for the 4-point superstring amplitude we now have instead
$$ (-ln¼w)^{-5} £ [-2¹i (c '+d)]^{-5} = i(2¹)^{-5} $$
The main difference in the analysis of divergences is that the lack of the 
$w^{-1}f^{-24}$ factor (before and after changes of variables) removes the divergences associated with the now-absent tachyon (see the argument at the end of subsection XIC4), so all open-string divergences cancel for gauge group SO(2$^{D/2}$) = SO(32) now.  (There is a $2^{N+1}=2^5$ in the final change of variables for the nonorientable loop.)  

The missing factor also means that now closed-string poles start with massless ones.  The $F^4$ factors (of the form $[tr(F^2)]^2$ for the nonplanar loop) now give spin to these poles, and for the massless poles the singlet currents $tr(FF)$ can be associated with the graviton, dilaton, and axion (4-form) coupling, as for the particle.  The latter couples to the Hodge dual of the dual of the 2-form, i.e., an $·$ tensor times a $(D-4)$-form $÷B$: in form notation,
$$ L_1 ¾ ÷B\wedge tr(F \wedge F) $$
(In $D=4$, $÷B$ would be the usual pseudoscalar axion.)  This preserves the Abelian gauge invariance of the $(D-4)$-form, since $tr(F \wedge F)$ is the curl of the Chern-Simons form (see subsection IIIC6):
$$ tr(F \wedge F) = dC,ââC = tr(üA\wedge dA +\f13 A\wedge A\wedge A) $$
We can see how this relates to the usual 2-form $B$ by a duality transformation:  Starting with a first-order form of the $(D-4)$-form action (ignoring the dilaton or string coupling),
$$ L_0 ¾ G\wedge d÷B -üG^2 $$
and varying with respect to $÷B$ in $L_0+L_1$, we find
$$ L_0+L_1 £ üG^2,ââG = dB +C $$
Gauge invariance of $G$ under Yang-Mills transformation implies an unusual transformation for $B$:
$$ ¶C ¾ tr( d \wedge dA)âÜâ¶B ¾ tr( \wedge dA) $$
(This is the same coupling and duality considered in $D=4$ in exercise XA3.1, and supersymmetrized in exercise XB5.1.)

Remarks similar to those about the open string apply to the closed-string amplitude (torus), where everything is replaced with $|¼|^2$'s again, but now including left and right-handed copies of the kinematic factor, with products of left and right-handed vector polarizations giving those of the massless states of the closed string.  The proof of modular invariance works as before, but with the changes listed above for the open superstring.  (For Type II strings one has only the torus, but Type I is nonorientable, and so has also the other graphs mentioned earlier.  Here we don't consider the heterotic string.)

There are several ways to see that this amplitude has no divergences:  (1) We already saw previously that such a divergence would be associated with a cosmological term.  This term, if expanded about flat space, would give a 4-point interaction with no derivatives.  But we have already extracted a kinematic factor that contains derivatives.  Furthermore, it would contribute to lower-point functions, which we saw vanish.  (2) Non-renormalization theorems in supersymmetric theories prevent generation of a cosmological term in loops.  (3) By duality, the limit where some of the $¨$'s get close would reveal this divergence as an intermediate dilaton state connecting a tree with a lower-point loop.  But we have already seen the lower-point loop graphs vanish.

Let's look at the last argument in more detail:  Applying this method in the same way as for open strings in section XIC4 and trees in subsection XIB6, we find
$$ A ¾ Ç_0^· d¨¼¨^{2n-1+üÝk_iÉk_j} $$
($1²n²N-1$) where again $n+1$ $¨$'s are taken close together, and the sum is over $i<j$ for those $n+1$ $k$'s, but now we get an exponent $2n$ because open-string $¨$'s are replaced with $|¨|^2$'s, and now
$$ 2Ýk_iÉk_j = -s_n +(n+1)M^2 $$
where $M^2=-8$ for the closed bosonic-string tachyons, but 0 for closed superstring ground states.  (As before, the $T$ integration converges because the fundamental region restricts $Re¼T>å3¹$.)  So for the bosonic case we get
$$ A_{bosonic} ¾ Ç_0^· d¨¼¨^{-s_n/4-3} ¾ {1\over s+8} $$
giving the usual closed-string tachyon pole (or divergence for no $s$).  However, the 4-point superstring amplitude is convergent (at $s=0$) for all $n$:
$$ A_{super} ¾ Ç_0^· d¨¼¨^{-s_n/4+2n-1} ¾ {1\over s-8}âfor¼n=1 $$
(i.e., 2 $k$'s).  As expected, there are no tachyons, and there is no contribution from massless poles because the 3-point loop that has been factored out vanishes for external massless states.  (Similar remarks apply for such singularities in the open-superstring loop.)

Ü7. Anomalies

There are no anomalies in odd dimensions, especially $D=11$.  Since string theories are equivalent to $D=11$ M-theory, they have no anomalies.  (This is similar to understanding cancellation of anomalies in the Standard Model by embedding it in a manifestly anomaly-free GUT.)  However, the manner of anomaly cancellation is unusual, and may have application beyond the present string theories.

Axial anomalies arise from massless (chiral) particles inside the loop.  Thus we will first analyze anomalies in such field theories, then consider how these fields appear in string theory.  The relevant field theories are 10D supersymmetric Yang-Mills coupled to N=1 supergravity, or N=(2,0) supergravity.  

In twice-odd dimensions (in particular D=10), unlike twice-even (like D=4), irreducible spinors are truly chiral (Weyl is not the same as Majorana):  In D=4, Yang-Mills can couple chirally because a Weyl spinor and its complex conjugate may be in different representations (i.e., complex), but in D=10 even a real group representation can give an anomaly.  Similar remarks apply to gravity:  In D=10 even gravity can couple chirally, because a chiral spinor need not be accompanied by its opposite chirality to construct an action.  Similar remarks apply to selfdual tensors (which also exist only in twice-odd dimensions), which can be generated from products of selfdual spinors.  Mass terms break chiral invariance, since they couple fields of opposite chirality, so only massless fields running around the loop contribute to the anomaly.

\x XIC7.1 Using the general results for irreducible spinors of subsection XC2, what are the simplest spinor actions for all dimensions and all signatures?  (Hint:  There is a simple correspondence with something in the big table.)  How many irreducible spinors are required to write an action in each case?  What happens when masses are introduced?

For the spinor in external Yang-Mills this calculation is standard (see subsections VIIIB2-3).  We are left to consider the group theory.  To mimic string theory, we use the 't Hooft double-line notation for the adjoint representation in terms of indices of the defining representation, as implied by the quark interpretation for Chan-Paton factors.  Thus each ``planar" propagator has added to it a twisted propagator, with the factor 0 or $à1$ as described in subsection XIC4.  (Equivalently, we could twist the vertices instead.)  In the planar and nonorientable (with respect to this notation) graphs, we get for the anomaly $F^{D/2}$ without traces:  For the planar case all vertices are on one side, and there is a factor of N from tracing the other side, while for the nonorientable case there is only one side.  However, there is only 1 planar graph, while for the nonorientable case there are $2^{D/2}$ (i.e., exactly half of all the graphs from twisting any of the $D/2+1$ propagators).  Thus this particular contribution to the anomaly is canceled for SO(2$^{D/2}$).

For the nonplanar graphs some $F$ vertices are on a different side than the $©_{-1}$ one, so there is a trace of those $F$'s.  For the case $D=10$, we already saw in subsection VIIIC4 that (using Bose symmetry to get an anticommutator) $tr(F^3)$ vanishes for SO(32).  Of course $tr(F)$ also vanishes, and for the same reason $tr(F^5)$ can't contribute because the anomaly itself is adjoint.  Finally, by adding a local counterterm to the action of the form $tr(AF)tr(AF^3)$ we can always convert the anomaly between $tr(F^2)F^3$ and $tr(F^4)F$.  (At this lowest order in the fields we can use the Abelian part of $F$ and of the gauge transformation.)  So for SO(32) in $D=10$ we can cancel all of the anomaly except a term $tr(F^4)F$.

At this point we remember that the nonplanar open superstring loop generates an unusual coupling in the 4-point amplitude between Yang-Mills and the axion (see subsection XIC6).  Using this fact, we can write a local counterterm that cancels the anomaly (at least at this order), namely
$$ B \wedge tr(F \wedge F \wedge F \wedge F) $$

Similar remarks apply to gravity anomalies, and mixed Yang-Mills/gravity anom\-alies.  Again the Yang-Mills generators are replaced with Lorentz generators.  But for pure gravity there are no Chan-Paton factors, so cancellations must be between different spins: spin 1/2, gravitino, and self-dual tensor.  It turns out that cancellations can be obtained in D=10 only for N=(2,0) supergravity or N=1 supergravity coupled to SO(32) or E$_8°$E$_8$ Yang-Mills (or some uninteresting non-semisimple groups).

Superstring theories generally violate (spacetime) parity because they use chiral spinors; the only exception is Type IIA, because the left-handed and right-handed modes have spinors of opposite parity, so spacetime parity includes worldsheet parity (switching left and right).

The calculation of the open-superstring anomaly is simpler in the covariant RNS formalism than in the lightcone.  It is very similar to the particle calculation of subsection VIIIB2, replacing $Öá$ with $Çd§¼ÆÉ»X$, $S_{ab}$ with $üÆ_{[a}Æ_{b]}$, $©_{-1}$ with its GSO analog, etc., since the background appears via $»X£»X(§)+A(x)¶(§)$, the $¶(§)$ putting $A$ on a boundary.  Integrating out the fermionic zero-modes gives the same kinetic factor, again leaving $m^2$ times the graph with external massless scalars, the internal mass$^2$ shifted by $m^2$, and no fermionic zero-modes.

Unfortunately, now that scalar graph is nontrivial, since there are always states with mass much greater than $m$, no matter how big $m$ gets.  This is related to the fact that this 6-point graph can be factored into a 4-point loop times a 4-point tree (or 5-point loop times 3-point tree) in an appropriate limit, while in the particle case we needed to consider only a 1PI graph.  The evaluation of this graph is similar to the one-loop superstring in the Green-Schwarz lightcone:  The factors of $f(w)$ again cancel, 10 being reduced to 8 for both the $X$ and $Æ$ oscillators by the ghosts.  

However, because of the shift in mass$^2$, there is now an extra factor of 
$w^{m^2/2}$ (i.e., the ground-state mass is now $m$).  As a result, the limit $m^2£¥$ is dominated by the region $w®1$ ($T®0$).  Therefore, we again transform coordinates, and again find the largest contribution comes from the (shifted) dilaton, whose propagator now gives a factor $1/m^2$ going into the vacuum.  (All other contributions die more rapidly, and vanish as $m£¥$ even after multiplying by the overall $m^2$.)  Again as for the 4-point amplitude, this dilaton contribution cancels between the planar and nonorientable graphs for SO(32).  However, for the nonplanar graph, this dilaton contribution vanishes, again as for the 4-point; we have seen the explanation in subsection XIC4.

\refs

£1 R. Hagedorn, ÓNuo. Cim.Õ É56A (1968) 1027:\\
	Hagedorn temperature (but not from string theory).
£2 K. Kikkawa, B. Sakita, and M.A. Virasoro, ÓPhys. Rev.Õ É184 (1969) 1701:\\
	first string loop calculation.
£3 J.A. Shapiro, \PRD 5 (1972) 1945:\\
	first closed string loop, and significance of modular invariance.
£4 M.B. Green, \PL 46B (1973) 392:\\
	first noticed divergence cancellation in RNS string with GSO projection
	(but not relation to supersymmetry).
£5 M.B. Green and J.H. Schwarz, \NP 198 (1982) 441:\\
	superstring loop with manifest lightcone supersymmetry.
£6 L. Alvarez-Gaum«e and E. Witten, \NP 234 (1983) 269;\\
	M.B. Green and J.H. Schwarz, \PL 149B (1984) 117:\\
	higher-dimensional field theory anomalies and their cancellation.
£7 M.B. Green and J.H. Schwarz, \NP 255 (1985) 93:\\
	superstring anomaly cancellation, using RNS formalism
	(as suggested by D. Friedan and S. Shenker).

\unrefs

ÚXII. MECHANICS

String theories describe particles of arbitrarily large spins:  So far in this
text we have concentrated on lower spins, but we can describe (at least)
free gauge-invariant actions for arbitrary spins based on quantum
mechanical BRST.

Gauge invariance is required in field theory to manifest Lorentz
invariance.  The basic problem is that a four-vector wave function cannot
have the obvious Minkowski inner product, since the time component
would have a minus sign in its normalization, resulting in negative
probability.  In the classical action there is a gauge invariance that allows
the time component to be dropped from the action.  However, such
gauges destroy manifest Lorentz invariance, since a three-vector cannot
represent Lorentz transformations in a local way.  More useful gauges
keep all components of the four-vector, while also introducing scalar
fermionic ``ghosts" to cancel the effects of the bad part of the
four-vector.  A certain symmetry between the bosonic and fermionic
unphysical degrees of freedom is needed to enforce this cancellation:  It is
the field theoretic version of the BRST symmetry discussed in section VIA.

Another complication is that gauge transformations do not allow the
elimination of traces in a simple way:  Although it is Lorentz covariant to
constrain a tensor to vanish when a pair of its vector indices is
contracted, this interferes with gauge invariance in interacting theories,
such as gravity.  A related complication is massive theories, which can't
always be described simply by adding mass terms to massless theories.

There is a simple solution to all these problems, which determines the
free part of the action for any theory.  (Interactions are a separate
problem.)  This method automatically introduces all the correct fields,
including ghosts, for any massless or massive theory.  It also gives a
simple universal expression for the BRST symmetry that cancels unphysical
modes, as well as providing a simple proof that these modes disappear in
the lightcone gauge.  The method is based on the idea of introducing
extra fermionic dimensions to spacetime that are unphysical (unlike
superspace for supersymmetry), which cancel unphysical degrees of
freedom associated with the time dimension. 

Although for most purposes the only spins of fundamental particles
relevant in field theory for are 0, 1/2, 1, 3/2 (maybe), and 2, and these
few cases can be studied separately, in this chapter we'll analyze all free
theories because:  
\item{(1)} The ultimate theory of particles may require them; 
\item{(2)} some of the theories presently under most active investigation (such
as strings and membranes) require them; 
\item{(3)} many observed, though
perhaps not fundamental, particles have higher spin; and 
\item{(4)} a better
understanding of field theory can be obtained by determining exactly
which properties all fields have in common as well as how they differ.

\sectskip\bookmark5{A. OSp(1,1\noexpand|2)}
	\secty{A. OSp(1,1\kern.1em\vrule width1.2ptÊ2)}
	\def\righthead{\rgbo{1 0 1}{A. OSp(1,1|2)}}

This construction involves the introduction of spacetime symmetries that
are not manifest on the physical coordinates.  An important analog is the
conformal group in D dimensions, which acts nonlinearly on the usual D
spacetime coordinates, but can be represented linearly on D+2
coordinates, since the group is SO(D,2).  As described in subsection IA6 for
spin 0, 1 space and 1 time coordinate can be eliminated, so that SO(D,2) is
still represented, but SO(D$-$1,1) is the largest orthogonal group that is
still manifest.  We have also seen that in the lightcone gauge this manifest
symmetry is reduced again in the same way, leaving SO(D$-$2).  In our
case the relevant group is OSp(D,2|2), the natural generalization of the
orthogonal group to D space, 2 time, and 2 anticommuting dimensions. 
This allows rotations between timelike and fermionic directions,
eventually resulting in their cancellation.

Ü1. Lightcone

We saw in subsection IIB3 how the single equation of motion
$S_a{}^b»_b+w»_a=0$, applied to field strengths, universally described all
spins in all dimensions, for free, massless particles.  (A possible exception
is the spinless case, where we need $õ=0$, which is redundant
otherwise.  However, we can use the universal field equation even in that
case if we use the vector field strength formulation of spin 0.)  One way
we solved this equation was to perform a unitary transformation.  We can
use the same unitary transformation, plus the constraints, to simplify the
Lorentz generators.  To further simplify matters, we can use the
constraint $õ=0$, solved for $»^-$, to choose the gauge $x^+=0$, which is
equivalent to working in the Schr¬odinger picture (no time dependence for
operators).  The procedure is thus:  (1) Start with the manifest
(antihermitian) representation of the Lorentz generators,
$$ J^{ab} = x^{[a}»^{b]} +S^{ab} $$
 (2) Apply the transformation
$$ J £ UJU^{-1},ââln¼U = S^{+i}{»^i\over »^+} $$
 which eliminates the only $S^{+i}$ term, in $J^{+i}$ (while complicating
$J^{-i}$).  (3) Finally, apply the constraints, which have already been
transformed by this same transformation,
$$ õ = 0â(ܼgauge¼x^+ = 0),ââ
	S^{ab}»_b +w»^a = 0 £ S^{+-} -w = S^{i-} = 0 $$

Our Lorentz generators are then
$$ J^{+i} = -x^i »^+,ââJ^{+-} = -x^- »^+ +w,ââ
	J^{ij} = x^{[i}»^{j]} +S^{ij} $$
$$ J^{i-} = -x^- »^i +{1\over »^+}[üx^i(»^j)^2 +S^{ij}»_j +w»^i] $$
 These generators satisfy the pseudo(anti)hermiticity condition
$$ J^{ab}ÿ(»^+)^{1-2w} = -(»^+)^{1-2w}J^{ab} $$
 This means that the Hilbert-space metric needs a factor of
$(»^+)^{1-2w}$.  This is related to the fact that the $w$ terms can be
eliminated by a ÓnonÕunitary transformation with the appropriate power
of $»^+$:  As part of step 2, we could have applied a second
transformation
$$ U_2 = (»^+)^{S^{-+}} $$
 with the result of eliminating all $S^{+-}$ terms, so the redundant
constraint $S^{+-}=w$ would not have been needed, so $w$ would not
appear.  
$U_2$ is in fact just the transformation that takes the surviving 
independent part of the field strength $F^{+...+i...j}$ to the 
lightcone gauge field $A^{i...j}$, taking us from the original 
constrained field strengths to unconstrained gauge fields.
In any case, we generally choose $w=0$ for bosons.  

We previously applied dimensional reduction to the field equations for the
field strengths, to obtain the equations for the massive, free theories
from the massless ones.  The same methods can be applied to the Lorentz
(or Poincar«e) generators.  (The Lorentz generators will be used later to
find the BRST operator, to obtain the field equations in terms of the gauge
fields, and the action.  For that purpose the dimensional reduction can be
performed at any stage in the derivation.)  We thus find the general result
$$ J^{+i} = -x^i »^+,ââJ^{+-} = -x^- »^+ +w,ââ
	J^{ij} = x^{[i}»^{j]} +S^{ij} $$
$$ J^{i-} = -x^- »^i +{1\over »^+}Óüx^i[(»^j)^2 -m^2] +S^{ij}»_j 
	+S^{i-1}im +w»^iÕ $$
 (The $-1$ is still an index, and should not be confused with an inverse.) 

Note that $S^{-i}$ and $S^{-+}$ were eliminated (after the unitary
transformation) by the constraints, and that $S^{+i}$ just dropped out. 
(In other words, $S^{+i}=0$ was the gauge choice for the constraint
$S^{-i}=0$.)  This leaves only $S^{ij}$ (and $S^{i-1}$ in the massive case),
whose representation is that of the highest-weight part of the original
field strength.  However, we can more simply choose the representation
of $S^{ij}$ as our starting point, since it is just the transverse part of the
gauge field; it defines the representation of the Poincar«e group.  We
therefore have an explicit construction of the generators of the Poincar«e
group, for arbitrary representations, defined on just the physical degrees
of freedom, given directly by the little group SO(D$-$2) spin generators
$S^{ij}$ (or SO(D$-$1) generators $S^{ij}$ and $S^{i-1}$ in the massive
case) that identify the representation.  For example, in D=4, SO(2) has just
one generator, the helicity, so for any state of a given helicity we know
the action of the Poincar«e generators. 

\x XIIA1.1 Check that this lightcone representation of the Lorentz
generators satisfies the correct commutation relations.

Classical free field theory is easy to define in the lightcone, since solving
the constraints in the lightcone formalism has picked out just the
physical components, so the only  remaining constraint is the
Klein-Gordon equation.  Thus, the kinetic term for ÓanyÕ massless bosonic
field is simply $-üÄ(üõ)Ä$, where $üõ=-»^+»^-+ü(»^i)^2$, and $»^-$ is
considered the time derivative.  (In general, the kinetic operator for a
massless boson is some second-order differential operator, which
reduces to $õ$ on the physical components.)  For fermions we have
instead $õ/»^+$, since we must then have an odd number of derivatives
to avoid getting a trivial result after integration by parts.  (For a boson,
$Ä»Ä=»(üÄ^2)$, for a fermion $Æ»»Æ=»(Æ»Æ)-(»Æ)^2=»(Æ»Æ)$ by
anticommutativity.  In general, the kinetic operator for a massless
fermion is some first-order differential operator, which reduces to
$õ/»^+$ after eliminating auxiliary fields.)

This quantum mechanical representation of the Lorentz generators has a
simple translation into classical field theory, in terms of field theory
Poisson brackets.  The definition of Poisson brackets in lightcone
quantum field theory follows directly from the action:  Defining as usual
the canonical momentum $¹$ as (minus, in our conventions) the variation
of the Lagrangian with respect to the time derivative $»_+$ ($=-»^-$) of
the variable $Ä$, we find the fundamental bracket for bosons
$$ ¹ = -»^+ ÄâÜâ[Ä(x^-,x^i),Ä(x'^-,x'^i)] = 
	i{1\over »^+}(2¹)^{D/2}¶(x^--x'^-)¶^{D-2}(x^i-x'^i) $$
 (Note that the $»^+$ was essential for the antisymmetry of the bracket. 
We can also evaluate its inverse as an integral, so
${1\over »^+}¶(x^--x'^-)=ü·(x^--x'^-)$.  For fermions we have instead
an anticommutator and no $1/»^+$.)  We then find that any quantum
mechanical group generator $J$ (including internal symmetries)  can be
represented in field theoretic form as
$$ \J = Ç{dx^- d^{D-2}x^i\over (2¹)^{D/2}}¼üÄ»^+JÄ = iüÒÄ|JÄÔ $$
 where we have used the relativistic inner product of subsection VB2, but
for a lightlike hypersurface:  For positive-energy solutions
$$ Ò1||2Ô = Ç{dx^- d^{D-2}x^i\over (2¹)^{D/2}}¼Æ_1* ü(-i)\onª»{}^+ Æ_2 $$
 Note that the free Poincar«e generators are local in this form, from
cancellation of $»^+$'s.  In interacting theories, the generator $\J^{-i}$, as
well as the translation generator $\P^-$, which is also the Hamiltonian,
gets additional terms higher-order in the fields.  In this manner,
relativistic quantum field theory can be quantized in a way that more
resembles nonrelativistic field theory than in non-lightcone methods,
since $õ$ is quadratic in the usual time derivative $»_0$.  We won't
consider lightcone quantum field theory further; however, in the
following sections we'll use this construction to derive free gauge theory
and its covariant quantization, in a way that we'll generalize
straightforwardly to interactions.  Thus, the same construction directly
gives the formulation of free representations of the Poincar«e group, from
field strengths to transverse fields to covariant gauge fields.

Ü2. Algebra

From the definition of the graded determinant in terms of Gaussian
integrals (see subsection IIC3), we see that anticommuting coordinates
act like negative dimensions:  For example, $sdet(kI)=k^{a-b}$ for $a$
commuting and $b$ anticommuting dimensions.  Thus, if we add equal
numbers of commuting and anticommuting dimensions, they effectively
cancel.  Here we'll do the same for theories with spin, which allows the
restoration of manifest Lorentz covariance to lightcone theories:  Adding
2 commuting and 2 anticommuting dimensions to SO(D$-$2) gives
OSp(D$-$1,1|2) (see also subsection IIC3), which has an SO(D$-$1,1)
subgroup.

We have seen that quantum field theory requires unphysical
anticommuting fields to cancel the commuting unphysical fields
introduced by using gauges that do not eliminate longitudinal
polarizations.  For example, the gauge field for electromagnetism has
only D$-$2 components in the lightcone gauge, but needs to keep all D
components to maintain manifest Lorentz covariance; this requires 2
``ghosts" to cancel the 2 extra components of the gauge field.  The
general result, at least for bosonic gauge fields, is to produce fields that
form representations of OSp(D$-$1,1|2), including gauge fields and
ghosts.  Furthermore, by adding 2 anticommuting dimensions the BRST
transformations that relate the ghosts to the longitudinal degrees of
freedom can be introduced in a natural way, as ÓtranslationsÕ in the new
coordinates.  The result is that OSp(D$-$1,1|2) multiplets are automatic,
and gauge fixing receives a geometric interpretation.  In this section we'll
see that an even more natural interpretation of these BRST
transformations is as ÓrotationsÕ of the anticommuting coordinates, and
that they not only make gauge fixing to the simplest Lorentz covariant
gauge trivial, but also give a simple derivation of the gauge ÓinvariantÕ
action itself.  (The two points of view are related in that translations can
be considered as part of ``conformal rotations", as we saw in subsection
IA6.)

The basic idea is very simple:  Take the lightcone representation of the
Poincar«e generators, found in the previous subsection, and extend the
SO(D$-$2) indices and representations to OSp(D$-$1,1|2) ones (including
appropriate signs for the grading).  Conversely, we can begin the
construction with the ``conformal" group OSp(D+1,3|2), find the equations
of motion for the ``Poincar«e" group OSp(D,2|2), and solve them for the
``lightcone group" OSp(D$-$1,1|2).  So we can use the same expressions
for the generators, but now the ``transverse" OSp(D$-$1,1|2) index is
$$ i = (a,Œ) $$
 where $a$ is a D-component index of SO(D$-$1,1) and $Œ$ is a
2-component index of Sp(2).  The OSp(D$-$1,1|2) metric is
$$ ú^{ij} = (ú^{ab},C^{Œº}) $$
 Furthermore, we divide up the full OSp(D,2|2) index as 
$$ (à,i) = (A,a),ââA=(à,Œ) $$
 We now interpret the SO(D$-$1,1) subgroup that acts on the $a$ index as
the usual physical one, since the generators take the usual covariant
form (because all the transverse generators are linear).  The orthogonal
subgroup OSp(1,1|2) that acts on the $A$ index, and leaves the
$a$ index alone, is then interpreted as a symmetry group of the
unphysical degrees of freedom, an extension of BRST (and ``antiBRST").  
(Note that OSp(1,1|2) is the group of coordinate transformations in 2 anticommuting dimensions; for later reference, IGL(1) is the same for 1 such dimension).
However, the
generators with $-$ indices are nonlinear, since the $à$ indices are no
longer independent from the rest.  (The + were gauged away, the $-$
were fixed by equations of motion.)  As a result, they act on transverse
indices in a nontrivial way. 

$$ \matrix{ \hbox{longitudinal, nonlinear SO(1,1):} & \hphantom{Ó} à \cr
		\hbox{transverse, manifest OSp(D$-$1,1|2): } i = & \hskip-.5em
			\leftÓ \matrix{ Œ \cr a \cr} \right. \cr}\hskip-1.5em
	\lower3pt
	\hbox{$\matrix{ \left. \vphantom{\matrix{ à \cr Œ \cr}} \rightÕ = A:
			\hbox{ghost OSp(1,1|2)} \cr
		\phantom{ÕâââöA}: \hbox{Lorentz SO(D$-$1,1)} \cr}$} $$

Since the OSp(1,1|2) generators act only in the unphysical directions, all
physical states should be singlets (with respect to the cohomology)
under this symmetry.  This is clear
from the original construction:  We started with linear generators for
OSp(D,2|2) (and translations and dilatations for (D,2|2) dimensions), applied
the equations of motion in terms of them, and now we apply the
OSp(1,1|2) singlet condition last.  If we had instead applied the OSp(1,1|2)
singlet condition first to the (D,2|2) dimensional space, we would have
gotten the usual (D$-$1,1|0) dimensional space, and finally applying the
equations of motion would have given us the lightcone results of 
subsection IIB3.  

$$ \matrix{ \hbox{manifest symmetry:} \hfill\cr \cr
	\searrow \hbox{field equations (fix $à$)} \hfill\cr
	\swarrow \hbox{BRST singlets (fix $A$)} \hfill\cr
	\nearrow \hbox{add 2+2 (extend $i£(a,Œ)$)} \hfill\cr}ââ
	\vcenter{
	\halign{\strut\hfil#\hfil&&\hfil#\hfil\cr
	& OSp(D,2|2) & \cr
	\hfill $\nearrow$\llap{\hbox{$\swarrow$}} && $\searrow$ \hfill\cr
	SO(D$-$1,1) & $\Longrightarrow$ & OSp(D$-$1,1|2) \cr
	field strengths && gauge fields/BRST \cr
	\hfill $\searrow$ && $\nearrow$\llap{\hbox{$\swarrow$}} \hfill \cr
	& SO(D$-$2) & \cr
	& lightcone & \cr}} $$

Explicitly, the OSp(1,1|2) generators are (choosing $w=0$)
$$ J^{+Œ} = -x^Œ »^+,ââJ^{-+} = x^- »^+,ââ
	J^{Œº} = x^{(Œ}»^{º)} +S^{Œº} $$
$$ J^{Œ-} = -x^- »^Œ 
	+{1\over »^+}[üx^Œ(õ-m^2+»^º »_º) 
	+S^{Œb}»_b +S^{Œ-1}im +S^{Œº}»_º] $$
 while the SO(D$-$1,1) generators take their usual manifest form
$$ J^{ab} = x^{[a}»^{b]} +S^{ab} $$

\x XIIA2.1 Write the general commutation relations of OSp(D$-$1,1|2). 
Specialize to the case OSp(1,1|2), in lightcone notation.  Show that this
representation satisfies them, paying special attention to signs.  (Use the
OSp(D$-$1,1|2) commutators for the $S$'s.)

We can also add  a ``nonminimal" part to the general ``minimal" part of the
OSp(1,1|2) algebra we have already derived, in the sense that the two
parts commute and separately satisfy the commutation relations:
$$ J^{AB} £ J^{AB} +÷S^{AB} $$
 (This is similar to adding spin pieces to orbital in the absence of
constraints relating them.)
The simplest choices are to choose this new part to be just quadratic in
new, ``nonminimal" coordinates and momenta.  It will prove convenient to
perform some transformations $J£UJU^{-1}$ that make the OSp(1,1|2)
generators more similar to what they were before adding the spin parts. 
We therefore make two consecutive transformations:
$$ J £ U_2 U_1 J U_1^{-1}U_2^{-1}:ââ
	U_1 = e^{÷S^{+Œ}»_Œ/»^+},ââU_2 = (»_+)^{÷S^{-+}} $$
 to return $J^{-+}$ and $J^{+Œ}$ to their previous forms.  In fact, these
are just the OSp(1,1|2) version of the same transformations we used in
the previous subsection to remove $S^{+i}$ and $S^{+-}$.  The result is
$$ J^{+Œ} = -x^Œ »^+,ââ
	J^{-+} = x^- »^+,ââJ^{Œº} = x^{(Œ}»^{º)} +öS^{Œº} $$
$$ J^{Œ-} = -x^- »^Œ 
	+{1\over »^+}[x^Œ(-K+ü»^º »_º) +\Q^Œ +öS^{Œº}»_º] $$
 Here we have
$$ \boxeq{ K = -ü(õ-m^2),âöS^{Œº} = S^{Œº} +÷S^{Œº},â
	\Q^Œ = ÷S^{Œ-} +S^{Œb}»_b +S^{Œ-1}im -÷S^{+Œ}K } $$
 but more generally we can satisfy the commutation relations by
requiring only that $K$, $\Q_Œ$, and $öS_{Œº}$ are independent of the
unphysical coordinates $x^-$ and $x^Œ$ and their momenta, and satisfy
that their only nontrivial commutators are
$$ Ó\Q^Œ,\Q^ºÕ = 2KöS^{Œº},ââ[öS^{Œº},\Q^©] = \Q^{(Œ}C^{º)©},ââ
	[öS_{Œº},öS^{©¶}] = ¶_{(Œ}^{(©}öS_{º)}{}^{¶)} $$

$U_2$ is nonunitary, which makes $÷S^{àŒ}$ hermitian 
(rather than antihermitian) after the transformation, 
requiring a modification of the usual representation for $÷S$. 
The usual representation can also be used by introducing an 
$i$ into the transformation, which gives $÷S^{àŒ}$ a factor of 
$ài$ in $\Q^Œ$. However, this $i$ can be removed by the 
same method used in subsection IIB4 to remove $i$'s 
associated with the index $-$1, only now it is applied to 
both the $+$ and $-$ indices.

Ü3. Action

We saw in the previous subsection that physical states are singlets under
the OSp(1,1|2) BRST symmetry.  It was introduced in a trivial way, but
became nontrivial after solving the equations of motion; on the
other hand, applying the singlet condition first reproduced the usual
lightcone analysis.  Reversing the order of applying the two conditions
has the advantage of allowing the physical state condition to be
expressed as a single equation, which can be derived from an action.  

There are two ways of doing this:  One is to use this algebra to generalize
to gauge fields the first-quantized BRST as applied to field theory in
subsection VIA3.  Because of their quantum mechanical origin, the
gauge-invariant $ìQì$ actions directly give a form suitable for choosing
the Fermi-Feynman gauge, where the kinetic operator is simply $õ-m^2$. 
However, this is somewhat unusual for fermions, whose simplest field
equation is first-order.  (But it is useful for supersymmetry, where
bosons and fermions are treated symmetrically.)  As a result, this
approach gives actions for fermions with an infinite number of auxiliary
and ghost fields.  The most convenient way to discover the usual
finite-component gauge-invariant first-order actions hidden there is by
performing an appropriate unitary transformation, after which this action
(for bosons or fermions) appears as the sum of three terms: the usual
gauge-invariant action, a term giving the usual second-quantized BRST
transformations, and a nonderivative term that would be considered
nonminimal under second-quantized BRST.  This approach will be
described in detail in the following sections.

The other way is to define a $¶$ function in the generators of the group
OSp(1,1|2), and use it as the kinetic operator for the action:
$$ S = -Çdx¼dx^- d^2 x^Œ¼\f14 ì »^+ ¶(J_{AB}) ì $$
 where the integration is over all the coordinates appearing in the
OSp(1,1|2) generators.  (The $dx$ part is the usual $d^D x/(2¹)^{D/2}$.) 
$»^+$ comes from the usual relativistic inner product; it is also a
``measure" factor, which is a consequence of our using generators
satisfying the pseudohermiticity condition
$$ Jÿ_{AB}»^+ = -»^+ J_{AB} $$
 Equivalently, we could redefine $ì£(»^+)^{-1/2}ì$,
$J£(»^+)^{-1/2}J(»^+)^{1/2}$ (assuming $»^+±0$, as usual in lightcone
formalisms) to eliminate it and restore hermiticity.  (This would only
affect $J^{-+}£J^{-+}-ü$, $J^{Œ-}£J^{Œ-}+»^Œ/2»^+$, making hermitian
the terms $üÓx^-,»^+Õ$ and $\f14Óx^Œ,»^º »_ºÕ/»^+$.)  Because of the
$¶$ function, this action has the gauge invariance
$$ ¶ì = üJ^{BA}ñ_{AB} $$
 Thus, the field equations and gauge invariance reduce $ì$ to states
in the OSp(1,1|2) cohomology.

More explicitly, the $¶$ function can be written as
$$ \li{»^+ ¶(J_{AB}) &= »^+ ¶(J_{Œº}{}^2)¶(J^{-+})¶^2(J^{+Œ})¶^2(J^{Œ-})\cr
	&= ¶(x^-)¶^2(x^Œ)¶(öS_{Œº}{}^2)»^{+2}J^{-Œ2} \cr} $$
 where we have used
$$ J^{-+}¶(J^{-+}) = ¶(J^{-+})J^{-+} = 0âÜâ
	¶(J^{-+}) = {1\over »^+}¶(x^-) $$
 (There is freedom in ordering of the original $¶$ functions:  Reordering of
any two $¶$'s produces terms that are killed by the other $¶$'s.)  The
$¶(öS_{Œº}{}^2)$ can be interpreted as a Kronecker $¶_{s0}$ in the Sp(2)
``spin" $s$:
$$ -üöS^{Œº}öS_{Œº} = 4s(s+1) $$
 (remember $öS^{Œº}$ is antihermitian, and $iöS^{¢\¢}$ is always integer
while $s$ can be half-integer).  The rest of the explicit $¶$'s are Dirac
$¶$'s in the unphysical coordinates, which can therefore be trivially
integrated out, leaving:
$$ \boxeq{ S = Çdx¼L_{gi},ââL_{gi} = üÄ K_{gi} Ä,ââ
	K_{gi} = ü(-õ +m^2 +ü\Q^Œ\Q_Œ) } $$
 where $Ä$ is $ì$ evaluated at $x^Œ=x^-=s=0$.  Furthermore, the
remaining gauge invariance is
$$ \boxeq{ ¶Ä = ¶_{s0}ü\Q^Œ ñ_Œ } $$
 from $J^{Œ-}$, since $J^{Œº}$, $J^{+Œ}$, and $J^{-+}$ have been used to
gauge to $s=x^Œ=x^-=0$, respectively.

\x XIIA3.1  Show explicitly that this action is invariant under the
OSp(1,1|2) gauge transformations.  (Hint:  Use the same method as
exercise XA2.1.)

Ü4. Spinors

As we saw in subsection VIA3, the BRST algebra for the (Dirac) spinor
requires nonminimal terms.  For the general case of fermions we add
these terms in the general way described in subsection XIIA2, choosing
them in terms of a (second) set of OSp(1,1|2) $©$ matrices:
$$ ÷S^{AB} = -ü [÷©^{A},÷©^{B}Õ,ââÓ÷©^{A},÷©^{B}] = -ú^{AB} $$
 where $÷©^A=(-û,-µ;÷Å,i÷½)$ in the notation of subsection VIA3.  In
particular, we find
$$ öS^{Œº} = S^{Œº} -ü÷©^{(Œ}÷©^{º)} $$
$$ \Q^Œ = S^{Œb}»_b +S^{Œ-1}im +÷©^Œ[÷©^- +÷©^+ ü(õ-m^2)] $$

The next step for general massless fermions (and similarly for the
massive case) is to apply
$$ \Q^2 = üS^{Œa}S_Œ{}^b »_a »_b  
	-(÷©^- -÷©^+ K)÷©^Œ S_Œ{}^a »_a  -üK $$
 where we have used 
$$ ÷©^Œ ÷©_Œ = C^{Œº}ü[÷©_º,÷©_Œ] = 1 $$
 At this point we note that the gauge invariance generated by $\Q^Œ$, for
gauge parameter $ñ_Œ=÷©_Œ ñ$, includes a term $÷©^- ñ$ that allows us
to choose the gauge
$$ ÷©^+ Ä = 0 $$
 One way to think of this is to treat $֩^+$ as an anticommuting
coordinate and $֩^-$ as its derivative; another way is to treat them as
2$ð$2 matrices.  Alternatively, we can unitarily transform the action to
contain just the $÷©^-$ term of the operator:  From the discussion of 2D $©$ matrices of subsection
VIIB5 we find, including that part of the spinor metric,
$$ ֍֩^- = -\tat1000 $$
 then acts as a projection operator.  Either way, the net result is to
reduce the action to, now restoring the mass,
$$ \boxeq{ S_f = Çdx¼L_{gi,f},ââL_{gi,f} = üöÄ K_{gi,f}öÄ,ââ
	K_{gi,f} = ü÷©^Œ (S_Œ{}^a »_a +S_{Œ-1}im) } $$
 where $öÄ$ is $Ä$ with the $÷©^à$-dependence eliminated (the top
component in the above matrix representation).  Thus, $öÄ$ differs from
the bosonic case in that it not only depends on $x^a$ and is a
representation of $S^{ij}$, but is also a representation of $÷©^Œ$, which
appears in $öS^{Œº}$ to define $s=0$.

The only type of representation we have missed in this analysis is
self-dual antisymmetric tensors.  In terms of field strengths, these
satisfy
$$ F_{a_1...a_{D/2}} = 
	à\f1{(D/2)!}·_{a_1...a_{D/2}b_1...b_{D/2}}F^{b_1...b_{D/2}} $$
 which is consistent, with Lorentz metric, if $D/2$ is odd (as seen from
applying the $·$ tensor twice).  A similar condition holds for the
gauge field in the lightcone gauge (with a (D$-$2)-dimensional $·$
tensor).  Because of the $·$ tensor, this condition can't be described by
adding extra dimensions to the lightcone.  However, the direct product of
two spinors contains all antisymmetric tensors, and the rank $D/2$ one
can be picked out by an appropriate OSp invariant constraint.  The
self-dual part of this tensor comes from the direct product of chiral
spinors.

\x XIIA4.1  We now consider this construction in more detail:
 ªa  Derive the generalization of $©_{-1}$ to OSp $©$ matrices,
anticommuting with both the fermionic and bosonic $©$'s,
$Ó©_{-1},©_AÕ=0$.  We can use the usual product for the fermionic $©$'s,
but obviously the bosonic ones will need something different.  (Hint:  For
each pair of fermionic or bosonic $©$'s there is a Klein factor, as in
exercise IA2.3e; for the fermionic $©$'s the exponential is equal to the
usual product.)
 ªb  In twice-odd dimensions, consider the direct product of two spinors
by representing the OSp spin operators as a sum in terms of the two
different sets of OSp $©$ matrices acting on the two different spinor
indices.  Define the U(1) (O(2)) symmetry that mixes the two $©$ matrices
by taking linear complex combinations of the $©$ matrices to form
fermionic creation and annihilation operators, so the OSp-invariant U(1)
generator is $aÿ^A a_A$.  Show that the eigenvalues of this generator pick
out the different Lorentz representations.  These can be made irreducible
by including $©_{-1}$ projections.  Using explicit U(1) and (both) $©_{-1}$
projectors in the action, show that self-dual tensors can be described. 
(Note:  This description contains an infinite number of auxiliary fields.)

Ü5. Examples

The OSp(1,1|2) method is thus an efficient method for finding
gauge-invariant actions (though not so useful for gauge fixing).  We begin
with examples of massless bosons, for which the gauge-invariant kinetic
operator is
$$ K_{gi} = ü(-õ +ü\Q^Œ\Q_Œ),ââ\Q^Œ = S^{Œa}»_a $$
 The scalar is a trivial example; the simplest nontrivial example is the
massless vector:  In terms of the basis $|{}^iÔ$ for an OSp(D$-$1,1|2)
vector (D-vector plus 2 ghosts), normalized to
$$ Ò{}^i|{}^jÔ = ú^{ij}âÜâÒ{}^a|{}^bÔ = ú^{ab},ââÒ{}^Œ|{}^ºÔ = C^{Œº} $$
 we can write the OSp(D$-$1,1|2) generators as
$$ S_{(1)}^{ij}=|{}^{[i}ÔÒ{}^{j)}| $$
 The Sp(2)-singlet field is then (dropping the $|{}^ŒÔ$ term)
$$ Ä = |{}^aÔA_a(x) $$
 We then have
$$ \Q^Œ = (|{}^ŒÔÒ{}^a| -|{}^aÔÒ{}^Œ|)»_aâÜâ
	\Q^2 ­ ü\Q^Œ\Q_Œ = |{}^aÔÒ{}^b|»_a »_b -ü|{}^ŒÔÒ{}_Œ|õ $$
$$ ÜâL_{gi(1)} = \f18 (F_{ab})^2 $$
 and for the gauge invariance
$$ ¶Ä = ¶_{s0}ü\Q^Œ ñ_ŒâÜâñ_Œ = |{}_ŒÔÂ(x)âÜâ¶A_a = »_a  $$

A more complicated example is the graviton (massless spin 2):  We write
the field, a graded symmetric, traceless OSp(D$-$1,1|2) tensor, in terms
of the direct product of two vectors, with basis $|{}^iÔ|{}^jÔ$.  The spin
operators are thus
$$ S^{ij} = S_{(1)}^{ij}°I_{(1)} +I_{(1)}°S_{(1)}^{ij} $$
 where the first factor in each term acts on the first factor in
$|{}^iÔ|{}^jÔ$, etc.; $I_{(1)}$ is the spin-1 identity.  The $s=0$ part of the
field is then
$$ Ä = |{}^iÔ|{}^jÔh_{ji},ââh^i{}_i = h^a{}_a +h^Œ{}_Œ = 0âÜâ
	Ä = (|{}^aÔ|{}^bÔ +ü|{}^ŒÔ|{}_ŒÔú^{ab})h_{ab} $$
 where $h_{ab}$ includes its trace.  The rest is straightforward algebra;
we use identities such as:
$$ \Q^2 = \Q^2_{(1)}°I_{(1)} +I_{(1)}°\Q^2_{(1)} +\Q^Œ_{(1)}°\Q_{(1)Œ} $$
$$ \Q^Œ_{(1)}°\Q_{(1)Œ} = 
	(|{}^ŒÔ|{}_ŒÔÒ{}^a|Ò{}^b| +|{}^aÔ|{}^bÔÒ{}^Œ|Ò{}_Œ|)»_a »_b $$
$$ (Ò{}^Œ|Ò{}_Œ|)(|{}^ºÔ|{}_ºÔ) = -Ò{}^Œ|{}^ºÔÒ{}_Œ|{}_ºÔ = -2 $$
 where $\Q^2_{(1)}$ was evaluated above, and in the last identity we used
the fact that $|{}^ŒÔ$ is anticommuting.  The final result is then
$$ L_{gi(2)} = -\f14[h^{ab}õh_{ab} +2(»^b h_{ab})^2
	-h^a{}_a õ h^b{}_b +2h^a{}_a »^b »^c h_{bc}] $$
 in agreement with subsection IXB1.  The original OSp(1,1|2) gauge
invariance reduces to
$$ ñ_Œ = (|{}_ŒÔ|{}^aÔ +|{}^aÔ|{}_ŒÔ)Â_aâÜâ¶h_{ab} = »_{(a}Â_{b)} $$

\x XIIA5.1  Consider a (D$-$2)-rank antisymmetric tensor (i.e., totally
antisymmetric in D$-$2 indices in D dimensions; see exercises IIB2.1
and VIIIA7.2, and subsection XA3). 
  ªa  Show from a lightcone analysis that it is equivalent to a scalar. 
Derive the gauge-invariant action using OSp methods.  Find the gauge
transformations and field strength.
 ªb  Find a first-order form for the action, (auxiliary field)${}^2$ +
(auxiliary field) $ð$ (field strength).  Show that eliminating the gauge
field as a Lagrange multipler results in the action for a scalar.  Show that
switching between scalar and antisymmetric tensor is equivalent to
switching field equation and constraint for the field strength.
 ªc  Find the description for the massive case by dimensional reduction.

\x XIIA5.2  Consider a tensor totally symmetric in its vector indices.  In
the lightcone gauge, the irreducible tensor is traceless.  Show that, upon
covariantization, the field appearing in the gauge-invariant action
satisfies a double-tracelessness condition (or equivalently the fields
appearing there are the totally traceless tensor and another totally
tracelsss tensor with two less indices).

For massless fermions we saw
$$ K_{gi,f} = ü÷©^Œ S_Œ{}^a »_a $$
 The next step is to use the fact that arbitrary fermionic representations
are constructed by taking the $©$-traceless piece of the direct product of
a (Dirac) spinor with an irreducible bosonic representation.  (Just as an
irreducible bosonic representation of an orthogonal group is found by
taking the direct product of vectors, choosing an appropriate symmetry,
as described by the Young tableau, and requiring the trace in any two
vector indices to vanish; here we also require that using a $©$ matrix to
contract the spinor index with any vector index also vanishes.  Of course,
simpler methods can be used for SO(3,1), but we need methods that apply
to all dimensions, so they can be applied to orthosymplectic groups.)  We
then can write
$$ S^{ij} = ×S^{ij} -ü[©^i,©^jÕâÜâöS^{Œº} = ×S^{Œº} -aÿ^{(Œ}a^{º)} $$
 where $×S^{ij}$ is the part of the spin acting on just the vector indices,
and we have combined $©^Œ$ and $÷©^Œ$ into creation and annihilation
operators, as in subsection VIA3:
$$ a^Œ = \f1{å2}(©^Œ +i÷©^Œ),âaÿ^Œ = \f1{å2}(©^Œ -i÷©^Œ);ââ
	[a_Œ,aÿ^º] = -¶_Œ^º $$

For spin 1/2 $×S=0$, $¶_{s0}$ projects to the ground-state of the
oscillators, and we immediately find
$$ S_Œ{}^a = -©_Œ ©^aâÜâL_{gi(1/2)} = \f14 öÄ ©^a i»_a öÄ $$
 where we have used
$$ ÷©^Œ ©_Œ = iü(aÿ^Œ a_Œ -a^Œ aÿ_Œ) = i(aÿ^Œ a_Œ -1) = -i(N+1) $$
 and this ``$N$" counts the $aÿ^Œ$ excitation level.  (Note that the
Hilbert-space inner product between the spinors includes the usual factor
of $©_0$.)  A less trivial case is spin 3/2:  Now
$$ Ä = |{}^iÔÄ_i,ââS^{Œa} = -©^Œ ©^a +|{}^{[Œ}ÔÒ{}^{a]}| $$
 where $Ä_i$ has an explicit vector index, and an implicit spinor index. 
Then from $©$-tracelessness (for irreducibility) we have for the Sp(2)
singlets
$$ ©^i Ä_i = 0âÜâÄ_Œ = -©_Œ ©^a Ä_aâÜâ
	a_Œ Ä_a = 0,âÄ_Œ = -\f1{å2}aÿ_Œ ©^a Ä_a $$
After a little algebra, using identities such as
$$ \f16 ©^{[a}©^{b}©^{c]} = ©^a ©^b ©^c +ü(ú^{b(a}©^{c)} -ú^{ac}©^b) $$
 we find
$$ L_{gi(3/2)} = -\f1{12}öÄ_a ©^{[a}©^{b}©^{c]} i»_b öÄ_c $$
 From inspection, or from $¶Ä=¶_{s0}ü\Q^Œ ñ_Œ$, we find the gauge
invariance
$$ ¶öÄ_a = »_a  $$

\x XIIA5.3  Let's now examine some massive examples:
 ªa  Find the gauge-invariant actions for massive spin 2 and spin 3/2 by
dimensional reduction of the massless cases.
 ªb  Note that for the spin-2 case the part of the mass term quadratic in
$h$ is proportional to $-h_{[a}{}^a h_{b]}{}^b$.  More generally, we might
have expected $(h_{ab})^2+k(h_a{}^a)^2$ for arbitrary constant $k$, since
the first part gives mass to the physical (transverse, traceless) part of
$h$, while the second term affects only the unphysical pieces.  Find the
St¬uckelberg terms generated from this generalized mass term by the
linearized gauge invariance.  Looking at just the terms quadratic in the
St¬uckelberg vector, what is special about $k=-1$, and why do other
values of $k$ give ghosts?  (Hint:  Compare gauge-fixed
electromagnetism.)

\refs

£1 K. Bardak'ci and M.B. Halpern, ÓPhys. Rev.Õ É176 (1968) 1686:\\
	lightcone representations of Poincar«e algebra.
 £2 M. Fierz and W. Pauli, ÓProc. Roy. Soc.Õ ÉA173 (1939) 211;\\
	S.J. Chang, ÓPhys. Rev.Õ É161 (1967) 1308;\\
	L.P.S. Singh and C.R. Hagen, \PRD 9 (1974) 898;\\
	C. Fronsdal, \PRD 18 (1978) 3624;\\
	T. Curtright, \PL 85B (1979) 219;\\
	B. deWit and D.Z. Freedman, \PRD 21 (1980) 358;\\
	T. Curtright and P.G.O. Freund, \NP 172 (1980) 413;\\
	T. Curtright, \PL 165B (1985) 304:\\
	covariant actions for totally symmetric representations of all (4D)
	spin.
 £3 W. Siegel and B. Zwiebach, \NP 282 (1987) 125:\\
	lightcone representations of Poincar«e algebra, and covariant actions,
	for arbitrary spin and dimension.
 £4 G. Parisi and N. Sourlas, \PR 43 (1979) 744:\\
	cancellation of extra dimensions.
 £5 R. Delbourgo and P.D. Jarvis, ÓJ. Phys.Õ ÉA15 (1982) 611;\\
	J. Thierry-Mieg, \NP 261 (1985) 55;\\
	J.A. Henderson and P.D. Jarvis, ÓClass. and Quant. Grav.Õ É3
	(1986) L61:\\
	ghosts from extra dimensions for indices.
 £6 W. Siegel, \NP 284 (1987) 632:\\
	1st-quantized BRST for fermions.
 £7 N. Berkovits, \xxxlink{hep-th/9607070}, \PL 388B (1996) 743:\\
	1st-quantized BRST for self-dual tensors.
 £8 Siegel, Óloc. cit.Õ
 £9 Fierz and Pauli, Óloc. cit.Õ (ref. 2 above):\\
	mass term for spin 2.

\unrefs

Û9 B. IGL(1)

Although the OSp(1,1|2) method is the simplest way to derive general free
gauge-invariant actions, it does not yield a simple method for gauge
fixing, even though the Hilbert space contains exactly the right set of
ghosts.  We now describe a related method that is slightly less useful for
finding gauge-invariant actions (it includes redundant auxiliary fields),
but allows gauges to be fixed easily.

Ü1. Algebra

For this method we use a subset of the OSp(1,1|2) constraints, and show
they are sufficient.  A simple analog is SU(2):  To find SU(2) singlets, it's
sufficient to look for states that are killed by both $T_3$ and the raising
operator $T_1+iT_2$.  This approach gives a formalism that turns out to
be easier to generalize to interacting theories, as well as allowing a
simple gauge-fixing procedure.  We first divide up the Sp(2) indices as
$Œ=(¢,\¢)$ (not to be confused with $à$).  We then make a similarity
transformation that simplifies some of the generators (while making
others more complicated):
$$ J £ UJU^{-1}:ââU =  (»^+)^{iJ^{¢\¢}} $$
 which changes the Hilbert-space metric (and corresponding hermiticity
conditions) to
$$ ç = UÿU = (-1)^{iJ^{¢\¢}},ââÒï|þÔ = Çïÿçþ $$
 This simplifies (looking at the massless case without $÷S^{AB}$ for
simplicity)
$$ J^{+\¢} £ {1\over »^+} J^{+\¢} = -x^{\¢} $$
$$ J^{-+} £  J^{-+} -iJ^{¢\¢},ââJ^{¢\¢} £ J^{¢\¢} $$
$$ J^{¢-} £ J^{¢-}»^+ +iJ^{¢\¢}»^¢ = 
	(-x^-»^+)»^¢ +üx^¢ õ +S^{¢a}»_a +S^{¢¢}»_¢ $$
 (We use the same conventions for raising and lowering Sp(2) indices as
for SU(2) in subsection IIA4 and SL(2,C) in subsection IIA5.)  These four
generators form the subgroup GL(1|1) of OSp(1,1|2) (=SL(1|2)):  We can
write the generators as $J_I{}^J$, where $I=(+,¢)$.  In subsection XIIB4
we'll see that the singlets of this GL(1|1) are the same as those of
OSp(1,1|2).

On the other hand, because of the simplified form of these generators,
it's easy to see how to reduce the group even further:
Applying some of the constraints on wave functions/fields to the right,
$$ J^{+\¢} = -x^{\¢} = 0,ââJ^{-+} +iJ^{¢\¢} = x^- »^+ = 0 $$
$$ÜâJ^{¢\¢} = -ix^¢ »_¢ +i +S^{¢\¢},ââ
	J^{¢-} = üx^¢ õ +S^{¢a}»_a +S^{¢¢}»_¢ $$
 (Of course, this further reduction could have been performed even
without the transformation.)  We are now left with the group IGL(1), with
just $J^{¢\¢}$ and $J^{¢-}$ as generators.  ($J^{¢-}$ acts as translations
for the GL(1) generator $J^{¢\¢}$.)  Also, we have reduced the unphysical
coordinates to just $x^¢$.  We now simplify notation by relabeling
$$ c = x^¢,ââb = »_¢,ââS^3 = iS^{¢\¢},ââJ = iJ^{¢\¢} +1,ââ
	Q = J^{¢-} $$
$$ Üâ\boxeq{ J = cb +S^3,ââQ = ücõ +S^{¢a}»_a +S^{¢¢}b } $$
 $J$ and $Q$ are versions of the ghost-number and BRST operators
introduced in subsection VIA1.  The net result for obtaining these IGL(1)
generators from the original OSp(1,1|2) generators can also be stated as
$$ J = iJ^{¢\¢}|_{»^¢=0,»^+=1},ââQ = J^{¢-}|_{»^¢=0,»^+=1} $$
 where $»^¢=0$ can be regarded as a gauge condition for the constraint
$x^{\¢}=0$, and $»^+=1$ for $x^-»^+$ ($¾»/»(ln¼»^+)$) $=0$.  The IGL(1)
algebra is
$$ [J,Q] = Q,ââQ^2 = üÓQ,QÕ = 0 $$

\x XIIB1.1 Show that any IGL(1) subgroup of OSp(1,1|2) ($J = iJ^{¢\¢}$, $Q
= J^{¢-}$) satisfies these commutation relations.  Check that this final
representation of the IGL(1) algebra satisfies them.

\x XIIB1.2 Using the results of subsection XIIA2,
 ªa Give expressions for $Q$ and $J$ in terms of ($c$, $b$,)
$K$, $\Q^Œ$, and $öS^{Œº}$.
 ªb Derive $Ó\Q^Œ,\Q^ºÕ$, assuming only $Q^2=0$ and the 
previous results for $[öS,öS]$ and $[öS,\Q]$.

Ü2. Inner product

The new inner product can be derived by the same steps:  Starting with
the lightcone inner product of subsection XIIA1, we add extra dimensions
to get the OSp(D,2|2) inner product.  We next drop dependence on $x^\¢$,
which will be eliminated in the IGL(1) formalism.  Then we perform the
transformation with $(»^+)^{iJ^{¢\¢}}=(»^+)^{J-1}$ used to simplify the
BRST operator.  This acts on both fields in the inner product; applying
integration by parts turns one such factor into $(-»^+)^{-J}$.  The net
effect is that it cancels the $»^+$ in the Hilbert-space metric, which
allows us to drop the $x^-$ integration, and it introduces a factor of
$(-1)^J$.

Rather than defining a Hilbert-space inner product, which is sesquilinear,
it is slightly more convenient to define a symplectic inner product,
replacing the Hermitian conjugate of the wave function/state on the left
with the transpose, in analogy to an ordinary vector inner product.  The
inner product is then
$$ \boxeq{ Òï|þÔ = -i(-1)^ï Çdx¼dc¼ï^T (x,c) (-1)^J þ(x,c) } $$
 A Hilbert-space inner product can then be defined simply as $Òï*|þÔ$
(where $(ï*)^T=ïÿ$).  We have included a sign factor corresponding to
what would be obtained if the $dc$ integration were moved to the
symmetric position between the two wave functions:  By $(-1)^ï$ we
mean take $ï=0$ in the exponent if $ï$ is bosonic and 1 if $ï$ is
fermionic.  We can make this manifest by defining
$$ ï(x,c) = Òx,c|ïÔâÜâI = -iÇdx¼dc¼|x,cÔ(-1)^J Òx,c| $$
 which allows the inner product to be evaluated between $Òï|$ and $|þÔ$
by inserting this form for the ``identity" $I$.

\x XIIB2.1  Work out the inner product for the vector field in terms of all
of the components (both expanding over $c$ and separating physical
and ghost parts of the OSp(D$-$1,1|2) vector).

As a result, any commuting or anticommuting constant factor ``$a$" can
be moved out of the inner product from the left or right in the usual way:
$$ Òï|þaÔ = Òï|þÔa,ââÒaï|þÔ = aÒï|þÔ $$
 As a consequence of the anticommutativity of the integration measure,
we have
$$ (-1)^{Òï|þÔ} = (-1)^{ï+þ+1} $$
 meaning that the statistics of the inner product is the opposite of an
ordinary product; we can think of ``$Ê|Ê$" as a fermion.

Because of the change in metric from the lightcone, the IGL(1) generators
now satisfy
$$ J^T = 1-J,ââQ^T = Q $$
 where the constant comes from dropping the extra coordinates, and the
transpose ``$Ê{}^TÊ$" indicates integration by parts (the usual transpose in
the infinite-matrix representation of operators):
$$ Çï\O þ = Ç(-1)^{ï\O}(\O^T ï)þ $$
 Before our transformations the generators all satisfied $G^T=-G$; now
they are pseudoantisymmetric with respect to the metric $(-1)^J$, up to
the constant:
$$ Òï|\O þÔ = Ò÷{\O}ï|þÔâÜâ÷{\O} = (-1)^J \O^T (-1)^J $$
$$ ÷J = 1-J,ââ÷Q = -Q $$
 From $J^T=1-J$ also follows the symmetry property of the inner product:
$$ Òï|þÔ = (-1)^{(ï+1)(þ+1)}Òþ|ïÔ $$
 This can be interpreted as antisymmetry once the anticommutativity of
the ``$Ê|Ê$" (metric) is taken into account.  

The hermiticity conditions that follow from the change from the
lightcone are
$$ Jÿ = 1-J,ââQÿ = Q $$
 Before all generators were antihermitian; now they are
pseudoantihermitian, up to a constant:
$$ ö{\O} = (-1)^J \Oÿ (-1)^J $$
$$ ÜâöJ = 1-J,ââöQ = -Q $$
 The factor of $i$ in the inner product compensates for the funny
hermiticity of $(-1)^J$.  We then find the usual hermiticity condition for a
vector inner product,
$$ Òï|þÔ* = Òþÿ|ïÿÔ $$

Ü3. Action

As explained in subsection VIA1, we are interested in states in the
cohomology of the BRST operator $Q$, which means states satisfying 
$$ Qì = 0,ââ¶ì = iQñ $$
 In particular, the physical states are states in the cohomology of $Q$ at
ghost number $J=0$.  However, now $Qì=0$ is the wave equation (as in
subsection VIA3), and $Qñ$ contains the usual gauge transformations (as
in the OSp(1,1|2) action of the previous section). 

The free gauge-invariant action for an arbitrary field theory is then
$$ S_0 = -(-1)^ìÇdx¼dc¼üì^T ¶_{J1}Qì = -(-1)^ìÇdx¼dc¼üì^T Q¶_{J0}ì$$
 for a real column-vector field $ì$.  (Complex fields can be decomposed
into their real and imaginary parts.  For relating to quantum mechanics,
we will usually consider the column-vector in our OSp Hilbert-space
notation.)  This action gives $Qì=0$ as the equations of motion, and has
$¶ì=Qñ$ as a gauge invariance, so the solutions are the cohomology of
$Q$.  The projector $¶_{J0}$ is a Kronecker $¶$ restricting $ì$ to vanishing
ghost-number (and thus $Qì$ to ghost number 1, since $[J,Q]=Q$).  As we'll
see in section XIIC, this projector is redundant:  The states in the
cohomology with nonvanishing momentum automatically have vanishing
ghost-number, and the states with nonvanishing ghost-number are
needed for gauge fixing.  The projection is useful only for eliminating
states that are redundant for discussing gauge invariance; we'll drop it
for the remainder of this section.  The ``complete" free action is then
$$ \boxeq{ S_0 = -(-1)^ìÇdx¼dc¼üì^T (-1)^{J-1}Qì = üÒì|iQìÔ = S_0ÿ } $$
 where we have included the inner-product metric.  This is just the
translation of the free BRST operator from first- to second-quantized
form, as for the lightcone in subsection XIIA1.

We now consider some simple examples, to see how this method
reproduces the usual results.  The simplest example is the scalar:  As
shown in subsection VIA3, 
$$ Q = cü(õ-m^2)âÜâ-Çdc¼üì(-1)^{J-1}Qì = -üÄü(õ-m^2)Ä $$
 without restrictions to vanishing ghost number, unitary transformations,
gauge fixing, etc.  Thus the scalar is in no way a gauge field:  The kinetic
operator follows from simple kinematic considerations.

The fundamental example of a gauge theory is a vector:  It is the defining
representation of the Lorentz group, and of the extended Lorentz group
we used to define the BRST operator.  In the rest of this chapter we will
see from its action most of the general properties of gauge theories:
ghosts, gauge invariance, BRST transformations of the fields, the
gauge-invariant action, gauge fixing, backgrounds, mass, etc.  In addition
to the equations given for this case in subsection XIIA5, we will use
$$ Ò{}^¢|{}^\¢Ô = -Ò{}^\¢|{}^¢Ô = i $$
 The BRST and ghost-number operators are
$$ Q = ücõ +(|{}^¢ÔÒ{}^a|-|{}^aÔÒ{}^¢|)»_a +2|{}^¢ÔÒ{}^¢|b,ââ
	J = cb +i(|{}^¢ÔÒ{}^\¢| +|{}^\¢ÔÒ{}^¢|) $$
 The field is real; it can't be called hermitian, since it is a column vector,
but each component of that vector is hermitian.  (This is the same as
reality, but for anticommuting objects reality includes extra signs that
are defined to be exactly those coming from hermitian conjugation.) 
Thus,
$$ ì = ì* =
	(|{}^aÔA_a -i|{}^\¢ÔC +i|{}^¢Ô÷C) -ic(|{}^aÔ×A_a -|{}^\¢Ô×{÷C} -|{}^¢Ô×C) $$
$$ ì^T = ìÿ =
	 (A_aÒ{}^a| +iCÒ{}^\¢| -i÷CÒ{}^¢|) +(×A_aÒ{}^a| -×{÷C}Ò{}^\¢| -×CÒ{}^¢|)ic $$
 where we denote the ``antifields" (those at order $c$) by a 
``$×{\phantom m}$".  

The BRST transformations of the fields can be found by comparing terms
in $ì$ and $Qì$:  If we define a second-quantized BRST operator $öQ$ such
that $Qì=öQì$, but $öQ$ acts only on the fields while $Q$ (as usual) acts on
$|{}^iÔ$ and the coordinates ($x$ and $c$), then
$$ ì = |{}^iÔ(Ä_i -icÆ_i)âÜâ
	Qì = öQì = |{}^iÔ[(-1)^i öQÄ_i -ic(-1)^{i+1}öQÆ_i] $$
 In other words, we compare terms in $ì$ and $Qì$, and throw in a minus
sign for transformations of fermions.  (So, e.g., ``$öQA_a$" is the
coefficient of $|{}^aÔ$ in $Qì$.)  Dropping the ``$Êö{\phantom n}Ê$" on $Q$,
the result is
$$ QA_a = -»_a C,âQC = 0,âQ÷C = -2i(×{÷C}-ü»ÉA) $$
$$ Q×A_a = -i(üõA_a -»_a×{÷C}),âQ×C = üõ÷C +»É×A,âQ×{÷C} = -üõC $$

Note that although $Q$ is hermitian, it is antihermitian with respect to the
inner-product metric $(-1)^J$, as expected from our convention of using
antihermitian generators for spacetime symmetries.  (The same extra
sign for hermiticity vs.¼pseudohermiticity, also because of ghosts
introduced by relativistic quantum mechanics, occurs for the spatial Dirac
matrices $©^i$: see subsection XC2.)  As a result, our transformations
agree with those of subsection VIA4.  However, while the first-quantized
Abelian transformations also agree with those of subsections VIA1-3, the
second-quantized nonabelian transformations will have the extra $i$
demonstrated in subsection VIA4, following from the $i$ introduced in the
inner-product metric in the previous subsection.  (This minor yet annoying
factor will be further discussed in section XIIC when we relate first- and
second-quantized BRST.)

The Lagrangian then can be expanded as (after some integration by parts)
$$ L_0 = -Çdc¼üì^T(-1)^{J-1}Qì = 
	-iü(AÉQ×A -CQ×C -÷CQ×{÷C} +×AÉQA -×CQC -×{÷C}Q÷C) $$
$$ = \f18 (F_{ab})^2 +(×{÷C} -ü»ÉA)^2 -i÷CüõC +i×AÉ»C $$
 where we have used the transpose of the field on the left of $ìQì$.  To
find the gauge-invariant action, we can evaluate it by keeping just the
(anti)fields with vanishing ghost number ($A_a$ and $×{÷C}$), and then
eliminate the remaining antifields by their equations of motion:
$$ L £ \f18 (F_{ab})^2 +(×{÷C} -ü»ÉA)^2âÜâ L_{gi} = \f18 (F_{ab})^2 $$

\x XIIB3.1  Consider the example of the second-rank antisymmetric
tensor (see exercises IIB2.1, VIIIA7.2, and XIIA5.1, and subsection XA3):
 ªa  Construct the states by direct product of two vectors.  Decompose
into fields plus antifields, physical plus ghost:  In particular, note the Sp(2)
representation of each SO(D$-$1,1) representation.
 ªb  Find the BRST transformations for all the (anti)fields.  In particular,
note that the tensor transforms into vector ghosts (as expected from the
gauge invariance), which themselves transform into scalar ghosts
(``ghosts for ghosts").  
 ªc  Graph all the states for $s$ (of the Sp(2) $S^{Œº}$)
vs.¼$J$, and indicate there how BRST relates them.
 ªd  Find the gauge-invariant action from $ìQì$.
 ªe  Generalize to arbitrary-rank antisymmetric tensors.  Compare the
results of exercise XIIA5.1a.

Ü4. Solution

The identity of the cohomology and the physical states can be proven
most easily by making a unitary (``gauge") transformation to the
``lightcone gauge":
$$ (Q,J) £ U(Q,J)U^{-1}:ââU =  e^{(S^{+i}»_i+S^{+¢}b)/»^+} $$
 which simplifies $Q$ while leaving $J$ unchanged:
$$ Q £ ücõ -S^{¢-}»^+,ââJ £ cb +S^3 $$
 These are the usual lightcone indices $à$ of any D-vector, not to be
confused with the $à$ used earlier when reducing from D+2 bosonic
dimensions.  Except for the extension to include the $¢$ index, this is the
same transformation used in subsection IIB3 (and XIIA1).  

This makes the generators separable, allowing us to treat the two terms
in $Q$ and $J$ independently.  Specifically, if we integrate the action over
$c$,
$$ ì = Ä-icÆâÜâ-Çdc¼üì^T(-1)^{J-1}Qì = 
	-\f14 Ä^T(-1)^{S^3}õÄ -iÆ^T(-1)^{S^3-1}S^{¢-}»^+ Ä $$
 Then $Æ$ is just a Lagrange multiplier enforcing the algebraic constraint
$S^{¢-}Ä=0$ (ignoring $»^+$, which we always assume is invertible in the
lightcone approach), leaving just the Klein-Gordon term for the part of
$Ä$ that satisfies the constraint.  We also have the gauge invariance
$$ ñ = Â+c,â¶ì = iQñâÜâ
	¶Ä = -iS^{¢-}»^+ Â,â¶Æ = -S^{¢-}»^+  -üõ $$
 so we can shoose the gauge where $Ä$ is restricted (algebraically) to be
in the cohomology of $S^{¢-}$.

To solve for the cohomology of $S^{¢-}$ it is sufficient to consider the
reducible representations formed by direct products of vectors (for
bosons), or the direct products of these with a single Dirac spinor (for
fermions), since by definition the OSp(D$-$1,1|2) generators $S^{ij}$ don't
mix different irreducible OSp(D$-$1,1|2) representations.  We'll show that
this cohomology restricts any reducible OSp(D$-$1,1|2) representation to
the corresponding reducible SO(D$-$2) lightcone representation, and
therefore restricts any irreducible OSp(D$-$1,1|2) representation to the
irreducible SO(D$-$2) representation from which it was derived.  (Also, the
irreducible representations in arbitrary dimensions are most conveniently
found by such a construction, where reduction is performed by
symmetrization and antisymmetrization and subtracting traces of vector
indices, and in the fermionic case also subtracting gamma-matrix traces
and using Majorana/Weyl projection.)

For bosons, we first consider the representation from which all the rest
are constructed, the vector.  Writing the basis for the vector states as
$|{}^iÔ$, where $S^{ij}|{}^kÔ=|{}^{[i}Ôú^{j)k}$, we find
$$ S^{¢-} = 0âÜânot¼|{}^+Ô, |{}^\¢Ô $$
$$ ¶ = S^{¢-}âÜânot¼|{}^-Ô, |{}^¢Ô $$
 This leaves only the transverse lightcone states, as advertised.  For the
direct product of an arbitrary number of vectors, we find the same
result:  The unphysical directions are eliminated from each vector in the
product.

We might worry that extra states in the cohomology would arise from a
cancellation of two terms, resulting from the action of the ``$¢$" and
``$-$" parts of $S^{¢-}$.  Specifically, this could happen if we could
separate out the supertraceless part of (the graded symmetric part of) the
product of two OSp(1,1|2) vectors.  (For example, for an SO(n) vector we
can separate the traceless part of a symmetric tensor as
$T_{ij}-\f1n ¶_{ij}T_{kk}$.)  However, this is not possible, since for
OSp(1,1|2)
$$ str(¶_A^B) = 2 -2 = 0 $$
 Explicitly, we can look at the two likely candidates for extra states in the
cohomology, 
$$ |{}^+Ô|{}^-Ô à i|{}^¢Ô|{}^\¢Ô $$
 (and their transposes).  But using
$$ S^{¢-}|{}^+Ô = -|{}^¢Ô,ââS^{¢-}|{}^\¢Ô=-i|{}^-Ô $$
 for these states we find
$$ S^{¢-} (|{}^+Ô|{}^-Ô +i|{}^¢Ô|{}^\¢Ô) = -2|{}^¢Ô|{}^-Ô $$
$$ S^{¢-}i|{}^+Ô|{}^\¢Ô  = |{}^+Ô|{}^-Ô -i|{}^¢Ô|{}^\¢Ô $$
 so neither state is in the cohomology.  Note that we take $S^{¢-}$ to
anticommute with $|{}^ŒÔ$; the states in the Hilbert space are assigned
statistics.  (This is the simplest way to allow a direct relation between
wave functions and fields.)

\x XIIB4.1  Check this analysis for spin 3.

Note that in the Lagrangian $-\f14 Ä^TõÄ -iÆ^TS^{¢-}»^+ Ä$ the fields in
$Ä$ that are nonzero when acted upon by $S^{¢-}$ are auxiliary, killed by
the Lagrange multiplier $Æ$.  On the other hand, the fields that are
$S^{¢-}$ on something are pure gauge, and do not appear in the
$ÆS^{¢-}Ä$ term because $S^{¢-}$ is nilpotent, while they drop out of
the $ÄõÄ$ term because the fields multiplying them there are exactly the
auxiliary ones that were killed by varying $Æ$.  This follows from the fact
that a field that is pure gauge with respect to $S^{¢-}$ has a
nonvanishing inner product only with an auxiliary field, since 
$Ä_1=S^{¢-}ÂÜÄ_2 Ä_1=Ä_2 S^{¢-}Â$.  Equivalently, a field redefinition
$Æ£Æ+AõÄ$ can cancel any terms in $ÄõÄ$ where one $Ä$ is $S^{¢-}$ on
something.

For the example of the vector, we have explicitly for the transformation
to the lightcone
$$ ln¼U = {1\over »^+}\left[(|{}^+ÔÒ{}^i| -|{}^iÔÒ{}^+|)»_i
	+(|{}^+ÔÒ{}^¢| -|{}^¢ÔÒ{}^+|)b\right] $$
 under which the Lagrangian becomes
$$ L £ L' = -\f14 AÉõA -i÷CüõC +×{÷C}»^+ A^- -i×A{}^- »^+ C $$
 The lightcone gauge transformations are
$$ ¶A^+ = -»^+ Â,ââ¶×{÷C} = -üõ $$
$$ ¶÷C = -»^+ ½,ââ¶×A^- = -üõ½ $$
$$ ¶×C = üõÅ^{\¢},ââ¶×A^+ = üõÅ^+,ââ¶×A^i = üõÅ^i $$
 where ``$i$" here refers to the transverse (D$-$2) components.

Ü5. Spinors

In general we can add nonminimal terms $÷S^{AB}$ of subsection XIIA2: 
The easiest way is to add them as the last step, remembering that $Q$
comes from $J^{¢-}$ and $J$ from $J^3$; this yields
$$ Q £ ücõ +S^{¢a}»_a +S^{¢¢}b +÷S^{¢-},ââJ £ cb +S^3 +÷S{}^3 $$
 This result can also be seen from first-quantization of spin 1/2
(subsection VIA3).  Alternatively, if we add $÷S$ at the beginning as in
subsection XIIA2, performing the transformations given there, followed
by the transformation 
$$ U = (»^+)^{iJ^{¢\¢}} $$
 of subsection XIIB1, where $J^{¢\¢}$ itself now contains $÷S$ terms, we
again find
$$ J^{+\¢} £ {1\over »^+} J^{+\¢} = -x^{\¢},ââ
	J^{-+} £  J^{-+} -iJ^{¢\¢} $$
$$ J^{¢\¢} £ J^{¢\¢},ââJ^{¢-} £ J^{¢-}»^+ +iJ^{¢\¢}»^¢ $$
 Adding a final transformation
$$ U = e^{-÷S^{+¢}b} $$
 (which actually undoes part of an earlier one), we again obtain the above
result.

As in the previous subsection, we can also transform to the lightcone
gauge, to find
$$ Q £ ücõ -S^{¢-}»^+ +÷S^{¢-} $$
 When analyzing the BRST cohomology in the lightcone gauge, the effect of
this nonminimal term is to replace (again ignoring the factor of $-»^+$)
$$ S^{AB} £ öS^{AB} = S^{AB} +÷S^{AB} $$
 (although actually only the $S^{¢-}$ and $S^3$ parts are used here).

As in subsection XIIA4, when treating fermions we choose the Dirac spinor
representation of OSp(1,1|2) for $÷S^{AB}$.  Thus, for the case of spin 1/2,
where $S^{AB}$ also is a Dirac spinor representation, $öS^{AB}$ is
represented by the direct product of two OSp(1,1|2) spinors.  We now use
the harmonic oscillator interpretation of the ghost coordinates used in
subsection XIIA5 (extending it trivially to the fermionic ones), which can
be applied to arbitrary OSp groups:
$$ S^{AB} = -ü[©^{A},©^{B}Õ,ââ÷S^{AB} = -ü[÷©^{A},÷©^{B}Õ $$
$$ a^A = \f1{å2}(©^A +i÷©^A),ââaÿ^A = \f1{å2}(©^A -i÷©^A) $$
$$ ÜâÓa^A,aÿ^B] = -ú^{AB},ââöS^{AB} = -aÿ^{[A}a^{B)} $$
 By expanding about the oscillator vacuum, we find this representation of
OSp(1,1|2) consists of the direct sum of totally (graded) antisymmetrized
tensors: $|0Ô, |{}^AÔ=aÿ^A|0Ô, |{}^{[AB)}Ô=aÿ^A aÿ^B|0Ô,$... .  But we have
already treated this case for the bosons, the result being that only the
singlet (vacuum) survives.  Of course, the complete spin representation is
given by the direct product of the representation of the unphysical
variables ($©^A$, $÷©^A$) and the physical ones, namely the transverse
lightcone gamma matrices.  Thus, the states in the cohomology are
given by the direct product of all the lightcone states with the vacuum
of the unphysical variables.  To treat arbitrary fermions, generalization to
direct products of the Dirac spinor with arbitrary numbers of vectors
works the same way, since the spinor looks like the direct sum of parts of
direct products of vectors as far as $öS^{AB}$ is concerned.

\x XIIB5.1  Work out the explicit $Q$ and Hilbert space for spin 3/2.

Ü6. Masses

 As usual masses can be added by dimensional reduction:  Our complete
result for application to massless and massive, bosons and fermions is
then
$$ \boxeq{ Q = üc(õ-m^2) +S^{¢a}»_a +S^{¢-1}im +S^{¢¢}bâ(+÷S^{¢-}),ââ
	J = cb +S^3â(+÷S{}^3) } $$
 with extra $i$'s introduced implicitly by the procedure given in
subsection IIB4.

For example, for the vector we have (the ``St¬uckelberg formalism")
$$ Q = üc(õ-m^2) +(|{}^¢ÔÒ{}^a|-|{}^aÔÒ{}^¢|)»_a 
	+(|{}^¢ÔÒ{}^{-1}|+|{}^{-1}ÔÒ{}^¢|)m +2|{}^¢ÔÒ{}^¢|b $$
$$ J = cb +i(|{}^¢ÔÒ{}^\¢| +|{}^\¢ÔÒ{}^¢|) $$
 Compared to the massless case treated in subsections XIIB3-4, the
corresponding field now has the extra terms
$$ ì £ ì +|^{-1}ÔÄ -ic|^{-1}Ô×Ä $$
 giving the action
$$ L_0 = L_{gi} +[×{÷C} -ü(»ÉA +mÄ)]^2 -i÷Cü(õ-m^2)C +iC(»É×A +m×Ä) $$
$$ L_{gi} = \f18 (F_{ab})^2 +\f14 (mA +Ȁ)^2 $$

For the spinor
$$ Q = üc(õ-m^2) -©^¢(Ö»-i\f{m}{å2}) -(©^¢)^2 b -÷©^¢ ÷©^- $$
$$ J = cb -i©^¢©^\¢ -i÷©^¢ ÷©^{\¢} $$
 where in the notation of subsection VIA3, 
$$ ©^¢ = Å,ââ©^\¢ = i½,ââ÷©^¢ = ÷Å,ââ÷©^\¢ = i÷½,ââ÷©^- = -µ $$

\x XIIB6.1  Use this method to work out the action and gauge 
transformations for massive spin 2.

Ü7. Background fields

The coupling of external fields can be treated by suitable modification of
the BRST operator.  In terms of self-interacting field theories, this
corresponds to writing the field as the sum of quantum and background
fields, and keeping in the action only the terms quadratic in the quantum
fields, as discussed for semiclassical expansions in subsection VA2 and for
the background field method in subsection VIB8.

One interesting case is the coupling of an external vector gauge field. 
Clearly the spacetime derivatives in $Q$ must be modified by the minimal
coupling prescription $»£á=»+iA$, but dimensional analysis and Lorentz
covariance also allow the addition of a nonminimal term proportional to
$F^{ab}S_{ab}$ to $õ$.  With the appropriate coefficient, the general result
is (see subsection VIIIA3)
$$ Q_I = üc(õ-iF^{ab}S_{ba}) +S^{¢a}á_a +S^{¢¢}b $$
 where $õ$ is now the covariant $á^2$.  ($J$ is unchanged.)

In the case of spin 0, this modification is trivial.  For spin 1/2, we
substitute the graded generalization of the Dirac matrices,
$S^{ij}=-ü[©^i,©^jÕ$, as discussed in subsection XIIA5.  We then find
$Q_I^2=0$ fixes the above coefficient of the nonminimal term, the
same as from squaring $©^a á_a$.  This follows from the simple
factorization of $S^{ij}$ in $Q_I$:
$$ Q_I = -c(©^a á_a)^2 -©^¢(©^a á_a) -(©^¢)^2 b
	= -(©^a á_a +©^¢b)c(©^b á_b +©^¢b)  $$
 where we have neglected the $÷S^{¢-}$ term of subsection XIIB5, and
used
$$ [©^a,c] = Ó©^¢,cÕ = 0 $$

In the spin-1 case, we find the interesting result that $Q_I^2=0$ requires
not only the above coefficient for the nonminimal term (as expected from
supersymmetry), but also that the background terms satisfy the
field equation $á_b F^{ab}=0$.  On the other hand, for spins >1, $Q_I^2=0$
implies $F_{ab}=0$, so these spins can't couple minimally (at least in flat
spaces).  Similar remarks apply to coupling gravity (spin 2) to spins >2.

\x XIIB7.1 Check these statements for spin 1.  Compare the analogous
result for background fields in the field theoretic approach from exercise
VIB8.2 for Yang-Mills for both the gauge transformations of the
gauge-invariant action and the Ófield-theoreticÕ BRST transformations of
the gauge-fixed action.

\x XIIB7.2  Show that electromagnetism can't couple minimally to 
(massless) higher spins (i.e., they can't have charge):
 ªa Show this for the graviton by considering $Q_I^2=0$ for spin 2
(symmetric traceless OSp tensor) in an external vector field.
 ªb Do the same for spin 3/2.

Another interesting feature of the spin-1 case is that we can define a
``vacuum" state
$$ |0Ô = |{}^\¢Ô $$
 which is in the free BRST cohomology of $Q$ only at zero momentum (constant field),
where $Q$ simplifies to $S^{¢¢}b$ without background.  However, this
state has ghost number $J=-1$.  In fact, it corresponds to the global
part of the gauge invariance of the theory:  Gauge parameters satisfying
$Qñ=0$ have no effect in the free theory (where $¶ì=iQñ$), but can act in
the interacting theory:  They do not contribute an inhomogeneous term
to gauge transformations.  However, gauge parameters of the form
$ñ+Qþ$ have the same effect as $ñ$, up to trivial transformations
proportional to the field equations.  Thus, while the BRST cohomology at
$J=0$ gives the physical states, that at $J=-1$ gives the global
invariances associated with the gauge field.  

Now the physical states can be derived by operating on the vacuum with
appropriate vertex operators:  If we expand $Q_I$ about the free BRST
operator $Q$,
$$ Q_I = Q + V $$
$$ Q_I^2 = 0ââÜââÓQ,VÕ +V^2 = 0 $$
$$ ¶Q_I = i[Q_I,Â]ââÜââ¶V = i[Q+V,Â] $$
 where $Â$ is the gauge parameter for the ÓbackgroundÕ field.  
The usual operator cohomology,
relevant for asymptotic states, follows from linearization:
$$ V £ V_0âÜâÓQ,V_0Õ = 0,â¶V_0 = i[Q,Â] $$
in the weak-coupling limit,
where $V_0$ is the part of $V$ linear in the background fields.
The asymptotic states in the cohomology of $Q$ are then given by
$$ ì = V_0|0Ô,âñ = Â|0ÔââÜââQì = 0,â¶ì = iQñ $$
 We can check this explicitly, as
$$ V = iüc(ÓA_a,»^aÕ +iA^2 -F^{ab}S_{ba}) +iA_a S^{¢a}âÜâ
	V_0|0Ô = A_a|{}^aÔ +icü(»^a A_a)|0Ô $$
 The second term gives $×{÷C}=ü»ÉA$, in agreement with the free field equations.

Ü8. Strings

Another interesting example is strings.  Since first-quantization is
essential in string S-matrix calculations, it's natural to associate string
field theory with quantum mechanical BRST.  As usual, for massive fields
this formalism automatically includes the St¬uckelberg fields that would
have been found by dimensional reduction, as well as all the ghosts. 
However, the explicit expression for the BRST operator does not explicitly
correspond to that obtained by dimensional reduction:  Although the spin
operators $S^{¢a}$, $S^{¢¢}$, and $S^3$ are quadratic in oscillators,
$S^{¢-1}$ is cubic (because $öP^-$, and thus $X^-$, is quadratic in the
lightcone gauge).  Nevertheless, the representation on any particular
irreducible Poincar«e representation contained among all the string states
is the same as obtained by dimensional reduction, as follows from the
generality of our analysis.  

As for any Poincar«e representation, reducible or not, all we need is the
lightcone spin operators, given for the general case in subsection XIIA1. 
In subsection XIIA2 we saw that the OSp(1,1|2) generators followed
immediately from just a change in notation.  The IGL(1) generators were
then found in subsection XIIB1 by a unitary transformation and solving
half the constraints of GL(1|1); the net result was equivalent to applying
``gauge conditions" to the original OSp(1,1|2) generators:
$$ J = iJ^{¢\¢}|_{»^¢=0,»^+=1},ââQ = J^{¢-}|_{»^¢=0,»^+=1} $$

For the string, the gauge condition $»^+=1$ simply removes the last
vestige of $X^+$, whose oscillator modes were already eliminated by the
string lightcone gauge.  On the other hand, the condition $»^¢=0$ makes
$X^¢=X^¢_{(+)}+X^¢_{(-)}$ the sum of two conformally covariant
objects:  With the elimination of the linear $ $ term in the expansion of
$X^¢$, $X^¢_{(à)}$ are periodic in their arguments, and have the usual
mode expansion in terms of exponentials only (no linear term).

As a result, the decomposition of $Q$ as obtained from the lightcone
becomes trivial ($J$ was easy anyway, since it's quadratic):  Relabeling
the result for the string's lightcone Lorentz generators from subsection
XIB1,
$$ J^{¢-} = iÇ{d§\over 2¹Œ'}Ê(X^¢ ÀX{}^- -X^- ÀX{}^¢) $$
 (where now the ``$i$" comes from using the antihermitian form of the
Lorentz spin), separating $ÀX{}^¢$ into its $(à)$ pieces, using their
``chirality" $ÀX{}^¢_{(à)}=àX'^¢_{(à)}$ to convert the $ $ derivative into a
$§$ derivative, integrating by parts, and applying the definition
$$ öP_{(à)} = \f1{å{2Œ'}}(ÀXàX') $$
 to $X^-$, we obtain
$$ Q = i å{2\over Œ'}Ç{d§\over 2¹}Ý_à X^¢_{(à)}öP^-_{(à)} $$
 In this form we can easily apply the Virasoro constraint, as solved in the
lightcone,
$$ öP^2 = 0,ââöP^+ = ûå{2Œ'}p^+âÜâöP^- = i{1\over 2ûå{2Œ'}}(öP^i)^2 $$
 where we have applied $»^+=1$.  Finally, we relabel
$$ X^¢_{(à)} = C_{(à)} $$
 for purposes of identification with the usual BRST procedure in terms of
ghosts $C$ and antighosts $B$ (see subsections XIB2 and XIB7); comparison of
the (equal-time) commutation relations then gives the further
identification
$$ [öP{}^i_{(à)}(1),öP{}^j_{(à)}(2)Õ = ¦iú^{ij}2¹¶'(2-1) $$
$$ Üâ[öP{}^i_{(à)}(1),X{}^j_{(à)}(2)Õ = -iå{\f{Œ'}2}ú^{ij}2¹¶(2-1) $$
$$ ÓB_{(à)}(1),C_{(à)}(2)Õ = 2¹¶(2-1) $$
$$ ÜâB_{(à)} = -å{\f2{Œ'}}öP^{\¢}_{(à)} $$
 The final result is then
$$ Q = {1\over û}Ý_à Ç{d§\over 2¹Œ'}¼
	C_{(à)}(-üöP_{(à)}^2 àiC'_{(à)}B_{(à)}) $$
 in agreement with direct first-quantization of this string in the
conformal gauge, in terms of the constraints $öP_{(à)}^2$.  The ghosts can
be separated into zero- and nonzero-modes as
$$ C_{(à)} = üc +å{\f{Œ'}2}Y^¢_{(à)},ââ
	B_{(à)} = 2ûb ¦å{\f2{Œ'}}Y'^{\¢}_{(à)} $$
 (For the closed string there is also an extra zero-mode in $C$ and $B$,
enforcing the constraint that the ``+" contributions to $M^2$ equal the
``$-$".)

\x XIIB8.1  By separating zero-modes in the string's $Q$, and comparing
with its generic expression
$$ Q = üc(õ-M^2) +S^{¢a}»_a +iS^{¢-1}M +S^{¢¢}b $$
 show that
$$ S^{ij} = iÝ_à àÇ{d§\over 2¹}¼Y^i_{(à)}Y'^j_{(à)} $$
$$ S^{i-1}M = {1\over 2ûå{2Œ'}}Ê
	iÝ_à Ç{d§\over 2¹}¼Y^i_{(à)}(Y'^j_{(à)})^2 $$
$$ M^2 = {1\over 2ûŒ'}Ý_à Ç{d§\over 2¹}¼(Y'^i_{(à)})^2 $$

Of particular interest is the massless level:  As mentioned in subsection
XIA4, using the fact that the Hilbert space of the closed string is the
direct product of the Hilbert spaces of open strings gives a simple
analysis of the massless states of any closed string, since the massless
states of any open string are given by a vector (multiplet) plus perhaps
some scalars in the nonsupersymmetric case.  To find the complete
off-shell structure, including auxiliary fields and ghosts, we can either
take the direct product of the two lightcone representations and then add
2+2 dimensions, or first add the 2+2 dimensions and then take the direct
product of the two OSp(D$-$1,1|2) representations.  In the
supersymmetric case the latter is more convenient, since the procedure
of adding dimensions to superspace is not yet understood, but
quantization of the vector multiplet is (at least for N=1, and probably for
N=2, in D=4).  

For example, for the bosonic closed string we just multiply two OSp
vectors, producing a tensor $t_{ij}$, which we can decompose into its
symmetric traceless part $h_{ij}$ (graviton plus ghosts), antisymmetric
part $B_{ij}$ (axion plus ghosts), and trace $$ (physical scalar): 
$$ t^{ij}|{}_jÔ°|{}_iÔâ£âh_{ij} = t_{(ij]} -\f2{D-2}ú_{ij}t^k{}_k,ââ
	B_{ij} = t_{[ij)},â⍠= t^i{}_i $$
 (See subsection XIIA5.)  However, we know that string theory prefers to
treat fields in the string gauge for Weyl invariance, where the action
(kinetic term) is not diagonalized in the fields (until coordinate
invariance is fixed also).  This is understood from this direct product
structure:  The ``natural" string fields are
$$ string¼gauge:â\cases{ t_{(ab)} & graviton \cr 
	t_{[ab]} & axion \cr
	t^Œ{}_Œ & dilaton (T-duality invariant) \cr} $$
 since duality affects only the $X$ modes, while the diagonal fields
(representations of OSp(D$-$1,1|2)) are
$$ normal¼gauge:â\cases{ 
	h_{ab} = t_{(ab)} - \f2{D-2}ú_{ab}(t^c{}_c +t^©{}_©) & graviton \cr
	B_{ab} = t_{[ab]} & axion \cr
	 = t^a{}_a +t^Œ{}_Œ & physical scalar \cr} $$

\x XIIB8.2  Show that the above OSp analysis is consistent with the
diagonalizing field redefinitions of the low-energy string action found in
exercise XIA6.2.  Explain the result in terms of the redefinitions
$$ ì^2{1\over å{-g}} £ e^{2Ä},ââì^2 g^{mn} £ å{-g}g^{mn} $$ 

In the heterotic case, as mentioned in subsection XIA4, we take the
product of the real prepotential plus two chiral ghosts of super
Yang-Mills with the usual vector plus two scalar ghosts of bosonic
Yang-Mills:
$$ ( V ¢ Ä_Œ ) ° ( A_a ¢ C_Œ ) = H_a ¢ ( V_Œ ¢ Ä_{aŒ} ) ¢ Ä ¢ Ä_{(Œº)} $$
 The result is a vector prepotential $H_a$ describing the physical
supergravity and tensor multiplets (in a string gauge, $G=1$), a chiral
scalar compensator (``superdilaton") $Ä$ appropriate for ``old minimal"
supergravity, first-generation ghosts (Sp(2) doublets) $V_Œ$ and
$Ä_{aŒ}$, and second-generation ghosts (an Sp(2) triplet, for the tensor
multiplet) $Ä_{(Œº)}$.  If the vector is accompanied by scalars, the closed
string also has additional vector multiplets:
$$ ( V ¢ Ä_Œ ) ° \Ä_I = V_I ¢ Ä_{IŒ} $$

We saw in subsection XIIB7 that for Yang-Mills it's natural to think of the (constant) ghost as the ``vacuum" state.  In open string theory, this happens automatically in conformal field theory.  Thus, we again have a direct relation between the operator BRST cohomology and state BRST cohomology, creating states from the BRST-invariant vacuum with operators $V=CW$, where $W$ has weight 1 and is constructed from $X$ (and not $C$ or $B$).

We can also relate more general $V$'s and $W$'s by
$$ [Q,{\textstyle È}WÕ = 0âÜâ[Q,WÕ = »VâÜâ[Q,VÕ = 0 $$
which defines a $V$ given a $W$.  
Thus $ÈV$ is the only part of $V$ in the cohomology, the
rest being a BRST variation:  Moving $V$ to a different value of $z$ is a
gauge transformation.  Thus $V(z)$ has explicit $z$ dependence, while $Q$
does not:
$$ Q_I = Q +V(z) $$
 corresponding to a vertex local in $z$ (as found in subsection XIB7).
Conversely,
$$ [Q,VÕ = 0,âW = [{\textstyle È}B,VÕâÜâ
	[Q,WÕ = [ÓQ,{\textstyle È}BÕ,VÕ = [{\textstyle È}T,VÕ = -»V $$
as long as $V$ has $w=0$, and generally under some weaker restrictions.  Thus this choice of gauge transformation
$$ ñ ¾ WâÜâ¶V =[Q,ñÕ ¾ »V $$
allows translation of $V(z)$ to arbitrary $z$.  We can see the relation of these 2 operators from the external field approach:
$$ Q £ Q+VâÜâ{\textstyle È}T £ {\textstyle È}T +W $$
so $V$ is a background term for $Q$, while $W$ is a background for the ``gauge-fixed Hamiltonian" $ÈT$.

The gauge can be generalized:
$$ H_I = H +W(z) = ÓQ_I,ñÕ = ÓQ,ñÕ +ÓV,ñÕ $$
 where the gauge-fixing operator is
$$ ñ = Ç(B +f),ââñ^2 = 0 $$
 the first term in $ñ$ giving the usual $ü(p^2+M^2)$ in $H$, while $f$ is
left arbitrary (as long as $ñ^2=0$ is preserved).  The inverse relation for $W£V$ is unchanged as long as $V$ transforms the same with respect to the new $T$.

\subsectskip\bookmark0{9. Relation to OSp(1,1\noexpand|2)}
	\subsecty{9. Relation to OSp(1,1|2)}

For comparison to OSp(1,1|2), we perform a unitarity (gauge)
transformation on the IGL(1) action.  (The most general OSp(1,1|2)
expressions are given at the end of subsection XIIA2.)
We first define an almost-inverse of
$öS^{¢¢}$:  Since $öS^{¢¢}$ annihilates states with $s^3=s$, where we
define
$$ -üöS^{Œº}öS_{Œº} = 4s(s+1),ââöS^3|sÔ = 2s^3|sÔ $$
 (so that $s$ and $s^3$ take their usual integer or half-integer values),
we can define $öS^{¢¢-1}$ such that
$$ öS^{¢¢-1}öS^{¢¢} = 1 -¶_{s^3,s},ââ
	öS^{¢¢}öS^{¢¢-1} = 1 -¶_{s^3,-s} $$
$$ öS^{¢¢}öS^{¢¢-1}öS^{¢¢} = öS^{¢¢},ââ
	öS^{¢¢-1}öS^{¢¢}öS^{¢¢-1} = öS^{¢¢-1} $$
 We then apply the transformation
$$ Q £ Q_{diag} = UQU^{-1} $$
$$ Q = -cK +\Q^¢ +öS^{¢¢}b,ââ
	ln¼U = c[ÓöS^{¢¢-1},\Q^¢Õ -öS^{¢¢-1}öS^{¢¢}\Q^¢öS^{¢¢-1}] $$
 The exponent of $U$ is nilpotent from the $c$, so it generates only
a linear term, $U=1+ln¼U$.  Using the commutation relations from
subsection XIIA2
$$ [öS^{¢¢},\Q^¢] = 0,ââ(\Q^¢)^2 = KöS^{¢¢} $$
 we find
$$ Q_{diag} = c¶_{s^3,s}(-K +\Q^¢ öS^{¢¢-1} \Q^¢)¶_{s^3,-s}
	+(cb\Q^¢ ¶_{s^3,s} +bc¶_{s^3,-s}\Q^¢) +öS^{¢¢}b $$
 Now we apply the identity
$$ [öS^3,A] = 0âÜâ¶_{s^3,s}A¶_{s^3,-s} = A¶_{s0} $$
 since any matrix element between $Òs,s^3|...|s',s'^3Ô$ gives
$s=s^3=s'^3=-s' £ s=s'=0$, as well as the facts
$$ öS^{¢¢}\Q^\¢ ¶_{s0} = [öS^{¢¢},\Q^\¢]¶_{s0} = 2i\Q^¢ ¶_{s0}âÜâ
	öS^{¢¢-1}\Q^¢ ¶_{s0} = -üi\Q^\¢ ¶_{s0} $$
$$ Ó\Q^Œ,\Q^ºÕ¶_{s0} = 0 $$
 This yields the final result
$$ \boxeq{ Q_{diag} = -c(K +\f14\Q^Œ \Q_Œ)¶_{s0}
	+(cb\Q^¢ ¶_{s^3,s} +bc¶_{s^3,-s}\Q^¢) +öS^{¢¢}b } $$

\x XIIB9.1  Use the commutation relations of $öS^{Œº}$, as well as
$-üöS^{Œº}öS_{Œº}=4s(s+1)$, to derive
$$ öS^{\¢\¢}öS^{¢¢} = 4[s(s+1) -s^3(s^3 -1)] $$
 from which follows the explicit expression
$$ öS^{¢¢-1} = {1\over öS^{\¢\¢}öS^{¢¢}}öS^{\¢\¢} =
	{1-¶_{s^3,s}\over 4(s-s^3)(s+s^3+1)}öS^{\¢\¢} $$

Integrating over $c$ in the action $S_{diag}=-Çüì^T(-1)^{J-1}Q_{diag}ì$
as in subsections XIIB3-4, we find the Lagrangian
$$ L_{diag} = üÄ^T(K +\f14\Q^Œ \Q_Œ)¶_{s0}Ä 
	+iÆ^T(-1)^{öS^3-1}¶_{s^3,-s}\Q^¢ Ä -üÆ^T(-1)^{öS^3}öS^{¢¢}Æ $$
 (Note that in a product of the form $Æ^T$ a state of eigenvalue $s^3$
multiplies one of eigenvalue $-s^3$, since $(öS^3)ÿ=(öS^3)^T=-öS^3$.)  We now
see that in this action only the $s=0$ (physical) part of $Ä$ appears in the
$ÄÄ$ term, while only the $s^3=-s$ (``minimal") part (including physical)
appears in the $ÄÆ$ term.  The only part of $Æ$ that appears in the $ÄÆ$
term is the $s^3=s$ (minimal) part of $Æ$ (the ``antifields" to the
corresponding ones in $Ä$), while all, but only, the remaining
(``nonminimal") part of $Æ$ appears in the $ÆÆ$ term.  In particular, the
$ÄÄ$ term is recognized as the OSp(1,1|2) action of subsection XIIA3.  The
terms involving $Æ$ can be eliminated by $Æ$'s gauge invariance and field
equation, and $Æ$ contains no propagating degrees of freedom (fields
with $õ$ equations of motion), as can be seen by the methods used to
analyze the cohomology in subsection XIIB4.  However, the auxiliary
fields $Æ$, and the ghosts in (the $s±0$ part of)
$Ä$ to which they couple, are useful in gauge fixing, as we'll see in the
next section.

Again looking at the example of the vector:
$$ öS^{¢¢} = 2|{}^¢ÔÒ{}^¢|âÜâ
	öS^{¢¢-1} = \f14 öS^{\¢\¢} = ü|{}^\¢ÔÒ{}^\¢| $$
$$ \Q^¢ = (|{}^¢ÔÒ{}^a|-|{}^aÔÒ{}^¢|)»_aâÜâ
	ln¼U = -iüc(|{}^\¢ÔÒ{}^a| +|{}^aÔÒ{}^\¢|)»_a $$
 The result of the transformation is then (cf.¼subsection XIIB3)
$$ Q_{diag} = üc(õ -|{}^aÔÒ{}^b|»_a »_b)¶_{s0} +cb|{}^¢ÔÒ{}^a|»_a
	-bc|{}^aÔÒ{}^¢|»_a +2|{}^¢ÔÒ{}^¢|b $$
$$ L_{diag} = \f18 (F_{ab})^2 +i×AÉ»C +×{÷C}{}^2 $$
 The BRST transformations now simplify to
$$ QA_a = -»_a C,ââQC = 0,ââQ÷C = -2i×{÷C} $$
$$ Q×A_a = -iü»^b F_{ba},ââQ×C = »É×A,ââQ×{÷C} = 0 $$

\x XIIB9.2  Find the ghosts and simplified BRST transformations for
massless spin 2.

\refs

£1 M. Kato and K. Ogawa, \NP 212 (1983) 443;\\
	S. Hwang, \PRD 28 (1983) 2614;\\
	K. Fujikawa, \PRD 25 (1982) 2584:\\
	1st-quantized BRST (for strings).
 £2 W. Siegel, \PL 149B (1984) 157, É151B (1985) 391:\\
	gauge-invariant actions from 1st-quantized BRST.
 £3 E. Witten, \NP 268 (1986) 253;\\
	A. Neveu, H. Nicolai, and P.C. West, \PL 167B (1986) 307:\\
	BRST operator as kinetic operator.

\unrefs

Û3 C. GAUGE FIXING

Although the quantum mechanical BRST operator is clearly useful for
gauge fixing, its relation to the second-quantized BRST we applied in
chapter VI is not obvious, since the latter BRST operator does not include
the gauge-invariant action.  Here we relate the two, and extend the
former to interacting field theories.  In particular, we show how the
$ìQì$ action leads directly to the gauge-fixed kinetic term as simply as
it led to the gauge-invariant one, without applying any transformations.

Ü1. Antibracket

In the usual Hamiltonian formalism we work in a phase space $(q,p)$ on
which is defined a Poisson bracket, useful for studying symmetry
properties and equations of motion in the classical theory, and for
relating to the commutator of the quantum theory (see subsections
IA1-2).  We want to interpret the present case of interest as an analogous
phase space, for which the fields $Ä$ (in $ì=Ä-icÆ$) correspond to $q$
and the antifields $Æ$ to $p$.  This automatically follows from the
lightcone commutator of subsection XIIA1, by the same steps used to
derive the IGL(1) algebra and inner product
$$ Òï|þÔ = -i(-1)^ï Çdx¼dc¼ï^T (x,c) (-1)^{J(c)} þ(x,c) $$
 in subsections XIIB1-2:  We thus define this generalization of the
Poisson bracket (see subsection IA2) in terms of the inner product as
$$ (f[ì],g[ì]) = f \circ g,ââ\circ =
	ú^{IJ}\leftÒÊ{\onÁ¶\over ¶ì^I}
	\mathrel{\hbox{$\left|\vbox to 22pt{}\right.$}}
	{¶\over ¶ì^J}\rightÔ $$
 where we have expanded the column vector $ì$ over a basis in the usual
way (see subsections IB1,5),
$$ ì = |{}^IÔì_I,ââì^T = ì^I Ò{}_I| $$
$$ Ò{}^I|{}^JÔ = ú^{IJ} = (-1)^I ú^{JI},ââ(ú^{IJ})* = ú^{JI},ââ
	ú^{IK}ú_{JK} = ¶_I^J $$
 etc., and the indices $I,J$ (not to be confused with the ghost-number
operator $J(c)$ appearing in the definition of the inner product above) run
over all indices on the field, which determine the statistics of the
corresponding component as $(-1)^I$.  (Thus $ì$ is always bosonic, the
statistics of the fields coming always from expansion over $|{}^IÔ$.)  This
generalized commutator ``$(¼,¼)$" is called an ``antibracket" because of
the unusual statistics associated with it, following from the same unusual
statistics of the inner product.  (The ordering of indices on $ú^{IJ}$ in the
antibracket is the opposite of usual to take into account the extra sign
factor from the intervention of the anticommuting ``$Ê|Ê$" between the
two $ì$'s.)  Plugging in the definition of the inner product, we have more
explicitly
$$ \circ = iÇdx¼dc¼{\onÁ¶\over ¶ì^I(x,c)}
	ú^{JI}(-1)^{J(c)}{¶\over ¶ì^J(x,c)} $$
 Note that while the inner product is defined between two functions of
the coordinates, the antibracket is defined between two functionals of
$ì$, $f$ and $g$, which don't depend explicitly on $(x,c)$ (although we can
specialize to cases where they depend on other values of the coordinates
$(x',c')$).  Thus the orbital term $cb$ in $J(c)$ acts only on the argument
$c$ of $ì^J(x,c)$ (and not on $g$), while the spin term $S^3$ acts only on
the index $J$ of $ì^J(x,c)$.

We have used the fact that $¶/¶ì$ is antihermitian, as follows from the
fact that the graded commutator between it and $ì$ is always a
commutator, since they always have opposite statistics:
$$ {¶\over ¶ì^I (x,c)}ì^J (x',c') = 
	\left( {¶\over ¶ì^I (x,c)}ì^J (x',c') \right)^\dagger $$
$$ = \left[ {¶\over ¶ì^I (x,c)}, ì^J (x',c') \right] = ¶_I^J ¶(x'-x)¶(c'-c) $$
$$ Üâ\left( {¶\over ¶ì} \right)^\dagger = \left( {¶\over ¶ì} \right)^T = 
	- {¶\over ¶ì} $$
 Thus,
$$ \left( {¶\over ¶ì}fÿ \right)^\dagger = 
	\left[ {¶\over ¶ì}, fÿ \rightÕ^\dagger = 
	\left[ f, -{¶\over ¶ì} \rightÕ = f{\onÁ¶\over ¶ì} $$

Other properties of the bracket follow directly from those of the inner
product:
$$ (-1)^{(f,g)} = (-1)^{f+g+1} $$
$$ (f,ga) = (f,g)a,ââ(af,g) = a(f,g) $$
$$ (f,g) = -(-1)^{(f+1)(g+1)}(g,f) $$
$$ (f,gh) = (f,g)h + (-1)^{(f+1)g}g(f,h) $$
$$ (-1)^{(f+1)(h+1)}(f,(g,h)) + cyc. = 0 $$
$$ (f,g)ÿ = (gÿ,fÿ) $$
 (for some commuting or anticommuting constant $a$).  Thus, the bracket
has the exact opposite symmetry as the inner product (as is the case with
the usual brackets):  It would be symmetric in its two arguments if not for
the $Çdc$ that sits effectively between the two arguments. Most of the
properties follow from this fact, and that the signs obtained from pushing
things around are determined by moving things naively while treating the
``$Ê,Ê$" in the middle as anticommuting.  Furthermore, the existence of a
bracket with these properties allows the definition of a Lie derivative,
$$ \L_A B ­ (A,B) $$

\x XIIC1.1  Find all the usual properties of this derivative (statistics,
linearity, distributivity, hermiticity, algebra, etc.), and relate to the usual
Lie derivative.

We also have functional identities such as
$$ {¶\over ¶ì^I (x,c)}Òì|ïÔ = i¶_I^J (-1)^{J(c)} ï_I(x,c) $$
$$ (ì^I (x,c),f[ì]) = -i(-1)^{J(c)} ú^{IJ}{¶\over ¶ì^J (x,c)}f[ì] $$
 from which follow
$$ (ì^I(x,c),ì^J(x',c')) = -i(-1)^{J(c)}ú^{IJ}¶(c'-c)¶(x'-x)
	= -i(-1)^{S^3}ú^{IJ}(c+c')¶(x-x') $$
 as well as
$$ (ì(x,c),Òì|ïÔ) = ï(x,c),ââ(Òï|ìÔ,Òì|þÔ) = Òï|þÔ $$
 Here $ï$ and $þ$ are wave functions in the same space as $ì$, but need
not be taken as bosonic (or real):  We can even take them as functionals
of $ì$ when applying the chain rule, using the above expressions for the
terms where the $¶/¶ì$'s don't act on them.

Expressions quadratic in $ì$ will be used to perform the
second-quantized (or just classical field theoretic) version of linear
first-quantized transformations:
$$ \O_A ­ üÒì|AìÔâÜâ(\O_A,\O_B) = \O_{[A,BÕ},ââAì = (ì,\O_A) $$
$$ df = -Çdx¼dc¼(-1)^I dì^I {¶\over ¶ì^I}fâÜâ
	(\O_A,f) = Çdx¼dc¼(-1)^I (Aì^I){¶\over ¶ì^I}f $$
 where $A$ and $B$ must satisfy
$$ [(-1)^{J(c)}A]^T = (-1)^{J(c)}AâÜâA = -(-1)^{J(c)} A^T (-1)^{J(c)} $$
 to give nontrivial contributions when appearing symmetrically between
the two factors of $ì$.  Corresponding group elements come from
exponentiating bosonic first-quantized generators, yielding fermionic
second-quantized generators:
$$ ¶f = (\O_A,f) = \L_{\O_A}fâÜâf' = e^{\L_{\O_A}}f $$
$$ (-1)^A = 1âÜâ(-1)^{\O_A} = -1âÜâ(\O_A,f) = -(f,\O_A) $$
 Clearly, the latter relations must hold when replacing $\O_A$ with their
nonlinear second-quantized generalizations.  We then find (also for
bosonic $A$)
$$ Aì = (ì,\O_A) = -(\O_A,ì) $$
 where the minus sign is the usual from translating first-quantized to
second-quantized language (see subsection IC1).

Expanding $(ì,ì)$ in $c$ as
$$ ì^I = Ä^I -icú^{IJ}(-1)^{S^3}×Ä_J $$
 we find
$$ (×Ä_I(x),Ä^J(x')) = -(Ä^J(x),×Ä_I(x')) = ¶_I^J ¶(x-x') $$
 This allows us to reexpress the antibracket as
$$ \circ = -Çdx¼(-1)^I \left( {\onÁ¶ \over ¶×Ä_I}{¶\over ¶Ä^I}
	+{\onÁ¶ \over ¶Ä^I}{¶\over ¶×Ä_I} \right) $$

For example, the antibrackets of the component fields for the vector are
$$ ì = |{}^iÔì_i = |{}^iÔ[ú_{ji}A^j -ic(-1)^{S^3}×A_i]
	= (|{}^aÔA_a -i|{}^\¢ÔC +i|{}^¢Ô÷C) -ic(|{}^aÔ×A_a -|{}^\¢Ô×{÷C} -|{}^¢Ô×C) $$
$$ (×A_i(x),A^j(x')) = -(A^j(x),×A_i(x')) = ¶_i^j ¶(x-x') $$
$$ Üâ(×A_a,A_b) = ú_{ab}¶,â(×C,C) = ¶,â(×{÷C},÷C) = ¶ $$
 where now ``$i$" refers to the OSp(D$-$1,1|2) index (and we use
$C=A^¢$, $÷C=A^{\¢}$, $×C=×A_¢$, $×{÷C}=×A_{\¢}$).

Ü2. ZJBV

To prove gauge independence of the path integral, it's useful to draw an
analogy of relativistic quantum mechanical BRST to second-quantized
BRST.  (We'll see below that this is not just an analogy, but an
equivalence.)  Translating the BRST quantization of subsection VIA2 into
path integral language, the general Lagrangian path integral for BRST
quantization in quantum physics is
$$ \A = ÇDq¼e^{-iS'},ââS' = S+ÓQ,ñÕ $$
 where $q$ is all coordinates, including ghosts.  While $S$ and $ñ$ depend
only on $q$, the BRST operator is linear in the conjugate momenta $p$ --- 
It generates a coordinate transformation:
$$ [Q,q^mÕ = -i¶_Q q^mâÜâQ = (¶_Q q^m)p_m $$
 Here the index ``$m$" includes all dependence of $q$, including time.  (We
saw a more explicit expression of this result in subsection VIA2, assuming
the constraints $G_i$ are themselves linear in the physical momenta.)  We
also have
$$ [Q,S] = 0 $$
 where $S$ can include not only the gauge-invariant action, but arbitrary
additional gauge-invariant pieces.  Such pieces can be used to construct
states in the BRST cohomology from the vacuum.  (This is the
path-integral translation of the operator construction given in subsection
VIA1.)  It can also include pure BRST variations, $ÓQ,ñ_0Õ$.  Thus, to prove
gauge independence of $\A$, we need only prove the vanishing of its
variation under infinitesimal change of $ñ$,
$$ ÇDq¼e^{-iS}ÓQ,¶ñÕ = 0 $$
 But this is trivial, since $Q$ acts as a total derivative, and $[Q,S]=0$. 
More generally, we require only
$$ 0 = 1É\onÁQ -i[Q,S] = -i(¶_Q q^m)\onÁ»_m -(¶_Q q^m)»_m S $$
 where $»_m=»/»q^m$, and ``$1É\onÁQÊ$" means the derivatives in $Q$ act
backwards onto the 1 (and itself), as found by integration by parts.  In
cases we have considered (and almost always), $1É\onÁQ$ and $[Q,S]$
separately vanish.  More generally, since there is a $1/\h$ multiplying
$S$ implicitly, nonvanishing values would require a ``quantum correction"
to $S$.  We can also write this condition as
$$ e^{-iS}\onÁQ = 0 $$
 Furthermore, we can write
$$ 1É\onÁQ = -i{»^2 Q\over »p_m »q^m} $$

These manipulations can be applied to the field theory expressions $\J$
for group generators, as found in subsection XIIA1 for the lightcone:  The
BRST operator as found from these generators is
$$ S = üÒì|iQìÔ,âQ^2=0âÛâ(S,S) = 0 $$
 Then a unitary transformation can be implemented by exponentiating the
infinitesimal transformation
$$ ¶Q = [G,Q]âÛâ¶S = üÒì|i(¶Q)ìÔ = (\G,S),ââ\G = üÒì|GìÔ $$
$$ Q' = e^G Q e^{-G}âÛâS' = üÒì|iQ'ìÔ = e^{\L_\G}S $$

We want to implement gauge fixing by performing a unitary
transformation on $Q$ and then evaluating $S$ at the antifields $Æ=0$:
$$ \A = ÇDļe^{-iS'}|_{Æ=0},ââS' = e^{\L_ñ}SâÜâ
	S_{gf} = (e^{\L_ñ}S)|_{Æ=0} $$
 By similar manipulations to the BRST case, we see that gauge
independence means
$$ 0 = ÇDļe^{-iS}(S,¶ñ) = iÇDļ(e^{-iS},¶ñ) $$
 where we evaluate this expression at $Æ=0$, and we have again included
arbitrary gauge-invariant pieces in $S$.  We thus obtain gauge
independence from $(S,S)=0$, or more generally (again using integration
by parts)
$$ 0 = Çdx¼(-1)^I {¶^2 \over ¶×Ä_I ¶Ä^I}Êe^{-iS}
	= Çdx¼(-1)^I {¶^2 S\over ¶×Ä_I ¶Ä^I} +iü(S,S) $$
 This is the approach to BRST of Zinn-Justin, Batalin, and Vilkovisky (ZJBV).

\x XIIC2.1  Find the unitary transformation, in ZJBV language, that
transforms the ``untransformed" action for the massive vector (that
which gives the gauge-invariant action upon dropping antifields) into the
action that has only the vector field (and not the scalar) upon dropping
antifields.

Writing the BRST transformations in this second-quantized ZJBV notation
will allow us to gauge fix interacting theories (found by adding
interaction terms to the free $S$) in a gauge-independent way.  In
particular, it proves the equivalence of the manifestly unitary lightcone
gauge (which has no ghosts, only physical degres of freedom) to the
manifestly Lorentz covariant Fermi-Feynman gauges (where the kinetic
operator is simply $õ-m^2$).  Specifically, the $ìQì$ action is already
unitarily transformed to the Fermi-Feynman gauge:  Keeping just the $Ä$
terms, we have for a bosonic theory
$$ S_{FF} = S|_{Æ=0} = -ÇdxÊdc¼üì^T(-1)^{J-1}cü(õ-m^2)ì 
	= -Çdx¼üÄ^T(-1)^{S^3}ü(õ-m^2)Ä $$
 In other words, the Fermi-Feynman kinetic term is just the sum over all
fields (but not antifields) of a $õ-m^2$ term (using the
OSp(D$-$1,1|2)-invariant inner product:  the $(-1)^{S^3}$ is just a
sign, and can be absorbed by a field redefinition).  This can also be seen
from the result for the complete $ìQì$ action after dropping the antifield
terms:  For example, for a vector the gauge-fixed (free) action is simply
(see subsection XIIB3)
$$ L_{FF} = -\f14 A^a õA_a -i÷CüõC $$
 (The Fermi-Feynman gauge for fermions also gives a $õ-m^2$ kinetic
term, but with an infinite number of ghosts; this may be useful for
supersymmetry.)

\x XIIC2.2  Let's again consider arbitrary-rank antisymmetric tensors (see
exercises\\ XIIA5.1 and XIIB3.1):
 ªa  Find the Fermi-Feynman actions.
 ªb  Do the same for the massive case.  (Note:  There are more fields.)

On the other hand, we saw in subsection XIIB9 that a unitary
transformation, and evaluation at $Æ=0$, gave the gauge-invariant
OSp(1,1|2) action in terms of just the physical fields:
$$ ñ_0 = üÒì|c[ÓS^{¢¢-1},\Q^¢Õ 
	-S^{¢¢-1}S^{¢¢}\Q^¢S^{¢¢-1}]ìÔâÜâS' = S_{diag} $$
$$ S_{diag} = Çdx¼[üÄ^T(K +\f14\Q^Œ \Q_Œ)¶_{s0}Ä 
	+iÆ^T(-1)^{S^3-1}¶_{s^3,-s}\Q^¢ Ä -üÆ^T(-1)^{S^3}S^{¢¢}Æ] $$
$$ S_{gi} = S_{diag}|_{Æ=0} = Çdx¼üÄ^T(K +\f14\Q^Œ \Q_Œ)¶_{s0}Ä $$
 In the usual gauge-fixing approach, we would start with $S_{diag}$ and
do the inverse transformation to obtain the Fermi-Feynman gauge:
$$ ñ = -ñ_0âÜâS_{FF} = (e^{-\L_{ñ_0}}S_{diag})|_{Æ=0} $$
 The same $ñ$ can be used in the interacting case, since the effect on the
quadratic piece of the action will be the same.  Thus, to apply the usual
ZJBV procedure we can either start with $S_{diag}$ and apply some $ñ±0$
sufficient to fix the gauge, or we can start with $S$ (the one we found
from quantum mechanical BRST) and apply some equivalent $ñ$, or no
$ñ$ at all (for Fermi-Feynman gauge).

To compare with the lightcone gauge, we start with
$$ S_{lc} = Çdx¼[-\f14 Ä(-1)^{S^3}õÄ -iÆ(-1)^{S^3-1}S^{¢-}»^+ Ä] $$
 which was itself obtained by unitary transformation from $S$ (see
subsection XIIB4), and make a further unitary transformation, of the form
$ñ=Çdx¼üÄAõÄ$, that has the effect $Æ£Æ+AõÄ$ for $A$ such that all
terms in $ÄõÄ$ containing auxiliary fields, and thus also pure-gauge
fields, are canceled.  For this transformed $S_{lc}$ we then have
$$ S_{lc,diag}|_{Æ=0} = -Çdx¼\f14 Äõ¶(S^{AB})Ä $$
 where the projection operator $¶(S^{AB})$ picks out the singlets of
$S^{AB}$ ($A=(à,Œ)$), i.e., the transverse (physical) degrees of the light
cone.  A similar procedure can be applied in the interacting case.

For the example of the vector:
$$ ln¼U = -iüc(|{}^\¢ÔÒ{}^+| +|{}^+ÔÒ{}^\¢|)(»^+)^{-1}õ $$
$$ Q_{lc} = ücõ -S^{¢-}»^+â£âQ_{lc,diag} = ücõ|{}^iÔÒ{}_i| -S^{¢-}»^+ $$
$$ L_{lc} £ L_{lc,diag} = -\f14 A_iõA_i +×{÷C}»^+ A^- -i×A{}^- »^+ C $$
 which has just the transverse (lightcone) degrees of freedom when the
antifields are dropped:
$$ L_{lc,gf} = -\f14 A_iõA_i $$

Ü3. BRST

In subsection VIA2, we saw that BRST could be used to gauge fix by adding
a BRST variation to the gauge-invariant Lagrangian.  In that case, physical
states are those that are not only in the BRST cohomology, but also
satisfy the equations of motion.

On the other hand, for relativistic mechanics we saw that the equations
of motion are rather redundant, since $ $ is unphysical, and so
$p^2(+m^2)=0$ is already included as a constraint, and contained in the
quantum mechanical BRST operator.  In fact, we have seen how in the
most general case of a free field the correct spectrum is specified by just
the cohomology of the quantum mechanical BRST operator.

We therefore want to identify the quantum mechanical BRST cohomology
condition with the combination of the second-quantized BRST cohomology
condition and the wave equation.  This essentially has been accomplished
in subsection XIIC2 by decomposing $Q_{diag}$ with respect to $c$, as
we'll now see by some further analysis.

From subsection XIIB9 we have
$$ Q_{diag} = -c(K +\f14\Q^Œ \Q_Œ)¶_{s0}
	+(cb\Q^¢ ¶_{s^3,s} +bc¶_{s^3,-s}\Q^¢) +S^{¢¢}b $$
 where the first term gives $L_{gi}$ in terms of the physical part ($s=0$)
of $Ä$, the second term gives the minimal BRST transformations in terms
of the minimal (anti)fields, and the last term adds the nonminimal stuff
needed for fixing to general gauges.  In particular, we see that the BRST
transformation of the physical fields are
$$ ¶(¶_{s0}Ä) ¾ ¶_{s0}\Q^¢ Ä = ¶_{s0}\Q^¢(¶_{s,1/2}¶_{s^3,-1/2}Ä) $$
 (Note that this differs somewhat from the expression found from $Q$,
since the transformation to $Q_{diag}$ is effectively a redefinition of $Æ$,
adding to it a piece proportional to an operator on $Ä$.)  

Thus the only occurrences of the physical fields in $Q_{diag}$ are in the
term that gives the gauge-invariant action and the term that gives their
BRST transformation; the remaining terms introduce nonminimal fields, as
well as account for the BRST transformations of the ghosts.  But this is the
definition of BRST:  Take the classical action in terms of physical fields,
construct the BRST transformation from the gauge transformation that
leaves the classical action invariant, add terms to the BRST operator that
insure its nilpotency on these ghosts, and add nonminimal terms to allow
gauge fixing.  We have just seen that $Q_{diag}$ is exactly of this
structure, where gauge fixing gives the desired Fermi-Feynman gauge by
unitary transformation to $Q$ (which becomes a canonical transformation
in second-quantized language, using the antibracket).

All that is left to see is how the ZJBV combination of the gauge-invariant
action with the BRST operator is equivalent to ordinary BRST.  Expanding in
antifields, ZJBV gives the gauge fixed action as
$$ S = S_0 +Æ_m(QÄ_m) +Æ_{nm}Æ_{nm}¼Ü¼
	S_{gf} = S_0 +(¶ñ/¶Ä_m)(QÄ_m) +(¶ñ/¶Ä_{nm})(¶ñ/¶Ä_{nm}) $$
 where we have used the fact that the three terms in $Q_{diag}$ contain
only the physical, minimal (``$m$"; including physical), and nonminimal
(``$nm$") fields, respectively.  In the usual ZJBV and BRST formalisms,
derived from BRST without antifields, there is no $Æ^2$ term, since this
generates a BRST transformation
$$ iQÄ_{nm} = (S,Ä_{nm}) ¾ Æ_{nm} $$
 One instead introduces further nonminimal fields, the
``Nakanishi-Lautrup fields", such that
$$ QÄ_{nm} = Ä_{NL} $$
 and use an extended gauge-fixing function
$$ öñ = ñ +Ä_{nm}Ä_{NL} $$
 Then the gauge fixed action is
$$ S_{gf} = S_0 +ÓQ,öñÕ
	= S_0 +(¶ñ/¶Ä_m)(QÄ_m) +(¶ñ/¶Ä_{nm})Ä_{NL} +Ä_{NL}^2 $$
 After eliminating the NL fields by their (algebraic) equations of motion,
we obtain the same result as found from ZJBV.  (Of course, the NL fields
can also be introduced directly into the ZJBV formalism, but are
redundant for purposes of finding Fermi-Feynman gauges.)

Consider the special case of Yang-Mills:  Generalizing our results for free
Yang-Mills to the interacting case, making use of the BRST
transformations of subsection VIA4, we have
$$ L_{ZJBV} = \f18 (F_{ab})^2 +×{÷C}{}^2 +i×AÉ[á,C] -×CC^2 $$
 The basic antibrackets are
$$ (×A_a,A_b) = ú_{ab}¶,ââ(×C,C) = ¶,ââ(×{÷C},÷C) = ¶ $$
 From the general relations we saw earlier, and the definition of $S$ in
terms of $Q$ for the free case, we have
$$ iQì = (ì,S) = (S,ì) $$
 Since in the above we have pulled out factors to the left of the fields, as
$$ ì = |{}^iÔì_i = |{}^iÔ[ú_{ji}A^j -ic(-1)^{S^3}×A_i] $$
 we pull them out of the left of the antibracket, to obtain
$$ i(QÄ)^I = (S,Ä^I),ââi(Q×Ä)^I = (S,×Ä^I) $$
 where as before $(QÄ)^I$, etc., means to evaluate the corresponding
component of $Qì$ and introduce the corresponding signs for effectively
pulling those factors to the left.  We then find the previous results for the
BRST transformations of the fields (by construction), but also those of the
antifields:
$$ QA_a = -[á_a,C],ââQC = iC^2,ââQ÷C = -2i×{÷C} $$
$$ Q×A_a = -iü[á^b,F_{ba}] +iÓC,×A_aÕ,ââQ×C = [áÉ,×A] +i[C,×C],ââQ×{÷C} = 0 $$
 (Remember that the funny signs of the antibracket come from its
symmetry, plus treating the comma in ``$(¼,¼)$" as anticommuting.  Note
the generic terms $i[C,¼Õ$.)

\x XIIC3.1  Generalize the above results for the action and BRST
transformations with antifields when Yang-Mills is coupled to matter.

BRST was described in a different way in subsection VIA4:  Here we apply
quantum mechanical BRST, and find it equivalent to applying the ZJBV
form of BRST to second-quantization.  The ZJBV action consists of the
gauge-invariant action, plus the antifields times the BRST
transformations of the fields, plus (antifield)${}^2$ terms.  The difference
between the BRST transformations obtained by the general methods of
subsection VIA1 as applied to second-quantization, and those found in
this chapter by applying OSp methods to first-quantization of relativistic
systems, is that the Nakanishi-Lautrup field is treated as a field in the
former approach and as an antifield in the latter.  The two give
equivalent results:  The latter uses fewer fields, but is slightly more
restricted in choices of gauge; however, this restriction is avoided in
practice.  (More ``nonminimal" fields can be added to allow more general
gauge choices in either case.)  For example, the ZJBV action for the
former treatment of Yang-Mills can be obtained from that for the latter
by the replacement
$$ ×{÷C}{}^2 £ ×{÷C}B $$
 The gauge-fixed action is the canonically transformed action (with
respect to the antibracket) evaluated at vanishing antifields:
$$ S_{gf} = e^{\L_ñ}S_{ZJBV}| $$
 Consider Yang-Mills in the most common type of gauge, where some
function of $A$ is fixed.  From the usual BRST approach (see subsection
VIA4), or the ZJBV approach with $B$, we find
$$ ñ = trÇü÷C[f(A) +üŒB]âÜâ
	L_{gf} = L_{gi} -üB[f(A) +üŒB] -üi÷C{»f\over »A}É[á,C] $$
 while in the ZJBV approach without $B$ we have
$$ ñ = trÇü÷Cf(A)âÜâL_{gf} = L_{gi} +\f14 f(A)^2 -üi÷C{»f\over »A}É[á,C] $$
 which is equivalent to the previous for positive $Œ$ (after elimination of
$B$).

\refs

£1 Siegel and Zwiebach, Óloc. cit.Õ (XIIA).
 £2 Siegel, Óloc. cit.Õ
 £3 J. Zinn-Justin, in ÓTrends in elementary particle theoryÕ, eds. H. Rollnik
	and K. Dietz (Springer-Verlag, 1975) p. 2:\\
	introduced antifields, antibracket, etc.
 £4 I.A. Batalin and G.A. Vilkovisky, \PL 102B (1983) 27, É120B (1983)
	166; \PRD 28 (1983) 2567, ÉD30 (1984) 508; \NP 234 (1984) 106;
	ÓJ. Math. Phys.Õ É26 (1985) 172:\\
	generalized Zinn-Justin's approach.
 £5 W. Siegel and B. Zwiebach, \NP 299 (1988) 206:\\
	2nd-quantized ZJBV from 1st-quantization.

\unrefs

Û0 AfterMath

\def\righthead{\hfil} \def\lefthead{\hfil}

\def\subsect#1\par{\par\nobreak\noindent{\sectfont{#1}}}

ÜConversions to other common conventions

$$ ú_{ab} £ -ú_{ab},ââ©_a £ \f1{å2}©_a,ââ©_{-1} £ -\f1{å2}i©_5,ââS £ -S $$
$$ g^2 £ {g^2\over 8¹^2}¼
	\left[{}viaâü(m^2 -õ) £ m^2 -õ,â
	{d^D x\over (2¹)^{D/2}} £ d^D x{}\right]
	¼or¼nonabelianâ{g^2\over 16¹^2} $$

ÜNatural (Planck) units

$$ c = \h = k = û = 1â(G = ¹) $$
$$ {1\over e^2} = {8¹^2\over e_{ft}^2} = {2¹\over e_m^2} = {2¹\over Œ}
	= 861.0225762(29),ââH^{-1} = 1.43(7) ð 10^{61} $$
$$ 1¼kg = 2.59225(20) ð 10^7,â
	1¼m = 1.096651(82) ð 10^{35},â1¼s = 3.28768(24) ð 10^{43} $$
$$ 1¼K = 3.98216(30) ð 10^{-33},ââ1¼GeV = 4.62109(34) ð 10^{-20} $$

ÜIndices

\\
$a,b,c,...$ --- (flat) vector\\
$i,j,k,...$ --- transverse (D$-$1 or D$-$2) vector or internal\\
$m,n,p,...$ --- (curved) vector or large summation\\
$A,B,C,...$ --- (flat) super or conformal vector\\
$I,J,K,...$ --- internal\\
$M,N,P,...$ --- (curved) super\\
$\A,\B,\C,...$ --- conformal spinor\\
$Œ,º,©,...;µ,Ã,¹,...$ --- spinor (usually 2-valued) or fermionic\\
$î,û$ --- internal\\
$0$ --- time\\
$-1$ --- mass (dimensional reduction)\\
$à;t,Ðt$ --- lightcone (longitudinal; transverse)\\
$¢,\¢$ --- spinor or spacecone reference line

ÜIntegration

$$ Çdx ­ Ç{d^D x\over (2¹)^{D/2}},ââÇdp ­ Ç{d^D p\over (2¹)^{D/2}} $$
$$ ¶(x-x') ­ (2¹)^{D/2}¶^D (x-x'),ââ¶(p-p') ­ (2¹)^{D/2}¶^D (p-p') $$
$$ Òx|x'Ô = ¶(x-x'),¼Òp|p'Ô = ¶(p-p');âÒx|pÔ = e^{ipÉx},¼Òp|xÔ = e^{-ipÉx};â
	p_a = -i»_a $$
$$ on-shell:ââÒp||p'Ô = {¶(p-p')\over 2¹¶[ü(p^2+m^2)]} $$

\newpage

ÜGROUP THEORY (I,X): Covering groups

\vskip.1in

\noindent
SO(2) = U(1),âSO(1,1) = GL(1)\\
SO(3) = SU(2) = SU*(2) = USp(2),âSO(2,1) = SU(1,1) = SL(2) = Sp(2)\\
SO(4) = SU(2)$°$SU(2),âSO(3,1) = SL(2,C) = Sp(2,C),â
SO(2,2) = SL(2)$°$SL(2)\\ 
SO(5) = USp(4),âSO(4,1) = USp(2,2),âSO(3,2) = Sp(4)\\
SO(6) = SU(4),âSO(5,1) = SU*(4),âSO(4,2) = SU(2,2),âSO(3,3) = SL(4)

\vskip.1in

\noindent
SO*(2) = U(1),â
	SO*(4) = SU(2)$°$SL(2),âSO*(6) = SU(3,1),âSO*(8) = SO(6,2)
	
\vskip.2in

ÜSpinors

\vskip.01in
$$ \vbox{\offinterlineskip
\hrule
\halign{ &\vrule#&\strut¼\hfil#\hfil¼\cr
height2pt&\omit&\omit&\omit&\hskip2pt\vrule&\omit&&
	\omit&&\omit&&\omit&\cr
&&\omit& $D_-$ &\hskip2pt\vrule& 0 && 1 && 2 && 3 &\cr
& $D$ &\omit&&\hskip2pt\vrule& Euclidean && Lorentz && 
	conformal &&&\cr 
height2pt&\omit&\omit&\omit&\hskip2pt\vrule&\omit&&
	\omit&&\omit&&\omit&\cr
\noalign{\hrule}
height2pt&\omit&\omit&\omit&\hskip2pt\vrule&\omit&&
	\omit&&\omit&&\omit&\cr
\noalign{\hrule}
height2pt&\omit&\omit&\omit&\hskip2pt\vrule&\omit&&
	\omit&&\omit&&\omit&\cr
&&\omit&&\hskip2pt\vrule& $Æ_Œ$ $Æ_{Œ'}$ && $Æ_Œ$ $Æ_{ÀŒ}$ &&
	$Æ_Œ$ $Æ_{Œ'}$ && $Æ_Œ$ $Æ_{ÀŒ}$ &\cr
& 0 &\omit&&\hskip2pt\vrule& $ú^{Œº}$ $ú_{ÀŒ}{}^º$ $ú^{ÀŒº}$ &&
	$ú^{Œº}$ && $ú^{Œº}$ $¯_{ÀŒ}{}^º$ $¯^{ÀŒº}$ && $ú^{Œº}$ &\cr
&&\omit&&\hskip2pt\vrule& $§_{Œº'}$ && $§_{ŒÀº}$ && $§_{Œº'}$ &&
	$§_{ŒÀº}$ &\cr 
height2pt&\omit&\omit&\omit&\hskip2pt\vrule&\omit&&
	\omit&&\omit&&\omit&\cr
\noalign{\hrule}
height2pt&\omit&\omit&\omit&\hskip2pt\vrule&\omit&&
	\omit&&\omit&&\omit&\cr
&&\omit&&\hskip2pt\vrule& $Æ_Œ$ && $Æ_Œ$ && $Æ_Œ$ && $Æ_Œ$ &\cr
& 1 &\omit&&\hskip2pt\vrule& $ú^{Œº}$ $ú_{ÀŒ}{}^º$ $ú^{ÀŒº}$ && 
	$ú^{Œº}$ $ú_{ÀŒ}{}^º$ $ú^{ÀŒº}$ && $ú^{Œº}$ $¯_{ÀŒ}{}^º$ $¯^{ÀŒº}$
	&& $ú^{Œº}$ $¯_{ÀŒ}{}^º$ $¯^{ÀŒº}$ &\cr
&&\omit&&\hskip2pt\vrule& $§_{(Œº )}$ && $§_{(Œº )}$ && $§_{(Œº )}$ 
	&& $§_{(Œº )}$ &\cr 
height2pt&\omit&\omit&\omit&\hskip2pt\vrule&\omit&&
	\omit&&\omit&&\omit&\cr
\noalign{\hrule}
height2pt&\omit&\omit&\omit&\hskip2pt\vrule&\omit&&
	\omit&&\omit&&\omit&\cr
&&\omit&&\hskip2pt\vrule& $Æ_Œ$ $Æ^Œ$ && $Æ_Œ$ $Æ^Œ$ && $Æ_Œ$
	$Æ^Œ$ && $Æ_Œ$ $Æ^Œ$ &\cr
& 2 &\omit&&\hskip2pt\vrule& $ú^{ÀŒº}$ && $ú_{ÀŒ}{}^º$ && 
	$¯^{ÀŒº}$ && $¯_{ÀŒ}{}^º$ &\cr
&&\omit&&\hskip2pt\vrule& $§_{(Œº )}$ $§^{(Œº )}$ && $§_{(Œº )}$
	$§^{(Œº )}$ &&$§_{(Œº )}$ $§^{(Œº )}$ && $§_{(Œº )}$ $§^{(Œº )}$ &\cr 
height2pt&\omit&\omit&\omit&\hskip2pt\vrule&\omit&&
	\omit&&\omit&&\omit&\cr
\noalign{\hrule}
height2pt&\omit&\omit&\omit&\hskip2pt\vrule&\omit&&
	\omit&&\omit&&\omit&\cr
&&\omit&&\hskip2pt\vrule& $Æ_Œ$ && $Æ_Œ$ && $Æ_Œ$ && $Æ_Œ$ &\cr
& 3 &\omit&&\hskip2pt\vrule& $¯^{Œº}$ $¯_{ÀŒ}{}^º$ $ú^{ÀŒº}$ &&
	$¯^{Œº}$ $ú_{ÀŒ}{}^º$ $¯^{ÀŒº}$ && $¯^{Œº}$ $ú_{ÀŒ}{}^º$ $¯^{ÀŒº}$ 
	&& $¯^{Œº}$ $¯_{ÀŒ}{}^º$ $ú^{ÀŒº}$ &\cr
&&\omit&&\hskip2pt\vrule& $§_{(Œº )}$ && $§_{(Œº )}$ && 
	$§_{(Œº )}$ && $§_{(Œº )}$ &\cr 
height2pt&\omit&\omit&\omit&\hskip2pt\vrule&\omit&&
	\omit&&\omit&&\omit&\cr
\noalign{\hrule}
height2pt&\omit&\omit&\omit&\hskip2pt\vrule&\omit&&
	\omit&&\omit&&\omit&\cr
&&\omit&&\hskip2pt\vrule& $Æ_Œ$ $Æ_{Œ'}$ && $Æ_Œ$ $Æ_{ÀŒ}$ &&
	$Æ_Œ$ $Æ_{Œ'}$ && $Æ_Œ$ $Æ_{ÀŒ}$ &\cr
& 4 &\omit&&\hskip2pt\vrule& $¯^{Œº}$ $¯_{ÀŒ}{}^º$ $ú^{ÀŒº}$ &&
	$¯^{Œº}$ && $¯^{Œº}$ $ú_{ÀŒ}{}^º$ $¯^{ÀŒº}$ && $¯^{Œº}$ &\cr
&&\omit&&\hskip2pt\vrule& $§_{Œº'}$ && $§_{ŒÀº}$ && $§_{Œº'}$ &&
	$§_{ŒÀº}$ &\cr 
height2pt&\omit&\omit&\omit&\hskip2pt\vrule&\omit&&
	\omit&&\omit&&\omit&\cr
\noalign{\hrule}
height2pt&\omit&\omit&\omit&\hskip2pt\vrule&\omit&&
	\omit&&\omit&&\omit&\cr
&&\omit&&\hskip2pt\vrule& $Æ_Œ$ && $Æ_Œ$ && $Æ_Œ$ && $Æ_Œ$ &\cr
& 5 &\omit&&\hskip2pt\vrule& $¯^{Œº}$ $¯_{ÀŒ}{}^º$ $ú^{ÀŒº}$ &&
	$¯^{Œº}$ $¯_{ÀŒ}{}^º$ $ú^{ÀŒº}$ && $¯^{Œº}$ $ú_{ÀŒ}{}^º$ $¯^{ÀŒº}$ 
	&& $¯^{Œº}$ $ú_{ÀŒ}{}^º$ $¯^{ÀŒº}$ &\cr
&&\omit&&\hskip2pt\vrule& $§_{[Œº ]}$ && $§_{[Œº ]}$ && $§_{[Œº ]}$ 
	&& $§_{[Œº ]}$ &\cr 
height2pt&\omit&\omit&\omit&\hskip2pt\vrule&\omit&&
	\omit&&\omit&&\omit&\cr
\noalign{\hrule}
height2pt&\omit&\omit&\omit&\hskip2pt\vrule&\omit&&
	\omit&&\omit&&\omit&\cr
&&\omit&&\hskip2pt\vrule& $Æ_Œ$ $Æ^Œ$ && $Æ_Œ$ $Æ^Œ$ && $Æ_Œ$
	$Æ^Œ$ && $Æ_Œ$ $Æ^Œ$ &\cr
& 6 &\omit&&\hskip2pt\vrule& $ú^{ÀŒº}$ && $¯_{ÀŒ}{}^º$ && $¯^{ÀŒº}$ 
	&& $ú_{ÀŒ}{}^º$ &\cr
&&\omit&&\hskip2pt\vrule& $§_{[Œº ]}$ $§^{[Œº ]}$ && $§_{[Œº ]}$
	$§^{[Œº ]}$ &&$§_{[Œº ]}$ $§^{[Œº ]}$ && $§_{[Œº ]}$ $§^{[Œº ]}$ &\cr 
height2pt&\omit&\omit&\omit&\hskip2pt\vrule&\omit&&
	\omit&&\omit&&\omit&\cr
\noalign{\hrule}
height2pt&\omit&\omit&\omit&\hskip2pt\vrule&\omit&&
	\omit&&\omit&&\omit&\cr
&&\omit&&\hskip2pt\vrule& $Æ_Œ$ && $Æ_Œ$ && $Æ_Œ$ && $Æ_Œ$ &\cr
& 7 &\omit&&\hskip2pt\vrule& $ú^{Œº}$ $ú_{ÀŒ}{}^º$ $ú^{ÀŒº}$ && 
	$ú^{Œº}$ $¯_{ÀŒ}{}^º$ $¯^{ÀŒº}$ && $ú^{Œº}$ $¯_{ÀŒ}{}^º$ $¯^{ÀŒº}$ 
	&& $ú^{Œº}$ $ú_{ÀŒ}{}^º$ $ú^{ÀŒº}$ &\cr
&&\omit&&\hskip2pt\vrule& $§_{[Œº ]}$ && $§_{[Œº ]}$ && $§_{[Œº ]}$ 
	&& $§_{[Œº ]}$ &\cr
height2pt&\omit&\omit&\omit&\hskip2pt\vrule&\omit&&
	\omit&&\omit&&\omit&\cr}
\hrule} $$

\newpage

ÜLORENTZ (I,II)

$$ -m^2 = p^2 = p^a p^b ú_{ab} = -(p^0)^2 +(p^1)^2 +(p^2)^2 +(p^3)^2 
	= -2p^+ p^- +2p^t Ðp^t $$
$$ E = p^0 = -p_0;ââp^à = \f1{å2}(p^0àp^1),âp^t = \f1{å2}(p^2 -ip^3) $$
$$ -ds^2 = dx^2 = dx^a dx^b ú_{ab},âp^a ds = mÊdx^a,âp^a d  = dx^a $$

$$ p^2 +m^2 = S_a{}^b p_b +S_{a,-1}m +wp_a = S_{-1}{}^a p_a +wm = 0 $$
	
Ü2-spinor

$$ C_{Œº} = -C^{Œº} = C_{ÀŒÀº} = \tat0i{-i}0;â
	p^{ŒÀº} = {\textstyle\left({p^+\atop p^t}¼{Ðp^t\atop p^-}\right)} $$
$$ Æ_Œ = Æ^º C_{ºŒ},âÆ_{ÀŒ} = Æ^{Àº}C_{ÀºÀŒ};ââ
	Æ^2 = üÆ^Œ Æ_Œ = iÆ^¢ Æ^\¢ = -iÆ_¢ Æ_\¢ $$
$$ ÐÆ^{ÀŒ} = (Æ^Œ)ÿâÜâ(Æ_Œ)ÿ = -ÐÆ_{ÀŒ},â
	(Æ^2)ÿ = ÐÆ^2 = üÐÆ^{ÀŒ}ÐÆ_{ÀŒ} = iÐÆ^{\rdt ¢}ÐÆ^{\rdt\¢} $$
$$ A_{[Œº]} = A_{Œº}-A_{ºŒ} = C_{Œº}C^{©¶}A_{©¶},âA_{[Œº©]} = 0;ââ
	V^2 = -2¼det¼V = V^{ŒÀº}V_{ŒÀº} $$
$$ ú_{ŒÀŒ,ºÀº} = C_{Œº}C_{ÀŒÀº},â·_{ŒÀŒ,ºÀº,©À©,¶À¶} 
	= i(C_{Œº}C_{©¶}C_{ÀŒÀ¶}C_{ÀºÀ©} -C_{Œ¶}C_{º©}C_{ÀŒÀº}C_{À©À¶}) $$
$$ ÒÆ| = Æ^ŒÒ{}_Œ|,â|ÆÔ = |{}^ŒÔÆ_Œ;â[Æ| = Æ^{ÀŒ}[{}_{ÀŒ}|,â|Æ] = |{}^{ÀŒ}]Æ_{ÀŒ} $$
$$ V = |{}^ŒÔV_Œ{}^{Àº}[{}_{Àº}|,âV* = -|{}^{ÀŒ}]V^º{}_{ÀŒ}Ò{}_º|;â
	f = |{}^ŒÔf_Œ{}^ºÒ{}_º|,âf* = |{}^{ÀŒ}]f_{ÀŒ}{}^{Àº}[{}_{Àº}| $$
$$ ÒÆÔ = ҍÆÔ = Æ^Œ _Œ,â[ƍ] = Æ^{ÀŒ}_{ÀŒ};âÒƍÔÿ = [ƍ] $$
$$ ÒÆ|V|] = Æ^Œ V_Œ{}^{Àº} _{Àº},âÒÆ|f|Ô = Æ^Œ f_Œ{}^º _º;âVW* +WV* = (VÉW)I $$
$$ ·_{0123} = -·^{0123} = 1,ââ·(V,W,X,Y) = i¼tr(VW*XY*-Y*XW*V) $$

Ü4-spinor

$$ ï = \pmatrix{ Æ_Œ \cr Ѝ_{ÀŒ} \cr},ââ
	Ðï = ïÿç = (^Œ¼ÐÆ^{ÀŒ});ââ-©^a ©^b = üú^{ab} +S^{ab} $$
$$ Öá = \pmatrix{ 0 & á_Œ{}^{Àº} \cr á^º{}_{ÀŒ} & 0 \cr},â
	ç = å2©_0 = \pmatrix{ 0 & ÐC^{ÀŒÀº} \cr C^{Œº} & 0 \cr},â
	©_{-1} = \f1{å2}\pmatrix{ -i¶_Œ^º & 0 \cr 0 & i¶_{ÀŒ}^{Àº} \cr} $$
$$ ©^a ©_a = -2,ââ©^a Öa ©_a = Öa,ââ©^a ÖaÖb ©_a = aÉb,ââ
	©^a ÖaÖbÖc ©_a = ÖcÖbÖa $$
$$ tr (I) = 4,ââtr (ÖaÖb) = -2aÉb,ââtr (ÖaÖbÖcÖd) = aÉb¼cÉd +aÉd¼bÉc -aÉc¼bÉd $$
$$ ï = |{}^ŒÔÆ_Œ +|{}^{ÀŒ}]Ѝ_{ÀŒ},ââÐï = ïÿ = ^ŒÒ{}_Œ| +ÐÆ^{ÀŒ}[{}_{ÀŒ}| $$
$$ ©_{ŒÀº} = -|{}_ŒÔ[{}_{Àº}| -|{}_{Àº}]Ò{}_Œ|;ââ
	¸_+ = |{}^ŒÔÒ{}_Œ|,ââ¸_- = |{}^{ÀŒ}ÔÒ{}_{ÀŒ}| $$

\newpage

ÜACTIONS (III-VI,XII)
$$ L = -üÀq{}^2 g(q) +ÀqA(q) +U(q),ââS = Çdt¼L $$
$$ L_H = -Àqp +H(q,p),ââS_H = Çdt¼H -dq¼p $$
$$ \A = ÇDļe^{-iS}ï;ââWick:â¼\A = ÇDļe^{-S}ï,âS³0 $$

ÜMechanics

\\
{\bf OSp(1,1|2):}
$$ K = -ü(õ-m^2),ââ
	\Q^Œ = S^{Œa}»_a +S^{Œ-1}imâ[+÷©^Œ (÷©^- -÷©^+ K)] $$
$$ öS^{Œº} = S^{Œº}â(-ü÷©^{(Œ}÷©^{º)}),ââ
	-üöS^{Œº}öS_{Œº} = 4s(s+1) $$
$$ S = Çdx¼L_{gi},ââL_{gi} = üÄ^T K_{gi}Ä,ââ
	K_{gi} = ü(-õ +m^2 +ü\Q^Œ \Q_Œ)ââ(öS^{Œº}Ä = 0) $$
$$ ¶Ä = ¶_{s0}ü\Q^Œ ñ_Œ $$
$$ S_f = Çdx¼L_{gi,f},ââL_{gi,f} = üöÄ K_{gi,f}öÄ,ââ
	K_{gi,f} = ü÷©^Œ (S_Œ{}^a »_a +S_{Œ-1}im) $$
	
\noindent{\bf IGL(1):}
$$ Q = üc(õ-m^2) +S^{¢a}»_a +S^{¢-1}im +S^{¢¢}bâ(+÷S^{¢-}),ââ
	J = cb +S^3â(+÷S{}^3) $$
$$ S = -(-1)^ìÇdx¼dc¼üì^T(-1)^{J-1}Qì = üÒì|iQìÔ,ââS_{gi} = S|_{Jì=0} $$
$$ S_{FF} = S|_{bì=0} = -Çdx¼üÄ^T (-1)^{S^3}ü(õ-m^2)Ä $$

ÜQuantum ChromoDynamics

$$ G_iÿ = G_i,ââ[G_i,G_j] = -if_{ij}{}^k G_k,ââ
	(G_i Æ)_A = (G_i)_A{}^B Æ_B $$
$$ á_a = »_a +iA_a = »_a +iA_a{}^i G_i,ââ-i[á_a,á_b] = F_{ab}
	= F_{ab}{}^i G_i = »_{[a}A_{b]} +i[A_a,A_b] $$
$$ L = \f1{8g^2}tr¼F^{ab}F_{ab} +L(á,Æ),ââ
	tr_D(G_i G_j) = ¶_{ij},âtr_A(G_i G_j) = 2N¶_{ij} $$
$$ QA_a = -[á_a,C],¼QC = iC^2,¼Q÷C = -iB,¼QB = 0,¼QÄ = iCļ(for¼¶Ä = iÂÄ) $$
$$ S_{gf} = S_{gi} -iQñ,¼ñ = trÇü÷C(f+üŒB)¼Ü¼
	L_{gf} = L_{gi} -üB(f+üŒB) +üi÷C(¶f)|_{Â=C} $$
$$ L_{Majorana} = Æ^Œ iá_Œ{}^{ÀŒ}ÐÆ_{ÀŒ} 
	+\f{m}{2å2}(Æ^Œ Æ_Œ +ÐÆ^{ÀŒ}ÐÆ_{ÀŒ})â£â
	-\f14 Æ^Œ(õ-m^2)Æ_Œ -üÆ^Œ f_Œ{}^º Æ_º $$
$$ L_{Dirac} = Ñï(iÖá+\f{m}{å2})ï 
	= (ÐÆ^{ÀŒ}iá^Œ{}_{ÀŒ}Æ_Œ +Ѝ^{ÀŒ}iá^Œ{}_{ÀŒ}_Œ)
	+\f{m}{å2}(Æ^Œ _Œ +ÐÆ^{ÀŒ}Ѝ_{ÀŒ}) $$

\newpage

ÜFEYNMAN (V)

$$ S = S_0 +S_I,ââS_0 = ÇüÄKÄ;ââ
	\A_N = Þ_{i=1}^N \left( ÇÆ_{Ni}{¶\over ¶Ä}\right) Z[Ä] $$
$$ Z[\Ä] = e^{-W[\Ä]}  = ÇDļe^{-(S_0[Ä] +S_I[Ä+\Ä])} =
	exp \left(Çü{¶\over ¶\Ä}{1\over K}{¶\over ¶\Ä}\right)e^{-S_I[\Ä]} $$

\noindent{\bf Effective action $ý[Ä]$ (unrenormalized):}
	(E.g., $L=-\f14 Ä(õ-m^2)Ä+\f16 gÄ^3$.)\\
 (A1) 1PI graphs only (plus $S_0$).  (For $W[Ä]$, connected graphs only.)\\
 (A2) Momenta: label consistently with conservation, with $Çdp$ for each
	loop.\\
 (A3) Propagators: $1/K$ for each internal line.  (E.g., $1/ü(p^2+m^2)$.)\\
 (A4) Vertices: read off of $-S_I$.  (E.g., $-g$)\\
 (A5) External lines: attach the appropriate (off-shell) fields and $Çdp$,
	 with $¶(Ýp)$.\\
 (A6) Statistics: 1/n! for n-fold symmetry of internal/external lines\\ 
	\phantom{(7) }(or keep just 1 of n! related graphs);
	$-$1 for fermionic loop; overall $-$1.

\noindent{\bf Vacuum:} (Renormalize before for minimal 
	subtraction/after for MOM.)\\
 (B1) Find the minimum of the effective potential (for scalars).\\
 (B2) Shift (scalar) fields to perturb about minimum; drop constant in
	potential.\\
 (B3) Find resulting masses; find wave function normalizations.

\noindent{\bf T-matrix:}\\
 (C1) Connected trees of (shifted, renormalized) $ý$: (A2-4) for L=0
	with $S£ý$.\\
 (C2) Amputate external $ý_0$-propagators.\\
 (C3) External lines: appropriate to $ý_0$ wave equation $÷KÆ=0$.  (E.g.,
	1.)\\
 (C4) External-line statistics: No symmetry factors; $-$1 for fermion
	permutation.

\smallskip
ÜProbabilities

$$ \S_{connected} = i¶\left(Ýp\right)T $$
$$ dP = |T_{fi}|^2 ¶^D \left(Ýp\right) Þ_{all}{(2¹)^{D/2}\over ¿}
	Þ_{out}{d^{D-1}p\over (2¹)^{D-1}},¼
	P = 2(Im¼T_{ii}) (2¹)^{-D/2} Þ_{in}{(2¹)^{D/2}\over ¿} $$
$$ {dP\over dt} = {2¼Im¼T_{ii}\over ¿},â
	{dP\over ds} = {2¼Im¼T_{ii}\over m} = -2¼Im¼M,â
	{dP\over d } = 2¼Im¼T_{ii} = - Im¼M^2 $$
$$ d§ = {dP\over v_{12}} = |T_{fi}|^2 ¶^D \left(Ýp\right) 
	{(2¹)^D\over Â_{12}} Þ_{out}{d^{D-1}p\over (2¹)^{D/2-1}¿},ââ
	§ =2(Im¼T_{ii}) {(2¹)^{D/2}\over Â_{12}} $$
$$ Â_{12}^2 = (p_1Ép_2)^2 -m_1^2 m_2^2 =
	\f14 [s-(m_1+m_2)^2][s-(m_1-m_2)^2] $$
$$ {d§\over dt} = ü(2¹)^3|T_{fi}|^2{1\over Â_{12}^2}â(4D);â
	s = -(p_1+p_2)^2,¼t = -(p_1+p_3)^2,¼u = -(p_1+p_4)^2 $$
$$ {d§\over d¯} = (2¹)^2|T_{fi}|^2
	{|\vec p_3|^{D-1} \over Â_{12}[ü(s-m_3^2-m_4^2)¿_3 -m_3^2 ¿_4]} =
	(2¹)^2|T_{fi}|^2{Â_{34}^{D-3}\over Â_{12}s^{D/2-1}}â(CoM) $$

\newpage

ÜGAUGES (II,VI):  Gervais-Neveu 

$$ L_A = -\f14 AÉõA -iA^a A^b »_b A_a -\f14 A^a A^b A_a A_b $$
$$ Yang-Mills:ââL = L_A +L_C,âL_C = -üi÷C(»+iA)^2 C -ü÷CC(»ÉA +iA^2) $$
$$ Gervais-Neveu:ââL = L_A +\f14 m^2 A^2 $$ 

ÜTwistors

$$ Òp| = p^ŒÒ{}_Œ|,â|pÔ = |{}^ŒÔp_Œ;â[p| = p^{ÀŒ}[{}_{ÀŒ}|,â|p] = |{}^{ÀŒ}]p_{ÀŒ} $$
$$ ÒpqÔ* = [qp] = -[pq],ââÒpqÔÒrsÔ +ÒqrÔÒpsÔ +ÒrpÔÒqsÔ = 0 $$
$$ P = |pÔ[p| = p^+ |+Ô[+| +p^- |-Ô[-| +p^t|-Ô[+| +Ðp^t|+Ô[-|,ââ-P* = |p]Òp| $$
$$ p^+ = Òp-Ô[-p],¼p^- = Ò+pÔ[p+],¼p^t = Ò+pÔ[-p],¼Ðp^t = Òp-Ô[p+],âÒ+-Ô = [-+] = 1 $$

ÜSpacecone

$$ \vbox{\offinterlineskip
\hrule
\halign{ &\vrule#&\strut¼\hfil$#$\hfil¼\cr
height2pt&\omit&\hskip2pt\vrule&\omit&&\omit&\cr
& axial\,\, gauges &\hskip2pt\vrule& non\hbox{-}null &
	& null\,\, (+\,\, auxiliary\,\, field\,\, eq.) &\cr 
height2pt&\omit&\hskip2pt\vrule&\omit&&\omit&\cr
\noalign{\hrule}
height2pt&\omit&\hskip2pt\vrule&\omit&&\omit&\cr
\noalign{\hrule}
height2pt&\omit&\hskip2pt\vrule&\omit&&\omit&\cr
& (partly)\ temporal &\hskip2pt\vrule& \hfill timelike:¼A^0 = 0 &
	& \hfill lightcone:¼A^+ = 0,¼¶/¶A^- &\cr 
& spacelike &\hskip2pt\vrule& \hfill Arnowitt\hbox{-}Fickler:¼A^1 = 0 &
	& \hfill spacecone:¼A^t = 0,¼¼¶/¶ÐA^tÊ &\cr
& scalar &\hskip2pt\vrule& \hfill unitary:¼Ä = Äÿ &
	& \hfill Gervais\hbox{-}Neveu:¼Ä = ÒÄÔ,¼¶/¶ÄÿÊ &\cr
height2pt&\omit&\hskip2pt\vrule&\omit&&\omit&\cr
}\hrule} $$
$$ nÉA = 0,ân = |+Ô[-|;ââtree¼¾¼Ò¼Ô^{2-E_+}[¼]^{2-E_-} $$
$$ L = L_2 +L_3 +L_4 $$
$$ L_2 = A^+ (-üP^2)A^- +Æ^+{-üP^2\over p}Æ^- $$
$$ L_3 = \left({p^¦\over p}A^à\right)([A^à,pA^¦] +ÓÆ^+,Æ^-Õ)
	+\left({p^¦\over p}Æ^à\right)[A^à,Æ^¦] $$
$$ L_4 = ([A^+,pA^-] +ÓÆ^+,Æ^-Õ){1\over p^2}([A^-,pA^+] +ÓÆ^+,Æ^-Õ)
	-[A^+,Æ^-]{1\over p}[A^-,Æ^+] $$
$$ A^+ = {[-p]\over Ò+pÔ},ââA^- = {Ò+pÔ\over [-p]};ââ
	Æ^+ = [-p],ââÆ^- = Ò+pÔ $$
$$ ref.¼lines:â{p^-\over p}A^+ = {p^+\over p}A^- =
	{p^-\over p}Æ^+ = {p^+\over p}Æ^- = 1 $$
$$ P_¢ = |-Ô[-|,âP_\¢ = |+Ô[+|;ââP_¢^a = ¶^a_-,âP_\¢^a = ¶^a_+ $$

ÜBackground-field

$$ Ä £ \Ä +Ä;ââá £ \D +iA,ââ
	F_{ab} £ \F_{ab} +\D_{[a}A_{b]} +i[A_a,A_b] $$
$$ »ÉA £ \DÉA,ââ÷C»ÉáC £ ÷C\D^2 C +÷C\DÉi[A,C] $$
$$ 1-loop:ââK = -ü(õ -i\F^{ab}S_{ba}) $$

\newpage

ÜSUPERSYMMETRY (II,IV,VI): Superspace

$$ q_Œ = -i\left({»\over »Ï^Œ} -üÐÏ^{Àº}p_{ŒÀº}\right),ââ
	Ðq_{ÀŒ} = -i\left({»\over »ÐÏ^{ÀŒ}} -üÏ^ºp_{ºÀŒ}\right);ââ
	(q^Œ)ÿ = Ðq^{ÀŒ} $$
$$ d_Œ = {»\over »Ï^Œ} +üÐÏ^{Àº}p_{ŒÀº},ââ
	Ðd_{ÀŒ} = {»\over »ÐÏ^{ÀŒ}} +üÏ^º p_{ºÀŒ};ââ(d^Œ)ÿ = -Ðd^{ÀŒ} $$
$$ Óq^Œ,Ðq^{Àº}Õ = Ód^Œ,Ðd^{Àº}Õ = p^{ŒÀº},ââ
	Çd^2 Ï = d^2 = üd^Œ d_Œ,ââÐd^2 d^2 Ðd^2 = üõÐd^2 $$
$$ »_M = (»_µ,»_{Àµ},»_m) = »/»z^M,ââz^M = (Ï^µ,ÐÏ^{Àµ},x^m) $$
$$ d_A = (d_Œ, Ðd_{ÀŒ}, »_{ŒÀŒ}) = E_A{}^M »_M;ââ[d_A,d_BÕ = T_{AB}{}^C d_C $$
$$ T_{ŒÀº}{}^{©À©} = T_{ÀºŒ}{}^{©À©} = -i¶_Œ^© ¶_{Àº}^{À©},ârest = 0 $$
	
ÜSuper Yang-Mills

$$ á_A = d_A +iA_A,ââ[á_A,á_BÕ = T_{AB}{}^C á_C +iF_{AB} $$
$$ Ñá_{ÀŒ}Ä = 0âÜâÓá_Œ,á_ºÕ = ÓÑá_{ÀŒ},Ñá_{Àº}Õ = 0;ââÓá_Œ,Ñá_{Àº}Õ = -iá_{ŒÀº} $$
$$ [á_Œ,á_{ºÀ©}] = C_{Œº}ÑW_{À©};ââÑá_{ÀŒ}W_º = 0,âá^Œ W_Œ + á^{ÀŒ}W_{ÀŒ} = 0 $$
$$ [á_{ŒÀŒ},á_{ºÀº}] = i(C_{Œº}Ðf_{ÀŒÀº}+C_{ÀŒÀº}f_{Œº}),ââf_{Œº} = üá_{(Œ}W_{º)} $$

ÜActions

$$ L_{N=1} = -\f1{g^2}tr Çd^2 ϼüW^Œ W_Œ +½Çd^4 ϼV
	-Çd^4 ϼÐÄe^V Ä + \left[ Çd^2 ϼf(Ä) +h.c.\right] $$
$$ L_{N=2} = -\f1{g^2}tr\left( Çd^2 ϼW^2 +Çd^4 ϼe^{-V}ÐÄe^V Ä \right)
	+Çd^4 ϼ½_0 V +\left( Çd^2 ϼ½_+ Ä +h.c. \right) $$
$$ -Çd^4 ϼÐÄ^{i'}(e^{V })_{i'}{}^{j'}Ä_{j'} +ü\left[Çd^2 ϼ ^{i'j'}Ä_{i'}(Ä+M)Ä_{j'} +h.c.\right] $$
$$ L_{N=4} = \f1{g^2}tr \left[ -Çd^2 ϼW^2 -Çd^4 ϼe^{-V}ÐÄ^I e^V Ä_I 
	+\left( Çd^2 ϼ\f16 ·^{IJK}Ä_I[Ä_J,Ä_K] +h.c. \right) \right] $$

ÜSupergraphs

\par\noindent
(A2$ü$) $Ï$'s:  one for each vertex, with an $Çd^4 Ï$.\\
(A3${}'$)  Propagators:  
$$ (VV,ÐÄÄ,ÄÄ,ÐÄÐÄ) :ââ\left( 1, -1, \f{m}{å2}{d^2\over -üp^2},
	\f{m}{å2}{Ðd^2\over -üp^2} \right) {1\over ü(p^2+m^2)}¶^4(Ï-Ï') $$
(A4$ü$)  Chiral vertex factors:  $Ðd^2$ on the $Ä$ end(s) of every chiral
propagator,\\
\phantom{(A4$ü$)} $d^2$ on the $ÐÄ$ end(s), but drop any one such factor
at a superpotential vertex.

\newpage

ÜLOOPS (VII,VIII):  Gamma function

$$ ý(z) = Ç_0^¥ d¼Â^{z-1}e^{-Â}
	= \f1z e^{-©z} Þ_{n=1}^¥ {1\over 1+{z\over n}}e^{z/n}
	= \f1z exp\left[-©z +Ý_{n=2}^¥ \f1n ½(n) (-z)^n\right] $$
$$ © = \lim_{n£¥}\left( -ln¼n +Ý_{m=1}^n \f1m \right) 
	= 0.5772156649...,ââ½(z) = Ý_{n=1}^¥ {1\over n^z} $$
$$ ý(z+1) = zý(z),ââý(z)ý(1-z) = ¹¼csc(¹z),ââ
	{ý(z)\over ý(2z)} = {2^{1-2z}å¹\over ý(z+ü)} $$
$$ ý(n+1) = n!,ââ
	ý(n+ü) = (n-ü)(n-\f32)...üå¹ = {(2n)!\over n!2^{2n}}å¹ $$
$$ \mathop{lim}_{z£¥}ý(z) ® å{2¹\over z}\left({z\over e}\right)^z,ââ
	\mathop{lim}_{z£¥}ý(az+b) ® å{2¹}(az)^{az+b-1/2}e^{-az} $$
$$ B(x,y) = Ç_0^1 dz¼z^{x-1}(1-z)^{y-1} = Ç_0^¥ d ¼ ^{x-1}(1+ )^{-x-y}
	= {ý(x)ý(y)\over ý(x+y)} $$

ÜRegularization

$$ {ý(a)\over [ü(p^2+m^2)]^a} = Ç_0^¥ d ¼ ^{a-1}e^{- (p^2+m^2)/2} $$
$$ 1 = Ç_0^¥ d¼¶\left(Â-Ý _i\right),ââ _i = Œ_i $$
$$ Çdk¼e^{-k^2/2} = 1,ââÇdk¼{k_a ... k_b\over (ük^2)^a} = 0 $$
$$ Çdk¼e^{ikÉx}{1\over ü(k+üp)^2 ü(k-üp)^2} = $$
$$ (üp^2)^{D/2-2}ý(\f{D}2-1)ý(2-\f{D}2) Ý_{n=0}^¥
	{ý(n+\f{D}2-1) \over n!ý(2n+D-2)}Ó\f14[p^2 x^2-(pÉx)^2]Õ^n $$
$$ Çdk¼{ý(a)\over (ük^2)^a}{ý(b)\over [ü(k+p)^2]^b}
	= {ý(a+b-\f{D}2)\over (üp^2)^{a+b-D/2}}B(\f{D}2-a,\f{D}2-b) $$

ÜSchemes

$$ {\rm MS}:â\h $$
$$ Ñ{\rm MS}:âý(\f{D}2)\h,âý(1-·)\h,âetc. $$
$$ {\rm G}:â{(-1)^{D_0/2}\over ·ý(2-\f{D}2)B(\f{D}2-1,\f{D}2-1)}\h,â
	{ý(1-2·)\over ý(1+·)[ý(1-·)]^2}\h,âetc. $$

ÜRunning coupling

$$ ý_{1,2g} ® tr Çdx¼\f18 F^{ab}º_1(ln¼õ -\f1·)F_{ab};â
	º_1 = üc_R(-1)^{2s}(4s^2 -\f13);¼Êc_{D+ÐD} = 2,¼c_A = 2N $$
$$ ý_M ® trÇdx¼\f18 F \left[ º_1 ln {õ\over M^2}
	 +{º_2\over º_1}ln\left( º_1 ln {õ\over M^2}\right)\right] F,ââ
	{µ^2\over M^2} = e^{1/º_1 g^2}(g^2)^{º_2/º_1^2} $$

\newpage

ÜGRAVITY (IX)

$$ [M_{ab},V_c] = V_{[a}ú_{b]c}âÜâ
	üÂ^{ab}[M_{ba},V_c] = Â_c{}^a V_a,ââ
	[M_{ab},M^{cd}] = -¶_{[a}^{[c}M_{b]}{}^{d]} $$
$$ üÂ^{ab}M_{ba} = üÂ^{Œº}M_{ºŒ} +üÂ^{ÀŒÀº}M_{ÀºÀŒ} $$
$$ [M_{Œº},Æ_©] = Æ_{(Œ}C_{º)©}âÜâ
	üÂ^{Œº}[M_{ºŒ},Æ_©] = Â_©{}^Œ Æ_Œ,ââ
	 [M_{Œº},M^{©¶}] = ¶_{(Œ}^{(©}M_{º)}{}^{¶)} $$
$$ á_a = e_a +ü¿_a{}^{bc}M_{cb},âe_a = e_a{}^m »_m;ââ
	[á_a,á_b] = T_{ab}{}^c á_c +üR_{ab}{}^{cd}M_{dc} $$
$$ -ds^2 = dx^m dx^n g_{mn},ââg_{mn} = e_m{}^a e_n{}^b ú_{ab} $$
$$ S = Ç dx¼{\bf e}^{-1}L,ââ{\bf e} = det¼e_a{}^mââ
	L = -\f14 R +L(á,Æ),ââR = R_{ab}{}^{ab} $$
$$ R_{ab} -üú_{ab}R = 2T_{ab},ââR_{ab} = R^c{}_{acb},ââ
	¶S_M = Çdx¼{\bf e}^{-1}(e_m{}^a ¶e^{bm})T_{ab} $$

ÜMethods

$$ \li{ (1): &ââD = 1âÜâá = » \cr
	(2): &ââds^2 = ds_1^2 +ds_2^2âÜâá = (á_1,á_2) \cr
	(3): &ââá_a(x) £ á_a(x') \cr
	(4): &ââds'^2 = ì^{-2}ds^2âÜâá'_a = ìá_a +(á^b ì)M_{ab}, \cr
		&ââR'_{ab}{}^{cd} = ì^2 R_{ab}{}^{cd} 
	+ì¶_{[a}^{[c}á_{b]}á^{d]}ì -¶_{[a}^c ¶_{b]}^d (áì)^2 \cr} $$
$$ [M_{12},V_2] = ú_{22}V_1,ââ[M_{12},M_{23}] = ú_{22}M_{13} $$
$$ [á_1,á_2] = [e_1+¿_1,e_2+¿_2] $$
$$ = Ó[e_1,e_2]+(e_1 ¿_2)M_2-(e_2 ¿_1)M_1Õ
	+Ó¿_1[M_1,á_2]-¿_2[M_2,á_1]-¿_1 ¿_2[M_1,M_2]Õ $$

ÜExamples

$$ L_G = -\f14 {\bf e}^{-1}R =
	\f1{16}·^{mnpq}·_{abcd}e_m{}^a e_n{}^b R_{pq}{}^{cd}
	= \f18 ·^{mnpq}e_m{}^a e_n{}^b ÷R_{pqab} $$
$$ = i\f18 ·^{mnpq}e_m{}^{ŒÀŒ}
	(ÐR_{npÀŒ}{}^{Àº} e_{qŒÀº} -R_{npŒ}{}^º e_{qºÀŒ}) $$

$$ S_G = 4\f{D-1}{D-2}Çdx¼{\bf e}^{-1}\f14 Ä(õ-\f14\f{D-2}{D-1}R)Ä $$

$$ -ds^2 = -\left(1-{2GM\over r}\right)dt^2
	+\left(1-{2GM\over r}\right)^{-1}dr^2 +r^2(dÏ^2+sin^2ϼdÄ^2) $$

\newpage

ÜSUPERGRAVITY (X)

$$ á_A = E_A{}^M »_M +ü¯_A{}^{º©}M_{©º} +ü¯_A{}^{ÀºÀ©}M_{À©Àº} +iA_A Y,
	ââ[Y,á_Œ] = -üá_Œ $$
$$ [á_A,á_BÕ = T_{AB}{}^C á_C +üR_{AB}{}^{©¶}M_{¶©}
	+üR_{AB}{}^{À©À¶}M_{À¶À©} +iF_{AB}Y $$
$$ ÓÑá_{ÀŒ},Ñá_{Àº}Õ = BÑM_{ÀŒÀº},ââÓá_Œ,Ñá_{Àº}Õ = -iá_{ŒÀº} $$
$$ [Ñá_{ÀŒ}, -iá_{ºÀº}] = C_{ÀºÀŒ}\W_º -ü(á_º B)ÑM_{ÀŒÀº},ââ
	[-iá_{ŒÀŒ},-iá_{ºÀº}] = C_{ÀºÀŒ}f_{Œº} -h.c. $$
$$ \li{ \W_Œ ={}& -Bá_Œ -G_Œ{}^{Àº}Ñá_{Àº} +ü(Ñá{}^{Àº}G_Œ{}^{À©})ÑM_{ÀºÀ©} 
		+üW_Œ{}^{º©}M_{©º} +iW_Œ Y +i\f16 W^º M_{ºŒ} \cr
	f_{Œº} ={}& iüG_{(Œ}{}^{À©}á_{º)À©} -ü(á_{(Œ}B +i\f13 W_{(Œ})á_{º)}
		+W_{Œº}{}^© á_© -ü(á_{(Œ}G_{º)}{}^{À©})Ñá_{À©} \cr
	& -(üÑá{}^2 ÐB +BÐB +\f1{12}iá^© W_©)M_{Œº} 
		-i\f1{16}[(á_{(Œ}{}^{À¶}G_{©)À¶})M_º{}^© +Œªº] \cr
	& +üW_{Œº}{}^{©¶}M_{¶©} 
		+\f14(á_{(Œ}Ñá{}^{À©}G_{º)}{}^{À¶})ÑM_{À¶À©} +iü(á_{(Œ}W_{º)})Y \cr
	W_{Œº©¶} ={}& \f1{4!}á_{(Œ}W_{º©¶)} \cr} $$
$$ G_a = ÐG_a,ââÑá_{ÀŒ}B = Ñá_{ÀŒ}W_Œ = Ñá_{ÀŒ}W_{Œº©} = 0,ââ
	Ñá{}^{ÀŒ}G_{ŒÀŒ} = á_Œ B -iW_Œ $$
$$ á^Œ W_{Œº©} -i\f13 á_{(º}W_{©)} = -iüá_{(º}{}^{ÀŒ}G_{©)ÀŒ},ââ
	á^Œ W_Œ +Ñá{}^{ÀŒ}ÑW_{ÀŒ} = 0 $$

ÜEctoplasm

$$ S = Çdx¼(-\f1{4!})·^{mnpq}e_q{}^D e_p{}^C e_n{}^B e_m{}^A L_{ABCD} $$
$$ L_{Œºcd} = ·_Œ{}^{ÀŒ}{}_{,ºÀŒ,cd}Ð{\L},ââ
	L_{Œbcd} = i·_{ŒÀŒ,bcd}Ñá^{ÀŒ}Ð{\L},ââ
	L_{abcd} = ·_{abcd}[(Ñá^2 +3B)Ð{\L} +h.c.] $$
$$ Ñá_{ÀŒ}\L = 0âÜâ\L = (Ñá^2 +B){\bf L} $$

ÜAction

$$ S_{SG,c} = 3Çdx¼d^4 ϼE^{-1}Ðìì $$

$$ L_{SG} = L_G +L_Æ +{\bf e}^{-1}L_a $$
$$ L_G = -\f14 {\bf e}^{-1}R,ââ
	L_Æ = ·^{mnpq}ÐÆ_{mÀŒ}üÓe_n{}^{ŒÀŒ},á_pÕÆ_{qŒ},ââ
	L_a = -\f38 (G_a)^2 +3ÐBB $$
$$ ¿_{mbc} = e_m{}^a[\on\circ{¿}_{abc}-ü(öT_{bca}-öT_{a[bc]})],ââ
	öT_{ab}{}^c = ·_{ab}{}^{cd}G_d +iÆ_{[a}{}^© ÐÆ_{b]}{}^{À©} $$
$$ ¶e_m{}^{ŒÀŒ} = -i(·^Œ ÐÆ_m{}^{ÀŒ} +з^{ÀŒ}Æ_m{}^Œ),ââ
	¶Æ_m{}^Œ = á_m ·^Œ +ie_m{}^{ºÀº}(·_º G^Œ{}_{Àº} -з_{Àº}¶_º^Œ B) $$
$$ ¶B = -\f23 ·^Œ t_{Œº}{}^º,ââ
	¶G_{ŒÀŒ} = -·^º(t_{ºŒÀŒ} +\f13 C_{ºŒ}t_{À©Àº}{}^{Àº}) +h.c. $$
$$ T_{ab}{}^© = -e_a{}^m e_b{}^n á_{[m}Æ_{n]}{}^©
	+i(Æ_{aº}G^©{}_{Àº} -ÐÆ_{aÀº}¶_º^© B -aªb)
	­ C_{Œº}t_{ÀŒÀº}{}^© +C_{ÀŒÀº}t_{Œº}{}^© $$

\newpage

ÜSTRINGS (XI)

$$ S_L = {1\over Œ'}
	Ç{d^2 §\over 2¹}å{-g}g^{mn}ü(»_m X^a)É(»_n X^b)ú_{ab} $$

$$ Œ(s) = ûŒ's +{1\over û},ââû = \leftÓ \matrix{ 1 & (open) \cr ü & (closed) \cr} \right. $$

$$ \li{ \A_4 = ¼& g^2 Ç_0^1 dz¼ z^{-Œ(s)-1} (1-z)^{-Œ(t)-1} \cr
	= ¼& g^2 B[-Œ(s),-Œ(t)] = g^2{ý[-Œ(s)]ý[-Œ(t)]\over ý[-Œ(s)-Œ(t)]} \cr
	= ¼& g^2 Ý_{j=0}^¥ \leftÓ {[Œ(s)+j][Œ(s)+j-1]ò[Œ(s)+1]\over j!} \rightÕ {1\over j-Œ(t)} \cr
	= ¼& \left( in \mathop{lim}_{\astop{s£-¥}{t¼fixed}} \right)¼g^2 ý[-Œ(t)][-Œ(s)]^{Œ(t)} \cr
	¾ ¼& \left( in \mathop{lim}_{\astop{s£-¥}{ϼfixed}} \right)¼e^{-f(cos¼Ï)Œ(s)};â
		cos¼Ï ® 1 +2{t\over s}, \cr
	&â f ® {t\over -s}ln \left( {-s\over t} \right) +{u\over -s}ln \left( {-s\over u} \right) \cr} $$

ÜConformal field theory

$$ L = Ѝ_{(1-w)}Ñá_{(w)}âÜâ
	T = Ѝü\onª» +(ü-w)»(Ѝ) $$
$$ L = \f14 ú^{ij}(áÄ_i)É(áÄ_j) +üRµ^i Ä_iâÜâ
	T = üú_{ij}(»Ä^i)(»Ä^j) -µ_i »^2 Ä^i $$
$$ Ò0|(z)Ѝ(z')|0Ô = {1\over z-z'} $$
$$ Ò0|Ä^i(z)Ä^j(z')|0Ô = -ú^{ij}ln(z-z') $$
$$ e^{iaÉÄ(z)} e^{i÷aÉÄ(z')} = 
	(z-z')^{aÉ÷a}e^{i[aÉÄ(z) +÷aÉÄ(z')]}
	+ ... $$
$$ -\left[ Ç{dz'\over 2¹i} Â(z') T(z'), _{(w)}(z)\right] =
	Â(z)(»_{(w)})(z) +w(»Â)(z)_{(w)}(z) $$
$$ -\left[ Ç{dz'\over 2¹i} Â(z') T(z'), Ä^i(z)\right] =
	Â(z)(»Ä^i)(z) +µ^i(»Â)(z) $$
$$ w\left( e^{iaÉÄ} \right) = üa^2 +iµÉa $$
$$ anomaly¼¾¼6(w-ü)^2 -ü $$

Û0 INDEX

\def\righthead{\hfil}

\parskip=0pt \parindent=0pt
\baselineskip=1.23em plus 1pt

\refrm 

Topics are listed by subsection (or section, etc.).

\vskip.15in

\frenchspacing

\twocolumn{

{\subsubsubtitlefont A}belianòIB2

ActionòIII-IV

Adjoint representationòIB2,5,C2

Advanced propagatoròVA3,B1-2

Affine parametrizationòIIIB2

Affine transformationòIB1

Analytic S-matrix theoryòXIA1

Anni'l'n op.òIC1,IIB3,C5,IIIA2,VIA1,3,XIB2

AnomalyòVIIIA7,B2-4,C1,XIB4,C7

AntibracketòXIIC1

AnticommutationòIA2

Anticommuting number/variableòIA2

AntifieldòVIA3,XIIB3

Anti-Gervais-Neveu gaugeòVIB4-5

AntighostòVIA2

AntiparticleòIA4-5,IIIB4-5

Antisym.tensoròIIB2,VIIIA7,XA3,XIIA5,B3

Arnowitt-Fickler gaugeòIIIC2

Asymptotic freedomòVIIIA3-4,C,XIA7

Atiyah-Drinfel'd-Hitchin-ManinòIIIC7

Auxiliary variableòIIIA1,5,B2,C2,4,IVC2

Axial anomalyòVIIIA7,B2-4,C1

Axial currentòIVA4,VIIIA7,B2-4

Axial gaugeòIIIC2,VIB6-8,IXB3

Axial vectoròIA5

AxionòVIIIC4,XIA4-7,C6-7,XIIB8

\vskip.2in

{\subsubsubtitlefont B}ackgr'nd fieldòVC1,3,VIIB2,XIA5-6,XIIB7

Background-field gaugeòVIB8,10,VIIIA3

BagòIVB1

Baryon numberòIVA4,B1

BaryonòIC4

Bhabha scatteringòVIC4

Bianchi id.òIIA7,IIIC1,IVC3,IXA2,XB2,C5

Big BangòIVA7,IXC3

Birkhoff's theoremòIXC5,7

Bjorken scaling variableòVIIIC3

Black holeòIXC7

Bohr-Sommerfeld quantizationòVA2

Borel summation/transformòVIIC3

BosonizationòVIIB5,XIA5-6,B7-8

Bnd.cond.òIIIC2,6-8,VA,C3,VIIIC5,XIA3,B1

}{

Bnd.stateòVC1,VIIB3,5,C2,VIIIB4,XIA1-2

BracketòIA1-2,XIIA1

BRSTòVIA-B,VIIA1,IXB1,XIB7,XII

\vskip.2in

{\subsubsubtitlefont C}abibbo-Kobayashi-Maskawa matrixòIVB3

Cartan metricòIB2

Cartan subalgebraòIB5

Casimir operatoròIB2

CausalityòVA4,C6,VIIA1

Center-of-mass frameòIA4,VC7,VIC4

Central chargeòIVC7,XC5

ChannelsòVIIB7,XIA1

Chan-Paton factorsòXIA4,C7

C. conj.òIA5,IIC5,IIIA4,IVA4,B1,XC1,XIA4

Chern-Simons formòIIIC6,IVC5,VIIIB2,XC5

ChiralityòIIA7,IVC1

Chiral projectoròIIA6-7

Chiral representationòIVC4-5

Chiral spinoròIIA7,IIIC4

Chiral superfieldòIIC5,IVC1-2,VIB7

Chiral superspaceòIVC1

Chiral symmetryòIVA4,VIIIB2-4,7

Chiral theoryòIIA7

ChromodynamicsòIC4,IVB1,VIIC4

Classical groupòIB4,IIC3

Classical limitòIA1-2,IIIA2-3,V

Clebsch-Gordan-Wigner coefficientòIB5,IIA4

Closed stringòIVB1,VIIC4,XI

Coherent stateòIA2

CohomologyòVIA1,3,XB2,XII

Coleman-Weinberg mechanismòVIIB2

Colinear divergence/particleòVIIA6

ColoròIC4,IVA4

Color decomposition/expansionòVC9,VIIC4

Color orderingòVC9,VIC1-3

Compact groupòIB2,4

CompactificationòIIB4,XIA4

CompensatoròIVA5,IXA7,XA3

Complex projectiveòIVA2,VIIB3

Complex representationòIB1-2

Component approachòXB4

Component transformationòIIC2,XB3

} \newpage \twocolumn{

Compton scatteringòVIC4

ConfinementòIVB1,VIIIB7

ConformalòIA6,IIB-C,IIIC5-7

Conformal anomalyòVIIIC1,XIB4

Conformal boostòIA6,IIB1,6-7

Conformal field theoryòXIB,XIIB8

Conformal gaugeòXIB1

Conformally flatòIXC2

Connected graphòVC2

ConservationòIA1,IIIB4,IXA6

Constituent quarkòIC4

ConstraintòIIIA5

Constructive quantum field th.òIIIC4,VIIC3

Continuum limitòVA1,VIIIB7,XIA7

Contour integralòIIA1

Contracted indicesòIA1

ContractionòIIC3

Coordinate representationòIC1

Coordinate transform.òIA,C2,IIIB1,3,IXA1

Coset spaceòIVA3

Cosmological red shiftòIVA7,IXC4

Cosmological termòIXA5-7,C2,XB1,6-7

CosmologyòIVA7,IXC3

CountertermòVIIA1

Covar.deriv.òIIC2,IIIA4,C1,IVC3,IXA2,XA1

Covariant momentumòIA5,IIIC1

Covering groupòIC5

CP(n) modelòIVA2,VIIB3

CPT theoremòIVB1,VC8

Creat'n op.òIC1,IIB3,C5,IIIA2,VIA1,3,XIB2

Critical dimensionòXIA7,B4

Crossing symmetryòVC8,VIC4

Cross productòIIA1

Cross sectionòVC7,VIC4

CurrentòIIA7,IIIA4,B4,VC3

Current quarkòIC4

CurvatureòIXA2

Curved indicesòIXA1

Cut propagatoròVC7

Cutting ruleòVC6

\vskip.2in

{\subsubsubtitlefont D}ecay rateòVC7,VIIA4,VIIIA3,B4

Deceleration parameteròIVA7,IXC3

De Donder gaugeòIXB1

Deep inelastic scatteringòVIIIC3

}{

Defining representationòIB4

Degree of freedomòIVC3-4,VB3,XC3,XII

DensityòIIIB1,IXA4,XIA6

Density parameteròIVA7,IXC3

De Sitter spaceòIXC2-3

DeterminantòIB3,C2,5,IIA1,5

Differential cross sectionòVC7

Differential formòIC2,IXA4,XB2

DilatationòIA6,IIB1,6

DilatonòIVA7,IXA7,XIA4-6,C4-7,XIIB8

Dim.redòIIB4,IIIC8,VIB6,VIIA1,VIIIA3,XC

Dimensional regularizationòVII

Dimensional transmutationòVIIB3

Dirac deltaòIIIA1,VA2

Dirac equationòIIB2,IIIA4

Dirac matrices/spinoròIC1,IIA6,XC1

Direct product/sumòIB2

Discrete symmetryòIA5

DivergenceòVII

Division algebraòIIC4

Double coveringòIIA3

Drell-Yan scatteringòVIIIC3

DualityòIIA7,IIIC4,VIIB5,XA3,B5-6,C6,XI

Dual modelòXIA1

Dual representation/spaceòIB1

Duffin-Kemmer matricesòIIB4

DustòIXB2,C3

Dynkin indexòIB2

\vskip.2in

{\subsubsubtitlefont E}ctoplasmòXB2

Effective actionòVC2

Effective potentialòVIIB2

Einstein-Hilbert actionòIXA5

Einstein summationòIA1

Elastic scatteringòVC7

ElectromagnetismòIA5,C4,IIA7,B5,B7,IVB2

EnergyòIA1,5,IIA5,B6,C1

Energy-momentum tensoròIIIA4,B4,IXA6

Equivalence principleòIXA4

Euclidean spaceòIA4,IIIC4,VB4

Euler numberòVIIC4,IXA7,XIA2

Euler-Mascheroni constantòVIIA2

Euler's theoremòVIIC4

Event horizonòIXC7

Evolution equationsòVIIIC3

} \newpage \twocolumn{

Exceptional groupòIB4,XC6

Exclusive scatteringòVIIIC3

Explicit supersymmetry breakingòIVC6

Explicit symmetry breakingòIVA

Extended supersymmetryòIVC7,XC3-6

External fieldòIIIA1,B3-5,VC

External lineòVC2

Extrinsic lengthòIIIB1

\vskip.2in

{\subsubsubtitlefont F}aceòVIIC4,XIA7

FactorizationòVIIIC3

FamilyòIC4,IVB3

Fayet-Iliopoulos termòIVC5-7

Fermi-Feynman gaugeòVIB2-3,9,IXB1,XII

FermionòIA2

Fermion doublingòVIIIB7

Feynman diagram/graphòVC2

Feynman parameteròVIIA2

Feynman tree theoremòVC6

Field strengthòIIA7,B3,5,IIIC1,XA2

Fine tuningòXB6

Finite theoryòVIIIA5-6

Finite transformationòIA3

1st-orderòIIIA5,C4,IVC5,IXA5,XB4-5,XIA3

1st-quantizationòVB1,VIIIB1,C5,XIB-C,XII

Flat indicesòIXA1

Flat superspaceòIVC3

FlavoròIC4,IVA-B

Flavor-changing neutral currentsòIVB3

Foldy-Wouthuysen transform.òIIB5,VIIIB6

Four-vectoròIA4

Fragmentation functionòVIIIC3

FunctionalòIC1,IIIA1

Functional integralòVA1

Functional derivativeòIIIA1

Fundamental representationòIB4

Furry's theoremòVIIA5,VIIIB3

\vskip.2in

{\subsubsubtitlefont G}alilean boost/groupòIA1

Gamma matrixòIC1,IIA6,XC1

GaugeòIIIB2,C1,VIB,IXB,XA4,XIB1

Gauge fieldòIIA7,III

Gauge invar.òIIA7,IIIA5,B1,C1,IXA1,XIA3

Gauge parameteròIIA7,III

GauginoòIVC5

}{

Gauss-Bonnet theoremòIXA7

GaussianòIB3,IIC3,VA2,5,B1,VIIA2,XIA7

General linear groupòIB4

General relativityòIX

General superfieldòIVC1

Generating functionalòVC1-2

GeneratoròIA1

GenusòXIA2

GeodesicòIXB2-4

Gervais-Neveu gaugeòVIB4-5

Gervais-Neveu modelòIVA6

GhostòVIA,XII

Ghost numberòVIA1,XIIB1

Glashow-Iliopoulos-Maiani mechanismòIVB3

Glashow-Salam-Weinberg modelòIVB2

GLDAP equationsòVIIIC3

Gliozzi-Scherk-Olive projectionòXIB5,C6

Global symmetry/transformationòI

GlueballòIC4,IVB1,VIIC4,VIIIB4

GluinoòIVC5

GluonòIC4,IIIC,IVB1,VIC,VIII

Goldstone boson/theoremòIVA

Goldstone fermionòIVC6,XB6

Graded commutationòIA2

Graded group/symmetryòIA2,IIC3,XII

Grand Unified TheoryòIVB4,VIIIA4

GraphsòVC2

Gravitational collapse/radiusòIXC7

Gravitational red shiftòIXC6

GravitonòIIIA4,C1,IVA1,IXB1,C1-2,XIA,C

GravityòIIIA4,IX

Green functionòVA-B,XIC3

Green-Schwarz superstringòXIB5,C6-7

Gribov ambiguityòVIB2

GroupòIA3-II

Group contractionòIIC3

Group metricòIB4

G-schemeòVIIA3

\vskip.2in

{\subsubsubtitlefont H}adronòIC4,IVA4,B1,VIIC4,VIIIC,XI

Hagedorn temperatureòXIC2

HamiltonianòIA1,IIIA,VA1,VIA1

Hamiltonian densityòIIIA3

Hamilton-Jacobi equationsòVA2

Hamilton's equationsòIA1

} \newpage \twocolumn{

HandleòVIIC4,XIA2

Hard particleòVIIIC3

Hardy-Ramanujan formulaòXIC2

Harmonic gaugeòIXB1

HelicityòIIB7,C5,IIIC2

Heterotic stringòXIA4-6,C6,XIIB8

Hidden matteròXB7

HiggsinoòIVC5

Higgs mechanismòIVA6

Higgs scalaròIC4,IVA6

Higher dimensionsòXC

Hilbert-space metricòIB4,VC5,XIIA1

HoleòVIIC4

Homogeneous spaceòIVA7,IXC3-4

Hubble constantòIVA7,IXC3

Hypermultiplet/hypersymmetryòIVC7

\vskip.2in

{\subsubsubtitlefont I}deal mixingòIC4

i$·$ prescriptionòVA3,B1

IGL(1)òXIIB-C

Improved perturbationòVIIC1

Inclusive scatteringòVIIIC3

Index notationòIB,IIA4-5

Induced metricòXIA3

Inelastic scatteringòVC7

Inhomogeneous groupòXIIB1

Infinitesimal transformationòIA1

InfinityòVII

Infrared divergenceòVIIA6,VIIIB

Infrared slaveryòIVB1

Inhomogeneous transformationòIA4,B1

Inner productòIB1,XIIB2

InstantonòIIIC6-7,VIIC2

Interaction pictureòVA4

Internal lineòVC2

Intrinsic lengthòIIIB1

InversionòIA6

Irreducible representationòIB2

IsospinòIVB2

Isotropic spaceòIVA7,IXC3-4

\vskip.2in

{\subsubsubtitlefont J}acobi identityòIA1,3

Jacobi theta functionòXIC2

JetòVIIIC2-3

JWKBòIA1,VA2,C3,VIB1,VIIIB1,6

}{


{\subsubsubtitlefont K}¬ahler manifold/potentialòXB6

Killing equation/vectoròIXA2,6,B2,C4,6

Kinoshita-Lee-NauenbergòVIIA6

Klein factor/transformationòIA2,VC5,XC1

Klein-Gordon equationòIIB1

Kronecker deltaòIA1

Kruskal-Szekeres coordinatesòIXC7

\vskip.2in

{\subsubsubtitlefont L}ab frameòVC7

Ladder diagramòXIA1

Lagrange multiplieròIIIA1,5

LagrangianòIIIA1

Lagrangian densityòIIIA3

Lamb shiftòVIIIB6

Landau equations/singularityòVC8

Landau gaugeòVIB2-3

Landau ghostòVIIC2

Laplace transformòVIIC3

LatticeòVIIIB7,XIA7

Legendre transformòVC3,XB5

LengthòIA4,IIIB1,IXB2,XIA3

LeptonòIC4,IVB

Levi-Civita tensoròIA5,B3

Lie algebra/bracket/derivative/groupòIA3

Light bendingòIXC6

LightconeòIA6

Lightcone basisòIA4

Lightcone coordinatesòIA4,6,XIIA1

Lightcone frameòIIB3,6

Lightcone gaugeòIIB3,IIIB2,C2

Lightcone spinoròXC4

Linear multipletòXA3

LinkòVC2,8,VIIIB7,XIB7

Little groupòIIB3-4

Local inertial frameòIXC4

LocalityòIIIA1,3

Local symmetry/transformationòIII

Loop graphòIIIB5,VC,VII-VIII,XIA2,C

Lorentz connectionòIXA2

Lorentz force lawòIXB2

Lorentz transformationòIA4-5,IIA5

Lorenz gaugeòVIB2,IXB1

Lowering operatoròIA6,B5,VIA1

\vskip.2in

{\subsubsubtitlefont M}agnetic chargeòIIIC8

} \newpage 
\twocolumn{

Magnetic momentòIIIC4,VIIA3,VIIIB5

Magnetic monopoleòIIIC8

Majorana spinoròIIA6,IIIA4,XC2

Mandelstam variablesòIA4,VC4,VIC4

Manifest symmetryòIB1

Maximal helicity violationòVIC1,3

Maxwell's equationsòIIA7,B2,IIIA4

MechanicsòIA,IIIB,VB,XII

Mellin transformòVIIC3

MesonòIC4,IVA4,B1,VIIC4,VIIIB4

Method of imagesòXIA3,B1-2,C3

MetricòIA4,B2,4,IIIA1,IXA2

Minimal couplingòIA5,IIB5,IIIA2,4,C1,IXA4

Minimal subtractionòVIIA3

Minkowski metric/spaceòIA4

Modified minimal subtractionòVIIA3

Modular invarianceòXIC5-6

M\o ller scatteringòVIC4

Momentum integralòVII-VIII

Momentum subtractionòVIIA3

MonopoleòIIIC8

\vskip.2in

{\subsubsubtitlefont N}akanishi-LautrupòVIA2,4,XIIC3

Natural unitsòIA1,4,IVA7

Negative energyòIA4-5,IIIB4,VB1

NeutrinoòIC4,IVB2-4

New minimal supergravityòXB5

Newtonian limitòIXB1,C6

Nielsen-Ninomiya theoremòVIIIB7

NonperturbativeòVIIC,XI

NonrelativityòIA1

Nonrenormalization theoremòVIC5

Normal orderingòVIA1,XIB2,6-7,C3

No-scale supergravityòXB7

Null basisòIA4

\vskip.2in

{\subsubsubtitlefont O}ctonionòIB4,IIC4

Off shellòIIB1,IVC1,3,7,XC3

Okubo-Zweig-Iizuka ruleòVIIC4

Old minimal supergravityòXB5

1/N expansionòVIIC4,XIB7

One-particle-irreducible graphòVC2

1.5-orderòXB4

On shellòIIB1,XC3

Open stringòIVB1,VIIC4,VIIIA3,XI

}{

Operator cohomologyòVIA1

Operator product expansionòVIIIC3

Optical theoremòVIIA4,6,VIIIC3

O'Raifeartaigh modelòIVC6

Orthonormal basisòIA4

Orthosymplectic groupòIIC3,XII

Orthogonal groupòIB4

OSp(1,1|2)òXIIA

Outer productòIIA1

Overlapping divergenceòVIIB8

\vskip.2in

{\subsubsubtitlefont P}air creationòIIIB5,VIIB4,6,VIIIA2-3,C2

Parallel transportòIXB4

ParityòIA5,IIA5,IVA4,B1,VIIA5,XIA4

Parke-TayloròVIC1-3

Partially conserved axial currentòIVA4

ParticlesòIA,IIIB,VB,XII

Partition functionòVIIIC5,XIC

PartonòVIIIC

Parton distribution/modelòVIIIC3

Path integralòVA1-2,C1-4

Path orderingòIIIC2,VIIIB7

Pati-Salam modelòIVB4

Pauli-Luba«nski vectoròIIB7

Pauli matricesòIIA1,IIIA2,XC1

Pauli-Villars regularizationòVIIIB2

Penrose transformòIIB6-7,C5

Perfect fluidòIXC5

Perihelion precessionòIXC6

PerturbationòV-VIII

PfaffianòIB3,C5

Phase spaceòIA5,VC7,VIIA5,B4,6,XIIC1

PhotinoòIVC5

Physical regionòVC7

PionòIC4,IVA4,VIIIB4

Pion decay constantòIVA4

Planar diagramòVC8-9,VIIC4

Planck unitsòIA1,4,IVA7

Plane waveòIIIC3,VB2,C7,IXC1

PlaquetòVIIIB7

Poincar«e groupòIA4,IIB

Poisson bracketòIA1

Polar vectoròIA5

Positive actionòIIIB1,VA5

Positive energyòIA4,IIC1,IIIA4,IVC2

} \newpage \twocolumn{

Power countingòVIIA5

PrepotentialòIVC1,4,XA1

Projective lightconeòIA6

Projective transformationòIA6,IVA2

PropagatoròVA-B

Propagator correctionòVIIB4-8,VIIIA

Proper Lorentz transformationòIA5

Proper timeòIA4

Pseudogoldstone bosonòIVA4,VIIIB4

PseudohermiticityòIB4

Pseudoreal representationòIB2

Pseudoscalar/pseudotensoròIA5

\vskip.2in

{\subsubsubtitlefont Q}CDòIVB1,VIC3,VIIIA3,C

Quantum-chromodynamic stringòXIB7

QuarkòIC4,IVA4,B

Quark-gluon plasmaòIVB1

QuaternionòIB4,IIA1,C4

Quenched approximationòVIIIB7

\vskip.2in

{\subsubsubtitlefont R}adial gaugeòVIB1,IXB4

RadiationòIVA7,IXC3-4

Raising operatoròIB5,VIA1

Ramond-Neveu-Schwarz stringòXIB4-5,C6

RapidityòIIA5

Real representationòIB2

RecursionòVIC2

Red shiftòIVA7,IXC4

Reducible representationòIB2

Reference momentum/vectoròVIB6

ReflectionòIA5,IIA3,5

Regge behavior/theory/trajectoryòXIA1

RegularizationòVIIA-B,VIIIB2

Relativistic Schr¬odinger equationòIIB1

RelativityòIA4,IIA5-B,IIIA4,IX

RenormalizabilityòVIIA1,5

Renormalizable gaugeòVIB3

RenormalizationòVIIA

Renormalization groupòVIIB3,7,C1

Renormalization mass scaleòVIIA3,B3

RenormalonòVIIC2

RepresentationòIB-C

Representation spaceòIB1

Residual gauge invar.òVIA1,B2,IXB3,XIB1-2

ResummationòVIIC

}{

Retarded propagatoròVA3,B1-2

Ricci scalar/tensoròIXA2

Riemann normal coordinatesòIXB4

Riemann zeta functionòVIIA2

Rindler coordinatesòIXB2

RotationòIA1,IIA2

R symmetryòXA1

Running couplingòVIIB7

\vskip.2in

{\subsubsubtitlefont S}calar fieldòIA1

Scalar multipletòIVC1

Scalar potentialòIIIA1

Scale weightòIA6,IIB1

Scaling variableòVIIIC3

Scattering matrixòVA4,C

Schwarzschild solutionòIXC5-7

Schwinger-Dyson equationsòVC3

Schwinger modelòVIIIA7

Schwinger parameteròVB1,VIIA2

S-dualityòIIA7,IIIC4,IXC1,XC6

Self-dualityòIIA7,IIIC4-8,IVC7,IXC1,XA3

Semiclassical expansionòIIIA4,VA2,C3

Semisimple groupòIB2

SesquilinearòIB4

Sigma matricesòIIA1,IIIA2,XC1

Sigma modelòIVA2-4

SignatureòIA4

Simple groupòIA6,B2

Simple supersymmetryòIVC7

SingularityòVC8,VIIC3,IXC7

SleptonòIVC5

SL(2,C)òIC5,IIA5

S-matrixòVA4,C

Soft divergence/particleòVIIA6

Sommerfeld-Watson transformòXIA1

SO(3)òIA3,IIA2

SourceòIIIA4,VC3

Sp(2)òXIB7,XIIA

SpaceconeòVIB6-7,C1-3

Spacetime latticeòVIIIB7

Special groupsòIB4

Special relativityòIA4,IIA5-B,IIIA4

SpinòII

Spin operatorsòIIB,XII

Spinor helicityòVIB2,6

} \newpage \twocolumn{

Spinor indexòII

Spinor notationòIB5,C5,IIA

Spinor representationòIC5,XC1-2

Spin-statistics theoremòVC5

Splitting functionòVIIIC3

Spontaneous symmetry breakdownòIVA1

SquarkòIVC5

S-supersymmetryòIIC4

Standard ModelòIC4,IVB

Stationary phaseòVA2,5

Steepest descentòVA5

Stirling approximationòVIIC2

Stokes' theoremòIC2

StringòIVB1,VIIC4,VIIIA3,C5,XI,XIIB8

String gaugeòIXB5,XA4,XIA6,C5,XIIB8

String tensionòXIA3

Strong interactionòIC4,IVB1

Structure constantsòIA3

Structure functionòIXA2

St¬uckelberg formalismòIIB4,IVA5

Subtraction schemeòVIIA1,3

Summation conventionòIA1

Superconformal groupòIIC4

SupercoordinateòIIC2,IIIB1

SuperdeterminantòIIC3

Superficial divergenceòVIIA1,5

SuperfieldòIIC2

SupergraphòVIC5

SupergravityòX,XIA4

SupergroupòIIC3

SuperhelicityòIIC5

SuperhiggsòXB6

SuperpotentialòIVC2

SuperrenormalizableòVIC5,VIIA5

SuperspaceòIIC2

SuperspaceconeòVIB7,C3

Superstrin'VIIIA3,XB6,XIA2,4,6,B4-5,C6-7

SupersymmetryòIIC,IVC,VIC3,5,VIIIA5-6,X

Supersymmetry breakingòIVC6

SupertraceòIA2,IIC3

SupertwistoròIIC5

Super Yang-MillsòIVC3-5

SU(2)òIA3,IIA2

SymplecticòIB4

SymmetryòI-IV

}{

Symmetry breakingòIVA,C6

\vskip.2in

{\subsubsubtitlefont T}adpoleòVIB8,C5,VIIB1

T-dualityòXIA5

Temporal gaugeòIIIC2

TensionòXIA3

Tensor multipletòXA3,B5,XIA6,XIIB8

Tensor notationòIB5

TetradòIXA2

'T Hooft ansatzòIIIC6

Three-vectoròIIA1

Time developmentòIA1

Time orderingòIIIA5,VA1

Time reversalòIA5,IIA5,XIA4

Time translationòIA1

T-matrixòVC4

Topological expansionòVIIC4

TorsionòIVC3,IXA2,XA1

Tree graphòVC2,XIB6

TrialityòXIB5

Triangle diagramòVIIIB2-5

TwistòVIIIC3

TwistoròIIB6-7,C5,IIIC5-7,VIB6-7,C1-3

Two-by-two matrixòIIA-B

\vskip.2in

{\subsubsubtitlefont U}ltraviolet divergenceòVII-VIII

UnitarityòIB4,VA4,C5-6,VIIA4

Unitary gaugeòIIIA5,B2,C2,IVA6,VIA4,B3

Unitary groupòIB4

U(1) problemòIC4

\vskip.2in

{\subsubsubtitlefont V}acuumòIC1,IIIB4,IVA,IXA5

Vacuum bubbleòVC7,VIB8,10,C5,VIIB1-2

Vacuum valueòIIC5,IVA1,VIIC3,IXA5

Van Vleck determinantòVA2

VectoròIA3-4,IIA1

Vector fieldòIC2

Vector multipletòIVC3-4

Vector potentialòIIA7

Vector representationòIC5

VelocityòIA1,IIIB1

VertexòVC2,VIIIB

Vertex oper.òVIIIC4-5,XIB2,5-7,C6,XIIB7-8

VielbeinòIVC3

VierbeinòIXA2

} \newpage \twocolumn{

Virasoro constraintsòXIA3,5,B1,7,XIIB8

Volume elementòIIIB1,VIIIC5,IXA2,XIC

\vskip.2in

{\subsubsubtitlefont W}ard-Takahashi identityòVIB8

Wave equationòIIB

Wave-function renorm.òVIB8,VIIIA3,5,B5

W bosonòIC4,IVB2

Weak interactionòIC4,IVB2-4

Weak mixing (Weinberg) angleòIVB2

Wess-Zumino gaugeòIVC4

Weyl scaleòIXA7,B5,XA3-4

Weyl spinoròIC1,IIA5

Weyl tensoròIXA3

White holeòIXC7

Wick rotationòIC5,VA5,B4

Wilson loopòVIIIB7

}{

WindowòXIA2

WinoòIVC5

WorldlineòIIIB1

WorldsheetòXI

Worldsheet latticeòXIC2

\vskip.2in

{\subsubsubtitlefont X}-rayòjust kidding

\vskip.2in

{\subsubsubtitlefont Y}ang-MillsòIIIC

Young tableauòIC3

Yukawa couplingòIVA4,B

\vskip.2in

{\subsubsubtitlefont Z}ero-modesòVIIIC5,XIB1

Zinn-Justin-Batalin-VilkoviskyòXIIC2

ZinoòIVC5

ZweibeinòXIA5,C3

}

\unrefs

%
%
%
%
%

\newpage

\parindent=72pt

\sectskip\bookmark0{Comments on Warren Siegel's Fields}
	\secty{Comments on Warren Siegel's {\subsubsubtitlefont Fields}:}

\def\righthead{\hfil} \def\lefthead{\hfil}

``The price is right."

``Oh, Warren, what have you done to your students now?"

``I can see you put a lot of work into it."

``That's nice, honey."

``You might want to add a reference to my paper..."

``It's different."

``Is this going to be on the exam?"

``I'll have a look at it when I get the time."

``Aren't there enough field theory books already?"

``So this is why you haven't written any papers lately."

``Where are the jokes?"

\bye